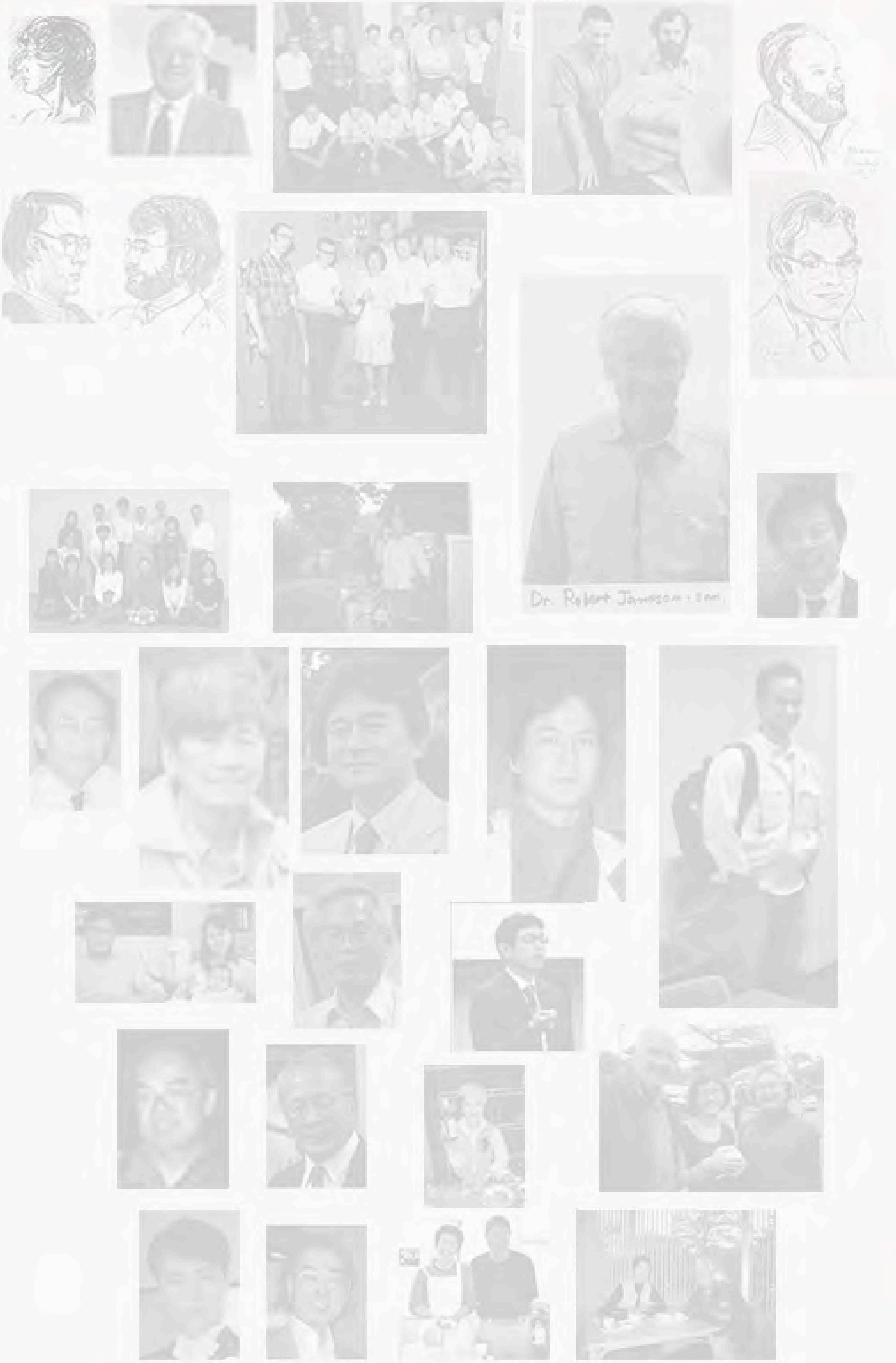
Dr. Robert Jameson · IBM

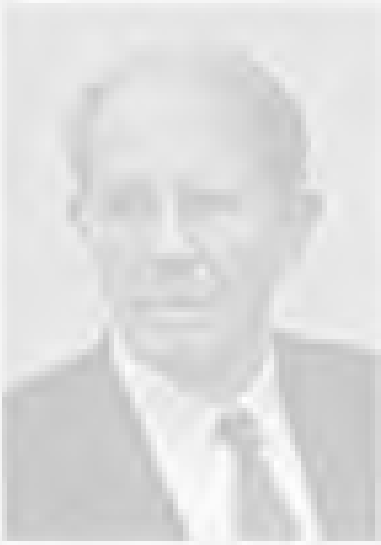
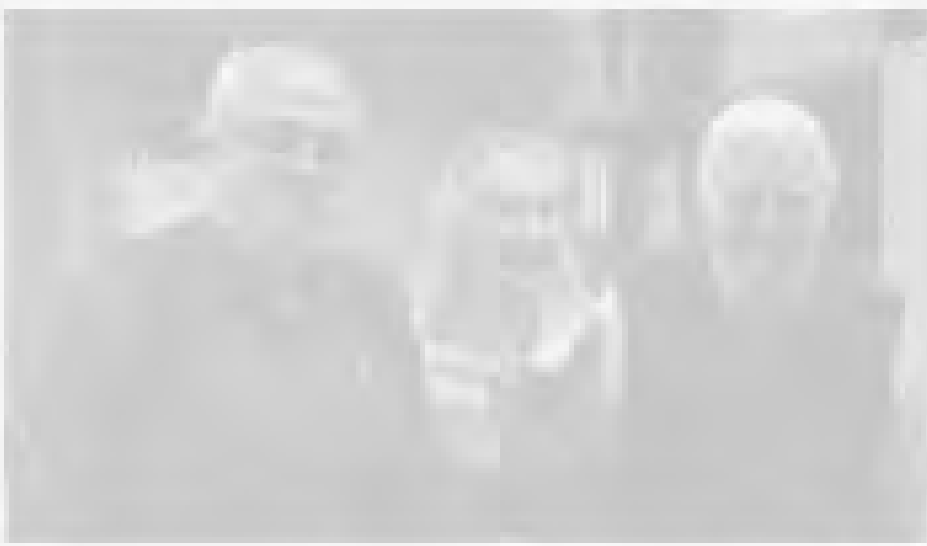
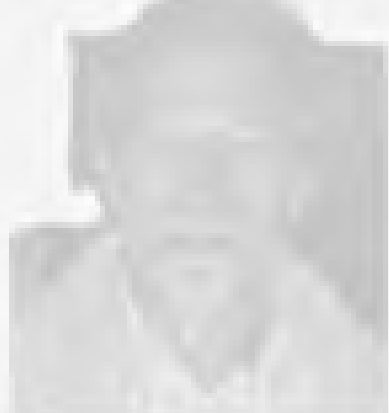
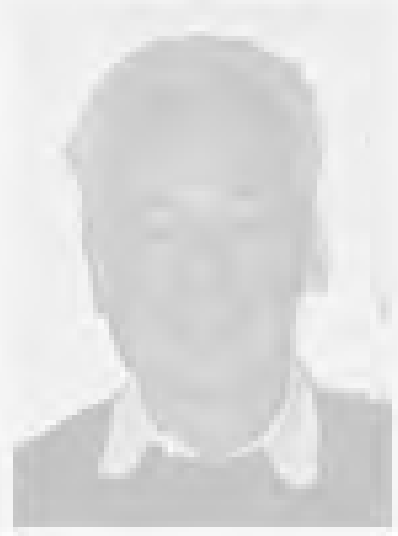
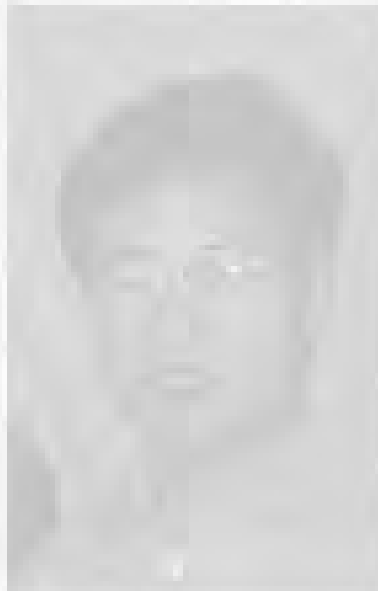
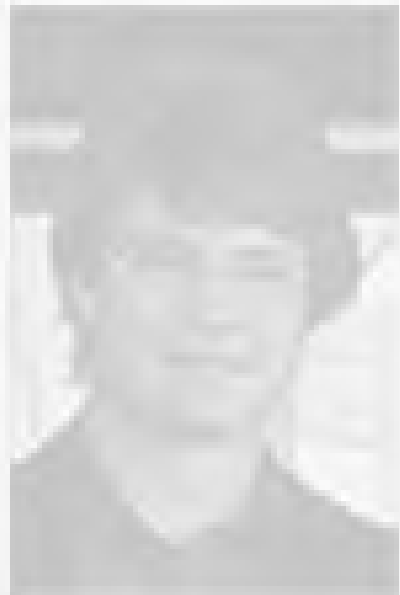
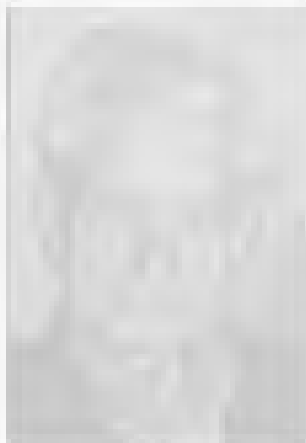
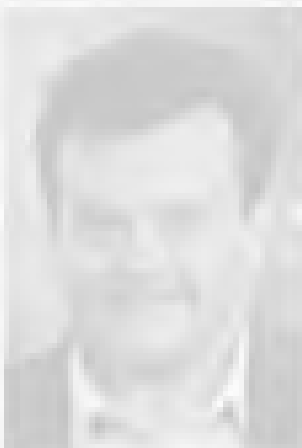
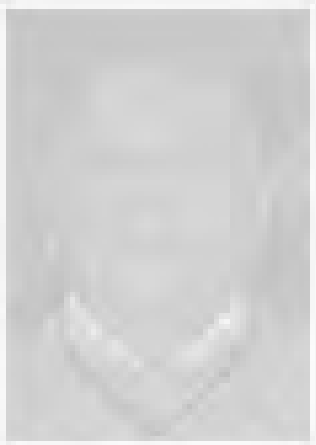
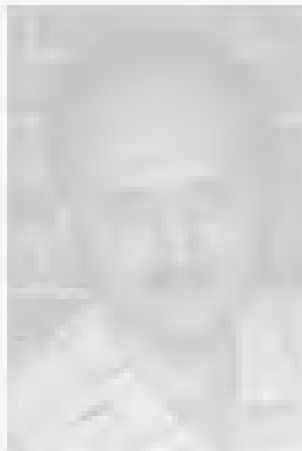
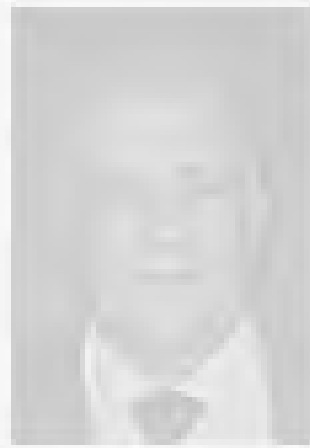
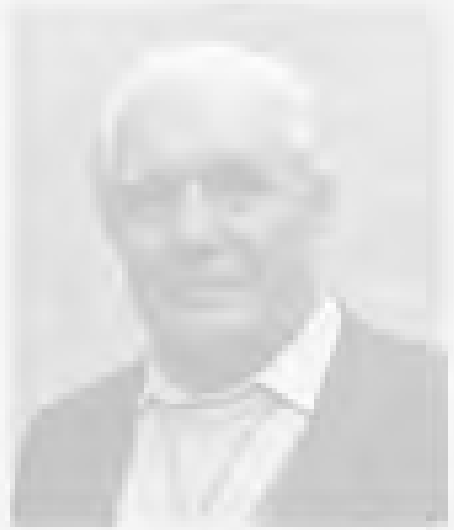
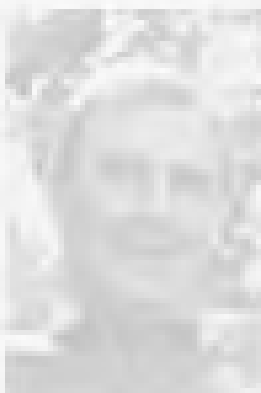
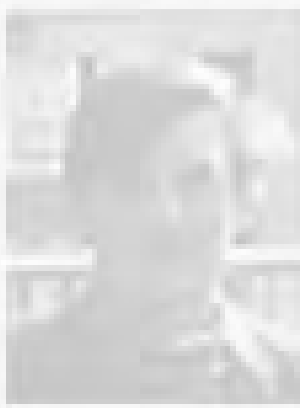
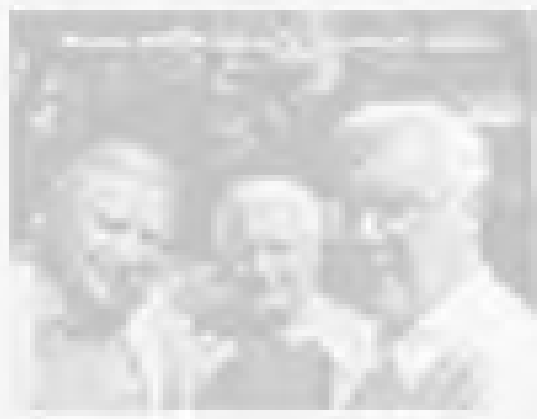
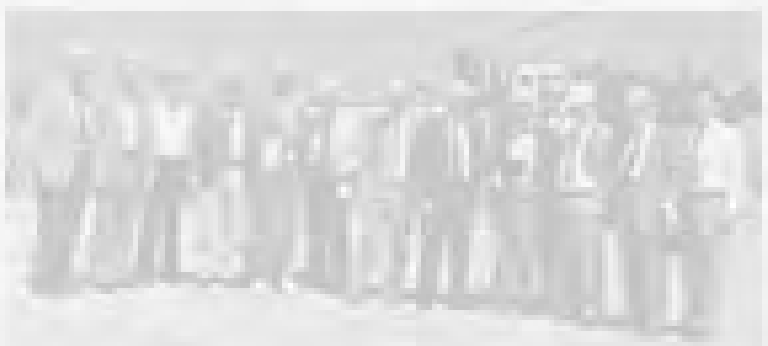
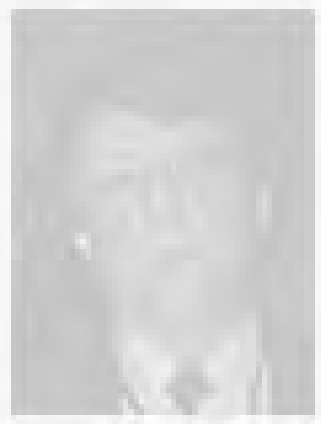
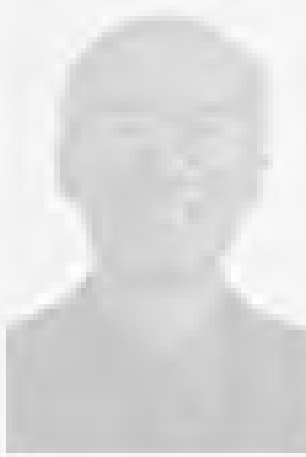
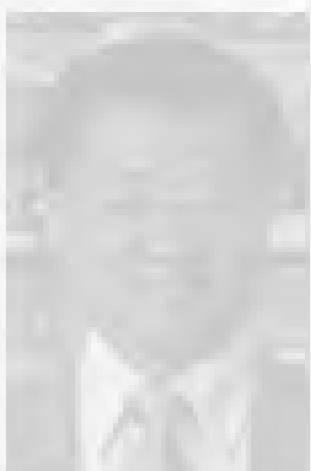
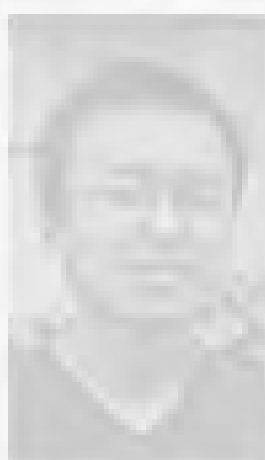
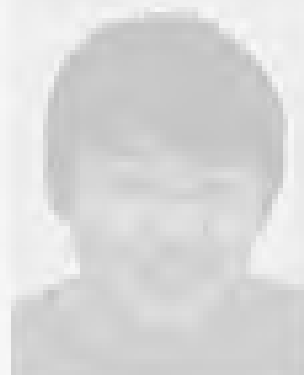
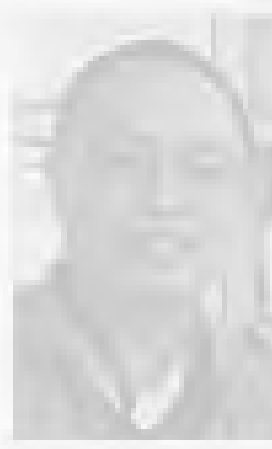
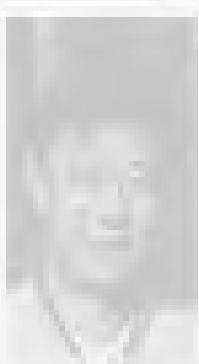
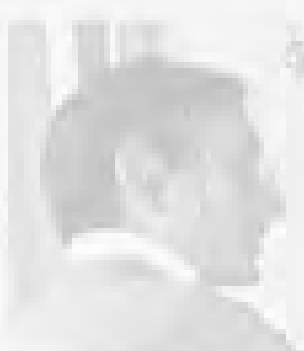
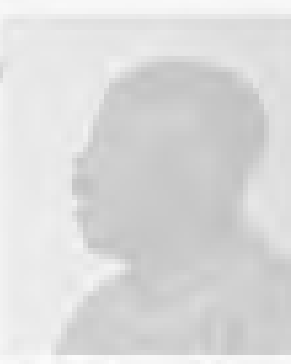
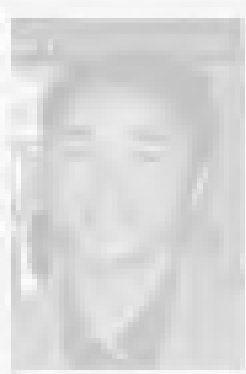
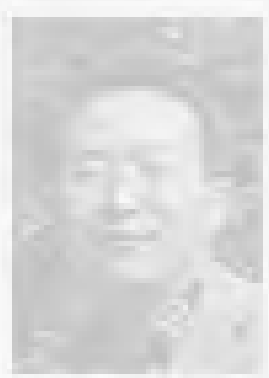
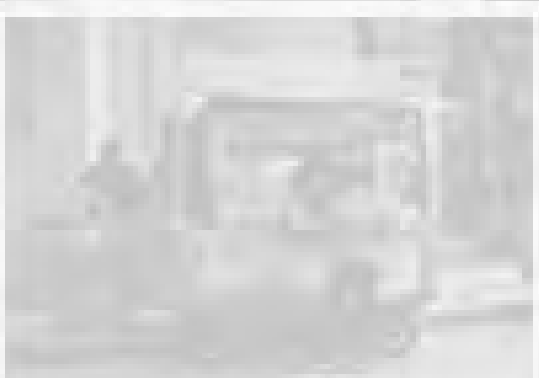
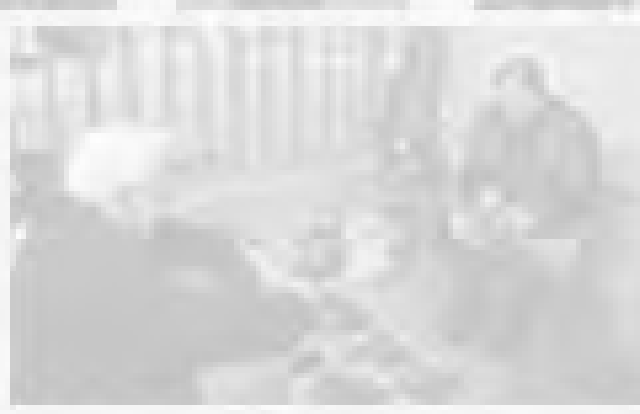

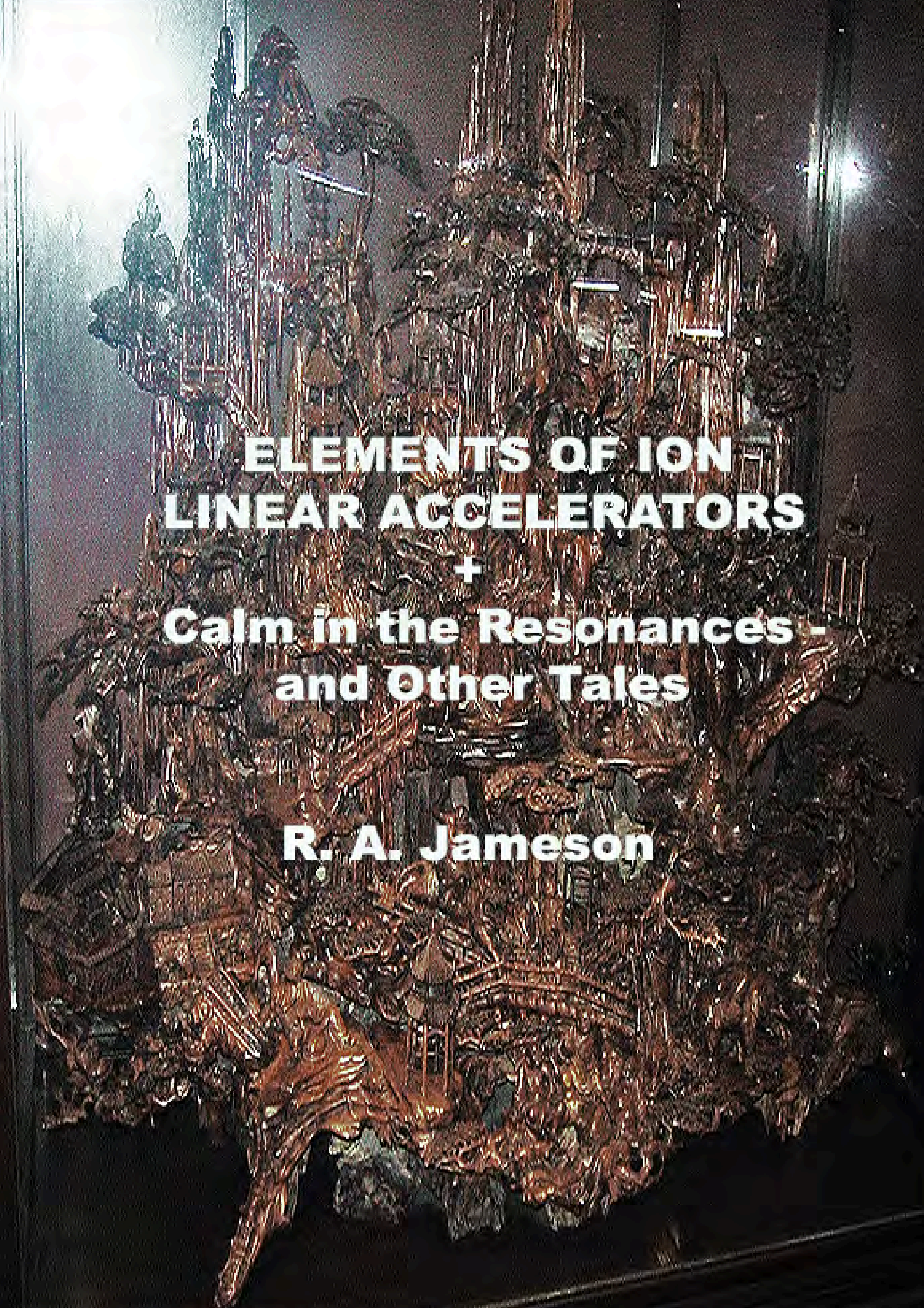

# ELEMENTS OF ION LINEAR ACCELERATORS
# +
# Calm in the Resonances - and Other Tales

R. A. Jameson

**(September 2023) The LINACS codes are released at no cost and, as always, with fully open-source coding.**

SEE - POSTED TO RESEARCHGATE:
**The LINACS Codes Release**
R.A. Jameson (retired), J.M. Maus (NTG, Germany), Bruce Yee-Rendon (J-Parc, Japan), M. Okamura (BNL, USA), P Jiang (IMP, China), C. Li (DESY, Germany)

DIRECT LINK TO THE FULL CODES PACKAGE:
https://ln5.sync.com/dl/01ce57220/v23cfj82-gsx74hws-runykxcf-btqmfe2a

**Abstract** — The main part of this book, "Elements of Linear Accelerators", outlines in Part 1 a framework for non-relativistic linear accelerator focusing and accelerating channel design, simulation, optimization and analysis where space charge is an important factor. Part 1 is the most important part of the book; grasping the framework is essential to fully understand and appreciate the elements within it, and the myriad application details of the following Parts. The treatment concentrates on all linacs, large or small, intended for high-intensity, very low beam loss, factory-type application. The Radio-Frequency-Quadrupole (RFQ) is especially developed as a representative and the most complicated linac form (from dc to bunched and accelerated beam), extending to practical design of long, high energy linacs, including space charge resonances and beam halo formation, and some challenges for future work. Also a practical method is presented for designing Alternating-Phase-Focused (APF) linacs with long sequences and high energy gain. Full open-source software is available. The following part, "Calm in the Resonances and Other Tales", contains eyewitness accounts of nearly 60 years of participation in accelerator technology.

The book is still in development, and may be updated from time to time.

A tutti coloro che avranno il coraggio di leggerla....

# Elements of Ion Linear Accelerators

## + Calm in the Resonances and other Tales

[GoToCalm]

R. A. Jameson
2024
rajameson@protonmail.com

Table of Contents [EITOC]

[eltoc]

NOTES

NOTES

# PART 1 — Elements of Ion Linear Accelerators

Once I read a technical book entitled "Elements of …", expecting as usual to find a maze of details, but was delighted that it in fact did present fundamental, essential and sufficient principles. It is like learning to dance – almost all courses concentrate on details, like the basic steps, awkward execution of memorized step patterns, etc., but never mention the **elements** – posture, balance, weight change, tempo, smooth transitions, how to put the basic steps into the background and improvise, maintaining or changing the instantaneous direction of the dance, foot closure at end of step, moving to accommodate the partner's movements, danceable music, etc.

I hope here to outline the crucial requirement for an overall **framework**, itself an **element**, in which to embed the **elements** of linear ion accelerators that constitute a simplified, compact, sufficient and practical method for ***ion linac beam dynamics design and simulation***, including optimization tools. Ion beams are the usual application, but more generally, non-relativistic particle beams.

In addition, many details are considered, and the additional myriad can be inserted as desired.

The emphasis is on the broad view, and this is approached in a different way than usually found in accelerator literature or the relatively few training opportunities that exist.

**The application of the work** concentrates on all linacs. large or small, with significant space charge effects, for factory-type application - which means for long and reliable life. **A comprehensive and self-consistent specification, plus the self-consistent big picture framework, is required for such applications**, which may not even top-rank low beam loss, but includes it in rank order relative to engineering, cost, RAMI and many other items, together adequate for the project lifetime including potential upgrades. The beam dynamics design and simulation are the primary thrust; the other items only when they relate but not in detail.

*It will be necessary to repeatedly remember the applications and specifications, to fully appreciate the framework and elements.*

*It will be necessary to repeatedly remember the applications, specifications, framework and elements for serious non-scientific, non-technical reasons.* Science is a human endeavor, and these aspects are interesting and even crucially important – considering not only the ultimate successes, but also some philosophy, and stumbling blocks or even failures that occur because of human aspects, not usually included in a technical book – and very necessary to the broad view.

One such aspect in particular has developed which must be addressed up front very early. A past history and current body of narrowly focused research has developed, that is relevant *but peripheral* to the framework for constructed linacs developed here. There is hardly consensus within this work, hardly effort to achieve consensus, and the literature is confusing. This is of course entirely fine, but extrapolations have started being made from it to linacs included in the applications, specifications and framework here. Such extrapolations, made without the framework and perspective required to address the applications concentrated on here, have sown and are sowing much confusion. Space is also given below to this problem. For two worst cases, see Sec. 1.7.4.

*These opening paragraphs are so important that they will be specifically expanded – Sec.B. Without fully understanding the ideas of a **framework** and **elements**, it would not have been possible to develop tools to explore the further details in this book, or to understand them – there would be just the usual disconnected details.*

The book is written specifically for younger, still curious, generations. I have been approached by many students in various lands who complained that they had been taught and shown many details, but that they could not see an overall view, approach, or sense, and were baffled. There was not time or opportunity enough to help them very specifically, except with the students with which I became closely involved; I hope that this book may be of help to the newcomers. Please make the admittedly nontrivial effort to understand the big picture here, and be very thoughtful in relation to the literature.

Before concentrating on beam dynamics design and simulation, I was fortunate to be the first, because of the enormous step of the LAMPF linac, to investigate other major aspects of linacs, such as the rf field phase and amplitude control, operation of megawatt rf amplifiers below saturation with phase and amplitude modulation, overall installation and bringing to design performance, alignment, operation and maintenance management – each of which was approached by formulating a broad view within a **framework** and basic **elements**.

My investigations have benefitted very much from the use of material and also *concepts* from fields outside the narrow field of "accelerator physics", in particular electrical, automatic control and other fields of engineering, also nonlinear systems and plasma physics. [1]

The progression here may also have a flavor of discovery, progression of the research, and philosophy, which is usually "not allowed" in what are considered to be "scientific publications". But these are essential to a big picture.

Exciting discovery and progression still occur – for example, now in 2021-2022, with new insights and synthesis presented in Ch. 23.7 and 28. There are plenty of subjects for the future.

It is not intended to provide exhaustive background for every topic [2]. Basic background in linear and ring accelerators is expected, and it is expected that serious readers could extend their background, for example exploring the modern aspects of nonlinear systems by reading outside the "accelerator literature" (which at present hardly addresses the subject). Problem sets are not developed. Unorthodox techniques such as repetition are used to hammer on the **framework** and most important **elements**. It is hoped that the expositions will be adequate to convey the **framework** and **elements** and enough detail to encourage the reader to use and further develop the required tools and techniques.

Therefore the views, presentations and conclusions are the results of my own fascination with an unexplored landscape, and there was and is no one "out there" with whom to review the whole. The reader is free to accept, question, add or reject.

The progression is my view, developed during and after the design and construction of the 800 MeV, 1 mA average, 17 mA peak current Los Alamos Meson Physics Facility (LAMPF) proton linear accelerator. This linac was a $10^4$ order advance over existing operating linacs in the US – a factor of 10 in energy, and factor of 1000 in average beam intensity. Progress was made on LAMPF that led to important insights into an overall view. It was possible to get advice on many details, but there was essentially no one to talk to about the big picture – it had to be developed from scratch. That has continued to be the case in my personal quest for understanding of the elements of high-intensity ion linear accelerators.

For almost thirty years after LAMPF began operation in 1972, there was little interest in ion linacs with significant beam current, but there were many outstanding questions that I wanted to understand. Two examples are 1) that LAMPF, and the subsequent linacs built the same way at FBAL, BNL, CERN, all had unexplained emittance growth, and 2) the infeasibility for significant beam current of using continuous-wave (CW) Cockcroft-Walton injectors to inject significant beam current into a drift tube linac, which led to development of the Radio Frequency Quadrupole (RFQ) in the West. The great success of the RFQ, even in its original very simple form, in transporting low energy, low-to-medium intensity (e.g. < ~50mA peak current protons) ion beams with small emittance growth, has satisfied most projects, but there were many questions about how to configure it for high beam intensity and very low beam loss, how to design it for desired beam space charge physics characteristics, how to negotiate design compromises between performance and cost, etc. Fortunately, I was able to participate in and influence the RFQ development and several projects with high-intensity beam requirements – the 1980 LASL AT-Division RFQ prototype and subsequent development, the 125 mA D+ cw beam current International Fusion Materials Facility (IFMIF) project for fusion-related neutron-resistant materials development; the J-Parc linac; and the Accelerator Transmutation of (radioactive) Waste (ATW) project. ATW required convincing evidence that even up to several hundred mA $H^+$ accelerators to 0.6-1.0 GeV energy were feasible, and the overall view provided this, leading to a worldwide effort (although current projects are at low beam current, for various reasons.).

---

[1] It is important to note that a concept can be used as a launching point, without having to fully understand its context, or to re-derive it from first principles for a journal article, as is often demanded by a constrained "reviewer".

[2] Although this seems to be a primary requirement for any paper hoping to pass "review", endlessly, boringly, and unnecessarily repeated.

The RFQ is in fact one of the hardest forms of linacs because of the beam transition from dc to bunched, and also incorporates the aspects of any linac. I decided to use the RFQ as a research platform, and most of this book involves the RFQ as the example. **But the big picture is the same for any linac.**

It was, and unfortunately still is, widely believed, in spite of clear demonstration to the contrary, that the internal workings of the RFQ are "too complicated" to be viewed in detail; only the RFQ output characteristics were used; and improvements in the design and simulation methods were not – and are still not - incorporated. In particular, many designs have been and are still based on the earliest approximate method that we developed at LASL [3] for the first RFQ built outside the Soviet Union. It *is not difficult* to understand the internal beam dynamics of any linac, including the RFQ. A powerful comprehensive analysis and design method is developed herein.

Later US projects, like linac designs for antimissile defense, tritium production and the LEDA RFQ, the Spallation Neutron Source, used the thirty-year-old design method of LAMPF plus the rudimentary RFQ design, unfortunately losing possible generations of advancement.

Very slowly, as several projects in the 21st century required beams of approximately the same peak beam current as LAMPF, the questions of how to produce them while minimizing beam loss to afford practical maintenance have begun to be investigated, but in most cases, restricted to the particular project and based mostly on "what has been done before". Some traces of the tools outlined herein are beginning to be seen. Unfortunately, the higher intensity SNS linac was not designed using the elements and space charge physics principles outlined herein, but only with earlier non-optimum procedures and some subsequent hand adjustment. Fortunately, the KEK/J-Parc linac, with similar intensity requirement, broke out and did use the concepts. Aspects of these decisions will be pursued.

[eltoc]

# A. The "Framework"

Desired **elements** of ion linac beam dynamics design and simulation are introduced in **bold** type. Overall, the elements should fit into a "**framework**", itself an **element**, for design, simulation and optimization. The importance of a **framework** is *paramount* for integration, coordination, and substantiation of the **elements** as a whole.

I use the example of a beautiful and intricate 3D sculpture in the root section of a large tree, Fig. A.1. The very many details could not properly be integrated without the "**framework**" of the tree.

[3] Los Alamos Scientific Laboratory. The change to Los Alamos National Laboratory (LANL) says a lot. At the time of the change, we noted that now we were on par with the Los Alamos National Bank…

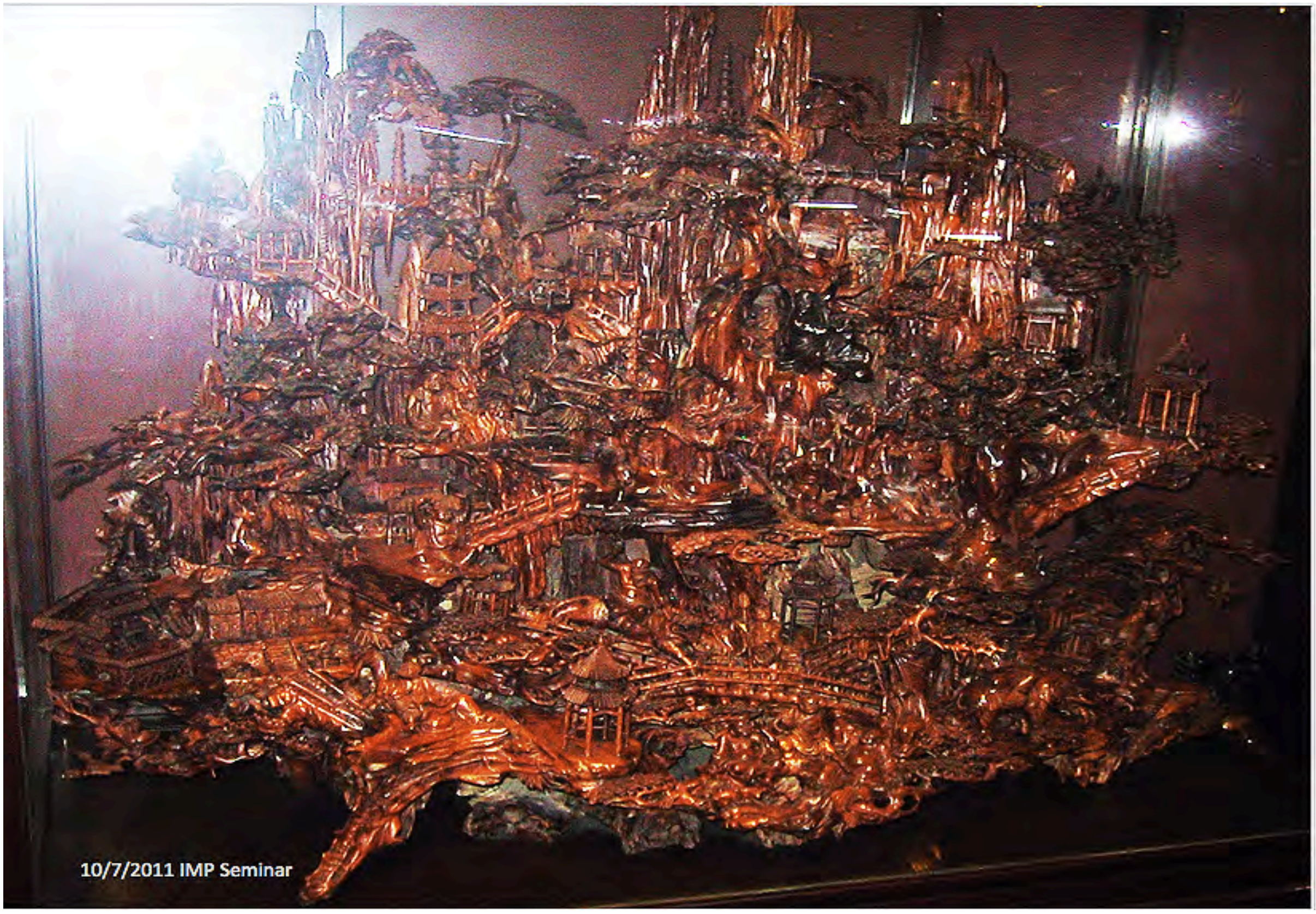

Fig. A.1 A "framework" supports all the many details…

Another beautiful analogy, representing the multidisciplinary nature of the tools used to carve the many details, is a thoughtful gift [4], the many-armed Bodhisattva, Fig. A.2. A different tool in each hand – that must be centrally coordinated.

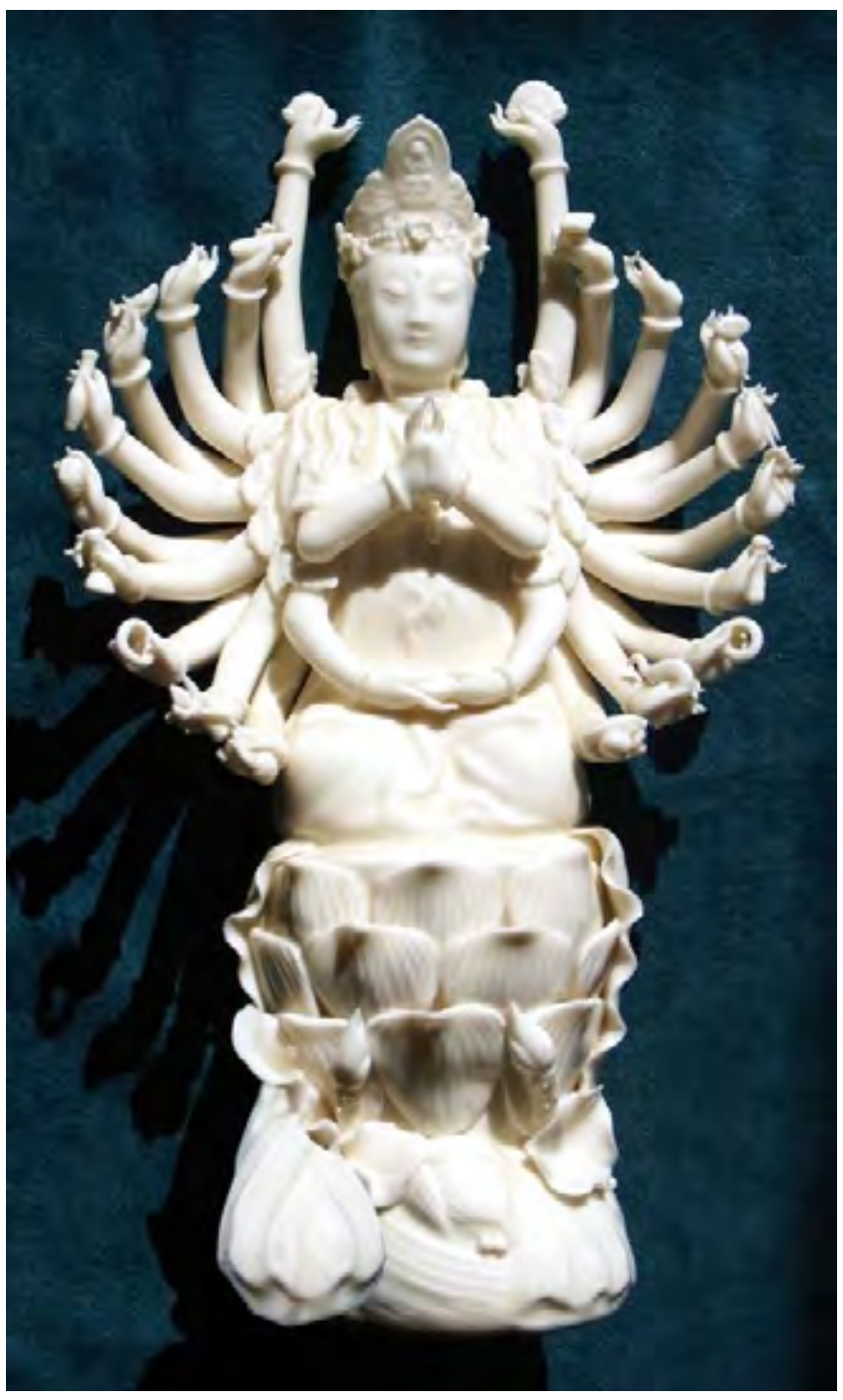

Fig. A.2 Multidisciplinary tools

[eltoc]

# B. Elements of the Framework

## B.1. The Philosophy of William of Ocham [5]

My preference evolved, when looking for the requirements of something, to try to find the simplest framework of a minimum set of necessary and sufficient elements. As usual, we stand on the shoulders of our predecessors.

William of Ockham (1285-1347) was an English Franciscan friar, scholastic philosopher, and theologian, who formulated a "principle of parsimony or the law of parsimony, the problem-solving principle that "entities should not be multiplied beyond necessity (not his words)"" – becoming known as "Occam's

---

4 Unfortunately, not sure from whom – think probably Prof. Chen Jia-erh on visit November 1992.

5 Assembled from Web, especially Wikipedia, Merrian-Webster. These should be carefully read – essential background material.

Razor" (also not his words), although the principle was expressed by numerous famous philosophers long before. – It "is interpreted as requiring that the simplest of competing theories be preferred to the more complex or that explanations of unknown phenomena be sought first in terms on known quantities.", Aristotle ("the more limited, if adequate, is always preferable"), Ptolemy ("we consider it a good principle to explain the phenomena by the simplest hypothesis possible"), Isaac Newton ("we are to admit no more causes of natural things than such as are both true and sufficient to explain their appearances").

A theory (framework) that is too simple can obstruct progress as much as one that is unnecessarily complicated. *The theory (framework) must be tested!*

"Occam's razor doesn't necessarily go with the simplest theory, whether it's right or wrong; it is not an example of simplicity for simplicity's sake. It merely tries to cut through the clutter to find the best theory based on the best scientific principles and knowledge at the time." "Similarly, in science, Occam's razor is used as an abductive heuristic in the development of theoretical models rather than as a rigorous arbiter between candidate models. In the scientific method, Occam's razor is not considered an irrefutable principle of logic or a scientific result; the preference for simplicity in the scientific method is based on the falsifiability criterion. For each accepted explanation of a phenomenon, there may be an extremely large, perhaps even incomprehensible, number of possible and more complex alternatives. *Since failing explanations can always be burdened with ad hoc hypotheses to prevent them from being falsified, simpler theories are preferable to more complex ones because they tend to be more testable.* "

The philosophy is extended by many, far beyond this simple précis. The Wikipedia coverage is necessary background to study - it includes important points concerning probability theory and statistics (including AI research and the free energy principle); difference in interpretation of scope and application, leading to controversial and diametric views ("controversial aspects", even humorous ones, as "ironically" presented by Galileo); (anti-razors).

**Without *study* of these Web articles, one will lack essential background for understanding this book.**

**You will shortly be introduced in Sec. 1.4, to three *time-dependent* nonlinear equations (1), (2) & (5) that are essentially the *only* equations presented in this book. Their exact, but sophisticated simultaneous solution as a function of time** (which is not at all difficult to obtain) **enable a powerful foundation – that of an equilibrium.**

**At this condition – called equipartition (EP) of the particle distribution's energies in the degrees of freedom – even instantaneously, but also held over the linac trajectory – the particle distribution has no unbalanced energy that could result in expansion of the distribution and possibly particle loss to the surrounding environment which makes access and maintenance difficult.** Minimizing any such loss is a highest priority for any linac with significant beam intensity.

**The guidance is that a rigorous EP design will be investigated first, against a rigorous specification** (including engineering, cost, RAMI, error tolerance)**.** If satisfied, this design will be sufficient. If there is any difficulty, adjustment of the specification must be considered. If still unsatisfied for a defensible reason, EP can be abandoned, the beam is just kept "matched" to the lattice, an ad-hoc or another programmed design technique explored, and often then also acceptable, in particular if the spec does not rank low beam loss as highest priority.

**This constitutes the Occam's Razor framework needed for ion linac beam dynamics design – this is the sufficient big picture for essentially all current routine design work – for both newcomers who have pleaded for such, and for practitioners. (It is augmented by additional essential elements and accurate simulation.)**

***If one does not grasp and understand Part 1.-A&B & Chapter1, the remainder becomes perhaps interesting, but inconsequent. Please accept until absorbing unbiased the further development. This material shows the success of the method and necessary tools via careful testing.***

An alternate approach - of interpreting from many details and unconnected examples without a framework - usually fails in testing [6].

The Occam's Razor approach developed here is purposely not complicated, but has been considered so, to what seems a somewhat astounding extent. Its purpose is to be directly useful both for new students and expert designers in the daily practice.

A very complicated framework approach may also be elegant, interesting and probably valuable, if it could be sufficiently tested and promulgated.

Garaschenko made the breakthrough in APF design, detailed in Part 2.B, using the modern control theory framework.

An elegant framework, based on the most extended concepts of modern control theory and artificial intelligence, has been pursued in Russia through a collaboration led by Prof. Dimitri Ovsyannikov of St. Petersburg University. This collaboration also included the powerful Moscow Radiotechnical Institute (MRTI) group that was led by Dr. Boris Bondarev (see "Calm in the Resonances"). Unfortunately the published record in English has been sparse, without much practical evidence and detailed comparison to other methods, and source code for very well developed software (e.g. the LIDOS and BDO codes).

## B.2. Elements of the Ion Linac Beam Dynamics Physics Framework

Ion linac beam dynamics physics is that of modern **nonlinear systems**, applied for ion acceleration below the stochastic limit. So it is the same as the physics of the universe but at faster time scale – and *not* at or above the stochastic limit, where real thermalization occurs.

- The ***most important element*** is that simulation, analysis, and aspects of the design practice always use the **instantaneous state of the beam**. There is no steady state in a linac.

This is essential to understand. The governing principles and equations are based on this. Because serious misconceptions have arisen, many have been confused. This **element** will be hammered on repeatedly.

The **essential theory, design and simulation procedures** require consideration of the **basic processes** – the **rms properties applied locally, and phase-space transport re-distribution, and single particle behavior** of the beam distribution. **The fundamental design basis is the rms distribution, obtained from the time-domain slow component, or smooth approximation, of the particle motion AND APPLIED AT EACH TIME INSTANT** (2018 – see Ch.29 & 30 on extension to remove the smooth approximation)**.** Simulation must go correctly beyond the rms.

- Important distinctions are necessary to correctly interpret the interplay between what happens in an actual machine (the ''experiment"), simulation, and theory. This **element** will be termed the "**Actual/ /Simulation/Theoretical Hierarchy**". It is essential to understand that the beam itself has no knowledge of any theory.

---

6 Although seldom in "reviews".

• The general design method should afford the designer **control over *all* parameters**, including not only the external fields, but also the space charge physics for high intensity linacs. Comprehensive, flexible design codes with control over all parameters are rare – and much rarer in open source code. The **space charge physics** is the key that affords **design from "inside out"**, ***from*** desired beam space charge physics characteristics ***to*** the external fields required to realize these characteristics. Extant methods set up external fields and then simulate to see what happens to the beam – an "outside in" approach (and typically do not even compute the relevant space charge physics parameters).

• Design and simulation codes require and should provide the **best physics and engineering models** affordable with modern desktop, laptop, main frame computers. The necessary **compromises should prefer the physics over computing aspects**. Other computational modes, for example with more approximate physics but faster running time, are desirable but the **compromises and relationship to the best models should be clear, documented, and tested**. Old codes contained many simplifications (holding quantities constant because easier to understand or implement in practice), approximations (theoretical, to address computer limitations, etc.) that needed to be discarded or updated.

• **Experimental methods should be used in testing, verifying, and also comparing, the physics and programming of codes, over wide parameter range** – anecdotal examples, theoretically simplified models of isolated aspects, etc. are insufficient and often misleading, and comparison to codes with known deficiencies is nonsense.

• ***Especially important*** is that **use of a design and simulation code must lead to the true design optimum**, and not a false result caused by the code itself. This insight and ever present concern from engineering design is hardly perceived in accelerator physics.

• It is an *essential* **element** that the development, use, verification and interactions between the design, simulation and optimization codes have an overall concept and be closely coordinated. This requires a comprehensive, multidisciplinary point of view, encompassing engineering, modern control and optimization, nonlinear systems, mathematics as well as what is termed accelerator physics.

An essential corollary is that **the design code physics must be kept as close as possible to the simulation code physics**.

Similar frameworks were crucial in helping me with other responsibilities at LAMPF, as embellished to some extent in "Calm in the Resonances and Other Tales", including:

• The first design and implementation of automatic rf field control for a linac, involving detailed understanding of all components of the control paths and the engineering discipline of modern automatic control systems. This included operating MW-class rf amplifiers out of saturation in order to modulate phase and amplitude, for the first time.

• Understanding and implementing the structure tuning method for the ~10,000 cells of the LAMPF room-temperature linac.

The frameworks for these two were simply that the situations had to be very well understood and demonstrated that full-power operation was as desired, and could be predicted to be stable over a long time with acceptable confidence.

• Managing the final construction, installation, commissioning, and initial turn-on of the LAMPF linac. This framework was basically very simple. One boss, and one (blueprint size) piece of paper [7].

• Managing the process of bringing the linac to design performance.

---

7 In stark contrast to the Harvard Business School of (Mis)Management concepts that have contributed so much to the decline of USA technical strength over the past ~30 years. In particular, "matrix management". When everyone has two bosses, loss of excellence is *guaranteed*. Especially when the grid is tipped the wrong way, with "project management" having the upper hand (vertical lines) over budget and schedule, with no responsibility for, or knowledge of, the technology - the usual case practiced. One technically competent boss is required, and project management is one of the tools. Matrix management is usually thought of as a 2D grid, but it is actually 3D, the third dimension being leakage of excellence.

• Managing a subsequent complete realignment of the linac and beamlines. This framework was again very simple – the survey was required to be presented on one piece of graph paper, with one origin. The initial many surveys were uncoordinated, each with its own origin; the realignment found very many incongruities, even up to 1 inch.

**FRAMEWORK**

**Nonlinear system**

**Instantaneous time state**
for all design, simulation, analysis

**Hierarchy**
Actual (Experiment) – "the truth"
Simulation (also is "actual")
Theoretical (beam has no knowledge of theory)

**Control over *all* parameters**
Including space charge physics

**Design from "*inside out*"**
Define desired beam physics, create fields to accomplish

**Basic processes**
Redistribution behavior (space charge mixing)
Rms behavior – fundamental design basis
Single particle behavior

**Best physics possible**
Essential – to max computer capabilities, no compromises or approximations.
Alternate running modes (e.g., faster execution) clear, documented, tested.

**Test codes with experimental methods**
Over wide parameter range (not anecdotal or with simplified models)

**Codes must result in true optima**

**Design code physics as close as possible to simulation code physics**

Fig. B.1 gives an outline of the framework for linac beam dynamics suggested above.

The goal of a framework is to establish, **as simple as possible**, the basis and controlling view of the big picture, upon which to build the details.

In this book, the goal is to find the simplest way to frame the handling of an ion beam in a linear accelerator that will function as desired in operation. The framework then includes design, simulation, theory, analysis and practical application.

It will be shown that the simplest form of the framework requires a complete understanding that the system must be described in the canonical time-domain, local, instantaneous system state. The design, simulation and analysis can be developed under three theoretical equations (one of which is optional). The simulation requires accurate solution for the external and internal forces by accurate, then also relatively fast, fully consistent Poisson solutions.

**Chapter 1** expands the definition of the framework. This is basic and general for any linac.

**Part 2.A. Chapters 2-5** introduce RFQ design from the "traditional" point-of-view, and show how this is extended to introduce the beam space charge physics in the design process.

The mostly RFQ-based exposition is then broken, to insert **Part 2.B. Chapters 6-12**, the discovery of a practical method for design of general alternating-phase-focused (APF) linacs. (The RFQ is a special smooth-focusing case of APF!) Although APF offers very significant cost reductions, design methods for general structures were previously very limited and tedious, which has strongly constrained their use. With this new practical method, the way is opened for a wide variety of possible applications. (2018 – see Chapters 27, 29 & 30, which also have APF connotations.)

It was clear since the mid-1990's that extant RFQ simulation still required significant improvement, but interest and support were lacking. **Part 3, Chapters 13-22,** on RFQ simulation code development, traces my efforts to understand and compare extant codes, and to develop a new code incorporating the best physics possible with the increasing computer power available. Deep and extensive collaboration, with Yuri Batygin and his *BEAMPATH* code, the MRTI *LIDOS* group (Boris Bondarev, Alexander Durkin and Stanislav Vinogradow), and the doctoral work of Johannes Maus IAP Frankfurt has resulted in new methods and a new code, *LINACS*, which incorporates the framework for any linac. The RFQ option, *LINACSrfq*, is now the standard for comparison.

Chapters **13 – 15** trace the considerations involved in comparing codes, as applied to the RFQ, leading to the evolution of the basis for developing a new simulation code, *LINACSrfqSIM*, in Chapter **16**. It was necessary to completely replace the approximations to the RFQ external and space charge fields with modern Poisson methods, made practical for presently available desktop/laptop computers - the doctoral thesis (**summa cum laude**) of Dr. Johannes Maus at IAP, Chapter **17**. Chapter **18** then outlines comparison to the well-developed Russian RFQ code *LIDOS*, whose authors were very open to extensive discussion and collaboration which was very helpful. Chapter **19** outlines the synthesis of the new code *LINACS* and *LINACSrfq*, including a simple graphical users interface (GUI) and a single output graphic which displays the information that is actually needed for RFQ design (in contrast to other extant codes). Detailed discussions cover the complex treatment of the RFQ injection and front end, input beam matching, estimation of the rf power requirement, the compatibility of the accurate Poisson and fast 2-term simulations, and combination of all the **elements** into the synthesis of the *LINACSrfqSIM* code. Chapter **20** compares the new code *LINACSrfq* to the *LIDOS* code, which has very carefully thought-out physics and programming [8].

**Part 4, Chapters 21-24**, are detailed investigations of several particularly important linac aspects. Chapter **21** discusses single particle behavior, and questions about the tune chart. Chapter **22** investigates the basic, underlying, always present driving force in channels with space charge – the space charge driven "space charge mixing" [9] which collectively affects the trajectory of each particle. Chapter **23** discusses RFQ design with arbitrary vane tip shapes. Presently included in *LINACSrfq* are the nearly standard circular vane tip cross-section, and sinusoidal, trapezoidal and 2-term-(not recommended) longitudinal modulation. Full Poisson simulation uses the exact geometry and is the accurate method. A multipole expansion for the fields is however very useful for *design* purposes, and several methods for finding the required coefficients are presented. Chapter **24** presents dated but still the most complete treatment of halo formation and phase space transport, which opened the way to extensive investigations by others. This area is still not complete, in terms of understanding or practical application, and must be pursued further for planned upgrades of existing facilities.

**Part 5** covers some specific linac investigations, interesting in themselves, but importantly to demonstrate the **elements**. **Chapters 25-26** are two design studies that indicate the power and flexibility of the "inside out" design technique. In Chapter **25**, the Japanese government issued a challenge to investigate the possibility of very high deuteron beam current in an RFQ [10]. 400mA D+ was chosen as the design operating current. A design found by a code with known deficiencies was checked in *LINACSrfq*, and found to be highly optimistic, as expected. A satisfactory new *LINACSrfq* preliminary design is presented. Chapter **26** required D+ and H+ RFQs for a small neutron source, with emphasis on very low injection energy, as well as the usual minimum cost, minimum length.

---

8 This comparison to a well done, open source code was very worthwhile. It is emphasized that comparison to codes which have known deficiencies (there are several) is pointless. This should be obvious, but is not – such "comparisons" appear so often in the "literature".

9 Even this terminology is not always understood by "accelerator community" "reviewers"…

Chapters 27-30 were recently added, therefore at the end. **Chapter 27** (2017) turned out to be multi-faceted, with discovery of a new kind of accelerating channel, that with further investigation (a worthy PhD topic), including connection to APF design, could have significant advantages for high-intensity linacs. In addition, the framework method for analysis is applied to analyze the compensation of existing linacs for H- intrabeam scattering. These *separated-function* linacs have significantly different focusing lattices from the simple lattices represented by the extant theoretical tools for locating dangerous space charge driven resonances in linacs and rings. Current research is directed towards extension of the theory and tools to afford deeper treatment of the very low beam loss requirement of high-intensity accelerators (both linacs and rings). The book should be interesting and challenging all the way to the end – Chapter **28** reviews and extends the **elements** of time dependence and *a priori* adjustments to improve the equipartitioning EP characteristic and ratios. Chapter **29** (2022) outlines the discovery of a new and powerful way to control the longitudinal beam emittance, resulting in a major improvement of the RFQ shaper strategy. Chapter **30** considers aspects of space charge influenced resonance codes and beam halo. Chapter **31** outlines still outstanding challenges for future high-intensity linacs and rings, and an approach that could be taken. Chapter **32** briefly discusses taking an expert system approach to the RFQ.

While many very detailed aspects are developed which have not been studied or presented elsewhere, the reader is urged, upon initial perusal and over and over again, to think in terms of the big picture – the **framework** and its **elements** – and how they interrelate to produce integrated tools with accurate physics and engineering fidelity – and without which the reliable and improved results demonstrated can not be produced.

Full-text versions of my referenced work are available at https://www.researchgate.net/ and https://arxiv.org/ .

The hospitality of many hosts is gratefully acknowledged:
Japan: Institute for Chemical Research (ICR) Kyoto University Uji campus
KEK High energy Accelerator Research Organization, Tsukuba
RIKEN, Wako campus, Eminent Scientist, Radiation Group, Nishina Center, RIBF
JAEA - Japan Atomic Energy Agency
(previously JAERI - Japan Atomic Energy Research Institute)
J-PARC Center, Tokai, Japan
KEK Tsukuba and KEK/J-Parc, Tokai-mura
National Institute for Radiological Science (NIRS), Chiba
Institute for Nuclear Science, Univ. Tokyo
Sumitomo Corporation
Germany: The Institute für Angewandte Physik, Goethe Uni Frankfurt, Frankfurt-am-Main
The Alexander von Humboldt Foundation
GSI Helmholtzzentrum für Schwerionenforschung, Darmstadt
Russia: Moscow Radiotechnical Institute (MRI)
St. Petersburg State University
Moscow Meson Physics Facility, Troitsk
China: Fudan University, Shanghai
Peking University, Beijing
Institute of Modern Physics (IMP), Chinese Academy of Science (SAS), China

[eltoc]

[10] Remarkable enough that such challenge should come from a government; formulated as a pure accelerator R&D problem, but directed toward possible ATW.

## B.3. My Critique of the Present Publishing System

You will have already noticed my concern especially about the present journal publication review practice. Because this will be an ever-recurring theme, its context needs to be outlined specifically.

A critique and criticism are not the same (Google this, e.g. the abacus.bates.edu page).

Appendix 5 of "Calm in the Resonances" summarizes and suggests that the existing framework could benefit from, but does not need, significant change, and that some non-emotional, non-egotistical, non-political improvements might be considered.

***The critique here is specifically limited to the narrow niche of particle accelerator beam dynamics –*** although it is a widespread concern.

"Reviewed" articles are not the same as good articles. Some quite adequate avenues have opened up and are probably the best for the future – especially "open publishing". Authors publish their work – it is the responsibility of the reader to accept and use or not.

My concern has focused on the problems created for young *researchers* (not MS-level PhD street-organ-handle grinders, who must "publish or perish"). Picking up the pieces and helping someone, who has done good work of value to the community, to continue after a *non-helpful*, ignorant, or insulting and demolishing "review" is a hard task.

Young researchers learn, among themselves and from their mostly negative experiences, as with how to pass a test in school, how to attempt to handle non-helpful or non-informed reviews. But after getting published, they regret that they often feel that what they wanted to say has been diluted, impact lost, and that they have wasted valuable time.

So the comments and critique herein should not be regarded as rank criticism. They highlight, in the immediate context of the passage, where the development is beyond the prevailing "state-of-the-art", where deeper thought and improvement are needed.

As developed in Appendix 5, the situation could be *split* into two essentially independent aspects – to have a recognized and well-used *world* mechanism for rapid communication and education, and the separate aspect of formal publishing.

[eltoc]

# Chapter 1 – Framework for Linac Focusing and Accelerating Channel Design, Simulation, Analysis

## 1.1 Definition of a Framework

Guidelines for linear accelerator focusing and accelerating channels are considered mature, but are mostly mechanical layouts of successive components, with simulation to see what happens. The fundamentals are not well understood, which is presently (2024) causing more and more confusion, unnecessary controversy, and "advice, given as general" from outside the context of a practical factory-grade high-intensity linac. A review of the general background and principles perhaps helps clear away a profusion of details and disconnected notions. Also there are sophisticated features that are still little explored, and many rich details and extensions for future research.

Concepts (repeat – concepts), presently mostly outside current "accelerator physics", from standard engineering practice (many areas), mathematics, nonlinear systems, non-neutral plasma physics can be useful for increased understanding. It is however dangerous to mention such because of almost impenetrable stove-piping of the accelerator [11], non-neutral plasma, and neutral plasma fields, there being seldom enough space (or patience) to completely re-derive and present all the background necessary to first re-prove the whole background, present language translations, then present the required very large number of examples asked for, etc., etc. Bridging languages is hard work, and to grasp concepts even harder. Background in nonlinear systems is very helpful, but even that is not necessary to be able to understand as presented here how a linac is working, and how that understanding is used for design, simulation and analysis. **It is simply that we find a method to characterize the beam at any time.** The method has long been shown to be effective in practice for simulated and operational linacs.

[11] Also within the accelerator field itself.

A primary challenge is to accelerate intense beams with very little beam loss, as loss causes radioactivity and consequent maintenance problems. The challenge is to balance the internal space charge forces, that cause the beam to expand, against the external focusing and accelerating forces, such that very few particles are lost in the linac channel. The system is highly nonlinear, and there is a tendency for particles to migrate far from the beam center into higher than Gaussian "tails" or "halos" which become lost.

Linac beam dynamics is basically a classical example of a nonlinear system, a field of great importance in recent years. During its development, different aspects have been emphasized. For a long time, the phenomenon of "chaos" observed in even very simple systems was emphasized, with frustration that theory based on infinitesimal perturbations of a Hamiltonian could give useful indicators, such as the Liapunov coefficient, but knowing that this basis was clearly false. Later development overcame this fundamental difficulty.

A key example of where a concept is useful in this book is the role of the Hamiltonian. Again, the Web articles are necessary background – Wikipedia, Britannica, others – and must be absorbed (as concepts, not necessarily in detail) – to go further here. "Inspired by, but distinct from, the Hamiltonian of classical mechanics, the Hamiltonian of optimal control theory was developed by Lev Pontryagin as part of his maximum principle. Pontryagin proved that a necessary condition for solving the optimal control problem is that the control should be chosen so as to optimize the Hamiltonian. (Wikipedia Hamiltonian (control theory)". [12] Crucial **elements** come from it. The most efficient beam trajectories follow, even to the extent of minimizing "free energy".

The Hamiltonian describes energy in terms of integrals in the time domain. The integral could integration across one time step of a simulation, i.e. at the local time state.

In 1979, Chirikov [13] wrote a classical review paper summarizing the extant situation. It included these unforgettable sentences: – *"Academician A.N. Kolmogorov has mentioned on an occasion that it is not so much important to be rigorous as to be right. A way to be convinced (and to convince the others!) of the rightness of a solution without a rigorous theory is a tried method of the science – the experiment."* .... *"It is obvious that those questions cannot be answered only by visual considerations of the type we used above. To answer them one needs either experiments, including numerical ones, or a rigorous theory. The first way has the common restrictions for experiment, for example, finite time interval over which one can be convinced (and convince the others!) of motion stability.*" [14]

---

[12] This seems particularly hard to grasp by the traditionally trained. Once I forgot to delete a sentence referring to this concept in an article, moving a "reviewer" to become very insulting. Part 2.B deals with a new practical method for design of Alternating Phase Focused (APF) linacs, where the enabling concept of very large synchronous phase swings was found by an optimal control theory researcher.

13 B.V. Chirikov, "A Universal Instability of Many-Dimensional Oscillator Systems", Physics Reports (Review Section of Physics Letters) 52, No. 5 (1979) 263-379, North-Holland Pub. Co.

[14] The philosophical comment "and to convince the others!" is a key part of making this observation unforgettable – one must work hard to convince one's self, but then it is even harder "to convince the others"…

## 1.2 The Instantaneous Time Beam State

***Working in the time-varying, local, instantaneous system state* is *the HIGHEST RANKING element in ion linacs beam dynamics design, simulation and analysis.*** This has to be stated and demonstrated over and over, because the prevailing view has been the steady-state preference of physics.

In the early days of the 1960's and before, when linacs were young, beam intensity was not a theme. Rings had low intensity or were in a relativistic range where space charge was not relevant. People were used to using the beam distribution position as it stepped along in beam transport lines or on ring orbits. When intensity became a major theme for the LASL/LAMPF linac, it was known that position coordinates (all particles at the same position) was not correct for evaluating space charge; the **element** of using the best physics was ignored - out of some consideration to stay with a nomenclature "that more are familiar with", it was decided to use position as the independent variable in developing simulation codes. Unfortunate even then, but the confusion this would eventually cause was not anticipated. It goes far beyond the space-charge issue, into the fundamentals.

All of the major theory for the equations governing high-intensity linac design, simulation and analysis are time based, with results that may be translated directly, at any time, into useful quantities, such as rms quantities, including energy states and rms energy balance. However, physicists trained in steady-state frameworks have insisted that the energy is that of a steady-state – in particular the thermodynamic state - and reject the system time-state. Clearly the major theorists, including Vlasov, Hamilton and others, knew about this and were not confused about the time-varying fundamentals, but this has been forgotten. That forgetting, or ignoring, is a source of major confusion, that has gotten steadily worse, and dangerous as hypotheses from peripheral investigations are heralded as "general for linacs".

In particular, in Sec. 1.5 the governing equations are presented, including one for an energy-balance equilibrium. The time-dependent integration involved in the derivation, that can be over a very short time such as one simulation step and therefore with no question as an instantaneous state, is forgotten, and it is vehemently promoted that this energy balance can only apply as a thermodynamic state, which is well known will never occur in nonlinear systems with the characteristics of the universe – that also apply to linacs.

Once the request was properly posed (how can the internal energy of the rms beam distribution itself be expressed?), Paul Channell provided that (Chapter 1 - Appendix 1), simply in terms of average energies over a particle distribution. That it was applicable instantaneously was immediately apparent. The use for design, simulation and analysis was immediately apparent.

Implications of this highest-ranking **element** will be repeatedly reinforced. The confusion, especially as so vehemently promoted, has made me wonder how this can occur in a "physics" environment, where it would seem that an equilibrium would be considered desirable, as it is in engineering. Some deliberations on this are thus inserted, questions with no obvious answers (like - is it built into the physics educational system?). But finally the promotion and the vehemence are addressed head on (e.g. Sec.1.7)

## 1.3 Hierarchy – The Actual (Experimental/Simulation/Theory) Hierarchy

For linacs, we are fortunate to have robust simulations, a rigorous but approximate rms theory yielding a design method [15] , and experimental proofs. A key **element** that helps the understanding of the big picture is that there is an *"actual (experiment)/simulation/theory hierarchy",* in that order.

---

15 R. A. Jameson, "Beam-Intensity Limitations in Linear Accelerators," (Invited), Proc. 1981 Particle Accelerator Conf., Washington, DC, March 11-13, 1981, lEEE Trans. Nucl. Sci. 28, p. 2408, June 1981; Los Alamos National Laboratory Report LA-UR-81-765, 9 March 1981. Correction, Jameson, R.A.; IEEE Trans. Nucl. Sci.; 1981; v.28, no.4, p.3665-3665

R. A. Jameson, "Equipartitioning in Linear Accelerators", Proc. of the 1981 Linear Accelerator Conf., Santa Fe, NM, October l9-23, 1981, Los Alamos National Laboratory Report LA-9234-C, p. 125, February 1982; Los Alamos National Laboratory Report LA-UR-81-3073, 19 October 1981.

Both actual devices and computer simulations are the experiments, which give information on ***the instantaneous time beam state***, which is analyzed and guided by the approximate theoretical methods. Or in other words, the actual machine (experiment) and simulations are the reality; the approximate theoretical methods assist in understanding and importantly in design. This point is basic. [16] [17]

For example, the LAMPF linac design strung components together and the first simulation program were written and gave some information about performance. Others followed the same way 30 years later . A linac or ring beam has no knowledge of "theory". With knowledge of accurate beam-in to beam-out component transfer functions, a machine is built, and simulated, with an appropriate sequence of components. The components can include many real details like errors, simplified or more accurate descriptions, etc. that are not possible to include in a theoretical model. The real machine, or this "map", is the reality that is being dealt with. The input/output characteristics of the beam distribution at each component - that is, the ***local, instantaneous characteristics*** - constitute the performance – it is not practical to measure all of them, but they are available in simulation. The beam itself has no knowledge of "Hamiltonian", "envelope equation", internal energy balance", "rms", etc. A "simulation", more or less accurate, can then reveal more information that cannot be practically measured in the real device. "Theory" then comes in as a way to reveal other information about the local, instantaneous state. The "theory" for beams with space charge is all approximate (mathematically limited to KV distribution as replaced by "equivalent rms" distribution, infinitesimal perturbations, improper interpretation of instantaneous/local theory as steady-state, etc., etc., etc.). It is not the reality - but it has been shown in actual linacs and rings and by comparison to accurate simulation results that theory can be used in this way to understand and evaluate performance.

It seems quite amazing to have to point out an additional **element**: If you are trying to understand experimental evidence from an as-built device, using simulation, you have to simulate the *actual* as-built, as operated, device. Comparing the experimental evidence to the more-perfect *design* device does not make sense. Seems completely obvious. But in 2018, interactions with two major labs (one inexperienced) revealed that significant effort had been expended on trying to correlate experimental evidence to simulation of the design machine, where in both cases it was well established that the actual machine deviated significantly (well beyond experimental error bars) from the ideal design conditions. For example, the known field distribution may differ significantly from the ideal designed field, or the operating level of superconducting cavities is adjusted according to sparking limit rather than to the ideal same level. In such cases, the first goal is to get agreement between experiment and simulation, so that work-arounds can be investigated, often requiring sophisticated optimization techniques.

---

[16] The statement here that the actual heirarchy places theory after reality (either experimental or a trustworthy simulation that clearly goes beyond the theory's capability) is intended here to state the case often found with respect to linacs, for example, concerning the oxymoron, or the definition of single particle phase advance spread being bounded by the zero-current tune (true) and the rms depressed tune. In both these cases, the evidence has long shown otherwise. Further discussion of this heirarchy otherwise is very complex – see "Out of my Later Years", A. Einstein, 1934-1955, *Science* section, and other deep thinkers.

[17] *(Aside – Here is an example of the kind of facetious difficulty we get into. We may be trying here to convince a rather "lay audience" to assimilate, dig deeper, and use this stuff. Therefore, guiding the reader by properly emphasizing would be good. I would tend to divide this paragraph, or other material, into some separate, single sentence paragraphs. This is an effective method in English writing to give emphasis.*

*But there has been objection to "single-sentence paragraphs" in some journals, e.g. by "hobby-reviewers" of PRSTAB! A lot of experience with students over frustration with PRSTAB "hobby-reviewers" indicates arrogance, comments that are insulting instead of constructive help, inadequate grasp of not only the relevant physics, but also of acceptable or even good English, not to mention engineering. Etc., etc.…*

*Another way of giving emphasis would be to use italics.*

*What do you think?)*

The same applies to comparing reliable experimental evidence to results of a simulation code which has known deficiencies – it doesn't make sense. The statement is: If confident of the experimental results and there is disagreement with the simulation, then improve the simulation.

## 1.4 The Particle Distribution Behavior

The framework for linac focusing and accelerating channel design involves consideration of the rms, single particle and plasma related behavior of a particle distribution.

The time-dependent evolution of the rms behavior is understood in terms of the external field evolution, its periodicity or lack thereof, its adiabaticity or lack thereof, and in terms of the mechanism of space charge induced redistribution of the particles and resonance interactions with the lattice and between the degrees of freedom. The theory is based on approximations, such as the smooth approximation for the beam envelope, or an assumed infinitely periodic system, or a continuous rather than bunched beam (4D vs. 6D), but provides useful measurements of the beam, which are then evaluated and applied locally. It is found that this approach tracks the actual full case very well, so it is then used to understand performance – and more: for design. The basic rms formalism is outlined in Sec. 1.5.

***It is essential to understand that it is the transient behavior – the local behavior – that is of interest, and not a steady-state condition.*** This essential element will be repeatedly emphasized.

***It is essential to get the rms design essentially right, in order to minimize beam loss.***
It is clear that design that has no direct control over the high-intensity dynamics is lacking.

***The impact of practical compromises*** (e.g., lack of strict periodicity, non-adiabaticity, matching considerations, lack of smoothness resulting with practical modularity, lack of beam equilibrium conditions, contribution of these to beam halo, and additionally cost and construction considerations and restraints) can then be clearly understood. These constitute a second significant challenge for every design. For example, low beam loss may not even be the most important criterion but many other constraints make design difficult, or a particular approach such as Alternating Phase Focusing (APF) introduce additional difficulties (such as finding the appropriate APF gap spacing or field generation).

***Single particle and plasma investigations*** require additional special consideration and tools.
The linac system has nonlinearities in both the external and space charge fields, and collective as well as individual particle behavior, with the result that there is a spread of behavior of the individual articles around the rms values. This spread must also be carefully observed and controlled as well as possible. It is possible that troublesome amounts of beam may be lost without much effect on the rms beam. In practical systems, compromises and exceptions may lead to larger size, low density beam components whose loss would be troublesome. It is popular to use the term "beam halo" for these components.

Ill-considered theory confuses, such as the definition of single particle phase advance spread being bounded by the zero-current tune (true) and the rms depressed tune.

There are many "definitions" of what constitutes a "halo" in focusing and accelerating channels. Definitions such as, e.g., "particles beyond the highest point of the beam space charge potential well", or "particles <u>starting in outlying areas of phase space</u> that move farther out (perhaps crossing a separatrix)" are not useful. A definition that seems correct is that "a halo consists of particles that occupy phase space other than the *expected* (perhaps *desired*) phase space". For example, it is known (but hardly, and therefore probably not expected) that particles starting well within the beam core, even at the transverse origin and at the quasi-synchronous phase, can later move far out in phase space, even crossing separatrices [18].

For example, mismatched injection into a channel causes the beam size to oscillate. Thus there is an expected increase in rms beam size, which does not constitute halo. But the nonlinear dynamics of the system also causes non-Gaussian "tails" to form. These are actually also known and must be expected, but are called "halo" because they can extend far enough that scraping can occur. The cure then has to be to minimize mismatch, or keep it below a tolerance, or to do intentional scraping. If the external fields are nonlinear, mis-steering errors in position and/or angle also cause "halo", although it must be expected and the cure is to minimize mis-steering. Separated function linacs with phase slip in the accelerating cells have a systematic driving term that might cause halo. Beams that are exactly rms matched and even rms equipartitioned also have individual particle tune spreads that characterize the total beam characteristics, and these may change along the channel. Again, these must be expected and kept within bounds. So another useful definition of "halo" is "a halo consists of particles that occupy *unwanted* phase space that may result in beam loss". A more rigorous theoretical definition is explored in Chapter 30. Chapter 24 presents a very detailed investigation of one particular source of halo, using specialized tools.

Exploring the actual form of these phenomena is a research topic of this book. It is essential to have this information, because in practical linacs the causes are to some extent unavoidable, and tolerances must be established so that the number of particles actually lost to the metal boundaries is low enough. A design tool called the tune chart (Hofmann Chart) is a useful result of this understanding, helpful in the design technique.

Background from the present status of nonlinear systems is necessary to have a basic understanding of the collective and single particle behavior. This is introduced by first presenting the rms theory and design procedure.

The design strategy presented has features, for control of both the rms and halo properties, that are outlined already decades ago in the theory, but have been exploited to date only by me.

One can still hope that the power of "inside out" design will become the state-of-the art, as it must. My explorations have been very absorbing, but there are many areas for further work, for example, with higher intensity (>~50mA peak H+ equivalent space charge) superconducting linacs, addition of an intrabeam scattering rule for H- linacs (now implemented at J-Parc and SNS) and design for compensation of it, or how to apply similar concepts to halo control (especially if it seems that non-equipartitioned design is somehow desired or to whatever extent unavoidable; halos have dynamics different from the core or rms).

[eltoc]

---

18 a. R.A. Jameson, ""Beam-Halo From Collective Core/Single-Particle Interactions", LA-UR-93-1209, Los Alamos National Laboratory, 31 March 1993.

b. R.A. Jameson, "Design for Low Beam Loss in Accelerators for Intense Neutron Source Applications - The Physics of Beam Halos", (Invited Plenary Session paper), 1993 Particle Accelerator Conference, Washington, D.C., 17-20 May 1993, IEEE Conference Proceedings, IEEE Cat. No. 93CH3279-7, 88-647453, ISBN 0-7803-1203-1. Los Alamos National Laboratory Report LA-UR-93-1816, 12 May 1993.

c. R.A. Jameson, "Self-Consistent Beam Halo Studies & Halo Diagnostic Development in a Continuous Linear Focusing Channel", LA-UR-94-3753, Los Alamos National Laboratory, 9 November 1994. AIP Proceedings of the 1994 Joint US-CERN-Japan International School on Frontiers of Accelerator Technology, , Maui, Hawaii, USA, 3-9 November 1994, World Scientific, ISBN 981-02-2537-7, pp.530-560.

d. C. Chen, R.C. Davidson, Q. Qian, R.A. Jameson, "Resonant and Chaotic Phenomena in a Periodically Focused Intense Charged-Particle Beam" (Invited paper for C. Chen), Proc. 10th Intl. Conf on High Power Particle Beams, NTIS, Springfield, VA 22151,(1994), 120-127.

e. C. Chen & R.A. Jameson, "Self-Consistent Simulation Studies of Periodically Focused Intense Charged-Particle Beams, Physical Review E, April 1995, PFC/JA-95-9 MIT Plasma Fusion Center

## 1.5 The Three Elemental Equations for RMS Design and its Analysis

***<u>The theory and design require satisfying only two, or optionally three, equations, which control the rms properties of the beam.</u>***

### *1.5.1 The transverse and longitudinal rms matching equations*

Two required design equations are the smoothed approximation transverse and longitudinal beam envelope equations, which can be derived for any linac type. These are called the rms matching equations, which must be satisfied ***<u>locally</u>***. These equations relate the rms beam emittances to the rms beam sizes and the rms smooth approximation phase advances:

$$\varepsilon_{tn} = \frac{a^2(\beta\gamma)\sigma_t}{L} \tag{1}$$

$$\varepsilon_{ln} = \frac{(\gamma b)^2(\beta\gamma)\sigma_l}{L} \tag{2}$$

where $t$ indicates transverse and $l$ indicates longitudinal; $\varepsilon_{ln}$ and $\varepsilon_{tn}$ are the normalized rms emittances (cm.rad), $a$ and $b$ the rms beam sizes (cm), $\beta$,$\boldsymbol{\gamma}$ the relativistic beta, gamma. $\sigma^t/L$ and $\sigma^l/L$ (*<u>same L for both</u>*) are the phase advances of the smooth betatron or synchrotron oscillation over a suitable unit length. (The division by a unit length L has been widely overseen, which is still (2024) causing confusion.)

The zero-beam-current smooth betatron or synchrotron oscillation periods are determined by the structure transverse and longitudinal periods, usually $n^t\,\boldsymbol{\beta}\lambda/2$ and $n^l\,\boldsymbol{\beta}\lambda/2$ resulting in cancellation of the $\boldsymbol{\beta}$'s. *Repeating, <u>same L for both planes in (1) and (2).</u>*

In simulation or experiment, σ's are computed directly from (1) & (2).

Expanding the phase advance (or "tune") terms:

$$\sigma^{t2} = \sigma_0^{t2} - \frac{I\lambda^3(1-ff)}{a^2(\gamma b)\gamma^2}k \tag{3}$$

$$\sigma_l^2 = \sigma_0^{l2} - \frac{2I\lambda^3 ff}{a^2(\gamma b)\gamma^2}k \tag{4}$$

where $\sigma_0$ is the zero current phase advance defined by the external field, $I$ is the beam current, $a$ and $b$ are the transverse and longitudinal rms beam radii, $ff$ is the geometry (form) factor ≈ $a/(3\gamma b)$, and

$$k = \frac{3}{8\pi}\frac{z_o q 10^{-6}}{mc^2}$$

<u>The equations are expressed relativistically, good for any linac.</u>

Averaging the x and y motions is usually sufficient for analysis and design with linac most structure types, but both the transverse and longitudinal should always be observed, and all three dimensions are also sometimes necessary.

The rms envelope equations describe a bunched beam of ellipsoidal form, so apply exactly after the beam is bunched enough to be described as an equivalent ellipsoid. They are derived from the time-dependent Vlasov equation by separating "fast" terms, that oscillate with the phase advance of the focusing period, from "slow terms", that oscillate with the phase advance of the betatron and synchrotron periods. An infinitely long periodic system and adiabaticity are assumed, so a' and b' are assumed to be zero, and also a" and b" are assumed to be zero. In reality, linacs are neither strictly adiabatic nor periodic, and there are a' and b' terms that can have effects, e.g. on transmission and beam loss, of several percent. These effects are discussed in Ch.16 & 28. However, for most practical purposes, the local use of the smooth approximation is satisfactory, with small enough steps in the design and simulation.

### *1.5.2 The element of simplifying theory and notation*

***An element, considered important for this book, is to keep the theory and notation as simple as possible.*** For example, in that sense, phase advance here is preferred over the definition of a somewhat unintuitive "frequency". In the broad sense, the reader will find essentially no further theory at all in this book. It is hoped that the reader will understand and assimilate the underlying **elements**, and can expand all details as necessary when needed.

### *1.5.3 The Contribution of Frank Sacherer [19] – Crucial and very early (1970)*

***Sacherer's work [20], showed that rms emittances can replace the original KV envelope equations, <u>and</u> that quantities in the envelope equations can be specified a priori – crucial <u>elements</u> of linear accelerators. (Paper also gives derivation of envelope equations.)***

The paper gives the fundamental reasoning in extraordinarily simple, clear and precise manner. Because its seminal nature is so important, the liberty is taken here to quote (bold added):

In his *1. Introduction*, Sacherer notes that Pierre Lapostolle [21] had already in the same year *"observed, in the course of computer experiments with continuous beams, the envelope oscillations of many different distributions followed very closely the KV equations. For one distribution (constant density in four dimensional phase space, axial symmetry in real space), he proved that the stationary or matched state is specified exactly by the K-V equation provided the beam boundary and emittance are defined by rms values. Following this, Gluckstern proved by considering moments of the Vlasov equation that the rms envelope motion of all continuous round beams rigorously satisfies the K-V equation, even for time dependent external forces. Thus the K-V equations, which were derived for the distribution that occupies the surface of a hyperellipsoid in four-dimensional phase space, actually describe any four-dimensional distribution that is round in real space. In this report these results are extended to continuous beams with elliptical symmetry as well as to bunched beams with ellipsoid form, and also to one-dimensional motion.*

*2. Moment equations for one dimension*

*The space-charge term in this equation has an interesting interpretation. .... .... The rms envelope equation depends only on the rms linear part of the forces. Or conversely, because (the average) $xF_S$ has very little dependence on the higher moments, the linear part of the self-force depends mainly on the rms size of the distribution and only slightly on its detailed form.*

*3. Moment equations for two and three dimensions*

*.... The linear external forces, including linear coupling terms, are included in $F_i$.*

*.... (the equations) are easily generalized to three dimensions.*

*4. Explicit forms for one-dimensional envelope equations*

---

[19] Both the original long internal CERN report and the conference paper are cited in [20].

The extremely widespread ignorance of the seminal nature and use of this work of Sacherer has been a mystery to me for decades. I have tried to teach it for decades, admittedly not in a book or formal courses. Why is it not an <u>element</u> in the (few) books and courses? How many have read the long version? Why has so much time been spent constrained by the "KV box", injecting beams into resonances when resonances are to be avoided in any high intensity accelerator, and summarizing the results by steady-state arguments without showing the phase-space transport – which time could have been spent on useful and still extant problems? Why is not the use of the envelope equations for analysis of the local, instantaneous state in simulations not immediately clear?

***20*** F. J. Sacherer, "Rms Envelope Equations With Space Charge", CERN/SI/Int. DL/70-12, 18.11.1970 (19p ); and PAC1971, pp. 1105-1107 (3p).

[21] Pierre Lapostolle was a real pioneer, a close friend and colleague whose work I followed closely.

*5. Envelope equations for continuous beams*
*6. Envelope equations for bunched beams*

*7 Conclusion*

*A rather surprising and useful result has been found for beams in free space, namely that the rms linear part of the self-field depends mainly on the rms size of the distribution and only very weakly on its exact form. Using this result, envelope equations for the rms beam size have been derived that are **exact** for continuous beams of elliptical symmetry, and **in practice also valid for bunched beams of ellipsoidal form**. The main restriction in applying these equations is that **the time dependence of the rms emittance must be known a priori.***

*Possible uses of the equations include the specification of stationary or matched states in the presence of space-charge. For example, the periodic solution of (47) for alternating-gradient structures, including radio frequency cavities, specifies the matched beam size (both longitudinal and transverse) as a function of rms emittances and intensity. The largest matched size attainable without exceeding aperture limits or bucket size determines a space-charge limit. For a beam matched in this way, envelope oscillations about the periodic solution are suppressed, although higher modes of oscillation (sextupole, octupole, etc.) may occur. Suppression of the higher modes will require constraints, as yet undetermined, on the higher moments of the distribution. Another use, already mentioned, is the design of low-energy beam transfer lines6).*

*Acknowledgement: I wish to thank P. Lapostolle for many helpful discussions. "*

[eltoc]

## 1.5.4 Equivalent rms

An essential feature of the use of rms quantities is Sacherer's proof that equivalent rms emittance represents any distribution; that is, that if the same rms value for any distribution is used in the envelope equations, the results will be very nearly the same, within a few percent - a seminal result that enables an "equivalent rms" design method.

## 1.5.5 Apriori Specification

The equations cannot be extended to account for emittance growth - the derivation requires the emittances to be either constant, or to have a functional form known *a priori*. This crucial, enabling statement appears to be completely unappreciated by anyone except me, in spite of decades long attempts to teach it. ***The ability to use an a priori form** – not only for the emittances, but **for any quantity in the envelope equations** – is an essential* **element** ***of this work and of the framework.*** It enables a widely expanded design space.

The remaining difference, between the equivalent rms and the detailed distribution, is, however, important, to achievement of the best beam-loss performance. A method to account for this difference in the design by purposely varying the form factor *a priori* is described in Ch.28, along with other possibilities.

## 1.5.6 Smooth matching

It is immediately obvious that the matched design beam radius and length should vary smoothly in order to avoid unwanted effects. Usually the derivative of the beam size is allowed to vary only slowly - approximately adiabatically in terms of the betatron and synchrotron oscillation wavelengths. This is a hard problem in many practical designs, especially at transitions between linac sections, which may involve different linac types or sections with instrumentation, choppers, etc.. It is best solved by built-in adjustments to the final cells of the preceding and initial cells of the following linac section, but often must be solved by an external "matching section". A typical consequent problem with the latter is that adjustment "knobs" are available, which later can lead to inconsistent tunes.

## 1.5.7 The equipartitioning equation

***The third equation describes the internal energy balance, called equipartitioning (EP), between the degrees of freedom of the beam. Balanced means in equilibrium and that the beam will be immune***

***from certain disturbances, such as a resonance corresponding to the EP parameters.*** The use of this equation in design is optional; if used, it also should be locally satisfied exactly (not in some approximation). The beam internal energy is balanced between the transverse (average of x and y) and longitudinal degrees of freedom if

$$\frac{e_{l_n}\sigma^{l}}{e_{t_n}\sigma^{t}} = 1 \qquad (5)$$

which also implies

$$\frac{\varepsilon_{ln}}{\varepsilon_{tn}} = \frac{\gamma b}{a} = \frac{\sigma^{t}}{\sigma^{l}} \qquad (6)$$

The simple and practical form of the equipartitioning Eqn. (5) was the breakthrough needed to realize that ***this equation can be <u>exactly, and easily</u>, solved simultaneously with the matching equations***, giving the possibility to design linac channels that maintain the beam in equilibrium, thus minimizing the potential for beam size and/or emittance growth that could lead to undesired beam loss causing radioactivity buildup and maintenance problems.

These equations deal with effective rms quantities – not Liouvillian. The concept of energy is also definable only at the rms level, and not at higher order, so the set of equations is consistent. The beam internal energy is derived in terms of local rms quantities, and not as a thermal quantity, although of course is also correct in the thermodynamic limit. Chapter 1 - Appendix 1 gives the derivation of the beam energy balance theorem.

## *1.5.8 Exact Simultaneous Solution*

It is very important **element** that the solution of the three equations be ***<u>exact</u>*** *at the local operating point*, and not some approximation that is not physically justified.

It is not difficult to get an exact solution – perceived difficulty is not an excuse.

***<u>ESPECIALLY NOT SOLUTION AT THE SPACE CHARGE LIMIT</u>***, as approximated and published in a book, which has misled many. The context of that book was based on electron beam physics, which except at very low velocities, does not apply to ion beam physics – but was applied to ion beam physics *without testing*, and in spite of long and serious attempts to discuss by myself and another acknowledged international expert in the field. The book concentrates on relativistic electron beam physics and thermal equilibrium; both do not apply to ion beam physics –the author prefers to give everything an expression in that frame, but does show the full formula.

The envelope equations (1)-(4) very clearly show that both the emittance and the space charge contribute to the beam sizes. The generic form of the tune depression vs. beam current from these equations (for any lattice) is highly nonlinear, indicating that an approximation at tune=0 is inappropriate (Fig.1.1).

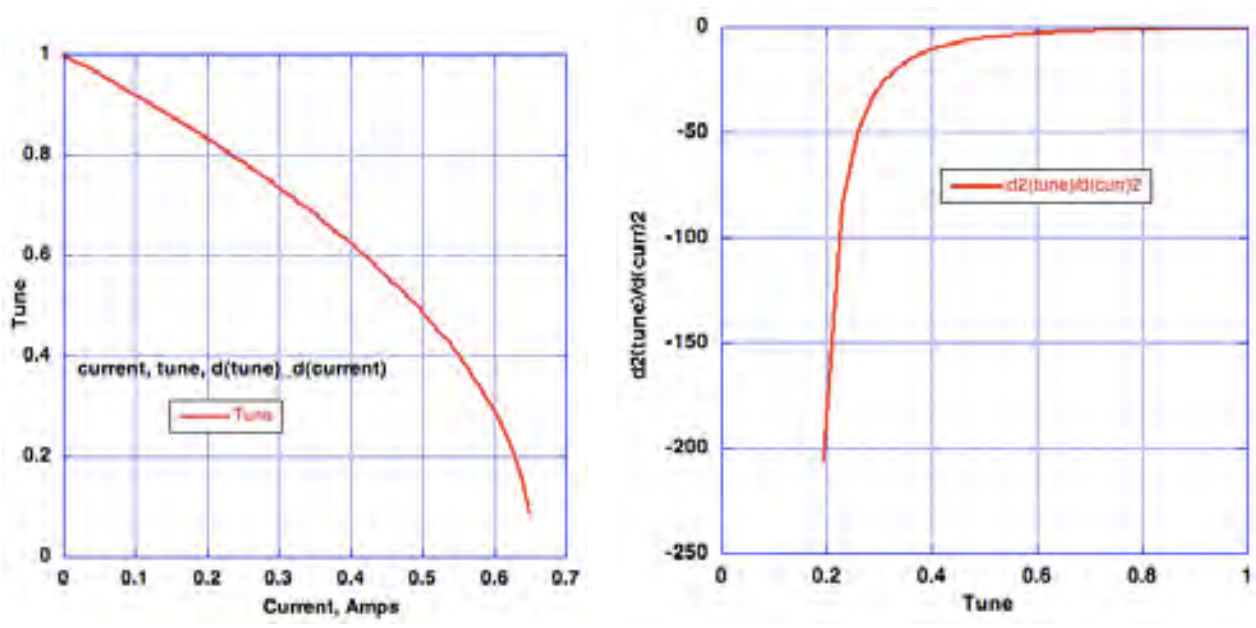

Fig.1.1 Nonlinearity of tune depression vs beam current for a linac example. Large beam current ~550mA reduces the tune to ≈ 0.4, where rmsr0i is ≈ rmsr0e and the curvature starts to become large. Tune = 0.8 is reached at ~230mA. With low beam current ≤~50mA there would be little space charge effect.

As a testing example, assume a 175 MHz proton linac, with nblt = # βλ in transverse focusing period = 1, and etnrmsgiven=0.0002, elnrnsgiven=0.0004, and the EP equation, as a function of sig0tr and the beam current curamp. Fig. 1.2 shows the b/a=rmsl/rmsr ratio and beam sizes for the full solution with EP = (elnrmsgiven/etnrmsgiven)*(rmsrestim/rmslestim) = 1, and for the zero emittance 0e approximation of this full solution. Significant error is apparent for small tune depressions down to ~0.4 (re Fig.1.1).

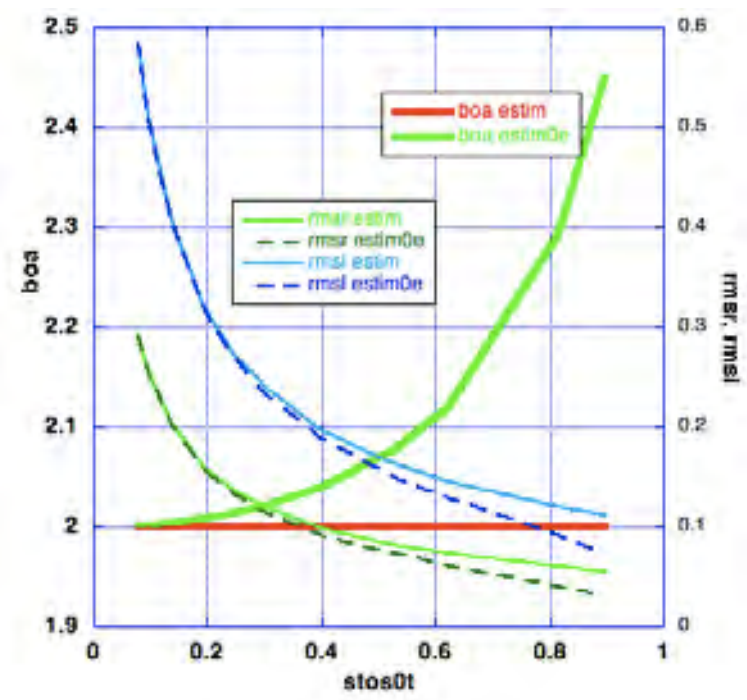

Fig.1.2 Simultaneous solutions under exact and and space-charge-limit conditions.

"Reviewers" are currently opining (2022) that:

- ion linacs are designed for transverse tune depression ≥≈0.8. *This is a grave restriction of the design space!!!* Not at all true. When working to a full specification, the designer has full freedom, and must not be restricted by uninformed "opinions". E.g., rf power and length are strongly influenced by the operating tune depression. Here, if the space-charge-limit approximation were to be used, the errors at transverse tune depression ≈ 0.8 are large (too large when dealing with deuterons for example, where beam loss has to be strictly kept very small at energies above ~2MeV).

### *1.5.9 Solution using Mathematical Approximation or Solver*

The simultaneous solution for the beam sizes at each working point can be approximated by analytical contraction mapping formulae that are accurate to within a few percent, or even with much better accuracy at higher order [22, 1st entry & (2nd entry, Appendix 4)]. From the envelope equations, the 0i terms are with zero beam current and describe the external field; the 0e terms are with zero emittance and describe the space charge effect.

```
!  Contraction mapping estimates for beam sizes:
! bmszest for dtl, ccl requires nblt
         rmsr0i=dsqrt(etnrmsgiven*xnblt*wavel/(gammaloc*sig0tr))
         rmsl0i=dsqrt(elnrmsgiven*xnblt*wavel/(gammaloc**3*sig0lr))
!
         xkhat=3.d0*z0*(dabs(chrg(1))/(partmass(1)*931.5016d0))/(1000000.d0*8.d0*pi)
         xifac=(curamp*chrg(1)*xkhat*xnblt**2*wavel**3/sig0lr**2)**(1.d0/3.d0)
         sig0fac=(sig0tr/sig0lr)**2 + (1.d0/2.d0)

!!@@      using total-to-rms beam size ratio of dsqrt(6)   .. is Sqrt[5] in M
         rmsr0e=(3.d0/2.d0)**(1.d0/3.d0)* &
             xifac*sig0fac**(-2.d0/3.d0)/dsqrt(6.d0)
         rmsl0e=(2.d0/3.d0)**(2.d0/3.d0)* &
             xifac*sig0fac**(1.d0/3.d0)/dsqrt(6.d0)
         rmsrestim=(rmsr0e**3 + rmsr0i**3)**(1.d0/3.d0)
         rmslestim=(rmsl0e**3 + rmsl0i**3)**(1.d0/3.d0)
```

<u>It is immediately seen that approximating to the space charge limit (the 0e terms dominate completely) is totally unnecessary – *it is trivial to use the full correct formulas.*</u>

---

22 N. Brown, et.al., Particle Accelerators, 1994, Vol. 43(4), pp231-233;
M.Reiser. "Theory and Design of Charged Particle Beams", 2nd Edition, 2008, Wiley ISBN: 978-3-527-40741-5;
Edward P. Lee, Reviewer of Reiser 1st Edition of above, Physics Today **48**, 6, 59 (1995);

Then why do "ion linac people" use the space charge limit approximation? First of all – *they do not think*. Perhaps second – they are not expert, or they would have read the reference, especially to the end (the Appendix is very short), where the full formulae are discussed (the same problem as with Sacherer's article concerning "*a priori*"). Third, they generalize on insufficient evidence or testing. They are not real designers.

Above all, cognizant ion linac designers would be aware of the book review ([22], 3rd entry), which notes that the 2nd entry needs to be used with great care, as I have warned many:

> Within context of full spec: "characteristic of most new applications that the quality of beam transport and manipulation is a concern at least as great as the technology and cost associated with acceleration". "The emphasis is clearly on theory." "It is not intended as a text on particle accelerators, which are presented as applications …" . "… some useful formulas …". "However, it is not a complete guide for the design of a beam-transport system … beam-dynamics design for real systems generally receives a detailed treatment with a variety of codes whose description is outside the scope of this book" . "…The book will best serve physicists (! Ed.) who are involved in a beam-line design and need to scope the beam parameters and phenomena at a preliminary stage.". "The derivations are sound and detailed, but there is some misleading discussion…".

Thus this analytical form could be used directly, as it was for the JAERI/J-Parc linac original design.

The contraction mapping is also very useful to get essentially exact starting conditions for the exact simultaneous solution of the three equations using an optimization program. Use of an optimization program package in the simulation program flexibly affords many applications. An **element** of working with any optimization program (whatever called, AI, neural net, steepest descent, many other names but all optimizers) is that the given starting conditions must be near enough to the optimum that the program is able to converge to the optimum.

In the *LINACS* design and simulation software, the exact solution is obtained with the aid of the optimization program *NPSOL* that is given an accurate mathematical starting point using the contraction mapping formulae. Using this method also enables other important conditions, such the rf coupling between planes, and other conditions simultaneously involving the beam sizes; the optimization functions can contain weighted sums of combinations of conditions, etc. (*a priori* decisions – see Ch.28).

### *1.5.10 Local, instantaneous system state*

It is very important to stress, as an **element**, that use of the rms quantities applies to the *instantaneous, local* situation, and not to a final state. Then analysis and design succeed. This crucial point is paramount and must be understood. (See throughout this book, also Ch. 16.1.3 and Ch.28).

### *1.5.11 Design using the (two) (or three) equations; "inside out" instead of "outside in" design*

"Design" with the rms equations aims at desired performance, with respect to emittance, or phase advances, or space charge behavior, all of which can be influenced "*a priori*".

#### 1.5.11.1 "Outside-in" Design

If only the two (transverse and longitudinal) rms envelope "matching" equations are used for design, they can be solved simultaneously for two quantities. Typically, as suggested by Eqns. 1-4, these are the transverse and longitudinal rms beam sizes.

Simultaneous satisfaction of the two (transverse and longitudinal) rms envelope "matching" equations in the degrees of freedom *locally (instantaneously)* along the complete channel is mandatory *in any case* for good performance of any design. Adequate emittance behavior and low loss design using only the matching equations is often possible if done appropriately.

In this case, then all the external field parameters (vane voltage, aperture, modulation and synchronous phase) must be given by rules as functions of z or β. The remaining unknowns are the transverse and longitudinal rms beam sizes, and can be found by simultaneous solution.

It is possible that these rules could contain *a priori* functions of the beam sizes in terms of β or z.

- If, as may be typical, the rules do not do this, then the beam is not directly involved in the design – the simultaneous solution just gives the corresponding uncontrolled beam sizes, and beam performance is just to simulate the design and see what happens. This is the prevalent method of the outdated "state-of-the-art" or "international expert level", termed "outside-in" design – just set up four rules, simulate and see what happens. There may be criteria (at least limits, as on peak surface field, modulation or synchronous phase), tuning considerations on whether to allow vane voltage variation, etc., but the form of the *a priori* rules given within those constraints are ad-hoc and just tested by simulation.
- For RFQs designed with *LINACS*, the synchronous phase is designed according to the Teplyakov criterion for controlling the charge density of the beam distribution – this does involve the beam physics to that extent in the design.
- Many possibilities have not been explored.

#### 1.5.11.2 "Inside-Out" Design

We are interested in the beam behavior (rms and total emittance, beam transmission, accelerated fraction, loss, beam loss location and energy). The form of the external fields is not interesting in itself. Nor are the rms beam sizes, when not controlled by beam-based physics.

Therefore it is desirable to find an **"inside out" design that specifies the desired beam behavior as influenced by space charge,** and determines the external fields that give that behavior. ***Other than the two matching equations (average transverse and longitudinal), which govern the transverse and longitudinal rms beam sizes, we have the possibility to add only one additional equation – the EP equation – no further equation with a binding physical basis is known.***

Therefore, only one of the four governing parameters can be made free in the design process for use in solving the three equations (perhaps including *a priori* conditions) simultaneously. The vane voltage has other constraints, the aperture is an easily visualized quantity, and the synchronous phase may have the additional Teplyakov constraint. The effect of vane modulation is not readily visualized and therefore is a logical choice to leave free.

Working to specify the beam physics and solving for the external fields required to make that happen is an engineering oriented approach, in contrast to the usual "outside in" practice of stringing together external field components and injecting a beam into a simulation to see what happens. [23]

#### 1.5.11.3 Characteristics of an EP Design

##### 1.5.11.3.1 There is not necessarily any disadvantage to a fully EP design.

It will be shown that there are sophisticated variations of EP design in conjunction with *a priori* conditions that give much design flexibility, especially for adjusting the RFQ length. There are misconceptions because of lack of testing, lack of thought (or other reasons) and "reviewed" papers that argue against EP design, in spite of the usually self-evident fact that an equilibrium is a desirable condition, certainly in engineering. This theme will be developed further throughout this book.

##### 1.5.11.3.2 Advantages of an EP design.

– *if EP is achieved, as verified by simulation, it is sufficient to insure a tight beam and low loss.*

---

[23] "Outside-in" linac designs have been used since the beginning. They are experience-based, that is – adhoc. Yet another "outside-in" design is MS-level work and definitely not a basis for a PhD degree, although many such have been, and are, being granted. Fully executed EP designs are proven for decades, are still rare, but also are not PhD work. There are just two kinds of designs – beam-based "inside-out", or "outside-in" – some ad-hoc variation of the latter is not worthy of being stated as a new, <u>general</u> design technique - although this has been, and is being done – see Sec. 1.7.4.

– *Insuring a tight beam seems to be generally not recognized – the equilibrium does also keep the 100% emittances tighter.*
– *Work at J-Parc, cited later, shows that the EP beam is also more robust against various errors.*

The experimental proofs are simulations and actual constructed devices, in which both the external and space charge fields are controlled – and the experiments work – the design rms conditions are achieved, the beam also remains "tight" without evolution of unexpected broadening or low level halo.

Thus it is clear that any project should develop a fully EP design, for comparison to any other imagined approach.

1.5.11.3.3 Addition of the EP condition is optional.

As with any design condition, EP has to be evaluated relative to the entire specification, that requires trade-offs and compromises for a particular project.

Direct control of the beam space charge physics is obviously more and more important as the beam intensity (beam current) requirement increases. As developed below, the intensity is indicated by the "tune depression" = ratio of (phase advance with beam current)/(phase advance with zero current). If this ratio is near 1, space charge, thus the EP condition, has little influence, and may be ignored. At ratio = 0, the space charge limit is reached – a very precarious condition. In between, the behavior is very nonlinear, and typical high intensity linacs operate in the tune depression region of high nonlinearity. This region is above the stochastic limit where all trajectories become chaotic. There can be advantages, e.g., rf power efficiency, to more tune depression.

There is much confusion in the literature, contending that requiring the EP condition constrains the design space and competes too much with economic criteria (e.g., length, but generally without engineering, cost or RAMI considerations). It has generally not been appreciated that sophisticated use of the three equations can, for example, also lead to shorter channels while maintaining an equilibrium beam [24]. The phase advances in Eqns. (1) and (2) contain all the parameters of the external field and a space charge term that includes the beam current, the beam sizes, and the space charge form factor, which is a function of the beam sizes. It is very important to realize that *all* quantities in Eqns. (1, 2 and 5) may be changed *a priori* along a design trajectory – including the emittances, the space charge form factor, and the equipartitioning ratios. This considerable opening of the design space, while gaining the advantages of equilibrium, has not been explored by others to date.

Non-EP designs will typically have interactions with resonances, formation of halo, and beam losses.

In any case, if a halo forms for whatever reason, the halo will have different phase advance characteristics from those of the rms distribution. If its source cannot be eliminated, a compromise matching between the rms core and the halo may help minimize beam loss. A change from rms matching will of course compromise the rms behavior also. Such compromise strategies have not been well researched yet (2022).

The guidance is: do an exact and thorough EP design first. Then if not satisfied, explore ad-hoc "outside-in", or program something else.

This is the essential theory and method for the basic design of any type of linac. Practically, of course there are many details. Compromises and exceptions must be made, for example for detailed matching between different linac types, the usual cost considerations, etc.

[eltoc]

24 "RFQ Designs and Beam-Loss Distributions for IFMIF", R.A. Jameson, Oak Ridge National Laboratory Report ORNL/TM-2007/001, January 2007.

## 1.6 Beam Particle Distribution Evolution Via Plasma Interactions [25]

The basic interaction, termed *space charge mixing*, is that particle redistribution inside the beam can occur in about (plasma period)/4. Space charge mixing *is continuously present* when there is space charge. This was first explained by O.A. Anderson [26], in terms of the time it takes for initially parallel trajectories to bend and first cross. Definitions of the plasma periods are:

$$\sigma_p^l = \sqrt{\left(\sigma_0^{l2} - \sigma^{l2}\right) ff}$$
$$\sigma_p^t = \sqrt{\left(\sigma_0^{t2} - \sigma^{t2}\right)(2/(1-ff)}. \qquad (7)$$

where ff = ~ a/(3γb) is the space charge form factor. Note the combination of transverse and longitudinal.

This is a collective effect – not a resonant effect - ***with much faster change than those caused by other driving forces like resonance interactions or mismatch, etc - and actually the driving mechanism for those interactions***. In terms of its rms definition, it will be observed in changes of the rms emittance and beam size, but changes in the detailed particle distribution, internal energy balance, detailed match to the local features of the channel, etc., are particularly important. Typical (plasma period)/4 for a tune shift of 0.4 or below is only a few cells of a linac.

Crucial **element**, using the *local, instantaneous* state:

This type of redistribution is clearly seen in simulations of the first few cells of machines with space charge when waterbag or other nonequilibrium distributions [27] are injected, the fast reaction of the distribution when the channel trajectory enters or leaves a resonance zone, the fast reaction when the equipartitioned condition is reached, even when in a resonance zone, or at transitions.

[eltoc]

## 1.7 Structure and Space Charge Resonances

### *1.7.1 Design Tool – the Hofmann Chart [28]*

Structure resonances from the inherent periodicity of most accelerator channels, and collective resonances caused by space charge, which broaden the stopbands of the structure resonances, are major driving terms for emittance growth and halo formation, which are realized through the mechanisms of space charge mixing (with period faster than the resonance growth periods as usually stated) and phase space transport. The study of such resonances has been in progress for decades (since 1960's or before). The analysis is mostly for a continuous focusing system – a very restricted set – with an infinitesimal space charge perturbation to the Vlasov equation, which correctly includes all relevant modes, or, for the usual charts, to the *rms* envelope equations, which are a restricted set, invoking the "smooth approximation", and the KV distribution which introduces spurious modes. The motivation and interpretation is from the perspective of rings, requiring some thought when related to linacs.

---

25 This very important point is presently (2022) not known, or sometimes falsely described, in the literature.

26 Oscar A. Anderson, "Internal Dynamics and Emittance Growth in Space-Charge-Dominated Beams", Particle Accelerators, 1987, Vol. 21, pp. 197-226;
"Emittance Growth in Intense Mismatched Beams", 1987 Particle Accelerator Conference, Washington, DC, IEEE Cat. No. 87CH2387-9, p. 1043.

27 G. Parisi, "Investigations on Particle Dynamics in a High Intensity Heavy on Linac for Inertial Fusion", PhD. Dissertation, Goethe Uni Frankfurt, 28 September 1999.

28 I. Hofmann, "Dynamical Aspects of Emittance Coupling in Intense Linac Beams", The 52nd ICFA Advanced Beam Dynamics Workshop on High-Intensity and High-Brightness Hadron Beams, HB2012, Beijing, September 17-21, 2012.

It must be emphasized that the chart is *rms* derived and usually used with *rms* quantities. Now occasionally it is presented as being used for "100% emittance". Then what is meant by "100% emittance should be clearly pointed out; whether this 100% is just an artificial technique of applying the rms ellipse at all particles, including the outermost, of a simulated distribution; or just application of the 100% to rms ratio for a theoretical distribution, or some other measure.

A resonant interaction coupling the transverse and longitudinal planes occurs at *every rational fraction* of $\sigma^l/\sigma^t$. The major ones for $\sigma^l/\sigma^t$ = 0.33, 0.5, 1, 2,… can cause significant change of the rms quantities, although continued interaction with minor resonances can also have a scattering effect producing approximately linear emittance growth.

A tune chart (named the Hofmann Chart), derived from the envelope equations, plots the growth rate, found from a specific simulation of a beam distribution involved long-term with the major resonances at a tune depression caused by beam space charge, in both transverse and longitudinal planes (which are more or less coupled depending on the type of accelerator structure), vs. the ratio of longitudinal to transverse tunes, as exemplified in Fig. 1.3.

The chart is generated by injecting a beam distribution directly into a resonance, a non-realistic situation in linacs. It adequately reveals the boundaries of the resonance bands, but not the growth rates relevant to the actual instantaneous system state. Thus it is a very valuable guide on which tunes to avoid or traverse carefully – but only such guide and not with growth rates useable for practical actual cases.

If the beam distribution is equipartitioned (EP), it has no free energy that can drive resonances – at that condition, and to a lesser extent in some area surrounding that condition, including tails of multiple resonances, where the EP conditions are not exactly satisfied but the free energy is still small, e.g. Fig. 1.4.

It needs to be noted that, for example, if eln/etn = $\sigma^t/\sigma^l$ = 2, <u>*the resonance*</u> at $\sigma^l/\sigma^t$ = 0.5 <u>*does not disappear*</u>. It is still there, but there is no *rms* energy to drive it, so that the peak, and its spreading, will be absent from the chart (Fig. 1.4). This includes the tails of overlapping resonances at that ratio.

The parameterization using ratios, resulting from a simplified model, was a very useful development, when correlated directly with local state analysis and design. With zero beam current, $\sigma^t/\sigma_0^t = \sigma^l/\sigma_0^l$ = 1.0 and the resonances are at single points along the top of the chart [29]. With increasing space charge, the resonance stopbands spread, and the growth rate of the interaction increases, as indicated by the deepness of the shading.

Parameterization to the tune depressions and the tune ratio requires that a different chart be used for each emittance ratio. This complicates practical use, again indicating the guidance aspect. More than one chart, or perhaps a video, will be needed for some cases, e.g., for *a priori* adjustment of the emittance ratio.

---

29 Single particle charts are of course incapable of indicating collective performance.

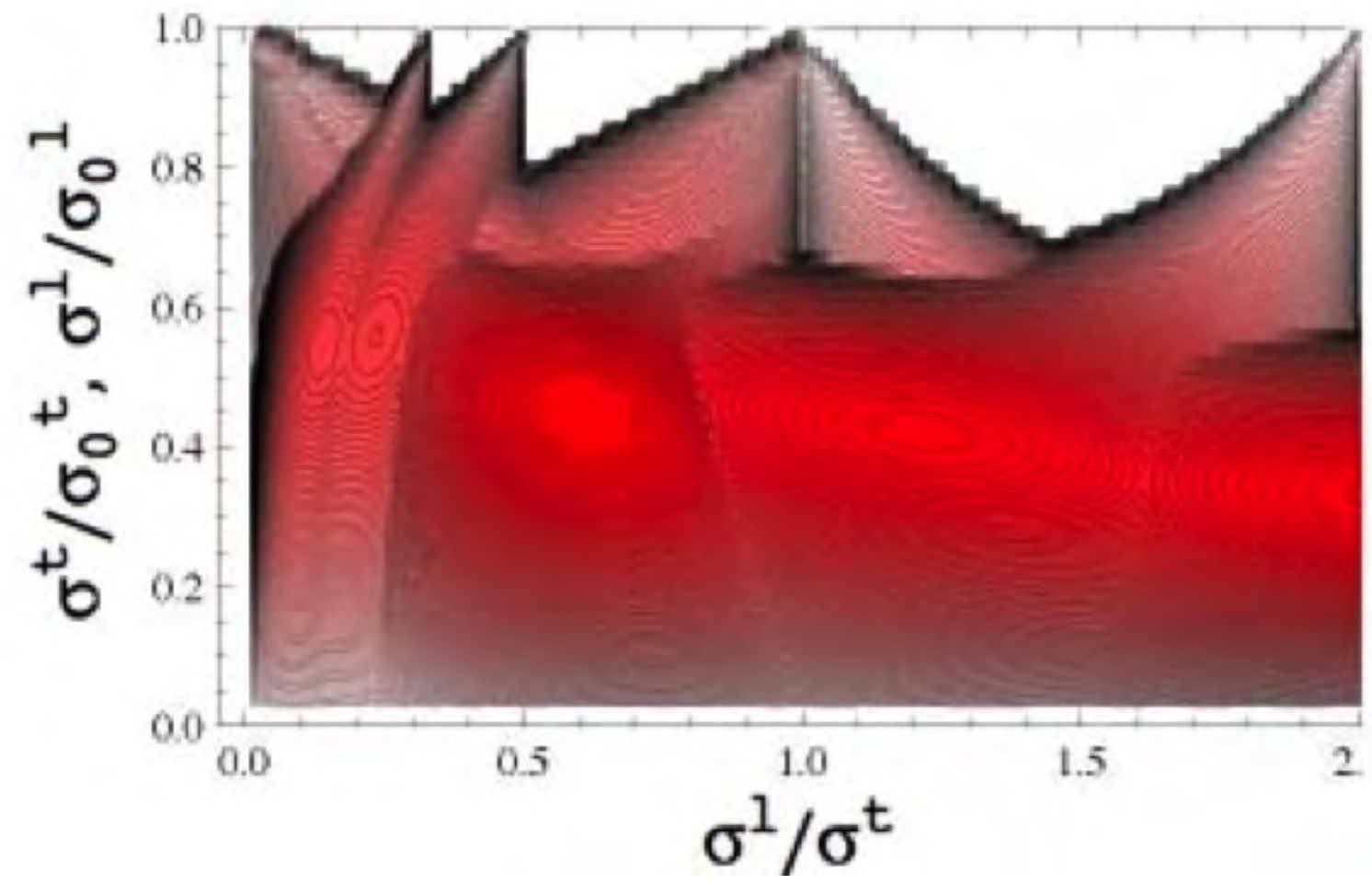

Fig. 1.3. Stability chart showing stop-bands for resonances at $\sigma^l/\sigma^t$ near 0 and at 0.33, 0.5, 1.0 and 2.0. (Chart is for rms emittance ratio eln/etn=10.) (($\varepsilon$ln, $\varepsilon$tn, $\sigma^t$, $\sigma^l$ and the zero-current phase advances $\sigma_0{}^t$ and $\sigma_0{}^l$ may be simplified for typing by etn, eln, st, sl, s0t and s0l).

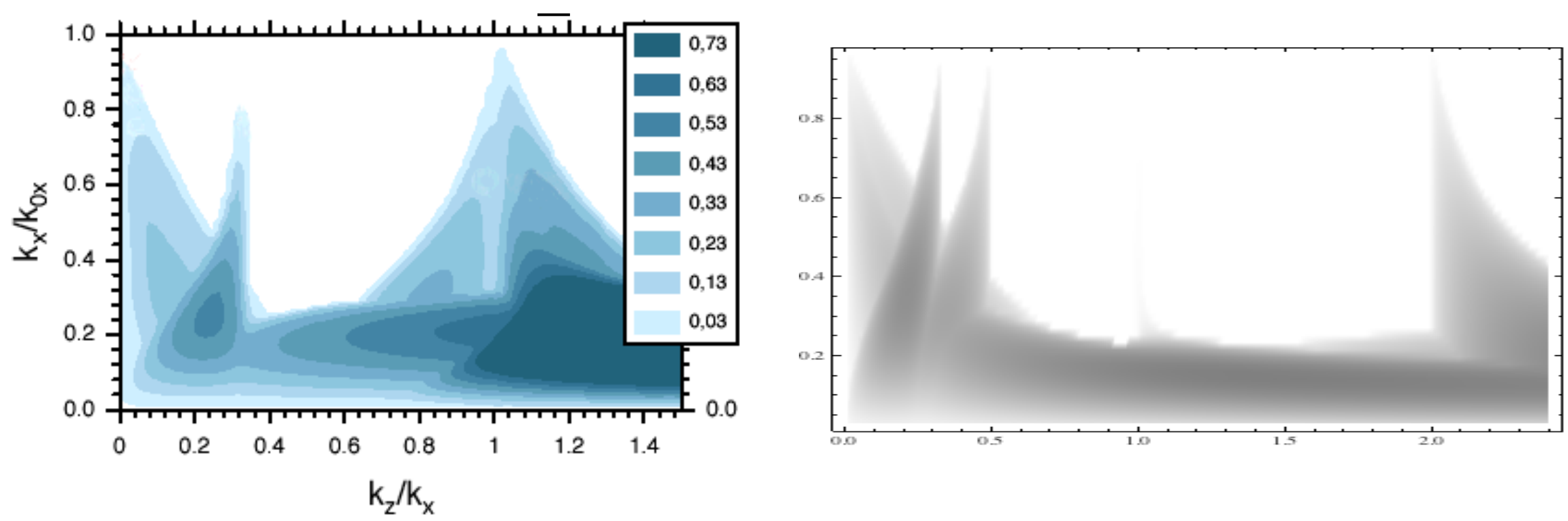

Fig. 1.4. Hofmann Chart (left) for longitudinal-to-transverse emittance ratio eln/etn = 2.0; thus kz/kx = 0.5 is the equipartitioned condition for this chart. (right) for eln/etn = 1.03. The shading represents a growth rate. The values are for a special case and should be ignored. The curious lack of response on one side of resonances is commented by Hofmann in Sec.21.6, where it it shown that there is actually a reaction on both sides.

The charts usually used by everyone are obtained from an algorithm provided by Hofmann. It is important to understand their derivation and use, including the terminology "growth rate" as presented.

The "growth rate" definition is not often repeated in every paper (!). It has to be related to a distance unit, and is "Contour lines indicate levels of growth rates of unstable modes in units of zero space charge betatron units in x". The unit is arbitrary and convertible, and may have useful relevance to rings.

The chart model is the smooth approximation of a transverse focusing channel without parameter change or acceleration. They are generated by injecting a beam distribution directly into a resonance and observing a result - after some arbitrary distance along the channel – of the rms emittance at this point, or represented by a growth rate in units as above. The number of zero-current betatron units used is probably ≥~100, and could be compared to the number in the linac being studied…

As the nonlinear dynamics are not thermodynamic, there is no steady-state, so this distance the distribution may indicate a "quasi-equilibrium", but may then significantly evolve further at larger distance – there is no clarification of this or the phase space transport that influenced the effects on the rms emittance and resulted in the chart data point.

These results, and the shading, represent "the growth rate of the resonance". When the beam distribution is not equipartitioned, it is seen in Fig. 1.3 that the growth rate of this special case may be significant already at tune depression of ~0.6.

*This is not the same as the effect on the instantaneous rms properties of a linac beam distribution along its trajectory.*

The thresholds have proven usefully accurate when used locally with acceleration and other local effects.

The high-intensity linac (and rings) design goal has always been to avoid beam loss. Injecting into a resonance on purpose, or allowing the trajectory to remain in a resonance zone, is thus clearly to be avoided. So it should have been clear from the beginning that the chart as presented is a guide, and this discrepancy and the numbers on the displayed growth rate bar are ignorable.

Considerable attention is paid to the time scale and effect of brief resonance interactions in the detailed chapters.

The small free energy in a zone around the exact EP condition increases the available "safe area" for design trajectories. One sees that there are "safe areas" where significant resonant coupling would be avoided:

- if the system has very small tune depressions ≥~0.8
- when the beam distribution is equipartitioned, or nearly so.

## *1.7.2 – Use of the Hofmann Chart, continued*

The cell-by-cell rms trajectories in both (or all three) planes will be plotted on this chart. Rules governing the design of linacs with intense beams were introduced into the cell-by-cell design procedures starting in 1981 [30,31,etc.]. It is apparent that a linac design should avoid the unstable zones of the main resonances, or use a purposeful strategy if they are purposely entered or crossed.

As different kinds of accelerator systems have different degrees of coupling between the longitudinal and transverse degrees of freedom, it is essential for understanding to plot both trajectories, although the literature mostly shows only the transverse one [32]. The accelerator external fields also contain some degree of nonlinearity and coupling, which mixes with the space charge nonlinearity. It is desirable to minimize these nonlinearities and couplings as much as possible, but not possible completely; an example is the nonlinearity caused by the non-ideal shape of RFQ vanes. The design equations applied locally account for these effects, and they are accurately included in the simulations.

The detailed derivation of the effects of these collective and structural resonances proceeds from a perturbation analysis of the (unphysical) KV shell distribution in the Vlasov equation, for channels assumed infinitely long and without parameter changes, with results mostly presented from a steady-state point of view. The usefulness of the mode identifications, however, is that although approximate, they can be used instantaneously, locally, and from the transient point of view, in terms of phase space transport, to afford powerful design techniques.

---

30 R. A. Jameson, "On Scaling and Optimization of High Intensity, Low-Beam-Loss RF Linacs for Neutron Source Drivers," AIP Conference Proceedings 279, ISBN 1-56396-191-1, DOE Conf-9206193 (1992) pp. 969-998, Proceedings of the Third Workshop on Advanced Accelerator Concepts, 14-20 June 1992, Port Jefferson, Long Island, New York, LA-UR-92-2474, Los Alamos National Laboratory, 28 July 1992.

31 R. A. Jameson, "An Approach to Fundamental Study of Beam Loss Minimization," AIP Conference Proceedings 480, "Space Charge Dominated Beam Physics for Heavy Ion Fusion," Saitama, Japan, December 1998, Y. K. Batygin, Editor. Workshop on Space Charge Dominated Beam Physics for Heavy Ion , 10-12 December 1998, Institute of Physical and Chemical Research (RIKEN), Wako-shi, Japan. Los Alamos National Laboratory Report LA-UR-99-129, 8 January 1999.

32 (This arose because previous charts were intended for rings, and in terms of instability thresholds of a very restricted model, and when one plane's trajectory went below a threshold, the other did also. The extent of the ignorance about this did not become fully clear until 2017-2018, when it was realized that the theorist had no idea about how actual linacs are laid out – see Ch. 28.)

Misunderstanding has resulted because the theory and simulations, for almost all studies, have been interpreted as for simple continuous and periodic channels, with no other components than the transverse focusing components, assumed infinitely long with no parameter changes, which is a very special case (that may be somewhat appropriate for a storage ring.). The interpretations of this special case have far too often been stated as general conclusions – a classic example of this error. There are at least several problems:

1.) Apparently because the perturbation analysis is formally possible using the KV distribution, a KV input distribution has been repeatedly used as input to simulation of a channel, although Sacherer's work outlined above showed that this only puffs up papers and only confuses. Use of the KV to match at the inside/outside beam boundary is required for the approximate theory, but there is no connection that requires the non-physical KV distribution to be the input to a simulation.

2.) Mainly, the simulations have concerned injecting a particle distribution directly into a resonance. Practically, resonances are to be avoided. So injecting a beam into a resonance is mainly academic. It could be useful for rings and educational however, if interpreted comprehensively:

3.) The simulations are continued through many periods, the distribution changes and moves in phase space, approaches a saturation as it leaves the resonance zone, and this saturated state (e.g. "final rms emittance") is reported. [33]. Often it is reported as having reached steady-state, although it is only a quasi- or local-quasi-steady state The settling times of linac dynamics are very long and longer runs would show more evolution. Also, there is no clarification of the phase-space transport that reveals *how* and *why* the distribution evolved.

This is a result of the very wide-spread lack of understanding that the instantaneous, local beam state is the important **element**. The essential aspect is also revealed in Appendix A of the classic reference [34], where the authors admit that their whole treatment is only indirectly related to the aspect of transport of the phase-space occupied by a collection of particles.

4.) Simulation *of simple transverse focusing only channels* over hundreds of cells has not indicated any evidence that modes with $m+n > 4$ play a role. This is not surprising if one keeps in mind that even in synchrotrons, resonances above fifth or sixth order are practically ignorable. The situation with more complex lattices has long been declared a worthy research topic, but only recently have investigations begun, for example in Chapter 27, in which other modes from rf gap kicks are seen, and Chapter 29.

5.) Considerable resources have been spent on "experiments to confirm the theory". To start with, the **element** is that "the approximate theory is checked against experimental (and simulation) results to determine if the approximation is useful", Sec. 1.2 – and this rests on the confidence that the experiment or simulation is "the truth", requiring very careful and accurate framing and procedure. Second, there seems to be widespread belief that only a "dedicated experiment" can provide this insight. It has been clear since 1981 and in very many other simulations and analyses of existing machines (e.g. the emittance growth observed in the LAMPF DTL class of DTL's of that generation) using the local method, that the theory is adequately useful, in this case to know which regions of tune space to avoid or use carefully.

Specific dedicated experimental confirmations were obtained in the heavy-ion fusion cesium beam demonstration accelerator at LBL (~1988), and a demonstration of the emittance exchange at the "main resonance" $kz/kx = 1$ was obtained recently at the GSI UNILAC.

---

33 Often see statements that "the resonance saturates". Resonances do not saturate.

34 "Regular & Stochastic Motion", Lichtenberg AJ &Lieberman MA, Appl. Math Sciences 38, Springer-Verlag 1983

Repeating, it is apparently believed that only such dedicated experiments are "valid" – but every machine built has been tested to some extent, and with proper analysis of such, and many detailed simulations; it is clear that the approximate theoretical results are valid and useful [35]

6.) Non-self-consistent methods can give useful information, but need to be correlated with self-consistent methods. E.g., test particles are often introduced (to make Poincare plots) – the initial placing of the test particles has no physical basis. The Poincare plot is a useful roadmap, but gives no information about how a real beam distribution travels on it.

These problems have existed since the beginning of these studies decades ago, and have proliferated. Clarification is complicated and difficult to cover in a single paper - [36] was a start of a planned series, and should be read as supplemental material here. At the same time however, with the understanding that the local, instantaneous, state should be used, it has been shown that the behavior seen in simulations can be explained well, and gives the basis for powerful design techniques.

### *1.7.3 The Relevant Dynamics for the RMS Beam Characteristics*

The linac channel particle dynamics is just a typical nonlinear system, whose behavior is a function of the strength of the nonlinearities in the system, as reviewed in the master paper by Chirikov [13]. With small beam current, the linac system operates with small tune depressions – at the top of Fig. 1.3. Space charge and nonlinear external fields increase the nonlinearity, and as the tune depression becomes larger, the degree of coupling between the degrees of freedom increases. The mechanism for the coupling and resulting instability is excitation of resonances. Fig. 1.3&4 indicate strong action for tune depressions in the region ~0.6 – 0.2, and an essentially totally coupled overlapped collection of many resonance in the tune depression region below ~0.3. Ion accelerators for many intense beam applications typically desire to operate with tune depressions reaching down into this region (for example because they would be perhaps shorter, use the rf more efficiently, and thus cheaper) [37].

With small nonlinearity, the resonance zones of attraction are independent, but with increasing nonlinearity, they overlap, as indicated by the resonance width and tails. The overlap results in a wider band of instability, and at a certain level of nonlinearity, called the stochastic limit, all resonances are broken and the whole system becomes a sea of chaotic behavior. In ion linac systems, it becomes very difficult to achieve desirable behavior, either of the rms quantities or outlying particles, when the tune depression is lower than ~0.3, even very temporarily.

Below tune depression of ~0.2, the growth rates become smaller again, and at the space charge limit, tune depression = 0, there is again a stable, but precarious, situation in which the beam is actually equipartitioned. Electron machines typically operate at the space charge limit when they become relativistic, where space charge has no collective effect.

Interpretations of *ion beam dynamic*s as if they were electrons and at the space charge limit, with thermodynamic properties, has produced considerable confusion and erroneous work.x

Above the stochastic limit, the system dynamics are suitably described by the term "thermalization", as used in classical dynamics. Mixing in phase space is fast, and the time required to reach the entropy end condition with equal space and time averages is short. However, the required beam quality needed to avoid beam loss problems in ion linacs cannot be achieved or preserved in this region – the beam is essentially out of control and will grow explosively. Ion accelerators must operate below the stochastic limit – and therefore the term "thermalization" is incorrect for the dynamics of ion accelerators. Below the stochastic limit, there can be local areas of chaotic behavior, but not overall chaos.

---

35 From which it follows that more such "dedicated experiments", and also rederiving, republishing, beating of the existing theory to death are essentially a waste of time, that is needed for extant problems…

36 "Structure resonances due to space charge in periodic focusing channels", Chao Li & R.A. Jameson, Phys. Rev. Accel. Beams 21, 024204 (2018)

37 The present tendency to conservatively limit project designs to tune depressions no less than ~0.8 reflect lack of knowledge about how to use beam-based design. Future requirements challenge.

Below the stochastic limit, the dynamics is described by interactions through resonances. Previously, interesting relations were clarified by perturbing the Hamiltonian, that could indicate whether a system was in chaos or not, but were very troubling because it was clear that the theory required the perturbation to be infinitely small, and such was not the reality. Most of the mathematical theory was not expressed in terms of resonances, as this was not a familiar language as it is for accelerator physicists. An important breakthrough came with an understanding of a geometrical description of chaos in nonlinear systems that clearly shows how the interactions below the stochastic limit are really the resonances.

A real example is the Japanese Pachinko board. If a ball is placed at the top of a slanted board, it will quickly roll to a stable position at the bottom. But if many obstacles are placed on the board, the ball's trajectory will be non-predictable and it may take a long time for it to reach the bottom. The obstacles could even be of a type that catches and holds the ball, and then perhaps releases it again later. In 3D, the obstacles can be placed either on the inside or the outside of a bottom-up bowl, with the same result. In the linac case, the obstacles are the resonances described above. [38]

Knowledge of such "transport" in phase space is important. How do particles get their starting conditions, and how do they move in phase space? The "road map" of Poincare intersections of orbits of particles with arbitrary initial conditions is informative but not sufficient. Investigations need to be self-consistent. Details of some types of transport in linacs, and tools for measuring, have been investigated [18].

An important way to get information about the rates of evolution is the "exit time" concept. For systems operating below the stochastic limit, the exit times characterize the time in which a particle may finally escape being trapped or strongly influenced by a resonance.

In general, the time that would be required for a system operating below the stochastic limit to reach full equilibrium is very long – much longer than the time required for a beam to traverse an accelerator. The use of the rms description is not that a thermal equilibrium would finally be reached, but that of a transient situation described locally as it evolves.

Linacs can have strong tune depressions and local chaos, but still have long evolution times. For ions, we do not want complete chaos, or the extremely delicate equilibrium implied by operation at the space charge limit!

Designs made at the space charge limit are not appropriate – the design equations need to be solved – exactly - at the real desired operating conditions. An optimal EP design can easily be done with exact simultaneous EP + envelope equations solution - without a non-physical approximation (difficulty of exact solution is claimed to be a reason for design at the space charge limit; another is the theorist's preference to promote his own theory [39]).

[38] Physics Today, November 2022 very interesting – "MemComputing – when (Long-Term) memory becomes a computing tool" – actually more than the title seems – with example of the Pachinko board (!!), the driving force in this case is gravity, equations of motion, solutions of which are bases of attraction (inside bowl paths always converging) are augmented by disturbance points which are saddle points – which deflect the trajectory, finally to path with fewer saddle points and then to the solution – claimed to better avoid local minima… Certainly applicable to particle trajectory among resonances?? – also have driving force(s). Modern nonlinear system solvers like NPSOL do "remember" previous improvements… Would be interesting to explore further in context of current MemComputing work.

A typical non-EP RFQ linac design example is shown in Fig. 1.5. *It is essential to realize that the actual EP condition and the three corresponding ratios need to be known at all points along the trajectory. At any point, if the EP condition does not = 1, and the three ratios are not equal, an EP tune chart at the local emittance ratio cannot be used directly.*

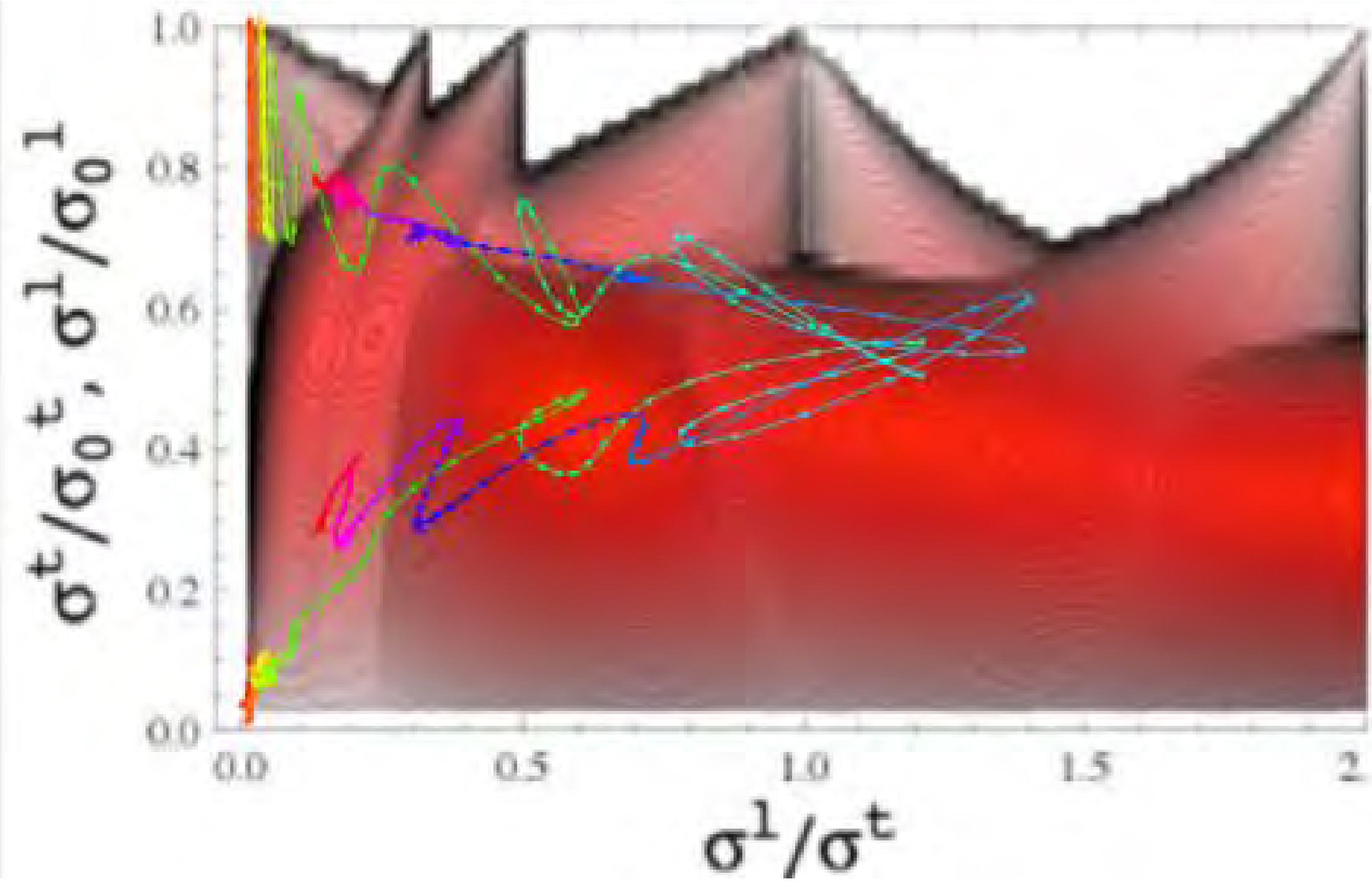

Fig. 1.5. Rms trajectories on Hofmann chart drawn for EP ratios eln/etn = b/a = sigt/sigl = 10. The trajectories progress from red at the RFQ input through orange, yellow, green, blue, magenta at the RFQ output. The chart shows stop-bands for resonances at $\sigma^l/\sigma^t$ near 0 and at 0.33, 0.5, 1.0 and 2.0. Note that the influence of the sigl/sigt=1 resonance is very broad and strong, with tails extending far away from 1:1.

In this particular case, the EP condition is never attained along the trajectory (for any extended distance, although perhaps instantaneously), so Fig. 1.3 is illustrative. At $\sigma^l/\ \sigma^t$ = 0.5, the (upper) transverse trajectory (here changing from yellow to green) makes a closed circle. This indicates that there was an actual trapping in this resonance for a short time (but then an exit). The (lower) longitudinal trajectory does not make a circle - it makes a zigzag - this indicates that is was strongly influenced by the resonance but not trapped in it. The time interval is short; the escape is aided by changing parameters and acceleration. If the beam distribution were actually equipartitioned at kz/kx=0.5 as in Fig. 1.4 (requires correct derivation and graphing the actually relevant information from the simulation – the EP condition, and the three ratios), the chart could be as shown with that resonance absent, as there would be no free energy in the beam to excite it, and the circle and zigzag would not be present. Similar things are happening in the main resonance around $\sigma^l/\ \sigma^t$ =1. Here the important points are how wide the resonance is when the trajectory is inside its growth zone, the tune shift, the growth rates, and how long the trajectory stays in the growth zone. Remember that here we are looking at the rms behavior and its evolution in a transient state, and also are interested in the transient total beam behavior.

[39] This grievous error has destructively spread and has been used for actual constructed apparatus and project designs. Once these wrong approaches are published in "reviewed" journals or books, they become accepted without thinking, become extremely hard to remove and are destructive to the state-of-the-art for a very long time!

An example of a well-satisfied EP RFQ design is included here as Fig. 1.6, from the IFMIF CDR RFQs; reference the widely distributed "RFQ Designs and Beam-Loss Distributions for IFMIF", R.A. Jameson, Oak Ridge National Laboratory Report ORNL/TM-2007/001, January 2007.

Fig. 1.6. The fully developed IFMIF Post-CDR RFQ, with *a priori* variation of the EP condition in the acceleration section.

" Figs. 6.2.-4,5,6 show the space-charge physics and emittance behavior. The ratios rise *a priori* as a function of beta, and the beam remains closely equipartitioned. A composite Hofmann Chart with eln/etn = 2 overlaid on eln/etn = 1.4 is shown, to convey the required change in the EP ratios from 1.6 at EOS to 2 at the end of the RFQ. There is no unplanned resonance growth.

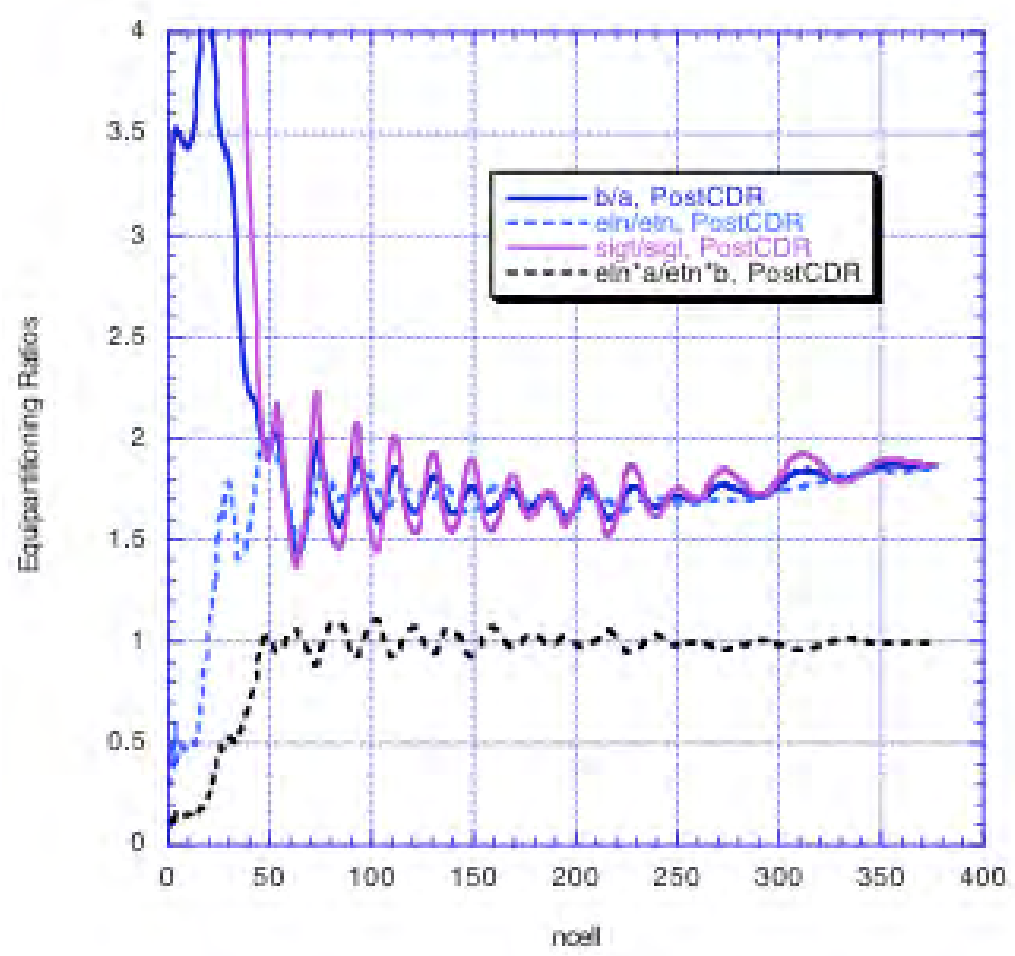

Fig. 6.2-4. Equipartitioning ratio, and corresponding beam size, emittance and tune ratios, Eqs. (7) and (8) for the Post-CDR equipartitioned RFQ using pteqHI including multipole and image-charge effects.

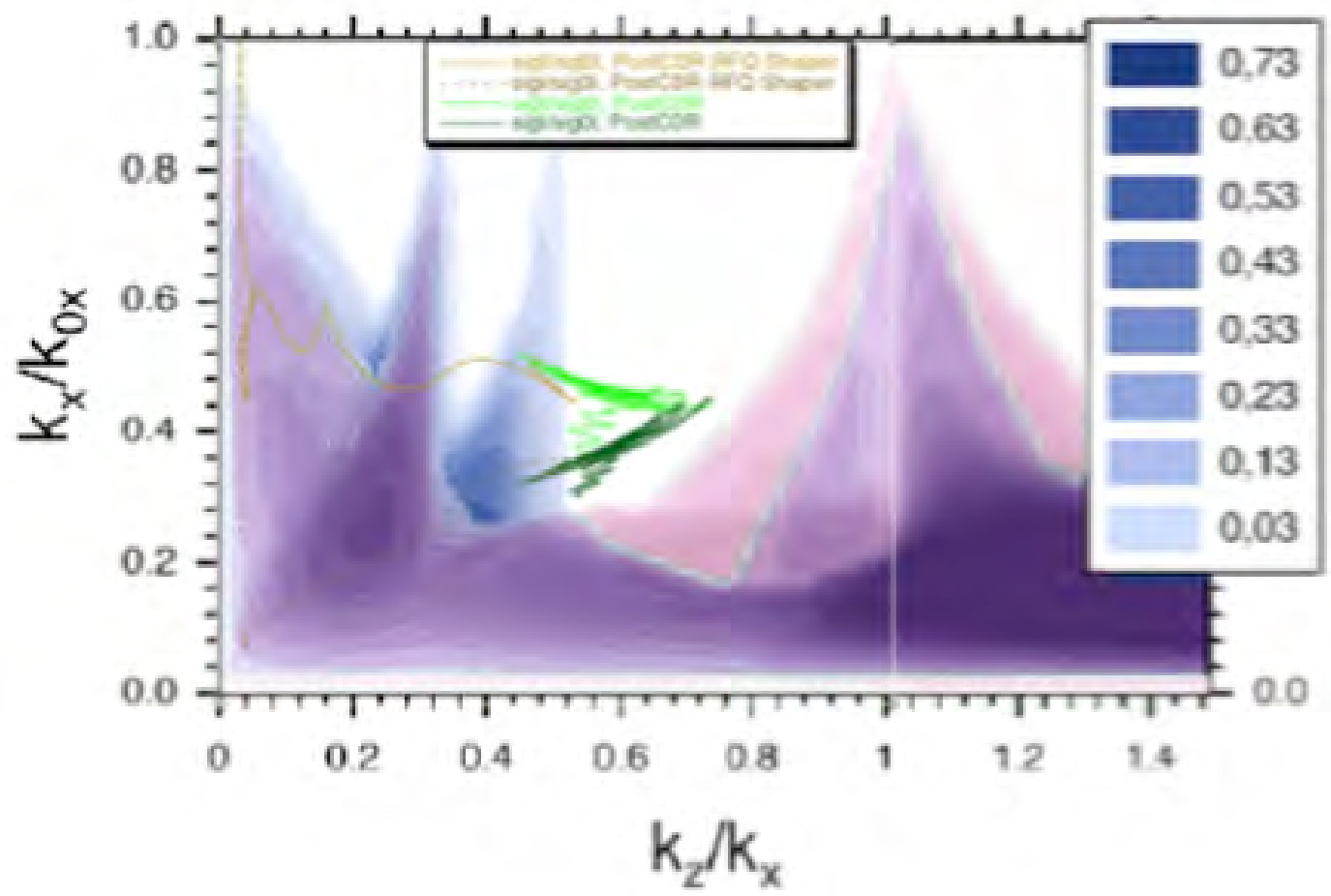

Fig. 6.2-5. Composite Hofmann Chart for eln/etn=1.4 (underlying blue-toned shadows) and eln/etn = 2 (overlying magenta toned shadows). The Post-CDR equipartitioned RFQ trajectories for the shaper and from the EOS to the output are shown, using pteqHI including multipole and image-charge effects.

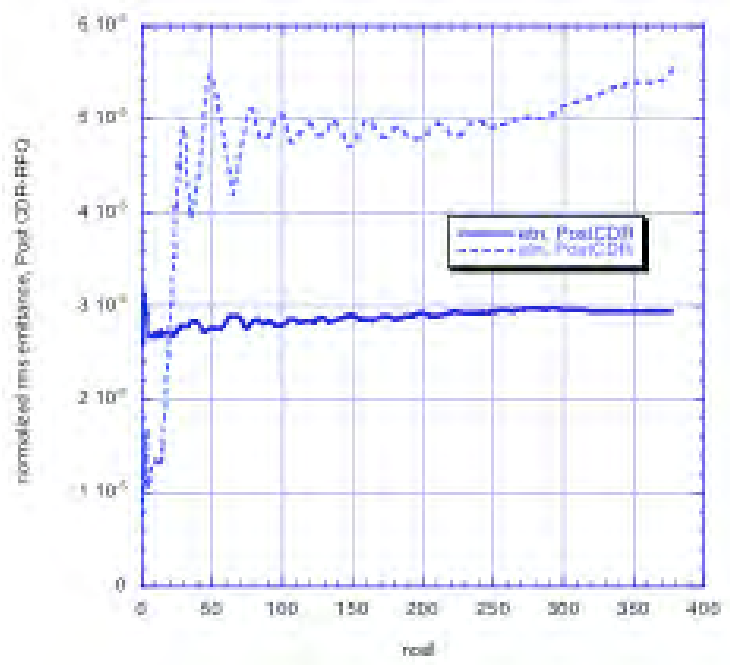

Fig. 6.2-6. Transverse and longitudinal rms normalized emittances for the Post-CDR equipartitioned RFQ, using pteqHI including multipole and image-charge effects. “

## *1.7.4 – BEWARE! “Safe areas”, the Oxymoron*

*Erroneous interpretation (or worse) of ‘safe areas” has led to egregious aberrations of the intentions here to establish a rigorous and self-consistent design framework. This has become serious, as promotional, “reviewed” but wrong papers, scientifically and technically, have been published, for some time and as recently as early 2022.*

Therefore, we must highlight and expand to this specific section 1.7.4 here to deal with this very specifically and directly.

Please review Sec.1.5.11 and footnote to Sec.1.5.11.2.

This addition has too much detail for this introductory chapter’s goal of establishing the framework; it was added later as the misleading aberrations began to appear – it does address the goals of the framework being developed here, and is valuable anyway, as it will give you an early glimpse into correct and incorrect usage of the framework, and into good scientific research practice for any researcher. Sometimes a very bad example can be a good learning tool – better if encountered in the correct context, so as not to be misled.

The framework being developed here is simple, logical, straightforward, and full. But it has appeared to be difficult to grasp. Thus there appears to be a background technical educational problem, which I drastically summarize as instantaneous system state vs. steady state. Other problems lie, e.g., in learning research methods.

But very clearly there are also problems in non-technical areas. Please allow a blunt, direct statement, confined here to the use of equipartitioning:

The precedence for the form and use of equipartitioning in the linac context is established. (Often, one may wonder “Why didn’t I think of that”!!)

<u>If used, it is a sufficient condition. Testable. Not falsifiable (Sec.B.1).</u>

If not used, the field is “outside-in” and wide open. Then, in order to have something to do, to publish, “outside-in” and peripheral issues may embellished. Academically, not out of order. But bluntly, if driven by human nature motives such as the illusory wish be to become famous, proposed as superior to a proven physically sufficient method, then such proposal has to be built on the scientific method and very well supported. The example here is probably the worst, maybe at the boundary of worst, of many observations of “reviewed”, published papers.

### <u>1.7.4.1 Why is a physical, flexible and readily solveable equilibrium avoided?</u>

This is the basic question. Not on scientific grounds.

<u>Every project should evauate if EP works with their spec – then decide.</u>

### <u>1.7.4.2 Safe areas without major resonances</u>

Figs. 1.3&4 show, for specific conditions, that there are areas (white), with no *major* resonances, at tune depressions above the areas where the resonance begin to drastically overlap, and above the the stochastic limit, where all resonance are overlapping. However, there are an infinite number of resonances on this chart – at every rational fraction $\sigma^l/\sigma^t$. These are discussed further later; they act like scattering centers and may become significant for long linacs.

**1.7.4.3 Safe areas because of the equipartitioning influence**

**Oxymoron** - An oxymoron is a figure of speech that juxtaposes elements that appear to be contradictory. Oxymorons appear in a variety of contexts, including inadvertent errors (such as "ground pilot") and literary oxymorons crafted to reveal a paradox. The most common form of oxymoron involves an adjective-noun combination of two words – examples "dark light", "crazy wisdom", "open secret", "controlled chaos".

In our case, lack of careful thought has created a sad oxymoron – and a real one, not a crafted literary trick, although even perhaps with aspects of that. *People are really believing, and have stated in the literature, that they "can ignore EP and just operate in the clear region around EP" !! And others are blindly following, or worse.*

Why a main approach to a design is proposed as avoiding the fundamental advantage of an equilibrium is a mystery, when its detailed evaluation is readily done. But as to the oxymoron:

Fig. 1.4 is for an equipartitioned (EP) particle distribution with a longitudinal-to-transverse emittance ratio eln/etn = (longitudinal-to-transverse) rms beam size ratio b/a = (transverse-to longitudinal) rms phase advance ratio = 2.0; thus kz/kx = 0.5 is the equipartitioned condition for this chart, and for eln/ etn = 1.03 (slightly high to avoid chart solver problems at eln/etn = 1.

It is clear that a large region free of *major* resonances has been achieved – and not just at the exact EP tune ratio where there is no "free energy", but also where the amount of available "free energy is small, where no resonant growth is seen until the tune depression falls to the stochastic limit. The tails of the major resonances are also affected. [40] "What is going on?" might be asked.

The charted resonant interaction areas and colorations are derived from injecting a distribution at a point *into* the resonance band - and observing its evolution as the resonance acts on it for an extended period, until a growth time can be determined. This gives only some information about where dangerous areas exist.

*1.7.3.1 Resonances with reference to high intensity, very low beam loss design*

An ion linac particle distribution evolving along a time-dependent trajectory is very unlikely to be purposely injected into a resonance, or left there to evolve with clearly unwanted results. The most important aspect is to understand that this chart is useful for consideration of the instantaneous system state. Points plotted on a trajectory are the (actual, simulated) rms quantities at that instant. If the trajectory is not EP at that point, a general mode tune chart should be used - it is not appropriate to use the exact EP Hchart there. The reaction at or in the vicinity of a resonance depends on the speed of the plasma space charge mixing at each point along the trajectory, and on the free energy available.

**1.7.4 ,4 Location of channel trajectories**

Now we reach the critical misuse and the oxymoron:

*It has been stated in the literature, that one "can ignore EP and just operate in the clear region around EP" !!*

A trajectory lying in a region with very little tune depression has hardly a space charge consideration in any case.

---

40 (The theoretical construct of the chart results in some resonance bands that are active only on the left side. e.g., of type $\sigma^l/\sigma^t = 0.33, 0.5$. As would be expected, and shown later, the actual resonances are two-sided.)

**But the greatly extended "clear region around EP" cannot exist if the EP condition is ignored.**

Lack of understanding has produced the above statement in the literature for linacs, originally from an author not very familiar with linacs, perhaps extrapolating from a ring situation, without proposing how a trajectory in the "clear region around EP" would be derived, or investigating its properties - but from a purely technical motivation and not personal intention. That is undesirable, but there is worse.

### 1.7.4.5 The non-moronic opposite of the oxymoron

To maintain orientation, keep in mind that the opposite of the oxymoron is not moronic – the EP=1 condition is a robust and desirable goal, and the resonances may be ignored.

### 1.7.4.6 Beam controlled design vs. uncontrolled design, general procedure vs. trial and error

Self-consistent linac *design* which includes control of the beam space charge physics is "inside-out", and a general procedure. It is recommended that a thorough EP *design*, with all the possible ramifications for controlling channel length, rms and total emittances, beam losses, error tolerance, etc., be understood as the initial step for any project.

If it then turns out that it seems that the (hoped for) specification cannot be met, then the direct use of EP can be abandoned, space charge physics ignored in the design, and "outside-in" designs with tradeoffs explored. Such design will almost certainly involve resonance interactions.

These are the two types of design – comprehensibly beam-based, or not comprehensively beam-based.

Only if and where the *design* trajectory attains the EP condition is the Hchart for the corresponding emittance ratio appropriate. Even then, the Hchart is only a guide as to where dangerous areas exist – note its derivation conditions. Otherwise, if the *local design* point is not EP, there is no "clear area", or maybe a smaller one, different growth rates, and a tune chart showing all the resonances is a better guide, as done in Fig.1.5.

In the RFQ, trajectory rules must be specified for all four of the controlling parameters: vane voltage, synchronous phase, aperture and modulation.

**Use of the beam-based EP condition, via vane modulation control to find the simultaneous solution of the matching and EP equations, is a *very clear, specific, general, sufficient* and *simple* "inside-out" way to include the attractive characteristic of an equilibrium. *Once the rules for the other three parameters are given, the EP trajectory is unique*.**

**An RFQ design with other *programmable* rules (an infinite number are possible) for all four of the controlling parameters might be possible. It would, at least, be *necessary*[41]:**

- ***to explain the reason for choosing the design condition.***
- ***to justify it* against the same specification as used for the above defined EP design, in order for the comparison to be fair.**
- ***to show how the design is performed*.**
- ***to show its advantage over the whole range of its defined design space, as it would probably not be unique*.**
- ***to show that such design could be a general, optimal design strategy over a wide range of specifications, or could not be*.**

**In other words, with understanding and rigorous scientific procedure – and scientific motivation.**

It would be refreshing to see such research, carried out with all three of these conditions.

---

41 e.g., "Beam dynamics design of an RFQ for a planned accelerator, which uses a direct plasma injection scheme", Zhang Zhou-Li, R.A. Jameson, et. al. NIMA 592 (2008) 197-200

Using the *a priori* **element**, it would be possible to free one parameter, such as again modulation, to add the EP condition as an *a priori* condition that is not = 1, and could vary along the trajectory. The corresponding tune charts would be at those conditions.

It would also be possible to program the condition of a corresponding EP ratio, such as eln/etn, or the rms emittances separately using *a priori* rules.

There has been some superficial use of the tune chart to see if major resonances are avoided, but with actual performance evaluation only from a simulation result, against a specification which probably only wants information at the machine output and does not concern itself with possible trouble if all is not as planned, or for eventual upgrade requirements, etc., etc.

What has mostly been done so far (for the RFQ) is just to fall back on giving rules for all four parameters – perhaps with a Teplyakov condition on the synchronous phase to control the longitudinal beam charge distribution, with non-space-charge physics consideration for the vane voltage, but ad-hoc and not beam-based for the aperture and modulation – "outside-in" - then plug into a simulation and see if the result satisfies the overall specification.

Serendipity might produce a useful design for a project specification. [42] *But that single case cannot be a basis for claiming a general procedure – it must be tested as above*. Ad-hoc, trial and error, "outside-in", is an effective method to find *one* acceptable design, however tedious, but is not beautiful, and is not general.

However, as stated above, exactly this has happened – without understanding [43], without use of proper scientific procedure, and ….??

A trial and error serendipitous design, for which it is always easy, after seeing desirable features in the simulation, to set up some (questionable) boundaries and a (questionable) analysis, is compared (inappropriately) to one other example, without independently testing it in the same framework, is declared to be a great, new (newly invented here), original procedure -- without declaration of its ad-hoc origin, without showing how the trajectories were actually obtained by design – even given a "new name" -- and stated to be the new general best procedure. Falseifiable?? (Sec.B.1)

In the particular case of early spring 2022 [44], a non-EP design is presented, claimed to be the new best procedure generally, and although "reviewed", without evidence of scientific procedure.

The simulation shows that the main σl/σt=1 resonance is crossed twice and that resonance trappings occur. Publications since at least 2008, when the concept of a "fair" EP vs. non-EP comparison was developed, have shown that this resonance crossing should be avoided in general if not EP.

The data is presented as representing "100% (or 99%, etc) of the distribution" (without definition of 100%). The remark on this above is emphasized by noting that it is essential that the rms properties of a design be sensible. If there is no physical control of the beam itself that emphasizes the desirability to maintain a tight distribution, there certainly is no control of the "100%".

The requirement is for σl/σt to stay within a restricted range, but more, that the emittance behavior will have a very specific behavior, obtained ad-hoc, that would not, in general, always be the desired specification, or in fact occur :

---

42 - that on checking may even actually be adequately EP! This has unwittingly been the case upon checking for some major projects. The EP condition and ratios should always be plotted, as they give fundamental information.

43 "Ich, Lichtenberg (Georg Christoph, ~1795)", pithy remarks, all still same (incompetent academics, etc., etc.) An aphorism: "Wer eine Wissenschaft noch nicht so innehat, dass er jeden Verstoss dagegen fühlt wie eine grammatikalischen Fehler in seiner Muttersprache, der hat noch viel zu lernen." -> "***Anyone who has not yet mastered a science in such a way that they feel every violation of it like a grammatical error in their mother tongue still has a lot to learn***." (DeepL Translate)

44 Also earlier

• stated that, in the entrance region, an equipartitioning occurs from longitudinal to transverse as the longitudinal emittance overshoots the transverse emittance and approaches the main resonance;

• then with eln decreasing and etn increasing and reaching eln/etn=1 in the middle of the tune crossover region where $\sigma l/\sigma t > 1$;

• after which the resonance is crossed again in the other direction as $\sigma l/\sigma t$ falls to <1, the longitudinal emittance grows again, the transverse emittance becomes smaller and ends approximately at the initial transverse emittance.

There is no evidence that such behavior of this nonlinear system could be programmed into a general design program and work over a full specification range [45] – whether if freeing even all four main parameters could be enough. Even possibly with the help of an optimization program. Unless this would be possible, such serendipitous result would have to be searched for, over a full specification range, by trial and error (also a valid and programmable optimization procedure), or by hand. Without invoking beam-based physics and with no design method presented, there are no grounds for claiming this, or some variation, could generally used as a new and best procedure.

There are very many errors in the exposition of this case. Another gross error concerns scientific procedure – an intent of the paper is to declare this case as generally superior to an EP design – and there is nothing at all shown as to the EP condition and ratios along the tune trajectory for this design.

Then comes something that is a far worse abuse of scientific procedure.

A comparison is intended to be made to an EP design, but very clearly without acknowledging first-hand knowledge of the published [46] ORNL Report and proven full possibilities and advantages of EP design, without understanding of it, and *WITHOUT TESTING of a reproduction of such with the same design, simulation and analysis environment for both*. The scientific method is totally ignored.

Instead, the results of a paper are quoted – and the use of this example reveals a great deal.

The comparison shows no understanding, no further investigation, no reproduction testing using the same techniques and tools as for the non-EP design, and the often serious problems of comparing anecdotal cases.

The EP design example paper simulation results shows and states that the EP condition is not met well over the entire acceleration profile, and it should have been immediately obvious that it is not suited to comparison with a fully EP design, and worse to fabricating a general conclusion.

The EP paper reports a very early design, done independently, and without later design refinements, the use of which, as stated, would have prevented trajectory oscillations that happen to be of the same order as reported for the non-EP case being compared to. Such oscillations are not desirable and avoidable for EP designs in general. The EP paper reports the as-built results of a particular design case, and was never intended to be taken as an optimal comparison basis. No contact was made for discussion.

Also, a very limited set of error tests were reported for the "new, general design", with inference that it is better than an EP design. Without understanding or awareness of how a design procedure can increase the tolerance to "off-design" conditions, or that, for example, it has been shown at J-Parc that an EP design is more resistant against various errors. [47].

Not only is there no generation of an EP case for comparison *with the same design, simulation and analysis environment for both* – there is even no graph showing the instantaneous state of the EP condition, and corresponding ratios, along the trajectory for both the non-EP and EP cases (or the cited EP case), as the basis for discussion and comparison of the supposed focus of the article - emittance transfer, emittance growth, etc.

---

45 For example, a different behavior, to attain minimum longitudinal output emittance, is required for the JAEA ADS linac, Chapter 29.

46 For example, see several examples herein, and especially the IFMIF CDR RFQs as shown in Sec.1.7.3, with reference "RFQ Designs and Beam-Loss Distributions for IFMIF", R.A. Jameson, Oak Ridge National Laboratory Report ORNL/TM-2007/001, January 2007, widely distributed, also with certainty to the 2022 author, and readily available to anyone staying abreast of this subject.

47 "Lattice and error studies for J-Parc Linac Upgrade to 50mA/400 MeV", Y. Liu, M. Ikegami, KEK/J-PARC, Tokai, Japan, Proceedings of IPAC2013, Shanghai, China. Note the "for any linac" aspect.

Repeating, the only supportable way to be sure about comparisons is to carefully do the cases within the same background, design and simulation framework – and with scientific procedure. Ad-hoc comparisons of anecdotal cases are always questionable.

Overall an egregious, totally misleading non-scientific effort to become famous for inventing something, exploiting an untested comparison of a coincidental non-EP case to an obviously not well-satisfied EP case, with totally unjustified conclusion. Basic scientific procedure is not followed, neither in the exposition nor in making a comparison. Clearly there is no understanding at all of the space charge physics and how to manipulate it. How the proposed design is performed is not presented, nor are the space-charge physics characteristics from the simulation presented. The content is based on a non-general premise. Comparison is made to an inappropriate example – instead of to an EP design generated in the same software. Unjustified and false conclusions are stated as general.

Throughout, it contains completely wrong statements and other statements that should have been immediately seen and questioned even by a "reviewer" unfamiliar with the overall framework and time-domain approach but supposedly cognizant of the scientific method and the elementary discipline. The commented .pdf is almost completely covered in yellow. Cursively the paper is adequate, apparently meeting the "usual" "reviewer" requirements (e.g. wanting excessive background, but not concerned about scientific aspects). It is known that some have fallen for it – whether impressed that it was "reviewed"?? – but the reader must beware. [48]

Repeating the point: **The greatly extended "clear region around EP" cannot exist if the beam distribution is not equipartitioned, or nearly so, at that point on the trajectory, i.e., that the distribution has no or little free energy.**

**The opposite of the oxymoron is a robust and desirable goal.**

### *1.7.5 Self-Oscillation*

It is worth remarking that, after many decades, there is no standard nomenclature for resonances that can be encountered in linacs, or in general for the space charge physics. This makes it hard not only for beginners. Again it is interesting to explore how other fields handle the nomenclature of resonances, including the excerpts below [49]:

*" Self-oscillation is the generation and maintenance of a periodic motion by a source of power that lacks a corresponding periodicity: the oscillation itself controls the phase with which the power source acts on it.*

*Self-oscillation is also known as "maintained," "sustained," "self-excited," "self-induced," "spontaneous," "autonomous,"*

*Physicists are very familiar with forced and parametric resonance, but usually not with self-oscillation, a property of certain dynamical systems that gives rise to a great variety of vibrations, both useful and destructive. In a self-oscillator, the driving force is controlled by the oscillation itself so that it acts in phase with the velocity, causing a negative damping that feeds energy into the vibration: no external rate needs to be adjusted to the resonant frequency.*

---

48 For this section, as when Ingo Hofmann told me was his problem [private communication] in his rebuttal of the misled argument that the EP equation only applies at a thermodynamic steady state, it took more time to remove polemic from the drafts than to state the bare facts.

***The confusion that is spread by a faulty "reviewed" paper, especially that is wrong to this magnitude, does huge disservice to the linac community, and it will be very difficult to remove the stain.***

49 "Self-oscillation", Alejandro Jenkins, arXiv:1109.6640v4 [physics.class-ph] 11 Dec 2012

*We review the general criterion that determines whether a linear system can self-oscillate. We then describe the limiting cycles of the simplest nonlinear self-oscillators,*

*Parametric resonance  q + w^2(t)q = 0, Hill's Eq.; w^2( t) = w0 ^2(1 + a cos( gt)) -> Matthieu Eq.*

*investigations in [82].)*

*Parametric resonance resembles self-oscillation in that the growth of the amplitude of small oscillations is exponential in time, as long as there is some initial perturbation away from the unstable equilibrium at q = 0. On the other hand, as in the case of forced resonance, the equation of motion for parametric resonance has an explicit time-dependence. A parametric resonator requires the tuning of g in Eq. (15) to 2w0/n , and it fails altogether for g = 0.*

*We showed that a linear system with more than one degree of freedom can self-oscillate, even if no single mode is negatively damped, as long as the couplings are not symmetric, which is possible only if the degrees of freedom describe perturbations about a non-stationary trajectory.*

*We mentioned how the approach of self-oscillators to a nonlinear limit cycle implies that they irreversibly erase information about their initial conditions, thus generating entropy. This explains why the negative damping of linearized self-oscillators near their equilibrium does not imply a reversal of the thermodynamic arrow of time.*

*Self-oscillation is both theoretically interesting and practically useful. Furthermore, it naturally connects with the mathematics and the history of control theory, since self-oscillation corresponds to the presence of positive feedback (and therefore of a dynamical instability). We see no excuse for the fact that the subject is hardly taught to physics students and that it remains, for most physicists, in the shadow of the notions of forced and parametric resonance. "*

### *1.7.6 Nonlinear Lattices, Intrinsic Nonlinearity, and a New Concept – Highly Oscillatory Designs*

**See Chapter 27.**

## 1.8 Single Particle Dynamics

The tune spread in a beam with space charge is very broad, and single particles may be locally affected by many resonances. Study of single particle behavior in terms of the Hofmann chart requires a definition of a *single-particle tune*, that must be sensible locally with respect to the balance between external and space charge fields acting on it. This definition must include the possibility of zero or negative local values, as a particle may be temporarily stopped or rotate backwards in phase space. Definitions appearing in the literature that single particle phase advance ranges from the rms to the zero current phase advances are wrong, for example as shown in mismatched beam studies [Ch.24]. On average over longer times, single particles will of course gather around the rms phase advances. The local single particle tunes thus extend dynamically over the whole tune depression range of the Hofmann Chart. Single particle phenomena are considered further in later chapters, and are particularly relevant to beam halo and very low beam loss.

## 1.9 Design and Simulation Codes and Design Optimization

### 1.9.1 The RFQ and the Alternating-Phase-Focused (APF) Linacs

Insights into how linac design can systematically include the full physics of space charge are particularly important. The principles of controlling the rms matching equations were already used at CERN in the 1970's, and from 1981 with the possibility of the full set of rms equations including equipartitioning. However, to date, still most linacs are not designed with desired space charge physics characteristics built in – but consist of sequences of parts strung together and then simulated to "see what happens", adjusted more or less by hand, re-simulated, and so on. This is termed

“outside in” design. Since the 1980’s, my goal has been to start from desired performance with respect to space charge, and set up external fields to make that happen – “inside out”, and the results are powerful.

The **elements** required for accurate design and simulation codes will be expanded in detail in subsequent chapters, using the RFQ and the Alternating-Phase-Focused (APF) linac as examples.

The RFQ is one of the hardest linacs to design and simulate, because special considerations are required as the beam changes from dc at the input to a fully bunched and accelerated beam. It is a good example also of a linac often followed by another linac with rf frequency change(s), where the beam phase length increases by the frequency ratio at the transition, requiring special measures.

Teplyakov’s invention of the RFQ was an epochal event in linac history, providing acceleration of high intensity ion beams with small emittance growth. The RFQ is so successful in practice that it almost removed further interest in the mechanisms of emittance growth. The author had the privilege of a close friendship with Prof. Teplyakov and was also closely involved with the RFQ development outside Russia. Some unique views of that history are given in the paralleling narrative.

The RFQ is used as an example – the **framework** and **elements** apply to any ion linear accelerator.

The Alternating-Phase-Focused (APF) Linac obtains both transverse and longitudinal focusing/ acceleration directly from the rf fields and does not need separate magnetic focusing (or could require less magnetic focusing), thus saving significant costs. Many forms have been investigated over many decades, but without much success in application because of difficulty in doing the design – with one dramatic exception – the RFQ. In 2012, a practical method for APF design was developed, aided significantly by the concepts of this framework [Part 2.B]. Now it is practical to consider APF as a design option for any linac. It is hoped that the reader might be excited by the many prospects now open for investigation and pursue them.

### 1.9.2 Framework for Design, Simulation and Optimization

The subsections below are all **elements** of the **Framework**.

Modernization of design and simulation began with detailed comparison of extant computer codes, and later development of the new code *LINACS*, with specialization to *LINACSrfqDES*, *LINACSrfqSIM* and later *LINACSapf* [50]). Before that could begin, it was necessary to consider how the tasks should be organized and executed.

#### *1.9.2.1 Source Code*

There are so many reasons why codes by different authors can differ that no comparison is meaningful between them unless the source code is open. It is also very helpful if scientific discussions are possible [51]. Comparison to “black box” codes is not meaningful, and although considerable effort was made on some such comparisons, which showed significant differences that could fairly be attributed to specific problems, no conclusions can be reached. It is strongly recommended that any serious project avoid the use of black box codes.

#### *1.9.2.2 Anecdotal Tests*

Comparison of an answer given by comparison of some specific project design is only an anecdotal result and can be misleading. It also gives no information or assurance about the optimality of a design.

---

[50] The superconducting linac is much easier in terms of the beam dynamics, and involves many engineering aspects. The *tracewin* sc linac work of Didier Uriot appears to be conceived in a comprehensive framework and is very effective for superconducting linac design and simulation.

[51] Unfortunately often not, because any question is interpreted as a criticism and/or competition.

*1.9.2.3 Tests Over Broad Parameter Range*

Codes should be able to handle a broad range of parameters. For design, this means that all parameters, including in this case the space charge physics of high intensity linac beams, must be available to the designer. For simulation, this means that off-design aspects can also be accurately tested.

*1.9.2.4 Experimental Approach for Comparison and Development*

An experimental point of view and methods should be employed. Comparison and development should always keep in mind the full range of possible code use for design and simulation. A key **element** of an understandable experiment is to vary only one parameter at a time.

*1.9.2.5 Finding True Optima*

The designer has to find a design that best satisfies the requirements subject to constraints that require compromises. It is quite possible that the design and/or simulation codes themselves can influence the optimum, via the physics or computing techniques used in the codes. It is important to know if this is happening, and how to compensate or accommodate it. The experimental approach is an essential **element** in determining if different codes reach the same optima.

When it has been demonstrated that an accurate optimum has been found, then alternative code operating modes, for example a faster but with known lesser accuracy mode, can be tested against that optimum.

*1.9.2.6 A Testing Code With Switches*

Different subroutines for testing an aspect can be inserted into a code that provides the rest of the framework, and tested by simply switching back and forth. This is also part of the experimental approach, helping insure that only the intended aspect is tested.

*1.9.2.7 Tests of Detailed Aspects*

Care in testing individual aspects is needed to insure that other aspects of the code are not unintentionally influencing the result. Modern codes are so complicated that this is an ever present concern.

## 1.10 Summary

The following chapters illustrate these **elements** in detail, as the RFQ design code *LINACSrfqDES*, APF design and simulation code *LINACSapf*, and RFQ simulation code *LINACSrfqSIM* are developed. The possibility for very detailed comparison to some existing RFQ codes was really essential, and the availability of source code and discussion is very gratefully acknowledged.

It is demonstrated by simulations that design via rms matching and equipartitioning is very effective, and sufficient to minimize rms emittance growth and to keep the beam distribution relatively tight.

It is necessary to satisfy the rms design as well as possible, given compromises. If EP is not used, then the design strategy is to have the rms trajectory cross resonances quickly. Keeping tune shifts small is also beneficial, but requires more magnet and rf strength for stronger focusing.

The design procedure affords the obvious advantage of an equilibrium beam distribution by the "inside out" [52] procedure of (exactly) solving the EP equation simultaneously with the matching

---

[52] October 2021! Finally, notice that someone else in accelerator field is using very sophisticated "inside out" design – termed "inverse design" - for laser driven "microchip accelerator" dielectric cavities!! And publishing elsewhere:

"Microchip accelerators", R.J. England, P. Hommelhoff, and R.L. Byer, Physics Today 74, 8, 42 (2021); doi: 10.1063/PT.3.4815: https://doi.org/10.1063/PT.3.4815

"Method for computationally efficient design of dielectric laser accelerator structures", T. Hughes, G. Veronis, K.P. Wooton, R.J. England, S. Fan, Vol. 25, No. 13 | 26 Jun 2017 | OPTICS EXPRESS 15414

equations. This is sufficient to get the best beam (little and/or controlled emittance growth, tight beam), but also requires compromises and is not necessary is all cases. Many designs coincidentally achieve equipartitioning in some part of the linac, perhaps only briefly, but this is also very beneficial. In room temperature linacs, where longitudinal focusing inevitably falls as velocity and cell length increase, EP requires that the transverse focusing must be reduced to match the longitudinal focusing, which may lead to larger aperture and/or raising the rf voltage to get more focusing, and more cost. In superconducting linacs, the longitudinal focusing can remain constant or rise as velocity increases, so the situation is quite different.

All the capabilities of the three equations should be considered by the designer – everything can be changed along the trajectory, including emittances and form factor.

It is recommended to always plot trajectories in all dimensions on the Hofmann Chart, to understand the space charge physics.

Very close coordination of the design and simulation programs affords experimental testing for a true design optimum and alternative code operating modes.

## 1.11 Conclusion

Linacs are robust – those presently built are based more on experience than on a comprehensive design strategy – but they work. Future higher intensities will require more powerful and self-consistent techniques. In any case, it is not a bad idea to have a comprehensive design strategy – for any linac… [53]

[eltoc]

53 "Linear Accelerators", Robert Jameson, Joseph Bisognano, and Pierre Lapostolle, Encyclopedia of Nuclear Physics and its Applications, 2013 Wiley-VCH Verlag GmbH & Co. KGaA. Published 2013 by Wiley-VCH Verlag GmbH & Co. KGaA.

***Chapter 1 - Appendix 1. Paul Channell's Derivation of the Equipartitioning Equilibrium Condition***, December 1980, LASL

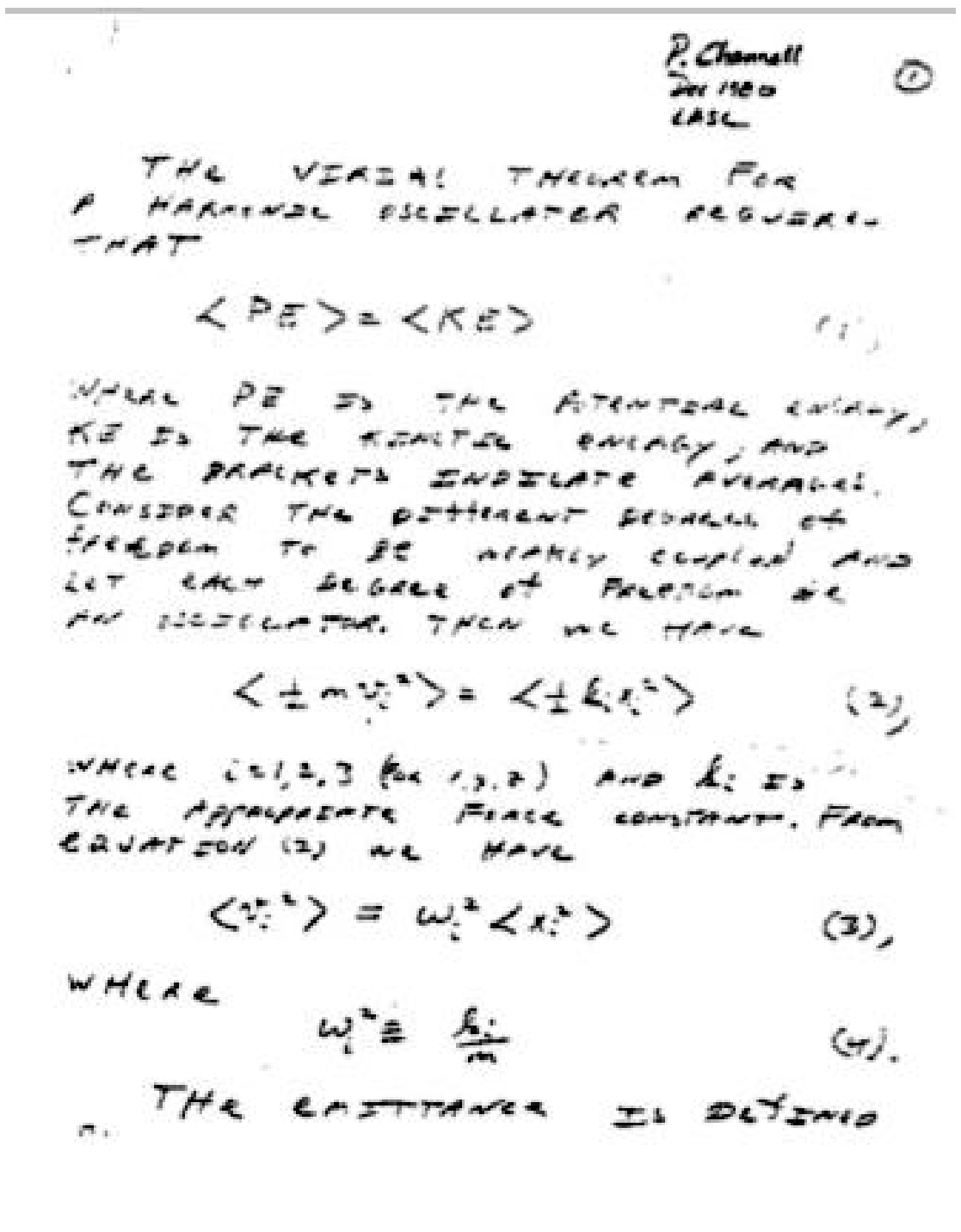

P. Channell Dec 1980 LASL ①

THE VIRIAL THEOREM FOR A HARMONIC OSCILLATOR REQUIRES THAT

$$\langle PE\rangle = \langle KE\rangle \quad (1)$$

WHERE PE IS THE POTENTIAL ENERGY, KE IS THE KINETIC ENERGY, AND THE BRACKETS INDICATE AVERAGES. CONSIDER THE DIFFERENT DEGREES OF FREEDOM TO BE WEAKLY COUPLED AND LET EACH DEGREE OF FREEDOM BE AN OSCILLATOR. THEN WE HAVE

$$\langle \tfrac{1}{2} m v_i^2\rangle = \langle \tfrac{1}{2} k_i x_i^2\rangle \quad (2),$$

WHERE i=1,2,3 (for x,y,z) AND $k_i$ IS THE APPROPRIATE FORCE CONSTANT. FROM EQUATION (2) WE HAVE

$$\langle v_i^2\rangle = \omega_i^2\langle x_i^2\rangle \quad (3),$$

WHERE

$$\omega_i^2 = \frac{k_i}{m} \quad (4).$$

THE EMITTANCE IS DEFINED AS

The enormous amount of written material intended to clarify in this area tends to be very confusing if its time-varying fundamentals are not firmly in mind - the focus is on a steady-state – a prevalent view of physics. Even then a cognizant statement may be embedded in a confusing context.

Virial Theorem Wikipedia is such, but states clearly:

"the virial theorem does not depend on the notion of temperature and holds even for systems that are not in thermal equilibrium.
en.wikipedia.org › wiki › Virial_theorem

Wikipedia on the Equipartition Theorem is good – if read to the end, and in the context of its assumptions. The 1st paragraph states that "the original idea" (assumed) thermal equilibrium", but goes on at length to many detailed expansions (often also with some internal conflict), including a statement about its relation to the virial theorem –always in the steady-state frame of physics, without considering how a system may have gotten to that state – the transient world with self-consistent initial conditions and transport. The article is clear enough to see that including the transient process is in no conflict with the assumed framework or vice versa.

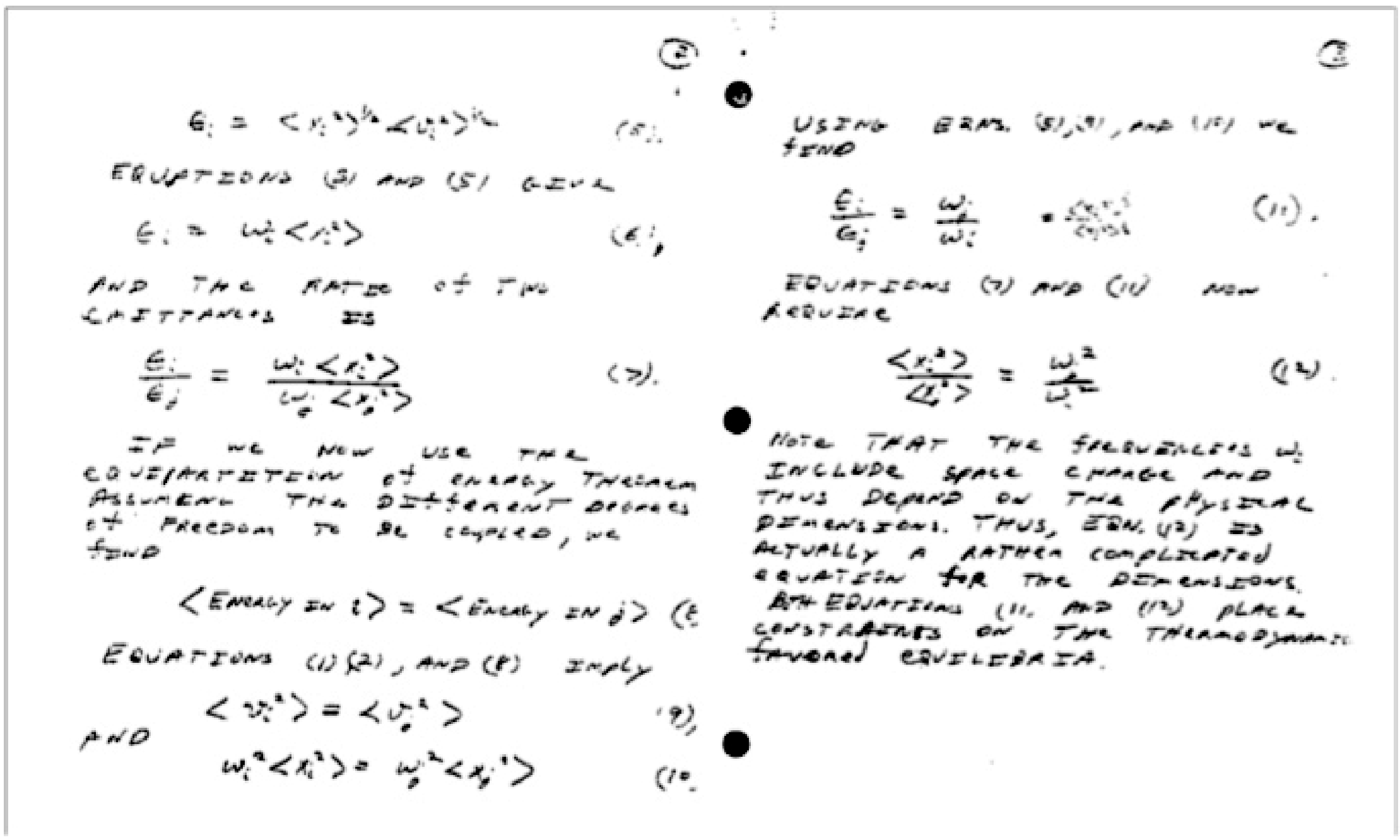

②

$$\epsilon_i = \langle x_i^2\rangle^{1/2}\langle v_i^2\rangle^{1/2} \quad (5).$$

EQUATIONS (3) AND (5) GIVE

$$\epsilon_i = \omega_i\langle x_i^2\rangle \quad (6),$$

AND THE RATIO OF TWO EMITTANCES IS

$$\frac{\epsilon_i}{\epsilon_j} = \frac{\omega_i\langle x_i^2\rangle}{\omega_j\langle x_j^2\rangle} \quad (7).$$

IF WE NOW USE THE EQUIPARTITION OF ENERGY THEOREM ASSUMING THE DIFFERENT DEGREES OF FREEDOM TO BE COUPLED, WE FIND

$$\langle \text{ENERGY IN } i\rangle = \langle \text{ENERGY IN } j\rangle \quad (8)$$

EQUATIONS (1), (2), AND (8) IMPLY

$$\langle v_i^2\rangle = \langle v_j^2\rangle \quad (9),$$

AND

$$\omega_i^2\langle x_i^2\rangle = \omega_j^2\langle x_j^2\rangle \quad (10).$$

③

USING EQNS. (5), (9), AND (10) WE FIND

$$\frac{\epsilon_i}{\epsilon_j} = \frac{\omega_j}{\omega_i} \quad (11).$$

EQUATIONS (7) AND (11) NOW REQUIRE

$$\frac{\langle x_i^2\rangle}{\langle x_j^2\rangle} = \frac{\omega_j^2}{\omega_i^2} \quad (12).$$

NOTE THAT THE FREQUENCIES $\omega_i$ INCLUDE SPACE CHARGE AND THUS DEPEND ON THE PHYSICAL DIMENSIONS. THUS, EQN. (12) IS ACTUALLY A RATHER COMPLICATED EQUATION FOR THE DIMENSIONS. BOTH EQUATIONS (11) AND (12) PLACE CONSTRAINTS ON THE THERMODYNAMICALLY FAVORED EQUILIBRIA.

Last sentence, add: "if thermodynamic equilibria were to be reached". The derivation is not restricted to that condition. Channell did not intend that, and knew that it is applicable to the system state at any time. We discussed when he gave me this, and he had no objection to how I proposed to try using it.

(1)

Let the linear Hamiltonian for the $i^{th}$ degree of freedom be

$$H_i = \frac{p_i^2 + \omega_i^2 q_i^2}{2}$$

The equipartition theorem says

$$\{H_i\} = \{H_j\} \quad \text{in equilibrium}$$

The virial theorem gives

NOTE $$\{p_i^2\} = \omega_i^2 \{q_i^2\}$$

The braces denote time averages, not distribution function averages.

---

**Conjecture**: Next level equipartitioning implies

$$\{H_i^2\} = \{H_j^2\}$$

(2)

If we write this out in detail we get

$$\{p_i^4 + 2\omega_i^2 q_i^2 p_i^2 + \omega_i^4 q_i^4\} =$$

$$\{p_j^4 + 2\omega_j^2 q_j^2 p_j^2 + \omega_j^4 q_j^4\}$$

However, First level equipartitioning and the virial theorem

$$\Rightarrow \{p_i^2\} = \{p_j^2\} = \omega_i^2\{q_i^2\} = \omega_j^2\{q_j^2\}$$

Harmonic motion then gives

$$\{p_i^4\} = \{p_j^4\} = \omega_i^4\{q_i^4\} = \omega_j^4\{q_j^4\}$$

Thus,

$$\boxed{\omega_i^2\{q_i^2 p_i^2\} = \omega_j^2\{q_j^2 p_j^2\}}$$

Average over distributions

(3)

$$\boxed{\omega_i^2 \langle q_i^2 p_i^2 \rangle = \omega_j^2 \langle q_j^2 p_j^2 \rangle}$$

__Note__ If the distribution function is of the form $e^{-\beta H_i} \propto e^{-\frac{p_i^2}{2} - \frac{\omega_i^2 q_i^2}{2}}$

Then $\langle q_i^2 p_i^2 \rangle = \langle q_i^2 \rangle \langle p_i^2 \rangle$

and the new condition on equipartitioning follows from the previous, so there is nothing new (This is how it should be.)

In general, though,

$$\langle q_i^2 p_i^2 \rangle \neq \langle q_i^2 \rangle \langle p_i^2 \rangle$$

and this is a new condition.

[eltoc]

# PART 2 — DESIGN CODE DEVELOPMENT

Systematic design code for linacs is altogether rare, and design within a comprehensive framework, and including space charge physics as a primary focus with "inside out" design is so far uniquely represented by *LINACS*. Therefore, heavy emphasis is placed here on design, followed later by simulation, where equal care and a framework approach are important. RFQ design is used as the example, being the hardest linac to design because of the beam evolution from dc injection to the formation of accelerated bunches.

Then the new practical method for APF linac design and simulation is presented. This somewhat breaks the flow of the mainly RFQ-oriented theme, but is presented in this leading chapter because its potential is too important to be relegated to the end of the book.

## 2.A. RFQ DESIGN

Reference [44 ] is a full development of systematic RFQ design incorporating space charge physics, circa 2007, and is an essential complement to this book. Sections 4&5 are presented in updated form here.

## Chapter 2 – External Field Quantities Defined by the RFQ Metal

The RFQ is an accelerating device with time-varying fields. The classical description of the RFQ external fields used the first two terms, or the first eight terms, of the field solution in the vicinity of the vane tips - called the '2-term potential' or '8-term potential" - to describe the transverse and longitudinal field. It is emphasized that this representation is only accurate near the beam axis. Whereas particles get lost on the vane tips, including areas around the tip, not just at the very tip.

A sinusoidal longitudinal modulation profile is instead mostly used now, which is easier to machine and gives somewhat more efficient acceleration. Other vane shapes are possible, and trapezoidal longitudinal modulation, one of the original shapes, has recently become popular again, giving further advantage in acceleration rate (Chapter 23). In practice, the transverse vane profile is also not machined to the exact 2-term potential shape; typically a circular radius at the vane tip is used. The 8-term multipole approximation was adequate near the axis for this transverse vane profile and sinusoidal longitudinal modulation, but trapezoidal longitudinal modulation, for example, would require many more (e.g. 40) terms, and 2-term modulation (not recommended) is worse.

For design, we need the physics, especially the space charge physics, to correspond closely to the actual physics as simulated accurately by the simulation code, but would also like a quick running design code so that many variations can be explored and even programmed for optimization. This is accomplished in a unique and particular way in the *LINACS* design codes.

The key point for design is that the 2 or 3 rms design equations are used. Therefore the key information must be accurate at the rms beam radii – which should occur well within the linac bore and bucket. And therefore the multi-term potential representation is adequate for design.

Older versions of simulation codes as developed by Crandall parameterized the potential terms over the design parameter space and generated look-up coefficient tables, for fast computer simulation. But as stated above, this approximation is not appropriate when low beam loss is required.

The external field and the space-charge field must both be found by direct Poisson solution with <u>the same</u> exact geometrical boundary conditions, which must extend far enough beyond the transverse vane tip to insure accurate fields in the vicinity of the vane tip. Part 3 below describes the unique *LINACSrfqSIM*, which affords full-featured accurate RFQ simulation much more efficiently than still cumbersome large electromagnetic solvers.

Image charge effects are important in RFQs with significant beam current. The direct Poisson solution automatically and fully includes the image charge effects. The clever analytical image method of Crandall was tested in the LINACSrfqDES code and found negligible in the envelope design process.

# Chapter 3 – Minimum Beam-Related Specification

The RFQ vane surface cell by cell is described by the geometrical shape, which is varied using four variables:

- either the minimum aperture a[rfq] or the average aperture r0[rfq]:
- the modulation em[rfq]
- the voltage between opposing vane tips v[rfq]
- the synchronous phase angle phis[rfq].

## 3.1 Teplyakov Synchronous Phase Rule

The RFQ generates both transverse and longitudinal focusing fields from the electric field only. It is an alternating-phase-focused (APF) device, but instead of requiring placement of rf gaps, its continuous fields required a unique insight, realized by Teplyakov.

The synchronous phase angle must be specified for each cell. There is a rigorous physical basis for this using the rule invented by Teplyakov [54] which was the key to the success of the RFQ: the charge density of the forming bunch should remain constant; i.e., the ratio between the accelerating bucket length and the beam length should be constant.

In "conventional" design procedures, this is the only place where a relationship between the beam and the structure is at least quasi-required at each cell (the rule is often approximated by a curve based on experience or an ad-hoc formula).

The charge density can be allowed to vary to some extent, and specific control is available in the exact cell-by-cell solution and more flexible design procedure available in *LINACS*. This control of the charge density is especially useful in controlling the length of the RFQ. See Ch.25.3.5 for more detailed discussion.

## 3.2 Beam Envelope Equations

As outlined above, two envelope equations describe the variation of the rms beam radius and length of a bunched ellipsoidal beam *in the smooth approximation* (rapid variation of the quadrupolar field and other possible variations within a focusing period are smoothed) of a focusing system, as a function of the external fields and the internal space-charge forces within the beam, which work to counteract the external fields. In addition, the optional equipartitioning relationship (also rms) can be added and the three equations solved simultaneously at each time step.

**These equations are commonly interpreted to a periodic focusing system of infinite length, or with implication of a long time, or thermodynamically, i.e. steady-state, but this is *wrong*. These views are correct, but only after a series of simplifying assumptions. Their much higher potential for use in relation to design and simulation is detailed herein – and repetitively because understanding and using this element is at the top of the list in this book. Ch.16 & 28 and elsewhere develop the crucially important point that they are time-dependent equations, and how they must be used in relation to design and simulation. They are indeed commonly applied locally, assuming (without checking) that the parameter variation is reasonably adiabatic, i.e., that the first derivative is zero. This condition is essentially satisfied in practical designs; however, there is evidence that in some cases the derivatives should be taken into account (here re use of the theoretical envelope equations, not self-consistent simulation, where the derivatives are accounted for).**

54 V. A. Teplyakov, "The First CW Accelerator in USSR and a Birth of Accelerating Field Focusing," EPAC 2006, Edinburgh, June 2006.

The physically accurate, self-consistent full Poisson simulation, with care in the selection of mesh density and step size, takes of the first derivatives. The envelope equations are used for design; the question of how to possibly account for non-negligible first derivatives in design is treated in Ch.28.

Equating the second derivatives of the beam sizes to zero means that the beam is "matched" to the focusing system, and this condition is sought as part of the design process.

### 3.3 Beam-Envelope Matching at Transitions

Transitions occur often in practical linacs; as energy increases, a different type of accelerator structure must be used, or engineering reasons require parameter changes, etc. The matching equations (1) and (2) immediately indicate how to maintain the matched condition at a transition, across which it would be desired that the emittances and beam sizes remain constant. Maintaining the matched conditions then requires that the phase advances per unit length, σ/(unit length), also remain constant.

### 3.4 An Historical Global Space-charge Rule

If the current is increased until the phase advances equal zero, the "space-charge limit" is reached. At some intermediate current, the matching equations (1) and (2) indicate that the emittances and the current will have equal effect on the beam radii. If the beam bunch is assumed to be spherical, this occurs when the "tune depressions" $\sigma/\sigma 0 = \sim 0.4$. The original procedure solved Eqs. (1) and (2) independently with $\sigma/\sigma 0 = 0.4$ to find "transverse and longitudinal current limits," which were brought to equality by parameter adjustments. Requiring that the desired operating current be ~half of this "current limit" was then used as a practical guide for the design input current and adequate transverse and longitudinal focusing if s/s0 were to equal 0.4 - not as a step-by-step design rule. It was quickly found wanting, as described next, although very unfortunately never developed further in the LANL codes.

[eltoc]

# Chapter 4 – "Conventional Design"

"Conventional" RFQ design, still used almost exclusively today, is exemplified by the design of RFQs as originally laid out by Los Alamos in 1978-1980, in which only a global beam-current-related criterion was invoked for guidance as to the practical maximum current that should be accelerated for given focusing parameters. That RFQ type is laid out in four sections – a radial matching section, a "gentle buncher", a "shaper", and the "accelerating section" as outlined below. Many parameters were held constant in the main accelerating section to ease manufacturing and tuning as understood at the time. Then beam acceleration is simulated and the resulting transmission to the output is observed as the measure of successful performance. The detailed pattern of beam loss has rarely been studied. It has been considered "too difficult" to understand in detail the cell-by-cell performance of the beam as it undergoes the very complicated process of forming a bunch from an initially dc beam and accelerating. This method is thus characterized as design "from the external field to the beam result," i.e., "from the outside in.

The rule that the "current limit" is twice the desired operating current guided the choice of rf frequency, which in turn governs the maximum allowable intervane voltage according to a "bravery factor" (typically ≤ 2) over the sparkdown criterion defined by Kilpatrick.

Parameters (such as the average aperture r0 between vanes, the voltage V between opposing vanes, the transverse focusing parameter B) were held constant along the vanes to ease manufacturing complexity. Later it was learned how the parameters could be varied, giving an additional **element** of flexibility to the design process.

The four RFQ sections begin with the "radial matching section", about 4-6 cells long, which brings the beam into the RFQ by ramping up the fields from zero to a chosen value and adapts the dc input beam to the time-varying rf.

The fields are then raised in a "shaper" section from (zero modulation and -90° synchronous phase) to the full transverse focusing ability available from the voltage and aperture chosen. Then the beam is slowly bunched and accelerated in a "gentle buncher" section, being careful not to create strong space-charge forces leading to beam blow-up and lost particles, either immediately or later downstream. Typically the gentle buncher accelerates to about ten times the injection energy. The end of this "shaper/gentle buncher" section is determined "by varying the synchronous phase reached at the end of the gentle buncher section". The reason for this was not well-stated in terms of the space charge physics, and the result is a shaper with undesirable properties.

The average aperture, r0, was held constant, to facilitate tuning the structure, and this also keeps the transverse beam size ~constant as the modulation increases. In terms of the space charge physics, if the emittance and transverse beam size are to be kept constant, then the transverse phase advance must also be held constant. These have an important consequence, because as the modulation, synchronous phase and beam velocity increase, the required minimum aperture decreases quickly (and finally very quickly) toward zero, as seen from the relation r0=(minimum aperture)(1+modulation)/2. Before this can occur, further increase of the modulation must be stopped. This point nominally defines the end of the "gentle buncher" section, and is considered the "choke-point" where the "current-limit" requirement applies. (In other words, the RFQ is basically a transverse focusing device, with a total focusing strength, so when longitudinal focusing is added, the available transverse focusing strength is reduced.)

An improvement for intense beams was to modify the shaper and gentle buncher to have a long initial "porch" section where the synchronous phase remains at -90° with no modulation to allow initial bunching with no acceleration, followed by the "gentle buncher".

The "acceleration section" then accelerates the beam to the final energy. The modulation is held constant at its end-of-gentle-buncher value. Therefore there is a sharp kink in the vane parameters at the end of the gentle buncher, which leads to poor beam loss performance, with which many groups have been unsatisfied and have pursued various smoothing techniques in this region. Within the constraint of the focusing available from the vane voltage such that r0 remains constant, the synchronous phase may be raised further, and then is clamped.

Major deficiencies of this design technique were quickly found to be that such RFQs tended to be very long, and that beam losses were higher than desired. Experienced designers quickly made hand modifications, and the situation was fully clarified with H. Takeda in 1993 [55]. Very unfortunately for many RFQs, the clarification was ignored and the "conventional" technique was and is used to the present, without improvement to the fully flexible and fully physics based technique developed herein.

It was also found that a lower B at the beginning of the shaper requires less convergence of the injected beam, making the space-charge effects at injection less, making it easier to achieve a good input match, and allowing if desired a lower vane voltage at the front end, lowering rf power.

The values at the end of the gentle buncher were then held constant through the remainder of the RFQ accelerating section. This means that the space charge physics is totally uncontrolled, and just reacts to whatever external field is present.

"constant r0, V" does make the overall RFQ cavity design and field tuning easier, and many simpler RFQs take advantage of this, although detailed design and tuning methods are now well known.

It has also been stated that an important advantage of the "constant r0, V" approach is that it is easier to find an "optimized" solution in the multi-parameter design space. In this sense, "optimization" is usually considered shortest length with acceptable beam loss. The optimization process using a flexible design process is still indeed tedious, but advantages of a more encompassing optimization are worth the extra work in the design phase. For example, the beam losses should be confined to occur at the lowest possible energy, in order to minimize buildup of radioactivity. More sophisticated parameter variation methods are described in **[56]**, and with present computers and the design method of *LINACSrfqDES* is also set up to be run under an optimization driver.

---

55 R.A. Jameson, Principal Investigator, B. Blind, G.P. Boicourt, R.W. Garnett, D.P. Rusthoi, H. Takeda, "Deuteron Linac Design Aspects for ESNIT", Study for the JAERI, LA-CP-93-5, Los Alamos National Laboratory, January 1993. (pp. 4-25; also openly circulated many times later.)

For an intense, factory-type linacs of IFMIF class, all aspects point to design sophistication beyond "constant r0, V."

[eltoc]

# Chapter 5 – Beam-Based Linear Accelerator Design Technique

This chapter describes accelerator design techniques that go beyond those commonly used, and which must be understood by project team members, as well as project leaders, reviewers and other interested parties.

Design of an RFQ is the focal point of this discussion, but it applies to linear accelerators of all types.

Repeating from Chapter 4: ""Conventional" RFQ design, still used almost exclusively today, is exemplified by the design of RFQs as originally laid out by Los Alamos in 1978-1980, in which only a global beam-current-related criterion was invoked for guidance as to the practical maximum current that should be accelerated for given focusing parameters. The RFQ is laid out in four sections – a radial matching section, a "gentle buncher", a "shaper", and the "accelerating section" as outlined below. Many parameters constant in the main accelerating section to ease manufacturing and tuning as understood at the time. Then beam acceleration is simulated and the resulting transmission to the output is observed as the measure of successful performance. The detailed pattern of beam loss has rarely been studied. It has been considered "too difficult" to understand in detail the cell-by-cell performance of the beam as it undergoes the very complicated process of forming a bunch from an initially dc beam and accelerating. This method is thus characterized as design "from the external field to the beam result," i.e., "from the outside in."

This method does involve the injected beam in the design, but not the detailed beam behavior inside the RFQ. For intense, factory-type linac facilities like the IFMIF, where continuous operation for a period of ~40 years is expected, and minimized radioactivity buildup is a key objective, a more rigorous design method was sought.

An inverse design procedure is more relevant to the problem of achieving low beam loss: a beam-based design procedure starting from the space-charge physics characteristics of the desired beam, and finding the external fields appropriate to confine it.

Therefore the author has derived a "from the inside out" beam-based approach, in which the desired space-charge physics of the beam is first specified at each cell, and then accelerator fields are derived for the desired conditions.

The beam-based method requires a practical formulation of the space-charge physics, understanding of the effect of accelerator structure resonances and their spreading by space-charge, phase-space transport mechanisms [57], controlled use of a beam internal energy equilibrium and the parameters related to it, and other factors.

*A major requirement of the beam-based method is that the desired design performance be very closely verified by the detailed beam simulation.* This was not lightly achieved, and required

**56** "RFQ Designs and Beam-Loss Distributions for IFMIF", R.A. Jameson, Oak Ridge National Laboratory Report ORNL/TM-2007/001, January 2007.

57 R. A. Jameson, "Self-Consistent Beam Halo Studies and Halo Diagnostic Development in a Continuous Linear Focusing Channel," LA-UR-94-3753, Los Alamos National Laboratory, 9 November 1994. AIP Proceedings of the 1994 Joint US-CERN-Japan International School on Frontiers of Accelerator Technology, Maui, Hawaii, USA, 3-9 November 1994, World Scientific, ISBN 981-02-2537-7, pp. 530-560.

extensive development of the design method to include all of the effects to be simulated, and of the simulation code itself.

Rules governing the space-charge physics of intense beams in linacs were introduced into the cell-by-cell design procedures starting in 1981. The evolution of the design methods, and their application, are discussed in the following paragraphs.

## 5.1 Extension of the ”Conventional” Procedure to Achieve Shorter RFQs

Through experience, it was quickly found that RFQs designed by the “conventional” procedure tended to be longer than desired. Other trial-and-error procedures were developed that produced shorter RFQs with nearly as good transmission; these methods remained largely private however.

A thorough investigation of the “traditional” gentle-buncher design procedure and the previously trial-and-error extension to shorter RFQs is given in [54]. Finding a satisfactory synchronous phase at the end of the gentle buncher corresponded to keeping the tune depressions there $\geq\sim 0.4$. It was found that the conventional approach was allowing the tune depression to be $<0.4$, and often even to go to zero (the space-charge limit), within the gentle buncher. The conclusion was that for intense beams, keeping constant beam size and using a single-point rule at the end of the gentle buncher were inadequate, and that the tune depression in the gentle buncher should be maintained above zero. For moderate to low beam intensities, the original procedure produces adequate results, although the RFQs are often considered long. In addition, the complication of the parameter space was shown, and how shorter RFQs are produced.

Other improvements were found empirically by experienced designers. [58,59,60,61,62] It was realized that lower focusing at injection eased input matching, and as a consequence that a lower vane voltage can be used in this region, saving rf power.

It was found that relaxing the transverse focusing in the accelerating section to keep the transverse phase advance more similar to the longitudinal phase advance gave better transmission. The reason for this was made clear by the author in the following amplification of the space-charge physics relations between the accelerator structure and a beam.

## 5.2 Space-charge Physics Relations Between the Accelerator Structure and a Beam

For a rigorous and practical beam-based design technique, the central requirement is to *use all available beam physics information in the design process*; the resulting design is then checked by a simulation code *with the same underlying physics*.

### *5.2.1 The Beam-Envelope Matching Equations*

In the “conventional” design, the beam-envelope matching equations are used to set a “current limit”. For a beam-based design, they are invoked at every cell.

The equations describe the variation of the rms beam radius and length, as a function of the external fields and the internal space-charge forces within the beam, which work to counteract the external fields. They describe a bunched beam of ellipsoidal form, so apply exactly after the beam is bunched enough to be described as an equivalent ellipsoid.

---

58 A. Schempp, Proceedings of EPAC88, 1989, p 464.

59 A. Schempp, Proceedings of PAC89, IEEE89, CH2669-0, 1989, p. 1093.

60 S. Yamada, “Buncher section optimization of heavy ion RFQ linac,” Proceedings of the 1981 Linear Accelerator Conference.

61 L. Young, “High Power Operations of LEDA,” LINAC 2000.

62 L. Young, “Operations of the LE[illegible] Resonantly Coupled RFQ,” WOAA004, PAC 2001, Snowmass.

Sacherer [63] showed, in his derivation of these equations, that the actual form of various particle distributions causes only a few percent difference in the solution, if the rms properties of the distributions are the same - a seminal result that enables an "equivalent rms" design method. The remaining difference is, however, important to achievement of the best beam-loss performance. The design strategy incorporated in *LINACS* can account for the variation in the form factor of the ellipsoid, and this is effective in reducing beam loss.

The equations cannot be extended to account for emittance growth - the derivation requires the emittances to be either constant, or to have a functional form known *a priori*. In fact, any of the parameters in the envelope equations can be varied *a priori*, including the form factor. The ability to use *a priori* forms is very valuable for certain design problems, and is a key **element** of the **framework**.

Cylindrical beam envelope equations can also be written. The beam transition in the RFQ from cylindrical to ellipsoidal form is however very complicated and a precise enough (less than ~2% error) analytical expression for the transition has not been found.

For design purposes (correct Poisson simulation is self-consistent), the matched beam radius and length should vary smoothly in order to avoid unwanted effects, so that the derivative of the beam size is allowed to vary only slowly - approximately adiabatically in terms of the betatron and synchrotron oscillation wavelengths, so that a' = da/dz is ~zero. When the second derivatives of the beam size are ~0, the beam is "matched" transversely and longitudinally:

Now in addition to the external field variables contained in the zero-current phase advances, four new variables appear - the transverse and longitudinal rms beam sizes and emittances.

Equations (1) and (2) may be solved for any two variables; the others must be prescribed. Usually, the matched rms beam sizes are solved for, with given rules for the external field variables.

Crucially important further discussion is given in Ch.28 – recommended to read and then return to here.

### *5.2.2 Beam Equilibrium - The Equipartitioned Condition*

One other space-charge physics relationship that can be employed for design is known, called the equipartitioning relationship, Eq. (5), which requires that the energy content within the beam be equal in the transverse and longitudinal degrees of freedom. When this condition is satisfied, there is no free energy within the beam that is available to drive a resonance condition.

The availability of an equilibrium, or equipartitioned (EP) condition for the beam is of course a powerful advantage. As explained below, the equilibrium requirement can be invoked at will by the designer; its utility varies according to detailed design requirements.

It is crucial to realize that the internal beam energy balance is transient, can be checked, can be controlled, step by step.

As the emittances can varied *a priori*, it is seen that the equipartitioned condition can be applied over a wide range of conditions, and can change within the accelerator.

The equilibrium condition can be required in addition to the matched conditions, and Eqs. (1), (2), and (5) solved simultaneously at each cell.

---

63 F. Sacherer, "RMS envelope equations with space charge, CERN Internal Report, SI-DL/70-12.

Now there are three equations, which may be solved for any three variables; the others must be prescribed. Typically, the beam transverse and longitudinal radii must be matched, and the EP condition forces a relationship between the beam radii. Thus two of the equations are effectively used for the beam radii, and the third equation to invoke the EP relationship through the use of one of the available RFQ external parameters, for example, the vane modulation.

### *5.2.3 Phase Advances - Resonances*

Eqs. (1), (2), and (5) represent a typical nonlinear system, with all the complex behavior [64,65] that today receives very much attention in the field called nonlinear dynamics, with which the particle accelerator community is in general not familiar. It is interesting that the field of nonlinear dynamics itself changed significantly in the past decade or so. Previously, it was attempted to explain "complex behavior" or "chaos" using theory stemming from a perturbed Hamiltonian. Some relations were found, for example, the Lyapunov Criterion which indicated whether a system was chaotic, which seemed to be useful even though the system was perturbed very far beyond the infinitesimal perturbation over which the theory was valid. More recently, a geometrically based theory of nonlinear behavior has been developed, which can handle large perturbations, and in which it becomes clear that the mechanism for phase-space transport is resonances - with which the accelerator community is very familiar. A combined point of view is useful [66].

With no beam (zero beam current), the structure resonances are defined completely by the local external fields, at all rational number combinations of the tune ratio $\sigma_o^l/\sigma_o^t$ = sig0l/sig0t. With beam current, these resonances are broadened by the collective effect of space-charge, depending on the rms $\sigma^l/\sigma^t$ = sigl/sigt and the internal tune spread of the beam particles.

All satisfactory RFQ and other linac designs should avoid the stronger resonances, either purposely or coincidentally to the design procedure used.

When the beam is in the equilibrium, equipartitioned, condition, there is no free energy available to be converted into changes in the particle distribution through resonant actions. In other words, the structure resonance still exists (it is not "cancelled"), but although the beam tunes may be in the resonance band, no action occurs because the beam is in equilibrium and there is no free energy.

This is the feature of a very useful tune chart for linear accelerators developed by Hofmann [67]. Fig. 5.2.3-1; the features depend on the emittance ratio between the degrees of freedom – in his case, between the x and y degrees of freedom on a continuous beam:

---

64 R. A. Jameson, 1993 Particle Accelerator Conference, Washington, D.C., 17-20 May 1993, IEEE Conference Proceedings, IEEE Cat. No. 93CH3279-7, 88-647453, ISBN 0-7803-1203-1.

65 R. A. Jameson, "Beam Halo from Collective Core/Single-Particle Interactions, Los Alamos National Laboratory Report LA-UR-93-1209, March 1993.

66 R. A. Jameson, "An Approach to Fundamental Study of Beam Loss Minimization," AIP Conference Proceedings 480, "Space Charge Dominated Beam Physics for Heavy Ion Fusion," Saitama, Japan, December 1998, Y. K. Batygin, Editor.

67 I. Hofmann,, "Emittance Growth of Beams Close to the Space Charge Limit," 1981 PAC, IEEE Trans. Nucl. Sci., Vol. NS-28, No. 3, June 1981, p. 2399, I. Hofmann and I. Bozsik, "Computer Simulation of Longitudinal-Transverse Space Charge Effects in Bunched Beams," 1981 Linac Conference, October 1981, LA-9234-C, Los Alamos National Laboratory, February 1982, p. 116.

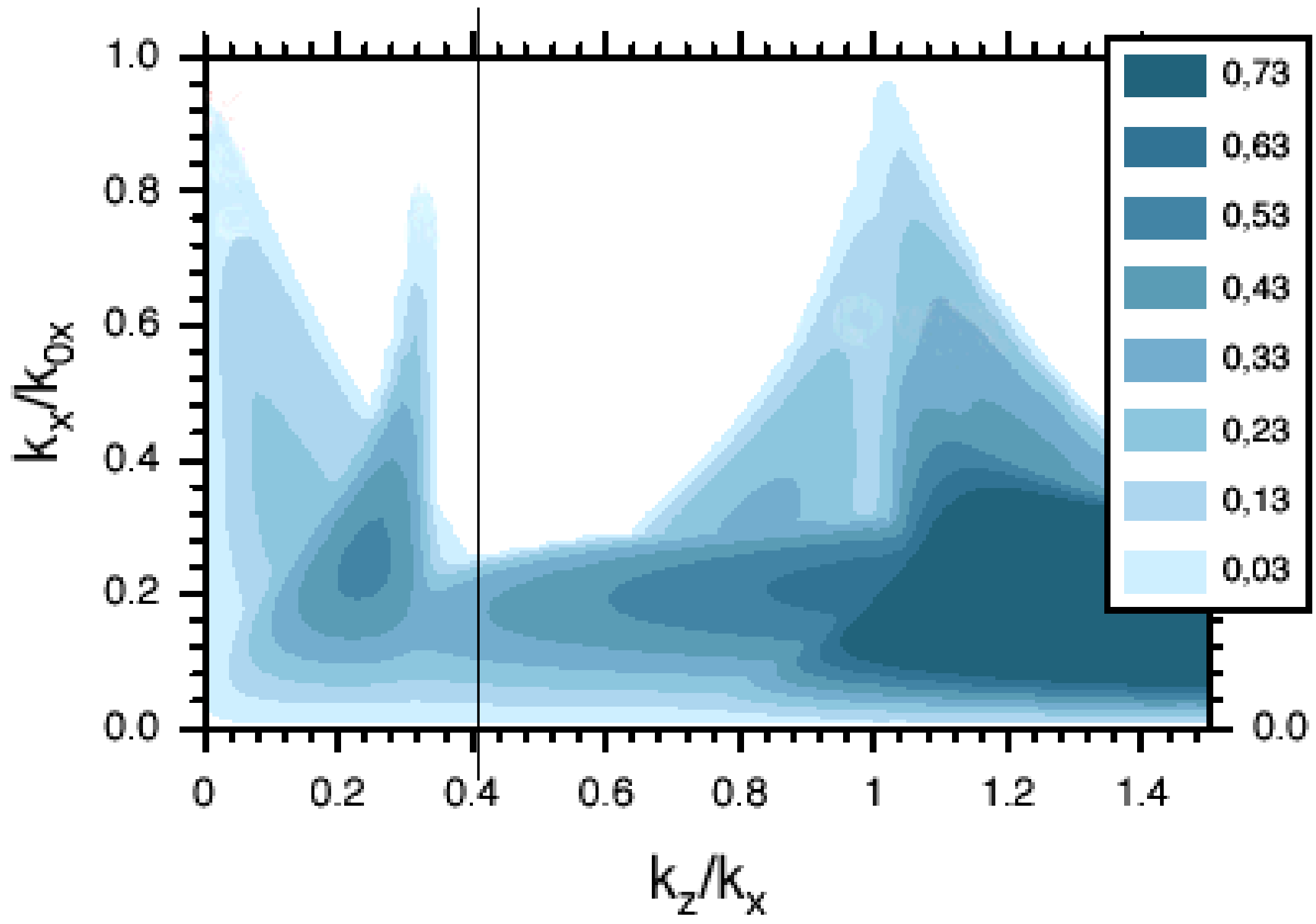

Fig. 5.2.3-1. Hofmann Chart for longitudinal-to-transverse emittance ratio eln/etn = 2.0; thus kz/kx = 0.5 is the equipartitioned condition for this chart.

I immediately checked the mode locations for an elliptical beam with acceleration, and found that the thresholds agreed very closely for the transverse and longitudinal degrees of freedom of the elliptical beam. (Sacherer had long before said that this would be the case …)

The abscissa is thus the ratio of longitudinal tune to transverse tune. The ordinate is the tune depression, here labeled kx/k0x signifying transverse, but in practice, the trajectories of the envelope tunes in the x,y average termed transverse, and z (longitudinal) are both plotted cell-by-cell on the chart. The coloring represents the growth rate of a space-charge broadened tune resonance. Only the \major resonances are shown. On this chart, the growth rates for kz/kx = 1, 1/3, and smaller fractions are seen. If the emittance ratio remains constant, one chart suffices, but the chart is very dependent on emittance ratio, so if it is changing, several charts must be used.

In the early days of the 1980's, it was very tedious to solve the resonance growth rate equations and only some thresholds for growth were available. However, it was found that if either the transverse or longitudinal tune trajectory went below the threshold, emittance growth in that trajectory would occur [68,69]. Later with Mathematica© the charts became practical to generate, and their use is slowly

68 R. A. Jameson, "Equipartitioning in Linear Accelerators," Proceedings of the 1981 Linear Accelerator Conference, Santa Fe, New Mexico, October l9-23, 1981, Los Alamos National Laboratory Report LA-9234-C, p. 125, February 1982; Los Alamos National Laboratory Report LA-UR-81-3073, 19 October 1981.

69 R. A. Jameson, "On Scaling and Optimization of High Intensity, Low-Beam-Loss RF Linacs for Neutron Source Drivers," AIP Conference Proceedings 279, ISBN 1-56396-191-1, DOE Conf-9206193 (1992) pp. 969-998, Proceedings of the Third Workshop on Advanced Accelerator Concepts, 14-20 June 1992, Port Jefferson, Long Island, New York, LA-UR-92-2474, Los Alamos National Laboratory, 28 July 1992.

being adopted [70,71,72], albeit mostly to check a design rather than to base a design trajectory at EP or on the desired location on the chart, and with some differences; for example, not generally realized are the usefulness of plotting the trajectories in both (or all) degrees of freedom, details about various placements of the trajectory on the chart, or the very detailed evidence of smaller resonances that can be detected.

As indicated above, resonances are present at every rational ratio of the tunes. In the design process, the major resonances should be either be avoided, or if traversed, then quickly so little growth in emittance occurs, or by requiring the beam to be in equilibrium in the vicinity of the resonance - so that even though the resonance (and its tails) is still there, there is no free energy and it is not excited.

Thus the eln/etn = 2/1 = 2 chart of Fig. 5.2.3-1 is equipartitioned at kz/kx = 1/2 = 0.5, and this is why no growth occurs in this region, until the tune depression reaches about 0.3 and less.

It is seen that there are areas in the tune chart where the trajectory can be placed without significant growth, especially around the equipartitioned region. It should be remembered, however, that there are tails, and also an infinite number of smaller resonances. A (non-equipartitioned) trajectory cutting across a region of kz/kx that is free from the major resonances shown will still exhibit a small, linearly increasing emittance as the beam is excited by these smaller resonances. The linear growth is typical of such a random scattering effect.

Thus, the EP condition need not be satisfied exactly, and good designs are possible without purposely using the equipartitioning relation. However, it is practical and easy to satisfy it exactly [73] after the bunch has been formed. The EP ratios eln/etn=b/a=σt/σl can even be varied along the trajectory if desired, always avoiding growth due to the corresponding resonance. That is, one could purposely traverse a major resonance without significant effect if the beam is maintained in equilibrium during the traverse.

The information represented by the Hofmann chart can also be represented in the typical phase-space plots of nonlinear systems, which show resonance islands, and which change depending on the degree of nonlinearity in the system. As the nonlinearity is increased, resonances begin to overlap. Eventually, the last free-standing resonance is overlapped and the system breaks into complete chaos. The degree of nonlinearity at this point is called the stochastic limit, and is seen to occur on the Hofmann Chart when the tune depression reaches ~0.3 and below, where there is a general area of growth.

There is a very significant change in the nature of the phase-space dynamics, phase-space transport, and growth rates or disturbance settling times when the stochastic limit is reached. Beyond the stochastic limit, the system is in chaos, with strong mixing and short settling time. In this region, the term temperature can be invoked. Below the stochastic limit, where a linac usually operates, the growth dynamics is completely different, and use of the terms "temperature" and "thermalization" are

---

70 R. A. Jameson, "An Approach to Fundamental Study of Beam Loss Minimization," AIP Conference Proceedings 480, "Space Charge Dominated Beam Physics for Heavy Ion Fusion," Saitama, Japan, December 1998, Y. K. Batygin, Editor. Workshop on Space Charge Dominated Beam Physics for Heavy Ion , 10-12 December 1998, Institute of Physical and Chemical Research (RIKEN), Wako-shi, Japan. Los Alamos National Laboratory Report LA-UR-99-129, 8 January 1999.

71 "Review of Beam Dynamics and Space Charge Resonances in High Intensity Linacs," I. Hofmann et al., EPAC 2002.

72 C. Zhang, Z. Y. Guo, A. Schempp, R. A. Jameson, J. E. Chen, and J. X. Fang, Low-beam-loss design of a compact, high-current deuteron radio frequency quadrupole accelerator, Phys. Rev. Special topics - Accelerators and Beams, Vol. 7, 100101 (2004).

73 Repeating, exact solution because typical trajectories are operating in the region of strong linearity. It is hard to imagine why/how an approximation at the space charge limit could have been scientifically proposed, and subsequently mislead many… not appropriate (although rather widely invoked in the accelerator community). The dynamics is characterized as "meta-stable," with very long settling times; there are areas of phase-space that are stable and others that are chaotic. Another way of saying this is that "simplification" by using the space-charge limit (tune depression = 0) is completely inappropriate for practical design work.

The Hofmann Chart will be used below to assist in defining and clarifying the beam dynamics of RFQs. As a reminder, the designer should remember that if the emittance ratio changes, a different chart is needed.

## 5.3 Beam-Based Design Procedure

### *5.3.1 LINACSrfq Design Interface*

The ingredients for beam-based design are now in hand - the two matching equations which should always be satisfied, and the third EP equation, which may be invoked if desired and solved simultaneously with the matching equations. Three RFQ parameters, typically the transverse and longitudinal beam sizes, and the modulation if EP is used, are used to satisfy the equations. At this time, the remaining (many) RFQ parameters must be chosen from other perspectives that do not involve beam/structure equations directly. (Design optimization is still an open issue, with many possibilities for more complex expression of complex optimzation goals with weighting, etc. and searches with AI, replacing the current work of an informed and experienced designer.)

Somehow, this concept has been hard to teach, so a fallback to simple algebra often helps. Our 2 or 3 equations are all f(a,b,aper,voltage, phis, modl), 6 variables. So if match only, 2 eqns, <u>any</u> two free, other four fixed. Adding EP, 3 eqns, <u>any</u> three free, other three fixed. **"<u>Any</u>" is the key**.

The default used in *LINACS* with clear reasons is (a,b,modl) free, (aper,voltage, phis) fixed. The reasons are that: the beam size needs to be rms matched, we can easily visualize the aperture and have some idea what its variation along the linac might mean, the voltage has strict limits from other considerations such as breakdown, and the synchronous phase is best controlled using a variation of Teplyakov's physics, whereas modulation may have a desired maximum but otherwise left free.

No one has ever tried (a,b,voltage) or (a,b,phis) or (a,b,aper) free – would be good MS research project, could write papers…

The *LINACSrfq* design code incorporates this approach (Part 3, Chapters 13-20). The designer has cell-by-cell control over all parameters, and over the space-charge physics behavior desired.

*It is important to note that now, as the beam itself is specified in terms of its rms emittances as well as current, the design and optimization process are now specific to this beam. This is why certain knowledge of the ion source and LEBT is important. In a factory-type application such as IFMIF, the characteristics of the ion source and LEBT should be reproducible within certain limits (for example, if the ion source is replaced with a spare).*

The beam-based design interface for the *LINACSrfq* code is next outlined to illustrate the design process.

## <u>(* Set up the general characteristics of the RFQ *)</u>

```
injenergy = 0.095; (* IFMIF EP RFQ, new spec *)
curamp = 0.130;
freq = 175.;
him = 2.0145;
q = 1;
endenergy = 5.; (*  final energy of RFQ, MeV  *) )
```

*(* IFMIF is a 4-vane type RFQ. Rf power factors are specified - see Chapter 19.6 *)*

```
rfqtype = 4vane;
powcuLEDAscaletoIFMIF = 7.94;
powcuFac = powcuLEDAscaletoIFMIF;
```

*(* Initialize V2TERM. VSINE or VTRAP(EZOIDAL) vane geometry functions for 4-vane RFQ. Turn multipoles on (mon) or off (moff), or (mon) with individual control of mpole terms (terms 1 = A01 and 3 = A10 are always 'on'. mpoleterms(A01, A03, A10, A12, A21, A23, A30, A32 *)*

```
geom = "VSINE";
mpoleswitch = "mon";
mpoleterms = {1, 1, 1, 1, 1, 1, 1, 1};
If[mpoleswitch == "moff," mpoleterms = {1,0,1,0,0,0,0,0}];
getvanegeomcoeffs; getimagecoeffs;
```

*(* Set RhooverR0 (could also be f(z) *)*

```
RhooverR0 = 0.75;
```

*(* The "field" is set to the desired Kilpatrick level KPlimit*KPfac. *)*

```
KPlimit = limkpfld[freq]; (* = 13.98 at 175 MHz *)
KPfac = 1.7;
```

*(* Enter the desired input normalized rms emittances (pi - cm - rad), *)*

```
etnrmsgiven = 0.000025;
elnrmsgiven = 0.000040;
```

*(* enter the rms emittances desired in the main RFQ after the shaper: *)*

```
etnrmsgivenmain=etnrmsgiven;
elnrmsgivenmain=elnrmsgiven;
```

*Here are some examples of a priori varying emittance:*
*(* etnrmsgivenmain := If[(z - zstart) < 1.0, etnrmsgiven*(1 - 0.2*(z - zstart)/1.0), 0.000016]; *)*
*(* elnrmsgivenmain := If[z < 1.2, elnrmsgiven, elnrmsgiven + 0.000025*((z - 1.2)/8)]; *)*
*(*elnrmsgivenmain := -5.3276*10^(-5) + 2.174*10^(-6)*(i/celldiv) - 1.9926*10^(-8)*(i/celldiv)^2 + 9.0126*10^(-11)*(i/celldiv)^3 - 1.9926*10^(-13)*(i/celldiv)^4 + 1.7246*10^(-16)*(i/celldiv)^5; *)*

## (* Set up the engineering choices and rules for the RFQ *)

### (* Shaper parameters : *)

*The design process starts at the end of the shaper section, where the beam is required to be equipartitioned. The aperture is entered as a fraction of the wavelength, as a comparison to the cell length, which is $\beta\lambda/2$. Transmission can be optimized by varying this aperture.*

```
EOSaperfac = 5.1;
apertgt = (100.*betalaminject/EOSaperfac);
```

*In conjunction with the main RFQ, the space charge form factor at the end of the shaper may be adjusted;*

```
shpr formfac adj
```

*The synchronous phase phis at the end of the shaper is near -90°, typically -88 to -84°. The RFQ will be shorter if phis is raised higher, but transmission may be lower.*

```
phistarget = -88.;
```

*Next the reduction in B at the beginning of the shaper is specified:*

```
bfraction = 0.55;
```

*Continuing to work backwards, specify the number of cells in the radial matching section.*

```
rmscells = 4;
```

*v[rfq] voltage rule for the shaper*

**`frontendvrule:=`**(e.g., the voltage found by the KP limit at the end of the shaper)

*Finally, the length of the shaper has to be specified. The program uses a rule involving the integrated zero-current longitudinal phase advance (i.e., the phase advance of a particle very near the synchronous particle), but the relation is not rigorous - siglint is only a number that controls the shaper length. The "porch" is the initial fraction of the shaper length where phis remains at -90°; phis is raised to phistarget in the remainder of the shaper. The shaper length and porch fraction are used as optimization variables to minimize the energy of lost particles.*

```
siglint = 270.;
porch = 75.;
```

**(* Main part of RFQ : *)**

*In a final design optimization step, the form factor is adjusted to conform to the actual distribution formed in the RFQ.*

(* Interpolating function for ffadj : *)

```
ffadjrule :=
```

*Rules are now specified for the main RFQ parameters. The general form is:*

- *Specify some rules*
- *Select the rule to be used in this design*

**(* phis[rfq] law in main rfq. *)**

*Specification of the Teplyakov-Kapschinsky Rule, here with upper limit at -20° if reached. The bucket-beam length ratio can be varied along the RFQ; lfacincr is the change in the ratio, lfacdist is the distance over which the change is made.*

```
lfacincr = 0.0;
lfacdist = 10.0;
c2 = lfacincr/lfacdist;
TKphisrule := (If[phisbw ≤ -20., phis[rfq] = phisbw,
                phis[rfq] = -20.];);
```

*Another rule, with cosine-like form:*

```
cossec = 0.4;
endphis = -20.;
Cosphisrule := (If[phizz ≤ -20., phis[rfq] = phizz,
                phis[rfq] = -20.];);
```

*A rule linear with position along the RFQ:*

```
philinear := -88. + 43.*((z - zstart)/(2.7 - zstart));
```

*Choose which rule to use in this run:*

```
mainrfqphisrule := TKphisrule;
```

*Another example of a rule which could be used:*

(* mainrfqphisrule:=If[philinear > phisbw, phis[rfq] = philinear, (philinear = -90.; TKphisrule)]; *)

**(* a[rfq] aperture rule for main RFQ : *)**

```
mainrfqaperrule := (endbeta = 0.073;
      c3 = 1.2;
      a[rfq] := atarget*(1 +
c3*((beta-betastart)/(endbeta-betastart))^1.0); );
```

**(* v[rfq] voltage rule for the main RFQ *)**

*(* mainrfq voltage definitions : *)*

```
vKP = KPlimit*KPfac*r0[rfq]/ckappa[rfq];
vlinear = vol0*(1.0 + (1.33* (z - zstart)/endz));
endz = 1.63;
```

*Choose rule:*

```
mainrfqvrule :=vKP;
```

**(* em[rfq] modulation law in main rfq. *)**

*(*1. For EP, em[rfq] is used to solve for the EP condition. Also used for Constant B strategy. *)*

```
mfree := (em[rfq] =.;);
```

*(* 2. Function for em[rfq] : *)*

```
mfunc := (em[rfq] = Min[emtarget*(1. + 1.*((z -
            zstart)/0.85)^1.0),3.0];);
```

*(* 3. em[rfq] is driven by a cosine function which starts using the em[rfq] slope at the end of the shaper, has length = cossecem, and ends with zero slope at the specified end value of em[rfq] (endmodl). *)*

```
cossecem = 1.75;
endmodl = 3.;
mCostype := emzz;
```

*(* Set mainrfqemrule to one of the laws *)*

```
mainrfqemrule := mfree;
```

**(* Choose strategy for main RFQ : *)**

(* 1. **matchonly** - *will use a rule for em[fq], e.g., mfunc, and the two matching equations.*

2. **matchEP** - *em[rfq] is a variable (use mainrfqemrule := mfree;), matching and EP equations are satisfied.*

3. **matchConstB** - *em[rfq] is a variable (use mainrfqemrule := mfree;), matching equations are satisfied and B is kept constant.*

4. **matchboaRatio** - *em[rfq] is a variable (use mainrfqemrule := mfree;), matching equations are satisfied and rmsl/ rmsr is kept constant* **boaratio = 1.2;**

5. *matchUser - user written criteria* *)

```
matchUser := ………;
```

*Select the strategy to be used for this run:*

```
mainRFQstrategy := matchEP;
```

*Everything has now been specified and the design program is ready to run. Many other variations could be specified; all the parameters are under control.*

**(* Run the subroutines that generate the shaper *)**

*The front end is generated from the end of the shaper (EOS) backwards. First the EP and matched condition at the EOS is found, then the conditions at the beginning of the shaper, then at the entrance of the RFQ. Working forward, the radial matching section and shaper are generated. The process is iterated using the energy found at the EOS on the first pass. Two passes suffice.*

```
nshprpasses = 2;
For[(j          =   1;emtarget = setbeginem;),j <= nshprpasses,
```

```
j++,
           (dum =x; shprpass=j; shprend;
           If[j==1, (energy = injenergy),
                (energy = energystart)];
           shprtgt; beginshprtgt; enterrms; rfqshramp; pass)];
```

**(* Run the subroutines that generate the main RFQ *)**

*Then the main RFQ is generated from the EOS to the final energy.*

```
      phisemrules; rfqaemvar; passmain;
```

The interface can be used in file form or from a GUI, which is programmed in Python/Kivy.

### *5.3.2 Design Strategy Discussion*

The RFQ has now three sections - radial matching, shaper (to bring the beam to EP), and acceleration.

Bringing the beam to the EP condition at the (end of the shaper) (EOS) has proven so robust and useful that it is the standard approach. It is easy to change the shaper rules, for example, to the constant r0 conventional method that exhibits the bottleneck, but tests of this method against the EP at EOS method have consistently shown better beam-loss performance for the EP at EOS method.

Complex rules in the main RFQ can be specified; for example:

- The CDR and Post-CDR IFMIF RFQs , typical of high-intensity cw RFQs, maintain EP from EOS to the end of the RFQ, and are no longer and have less beam loss than a "conventional" design.
- Bringing the RFQ to EP at the EOS, and then relaxing the EP condition downstream while taking care to not let the tune depressions become too low, and if a major resonance is crossed, then to cross it quickly, has been an effective method for some applications.

The original goal of this approach was first to bring all of the RFQ parameters and beam physics (the matching and EP envelope equations, plus some detailed extensions, which are all we know at this time) under complete and flexible control. This is now achieved. It affords the possibility to investigate optimization.

Further details of the design code *LINACSrfqDES* are given in Chapter 19.

### *5.3.3 Design Optimization Discussion*

There are so many parameters, and so many different views of what an optimum constitutes, that optimization is still an open and very interesting research topic. As computer power continues to increase, it becomes more feasible, and modern optimization methods, which can handle many variables and strategies, should be applied to the accelerator design problem. *LINACS* already incorporates a powerful optimization package, which can handle nonlinear as well as linear systems with constrained as well as unconstrained variables.

Experience with *LINACS* has resulted in an effective, but tedious optimization procedure; after each design iteration, *LINACSsim* has to be run to check transmission, energy at which losses occur, etc. This can be invoked through the GUI.

1. Using the 2-term fast option, get a rough design using the desired design strategy and default values that results in a solution.
2. Full Poisson simulation will soon become necessary to go beyond the rough design.
3. Find a reasonable input beam ellipse alpha and beta.
4. Optimize transmission varying EOSaperfac.
5. Minimize beam loss above ~10 times the input energy (e.g., above 1 MeV) by varying the shaper parameters siglint and porch. The length of the shaper and porch have influence on the transmission and also on the maximum energy of the bulk of the lost particles.

6. Find the input match for best transmission using the matching options in *LINACS* (most accurate is the transmission matrix option). The input match must be checked occasionally by simulation as the design develops.

All of the main RFQ specifications are strongly influenced by the coefficients used in the Teplyakov synchronous phase rule, the voltage and aperture rules, and by the equipartitioning and emittance rules selected.

7. The aperture must open enough as a function of beta to keep radial beam loss low; this raises the length and/or the voltage.
8. Reduction of the RFQ length is possible using a negative lfacincr to allow the beam length to grow relative to the bucket length.
9. Length is also reduced for lower EP ratio. A varying EP strategy may be useful.
10. For lower current designs, abandoning EP after the shaper may be useful.
11. Iterate, iterate. Tedious.

When the design has been optimized, a final optimization step almost always produces another significant improvement in transmission - an adjustment in the design code of the space-charge form factor, sometimes augmented by an *a priori* adjustment of the transverse and longitudinal rms emittances.

[eltoc]

# 2.B APF DESIGN and Simulation

## "Practical design of alternating-phase-focused linacs"

R. A. Jameson
2013

Reference [74] covers the invention and development of a practical method for systematic APF linac design and simulation, and is an essential complement to this book, presented and updated here.

Conventional magnetic transverse focusing in conventional linacs represents a high fraction of their cost and complexity. Both transverse and longitudinal focusing can be obtained from the rf field by using the strong-focusing effect of alternating patterns (sequences) of gap phases and amplitudes – known as Alternating-Phase-Focusing (APF). Simple schemes have small acceptances and have made APF seem impractical. However, sophisticated schemes have produced short sequence APFs with good acceptances and acceleration rates that are now used in a number of practical applications. Although studied for decades, the design of suitable sequences has been difficult, without direct theoretical support, inhibiting APF adoption. Synthesis of reported details and new physics and technique result in a new, general method for designing practical APF linacs. Very long sequences with high energy gain factors are demonstrated, motivated by the desire to accelerate very cold muons from ~0.340-200MeV (factor 600). An H+ example from 2-1000MeV (factor 500) is also given (the problem scales with the charge-to-mass ratio). The method is demonstrated with simple dynamics and no space charge – incorporation of space-charge and more accurate models is straight-forward. APF linacs can now be another practical approach in the linac designer's repertoire. APF can be used in addition to conventional magnetic focusing, and could be useful in minimizing the amount of additional magnetic focusing needed to handle a desired amount of beam current.

[eltoc]

# Chapter 6 – Introduction – Alternating-Phase-Focusing

APF acceleration has been studied for decades as it offered the possibility to avoid the high cost of magnetic focusing in the linac. Particles exposed to an rf field in a gap may receive focusing or defocusing forces in the transverse and longitudinal directions, according to the phase and amplitude of the gap field. Arranging a sequence of gaps in some particular manner, termed here an "APF sequence", can provide large simultaneous transverse and longitudinal acceptances with high acceleration rates and good emittance preservation, even better in some cases than the well-known RFQ or linacs with magnetic transverse focusing.

Typical transverse focusing and energy gain per unit length (dW/dz) forces over the full 360° range of rf phase (phi) are shown in Fig. 1.

$$xvsefoc = \pi\, qomE_oTsin(phis)\lambda nblt\hat{}2/(\beta\gamma\hat{}3),$$

$$dWdz = E_oqTcos(phis)$$

where *qom* is the charge-to-mass ratio, $E_o$ is the average accelerating gradient, *T* is the transit-time-factor, *phis* is the rf synchronous phase at the gap center, λ the rf wavelength, *nblt* is the nominal number of βλ in one cell (depending on the structure), β and γ are the relativistic factors.

74 "Practical design of alternating-phase-focused linacs", R. A. Jameson, 2013, arXiv identifier 1404.5176, available at:http://arxiv.org/abs/1404.5176

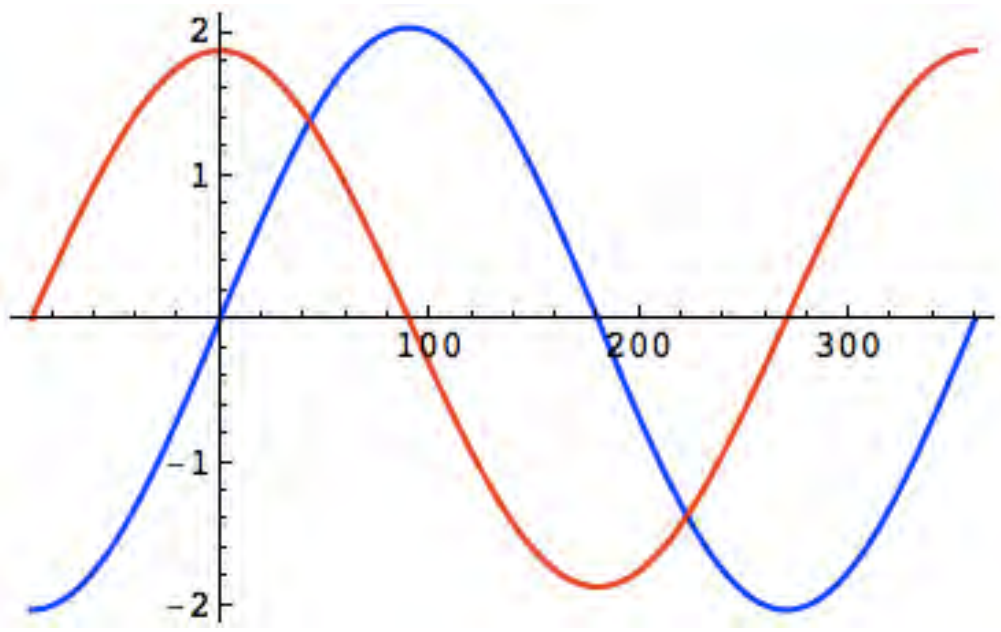

Fig. 6.1. Typical longitudinal acceleration (red) and transverse focusing forces (blue) over the full 360° range of phi.

There has been no adequate general theory or convenient method for determining an APF sequence with large energy gain. There are some APF linacs in operation, but their designs were laboriously produced by hand, and the sequences are short. Further APF development has also been hindered by a misconception that APF acceptances are necessarily small.

The purpose here is to demonstrate a practical method for designing and optimizing APF linacs. Chapter 7 shows the most successful present designs, and summarizes, from the detailed research, important factors needed to synthesize a new general method. Chapter 8 outlines the method. In Chapter 9, the initial design and optimization steps for a 0.34-20 MeV muon APF linac are outlined, as an example. The development continues in Chapter 10 using a proton APF linac as the second example, just to demonstrate that the problem scales with charge/mass ratio, and reaches the conclusion that a more sophisticated optimization procedure is required. Chapter 11 gives a new optimization procedure based on intelligence derived from APF physics and modern control theory, and applies the procedure to the H+ and muon APF linacs. Next steps for further development are outlined in Chapter 12, and Chapter 13 concludes the development.

[eltoc]

# Chapter 7 – The APF sequence, and summarization of important factors

The most sophisticated work was realized in the USSR during the 1960's-1980's [75,76, and culminated in the "Garaschenko Sequence" [77], found using modern control techniques. Simple schemes using modulation of the synchronous phase by ±Δphis had led to the realization that large swings could work (e.g. ±60°, and extended variations of these schemes, including offsets of the average phis), and it had been realized that smooth modulations over the full range of ±90°, with appropriate variations as the particle velocity increases, were possible.

75 "Efficiency of ion focusing by the field of a traveling wave in a linear accelerator", V.K. Baev & S. A. Mineav, Sov. Phys. Tech. Phys, 26(11), November 1981, p 1360.

76 "Linear resonance ion accelerators with a focusing axisymmetric accelerating field", V.K. Baev, N.M. Gavrilov, S.A. Minaev & A.V. Shal'nov, Sov. Phys. Tech. Phys. 28(7), July 1983, p.788.

77 F.G. Garaschenko, V.V. Kushin, S.V. Plotnikov, L.S. Sokolov, I.V. Strashnov, P.A. Fedotov, I.I. Kharchenko, A.V. Tsulaya, "Optimal regimes of heavy-ion acceleration in a linear accelerator with asymmetric variable-phase focusing", Zh. Tekh. Fiz. 52, 460-464 (March 1982)

## 7.1. Garaschenko APF sequence

Garaschenko's 51-cell synchronous phase sequence for a 0.0147-1.0 MeV/u (factor 68), $_{238}U^{7+}$ (qom=1/24), 6.8m, 25 MHz uranium-ion accelerator is shown in Fig. 7.1. It is seen that it has quasi-sinusoidal features.

It is the result of a complicated nonlinear optimization procedure, grounded in modern control theory, to find the maximum longitudinal and transverse acceptances. The longitudinal acceptance was larger than that of a typical RFQ.

Typically, a nonlinear optimization program must be given good enough starting points that it can converge to the correct optimum. Here the preliminary sequence was based on the extant schemes and a successfully operating APF linac at Dubna, which set out the general properties of the sequence in six separate focusing periods of 6-13 gaps and different spacings in each.

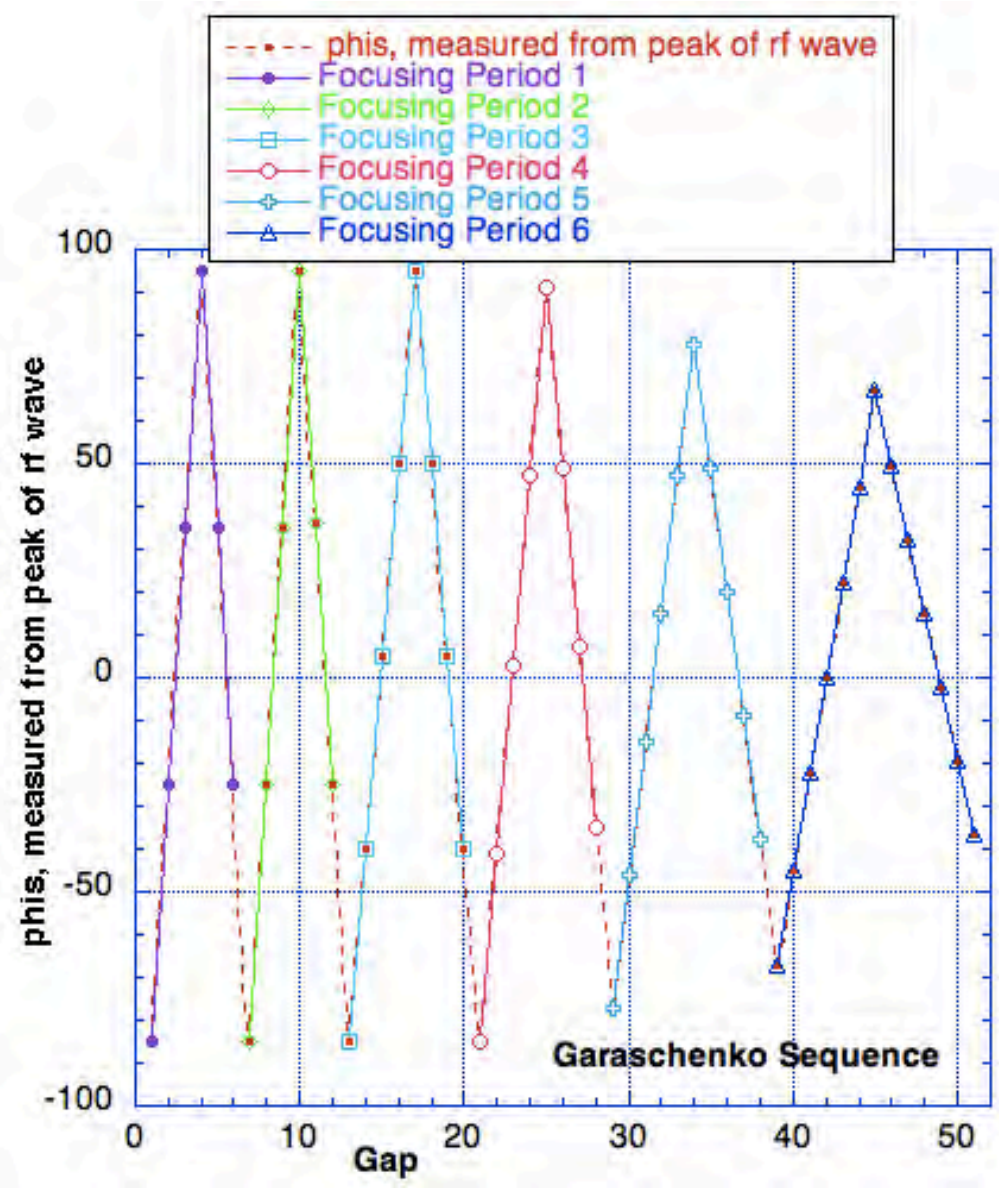

Fig. 7.1 Garaschenko APF sequence. Phis is measured from the peak of the wave.

The great value of this paper is not that it helps find a good initial sequence, but that *it shows the form of an optimized sequence*. From this, we can draw very useful general conclusions.

## 7.2. NIRS APF sequence

The IH linacs now in operation at NIRS, U. Gunma, and similar therapy machines under construction in Japan [78], also have an APF synchronous phase pattern very similar to that of Garaschenko. A general sequence function for the $C^{4+}$(qom=1/3), 0.608-4.0 MeV/u (factor ~6), 3.44m, 200 MHz NIRS linac was written as:

$f_s(n)=f_0\, exp(-an\,)sin((n-n_0)/(b\; exp(cn))$
where $n$ is the cell number.

78 Y. Iwata, T. Furukawa, T. Kanai, M. Kanazawa, N. Kanematsu, M. Komori, S. Minohara, T. Murakami, K. Noda, S. Shibuya, M. Torikoshi, S. Yamada, V. Kapin, Proc. APAC 2004, p.423-425; Y. Iwata, T. Fujisawa, T. Furukawa, T. Kanai, M. Kanazawa, M. Kanematsu, M. Komori, S. Minohara, T. Murakami, M. Muramatsu, K. Noda, M. Torikoshi, S. Yamada, H. Ogawa, T. Fujimoto, S. Shibuya, T. Mitsumoto, H. Tsutsui, Y. Fujii, V. Kapin, Proc. EPAC 2004, p.2631-2633; Y. Iwata,, S. Yamada, T. Murakami, T. Fujimoto, T. Fujisawa, H. Ogawa, N. Miyahara, K. Yamamoto, S. Hojo, Y. Sakamoto, M. Muramatsu, T. Takeuchi, T. Mitsumoto, H. Tsutsui, T. Watanabe, T. Ueda, NIM A 567 (2006) p.685-696

The five parameters were searched, and the sequence then optimized for small output energy spread and output matching to the following section. The 72-cell sequence, closely resembling Garaschenko's, is:

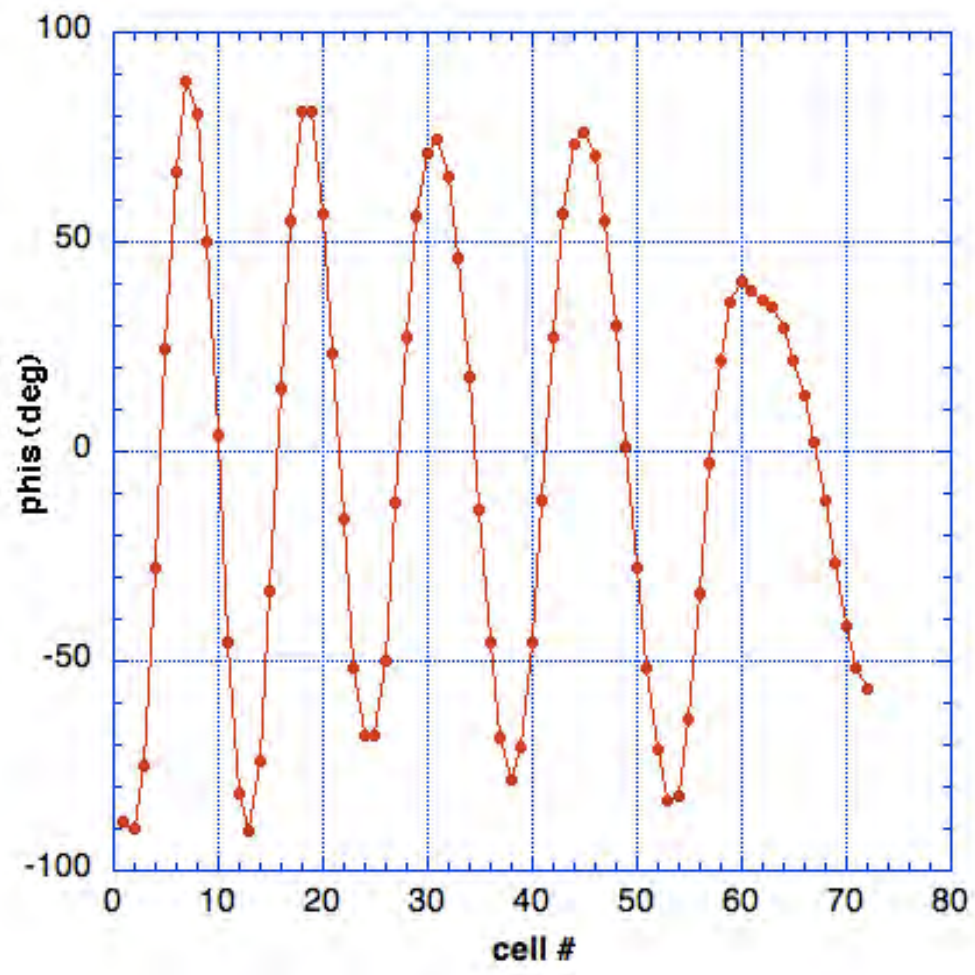

Fig. 7.2 The NIRS APF sequence. (Courtesy of NIRS.)

A similar formula (see Eqn. 1 below) can be fitted to the Garaschenko sequence.

## 7.3. Collection of important APF research details

Overall, many different sequence schemes have been developed, and important details discovered. Summarizing these leads to a practical design method.

In 1988 during research at KEK, the author explored the effect of superimposing various field and synchronous phase error patterns on a long 1 GeV magnetically-focused proton linac for waste transmutation, and found that certain patterns of sinusoidal modulation produced an APF, or anti-APF effect. A generalized error function with a number of parameters was used, and searches made over the parameter space.

The author had gathered a large collection of APF papers (not so easy at that time), and during the second half of that year at the Keage Accelerator Institute of Kyoto University, under the leadership of Prof. H. Takekoshi, his student H. Okamoto became interested in APF and used the smooth approximation method to characterize the stability region, acceptances, and other features of the sequence [79]. This well-known method leads to the Mathieu-Hill equations, for which suitable sequence trajectories can be investigated on a transverse stability chart. The method is useful for understanding the general properties of APF sequences but limited because of the assumptions of periodicity, small phase advances, and no acceleration and thus is not well suited to determining an actual sequence with acceleration. However, as shown below, it has an important utility in the synthesis of the new method.

During the next decade, work in Russia continued, especially by V.V. Kushin, V.K. Baev, and S. Minaev, who was a leading practitioner of actual APF designs until his death in 2010 and who influenced many extant APF designs.

---

79 "Beam dynamics of Alternating Phase Focused Linacs", Hiromi Okamoto, Nuclear Instruments and Methods in Physics Research A284 (1989) 233-247, North-Holland, Amsterdam

However, Minaev noted in [80] that "there is no theory for optimization of drift tube array so far", and this is still the situation at this date. In this paper, Minaev shows that he uses a quasi-sinusoidal sequence progression similar to that shown by Garaschenko, where the period of the sinusoid lengthens as the beam accelerates, but the three succeeding sections are optimized in turn by hand.

The difficulty of deriving a sequence is a main reason why previous APF designs have been for short sequences. Another is the misconception that APF acceptances are necessarily small. However, even the short sequence examples above show the possibility of high energy gain per meter, and high energy gain ratios (x68 for the Dubna linac). A motivation for this paper is to facilitate design and optimization of long sequences with high energy gain ratios.

The extensive literature contains details that are important for the synthesis of a new method.

Different spatial harmonics of an accelerating structure can be used for the transverse focusing and the longitudinal accelerations [81,82]. The latter paper also points out that in the Alvarez $E_{010}$-mode structure, the stored energy is distributed uniformly, and therefore the rf voltage applied at each gap is automatically proportional to the cell length, which is advantageous for APF design and practice. It was shown [83] that alternating both the field phase and amplitude allows small transverse emittance growth by aligning the sequence more along a line of constant phase advance on the stability chart.

It is noted in [83] and also in [77] that there should be an offset of 5°-10° in the initial average value of the synchronous phase, and [83] also notes that it is helpful to also have a tilt, so that the average synchronous phase tends from ~+5° to ~-5°. Such adjustment will increase the acceptance in one plane, while decreasing it in the other. Okamoto [84] adds valuable insight into the value of a tilt by pointing out that it helps avoid the danger of emittance equipartitioning [85] via a synchrobetatron coupling resonance.

As in conventional linacs, as acceleration occurs and the velocity beta increases, the focusing strength can be maintained by lengthening the focusing period.

---

80 "APF or KONUS Drift tube Structures for Medical Synchrotron Injectors – A comparison", S.Minaev, 1999 PAC, p. 3555-3557.

81 "Ion Beam Acceleration and Focusing in the Polyharmonical Drift Tube Systems", S. A. Minaev, EPAC 1990, p. 1744-1746.

82 "Spatially Periodic RFQ Accelerating Structure Based on e010 Mode Cavity", "Spatially Periodic RFQ Accelerating Structure Based on e010 Mode Cavity", S. Minaev, D. Kashinskiy, V.Pershin, B.Sharkov, T.Tretiakova, R.Vengrov, S.Yaramyshev, LINAC 2002, pp. 222-224., LINAC 2002, pp. 222-224.

83 Alternating Phase Focusing (APF) Linacs Developments and their Possible Applications", V.V. Kushin and S.V. Plotnikov, EPAC 1994, pp2661-2663.

84 W.H. Cheng, R.L. Gluckstern, and Hiromi Okamoto, "Synchrobetatron-coupling effects in alternating-phase-focusing linacs", Phys. Rev. E, Vol. 48, No. 6, December 1993, p. 4689-4698.

85 R. A. Jameson, "Beam-Intensity Limitations in Linear Accelerators," (Invited), Proc. 1981 Particle Accelerator Conf., Washington, DC, March 11-13, 1981, IEEE Trans. Nucl. Sci. 28, p. 2408, June 1981; Los Alamos National Laboratory Report LA-UR-81-765, 9 March 1981; R. A. Jameson, "Equipartitioning in Linear Accelerators", Proc. of the 1981 Linear Accelerator Conf., Santa Fe, NM, October l9-23, 1981, Los Alamos National Laboratory Report LA-9234-C, p. 125, February 1982; Los Alamos National Laboratory Report LA-UR-81-3073, 19 October 1981; and subsequent. Coupling resonances are the mechanism for emittance transfer, with or without significant space-charge.

As seen in Figs. 7.1&2, the full range of longitudinal focusing is being exploited by the Garaschenko and NIRS sequences. An APF linac employing the range -90° < phis < +90° accelerates efficiently and is similar to a conventional drift-tube-linac (DTL) in length.

APF linacs in operation so far are apparently for small beam currents, for which there are many applications. It is of interest to explore APF capabilities for higher beam power as well, because of the potential for large cost savings. Acceptances with space charge will be within the bounds of good small-(zero)-current transverse and longitudinal acceptances, so design at zero current should be done first, and will be faster to execute. This paper concentrates on zero-current design; other effects, such as space charge or combination with magnetic focusing, are straight-forward to add later.

The smooth approximation theory has been extended to include acceleration and amplitude modulation [86, 87], both again using a sequence of sinusoidal form. The first uses six parameters to describe the sequence: equilibrium synchronous phase, phase modulation amplitude, length of APF period, incremental energy gain, plus two additional parameters to include simultaneous modulation of the accelerating field amplitude - the relative phase between the amplitude and phase modulation and magnitude of the amplitude modulation. Stability boundaries are shown, but practical examples are not explored. The latter gives an example for an a superconducting accelerator structure period of one solenoid and two spoke-resonator cavities, one focusing and the other defocusing, demonstrating good transmission of up to 50 mA over an energy gain of ~x4.

## 7.4. Sequence optimization

However, it is apparent in Figs. 2 & 3 that the optimized sequences are only quasi-sinusoidal, and therefore a method for determining a preliminary sequence is needed, to be followed a procedure for subsequent optimization by a typical nonlinear constrained optimization program for required characteristics like maximum transmission, smallest energy spread, minimum emittance growth, etc. The significant deviations from the formulas in the Garaschenko and NIRS sequences are the result of careful final optimizations, using a nonlinear optimization technique by Garaschenko, and a Monte Carlo method by Tsutsui at NIRS.

A similar method is outlined in [88] for design and optimization of a short 26-cell sequence, which resembles Figs. 2 & 3, to accelerate $_{238}U^{34+}$ (qom=1/7) from 0.140 MeV/u to 0.542 MeV/u (factor~4) in 2m (~2.6MV/m) at 52 MHz. The optimization minimizes a desired objective function by adjusting the 26 gap phases directly.

Using the gap phases as the optimization parameters directly can work for short sequences, but even then may have difficulty converging. If the gap amplitudes are to be optimized as well, the number of variables doubles, and for long sequences, the difficulty compounds further. Therefore another strong motivation for this paper is to provide more information to the optimization procedure for quicker and more accurate convergence.

[eltoc]

---

86 L. Sagalovsku, J.R. Delayen, "Alternating-Phase Focusing With Amplitude Modulation", 1993 Particle Accelerator Conference.

87 J. Qiang, R.W. Garnett, "Smooth Approximation with acceleration in an alternating phase focusing superconducting linac", NIM Phys. Res. A 496 (2003) 33-43

88 Z-J Wang, Y. He, W. Wu, Y-Q Yang, C. Xiao, "Design optimization of the APF DTL in the SSC-linac", 2010 Chinese Phys. C 34 1639 (http://iopscience.iop.org/1674-1137/34/10/017)

# Chapter 8 – Synthesis of a practical APF design, simulation, and optimization method

The present situation is that practical methods have seemed cumbersome and limited to short sequences. As there seems to be no inherent limitation to the beam energies for which APF can be applied, it is of interest to synthesize a general method to explore long sequences with large energy gain, and efficient optimization. If reasonable zero-current transverse and longitudinal acceptances can be found, then APF-focused beams with space-charge up to some limit will also be possible.

By considering the general principles underlying the details of a large amount of previous work as outlined above, a generally practical and straight-forward method is synthesized for generating and optimizing APF linacs, as follows:

1. An optimal sequence can be based on a general sinusoidal form for modulation of the gap synchronous phase *phis*, with a 7-parameter function of the cell number *ncell* (*radian (=π/180°) indicates that the corresponding parameters are in degrees):

```
phis = phioffset*radian - phitilt*radian*ncell + phiampl*radian*Exp[-phiatten*ncell]*Sin[2.*Pi*ncell/
(phiperiod*Exp[peratten*ncell]) + phistart*radian]          (1)
```

As indicated above, it may be desired for further reduction of transverse emittance growth to apply sequences to both the gap phase and amplitude. The method can be extended for this. The form of the attenuations can be changed. The APF linac is then designed by simulation.

2. A 7-dimensional grid search over the seven parameters can be performed quickly for zero beam current with a fast multi-particle simulation code, with finer and finer grids.

3. When parameters are found which give an initial adequate transmission, the sequence is optimized, using a new strategy, for the desired beam properties using nonlinear optimization techniques.

4. The modeling can then be refined, with more accurate modeling, with space-charge, etc., and the process repeated.

## 8.1. *LINACSapf* dynamics method

A new simulation code, *LINACSapf*, was written for APF linac design and simulation.

A simple chain matrix representing each cell as a drift and an rf gap can be used. The linac is designed sequentially cell by cell, and thus acceleration is taken into account in fitting the phase sequence. The cell lengths are irregular, as determined by the local phase difference across the cell from the APF phase sequence, and the local velocity β. The Trace3D code rf gap transformation is used at the end of the cell.

The method is also well-suited for simulation of sequences with more accurate gap models (e.g. also including underlying magnetic focusing), and for beams with space-charge.

It is useful for preliminary design work not to insist on completely realistic conditions. The gap voltage should be realistic, as determined from the Kilpatrick Criterion and structure peak field characteristics. However, at first the aperture should not be a restriction. Zero beam current and small input beam emittances are useful when searching for workable sequences. In this paper, the beams are input with upright ellipses (α's = 0). The program can easily compute the acceptances, and this matching information can be used at any time to correct the input distribution.

The linac generation and dynamics are embedded in a driver, and execution is very fast, enabling parameter scans with thousands of runs to be performed quickly.

The beam performance is sensitive to the APF sequence parameters, and this made the initial programming for setting up various traps and debugging difficult *.
[eltoc]

# Chapter 9 – Initial design of a 0.34-20 MeV muon APF linac

The KEK Muon Project [89] requires acceleration of a small number of very cold muons (qom = 1/ 0.11272122345 = 8.87) over the range of ~0.34-200 MeV, with small transverse emittance growth and minimal output energy spread. The proposed muon beam current is very small and space charge can be neglected. The APF linac is an attractive candidate, because significantly lower cost is realized when magnetic focusing components are not required.

RFQ designs to bring the muon beam to 0.340 MeV have been developed [90]. Case 1.3 of that study has an output beam with transverse normalized rms emittance = 0.32 mm.mrad, a phase spread of ±22°, and an energy spread of ±2.1%. This beam is input to a 324 MHz APF linac, for which 1.8*(Kilpatrick limit)=0.32 MV/cm. A flat field and a peak-to-gap field ratio of ~6 are used to determine the working Eo ~ 0.055 MV/cm. The initial bore radius is 0.5 cm, with linear expansion to ~1.5 cm after the ~23 cells (~6.5 m) required to reach a final energy of ~22 MeV, where β=~0.56. Particles are considered "accelerated" if their final energy is within 3% of the final synchronous energy.

For the first stage of the study, a design for 0.34-20 MeV is sought. Only ~23 cells are required – a short sequence, for which the simulation is extremely fast.

## 9.1. Initial parameter search

Fig. 9.1 shows three successive searches over the six parameters of Eqn.(1) with *phiampl* = 90°.. The first search showed a tendency toward an optimum transmission, which was then found in the second search with narrower parameter ranges.

The case for best transmission showed transmission of 99.75%, but accelerated fraction of only 67.3%. The parameters are shown in Table 1.

---

* For example, the Trace3D gap transformation starts with the standard formula for the new energy after the gap as the previous cell's synchronous energy plus the energy gain across the gap, W(n+1) = Ws(n) + qEoTLcos(phi), where Ws(n) is the present cell's synchronous energy, q is charge, Eo is electric field, T is transit-time-factor, L is cell length, phi is the rf phase at the gap center. The gap transformation then finds the new energy again as the new cell's synchronous energy Ws(n+1)plus an energy offset Δw. The two new energies are usually very close, but for APF sequences with bad performance, there could be significant difference, which contributed to initial confusion. With good APF sequences, the difference is very small – the Trace3D transformation is correct.

89 Conceptual Design Report for The Measurement of the Muon Anomalous Magnetic Moment g – 2 and Electric Dipole Moment at J-PARC, December 13, 2011.

90 "Acceleration of Cold Muons", R.A. Jameson, KEK Report 2013-2, April 2013. (Includes method for practical design of alternating-phase-focused linacs).

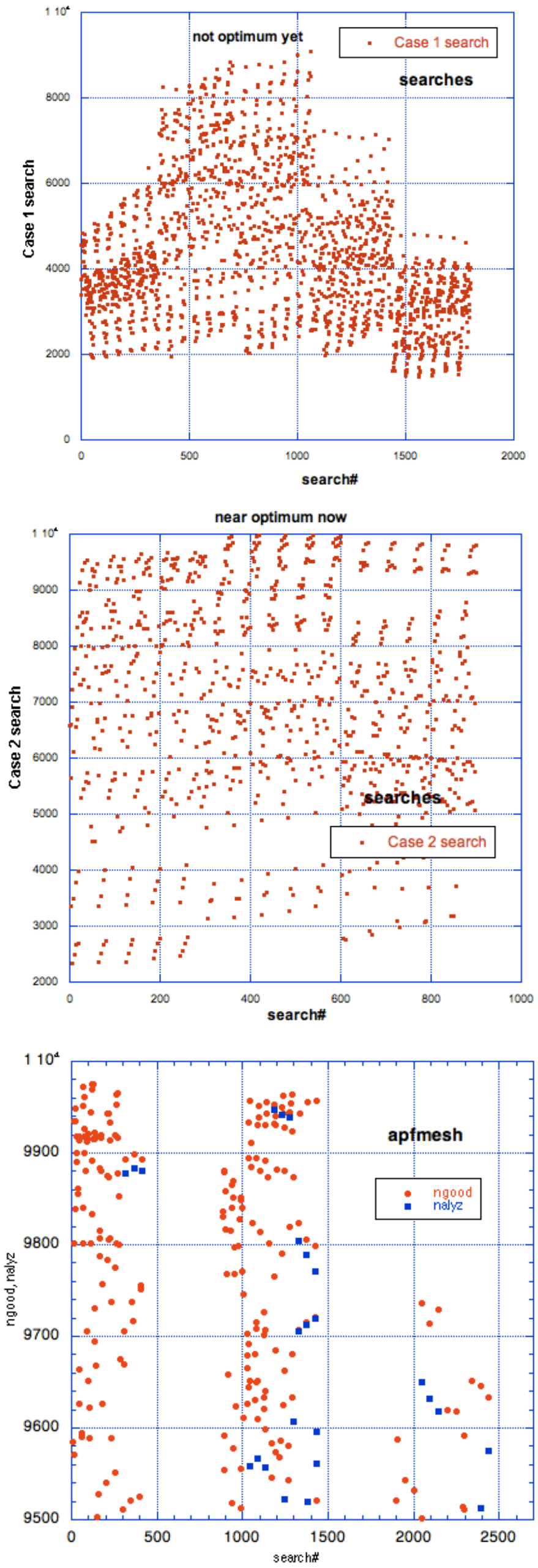

Fig. 9.1 Results of three successive searches over the six parameters of Eqn.(1). The vertical axis is the number of particles transmitted (ngood – red) and the number of particles accelerated to within 3% of the final synchronous energy (nalyz – blue).

| | |
|---|---|
| phioffset | 20.0 |
| phitilt | 0.00 |
| phiampl | 90.0 |
| phiatten | 0.01 |
| phiperiod | 2.3 |
| peratten | 0.008 |
| phistart | 50.0 |

Table 1. Initial parameter search result for best transmission.

The third search looked for best accelerated fraction; the third graph in Fig. 9.1 shows that this is reached at different parameters. The transmission is 99.53%, with accelerated fraction 99.47%, and the parameters are shown in Table 2:

| | |
|---|---|
| phioffset | 25.0 |
| phitilt | 1.00 |
| phiampl | 90.0 |
| phiatten | 0.008 |
| phiperiod | 2.3 |
| peratten | 0.009 |
| phistart | 55.0 |

Table 2. Initial parameter search result for best accelerated fraction.

Fig. 9.2 shows the two phase sequences, and it is seen that they are quite different. It is interesting to note that the tilt parameter is invoked for best accelerated fraction.

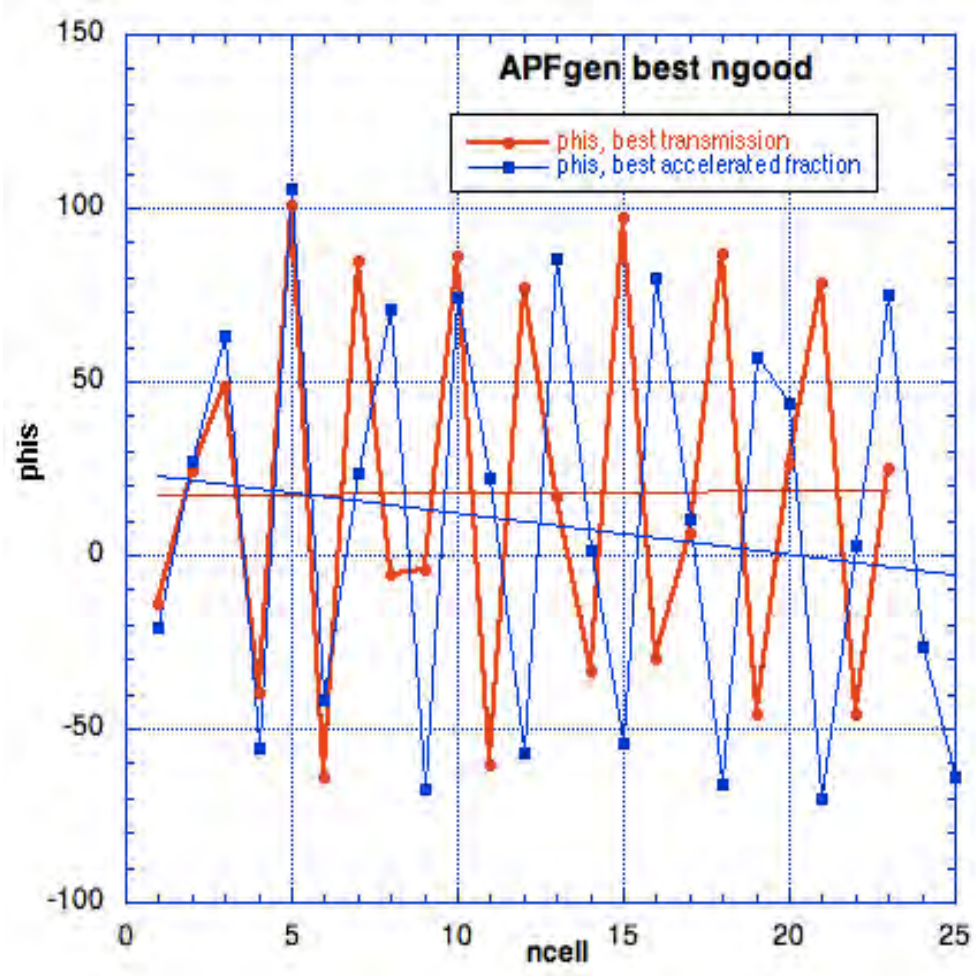

Fig. 9.2 The APF gap phase sequences for the best transmission (red) and best accelerated fraction (blue) cases. The linear fits show the effect of the tilt parameter.

The output x-xp and dphi-dw phase spaces for the two cases are shown in Fig. 9.3. As there is zero current, the beam remains a filament, as evidenced in the longitudinal phase space. Although almost all of the input muon beam has been captured and accelerated, smaller final transverse emittance and energy spread are to be sought.

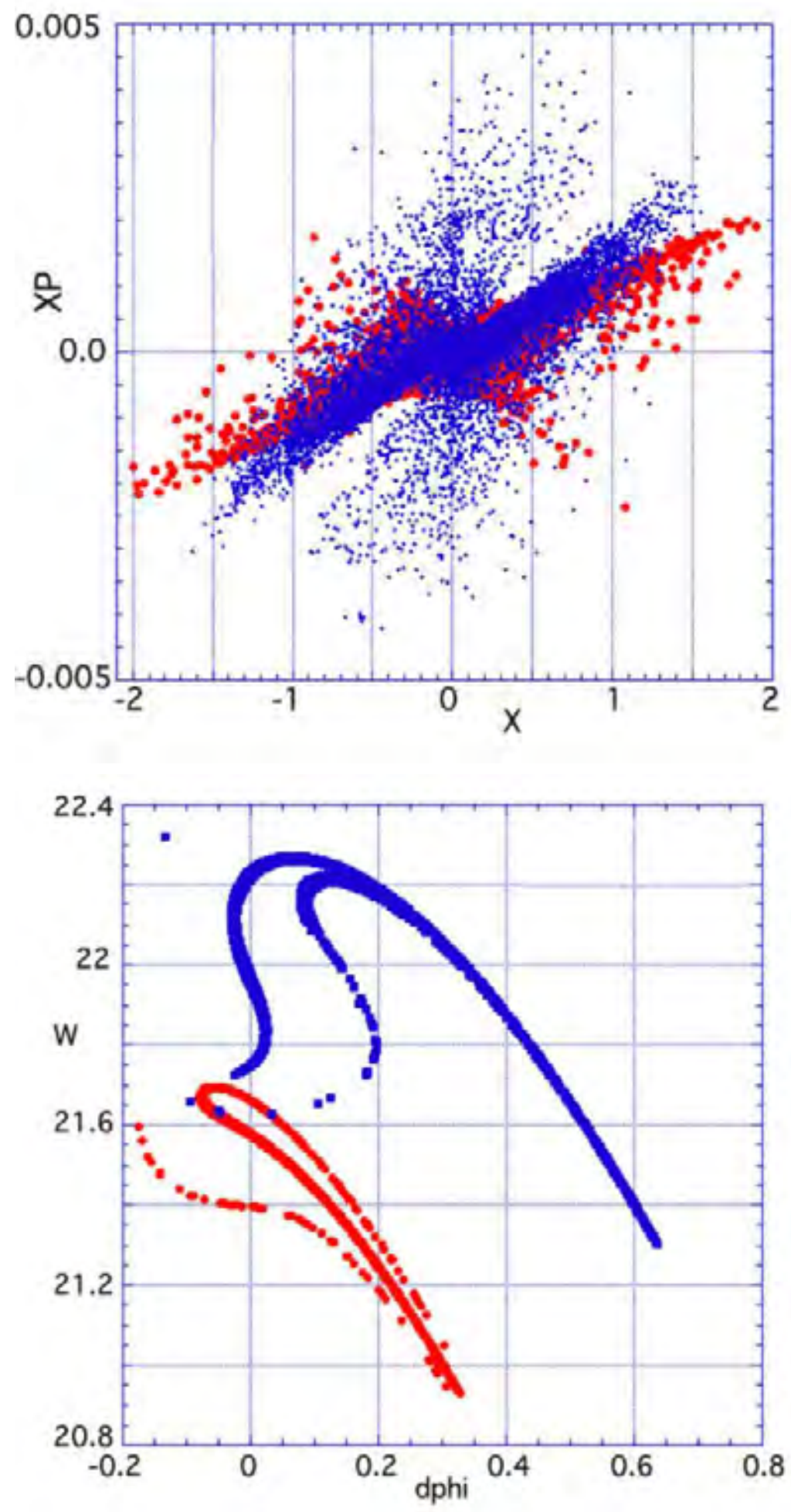

Fig. 9.3 Above: x-xp (red), y-yp blue)). Below: dphi-W phases spaces for the best transmission case (red) and the best accelerated fraction case (blue).

## 9.2. Find minimum output transverse emittance and energy spread

The output transverse emittance and energy spread are computed and the smallest emittance growth and smallest energy spread (%) are sought.

Fig. 9.4 shows search results on finer meshes. 10000 particles were computed, and the run kept only if the number of accelerated particles was ≥ 9900 (99%). For each of the six varied parameters, the number of transmitted particles (ngood), the number of particles accelerated to within 97% of the final synchronous energy (nalyz), the transverse normalized rms emittance growth (etngr), and the energy spread (dw,%) are shown, giving information on the optimum and its sensitivity.

The last graph shows a correlation between the transverse rms emittance growth and the output energy spread. Lower energy spread is obtained at the expense of more transverse emittance growth.

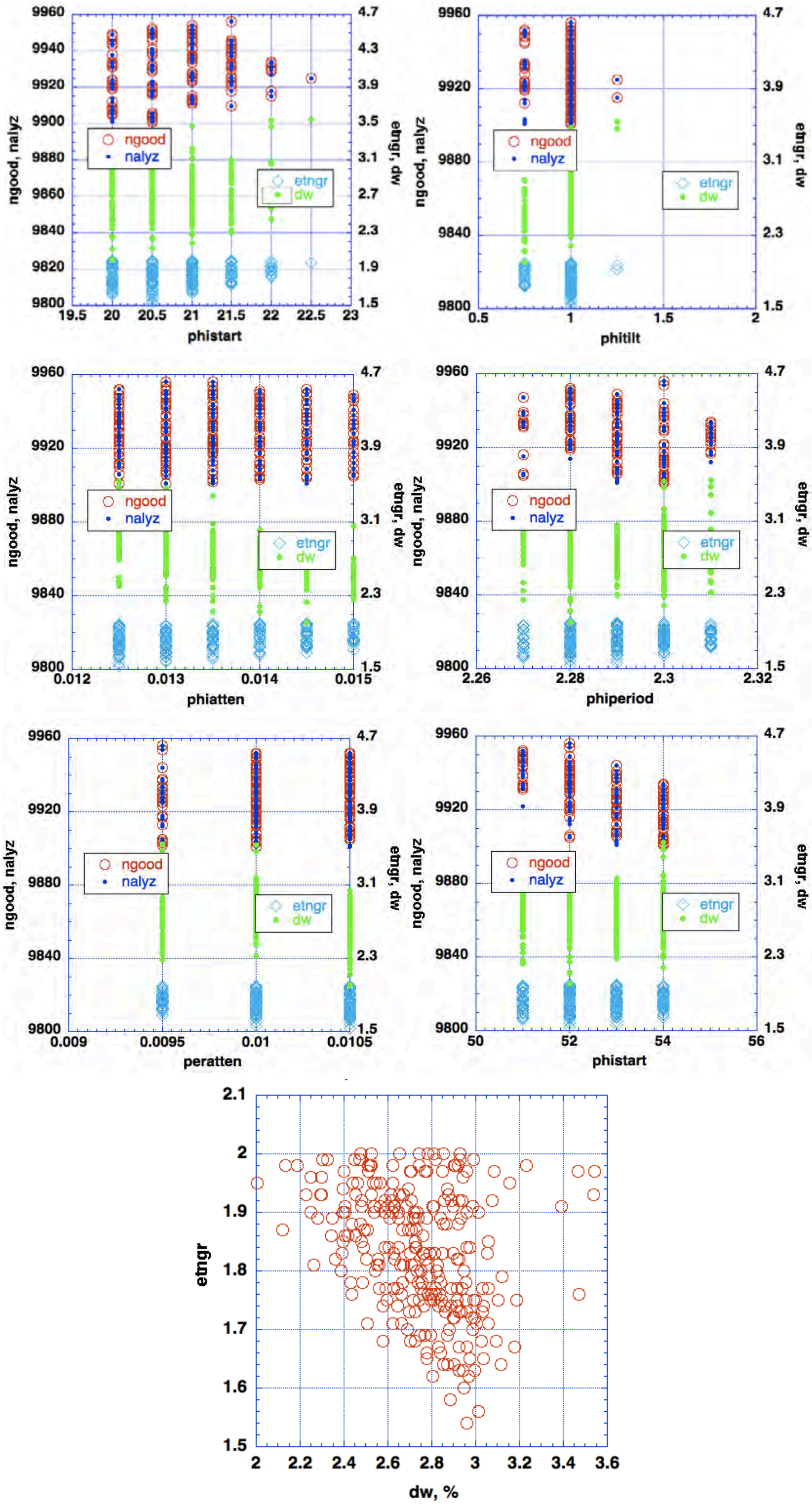

Fig. 9.4 Results of parameter search on finer grids.

The parameters and output characteristics which gave minimum energy spread (Case 430), are shown in Table 3. and in Fig. 9.5:

| | | | |
|---|---|---|---|
| phisoffset | 10.0 | #cells | 23 |
| phitilt | 0.75 | length,m | 7.14 |
| phiampl | 90.0 | Xmsn,% | 99.24 |
| phiatten | 0.0145 | Accel,% | 99.24 |
| phiperiod | 2.28 | etn growth | 1.95 |
| peratten | 0.0105 | output energy spread,% | 2.0 |
| phistart | 52.0 | | |

Table 3. Parameters and output characteristics for minimum energy spread.

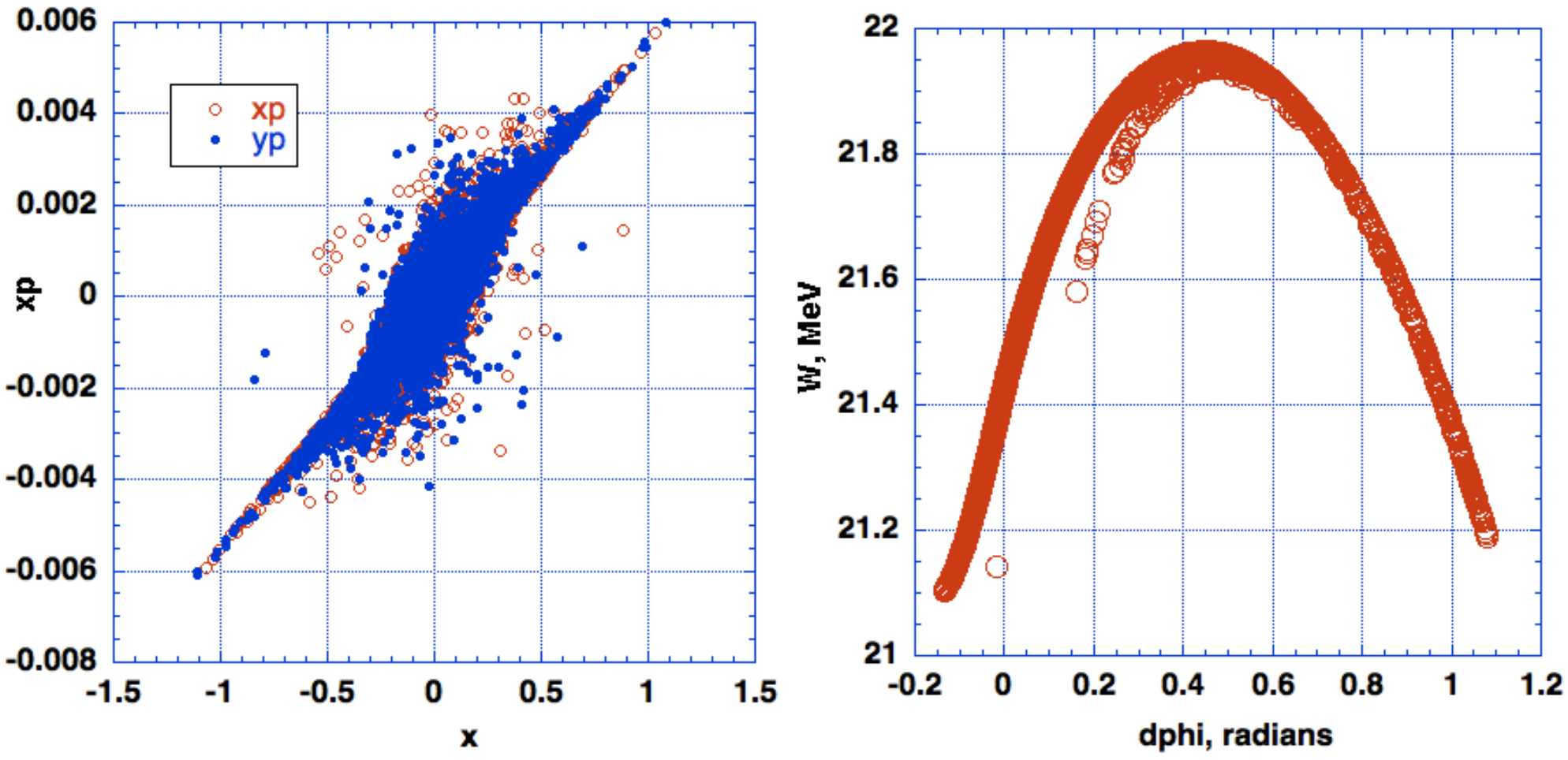

Fig. 9.5. Output x-xp, y-yp and dphi-W phase spaces for minimum energy spread case.

The parameters and output characteristics which gave minimum transverse emittance growth (Case 2926) are shown in Table 4. and in Fig. 9.6:

| | | | |
|---|---|---|---|
| phisoffset | 20.5 | #cells | 23 |
| phitilt | 1.0 | length,m | 7.05 |
| phiampl | 90.0 | Xmsn,% | 99.29 |
| phiatten | 0.0125 | Accel,% | 99.29 |
| phiperiod | 2.28 | etn growth | 1.54 |
| peratten | 0.0105 | output energy spread,% | 2.96 |
| phistart | 52.0 | | |

Table 4. Parameters and output characteristics for minimum transverse rms emittance growth.

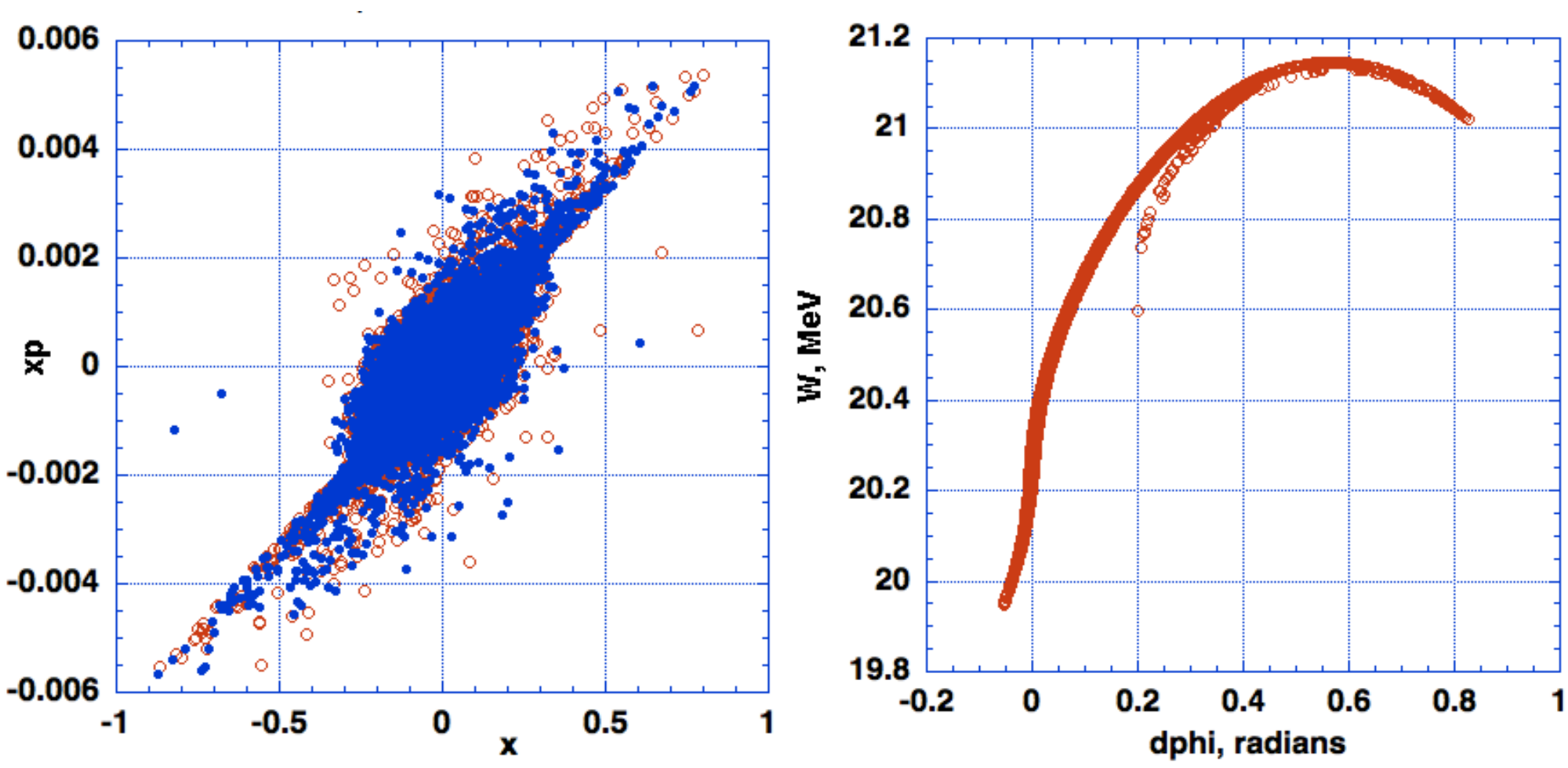

Fig. 9.6. Output x-xp, y-yp and dphi-W phase spaces for minimum transverse rms emittance growth case.

The difference in energy spread is a bucket rotation effect.

Fig. 9.7 shows the differences between the phis sequences for these two cases.

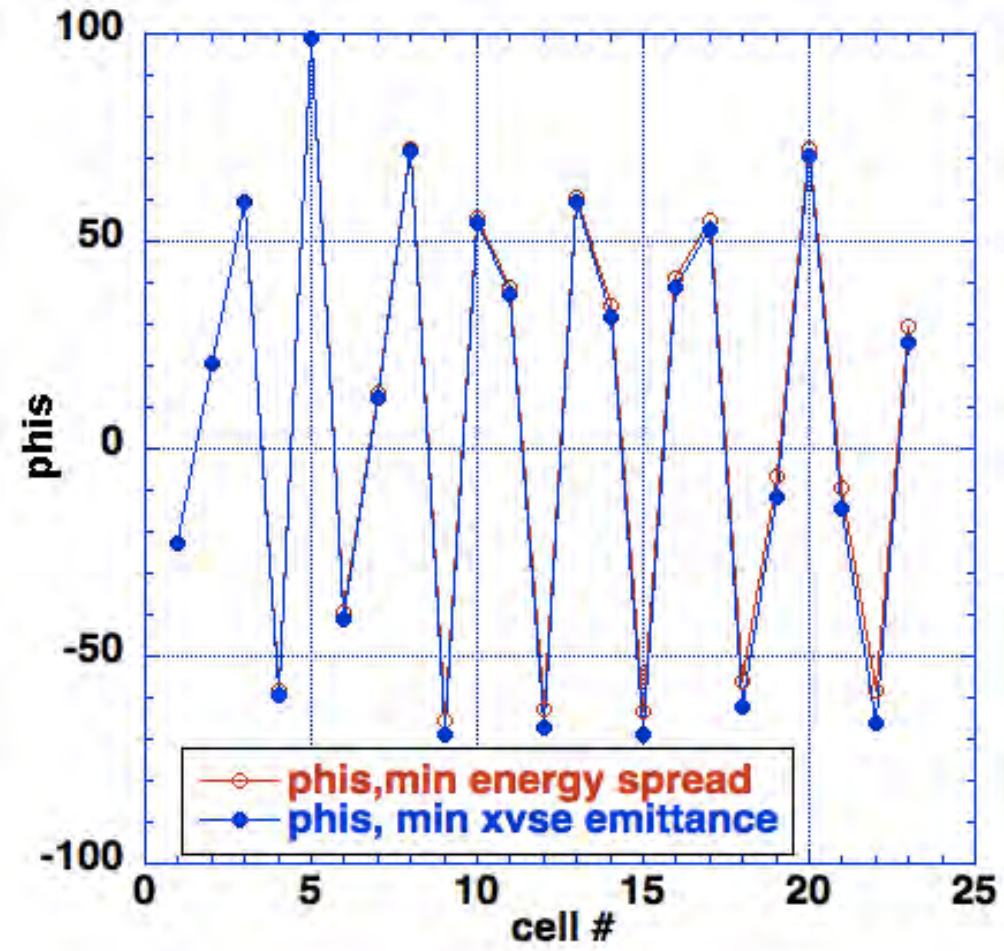

Fig. 9.7. The phis sequences for the minimum energy spread and the minimum transverse normalized rms emittance cases.

## 9.3. Search using a constrained nonlinear optimization program.

### 9.3.1. Optimization on the phis's of the sequence

The individual phis values at the 23 cells (23 variables) are optimized by the constrained nonlinear optimization solver NPSOL [91], for various objective functions:

1. Best transmission
2. Best accelerated beam fraction
3. Minimum transverse normalized emittance (etn) growth
4. Minimum output energy spread (dw), %
5. Minimize the sum of etn growth and dw/2. dw is divided by 2 for good weighting against the etn growth.

91 User's Guide for NPSOL 5.0: A FORTRAN Package for Nonlinear Programming, P E. Gill, W. Murray, M. A. Saunders, M. H. Wright, Technical Report SOL 86-6_, Revised June 4, 2001

6. Minimize the sum of etn growth, dw/2, and (npoints - accelerated points)/20. npoints = 10000, and the divide by 20 is for proper weighting.

As is generally the case with nonlinear optimizers, settings for the constrained range on each of the working variables, and on various features of the optimizer such as the way it searches for and handles derivatives, must be explored. Also the starting solution must be close enough to the final solution that the optimizer does not converge to a false local minimum. In this case, bounds of ±5° to ±10° on the phases were used, along with appropriate settings for the difference intervals.

A starting sequence was selected midway along the lower correlation boundary of the last graph in Fig. 9.4. As would be expected, the initial fine mesh search gives a good starting sequence, and not much improvement is found by the optimization, but it does succeed in finding the desired improvements (Table 5).

| *Objective function* | *Xmsn, Accel, %* | *etn growth* | *energy spread* |
|---|---|---|---|
| Starting sequence | 99.06 | 1.683 | 2.681 |
| 1. max xmsn | 99.27 | 3.534 | 2.773 |
| 2. max accelerated | 99.40 | 2.187 | 2.916 |
| 3. min etn growth | 95.05 | 1.674 | 2.679 |
| 4. min energy spread dW, % | 95.39 | 2.137 | 1.566 |
| 5. min (etngrowth, dW/2) | 99.18 | 1.230 | 1.714 |
| 6. min (etngrowth, dW/2,(npoints-naccel)/20) | 99.28 | 1.401 | 1.988 |

Table 5. Results of optimization of the 23 phis points directly.

**9.3.2. Comparison of starting sequence and optimized sequence**

The starting sequence was generated by the general 7-parameter sequence formula (1). The parameters of the initial (starting) sequence) were:

```
phisapf /. {phioffset->20.5,
phitilt->1.00, phiampl->90.0,
phiatten->0.0145, phiperiod->2.28, peratten->.0105, phistart->52.}
```

Mathematica FindFit can fit the parameters to the optimized sequence. The unconstrained natural form for Case 6 in Table 1 gives:

```
Gfit = FindFit[sequence,
 phisapf, {{phioffset,20.5},
          {phitilt,1.00},      {phiatten,0.0145},
```

```
{phiperiod,2.28},{peratten,.0105},
{phistart,52.}},ncell]
```

where the starting values are those of the parameters of the starting sequence. The result is:

```
{phioffset->17.0621, phitilt-> 0.685082, phiatten->0.0161262, phiperiod->2.28007,
peratten -> 0.0105318, phistart->52.2236}
```

The differences from the initial sequence are shown in Fig. 9.8:

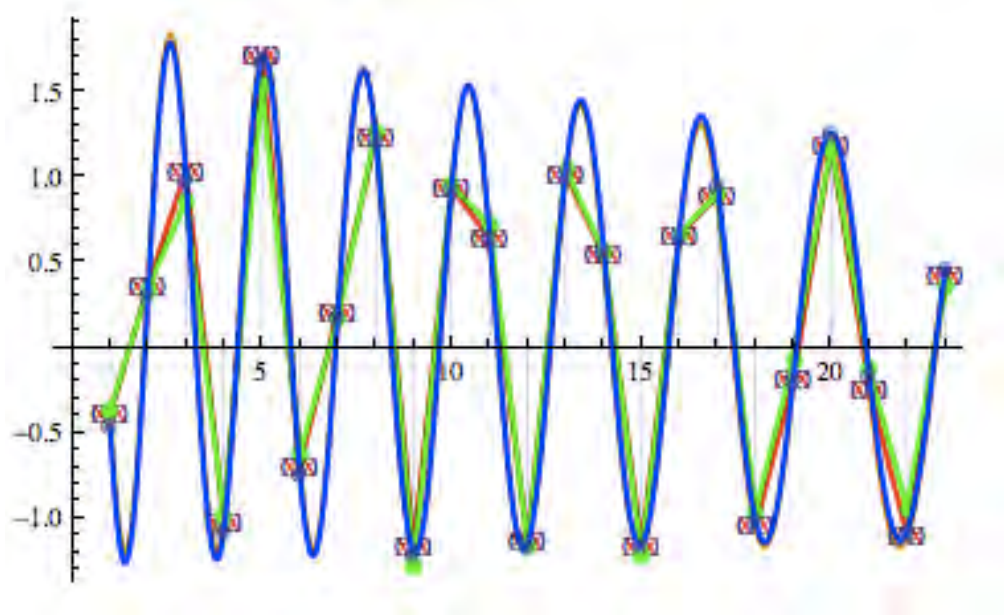

Fig. 9.8. The starting sequence curve from the general 7-parameter sequence formula (orange), and the synchronous phases at the 23 cell gaps (red). The ordinate is radians. Plotted over these are the optimized sequence curve from FindFit to the general 7-parameter sequence formula (blue), and the synchronous phases at the 23 cell gaps (green).

## 9.3.3. Optimization on the seven parameters of the sequence

Optimization on the seven parameters of the sequence was rather sensitive to the bounds and the optimizer settings, probably because there is less information to work with. Optimization on the gaps is recommended.

## 9.3.4. Optimization on the phis's and the (gaplength/bl)'s "gapobl" of the sequence

It was noted above that a tip in the references indicated that less transverse rms emittance growth might be obtained by varying both the synchronous phase and the accelerating field amplitude at the sequence gaps.

The energy gain per cell = VTcos(phis), where the transit-time-factor T = sin(gap length/βλ)/ (gap length/βλ) = sin(gapobl)/gapobl. The Trace3D convention is that "gap voltage" = V = E0*cell length because E0 is integrated over the cell length. The cell lengths are already determined from the phis sequence.

As no information exists on the possible or optimum form of a gapobl sequence, whether in conjunction with a phis sequence or not, the constrained nonlinear optimization procedure is very useful.

Optimization was performed on the (gap length/ βλ) and phis at each cell, with ±5° bounds on phis and 0.13 – 0.25 on gapobl. The starting sequence is the result of Table 5 Case 6. The objective function was set as a weighted combination of the emittance growth, the output energy spread, and desiring a large accelerated fraction:

```
objfunc=2.d0*etngrowth+energysprd+
dble(npoint-nalyz)/100.d0
if(objfunc.lt.0.d0)objfunc=1000.d0
```

The results: 97.31% accelerated, etn growth = 1.314, output energy spread 1.016%.

In comparison with Table 5 Case 6, there is less accelerated fraction, but adjusting also the gap lengths appears to indeed produce somewhat less transverse emittance growth and also less output energy spread.

The sequences are shown in Fig. 9.9:

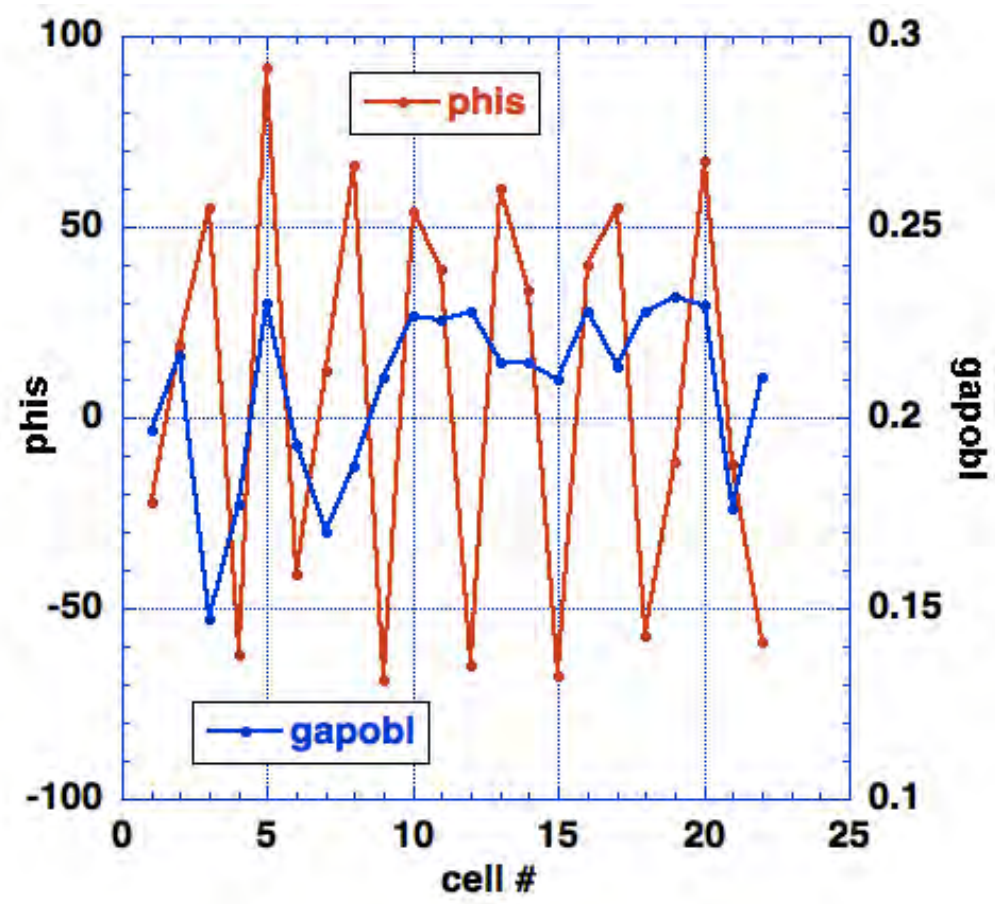

Fig. 9.9. Sequences resulting from optimization on the phis's and the gap lengths.

The output phase-spaces are shown in Fig. 13.

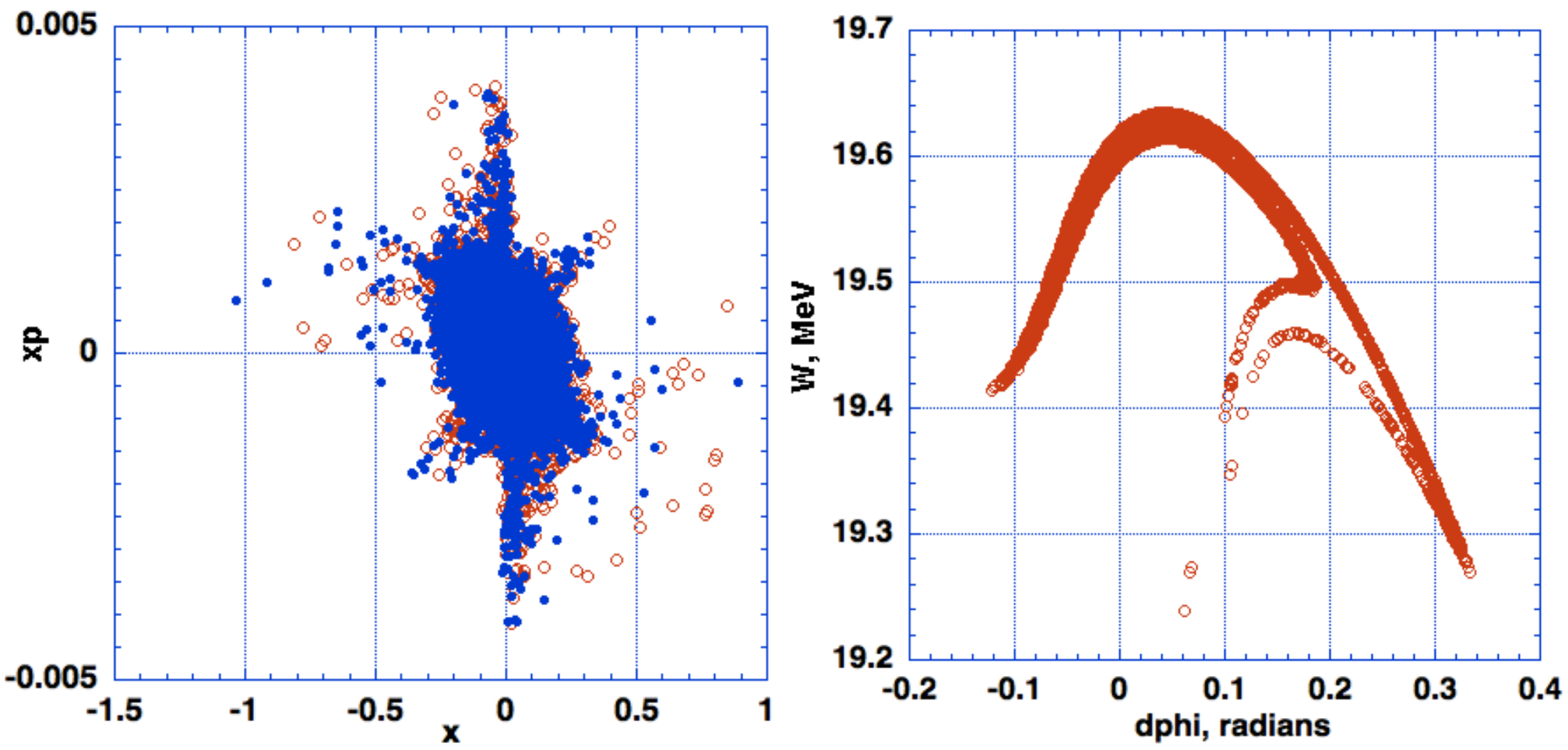

Fig. 9.10. Output x-xp, y-yp, and dphi-W phase space for optimization of both the phis's and the gapobl's.

The other objective functions used in Table 5 did not work well, tending to much reduced accelerated fraction.

### 9.3.5. Reduce the aperture expansion, try other apertures

As stated, the APF linac above had initial aperture = 0.5 cm radius, and final aperture = ~1.5cm radius.

At different aperture expansions, the beam loss changes (Table 6):

| Final Aperture | Accel fraction, % | etn growth | dw |
|---|---|---|---|
| 2 | 99.22 | 1.344 | 1.016 |
| 1.5 | 97.31 | 1.314 | 1.016 |
| 1 | 89.77 | 1.164 | 1.016 |
| 0.5 | 69.25 | 0.856 | 1.016 |

Table 6. Case optimized on both phis's and gapobl's. Initial aperture is 0.5 cm. Effect of linear aperture expansion.

Reoptimizing with constant aperture = 0.5 gave 76.66% accelerated, etn growth = 1.381, dw = 0.851.

The rms transverse emittance rms ellipse (betax+betay)/200 was added to the optimization function, yielding very small improvement in the accelerating fraction.

The beam loss occurs early with 0.5 cm initial aperture and 1.5 cm final aperture. Opening the initial aperture and keeping it constant gives good accelerated fraction (Table 7). The table shows improvement of accelerated beam fraction with larger, constant apertures.

| Aper, cm | Aper coeff | Accel frac, % | etn growth | dw |
|---|---|---|---|---|
| 0.75 | 0.0 | 92.77 | 1.260 | 1.074 |
| 1.0 | 0.0 | 99.38 | 1.436 | 1.074 |
| 1.5 | 0.0 | 100.00 | 1.490 | 1.074 |

Table 7. Beam loss vs. cell number for 0.5 cm initial aperture and 1.5 cm final aperture.

[eltoc]

# Chapter 10 – Initial result for 2-200MeV H+ APF linac

The research then proceeded with protons, just to illustrate the same procedure with a different particle (the problem scales with the charge-to-mass ratio). A good design for H+ at 324 MHz from 2-20MeV similar to the above muon design was quickly found, so a longer sequence for 2-200 MeV (factor 100) was explored next.

It was learned that the mesh search over the sequence parameters must be done with both fine enough resolution and with wide enough range, because there can be several areas of parameter space where a significant percentage of accelerated particles is obtained, but one area will be best. It is useful to plot the formula. The phis sequence swing should be kept large over the desired energy range to maximize the acceleration rate. The period attenuation is sensitive and if too large can overwhelm the sinusoid.

For easier control of the attenuation, the sequence formula was changed to linear attenuation of the period:

```
phis =  phioffset*radian - phitilt*radian*ncell +
phiampl*radian*Exp(phiatten*ncell)*Sin[2.*Pi*ncell/
(phiperiod*(1.d0-peratten*ncell)) + phistart*radian]        (2)
```

~164 cells are required, with total length ~45 m. The best mesh search result was with parameters:

```
{phioffset->22.°, phitilt->0.2,
phiampl->90°, phiatten->0.006,
phiperiod->4.4, peratten->0.005,
phistart->56°},
```

with transmission of only 39.23% and accelerated fraction of 36.47%.

Now optimization to raise the accelerated fraction became difficult.

Several methods were ineffective and difficult to make converge:
- optimization on all sequence points directly.
- optimize successively each sequence point individually.
- optimize from each cell to the end successively.

In each case, the objective function is computed over the whole linac.

It was necessary to find a better optimization strategy.

# Chapter 11 – Intelligence for a new optimization procedure from APF physics and modern control theory

The optimization strategy from a given local point to the end successively is based on the principle from modern control theory (and for any journey) that it is most effective to reoptimize from each present state to the end. Even so, enough information must be available on sensible paths for convergence; otherwise a solution may not be found, or found only with a large amount of unnecessary work. As indicated above, adjusting the whole matrix of remaining cell phases and gap lengths may eventually converge with an extreme amount of computation, or may not[92].

Applying this principle, new information is synthesized from the detailed APF development outlined above.

Recall that the Garaschenko [77] and Minaev [80] sequences use adjoining sequences of varying period length (N); Garaschenko uses six sequences with N = {6,6,8,8,10,13}. The NIRS or Eqn. (1 or 2) sequences use a sinusoid with period `(phiperiod*(period attenuation function))`, where the attenuation function is of some form like `Exp[`*`peratten`*`*ncell]` or `(1-`*`peratten`*`*ncell)`.

**The local N at each cell is obtained by applying the sequence formula (1) or (2) at that cell, and computing the ncell ahead for which the period accumulates by 1. (phase advance of 2*π)** [93]**.**

The new optimization strategy is then, at each cell sequentially, to optimize over the local period length N, with the objective function computed for the whole linac.

The transverse and longitudinal phase advances from the smooth approximation for the focusing characteristics of a beamline contain information about both the external fields of the linac or beam line, and also about the beam itself, in terms of the space-charge forces or other effects. When the parameters of the linac or beam line change slowly with respect to the betatron (transverse) or synchrotron (longitudinal) oscillation wavelengths of the phase advances, the smooth approximation

---

[92] This is directly analogous to the difficulties existing in correcting beam orbits, where the “experts” gathered data from all instrumentation (such as beam position monitors) at once and attempted to find correcting information to the beam control devices (such as beam steerers) by inverting the data matrix. This resulted in corrections to all of the control devices. In contrast, in the early-1980’s the subject of beam-based control was emerging, and new tools were contemplated from new fields such as “artificial intelligence (AI)” and “expert systems”. I dispatched S.H. Clearwater from the Los Alamos AT-Division as a Post-Doc to SLAC to work with M.H. Lee, and a full system of programs (COMFORT, ABLE) was developed and implemented on the SLC control system, which could efficiently identify where a beam orbit error, or beam instrumentation error, actually occurred and then made local corrections. (The experienced reader will wonder how it turned out, and will with understanding nod to learn that the “matrix-inversion” school nevertheless won, and the AI work was forgotten…)
“Prototype Development of a Beam Line Expert System”, S.H. Clearwater & M.J. Lee, 1987 PAC, p. 532; “Error-Finding and Error-Correcting Methods for the Startup of the SLC”, M.J. Lee, S.H. Clearwater, S.D. Kleban, L.J. Selig, 1987 PAC, p. 1334; “Modern Approaches to Accelerator Simulation and On-Line Control”, M. Lee, S. Clearwater, V. Paxson, E. Theil, 1987 PAC, p. 611.

**93** Using the concept of phase advance from the rms envelope equations of our standard Framework led to this adaptation for APF.

formulae can be applied locally at each cell, and are in principle sufficient for "beam-based" ("inside-out") design of any linac [94], although not, as seen here, for specifically optimized APF.

In the cited references, the smooth approximation phase advances are derived for a general APF sequence. Instead, in *LINACSapf,* the zero-current transverse and longitudinal phase advances s0t and s0l are computed locally at each cell from the rms envelope equations. (These quantities are often negative, so the absolute value is used to present the result in interpretive form.).

## 11.1. Results for H+ APF linacs, continued

Successively optimizing the gap synchronous phases (phis's) over each N for maximum accelerated fraction, an immediate dramatic improvement resulted, of transmission and accelerated fraction to ~85%, transverse emittance (etn) growth 2.7, output energy spread (dw) ~±1.0%

Further experimentation with the bounds on the phis's and gapobl's, and some iteration, resulted in transmission and accelerated fractions of >94%, etn growth ~2, dw <±1% (aperture 1.5-3.0 cm)

The changes to the H+ 324MHz 2-200MeV original sequence (from the mesh search) by optimization on phis only with ±5° bounds are shown in Fig. 11.1.1.

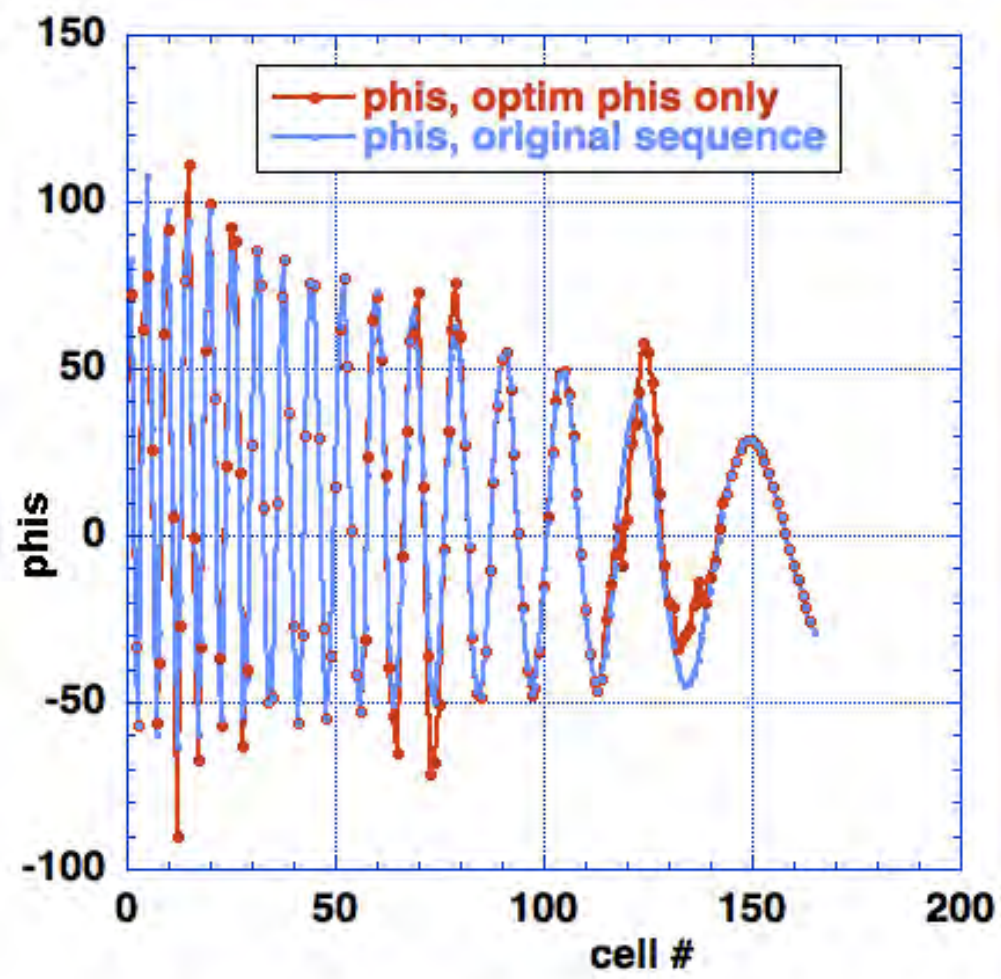

Fig. 11.1.1. Changes to original sequence by optimization on phis only, ±5° bounds.

Fig. 11.1.2 shows the changes to original sequence by optimization on both phis with ±10° bounds and on gapobl with bounds 0.15-0.25.

Random errors in the phis's up to ±2% had only small effect, and even improved the results for some random number sets.

The real, rms transverse acceptance is 2 π.cm.rad (Fig. 10.3.a). The longitudinal phase capture measured at the upper cusp of Fig. 16b is ~±17° with rms value ~8°, and energy capture +0.010 to -0.045 MeV.

---

94 R. A. Jameson, Principal Investigator, et. al., "Scaling and Optimization in High-Intensity Linear Accelerators", LA-UR-07-0875, Los Alamos National Laboratory, 2/8/2007 (introduction of *LINACS* design code), re-publish of LA-CP-91-272, Los Alamos National Laboratory, July 1991; R.A. Jameson, "A Discussion of RFQ Linac Simulation", Los Alamos National Laboratory Report LA-CP-97-54, September 1997. Re-published as LA-UR-07-0876, 2/8/07; "RFQ Designs and Beam-Loss Distributions for IFMIF", R.A. Jameson, Oak Ridge National Laboratory Report ORNL/TM-2007/001, January 2007; "*LINACS* Design and Simulation Framework", R.A. Jameson, KEK/J-PARC Seminar, 6 March 2012.

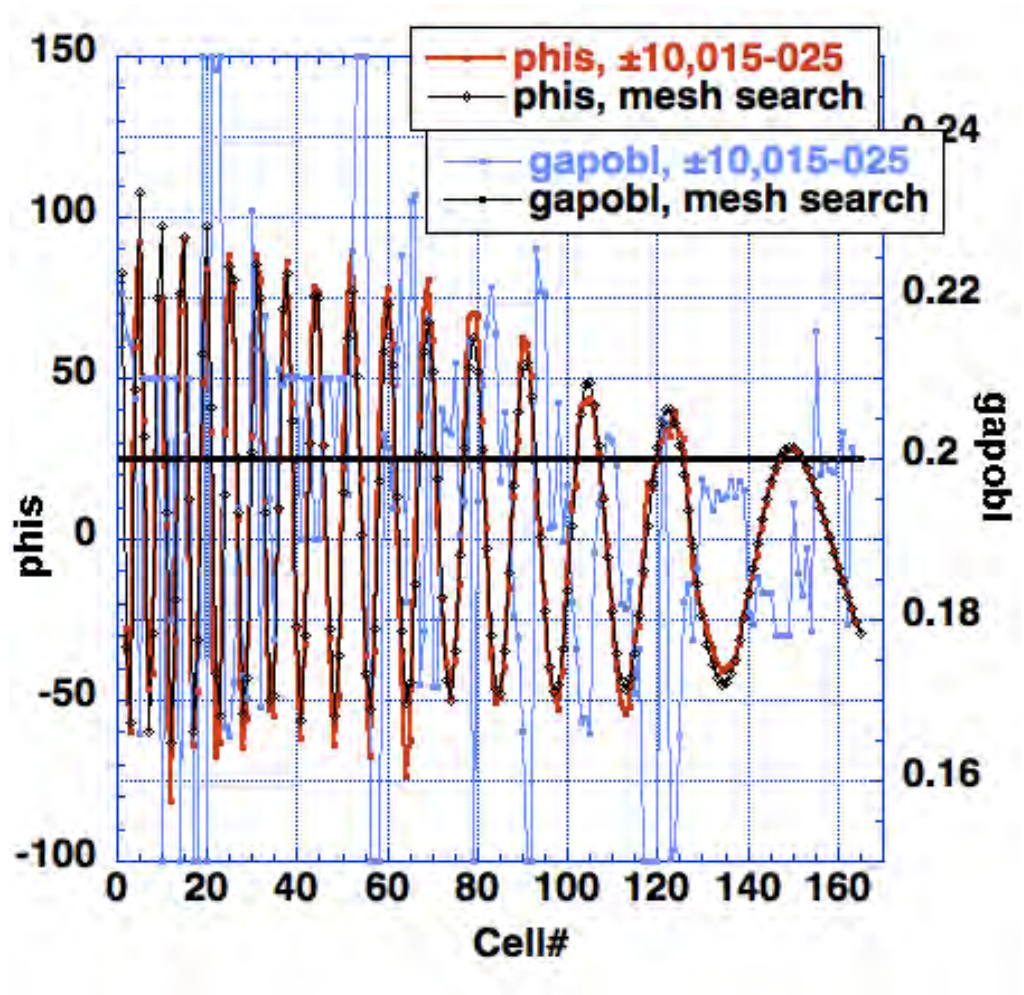

Fig. 11.1.2. Change to original sequence by optimization on both phis with ±10° bounds and on gapobl with bounds 0.15-0.25

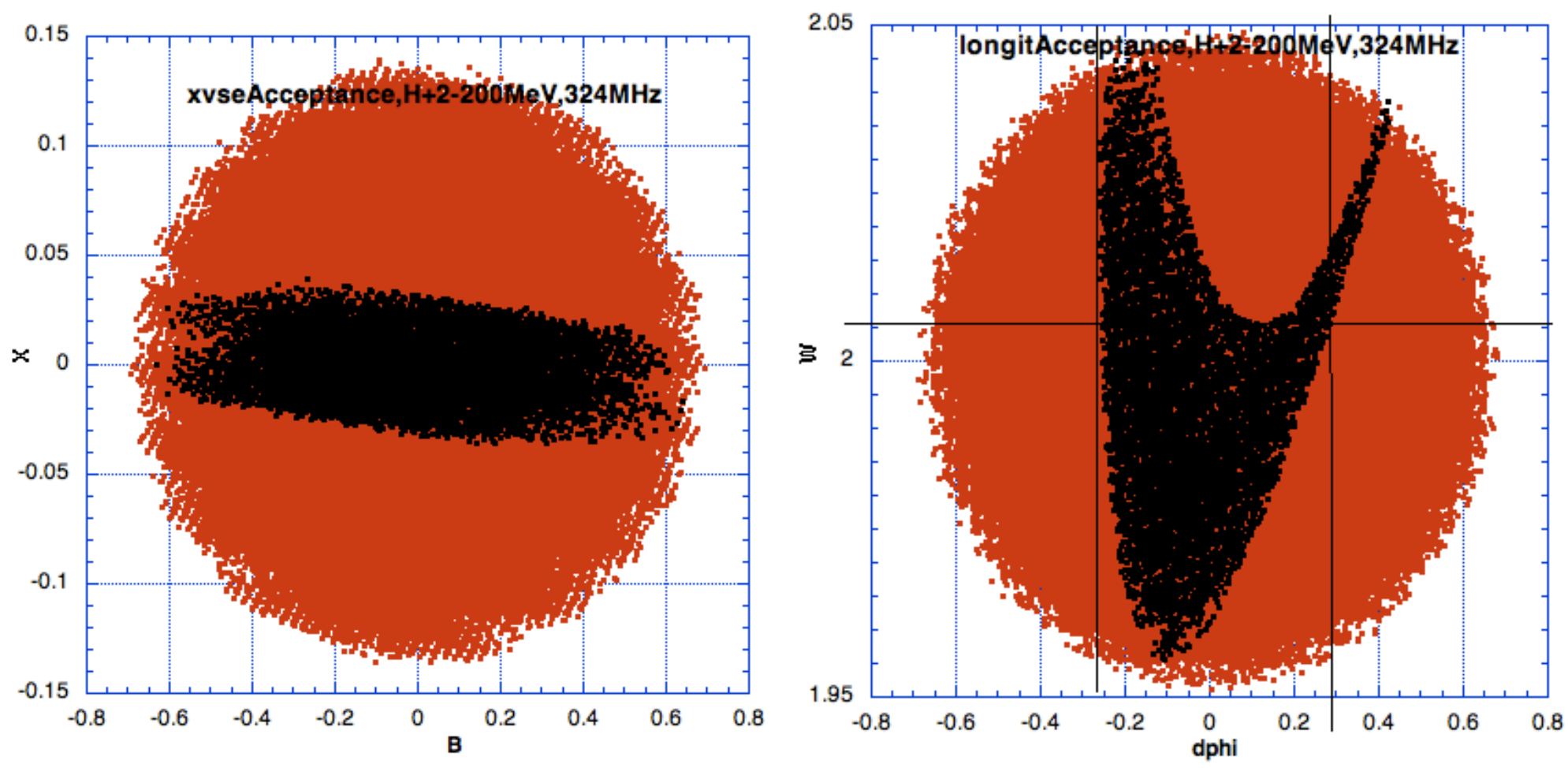

Fig. 11.1.3. a. Transverse acceptance of a 2-200MeV, 324 MHz H+ APF linac. b. Longitudinal acceptance.

A 200-1000MeV, 972MHz APF linac was then attempted, with the same Kilpatrick factor of 1.6, being careful to keep phis swings high, and not letting the attenuation factor overwhelm the sequence. The mesh search resulted in an input N that matched the output N of the 2-200 MeV linac. The total length is ~174 m, with 771 cells. With no aperture restriction and tiny ±22°, ±0.007MeV bunch. the mesh search yielded a starting sequence and optimization resulted in ~94% accelerated fraction. Acceptance matching gave improvement to >99%, with longitudinal acceptance region ~±20°, ±0.1MeV, Fig. 11.1.4.

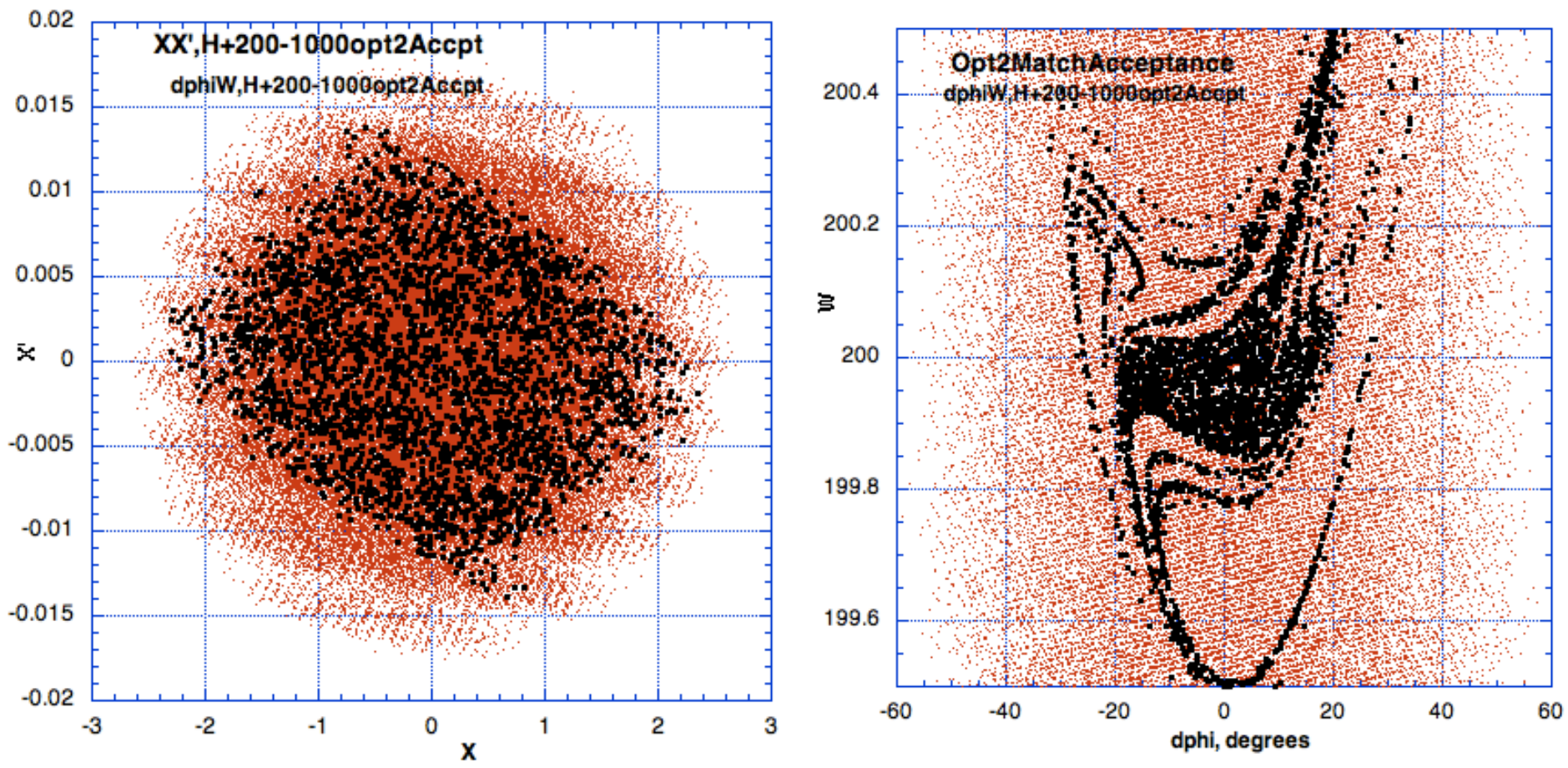

Fig. 11.1.4. a. Transverse acceptance of a 200-1000MeV, 972 MHz H+ APF linac. b. Longitudinal acceptance.

A reality check with a ±72° (at 972 MHz), ±1.9MeV beam output from the H+ 2-200MeV design yielded only 8% accelerated.

Some further sequence adjustments or reality factors might raise or lower the acceptances. The acceptance with beam current will be less, but some amount of beam current could be accelerated. *The point is that the design and optimization procedure works even for long sequences and large energy gain factors.*

## 11.2. Results for Muon APF Linacs, continued

Optimizing the 0.340-20MeV, 324 MHZ , 0.5 cm initial aperture and 1.5 cm final aperture muon linac design with the period N procedure on the phis's resulted in about the same performance, ~98% transmission, 98% accelerated, as the direct optimization of the 23 phases. The transverse and longitudinal gap impulses are opposite in sign and the longitudinal impulse is stronger. The phase advances s0t and s0l, Fig. 11.5.a., show that s0t lies essentially under 90° as necessary to avoid the resonance. s0l is above 90° initially and then falls below; this may have a negative effect on the acceptances and needs to be investigated.

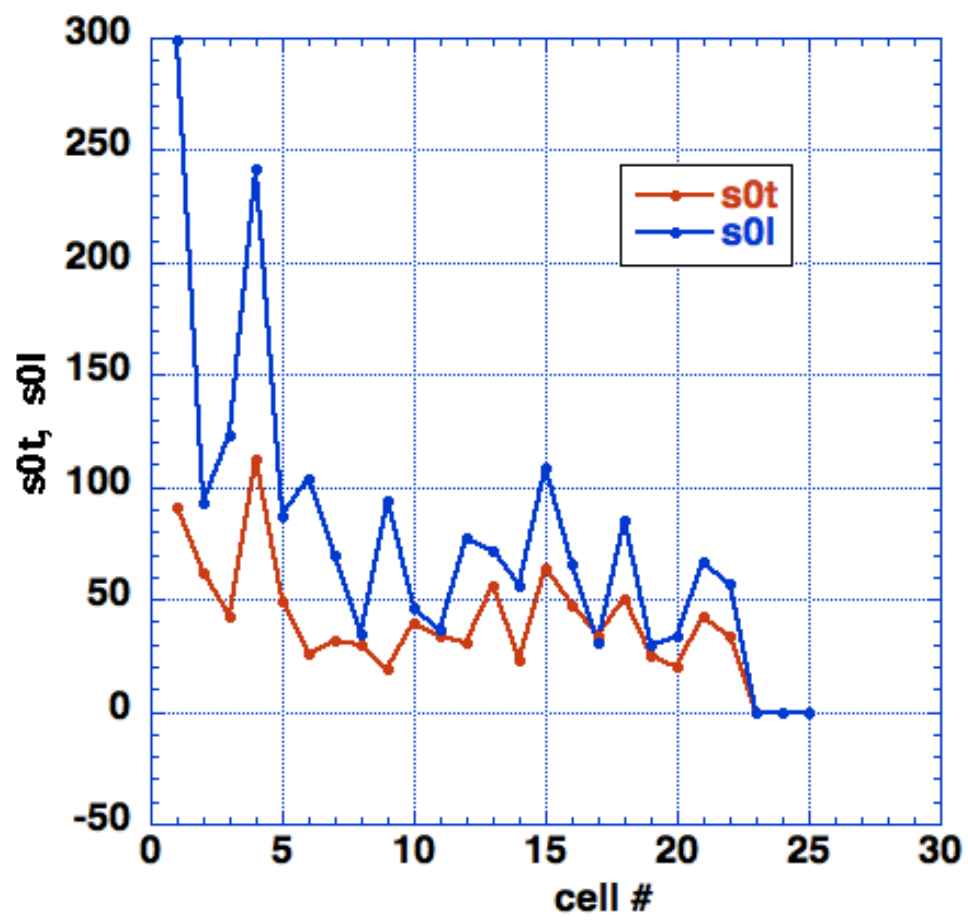

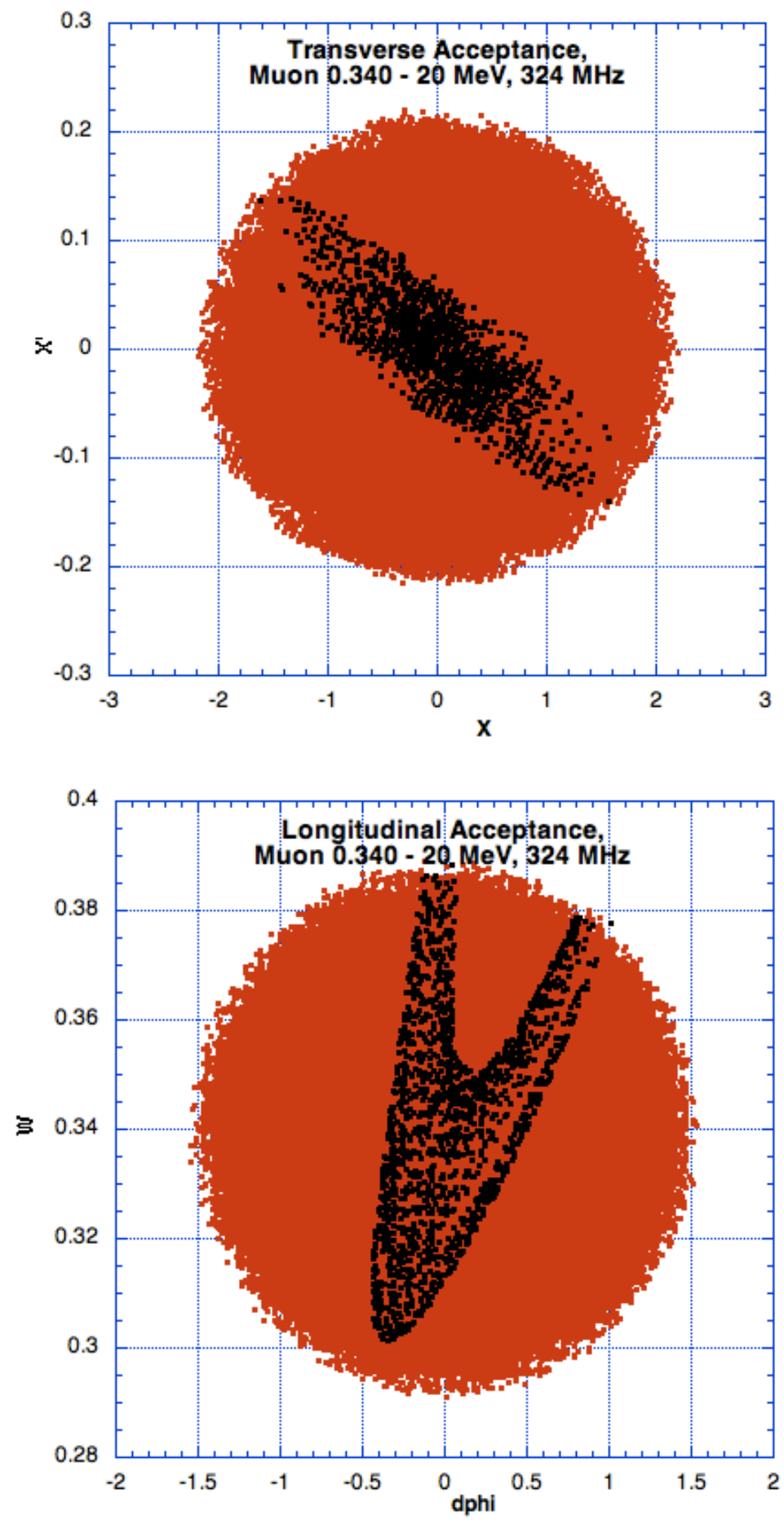

Fig. 11.5. a. Zero-current phase advances for muon 0.340-20MeV, 324MHz APF linac. b. & c. Transverse and longitudinal acceptances.

A 20-200MeV, 972MHz muon APF linac was designed using a linear *peratten* rule. The real transverse emittance input from the 0.034-20 MeV output (where β=0.542) = 0.00028. The dphi output ~±22° is multiplied by 3 = ±66 at 972 MHz, dw=0.2 MeV. The linac has 188 cells, length ~47 m. Iterations with the final one on both phis's and gapobl's resulted in 100% transmission and acceleration with no aperture restriction, and 55% with a constant aperture of 1 cm. Fig. 11.6 shows the phase advance and acceptances.

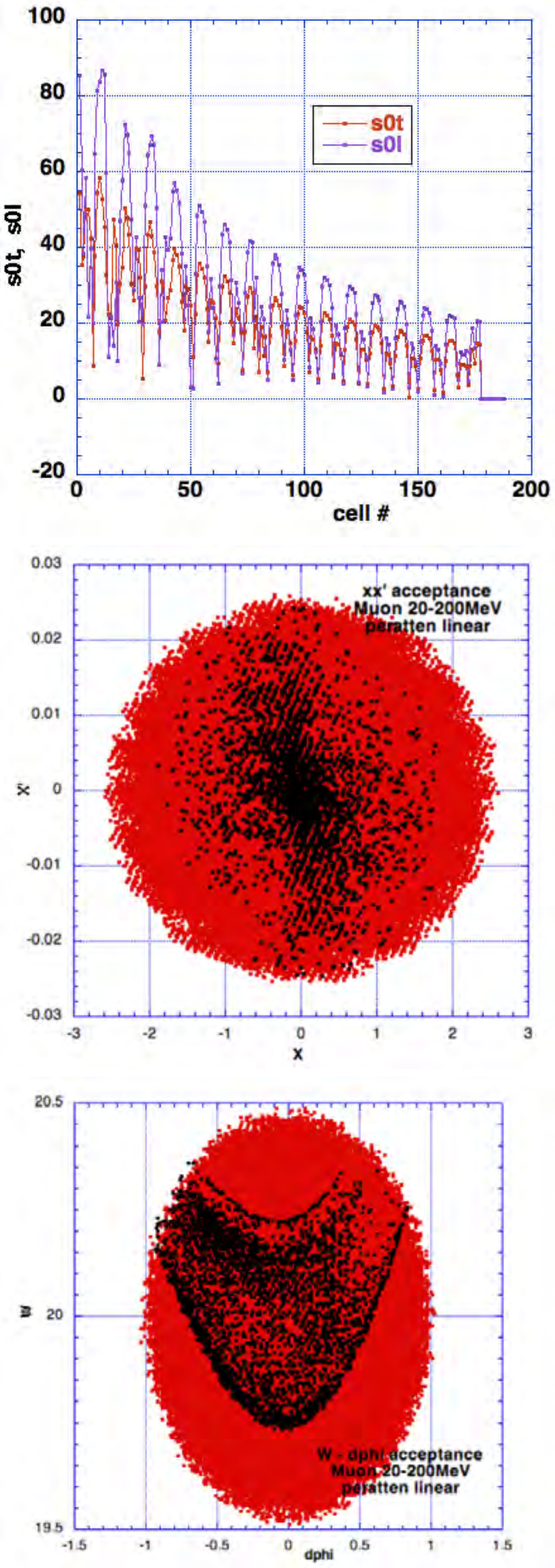

Fig. 11.6. a. Zero-current phase advances for muon 20-200 MeV, 972MHz APF linac. b. & c. Transverse and longitudinal acceptances.

A 0.340-200MeV, 972MHz muon APF linac, 251 cells, ~54 m, resulted in ~65% transmission and acceleration, Fig. 11.7.

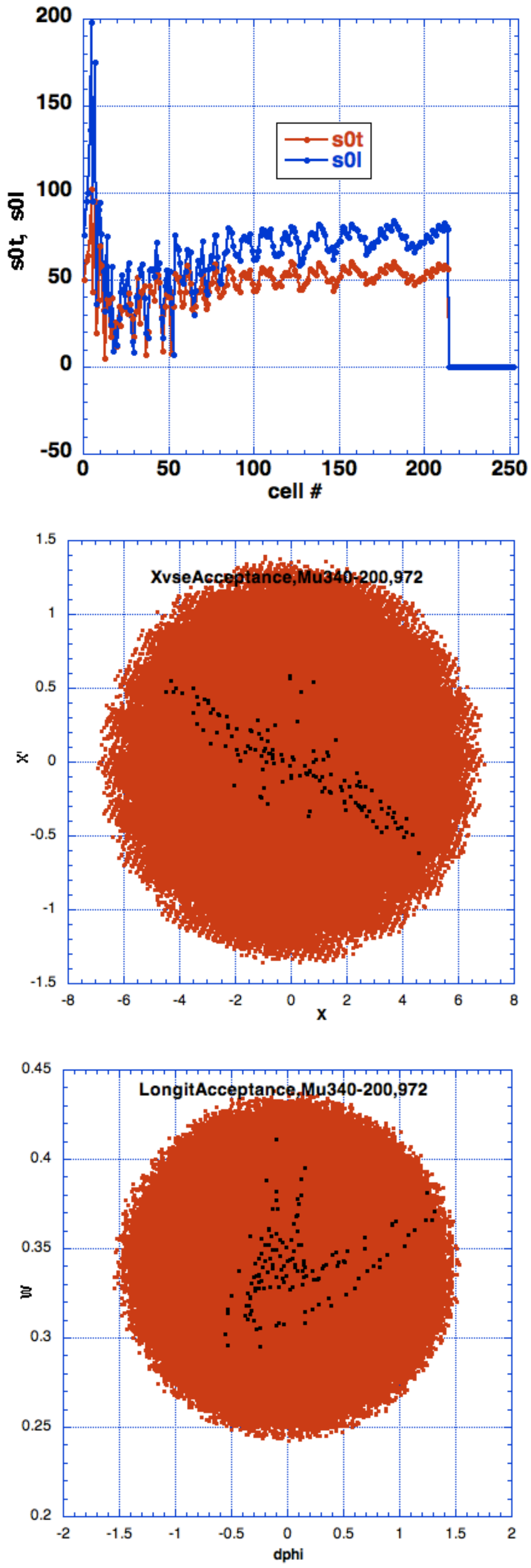

Fig. 11.7. a. Zero-current phase advances for muon 0.340-200 MeV, 972MHz APF linac. b. & c. Transverse and longitudinal acceptances.

Again it is seen that designs can be efficiently found and optimized for long sequences with large energy gain factors, upon which realistic practicalities can be investigated.

[eltoc]

# Chapter 12 – Model Enhancement, Conclusions, Acknowledgements

Model enhancement

The simple but efficient tools now implemented in *LINACSapf* can readily be extended. Some steps would be to:

- Extend *LINACSapf* to have the option for a more accurate gap models, such as an analytical one, or a field map.
- Add space-charge. The preliminary design procedure would use the rms envelope method of *LINACS*, and afford "inside-out" design from desired beam space-charge physics considerations, including the possibility of using equipartitioning if desired. Final simulation could then incorporate standard space-charge routines.
- Investigate actual construction. The 'fish filet' structure suggested at TIT seems an ideal candidate.
- The above examples were designed with an average longitudinal electric field Eo = (1.6 to 1.8)*Kilpatrick factor at the designated frequencies. Thus they represent approximately the maximum strength fields that might be obtained throughout the linac. Many practical aspects, such as actual design of the structure and determination of possible peak surface fields, other phase advance strategies, other sequence and optimization strategies, etc. can now be investigated.
- APF can be used in addition to conventional magnetic focusing, and could be useful in minimizing the amount of additional magnetic focusing needed to handle the desired amount of beam current.

APF linacs can now be another practical approach in the linac designer's repertoire, and can be considered as a candidate for any application, either alone or in conjunction with magnetic focusing.

Conclusions

Working tools for the generation and beam dynamics simulation of APF linacs using a general synchronous phase sequence have been developed in the program *LINACSapf*. A driver framework allows quick searching over the parameter space of an 7-parameter general APF phase sequence, to look for desired characteristics such as maximum transmission, maximum accelerated beam fraction, low transverse emittance growth, low output phase of energy spread, etc.

Optimization from the parameter mesh search result is then done with a constrained nonlinear optimization program, using essential accelerator physics information about the period of the smooth approximation phase advance for convergence and efficiency.

The program was used to demonstrate that very long APF linacs, with high energy gain factors, are possible, for the first time. Examples for zero-current muon and H+ APF linacs were demonstrated. Future steps to incorporate space-charge and more accurate components are straight-forward.

Acknowledgements

This synthesis and extension of much foregoing detailed research in APF accelerator physics and modern control theory optimization techniques is based in particular on those authors cited in the references, particularly in Russia and Japan, and on pioneering work in optimal control of particle accelerator dynamics and Beam Dynamics and Optimization (BDO) Workshops at the Institute of Prof. D. Ovsyannikov, St.Petersburg State University [95].

# Chapter 12 – Update 2016

In 2016, working with Dr. Johannes Maus, a short sequence C4+ APF linac was investigated. The simple dynamics code described above was overhauled and improved, definitions clarified, space charge added (using an improved scheff r-z PIC subroutine), and extensively compared to the successful APF C4+ linac of the National Institute for Radiological Science at Chiba, Japan. Very

---

95 "Design and Simulation of Practical Alternating-Phase-Focused (APF) Linacs – Synthesis and Extension in Tribute to Pioneering Russian AFP Research", R.A. Jameson, Proc. RUPAC2012, St-Petersburg, Russia, MOACH02 & MOACH02-talk many possible sequence combinations are easily found, and many variations in strategy and optimization have been investigated.

An option for full Poisson solution of the external and space charge fields was added by Maus. As the Poisson mesh must be periodic, this involved careful consideration. For the Poisson solver a different definition of a cell must be applied in order to get the longitudinal boundary condition right. The solver reflects the geometry at the longitudinal boundaries. Therefore, the boundary are placed at the center of the drift tubes. With this definition the cell starts at a center of the drift tube, the center of the gap is "somewhere" in the middle of the Poison cell, because of possible phase shift, the cell length is usually not beta*lambda/2 in length, and it ends at the center of the next drift tube. Cell lengths different from beta*lambda/2 are not represented very precisely in the simple dynamic code due to the assumption made for the calculation of the transit time factor – this needs further work.

Therefore, one observes differences between the two approaches.

In order to make use of the powerful optimization tool, a shorter simulation time was needed and could be achieved by setting up a multi-processor version of the Poisson solver and the dynamic routine. A given sequence can be optimized within a reasonable time frame of several hours.
Some variations are implemented to allow for basic error studies of a given sequence in terms of voltage and alignment deviation.

The quality of the results of the optimizer depends on how well the arguments of the objective function are chosen. For the energy spread an option was implemented to search for certain percentages of the all particles. So, the energy spread e.g. for 95% of the particles can be used as summand for the objective function. Similar features can be added for other parameters as well.

As expected, the geometrical configuration of the drift tubes then influences the results. There are advanced ideas about how to configure the drift tube noses. Investigations are still in progress.

*LINACSapf* with the simple dynamics ~~has been and is available for purchase (2500 Euro) from the author~~ (2023 posted on Researchgate at no-cost). The full Poisson version is proprietary at this date.

[eltoc]

[eltoc]

# PART 3 — RFQ SIMULATION CODE DEVELOPMENT

# Chapter 13 – Introduction to Comparison and Development of RFQ Simulation Codes

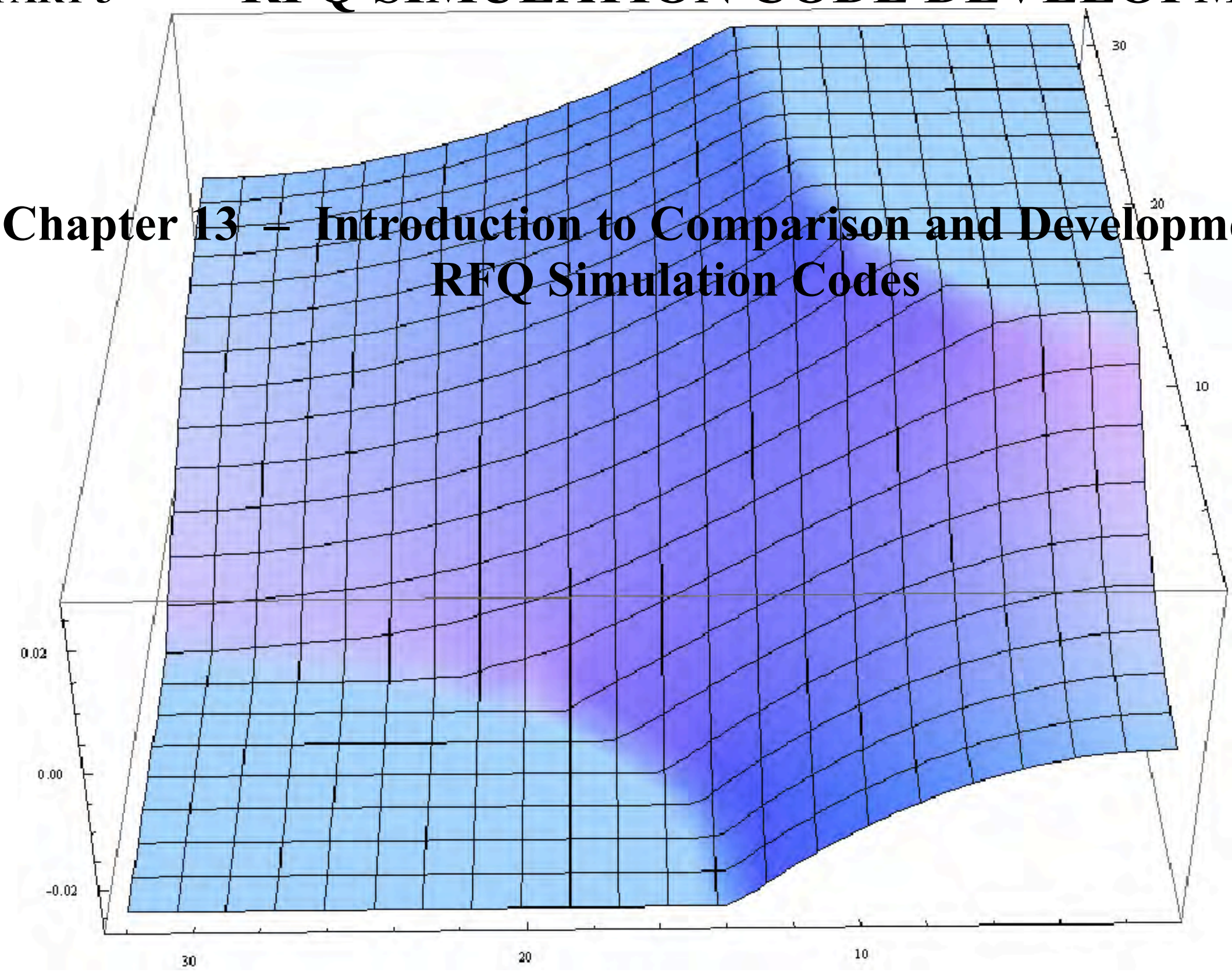

## 13.1 Dedication

This quest to improve the physics fidelity of general linac and particularly RFQ design and simulation to reflect the computational capabilities of today's desktop and laptop computers is dedicated to Kenneth R. Crandall (KRC), whose superb physical insight and numerical computational abilities produced the world-standard linac simulation codes of the 1970's-~1996 era.

## 13.2 Introduction

I and several students were engaged for several years at the Institute für Angewandte Physik (Institute for Applied Physics) IAP, Goethe Universität, Frankfurt-am-Main, Germany, previously under the direction of Prof. Dr. Horst Klein†) in research on the physics fidelity of general linac and particularly RFQ design, simulation and optimization, and the correspondence that our findings would have with the actual working RFQ performance.

Motivations include:

• Design methods have been limited by lack of direct control over all parameters – not only over all external field parameters, but also over the desired space-charge physics. For small beam currents, "outside-in" design using external field parameters is appropriate, but for intense beams where minimizing beam loss is crucial, "inside-out", "beam-based" design where the external fields are determined from the desired space-charge physics behavior is preferred – and is possible.

• The performance of actual RFQs has generally not agreed so well with the results expected from simulation codes; in particular, beam transmission has generally been lower than predicted, of order 10% or more. Also, some simulations for very intense beams have predicted very high beam transmissions (even exceeding 99%), and it is of interest to know if such predictions are justified.

• Simulation methods have contained various approximations in the past, some from simplified theoretical models, some from computer limitations. Today's computers, especially desktop and laptop computers, allow better models. Improving the physics modeling, and testing them has been the top priority of our research. Comparison of older methods with the improved models is expected to clarify many questions apparent with the older models, that were not possible to explain until greater computational power was available.

- Design and simulation tools, in particular for RFQs, have become more and more "black boxes", with no access to source code, inadequate support, and even lack of willingness to entertain questions. Open source programs, extensive testing, written reports with clear decisions made on the basis of the testing, and openness for collaboration are needed again.
- Computer limitations will still exist., although not with significant regard to the number of particles handled. The number of particles in a high current particle beam is of order $5x10^9$. This number has been investigated with parallel computing some time ago; there is some increase in beam size, but the effect is relatively small. This becomes clear when it is realized that collective effects dominate and not single particle effects. So the challenge is not that, nor alluring graphics. Once the top priority goal of modeling and testing improved physics has been met, decisions can be made with respect to code running time, and which approximations may be justified under which conditions in order to have faster run times. Then the main job requiring large computer power can be undertaken – optimization, checking with many particles.
- Optimization has been the underlying ultimate goal for many years, but has been postponed as a formal subject of modern control practice until the many questions about underlying physics and simulation could be better addressed. The work presented here will be summarized with decisions and recommendations for future work on optimization. An insight will be presented for the RFQ problem, explaining how the accurate but slow physics result from full Poisson simulation can be approximated by a fast result without compromising correct determination of the optimum.

This report concentrates on a "working designer's" comparison of simulation codes for RFQs, and development of a more accurate and capable new code package.. The designer should, however, be constantly aware that all of the aspects, tools, methods, and results are fully applicable to any linear accelerator. For example, it is expected that questions regarding subtle beam losses observed in the newer SNS linac (a design essentially the same as that of LAMPF except for the superconducting section) will not be answered until a comprehensive view along the lines outlined herein has been developed.

The "working designer" starts with an overall specification, and gradually works toward a design that is optimum for his application. During this process, the simulation code will be subjected to a wide variety of off-optimum situations. The designer needs to have knowledge of how his simulation code handles off-optimum conditions, *and whether it will find the correct optimum*. This means that a code needs to be well tested, and one way to enforce this is to compare it to other codes.

We know of no other linac or RFQ code comparison project of the breadth or depth we are attempting[96] [97]. Some code comparisons can be found in the literature; on examination, they are anecdotal comparisons of a single design, or comparisons of codes that use essentially the same simulation techniques, or tests of a theoretical or highly simplified model, such as a fixed distribution to compare space-charge forces. Extensive study had shown that simulation of anecdotal cases designs (across a wide variety of parameters, e.g., IFMIF (International Fusion Materials Irradiation Facility), heavy ion injectors, cancer-therapy injectors, etc.) by different codes gave different results, and that simulation of the same cases by a particular code not only also gave different results, but also a different pattern of results. And in any case, comparison of an advanced code to a code that has known deficiencies can only qualitatively make sense.

---

96 See reference for a good comparison of various programs for a simple, analytically evaluable lattice, including off-optimum conditions – showing agreement at the central design momentum but substantial differences for non-zero momentum deviations.

[97]. "A comparison of several lattice tools for computation of orbit functions of an accelerator", E.D. Courant, et.al., Proc. 2003 Particle Accelerator Conference, p3485.

*Code comparison over a wide range of parameters is essential, and this is only possible if a controlled simulation experiment is set up. It cannot be done by looking at a number of different anecdotal examples.*

In usual experimental practice, a good controlled experiment is one in which only one parameter of a system is varied. The RFQ design code *LINACSrfqDES* [98,99] is unique in that it affords control of literally all RFQ parameters, including space-charge rules, and therefore can be used to set up controlled comparison studies. Repeating for emphasis, because the design process for RFQs is equally or even more important than simulation, the RFQ design code *LINACSrfq* is *unique* in that it affords control of literally all RFQ parameters, including space-charge rules.

Using *LINACSrfq*, a family (the *aperfac* family) of thirteen high-current RFQs was designed, for which only one parameter is changed – the minimum aperture at the end of a shaper [Chapter 13 - Appendix 1]. All other parameters, including space-charge rules, remain unchanged or automatically follow the consequence of the aperture change (e.g., the vane voltage changes because a Kilpatrick criterion is used). The minimum aperture changes from too large to too small; the accelerated beam fraction is low at either extreme, with an optimum at an intermediate aperture, as indicated in Fig. 13.1.

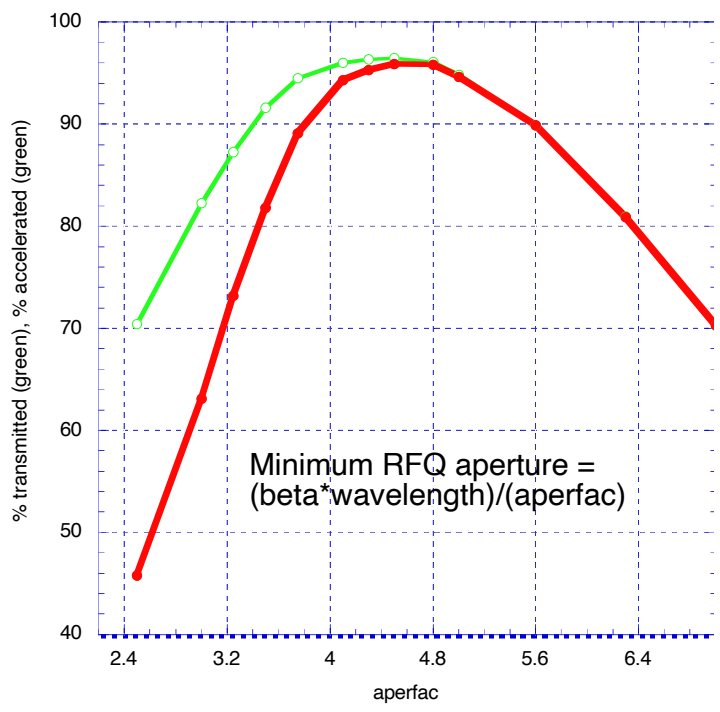

Fig. 13.1 Typical simulation result for the percentage of transmitted (green) and accelerated (red) beam for an *aperfac* family of 13 RFQs designed by changing one parameter.

These are simple RFQs, with a radial matching section, shaper section, and the main section where most of the bunching and acceleration occurs. Full RFQs include other features, such as output radial matching section, transition cell, zero-modulation section, resonant coupling gaps between longitudinal vane sections. Most RFQ codes do not have these features, and the interest here is on the dynamics and space-charge treatments.

The corresponding simulation code at the time, *pteqHI* [100], was a highly developed improvement of the original PARMTEQ, and a development requirement was that the *pteqHI* result have high fidelity to the *LINACSrfq* design requirements, at least for favorable conditions such as near the optimum of Fig. 13-1.

---

[98] "Scaling And Optimization In High Intensity Linear Accelerators", R. A. Jameson, LA-CP-91-272, Los Alamos National Laboratory, July 1991, re-published as LA-UR-07-0875, 2/8/07 – introduction of *LINACS* design program, and *LINACSrfq* subprogram.

[99] "RFQ Designs and Beam-Loss Distributions for IFMIF", R.A. Jameson, Oak Ridge National Laboratory Report ORNL/TM-2007/001, January 2007.

[100]"A Discussion of RFQ Linac Simulation", R.A. Jameson, Los Alamos National Laboratory Report LA-CP-97-54, September 1997. Re-published as LA-UR-07-0876, 2/8/07.

We have been comparing the results for the 13-RFQ family from *pteqHI*, *PARMTEQM* [101] (developed at LANL, with no essential changes since ~1996), *LIDOS* (developed at MRTI, Moscow [102]), and *BEAMPATH* (developed by Y. Batygin [103]). These comparisons have been fruitful because source code has been available, or because the authors have intensively collaborated (*LIDOS, BEAMPATH*).

Comparisons to other black box codes have raised so many questions[104] that we can only recommend that no serious project use simulation codes for which the source code or intensive collaboration with the author is not available.

We immediately found (Fig. 13.2) that different codes produce not only different accelerated beam fractions for the 13-RFQ family, *but that the optimum aperture may also be different* – this is a situation that should be avoided for a serious project [105].

---

[101] "RFQ Design Codes", K.R. Crandall, et. al., RFQCodes.doc, PARMTEQM Reference Manual, LA-UR-96-1836, Rev. Aug 21, 1998, Los Alamos National Laboratory.

[102] "Code Package for RFQ Designing", Boris Bondarev, Alexander Durkin, Yuri Ivanov, Igor Shumakov, Stanilav Vinogradov, Moscow Radiotechnical Institute, Proc. 2nd Asian Particle Accelerator Conference, Beijing, China, 2001.

[103]. Yuri K. Batygin, "Particle-in-cell code BEAMPATH for beam dynamics simulations in linear accelerators and beamlines", Nuclear Instruments and Methods in Physics Research A 539 (2005) 455–489

104 One example is discussed in Appendix 13A3, on integrator stability.

105 Jameson, R.A., "RFQ Design Studies – Investigation (incomplete) of Dependence of Optimization on Codes", Injector/RFQ Working Group, 1st EU=JA Workshop on IFMIF-EVEDA Accelerator, 7-9 March 2007, Paris. (RFQ Workshop Paris.ppt, March 8, 2007)

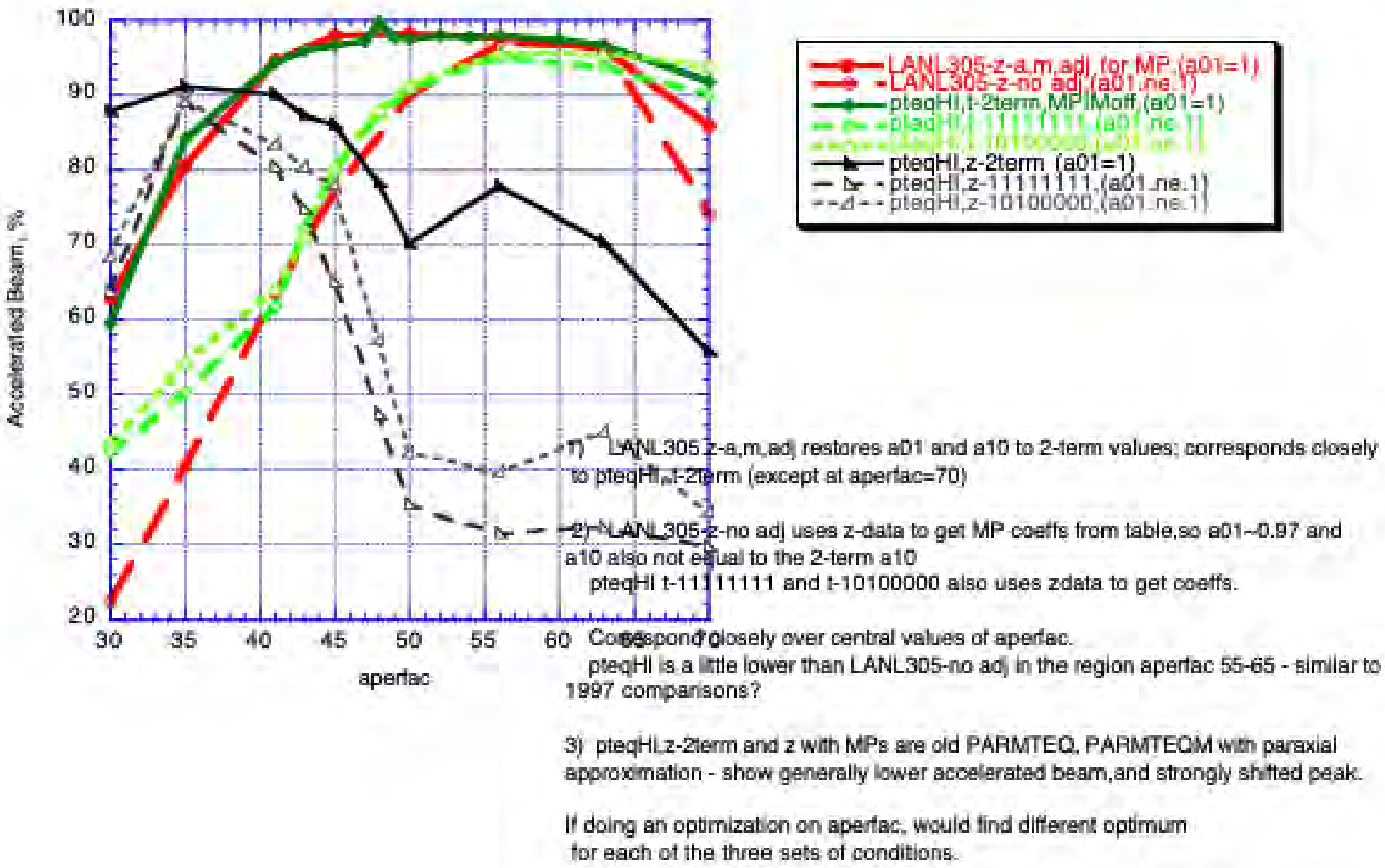

Fig. 13.2 Percent of accelerated beam using different simulation codes on the *aperfac* family of RFQs. This is a very early comparison in 2007. Other codes were also tested and gave other different curves, but were not discussable so are not shown. The general pattern of disagreement is what is important. here; the legend gives some details, e.g. the 8-figure numerical sequences show the multipole terms turned on (1) or off(0).

The importance of differences is in the eye of the beholder, but here were seen differences ranging from a few percent to 10% or more, and this was worrisome.

There are dozens of reasons why different codes can have different results. The physics used in the codes must have the same fundamental basis, but can be implemented in different ways. Many decisions have to be made about the internal workings of the simulation, starting from the choice of methods for simulating the external and space charge field, and then many details such as mesh sizes, starting conditions, boundary conditions, etc. It would seem apparent that some differences in results should be expected – the question is, how large, and of what consequence.

Part of the motivation for this study was from hearing strong statements that designs compared in two codes, known to have differences in approach and technique [106], gave "transmission agreement to four significant figures", gave transmission that seemed unrealistically high compared to known practical experience, and were supported by no available evidence from either the codes or the underlying simulation conditions. Using the 13-RFQ family and looking only at transmission, we found that these two codes have different curves, which cross at a certain aperture, so it is possible, although unlikely, to have transmission agreement to four significant figures for some anecdotal design.

The comparisons between *pteqHI*, *PARMTEQ*, *PARMTEQM, LIDOS* and *BEAMPATH* have shown that the 13-RFQ family comparison is a strong test. Of course many other controlled experiments of this sort could be proposed, with analysis of characteristics beyond simple transmission, especially the phase-space ellipse and total emittance patterns. But the *aperfac* family test raised many puzzles, and addressing these has required intensive investigation.

106 One code with known severe approximations, the other with Poisson field solutions.

It is clear that Poisson solvers are necessary to accurately compute the RFQ external and space charge fields. This is done in *LIDOS*, and in a limited form for space charge in *BEAMPATH*. Development of Poisson solvers for *pteqHI* was done by J. Maus at IAP for his doctoral dissertation, with the initial goal that the physical correctness be apparent, explainable and open. This is because the same kinds of decisions about the programming have to be made (meshing, boundary conditions, convergence criteria, etc.), and each has its consequences. From this basis, further development (faster execution, etc.) can systematically follow.

The final goal has been to build a new, open-source RFQ code incorporating the best physics that present-day computers can handle, a flexible library of analysis capabilities, and a variety of options that allow the designer to choose between accuracy and execution speed with an understanding of the consequences – realized in the code *LINACS*.

This report will compare *BEAMPATH*, *PARMTEQM*, *LIDOS,* the earlier version of *LINACS* that was named *pteqHI*, and *LINACS*. Each section will introduce increasingly complex aspects and concentrate on different codes; however, as the complexity is indeed high, each section will pave the way by including some points regarding the codes to be studied in detail in later sections. Chapter **14** compares *pteqHI* and *BEAMPATH* for the simplified 2-term beam dynamics representation but different space-charge routines and introducing the subject of image charge. In Chapter **15**, *PARMTEQM* is added. The effects of z-code representation, paraxial approximation, external field representation using a multipole expansion of the RFQ electric fields, and a representation of image charges are discussed, and comparison is made to the time based (t-code) *pteqHI.* In Chapter **16**, it is shown why the approximations of the earlier codes should be removed, as is now possible with extant desktop and laptop computers, and the evolution of the physics basis for a new simulation code *LINACSrfqSIM* is developed. Chapter **17** discusses in detail a new Poisson-solver method for *LINACSrfqSIM,* developed and written by Johannes Maus, IAP, Goethe Uni Frankfurt. It is important to note that it IS (easily) possible to turn image charge on and off for checking, although it has been strongly claimed that the Poisson method does not allow this (!!) "Turning the image charge on and off" just means using different boundary conditions for the space charge mesh – "on" meaning here the exact quadrupolar RFQ boundaries, "off" meaning using a cylindrical open or closed boundary condition. Chapter **18** then extends to *LIDOS*, in which both external and space-charge fields are found using a Poisson solver. The synthesis of the new *LINACSrfqSIM* code is outlined in Chapter **19**, including a simple graphical interface unit (GUI) and graphic containing information that is really pertinent to the space charge physics in a linac. A comparison to *LIDOS* follows in Chapter **20**.

Appendices give details of: 13A1. Analysis of the Aperfac RFQ Family; 13A2. Synopsis of Reference [2]; 13A3. Integration of the Equations of Motion, Difference Equation Integrator Stability

[eltoc]

## *Chapter 13 - Appendix 1. Analysis of the aperfac RFQ Family*

IAP Memo RAJ-2009/09/15

# ANALYSIS OF THE APERFAC RFQ FAMILY

R. A. Jameson
Institüt für Angewandte Physik
Johann-Wolfgang-Goethe Universität
Frankfurt-am-Main

### INTRODUCTION

We have been engaged for the past almost three years at IAP in a "working designer's" comparison of simulation codes for RFQs. Using the RFQ design code *LINACSrfq* [1], which affords control of all RFQ parameters, including space-charge rules, we designed a family of thirteen RFQs (the "*aperfac* family") for which only one parameter is changed – the minimum aperture at the end of a shaper. All other parameters, including space-charge rules, remain unchanged and automatically follow the consequence of the aperture change (e.g., the vane voltage changes because a Kilpatrick criterion is used). The minimum aperture changes from too large to too small; the accelerated beam fraction is low at either extreme, with an optimum at an intermediate aperture, as indicated in Fig. 13A1.1.

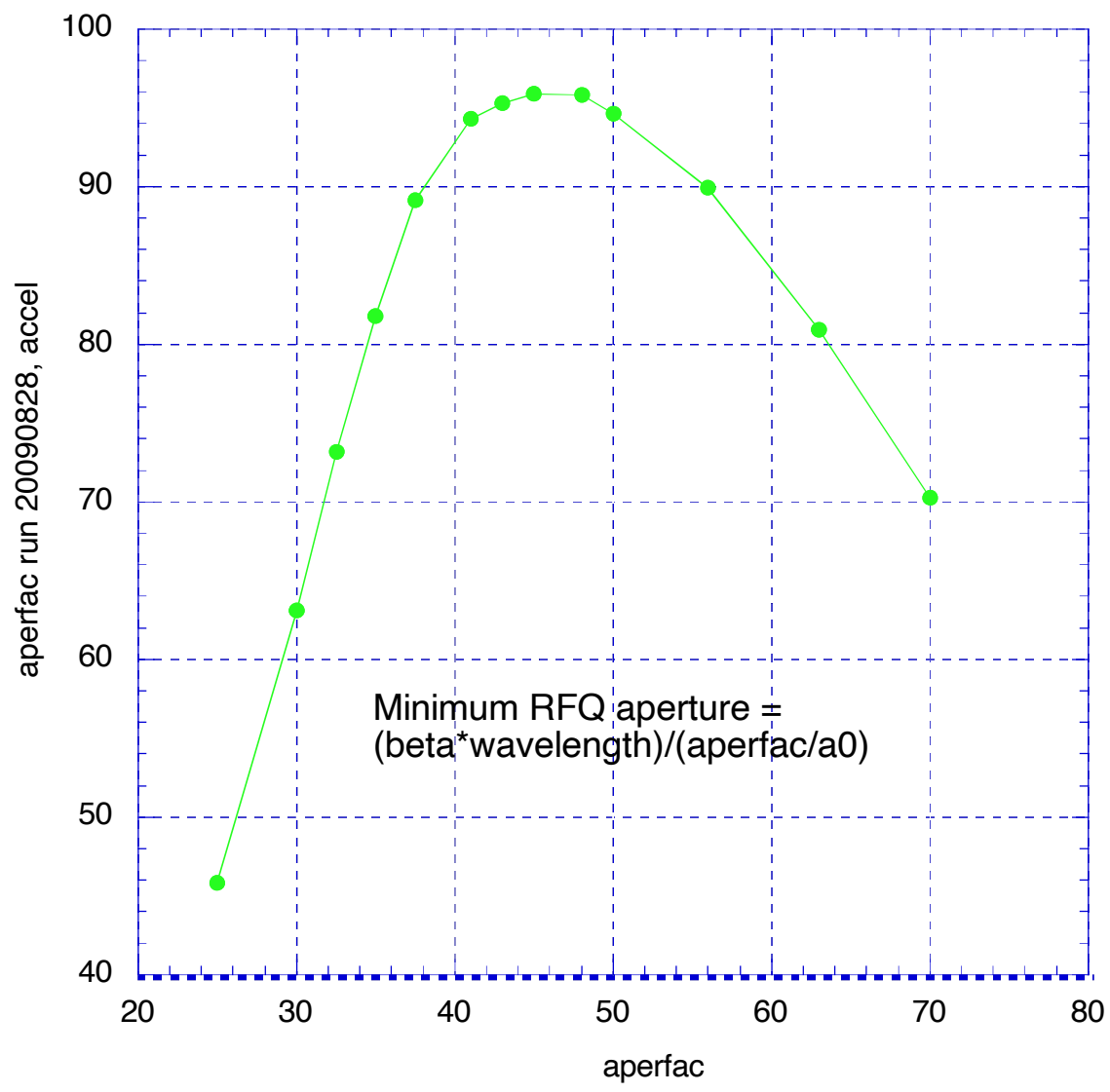

Fig. 13A1.1 Typical simulation result for the percentage of accelerated beam for a family of 13 RFQs designed by changing one parameter.

The corresponding simulation code, *pteqHI* [2], is a highly developed improvement of the original PARMTEQ, and a development requirement is that the *pteqHI* result has high fidelity to the *LINACSrfq* design requirements.

We have been comparing the results for the 13-RFQ family from *pteqHI*, *PARMTEQM* (developed at LANL, with no essential changes since ~1993), *LIDOS* (developed at MRTI, Moscow [3]), and *BEAMPATH* (developed by Y. Batygin [4]). These comparisons have been fruitful because source code has been available, or because the authors have intensively collaborated (*LIDOS, BEAMPATH*). Comparisons to other black box codes have raised so many questions that we can only recommend that no serious project use simulation codes for which the source code or intensive collaboration is not available.

We know of no other code comparison project of the breadth or depth we are attempting. Some code comparisons can be found in the literature; on examination, they are anecdotal comparisons of a single design, or comparisons of codes that use essentially the same physics techniques, or tests of a theoretical or highly simplified model, such as a fixed distribution to compare space-charge forces. The "working designer" starts with an overall specification, and gradually works toward a design that is optimum for his application. During this process, the simulation code will be subjected to a wide variety of off-optimum situations. We immediately found that different codes produce not only different accelerated beam fractions for the 13-RFQ family, but that the optimum aperture may also be different – this is a situation that should be avoided for a serious project.

There are dozens of reasons why different codes can have different results. The physics used in the codes must have the same fundamental basis, but can be implemented in different ways. Many decisions have to be made about the internal workings of the simulation, for example about mesh sizes, starting conditions, boundary conditions, etc. It would seem apparent that some differences in results should be expected – the question is, how large, and of what consequence.

Part of the motivation for this study was from hearing strong statements that designs compared in two codes, known to have differences in approach and technique, gave "transmission agreement to four significant figures", gave transmission that seemed unrealistically high for known practical experience, and were supported by no available evidence from either the codes or the underlying simulation conditions. Using the 13-RFQ family and looking only at transmission, we found that these two codes have different curves, which cross at a certain aperture, so it is possible, although unlikely, to have transmission agreement to four significant figures for some anecdotal design.

The comparisons between *pteqHI, PARMTEQ, LIDOS* and *BEAMPATH* [5] have shown that the 13-RFQ family comparison is a strong test. Of course many other controlled experiments of this sort could be proposed, with analysis of characteristics beyond simple transmission. But the *aperfac* family test raised many puzzles, and addressing these has required intensive investigation.

It is clear that Poisson solvers are necessary to accurately compute the RFQ external and space charge fields; development of such is in progress at IAP, with the initial goal that the physical correctness be apparent, explainable and open. This is because the same kinds of decisions about the programming have to be made (meshing, boundary conditions, convergence criteria, etc.), and each has its consequences. From this basis, further development (faster execution, etc.) can systematically follow.

Work to document all the findings is in progress. The final goal is to build a new, open source RFQ code incorporating the best physics that present-day computers can handle, a flexible library of analysis capabilities, and a variety of options that allow the designer to choose between accuracy and execution speed with an understanding of the consequences – still a large undertaking.

In this note, the characteristics of the 13-RFQ *aperfac* family are outlined.

## *LINACSrfq* DESIGN

*LINACSrfq* (open-source) is the RFQ implementation of the general *LINACS* design code [1,6]. It is unique, in that it gives the designer complete control of literally every RFQ parameter, including both the external fields and the desired space charge physics. The space-charge physics is defined by the transverse and longitudinal rms envelope matching equations, and optionally by the beam equilibrium, or equipartitioning, condition.

The *LINACSrfq* design summary table for each RFQ in the *aperfac* family has the following form, with short explanations in italics. Only the minimum aperture at the end of the shaper is varied.

**rfqtype=4 vaneWin=0.095 MeV Wout= 5. MeV**
**I=0.13 him=2.0145** *(deuteron)* **q=1** *(charge)* **freq=175.** *(MHz)*
**RhooverR0=0.75** *(ratio of vane tip radius to r0)*
**vane geometry=VSINE** *(vane modulation is sinusoidal)*
**mpole=1 mpoleterms=11111111 image=**0 *(Multipoles are included in the design procedure, both multipole and image terms are output for input to simulation codes.)*
**KPfac=1.7** *(Kilpatrick bravery factor)*
**etnrmsgiven=0.000025** cm.rad **elnrmsgiven=0.00004** cm.rad
*(Normalized rms transverse and longitudinal emittances, both in cm.rad. Ratio of longitudinal to transverse emittances is 1.6)*
**etnrmsgivenmain:=etnrmsgiven elnrmsgivenmain:=elnrmsgiven**
*(The RFQs have two sections – a shaper, and the main RFQ. Here the rms emittances are to remain constant through the entire RFQ. (It is possible to have a-priori varying emittances.))*
**Shaper parameters:**
**EOSaperfac=(VARIED)** *(The minimum aperture in the RFQ will occur at the end of the shaper, and is entered as (beta*wavelength)/EOSaperfac, where EOSaperfac varies from 2.5 to 7.0)*
**phistgt=-88. bfrac=0.6**
**rmscells=4 siglint=180. porch=60. celldiv=10**
*(A radial matching section of 4 cells is used to bring the transverse focusing parameter B to 0.6 times the value that B will have at the end of the shaper. The shaper brings the injected beam to*

*an equipartitioned, equilibrium condition at the end of the shaper, where the synchronous phase has increased to -88°. Siglint controls the length of the shaper, and is the zero-current longitudinal phase advance accumulated through the shaper. A "porch" with zero modulation is maintained for 60% of the shaper length. Each cell is divided into ten steps.)*

**ffadjrule:=ffadj:=1.** **ffadj0=1**. *(The design can be refined by an adjustment of the beam form factor – this is not done in this case)*

**phis rule: lfacincr=-0.5 lfacdist=12.**

**TKphisrule:=(If[phisbw<=-20.,phis[rfq]=phisbw,phis[rfq]=-20.];)**

**mainrfqphisrule:=TKphisrule**

*(The original Teplyakov concept for keeping the charge density in the bucket constant is used, but a controlled change in the density can also be used; in this case the ratio of bucket length to beam bunch length is allowed to grow by 0.5 over a length of 12 meters. If the resulting synchronous phase reaches -20°, it is then held constant for the rest of the RFQ. This does not occur in the aperfac family.)*

**mainrfqaperrule:=(endbeta=0.073;c3=1.;**

**a[rfq]:=aEOS (1+c3 ((beta-betaEOS)/(endbeta-betaEOS))^1.);)**

*(A rule is also used for the aperture in the main part of the RFQ, starting from the end of the shaper. Here the aperture is programmed to increase as a function of beta.)*

**frontendvrule:=(v[rfq]=vtarget;)**

**mainrfqvrule:=(v[rfq]=(KPlimit KPfac r0[rfq])/ckappa[rfq]/.{em[rfq]->mstart,a[rfq]->astart};)**

*(Here the vane voltage rule uses the Kilpatrick limit at the end of the shaper for the entire shaper, and in the main part of the RFQ, the vane voltage is determined by the Kilpatrick limit, the bravery factor, r0, and the peak surface field on the vane.)*

**mainrfqemrule:=mfree** *(The RFQ will be required to maintain the equipartitioned, equilibrium beam condition from the end of the shaper to the end of the RFQ. This condition plus the requirement that transverse and longitudinal match are maintained requires three variables. The transverse and longitudinal rms beam sizes and the modulation are used.)*

**mainRFQstrategy:=matchEP** *(This rule tells the program that the RFQ will be required to maintain the equipartitioned, equilibrium beam condition from the end of the shaper to the end of the RFQ.)*

## *pteqHI* SIMULATION

*pteqHI* is an open-source RFQ simulation code derived from the original *PARMTEQ* code developed at Los Alamos for the first RFQ proof-of-principle test outside of the then Soviet Union, in 1980. It uses time as the independent variable in order that the space charge forces are correctly calculated, handles multiple inputs with arbitrary mass and charge, has no paraxial approximations, has a large repertoire of analysis capabilities, and served as a test bed for the code comparisons. Available source-code subroutines from other codes were installed in the *pteqHI* framework, affording direct comparisons.

To understand the dynamics throughout the RFQ in terms of available theory, only the particles that are successfully transmitted through the entire RFQ are analyzed. Particles are designated as lost only if they intercept the actual vane surface. Thus, two runs are required. The transmitted particles are flagged on the first run. The second run is with identical dynamics, with cell-by-cell analysis of only the flagged particles. In this way, the rms quantities are readily understandable and can be studied in comparison with the characteristics that were requested in the *LINACS* design program. (Now changed to one run, with saving of particle coordinates along the trajectory for the analysis.)

## EXTERNAL FIELD QUANTITIES

The phase advances per transverse focusing period (in RFQ $\beta\lambda$ (2cells)) are designated as sig0t = transverse zero-current rms phase advance; sig0l =longitudinal zero-current rms phase advance; sigt = transverse rms phase advance with current, sigl = longitudinal rms phase advance with current.

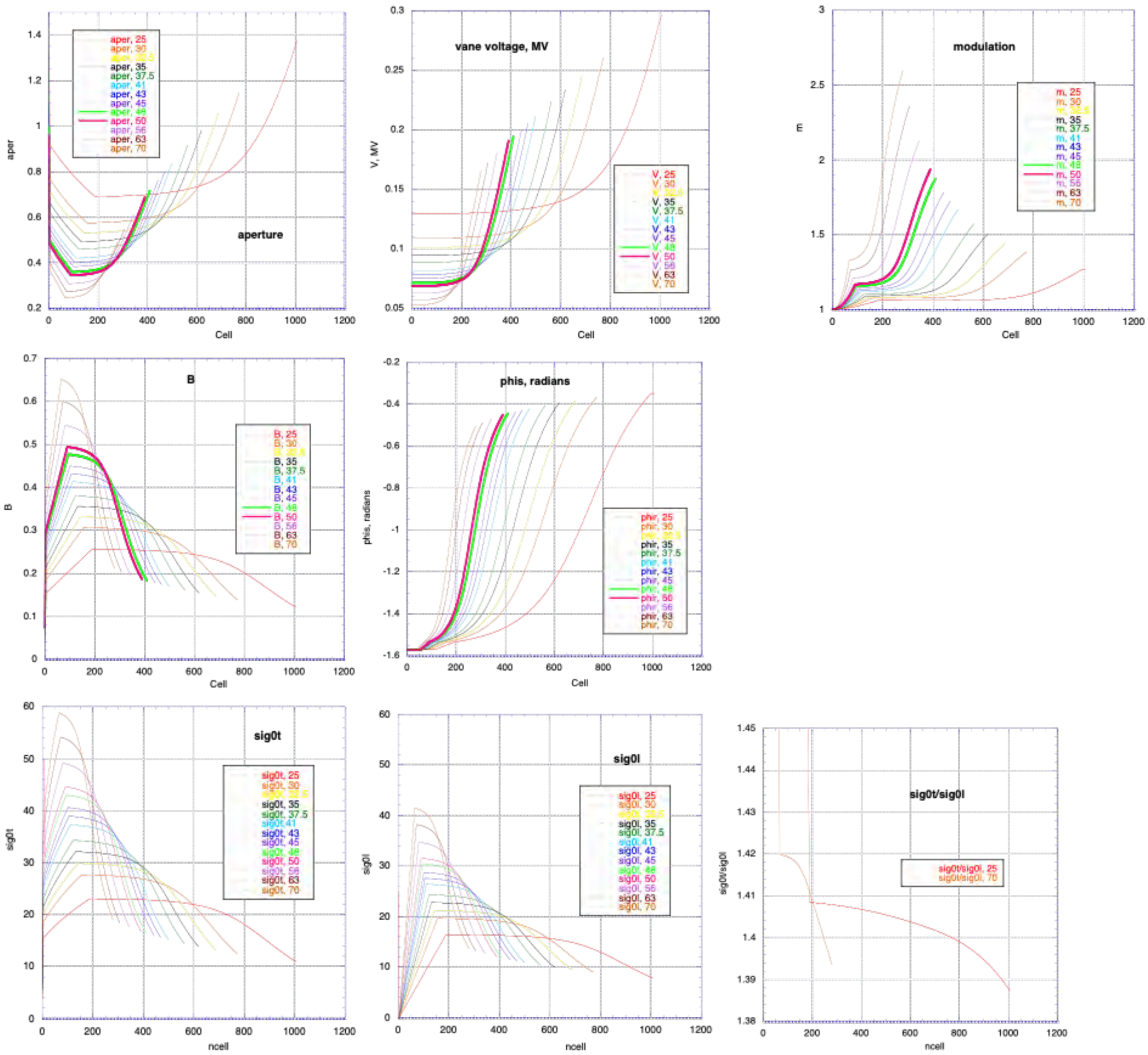

Fig. 13A1.2. External field parameters and phase advances for the *aperfac* family of RFQs.

o

The multipole characteristics (pure multipole coefficient, not multiplied by Bessel function) are shown in Fig. 13A1.3.

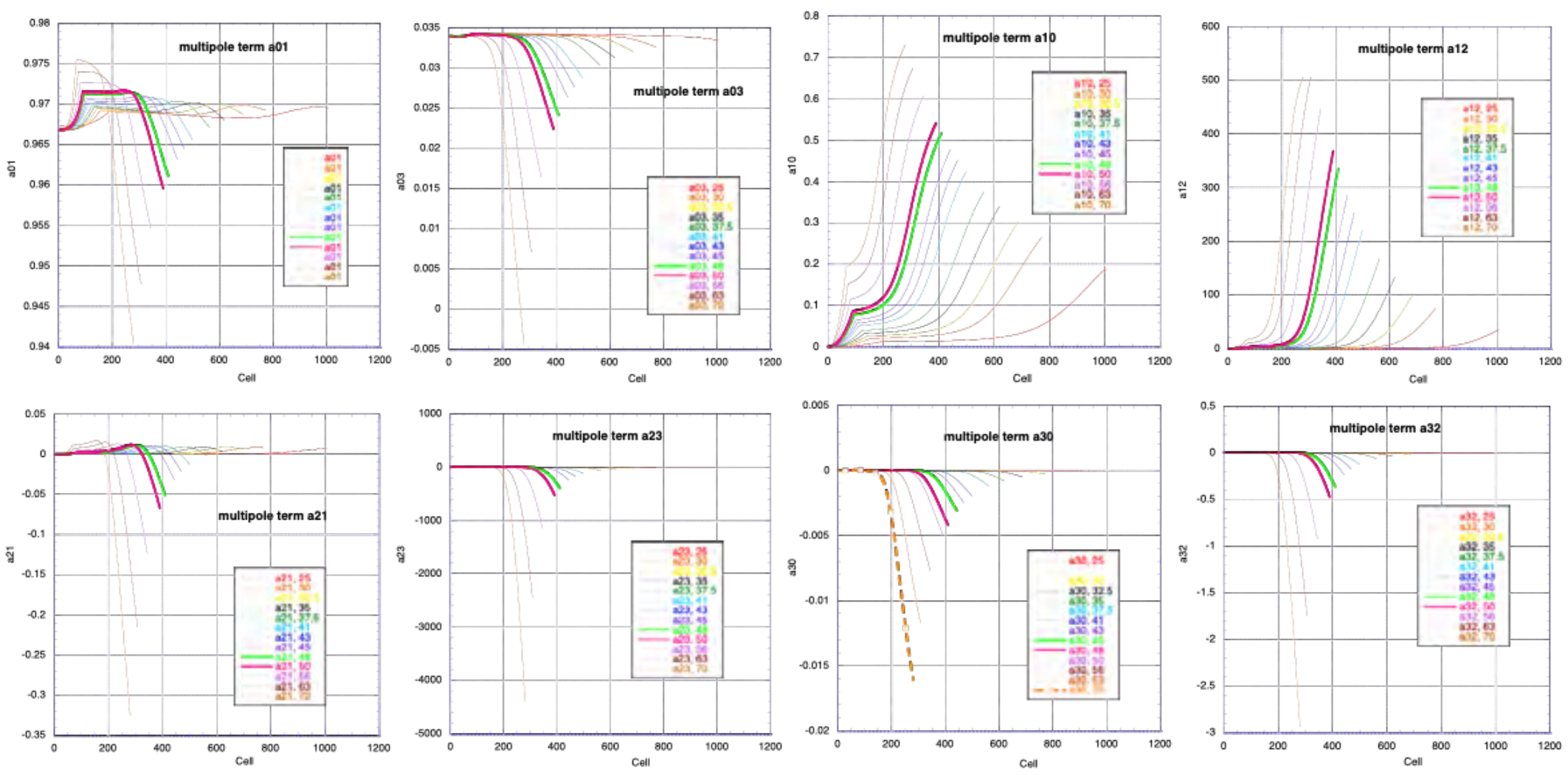

Fig. 13A1.3. Multipole coefficients.

## TRANSMITTED AND ACCELERATED BEAM

pteqHI results are shown for simulations with 10000 particles, with transverse waterbag and longitudinal uniform initial distributions. Fig. 13A1.1 shows the percent of beam accelerated within 10% of the synchronous energy at the end of the RFQs.

Fig. 13A1.4 (left) shows the transmitted particles at each cell – particles are designated lost only if they intercept the vane boundary, which is exactly described at each step, or a radius 1.25*(modulation*aperture). Fig. 13A1.4 (right) shows the number of particles in an energy band within 10% of the synchronous energy at each cell. The initial dip in the in-bucket curves is explained in the next figure.

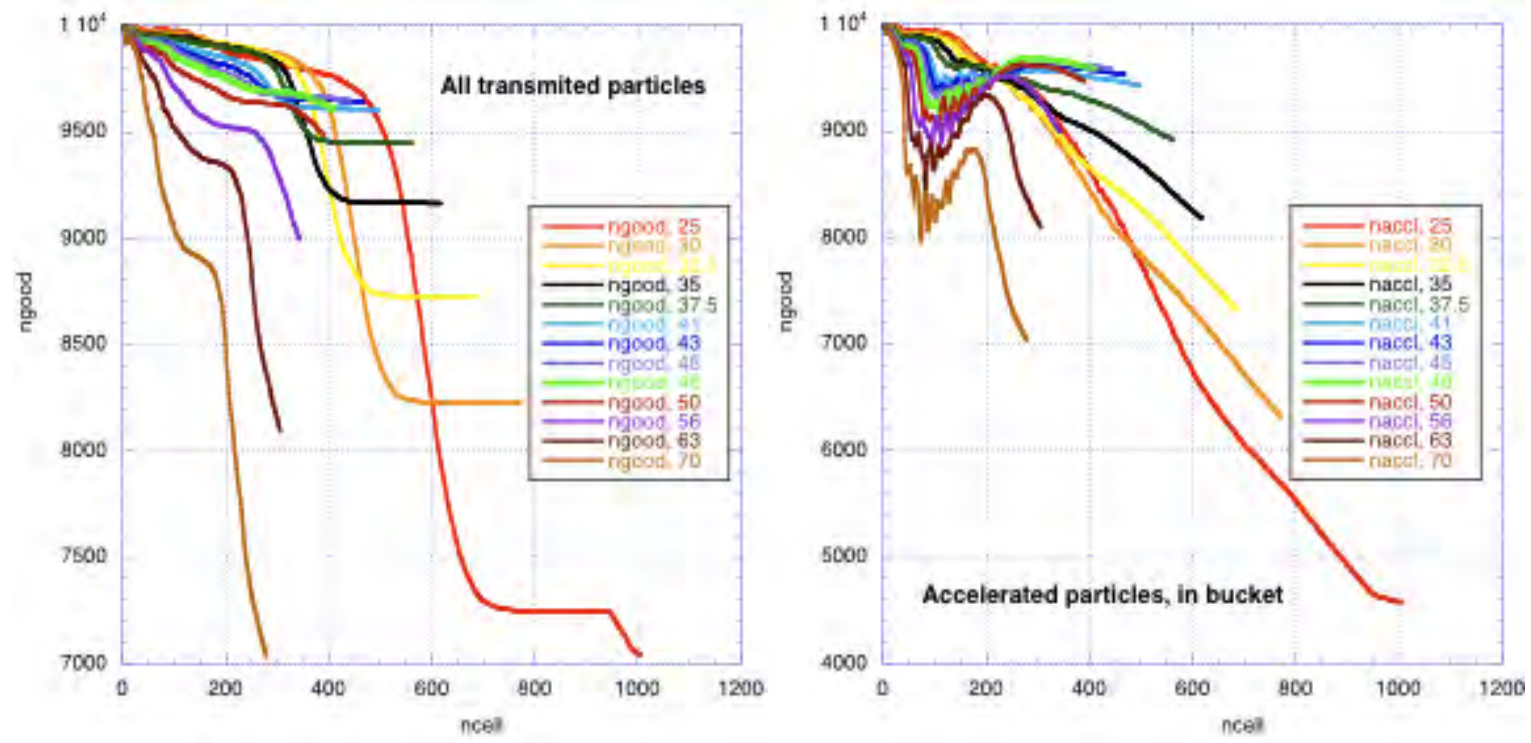

Fig. 13A1.4. Transmitted and accelerated particles. 10000 prticles injected.

It is important to know what kind of computational mesh a code uses, and how it handles particles that may get outside the mesh during the computation.

For computer storage reasons, most simulation codes use a moving external field mesh, for RFQs the mesh is typically two cells (βλ) long. (*LIDOS* is an exception, with one mesh for the whole RFQ.)

*PARMTEQM/pteqHI* provide special treatment for particles that get ahead or behind the external field mesh. If ahead, the particle "sits out" of the external field integration until the moving mesh reaches it, and then drops back into the mesh (Fig. 13A1.5). This is accurate and there is no loss of information, e.g., space-charge kicks. The accelerated particle curves in Fig. 13A1.4 do not include the particles temporarily ahead of the computational mesh.

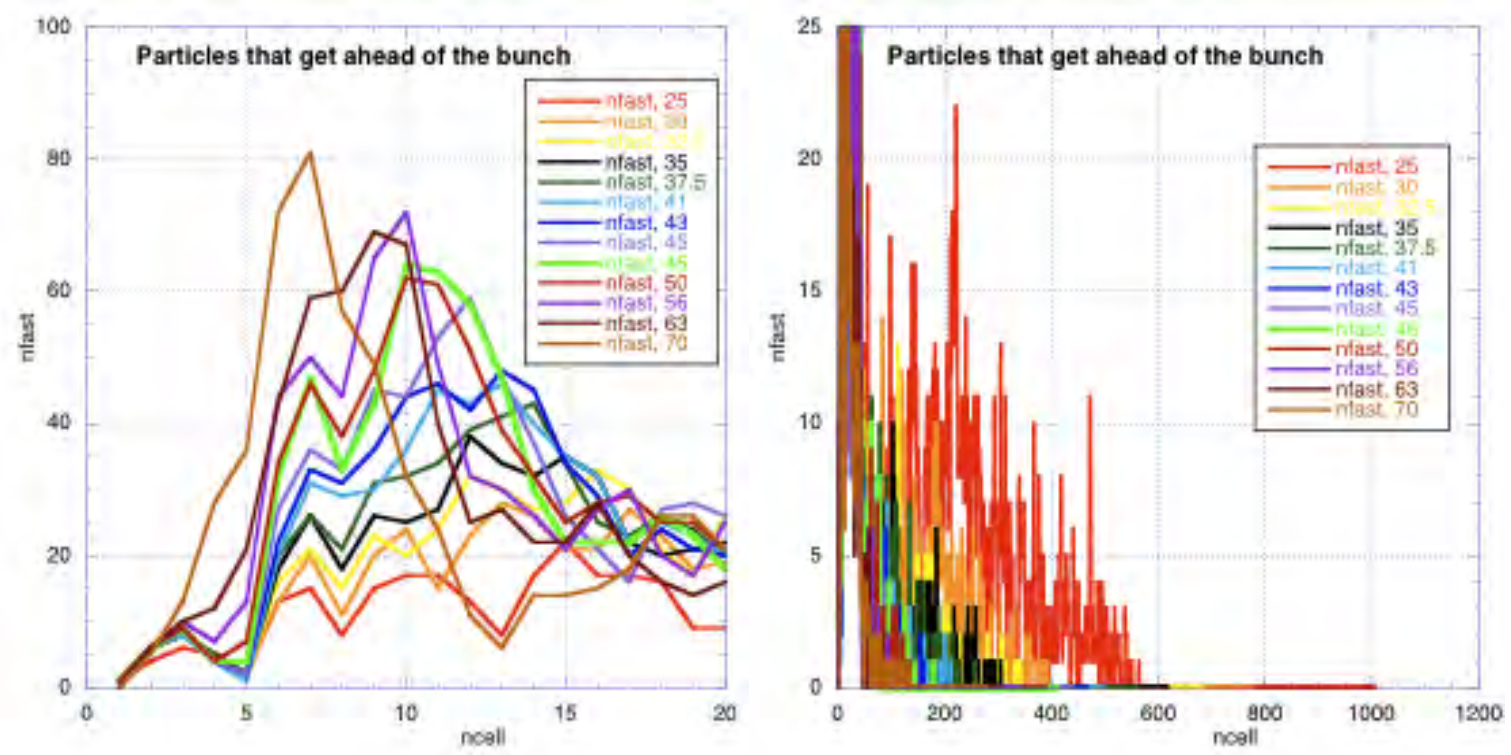

Fig. 13A1.5. Particles which temporarily get ahead of the computational mesh.

Particles that fall behind the mesh are separately re-advanced into the mesh at each step, by applying extra external field kicks and space charge kick based on point-to-point interaction of the single particle and the charge of the main bunch at its centroid. This method is also used in *pteqHI*. (e.g., *BEAMPATH* does not reposition particles (and does not compute space charge forces on particles outside the main mesh)).

Fig. 13A1.6 shows the radial losses and the number of particles outside the acceleration band at each cell.

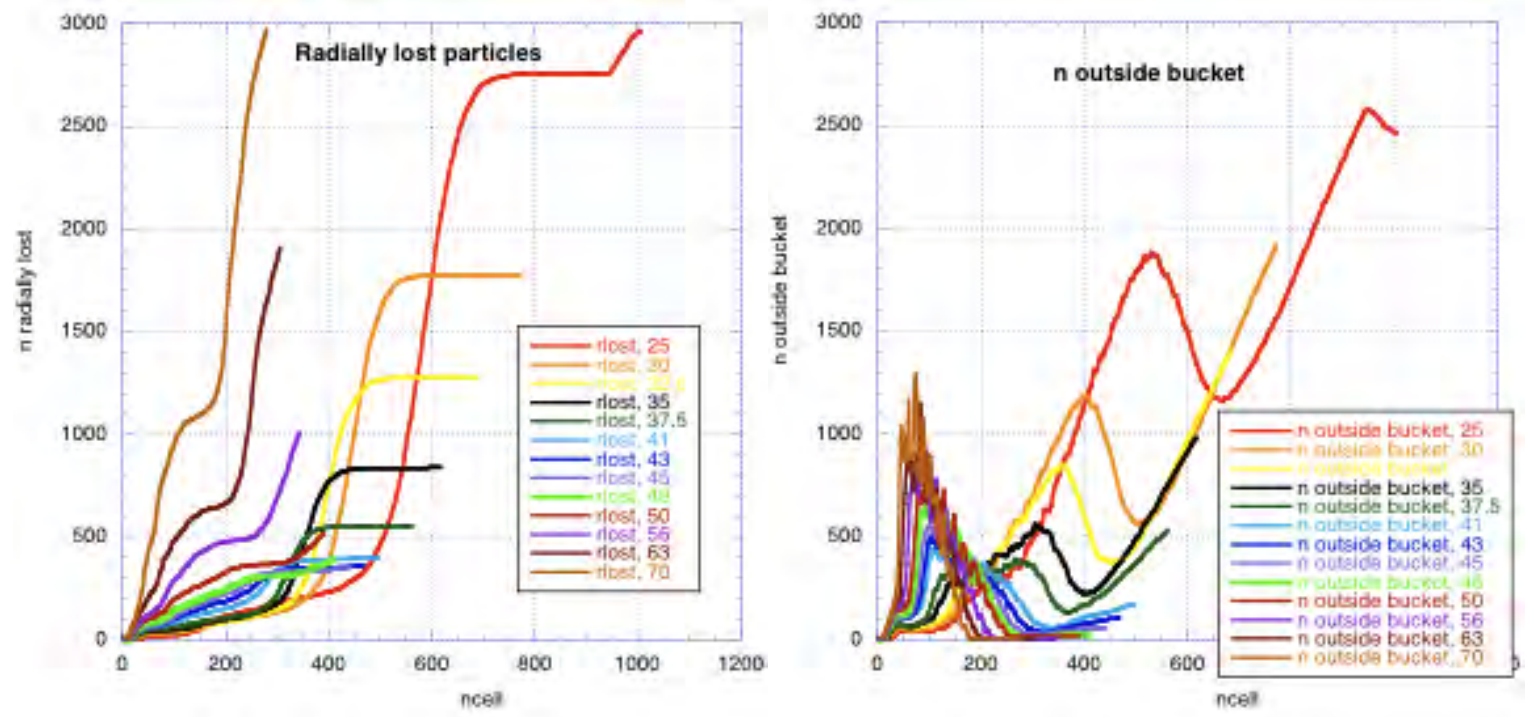

Fig. 13A1.6 Radial losses and particles outside the acceleration band at each cell.

Figure 13A1.7 shows typical x-y characteristics of an output beam, indicating the radial loss criteria.

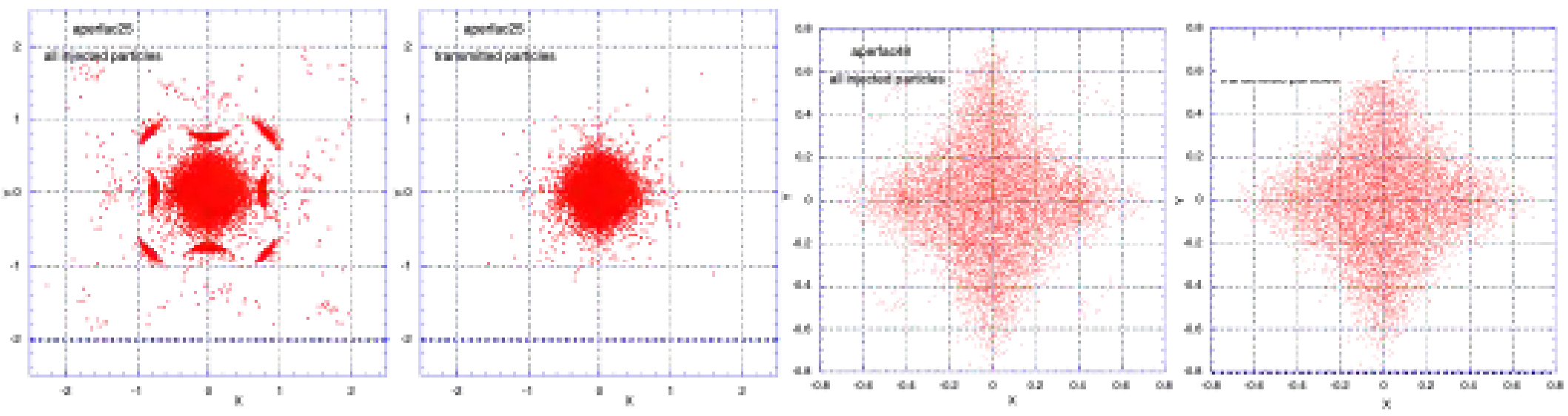

Fig. 13A1.7. x-y plots of particles at the end of the *aperfac*25 RFQ (left) and *aperfac*48 (right), showing all injected particles, and particles transmitted to the end.

## SPACE CHARGE PHYSICS CHARACTERISTICS

The accelerated beam fraction at the output is usually the only judgment made of RFQ performance, largely because it has been deemed too difficult to analyze what is happening cell-by-cell along the RFQ. Also, other design programs do not afford the degree of space-charge physics control possible with *LINACSrfq*, in which arbitrary rules for phase advances, *a priori* emittance growth, equipartitioned beam, etc., can be required. Then the ability of the actual RFQ to reproduce the design conditions needs to be confirmed, and the beam loss pattern is also of interest to determine shielding requirements, etc. The *LINACSrfq* and *pteqHI* methods outlined above afford such analysis [6,8,9], and it is interesting to investigate the underlying characteristics of the *aperfac* family to understand the Fig. 13A1.1 transmission result.

The *aperfac* RFQs are all designed by *LINACSrfq* to reach an equipartitioned condition at the end of a shaper section, where the ratios eln/etn-b/a=sigt/sigl = 1.6. The rms envelope equations do achieve this requirement in *LINACSrfq*. However, it is not guaranteed that the actual (simulated) RFQ will succeed to meet all the design requirements – this requires design optimization. Only the RFQs near the optimum transmission in Fig. 13A1.1 actually have equipartitioned performance, as shown in the following figures (Fig. 13A1.8).

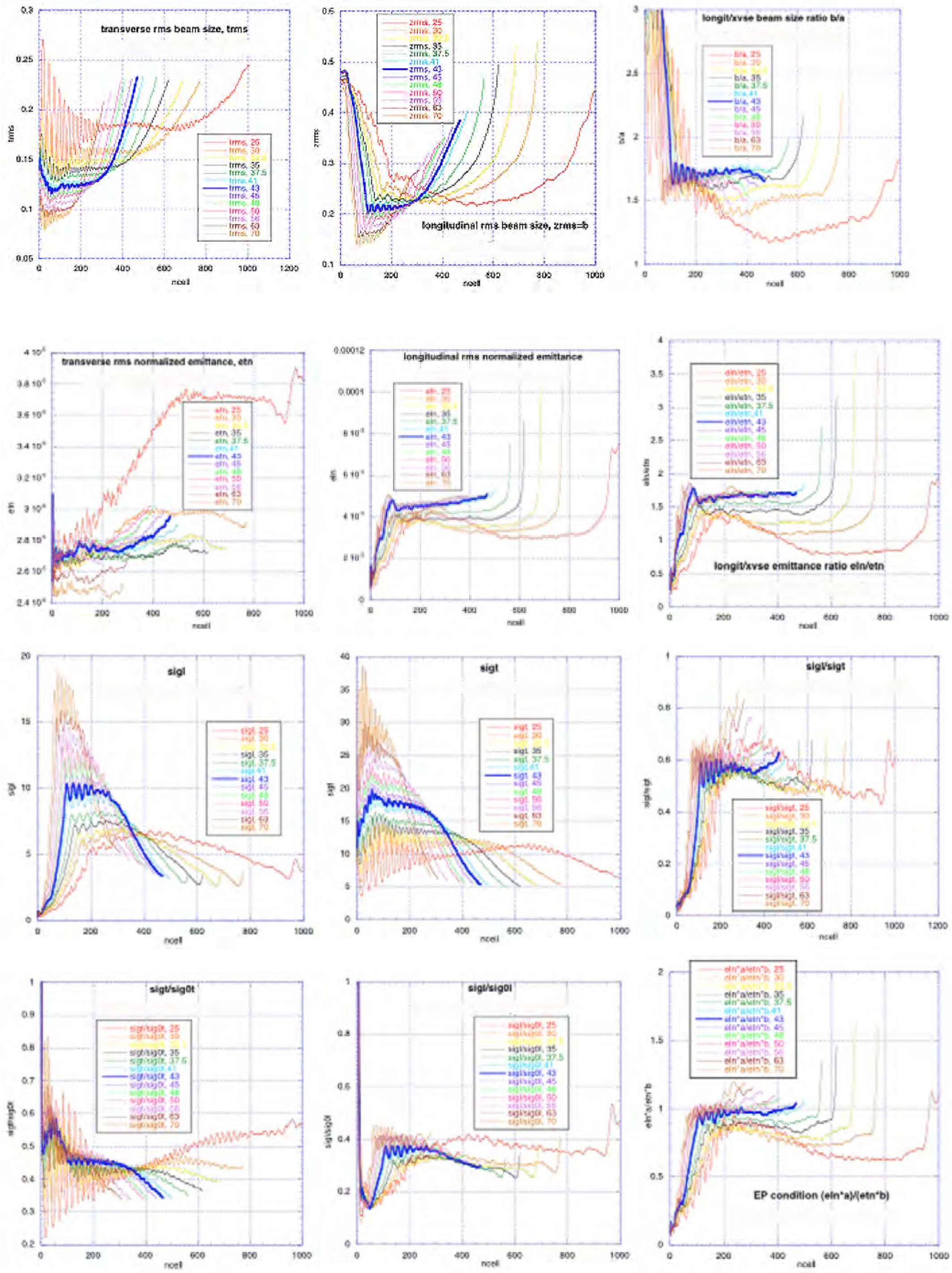

Fig. 13A1.8. The first three rows show the rms beam sizes, rms normalized emittances, the rms phase advances per βλ, and the associated equipartitioning ratios. The fourth row shows the transverse and longitudinal tune depressions and the equipartitioning (EP) condition. Only the RFQs near the optimum of Fig. 1 can actually attain the EP condition throughout the RFQ.

The space charge physics occurring along the RFQ is primarily concerned with resonance interactions and particle loss to the aperture, and is clarified using the Hofmann Chart [7,6] and other information, here for the bad RFQs at *aperfac*25 and *aperfac*70, and for an optimal one at *aperfac*48.

*Aperfac*25 RFQ

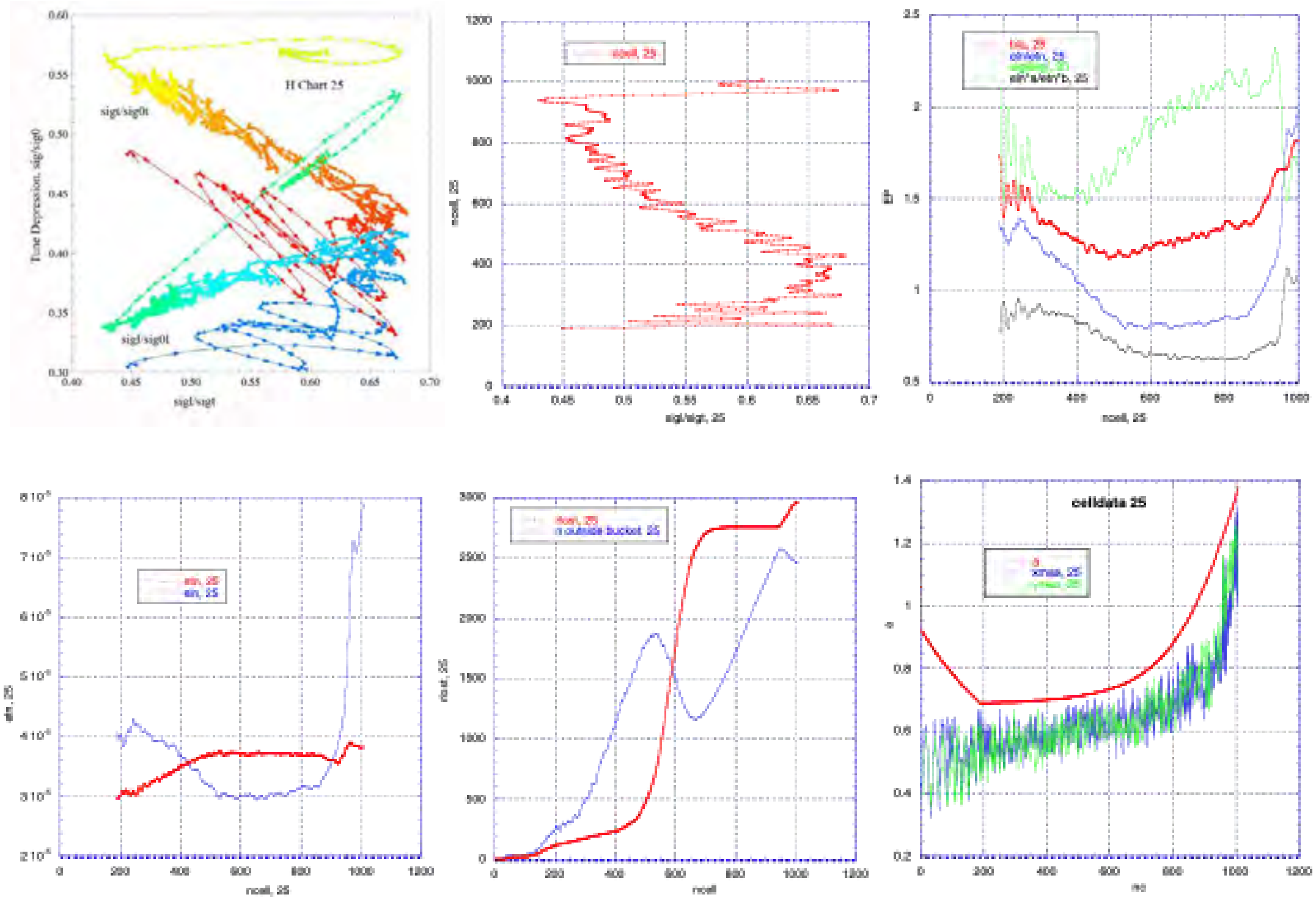

Fig. 13A1.9. Space charge physics characteristics for *aperfac*25 RFQ. First graph is the Hofmann Chart, with trajectories from the shaper end to the RFQ end. The sigt/sig0t trajectory starts with red, and sigl/sig0l starts with blue. Second graph plots ncell vs. sigl/sigt to aid the eye with the Hofmann Chart. Third graph shows the equipartitioning ratios and the equipartitioning condition (eln*b)/(etn*a). Fourth graph is rms emittances, fifth shows radial losses and the number of particles that have fallen out of the bucket but are still being transmitted. Sixth graph shows the aperture and the x- and y- maximum beam sizes.

The *aperfac*25 RFQ has heavy losses both radially, and longitudinally from the bucket. The minimum aperture is extremely large, at βλ/2.5 almost as large as the cell length βλ/2. In addition, this RFQ has been given a poor input match (Fig. 13A1.8) – this introduces another code comparison aspect. (Bad match removed in more recent runs.).

The shaper end is at cell 188; there are essentially no losses in the shaper.

In Fig. 13A1.9, the third graph shows that equipartition, where (eln*b)/(etn*a)=1, is never attained, and therefore all the resonances on the Hofmann Chart are active. The initial tune depressions are strong, with sigt/sig0t~0.5 and sigl/sig0l~0.3. st/s0t initially decreases and sl/s0l increases, and the 0.5 resonance is strongly involved from the end of the shaper at cell 188 to ~cell 450. (A loop in the Hofmann Chart trajectory indicates temporary trapping in a resonance; a kink back and forth indicates strong attraction by the resonance.) The transverse emittance is growing, equipartitioning with the longitudinal emittance is occurring, and particles are being lost from the bucket.

The acceleration rate rises from ~cell 450. Many resonances are involved as sigl/sigt decreases from ~0.65 to ~0.45. Radial losses increase, including particles that have left the bucket. The aperture acts as a strong scraper and limits the rms emittance of the analyzed particles (that successfully reach the end of the RFQ); the maximum x,y extent of these particles is smaller than the aperture. Between ~cells 450-950, as the cells get longer, st/s0t increases, space charge resonance spreading becomes weaker in this plane, and the radial loss levels off. sl/s0l decreases, and the trajectory enters the strongly resonance overlap region below s/s0~0.4. Particles are again lost from the bucket. From ~cell 950, eln increases very rapidly and sigl/sigt grows. A resonance at ~5/8=0.625 causes immediate emittance and radial beam loss reactions.

*Aperfac*48 RFQ

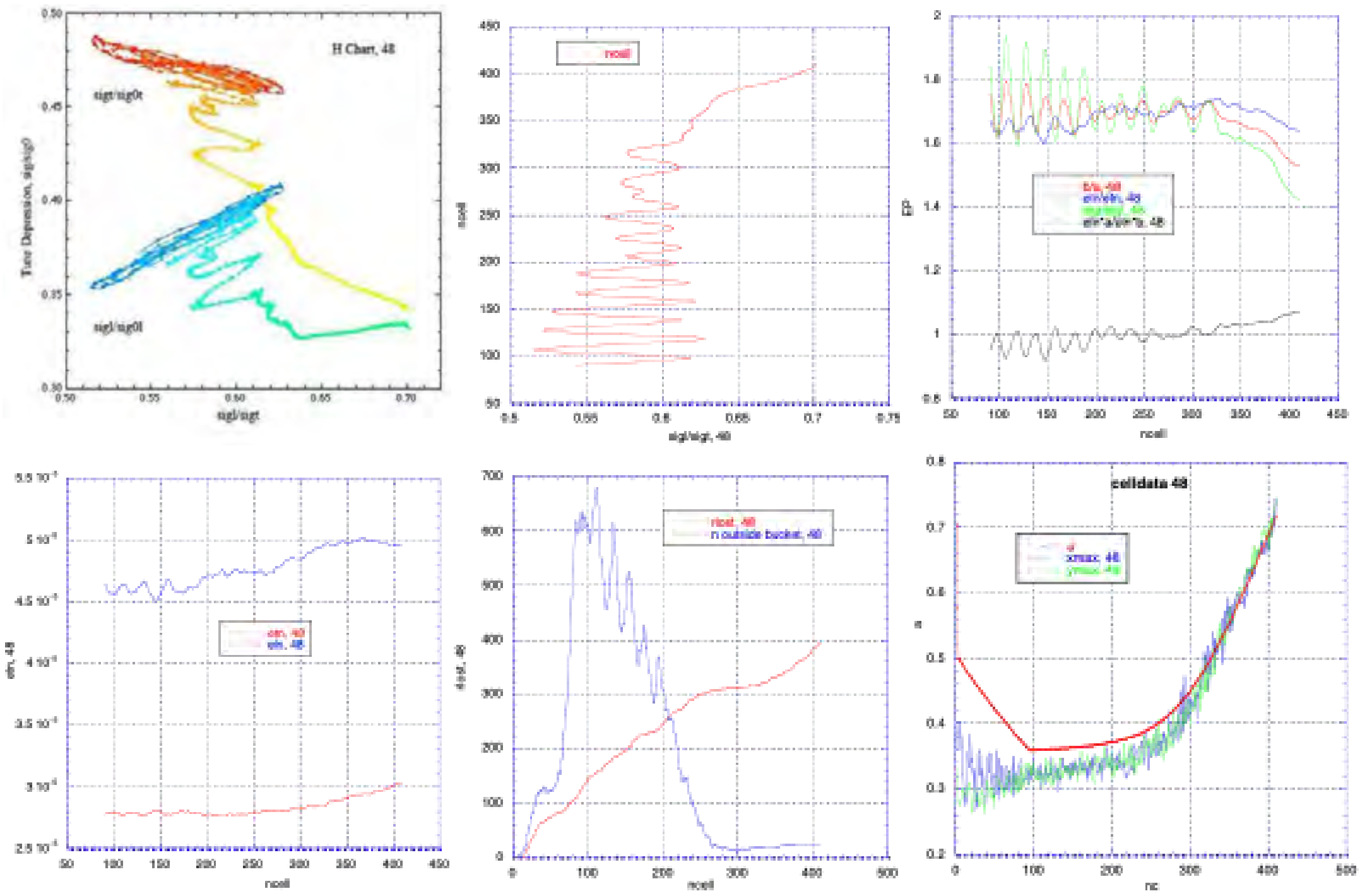

Fig. 13A1.10. Space charge physics characteristics for *aperfac*48 RFQ

The *aperfac*48 RFQ end of shaper aperture is near the optimum accelerated beam fraction (Fig. 13A1.1). It comes close to meeting the design objective of a fully equipartitioned RFQ. The shaper end is at cell 90; some particles have left the bucket in the shaper and are then lost radially. EP is attained at a ratio of ~1/1.7 = 0.588 from the shaper end to ~cell 300. In this zone, the rest of the particles leaving the bucket in the shaper are lost radially. The 0.5 resonance is crossed repeatedly in both directions, but there is little free energy for resonance excitations and the emittances remain nearly constant. After ~cell 300, the EP condition departs to EP~1.08, which would still be relatively safe with less tune depression. However, the tune depressions become large, down to 0.3, in the strongly resonance overlap zone, causing some emittance growth and increase in loss. Successfully transmitted particles are analyzed at mid-cell, where the vane tip is at r0, and there can have maximum x,y extent somewhat beyond the minimum aperture.

*Aperfac*70 RFQ

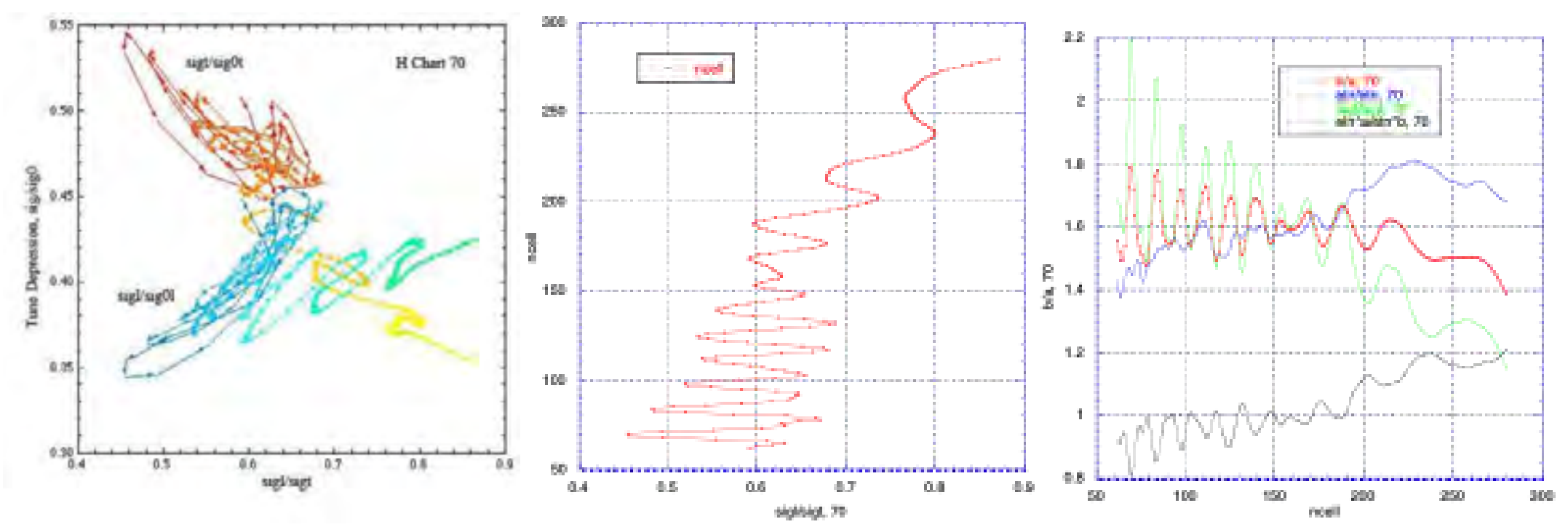

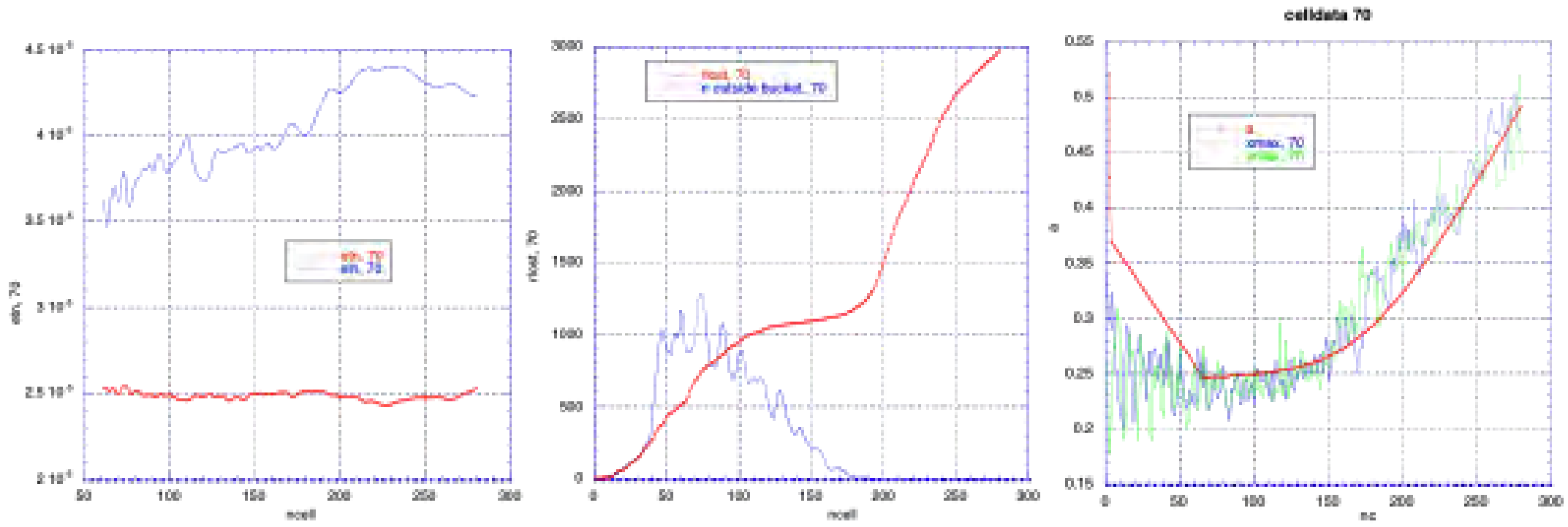

Fig. 13A1.11. Space charge physics characteristics for *aperfac*70 RFQ

The *aperfac*70 RFQ is mostly weak radially. The tune depressions of all the RFQs in the family are similar, but here the aperture is too small and the radial losses are large. The shaper end is at cell 61. The design EP goal is not attained there as the longitudinal focusing and space charge forces are not appropriate to give the mixing required for eln to equal 1.6*etn. Particles lost from the bucket in the shaper are immediately lost radially. The 0.5 resonance is involved from cells 61-100, particles are lost from the bucket. The design EP goal is partially attained between cells100-180, with no transverse emittance growth. Particles lost from the bucket earlier are being lost radially. Then no more particles are lost from bucket. There is departure from EP after ~cell 150; sigt/sig0t decreases down to 0.35, into the strong resonance region. Attraction to resonances around 0.6, 0.7, and 0.8 is evident, with emittance growth and radial losses. Successfully transmitted particles are analyzed at mid-cell, where the vane tip is at r0, and there can have maximum x,y extent somewhat beyond the minimum aperture.

<u>Beam Compactness</u>

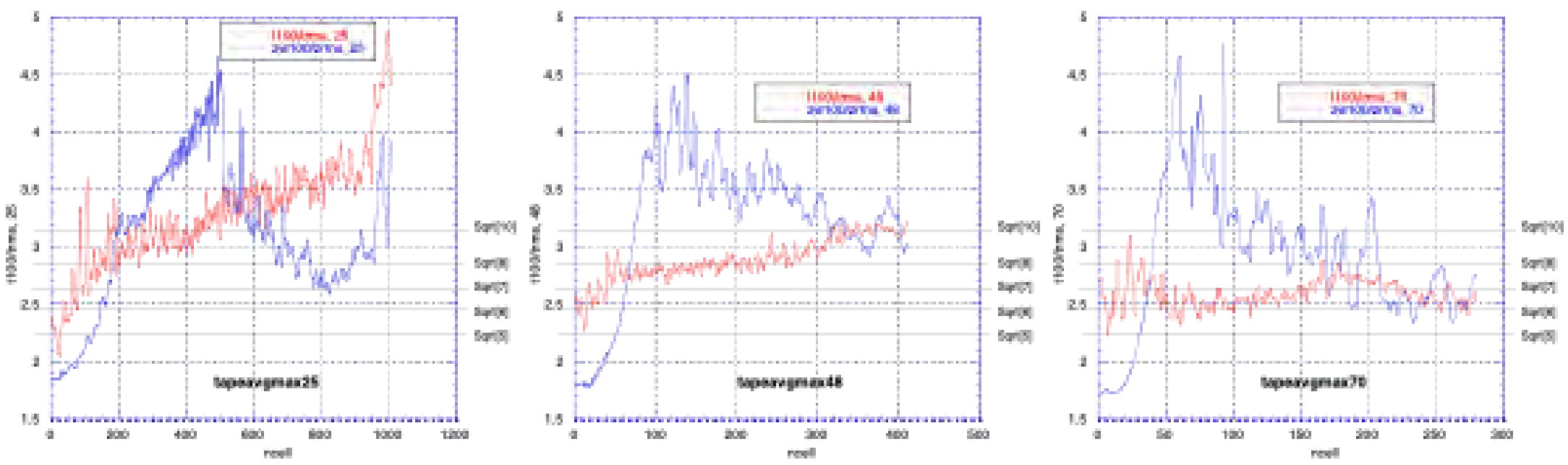

Fig. 13A1.11. Ratios of the transmitted particles' maximum transverse and longitudinal beam extent to the rms extent, t100/trms and zw100/zrms.

The equipartitioned beam strategy, in which there is no free energy to drive parametric resonances, results in little emittance growth and halo – the beam remains more compact. A uniform beam has a Sqrt[5] ratio of total to rms size, a waterbag Sqrt[6], and a Gaussian distribution cut off at three standard deviations has a ratio of ~Sqrt[10]. Here 10K particles were injected with a waterbag transverse distribution and a uniform longitudinal distribution.

The transmitted particles from the *aperfac*70 RFQ have highly spread distributions. The tails are actually higher than Gaussian, intrinsic to the dynamics, which is characterized by resonance phenomena and not by long-term thermalization (i.e., local, but not overall, chaos in the phase space).

The *aperfac*48 RFQ beam has initial transverse spreading in the shaper as the longitudinal bunch is formed, and the longitudinal bunch has a long tail until most of these particles are lost radially. A slow transverse widening then occurs, typical of multiple small resonance interactions, and this remains because there is enough free aperture and only small radial loss. This RFQ design could be further optimized with *LINACSrfq* to achieve fully EP performance by accounting for the change in the space charge form factor.

The *aperfac*70 RFQ transverse distribution remains near waterbag because of heavy losses to the aperture.

## SUMMARY

The *aperfac* RFQ family is generated by changing one RFQ parameter, the minimum aperture at the end of the shaper, with consequent effect on the other completely specified design rules. This affords a consistent experimental framework for simulation code comparison, in contrast to a patchwork of anecdotal cases. The accelerated beam fraction has a clear optimum (Fig. 13A1.1). The work of the designer is to find such an optimum, and it is important to have confidence that the physically correct optimum is found. It is known that different codes find different optima, and it is useful to investigate why. For code comparison study, the simplest and mostly only used criterion is the percentage of accelerated beam, Fig. 13A1.1. Determining the reasons why different codes produce different curves for the *aperfac* family using only this criterion, reported elsewhere, has been very intricate. Other comparison criteria can readily be imagined, and could also be possible in a controlled experiment, using the capabilities of the design code *LINACSrfq*, where all parameters, including the desired space charge physics, are under the control of the designer.

**References for this appendix**

## *Chapter 13 - Appendix 2. Synopsis of Ref. [2]*

**Jameson, Robert A., LA-UR-07-0876, A Discussion of RFQ Linac Simulation, Los Alamos National Laboratory, 10/2009 (re-publish of LA-CP-97-54, September 1997).**

- z-code dynamics subroutines rfqdyn.f, tcord.f, and calctm.f

*Abstract*

Intended studies of new linac design techniques, using the RFQ as an example, required a computer simulation code that is fast but yet of sufficient physical accuracy. Source code had to be available, as continuing modifications were planned, and results had to be open to IFMIF project members. Extensive investigation of the original PARMTEQ codes with time or position as independent variable (t-code or z-code) was required to fully substantiate the requirements and to produce working tools. While beam transmission may be similar, differences in the space-charge treatment in the z- vs. t-codes result in somewhat different rms ellipse parameters and beam emittance along the RFQ, and in different beam loss patterns. This affects the validity of the simulation for detailed purposes, e.g., the study of the space-charge physics in the channel, design of the channel to have special properties such as equipartitioning, and beam loss patterns for residual activation and shielding estimates. The conclusion is that a time-domain simulation should be used. Use of a fast r-z mesh method for space-charge is sufficient for many purposes, generally agreeing qualitatively and reasonably well quantitatively with slower 3D methods.

Detailed effects - multipoles, images, end-cell details, vane-shape details, transition cells, etc., were not investigated. The discussion includes comment on comparison to experimental results, where apparently overestimated transmission by the z-code had led to concern about the simulation veracity. The material studied here indicates that the z-code tends to overestimate transmission by a few percent compared to the t-code, but both the PTEQ simple t-code and the powerful RFQTRAK indicate high transmission compared to experiment. It appears that the challenge passes to a better simulation of the complete experiment, including, for example, a complete model of beam neutralization. The characteristics of bucket-filling in the RFQ bunching/accelerating channel are discussed. The general fundamental physics conclusion concerning time-domain simulation applies to linac dynamics simulation in general, and may argue for the use of t-code procedures in bunchers and in higher-energy, high-intensity linacs. The conclusion is critical to estimation of particle losses. It may be important up to quite high energies, because although the effects decrease with beam bunching and higher velocity, beam current loss tolerances vary inversely with energy.

*Introduction and Overview*

In early 1995 at the Institute für Angewandte Physik of the Johann-Wolfgang-Goethe Universität in Frankfurt-am-Main, Germany, a visiting Chinese graduate student, Li Deshan, was doing a doctoral dissertation on the RFQ, and requested that I involve him if possible in some advanced problems of the RFQ. At the time, interested in fundamental questions of energy balance in multidimensional beam dynamics systems and the avoidance of beam halos, I felt that the use of simpler systems using already bunched beams would be easier than working with the full dc-to-bunched beam conversion and subsequent acceleration of the full RFQ. However, Li needed help, and it was very unusual and nice for me to have a helper, so we started to explore using the RFQ vane modulation recipes to produce equipartitioned beams.

After Li's graduation and necessary departure from physics to the business arena, given present economic conditions in China, I decided the RFQ is in fact a very interesting vehicle for fundamental studies, in that it does contain all the processes, and also because it is the same problem as when an pre-linac injects into a subsequent linac operating at a higher harmonic, where the beam will start out expanded again in phase and usually needs to be rebunched. So work began in earnest on the RFQ vane parameter generation program and then on simulation runs of RFQ's generated to satisfy certain rules regarding equipartitioning. The result was the IFMIF RFQ conceptual design, in which the beam is equipartitioned after the shaper.

The code used was PARMTEQB, a "standard" version widely distributed by the AT-Division at Los Alamos, which uses z-position as the independent variable (we will name this a "z-code") along with some adjustment for the space-charge computation as explained below. It does not contain multipole, image charge, or several other effects incorporated into more modern versions of PARMTEQ. The source code for these newer versions is unavailable. PARMTEQB uses an r-z mesh for space-charge computation and is fast compared to the more complex versions. A fast code was desired because the main interest was in r-z physics and in optimization, requiring many runs.

Close observation of various emittance patterns and other diagnostics of PARMTEQB raised concerns about the space-charge computation. Investigation of these questions has necessitated a lengthy detour from the main research path, which is to be continued and reported elsewhere.

To check PARMTEQB, it was necessary to use a code with similar characteristics (fast, space-charge subroutine, etc.) but with time as the independent variable (t-code). A sister code, PARMTEQC, was available from the early RFQ work at Los Alamos. PARMTEQB and PARMTEQC were combined into a code named PTEQ for flexible operation and comparisons.

The source code and numerous questions and bugs encountered were communicated to the IFMIF[107] Accelerator Team. Surprise was expressed by some partners at the simplicity of the dynamics formulation in these codes. At this point, it was decided to initiate a worldwide collaboration of IFMIF accelerator team members and others to explore all aspects of RFQ dynamics and simulation, the first such effort. At present, the main codes in use are either based on PARMTEQ or Russian codes programmed independently from PARMTEQ, using the formulations of Kapchinsky and Teplyakov. A meeting of this team was hosted by Saclay under the auspices of the IFMIF Accelerator Team from 26-30 May 1997 in Paris. RFQ simulation was the primary topic; results from this meeting are included herein at appropriate points.

The report is written to take both the general reader interested in the beam dynamics physics issues and the RFQ simulation code expert through the rather complex and interrelated aspects. The first section addresses several basic questions that arose from the use of the z-code PARMTEQB. These lead to a discussion of the several coordinate systems that can be used, and to the necessity to use a code with time as the independent variable to compare with PARMTEQB. The PARMTEQB z-code and PARMTEQC t-code, essentially the same in structure, are consolidated into one code called PTEQ.

PTEQ is then correlated with some other PARMTEQ versions. A time-domain code that had evolved from PARMTEQC via CRNL and SSCL was very useful in elucidating further aspects of synchronous particle diagnostics, transverse loss criteria, and the space-charge treatment of slow/fast particles that stray outside the moving frame used in the dynamics computation.

Input matching is discussed next. A problem in scrmstm.f, the t-code space-charge subroutine for the radial matching section, caused confusion for a long time, but then, as expected, the input match parameters for the t- and z- codes are nearly the same. A new nonlinear optimization technique for matching using the full beam produces a better match than earlier procedures using TRACE-3D and RFQUIK.

Results from the t-code are presented and compared with the z-code. The t-code exact space-charge treatment, compared to an approximation in the z-code, results in somewhat different rms ellipse parameters and emittance along the RFQ, and in different beam loss patterns.

The conclusion is that a time-based code should be used.

It has been noted at several laboratories that the "original" PARMTEQB z-code apparently overestimated transmission compared to experimental results. Multipole and image charge effects were introduced; it was found that transmission of some designs was lowered. Modified design procedure raised the vane voltage or the modulation until the quadrupole focusing again equaled the strength it had before multipole or image charge effects diminished it. It was then observed the transmission results were very similar when the simulation was run either with or without these effects turned on. Some comparisons were made to the PARMTEQC code in several incarnations; results, usually in terms of transmission, were reported to be "similar".

However, discussion of the simulation code veracity continued. For this study, design files were obtained on a number of RFQs that have been tested experimentally. These files include the input match conditions found by the designers, typically from a z-code based procedure. The transmissions predicted by the PTEQ z- and t-routines with that matching condition are compared in the next section, and to the published simulation and experimental results. The transmission predicted by the PTEQ t-code is consistently lower by a few percent than that predicted by the z-code, but both the PTEQ simple t-code and the powerful RFQTRAK indicate high transmission compared to experiment. It appears that the challenge passes to a better simulation of the complete experiment, including, for example, a complete model of beam neutralization.

The discussion is extended to the effects of bunching and acceleration, and the bucket formation and filling from the bunching through the accelerating stages in an RFQ.

Finally, the report is summarized, indicating that time should be used as the independent variable in RFQ simulations, in order to do the space-charge computations correctly, and that it is important to do input matching well. These conclusions are important to the problem of beam loss assessment and consideration of new design techniques.

The general fundamental physics conclusion concerning time-domain simulation applies to linac dynamics simulation in general, and may argue for the use of t-code procedures in bunchers and in higher-energy, high-intensity linacs. The conclusion is critical to estimation of particle losses. It may be important up to quite high energies, because although the effects decrease with beam bunching and higher velocity, beam current loss tolerances vary inversely with energy.

Appendices give details of the code conversion and development, a flow-chart for the PTEQ t-code, input and output files for the IFMIF RFQ, input files for other RFQs tested, and the PTEQ source code for the main dynamics subroutines.

[eltoc]

## *Chapter 13 - Appendix 3. Integration of the Equations of Motion*

IAP Memo RAJ-20091125

### Difference Equation Integrator Stability

Particles moving with Newton's laws of motion are approximated by discretization, which needs to be evaluated carefully [1]. Consistency, accuracy, and stability are considered here.

**The "leapfrog scheme"** is presented as

$$(x^{n+1} - x^n)/dt = v^{n+1/2} \qquad (4\text{-}3)$$
$$(v^{n+1/2} - v^{n-1/2})/dt = F(x^n)/m \qquad (4\text{-}4)$$

Eliminating velocity using $(x^n - x^{n-1})/dt = v^{n-1/2}$

$$x^{n+1} - 2x^n + x^{n-1} = (F(x^n)/m)dt^2 \qquad (4\text{-}14)$$

In general: $a_{k-}\sum_{i=0}^{k} {}_i x^{n+k-i} = (dt^2/m)\, b_{k-i} F^{n+k-}\sum_{i=0}^{k} {}^i$, where k=1 for leapfrog.

If $b_k = b_1$ is zero, the scheme is termed explicit, because $F^{n+k}$ does not enter the right=hand side of the equation, and $x^n$ is found directly from known quantities. If $b_k \neq 0$, the scheme is implicit because $x^n$ must be found iteratively unless F is a simple function of x. The leapfrog scheme (4-3,4-4) is explicit.

It is also "consistent" in that the differential equation it represents is time-reversible, and that requires that the time-reversible derivatives are obtained from time-centered derivatives. $(x^{n+1}-x^n)$ is centered about $t^{n+1/2}$ and the acceleration term $(v^{n+1/2} + v^{n-1/2})$ is centered about $t^n$.

Truncation errors occur when continuous quantities are represented by discrete sets of values; following [1], the leapfrog integrator is accurate to 2nd order $dt^2$.

Integrator stability:

The integrator must be stable, in that the truncation errors (or round-off errors, usually much smaller) do not grow enough to significantly affect the numerical result. Following the root-locus method in [1] (also familiar from modern control theory), the error propagation equation is

$$\varepsilon^{n+1} - 2\,\varepsilon^n + \varepsilon^{n-1} = -X^2\,dt^2\,\varepsilon^n$$

The characteristic equation is

$$\lambda^2 - 2\,\lambda + 1 = -X^2\,dt^2\,\lambda$$

Solving for the roots λ and plotting vs. X, Fig. 13A3.1 shows the leapfrog root-locus diagram:

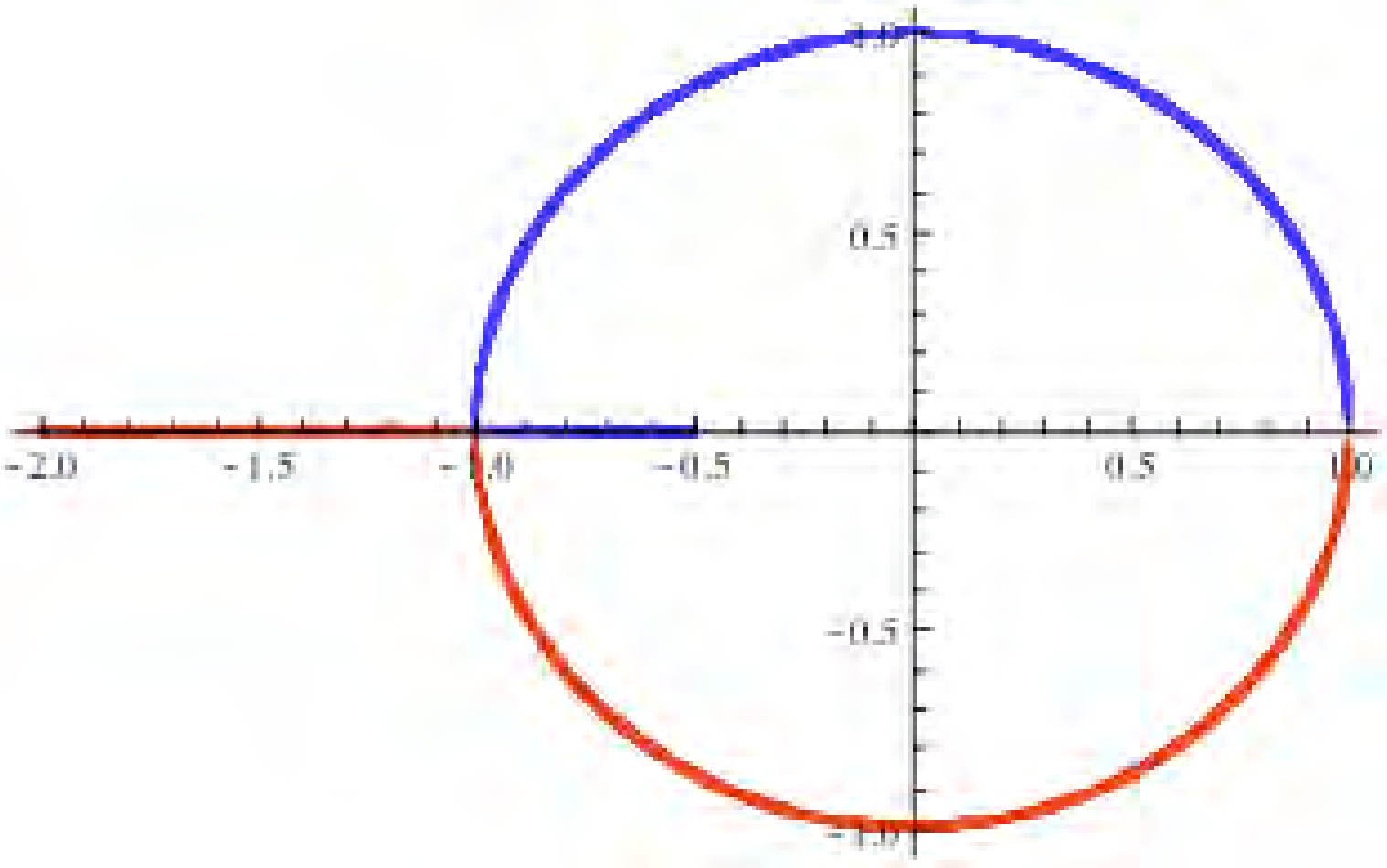

Fig. 13A3.1 Root-locus diagram for stable leapfrog integrator.

Starting at λ =1 for X=0, the roots traverse the unit circle until λ =-1 at X=4, and then become real for larger X. As long as the roots are on the unit circle (as they are for suitably small step size), the integrator is stable.

Plotting the error propagation equation for initial round-off error estimate of $10^{-20}$ using a step size of 1/20 cycle, the error oscillates with the initial condition amplitude (Fig. 13A3.2):

```
e0=10^(-20);e1=1.001*10^(-20);dt2=4*(1/20)^2;
For[{en=e1,enm1=e0,i=1,estep={}},i<1001,i++,
        enp1=2*en-enm1-dt2*en;
        If[i==1000,Print[enp1/en]];
        estep=Append[estep,enp1];
        enm1=en;
        en=enp1; ];
ListPlot[estep,Joined->True,PlotStyle->Thick,PlotRange->All]
1.05613
```

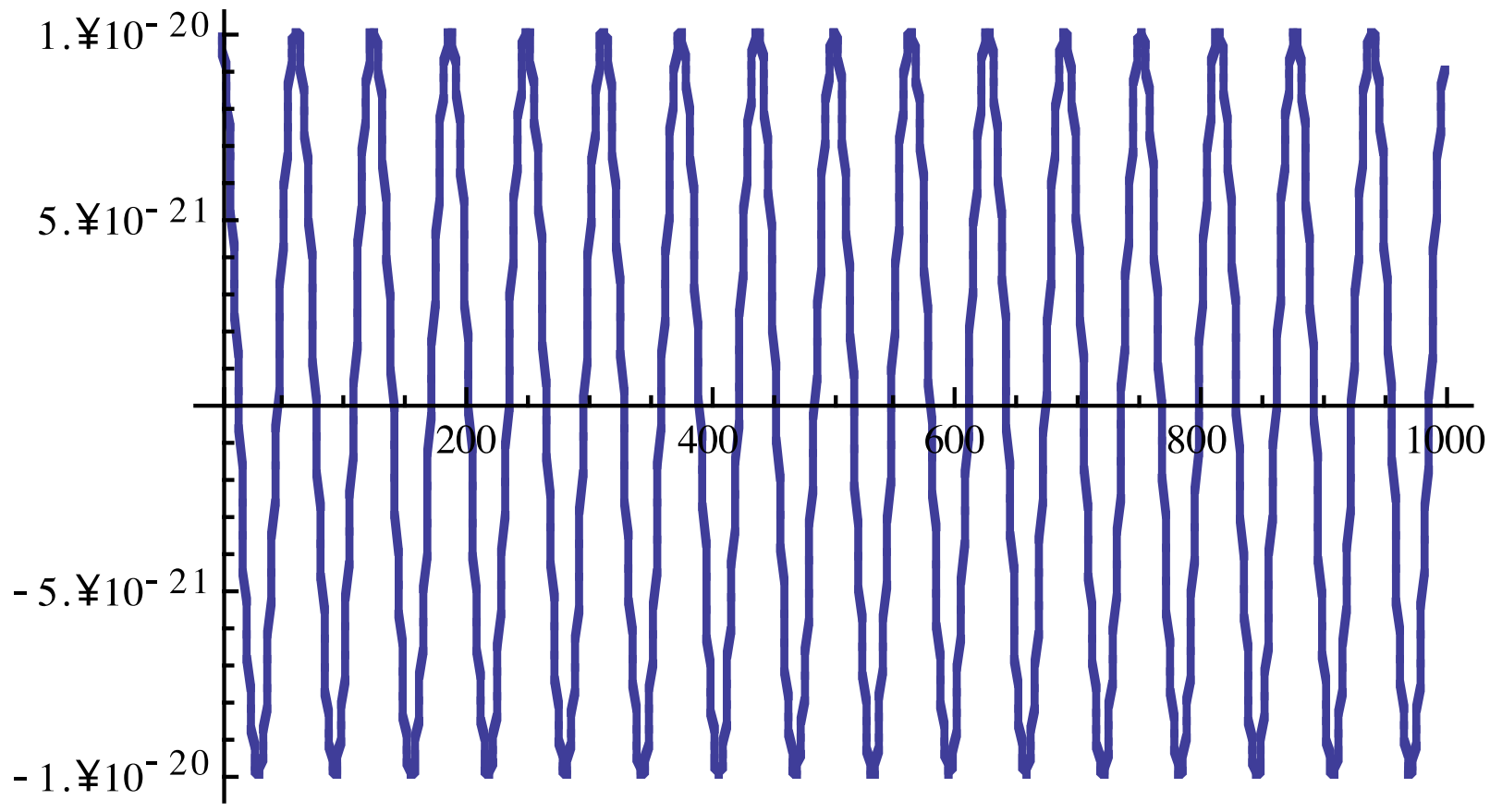

Fig. 13A3.2 Error propagation for stable leapfrog integrator.

***pteqHI/PARMTEQM*** **integrator** implements the leapfrog as follows:

Using $v^{n-1/2}$, x is advanced by ½ step, where a new acceleration term is computed and applied to the velocity over a full step, and then x is again advanced by ½ step using the new velocity:

$$x^{n+1/2} = x^n + (dt/2)v^{n-1/2} \quad -> \quad v^{n-1/2} = (x^{n+1/2} - x^n)/(dt/2)$$
$$v^{n+1/2} = v^{n-1/2} + (F(x^n)/m)dt \qquad (1)$$
$$x^{n+1} = x^{n+1/2} + (dt/2)v^{n+1/2} \qquad v^{n+1/2} = (x^{n+1} - x^{n+1/2})/(dt/2)$$

yielding $x^{n+1} - 2x^{n+1/2} + x^n = (F(x^n)/m)(dt^2/2)$.

*pteqHI/PARMTEQM* integrator stability:

The error propagation and characteristic equations equations are:

$$\varepsilon^{n+1} - 2\,\varepsilon^{n+1/2} + \varepsilon^n = -X^2\,(dt^2/2)\,\varepsilon^{n+1/2}$$
$$\lambda - 2\,\lambda^{1/2} + 1 = -X^2\,(dt^2/2)\,\lambda^{1/2}$$

Solving for the roots m gives the *pteqHI/PARMTEQM* leapfrog root-locus diagram (Fig. 13A3.3):

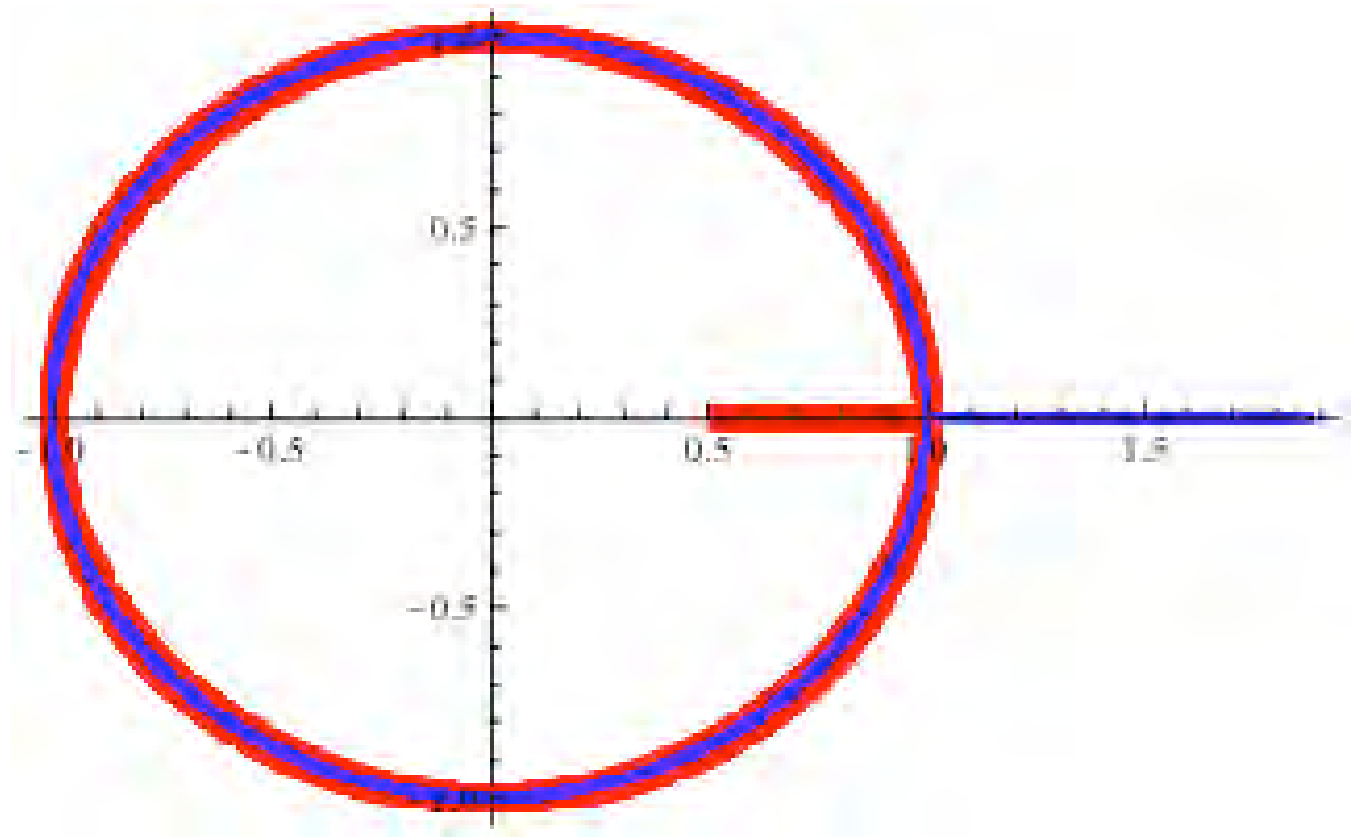

Fig. 13A3.3 Root locus diagram for stable *PARMTEQ/pteqHI/LINACSrfqSIM* integrator.

The roots traverse the unit circle in opposite directions from $\lambda = 1$ at $X^2\,(dt^2/2) = 0$ back to to $\lambda = 1$ at $X^2\,(dt^2/2) = 4$ and the integrator is stable for dt^2 in that range. At $X^2\,(dt^2/2) = 4$, the sinusoidal solution is sampled only ~twice per cycle. When $X^2\,(dt^2/2) > 4$, the roots are on the real axis and errors overwhelm the solution.

The error propagation plot is the same as for the leapfrog above.

**The integrator used in Batygin's *BEAMPATH* simulation code**, Ref.[2], which also refers to [1], is now very helpful, containing a detailed description of particle motion in an electric field and aspects of the integration of difference equations[108].

The *BEAMPATH* integrator for use with an RFQ is an implicit scheme with n+1 on both sides of the equations, requiring that iteration be applied in the algorithm. The iterated form was tested separately in *pteqHI* during the *BEAMPATH* comparison study and presented no problems.

$$\vec{p}_{n+1} = \vec{p}_n + \tau \vec{E}_n, \quad \vec{x}_{n+1} = \vec{x}_n + \tau \frac{\vec{p}_{n+1}}{\gamma_{n+1}}. \qquad (3.14)$$

where $\tau$ is dt. The algorithm is properly time-centered, and symplectic, meaning that the phase space is conserved (the Jacobian is equal to unity[109]). But it, as also the *pteqHI* leapfrog integrator , conserves the total energy of the particles only to order of dt. The difference is negligible if dt is small enough. Usually this might be of order 0.01 to 0.001*(rf or plasma period), and the code must be tested accordingly.

The stability in terms of error propagation are not discussed in [2], but would be stable, the same as for the *pteqHI* integrator.

**<u>Another integrator</u> has been suggested:**

$$x^{n+1} = x^n + dt\, v^n + (dt^2/2)(F(x^n)/m)$$
$$v^{n+1} = v^n + (dt/2)(F(x^n) + F(x^{n+1}))/m \qquad (4)$$

are consistent in that the differential equation is recovered at the limit dt->0, but are not time-centered, similar to the example (4-6) in [1]. The integrator is also implicit, as the index (n+1) appears on both sides. It may be possible to iteratively solve for the values at time (n+1) with acceptable speed.

The velocity equation advances by ½ time step using the acceleration found at step n and then ½ time step using the value at time n+1, while the position is advanced by a whole time step. This implementation is opposite to that of the *pteqHI* integrator, but is equivalent (the *BEAMPATH* integrator is also like this).

The Jacobian of this integrator is formally equal to one, so it is symplectic.

The 2nd order term in Eq.(4) implies accuracy of the integrator to higher order in dt.

A direct implementation as indicated in (4) without interation was attempted, but some tearing and folding of the longitudinal phase space was noted in the long RFQs with very large apertures of the *aperfac* family of test RFQs, so this attempt was abandoned.

Then (4) was implemented as leapfrog in the same order of steps as the *pteqHI* integrator as:

$$x^{n+1/2} = x^n + (dt/2)v^{n-1/2} + (1/2)(dt/2)^2 F(x^{n-1/2})/m$$
$$v^{n+1/2} = v^{n-1/2} + dt\, F(x^{n+1/2})/m \qquad (6)$$
$$x^{n+1} = x^{n+1/2} + (dt/2)v^{n+1/2} + (1/2)(dt/2)^2 F(x^{n+1/2})/m$$

---

108 Another reason that position, z, is not an appropriate independent variable is also pointed out. A main reason is that space-charge forces cannot be computed properly when all particles are at the same position – they must all be at the same time. But here, Batygin points out that in the z=representation, the right-hand sides of the equation of motion must be divided by the square of the velocity vz2, which is a dependent variable and therefore defined with some error, thus making the integration of the equations of motion less accurate.

109 Formally, the requirement that the Jacobian=1 applies to 1D, and is not enough for 6D. Other variables also have to be preserved. (Alex Dragt – private communication.)

A small but significant, steady and essentially linear growth in transverse emittance was observed, that is not present in (1) which does not have the 2nd order term; therefore this integrator was abandoned.  It is of interest to investigate the root-locus diagram of this integrator; continuing from (6):

$$x^{n+1} = x^n + (dt/2)(v^{n-1/2} + v^{n+1/2}) + (1/2)(dt/2)^2 (F(x^{n-1/2})/m + F(x^{n+1/2})/m)$$

Then using $(x^{n+1/2} - x^{n-1/2})/dt = (v^{n+1/2}) + v^{n-1/2})/2$

$$x^{n+1} = x^n + x^{n+1/2} - x^{n-1/2} + (1/2)(dt/2)^2 (F(x^{n-1/2})/m + F(x^{n+1/2})/m)$$

$$\varepsilon^{n+1} - \varepsilon^{n+1/2} - \varepsilon^n + \varepsilon^{n-1/2} = (1/2)(dt/2)^2 (F(x^{n-1/2})/m + F(x^{n+1/2})/m)$$

And the characteristic root-locus equation is

$$\lambda^{3/2} - \lambda - \lambda^{1/2} + 1 = -X(\lambda + 1)$$

giving the root-locus diagram Fig. 13A3.4:

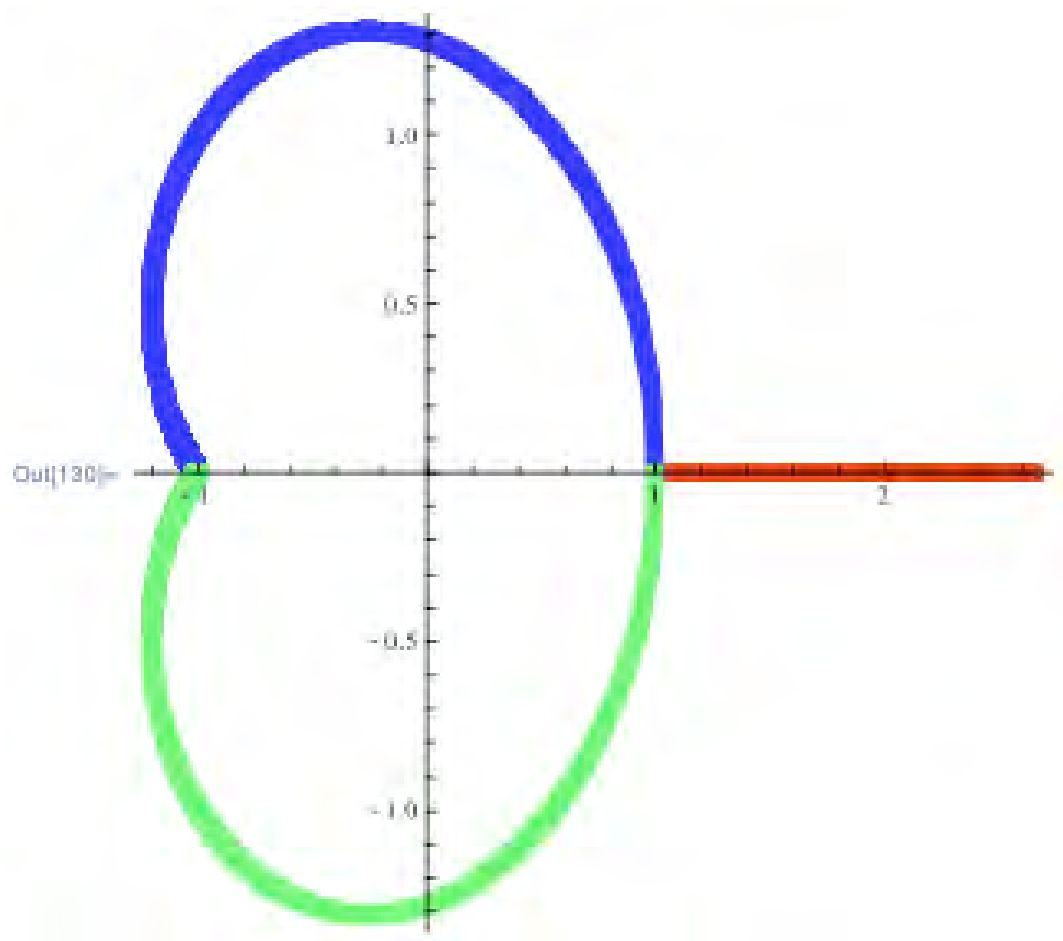

Fig. 13A3.4  Root locus diagram of an unstable integrator.

The extra, parasitic (red) root indicates growth for all dt>0; this integrator, although accurate and symplectic,  is unstable and any truncation or round-off error, however small, with propagate; in the similar example, by more than four orders of magnitude (Fig. 13A3.5).

```
e0=10^(-10);e1=1.001*10^(-10);e1=1.002*10^(-10);dt2=4*(1/20)^2;
For[{enm12=e0,en=e1,enp12=e1,i=1,estep={}},i<10001,i++,
        enp1=enp12+en+enm12 -dt2*(enm12+enp12);
         If[i==10000,Print[enp1/en]];
         estep=Append[estep,enp1];
         enm1=en;
         en=enp1; ];
ListPlot[estep,PlotStyle->Thick,Joined->True,PlotRange->All]
1.0001
```

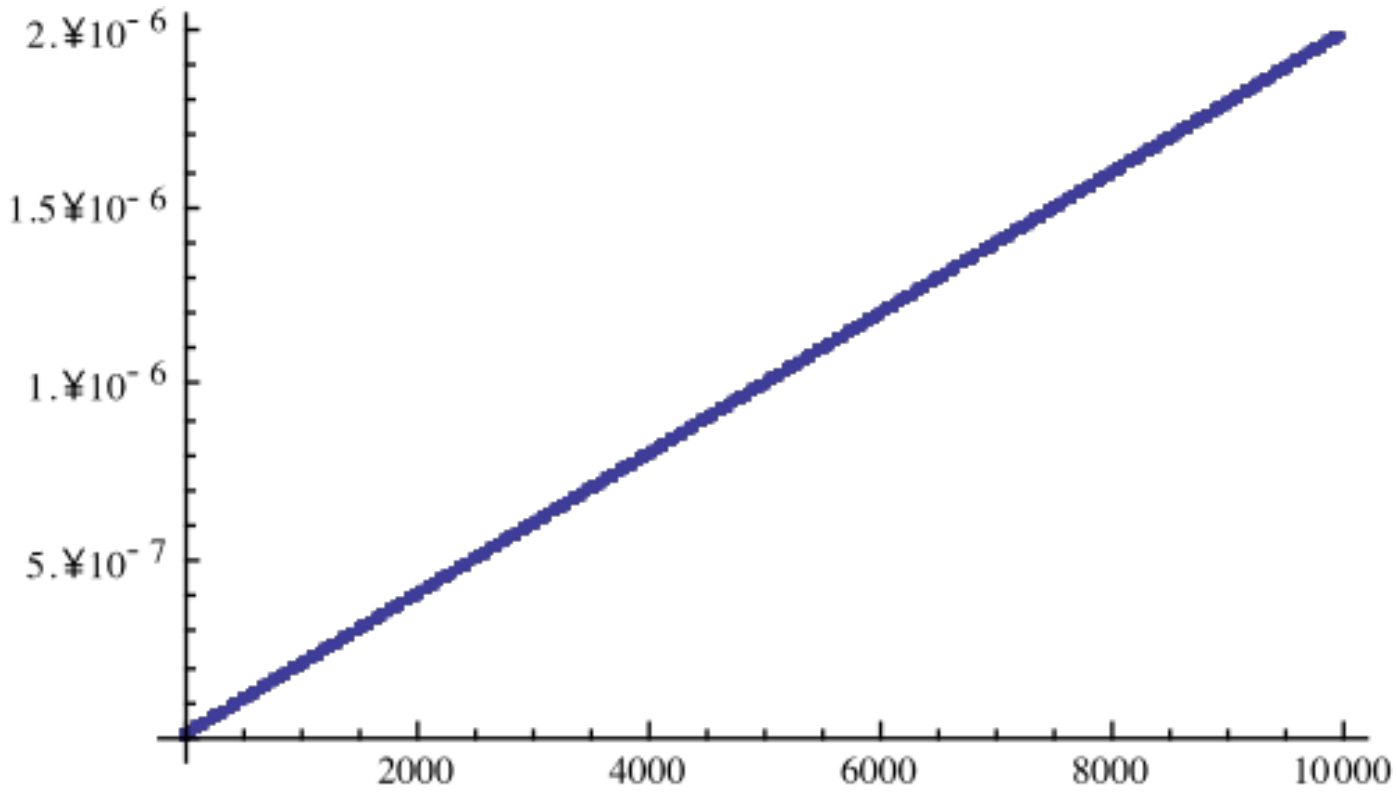

Fig. 13A3.5 Error propagation of an unstable integrator.

**Conclusion** – Stay with the *pteqHI/PARMTEQM* leapfrog integrator.

[eltoc]

# Chapter 14 – Comparison of *BEAMPATH* and *pteqHI* 2-term RFQ Simulations

We have been privileged to have the opportunity to collaborate with Dr. Yuri Batygin and to subject his code *BEAMPATH* to the 13-RFQ family test. *BEAMPATH* is written to handle only RFQs based on the simplified two-term (2-term) A01 and A10 potential description developed by Teplyakov and Kapchinsky for the original RFQ development (Fig. 14.1). *pteqHI*, *PARMTEQM*, *LIDOS* and *LINACSrfqSIM* also have a 2-term mode.

$$U(r,\theta,z) = \frac{V}{2}\left(\sum_{m=1}^{\infty} A_{om} r^{2m} \cos 2m\theta + \sum_{m=0}^{\infty}\sum_{n=1}^{\infty} A_{nm} I_{2m}(nkr) \cos 2m\theta \cos nkz\right)$$

Fig. 14.1 Multipole expansion of the RFQ potential.

The 2-term description does not allow for fully accurate description of the electric fields, which have multipole components from the actual vane shape, and must eventually be simulated by a modern Poisson solver. Comparison on the 2-term basis is a good starting point, because the theory and dynamics simulation are simple, and the effect of programmer decisions, off-optimum conditions and space charge can be initially explored.

A primary comparison requirement of this 2-term case was that the zero-current results agree very closely, because the main interest could then focus where real differences might occur, in this case in the treatment of space charge. Here the bar was also set very high – it was desired to get agreement, or to understand disagreement, to the order of 1% if possible. That this succeeded forms an important basis for further comparisons with other codes.

Although one might think that the 2-term comparison would be easy, it took a long time to finally find, understand, and correct the difference shown in Fig. 14.2. There were many confusing issues, and a few comments about steps along the way may be instructive, at least in terms of how difficult code comparisons can be[110]. It was essential that both source codes were available, that the time be spent to master each essentially line-by-line, and often that the intentions of the author be clarified. The steps proceeded from the basic setups, to the external fields, to the space charge computations, then to a large number of iterations, exacerbated by lack of cleverness in working out questions in the right order, etc.

The best procedure may be to start with very rigorous comparison of input beam, zero-current, input conditions timing offsets, etc., followed by the zero-current behavior, and then with space charge. The description in the following sections will proceed in this order. It is natural, however, to explore for large differences first, (clearly with space charge between *BEAMPATH* and *pteqHI*…), then later to clean up small differences, and this is how the work actually proceeded. Confusing nomenclature, use of different representations of the basic formulae, etc. led to time-consuming diagnostics – which could often show differences seemingly related to basic numbers (pi, pi/2, 4, sqrt(6), swt ("sine omega t" ?) = sin(omega t – pi/2), …) that wink in and out of the formulae like virtual particles. Numerical computation traps can also occur – occasionally really bizarre (will be described, even if not completely solved!).

## 14.1. Initial and final Comparisons

Fig. 3.2 Left shows the initial comparison between *pteqHI* and *BEAMPATH* running independently. Significant differences are seen – both in magnitude and in the location of an apparent optimum.

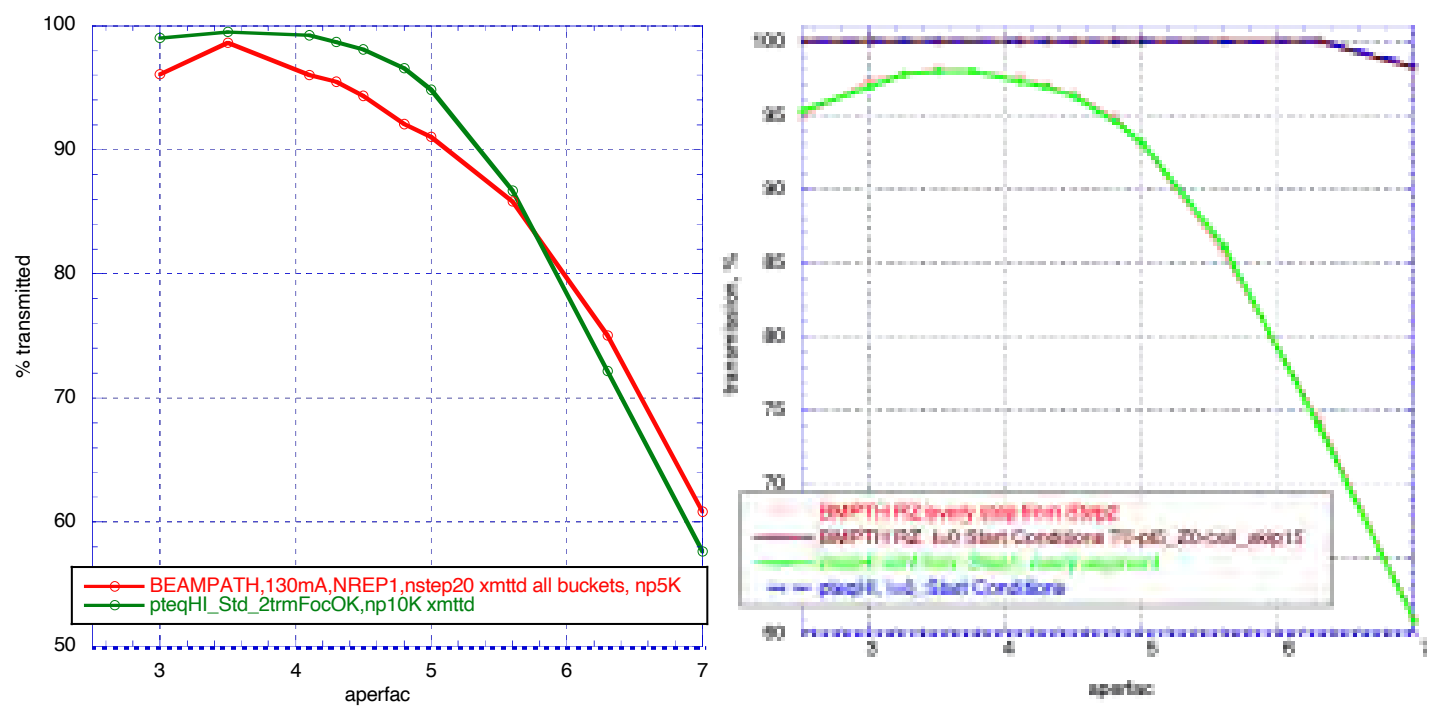

Fig. 14.2 Left - Initial comparison of accelerated beam percentage with 130mA beam current. Right - Final comparison of *BEAMPATH* and *pteqHI* transmitted percentage, for 130mA current and for zero current.

Final comparison results are shown in Fig. 14.2 Right. There is complete agreement with zero current – a requirement, although zero current is a very forgiving condition. The agreement with

110 Just the *BEAMPATH*/*pteqHI* 2-term and space-charge comparisons took an unexpectedly large amount of time over a period of one year. In large measure, the author's long bouts of uncertainty would have quickly been overcome by short personal meetings – the e-mail exchange was very prompt and good and eventually provided the right answers or clues, but of course could not replace personal contact. The old rule-of-thumb that it takes about six months to thoroughly master a new source code seemed confirmed. *pteqHI* was of course already known in detail, but because the basic position is that no "trust" be given *a priori* to either object of a comparison, it too was re-scrutinized in every detail.

Such effort in today's "managed" environment is "not affordable", except perhaps to some extent in a university environment with a long-term point-of-view. It is somewhat painful to see, on apparently important projects, the mechanical use of an executable code with unquestioned acceptance of clearly questionable manifestations, but this is now common.

space charge is also complete – given the underlying decisions, as elaborated in the following sections.

Achieving agreement involved mainly:

- understanding the intention of *BEAMPATH* RZ space-charge boundary conditions and then using them in *pteqHI* schefftm.
- Making the injection and first cell conditions match.

Very careful graphical and tabular checks were necessary to be sure that the RFQ parameters all matched exactly, and that all computing conditions, such as the dynamics and space charge meshes and their use, were the same. A number of smaller considerations of order 1-2% in transmission were also necessary

If source code for both programs had not been available, correspondence would doubtless not have been achieved.

## 14.2. Setup conditions and Basic Procedures

### 14.2.1 *BEAMPATH* Setup

Source code was made available.
Minor changes were needed to change computing platform – *BEAMPATH* is portable.
Documentation is adequate, and examples were provided.

### 14.2.2 Setup of the *Aperfac* Family RFQs for 2-term Comparison With *BEAMPATH*

The aperture, modulation, phis, and vane voltage are fixed for the *aperfac* family; a translator was written to create *BEAMPATH* input files. It is important to use enough significant figures for double precision.

The *pteqHI* rfqgen.f subroutine option for generating the required 2-term focusing and r0 terms for given cell-by-cell aperture, modulation, phis, and vane voltage was used.

### 14.2.3 "Primary" vs. "Secondary" Parameters

The choice of input parameters (aperture, modulation, etc.), or the handling of them, turned out to be more important than might be expected, and is discussed in the section on external fields.

### 14.2.4 Representation of the First Cell

Some extra handling is needed to represent the first cell in *BEAMPATH*; this is also discussed in the section on external fields.

### 14.2.5 Definition of Constants

A variety of definitions for the speed of light and mass are found in the various codes. The differences have little effect.

*14.2.5.1 Speed of Light or Wavelength*

*pteqHI, PARMTEQ, PARMTEQM*: c=29979.2458 cm/sec
*LIDOS* requires entry of the rf wavelength, and intrinsically uses speed of light = c =30000 cm/sec.
*BEAMPATH* requires entry of rf wavelength, and intrinsically uses speed of light = c = 29979.2458 cm/sec.

*14.5.2.2 Rest Mass*

*pteqHI*, *PARMTEQM*, *LINACSrfq*:
rest mass = 931.5016
amu proton = 1 (-> proton rest mass = 931.5016)
amu deuteron = 2.0145 (-> deuteron rest mass = 1876.51)

*Rfquik*: amumass=931.5016d0

*PARMTEQM*: MeVperAmu=931.494013d0
AmuProton=1.00727646688d0
AmuHminus=1.00837361135d0
AmuDeuteron=2.01355321270d0
AmuDminus=2.01465035717d0
AmuElectron=0.00054857990945d0
(MeVperAmu*AmuProton = 938.27199833451
(MeVperAmu*AmuDeuteron = 1875.6128

*BEAMPATH*: DATA PROTON /938280/ , A = 2.0145 for deuteron

## 14.2.6 Input Particle distribution

*14.2.6.1 Input Beam Emittance and Ellipse Parameters*

*LINACSrfq* enters normalized transverse and longitudinal emittances in cm.rad, and assumes a uniform distribution (total emittance = 5*rms emittance) but can use other ratios if desired.

*pteqHI* and *PARMTEQM* enter total real transverse emittance, and longitudinal emittance depending on the chosen input distribution. The usual distribution, named “Type 6”, creates “a four dimensional transverse hyperspace with uniform phase and no energy spread”. The transverse phase-space is a waterbag distribution with the requested number of particles placed randomly in “a uniform transverse radial distribution”. The ratio of total to rms transverse emittances for the waterbag distribution is six.
The ellipse alphas and betas for the total real transverse emittance are entered.

*BEAMPATH* enters total normalized transverse emittance – based on the rms normalized emittance of a KV distribution. E.g., rms normalized emittance = 0.025 cm.mrad * (totalrms emittance ratio for KV distribution = 4) = 0.1 cm.mrad. If a waterbag distribution is desired, ITYPE=WB and the program multiplies the area by 1.5.
The area of this *BEAMPATH* distribution matches that of the *pteqHI/PARMTEQM* Type 6 distribution. However generation of the *BEAMPATH* ellipse parameters is tricky.
*BEAMPATH* enters RX (x) = Sqrt[beta*emittance] and DRXDZ (xp) = -alpha/Sqrt[emittance/beta]. However, the emittance here is for the total *real* KV transverse emittance, not the total normalized KV emittance.

*LIDOS* enters total normalized transverse emittance, with ellipse parameters X and XP based on real total emittance.

Any code should have options to read in an external distribution and to output its source distribution.

Another good check is to start with a distribution from one code, translate it dimensionally to the units of the other code, and read it into the other code.

It is really necessary to make a graphical comparison.

*14.2.6.2 Subtle Point on Comparison of Time-Based (t) Codes and/or Position (z)-Based Codes, or in Using Emittance Measurements*

The input distributions for time-based codes to be compared would have the same position {x,y,z} coordinates, and may have velocity or (normalized) momentum coordinates.

Use of position as the independent variable is not recommended because of inherent difficulties in computing space charge forces, where the particles must obviously be at the same time. All modern RFQ codes use time as the independent variable for this reason. However, position was chosen as the independent variable for the widely distributed LANL *PARMTEQM* (and later versions to the present) RFQ code; this was done after some comparison of t- and z-versions, because it seemed more natural in terms of the handling of input and output transport lines. At points where space charge is calculated, *PARMTEQM* makes a rough conversion (accounting only for quadrupole focusing and neglecting rf and space charge defocusing) of the particle transverse coordinates to bring them to the same time, compute the space charge forces, then converts back to z-coordinates.

Similarly, in the original LANL z- vs. t- comparisons, the input distribution for the z-version was generated first {x,dx/dz,y,dy/dz,phase, energy} with the corresponding emittances and ellipse parameters. Then the t-version input distribution transverse coordinates were converted to move the particles "to the same time": {1,3}->{x,y}, {2,4}->{dx/dz, dy/dz}, (5,6)->{z,dz/dt}:

```
cord(3,np)=cord(3,np)+cord(4,np)*cord(5,np)/cord(6,np)
cord(1,np)=cord(1,np)+cord(2,np)*cord(5,np)/cord(6,np)
```

This results, in the time domain, in considerable skewing of the transverse ellipses (different ellipse parameters) and distortion at the ends, as indicated in Fig. 14.3 This conversion resided in the original LANL *parmteqc.for* t-code, and by inheritance in *pteqHI*. It must be removed in *pteqHI* for direct t-code comparisons.

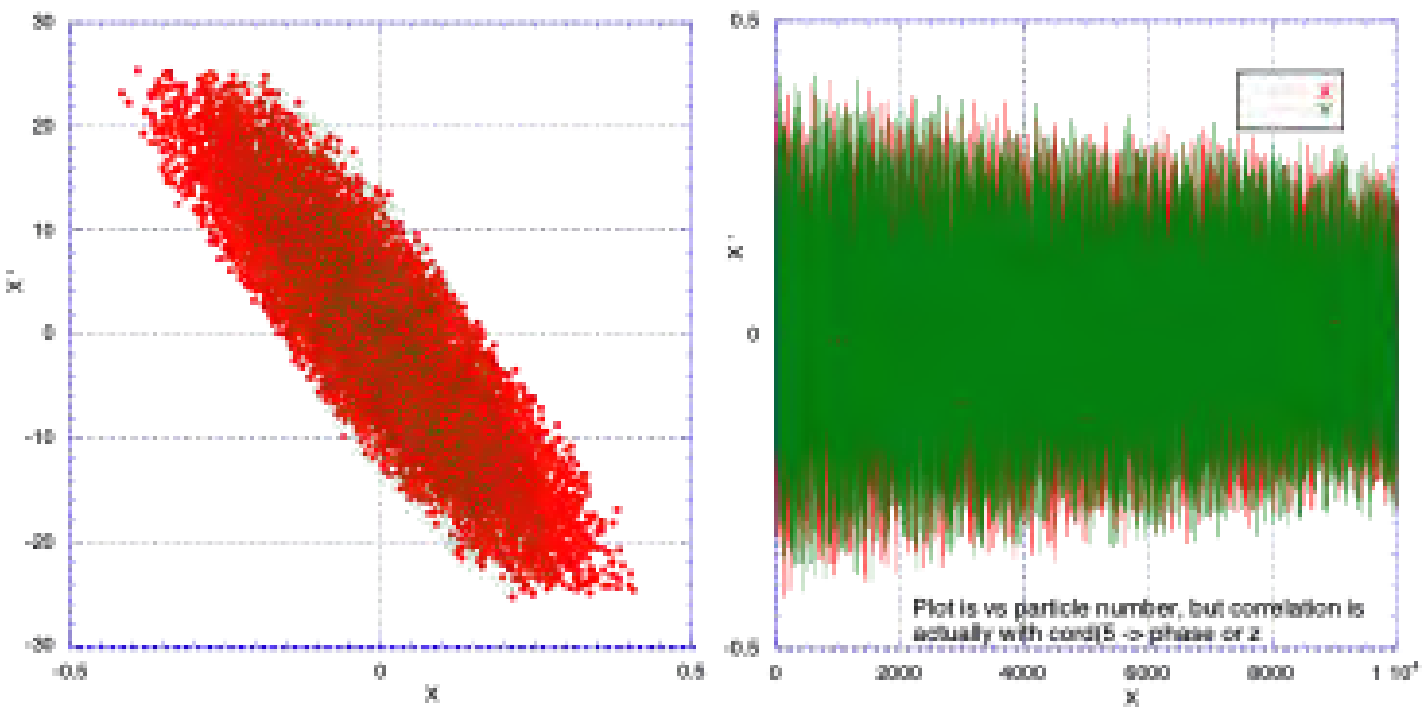

Fig. 14.3 Left – green is input distribution with no x conversion. Red is with conversion. Right shows converted x and y correlation with z.

*It is important to note* that this point has probably been largely overlooked when using data from emittance measurements to generate an input distribution. The emittance measurements are for particles at a certain position – therefore this conversion for use in a t-code would be appropriate. This may be one factor in the common disagreement of measured vs. simulated input matching values.

### 14.2.7 Synchronous Particle

There is no particular synchronous particle reference in a time-based code, as is necessary in a z-based code. However, including one is a particularly important diagnostic.

*BEAMPATH* indicates that particle #1 in the distribution is a synchronous particle, with initial transverse coordinates set to zero. However, this particle was usually not accelerated to full energy in this study (with full beam current). In the space charge dominated regime, the synchronous particle is shifted with respect to zero-current regime. It is also possible that space charge or mismatch interactions can cause an on-axis, synchronous particle to be kicked off-axis and out of exact synchronism [111]

*LIDOS* also occasionally "loses" its synchronous particle.

In *pteqHI*, a synchronous on-axis particle initially positioned corresponding to the specified synchronous phase (-90°) is added, and is not affected by the space-charge routine.

Observation of this diagnostic is important to eliminate programming bugs, for example, that sometimes cause too-many or too-few steps per cell. This problem requires round-off tolerances to be specified in the dynamics for cell boundary transitions, used in *pteqHI*. When testing *PARMTEQM* subroutines embedded in *pteqHI*, this diagnostic revealed that round-off tolerances are needed there also. The same is true for *BEAMPATH*, and were added for this study.

### 14.2.8 Starting Conditions

*BEAMPATH* populates z randomly in the initial distribution. *pteqHI* populates z uniformly according to particle number. This accounted for different appearance of some output analyses discussed below.

The *pteqHI/PARMTEQM* default Type 6 RFQ dc distribution positions the input bunch with its head at z=0. Offsets can be entered in the input file, but were not.

*BEAMPATH* generates the RFQ dc input beam ± 1 cell around z=0, and allows a position offset to be entered in the input file as Z0, which was set as (minus length of first cell) in order to correspond to *pteqHI*.

The default starting value for time in *pteqHI* is (-3pi/2)*period. The working time term in the dynamics expressions is swt=sin(twopi*freq*(t+0.5d0*dt)).

The corresponding time start value in the *BEAMPATH* input file is T0=-0.50. The working time term in the dynamics expressions is
COST=-cos(2.*pi*(t+dt/2)+pi/2).

This is one of the places where nomenclature and pi/2's were confusing. There is also a one-step difference between *BEAMPATH* and *pteqHI* whose origin has not been discovered. As a result, careful diagnostic outputs and graphics were used to be sure that the starting conditions were actually the same.

### 14.2.9 Dynamics & Space-Charge Mesh, Handling of Out-of-Mesh Particles

For this comparison, both *BEAMPATH* and *pteqHI* codes use a basic dynamics mesh and a space-charge mesh of ± 1 cell around a synchronous particle. (Other meshes can be selected in *pteqHI*, and another is the preferred standard.)

*BEAMPATH* routine *RZ* uses a Fourier Poisson solver to compute space-charge forces on an rz axially symmetric mesh. *pteqHI* uses an r-z mesh with charges represented as rings. Thus both are restricted to cylindrical symmetry. Both proportion the particle charge to the nearest four mesh corners. Space charge simulation is discussed further below, including aspects such as the number of mesh points, particle population, boundary conditions, etc..

---

[111]. R.A. Jameson, ""Beam-Halo From Collective Core/Single-Particle Interactions", LA-UR-93-1209, Los Alamos National Laboratory, 31 March 1993.

R.A. Jameson, "Self-Consistent Beam Halo Studies & Halo Diagnostic Development in a Continuous Linear Focusing Channel", LA-UR-94-3753, Los Alamos National Laboratory, 9 November 1994. AIP Proceedings of the 1994 Joint US-CERN-Japan International School on Frontiers of Accelerator Technology, , Maui, Hawaii, USA, 3-9 November 1994, World Scientific, ISBN 981-02-2537-7, pp.530-560.

*BEAMPATH* continues to track particles that leave the mesh, but space-charge is not computed for these particles. Thus typically the main bunch plus leading and trailing bunches are seen.

*pteqHI/PARMTEQM* usually employs a methods to bring fast or slow particles back into the mesh, but these methods were disabled for this comparison.

*pteqHI/PARMTEQM* also computes the space-charge for particles outside the mesh, by applying point-charge forces between the particle and a point-charge representation of all the particles within the mesh; this was also disabled.

#### 14.2.10 Lost Beam & Accelerated Beam Criteria

*14.2.10.1 Radial Loss:*

*BEAMPATH* uses the minimum cell aperture (at the end of each cell) and sets XAPER=YAPER, so particles exceeding this square boundary are declared lost.

*pteqHI* has this mode and was set accordingly.

*14.2.10.2 Longitudinal Loss:*

No losses are declared longitudinally.

*14.2.10.3 Transported, Accelerated Beam:*

*BEAMPATH* presents the transmission efficiency (Eff. Trans) as all particles not lost radially, and terms the accelerated efficiency (Eff.accel) as the particles in the main bunch that are within one cell of the synchronous particle.

*pteqHI* defines transmission as all particles not radially lost, and accelerated beam as all particles near the synchronous energy.

We use all particles not lost radially for the comparisons.

#### 14.2.11 Number of Particles, Step Size

Comparison runs were made with 10000 particles, and 10 steps per cell. (For these 2-term simulations, there was $<\sim 1\%$ difference if 20 steps per cell were used.) See also further discussion in Sec. 14.4.5.

#### 14.2.12 Subroutines

*pteqHI* subroutines could be embedded in *BEAMPATH*, and vice-versa, to enable active/passive comparisons in the same framework. For example, the *pteqHI* space charge subroutine *schefftm.f* was inserted in *BEAMPATH*, and the *BEAMPATH* subroutine *RZ* and its associated subroutines were inserted in *pteqHI*, and used in active/passive modes.

#### 14.2.13 Driver and Physics/Mathematical Frameworks

The driver frameworks of *pteqHI* and *BEAMPATH* are completely different, and one has to learn the flow. The form of equations is also different. For example, use of normalized equations have always been a source of confusion for this author, because the physical meaning becomes lost and comparison with physical expressions becomes difficult.

## 14.3. EXTERNAL FIELDS

Each code lent some confusion to making sure that the simple 2-term potential equation were actually the same in each, which they are.

For example, a factor of (4/Pi) shows up in *BEAMPATH*, but not in *pteqHI/PARMTEQ*. This turned out to be that some authors, including the original RFQ theory paper by Kapchinsky [112], include the

[112]. "The Linear Accelerator Structures with Space-Uniform Quadrupole Focusing", I.M. Kapchinsky and N.V. Lazarev, IEEE Transactions on Nuclear Science, Vol. NS-26, No. 3, June 1979

transit-time factor for an RFQ cell in the equations, but then have to factor it out again for the dynamics, because the transit-time factor is obtained as the result of averaging (integrating) over one cell or period, whereas in the dynamics, the integration is done step-by-step.

Another example involved the apparent disappearance of a 1/(particle radius) term in the *pteqHI* Bessel terms, compared to the corresponding *BEAMPATH* terms. This turned out to be a kind of nomenclature problem. The formula for *pteqHI* xi0kr contains (0.5*cay*r)^2, which suggests BesselI[0,(0.5*cay*r)^2] but it isn't. It is BesselI[0,cay*r], which is what the nomenclature xi0kr suggests - and is the series expansion to 3rd order (r^6). The term xi1kr is worse. The formula again contains (0.5*cay*r)^2, the nomenclature looks like BesselI[1,cay*r], but it is actually the series expansion of BesselI[1,cay*r]/r, so the particle radius had not disappeared after all.

But *BEAMPATH* and *pteqHI* do compute the same 2-term external fields (Fig. 14.2)

#### 14.3.1 Representation of the Injection and the First Cell

The RFQ classically uses an input "radial matching section" in which the transverse focusing is raised linearly from ~0 to a final value. Performance is sensitive to this region. *pteqHI/PARMTEQM* and *BEAMPATH* both store the value of cell parameters at the end of each cell. *pteqHI* also stores and uses values at the beginning of the first RFQ cell. *BEAMPATH* does not, so a bit of programming was added to make a radial matching section in *BEAMPATH* (see also next section).

The above description of "starting conditions" indicates that the beam head is initially placed at z=0. *pteqHI* applies a drift at each step for each particle until it reaches z=0, and then starts applying the external field forces. Thus the starting point for the synchronous particle is not at the beginning of Cell 1, but one cell back – a realistic representation of a beam coming in through the RFQ end wall.

#### 14.3.2 "Primary" vs. "Secondary" Parameters

The programs use different parameters in the input files. The aperture, modulation, and vane voltage are primary parameters with direct physical meaning. A fourth parameter is needed. Synchronous phase seems preferable, because it can be generated according to the original idea of the RFQ, to keep the longitudinal focusing uniform by keeping the ratio of beam length to bucket length constant (or to a prescribed variation). Other parameters, such as beta or cell length, can be derived in various ways to define an acceleration schedule, but in themselves have little physical meaning.

This should not be important, only care should be taken to use enough significant figures for double precision.

A peculiarity was noted when the addition of a radial matching section (RMS) to *BEAMPATH* resulted in a curious (and unexplained) phenomenon. *BEAMPATH* enters aperture as a primary variable, and uses the aperture directly in its 2-term formulas for the electric fields. The first RMS programming just included a large aperture at the beginning of cell 1 and used linear interpolation for the aperture at each step. The result was a disastrous change of the particle phase space into separate distributions at each step. The curious thing was that by simply changing the step interval from linear to a tiny bit nonlinear (Using exponent **0.9 or **1.1), or by adjusting the step sizes by a small random number, the gross distortion disappeared. The matter was not pursued further, because it was realized that the KRC's intent [113] for the RMS was to linearly increase the transverse focusing force, which is proportional to 1/(aperture**2). The *BEAMPATH* RMS strategy was simply changed to find the desired focusing forces at the beginning and end of the cell, and linearly interpolate.

Indeed, *PARMTEQM/pteqHI* formulas for the external fields in the whole RFQ are written in terms of the focusing forces, linearly interpolated from stored arrays for each cell.

---

[113]. "Proposal for a new radial matching section for rfq linacs", K.R. Crandall, LASL memo at-1:83-3, 830131

### 14.3.3 Dynamics Integrator

Both *BEAMPATH* and *pteqHI* use the simple leapfrog symplectic integrator, although in opposite order. *BEAMPATH* updates using the old momenta over ½ step, then the physical coordinates for a whole step, then the updated momenta over the 2nd ½ step. *pteqHI* updates the other way around. This difference required special care in interpolating parameter values.

There is no paraxial approximation in either code.

*pteqHI* had previously explored the use of an integrator with an added second-order term. Its use was also explored in the standard and usual leapfrog manner, but also as indicated by given formulae in a stepwise manner. The stepwise implementation resulted in tearing and folding of the longitudinal phase-space in some, but not all, conditions of the RFQs being simulated – this did not occur for leapfrog implementation. However, the second-order term produced a low, almost linear, persistent increase in the transverse emittances, which disappeared when the second-order term was simply omitted. Chapter 13 - Appendix 1 discusses integrator requirements further. Beyond symplecticity (conservation of phase-space), the integrator must also be consistent and accurate, and also stable.

## 14.4. SPACE CHARGE

Finally we come to the most interesting part. It was immediately apparent that some difference in the *BEAMPATH* and *pteqHI* space charge methods was causing most of the difference in Fig. 14.2 Plotting the sum of the absolute values of the transverse and longitudinal impulses, Fig. 14.4 indicated that the transverse kicks were similar but that the longitudinal kicks were very different.

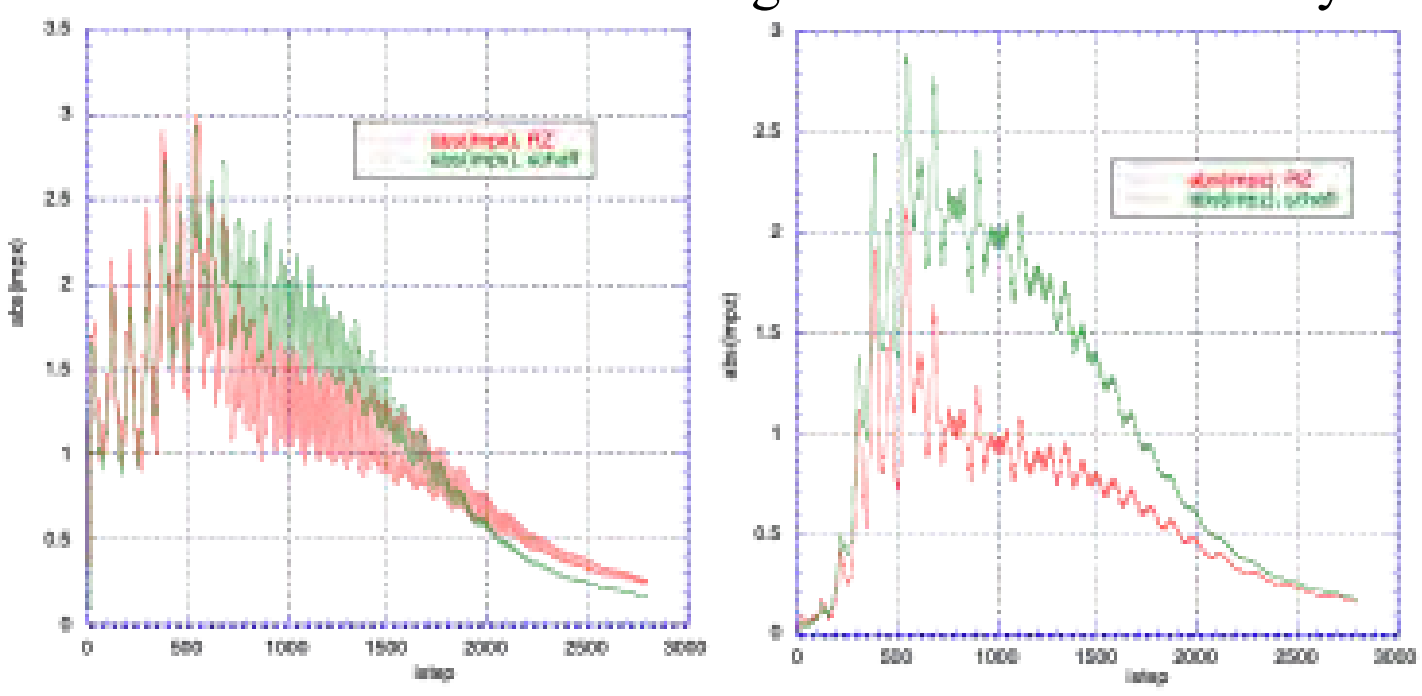

Fig. 14.4 Sum of the absolute values of the transverse and longitudinal impulses, red – *BEAMPATH* subroutine RZ, green – *pteqHI subroutine schefftm,*

To make a very long story very short, the difference was due to not understanding the intent of the *BEAMPATH* method, in terms of its boundary conditions – a clear case where personal meetings would have cleared everything up very quickly. The learning experience was valuable, however, and is outlined here.

### 14.4.1 Basic methods for computing space charge

*BEAMPATH* and *pteqHI* are both time-based codes, so space charge is easily computed directly from the coordinates at any point in the computation.

Space-charge is computed from the first step. For Fig. 14.2, *pteqHI* space charge was computed at each step. Often *pteqHI/PARMTEQM* are run with space charge computed once per cell at cell center to save run time; the difference is of order 1-2%. These considerations are shown in Fig. 14.5

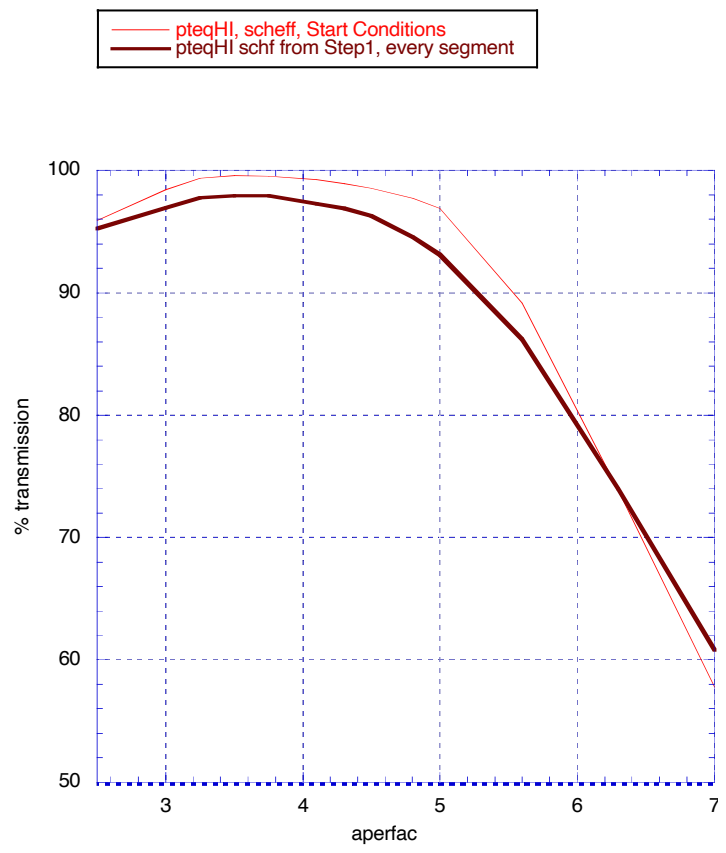

Fig. 14.5 *pteqHI schefftm* space charge computation variations. Red – computed at each step in the radial matching section and thereafter once per cell when the synchronous particle is at cell middle. Dark red – computed at each step, starting at first step.

*14.4.1.1 Theoretically Exact pteqHI/PARMTEQM scheff RZ Rings-On-Rings and Point-to-Point Methods*

*pteqHI* subroutine *schefftm.f* allocates the particle charge proportionally to the corners of an RZ mesh. The accumulated charge at each mesh point is represented as a ring of charge, so there is strict cylindrical symmetry. The forces of every other ring on each ring are then computed.

*IT HAS NOT BEEN POINTED OUT IN ANY KNOWN REFERENCE that, with a given charge distribution on the r-z mesh points, this ring-on-ring method is a theoretical model with exact solution, and therefore gives exactly known space-charge forces for any particle distribution. It does not depend on any boundary conditions. It gives no information outside the particle distribution (the basis function has no tails; it is an open boundary condition.).*

Although turned off in this study to match *BEAMPATH*, *pteqHI/PARMTEQM* also account for space charge of a particle which has fallen behind the mesh, by applying an impulse calculated as point-to-point between the particle and the bunch center of gravity. Each particle is also adjusted for multiple adjacent bunches on both sides of the mesh (typically 20 on each side) using the point-to-point method. (For comparison with *BEAMPATH*, the neighboring bunch effect was applied only to particles within the space charge mesh.) . These are also theoretically exact impulses.

The knowledge that these are exact affords a great advantage in comparing space charge methods having the same symmetry conditions. [114]

[114] (The literature contains “comparisons” using, e.g., unsatisfying “semi-analytical” distributions, etc. (and with no conclusions drawn), whereas more exact comparisons might afford applicable conclusions…)

*14.4.1.2 BEAMPATH Space Charge Methods [115,116,117,118]*

*BEAMPATH* has three methods that could be used. The *RZ* method assigns the particle charge proportionally to the corners of an RZ mesh. A Fourier solver is used to find the longitudinal space charge potential and a three-diagonal matrix is used for the transverse. The *QZ* method is fully 3D. The *RZ2D* method uses a Green's Function space charge calculation for an axially-symmetric beam.

**14.4.2 Separate Tests with Ball and Cylinder**

Extensive tests were made for the theoretically known cases of a ball or cylinder.

For the ball, the longitudinal boundary condition for *BEAMPATH* and the neighboring bunch contributions in *pteqHI* were switched off. The *BEAMPATH* radial mesh size was set to twice the ball radius, in order to check the field outside. For both ball and cylinder, the correct field shape and magnitude were obtained for all four subroutines (*schefftm, RZ, QZ, RZ2D*) with fidelity depending on the number of mesh points and number of particles. The *RZ2D* method requires the number of harmonic terms be specified; fidelity was good with 40 or more harmonics, but is very slow. *QZ* is also slower. *RZ* corresponds to *schefftm* and the comparison proceeded using these.

**14.4.3 Closed or Open Radial Boundaries for Fourier Solver in the RFQ**

*BEAMPATH* sets the radial mesh boundary at the minimum aperture in a cell (XAPER). It was then noted that if instead the radial mesh boundary was set at twice or more times XAPER, the sum of the absolute values of the transverse and longitudinal impulses (Fig. 14.4) substantially agreed with those of *schefftm*, and the *BEAMPATH* transmission curve also agreed much more closely. It took some time to figure this out, because the problem seemed to be with the longitudinal plane somehow, but actually was fixed by changing the radial boundary. The particle distributions, potential and electric fields on the grids were explored with extensive graphics.

Correspondence with Batygin then quickly pointed out that the whole intent of the *BEAMPATH* boundary had been missed. Implicit in the solver is that the longitudinal boundary is periodic, and is solved via an expansion of space charge and potential in a discrete Fourier series containing both sine and cosine functions. But the radial boundary is intended to also account for image forces, and therefore the edge of the radial mesh is a conducting pipe at zero potential. The transform is performed using only sine functions which are zero at the pipe - automatically satisfying zero potential condition at the aperture. It was easy to see that the r-boundary was set to zero in the code, as it would be also for an open boundary condition at suitable extent, but that sine Fourier was used was not so obvious.

By changing the radial boundary condition to 2*XAPER (or greater), the situation becomes more and more like an open boundary, with no image effect, and thus comes into agreement with *schefftm*. There is little change beyond 2*XAPER, which was adopted for further studies. Methods exist to more formally deal with the open boundary in suitably symmetric systems; by manipulation of the Green's function and at the expense of a two-fold increase in grid size in each dimension, the open

---

[115] . Yuri K. Batygin, "Accuracy and Efficiency of 2D and 3D Fast Poisson's Solvers for Space Charge Field Calculation of Intense Beam", EPAC1998 ThPo5H.pdf

[116]. Yuri K. Batygin, "Test Problems for Validation of Space Charge Codes", PAC1999, pp.1740-1742.

[117]. Yuri K. Batygin, "Stationary Relativistic Bunch in Space-Charge-Dominated Regime", PAC2001, pp.3084-3086.

[118 ]. Yuri K. Batygin, "Spectral Method for 3-Dimensional Poisson's Equation in Cylindrical Coordinates With Regular Boundaries", PAC2001, pp.3087-3089.

boundary can be analytically extended to infinity [119,120]. The RFQ is not cylindrically symmetric; the *BEAMPATH* methods have to be extended, but provide some valuable guidance as the intricacy of more accurate methods for external, space charge and image fields is explored in increasing detail in the following sections.

The transmission difference between the conducting pipe boundary at XAPER, and the open boundary represented by the 2*XAPER radial boundary is shown in Fig. 14.6

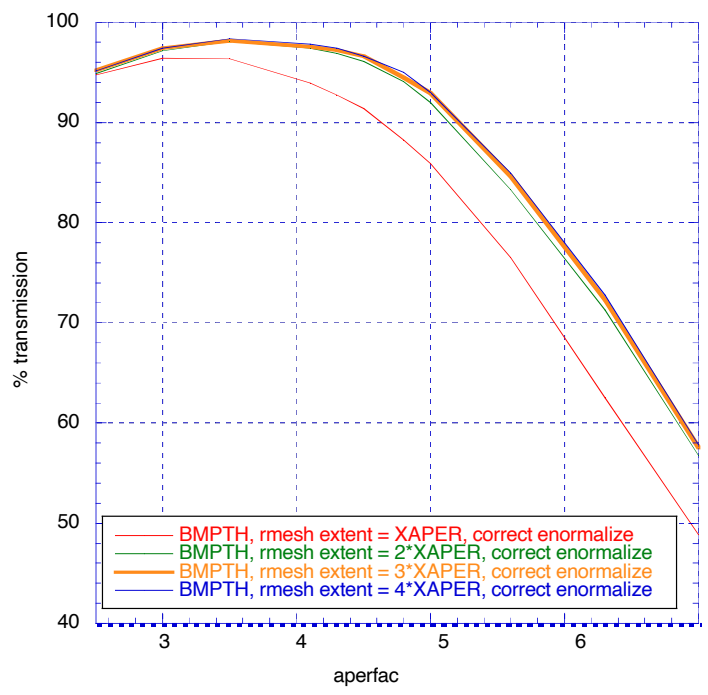

Fig. 14.6 Transmission curves for different *BEAMPATH* RZ space charge radial mesh boundaries.

### 14.4.4 Longitudinal Boundary Condition

The periodic condition at the longitudinal mesh boundary in the Fourier method is also known to be problematic, and the much studied problems with the Fourier method start to become apparent – the actual RFQ situation is not well suited, neither in terms of symmetry nor in periodicity. Batygin noted:

> "The problem with periodicity in Poison's solvers was a long-term subject in our study long time ago. One method assumed inclusion of particle which left one longitudinal boundary (let say at z) through the next boundary of z+ beta*lambda. It approximates situation when particles are moving from one bunch to another one. In my experience, it does not give good results. Finally, I came to situation when particles which left the period of beta*lambda are just removed from space charge calculations. it overestimates longitudinal space charge forces, but at least you know that, otherwise there is no info about degree of space-charge approximation".

Tests were made in *BEAMPATH RZ* by turning off the longitudinal periodicity, applying zero boundary at the ends of the ±1cell mesh, and applying the neighboring bunch effect. The results indicated that:

- *schefftm* with neighboring bunches and *BEAMPATH* with standard longitudinal Fourier boundary periodic condition are essentially equivalent.
- Neighbor bunches on non-periodic Fourier is not as strong as periodic Fourier.

The *PARMTEQM/pteqHI* method of returning particles to the mesh with accompanying point-to-point impulses is a straight-forward improvement that could be added to *BEAMPATH*.

*PARMTEQM* uses the separatrix length for the z-mesh length. Previous *pteqHI* work with full *schefftm* indicated that the particle return method gave somewhat higher transmission if the z-mesh

---

[119]. "Wavelet-space solution of the Poisson equation: An algorithm for use in particle-in-cell simulation", B. Sprague, MS Thesis, Northern Illinois University, 2009 (in progress). (internal file 'ben_thesis_final_20080818.pdf'. Also internal file 'Green's Function Notes.docx.

[120]. "Green's functions", G.E. Stedman, (1968),Contemporary Physics,9:1,49 — 69

[67,68] are rare examples of clear explanations in terms of what is happening physically in engineering terms, without the use of irrelevant physicists' names! Same as the problem with the use of Dirichlet and Neumann for boundary conditions…)

was slightly longer than the separatrix, because of complex behavior at the separatrix and consideration of simultaneous H+ and H- beams; and 1.3*(separatrix length) was chosen. Other z-mesh rules can be used, such as the common ±1 cell, or 3.5*(rms longitudinal beam length included in the separatrix), or a rule that switches to this from 2-cells when the rms measure is some percentage of the 2-cell length. The neighboring bunch and return of particles to the mesh account for the periodicity.

This subject will be revisited in in the following sections.

### 14.4.5 Mesh Basis, Mesh Size, # Mesh Points and Particle Population.

Subtle interactions between these variables affects the transmission curve. Extensive tests were performed.

For example, *schefftm* was tested with its full capabilities for returning particles to the mesh and full neighboring bunch effects. Using *schefftm* with given number of particles and mesh points, it was noted that:

- transmission falls order 2-5% as the r-mesh size is extended from XAPER to 4*XAPER.
- The 13-RFQ *aperfac* family RFQs' transmission characteristic for an r-mesh extent based on the beam size is qualitatively different than that for an aperture-based r-mesh. The peak of the beam-based would be interpreted as to the left of the peak for the aperture-based.
- Beam-based r-meshes for 1.05*(maximum particle radius) and 4*(rms transverse beam size) have ~same transmission

#### *14.4.5.1 Number of particles relative to mesh dimensions*

PIC methods often use a maximum particle to mesh dimension ratio of about 10 – for 10K particles in these runs, the scheff r-z mesh is 20x40 = 800. The *BEAMPATH RZ* equivalent needs to be at least 32x128=4096 to be accurate, standard is 64x256=16384.

For the same number of grid points, as the extent increases, the particle distribution becomes more concentrated into the inner cells. Both codes use cloud-in-cell spreading of charge to the nearest corners of a grid; space-charge kicks and transmission fall. *BEAMPATH* needs more cells for accuracy, and a larger radial extent for accurate (open) boundary conditions; so a mesh extent of 3*XAPER has about the same resolution as *schefftm* with a mesh extent of XAPER.

#### *14.4.5.2 Required size of r-meshes*

Repeating from above, the *schefftm* method represents each charge as a ring of charge, and directly computes the effect on each ring by every other ring – the method is restricted to cylindrical symmetry. Particles outside the mesh are treated using point-to-point forces. Both are theoretically exact for any particle distribution.

This observation has great practical application in code comparison.

In terms of the required size of the *schefftm* r-mesh, it means that the maximum extent can be the maximum particle radius.

If the extent of the *schefftm* r-mesh is greater than the maximum particle radius, cloud-in-cell smoothing will immediately affect the result by smoothing the charge at the maximum particle radius out to the next mesh box corners. The smoothing over the whole mesh depends on the mesh coarseness (fineness).

Fig. 14.7 helps explain why different transmission results from the choice of a beam-based vs. geometry-based r-mesh extent.

Fig. 14.7 Left – *aperfac*45 case, right – *aperfac*70 case. Four boundaries are shown: Red – XAPER, the minimum vane aperture. Green – 2*XAPER (also the maximum aperture when the modulation = 2). Blue – 1.05*(maximum particle radius (of particles successfully transmitted to the end of the RFQ). Yellow – Four times the transverse rms radius of particles successfully transmitted to the end of the RFQ.

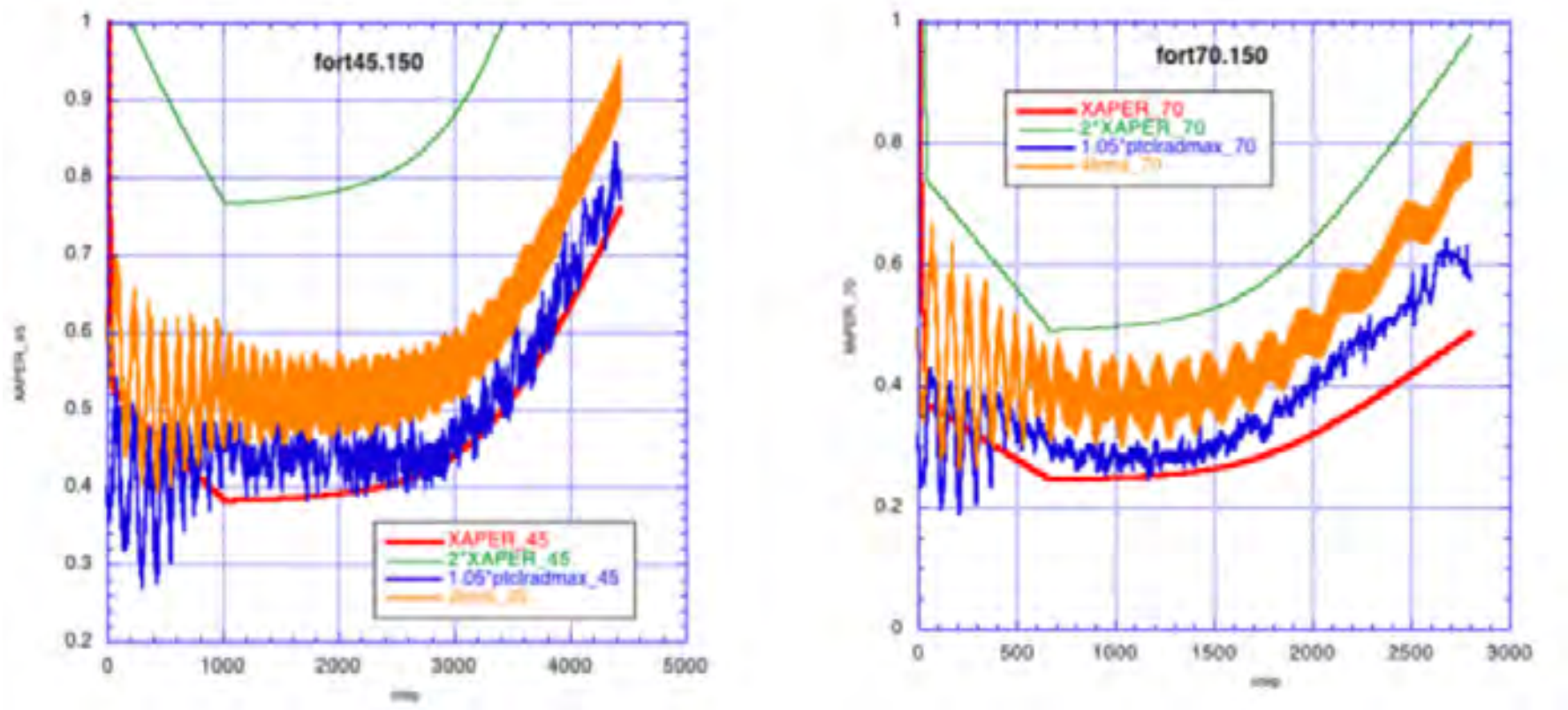

The smoothing in the critical shaper region would be quite different for an aperture-based or a max-particle-radius–based mesh. Both the max-particle-radius and 4*rms-transverse-beam-size are considerably less that XAPER some of the time, there the XAPER-mesh would smooth more. As there is no basis for including empty space beyond the beam boundary with the rings-on-rings method, based on a tighter fit being better, we would choose the max-particle-radius–based mesh (have chosen slightly larger, *1.05).

The minimum aperture is less that the max-particle-radius, as the beam experiences the alternating-gradient focusing through the cell, and also might travel in the region somewhat to the sides of the vane tip. This means (if the space charge r-mesh extent is XAPER and point-to-point space-charge is not applied to outer particles with radius outside the mesh) that the XAPER criterion would result in higher transmission (Fig. 14.8 Left.). Thus the max-particle radius mesh is preferred in the acceleration region also. (Reminder – the discussion here is for full *pteqHI*. The formal comparison to *BEAMPATH* above eliminates all particles with extent greater than XAPER *at each step*, which is too stringent.)

*BEAMPATH RZ* showed the same characteristics as *schefftm*, so the max-particle-radius rule could be chosen for both, with proper boundary consideration for *BEAMPATH*, as indicated in Fig. 14.8 Right, suggesting the conclusion that rmesh extent of 1.05 *(max particle radius) for *schefftm*, and 2 *(max particle radius) for *BEAMPATH*, would be realistic criteria.

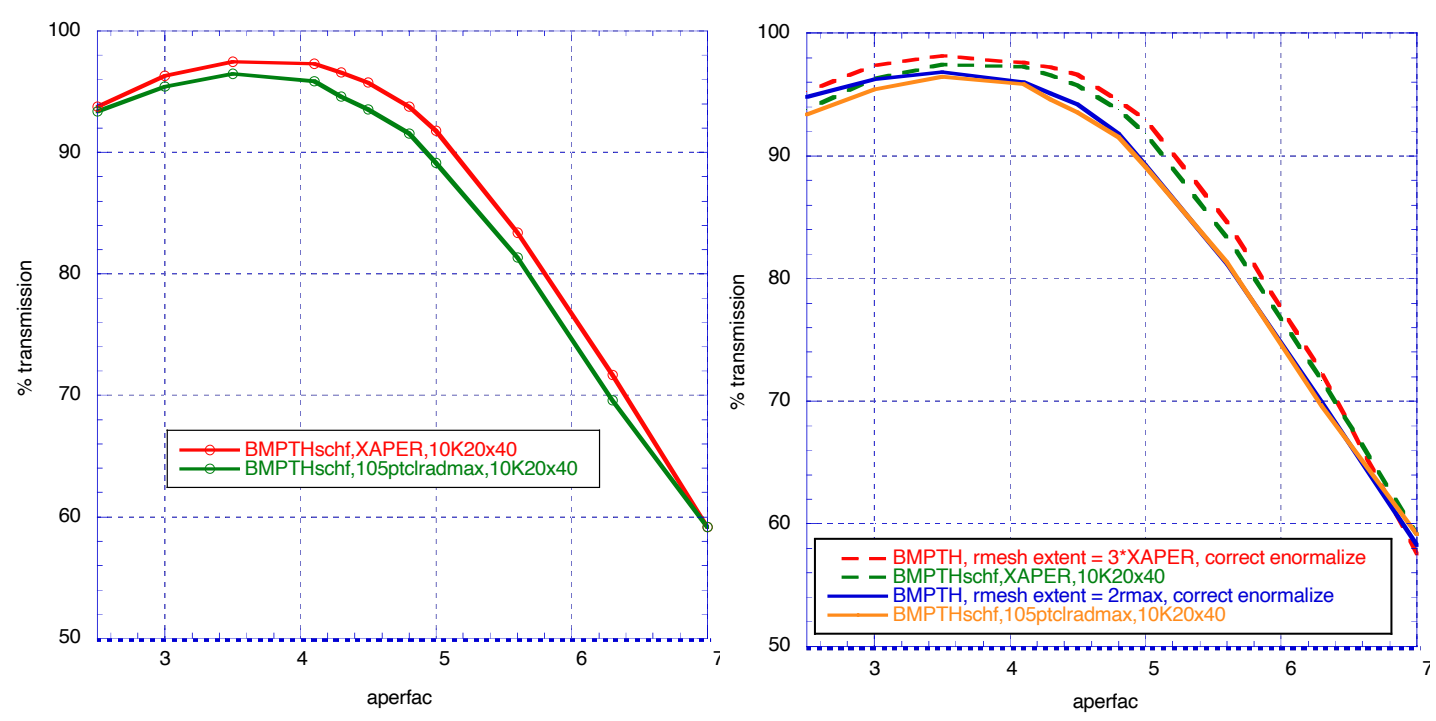

Fig. 14.8 Left - Demonstration of how the choice of radial mesh size can influence transmission. Right - Equivalent r-mesh extents for *BEAMPATH RZ* and *pteqHI schefftm*.

### 14.4.6 *Schefftm* beam shape ellipticity factor

The beam transverse shape is round only at the cell centers. *PARMTEQM/pteqHI schefftm* have an ellipticity correction feature that transform the beam to the circular frame for performing the space charge computation, then transforming it back. The difference is <1% between this and just using the coordinates as they are. Fig. 14.9 shows the difference in x-impulses given to a typical particle.

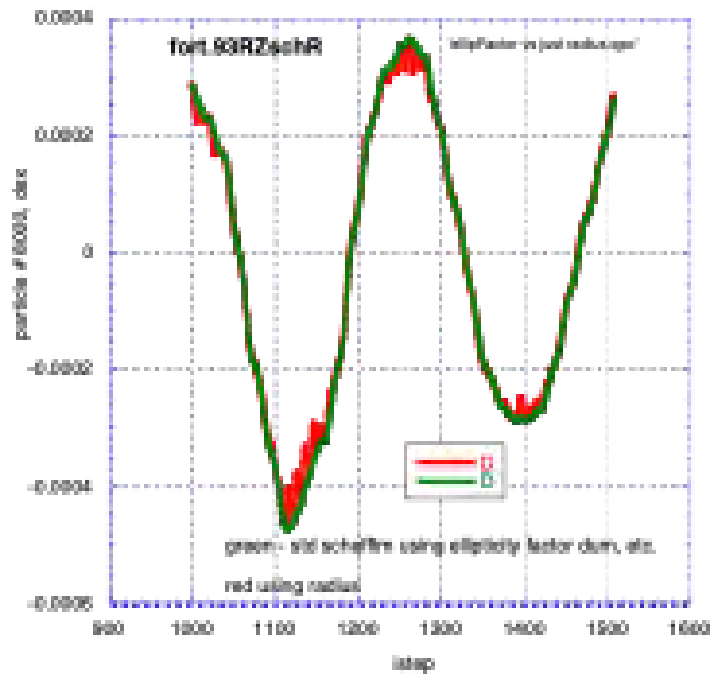

Fig. 14.9 X-impulses to a typical particle with (green) and without (red) ellipticity correction before computing space charge forces.

## 14.5. Conclusions – *BEAMPATH* and *pteqHI* 2-term Comparison

With an open r-mesh boundary condition in *BEAMPATH*, and with all the other conditions and decisions outlined, the two codes agree for the 2-term dynamics and space charge to within 1%.

The very fast running time of the *BEAMPATH* space charge method is attractive, as it would be fast enough to make optimization programming feasible. The method would be improved by adding the catchup provision for particles that leave the space charge mesh. However, correct treatment of image charges is crucial, but a cylindrically symmetric transverse boundary is required for the *BEAMPATH* method. Whether some such boundary could be adequate cannot be judged until there is the possibility for comparison to a fully accurate solution. A full Poisson solver enables this, and some important, rather surprising, conclusions await the reader below.

The Appendix given here as Ch.13 Appendix 3 Integration of the Equations of Motion – was included in original memo on this Chapter . This Appendix is important.

[eltoc]

# Chapter 15. Basis of *PARMTEQM*, Comparison of *PARMTEQM* & Basic *pteqHI*

*PARMTEQM* and *pteqHI* evolved from the same roots – the original first RFQ simulation codes outside the then USSR developed at LASL during the period ~1978-1980's. The early RFQ development was an open scientific undertaking, with many visitors and open source code and collaboration. When "management" descended on Los Alamos, morale and productivity plummeted, especially and very quickly starting in 1986, and the staff had no doubt that "the survival unit is now one", with the swift result of vanishing staff. Unfortunately, in an effort to maintain some identity, the group responsible for the RFQ simulation codes stopped open-source distribution and started marketing them as black box. After about 1996, no further improvements of the underlying physics of the codes were made.

Large transmission differences were noted (also by other designers) between designs run with the LANL2 or the LANL305 versions, with LANL305 transmission significantly higher (order 10%). A vane shape description for radial loss criterion added to *PARMTEQM* in 2001, which would explain at least part of this effect. LANL307 has no change to the handling of dynamics or space charge. The Changes.doc for *PARMTEQM* v307 notes no changes to the dynamics or space charge subroutines since 1996.

*PARMTEQM* was and is still used widely, but many questions arose, for example, why experimental results generally gave lower beam transmission, and concerning the effect of known approximations in *PARMTEQM*. With the absence of collaboration, no open source code, and diminishing (and finally no) support for responding to user questions, people began to develop new codes – which of course raised new questions.

The challenge of the IFMIF (International Fusion Materials Irradiation Facility) project required the most powerful RFQ yet envisioned - 125mA continuous wave deuteron current - and in earlier phases generated intense interest in improved design and simulation tools. *Pteq,* later *pteqHI*, with the yet present hope of prolonging the life of the earlier codes, began by fully exploring the choice of independent variable. Methods to understand the internal RFQ dynamics were developed, advanced computing power allowed removal of earlier approximations, and the code was configured as an open research tool. In 1997, the work with 2-term dynamics and very detailed comparisons with *PARMTEQ* were described in [58]; a synopsis is given in Chapter 13 Appendix 2. The work demonstrated in detail that time should be used as independent variable, as all known newer codes do.

Fortunately, KRC had written technical reports concerning the essential physics features [121,122,123,124,125], and subroutines could be prepared for the initial comparisons. The next steps were extension of *Pteq* to the full multipole field and image representation, following KRC, and then to *pteqHI*, the first RFQ code to handle multiple inputs of heavy ions of arbitrary charge and mass and to explore the problem of a strong dc field at the RFQ entrance, which was used in continued IFMIF research in its earlier phases and for laser ion source (LIS) development.

In January 2007, a full (143 pages) report "RFQ Designs and Beam-Loss Distributions for IFMIF" [126] was issued and widely distributed, summarizing the work of the IFMIF project. The Abstract is:

> The IFMIF 125 mA cw 40 MeV accelerators will set an intensity record. Minimization of particle loss along the accelerator is a top-level requirement and requires sophisticated design intimately relating the

121 "Effects of Vane-Tip Geometry on the Electric Fields in Radio-Frequency Quadruple Linac", K. R. Crandall, LA-9695-MS, UC-28, Issued April 1983.

122 "Radio-Frequency Quadrupole Vane-Tip Geometries", K. R. Crandall, et. al., IEEE Trans. Nucl. Sci., Vol. NS-30, No. 4 August 1983.

123 "RFQ Radial Matching Section and Fringe Fields", K. R. Crandall, Proc. 1984 Linac Conf. GSI-84-11

124 "PARMTEQ With Image charges", K. Crandall, LANL Memo AT-1-92-213, October 1991.

125 "Ending RFQ Vanetips With Quadrupole symmetry", K. R. Crandall, Proc. 1994 LINAC Conf.

126 "RFQ Designs and Beam-Loss Distributions for IFMIF", R.A. Jameson, Oak Ridge National Laboratory Report ORNL/TM-2007/001, January 2007. (Available on ResearchGate.)

accelerated beam and the accelerator structure. Such design technique, based on the space-charge physics of linear accelerators (linacs), is used in this report in the development of conceptual designs for the Radio-Frequency-Quadrupole (RFQ) section of the IFMIF accelerators. Design comparisons are given for the IFMIF CDR Equipartitioned RFQ, a CDR Alternative RFQ, and new IFMIF Post-CDR Equipartitioned RFQ designs. Design strategies are illustrated for combining several desirable characteristics, prioritized as minimum beam loss at energies above ~ 1 MeV, low rf power, low peak field, short length, high percentage of accelerated particles. The CDR design has ~0.073% losses above 1 MeV, requires ~1.1 MW rf structure power, has KP factor 1.7,is 12.3 m long, and accelerates ~89.6% of the input beam. A new Post-CDR design has ~0.077% losses above 1 MeV, requires ~1.1 MW rf structure power, has KP factor 1.7 and ~8 m length, and accelerates ~97% of the input beam. A complete background for the designs is given, and comparisons are made. Beam-loss distributions are used as input for nuclear physics simulations of radioactivity effects in the IFMIF accelerator hall, to give information for shielding, radiation safety and maintenance design. Beam-loss distributions resulting from a ~1M particle input distribution representative of the IFMIF ECR ion source are presented. The simulations reported were performed with a consistent family of codes. Relevant comparison with other codes has not been possible as their source code is not available. Certain differences have been noted but are not consistent over a broad range of designs and parameter range. The exact transmission found by any of these codes should be treated as indicative, as each has various sensitivities in its internal methods. Continued work to compare results between different codes more broadly and deeply than heretofore is highly recommended - this requires comparison at source code level and devising of appropriate tests. It is strongly recommended that the project obtain source code for all important simulation work.

In early 2007, IFMIF entered a prototype phase and it seemed necessary to emphasize that different codes were giving different results, over a broad range of parameter space, and that it seemed time to make a real comparison and decide on suitable design and simulation programs for this challenging project. The *aperfac* experiment used in this report was devised, and the first results from different programs showed widely varying results in the transmission and accelerated beam fraction curves, and worse, in the location of the optimum aperture. With regard to this section, comparing the available executable version of *PARMTEQM* with *pteqHI*, the latter gave significantly lower transmission and optimized at a larger aperture. Curves of the early results are not shown because the final results presented here confirm these characteristics. The detailed code comparison of this report was then started, and development of Poisson solvers at IAP in 2008.

Until 2009, *PARMTEQM* was investigated as a black box. Programming the KRC techniques within the common *pteqHI* platform afforded features to be individually switched on and off. The approach was aimed at understanding what *PARMTEQM* was doing, in terms of possible improvements to the *PARMTEQM* methods. Many interrelated and confusing issues were investigated. In 2009, access to *PARMTEQM* source code was afforded to me, which confirmed many hypotheses about its techniques and put the comparisons on a firm basis.

At this point, a comparison of *PARMTEQM* to *pteqHI* with the same basic procedures (loss criteria, etc.) could be made. The only difference from the 1997 study was the addition of the 8-term multipole representation of the external field in both codes, so it was expected that the comparison would be very similar to that result, and it is. The difference is again found to result from the general tendency of the t-code to give ~2% lower transmission than the z-code, and from a small effect of the residual paraxial approximation in *PARMTEQM*, as shown in Fig. 15.1. In addition to that result as found in 1997, now it is seen that there is also a shift in the optimum aperture.

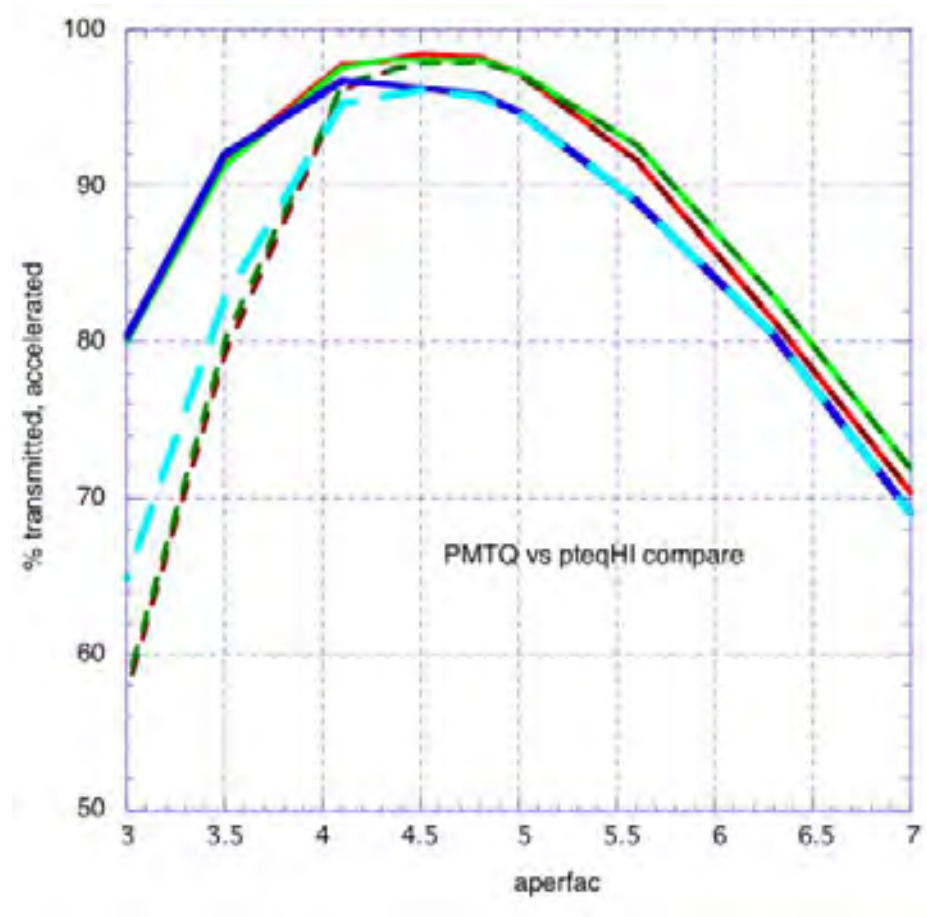

Fig. 15.1 *Aperfac* comparison of *PARMTEQM* and *pteqHI* with same basic procedures except for the independent variable and residual paraxial approximation in *PARMTEQM*. Red- *PARMTEQM* with individual particle paraxial approximation in dynamics. Green *PARMTEQM* with no paraxial approximation. Blue and light blue - *pteqHI*. Solid lines are transmission, dashed are accelerated fraction.

The *PARMTEQM* comparison is not intended to bring it into agreement with another code, because it is known that problems with various aspects require that either *PARMTEQM* be heavily improved eventually, or that newer codes be used. Of course, comparison to a code with known defects is nonsense. However, as *PARMTEQM* are still rather widely used, it is important to review it in detail.

*PARMTEQM* has useful capabilities beyond just the dynamics, e.g., for preparing detailed machining instructions. It (was) hoped that the best aspects of *PARMTEQM, pteqHI* and other codes (could) eventually be combined into a more physically accurate, open-source design and simulation package (which independently became *LINACS*).

## 15.1. *PARMTEQM/pteqHI* ISSUES

### 15.1.1 *PARMTEQM* Setup

*PARMTEQM* has a companion program *PARI*, which generates an RFQ from user input data. The design method is not discussed here; it is a very limited approach [127]. The user is asked during execution if certain "adjustments" are desired, for example, to adjust the aperture and/or modulation, given as input, so that the quadrupole multipole coefficient is returned to the 2-term value (=1), or the acceleration rate be adjusted to have the RFQ end at an exact energy or length. These adjustments are not transparent and result in a different RFQ than was input. They were strictly not used herein.

Even so, the *PARI* RFQ generation program for *PARMTEQM* is confusing. Some columns of its two output tables do not agree with each other, and as a black box, it is not possible to produce diagnostics. Input data from *LINACSrfq* were input to *PARI*, choosing no adjustments, with directly following *PARMTEQM* run; the *PARMTEQM* output listing for aperture, modulation and multipole data did not agree with the input.

*PARMTEQM* v.305 and 307 do not use zdata from an input file, but read in cell data from files generated by *PARI*. Earlier versions that read zdata use the transverse focusing term B as input, from which the aperture is computed internally, using the 2-term field formulae. However, sinusoidal longitudinal modulation is often used, which results in different relation between B and the aperture. This problem was only clarified by studying the source code.

127 ""RFQ Optimization Study for ESNIT," H. Takeda and R. A. Jameson, Los Alamos National Laboratory Report LA-CP-93-5, "Deuteron Linac Design Aspects for ESNIT," pp. 4-25, January 1993.

Input files for *PARMTEQM* must be in a proprietary binary format which is produced by *PARI*, so it is not possible to use *PARMTEQM* for an independently produced RFQ unless this format is known and can be reproduced. (We were successful in hacking it.)

These problems have been noted by a number of designers.

With *PARMTEQM* techniques programmed into the *pteqHI* framework, then it could be insured that exactly the same cell quantities were being used by both programs.

The same recalculated multipole coefficient tables are used by *PARI* and *pteqHI/LINACSrfq*. *PARMTEQM* multipole coefficient interpolation from the table was observed to be ragged and produced different values from the *LINACSrfq* Mathematica Interpolation function, which contributes to the disagreement (Fig. 15.2).

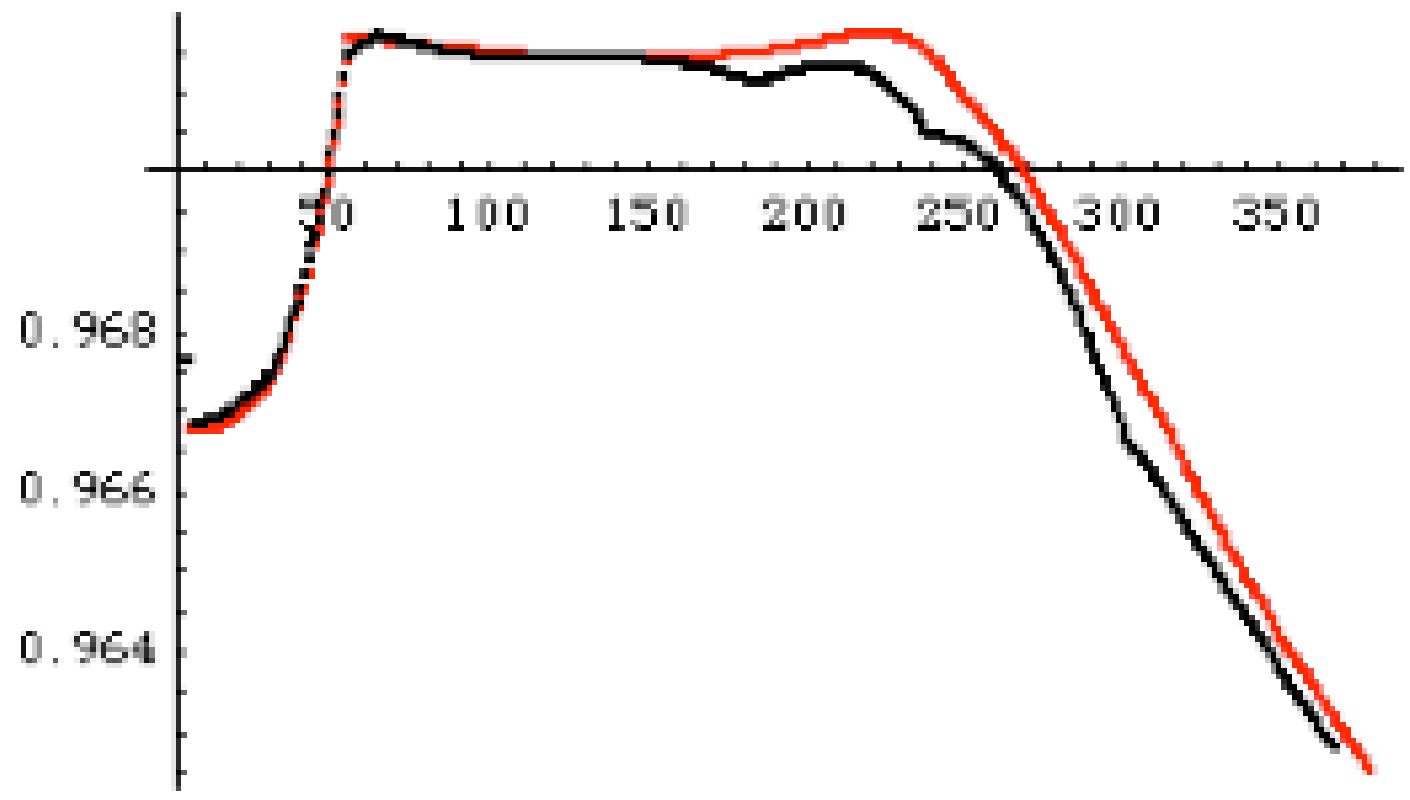

Fig. 15.2. A01 coefficient along an RFQ. Black – *PARI/PARMTEQM*. Red – *LINACSrfq* Mathematica Interpolation

Tests in another code using *PARI* output are possible using a high precision ASCII file of the cell data that *PARI* provides; the format has to be deduced. The high precision is necessary for comparison, it is not sufficient just to copy the limited resolution columns from the *PARI* output files.

As usual, many settings must be made and it takes time to learn them. For example, the rule for the relation between vane tip and r0 radii is set in the input, say as RhoCoeff 0.75, 0, or the lines in the LANL.INI file: [Pari]; RhoCoefficients = 0.75 0. For these settings, the RFQ vane tip would have a constant ratio $\rho/r_0 = 0.75$. Notice the importance of the zero on the RhoCoeff or RhoCoefficients line. If the zero were missing, then the vane tips would have a constant $\rho = 0.75$.

As there was no intention to alter *PARMTEQM*, the z-version multipole, space charge and image charge dynamics were incorporated as separate subroutines into the *pteqHI* platform, with switches allowing active/passive comparisons.

**15.1.2 Representation of the First Cell**

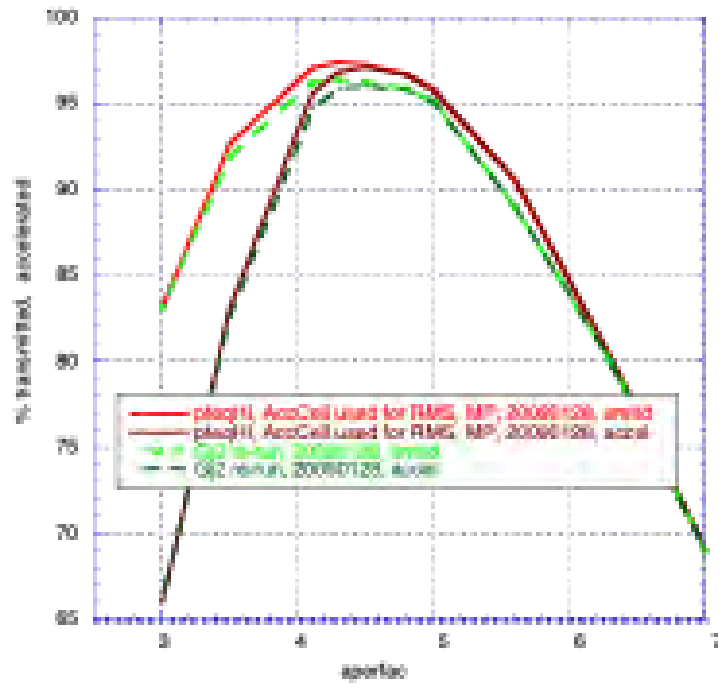

Fig. 15.3. Red, dark red – % transmitted, accelerated using accelerating cell formulae in radial matching section. Green, dark green – standard *PARMTEQM*.

Both codes store parameter values at the start of the first cell. *PARMTEQM* uses some approximations for the radial matching section. If the accelerating cell multipole field expressions are also used in the radial matching section (*pteqHI*), ~1% difference is observed (Fig. 15.3).

**15.1.3 Input Particle distribution**

As discussed above, the *PARMTEQM* input distribution is for all particles "at the same position". The *PARMTEQM* dynamics, space charge and image subroutines transplanted to *pteqHI* continue to use position as the independent variable. The input distribution for *pteqHI* is converted by moving the particles "to the same time", as discussed above.

**15.1.4 Synchronous Particle**

The synchronous particle diagnostic was added to the *PARMTEQM* subroutines embedded in *pteqHI*, and revealed problems, including that round-off tolerances for cell boundary crossings, etc., are needed in the *PARMTEQM* subroutines also.

**15.1.5 Starting Conditions**

In both codes, the default Type 6 RFQ dc distribution was positioned with the head of the input bunch at the input to the first cell, defined as z=0. Offsets can be entered in the input file, but were not.

**15.1.6 Dynamics & Space-Charge Mesh, Handling of Out-of-Mesh Particles**

For this comparison, both *PARMTEQM* and *pteqHI* codes use a basic dynamics mesh and a space-charge mesh of ±1 cell around a synchronous particle. (Other meshes can be selected in *pteqHI*, and another is the preferred standard.)

*pteqHI/PARMTEQM* employ the same methods to bring fast or slow particles back into the mesh, and for computing space-charge for particles outside the mesh, by applying point-charge forces between the particle and a point-charge representation of all the particles within the mesh. (This is a major difference from the *BEAMPATH* comparison.)

**15.1.7 Lost Beam & Accelerated Beam Criteria**

*15.1.7.1 Radial Loss:*

Radial losses were declared using a *pteqHI* subroutine independent of which dynamics, space charge or image force subroutines were active. Instead of the too-restrictive smallest-aperture rectangle used in Sec. 3 and in the executable *PARMTEQM*, an exact geometrical representation of the vane surface is used at each step, and a particle is declared radially lost if it hits this surface. A significant number of particles traveling near and around the vane tip are transmitted successfully.

*15.1.7.2 Longitudinal Loss:*

No losses are declared longitudinally.

*15.1.7.3 Transported, Accelerated Beam:*

Transmission is defined as all particles not radially lost, and accelerated beam as all particles near the synchronous energy. The tolerance for accelerated particles is set in the input data, and was typically ±10% of the synchronous particle energy.

**<u>15.1.8 Number of Particles, Step Size</u>**

Comparison runs were usually made with 10000 particles, and 10 steps per cell, but various tests were made with 5K-1M particles, and up to 1000 steps per cell.

The nonlinearity of the longitudinal dynamics causes an unavoidable dependence on the number of integration steps per cell. Fig. 15.4 indicates a typical dependence for a particular RFQ. This is not a particular feature of *PARMTEQM* or *pteqHI* but will occur with any linac simulation code using a similar integration technique[128]. A large number of steps must be investigated as a final step in a detailed design procedure.

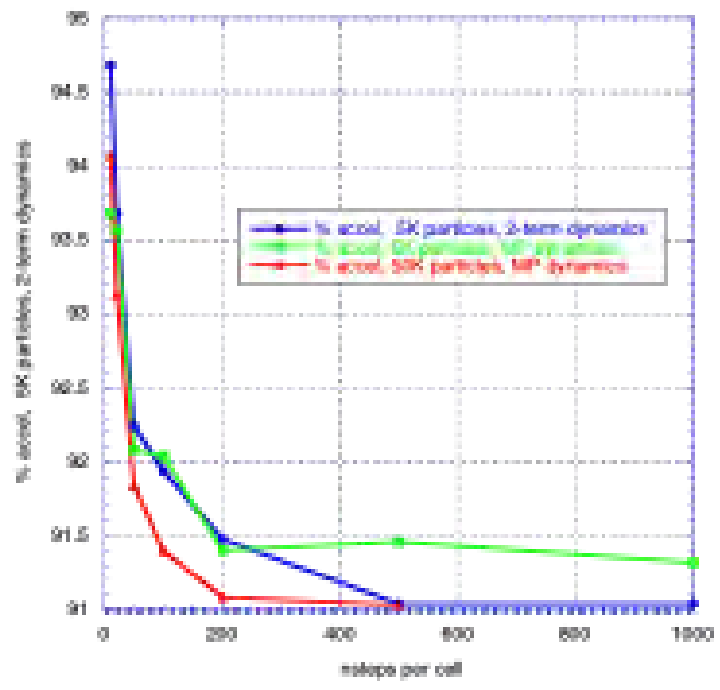

Fig. 15.4. Dependence of the percentage of accelerated particles on number of integration steps per cell for a typical RFQ. Blue – 5K particles, 2-term dynamics. Green – 5K particles, 8-term multipole dynamics. Red – 50K particles, 8-term multipole dynamics.

**<u>15.1.9 Subroutines</u>**

As noted, *PARMTEQM* subroutines were embedded in *pteqHI*, to enable active/passive comparisons in the same framework.

**<u>15.1.10 Driver and Physics/Mathematical Frameworks</u>**

The driver frameworks of *pteqHI* and *PARMTEQM* are essentially the same; *pteqHI* was used as the platform.

Both use the standard leapfrog symplectic, stable, integrator. Care must be exercised in selecting the integrator for the discrete difference equations representing the underlying differential equations. It is entirely possible that an integrator is consistent, accurate, and symplectic but still unstable with respect to truncation or round-off (usually smaller) errors [129, Chapter 13 - Appendix 1].

[eltoc]

# Chapter 16 – Evolution of the Physics Basis of *LINACSrfqSIM*

In this Chapter, the evolution of the physics basis and code details for RFQ simulation from *pteqHI* to a new RFQ code named *LINAC*S, with RFQ design *LINACSrfqDE*S and RFQ simulation *LINACSrfqSIM* are discussed. The approximations of computing the external field from the cylindrically symmetric 8-term multipole coefficient expansion, the space charge fields from the cylindrically symmetric open boundary RZ PIC method, and the image effect from an approximate method are removed, and replaced by solution of the Poisson equation. Interrelationships between design and simulation are of particular importance.

---

128 e.g. LIDOS, B. Bondarev, private communication.

129 "Computer Simulation Using Particles", R.W. Hockney & J.W. Eastwood (H&E), ISBN 0-85274-392-0, 1988, Chapter 4.

## 16.1. EXTERNAL FIELD

### 16.1.1 The 8-Term Multipole Representation of the External Field

Both *PARMTEQM* and *pteqHI* codes compute the transverse and longitudinal external field forces on each particle using the 8-term multipole expansion (Fig.14.1, A01,A03,A10,A12,A21,A22,A30,A32). The expression is cylindrically symmetric, and relies also on other symmetries. It is accurate only out to a transverse radius somewhat less than the minimum aperture of the RFQ cell.

The capability of the *LIDOS* code are introduced now – the potential at the beginning of a cell near the end of an RFQ was computed using the *LIDOS* Poisson solver, as shown in the upper left of Fig. 16.1.

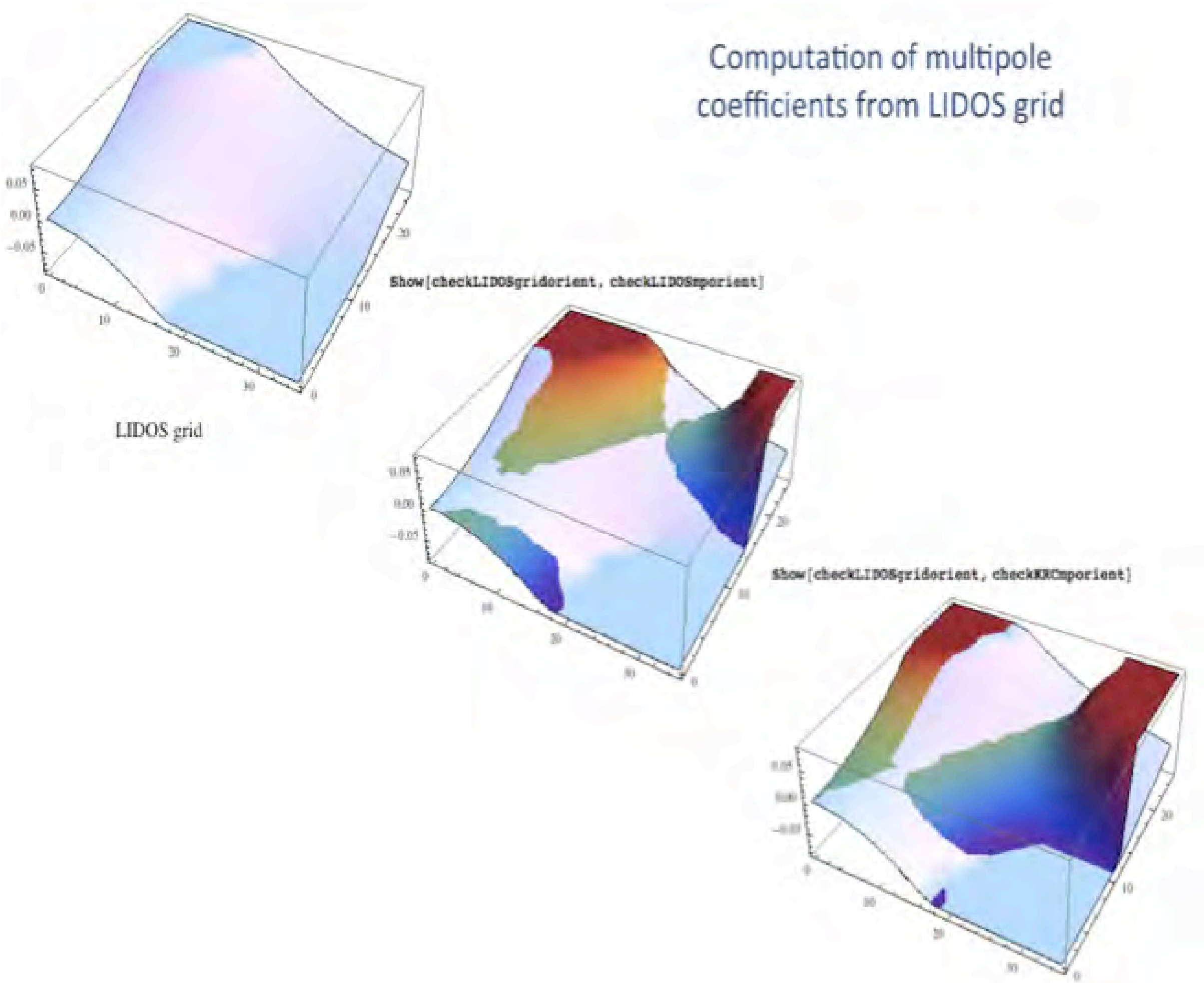

Fig. 16.1 Comparison of potential surfaces generated by a Poisson solver and by the 8-term potential.

Multipole coefficients were computed from the *LIDOS* grid, and the center figure is the resulting potential surface, superimposed on the Poisson surface,

The corresponding potential surface generated from the coefficient tables in *PARMTEQM/pteqHI*, superimposed on the *LIDOS* surface, is shown at the lower right.

Fig. 16.2 compares the coefficients computed from an arc slightly larger than the sinusoidal r0 radius, halfway between the minimum and maximum apertures of the cell, at the start, middle, and end of the cell. The lower line of the graphic is at the cell center, the others at the cell beginning and end.

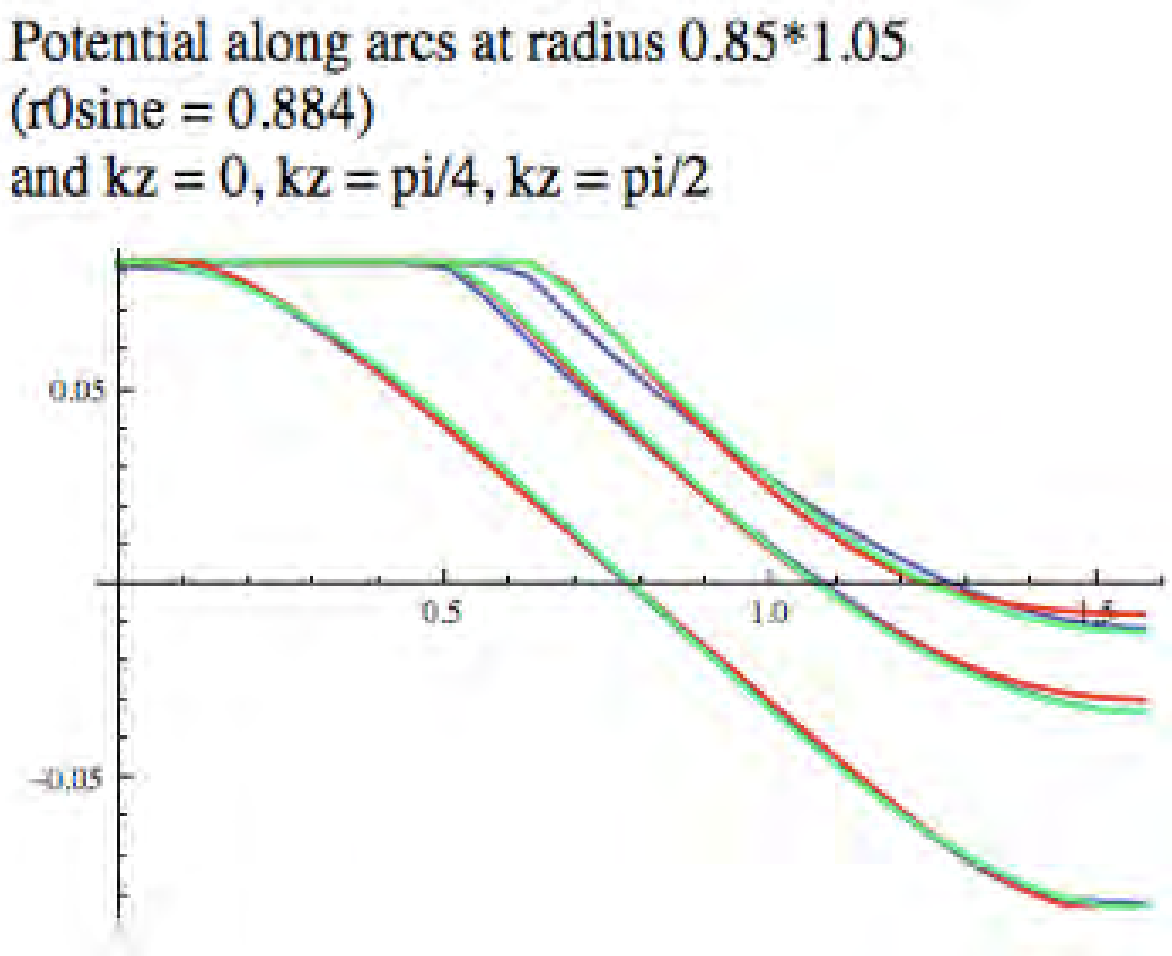

Fig. 16.2. Comparison of coefficients computed from the *LIDOS* grid and from the *PARMTEQM* coefficient tables.

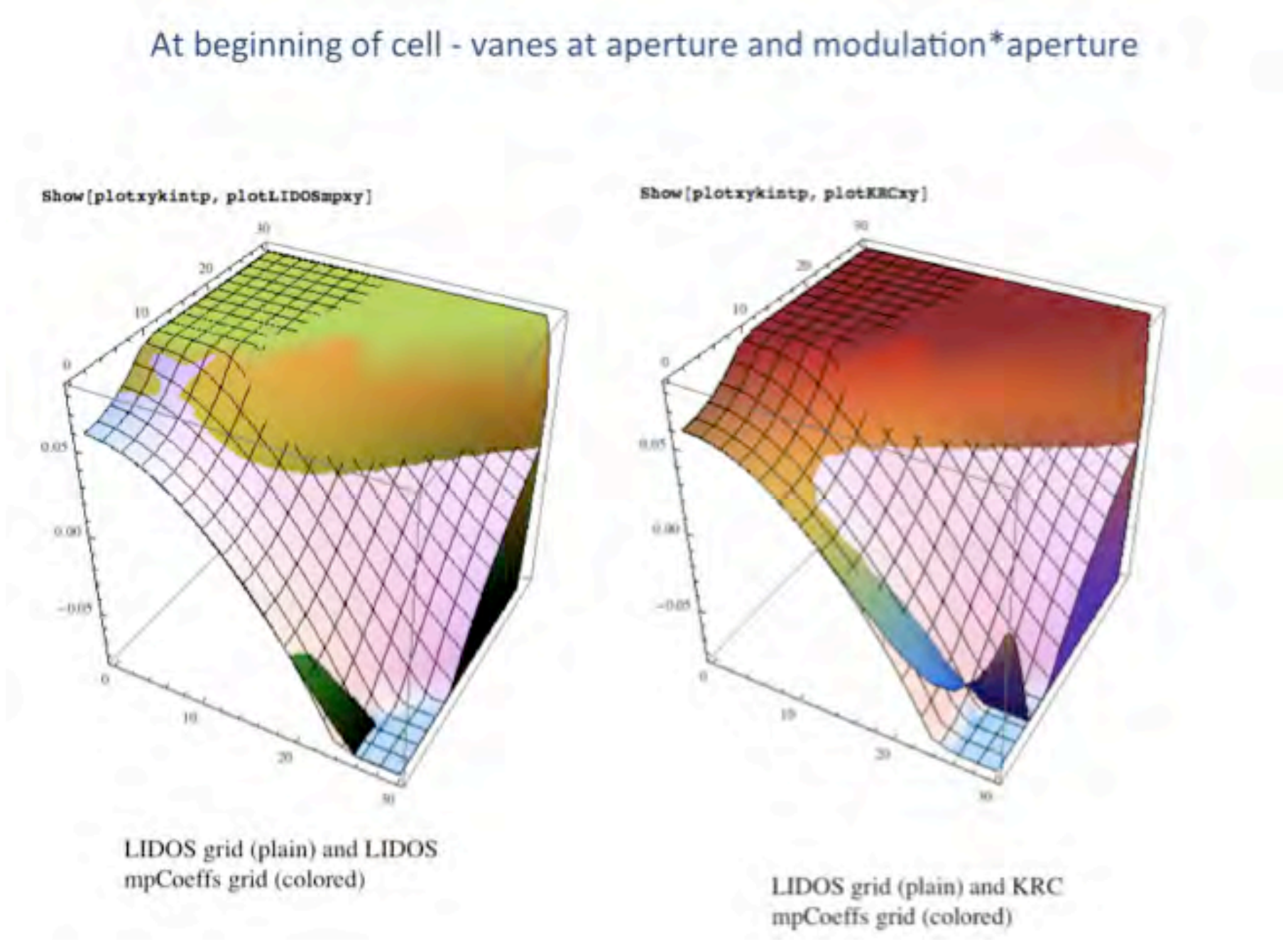

Fig. 16.3. Potential comparison at cell beginning.

Figs. 16.3 shows the cell beginning superposition in more detail. Fig. 16.4 illustrates definitively the problem with the multipole representation – it is reasonably accurate at the cell middle, but does not represent the potential well at radii greater that the minimum aperture and thus is in error over most of the cell, and in the important regions where particles are approaching the vane surface and may be lost.

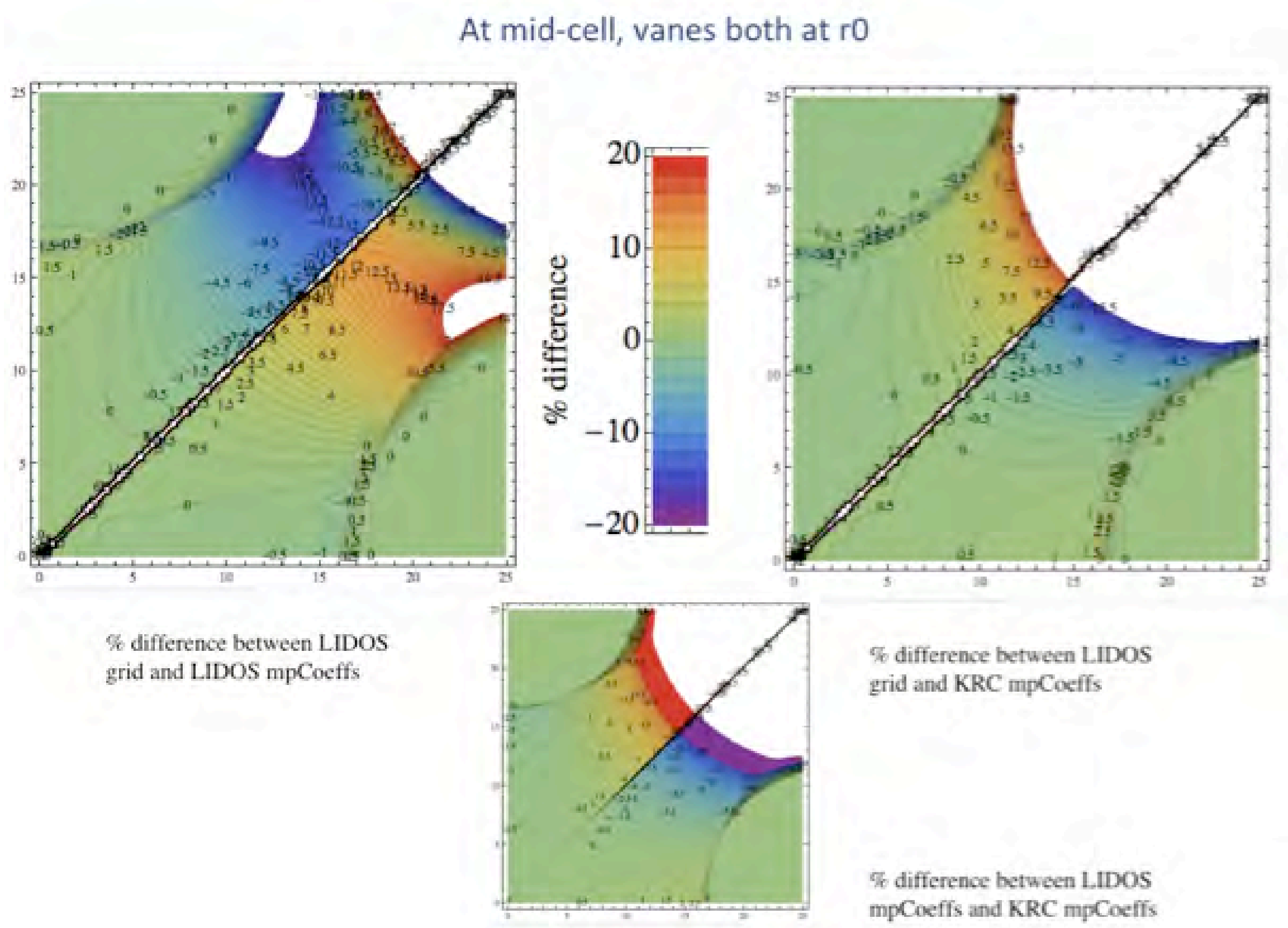

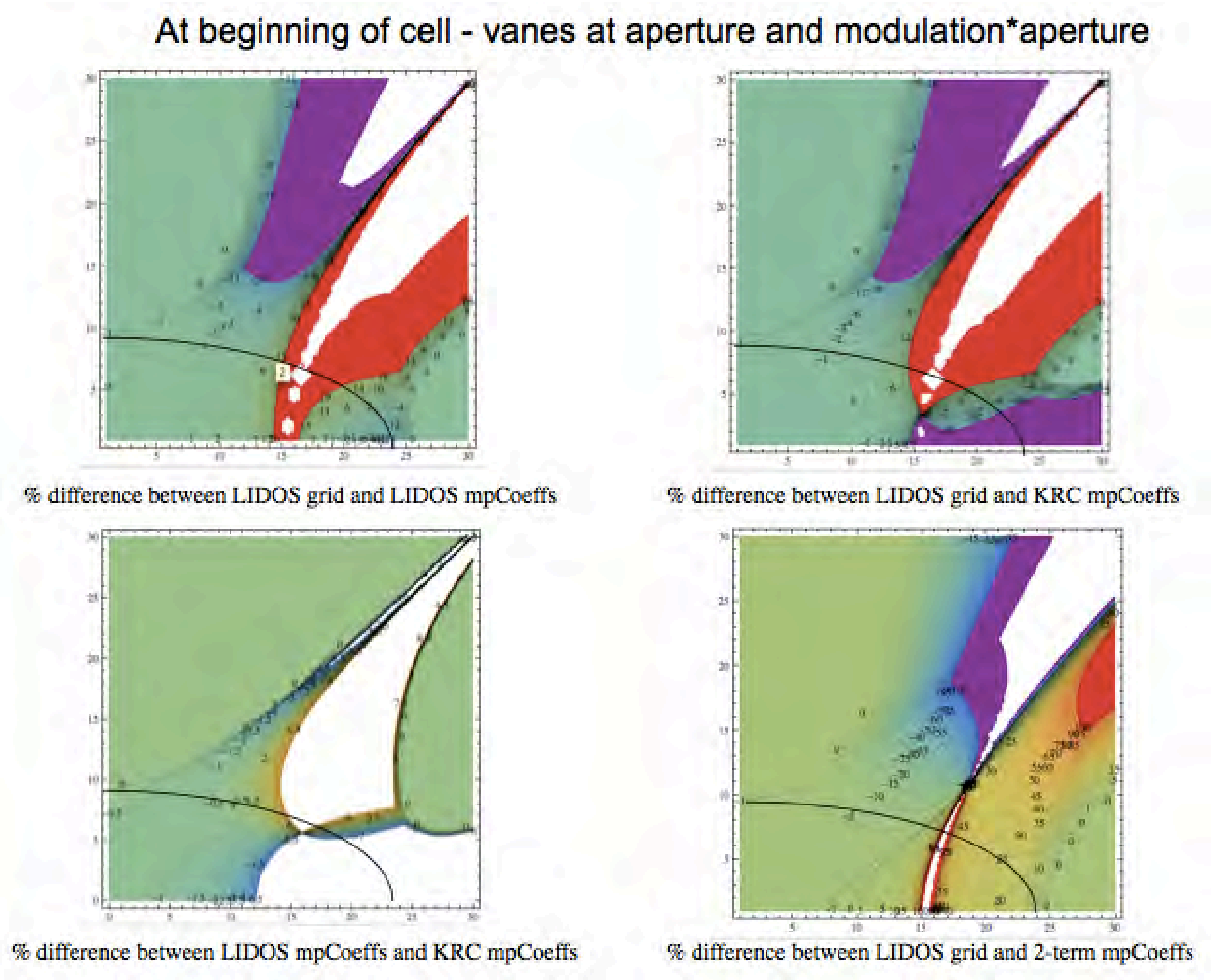

Fig. 16.4 Contour plots of the potential difference between Poisson and multipole solutions.

### 16.1.2. Removal of Paraxial Approximation

*PteqHI/LINACSrfqSIM* has no paraxial approximations. *PARMTEQM* uses the individual particle velocity in z as beta, instead of the full beta – apparently a residual left-over from very early versions when computer power was limited and the average velocity of the entire bunch was used for each particle. The effect is small, as indicated in Fig. 16.5.

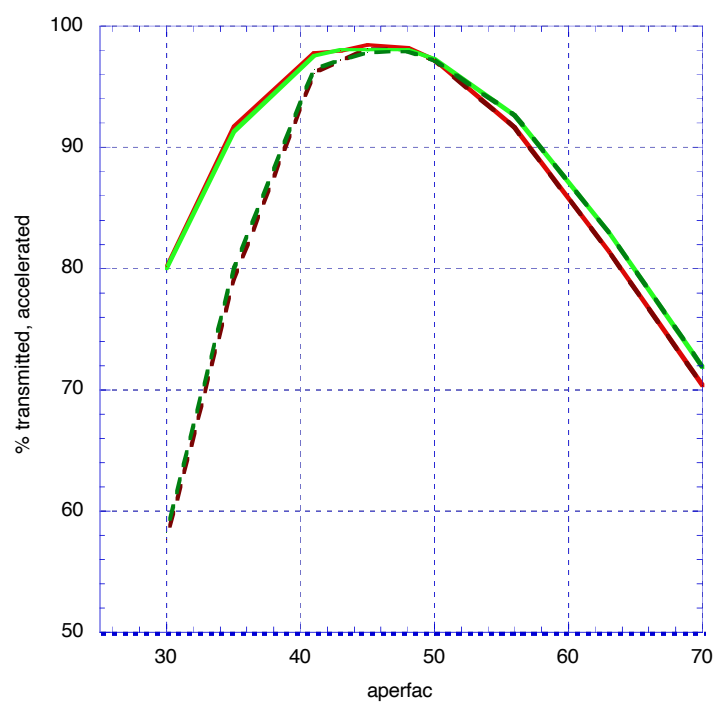

Fig. 16.5. Effect of paraxial approximation in *PARMTEQM*.

### 16.1.3. Missing Acceleration and dq/dt Terms

Please refer also to Ch.28, for review and some extension this very important and key discussion.

Following the derivation of the equations of motion, it is usually assumed that the changes in transverse and longitudinal focusing over a step are small enough to ignore. Typical RFQs are essentially adiabatic – experience shows that attempts to push too hard, e.g., for shorter length, result in inacceptable emittance and beam loss performance. Atypical RFQs with extremely rapid parameter changes are also occasionally encountered in the literature.

16.1.3.1 On Rms Envelope Equations of Motion "With Acceleration"

R.A. Jameson – November 2021

The theoretical development of envelope equations of motion in a beam transport system is very old. They are derived from the *time-dependent* Vlasov equation, averaging over a distance of interest and discarding detailed motion within that distance – the "smooth approximation" – using "the slow part" of the motion.

This was done first for a continuous smooth transverse focusing channel with no parameter variation, and later extended to periodic channels (also with no parameter variation), with averaging over the period. By requiring the second derivative of the envelope axes with respect to time to equal zero, the motion will repeat exactly over each period and the channel is called "matched."

It was shown for this restricted model that destructive resonances could occur. The location of these in the tune space was found, and the usefulness of this knowledge verified in (many) operating machines [130].

This construct was clearly not exact for real practice, where lattices are not strictly periodic and may contain additional components that influence not only the transverse motion (e.g. higher multipole magnets – not in every "period", perhaps in "superperiods"), but also the longitudinal motion (e.g. acceleration, bunchers, wigglers etc.). Then the system is no longer Hamiltonian, but involves dynamics that involve dissipation and becomes extremely difficult to treat theoretically. At this point, a theory with simplifying assumptions can be helpful in understanding, but must be augmented – in order to be useful – by numerical simulation.

A key **element** is that a theory, or at least an underlying theory with numerical augmentation, needs to be *useful in practice*. In the following, some background is reviewed on where envelope equation theory usefully influenced practice, and how it is extended to include the effect of acceleration.

16.1.3.1.1 Historical linacs

The LAMPF 800MeV, 1 mA average current proton linac designed and built in the 1960's, was a huge step. It used only newly developed simulation codes with components setting up "outside-in" external fields, inserting a particle distribution with only a matched initial injection condition, and seeing what happened. After commissioning, many investigations were made on its performance and resulted in its improvement, but not completed as it became a purely operational facility. Nevertheless, from that point, it has operated well in that condition for 50 years.

Lower final energy linacs designed in the same way were built at Fermilab, BNL and CERN. An undesirable characteristic of these and LAMPF was transverse emittance growth and beam loss.

In the 1970's, the ideas of matched envelope equations, and (Lapostolle & Sacherer)'s escape from the confusing KV constraint, motivated CERN to build a "New Linac" designed to keep the beam rms matched throughout. As related elsewhere in this book, it had the same emittance growth problem. Now not surprising – an "outside-in" design, the solutions for matched beam size academic – can just simulate to see what happens.

---

130 It has been claimed in recent years that an accelerator (linac, ring,) theory can only be tested by a "dedicated experiment", and that experiments and analyses on all the machines in operation do not count… (!!!)…

In the 1980's, it was shown how a transported and accelerated 3D elliptical beam could be kept in rms equilibrium, and how the matched transverse and longitudinal rms envelope equations, plus possible other equations governing the beam, are used in their true time-dependent state to locally determine – from the desired beam space charge physics "inside-out" point of view – the external fields to achieve equilibrium transport and avoid the dangerous resonances. This is done in conjunction with numerical techniques [131].

From the 1990's, this practice has been used to extensively develop the design and simulation practice for the breakthrough RFQ accelerator and a number have been built and tested. In addition, the major J-Parc accelerator was designed with this practice, and demonstrates very good performance, with better tolerance of machine errors than "matched only" design, and also good tolerance to a major departure from the underlying equipartitioned design in which the entire transverse focusing profile is weakened to make the transverse beam dimension larger in order to avoid the effect of intrabeam scattering. Other new projects are showing evidence that this practice is at least partially and slowly being established as the state-of-the-art.

16.1.3.1.2. Removal of the Smooth Approximation Theoretical Constraint

The "smooth approximation" is of course no constraint in actual computer design and simulation practice. Chao Li showed (2018) how this theoretical constraint can be removed, with the aid of numerical techniques. This has been usefully applied to show the influence of transverse lattice details within the focusing period on resonance bandwidths. The approach is potentially useful as an aid to understanding longitudinal and coupled transverse/longitudinal effects as well [132].

16.1.3.1.3. Rms Envelope Equations of Motion "With Acceleration"

Studies since the 1950-1960's have considered theoretically including the effect of acceleration and parameter variations in the rms envelope equations. My ~172 page detailed diary goes back to 1969, 1981-1985, with very extensive work in 1994-1998, collaborating with theorists including P. Lapostolle, O. Anderson, E.P. Lee, B. Bondarev, K. Mittag, R. Gluckstern, and especially in 1998 with R. Ryne and the work of M. Ikegami and myself to re-derive and extend Ryne, and extend in application to practice [133]. These studies were plagued with many questions, (e.g., I found that the assignment of expanded terms to the potential and kinetic parts of the Lagrangian needed correction, and questioned why some derivatives would be included and many others also varying with time ignored, etc.).

In 1998, by considering the first derivatives of the beam envelope ellipse radii (a & b) in the framework of instantaneous system state and inside-out design, compact modifications of the beam envelope equations to account for changing external field parameters and space charge were found, programmed and simulated in the *LINACS* code, as outlined below.

After more than 20 years, the question is resurfacing.

---

[131] It has also been recently claimed that it is senseless to design a machine to a "design beam current" – under any circumstance - apparently in the belief that everything works only from "outside-in" in any case… ( But placing a design trajectory "in the resonance-free areas between major modes" is allowed – which requires specification of a design current…? And ignores the often present tune oscillations that may cross the local (instantaneous) EP with no excitation of cancelled modes there… Modern inside-out practice also designs to a beam current somewhat higher than the desired operating current – in order to have a tolerance and showing better performance at the operating and lower currents. Sensitivity testing to various off-design features is also part of standard practice.

[132] The stronger damping effects of human factors has unfortunately stalled this work.

[133] The works of Etienne Forest on this and many other aspects are very highly recommended – they are very well thought out in terms of giving a framework and elements very much in parallel to this book, as well as with lots of theory. See reference Ch. 27, ResearchGate, etc.

*16.1.3.1.3.1 Simulation:*

Simulation with correct Poisson solution for external and space charge fields is self-consistent, has grid with the changing cell dimensions as acceleration proceeds, and needs no "correction" for acceleration or other first derivatives. The fields acting on the particles are found on a mesh that has all the geometrical features, the full symmetries, boundary conditions, the potentials, the variations along the mesh. The simulation has no knowledge of envelope equations or any theory at all.

*16.1.3.1.3.2 What is the intended use of rms envelope equations "with acceleration"*

Codes to track envelopes are very old, known to be only approximate and cannot give beam transmission and loss information. Accurate tracking with simulation as above is reasonably fast with present computers. My interest in the usefulness of extended rms equations was then, as now, to possibly have a more accurate *design*, so let us review some aspects of that, using the RFQ as the example.

*16.1.3.1.3.3 Instantaneous beam state and inside-out design:*

**It is necessary to start with, accept and be comfortable with,** <u>two highest ranking **elements**</u>:

**1.)** **<u>Local transient beam state.</u>** There is no steady-state in a particle accelerator. The rms envelope and equipartitioning equations are time-dependent equations derived from the time-dependent Vlasov/Lagrangian/Hamiltonian roots. Their usual expression with the $2^{nd}$ derivatives of the transverse and longitudinal rms ellipse dimensions = 0 then requires the $1^{st}$ derivatives to be constant or zero (equal zero is the usual expression, assuming the non-changing continuous channel), leaving what seem to be time independent equations (but they are still time-dependent and in the Hamiltonian frame), and can be thought of as being for an infinitely long periodic lattice. This steady-state, physics accented, point of view is correct, but only a minor part of the deeper meaning and far too limiting. This chimney, or box, was broken out of by 1980, but the steady-state view still by far prevails.

***The crucial point is that at any particular time or position, the solutions are a "time (or position) snapshot (a "selfie"!), giving the instantaneous local state of the beam distribution.***

***The desired local state is <u>specified</u> by the designer at every point as the beam proceeds. <u>At least, rms matching is desired at every point</u>, and the lattice is constructed; this is what the practical designer wants to control.***

**2.)** **<u>"inside-out".</u>** The desired beam performance should drive the design of the external fields.

*16.1.3.1.3.4 Design:*

Sacherer wrote only that the mathematical difficulty requires the assumption of constant or *a priori* emittance – not pointing out that this has an important *practical* consequence:

***It is interesting, and crucially important, that constant or a priori emittance (preserved emittance, no undesired growth) is exactly what everyone always wanted and dreamed of !!! A design condition, regardless of the mathematics.***

The question is: Could the *design* process, using some modification of the rms envelope equations, specifically, to include first derivative terms, give a "more real" design to which the simulation and constructed device agree more exactly. (Close design-to-simulation and constructed device correlation is an **element** of the desired foundation.)

The rms envelope equations require that the beam ellipsoid dimensions are such that the $2^{nd}$ derivatives of the transverse (x and y, or their average (called "a")) and longitudinal radii (γb) = 0; the beam is "matched. This condition must always be satisfied – point-by-point – step-by-step. The equations are "used up" by simultaneous solution for the radii. If $1^{st}$ derivatives are included, their coefficients will be fixed at every point. Higher order terms such as emittance, = aa', are held constant or varied *a priori*.

In the various types of linacs, there is more or less coupling of the external fields between the transverse and longitudinal planes, and also the coupling via the space charge fields. The design method involves the *simultaneous solution, at each step,* of the coupled rms envelope equations and the EP or other equation(s) that specify the desired beam physics requirements, to determine the external field parameters. Therefore, *any modification, such as the inclusion of 1st derivatives, is immediately coupled and influences the simultaneous solution*.

The form of expression used may emphasize or obscure this:

The form (emittance = right side) says that if you make a new right side, like adding an "acceleration" term, the new right side still has to = emittance. I.e., the original (constant or *a priori* varying) part of the new right side has to change.

The equations are often squared, for expanding with detailed (phase advance)^2, and then written (left side - right side) = 0, obscuring the above.

The *LINACS* design process proceeds to obtain the desired local "inside-out" system state step-by-step:

- The number of steps per cell is set large enough that the performance change with a larger number of steps is considered small enough. The step size is the "dt" of any derivative.
- The coupled transverse and longitudinal envelope (plus optional EP or other) equations are solved simultaneously *and exactly* at each step. These equations contain additional parameters that are functions of time. The (2 or 3) equations allow (2 or 3) quantities to be solved for (usually the beam ellipsoid radii using the two matching equations) and additionally the vane modulation for the EP condition if used.
- With only 2 equations used to solve for the beam ellipsoid radii giving a matched beam, there are no further equations that could be used to control the beam; all the external parameters (V, phis, aperture & modulation) (the above mentioned extra parameters) have to be specified with rules as functions of time.
- The synchronous phase is best ruled by the Teplyakov requirement to keep the phase density of the bunch under control. Thus the phase advances are functions of the envelope radii.
- The derivative(s) of these rules would be available, but we show that these detailed derivatives are not needed.

The method for obtaining "derivatives" across the step has to be specified, as *they are part of the solution and not a priori available*. The most obvious programming would involve finding the derivative as ((value at current search)-(value at last step))/stepsize*, involving two instantaneous points.*

16.1.3.1.4. Questions Concerning the Theoretical Addition of "Acceleration"

At a fundamental level, if "acceleration" or other derivatives are added to the Hamiltonian-based rms envelope equations, the equations are no longer Hamiltonian. The "acceleration" term is a dissipative (damping) effect, but there is no diffusion. As in the case of an oscillator with damping, the system becomes non-integrable, non-linear, and non-Hamiltonian.

The biggest questions seemed to be "which derivatives to include in the theory", and, "is ignoring other derivatives rigorous or a theoretical maneuver".

This became incredibly complicated in published and internal papers at the time. Theoretical assumptions turned everything into special cases, and worse, concatenated special cases. Sometimes theoretical tricks like re-normalizing or transforming Hamiltonians removed the exposition very far away from any conventional terminology or practical point of view. Each author used his own nomenclature, some working in the inappropriate z-domain, there was no re-generalization of the specializations.

Using the local system state point of view, it was realized that if the envelope equations were expressed simply in terms of the ellipse radii {a,a',b,b'}, the expression and computation become very simple, just requiring the local {a,b} state at the new step, where the new {a,b} are being sought by exact solution of the simultaneous equations, and the {a,b} at the previous step, for finding {a',b'}. The derivatives of all the internal details of both the external fields and the space charge fields are thus included in {a',b'}, and do not have to be worried about separately.

This form was found to be:

$$etn^2 = (a)^3(\beta\gamma)^2[(a)'' + (\frac{ez}{2T})(a)'] + (a)^4\frac{\gamma^2\sigma_t^2}{(\beta\lambda)^2}$$

$$eln^2 = (\gamma b)^3(\beta\gamma)^2[(\gamma b)'' + (\frac{ez}{2T})(\gamma b)'] + (\gamma b)^4\frac{\gamma^2\sigma_t^2}{(\beta\lambda)^2}$$

where {(a) ″,(γb) ″} = 0 for matching, T is the beam energy, ez is d(beam energy)/dt. With {etn,eln} constant or specified *a priori*, the added "acceleration" terms result in a required weakening of the local state phase advance term at the point being computed.

A recent example gave in detail:

| | | | | |
|---|---|---|---|---|
| No 1st derivatives: | Transmission | 98.48% | Accelerated beam fraction | 97.68% |
| 1st derivatives in DESIGN: | | 98.62 | | 97.82 |

and also insignificant differences in other quantities.

"Typical" here means with parameters that change smoothly and not too fast – experience quickly shows how fast the bunching and acceleration schedule can be pushed before unacceptable emittance performance and beam loss occurs. In the RFQ, the schedule is then typically adiabatic.

16.1.3.1.5 Conclusion

In the framework for any linac type of inside-out design and instantaneous system state, addition of the first derivatives of the beam rms ellipse radii {a,γb} is readily evaluated step-by-step, and requires no further derivatives.

Comparing designs with and without 1st derivatives indicates how adiabatic the design is.

A non-adiabatic ion linac design would probably be a rarity, given that rms matching is always required and phase advances should be as smooth as possible, also over transitions in lattice types, because non-adiabatic variations will probably compromise the desired emittance and beam loss performance. But maybe in a superconducting linac at higher energies? Semi-relativistic electrons can be handled non-adiabatically, as in the FEL injector.

## 16.2. SPACE CHARGE

### 16.2.1. Choice of Independent Variable z vs. t

*PteqHI/ LINACSrfqSIM* uses canonical variables in the time domain as the independent variables, which makes the space-charge computation correct and natural; whereas *PARMTEQM* uses the z-position along the RFQ as the independent variable. The resulting *PARMTEQM* coordinates (x, dx/dz, y, dy/dz, phase, energy) are not canonical, and a transformation must be made to time coordinates for the space-charge computation. The transformation from z-to-t and back in *PARMTEQM* is approximate (only the quadrupole focusing is considered, rf defocusing and space charge are neglected) and leads to inaccuracy in the location of particle losses along the RFQ, even though the bulk transmission may appear similar. For high-intensity factory accelerators, the location, as well as the magnitude, of the losses is important. Typically *PARMTEQM* computed space charge once per cell, at the time when the synchronous particle reaches the cell middle; then the rf field is zero, the beam is essentially round, and the ellipticity correction described above is used.

This was the standard KRC method until sometime in the 1990's, when an undocumented change was made in the LANL305 source code, probably because of the choice of others at that time to write new codes with time as the independent variable. The space charge computation for cells after the radial matching section was changed to be done at each step instead of once per cell, and the z->t->z conversion was no longer performed. As the difference between z-based and canonical coordinates decreases with step size, this did reduce the error significantly, at the expense of computing time. The *PARMTEQM* results use this method.

*pteqHI* applies space charge at every segment in the radial matching section, but after that it is usually sufficient to apply once at the middle of each cell, as the difference in result is insignificant and the run time is reduced.

### 16.2.2. Starting condition for, and frequency of space charge application

Both *PARMTEQM* and *pteqHI/LINACSrfqSIM* position the starting beam segment with its head at the start of the first cell, defined z=0. The particles are drifted with no focusing until their z is >0. *PARMTEQM* applies space charge in the radial matching section when the synchronous particle reaches the center of each cell, a delay in space charge application of 1.5 times the number of steps per cell. However, this procedure seems unrealistic, as space-charge acting on the real beam in the injection region has a large effect. *pteqHI/LINACSrfqSIM* applies space charge at every segment in the radial matching section, but after that it is usually sufficient to apply once at the middle of each cell, as the difference in result is insignificant and the run time is reduced. The change in the injection physics reduced transmission significantly, as indicated in Fig. 16.7.

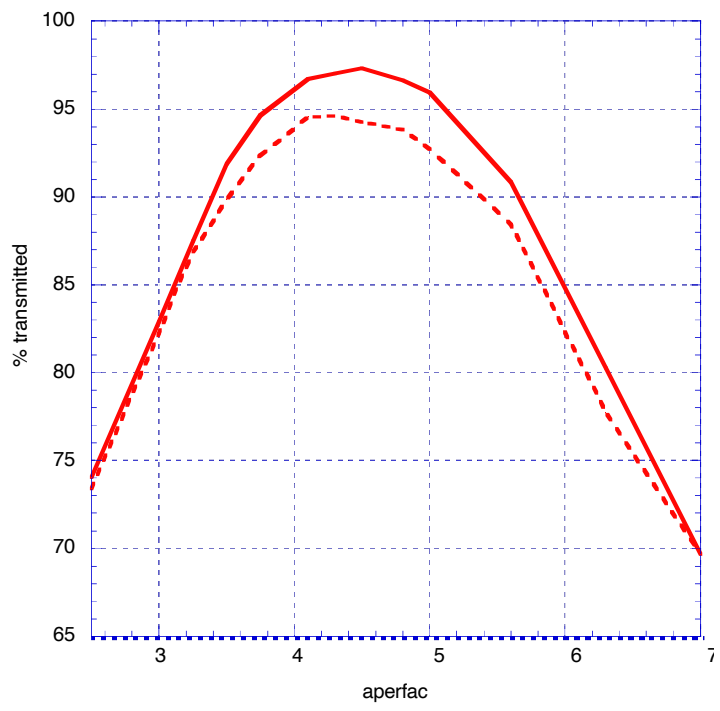

Fig. 16.7 Red solid – Delay of 1.5 times number of steps per cell before first application of space charge. Red-dashed – Application of space charge starting with Step 1.

### 16.2.3 The *scheff* Method

*Scheff* begins by placing the transverse radius and z-coordinate of each particle into an rz-mesh, with the charge distributed proportionally to the four corners of the mesh box, which results in a smoothing and avoidance of the effect of particles approaching each other very closely. The accumulated charge at each mesh corner is assumed to be a transverse ring of charge (cylindrically symmetric), the space-charge force of each ring on every other ring is computed, and the forces are then applied to each particle.

As pointed out above, with a given charge distribution on the r-z mesh points, this ring-on-ring method is a theoretical model with exact solution, and therefore gives exactly known space-charge forces for any particle distribution. It does not depend on any boundary conditions. It gives no information outside the particle distribution (the basis function has no tails; it is an open boundary condition.).

*pteqHI/LINACSrfqSIM/PARMTEQM* account for space charge when transporting a particle, which has fallen behind the mesh, back into the mesh, by applying an impulse calculated as point-to-point between the particle and the bunch center of gravity. Each particle is also adjusted for multiple adjacent bunches on both sides of the mesh (typically 20 on each side) using the point-to-point method. These are also theoretically exact impulses.

The knowledge that these are exact affords a great advantage in comparing space charge methods having the same symmetry conditions, and also in investigating other aspects, as discussed next.

### 16.2.4 Extent of Space Charge Mesh

*Scheff* allows study of the number of mesh points needed in the grid without influence of boundary conditions.

Discussion of the radial mesh size was given in Chapter 14.

Early concern that other codes gave very high transmissions compared to *pteqHI* led to tending to look for settings for the longitudinal mesh extent in *pteqHI* that would give higher transmission. A "standard" longitudinal mesh extent (*BEAMPATH, LIDOS*) is ±1 cell around the synchronous particle, naturally selected because of the boundary conditions for a Fourier or Poisson type solution. *PARMTEQM* uses the separatrix length.

Fig. 16.8 shows a study of the maximum longitudinal beam extent with respect to various boundaries. A z=mesh extent of 1.2*cellgth encloses the beam for all cases. This rule gives a higher transmission, and might be justified considering the complex dynamics that occurs at the longitudinal separatrix.

On the other hand, a rule based solely on a multiple of the rms beam length would result in a coarse grid in the shaper section and a fine grid in the acceleration section, with particles outside being handled by the point-to-point technique.

A combined rule, using ±cellgth when the separatrix is long and a multiple of zrms when the beam becomes well bunched, keeps the longitudinal space charge grid adapted to the beam length.

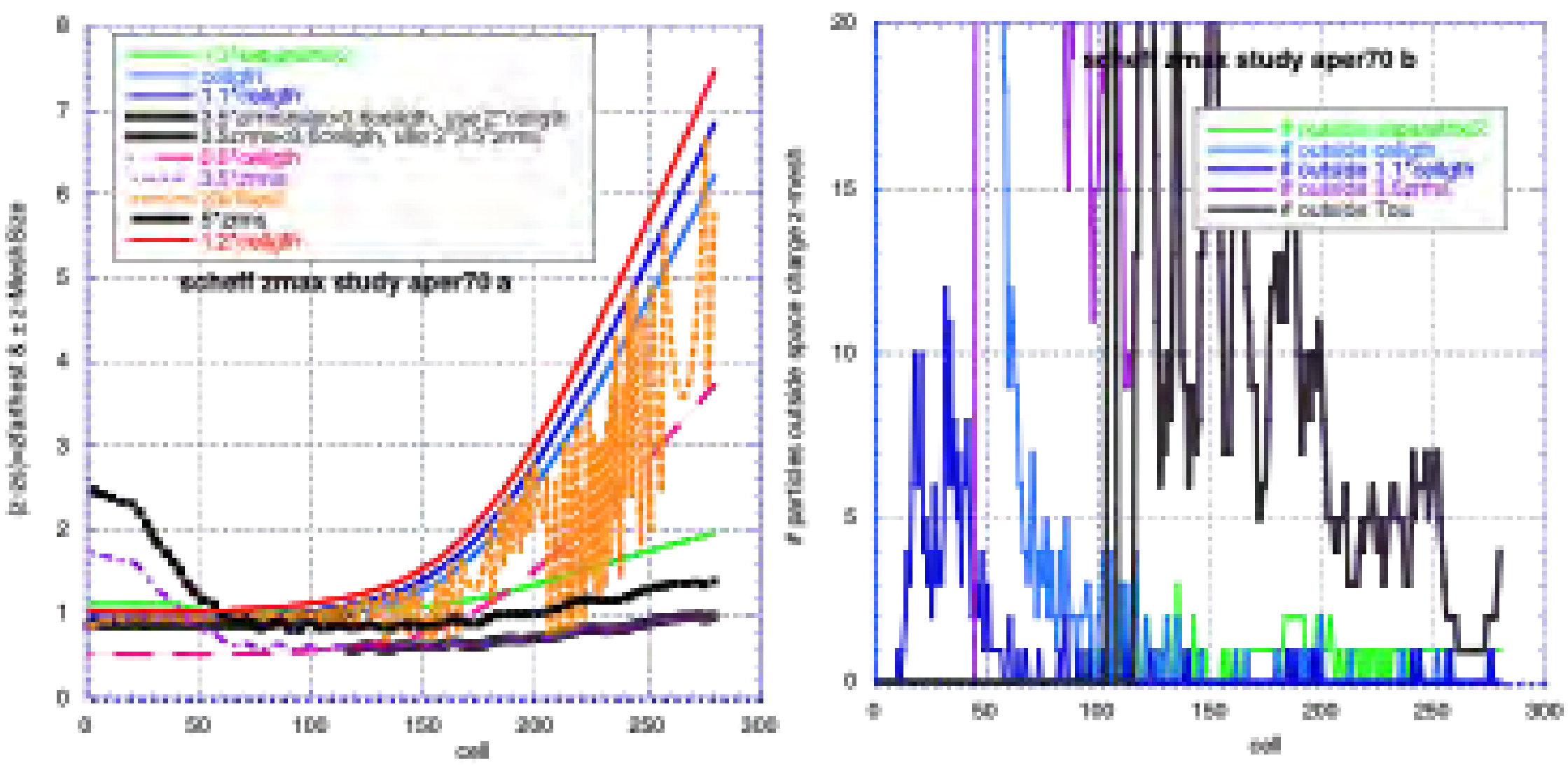

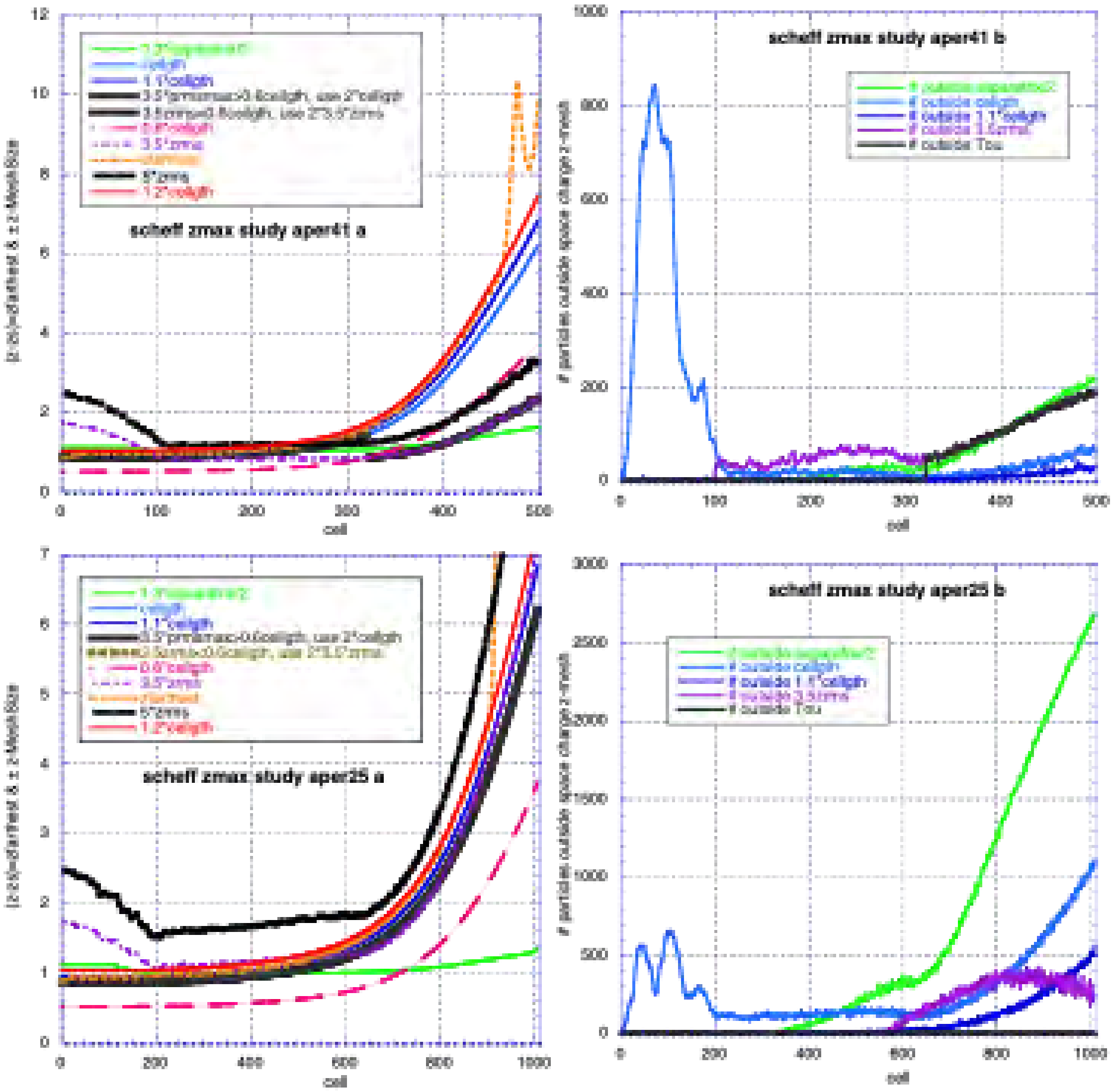

Fig. 16.8 Cell and beam length boundaries compared to the maximum z-extent from the synchronous particle (zfarthest), of the beam as it transits the RFQ, for the *aperfac* cases 70, 41 and 25. The left-side figures (a) show 1.3*(separatrix length)/2; 0.6, 1.0, 1.1 and 1.2 times the cell length (cellgth); 3.5 and 5 times the rms beam half-length (zrms); and the location of a z-mesh-extent rule (Tou) that uses ± cellgth when 3.5*zrms > 0.6*cellgth but ±3.5*zrms when 3.5*zrms < 0.6*cellgth. The right-side figures (b) show the number of particles (from 10000 particles injected) which travel outside the various regions All cases are with a 20x40 RZ mesh (i.e. 20 mesh points per beam half-length).

It was found that the long z-mesh-extent rules were sensitive to the number of z-mesh points, the number of particles, and the number of steps; but that the rule adapted to beam length is insensitive to these variations. Fig. 16.9 shows (red) the *aperfac* curves for z-mesh-extent of ±1.2*cellgth, which show higher transmission particularly for the smaller apertures than the curves for z-mesh-extent following the adapted rule. The adapted rule shows very little sensitivity to various number of grid points, number of particles number of integration steps per cell, and number of space charge applications per cell.

This is an important finding. It indicates one of the reasons why *PARMTEQM* transmission results are optimistic.

*pteqHI/LINACSrfqSIM* uses ± cellgth when 3.5*zrms > 0.6*cellgth but ±3.5*zrms when 3.5*zrms < 0.6*cellgth.

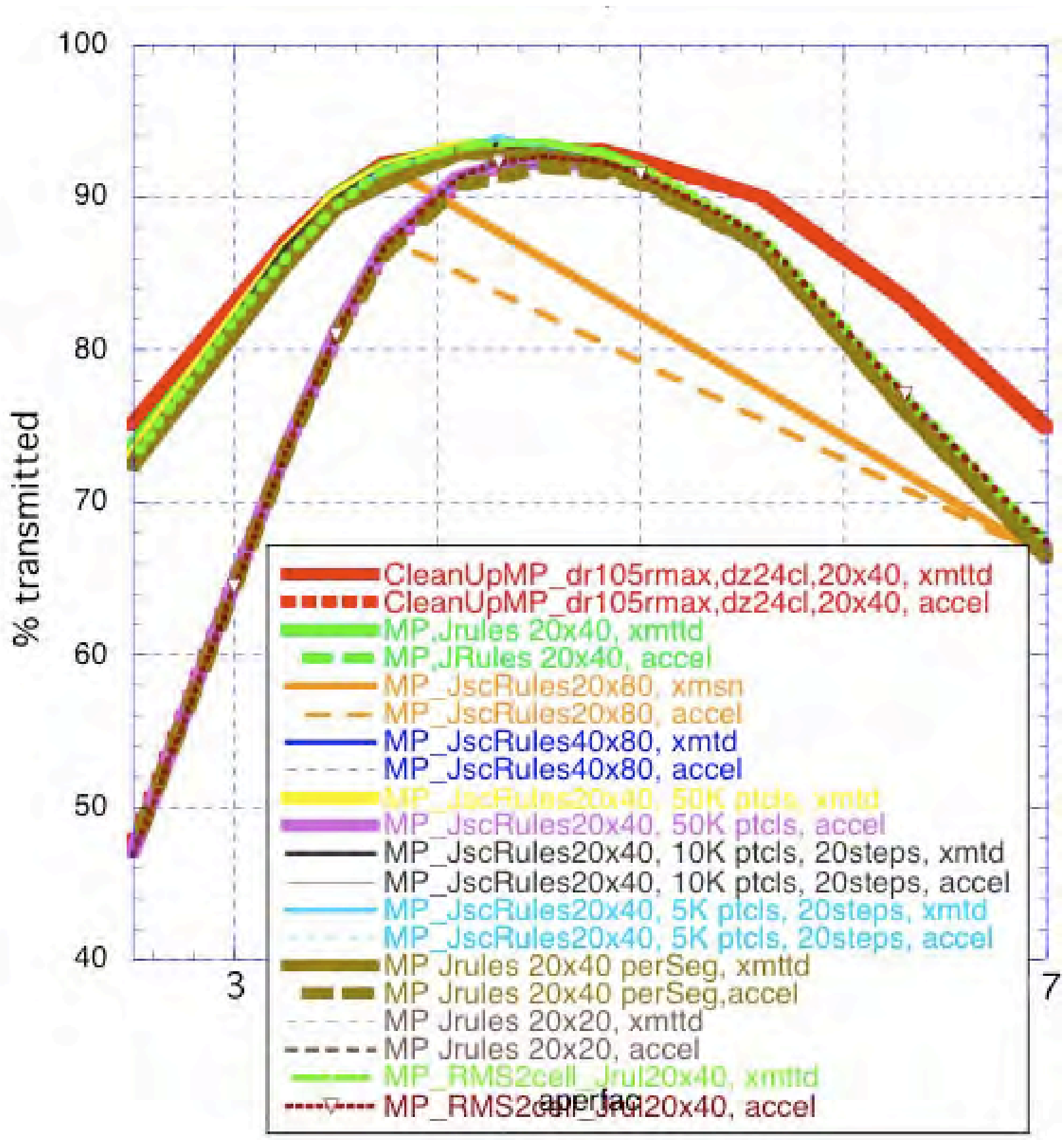

Fig. 16.9 Dependence of transmission and accelerated beam fraction curves on the scheff z-mesh-extent for different rules and different number of grid points, number of particles, number of integration steps per cell, and number of space charge applications per cell.

### <u>16.2.5 Comparison to 3D *PICNIC*</u>

The actual RFQ beam with significant space charge forces has a diamond shape transversely at the beginning and end of a cell, more circular at the cell center. The cylindrically symmetric *scheff* was checked using the 3D space-charge routine *PICNIC* [134], a PIC code in which the particles are located in x,y,z, cubes. The mesh size was selected to correspond to scheff, with ±3.5*zrms (the default setting), and the charge assignment mode was set to allocate particle charge to the nearest corner of the cube containing it. Fig. 16.10 indicates that the *scheff* method corresponds very well to the full-3D method.

134 PICNIC – courtesy of Nicholas Pichoff and Didier Uriot, private communication.

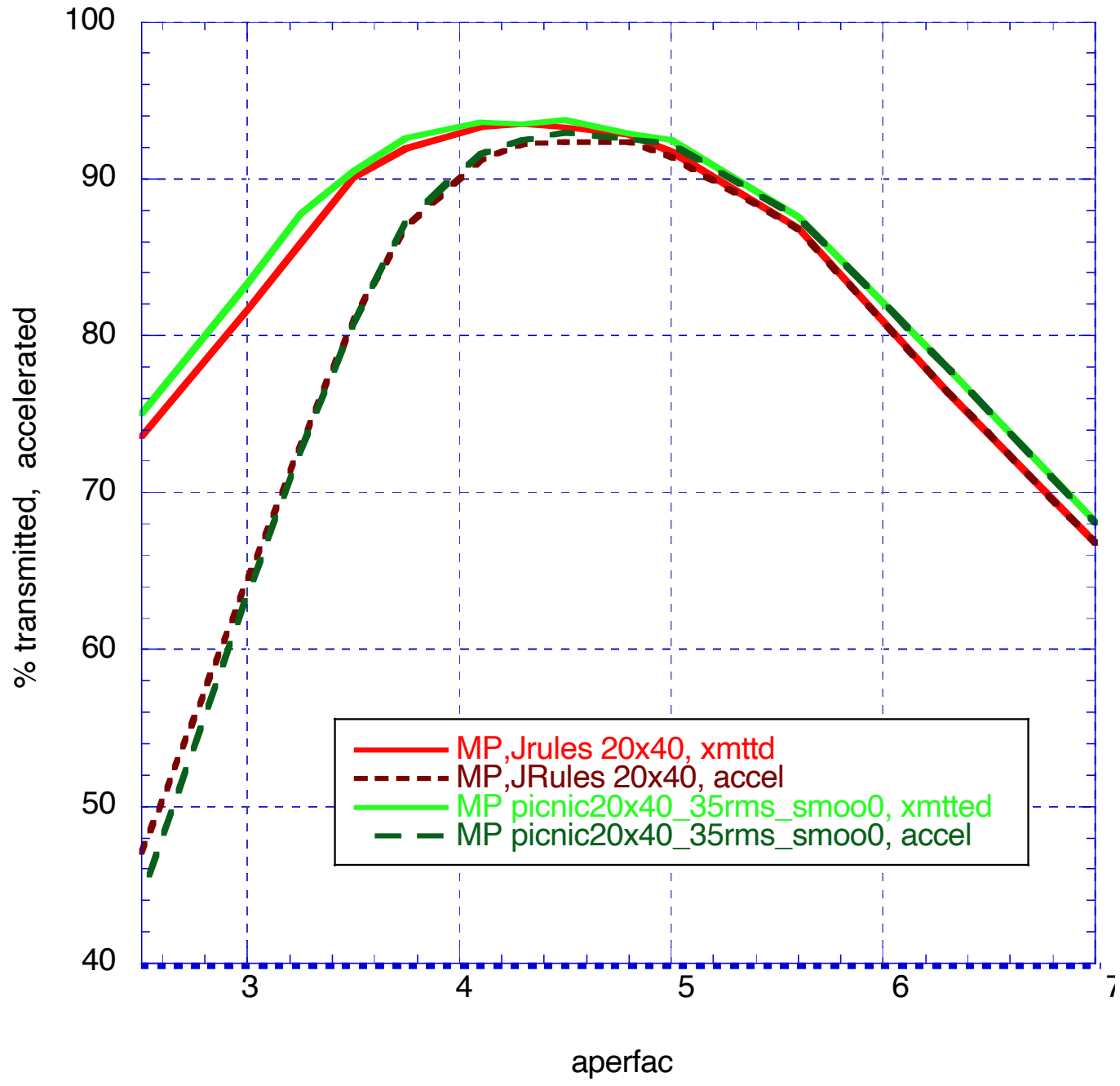

Fig. 16.10 Comparison of RZ *scheff* and 3D *PICNIC* space charge methods in the multipole coefficient external field.

## 16.3. IMAGE CHARGE

On order to approximate image forces in an RFQ, KRC used the general solution of the Laplace equation in cylindrical coordinates at the vane tip and eliminated multipole coefficients not having the correct symmetries. The beam is cylindrical at the beginning, and an ellipsoid when bunched. At an intermediate position, it is represented as a fraction that is cylindrical and can be represented by a line charge on the axis. The remaining fraction is assumed spherical, and is represented as a point charge on the axis, and image charges are calculated when the point charge is at the beginning and at the center of each cell.

Anecdotal results indicated that the image charge effect of this method might have a small, order 1-2%, effect on transmission, which appeared reasonable; however the systematic *aperfac* test reveals that anecdotal results can be seriously misleading. Fig. 16.11 shows the results with and without the KRC image effect. For the smaller apertures, the image effect lowers transmission and acceleration significantly. With the image effect turned on, the optimum is shifted quite strongly toward a larger aperture, which is intuitive.

However, for even larger apertures, transmission and acceleration are significantly increased. This effect, and the magnitude of it, was questionable – and a major motivation for pursuing an accurate Poisson solver.

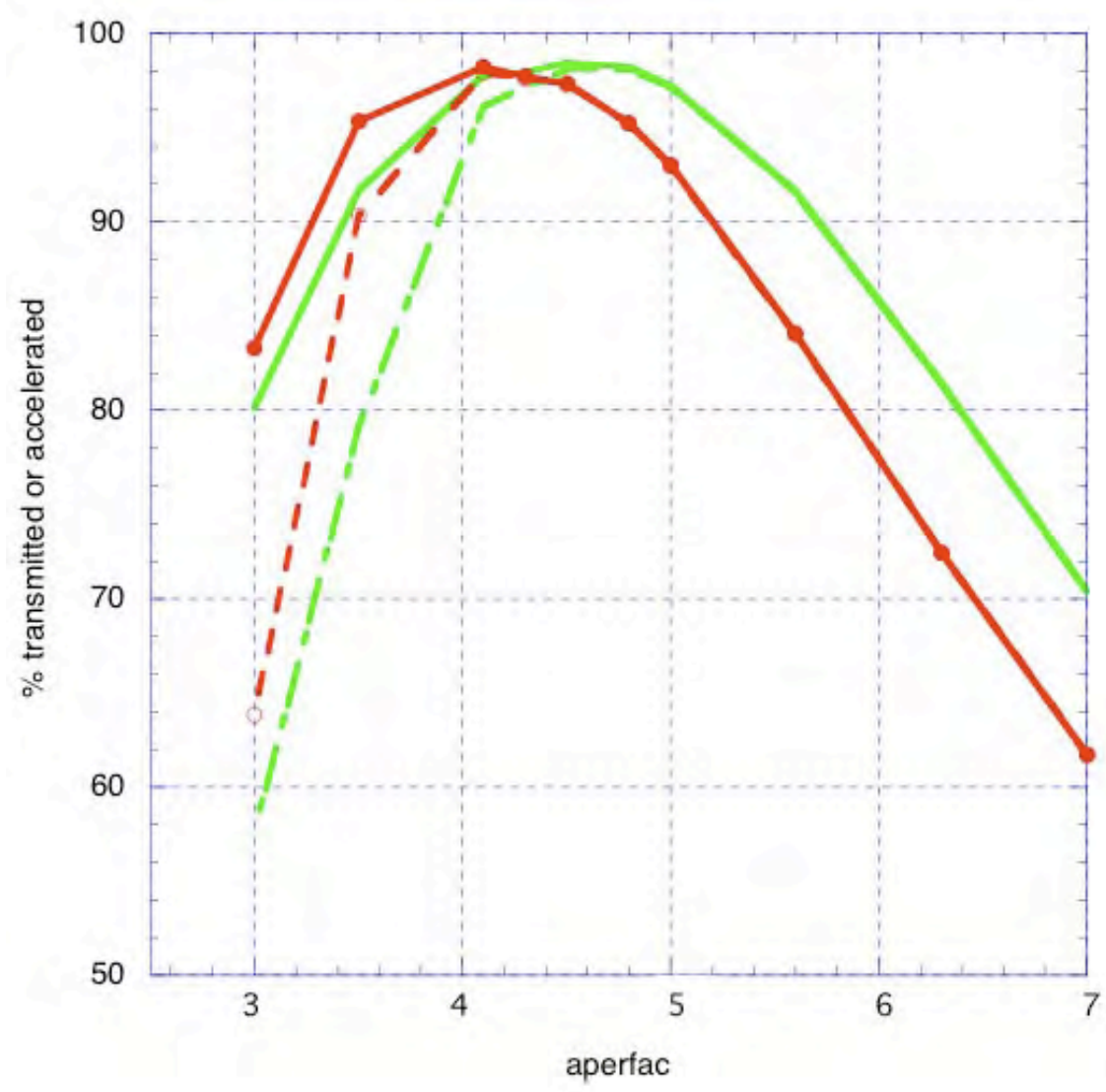

Fig. 16.11 Green – *PARMTEQM* result with multipole external field representation, *scheff* space charge, and no image effect. Red – with KRC image effect.

The image effect in the RFQ is defocusing in the transverse direction, but focusing in the longitudinal directions, as indicated by Fig. 16.12 [135]. Particles ahead or behind the bunch centroid tend to see an image of the bunch centroid and are thus attracted back longitudinally toward the center of the bunch.

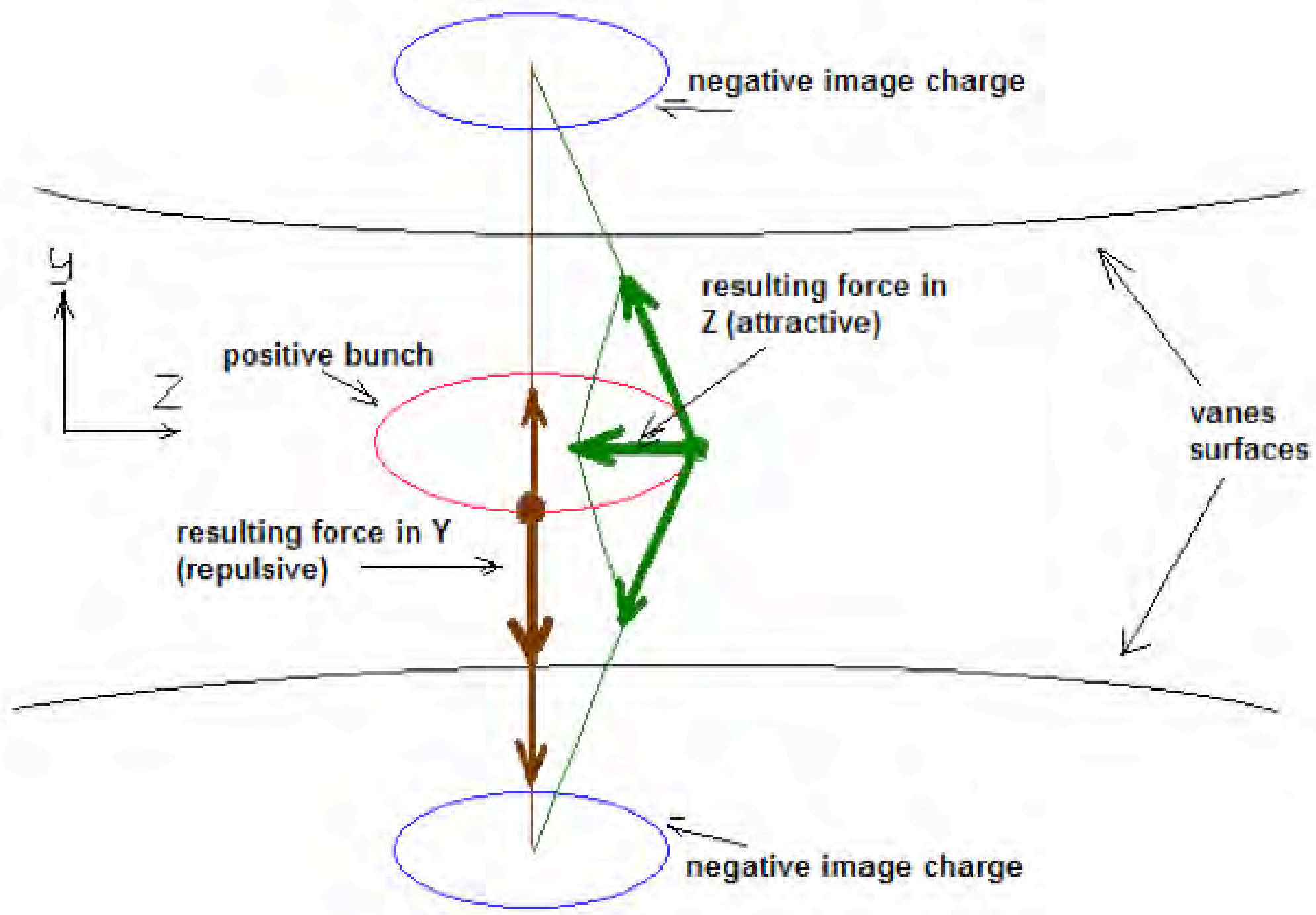

Fig. 16.12 Schematic of image forces acting in the RFQ.

Fig. 16.11 can be interpreted as that as the aperture increases, the transverse image effect lessens, while the longitudinal image effect remains as strong, or stronger, for a longer beam.

135 Stanislav Vinogradov, private communication.

The sign and relative magnitude of the external field, space charge and KRC image forces were checked by running with space charge only up to the beginning of Cell 200 in an RFQ. Then test particles were placed from the beam centroid, transversely in x and y from the beam axis out to 99% of the aperture, and longitudinally over ±1 cell. The forces acting on these test particles over two cells are indicated in Fig. 16.13. Transversely, space charge and image forces are defocusing, and it is noted that the image force near the vanes is of comparable magnitude to the maximum space charge force. Longitudinally, space charge is defocusing and the image charge is focusing.

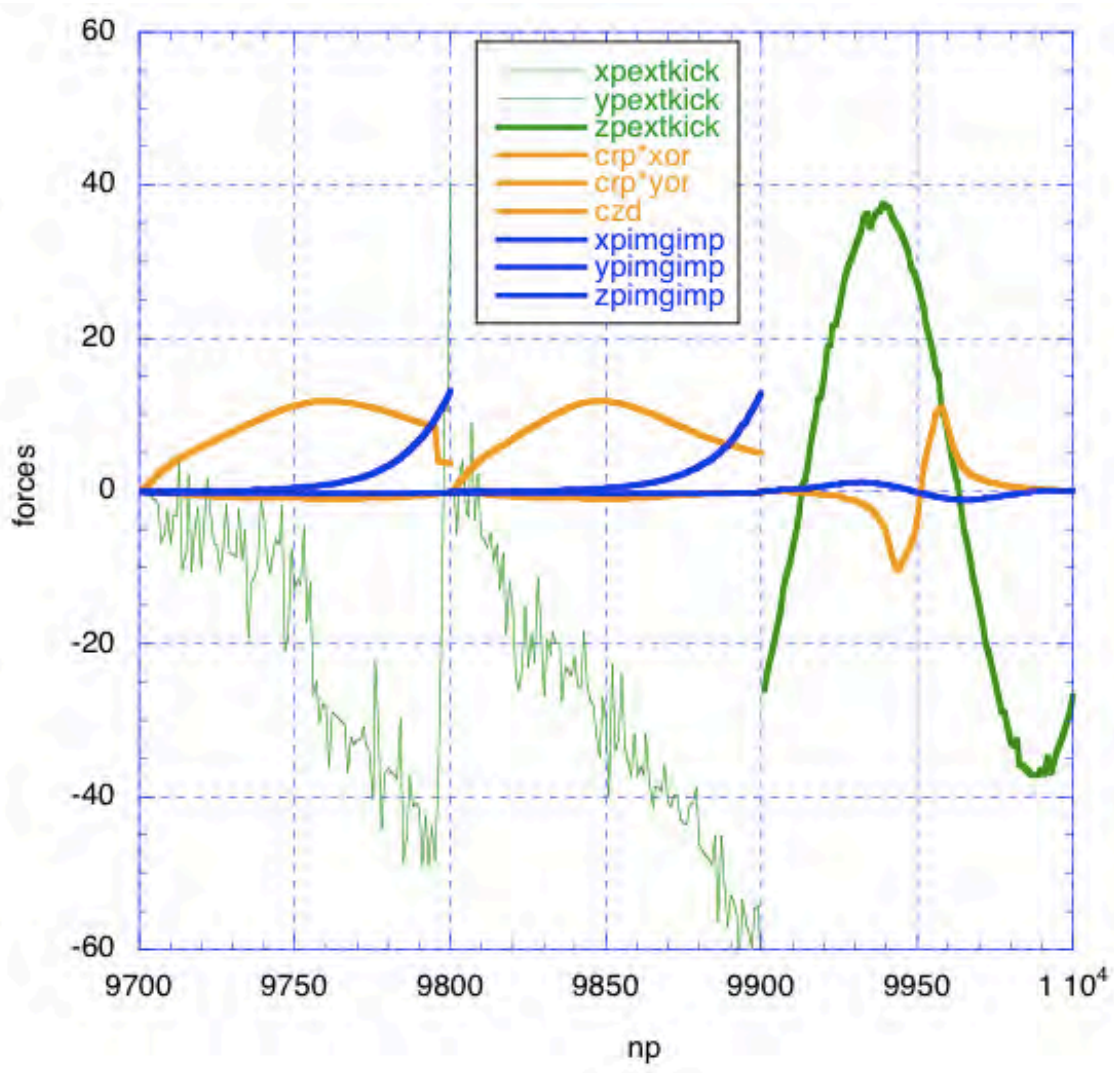

Fig. 16.13 Forces acting on test particles over one transverse focusing period (2 cells). Particles 9701-9800 are on x-axis, 9801-9900 on y-axis, 9901-10000 on z-axiz.

It is very important to test a code with image forces on and off (with actual and other boundary conditions) - and this is easy to do.[136]

## 16.4 Summary

Approximations made in *PARMTEQM* for the external, space charge and image effect fields can be removed using a full Poisson solver, achievable on present laptop and desktop computers. Starting conditions and space charge handling for the input beam are also modified to be compatible with the realistic conditions expected with the Poisson solver.

[eltoc]

# Chapter 17. IAP Poisson Solver [137] for *LINACSrfqSIM (J. Maus)*

136 It has been vociferously claimed that this is not possible, especially in a Poisson code. Obviously ridiculous – the claim was actually made to avoid discussion of questionable evidence…

[137] PhD. Thesis, summa cum laude, Johannes Maus, Institute for Applied Physics, Goethe University Frankfurt, October 2010.

Papers at LINAC2010, elsewhere…

Poisson solvers for *LINACSrfqSIM* have been developed at IAP by Johannes Maus for his PhD Thesis, and incorporated into *LINACSrfqSIM* by R.A. Jameson [138]. The primary goal was that the physical correctness be apparent, explainable and open. This is because the same kinds of decisions about the programming have to be made (meshing, boundary conditions, convergence criteria, etc.), and each has its consequences. From this basis, further development (faster execution, etc.) can systematically follow.

For accurate beam dynamic simulation, the electric field of the useful region inside the RFQ has to be known. Therefore the Poisson equation has to be solved. One method which can be used to solve the Poisson equation is the finite difference method. The differential equation is discretized on a suitable grid (or mesh). The simplest type of grid for calculation of the field inside an RFQ is an isotropic Cartesian grid in 3D. Generally, it is not necessary to have the same grid spacing in the longitudinal and in the transverse plane, so an anisotropic Cartesian grid would be a reasonable choice. An anisotropic grid has the disadvantage that the Poisson solver becomes more complex (and slower).

The simplest finite difference method is the Jacobi iteration solver where new values are calculated for all grid points using the old values. A damped Jacobi relaxation is obtained by introducing a relaxation parameter. A second type of iterative solver is the Gauss Seidel method, which uses the value of the new grid point as soon as it is calculated. A good choice for the overrelaxation parameter will lead to a good reduction of the error and therefore a good convergence.

The following characteristic of either one of the two solvers are very important for the multigrid idea: after applying a few iteration steps, the error of the approximation does not necessary become much smaller, but it becomes smooth. The smoothing capability of the solver is equivalent to the reduction of the high frequency part of the error of the approximation. The low frequency part of the error is much more difficult to reduce. This is illustrated in Figure 17.1. The initial values are randomly spread between 0 and 1. After a few cycles the high frequencies are damped and only low frequency variations on the grid remain. The smoothing factor of the iterative smoother strongly depends on the overrelaxation parameter; A good choice lies between 4/5 and 1.

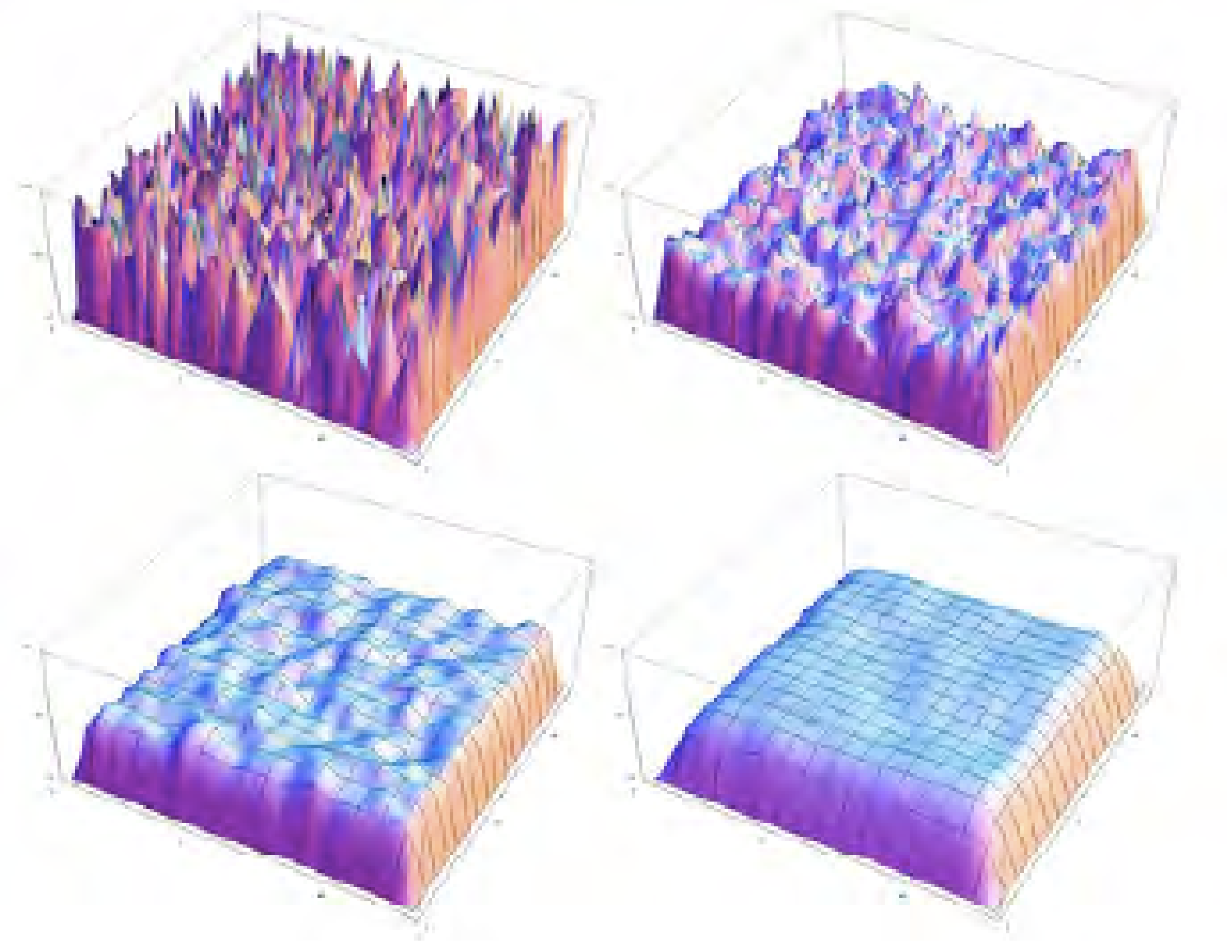

138 Very extensive collaboration, during the preparation of the Poisson solvers and integration into LINACs, from colleagues, especially Stanislaus Vinogradov and Sasha Durkin, at the Moscow Radiotechnical Insitiue (MRTI) is gratefully and fully acknowledged. The collection of emails plus comments is contained in 'Poisson Notes200809.doc'

Fig. 17.1: Initial values and approximation after 1, 3, and 10 cycles. The high frequency components are sufficiently reduced, whereas the low frequency components remain.

## 17.1 Multigrid Solver

A classical reference on Poisson equation solution by grid methods is [139]. The best modern method is to use multiple grids - the multigrid method - as presented in [140], which includes a good summary of how the multigrid method absorbs and extends the earlier methods.

The first concept of the multigrid idea is that an iterative solver such as the Gauss-Seidel solver smoothes the error of an approximation within a few iterations and can therefore be used as a smoother.

The second principle is the so-called coarse grid principle: if an error is well smoothed it can be approximated on a (much) coarser grid without losing information. The low frequency components of the fine grid are transferred to high frequency components on the coarse grid. The error can be further reduced on the coarser grid with less computational effort, since the number of grid points is reduced.

The error cannot be calculated directly, since the solution of the Poisson equation is not known at any time. Therefore it is useful to define the defect or residual of the approximate solution. The defect is a measure of how much the Laplacian of a given approximation differs from the source term of the Poisson equation. It can therefore be used to determine the quality of the solver and its ability to converge. The defect equation is equivalent to the definition of the error.

The discrete eigenfunctions of the discrete Laplace operator can be split into two components: the high frequency part, and the low frequency part. The ability of the smoother to smooth the error within a few iterations means that the high frequency components of error are reduced, whereas the low frequency components remain comparably unchanged.

Assume that we have two isotropic grids, a fine one, and a coarse second one with twice the grid spacing of the fine one The low frequencies on the fine grid are also visible on the coarse one, where they represent high frequencies. Whereas the high frequencies on the fine grid vanish on the coarse one, Fig. 17.2.

---

[139] "Computer Simulation Using Particles", Hockney, R.W., Eastwood, J.W., Bristol, Hilger, 1988.

[140] "Multigrid", Trottenberg, U.; Oosterlee, C.; Schüller, A; Academic Press, 2001.

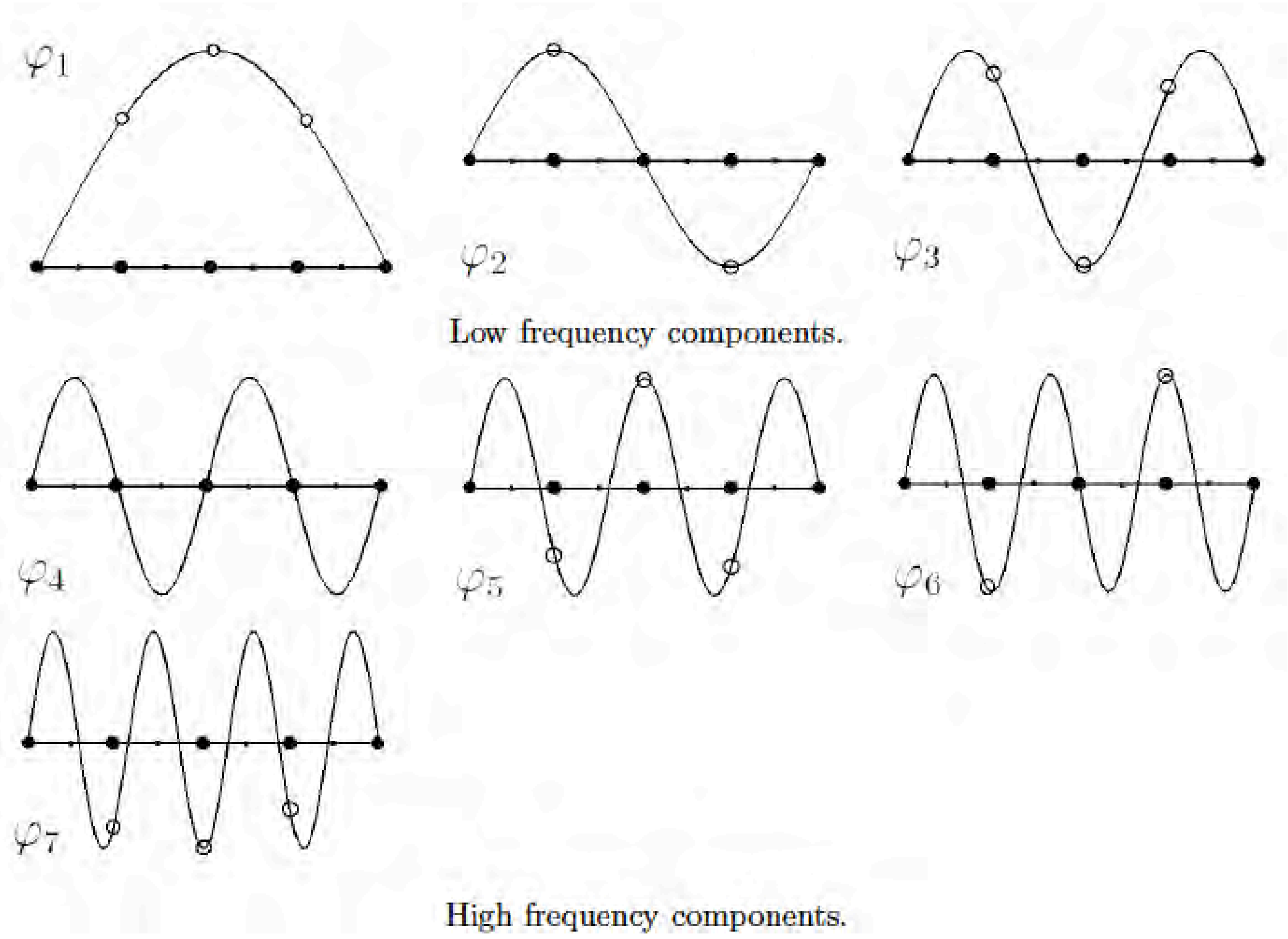

Fig. 17.2 Low and high frequency components of the error on a coarse grid with grid point spacing (dark points) twice that of a fine grid (light points), for a simple 1D example. Low frequency components are visible on the coarse grid (first row), whereas high frequency cannot be represented.

This means that the low frequency part of the error on the fine grid, which can not be damped easily with, for example, a Gauss-Seidel smoother, can be converted to a high frequency error on the coarse grid, which can then be damped easily using one of the described smoothers. This is the basic concept of a (2-grid) multigrid solver.

## 17.2 Ingredients of Multigrid Cycles

Generally, a multigrid iteration starts on the finest grid by applying some smoothing cycles to reduce the high frequency error. Then the defect is calculated and restricted to the coarser grid by a restriction schema.

The corresponding Laplacian is then solved on the coarse grid, either recursively with another multigrid approach since the equation has the same form as the initial Poisson equation, or by a fast iterative solver.

The defect equation does not need to be solved exactly. A suitable approximation will work as well without essential loss of convergence speed. After this approximation is computed, it will be interpolated to the fine grid, where a new approximation is found. Finally, some post smoothing steps will be performed.

In the following, different structures of one multigrid iteration consisting of four grids are discussed. For more or less grids, similar structures are feasible. All structures start on the finest grid and perform a specified number cycles to smooth the approximation. Then the defect is restricted to the next coarser grid where again smoothing cycles are performed. This procedure is repeated until the coarsest grid is reached. The coarsest grid is solved using an iterative solver. From this point, the different cycle schema then differ from each other.

The V-cycle illustrated in Fig. 17.3a solves the coarsest grid once and then prolongates back stepwise to the finest grid, performing smoothing cycles on the error of every grid level. The W-cycle shown in Fig. 17.3b is a more complicated structure. The coarsest grid is now solved more than once and the computational effort for one full W-cycle is higher than for one V-cycle, but the defect will be reduced stronger than for one V-cycle iteration. So the total time versus reduction of the defect might be better for the W-cycle. A last type of multigrid structure is the F-cycle shown in Fig. 17.3c. The selection of the type of multigrid structure and how many smoothing steps are used at each level that result in an effective solver strongly depend on the type of problem.

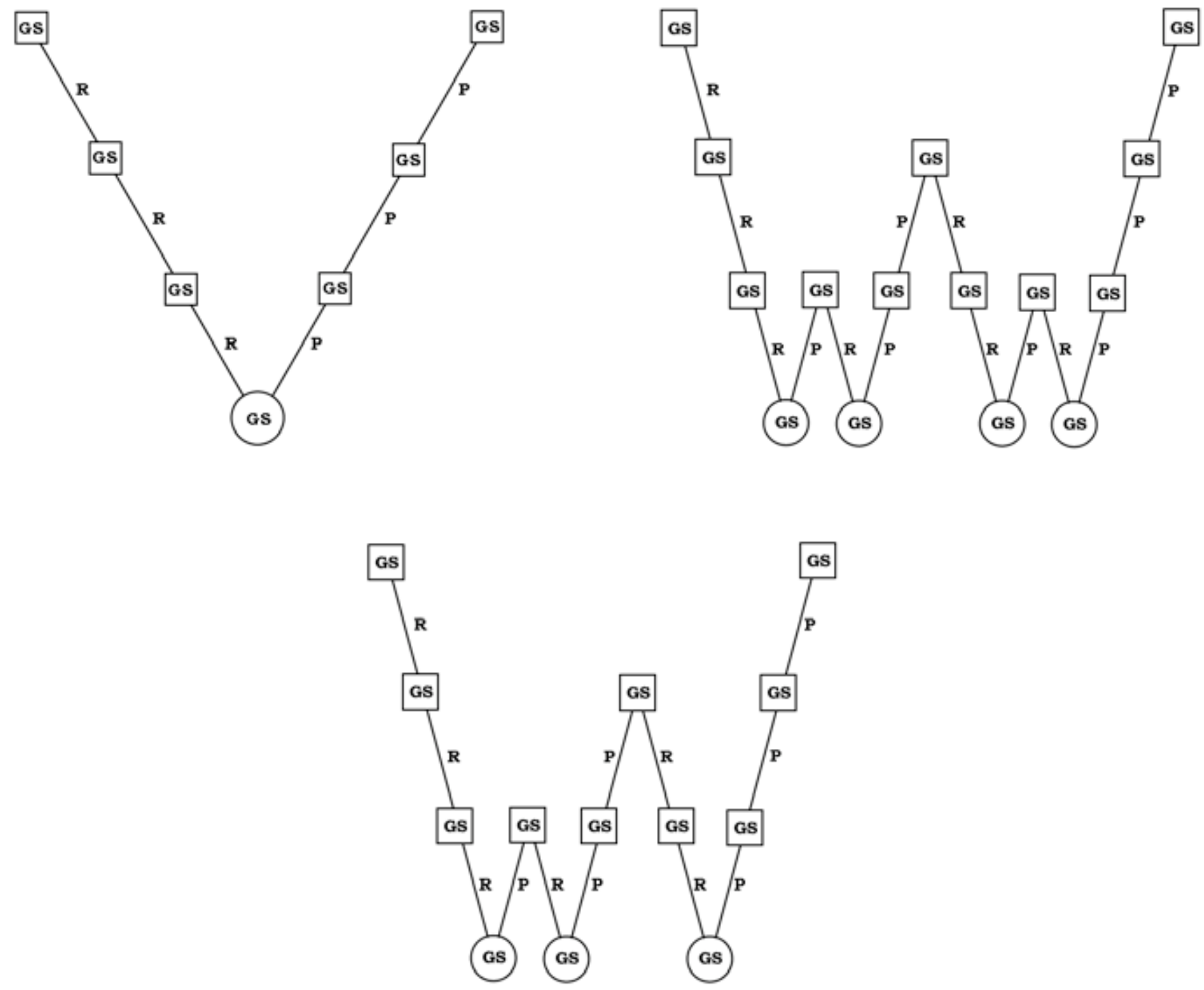

Fig. 17.3 a. V-cycle, b. W-cycle, c. F-cycle.

### 17.2.1 Restriction and Prolongation Operators

The purpose of the restriction operator is to map grid functions on the fine grid to grid functions on the next coarsest grid. The simplest restriction operator is the injection operator that connects coinciding grid points directly. Grid points on the fine grid which vanish on the coarse grid will then not be considered. More complex restriction operators are the half weighting (HW) operator and the full weighting (FW) operator, which assign a combination of the neighboring points to the center grid point. The choice of the restriction operator depends on the actual problem, and it cannot be said in general that any restriction operator is superior. For the case of the field inside an RFQ, the injection operator performed the best.

The purpose of the prolongation operator is to map grid functions on the coarse to grid functions on the next finest grid. This can done by a trilinear interpolation that is illustrated in Fig. 17.4.

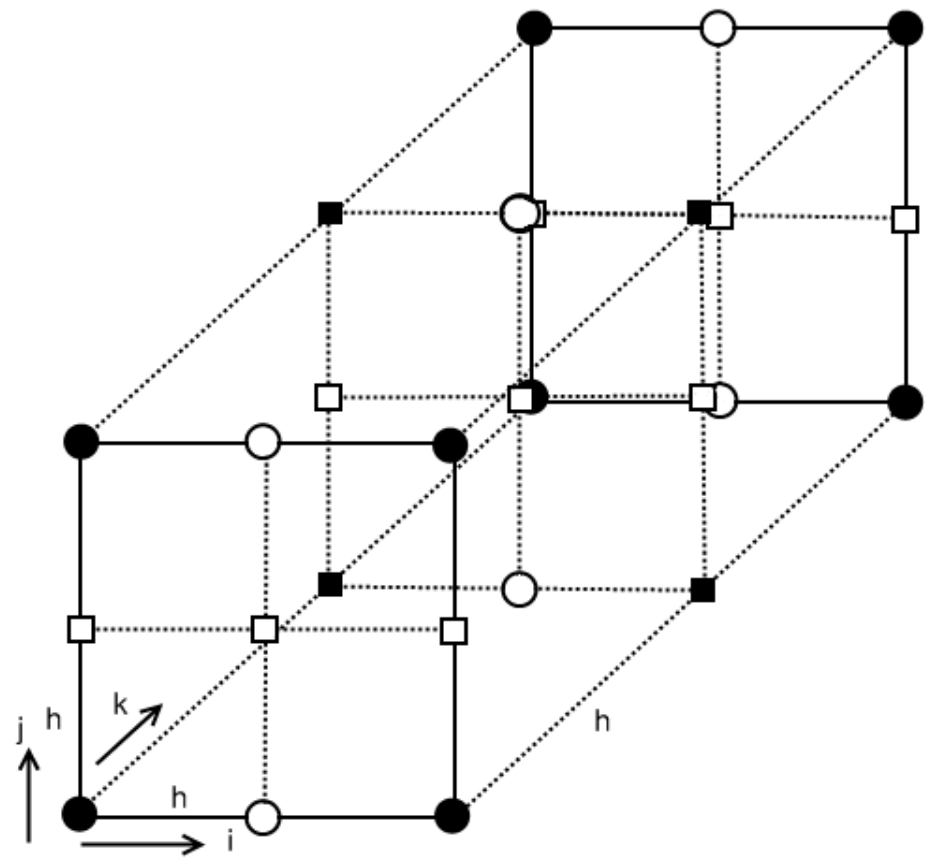

Fig. 17.4  Trilinear interpolation in 3D.

## 17.3 Verification of the Multigrid Solver

Since the solver will be used to calculate the RFQ external field as well as the internal space charge field, two different types of test problems will be considered: one with a charge density introducing the field (Poisson equation) and one without charges, where the field is generated by the boundary conditions (Laplace equation).

### 17.3.1 Laplace Equation

For the following test problems no charges are placed on the grid. The first test problem is a simple box in three dimensions with potential = 1 on its surface and starting values of zero on the grid. The solver should be able to increase the values on the grid to their exact answer of 1 over the entire grid. Figure 17.5 shows the initial potential and the potential after one and three multigrid iterations. It can be seen that the approximation is already within 20% of the exact answer after one iteration. After six iterations the difference is less than $10^{-6}$. The oscillations on the right hand side of the box are due to the running direction of the Gauss-Seidel-Smoother, which updates those grid points which are closer to the origin first. The corresponding defects are shown in Fig. 17.6. The absolute values of the defects depend on the grid spacing and are therefore of no interest, instead the change of the magnitude is the important measure. The defect was reduced by a factor of $10^5$ within six iterations. The shape of the defect coincides with the approximations minus the solutions which are 1 in this case.

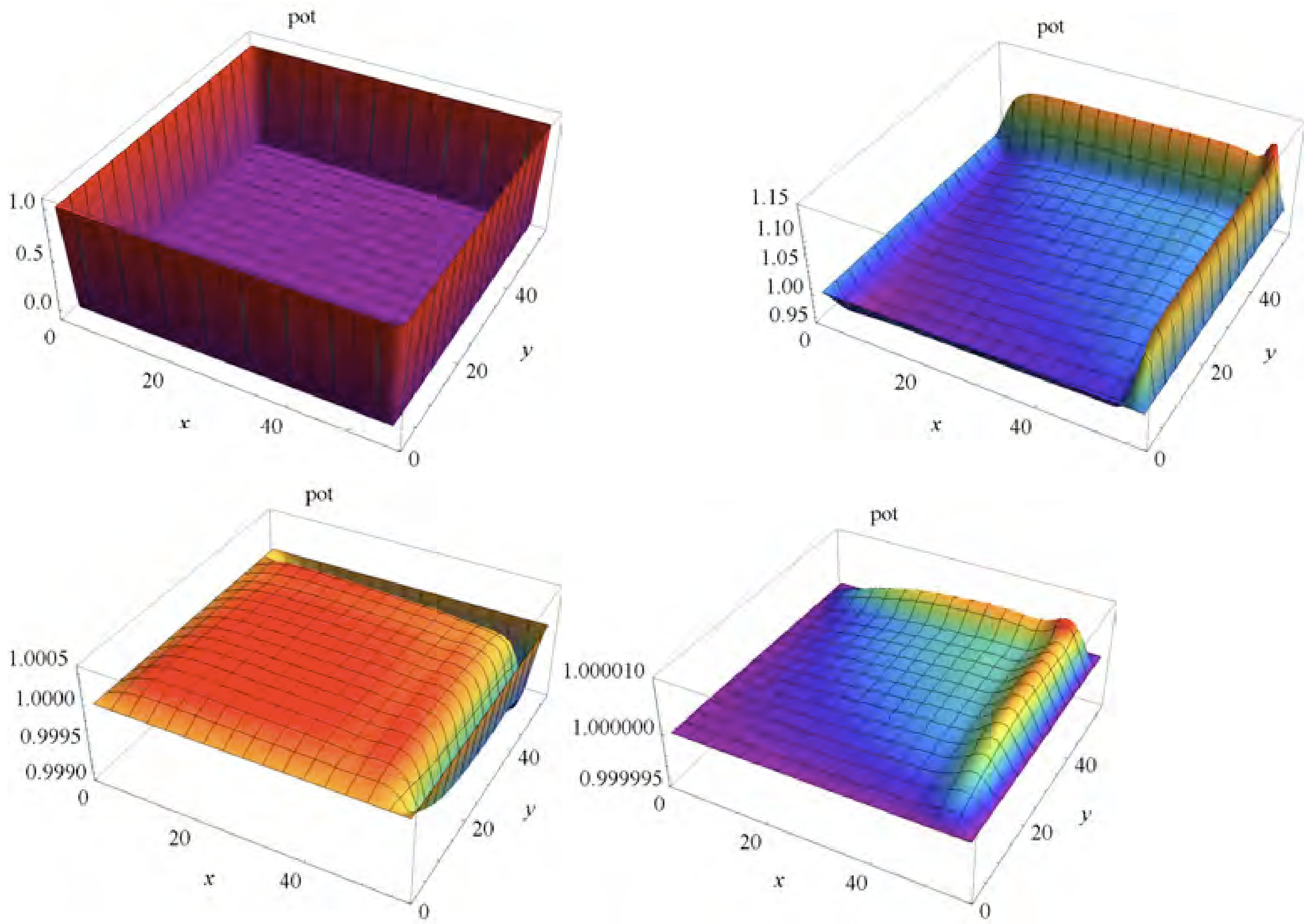

Fig. 17.5  Initial potential and potential after 1, 3, and 6 multigrid iterations of a box with potential = 1 on the surface

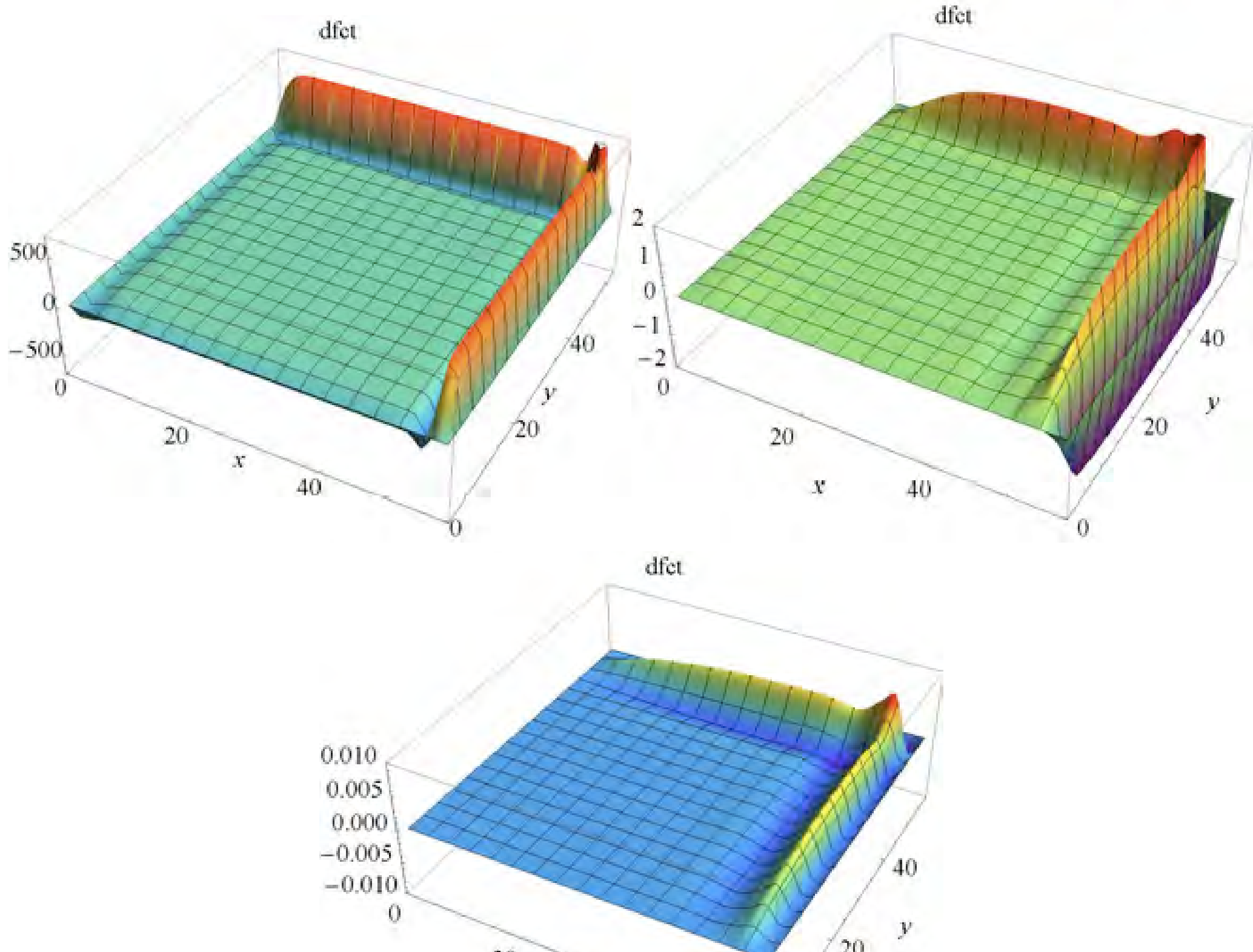

Fig. 17.6  Defect after 1, 3, and 6 multigrid iterations of a box with potential = 1 on the surface.

As a second test problem a conducting ball with radius of 50 grid points was chosen having zero potential on the surface. In its center the potential is forced to equal 1. The cut through the initial potential and the approximation of the potential after one and ten multigrid iterations at z = 0 are shown in Fig. 17.7. After one iteration, the shape of the approximation is already close to the one after ten iterations, so that just minor changes were applied. This demonstrates the advantage of the multigrid method: the ability to reduce the error within a few iterations. The corresponding defects are shown in Fig. 17.8. Their maxima are located around the center where the potential equals 1 and by the reflecting walls at x = 0 and y = 0. The defect is reduced by a factor of $10^{12}$ within 10 iterations.

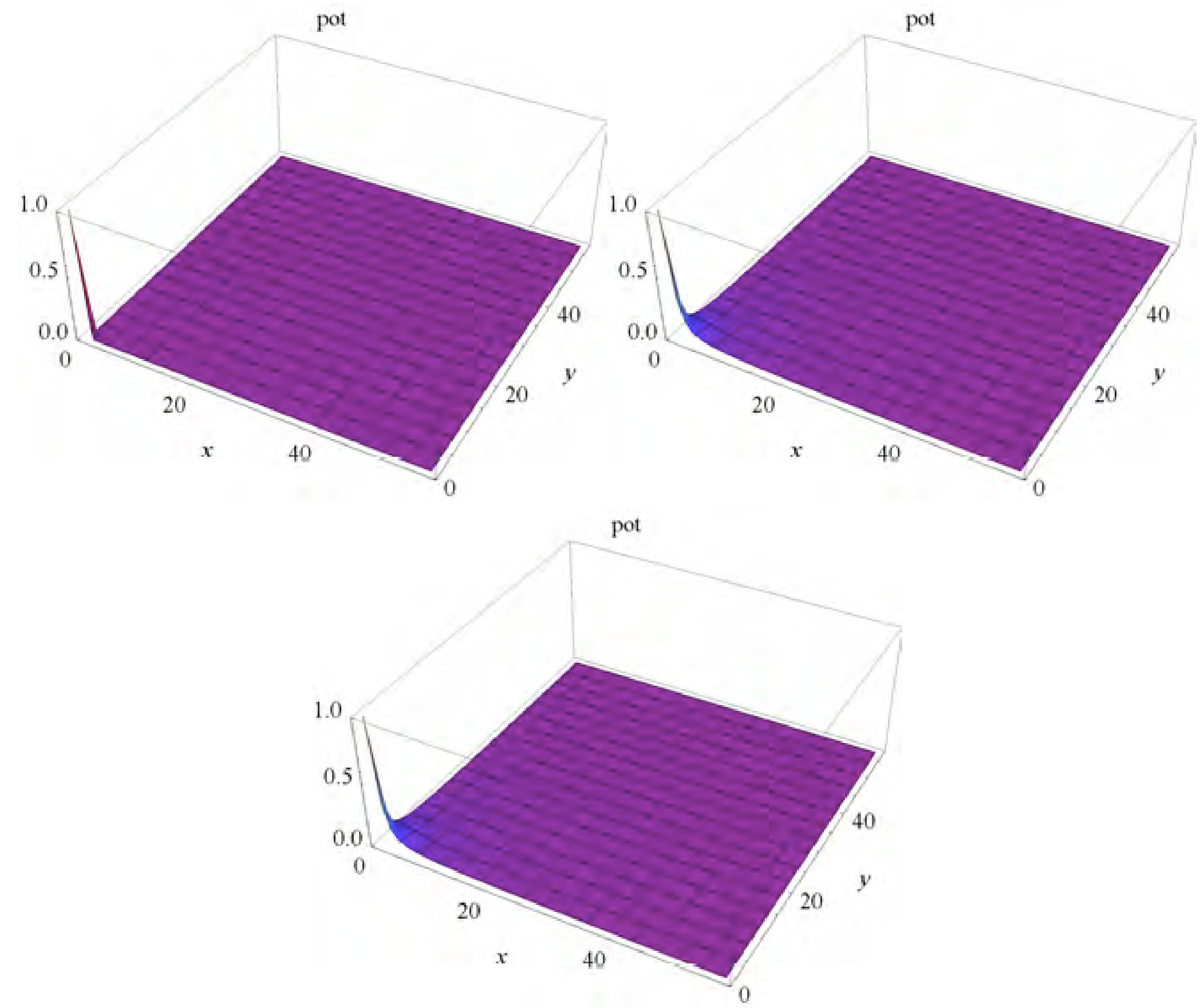

Fig. 17.7 Initial potential and potential after 1 and 10 multigrid iterations of a ball with vanishing potential on the boundary and potential equal 1 at the center.

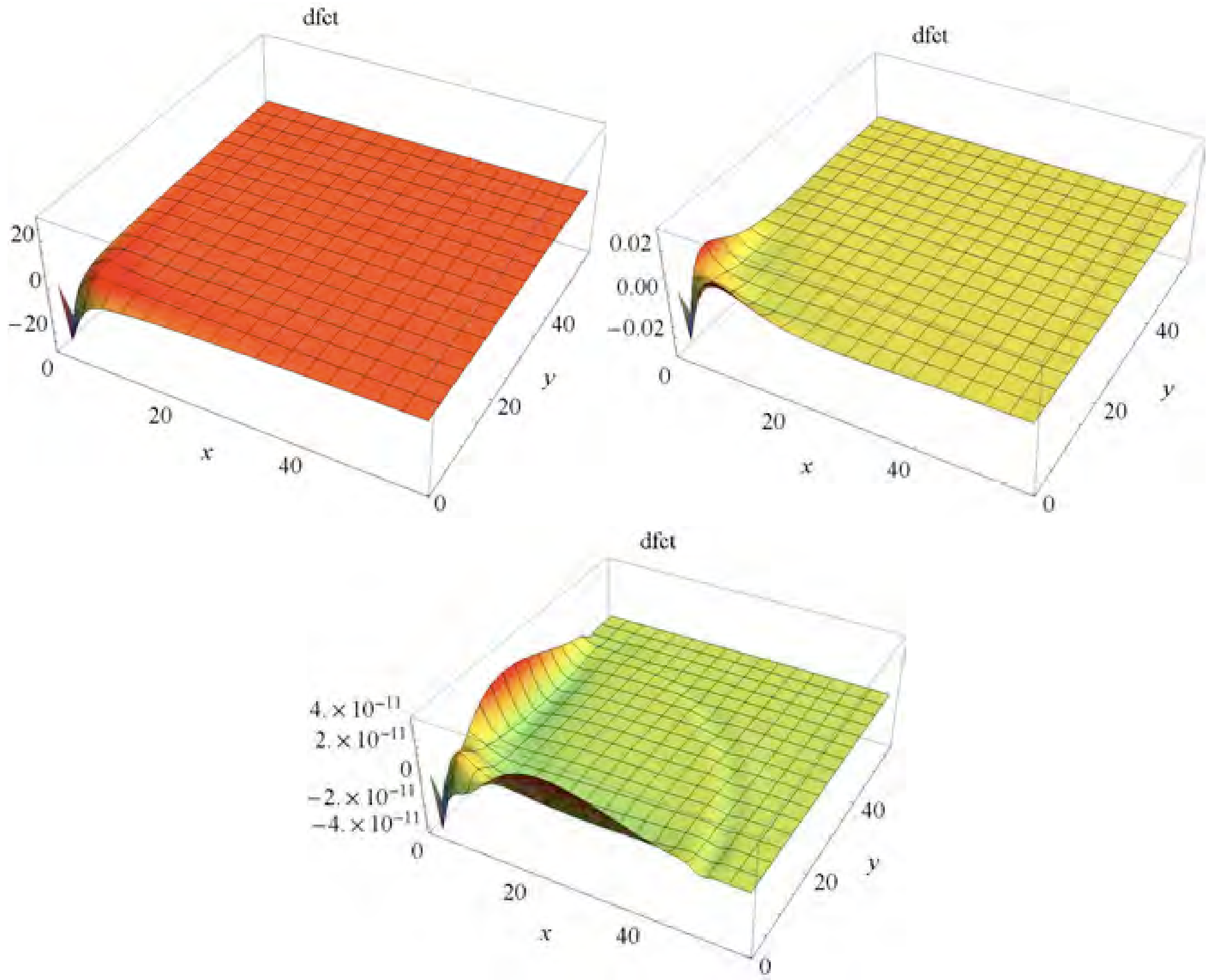

Fig. 17.8 Defect after 1, 3, and 10 multigrid iterations of a ball with radius of 50 grid points, with vanishing potential on the boundary and potential equal 1 at the center.

The last test problem is a parallel-plate capacitor placed inside a grounded box. The initial potential and the approximations after one and ten multigrid iterations are shown in Fig. 17.9. On the initial approximation, only the two plates are of opposite potential and the remaining grid points have zero potential. After one iteration the expected shape of the potential was observed. After ten iterations the shape of the approximation has changed a little. This differs when looking at the corresponding defects shown in Fig. 17.10. The defect is further reduced after ten iterations by approximately the same factor as in the other test problems above. If the width of the plate is reduced, the convergence of the solver becomes worse. This is due to too fine a structure on the finest grid, which might vanish on a coarser grid.

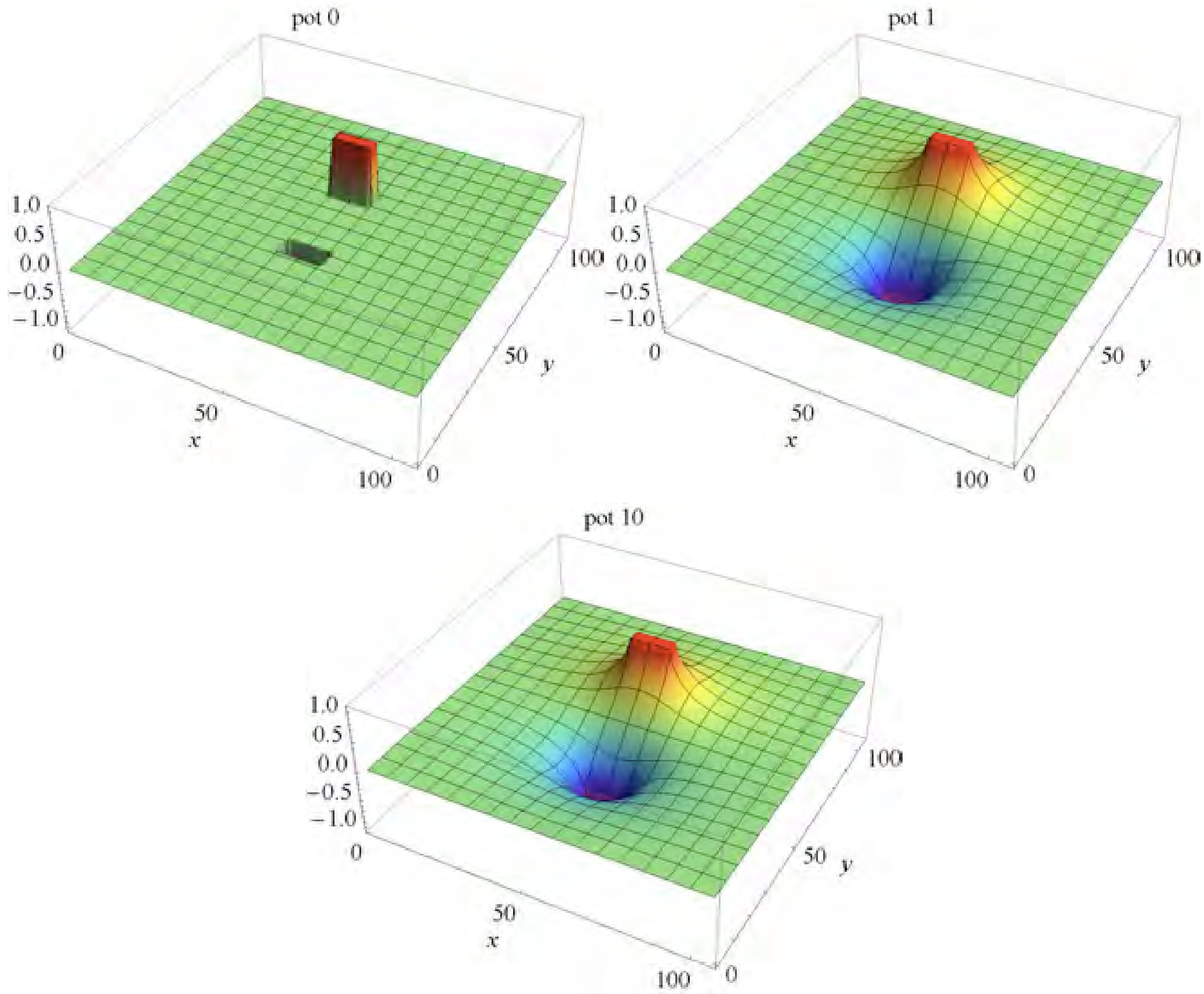

Fig. 17.9 Initial potential and potential after 1 and 10 multigrid iterations of a parallel-plate capacitor in a grounded box.

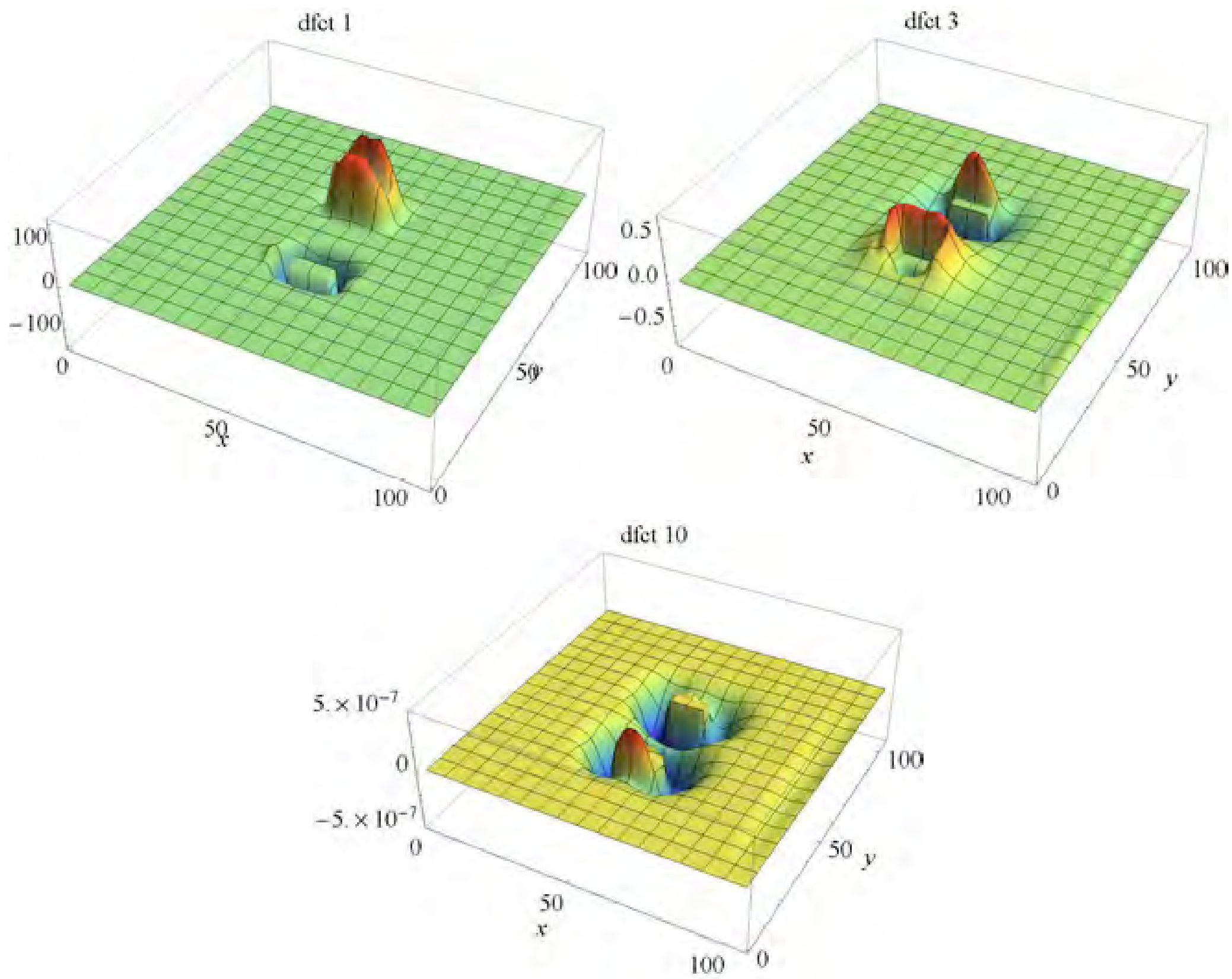

Fig. 17.10 Defect after 1, 3, and 10 multigrid iterations of a parallel-plate capacitor in a grounded box.

### 17.3.2 Poisson Equation

This section covers test problems involving nonzero charge densities and zero potential on the boundaries. These types of problems will occur when calculating space charge effects inside the RFQ. The boundaries are then forced to have zero potential, so that effect of image charges is realizable as well. The principle of superposition allows computation the external and internal fields separately. In most applications a dc beam having the shape of a cylinder two cells long is injected into the RFQ. Further down in the accelerator the beam is bunched and of elliptic shape. It will be surrounded by either the electrodes for the cases considering the image effect, or by the cylinder with zero potential and a certain distance to the beam axis for calculations neglecting the image charges.

*17.3.2.1 No Image Effect*

The test problems described below have a cylindrical boundary with zero potential in which the charge density is placed. For space charge simulations in an RFQ the radius of the cylinder needs to big enough to avoid particles coming too close to the cylinder. Twice the maximum aperture (2ma) is normally sufficient.

The first case discussed here is a charge density with the shape of a cylinder placed inside a grounded cylinder. This is the situation of a DC beam injected into the RFQ neglecting the image effect. The ratio of the grounded cylinder radius to the beam radius is 5 to 1. A cut at y = 0 of the charge density as well as the approximation of the potential is shown in Figure 17.11 and the corresponding defects are shown in Figure 17.12. The potential has its maximum at the center of the cylinder and falls off to the grounded cylinder. It does not show any longitudinal dependency as expected so that no further steps are necessary to include neighboring bunches. The defect is reduced by a factor of $10^8$ within 10 iterations.

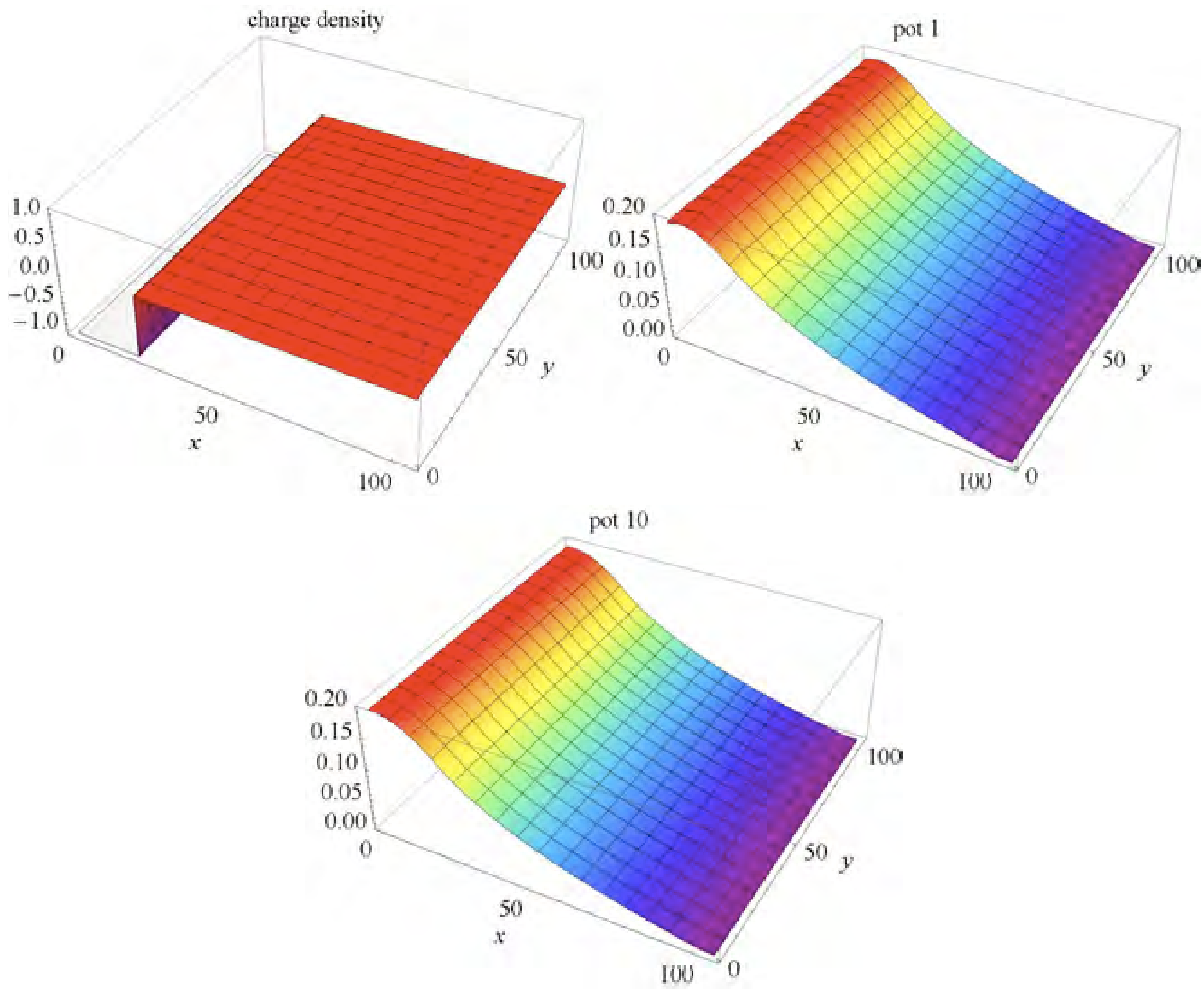

Fig. 17.11 Charge density and approximation of the potential after 1 and 10 multigrid iterations of a charge cylinder placed inside a grounded cylinder. X corresponds to the radius and y along the axis of the cylinder

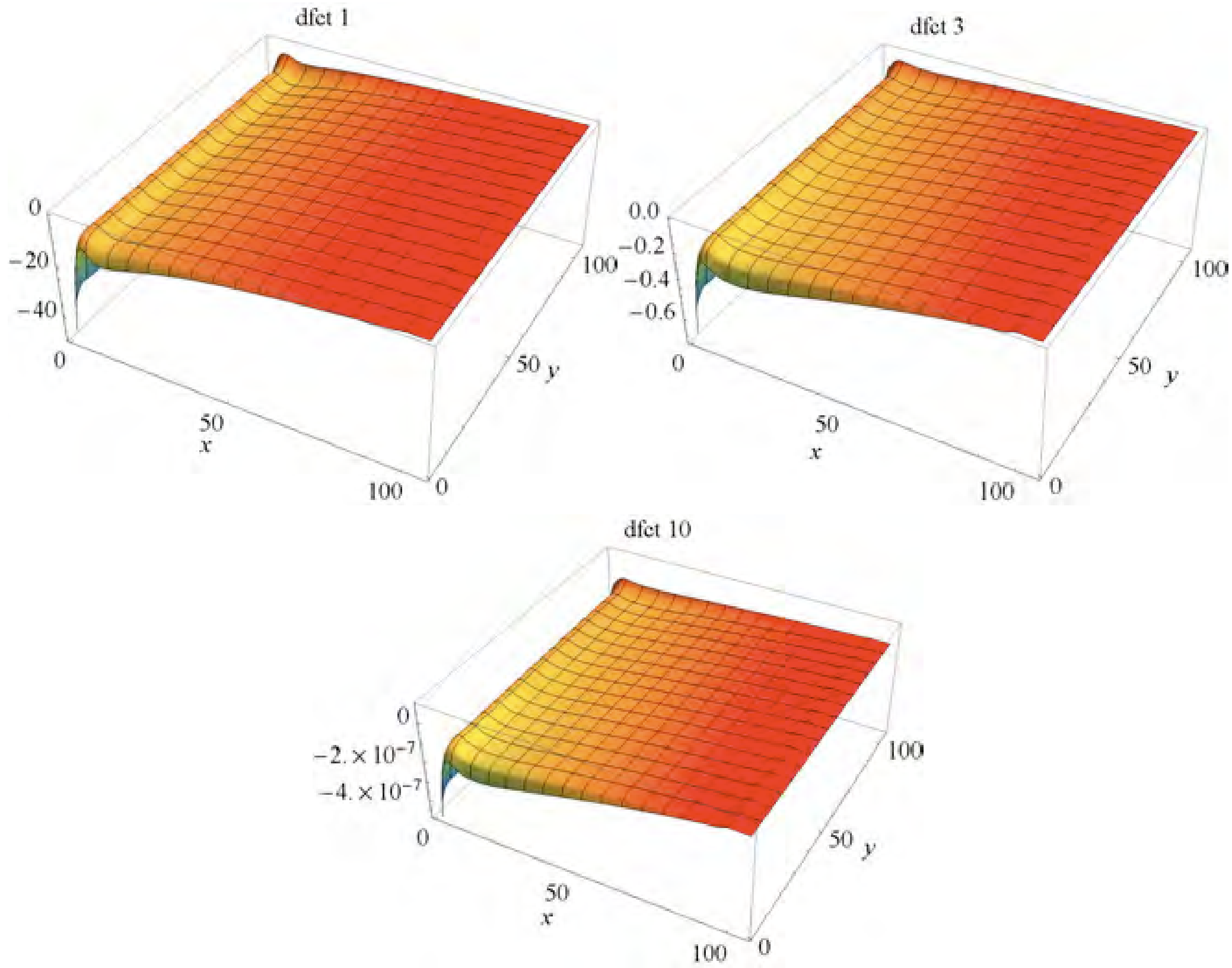

Fig. 17.12 Defect after 1, 3 and 10 multigrid iterations of a charge cylinder placed inside a grounded cylinder.

A cut at y = 0 of the approximation of the potential as well as of the charge density of a charged ball placed inside a grounded cylinder is shown in Fig. 17.13. The grid is periodic in the longitudinal direction, so that neighboring bunches are fully taken into account. The charged ball has a radius of 20 grid points, and the grounded cylinder radius is 100 grid points. The potential has its maximum at the center of the charged ball and falls of in all directions. In the transverse direction the potential decreases to zero whereas in the longitudinal directions it does not fall down to zero, since the neighboring charge is taken into account and no conducting boundary is present there. The solver reduces the defects shown in Fig. 17.14 strongly by a factor of $10^8$ within ten multigrid iterations and is therefore sufficient for computing space charge effects as well. It has to be noted that the maximum of the defect after some iterations is not in the region where the charge is located, but at the edges of the grid where grid points are reflected. Fortunately this is just a minor effect and can be neglected.

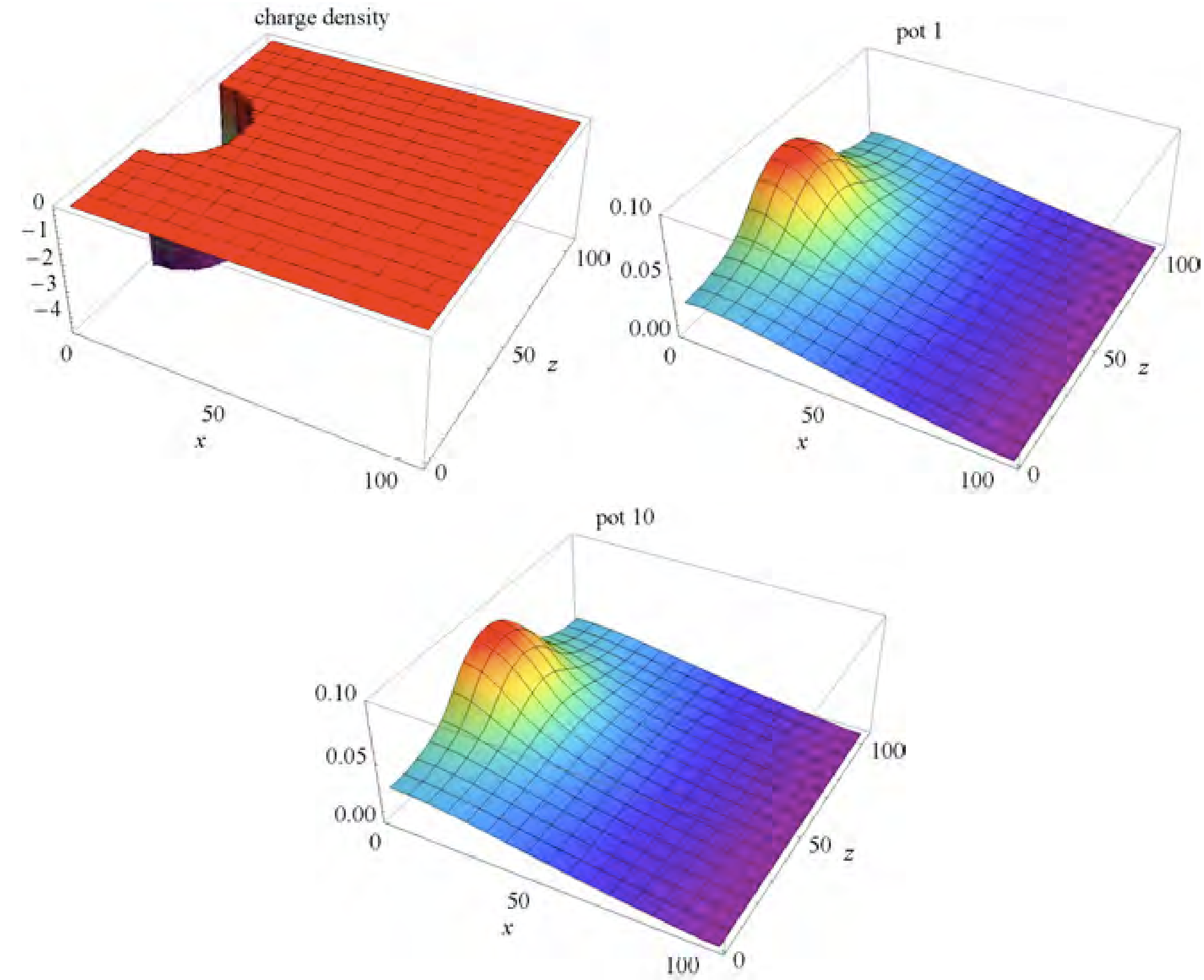

Fig. 17.13 Charge density and approximation of the potential after 1 and 10 multigrid iterations of a charged ball placed inside a grounded cylinder.

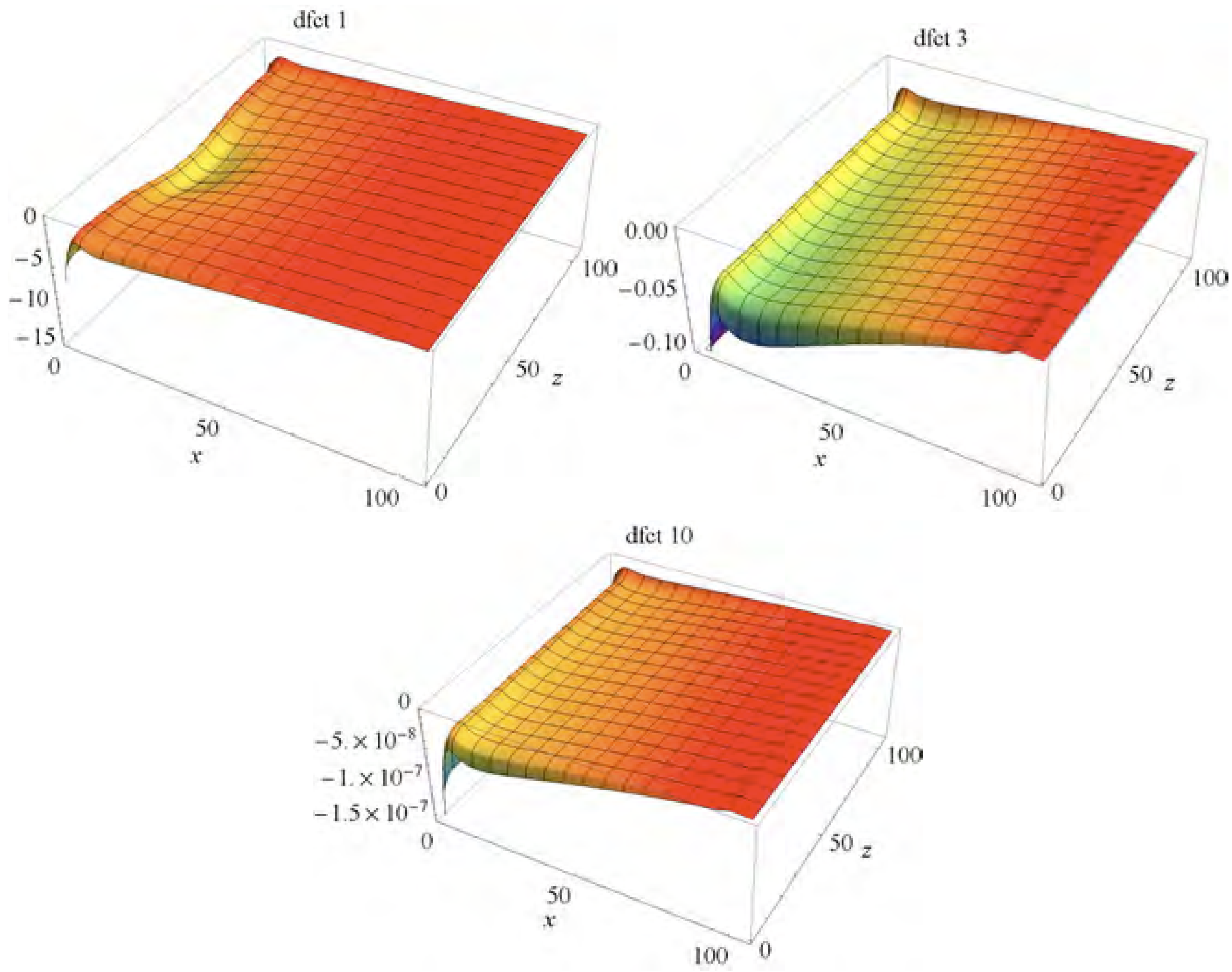

Fig. 17.14 Defect after 1, 3 and 10 multigrid iterations of a charged ball placed inside a grounded cylinder.

*17.3.2.2 Image Charge Effect*

In the following, test problems are described in which the image charge effect is not neglected. The boundary therefore has the shape of the electrodes of an RFQ and has zero potential, so that the external and the internal fields can be calculated separately making use of the superposition principle.

A simple example of an image charge is the potential of a point charge placed in front of a conducting plate at a distance d. It is equivalent to the situation of two point charges of opposite charge with a distance of 2d between them. In Fig. 17.15 the potential at a perpendicular line to the plate crossing the charge is shown, as well as the potential of a two charges with twice the distance. Both potentials have the same shape as expected, so the solver can calculate the image effect correctly as well.

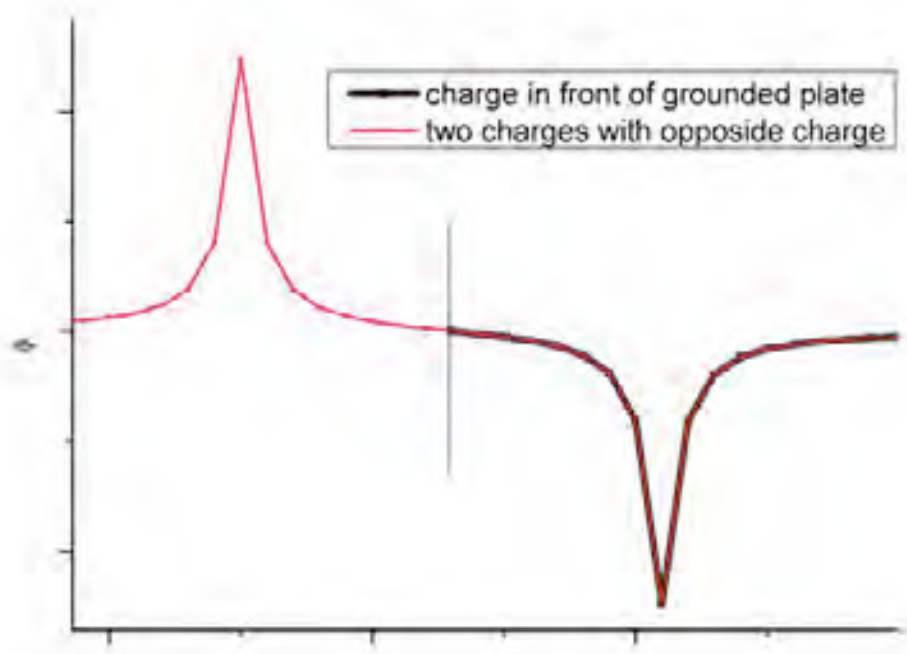

Fig. 17.15 Potential of a charge placed in front of a grounded plate and potential of two charges of opposite charge to demonstrate the image charge effect.

The charge densities used above are now placed inside a grounded quadrupole channel with different apertures. Fig. 17.16 shows the potential for the charged cylinder. The maximum of the potential on the beam axis depends on the aperture; with a bigger aperture the maximum increases. Both potentials fall off to zero at the vane tip but in different shape (more stretched for the bigger aperture) and will therefore result in different electric fields. A dependence on the longitudinal position cannot be seen, as expected from the reflected longitudinal boundary conditions.

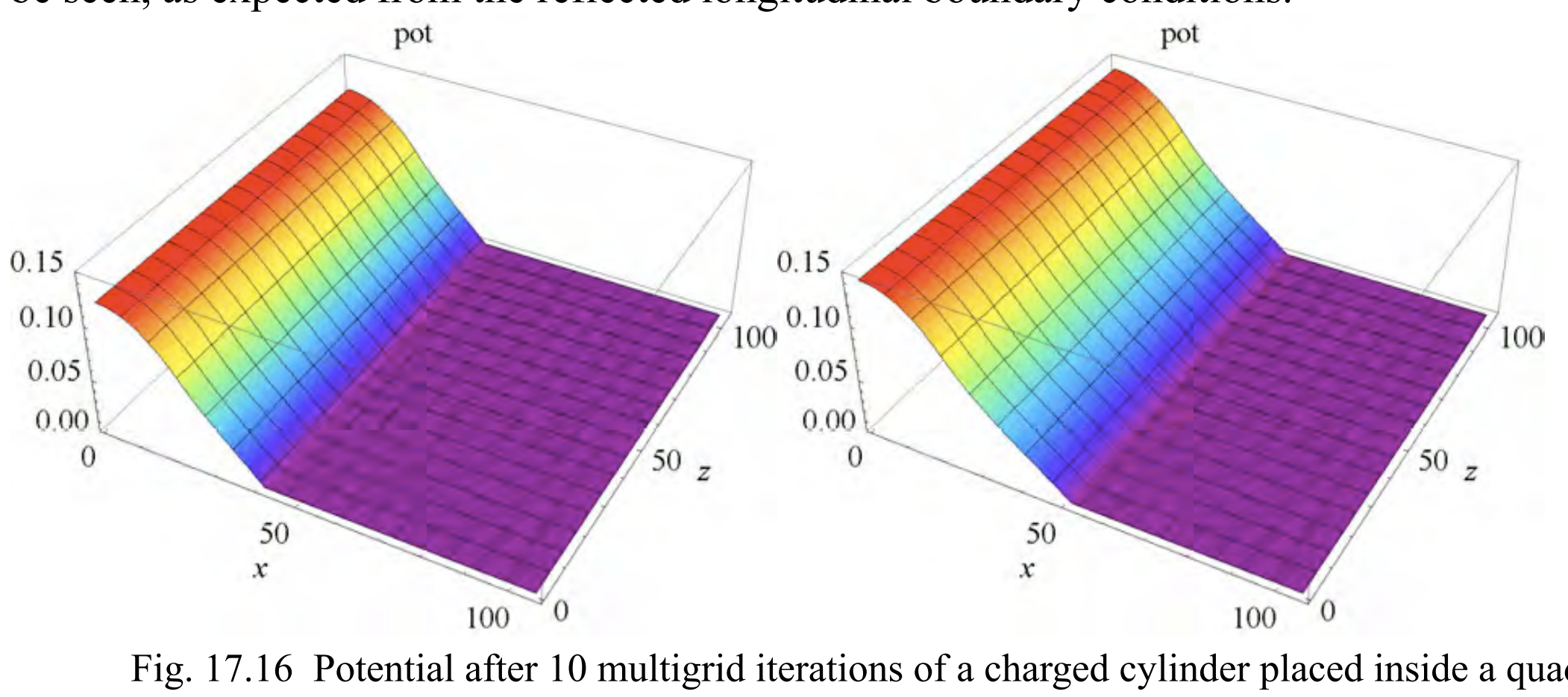

Fig. 17.16 Potential after 10 multigrid iterations of a charged cylinder placed inside a quadrupole channel with aperture of 0.4 cm and 0.5 cm.

Figure 17.17 shows the potential of a charged ball placed inside a grounded quadrupole channel for different apertures (0.2 cm – 0.8 cm). Just like the charged cylinder, the maximum of the potential increases with increasing aperture. The value of potential at the beginning and at the end of the cell also depends on the aperture in the same manner.

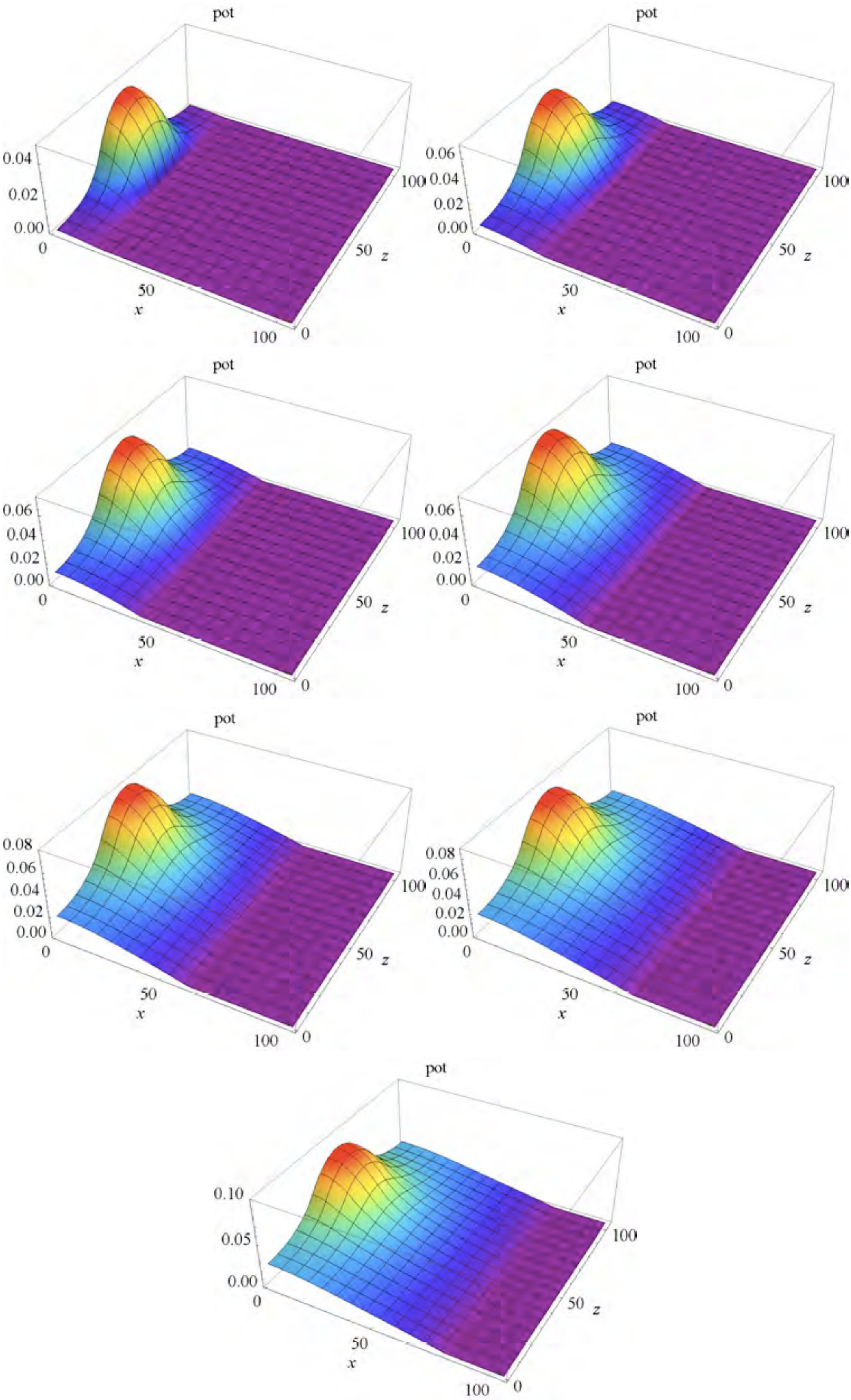

Fig. 17.17 Potential after 10 multigrid iterations of a 0.2 cm radius charged ball placed inside a quadrupole channel with aperture from 0.2 cm to 0.8 cm, increments 0.1 cm.

The shape of the potential on the beam axis is also shown in Figure 17.18 to illustrate the influence of the aperture on the potential. The curves for the smaller apertures are shifted to lower values compared to the curves with bigger apertures. The resulting longitudinal field components will therefore be somewhat close for the different apertures, since the difference between the curves is mainly a constant, which holds only for the longitudinal field. The black curve (aperture and the radius of the ball are both 0.2 cm) differs in its shape a little from the other curves, since the vane surface is placed directly in front of the charged ball and the image effect is therefore stronger than in the cases having a bigger aperture. If the vane surface is far enough away from the charge density the influence of the image effect is weaker and differences between axial potentials decrease.

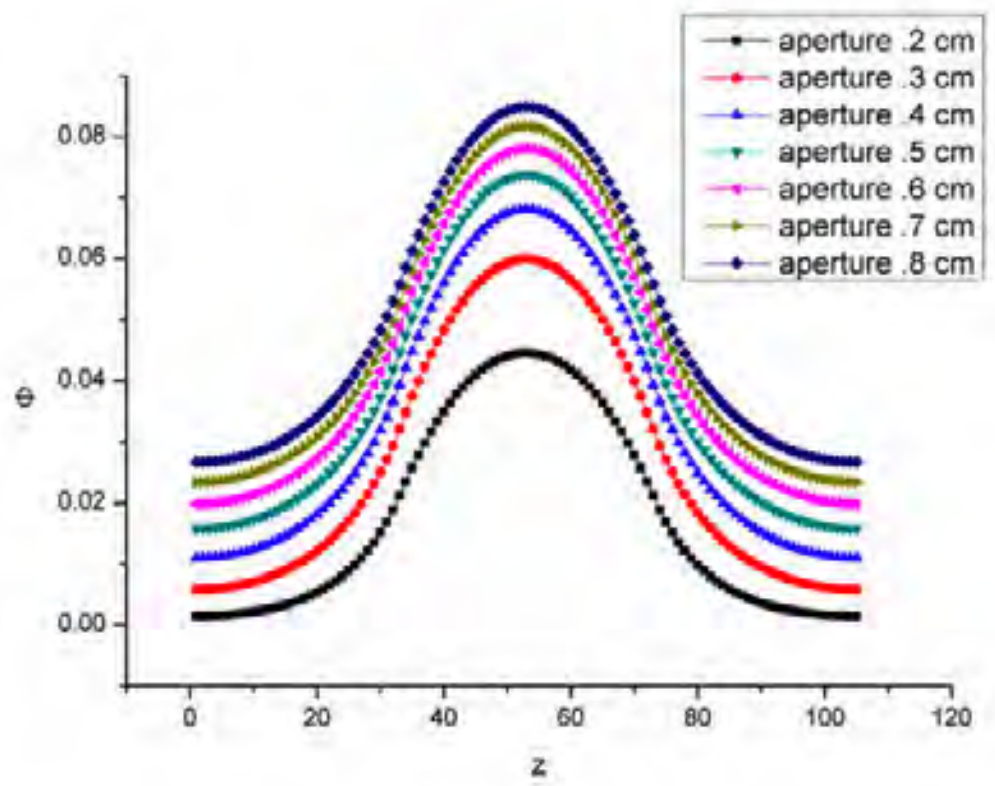

Fig. 17.18 Potential along the beam axis of a 0.2 cm radius charged ball placed inside a grounded quadrupole channel with aperture from 0.2 cm to 0.8 cm.

The transverse field (Fig. 1719), which is the derivative of the curves shown in Fig. 17.18, inside the charged ball is quite independent of the aperture, because the potentials (from the center to grid point 20) have the same shape with a constant offset. The field outside the charge density then depends on the aperture. For small apertures the potential falls off steeper than for big apertures. This results in higher transverse field component for small apertures, because the boundary forces the potential to zero, so it will drop down more rapidly.

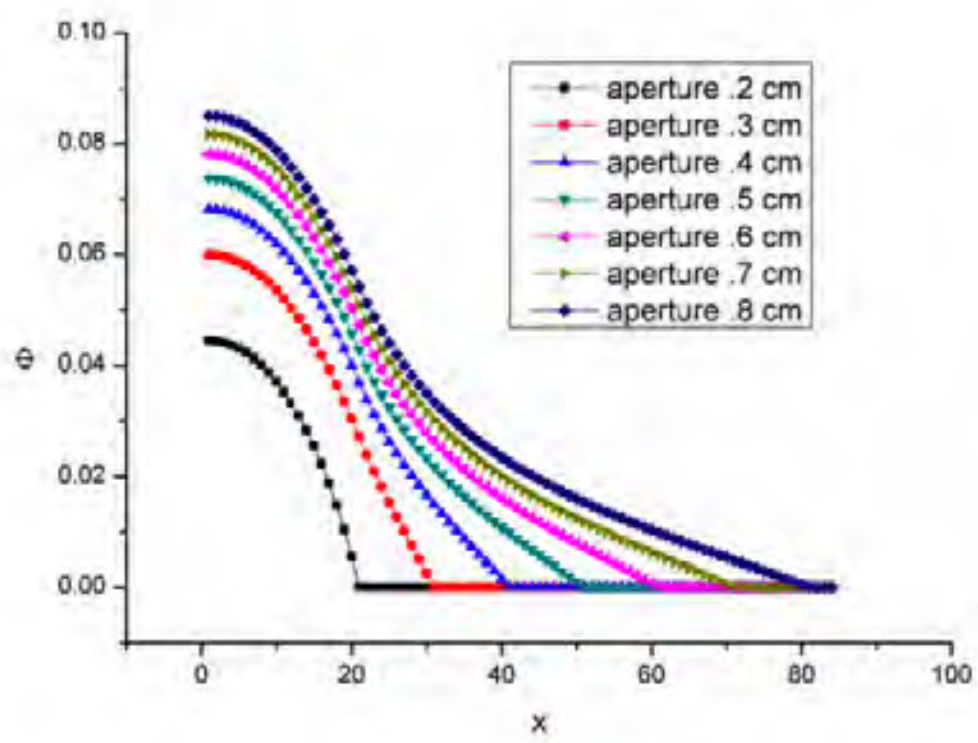

Fig. 17.19 Potential on a line from the center of the 0.2 cm radius charged ball to the vane surface.

## 17.4 Application of the Multigrid Poisson Solver to RFQ

### 17.4.1 External Field Grid

*17.4.1.1 Computer File Size Restriction – Overlap Method*

External field is computed in blocks of 15 cells, of which the middle 13 cells are used. Extra cell at front and back insures smooth transition to end boundaries, which are periodic.

*17.4.1.2 Transverse Mesh Boundary Conditions, One Quadrant Computation*

At present, one quadrant is computed and reflected to the other quadrants.

The transverse boundary conditions are the actual vane surface, and an arc between the vanes at a radius xdim = 1.d0*sa*(sm+1.5d0), i.e., the maximum aperture of the local cell plus 1.5 times the minimum aperture of the cell – far enough to give an open boundary. Similar considerations are used for the entrance and exit sections consisting of a tank wall and space between the wall inner surface and the vane ends. Mesh points at the boundary are adjusted to be exactly on the boundary.

*17.4.1.3 Selection of Grid Resolution for Transverse and Longitudinal External Field Meshes*

It is of primary importance that the transverse grid has good resolution near the vane tips, as usually particle losses must be minimized and therefore must be simulated accurately. This resolution must extend farther than the vane tip, as particles are not just lost at the vane tip, but also on the vane away from the tip. The grid resolution must also be chosen small enough that the change in result is considered small enough, and running time is involved in the grid resolution decision.

The distances between two neighboring grid points in x,y and z are determined transversely as a fraction (default 1/20) of the minimum aperture radius in the current 15-cell block, and longitudinally as a fraction (default 1/40) of the first cell length (the shortest cell) in the current 15-cell block, with adjustments to prevent the longitudinal to transverse aspect ratio from becoming too different. The number of grid points is then increased as necessary to have an even number divisible by four for the Poisson solver.

In Fig. 17.20 the influence of the denominator of the fractions on the number of accelerated particles are shown. When the grid is too coarse (black and red curve) the shape of the curve differs from the one with a finer grid. The values obtained are too low. Once the grid is fine enough the number of accelerated particles does not change when the grid resolution is further increased.

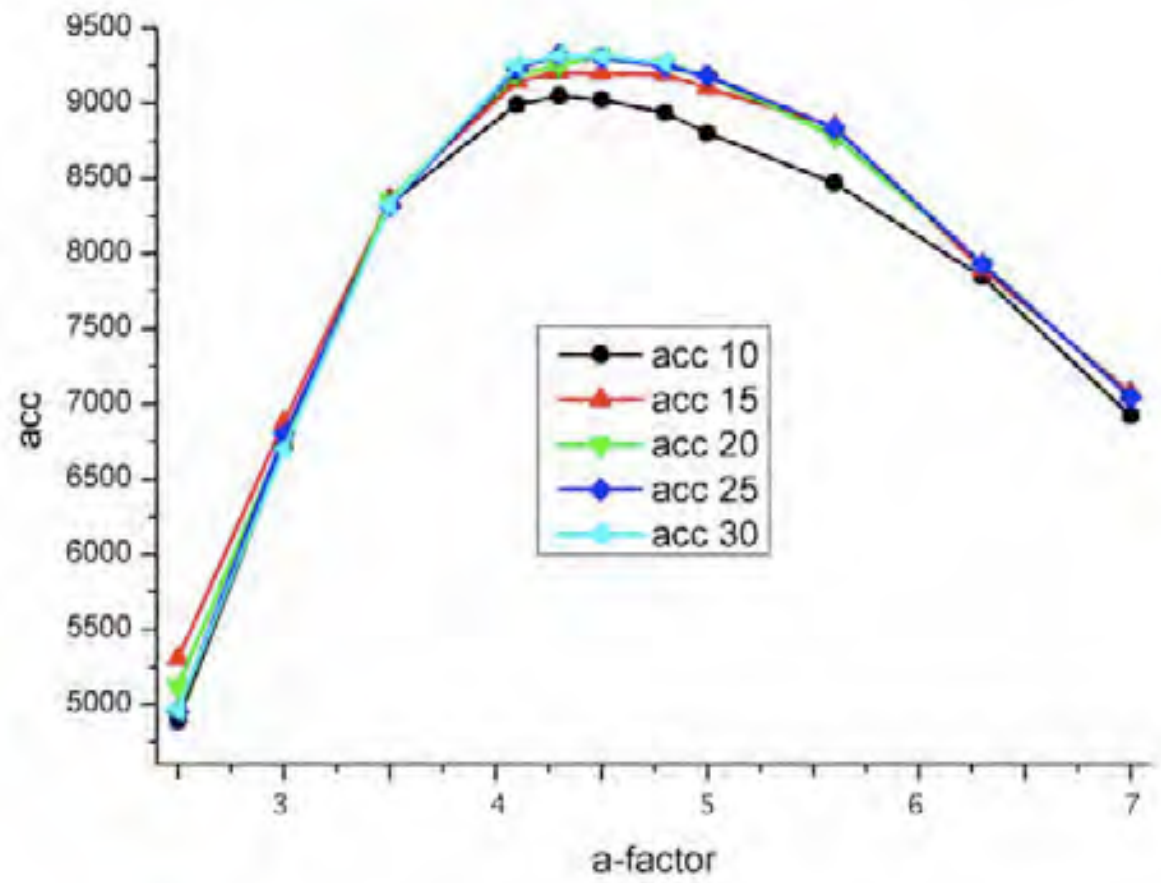

Fig. 17.20: Number of accelerated particles as a function of the external field grid resolution (dr=aperture/x and dz=cell length/x).

*17.4.1.4 Selection of Number of Multigrid Iterations and Smoothing Cycles for External Field Solution*

A three-level W-cycle is chosen for the RFQ simulation.

The sensitivity of the multigrid Poisson solver is analyzed. There are two different types of parameters to be chosen. The first are the number of multigrid iterations being executed and the number of smoothing cycles on every grid. Fig. 17.21 shows the influence of these two parameters on the number of accelerated particles for the set of test-RFQs. All curves are very similar except the black curve (3 multigrid iterations with 4 smoothing cycles) which is the roughest setting used in this comparison. As the standard, settings 5 multigrid iterations and 4 smoothing cycles on every grid (except when the ratio of dr and dz exceeds 4 (then 7)) are used; this is the green curve in the plot. A further increase of the number of multigrid iterations does not change the transmission.

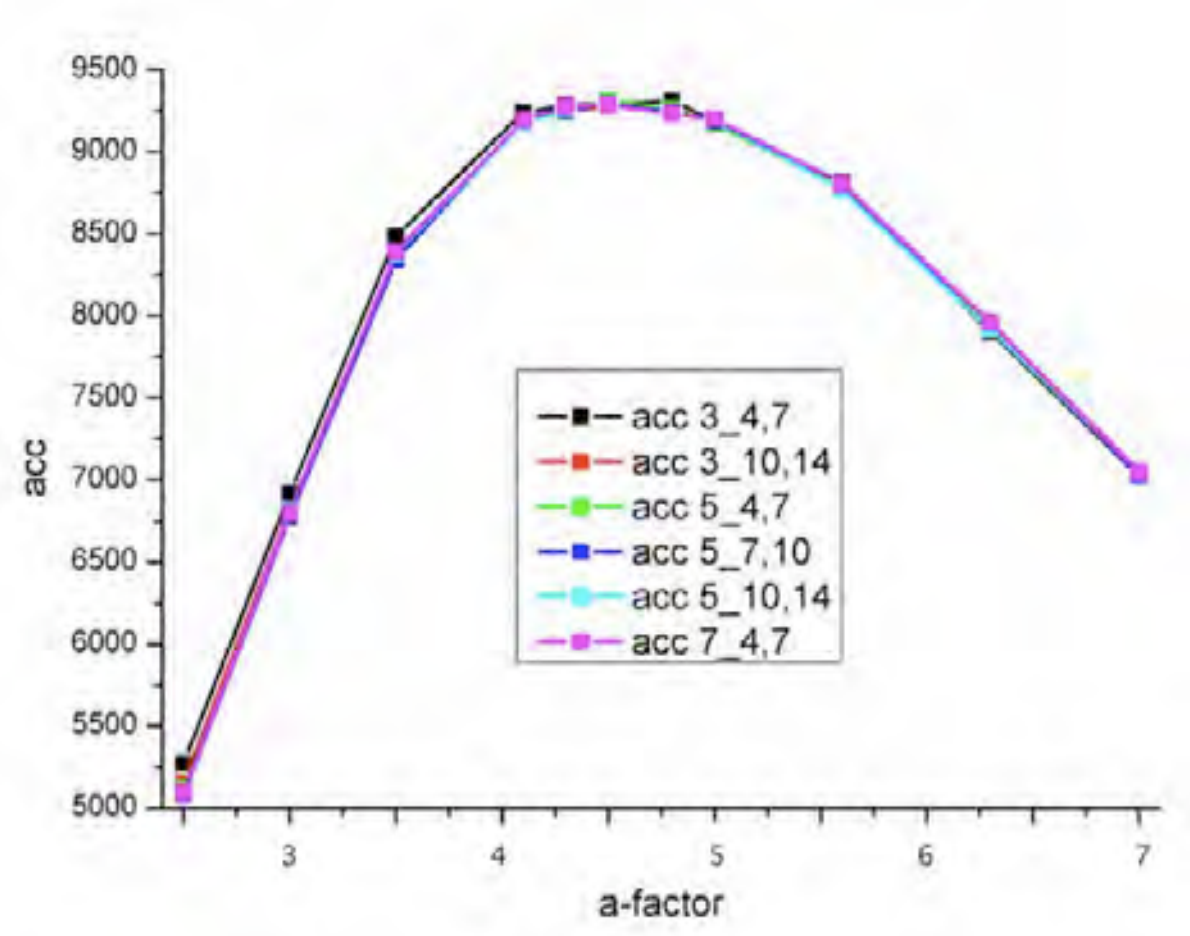

Fig, 17.21. External field grid: Influence of the number of multigrid iterations (first number) and smoothing cycles (second number and third for ratios of hx/dz greater than 4) on the number of accelerated particles.

## 17.4.2 Space Charge and Image Effect Grid

### *17.4.2.1 Superposition*

The external and space charge fields are computed separately, for computational efficiency. Using exactly the same grid boundaries insures accuracy of the superposition. The number of space charge computations per cell can be changed. The standard setting is to compute space charge every step in the radial matching section, where the beam is changing rapidly. Once per cell thereafter is adequate for most purposes.

If a particle falls behind or accelerates ahead of the moving mesh, it is shifted by $\pi$ to return it into the mesh

Description of several extant codes reveals that the space charge is computed with cylindrical boundary conditions (intrinsic to Fourier methods, etc.), so the physics cannot be accurate for RFQ simulation. Other disjoints between the external and internal field solutions may also be observed in extant codes, for example, different meshes and different boundary conditions for the external and internal fields, with approximate joining of the meshes (for example a space charge mesh that is confined to the RFQ minimum aperture or less) With black box codes, only comparing runs is possible, particularly over the RFQ aperture family; this has indicated severe differences, unfortunately sometimes not discussable.

### *17.4.2.2 Image Effect Check – Easy to Turn On/Off*

Using the same boundaries for both external and internal fields insures that image effects are correct. Other boundary conditions for the space charge are easily studied if desired, by changing the space charge grid boundary (Sec. 17.3).

### *17.4.2.3 Selection of Grid Resolution for Transverse and Longitudinal Space Charge Meshes*

As for the external field grid, it is of primary importance that the space charge grid has good resolution near the vane tips, as usually particle losses must be minimized and therefore must be simulated accurately. The resolution in the beam core should be correspondingly good, to insure that particle motion in phase space which might push particles into the vane tip region is accurate. The number of grid points is determined with the same procedures as for the external field.

The influence of the grid resolution on the number of accelerated particles is shown in Fig. 17.22. It is stronger than the influence of the number of multigrid iterations. With increasing grid resolution (from black to blue) the number of accelerated particles decreases on both sides of the curve (small and large apertures) and at the optimum for medium apertures the effect is much smaller and the values are similar for the different settings. The red curve (dr=aperture/20; dz=cell length/40) is a good compromise between running time and accuracy and is therefore chosen as the standard settings for the solver.

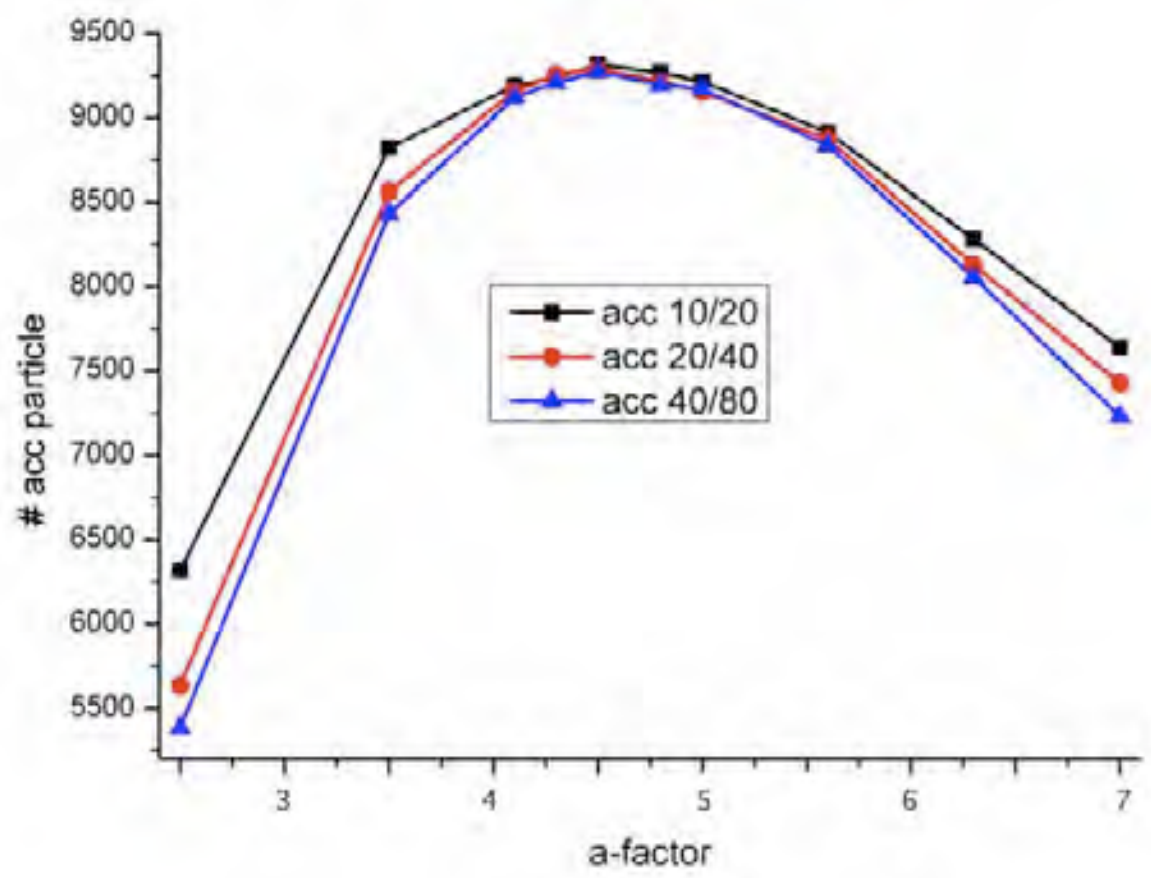

Figure 17.22: Space charge grid, no images: Number of accelerated particles as a function of the grid resolution (dr=aperture/x and dz=cell length/y). The external field grid uses the default settings of dr=aperture/20 and dz=celllength/40.

A note from pre-Poisson extensive studies with schefftm: – schefftm allows study of the length of the grid and the number of mesh points needed in the grid without influence of boundary conditions. Detailed investigations showed that a multiple of the rms beam length was a better mesh length than ±1cell, with grid spacing also based on a fraction of the rms beam length. Then in the Poisson space charge subroutine, where boundary conditions require ±1cell mesh length, the mesh spacing could be based on a multiple of the rms beam length, adjusted for integer number of mesh points, and divisible by four for coarser grid, etc.

*17.4.2.4 Selection of Number of Multigrid Iterations and Smoothing Cycles for Space Charge Field Solution*

A three-level W-cycle is chosen for the space charge Poisson solution.

The sensitivity of the Poisson solver has already been analyzed for the external field which does not include a charge density on the grid. Because a separate grid with exactly the same boundary conditions is used for the internal field calculation, the sensitivity of the Poisson solver is analyzed again, with a charge density introducing the field. Fig. 17.23 illustrates the effect of the number of multigrid iterations and the number of smoothing cycles on the number of accelerated particles. All used settings produce similar curves which lie within 1%. The standard settings are 5 multigrid iterations and 7 (10 for the ratio of dr/dz greater than 4) smoothing cycles per grid (blue curve).

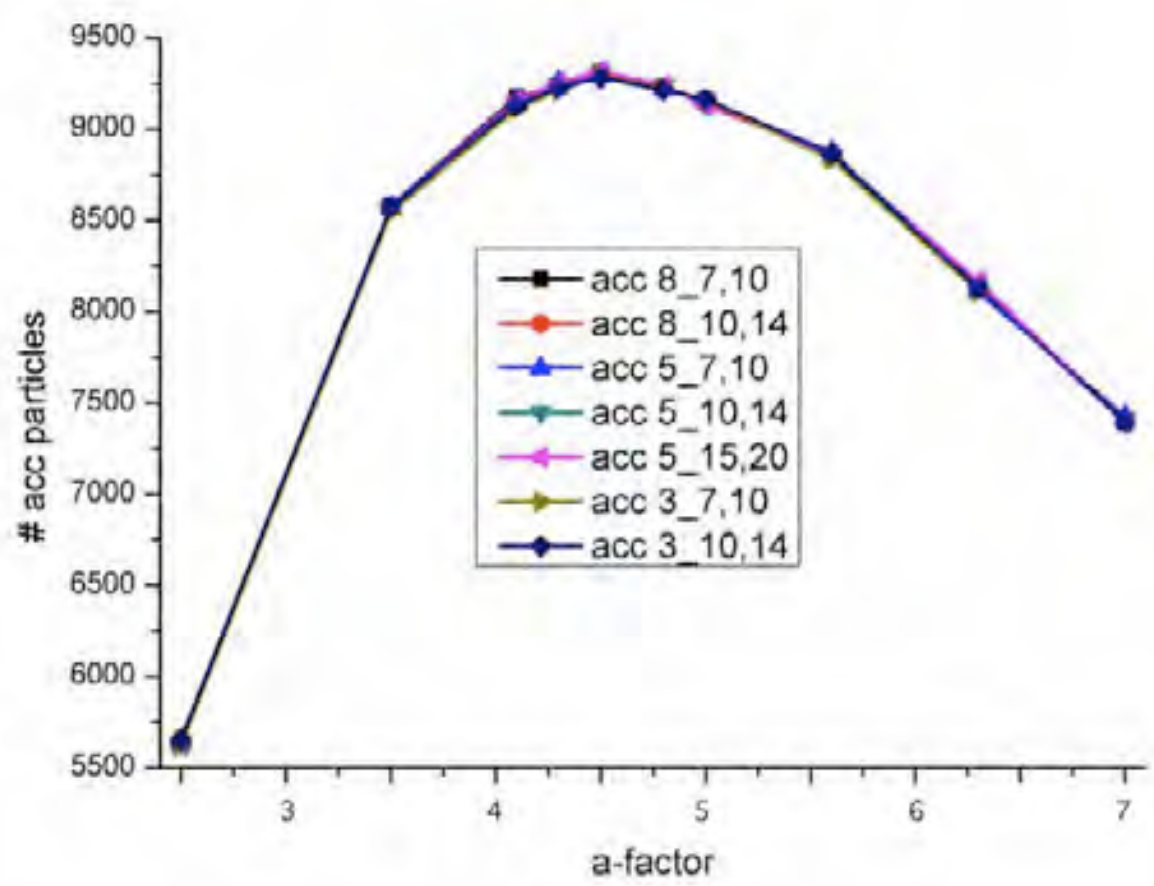

Fig. 17.23: Space charge grid, no images: Influence of the number of multigrid iterations (first number) and smoothing cycles (second number and third for ratios of hx/dz greater than 4) on the number of accelerated particles. The external field grid uses the default settings of dr=aperture/20 and dz=celllength/20.

*17.4.2.7 Differentiability of Solution; Use of splines*

The following is paraphrased from an email discussion [141].

Some questions beyond the accuracy of the Poisson solver arise concerning use of the results in dynamics codes.

Two types of solvers can be considered:

1. Fixed grid: A principal difficulty is that the solution is not differentiable, although it is usually continuous. When dynamics equations have non-differentiable right-hand sides, they either do not have solutions, or do not have unique solutions (the usual case). Thus there is no reason to trust the results. It should be possible to overcome this difficulty if higher order spline fits are used (one should probably assure at least four derivatives for theoretical reasons (KAM theorem)), but the computational and memory cost and programming complexity of the solution mean that few have tried to do this. There is anecdotal evidence that a few tests for an MHD code were tried and that the smoother solutions did change the behavior and make it better.
2. 2. Adaptive grid: The fixed grid difficulties continue to hold but now discontinuities are introduced in time as well as in space, and there are apparently no methods to correct this.

Discontinuities that introduce numerical chaos in the solutions are a particularly insidious type of error. Recall that orbital chaos (orbital instability) is structurally stable (i.e. persists under changes of system parameters - Anosov theorem) so that refining a mesh may give the same results but only be evidence that one is observing numerical instability, not genuine dynamics. Thus the usual argument that one cut the grid spacing in half and one still got the same answer may actually be evidence that the answer is wrong, not right.

However, while numerical instabilities may exist which cannot be suppressed by a simple diminishing of spatial and time steps, cutting of grid spacing is still a really very good and necessary check for the scheme quality regarding other parameters. Unfortunately, in the 3D case this cutting cannot be made fine enough even on modern computers, so other checks as mentioned above must be done also.

Other practical considerations also apply. The Poisson solver calculates the values of the potential at grid nodes (in a practical sense - accurately). Then (as in *LIDOS*, but a very common scheme) it defines mean values of Ex, Ey and Ez at the centers between nodes, and these values are also considered accurate. Then the program can calculate values of Ex, Ey and Ez for any other point, for example as linear interpolation from nearest 8 points with known accurate values. As result Ex, Ey and Ez along any straight line represent broken lines - left and right derivatives are not equal for some points –the fields are not differentiable, but they are piecewise differentiable and continuous. Or spline fits, with continuous derivatives to a certain order, may be used.

We applied spline fits to both the field solutions and to the dynamics, with little penalty in computing time, and saw no significant change in the results of the *aperfac* curves.

The discussion continues using a very simplified model. If there exists a finite solution of a dynamic equation (e.g., orbit of a planet around the Sun), and this solution is a subject of small perturbations (influence of third body from far away) and these perturbations are "bad enough" (include resonant frequencies), then finally (after very long time - this is important for further discussion) the solution (planet orbit) may (but not necessarily) change dramatically. And diminishing of the perturbations amplitude cannot prevent this, only increase the time.

---

[141] Discussion with P. Channell and Stanislav Vinogradov, January-February 2008, private communicatons.

Surely this has direct relationship to transverse motion in RFQ. It is finite, differences between true field and its piecewise differentiable approximation can be considered as perturbations, and in this case these perturbations are really "bad" in the sense of the KAM theorem. One important note - the perturbations, which are "seen" by particles through the time-integration scheme are different and seem to be "better", because this scheme takes only the mean value of the force during the time step and is insensitive to such things as field differentiability inside the step. Nevertheless one is not sure they become "good". So if there would be the possibility to keep particles inside RFQ for an infinite time, there is a probability that essential differences will be seen between trajectories of some particles in nature and in the numerical experiment.

It is also possible to make perturbations "not bad" using interpolations of higher order.

But it must noted that above conclusions are essential only for investigation of long-time particle behavior – that is, investigations of equilibrium distributions in different fields. For short-time behavior (like in RFQ – there would be only rare cases involving more than 200-400 oscillation periods), the influence of perturbations caused by different approximations may be easily suppressed by simply diminishing of time and spatial steps and proving the final accuracy strictly enough. The proof is to compare the results of numerical experiments with analytically derived results for cases when that is possible. For example, the radial oscillations of uniform electron ball inside an unmovable larger ion ball could be watched, comparing the oscillation frequency with the theoretical value and checking the absence of numerical instabilities for, say, 1000 periods. Or, particularly concerning the RFQ, the RFQ can be simulated with the ideal form of vanes, and results compared with simulations where the field from vanes was defined not with Poisson solver, but analytically through the 2-term potential. And if results are close in this case, it is difficult to suppose that for real vanes (which actually are rather similar to ideal) one can get significant errors. (This possibility remains in *LIDOS*, till now as an undocumented one. To simulate ideal vanes through Poisson solver, it is necessary to change the RealForm parameter in the real.prm file to "n" (this parameter is also used to change modulation from 2-term to sinusoidal). To simulate ideal vanes through 2-term potential, it is necessary to choose "ideal" form from the shell (this is documented)).

Such comparisons to time-independent analytical models are made in Sec. 17.3.

## 17.5 RFQ Simulation Results Using the Multigrid Poisson Solver

In this chapter the influence of the multigrid Poisson solver on the simulations of the particle dynamics in RFQs will be described. As the test case, the *aperfac* set of RFQs is used.

### 17.5.1 External Field

In this section the influence of the multigrid solver on the external field of the RFQ will be analyzed and compared to the old multipole expansion method. First the field will be illustrated, then the effects on the dynamics of a single particle will be considered and finally, the effect on the particle ensemble will be described. All the corresponding simulations were either done with zero current or with the same space charge subroutine to make sure that the effect of space charge is treated in the same way for multigrid and for multipole expansion runs.

#### *17.5.1.1 Illustration of the Field*

For illustrating the external field it does not matter which RFQ to choose, since the generation of the mesh and the solving for the potential is always done in the same way. The RFQ with an aperture factor of 48 was chosen, because it is somewhere in the middle.

In the following the potential as well as the corresponding electric field components are shown for different cells along the RFQ for two different planes. In Fig. 17.24 the potential and the electric field for cell number 10 in the xy-plane are shown. On the picture of the potential the electrodes at different potentials can be seen. The beam axis has a zero potential, since the modulation at this position of the RFQ is very small. The breakout angle of 10 degrees of the electrodes is also visible. The second picture illustrates the x-component of the electric field ($E_x$). For the region of the electrodes the electric field is equal to zero. At the beam axis $E_x$ is equal to zero and it rises linearly up to its maximum value by approaching the x-electrode. The transition from vacuum to electrode seems to be somewhat rough. There are two reasons for this kind of behavior.

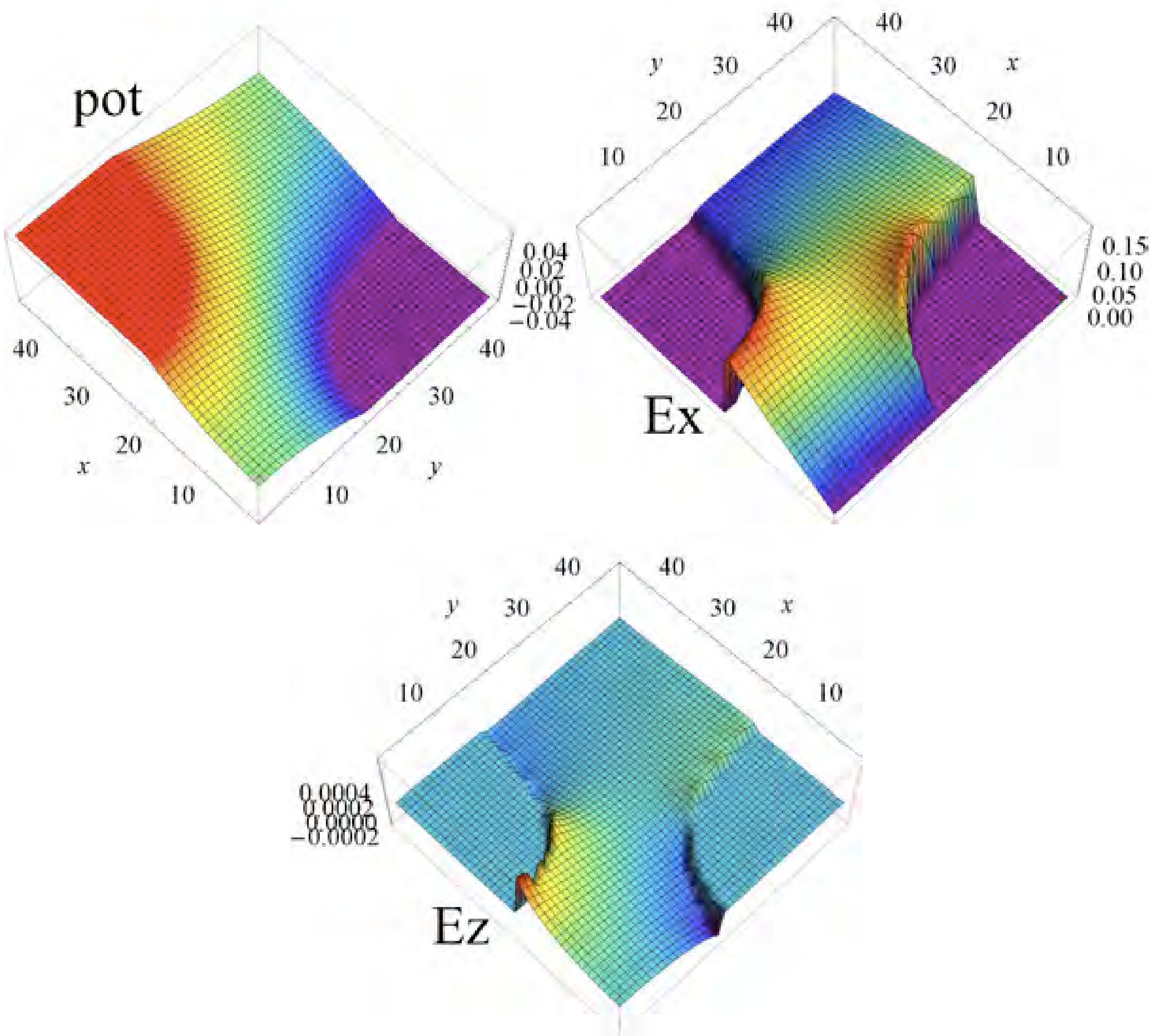

Fig. 17.24 XY-plane of the potential, $E_x$ and $E_z$ at cell 10.

For finding each field component for one regular grid point, the two-sided derivative schema is used and therefore the potential of each neighboring grid point has to be known. When the center grid point lies on the surface of the electrode (shifted grid point), the value of one of the neighboring grid points is equal to the center grid point and therefore the one-sided derivative schema is used. The necessary change of the schema is not perfectly smooth.

The second reason is due to shifting the last grid point to make sure that there is always a (shifted) grid point lying on the surface of the electrode. This effect is not taken into account while plotting the potential and field components. So the surface plots look rougher than they actually are. This is the dominant effect.

The longitudinal field component $E_z$ indicates a small longitudinal field at the region close to the electrodes, with the amplitude lower than the transverse field by a factor of 400. Again, there is only a very small modulation at this point. Fig. 17.25 shows the potential and the electric field components for the same cell, but in the xz-plane. On the potential plot the electrode can be seen with its constant potential. A little effect of the small modulation can be noticed. The $E_x$-component of the field increases linearly from the beam axis to its maximum at the electrodes. The longitudinal field component is very small around the beam axis, but it increases a little at the vanes due to a very small modulation. Along the beam axis as well as on any line parallel to the beam axis the longitudinal field has a sinusoidal shape.

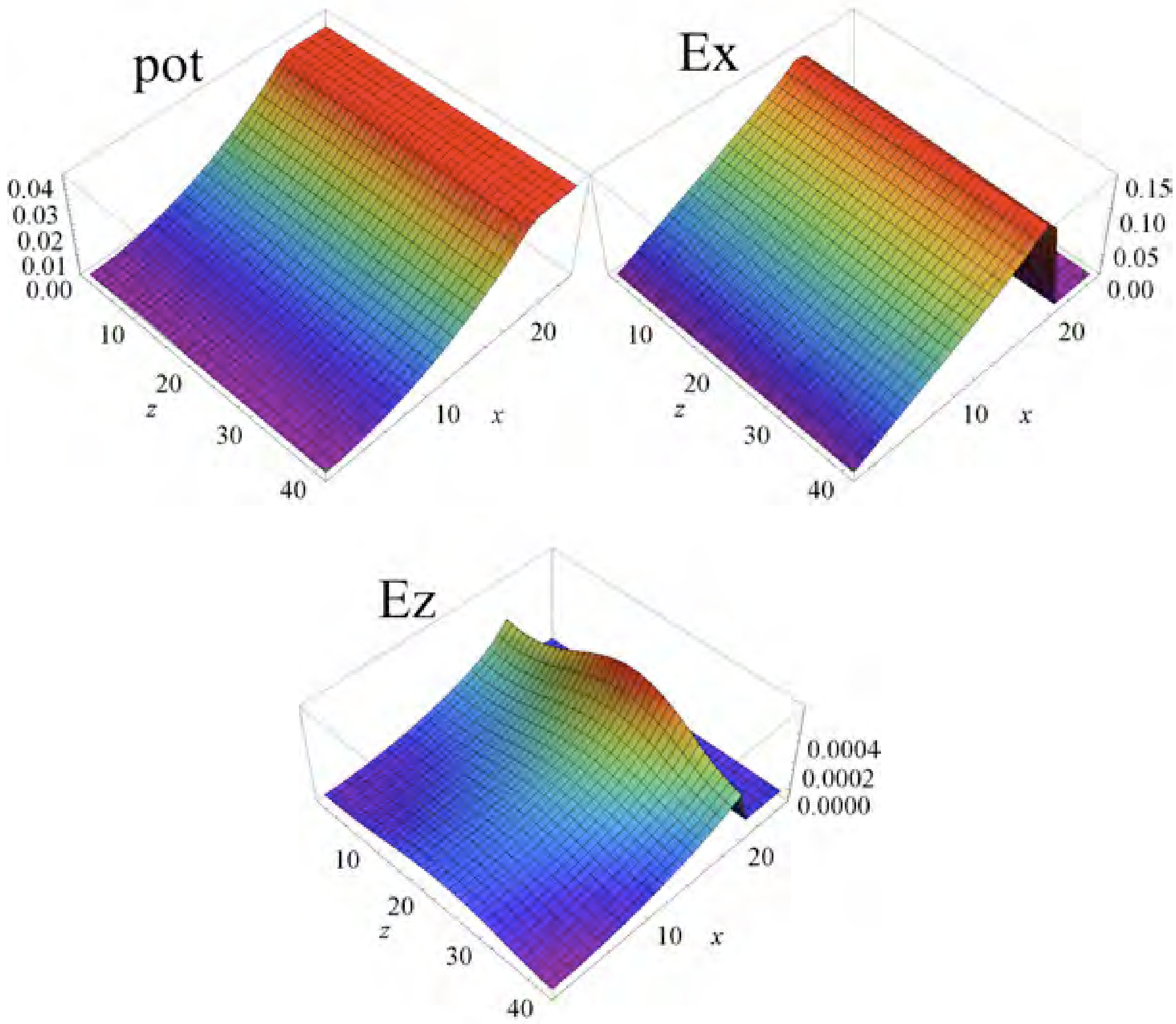

Fig. 17.25 XZ-plane of the potential, $E_x$ and $E_z$ at cell 10.

The potential and field components for cell 50 for the xy- and xz-plane are shown in Figs. 17.26 and 17.27. The longitudinal field components have increased compared to cell 10 as expected. Particles on the beam axis see some bunching forces, which increase smoothly up to the electrodes with some remains of the shifted grid point effect. The transverse field component has not changed much. On the xz-plane (Fig. 17.27) the influence of the modulation on the potential and the longitudinal field component can be seen. The potential at the beam axis has a sinusoidal change in its amplitude, which results in a non-zero longitudinal field component. The maximum is again at the region close to the electrodes, but it is quite flat up to 2/3 of the aperture.

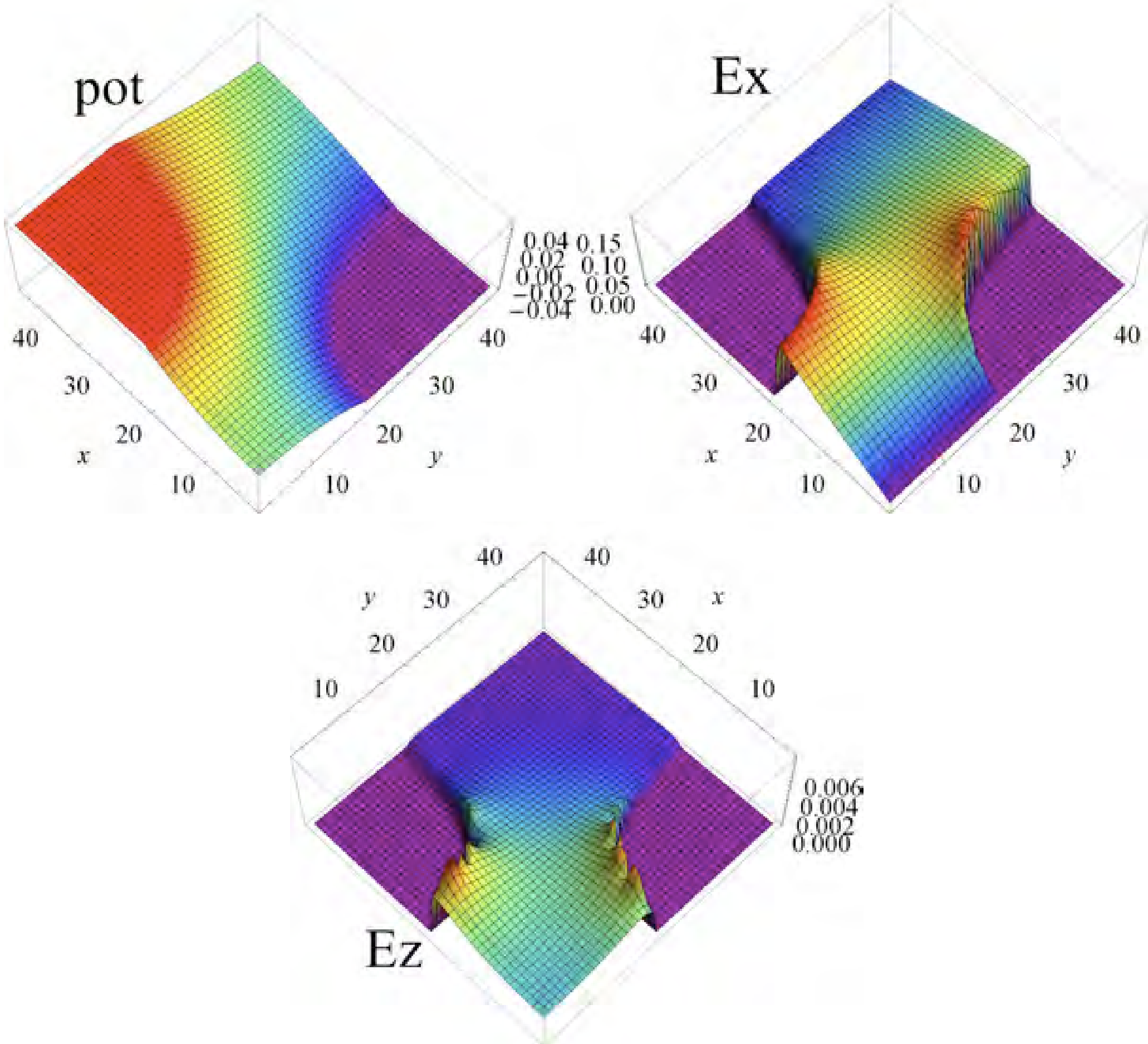

Fig. 17.26 XY-plane of the potential, Ex and Ez at cell 50.

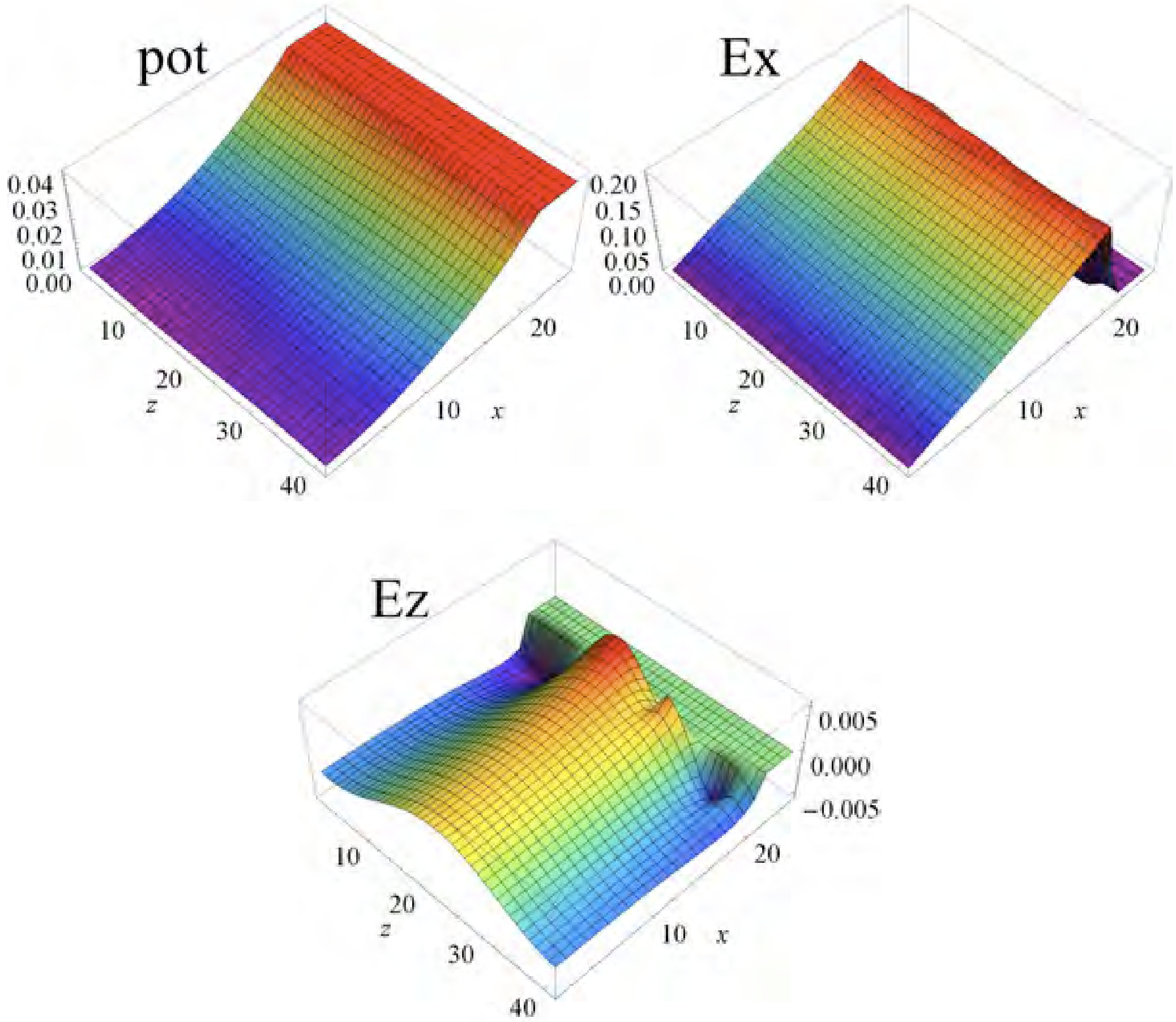

Fig. 17.27 XZ-plane of the potential, $E_x$ and $E_z$ at cell 50.

The potential and the two components of the electric field for accelerating cell 400 with modulation = 1.8 are shown in Fig. 17.28 for the xy-plane and in Fig. 17.29 for the xz-plane. In the transverse plane the potential at the beam axis is not zero any more, but pushed to lower values by the modulation. The displacement of the horizontal and vertical electrode is quite different, changing the shape of the potential. The shape of Ex is basically the same as in cell 50: linear increase towards the x-electrode. The potential along the beam axis has a sinusoidal shape, corresponding to the modulation on the vanes. The transverse field has also remained unchanged (besides the modulation of the vanes), but the longitudinal field has increased its strength compared to the previously shown cell.

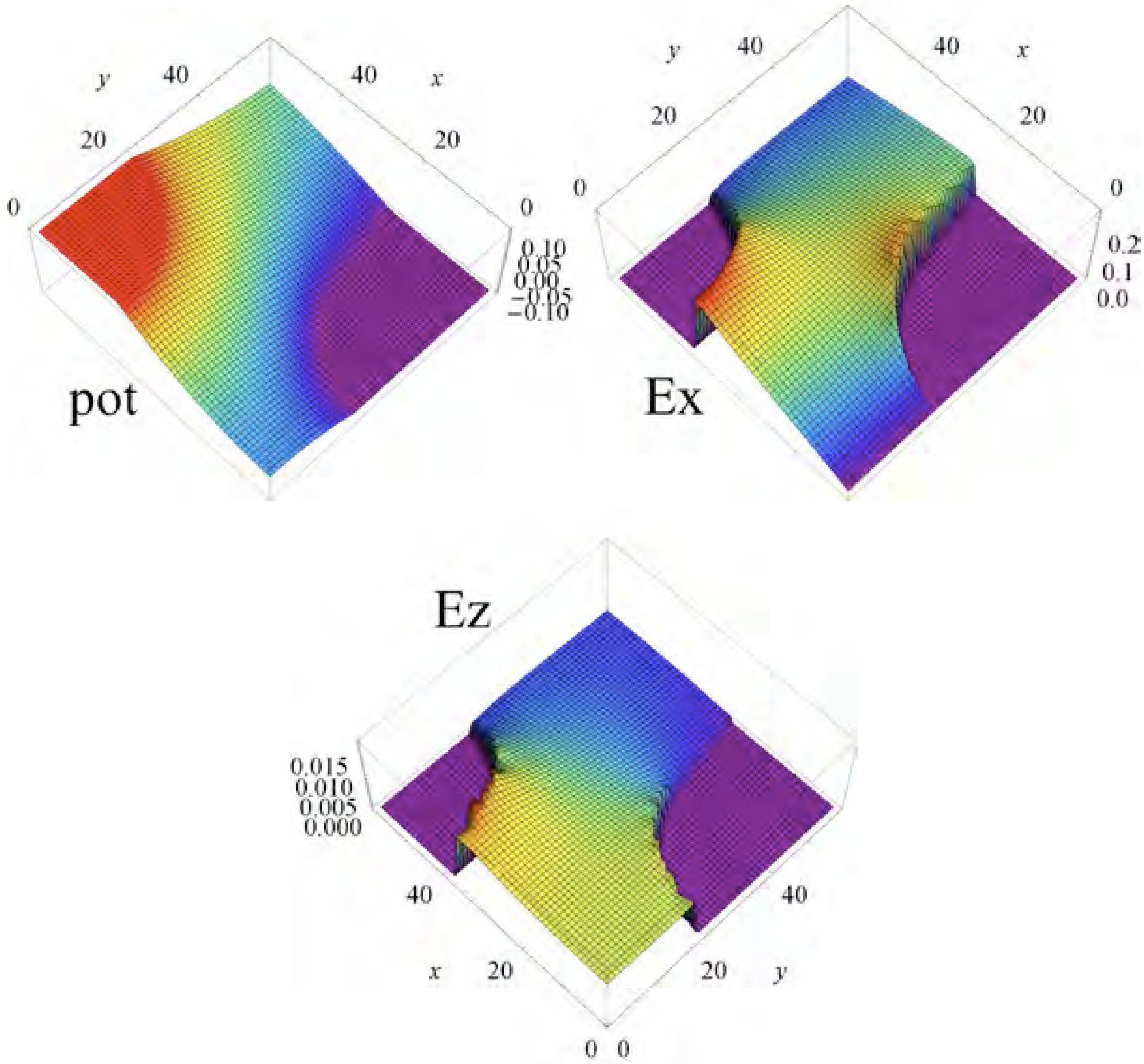

Fig. 17.28 XY-plane of the potential, Ex and Ez at the center of cell 400.

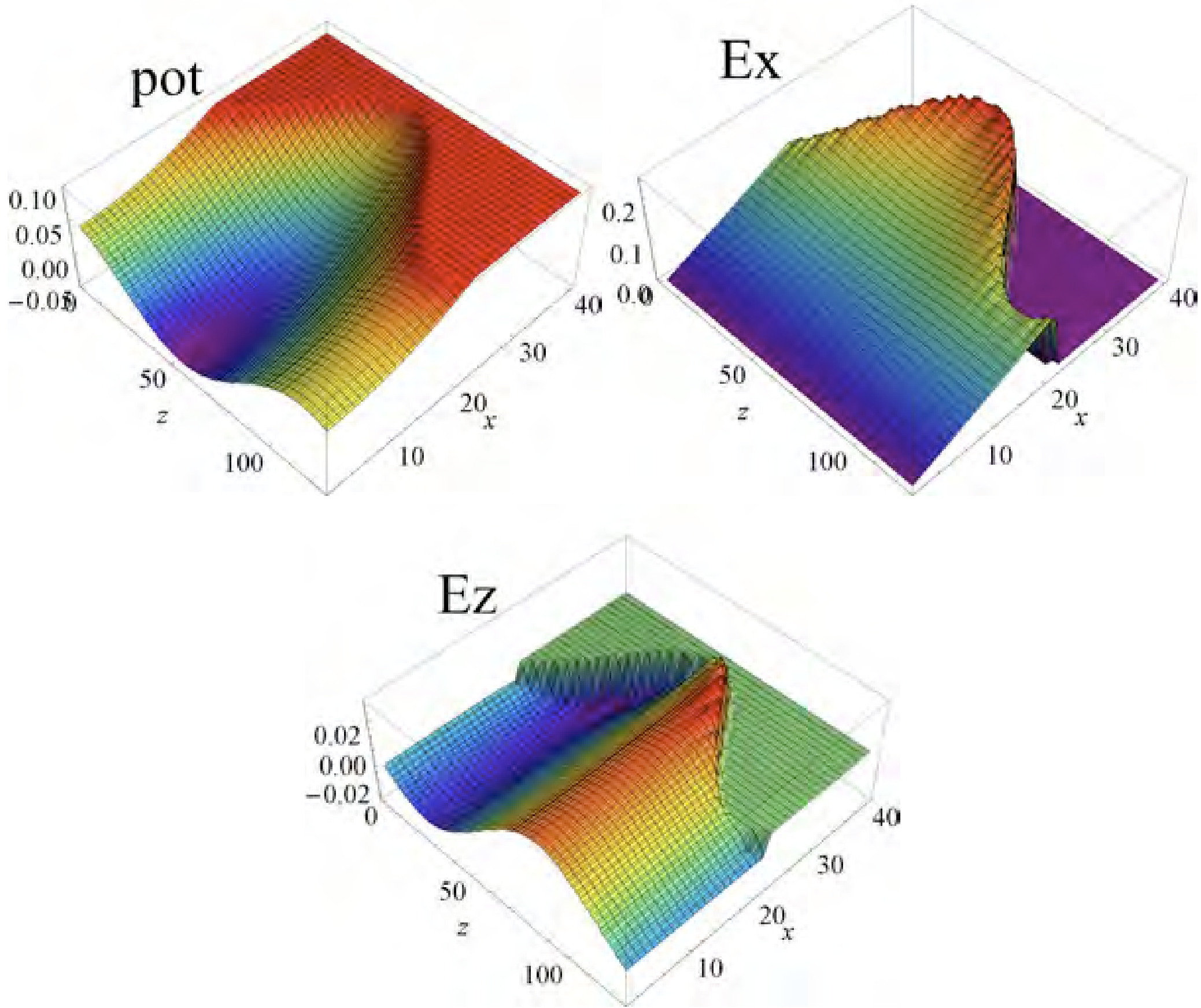

Fig. 17.29 XZ-plane of the potential, Ex and Ez at cell 400.

*17.5.1.2 Comparison to Potential of Multipole Expansion Method*

In this section the external field from the multigrid Poisson solver will be compared to the field from the multipole expansion method. One reason for implementing a Poisson solver was that the multipole expansion is only accurate within the region of the minimum aperture, since it derives from the cylindrically symmetric approximation, and its coefficients are calculated by integrating along an arc with a radius near the minimum aperture on a field map found by another Poisson solver. Therefore a difference in the region beyond the minimum aperture is expected. Figure 17.30 shows the 8-term multipole potential in the xy-plane at the center of a cell. The white areas at the top and bottom indicate regions where the potential is higher than the positive vane voltage or lower than the negative voltage. Apparently, there seems to be a second positive and negative electrode, which is due to the inaccuracy of this description of the potential. At the region of the beam axis the potential has the expected shape.

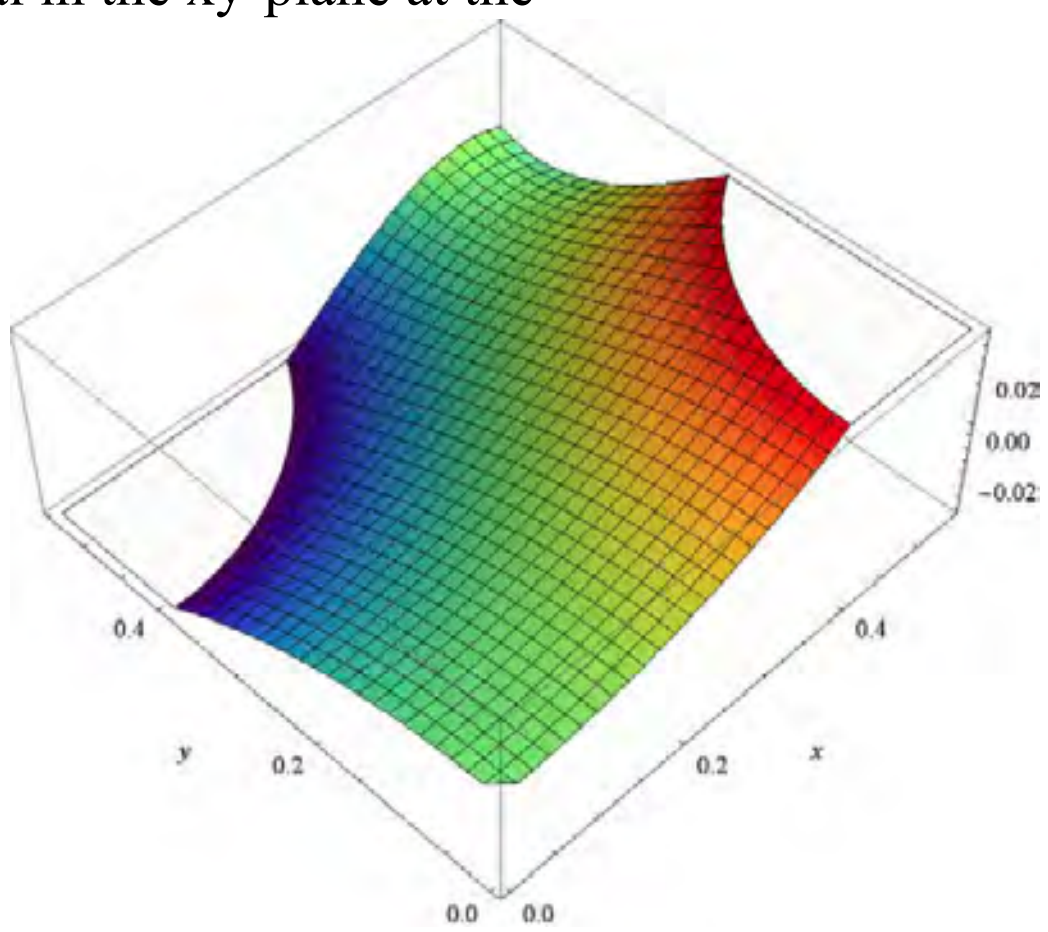

Fig. 17.30 Potential from the multipole expansion method at the center of cell 150 as a function of x and y.

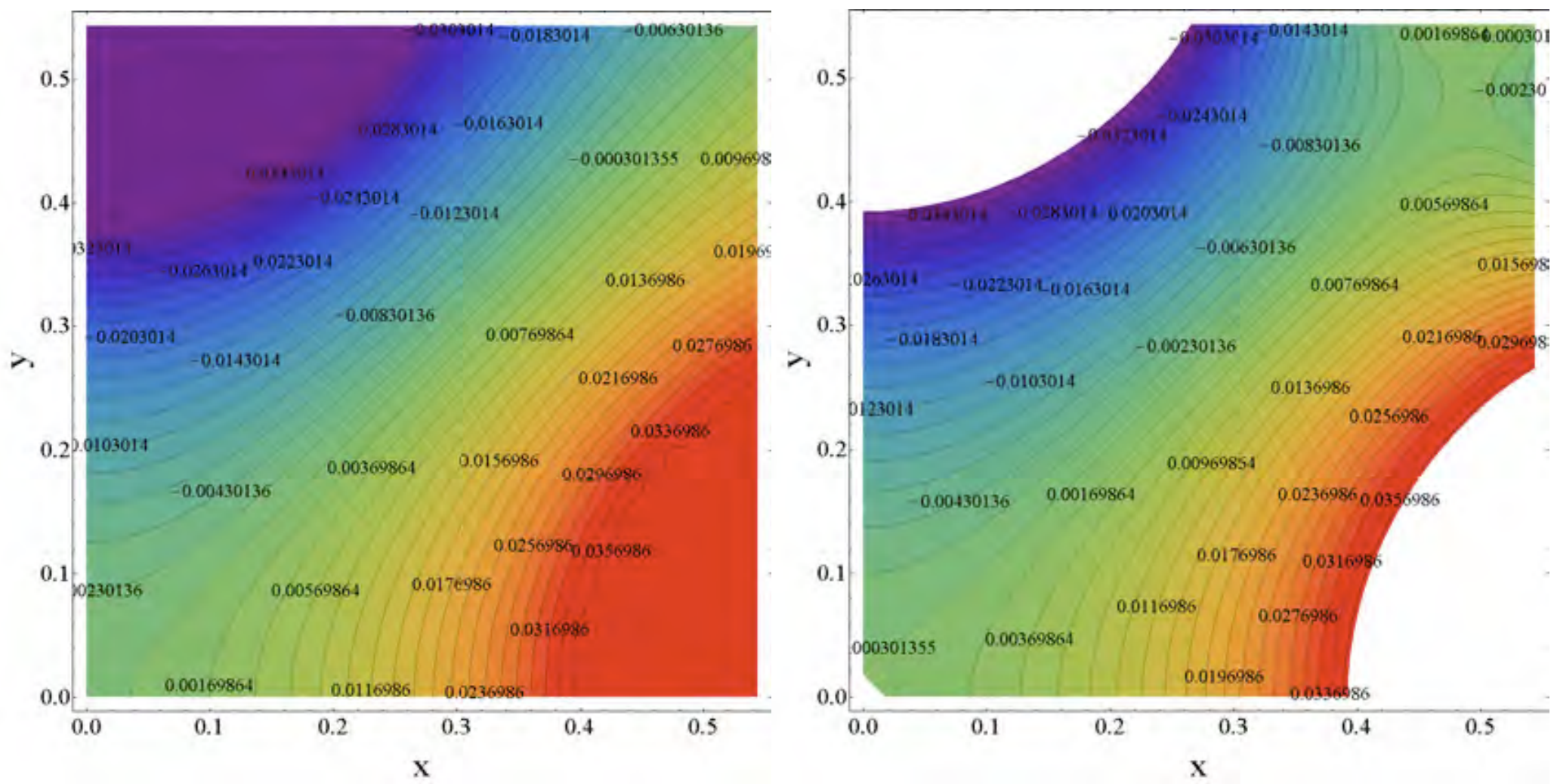

Fig. 17.31 XY-plane of the potential from the multigrid Poisson solver and from the multipole expansion method at the center of cell 150.

Fig. 17.31 shows the potentials from the two different methods at the same position. In the useful zone at the beam axis, the two potentials compare quite well. Only further out the there are major differences. Most of the particles will never go so far and therefore both methods can be reasonably be used. It has to be noted that the modulation at this position is quite small (m = 1.16) and the potential at the middle of the cell is shown where the displacement of the horizontal and the vertical electrode is the same. So it is neither surprising nor significant that the two methods give similar potentials. It suggests that the Poisson solver finds a potential that compares quite well to the MP-potential (potential from the multipole expansion method) in a situation where the multipole expansion method is a accurate description of the external field.

The absolute differences of the two potentials are shown in Figure 17.32. The z range of the plot is ±2% of the vane voltage. It can be seen that there is a offset of around 1.5% of the vane voltage. Beside that the difference is relative smooth and small. The point at the beam axis is missing, because the multipole expansion method gives its potential normally in cylindrical coordinates and the transformation to Cartesian coordinates involves an arctangent.

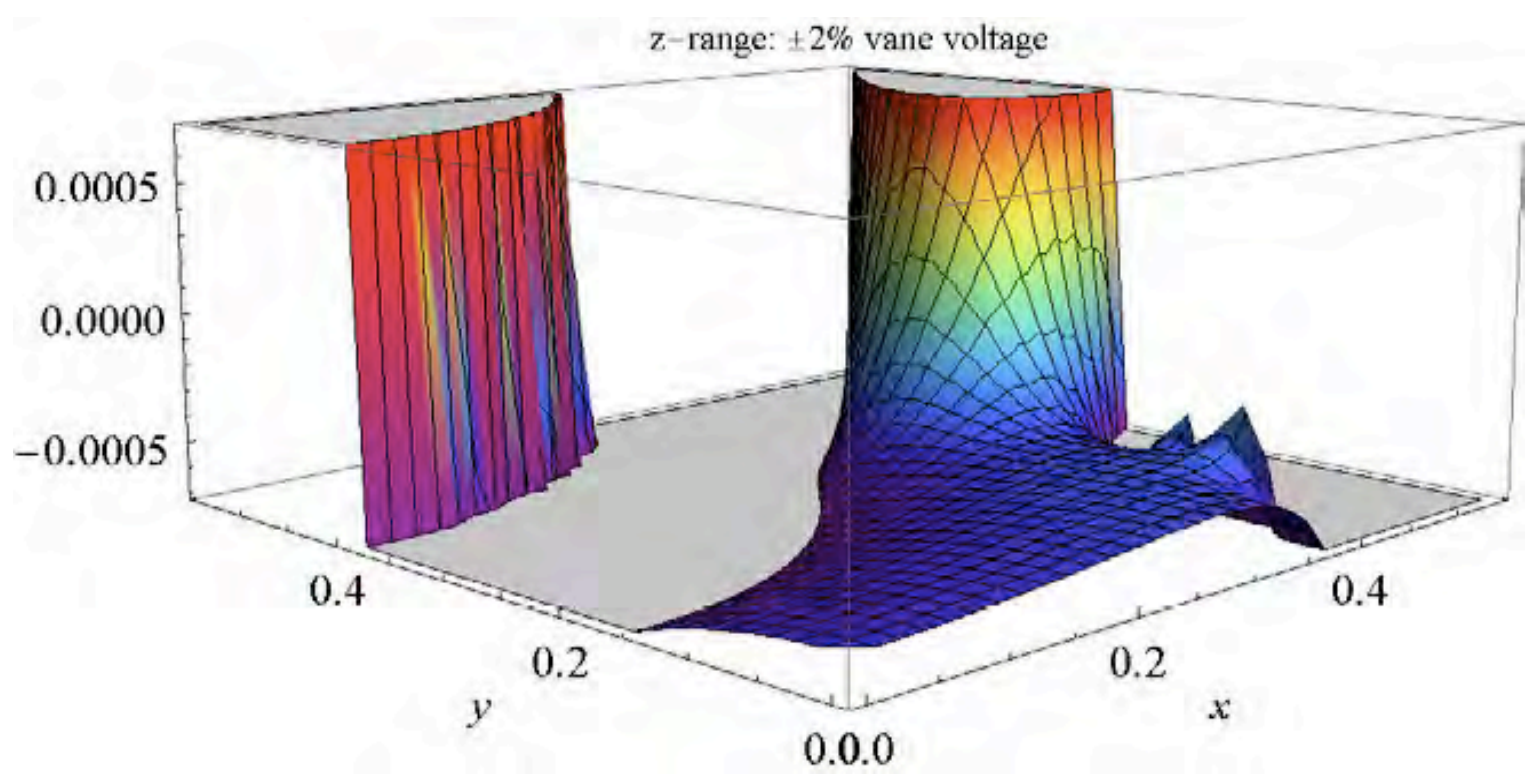

Fig. 17.32 Absolute difference of the potentials from the multipole expansion method and from the multigrid Poisson solver at the center of cell 150.

The last picture of cell 150 is the relative difference shown in Figure 17.33. The problem with relative differences is that the potential crosses through zero and there the relative values become big. That can be wonderfully seen and should not be taken seriously. Beside this effect the relative difference at the beam axis is less than 10%.

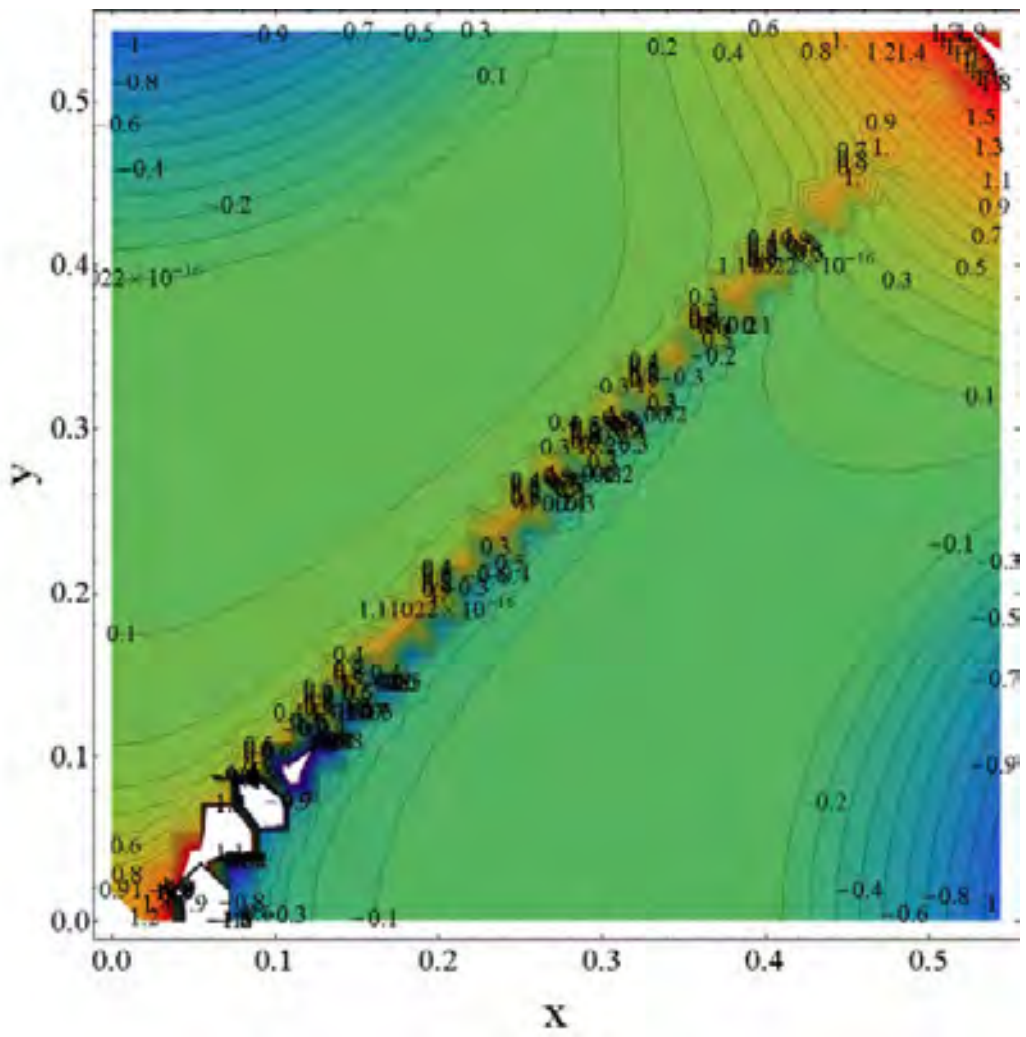

Fig. 17.33 Relative difference (MG-MP)/MG of the potentials from the multipole expansion method and from the multigrid Poisson solver at the center of cell 150.

Now, we consider a situation in which the limitation of the multipole expansion method is revealed. The potential at the end of a cell with a modulation of m = 2.3 for the two different methods is shown in Fig. 17.34. Potentials that are greater than the vane voltage have been cut off.

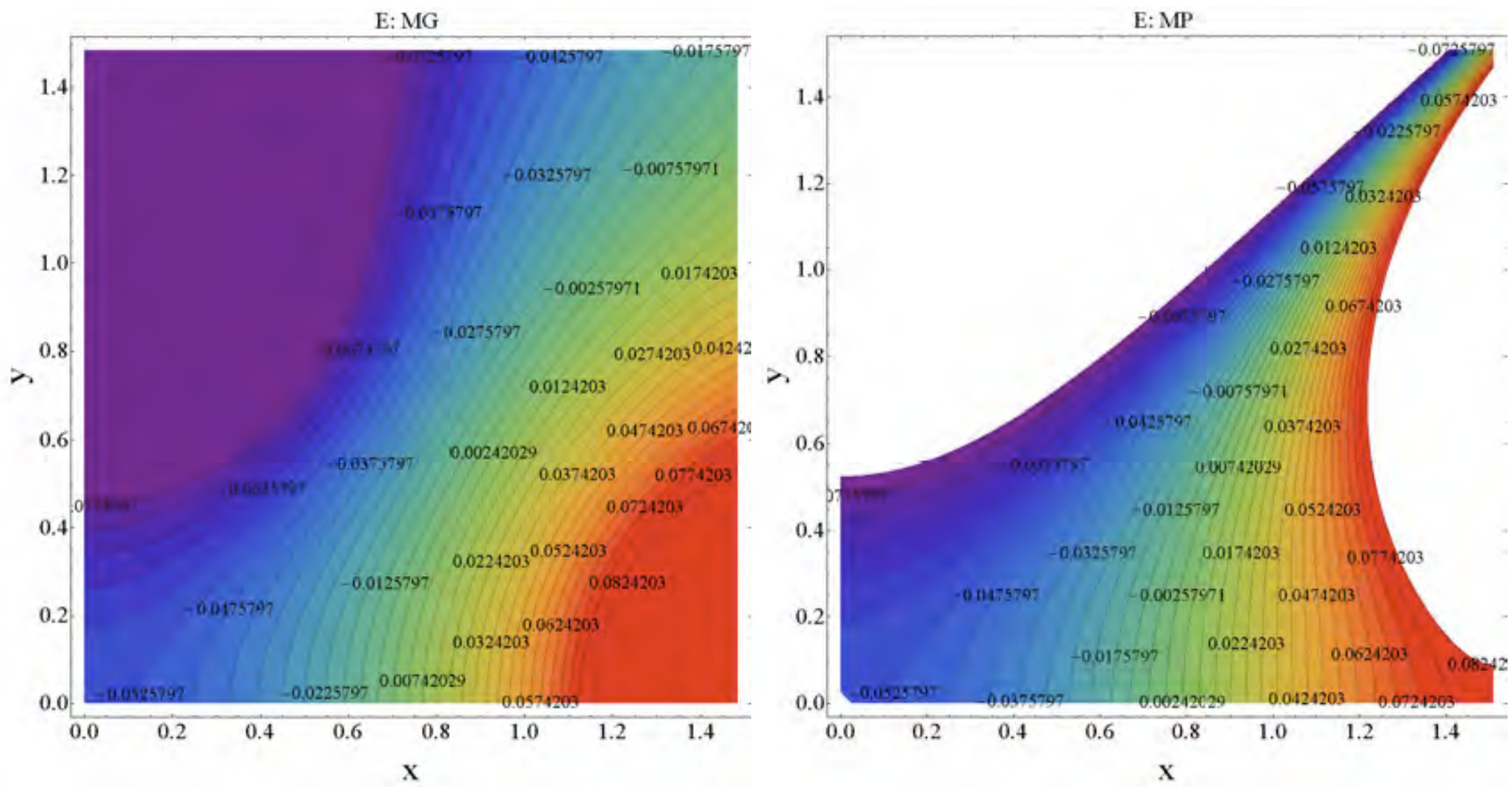

Fig. 17.34 Potential from the multigrid Poisson solver and from the multipole expansion method at the end of cell 300.

Obviously, the shape of the electrodes (white area) of the MP-potential is not even close to the shape of the vanes. Close to the beam axis the two method give similar potentials, but with increasing displacement from the beam axis, differences increase as well. In between the electrodes, the potential has to change from plus to minus the vane voltage. For the MP-potential the distance between the electrodes has become very small compared to the actual shape of the vanes used in the MG-potential (potential from the multigrid Poisson solver), therefore the electric field calculated from the MP-potential will be higher than it essentially is. Also the position of the horizontal vane in the MP-potential is too far away from the axes and therefore the corresponding electric field is too low. In Fig. 17.35, the relative differences of the potentials are shown. At the beam axis the difference is again less than 10%, but as soon as the displacement from the axis increases the difference become easily bigger than 40%. Towards the horizontal vane the error of the MP-potential becomes even bigger.

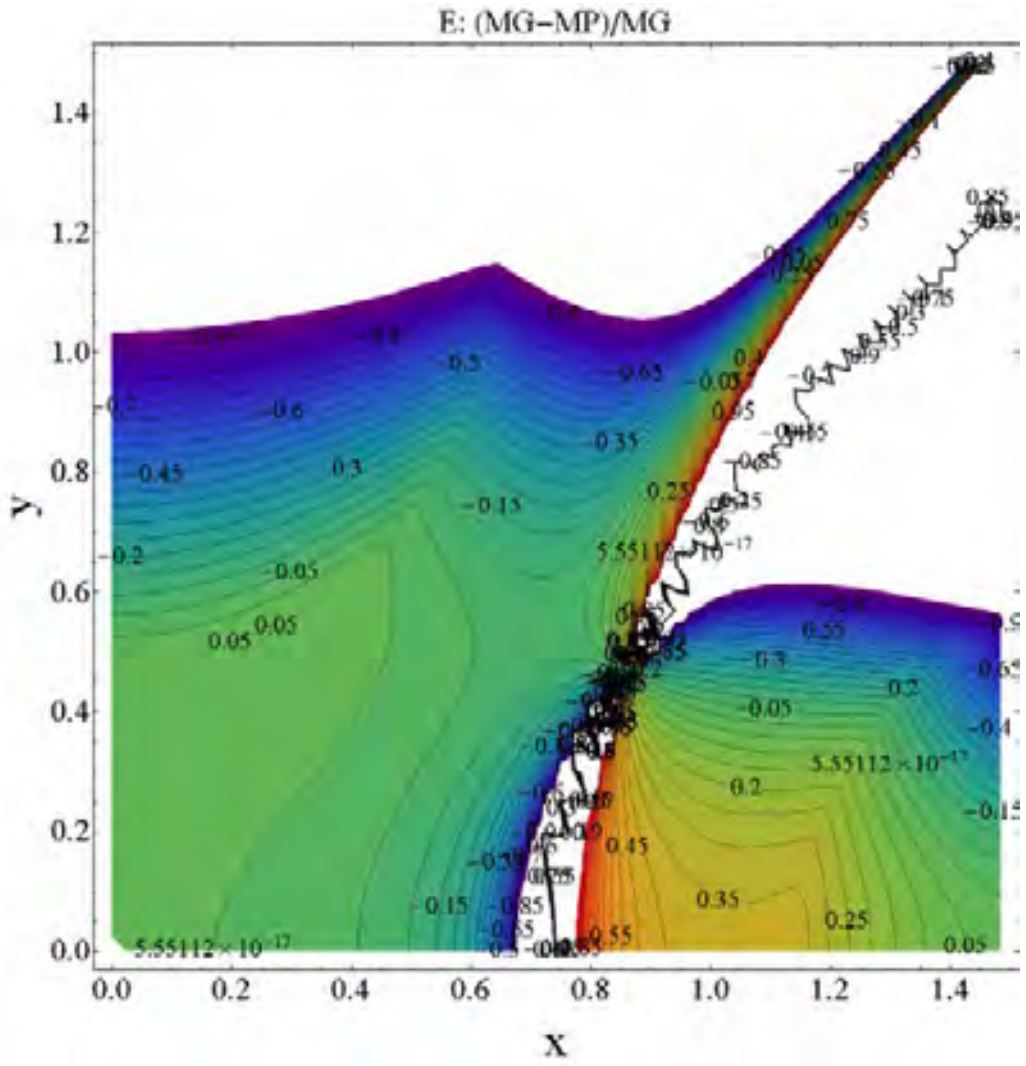

Fig. 17.35 Relative difference (MG-MP)/MG of the potentials from the multipole expansion method and from the multigrid Poisson solver at the end of cell 300.

### *17.5.1.3 Comparison of Ex*

Different potentials as shown in the previous section will lead to different electric fields. The x-component of the electric field in the xy-plane for the multipole expansion method and for the multigrid Poisson solver is shown in Fig. 17.36.

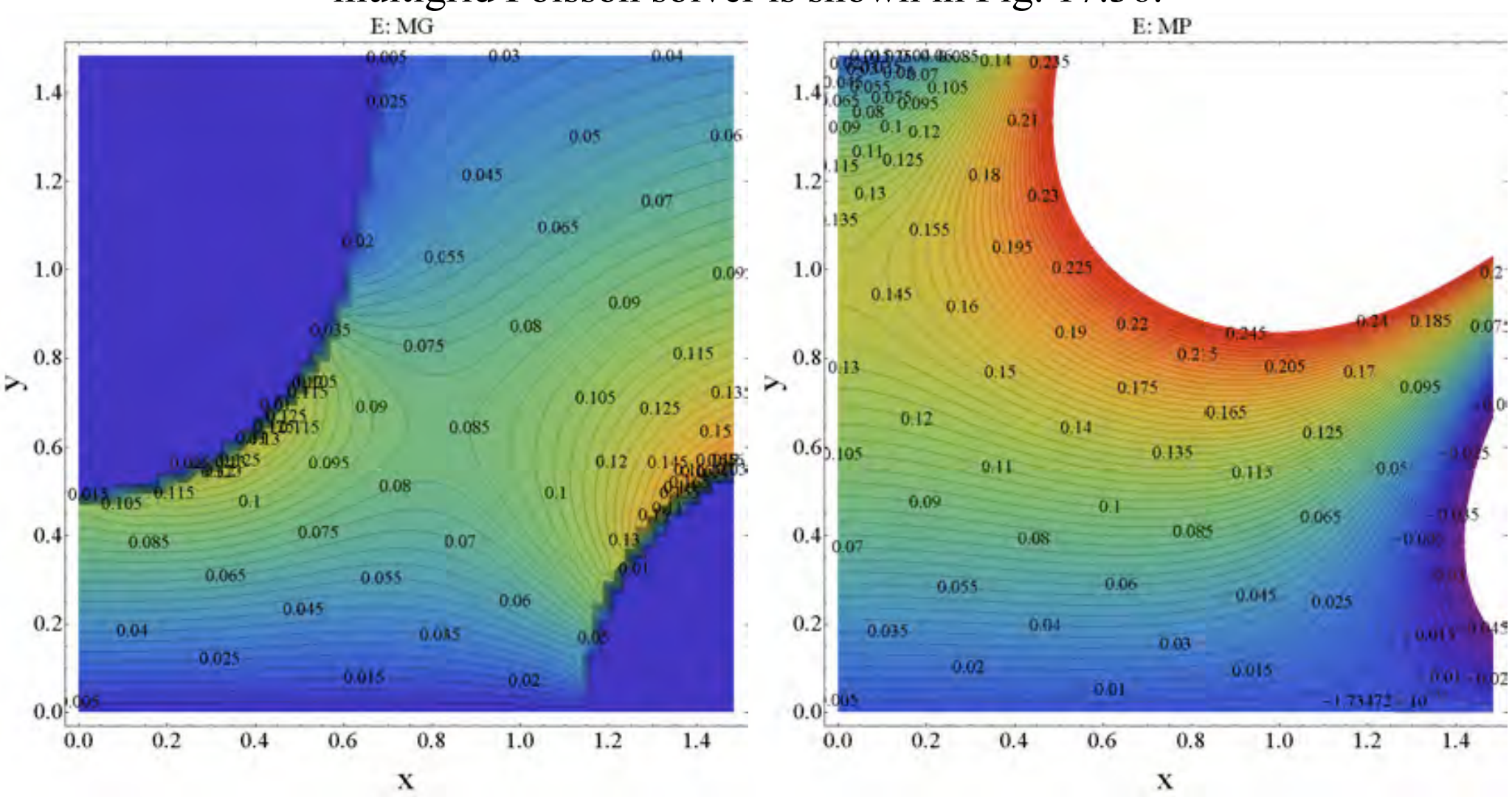

Fig. 17.36 X-component of the electric field (Ex) in the xy-plane at the end of a accelerating cell.

In the left picture (MG-field) the vanes are visible as those regions with a flat zero electric field. The multipole expansion method does not have any information about the vanes. Starting from the x-axis in both plots the field components increases linearly with equidistant equipotential lines. In the MG-case this incline stops at the vertical electrode whereas it continues for the MP-case. This is not a problem, since particles which touch the electrodes are considered transversely lost. In between the electrodes the field component increases further for the MP-case. In the MG plot there are some local maxima at the vanes, which are not present in MP plot. The relative difference of the two plots is shown in Fig. 17.37; at the useful zone around the beam axis the difference is less than 15%. Starting from in between the vanes the field starts to look quite different. Especially in front of the electrodes there are effects that cannot be represented by the multipole expansion method.

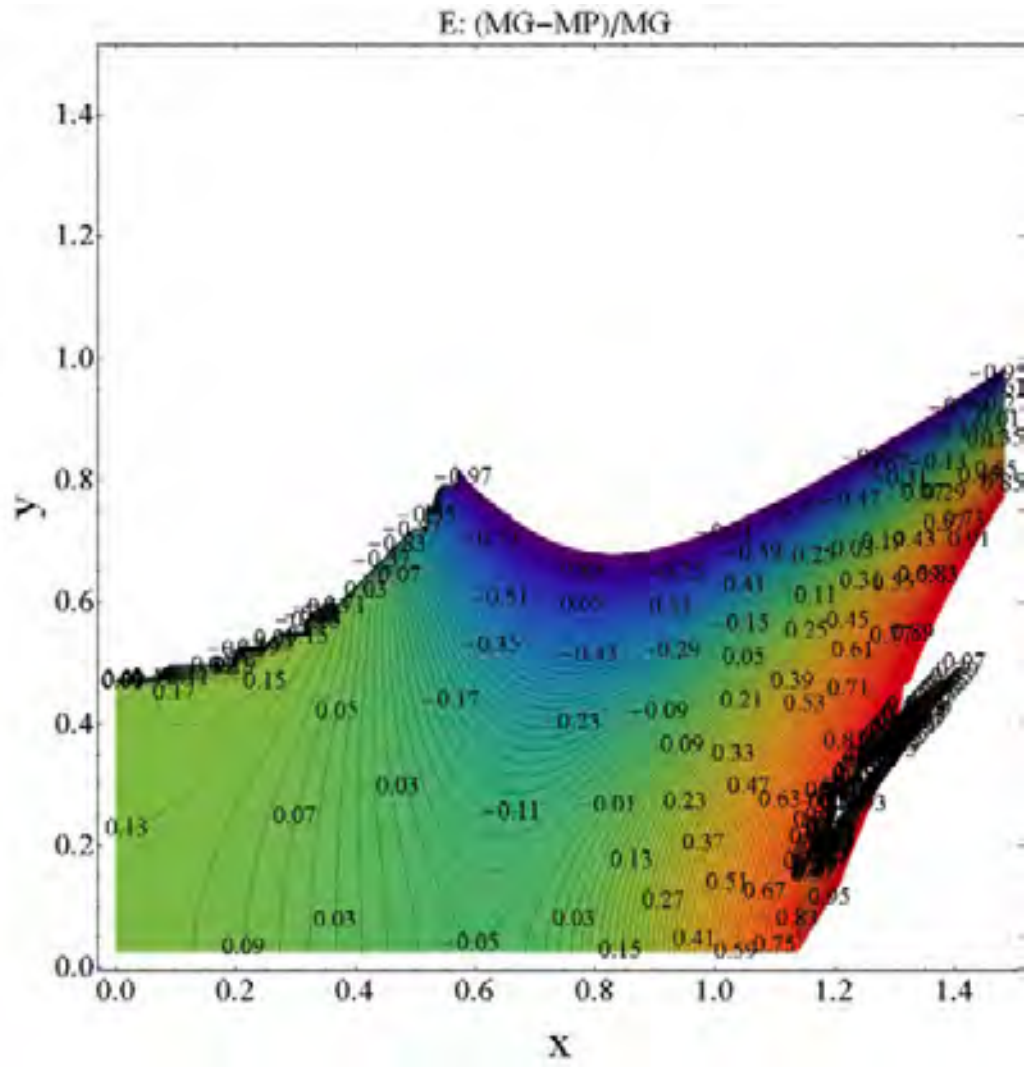

Fig. 17.37 Relative difference (MG-MP)/MG of the x-component of the electric field ($E_x$) in the xy-plane at the end of an accelerating cell.

The absolute value of the electric field at the same position is show in Fig 17.38. The maximum electric field is in the region close to the vanes and does not appear on the field plot from the multipole expansion method. At bigger distances away from the beam axis (lower left corner) the MP-field increases to its maximum whereas the MG-field tends to decrease which is reasonable since the vane voltage remains constant and the distance between the electrodes increase. Therefore a big relative difference is expected with a minimum at the beam axis. Fig. 17.39 showing the relative difference of the absolute values of the field confirms this.

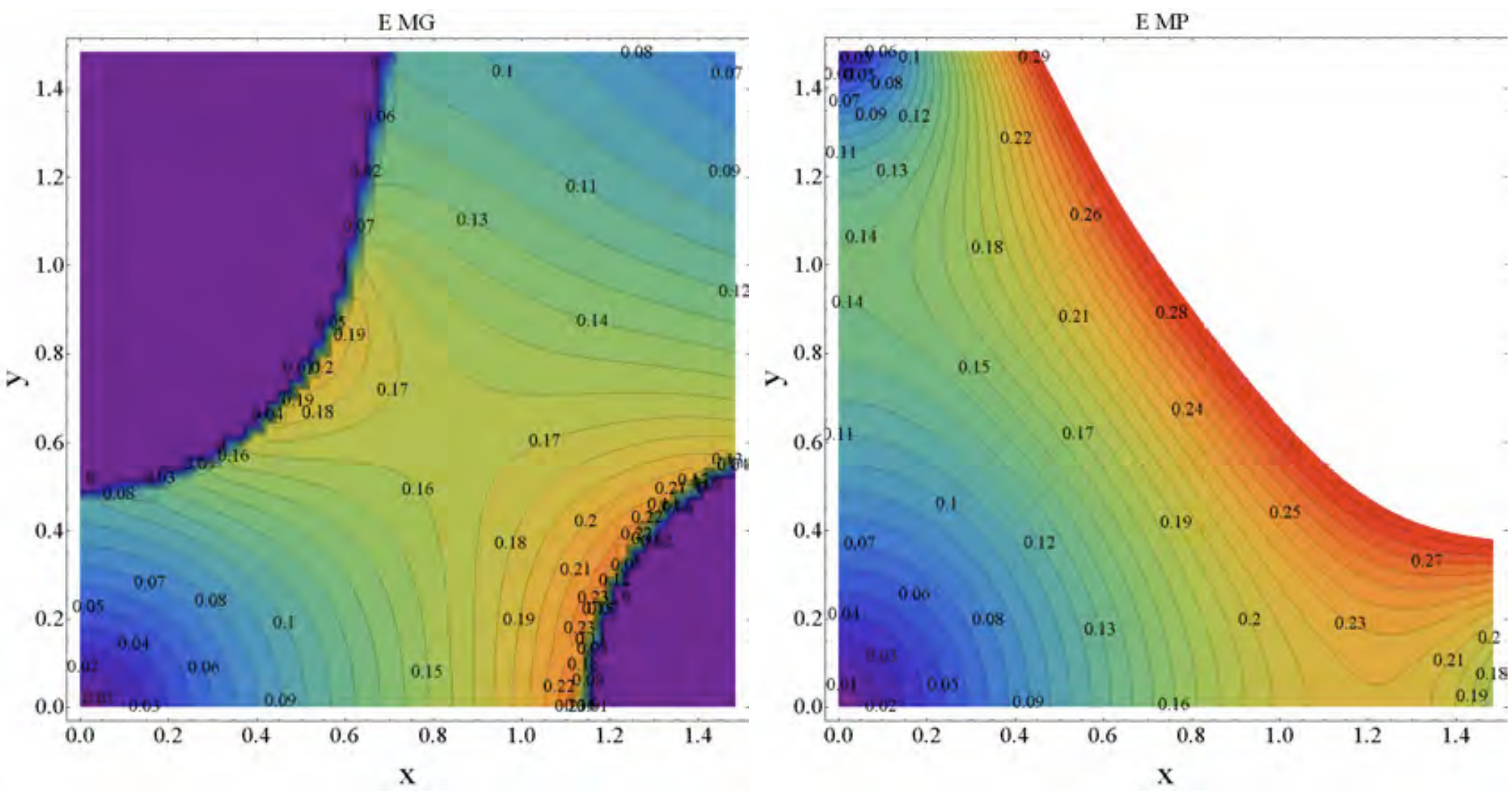

Fig. 17.38 Absolute value of the electric field (E) in the xy-plane at the end of a accelerating cell for multigrid Poisson solver and multipole expansion method.

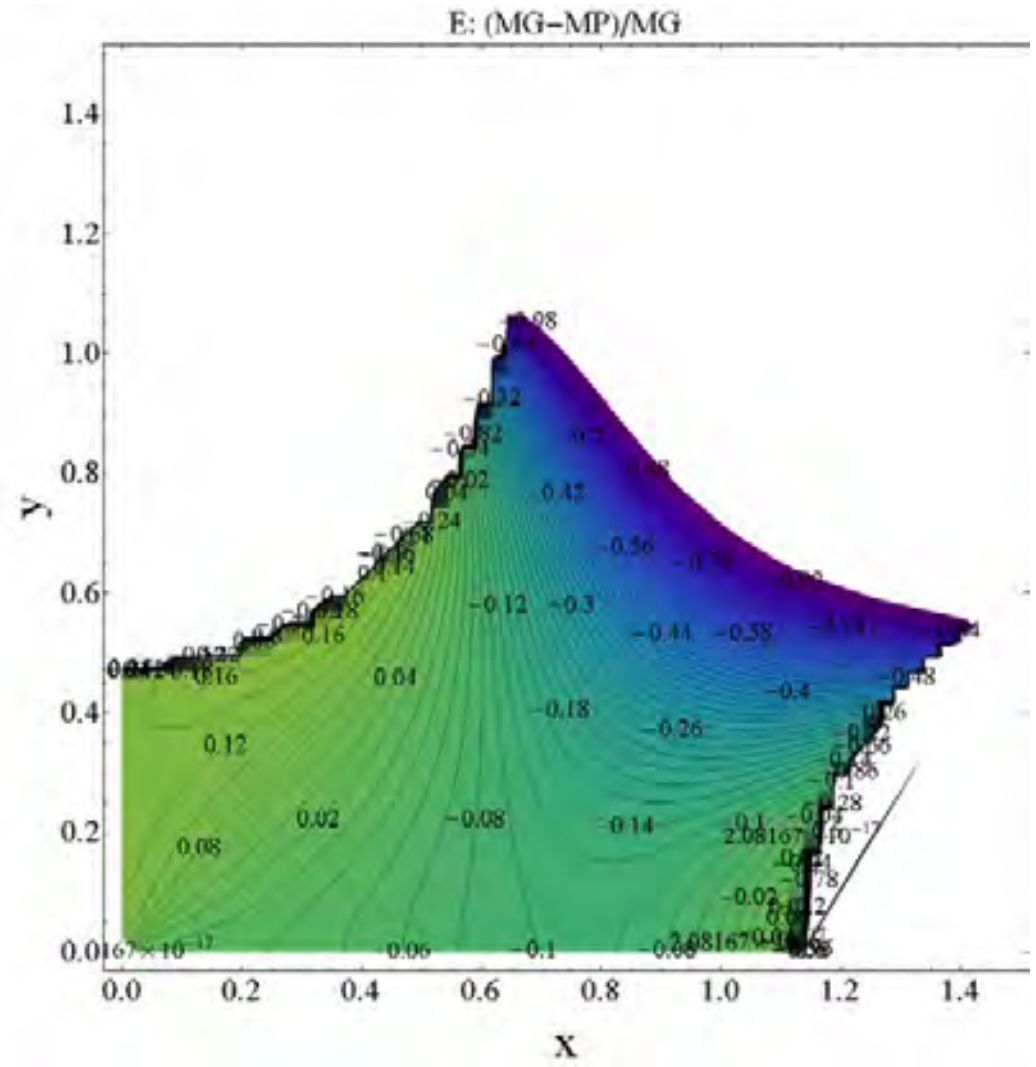

Fig. 17.39 Relative difference (MG-MP)/MG of absolute value of the electric field (E) in the xy-plane at the end of a accelerating cell.

### *17.5.1.4 Influence on the Single Particle Dynamics*

After comparing the differences of the method on the basis of potentials and fields, the influences on single particle dynamic is now analyzed. The first particle analyzed is the synchronous particle shown in Fig. 17.40.

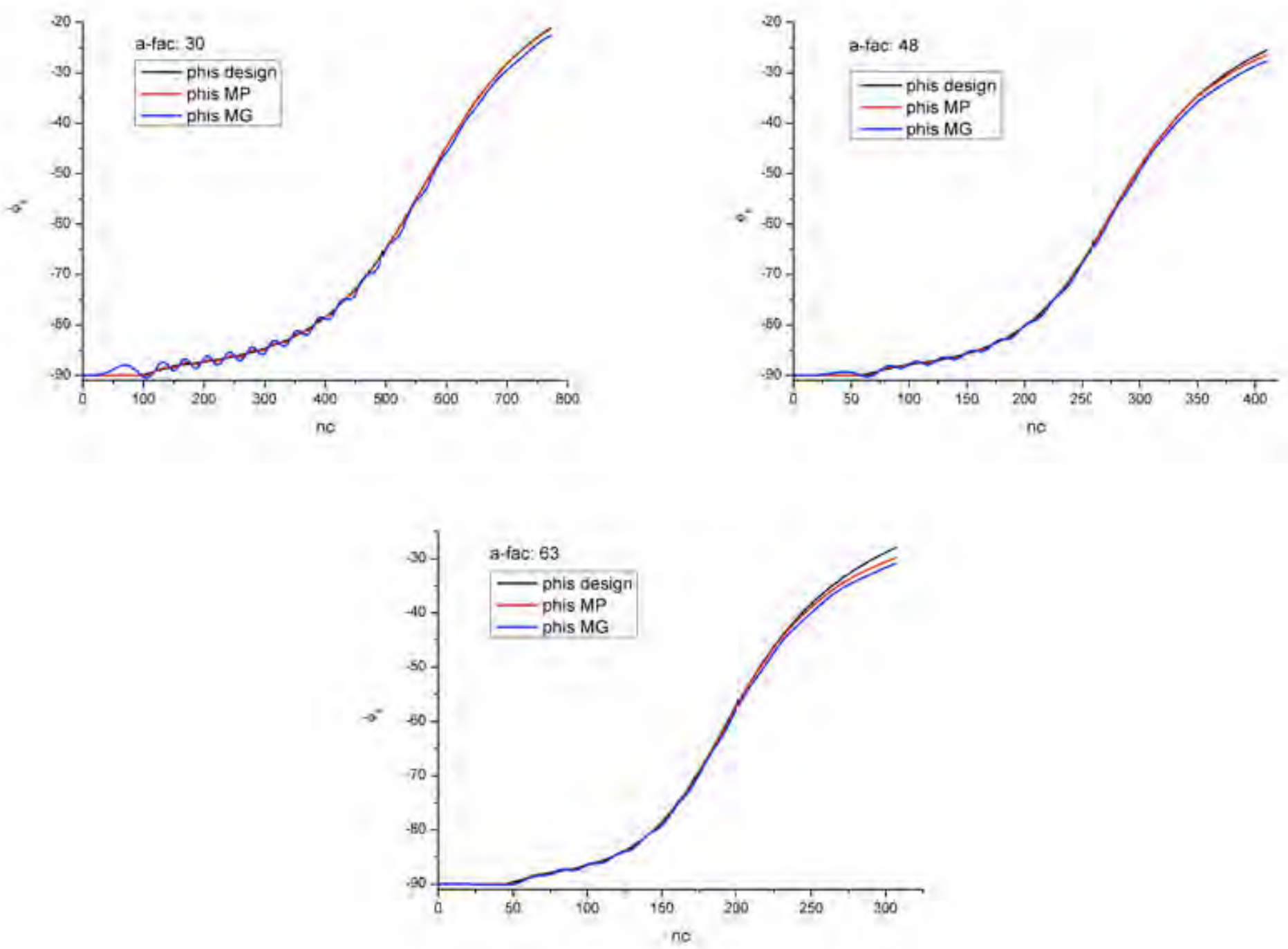

Fig. 17.40 Synchronous phase as function of the cell number for different RFQs (30,48 ,63) and different method for the external field.

As there is no true synchronous particle in the time code, a particle that remains on-axis transversely, receives only the accelerating force, and is not affected by space charge is added for diagnostics. Here the phase of that particle is plotted as function of the cell number. The black curve is the design synchronous phase, the red curve comes from a run using the multipole expansion method for representing the electric field and the blue curve is the one from external field of the Poisson solver. The behavior of the MP-phase (multipole expansion method) compares very well to the design phase. This is due to the usage of the multipole expansion method in the design procedure. So the synchronous particle is excited to the same external field, ideally, with which the RFQ was designed.

When the Poisson solver is used to calculate the external field the behavior of the synchronous particle changes. For small apertures (*aperfac*=63) the difference is small and increases with decreasing apertures. For big apertures the phase of the synchronous particle oscillates around the design phase in the beginning of the structure. With increasing phase this oscillation is damped. This graph clarifies that the dynamics of the particles depend on the chosen method for the external field.

Secondly, an arbitrary particle was chosen and the two different approaches were set up to determine the longitudinal kicks (change in momenta) of that particle along the RFQ, but only one method was driving the particle, to make sure that the positions for the two setups remain the same. The transverse kicks along the complete structure are shown in Fig. 17.41.

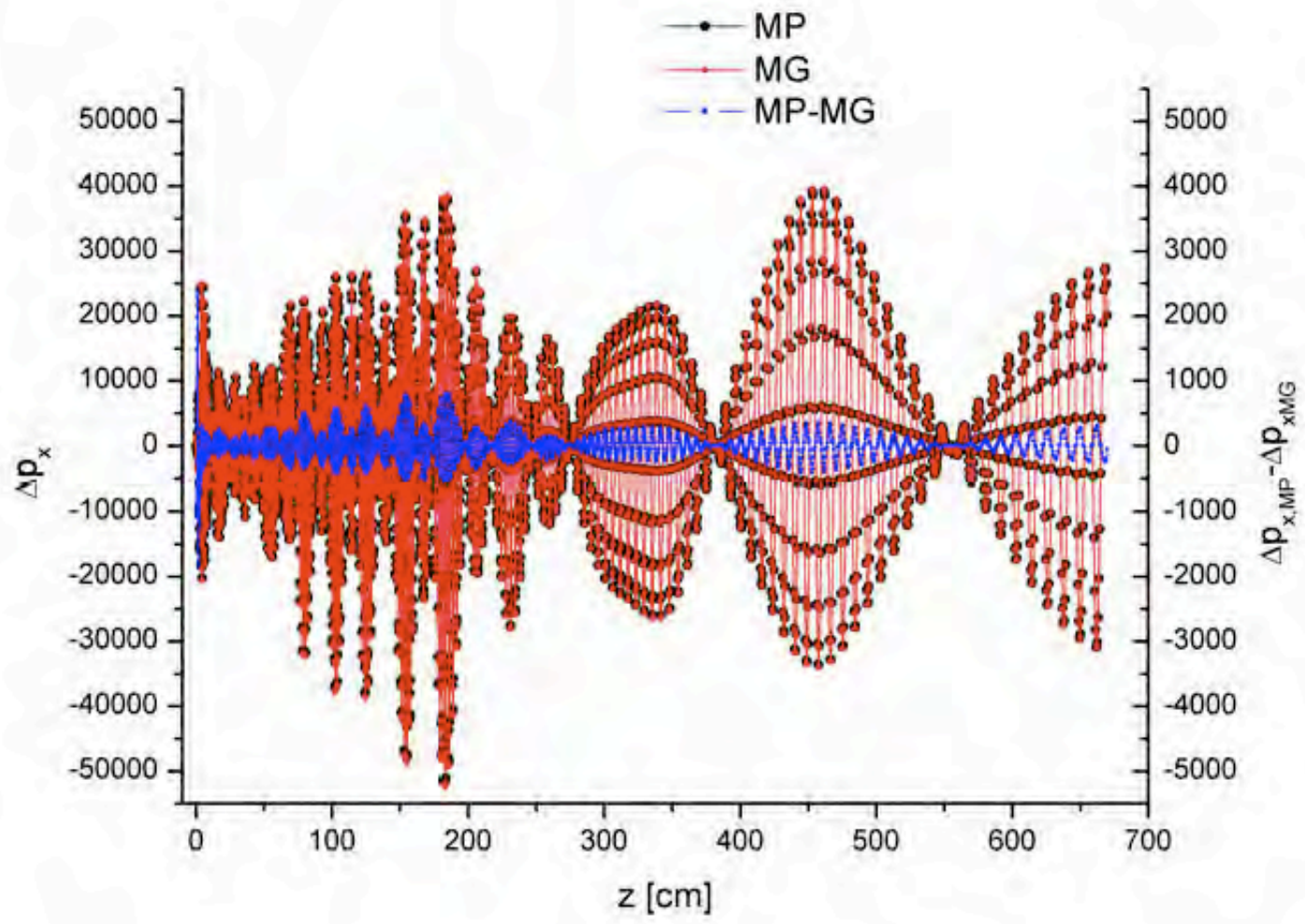

Fig. 17.41 Transverse kicks from multipole expansion method (MP) and multigrid Poisson solver (MG) on a certain particle along the RFQ, and the difference between MP and MG .

The blue curve indicates the difference of the kicks from multipole expansion method and Poisson solver. The oscillation of the RF (high frequency) as well as the betatron oscillation of the single particle through the bunch (low frequency oscillation) can be seen. The betatron oscillation frequency decreases along the RF due to the energy gain of the particle. In general the amplitude of the MP-kicks is bigger than the MG-kicks, but the difference is not very big in the main part of the RFQ. In the beginning of the RFQ the difference it a little bigger than in the rest of the structure and hence the transverse kicks for the first 20 cm are shown in Fig. 17.42.

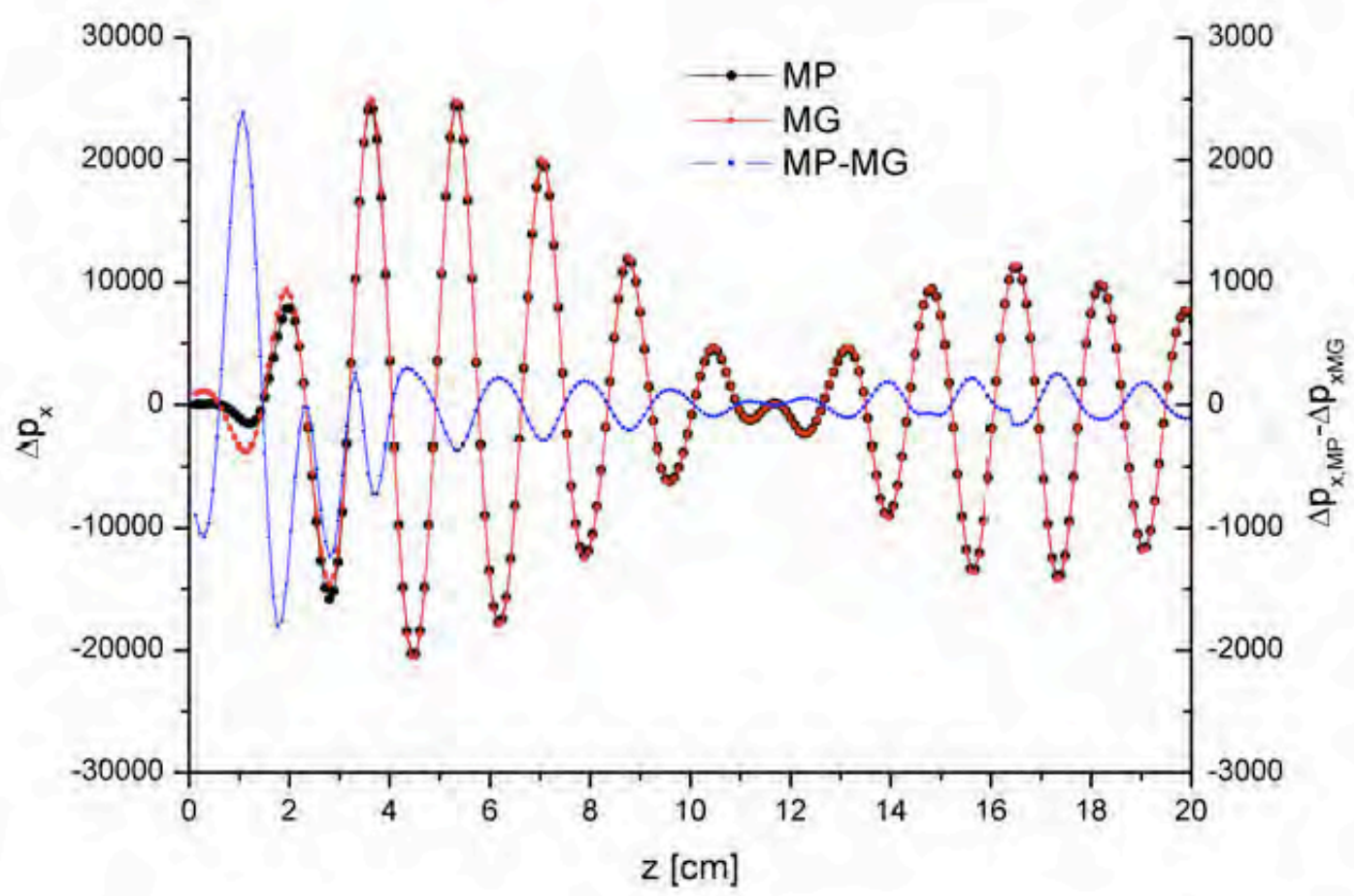

Fig. 17.42 Transverse kicks from multipole expansion method and multigrid Poisson solver on a certain particle along the beginning of the RFQ.

The multipole expansion method does not have an accurate representation of the fringe field region and it starts in our case directly with the first cell of the radial matching section, whereas the multigrid Poisson solver starts with a grounded plate (representing the tank wall) followed by an empty cell and then the electrodes start. This gives a more realistic picture of the real machine.

For further studies a more precise map of this region can be drawn with bore hole and a thick tank wall. Thus, the rise of the field is different from MP to MG. The MP-kicks start with zero, since there is no fringe field effect taken into account, but the multigrid Poisson solver assigns some non-zero kicks to the particle. In the radial matching section the aperture changes quite rapidly and hence the multipole coefficients need to be changed as well, which does not happen. The Poisson solver on the other hand finds the potential introduced by the actual shape of the vanes giving a more precise description of a real machine.

In Fig. 17.43 the evolution of the longitudinal kicks are shown for the same particle. As for the transverse kicks, two different types of oscillations take place. One is the oscillation of the RF and the other is the synchrotron oscillation of the particle passing longitudinally through the bunch. In the beginning of the structure the kicks oscillate about zero and the energy of the particle remains unchanged (only some bunching takes place). After some distance inside the RFQ the longitudinal kicks oscillate about a positive value and the particle gets accelerated. The shape of the curves from the multigrid Poisson solver and from the multipole expansion method are again quite similar in the main part of the RFQ.

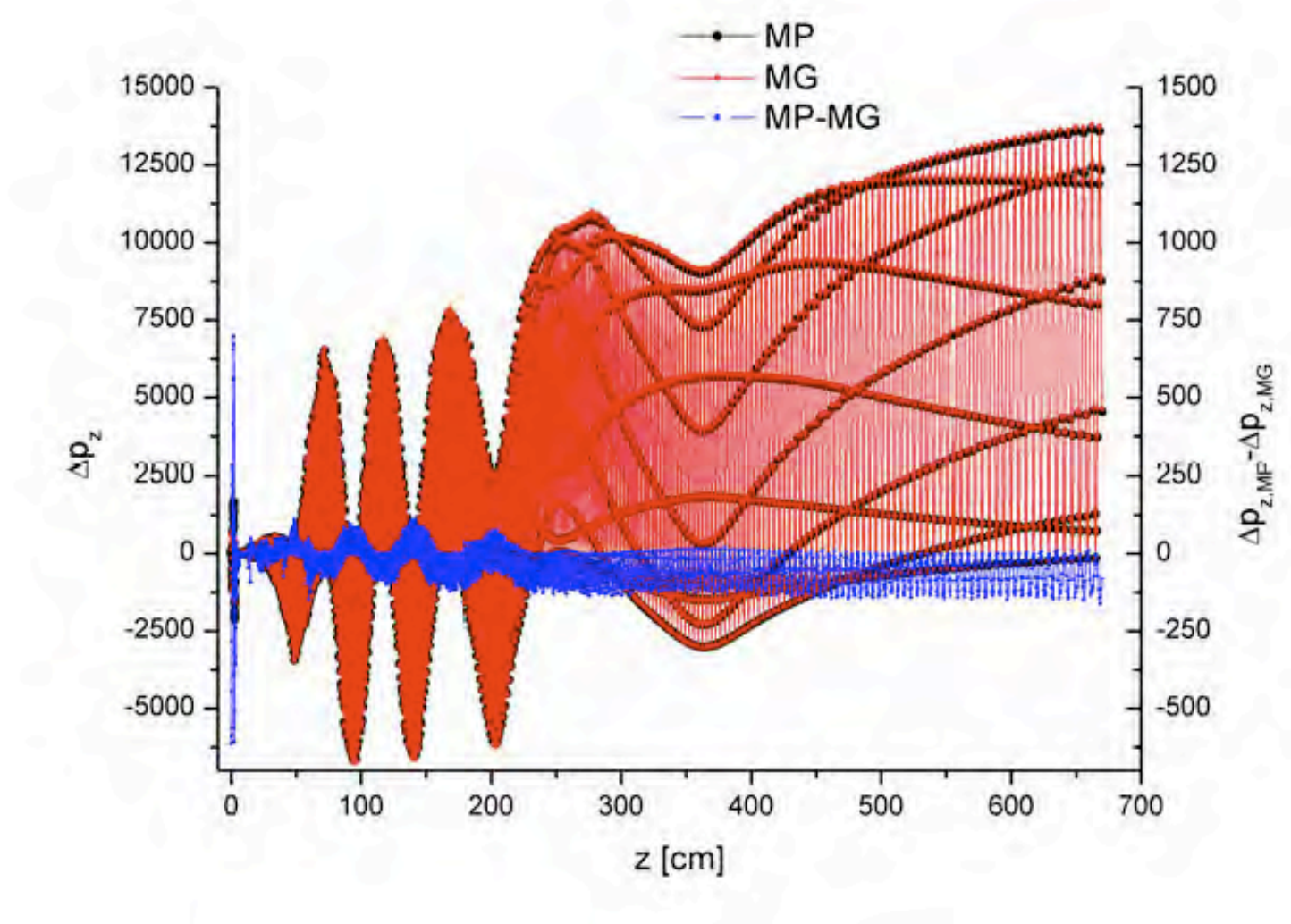

Fig. 17.43 Longitudinal kicks from multipole expansion method and multigrid Poisson solver on a certain particle along the RFQ.

In the beginning of the structure they differ the most (Fig. 17.44). After the radial matching section the longitudinal kicks are very small and they increase slowly to form the bunch. In the radial matching section however the longitudinal electric field is not equal to zero, because the quadrupole geometry is perturbated by the change of the aperture. The multipole expansion method finds here a different field than the multigrid Poisson solver which takes the actual shape of the vanes into account.

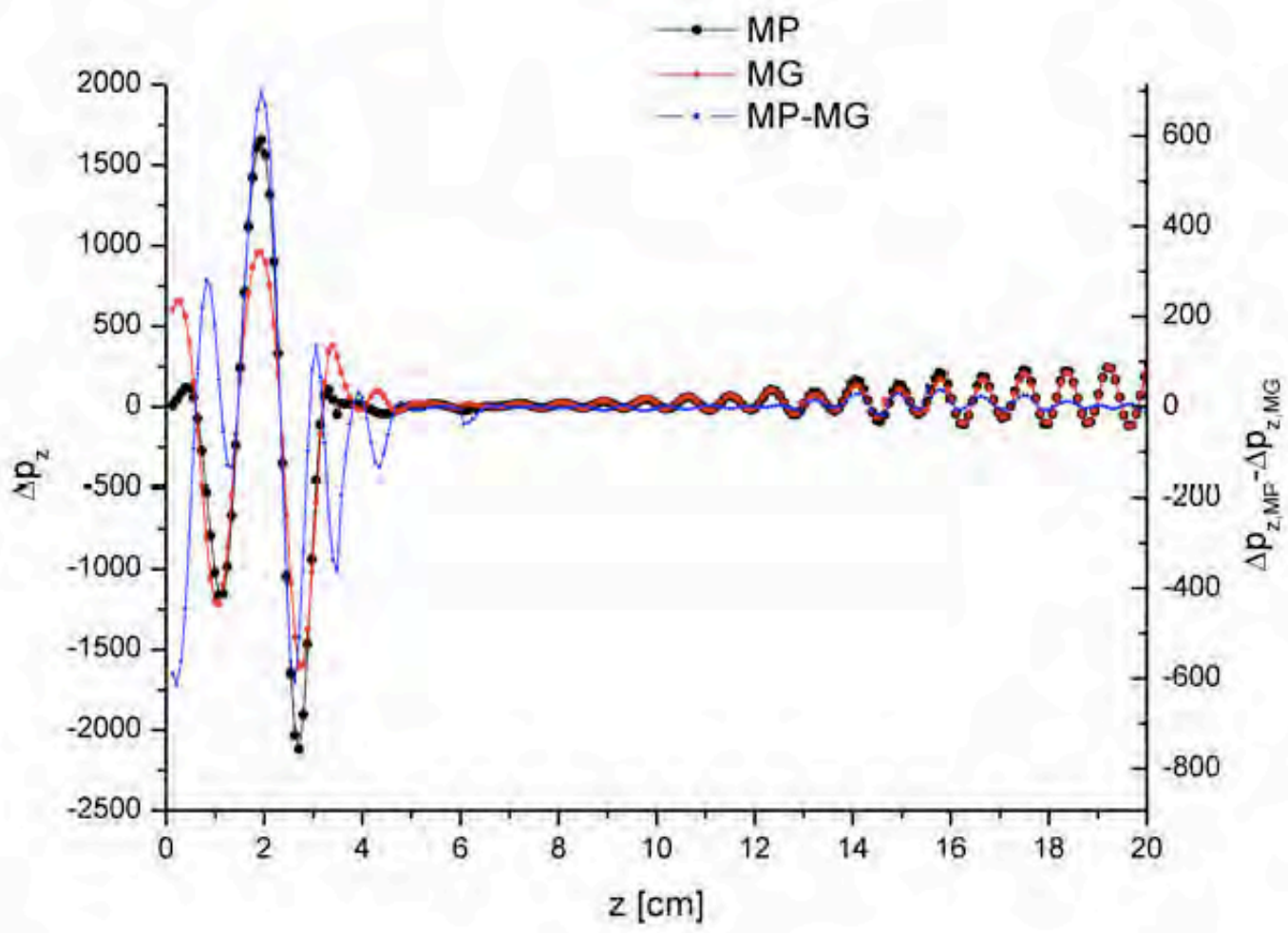

Fig. 17.44 Longitudinal kicks from multipole expansion method and multigrid Poisson solver on a certain particle along the beginning of the RFQ.

*17.5.1.5 Collective Effects*

The influence of the different descriptions of the external field on the transmission are discussed in this section. The runs used here were both done with the same space charge routine (SCHEFFTM) and different options for the external field. A zero current run contains little information due to the very high transmission for all the RFQs. Fig. 17.45 shows the transmission and the fraction of accelerated particles as a function of the aperture factor.

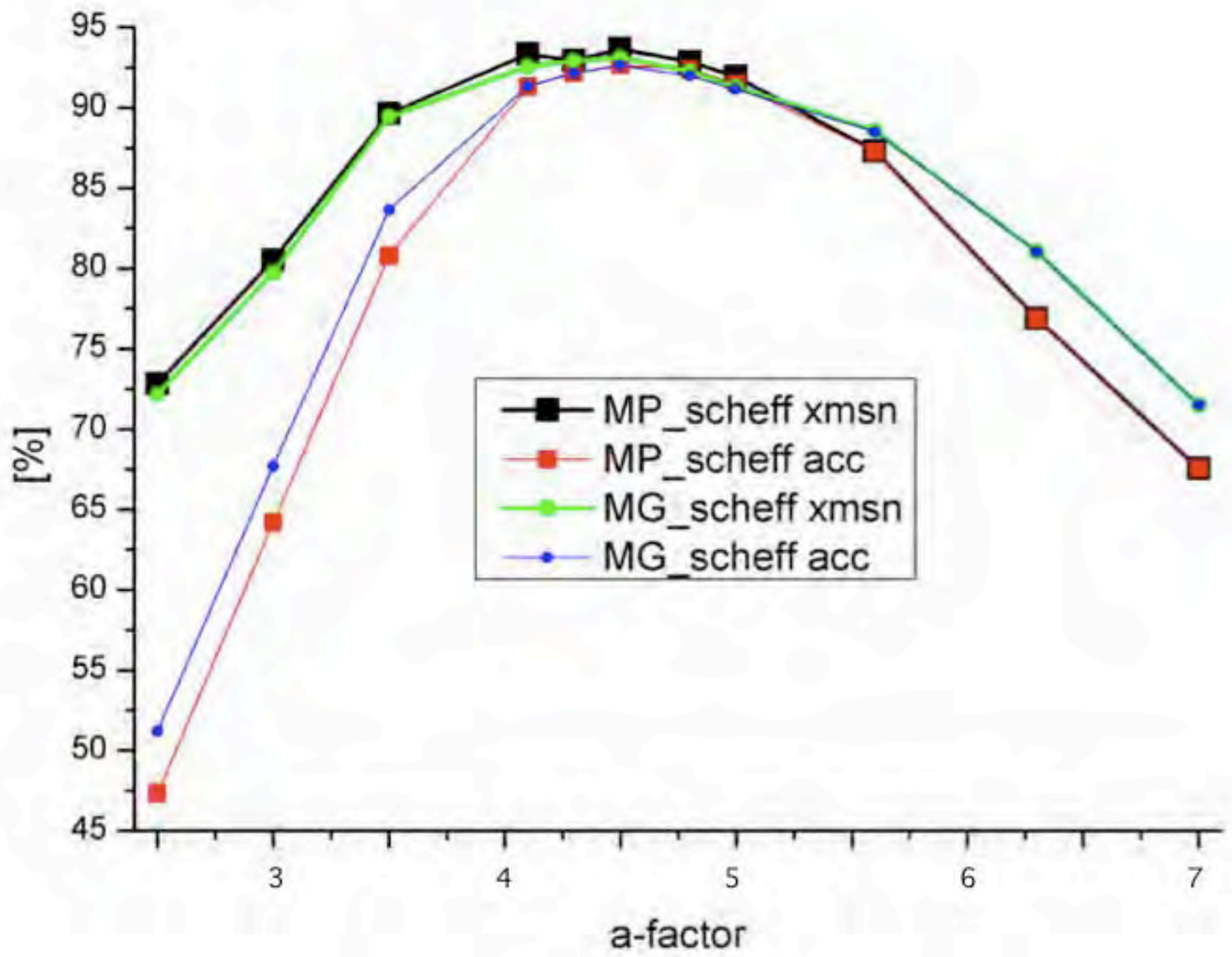

Fig. 17.45 Transmission and percentage of accelerated beam for different external fields.

The a-factor (= aperture factor) indicates the minimum aperture (βλ/aperfac; Sec. 5.3.1) at the end of the shaper section. On the left hand side the aperture of the RFQs is big and becomes smaller with increasing *aperfac*. The transmission curves for the multipole expansion method and for the multigrid Poisson solver are very close for the RFQs with a medium and big aperture. The optimum aperture remains the same. Once the aperture has become small enough, the results from the two different methods start to diverge. The multigrid Poisson solver gives higher results. For the large aperture (aperfac 25, 30,35) RFQs the fraction of accelerated is higher for the MG cases than for the multipole expansion method. This behavior of the transmission curve fits quite well with the considerations so far. For big apertures the multipole expansion method agrees more closely with the Poisson solver and radial losses are not as large an effect. For smaller apertures the effects at the aperture and therefore at the edge of the area of validity of the multipole expansion method become important and the multigrid Poisson solver is a more accurate description of the external field. There is a difference between the multipole expansion method and the multigrid Poisson solver, and a Poisson solver should be used for precise simulations.

### 17.5.2 Internal Field

Having discussed two different methods for the external field, the internal field is now in focus. First, space charge field routines are compared that do not take the influence of image charges on the vanes into account.

The first routine is *schefftm*, in which the charged beam is represented by charged rings and the effect of charged rings on rings can be calculated exactly analytically. The crux of *schefftm* is how well a three dimensional charge density can be mapped onto a two dimensional grid.

The second routine used for comparison is *PICNIC*, a particle-in-cell approach based on numerical calculation of the interaction between cubes.

The last routine is the multigrid Poisson solver. The boundary of the mesh is a conducting cylinder with a radius of twice the maximum aperture (m*a) and with zero potential.

Calculation time is quite bad for *PICNIC* (5 h), it becomes better with the multigrid Poisson solver (30 min), and it is best for *schefftm* (10 min).

Fig. 17.46 shows the effect of the size of the grounded cylinder on the transmission for the multigrid Poisson method.

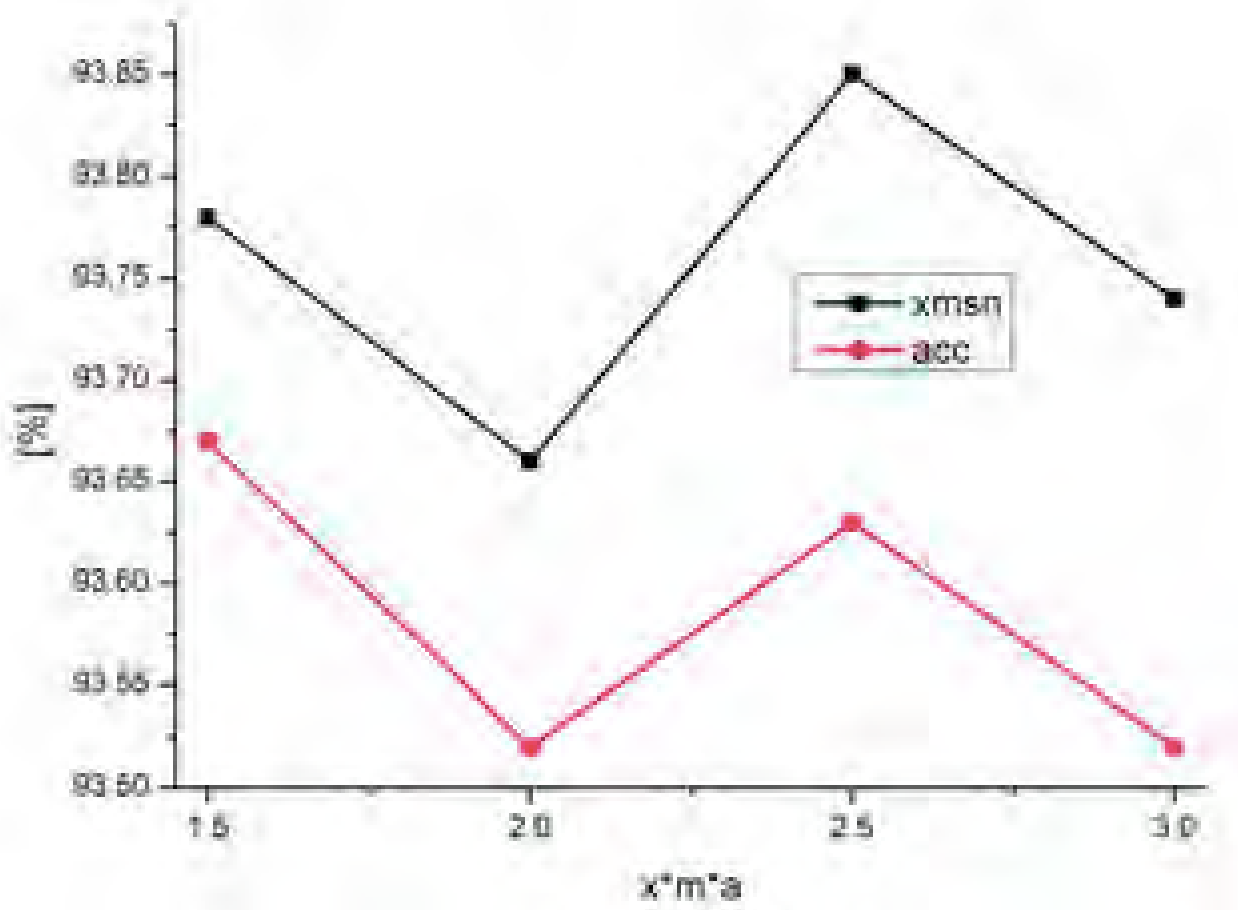

Fig. 17.46 Influence of the radius of the zero potential cylinder for the multigrid Poisson solver as the space charge routine on the transmission.

There is a little influence, within the range of 0.2%, as long as the grounded cylinder radius is > 1.5*(m*a).

In the multigrid Poisson method, it is easy to change only the boundary conditions to have the shape of the vanes in order to take the effect of image charges on the vanes into account.

The *schefftm* and *PICNIC* space charge routines only consider a single bunch of particles. The effect of neighboring bunches is added later by superposition, with the neighboring particles treated as point charges acted on by the centroid of the main bunch represented as a stronger point charge. Up to 20 leading and trailing bunches are taken into account in order to fully represent this effect. The multigrid Poisson solver uses its boundary conditions to take the neighboring bunches fully into account except for the slight effect of changing vane parameters.

In *LIDOS*, the conditions Fi(left)=Fi(right) *and dFi/dz(left)=dFi/dz(right)* are used on each iteration step. So *LIDOS* calculates fields not for a single bunch, but for a periodical sequence of similar bunches. Moreover, *LIDOS* has an opportunity to take into account the violation of periodical conditions during the acceleration, because it can include several bunches in the movable grid, but understanding results in this case demands a manual work with output files.

The *LINACS* MG SC grid also has n+1=1. Can't reflect left and right independently – bunch shape wouldn't be right, and vane shape would abruptly kink. Vane shape is imposed on the mesh, so when boundary (right side=left side) is applied, there will be a little step (down for increasing m) at the right boundary. Bunch centroid is at center of sc mesh.

*17.5.2.1 Influence on Single Particle Dynamics*

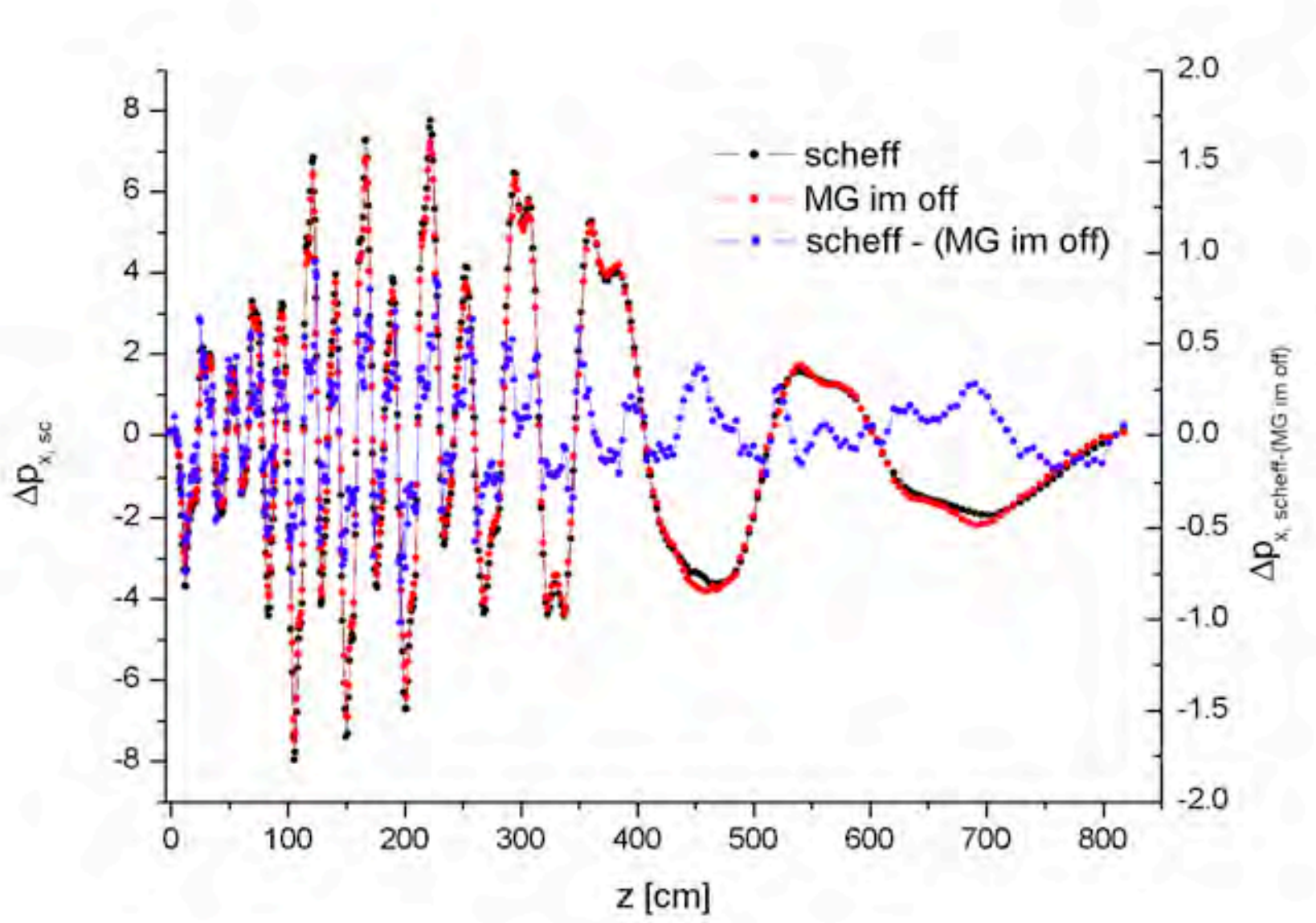

Fig. 17.47 Transverse space charge kicks from *scheff* and the multigrid Poisson solver.

As for the external field, an arbitrary particle was chosen and the space charge kicks on the particle from different space charge routines were logged as it travels through the quadrupole channel. In Figure 17.47 the evolution of the transverse space charge kicks from *scheff* and the multigrid Poisson solver with and without image charges is shown. The particle oscillates with its transverse phase advance through the bunch, the resulting kicks show this kind of oscillation as well. The phase advance is defined per focusing period and the geometric length of this focusing period increases with the energy of the particle. Hence the oscillation seems to be damped when plotting against the longitudinal z position. The shape of the curves from *scheff* and from the Poisson solver are very close, but the extrema are different. In the beginning (shown in Fig. 17.48 in more detail) *scheff* gives somewhat higher kicks than the Poisson solver does. At the end of the structure it is the other way around and the Poisson solver gives higher kicks. The microstructure of the curve, when the particle is at the edge of the bunch, is also very similar in the two curves. For this particle, which does not leave the bunch too far, the effect of images charges is minor.

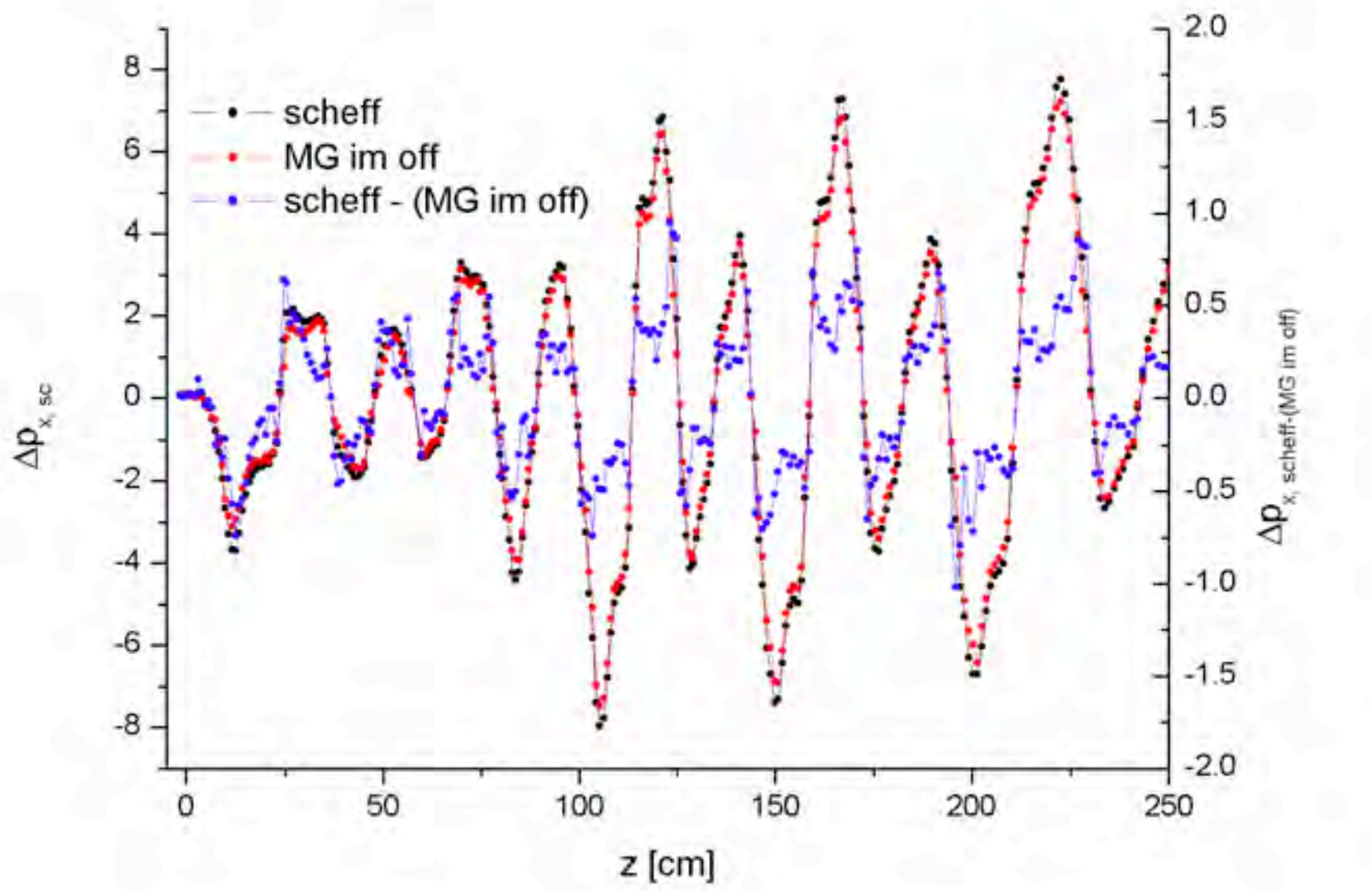

Fig. 17.48  Transverse space charge kicks from *scheff* and multigrid Poisson solver for the first 2:5 m.

The picture for the longitudinal kicks ($\Delta p_z$) is quite similar. The longitudinal kicks are shown in Fig. 17.49. Again, the shape of the curves is quite the same, but the values from *scheff* are higher for most of the structure. In the beginning of the RFQ the effect of space charge on the longitudinal dynamic of the beam is small, because the input beam is DC and longitudinal forces cancel out. This changes when the beam becomes bunched. It has to be noted that behavior for other particles can differ a lot from the particle shown.

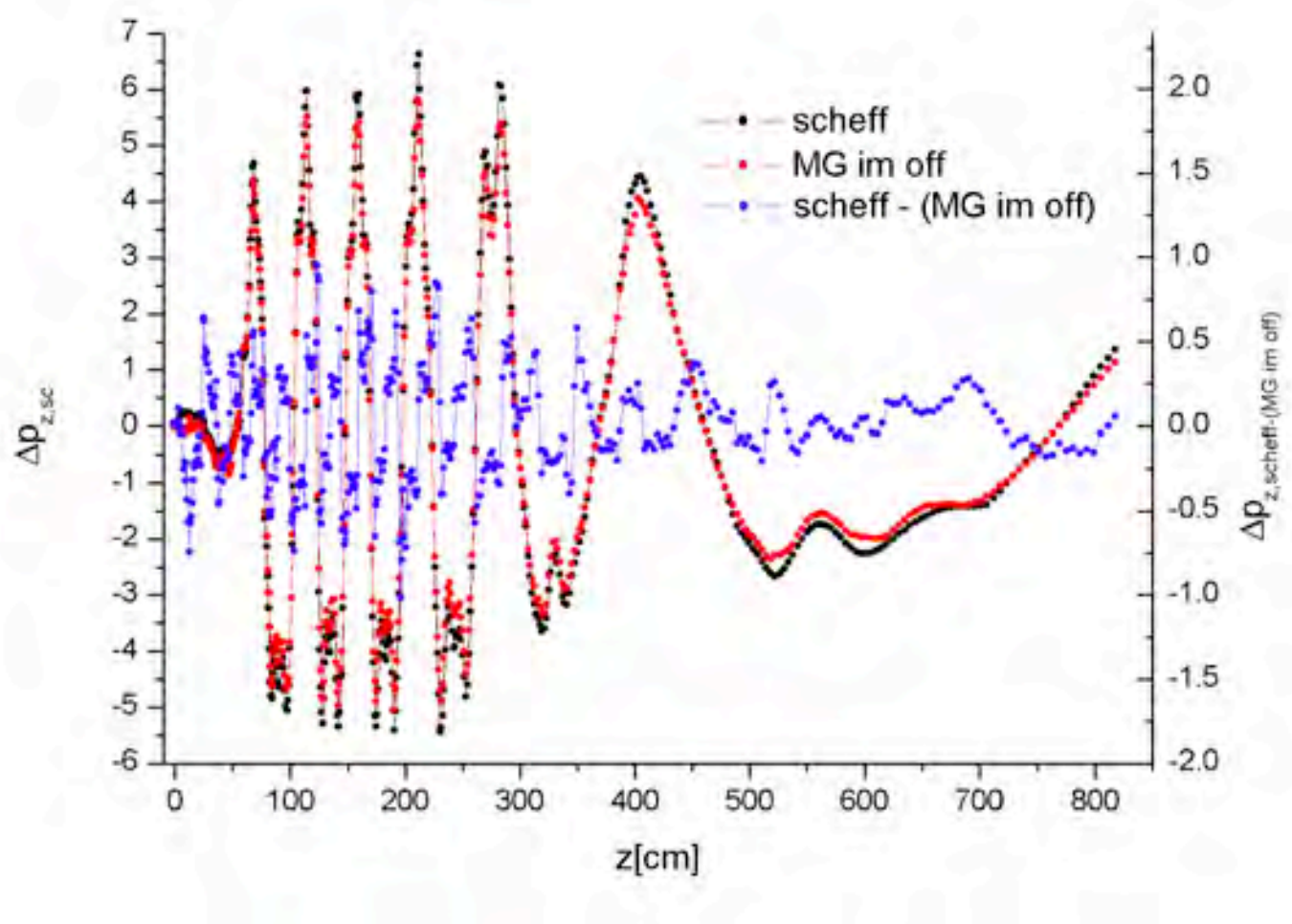

Fig. 17.49  Longitudinal space charge kicks from *scheff* and multigrid Poisson solver

### *17.5.2.2  Collective Effects*

To compare the influence of the routines used to calculate the space charge effect on the transmission, the set of RFQs was simulated using the same MG-routine for the external field.  The settings for the mesh in terms of the resolution of the mesh for *scheff* and for the Poisson solver were chosen equal, and with regard to the discussion above.  For *PICNIC* two different settings were used (one with the longitudinal grid extent set at ±3.5 times the longitudinal rms beam size including smoothing, and with ±5 times excluding smoothing).  The transmission curves for different space charge routines as a function of the aperture-factor are shown in Fig. 17.50.

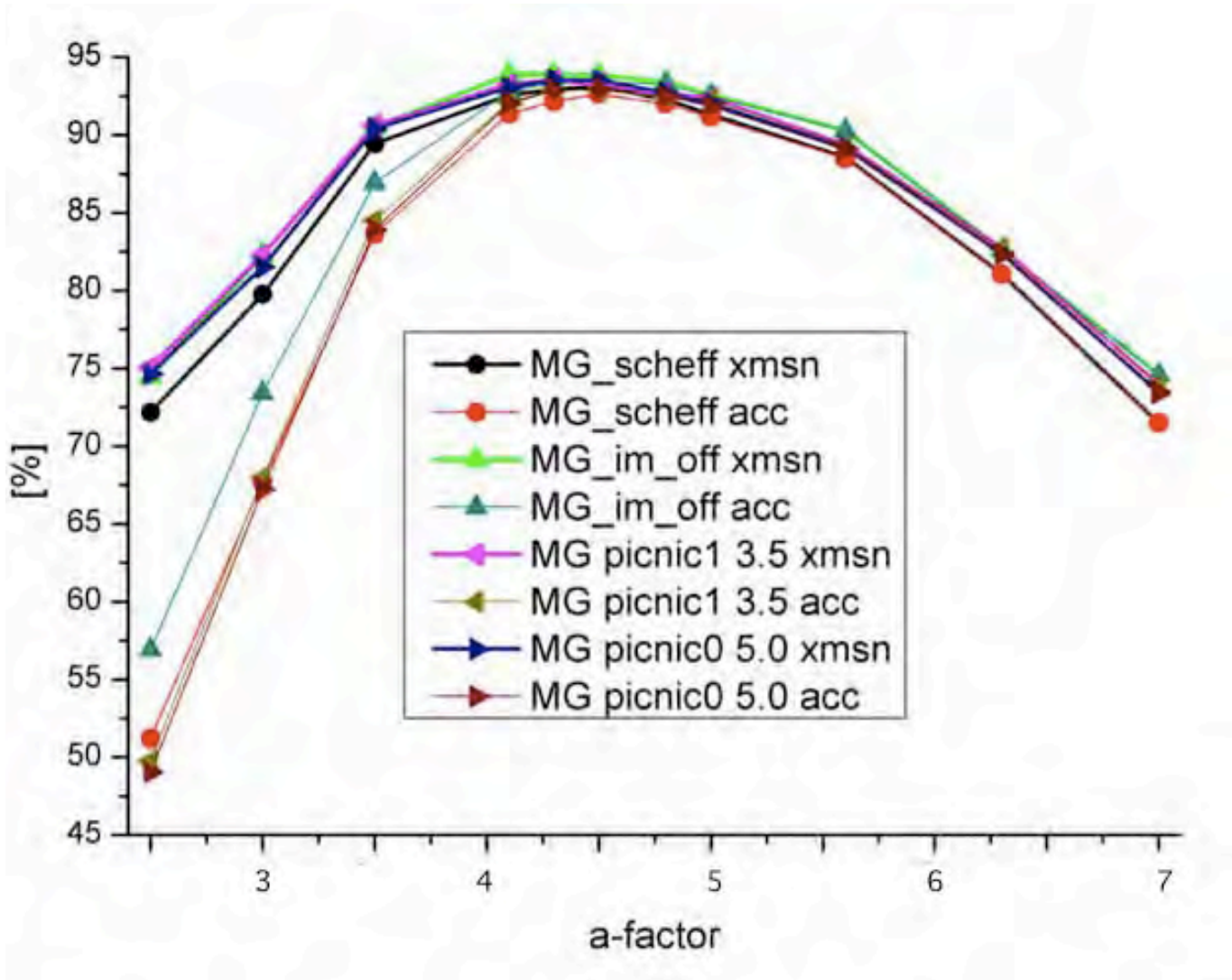

Fig. 17.50 Transmission and percentage of accelerated beam for different routines for the internal field without image charges, in the multigrid Poisson external field.

Roughly, the curves are similar. Over all SCHEFF gives the lowest transmission for every RFQ, but the difference from the other routines become smaller at the optimum aperture. The transmission results for *PICNIC* with both settings and the multigrid Poisson solver are very close; SCHEFF agrees well near the optimum and is ~2.5% lower for both larger and smaller apertures. The multigrid Poisson solver gives the highest transmission results and stays high longer than the other curves leading to a flatter plateau – the optimum is shifted somewhat toward a larger aperture. The percentages of accelerated beam for SCHEFF and PICNIC are very close over whole range, while the Poisson solver gives higher values for RFQs with big apertures.

In the multipole coefficient external field, PICNIC gave slightly higher transmission and nearly the same accelerated fraction as *scheff*; this tendency is repeated using the Poisson external field. Fig. 17.45, using the same space charge method, indicates nearly the same (for larger apertures) or higher transmission (for smaller apertures) for the Poisson external field; the Poisson external field accelerated fraction is higher for large apertures and nearly the same for small apertures. The Poisson external field is more accurate and some difference in the *aperfac* curves is to be expected. These features are fully apparent in Fig. 17.50, and it is concluded that the full Poisson solution for both external and space charge fields is more accurate, giving somewhat higher transmission, and higher accelerated fraction especially for the larger apertures.

*17.5.2.3 Space Charge With Image Effect*

Now the influence of image charges on vanes introduced by the beam will be determined.

The space charge tests with the grounded cylinder at large enough radius (essentially open boundary, and essentially equivalent to the boundary conditions of *scheff* and *PICNIC*, for which no boundary definitions are necessary) are a vitally important part of the development of the Poisson solver, as they prove that the Poisson solver is working correctly. Then to take image charges on the RFQ vanes into account, it is only necessary to impose the vane boundary condition on the same mesh[142]. The Poisson solution then naturally includes the image charge effect, as the images force the potential on the vane surface to equal zero.

The potential with and without image charges in the xz-plane of a certain cell in an RFQ is shown in Fig. 17.51. On the left plot the position of the vane can be seen as the region where the potential is

142 It has been claimed that it is impossible to turn the image effect on and off in an RFQ Poisson solver; however, it is actually very easy, and very important to do so.

equal to zero. The potential on the other plot falls off smoothly toward the outer x grid boundary, which is at 40 on the scale. The shape of the two potentials along the beam axis are quite similar.

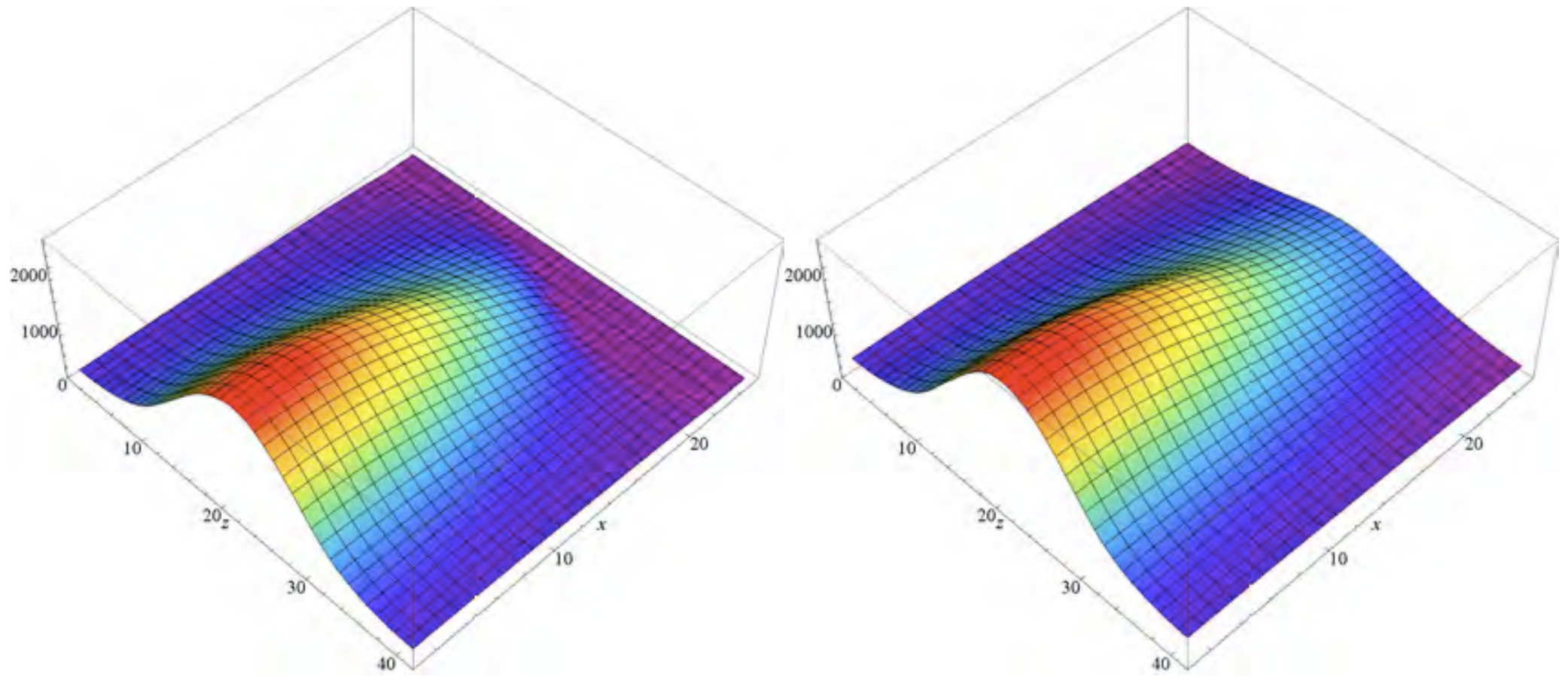

(a) with image charges (b) without image charges

Fig. 17.51 Space charge potential at a xz-plane with and without image charges.

The x-component of the field is shown in Fig. 17.52. On the axis the transverse electric field vanishes as expected. Without the image effect (right picture), at the bunch center (at the middle of the z-axis), $E_x$ increases to its maximum and than falls off, whereas with image charges, $E_x$ increases further after a small plateau to maximum in front of the vane. Note that the maximum field from the image effect is larger than the maximum space charge field. The picture indicates that the effect of image charges on the vanes can be quite strong for particles leaving the core.

The field in front of the vane is not very smooth and some steps are present. This is due to the method of shifting grid points at the boundary. Shifted grid points are not shifted for plotting and the effect is therefore smaller than it appears, so the roughness there does not have any influence on the beam dynamics, furthermore particle entering the electrodes are considered lost.

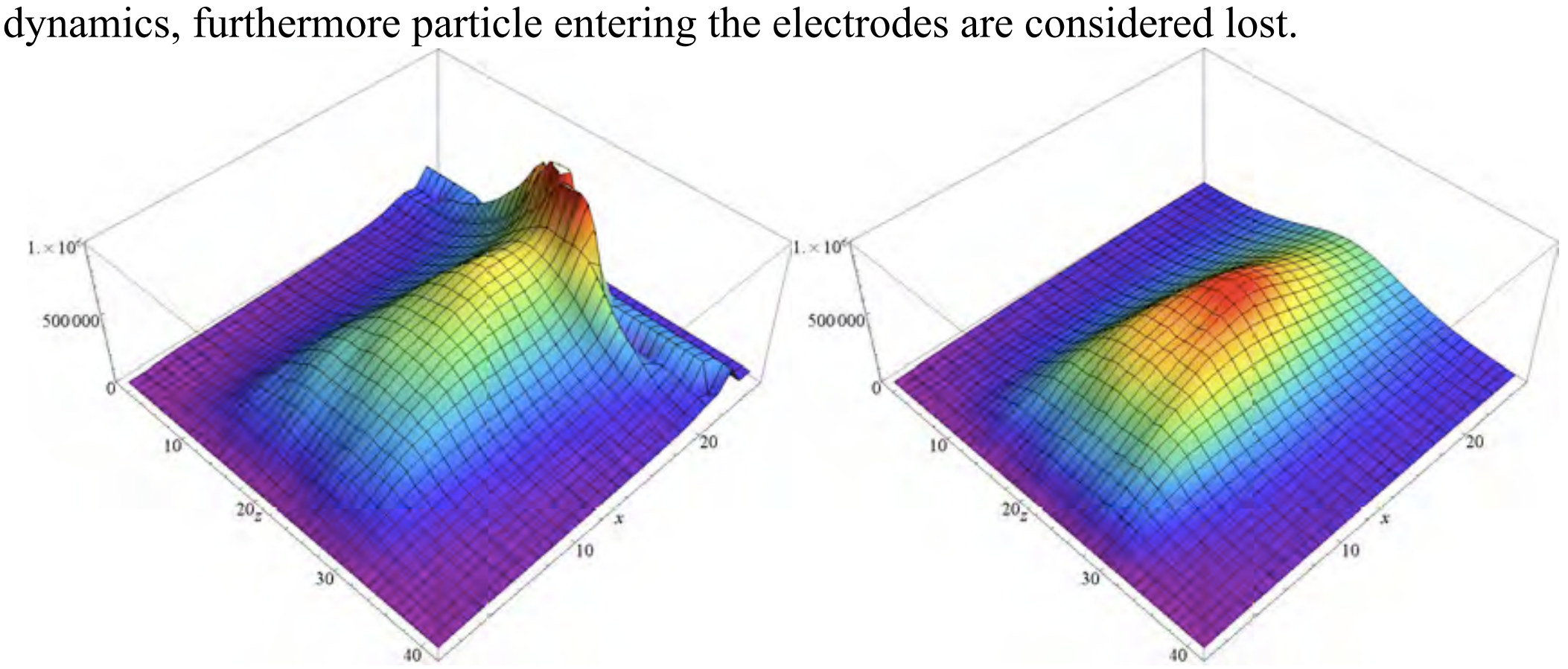

(a) with image charges (b) without image charges

Fig. 17.52 Transverse field $E_x$ from space charge at a xz-plane with and without image charges.

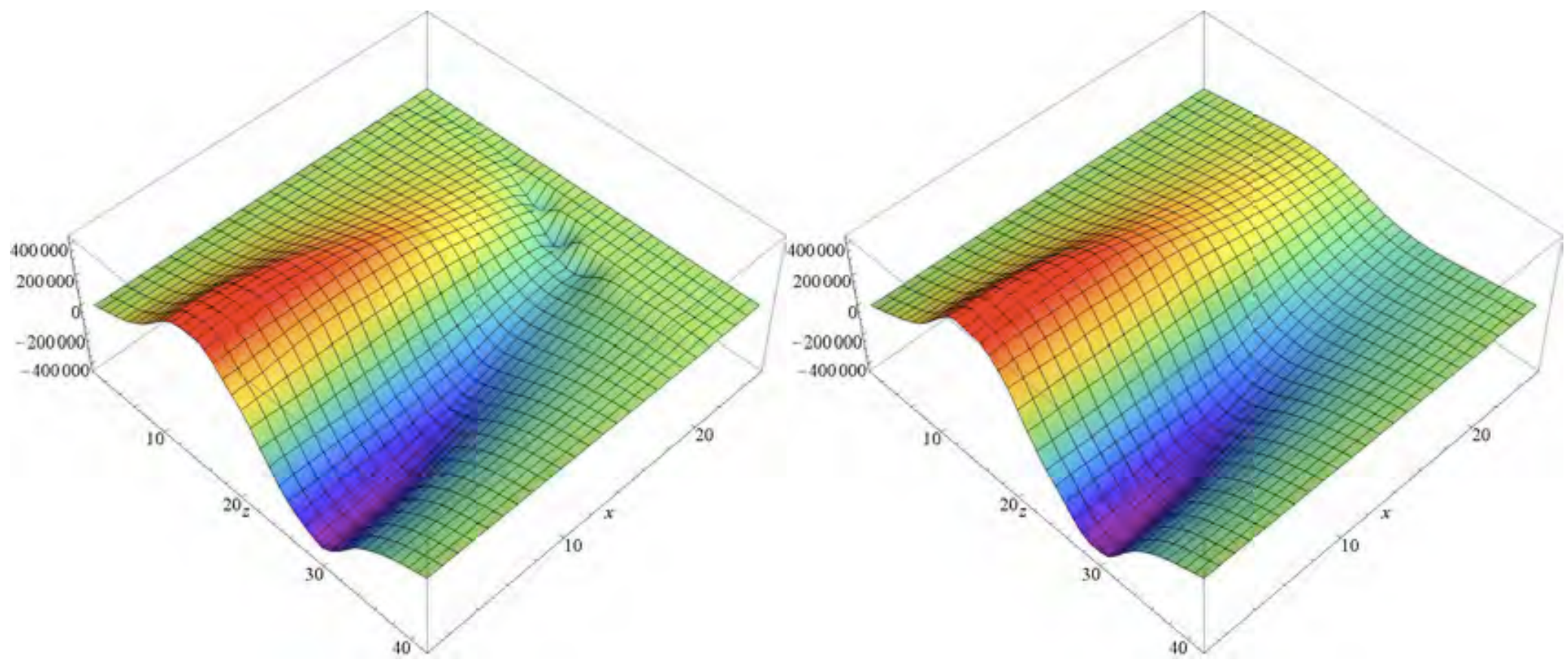

(a) with image charges (b) without image charges

Fig. 17.53 Longitudinal field Ez from space charge at a xz-plane with and without image charges.

In Fig. 17.53 the corresponding longitudinal field Ez is shown. Again, the transition from vacuum to the vane looks more rough and stepped than it actually is, because the plotting routine does not take the shift of the grid points inside the vanes into account.

The relative difference between the longitudinal field Ez with and without image charges on the vanes is shown in Fig. 17.54. (At the center of the bunch Ez is zero and the relative values can become quite big.) On the axis, (IMON-IMOFF)/IMOFF is ~(-10)% negative on both sides of the bunch centroid, indicating that the image effect is focusing. In front of the vanes the relative difference is large.

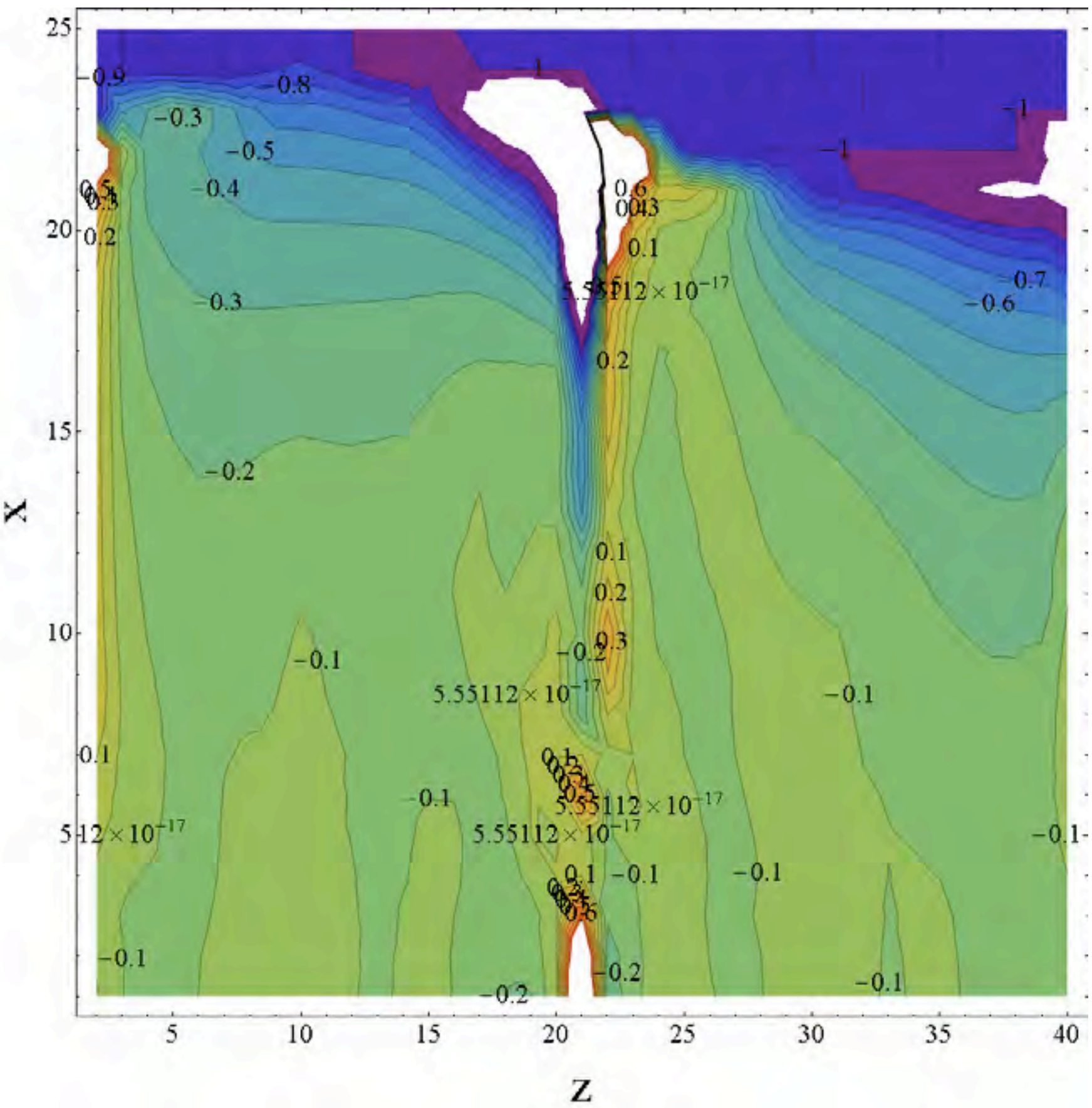

Fig. 17.54 Relative difference (IMON-IMOFF)/IMOFF of the longitudinal field Ez with and without image charges in the xz-plane.

The influence of image charges in terms of transmission and fraction of accelerated particles for an *aperfac* set of RFQs is shown in Fig. 17.55. The ochre curves are a result using the full multigrid Poisson solver without image charges (grounded cylinder with a radius of twice the maximum aperture). The same settings are used for the runs with image charges (red), with the boundary conditions of the space charge Poisson solver changed from the cylinder to the vane shape.

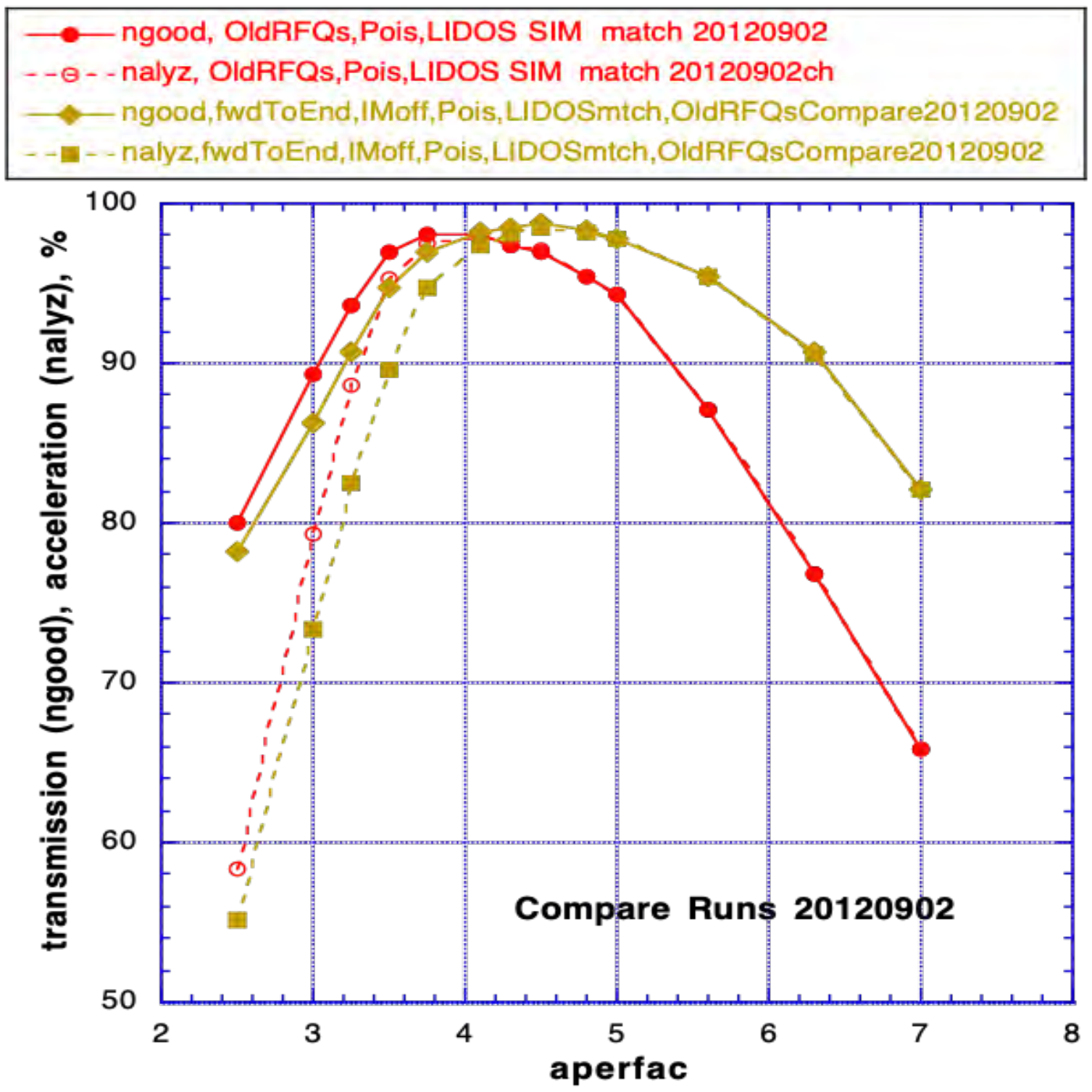

Fig. 17.55 Transmission and percentage of accelerated beam illustrating the effect of image charges on the vanes.

The optimum with image charges is shifted. There is enhancement for larger apertures, and a strong negative effect for smaller apertures.

To motivate why the effect is stronger for smaller apertures, the number of particles are plotted in Fig. 17.56 as a function of their minimum distance to the vanes along the structure.

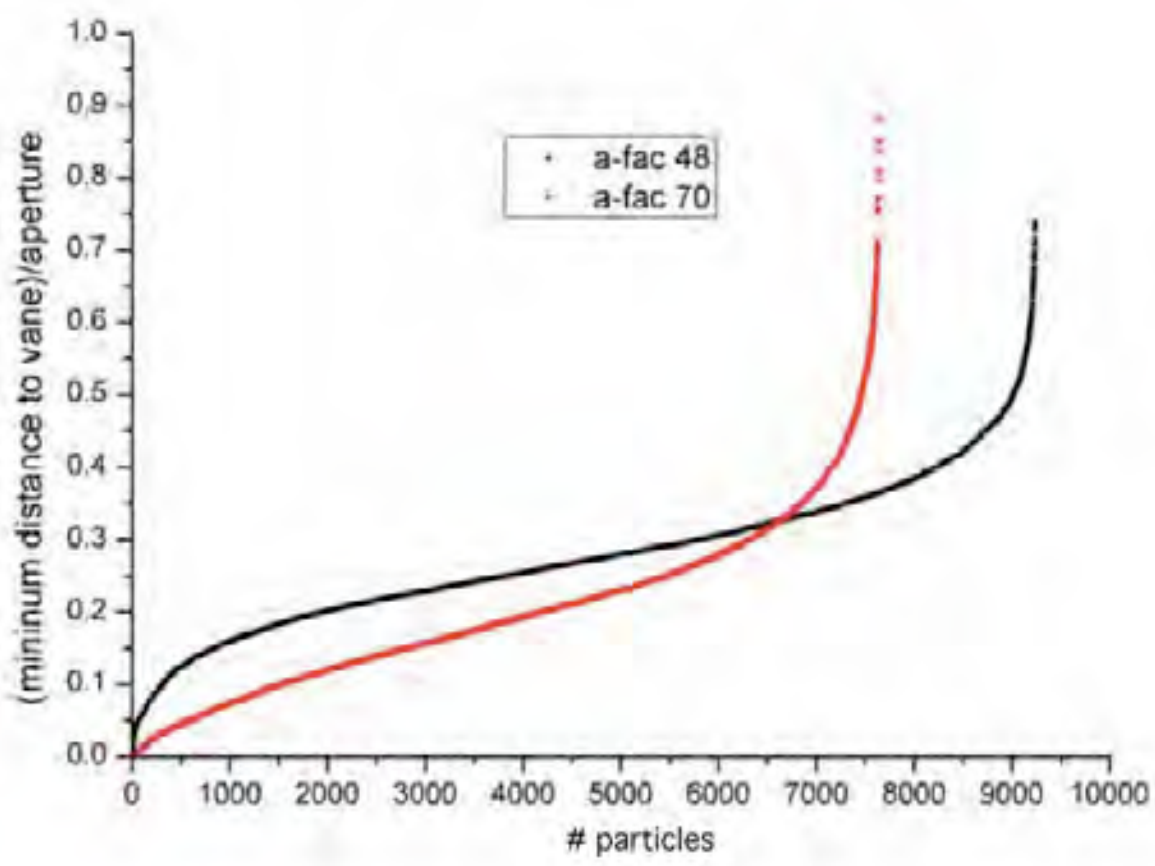

Fig. 17.56 Number of particles as a function of their minimum distance to the vanes along the structure.

In the RFQ with a medium aperture (black curve) only a few percent of the transmitted particles come closer to the vanes than 10% of the aperture. For an RFQ with a smaller aperture, 15% of the particles come closer to the vanes than 10% of the aperture. The fraction approaching the vane surface to within 20% of the aperture is 20% for the medium aperture (a-fac 48), and more than 40% for the small aperture (a-fac 70). For both RFQs the number of particles staying more than 40% of the aperture away from the vanes is small. So the number of particles entering the region where the effect of image charges has a strong influence on the field increases with smaller apertures. Also the image charge effect is stronger in those cases, since the distance between the beam introducing the image charges and the surface of the vanes is small. Because of the opposite charge of the image charges, the image charge effect is always defocusing in the transverse plane, but has a longitudinal focusing effect.

Summing up, a small aperture will increase the strength of the image charge effect compared to a big aperture in two ways and correct image charge computation is important. For larger apertures, the longitudinal image effect can enhance the transmission and accelerated beam fraction

#### *17.5.2.4 Discussion of Approximate Image Effect Models*

##### *17.5.2.4.1 KRC Image Method*

Conclusions regarding the approximate KRC method are given above. The beam is cylindrical at the beginning of the RFQ, and an ellipsoid when bunched. The method is to represent the beam as a fraction that is cylindrical and can be represented by a line charge on the axis. The remaining fraction is assumed spherical, and represented as a point charge on the axis, and image charges on the vanes are calculated for the line charge and for the point charge when it is at the beginning and at the center of each cell. The vane boundary is represented by the general solution of Laplace's equation in cylindrical coordinates, as in the multipole coefficient representation of the external field, as restricted by the symmetries imposed on the beam and on the vane surfaces. The qualitative effect on the *aperfac* curves is a significantly higher transmission for the larger apertures, changing to significantly lower transmission for the smaller apertures (Fig. 16.11). It is seen that the clever KRC method is actually quite good.

##### *17.5.2.4.2 Very Simple Point Model*

A very simple model based on Fig. 16.12 was programmed, in two versions, either by representing the bunch centroid and its images in each vane with the full charge and singly-charged test particles, or with every particle seeing its own image in each vane. The resulting forces on the test particles agreed qualitatively with Fig. 16.13 and the qualitative effect on the *aperfac* curves was the same as for the KRC method, Fig. 16.11, but smaller. A serious shortcoming of this model is that the multiple reflections of images on images are not included.

##### *17.5.2.4.3 BEAMPATH Model - Conducting Cylinder*

The *BEAMPATH* method for space charge and images is discussed in Chapter 14 – the boundary condition is a conducting cylinder with radius equal to the minimum aperture of the RFQ cell. This boundary could be changed to an effectively open boundary by increasing the radius to twice or more times the minimum aperture (there are also formal methods for doing this which can effectively move the boundary to infinity [143]). Fig. 17.57 shows the aperfac curves using the *pteqHI* multipole coefficient method for the external field, with 1) *scheff* alone for no image effect (red), 2) *scheff* plus the KRC image effect method (brown), 3) *BEAMPATH* space charge alone using a boundary of twice the transverse minimum aperture radius (dark green – 2*XAPER), and 4) *BEAMPATH* space charge plus images using the conducting boundary at the minimum aperture radius (light green - XAPER). The KRC method and the *BEAMPATH* conducting cylindrical boundaries give nearly the same result.; i.e., the KRC method boundary condition is equivalent to the conducting cylinder.

---

[143] R. Ryne, private communication.

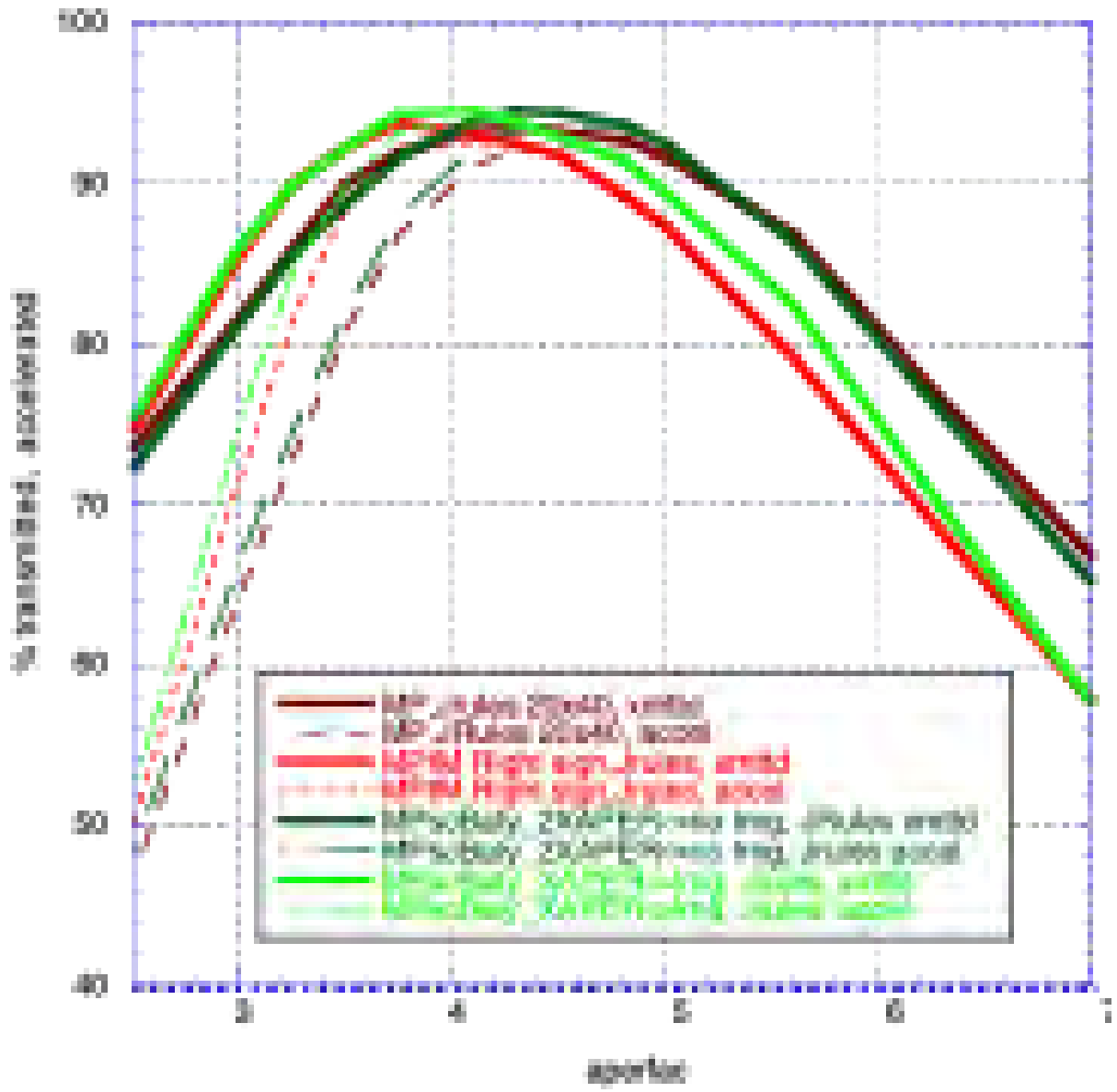

Fig. 17.57 Comparison of *pteqHI* KRC and *BEAMPATH* space charge and image methods.

A good test of the full multigrid method is then to check it with the same *BEAMPATH* boundaries. Fig. 17.58 shows the same qualitative characteristic.

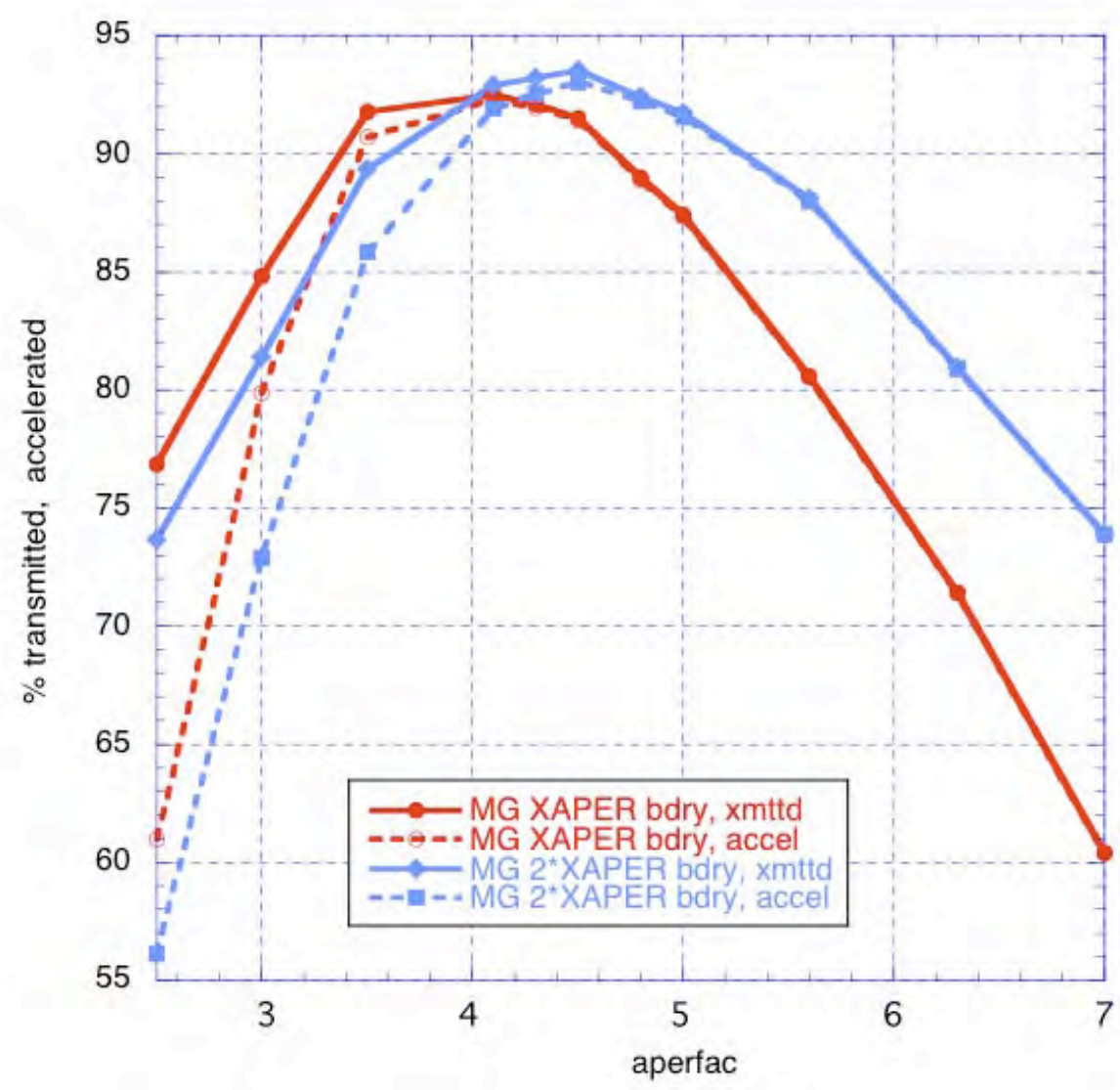

Fig. 17.58 Multigrid result using *BEAMPATH* conducting cylindrical pipe boundary at the minimum aperture ( red - XAPER) compared to boundary at twice the minimum aperture (light blue - 2*XAPER)

### *17.5.2.4.4 Point charge Exterior to Conducting Cylinder*

Others have noted that previous RFQ code predictions are often higher than experimental results, and have sought to model the image charge effect using an approximate method [144]. Space charge was computed accurately using 3D point-to-point. Images were then treated approximately by giving every particle a primary image in each vane. The Green's function was computed at a test point for the potential caused by the image of a point charge exterior to an infinitely long conducting cylinder representing an RFQ cell. Changes in the cell geometry were approximated by using several radii steps during the calculation. The electric force on the test particle was found by differentiating and interpolating the potential grid, summing over all particles in the bunch and over the corresponding particles in neighboring bunches, for each of the four vanes. Particles in as many as five leading and trailing bunches could be included. With no image charges, two or three neighboring bunches gave sufficient accuracy. With image charges, it was found that one leading and trailing neighboring bunch was sufficient, attributed to the fact that the effect of neighboring bunches is reduced by image charge because the total charge is zero when integrated over all the actual particles and their images.

[144] "Three-Dimensional Space Charge and Image charge Effects in Radio-Frequency-Quadrupole Accelerators", F.W. Guy, PAC1991, pp. 3032-3034.

Simulation results were obtained for aligned beams in three actual RFQs and two designs. One actual RFQ has very large aperture until just at the end, where there is a sharp restriction. The simulation showed essentially no image effect. The other two actual ~100mA proton RFQs showed reduced transmission by 3.7% and 12.1%, and the simulation with images agreed much better with the experimental results. The design cases with images showed reductions with images of 2.3% and 13.9%.

The simulation input files for several of these cases were also available for the work. Considerably more analysis would be required to make a conclusive comparison of these anecdotes to the *aperfac* family. The *aperfac* family is a controlled experiment with aperfac as a controlled parameter, whereas the anecdotes would have to be analyzed including comparison of actual focusing strength to beam size ratios, etc.

The observed results of the referenced paper are not claimed to be conclusive, but the main result is general significant lowering of transmission and better agreement with experiment.

The boundary condition for this method is also a conducting cylinder, as for the KRC and *BEAMPATH* methods.

*17.5.2.4.5 Discussion of Models Attempted in Russia*

Russian colleagues told that many attempts were made to approximate image charges in Russia, but none was considered adequate. *BEAMPATH* does not include an image charge effect for this reason.

## 17.6 Special Cells

The injection and output sections of the RFQ are also included in the Poisson solution. The input section has an end wall of specified thickness, a beam hole, rounding of the inner wall at the hole, and a drift space before the vanes start. The dc biased “straw” for injecting the plasma from a laser ion source (LIS) may also be included. After the last normally modulated cell, several types of end conditions can be added: a Crandall transition cell, cells with zero modulation, an output radial matching section, a drift space, and an end wall of specified thickness with a beam hole that can be rounded of the inner wall.

Vane surface and open or closed boundary conditions are used. As the “cell number” is a main index, special measures are taken to afford the indexing of the regions with lengths that are not $\beta\lambda/2$, but specified in cm.

## 17.6 Final Note on the Poisson Method

As of this writing (late 2022), *LINACS* is still the only code with correct dynamics using full Poisson solutions for the external and space charge fields in an RFQ, justified by thorough testing and open source code and documentation.

Intricate interaction of mesh size, extent, boundary conditions, number of particles, number of steps, …, plus Poisson features of convergence criteria, acceleration factor – requires extensive testing in more comprehensive framework than just an anecdotal case or two. The *Aperfac* family test is a good start as a working designer’s **framework**.

[eltoc]

# Chapter 18. Investigation of the LIDOS Simulation Code

We were very fortunate to have the executable version of *LIDOS* and the very cooperative support of the *LIDOS* team since the beginning of this study [145]. *LIDOS* uses a Poisson solver for both the external and space charge fields of an RFQ. Working with that data file for external field, we learned about what to expect from a Poisson solver, how to generate multipole coefficients from the grid, and had a checkpoint for *LINACSrfqSIM* development.

The *LIDOS* result for the aperfac family indicated very high transmission, as indicated in Fig. 17.1. As *pteqHI* was indicating lower transmission and a shifted curve, this was strong motivation for the development of Poisson solvers for *pteqHI*. Final comparison of these two codes, with the same physics approach and correct Poisson handling of image charges, was not possible until the *LINACSrfqSIM* development was completed, Chapter 20.

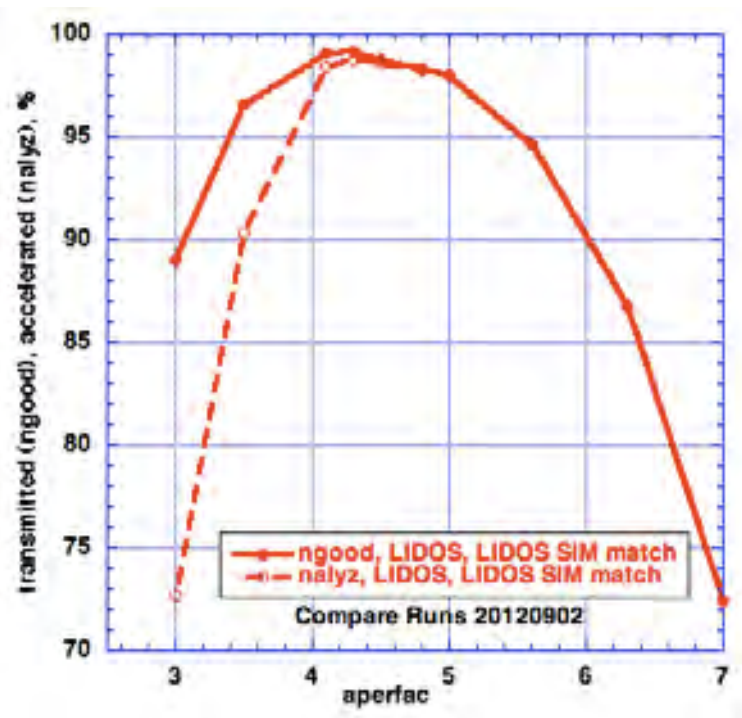

Fig. 18.1. *LIDOS* result for aperfac family.

## 18.1 *LIDOS* Framework

### 18.1.1 Features & Availability

*LIDOS* has features for RFQ design, simulation, and processing of files for vane machining. The simulation feature was used for this report.

Executable and also source code versions of *LIDOS* are available commercially through AccelSoft Inc., P.O. Box 2813, Del Mar, CA, 92014 USA. [146]

### 18.1.2 *LIDOS* Setup

*18.1.2.1 Cell Parameters Input File*

A format is specified for a user-defined RFQ input file; the *LIDOS* input files for the *aperfac* family are written, with sufficient precision, directly from *pteqHI/LINACSrfqSIM*.

The cell parameters of the *aperfac* family were read in, and a "standard" vane geometry as provided by the "Vane Real Shape" feature was used with the same geometry (Rho/r0 circular vane tip, 10° vane breakout angle, sinusoidal modulation) as used in *pteqHI* and *LINACSrfqSIM*.

*18.1.2.2 Lost Beam & Accelerated Criteria*

145 "Code Package for RFQ Designing", B. Bondarev, A. Durkin, Y. Ivanov, I. Shumakov, S. Vinogradov, Proc. 2nd Asian Particle Accelerator Conf.., Beijing, China, 2001.

146 "LIDOS.RFQ.Designer", User's Guide, publ. by AccelSoft, Inc.

Particles are declared lost only if they strike the vane surface. Particles are declared accelerated if their ratio of their velocity to the synchronous particle velocity is greater than a certain value; the default is 0.98, a change of this value would require correspondence with the *LIDOS* team.

*18.1.2.3 Longitudinal Modulation Sine vs 2term*

Either 2-term or sinusoidal longitudinal modulation can be specified. The default through the user interface is 2-term. For sinusoidal modulation, the file "real.prm" in the /RFQ directory must be modified, changing the very first parameter "RealForm" from "c" to "y".

*18.1.2.4 Input and Output Particle Coordinates*

To generate input and output particle coordinate files, the "inj_out.exe" and "inj.dat" files and instructions must be obtained from the *LIDOS* team. Also, if it is desired to start from a given particle coordinate array, file convertors are available from the *LIDOS* team.

### 18.1.3 *LIDOS* Representation of the RFQ

*18.1.3.1 Representation of the RFQ Front End*

Using the "Vane Real Shape" interface, the RFQ front end starts as an end-wall with a specified thickness and entrance bore hole. This is followed by a cylindrical section with specified diameter. The sum of the lengths of the end-wall and wall cylinder must be arranged to be an integral number of period lengths.

An RFQ output section of similar form can also be specified (but was not for this study).

The vane shape is specified using the interface, specifying the Rho/r0 ratio and the vane breakout angle. Execution shows a graphic of the entrance geometry and the vane shape.

*18.1.3.2 Beam Starting Conditions, Input Matching*

Running the generated "Vane Real Shape", the input matching condition is found for a beam starting with its head positioned at the outside of the entrance end-wall. The matching procedure is sophisticated, has a short running time, and produces a better match than has been available in other codes. The ideal emittance parameters of ~360 2° beam slices are computed "at the beginning of the regular RFQ part", and each slice is transported backward to the entrance, where there would be one common ellipse for a perfect radial matching section, but typically there will be some scatter. Then the best fit common ellipse is found. The *LIDOS* match is gives only slightly lower transmission than from a transmission match grid search. Matching is discussed further in Ch.19 and the *LIDOS* match in Ch.19.4.8

## 18.2 *LIDOS* Dynamics Method

The particles are advanced through the external field using the standard leapfrog integrator.

## 18.3 *LIDOS* Representation of the RFQ for Poisson Solution of the Fields

### 18.3.1 External Field

*LIDOS* makes a full 3D grid for the whole RFQ before starting the dynamics. The vane surface is located exactly by "cutting" the mesh parallelepipeds. The outer boundary between the vanes is located on a radius far enough from the axis (default is 1.5 times the maximum aperture of the whole RFQ, i.e., 1.5*(modulation*aperture), or has also been observed to be 4*(minimum aperture at the end of the RFQ)). The potential on this radius is varied linearly between the vanes, which is exact for a wedge.

The specification of the mesh resolution is given using the user interface. There is an interface problem with specification of the x,y mesh increment. The increment entered will actually be doubled in the code, because the mesh is generated across ±x and ±y.

The *LIDOS* team has recommended that the transverse external field mesh should have at least 10 mesh points across the minimum aperture *diameter*. This study, however, indicates that at least 20 mesh points across the minimum aperture *radius* are required.

Similarly, about 20 mesh points per cell length are recommended longitudinally.

The specified mesh resolution is then used for the whole RFQ.

However, because *LIDOS* generates a 3D mesh for the whole RFQ, the mesh file size using the recommended resolution becomes too large for modern laptop or desktop computers, which are limited at a file size of order 2.5GB.

One problem is that modern RFQs may have an increasing aperture for the main acceleration section, for example in the *aperfac* family to ~twice the end-of-shaper aperture. Early RFQs (and many still) have constant aperture in the accelerating section, but this strategy can be subject to resonant interactions causing emittance growth, halo production and beam loss. The growing aperture then requires many mesh points. Also, for enough longitudinal mesh points in the input cells, many mesh points result as the cell length increases.

The consequences of this aspect are investigated below.

#### 18.3.2 Space Charge Field

The space charge has the same transverse dimension as the external field mesh, and longitudinally covers ±1 cell (for periodicity of the Poisson solver), centered on the synchronous particle. The resolution can be specified separately from the external mesh resolution in the user interface; 30x30 is the default, but a finer mesh can be used, as this mesh is not so large and there is no computer restriction.
The outer open boundary between the vanes is an arc located at a radius far enough from the axis to have little influence on the field at the axis. The vane surface and the outer boundary are at zero potential. Space charge is calculated at each step.

#### 18.3.3 Solution of the Poisson Equation

The potentials on the external and space charge grids are found independently, using a single grid Poisson solver with Chebychev acceleration. Typically, of order 300 iterations are required to achieve ~$10^{-4}$ accuracy of the external fields for the aperfac RFQs. The electric fields are found by differentiating the potential grid, and applied to the particles by superposition.

## 18.4 Effect of Mesh Sizes on Simulation Result

The effect of the number of mesh points in the radius of the transverse mesh (nXext/2) on transmission was tested for an RFQ having nearly optimum transmission, Fig. 18.2 red curve. At least 60 mesh points across the full transverse extent are required.

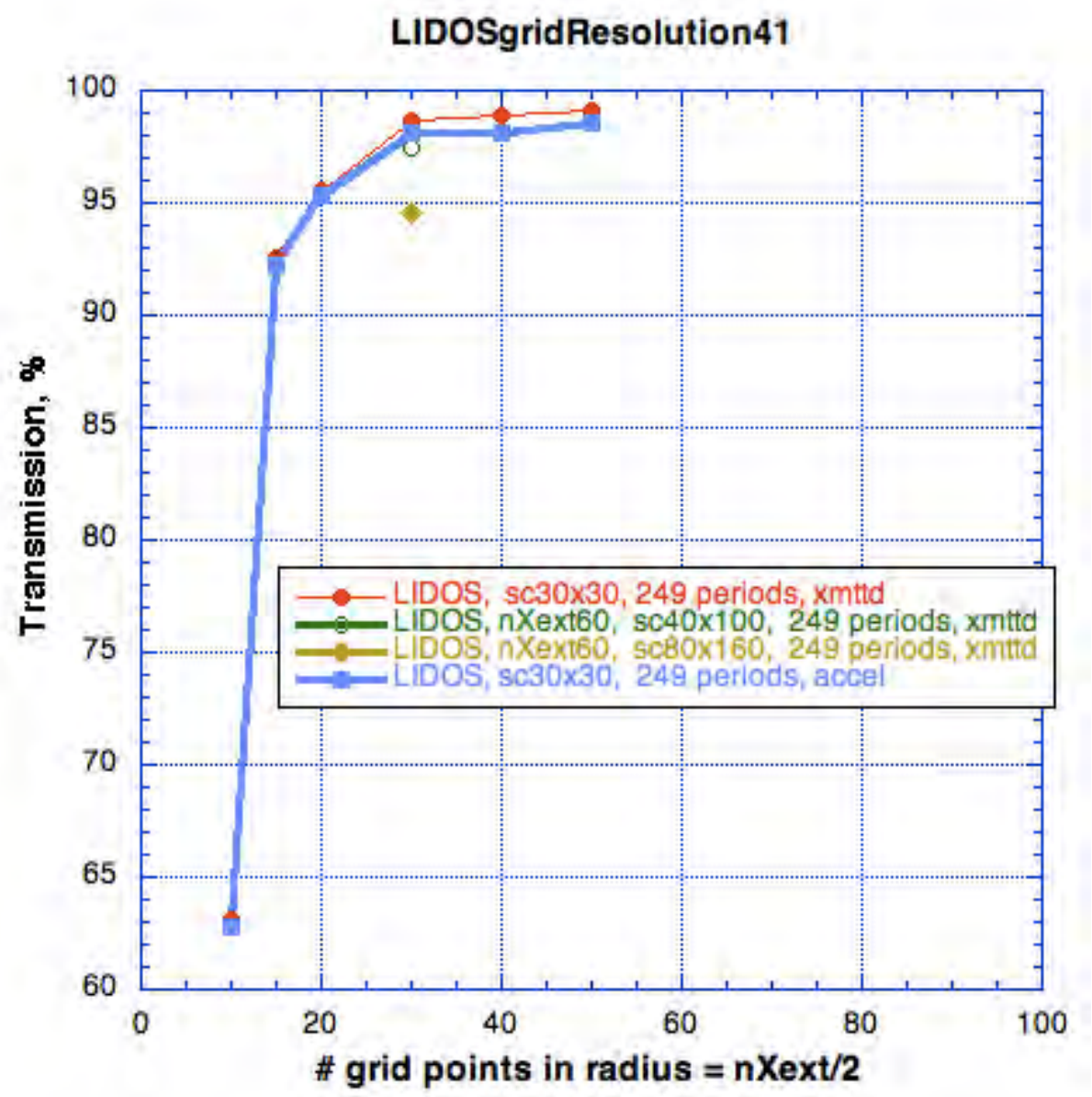

Fig. 18.2. Effect of the number of mesh points in the radius (nXext)/2 on transmission.

Then, keeping nXest=60, the number of space charge mesh points was varied, from 30x30 in transverse and longitudinal to 80x160. Transmission is sensitive, with significant reduction as the number of space charge mesh points is increased.

Fig. 18.3 shows the result for the whole family of RFQs. Using a fast desktop PC, the mesh sizes were pushed to the limit of the ~2.5 GB file size limit, with nXext≥60. With a coarse longitudinal mesh of only ~4-5 mesh increments per cell in the early cells, the maximum transverse resolution at the minimum aperture over the aperfac family was only ~6-7 mesh increments. The number of space charge mesh points was set to 80x160 in transverse and longitudinal. Transmission is sensitive, with significant reduction as the number of space charge mesh points is increased. The two red curves in this figure will be shown to bracket the *LINACSrfqSIM* result in Chapter 20.

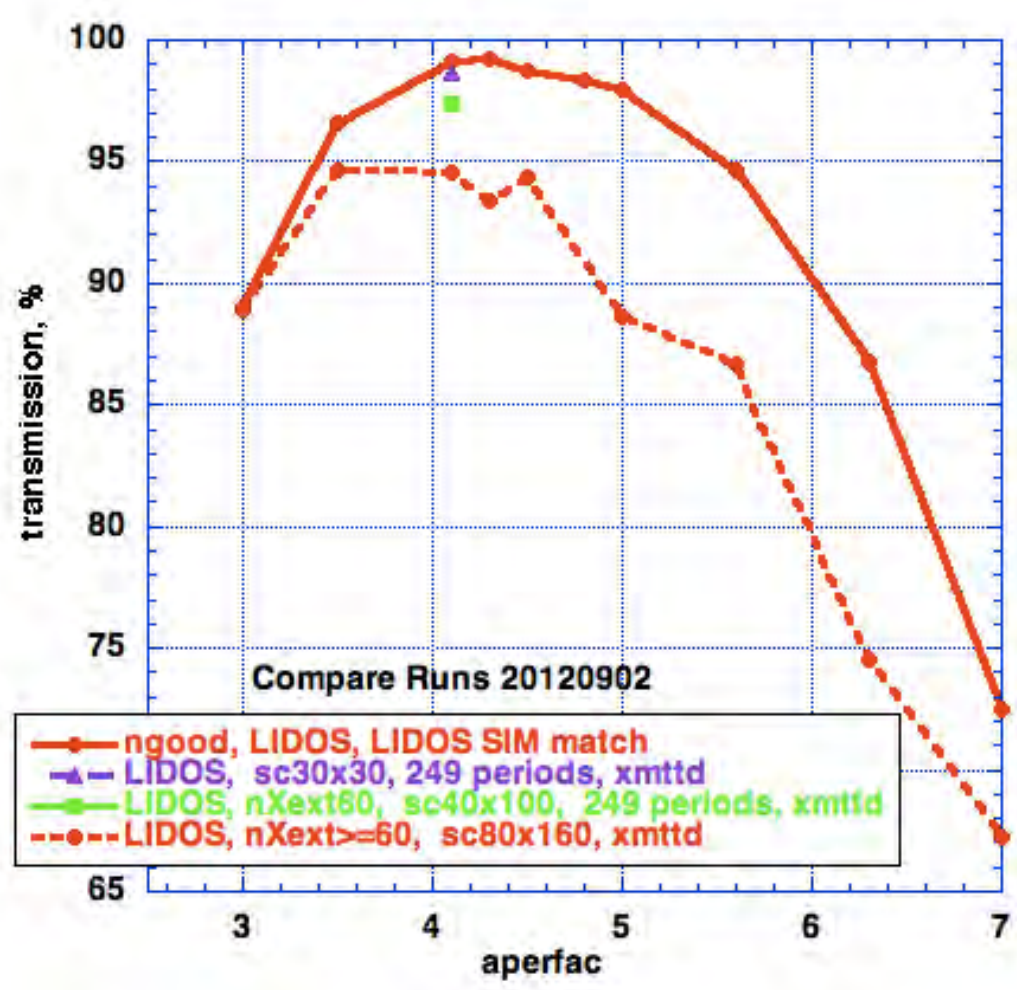

Fig. 18.3. Sensitivity of *LIDOS* results to number of space charge mesh points.

[eltoc]

# Chapter 19.  Synthesis of the New *LINACS* Code

The final goal was to build a new, open-source RFQ code, named *LINACSrfq*, for RFQ design and simulation, incorporating the best physics that present-day computers can handle, options that allow the designer to choose between accuracy and execution speed with an understanding of the consequences, and a flexible library of analysis capabilities.

## 19.1 *LINACSrfqDES* Framework

The beam-based linac design is described in Chapter 5, with the code interface in Ch. 5.3.  A simple GUI is shown in Fig. 19.1 [147]

The RFQ GUI is developed (2018); as *LINACS* is for any linac, placeholders are included for other forms.

The grey buttons are popup menus.  Buttons for saving and loading GUI's were added in 2018 by Peiyong Jiang.

The code can be run from the GUI or directly. The GUI is parsed by writing a subroutine, which is recompiled with the rest of the program by a command file.  This allows the optimization driver for the whole program to be set up.

The design program produces a cell-by-cell design to a file 'celltables/txt'. The RFQ cell parameters can be generated using the 2-term potential description or from 8-term multipole coefficient tables applied at the rms beam radii.

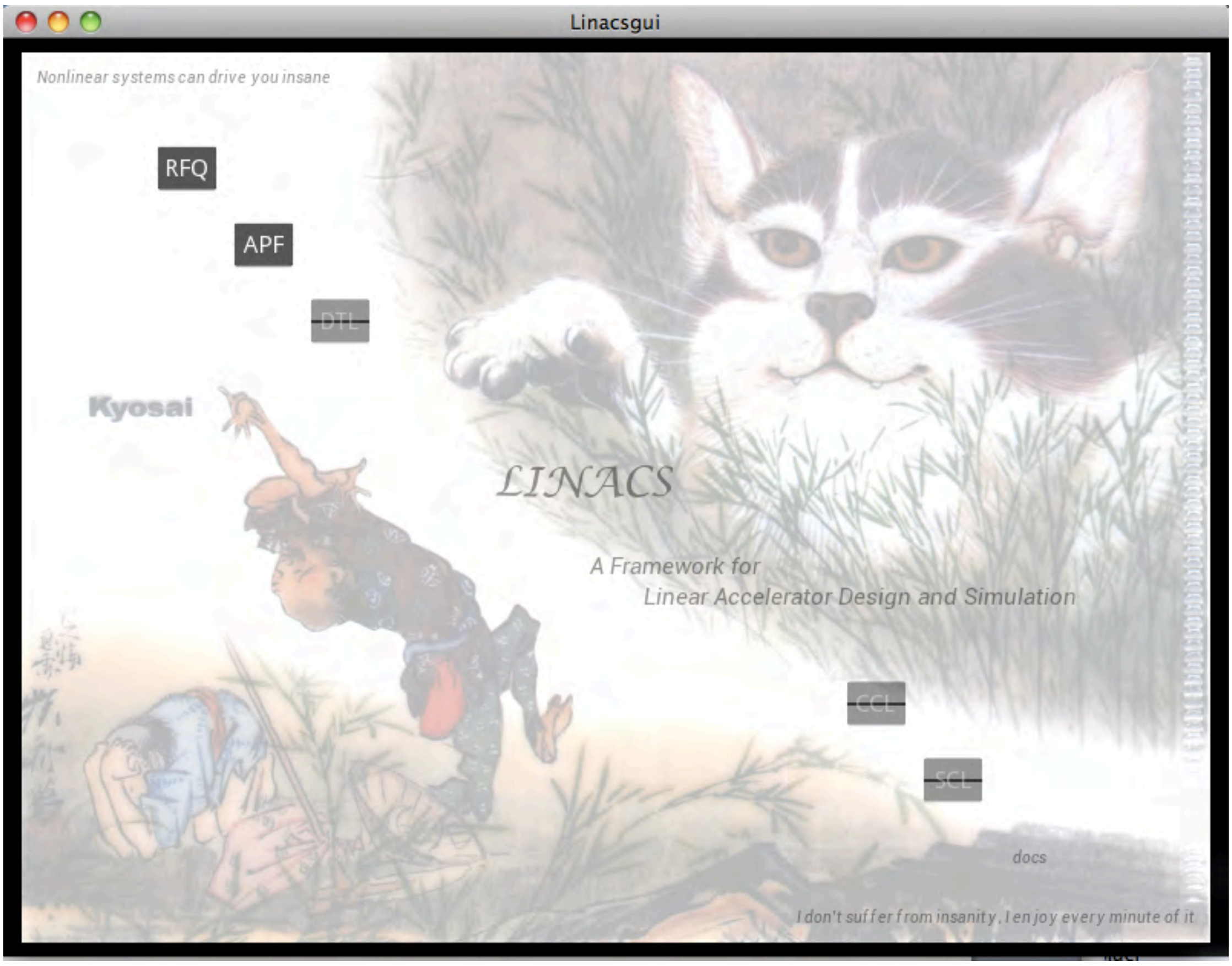

---

[147] Kyosai is one of my favorite artists – it provokes to comment that he is more interesting that either Dali or Picasso.  Kyosai (1831-1889) and Lewis Carroll (1832-1898) were contemporaries.  The grinning (Cheshire) cat was a motive in both English and Japanese cultures at the time.

Fig. 19.1  GUI opening window

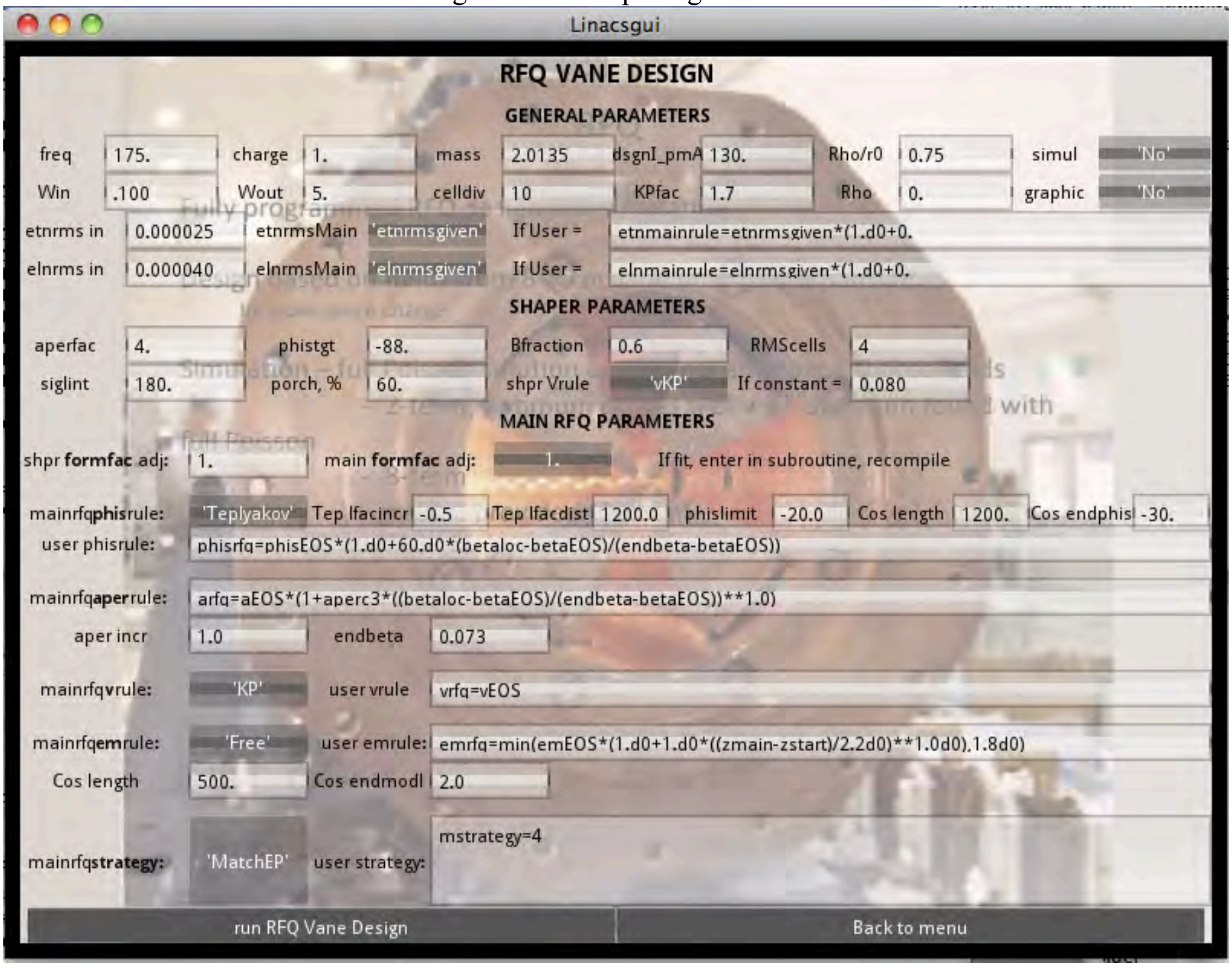

Fig. 19.2.  GUI RFQ window.

As the design process could initially span a very wide parameter space and then home in on a detailed specification, the process is not automatic, and requires experience.  The availability of all parameters, including beam space charge physics requirements, gives the designer complete control; this makes the job harder, but affords better designs.

For example, achieving an EP design is not always successful, if the accompanying parameters are not suitable, and at present there is not much particular guidance.

Anecdotally, it has been noted that in some cases, such as higher frequencies, e.g., 400-600 MHz, eln may drive up to only a fraction of etn, or to <2 when 2 is requested.   Anecdotal cases included 400 and 600 MHz; an EP RFQ could be obtained by lowering the eln/etn requirement, for 600 MHz down to nearly eln/etn=1.   Similarly for 400 MHz, eln/etn=2 did not work, as well as 7/5=0.000035/0.000025, Fig. 19.3.

There is often a seeming obsession for minimum length, often accompanied by no clear economic estimate of the cost/meter, or of other cost factors such as the cost/watt of rf power, detailed structure design, machining, tuning, etc.

The high benefit of equipartitioning for low beam loss should not be lightly discarded, as the EP equilibrium also insures a tighter beam (smaller 100% emittance).  The *a priori* variation of EP can often give a shorter length while keeping the EP condition, for example with ratios = 2 at the end of the shaper, decreasing in the main part of the RFQ down to ratios approaching 1 .

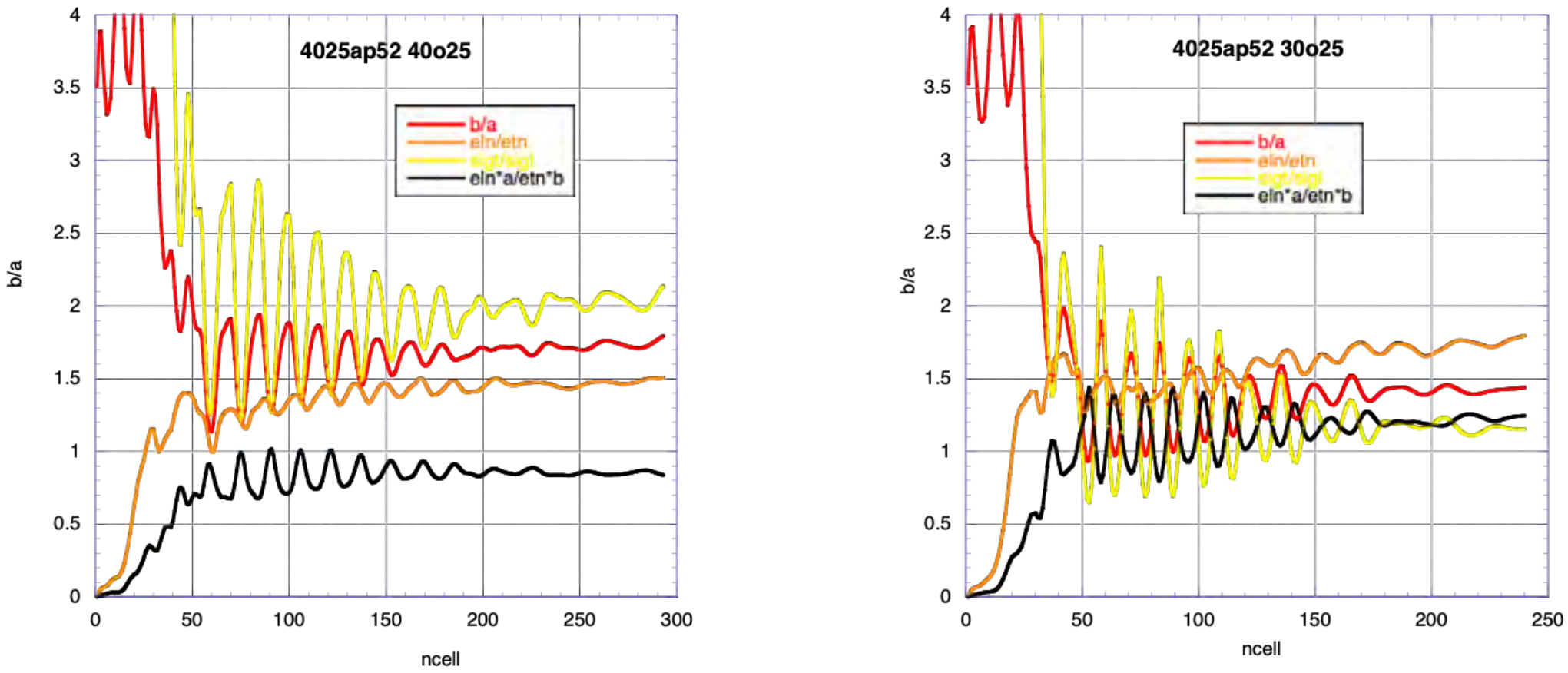

Fig. 19.3 The EP characteristics of an anecdotal 400 MHz preliminary design search. (The ripple like this is usually improved by better input transverse matching.)

Fig. 19.3 is an example where EP has been nominally achieved but the corresponding ratios are not equal, on the left example to 1.6, and in the right to 1.2. Often when the design is already this good, closer EP and ratio equality can be achieved by an *a priori* adjustment of the form factor in the space charge terms (Ch.28). The b/a ratio is obtained along the RFQ trajectory from the run shown, normalized to the design b/a, and fit to a polynomial in z (or cell number or beta), which is then incorporated in the design as an *a priori* adjustment of the form factor. Attempts to find a physical formula for the complicated form factor variation along the RFQ, based on the Debye length or other measures, were not successful; this simple fitting and *a priori* use is justified and works well as a final design step.

Continuing on the theme of having a clear view of the cost of various factors, especially for small beam currents (≤ ~50mA H+ equivalent), shorter RFQ length with still satisfactory beam loss can often be achieved without EP, as long as resonance interactions are avoided. Even in this case, and often obliviously, the design may be locally EP for points along the tune trajectory – thus having the benefits of local EP and operation near EP (the oxymoron!)

There have been many specifications that want everything – short length, very short length to fit into an existing room, little rf power, no beam loss, low injection energy, preservation of a non-realistic input emittance, or just emittance preservation, low longitudinal output emittance, very low cost, perhaps even more – all at the same time. In one case, such specification was given a student for his PhD thesis, and it would have destroyed him without help. In some of these cases, using the full power of *LINACSrfq*, it has been possible to find a design that meets the spec, usually without full EP. Or, a little negotiation with the specifier will be necessary.

It is actually harder to design for low space charge than for significant space charge, because the emittance dominates for the former, and for the latter, the space charge mixing affords a counteraction to the emittance. ***(NOTE!: This finally has some alternatives! – Ch.29)***

Some material on finding a proper regime was outlined in some detail in [148]. Part 4, Ch.22 and Part 5, Ch. 25 & 26 give other examples of RFQ design problems and their solution.

---

148 R.A. Jameson, Principal Investigator, P.J. Tallerico, W.E. Fox, N. Bultman, T.H. Larkin, R.L. Martineau, S.J. Black, "Scaling and Optimization in High-Intensity Linear Accelerators", Study for the JAERI, LA-CP-91-272, Los Alamos National Laboratory, July 1991. Republished as LA-UR-07-0875, Los Alamos National Laboratory, 2/8/2007.

## 19.2 *LINACSrfqSIM* Framework

### 19.2.1 Overall Summary of Physics

A number of approximations, made in older codes largely because of computer limitations, have been removed, and other improvements have been made.

- Laptop computer capability since circa 2010 now allow external and space charge fields in the RFQ to be computed with full Poisson solution fidelity to the quadrupole symmetry and vane geometry, including the effect of image charges.
- Space charge treatment at injection is greatly improved.
- Space charge computation is accurate using time as the independent variable.
- There are no paraxial approximations.
- Correlation between design and simulation is close, using the 8-term multipole potential in the design process. This also makes the fast 2-term simulation mode more accurate, although this mode is only useful during initial, very preliminary design search.

Each of these improvements showed some reduction of transmission and accelerated beam fraction from the optimistic results of older codes, which many experimental results from operating RFQs had consistently shown to be too high.

### 19.2.2 *LINACSrfqSIM* Features and Setup

Many options are available in *LINACS*.

Analysis: Analysis of the cell-by-cell rms properties of the beam inside the RFQ is nonsense if particles are included that are already lost or in the process of being lost within the RFQ. *LINACS* saves the particle coordinates at each cell through the simulation. Particles lost within the RFQ are identified and flagged. In a following analysis procedure, the rms local analysis is performed only on the particles that were successfully transmitted to the RFQ output. This method has made it possible to reliably understand the space-charge physics within the RFQ and to develop the design code in exact correspondence. Multiple analysis options for total beam, single particle, halo are available or easily added.

The code is structured to handle simultaneous distributions of ions; each distribution carries its own number of particles, input emittance data, charge, mass, current, and input energy.

The input distributions can be generated as usual from the input parameters in a number of analytical forms; typically a transverse waterbag and dc longitudinal for the RFQ, or input particle distribution tables can be read in.

Matrices of runs can be executed in batch mode, varying (for the first distribution listed) the input emittance alpha, beta or emittance, current, injection energy, and the RFQ vane voltage level by a vfac multiplying coefficient; in addition the beam current of two additional distributions can be varied.

The powerful linear/nonlinear, constrained/unconstrained optimizer program NPSOL is built in, and used in many ways. An NPSOL overall driver for the entire design and simulation process is also available, so that NPSOL or other optimization techniques can be used to study RFQ optimization ideas.

Input and output beam transport lines can be added. For example, an output transport line to an experimental line or for matching section to the next linac stage.

Other linac stages can be added. At present (2018), a simple DTL is available. An initially simple matrix chain superconducting linac would be easy to add, then upgraded to a full Poisson superconducting linac for study of very low beam losses.

There are many parameters to set up and options to choose, so the structure will be shown in terms of the input file ‘tapeinput.txt’ and corresponding instructions.

### 19.2.3 *LINACSrfqSIM* Setup

The code uses a dictionary to look for the various lines possible in the tapeinput.txt file:

```
/'run','linac','tank','null','rfqout', &
'title','start','stop','elimit','optcon','input','output', &
'error','scheff','end','adjust','begin','trans1','trans2', &
'tilt','trance','exitm0','exirms','lisstw'/
```

The lines are called "cards", in deference to the very old days when programming was done on punched cards, stored in long metal trays , and toted to the computer center for the one run per day – which made one think very carefully; now we can recompile and rerun so fast that much less thinking is required between runs…

***A typical tapeinput.txt file is shown here:***

The simulation requires a 'celltables/txt' file, giving the cell by cell parameters – this is the design RFQ. When this is available, the simulation can be run independently. The main design parameters (frequency, design particle charge and mass, input and output energies) would usually agree with those used to make the "celltables.txt' (either the GUI directly, or the corresponding 'rfqfileGUI.f90' produced by the GUI).

*However, the simulation is independent, and the design RFQ may be simulated at off-design conditions.*

```
run 1 0 0 0 1 0 0 0 0 2.0 0 0 0 0 0 0 0 0 0
title
 RFQ III:f=324Mhz,q=1.0,Wi=0.050,Wf=3.00,amu=1.008373582,ermsn=0.20pi wb,I=60mA
linac 1 0.050 324. 1.008373582 1
   0 0.0 0.0 0.75 10. 2 0.6 0  2  1.5 0.6  62.0  0.6  0.5  0.3
   1  50.0  1.5
trance
exirms 1
tank 1 3.0  -90 0.1 0 1.0 0 1.0 0 0 1.00 10 1 36 0.0 0.0
rfqout 1
start 1
stop -1
elimit 0.25 0.25
input -6 -10000 2.988 14.345 0.011629954 2.988 14.345 0.011629954
  180. 0.0 0.0
 0.0 0.0 0.0 0.0 0.0 0.0 1. 1.008373582 50.0 0.050
output 2 -1 2 1 1 1 1 2000 1 0 0 0 0 0 0 0 0 0 0 0
scheff 50.0 0.026162950 -0.023866890 20 40  5  10  4
optcon  120000   7 0.8 2.0 .2    6  3. 8. 1.0
 7 0.0105 0.0195 .00105
 6  0  50 10   0 0 0 0    5 1.0 1.2 0.05
begin
end
```

The instruction file is shown next. It is very complicated and will not be elaborated further – the reader should get an overview of the extensive features available. The full code is open-source, so the user can see all and is free to add other features. The extensive use of switching options is an important **element**, enabling various versions of subroutines to be compared, directly embedded in the same f**ramework**.

***LINACSrfqSIM card instr's.txt:***

```
!       LINACSrfqSIM calculates phase and radial motion in RFQ linacs.
!         Written by R. A. Jameson, adapting from original Los Alamos AT-Division
!       1980-1982 versions of PARMTEQ, later through PARMTEQM 1996 and v305, v307,
!       and RAJ pteq(~1994), pteqHI(~2005), pteqHIpoisson(~2010), LINACSrfqSIM(2010),
!       LINACSrfq(design + simulation, input GUI, output graphic 2013)
!
!  NOTE!!
!   NOTE!! - LINES IN TAPEINPUT.TXT MUST BE ≤ 80 COLUMNS.   CONTINUE ON NEXT
LINES IF LONGER.
!  NOTE!!
!
```

```
!     FUNCTIONS OF RUN CARD:   vv() -> irunopt()
!       vv(1)=run number
!            RUN 1 - Input distribution original z-coordinates written to 'rdinptOrigz.txt'
!                 and t-coordinates to 'rdinpt.txt'
!                  - Output t-coordinates written to 'rdoutpt.txt'
!                  - Particle status indices are found at NSTOP and stored in
!                  (13,file='tapegoodps'), and their initial coordinates cor(1-6) are
!                  stored in (93,file='tapeACCEPTED.txt").
!                  - Radially lost particle initial coordinates cor(1-6) are
!                  stored in (92,file='tapeINPUTLOST.txt").
!                  - xx', yy' and dpdw acceptances
!                  (dynamic apertures) written to tapeoutstd.txt.
!            RUN 2 - 2nd moments are found, stored on tapemoms1
!                 - also used for matching runs.
!            RUN 3 - 4th moments are found, stored on tapemoms1 and tapemoms2
!
!            irun options, vv(1)-vv(20) -> irunopt(1)->irunopt(20)
!
!            vv(2) source and type of input distribution
!                 =0, use input cards for distribution. (see rfqdyntm)
!                 =1, read z-coord input from file 'emitdist.txt'
!                 =2, read t-coord input from file 'emitdist.txt'
!                   If>0, usual INPUT card is still necessary for info on design and
!                   additional distributions.  The external distribution overwrites
!                   the cord array, which has been set up by usual input cards.

!                   The number of particles in the
!                   external file should match the INPUT cards.
!
!            vv(3)  placement of input beam
!                  =0, Stas Input is applied to input distributions, if ibeam≠0
!                  =1, Input distributions are used directly.
!
!            vv(4) =0, read in celltables.txt
!                 =1, read in existing RFQ cell info from rfqtable.txt
!
!            vv(5) choice of dynamics and space charge method
!                  =0 call 2-term dynamics and Batygin Fourier space charge method
!                  =1 call Multigrid Poisson
!                  =2 call 8-term multipole dynamics rfqdyntm, image effect OFF
!                  =3 call 8-term multipole dynamics rfqdyntm, image effect ON
!
!            vv(6)  =1 will find input ellipse alpha and beta match by running a model
!                    beam from the end of the shaper backward to the RFQ input.
!                    THIS ONLY WORKS IN 2-TERM MODE, SO VV(5) MUST = 0
```

```
!                  - result is given in tapeoutstd.txt, e.g.
!                        TALPHA,TBETA,TEMITRMSREAL  2.55845442757804  14.1743825685736
1.404044315827824E-002
!                        end-of-run        1
!
!          vv(7) =0, GUI not used; existing celltables.txt read in
!              >0, use rfqdesignGUI -> uses rfqfileGUI.f90 -> then reads in resulting celltables.txt
!                   If there is a typo in the GUI -> rfqfileGui, program will execute with previous files
!                 - e.g., with simul='Yes' - setting simul-'No' will not work until the typo is fixed.
!                   Or sometimes balks (runs with old files) even when new rfqfileGUI is ok -
!                 try just recompiling main program and try again...
!
!          vv(8) = 0, standard run
!               > 0, series of runs, use with OPTCON card - see below
!               =2_, (two digits: 2_) make contour map of objective function, using one of these series.
!                  The second digit directs to which matching routine in outmatch.f90.
!                  output card will normally be output 2 -1 2 1 1 1 1 2000 1 0 0 0 0 0 0 0 0 0 0 0
!            RUN GRIDS: use existing celltables,
!                   irunopt(8)=1 with irunopt(7)=1 & irunopt(10)=0
!                   SET STOP < LAST CELL, NOT = -1   -  Some kind of bug - CHECK...
!                   run from GUI or LINACS.command alias or executable directly.
!
!          vv(9) =0 for normal single run
!              =n,  the input currents for 3 distributions (distcurrent(1, 2 and 3))
!               can be overridden for a series of n runs.  The n values of the x variable
!               (presently not used for anything) and the three currents for each run
!               are entered in the file "tapeslice"; e.g., for 3 runs:
!                  .5  29.5 21.4 5.
!                  .75  43.33  38.4  12.5
!                  1.0  41.83 44. 40.
!               Can be embedded in the series using vv(8)
!
!           vv(10)= 0 Graphics is not run
!               ≠0 = EP ratio for graphics.  (0.1 -> 2.0 in increment 0.1 (e.g.,1.6),
!                     2.5, 3.0, 4.0,5.0,10.0)
!         (1,2,2.5,3.0,4.0.5.0 are actually that value + 0.03).  Graphics is run.
!
!           vv(11) = 0, velocity limits for analysis. vlalyz1 sets lower limit below
!                synchronous velocity, vlalyz2sets upper limit above synchronous
!                = 2, PARMTEQM elimits
!
!           vv(12) - not used - still present in rfqdyntm, which is no longer active
!
!           vv(13) - not used - still present in rfqdyntm, which is no longer active
!
!           vv(14) - not used
!
!           vv(15) is for trans2
!             TRANSPORT BEAM THROUGH TRANSPORT SECTION 'trans2',
!
!             if(ntr2.ne.0.and.nstop.ge.nctotl) then
!               write(6,*)'DO TRANS2'
!
!               = 0  RFQ has just been run, and want to continue to trans2 - file rdoutpt.txt is created
!
!               = 1  trans2 is run directly.  RFQ has been run previously and has generated an input file
!                 rdoutpt.txt file;
!
```

```
!                 = 2  trans2 is run directly.  A previously generated rdoutpt.txt is needed, just for dummy.
!                      A transverse distribution is created from the cor(1-6,np) array,
!                     which is created from the input card first distribution;
!                           (change the input ax,bx,ay,by,real emittances, input distribution injection energy
(different beta), etc
!                     (if e.g. starting from DTL - don't forget to change back!)
!                      the cord(1-7,np) array is overwritten.
!
!                        .ge.10.and.lt.20  A matching procedure is run, using input distribution as above in
second digit,
!                      as set up in pteqjob.f90 at "DO TRANS2"
!
!
!      -----------------------------------------------------------------
!                 for special RIKEN RFQ
!           vv(16) = 0, stop at end of RFQ section.
!                 > 0 and STOP = -1, stop at end of RFQ section.
!                 > 0 and STOP = 144, run RIKEN RFQ end section.
!                 =1  No special output files are produced.
!                 =2  tape118tm & tape 118z  produced at end of RIKEN RFQ section
!                    (Cell 118): particles coordinates x,px,y,py,z,pz in time domain
!                    (tape118tm), x,x',y,x',dphi,dw (tape118z).
!                 =3  tape118tm & tape 118z produced, and tape exittm, tapeexitz at end wall
!                   of RIKEN RFQ tank, coordinates as above.
!                 =4  tape exittm, tapeexitz produced.
!
!               (          old RIKEN work did not have zero-modulation section built in.  If something
!                         from old RIKEN were to be investigated again, probably best to regenerate using
!                         EXITm0 in tapeinput.  Old work used trans2:
!                 = cell # at end of RFQ as generated by rfqgen;
!                      rfqdyntm has not been run. So zero-modulation cells in RIKEN RFQ case
!                      are not set. Input will be read from trans2rfqinpt (unit 42).    )
!

!      -----------------------------------------------------------------
!
!
!             vv(max) = vv(30) dimensioned
!
!
!               TRANS1 runs first, so no special control cards
!
!               TRANS2
!                 ! GENERATE THE RFQ BEFORE RUNNING TRANS2 !
!                 Two control parameters are used with TRANS2 –
!                 irunopt(15) and vv(3)=opt3 on the OUTPUT card.
!
!          in pteqjob, AFTER BEGIN, concerning TRANS1 and TRANS2:
!
!          TRANS1 runs first, if cards are given:
!
!          If (opt3.eq.1.or.opt3.eq.2)  RFQDYNTM is run.
!            The cord array is ready for TRANS2.
!            The file trans2rfqinpt is generated, to be read in as input when
!            TRANS2 is run, if desired.
!
!          TRANS2 runs if cards are given:
!
```

```
!         if(irunopt(15).eq.0.and.opt3.eq.0) then
!           RFQ is not being used at all - input for trans2 is generated
!           from input card.
!            cord has been loaded from cor above.  If RFQ output is desired:
!
!         if((irunopt(15).ne.0).and.(opt3.eq.0)) then
!           RFQ has not been run on this run and want to run trans2
!           from RFQ output file trans2rfqinpt;
!            read input array from trans2rfqinpt, unit 94.
!            cord is in time coordinates.
!
!         if(irunopt(15).eq.0.and.opt3.ne.0)) then RFQ has been run,
!            want to continue to trans2 - the cord array is ready to be
!            converted to z domain.

!         output:
!           File 'trans2zinpt" is generated, in z coordinates with absolute phi and w –
!           no subtraction from phis or ws.
!
!           File 'trans2outpt" is generated, in z coordinates with absolute phi and w –
!           no subtraction from phis or ws.
!
!
!
!
!____________________________________________________________
!   TITLE card    (do not write a title on this line! )
!    write a title on this next line if desired
!
!____________________________________________________________
!   LINAC card:

! NOTE: vv(21) sets vane modulation type: =1 sinusoidal, =2 trapezoidal.  MUST BE SET !!
!   Example:
!       linac 1 0.240 200. 12 6
!         0 0.0 0.2 0.00 10.   2 0.6 0  2  1.0 0.2  1.0  0.6  1.0  0.2  1
!
!    vv(1) = number of last tank = 1 for RFQ
!    vv(2) = injection energy (total), MeV
!    vv(3) = frequency, MHz
!    vv(4) = Design particle mass in amu
!        WARNING - if running distribution of design particles with INPUT card,
!               enter same number there!
!    vv(5) = Design particle charge, stored as qstate
!       NOTE:  It is assumed that the RFQ is designed for a + charge
!
!   Vane data
!    vv(6) = vaneshape =0 for vane with breakoutangle; vane tip is then fitted smoothly
!                    to the breakout angle.
!                    Then AS_elec=.FALSE.
!                 =1 for electrode with same thickness
!                    Then AS_elec=.TRUE.
!            The breakout angle vv(10) is still used for either vane shape, to fit to
!            a vane tip radius = constant Rho or = Rho/r0*r0(cell#);  Therefore vv(6) could be
!                eliminated, but it helps to remember to make the desired setting for the breakout angle
vv(10)
!    vv(7) = relec  ! vane radius, i.e., half-thickness of vane, e.g.,
!          for 0.65 cm thick vane, = .375, and if constant Rho, then Rho should also = 0.375
```

```
!    vv(8) = Rho; vane tip radius, cm;  If ≠0, means fixed Rho (then Rho/r0 vv(9) must =0);
!            Must=0 for fixed Rho/r0.
!            The breakout angle vv(10) is still used to fit to this vane tip radius = Rho;
!           Therefore if a constant thickness vane is desired, set vv(10)breakout angle = 0.001
!    vv(9) = Rho/r0.   If ≠ 0, means fixed ratio Rho/r0;
!            Must=0 for constant Rho.  One of vv(8) or vv(9) must be ≠0, and the other one must be =0.
!    vv(10) = breakout angle of vane, degrees.  Default=10. degrees.  MUST NEVER = 0.
!
!   Entrance section - tank wall + gap = L1+L2 = even number of bl/2
!     vv(11) = nextra1 ! # extra cells before the electrodes (tank wall + gap)   !e.g. = 2
!
!     vv(12) = delL1  !tank entrance wall thickness, cm   !e.g.= 0.6
!      vv(13) = nblL1   !Extra thick wall section, in addition to delL1. Defined as an integer number of
bl/2   !e.g. = 0
!       L1=nblL1*btazro*wavel/2.d0 + delL1   ! Thickness of tank entrance wall   !e.g. = 0.6 cm
!
!     vv(14) = nblL2   !number of bl/2 in L2, e.g. 2
!       L2=nblL2*btazro*wavel/2.d0 - delL1   !gap length
!           L1+L2 must be an even number of bl/2
!
!     vv(15) = R1  !Radius of entrance hole in tank entrance wall, cm  !e.g. 1.5
!      vv(16) = R2  !Radius of round-off of inner side of entrance hole in tank entrance wall, cm  !e.g.
0.6
!
!   Exit section - gap + tank wall = L3+L4
!     vv(17) = L3  !distance, cm, between vane end and inside of tank end wall
!          vv(18) = L4   !tank exit wall thickness, cm.   CAUTION - must not be too large or mesh
generation
!       will run away.  Rule-of-thumb ≤ (end cell betalambda)/3.5
!    vv(19) = R3  !Radius of exit hole in tank exit wall, cm  !e.g. 1.5
!     vv(20) = R4  !Radius of round-off of inner side of exit hole in tank exit wall, cm  !e.g. 0.6
!      L3 will be slightly adjusted for external mesh correspondence and will print new value
!
!    Set type of vane modulation  (vanemodl)
!     vv(21) = 1 sinusoidal vane modulation
!       if((vanemodl.eq.1).or.(emrfq.le.altgeommbdry)) then      ! sinusoidal modulation
!     vv(21) = 2 trapezoidal vane modulation
!        if((vanemodl.eq.2).and.(emrfq.gt.altgeommbdry)) then   ! trapezoidal modulation
!          dsgncell='trapezoidal'
!        endif
!     vv(22) = slopepercent; percent of cell that has trapezoidal slope
!     vv(23) = altgeommbdry, modulation below which sine modulation will be used, and above which
!            trapezoidal modulation will be used.
!
!____________________________________________________________
!    lisstw (LISstraw) card:
!
!    adds LISstraw on inner wall of entrance tank wall
!    vv(1) = straw inner radius, cm
!    vv(2) = straw thickness, cm
!    vv(3) = straw dc voltage, MV

!____________________________________________________________
!    TRANCE card:
!
!    adds Crandall transition cell.
!
!    must be entered after LINAC and before EXITM0, EXIRMS,TANK
```

```
!
!
!______________________________________________________
!   EXITM0 card:
!
!   adds exit section with zero modulation.
!   vv(1) = length of exit zero-modulation section, cm
!
!   must be entered after LINAC, TRANCE, and before EXIRMS, TANK
!   DO NOT USE WITHOUT TRANCE - bad transition to zero modulation
!
!
!______________________________________________________
!   EXIRMS card:
!
!   adds exit radial matching section.
!   vv(1) = number of cells in exit radial matching section
!
!   must be entered after LINAC, TRANCE, EXITMO and before TANK
!   DO NOT USE WITHOUT TRANCE - bad transition to zero modulation
!
!
!______________________________________________________
!   TANK card:
!
!   Example: tank 1 59.5  -90 0.1 0 1.0 0 1.0 0 0 1.0  20 1 36 0.0 0.0
!
!    vv(1) = number of tanks. = 1 for RFQ
!    vv(2) = final total energy of RFQ, MeV
!           For heavy ions, (final energy per nucleon)*(Design particle mass in amu (LINAC vv(4))
!    vv(3) = Design initial synchronous phase angle, degrees
!    vv(4) = 0.1  (A default initial value for vane voltage)
!    vv(5) = 0
!    vv(6) = initial modulation = 1.0
!    vv(7) = 0
!    vv(8) = 1.0  (modulation scaling factor, active in rfqgen.f90
!    vv(9) = 0
!    vv(10) = 0
!    vv(11) = vfac, for running at vfac*(design vane voltage)
!    vv(12) = number of segments calculated per cell
!    vv(13) = qstate is stored here later after reading from LINAC card
!       NOTE:  It is assumed that the RFQ is designed for a + charge
!    vv(14) and following - not used (are used in PARTMEQM)
!
!
!
!______________________________________________________
!   RFQOUT card:
!
!   rfqout 1 prints cell parameters every cell.  rfqout 2 every 2 cells, etc.
!
!
!
!______________________________________________________
!   START card:
!
!   starting cell for simulation.  Usually = 1
!
```

```
!
!
!_______________________________________________________________
!   STOP card:
!
!   cell number at which the simulation is stopped.  = -1 to run to the end of the RFQ.
!
!     Special for RIKEN RFQ:
!           vv(16) = 0, stop at end of RFQ section.
!                 > 0 and STOP = -1, stop at end of RFQ section.
!                 > 0 and STOP = 144, run RIKEN RFQ end section.
!
!
!
!_______________________________________________________________
!   ELIMIT card:
!
!     Old version using elimit in MeV (c6=dmax1(c6-elimit,0.d0),
!  settable by elimit card in input deck,
!  elimit is defined as a decimal fraction; this fraction of the
!  synchronous energy defines the band below the synchronous energy
!  within which particles are accepted, particles with energy below
!  this band are considered lost.
!
!     Example:  elimit = 0.1; particles with energy less than 0.9*Ws are considered lost.
!               elimit = 0.9;    "      "    "     "    "  0.1*Ws  "    "        "
!
!     This will then be computed at every step, and also takes the
!  particle mass into account.
!     Similarly, compute a velocity limit at every step, using already
!  computed lower bound energy.
!
!     Problem with slow low energy particles needing nextra steps,
!   but never catch up - only oscillate.
!     If using vv(11) =0 new elimit=%Ws on the RUN card, can set elimit to correspond to
!   an energy somewhat below the injection energy of lowest energy
!   species; e.g. if the final energy is 2 MeV, and the lowest total
!   injection energy of the different species is 0.1 MeV, set elimit
!   to about 0.97.
!     Or use vv(11) = 2 on the RUN card, old elimit (as in PARMTEQ)=_MeV, vlimit.
!   This means that, for example, elimit = .3 will eliminate any particle with
!   energy less than 0.3 MeV below the synchronous particle energy.  A velocity
!   limit is computed in the code corresponding to this elimit.
!
!    velocity limits for analysis. elalyz1 is fractional band below ws that is
!    accepted for analysis; elalyz2 is above same above ws.
!    valyz1 sets lower limit below synchronous velocity, valyz2 sets upper limit
!    above synchronous
!        ebdry=esynch*(1.d0-elalyz1)
!        if(ebdry.lt.0.d0)ebdry=0.d0
!        valyz1=sqrt(ebdry*(2.d0*erest+ebdry))/(erest+ebdry)*29979.2458d0
!
!        ebdryalz2=esynch*(1.d0+elalyz2)
!        if(ebdryalz2.lt.0.d0)ebdryalz2=0.d0
!        vlalyz2=sqrt(ebdryalz2*(2.d0*erest+ebdryalz2))/(erest+ebdryalz2)*29979.2458d0
!
!    particle sorting for analysis then, e.g. for tapezlost:
!
```

```
!________________________________________________________________
!  INPUT card:  Generalized input subroutine
!
!  vv(1) = type lmn
!
!       type 1, np particles are generated randomly in three
!             phase plane ellipses
!       type 2, np particles are generated randomly in a six
!             dimensional ellipse
!       type 3, np/3 particles are generated uniformly on the
!             perimeters of the phase plane ellipses
!       type 4, particle number np is positioned at six specified
!             coordinates.  np is relative to the end of the last
!             distribution entered, and npoint is incremented by
!             np.  e.g.-if you want 2 particles at locations 1-2,
!             enter input 4 1 .., input 104 1 ..; final npoint=2.
!             but input 4 106 .., input 104 7 .. puts a particle
!             at location 106 and one at location 113; final
!             npoint=113.  be careful.
!       type 5, np particles are generated randomly in a four
!             dimensional transverse hyperspace with random
!             phase and energy spread within an ellipse.
!       type 6, np particles are generated randomly in a four
!             dimensional transverse hyperspace with uniform
!             phase and no energy spread.
!       type 7, np particles are generated in a Kapchinskij-
!             Vladimerskij distribution transversely with
!             random phase and energy spread  within an ellipse.
!       type 8, np particles are generated randomly with uniform
!             distribution in real space (x,y,z), then xp, yp,
!             and zp are chosen from within each phase-plane
!             ellipse.  z,zp are then converted to phi,w.
!               (((Following NOT RECOMMENDED FOR USE, E.G., WITH RFQ:
!             If longitudinal is entered as dphi, dw, 0.0
!             a uniform transverse radial distribution with d!
!             longitudinal will be generated. Type -8 is not used
!             (reset to Type 8) to preserve uniformity; (also the
!             proper correction dimension has not been derived.))))
!       type 9, rectangular array in one of the three phase planes.
!             ntype,kind,hl,hr,nh,vb,vt,nv.  kind(1,2,3->x,y,z).
!             Grid is located at horizontal -hl and +hr, vertical
!             -vb and +vt, from 0,0 in transverse planes, or phis,
!             es in longitudinal.  nh and nv are number of grid points
!             in each direction.
!               NOTE:  vv(9, 10)=0.0!!
!                  NOTE: input 9 MUST HAVE 0.0 phase spread and zero
!                      currents on both input and scheff cards.
!     type 10*n, as type n, but coordinates (x, xp, y, yp) are
!             transformed from their input values at phis +
!             dphi to their values at each particles phase,
!             assuming rf quad focussing.   (NOT USED NOW - 10's DIGIT SHOULD = 0)
!      type lmn, l=1 will, each time used, add a type mn distribution
!             to a previously generated distribution and increment
!             npoint by np.
!              Overflow of the coordinate array gives an error message
!             and the program stops.
!              lflag, nodist(lflag) and nbegin are carried in common and =
!             number of separate distributions and number of particles
```

```
!               in each distribution.  subsequent subroutines can thus
!               be made to distinguish the 1st input from all following.
!       type -lmn, will correct the standard type (m=0,1,2,3,4)(n=1,2,5,
!               8,and 9(longit. only)), random distribution to have z
!               zero means and the correct rms ellipse alpha, beta, and
!               area.  The total area will probably be different from
!               that specified on the input card, but the rms area will
!               be correct, and the total area to rms area ratio will
!               be exact.  The correction is applied as required by the
!               type spec; e.g. type 1 corrects each plane independently,
!               type 2 corrects 6d, type 5 corrects 4d xverse and 2d longit.
!
!     for types 1, 2, 3, 5, 6, 7, and 8.
!
!       vv(2)=no of particles or particle no
!         If -vv(2), random number generator reset to beginning
!         When using multiple distributions, vv(2) can be negative for the
!         first (design) distribution, but should be positive for the
!         added distributions, to avoid placing more than one particle
!         at the same point.
!
!       vv(3-8)=transverse ellipse rms parameters: ax, bx, ex, ay, by, ey
!             emittances are real, in cm.rad
!       vv(9-10)=phase and energy spread halfwidths, degrees and MeV
!       vv(9-11) are longitudinal ellipse parameters for some types of distributions
!       vv(11-16 or 12-17)=displacements of input beam. start at vv12 for type 8
!       phase spread or delta phase is entered in degrees
!RIKEN
!       vv(18-21)=charge, mass, PARTICLE current (mA) = mA/charge, (NOT total currrent),
!               TOTAL injection energy (MeV) (NOT per nucleon)
!    (Sorry - this inconsistency should probably be resolved to either all total or per nucleon.)
!               for ions in the particle distribution
!              WARNING - if running a distribution of design particles, enter mass same as LINAC card
vv(4)
!          Example - check charge-to-mass scaling:
!            Design for 238/34=7/1.
!                 linac 1 0.833 52.0 238.00 34. 0.0 0.75 10.
!                 tank  1 34.034 -90 0.1 0 1.0 0 1.0 0 0 1.0 10 34. 1 0.0 0.0
!                 input vv(18-21) design:   34.0 238.0 0.0147 0.833
!                    particle current = 0.0147 -> total current = 0.5
!                    total injection energy = 0.833 ->
!             To check 7/1 scaling, the LINAC and TANK cards are NOT changed - their
!           function is to generate the RFQ.  For example, scale the input vv(18-21) parameters:
!                  1.0  7.0 0.50 0.0245  or
!                  2.0 14.0 0.25 0.049
!             The total injection energy also has to be scaled to match the input velocity
!           based on the charge per particle.
!
!

!       21 NUMBERS OR ZEROS ARE REQUIRED FOR EACH DISTRIBUTION!  The distribution
!       charge, mass and current are read from the INPUT card.  The design particle
!       values should be entered on the LINAC and SCHEFF cards.
!
!
!
!____________________________________________________________
!     FUNCTIONS OF OUTPUT CARD:  -> common/outopts/ opt2,opt3,opt4,...,opt20
!
```

```
!        vv(1) -> cell or tank location of output;  As traditional:
!              = 1, output at cells vv(7), vv(8), vv(9),  -  -  vv(n).
!              = 2, output at vv(7) through vv(8) in steps of vv(9).
!              ≥ 2, output at entr. or exit rf surf. of tanks vv(7),vv(8)..
!
!
!        vv(2) -> location of output in the cell -> opt2:
!              = -1, output at cell center before space-charge
!              = -2, output at cell center after space-charge
!       If multipole cell tables are read in from LINACS, energy at
!       end of cell will be stored in cell(29 and used here.
!
!              >0  , output at cell end
!
!              also -> cell(10,ncell)
!
!              Some routines sense if sign of vv(2) is (-) in order to select phase,
!              energy at center of end of cell; e.g. out10, outmatch,
!              outmultm, outmoms.  Note - If multipole cell tables are read
!                   in from LINACS, energy at end of cell will be stored in
!                   cell(29 and used even if vv(2) is (-).
!
!
!        vv(3) -> JOB CHOICE - different jobs can be specified -> opt3:
!              = 0, no rfq is run - can run trans1 and trans2.  (See vv(7)).
!              = 1, TRADITIONAL OUTPUT.  Uses outstd.f
!              = 2, MOMENTS.  See vv(4).  Uses outmoms.f
!              = 3, MATCHING.  See vv(4).  Uses getmatch.f, confnmatch.f,
!                  objfnmatch.f, NPOSL.f, outmatch.f
!                    Use with RUN 2, and use a tapegoodps file with all zeros
!                  so that all particles will be used in finding the match.
!                  fort.27 contains NPSOL results, tapematch.txt is produced
!              =12, both traditional and moments
!
!
!        vv(4) -> JOB CONTROL -> opt4:
!
!           TRADITIONAL OUTPUT; vv(3) = 1, no control.
!
!           MOMENTS: vv(3) = 2, Find moments of the particle distribution
!                   at cell(s) specified by vv(1) and vv(2)
!              Controlled by RUN card:
!              Two runs are required:  RUN 1 and either RUN 2 or 3:
!
!              RUN 1
!              Good particle indices are found at NSTOP and stored on
!              (4,file='tapegoodps')
!              RUN 2 - 2nd moments are found, stored on tapemoms1
!                  opt4=0 causes ncell,avex,avexp,avey,aveyp,avez,avezp,xmax,ymax,
!                zmax,zmin,t100,zw100 to be written to tapeavgmax
!                  opt4>0 causes size spreads of various %'s of beam to be computed,
!                and tapeavgmax written with ncell,xmax,ymax,zmax,t100/trms,t995/trms,
!                t99/trms,zw100/zrms,zw995/zrms,zw99/zrms,ff100,ff995,ff99,ffb100,
!                ffb995,ffb99,gnb
!              RUN 3 - 4th moments are found, stored on tapemoms1 and tapemoms2
!
!           MATCHING JOB
!              vv(3) = 3
```

```
!              Use with RUN 2, and use a tapegoodps file with all zeros
!            so that all particles will be used in finding the match.
!
!           vv(2) is ignored - outmatch or outmultm called at end of cell
!
!           vv(4) controls the type of matching:
!
!                = 1 -> OUTPUT 1 -1 3 1 0 0  (cell#1st-point) (cell#2nd-point)
!                Local match across vv(7)-vv(8).
!                   gets input particle distribution alpha and beta in x and y
!                   that give a match across two points downstream
!
!                (see vv(5)->opt5: =0, 2 variable ax=ay, bx=by matching, from RFQ input; for one beam
!                             =1, 4 variable match, ax,bx,ay,by; for one beam
!                             >1, 2 variable ax=ay, bx=by matching from
!                                RFQ input for multiple beams)
!


          (7)                                    (8)
!                = 2 -> OUTPUT 1 -1 3 2 0 0  (cell#1st-point) (cell#2nd-point)
!                Local match across vv(7)-vv(8).
!                   gets input particle distribution alpha and beta in x and y
!                   that give a match across two points downstream, but to specified
!                   values for alphas and betas at (cell#2nd-point).  Must enter these
!                   values into outmatch.f or outmultm.f by hand and recompile.
!                   (as above regarding opt5 )
!
!                = 4 -> OUTPUT 1 -1 3 4 0 0 cell#(usually nstop)
!                match on transmission ( minimization of lost particles)
!
!                = 5 -> OUTPUT 2 -1 3 5 0 0  (cell#1st-point) (cell#2nd-point)
!                minimization of standard deviation of beam size from (cell#1st-
!                point - usually 1) to (cell#-2nd-point - e.g., end of shaper) +
!                1/angood (weighting of one probably ok, but check - weight of
!                1/angood should be ~ or some less than weight of std dev.).
!                nstop should be one cell or more beyond (cell#2nd-point).
!                Check that desired tapegoodps is used.
!
!                =7 alpha match
!
!                =9 phase wobble
!
!
!
!!      if(opt3.eq.3) then
!!        if((opt5.le.1).and.(lflag.eq.1)) call getmatch
!!        if((opt5.le.1).and.(lflag.gt.1)) then
!!          write(6,*)'getmatch called with #dists gt 1 - STOP'
!!          stop
!!        endif
!!      endif
!!      if(opt3.eq.3) then
!!        if((opt5.eq.2).and.(lflag.eq.2)) call getmultm
!!        if((opt5.eq.2).and.(lflag.ne.2)) then
!!          write(6,*)'getmultm called with #dists ne 2 - STOP'
!!          stop
!!        endif
```

```
!!     endif
!!
!             Subroutines getmatch.f, confndum.f, objfnmatch.f and outmatch.f used for single
!             input distribution; Subroutines getmmultm.f, confndum.f, objfnmultm.f and
!             outmultm.f used for multiple distributions.
!
!             Main program pteqjob modified at 'AFTER INPUT' to put input card
!             vv(3->20) data into emdata(1-18).
!             emdata(20-23) will be used for ax,bx,ay,by at the first match point,
!             emdata(24-27) will be used for ax,bx,ay,by at the second match point.
!
!             inputt.f modified to write input ellipse parameters to common/multdist3/
!             rfqdyntm.f modified for call outmultm
!
!
!       vv(5) -> output CONTROL -> opt5:
!             vv(5) controls # of variables used for matching, and whether for a single
!                input distribution or for multiple input distributions
!                =0, 2 variable ax=ay, bx=by matching, from RFQ input; for one beam
!                =1, 4 variable match, ax,bx,ay,by; for one beam
!                >1, 2 variable ax=ay, bx=by matching from RFQ input for each of multiple beams -
!                   (4-variable match). Now set up for only 2 beams (2 input distributions).
!
!
!       vv(6) -> output CONTROL -> opt6:
!             OLD!!  LINES ARE STILL IN OUTMOMS, COMMENTED
!                  vv(6).ne.0 and Run 2 writes  ncell beta totlgth sig0t sig0l trms zrms etn eln
!                    sigt sigl sigt/sig0t sigl/sig0l sigl/sigt  to file tapehamil1
!                    writes  b/a eln/etn sigt/sigl eln*a/etn*b a' b' Htk Hlk
!                    Htp Hlp Ht Hl Hk Hp H)  to file tapehamil2
!             CURRENT::
!                  vv(6).ne.0 and Run 2 writes ncell,betacell,totlgth,sig0t,sig0l,arms,brms,etn,eln,
!                    sigt,sigl,sigt/sig0t,sigl/sig0l,sigl/sigt,brms/arms,eln/etn,sigt/sigl,
!                    (eln*arms)/(etn*brms),aprime,bprime,zmax,zmin  to tapedist1 .txt
!                    (and tapedistn .txt for multiple distributions)

!             vv(6)=2 writes 'trans2zinpt', particle coordinates at input to TRANS2,
!                   in z coordinates with absolute phi and w – no subtraction from phis or ws.
!             opt6 may also be used for weighting or other control in outmatch.f - see subroutine.
!
!       vv(7) - ... can be used for specifying cell or tank outputs
!
!       vv(15)-vv(20) -> opt16-opt20         CALL PHASE ADVANCE ROUTINES
!          opt15=0 make "tapephadv.txt(numtape).txt" files
!             =1 make resonances count "Resphadv__.txt" files
!          opt16=0 call phadv1 transfer matrix
!             =1 call phadv2 (in,in') (out,out')
!          opt17 : startcell
!          opt18 : endcell
!          opt19 : =0 filter on zacc transmitted particles; =1 filter on present c7
!          opt20 : =0 summed phase advances;  ; =1 individual periods
!
!____________________________________________________________
!       OPTCON card
!
!  For series of runs, on RUN card: vv(1) = 1;  vv(8) > 0
!
!  This series changes input variables ONLY FOR THE FIRST INPUT DISTRIBUTION.
```

```
! You can still have additional distributions, but their variables are not changed.
!
!   However, within these loops, you can also run another series based on a table
! of input values (in file 'tapeslice', specific 4f10.4 format required,
! xvar(20),yvar(3,20)).  Presently this is set up to vary only the input current of
! the input distributions, e.g., for time slices across the whole pulse.
!   In this case, vv(9) on the RUN card = # of runs(e.g., # of time slices)
!   If you want to make only this series of runs (without the outer loops, set
!       optcon 0 0 0 0  0 0 0 0  0 0 0 0  0 0 0 0
!            0 0 0 0  0 0 0 0
!
!       As vv(1)=1, the series is a run 1 series only - no filtering of lost particles,
! useful for transmission comparisons using tapengood.
!
!
!                 on OPTCON card:
!           vv(1) = 1 for series of runs varying input ellipse alpha
!                 2 for series of runs varying input ellipse beta
!                 3 for series of runs varying input ellipse emittance
!                 4 for series of runs varying input current
!                 5 for series of runs varying input energy
!                 6 for series of runs varying vfac
!           vv(2) = _, number of runs in the series
!           vv(3) = min value of variable for series
!           vv(4) = max  value of variable for series
!           vv(5) = step size
!
!
! For series of runs, on RUN card: vv(1) = 1;  vv(8) > 0
!
!                on OPTCON card:
!           vv(1) = 100000 for series of runs varying input ellipse alpha, called variable 1
!                 20000 for series of runs varying input ellipse beta, called variable 2
!                 3000 for series of runs varying input ellipse emittance, called variable 3
!                 400 for series of runs varying input current, called variable 4
!                 50 for series of runs varying input energy, called variable 5
!                 6 for series of runs varying vfac, called variable 6
!           vv(1) = six or fewer digit number; 123456 cycles all six variables, 123000 cycles
!                 variables 1,2,3;  100006 cycles variables 1,6.  50 cycles only variable 5.
!
!         Lowest number variable is deepest in nest.
!
!         You can enter all of the following - if vv(1)=0 for a variable, these aren't used.
!
!           vv(2) = number of runs in the series 1
!           vv(3) = min value of variable for series 1
!           vv(4) = max  value of variable for series 1
!           vv(5) = step size for series 1
!
!           vv(6) = number of runs in the series 20
!           vv(7) = min value of variable for series 20
!           vv(8) = max  value of variable for series 20
!           vv(9) = step size for series 20
!
!           vv(10) = number of runs in the series 300
!           vv(11) = min value of variable for series 300
!           vv(12) = max  value of variable for series 300
!           vv(13) = step size for series 300
```

```
!
!            vv(14) = number of runs in the series 4000
!            vv(15) = min value of variable for series 4000
!            vv(16) = max  value of variable for series 4000
!            vv(17) = step size for series 4000
!
!            vv(18) = number of runs in the series 50000
!            vv(19) = min value of variable for series 50000
!            vv(20) = max  value of variable for series 50000
!            vv(21) = step size for series 50000
!
!            vv(22) = number of runs in the series 600000
!            vv(23) = min value of variable for series 600000
!            vv(24) = max  value of variable for series 600000
!            vv(25) = step size for series 600000
!
!                  on OUTPUT card:
!            set e.g.: output 1 -1 2 0 0 0 (endcell) ;  on Run 1,
!            this will print out ngood at the endcell. (you can
!            set endcell for any cell - you can also set stop to
!            correspond)
!
!                  File produced:
!            tapemap is produced, with
!            alpha beta emittance current input_energy vfac ngood(1-lflag)
!
!
!
!
!
!______________________________________________________________________
!  SCHEFF card:
!
!       sce(1)=beam current switch.  Only used as switch to turn space-charge
!            computation on (>0) or off (=0)
!              beam current is read from INPUT card.
!       sce(2)=radial mesh interval in cm.
!       sce(3)=longitudinal mesh interval in cm.
!       sce(4)=no. of radial mesh intervals (le 10)    ???le??? should be ge ??
!       sce(5)=no. of longitudinal mesh intervals (le 20)   ???le??? should be ge ??
```

These are not used in Poisson but scheff is still used in trans1 or trans2

```
!       sce(6)=no. of adjacent bunches
!       sce(7)=update mesh every _ cells.  (Should be 1 for simultaneous +,- beams)
!       sce(8)=no. of cells in radial matching section
!
!           for picnic:
!       sce(10)=particle charge spreading in the 8 closest nodes
!            (0:OFF(particle full charge to the closest node), 1:ON)
!       sce(11)=Mesh size (+/-Number of RMS size). Default:3.5
!       sce(12)=Number of skipped particles for field calculation
!            (ex:10 means that one takes one particle every 10)
!            default : 1.
!
```

```
!______________________________________________________________
!   TRANS card
!
!  NOTE:  ++++++++++++++++++
!         if(irunopt(15).eq.0) and trans1 or trans2 card(s) are present, does transport.
!         If(irunopt(15).ne.0) does a simple matrix model dtl design and simulation
!   ++++++++++++++++++++++++
!
! WHERE SHOULD THE TRANS1 OR TRANS2 CARDS BE LOCATED IN THE INPUT
FILE?????
!   after rfqout
!
!TRANS - NOTE THAT SCE(2) RADIAL MESH INTERVAL (CM) AND
!TRANS   SCE(3) LONGIT MESH INTERVAL (CM) MUST BE SET CORRECTLY
!TRANS   FOR TRANS.
!
!=============================================================
! trans1                 called before the rfq. (max elements = 80)
!                          transX(i,vv(1))=vv(i+1)
!        1             element sequence number
!                     the largest must be read last
!                     (zero parameters in any unused mid sequence
!                                               positions)
!        2             element type, defined below
!        3-6            defined below
!        7             if .ne.0 write output to disk at end of element.
!                     written to disk under cell number "zero",but
!                     ask DTLPROC for output at cell "one".
!        8             stored in cell(17,1) on disk
!                     if .ne.0 print emittance
!        9             stored in cell(18,1) on disk ,may be used below
!        10             stored in cell(19,1) on disk ,may be used below
!        11             stored in cell(20,1) on disk ,may be used below
!
!
!   -----------------------------------------------------
!        2             =1 drift
!        3    cm        drift length
!        4    cm        radius of aperture
!        5             number of space charge impulses
!
!
!   -----------------------------------------------------
!        2             =2 Buncher Cavity
!                        (no space charge)
!        3    MV        voltage, E0*T*L
!        4    deg       phase of RF seen by a particle at the
!                     DTL synchronous phase
!        5             harmonic factor relative to linac freq.
!                          (non integers are ok)
!                          (buncher freq. / DTL freq.)
!        6    cm        radius of aperture
!
!
!   -----------------------------------------------------
!        2             =3, Quadrupole Magnet
!                        (with space charge)
!        3    G/cm      gradient
```

```
!       4     cm       length
!       5     cm       bore radius
!       6              error flag, >0 is on.  (see code for errors created)
!
!
!  ---------------------------------------------------
!       2              =4, Horizontal Bend (x axis, rotation about Y axis?)
!                        (with space charge)
!
!    Define losses in the bending magnet by rectangular aperture
!    cards before and after the bend.
!
!       3     deg      angle of bend alpha
!       4     cm       radius of curvature rho
!       5     deg      entrance edge angle beta1
!       6     deg      exit edge angle beta2
!       7              number of space charge impulses
!
!                  note on sign convention: alpha and rho have
!                  same sign; alpha positive bends protons to the
!                  (neg x axis, which is right or left?);
!                  for vertical focusing beta1 and beta2
!                  have the same sign as alpha.
!
!
!  ---------------------------------------------------
!       2              =5, space charge effects,
!                        changes space charge parameters
!       3     mA        beam current
!       3-11           sce(1-9);    see the "scheff" line
!
!
!  ---------------------------------------------------
!       2              =6, circular aperture
!                        (no space charge)
!       3     cm       aperture radius
!
!
!  ---------------------------------------------------
!       2              =7, rectangular aperture
!                        (no space charge)
!       3     cm       X half width
!       4     cm       Y half width
!
!
!  ---------------------------------------------------
!       2              =8, thin lens
!                        (no space charge)
!       3     cm       focal length
!
!
!  ---------------------------------------------------
!       2              =9, alignment transformation (no space charge)
!                if an element is displaced, offset it at input with one card,
!                then remove the offset with other sign of dx and dy at output
!                of the element.
!       3     mils     X displacement
!       4     mils     Y displacement
```

```
!
!
!  -----------------------------------------------------
!        2            =10, solenoid lens
!                        (with space charge)
!               define losses in the solenoid by circular aperture
!               cards before and after the solenoid.
!        3     Gauss?   field
!        4     cm?      length
!        5     cm       aperture
!        6              number of space charge impulses
!
!
!  -----------------------------------------------------
!        2            =11, special
!                    user defined code
!                        subroutine special (element,n) set up now to
!                        rotate beam, n is angle in degrees
!
!
!  -----------------------------------------------------
!        2            =12, steering magnet
!                        (no space charge)
!        3*4           X steering, Gauss cm.
!        5*6           Y steering, Gauss cm.
!
!
!  -----------------------------------------------------
!        2            =13, fringe fields for bends
!                    (use with type 4)
!                        (no space charge)
!        3     cm       half vertical gap
!        4              ck1  default=0.450
!        5              ck2  default=2.80
!        6     cm       half horizontal gap
!     The vertical and horizontal half-gap distances in cm
!     will be used for loss calculations and are required on the card.
!
!
!  -----------------------------------------------------
!        2            =14, Electric Quadrupole
!                        (with space charge)
!        3     MV       Vx-Vy
!                    Voltage x vane - voltage y vane.
!                    Vx and Vy will usually have different signs.
!                    vv(3) positive means x-vanes (horizontal)
!                         are focusing for positive particles.
!                    vv(3) negative means x-vanes (horizontal)
!                         are focusing for negative particles.
!        4     cm       length
!        5     cm       bore radius
!                    also determines vane spacing, + or - vv(5)
!        6              error flag, >0 is on.  (see code for errors created)
!        12    MV       (Vx+Vy)/2           DC offset from Vx.ne.Vy.
!                    Vx and Vy will usually have different signs.
!
!=====================================================
! trans2
```

```
!               parameters are the same as trans1.
!
!
!!!  !!!!!!!!!(below from parmila.doc - not necessarily all correct for pteqHI, where
!!!  multispecies, multi-charge states are possible.!!!!!!!!!
!!!
!!!    trans notes, I (Takeda or someone) recommend never using the "tank"
!!!              line to extend a quad beyond the edge
!!!              of a tank (1st or last tank).
!!!              Do it with "trans" instead.
!!!         you must generate a dtl to use "trans", even
!!!              if you are not going to look at any
!!!              beam dynamics in the dtl.
!!!
!!!         suppose: You have a 1.9 MeV beam.
!!!                In a transport line before the DTL you
!!!                have a buncher that increases the beam
!!!                energy to 2.0 MeV.
!!!                You would have a "linac line" with the
!!!                value 2.0 MeV and a "input line" with the
!!!                value -0.1 MeV.
!!!                (if your buncher is set wrong then the stored synchronous
!!!                beam energy & the real beam energy will be inconsistent
!!!                at the DTL entrance.) Dyn. synchronous energy (DTL) comes
!!!                from the linac line, not from the beam or partran.
!!!                Dyn. synchronous energy (partran) comes from linac line plus
!!!                input line. (-0.1 MeV shift does not keep longitudinal emittance
!!!                correct)
!!!
!!!
!!!trans2                 called after the dtl. (max elements = 40)
!!!                     parameters are the same as trans1.
!!!                     information written to disk is under cell
!!!                          number "last" instead of "zero/one"
!!!=================================================================
```

## 19.3 Beam Starting Conditions - On the Longitudinal Potential and Field in an RFQ [149],[150]

A distribution created in z-coordinates can be converted directly to t-coordinates, because the input beam is dc.

At present, the fields are found for one quadrant and symmetry applied.

The RFQ parameters change rapidly, even over a length of two cells. To accommodate this for the external fields, we find the Poisson solution for the external potential in a section that is a mesh n cells long, but use only cells 2 to n-1 for the dynamics. The external potential in this section needs to be computed only once. The initial and final sections of the RFQ require special handling. A particle

[149] File references and procedures – see Appendix 1

[150] From: On the Longitudinal Potential in an RFQ - revised 20111116, On the Longitudinal Potential in an RFQ – revised20111115_commentedJMM20111123

distribution, centered and within ±1 cell length of a synchronous particle, is moved dynamically through the external fields found from the potential.

In contrast, the space charge fields are dynamic and have to be computed at each space charge call. For reasonable computation times, it is strongly desired to use a beam slice of minimum length, which the RFQ geometry dictates as the transverse focusing period of 2 cells (after bunching, there is a bunch in every 2nd cell). Then several assumptions are made:

– The mesh length is ±1 cell centered on the beam centroid; the length is twice the length of the cell that the beam centroid is in; so this is slightly approximate.

– When loading the charge, in the first and last mesh cells, particles are assigned left and right to mesh transverse planes 0_1 and k-1_k. Transverse planes 0 and k get charge only from the right or left, respectively. Interior mesh planes are getting particles from both sides, so they have essentially twice as much charge as do planes 0 &k. The periodic boundary condition must still apply, so an apparent charge from cell -1 needs to be assigned to plane 0, and from cell k+1 to plane k. The periodic boundary condition then requires the charge on plane k to be added to that on plane 0, and vice versa, so both planes 0 and k have the sum of the initial loading.

– In solving for the longitudinal space charge potential, periodical longitudinal boundary conditions are required to account for the effects of neighboring bunches, and are given by:

```
IF (k.eq.0) THEN
  Uzp = SC(i,j,1)%Phi
  Uzm = SC(i,j,k-1)%Phi
ELSE IF (k.eq.k) THEN
  Uzm = SC(i,j,k-1)%Phi
  Uzp = SC(i,j, 1)%Phi ,
```

where 0 and k are the left and right mesh boundaries. With changing focusing parameters over the mesh length, and because the particle velocities of the neighboring bunches and the external fields and cell lengths affecting them are actually different, this is clearly also approximate.

- In similar manner, the longitudinal electric field, found by differentiating the potential, requires a boundary condition at the left and right mesh boundaries, which must also be assumed to be periodic.

- As the cells get longer and the beam bunches, the fraction of the mesh that the bunch occupies becomes much smaller. The total mesh length must remain 2 cells for periodicity. We need to keep the number of mesh cells that the bunch occupies about constant, so we base the resolution (mesh cell size) on the rms beam length, resulting in a significant increase in the number of mesh cells and computation time along the RFQ.

The resulting behavior of the injected dc beam is of critical importance, because here these longitudinal boundary effects will have the most influence. This behavior is outlined next. As the beam becomes bunched, the ends of the longitudinal mesh become less important.

### 19.3.1 Tests of Beam DC Behavior at RFQ Input

#### *19.3.1.1 On-axis beam:*

A channel was set up as a grounded 1.5 cm radius conducting pipe. A standard (Type 6) dc input beam, with on-axis distribution was placed in the pipe, and the initial transverse coordinates set to zero. 10,000 particles are uniformly distributed in z. Space charge is computed after ½ time step.

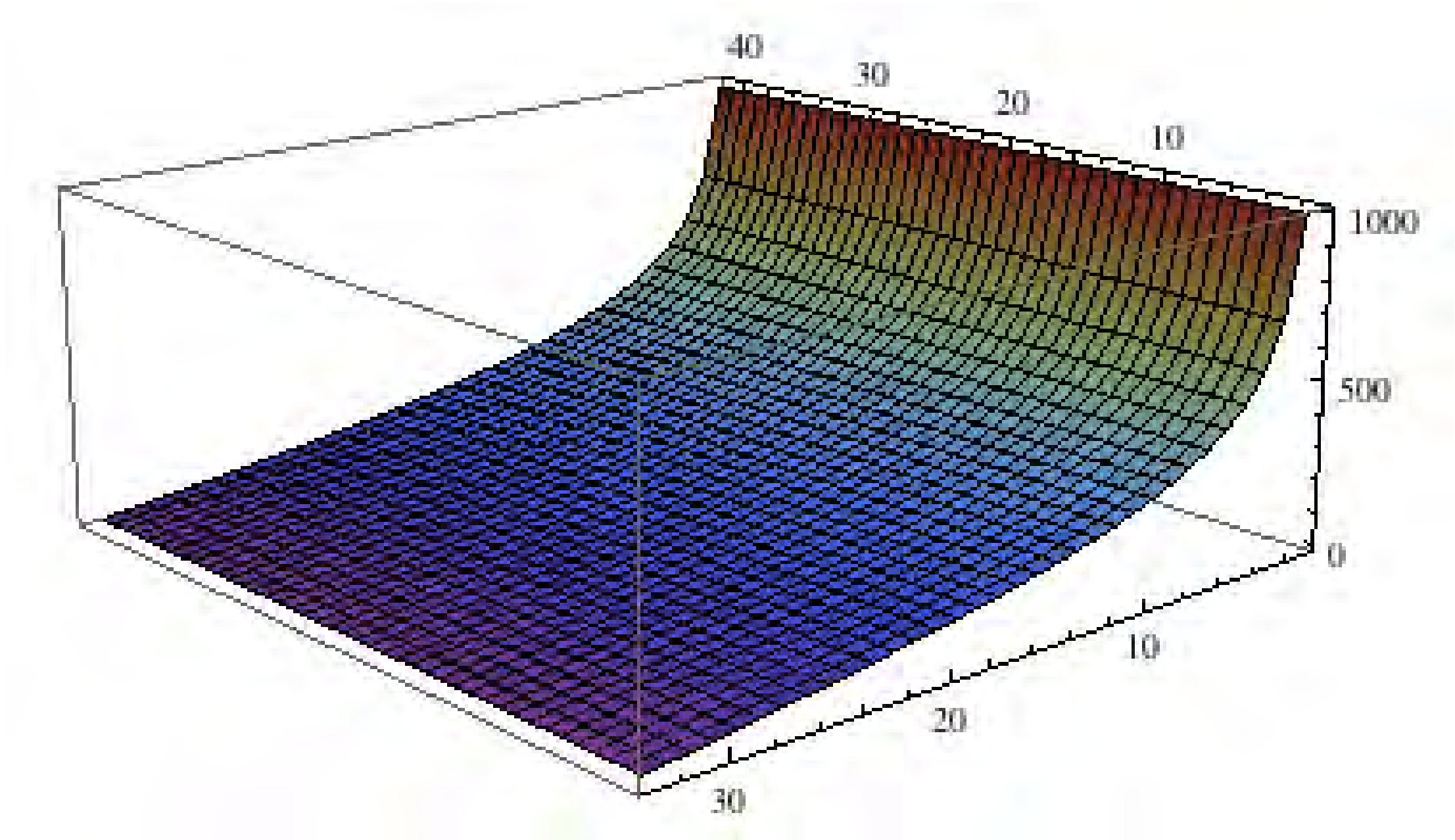

Fig. 19.3  Space charge x vs. z potential of on-axis beam.

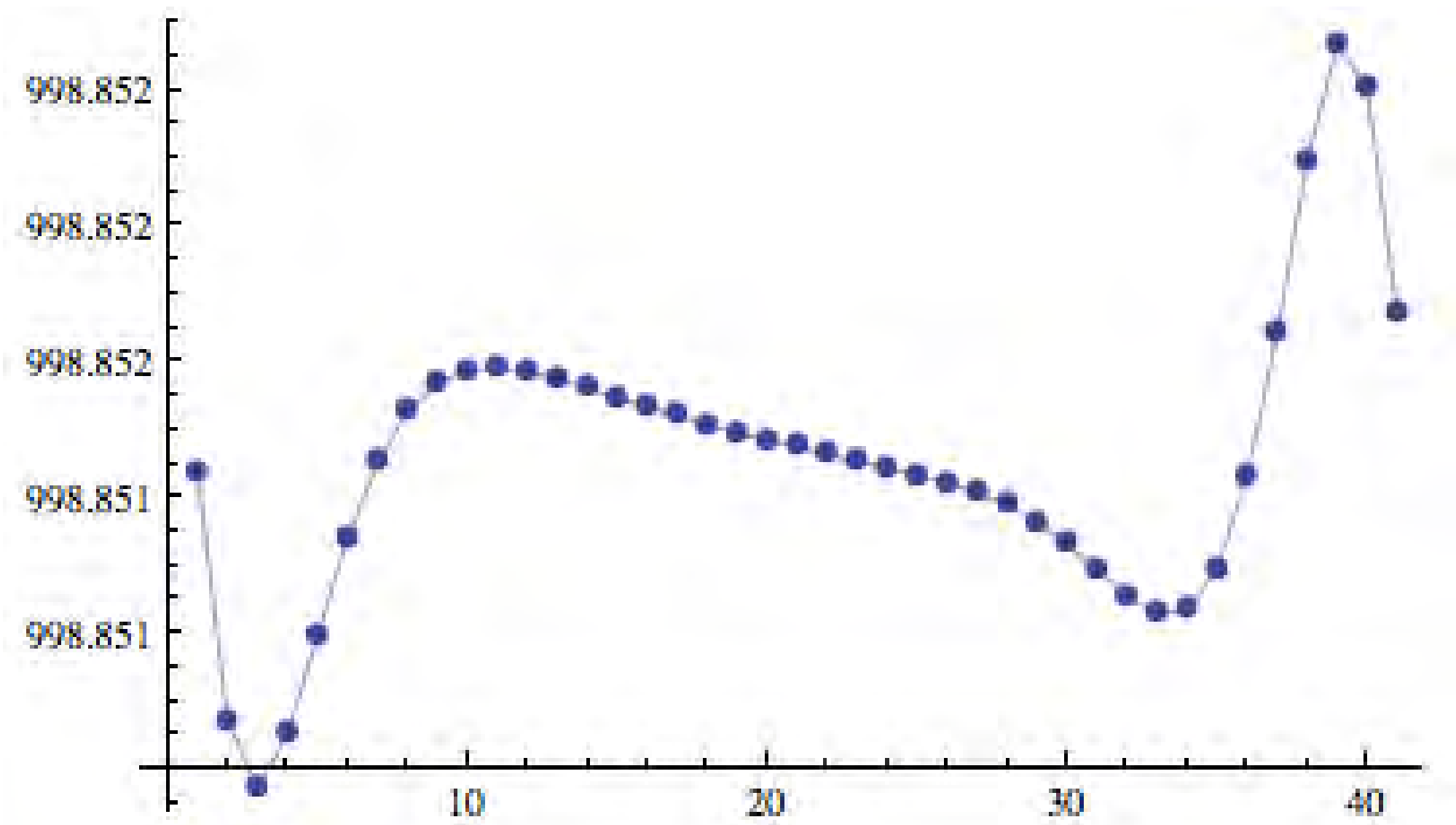

Fig. 19.5  Potential on the axis for on-axis beam, for Poisson cycle variables the standard m=5, n1 = n2 = 7.  The difference between the maximum and minimum is 0.00011%.
The defect converged as:
SC defects : full, sum20, sum10  2  3.39566E+06  1.02377E+07  3.80040E+07  1.17838E+00
SC defects : full, sum20, sum10  2  1.39171E+06  3.37411E+06  6.50810E+06  1.17838E+00
SC defects : full, sum20, sum10  2  2.25919E+05  5.26786E+05  8.77859E+05  1.17838E+00
SC defects : full, sum20, sum10  2  4.01037E+04  9.01625E+04  1.37122E+05  1.17838E+00
SC defects : full, sum20, sum10  2  5.09035E+03  1.08206E+04  1.55324E+04  1.17838E+00

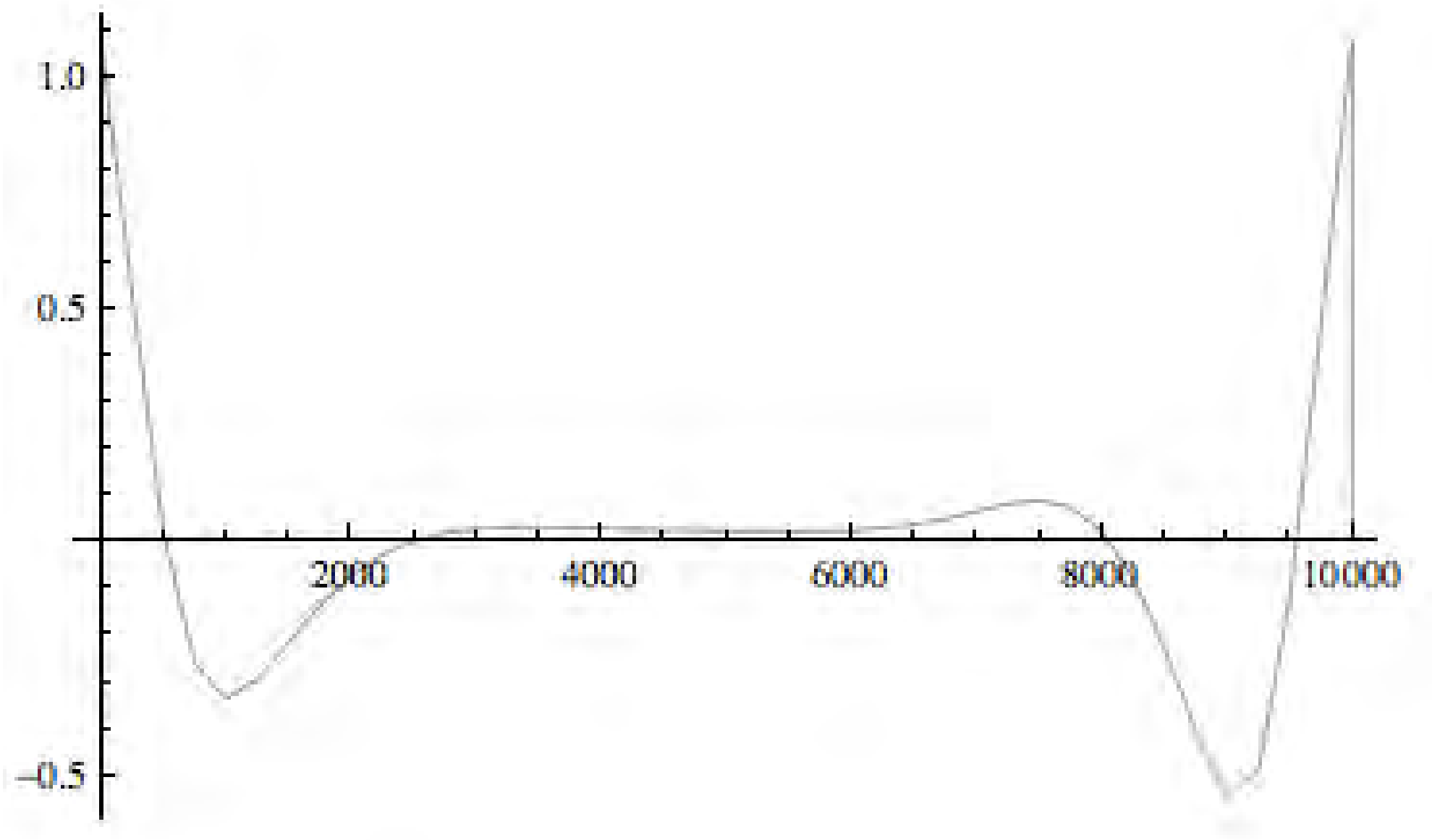

Fig. 19.6 Ez longitudinal field for on-axis beam on each of the 10000 particles. m=5, n1 = n2 = 7.

Using m=10, n1 = n2 = 14 doubled the running time but improved considerably for the on-axis beam. It is concluded that the standard accuracy is sufficient:

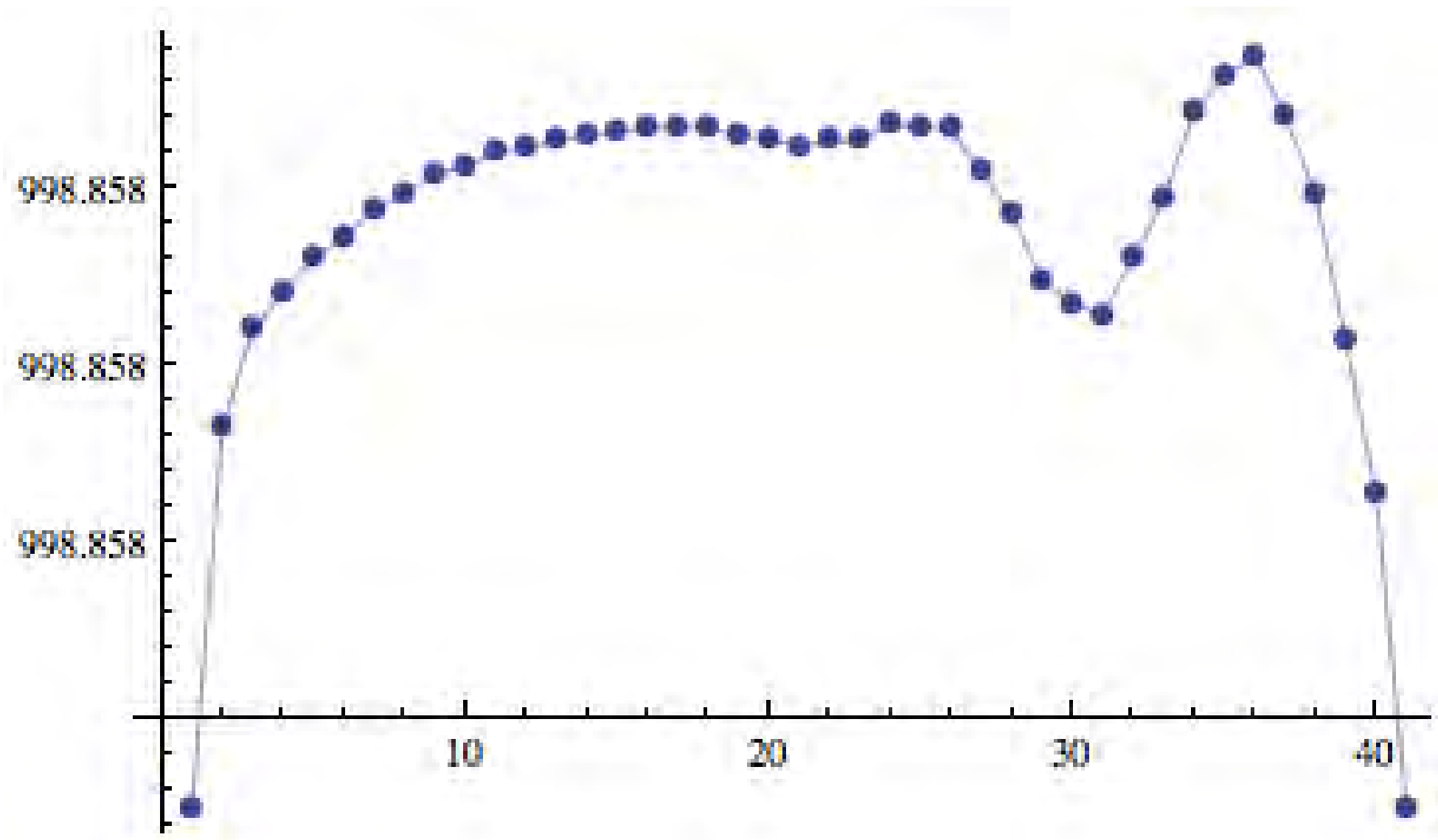

Fig. 19.7 Potential on the axis for on-axis beam, for Poisson cycle variables m=10, n1 = n2 = 14. The difference between the maximum and minimum is $4.3x10^{-8}$%.

The defect converged as:

SC defects : full, sum20, sum10 2 3.04973E+06 9.19455E+06 3.34880E+07 1.17838E+00
SC defects : full, sum20, sum10 2 5.11222E+05 1.21090E+06 1.81584E+06 1.17838E+00
SC defects : full, sum20, sum10 2 3.59147E+04 8.25521E+04 1.18300E+05 1.17838E+00
SC defects : full, sum20, sum10 2 1.72300E+03 3.42446E+03 4.67129E+03 1.17838E+00
SC defects : full, sum20, sum10 2 9.23686E+01 1.62992E+02 2.52228E+02 1.17838E+00
SC defects : full, sum20, sum10 2 2.76556E+00 4.34796E+00 8.61916E+00 1.17838E+00
SC defects : full, sum20, sum10 2 1.25478E-01 2.52033E-01 4.22201E-01 1.17838E+00
SC defects : full, sum20, sum10 2 9.30121E-03 2.24175E-02 3.16156E-02 1.17838E+00
SC defects : full, sum20, sum10 2 5.93091E-04 1.28049E-03 1.73229E-03 1.17838E+00
SC defects : full, sum20, sum10 2 3.04236E-05 5.70975E-05 8.11621E-05 1.17838E+00

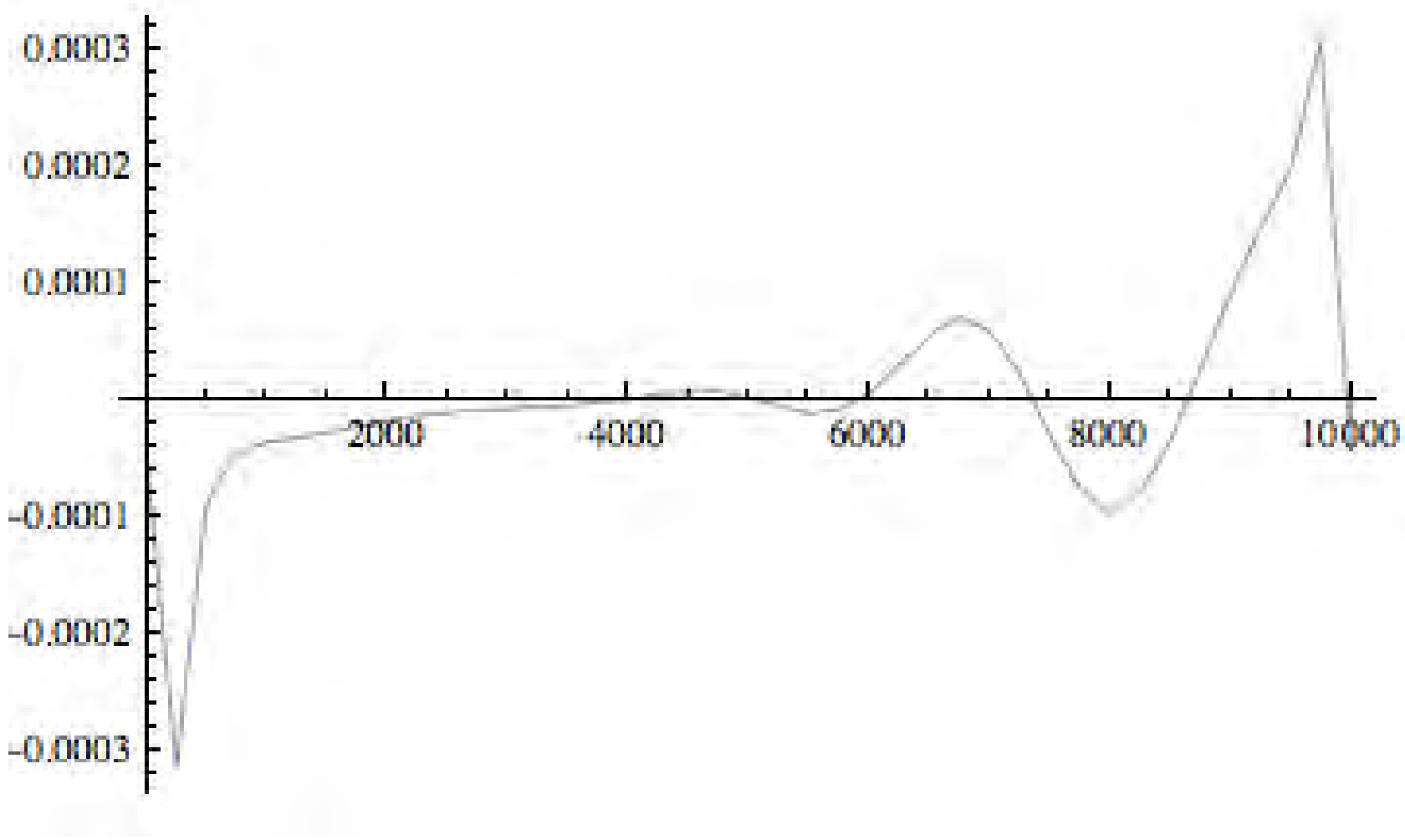

Fig. 19.6 Ez longitudinal field for on-axis beam on each of the 10000 particles. m=10, n1 = n2 = 14.

*19.3.1.2 Standard Beam*

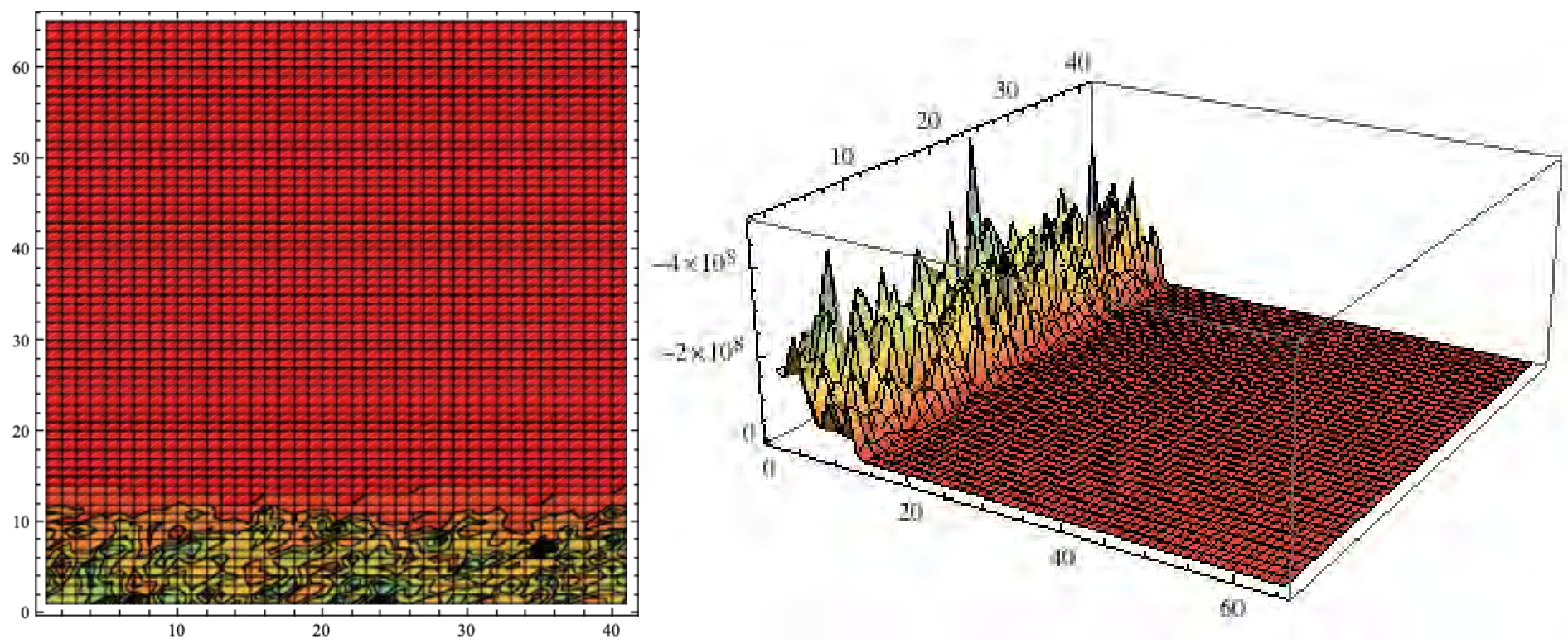

Fig. 19.9 Charge distribution for 10,000 particles.

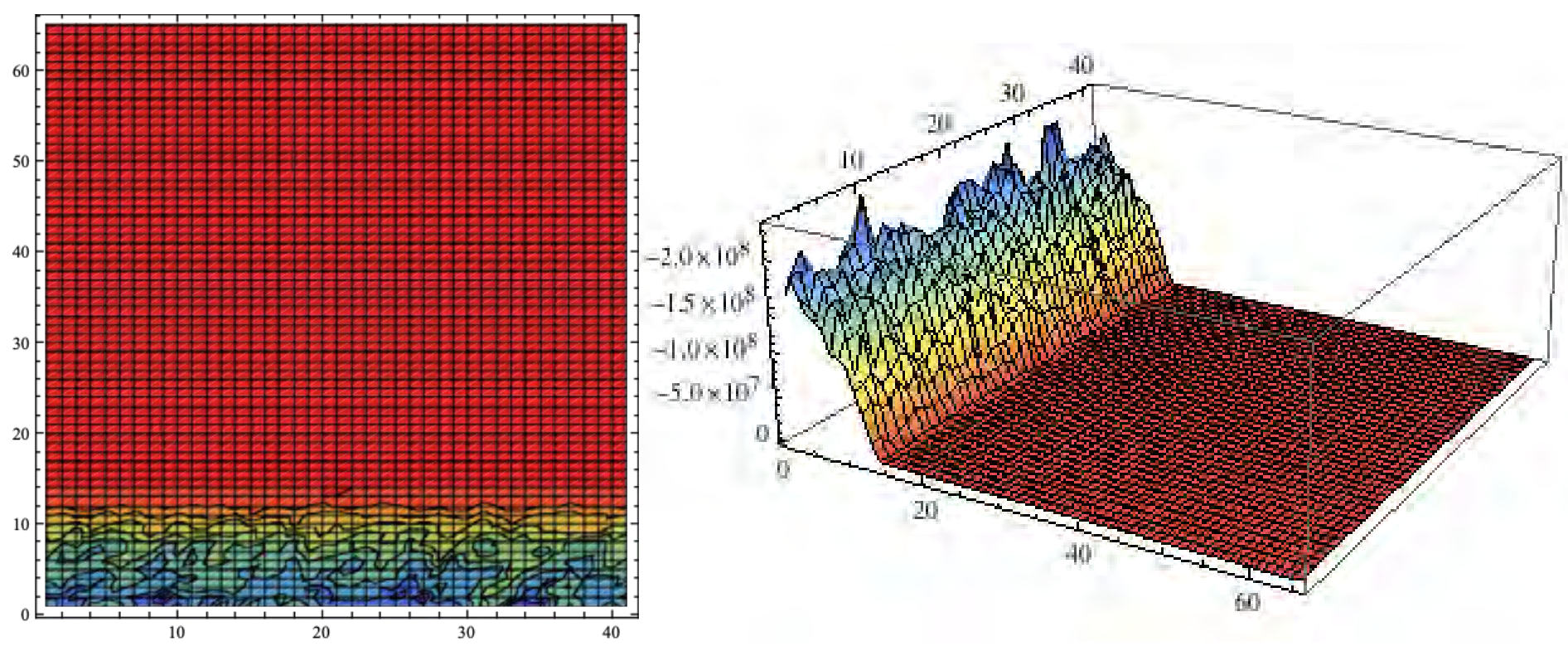

Fig. 19.10 Charge distribution for 100,000 particles.

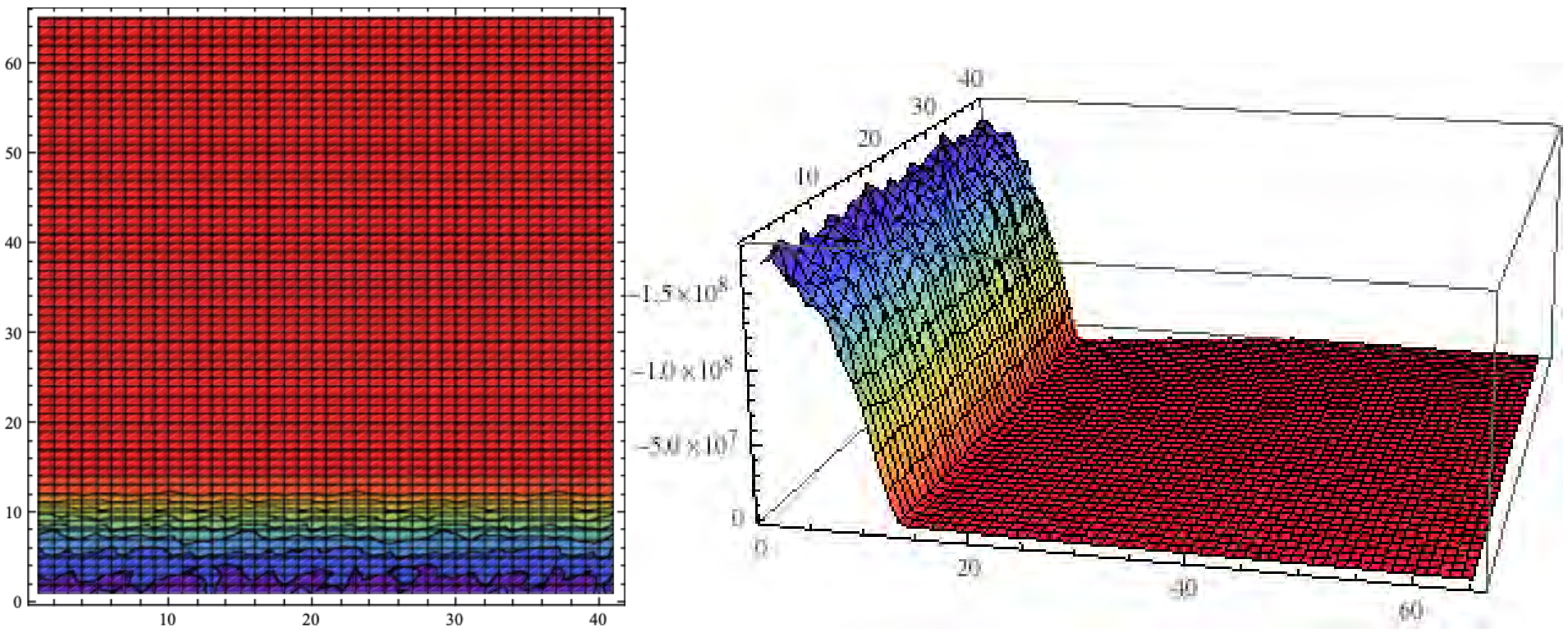

Fig. 19.11 Charge distribution for 1M particles.

A channel was set up as a grounded 1.5 cm radius conducting pipe. A standard (Type 6) dc input beam, with transverse waterbag or on-axis distribution was placed in the pipe, and space charge computed after ½ time step.

In Figs. 19.9-11, 10K, 100K, and 1M particles are uniformly distributed in z. The potential found by the Poisson space charge solver will reflect the coarseness of the charge distribution.

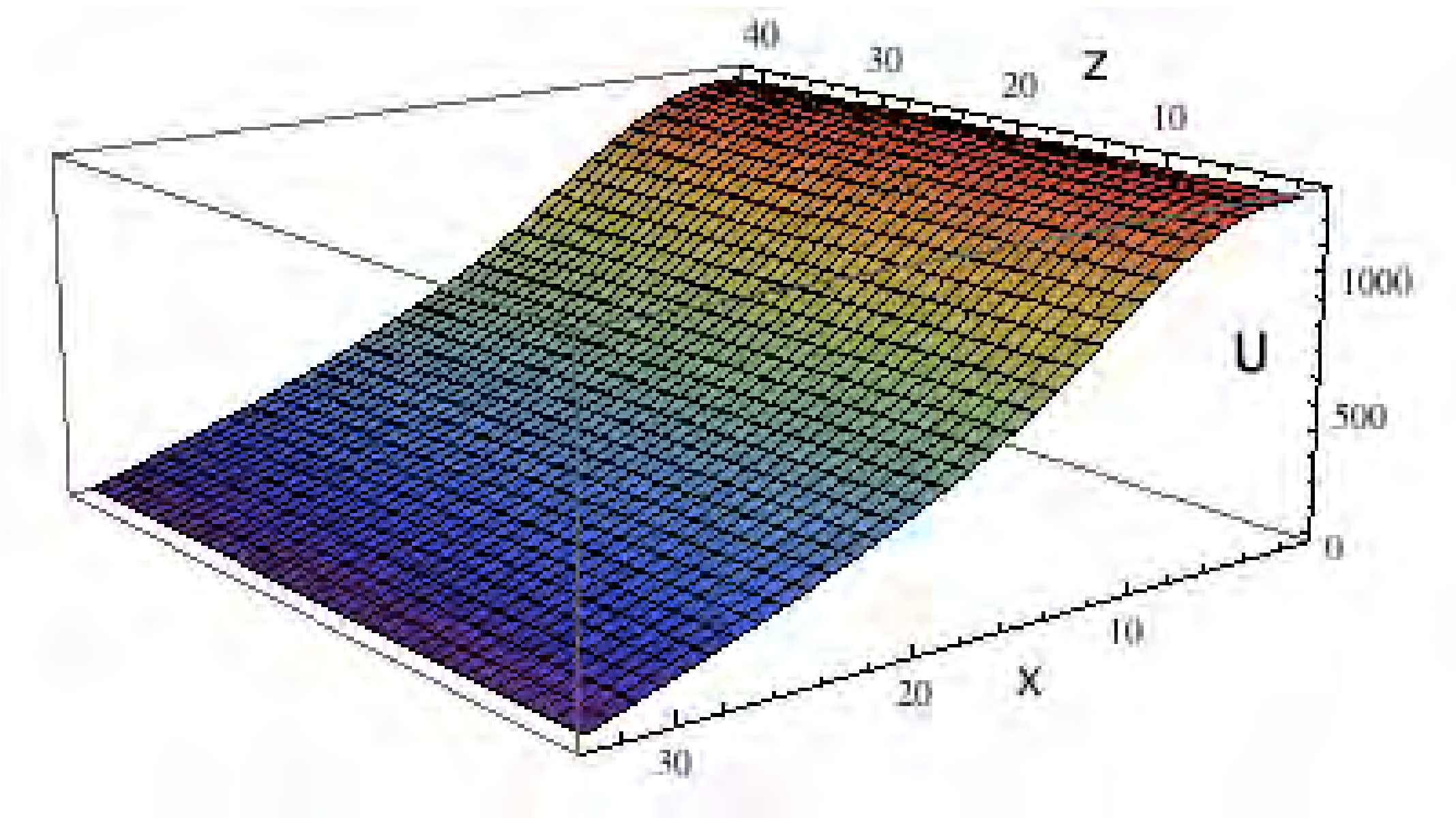

Fig. 19.13 Space charge x vs. z potential of standard beam.

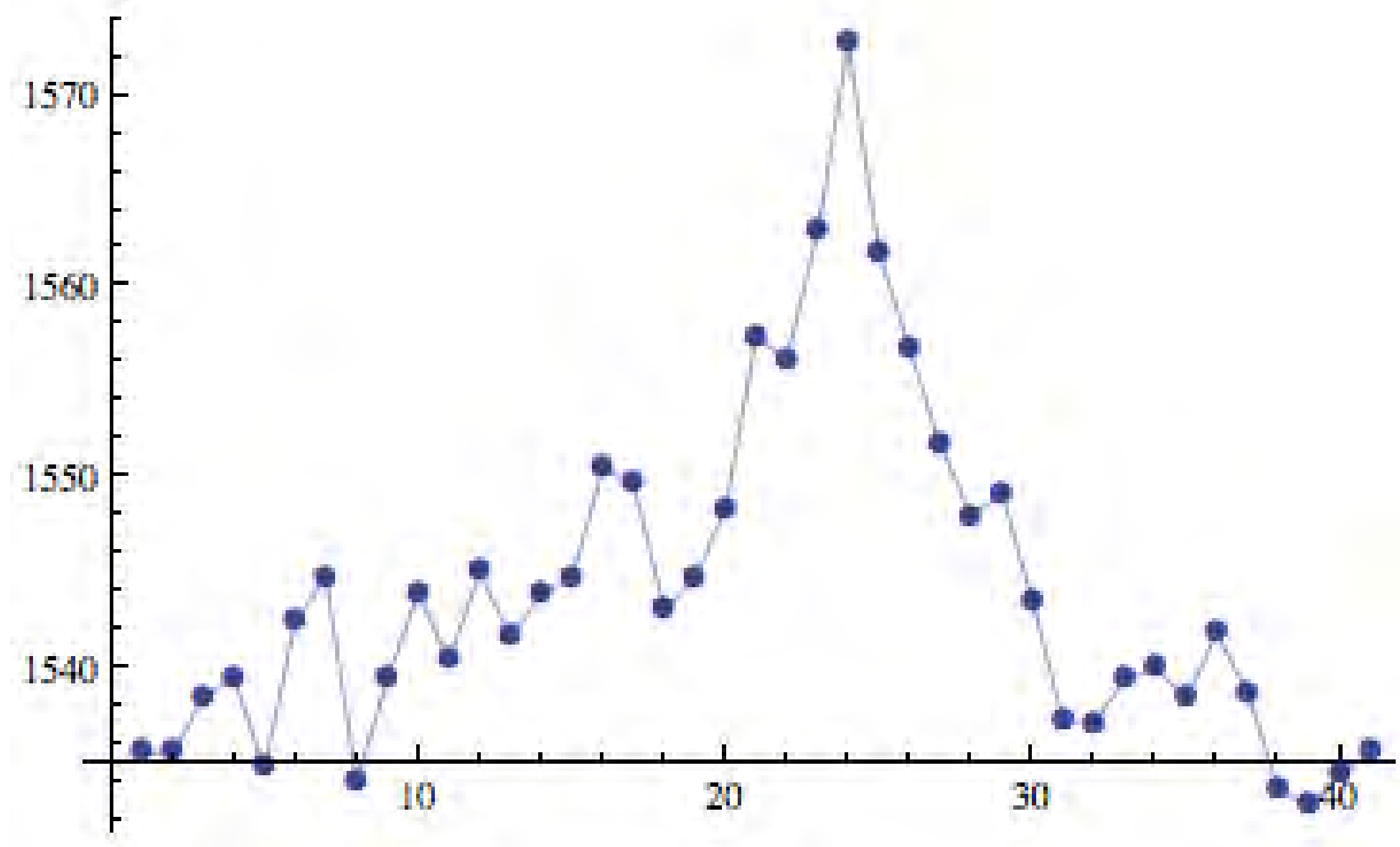

Fig. 19.14 Potential on the axis for standard beam with 10,000 particles. The difference between the maximum and minimum is 2.53%. The Poisson cycle variables are the standard m=5, n1 = n2 = 7. Using m=10, n1 = n2 = 14 doubled the running time (for a 10 cell run) but made no discernible difference in the potential on axis.

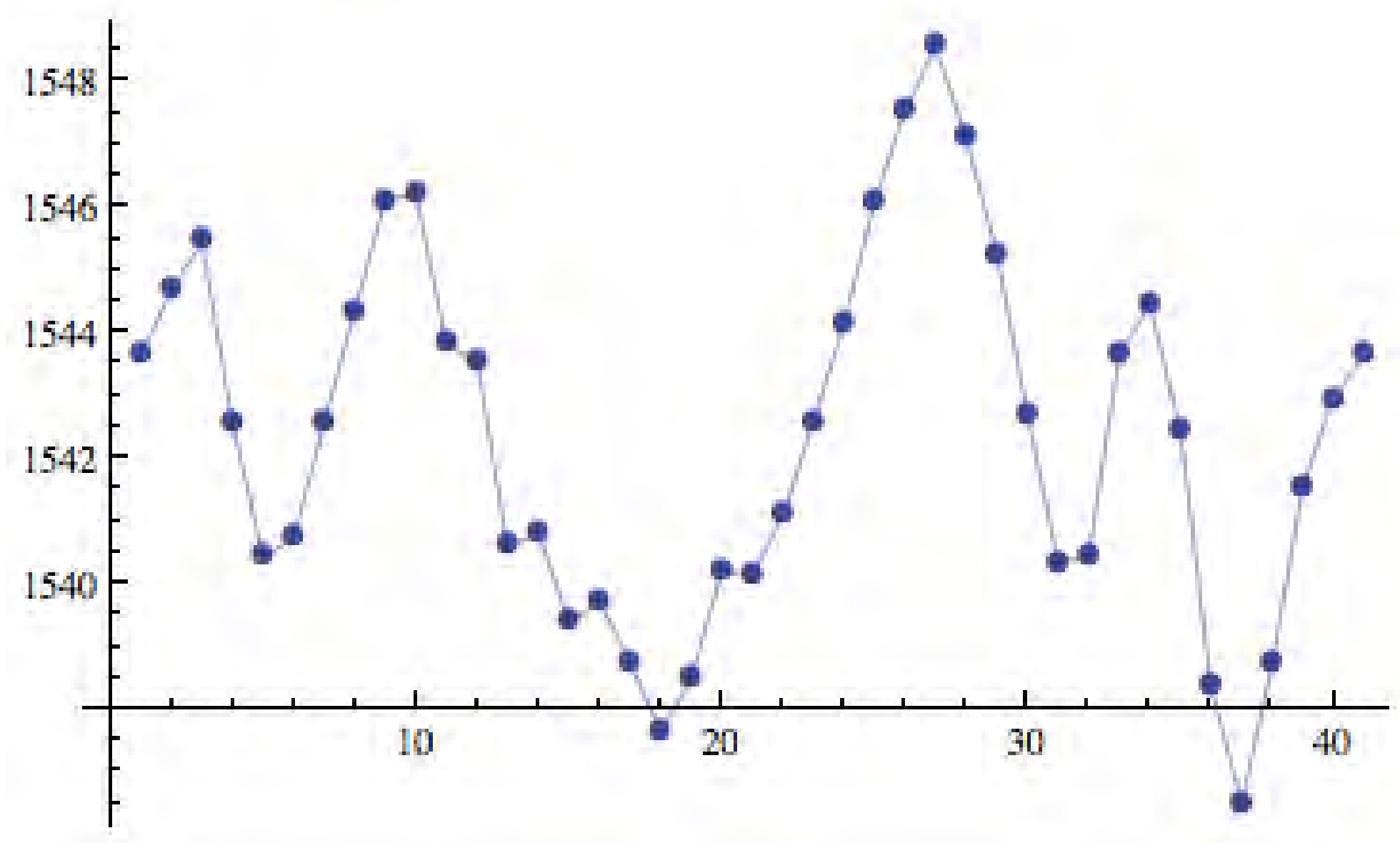

Fig. 19.15 Potential on the axis for standard beam with 100,000 particles. The difference between the maximum and minimum is 0.78%.

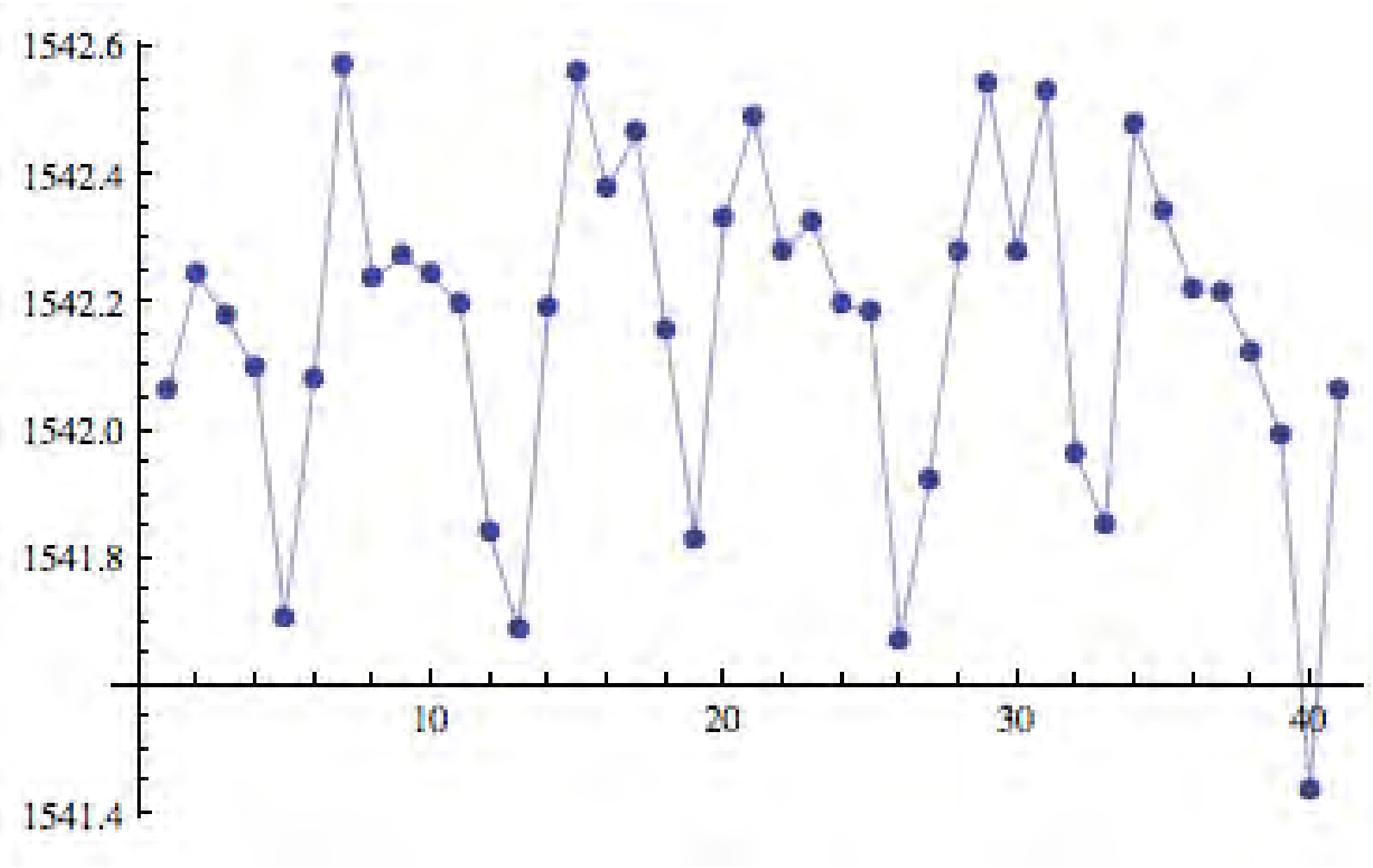

Fig. 19.16 Potential on the axis for standard beam with 1,000,000 particles. The difference between the maximum and minimum is 0.074%.

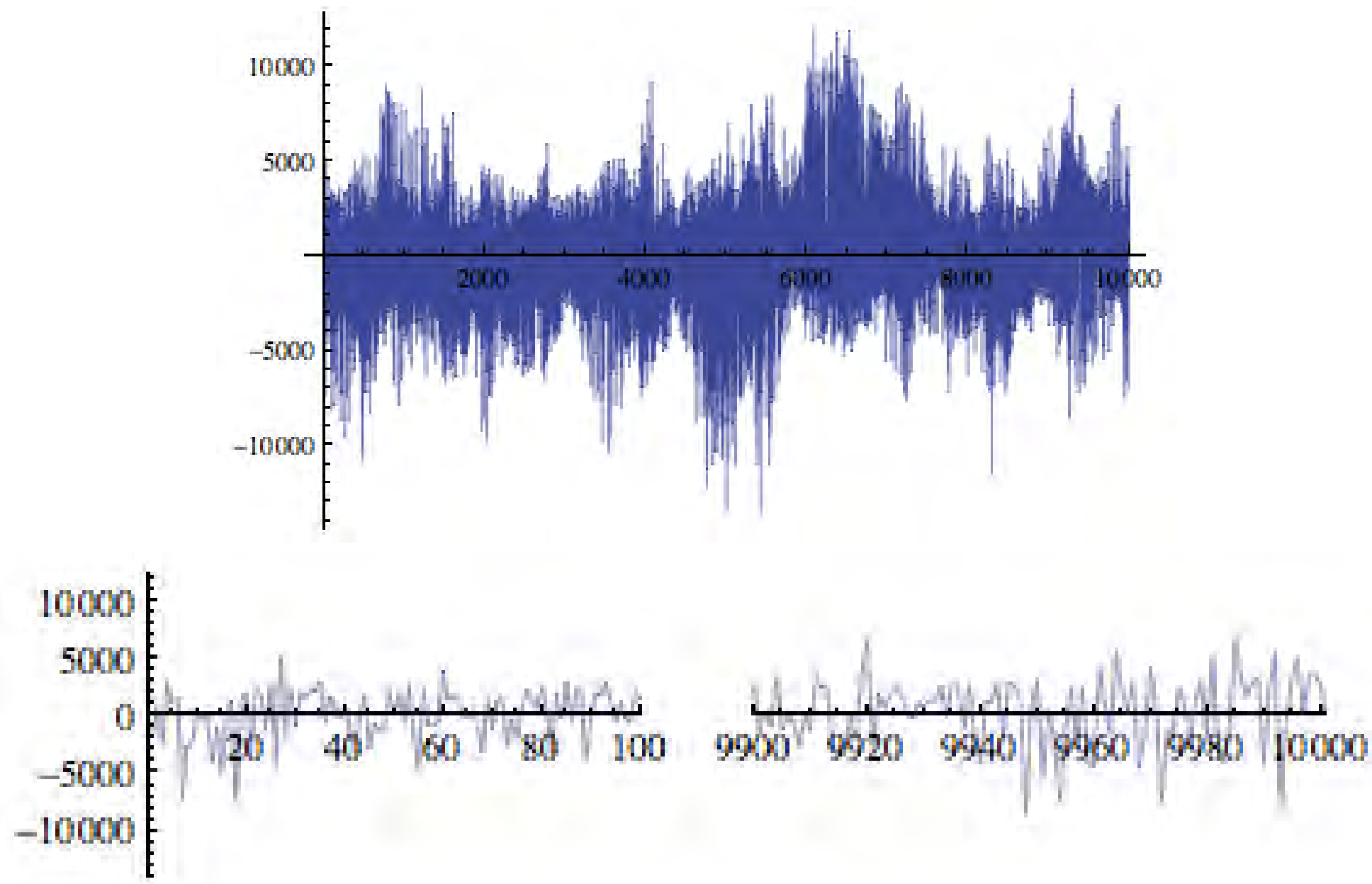

Fig. 19.17 Ez longitudinal field for standard beam on each of 10,000 particles, with detail for the ends.

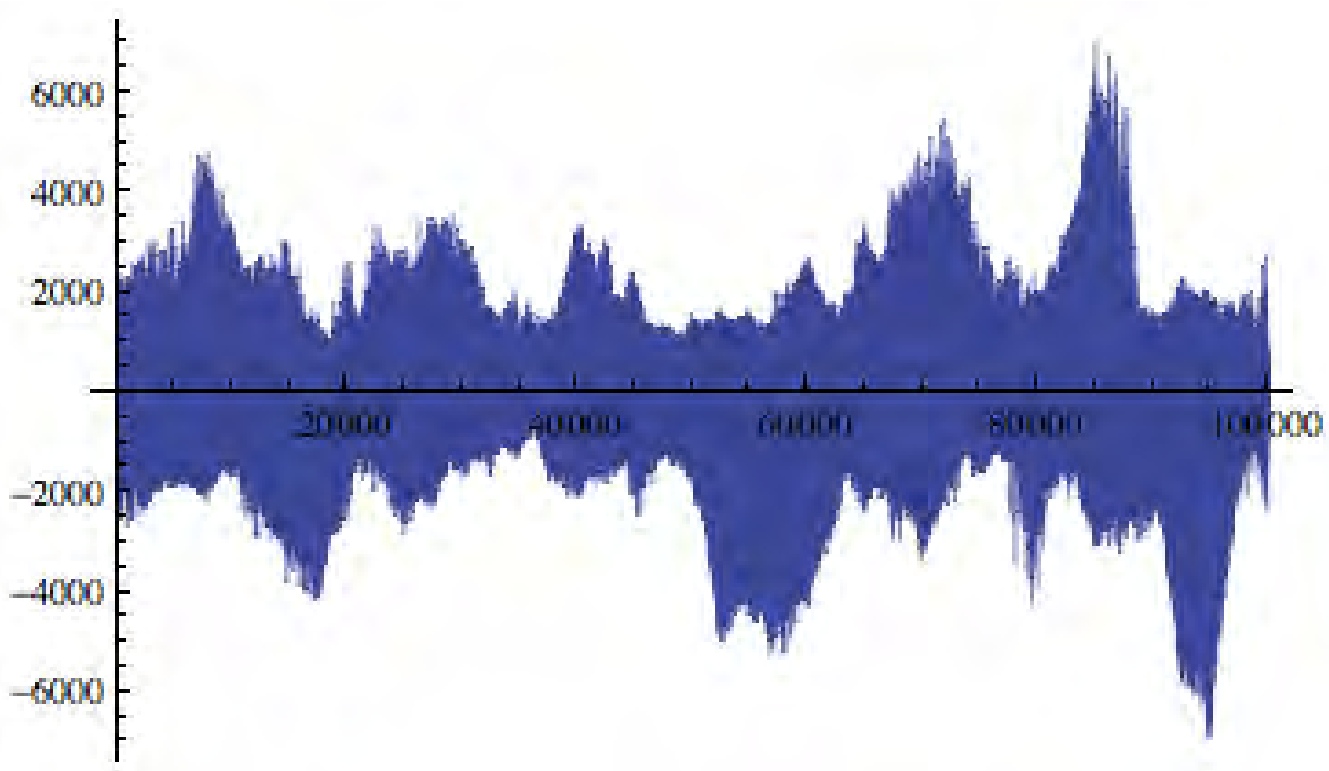

Fig. 19.18 Ez longitudinal field for standard beam on each of 100,000 particles

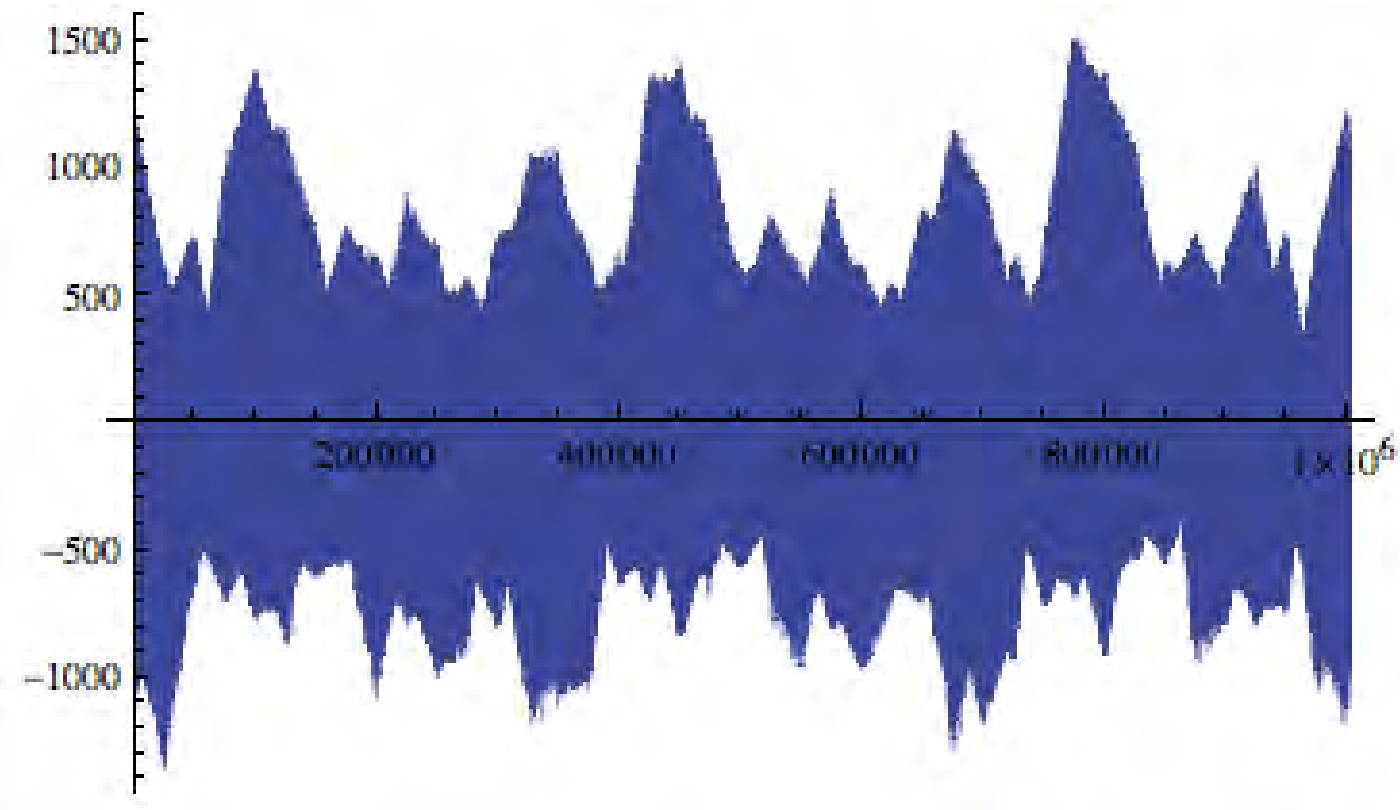

Fig. 19.29 Ez longitudinal field for standard beam on each of 1,000,000 particles.

### 19.3.2 Conclusion – Poisson DC Test

The Poisson space charge computation has very good performance for a dc beam.

The actual RFQ cells, with full vane voltage, image forces, and changing parameters produce much larger potentials and fields. This dc beam test assures that the Poisson solvers are functioning correctly.

Further refinement would require removing the approximations outlined above. That would require that many bunches be included in a long space charge mesh. In older methods, neighboring bunches were treated with a point-to-point charge method, and a minimum number of about 5, but a typical number of 20, bunches on each side of the main bunch were needed. So a factor of ~100 more computational speed would be required to run a simulation with approximately the same run time as at present.

### 19.3.3 Injection Study Stas

Space-charge beam end treatment

Age-old discussion (1997 meeting reference). With faster computation, now time to develop more accurate method.

Now have dynamic problem.

Poisson test (ref J Poisson verification section) with cylinder in cylinder was static, and longitudinal boundary condition seemed satisfied (check with high resolution).

Now have particles. Are transported step by step; then space charge computed and divergences updated (each step in rms, later once per cell is adequate).

Space charge effect on particles:

Consider single slice +-2 cells. Space charge unbalanced at beam ends – trailing particles pushed out of mesh behind – have lost energy. Leading particles pushed out ahead of mesh – have gained energy.

Particles further inside may also get this effect depending on distance from beam center – introducing tilt (check for single beam bunch)

Compensation method(s) may not be perfect, so some of this effect may remain…

Beam is continuous (periodic), so need periodic boundary condition:

Schefftm compensated for this by computing effect of neighboring bunches at point-to-point, and by returning particles to mesh

*BEAMPATH* has periodic longitudinal boundary condition for Fourier solver. But gave up on particles leaving the mesh – if particle leaves mesh, is not included in any further space charge computation. Therefore beam remaining in the mesh looks “dc”. Actually escaped particles go to a leading and trailing bunch, which look like bunches with no space charge.

KRC method – fast wait and slow particle catchup.

Some change by 2pi. E.g., *LIDOS*. But when exactly do they do the 2pi??

*LIDOS* Stas method – two bunches are injected, forward one discarded after first period. But does not compensate for particles that leave the trailing edge.

All of these have some unbalance at the beam ends. Space-charge pushes particles out of the mesh. So they have less or more energy than they should if it were pure dc beam. And energy stays wrong with re-insertion procedure.

Original Message ----- AD> From: <jameson@riken.jp>
To: "Alexandr Durkin" <apdurkin@mtu-net.ru>; <dovs45@mail.ru>
Cc: <maus@iap.uni-frankfurt.de>
Sent: Monday, February 14, 2011 6:09 PM
Subject: injected beam

Dear Sasha,

Thanks very much for your quick reply about real vs. ideal fields. I learned from that, and now am anxious to learn about the following:

Please tell me exact details about how you treat the injected beam in *LIDOS*:

The Vanes Real Shape part makes an input section - for our examples, the outer wall is 6mm, plus the space to vanes beginning equals one period in length. The mesh starts at the upstream outer wall; z=0 here.

The input distribution ranges from 0 degrees to +360 degrees.

How is the input beam positioned in z with respect to the input section mesh? (Is 0 degrees positioned at z=0?)

Do you apply space charge at every step, starting with the first step?

Etc.

When you compute the input match, the paper says you know what the matched slices are and then compute "back". Please send much more detail - for example, how do you know what the matched slices are? Do you compute particles back, or ellipses? How far back do you compute? Etc.

I look forward very much to your reply. I think these aspects may be significant to explain the apparent differences we have been struggling to explain, and I want to change several aspects of our code to be sure they are equivalent to yours. It will be a great relief if this turns out to be true.
With best regards, ... Bob

Stanislav Vinogradov, 2/17/11 12:29 AM +0300, Re: injected beam 1
From: "Stanislav Vinogradov" <vinogradov@mtu-net.ru>
To: "Alexandr Durkin" <apdurkin@mtu-net.ru>,
"Bob Jameson" <jameson@riken.jp>
Subject: Re: injected beam
Date: Thu, 17 Feb 2011 00:29:10 +0300
Dear Bob,

The procedure of beam injection is rather complex. During the very first period of time T the movable (space charge) mesh doesn't move, it begins to move with the beam only when t>T. Initially (at time t=0) this mesh is filled with so-called preliminary particles (these particles are not shown in cartoons), with input distribution in x and y and uniformly in z. These preliminary particles leave the mesh and the system during the period 0<t<T, because the phase jump in 2pi is not applied to particles if t<T. But these particles participate in space charge field creation, they help to avoid large z-field and lack of x and y fields near the head edge of the injected beam. So during the period 0<t<T preliminary particles gradually leave the mesh, while new injected particles gradually fill it. New particles are injected with input distribution in x and y and uniformly in time, and they are influenced by space charge fields from the moment of the injection. So at moment t=T we get movable mesh properly filled with particles, and the mesh begin to move with them.

The matched slices are ellipses, not particles. They are computed back from the beginning *of regular part of the linac (This was an important clue, but not understood until meeting in Moscow 2012!!)*, after input section. It is Sasha's code, probably he'll describe it in more details.

Best wishes,   Stas

This injection method was adopted for *LINACSrfq*.

## 19.4 Input Matching [151]

### 19.4.1 Overview

At the RFQ input, a beam coming from a time-independent focusing system has to be "matched" into the time-dependent focusing fields of the RFQ.

The dc input beam arrives with transverse phase-space orientation that is independent of time. Imagine 1° segments injected sequentially over one rf cycle. The RFQ acceptance varies with the phase of the rf voltage. Direct injection into full field would be mostly mismatched and most particles would not be accepted. Although time varying, the RFQ is still a standard focusing channel and is characterized by its betatron and synchrotron motion. The standard smooth approximation method separating the equations of motion into a "slow part" and a "fast part" gives the quantities needed for the channel design. The "slow part' gives the betatron and synchrotron oscillation phase advances, with or without beam current. The "fast part" describes the parameters during one rf cycle. An RFQ input section is needed that brings all phases of the injected beam into a common "slow part" ellipse.

The miracle of Crandall's solution to this problem is that a very simple "radial matching section (RMS)" accomplishes this, basically by raising the transverse focusing strength linearly from zero to its full value in a few (usually 4 or 6) cells. However, the xx' and yy' ellipse parameters of the entering beam have to be found, for the RFQ simulation program, and also for real operation.

### 19.4.2 Matching Methods

Usually, the RFQ input beam ellipses are assigned the same alpha and beta for both xx' and yy' phase spaces, although somewhat better transmission and/or accelerated beam fraction performance may be obtained for separate assignments.

Assume that the maximum transmission and/or the maximum accelerated beam fraction are the criteria for the input ellipse parameters.

#### *19.4.2.1 Transmission Matching Method*

The most reliable way to find the best transmission alpha and beta is to make a matrix of transmission results over a grid of input ellipse alphas and betas [152], make a contour plot and decide on the best alpha and beta by eye. The transmission surface may be bumpy for the initial choice of grid; in this case, a finer grid should be used – alpha steps of 0.05 and beta steps of 0.5 are usually satisfactory. This match will be termed the "transmission match". A disadvantage is that a long time is required to make so many full runs.
The transmission optimum can also be found using a nonlinear optimization program such as NPSOL [153], but the search time can be comparable to the time needed to make a map, and the method is subject to the vagaries of nonlinear optimization, so making the map is recommended.

#### *19.4.2.2 Front-End Matching Methods – Existing Models*

151 RFQ INPUT MATCHING METHODS STUDY, R.A. Jameson, August –September 2012

152 This is done conveniently using irunopt(8) in LINACSrfqSIM.

153 User's Guide for NPSOL 5.0: A FORTRAN Package for Nonlinear Programming, P E. Gill, W. Murray, M. A. Saunders, M. H. Wright, Technical Report SOL 86-6_, Revised June 4, 2001

A way to quickly determine an input match has long been sought. Many methods using some cells just after the end of the RMS have been programmed[154].

*19.4.2.2.1 Quasi-Periodic Matching Methods*

For example, by trying to have the same ellipse parameters across one or more focusing periods just after the RMS, as if in a non-varying periodic focusing system. The problem with this is that the RFQ input is highly non-periodic, not adiabatic and contains the tank wall and drift region.

*19.4.2.2.2 Phase Wobble Matching Method*

A logical approach is to minimize the scatter in the segments described above. The term "phase wobble matching" is coined to describe this approach.

One method is to use as input the design parameters of a match at the end of the RMS, as a function of time at, say, 1° intervals, and to compute these backwards through the RMS – adjusting the input parameters until the scatter at the input is minimized.

*LIDOS* computes the ideal emittance parameters of ~360 2° beam slices "at the end of the radial matching section" [155] (*See important footnote.*), and transports each slice backward to the entrance, where there would be one common ellipse for a perfect radial matching section, but typically there will be some scatter. Then the best fit common ellipse is found. The *LIDOS* match is good, and usually yields transmission near that of the transmission map.

*PARI/PARMTEQM* uses (separately) the program 'rfquick' in the same manner. The smooth approximation formulae for a uniform charge distribution in a cylinder, including space charge, are used at the end of the radial matching section (RMS) to find the total, minimum and maximum beam sizes. Many points across this interval are computed backwards with many small steps to the RFQ entrance using a 2x2 matrix transformation, (time is reversed, but true time reversal is not used (the r matrix is not inverted)), and the average alpha and beta constitute the input match. The 'rfquick' procedure is a good use of approximate design equations for a cylindrical beam, is fast, and gives a reasonable match, but again it typically yields lower transmission than the transmission match, and generally not as good as the methods of Sec. 19.4.4.

*19.4.2.2.3 Beam Behavior Observation Matching Method*

Attempts to minimize the betatron oscillation by observing the simulated beam behavior were also not successful and not amenable to an optimization program; this is because observing the beam alone has no firm basis upon which to minimize.

*19.4.2.2.4 Summary*

None of these methods gives as high a transmission as the result of the transmission matrix. The methods can give reasonable estimates, which can be used to position the alpha-beta matrix for a transmission search.

---

154 By many researchers. *LINACSrfqSIM* is set up to make it easy to explore different methods, either by making a contour map of an objective function as a function of the variables, or by using a constrained nonlinear optimization program to seek an objective function minimum.

155 Important footnote: here is written "at the end of the radial matching section" – upon reading further, it will be seen that that Durkin wrote "at the end of the radial matching section, but Vinogradov wrote "at the beginning *of regular part of the linac"*. Matching to the end of the rms has never worked well, and the reasons are developed below. In 2012, it became clear that matching to the beginning of the regular part of the linac, i.e. for the RFQ at the end of the shaper, where the beam is finally well formed, is actually what LIDOS does – and which now makes sense and works well.

### 19.4.3 Matching Method Criteria

Desired criteria for a matching method are:

1) Should work for any linac (not only an RFQ, although this is perhaps the hardest because it is highly non-periodic.

2) Should require use of only the early part of the channel – not a simulation through the whole linac.

3) Should move the actual beam forward or backward, from the input over this early part of the channel.

4) Should be available in the simulation program at all levels (e.g., in RFQ simulation at full Poisson level (slow running time, RFQ entrance and vanes sections are simulated with accurate geometrical, space-charge and image effects), or approximate but fast 2-term level), but accuracy using fast version should be good compared to slow version.

5) Should be relatively easy to use; programmed as a standard program result or on call as a separate run, without requirement for additional interim input.

6) should be inexpensive – fast running time, minimum number of channel computations.

### 19.4.4 Matching Method Development

Four new methods are developed, particularly for the RFQ, which is the hardest situation:

- using description of the dynamic aperture of the RFQ
- using a forward form of the phase wobble matching method
- using a complete description of beam performance based on the design physics.
- using a backward simulation method, based on beam performance and the design physics.

The development is progressive – at each step, new information is obtained which is useful in the next step.

### 19.4.5 Dynamic Aperture Matching Method

#### *19.4.5.1 Acceptance*

Acceptance means the acceptance area of particles from the input transverse space, which are transmitted through the whole RFQ with only the external fields and zero current.

##### *19.4.5.1.1 Classical acceptance, all particles at synchronous phase and energy*

The classical transverse acceptance tool in PARMILA, *PARMTEQM, pteqHI, LINACSrfqSIM* [156] is to use a "type 9" input particle distribution over an xx' or yy' grid, all at the synchronous phase and energy, and zero beam current. In *LINACSrfqSIM*, plotting "rdinptOrigz.txt" and "tapeACCEPTED.txt" gives a result like Fig. 19.30. The size of the grid must be adjusted to get good resolution for the number of particles used.

156 R. A. Jameson, Principal Investigator, et. al., "Scaling and Optimization in High-Intensity Linear Accelerators", LA-UR-07-0875, Los Alamos National Laboratory, 2/8/2007 (introduction of *LINACS* design code), re-publish of LA-CP-91-272, Los Alamos National Laboratory, July 1991; R.A. Jameson, "A Discussion of RFQ Linac Simulation", Los Alamos National Laboratory Report LA-CP-97-54, September 1997. Re-published as LA-UR-07-0876, 2/8/07; "RFQ Designs and Beam-Loss Distributions for IFMIF", R.A. Jameson, Oak Ridge National Laboratory Report ORNL/TM-2007/001, January 2007; "*LINACS* Design and Simulation Framework", R.A. Jameson, KEK/J-PARC Seminar, 6 March 2012.

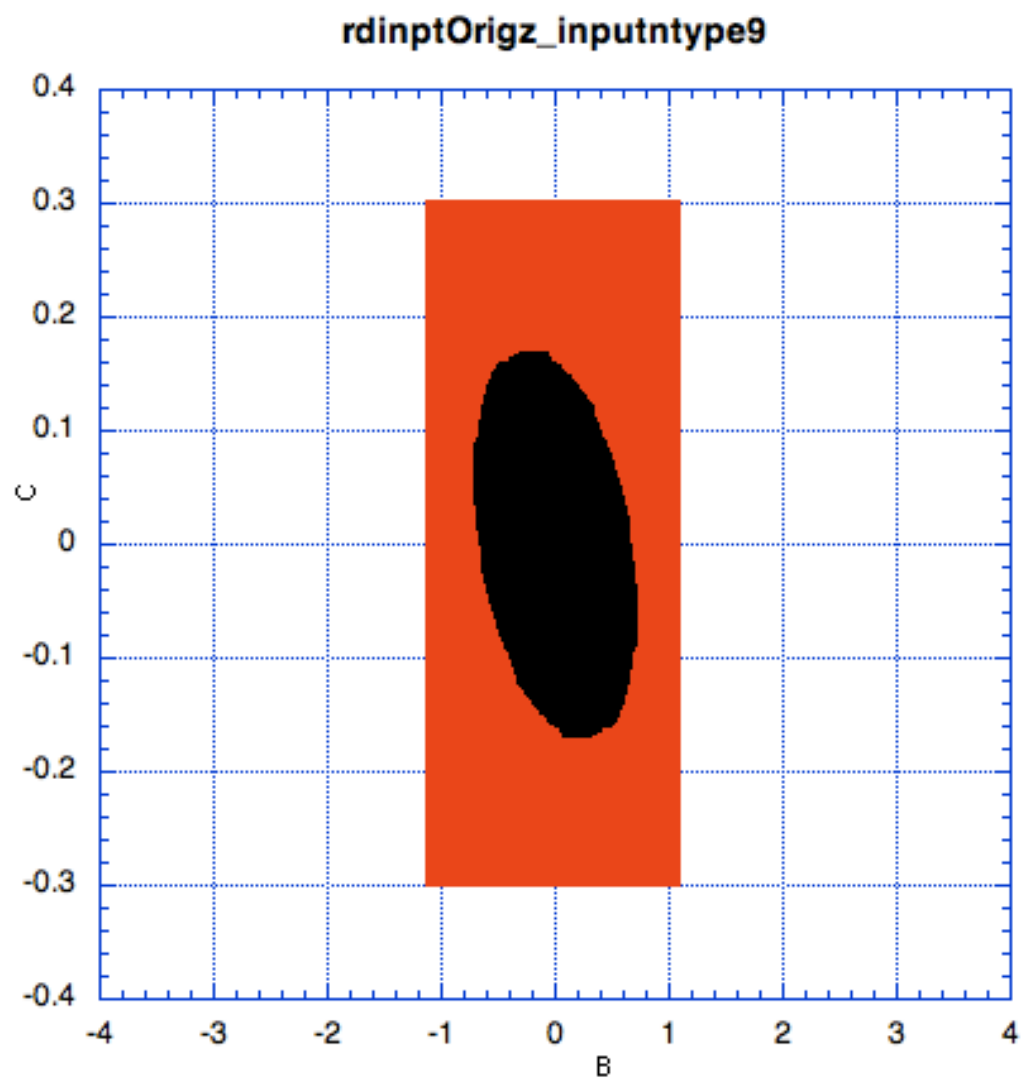

Fig. 19.30 Black - xx' acceptance (zero current, type 9 input);
Red – xx' input particle grid.

```
input 9 1 -1.1 1.1 200 -0.3 0.3 200
  0. 0.0 0.0
 0.0 0.0 0.0 0.0 0.0 0.0 34. 238.051 0.0 0.88726418
 TRANSMISSION ACCEPTANCE X-XP:
alphax   betax  exrmsreal extotalreal   xmax    xpmax
0.3677   4.3401  0.0272     0.1090        0.6910  0.1673
```

*19.4.5.1.2 Acceptance for full dc beam*

Placement of all the particles at the synchronous phase and energy is a rather special case more appropriate for investigating the acceptance of a transport or accelerator system with a well-bunched input beam. The RFQ has a dc beam input, and particles are accepted in transverse over a wide range of longitudinal phases. This more meaningful acceptance can be explored using a normal "type 6" dc input – zero current, and set the transverse alpha, beta and total emittance so that the accepted particles fall within the initial distribution. Fig. 19.31 shows an example with alpha=0.

```
input -6 -100000 0. 10. 1.0 0. 10. 1.0
   180. 0.0 0.0
 0.0 0.0 0.0 0.0 0.0 0.0 34. 238.051 0.0 0.88726418
 TRANSMISSION ACCEPTANCE X-XP:
alphax   betax  exrmsreal extotalreal   xmax    xpmax
0.8293   4.9615  0.0333     0.1332        1.0422  0.2668
```

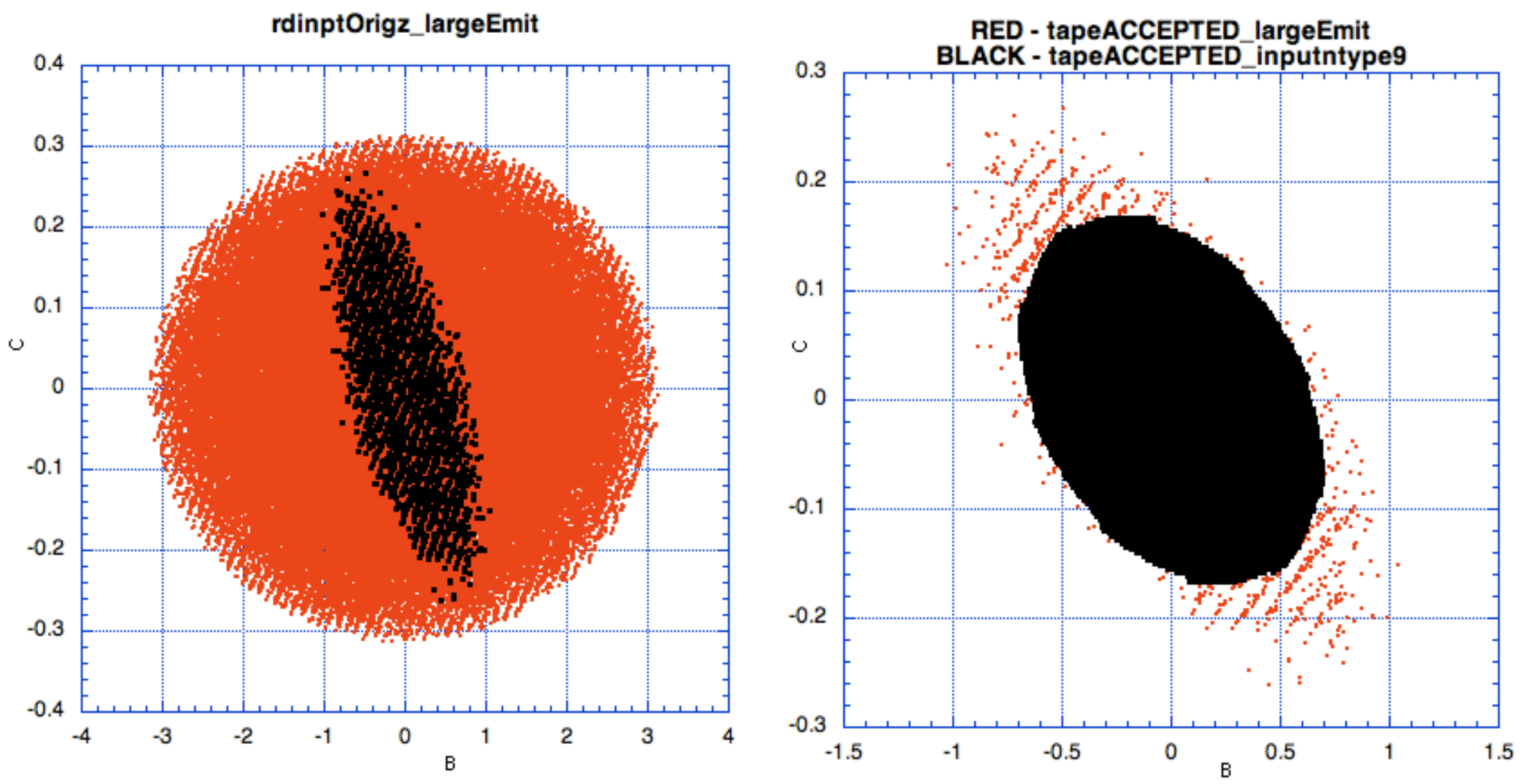

Fig.19.31a. Typical pattern. Red - normal Type 6 input beam, alpha=0, emittance and beta large enough to encompass all the accepted particles. Black – accepted particles. Fig. 19.31b Comparison of transverse acceptance:for particles all injected at synchronous phase and energy

| alphax | betax | exrmsreal | extotalreal |
|---|---|---|---|
| 0.3677 | 4.3401 | 0.0272 | 0.1090 |
| vs. over full dc input, zero current | | | |
| 0.8293 | 4.9615 | 0.0333 | 0.1332 |

*19.4.5.1.3 Dynamic Aperture*

Acceptance can also be measured the same way with current, and constitutes the “dynamic aperture”.

*19.4.5.1.4 Transverse Dynamic Aperture*

Again, an input ellipse is specified with parameters that will give an ellipse somewhat larger than the accepted particles. Fig. 19.32 shows the accepted particle distributions vs. input current.

```
input -6 -100000 1.2 7. 0.5 1.2 7. 0.5
 0. 0.0 0.0
 0.0 0.0 0.0 0.0 0.0 0.0 34. 238.051 ParticleCurrent,mA    0.88726418

 TRANSMISSION ACCEPTANCE X-XP:
 Current  alphax     betax      exrmsreal   extotalreal
  0 mA    0.3669     4.3340       0.0308      0.1233
 10 mA    0.4906     5.4258       0.0210      0.0840
 50 mA    1.0743     7.6478       0.0230      0.0921
100 mA    1.3614     9.9417       0.0233      0.0932
```

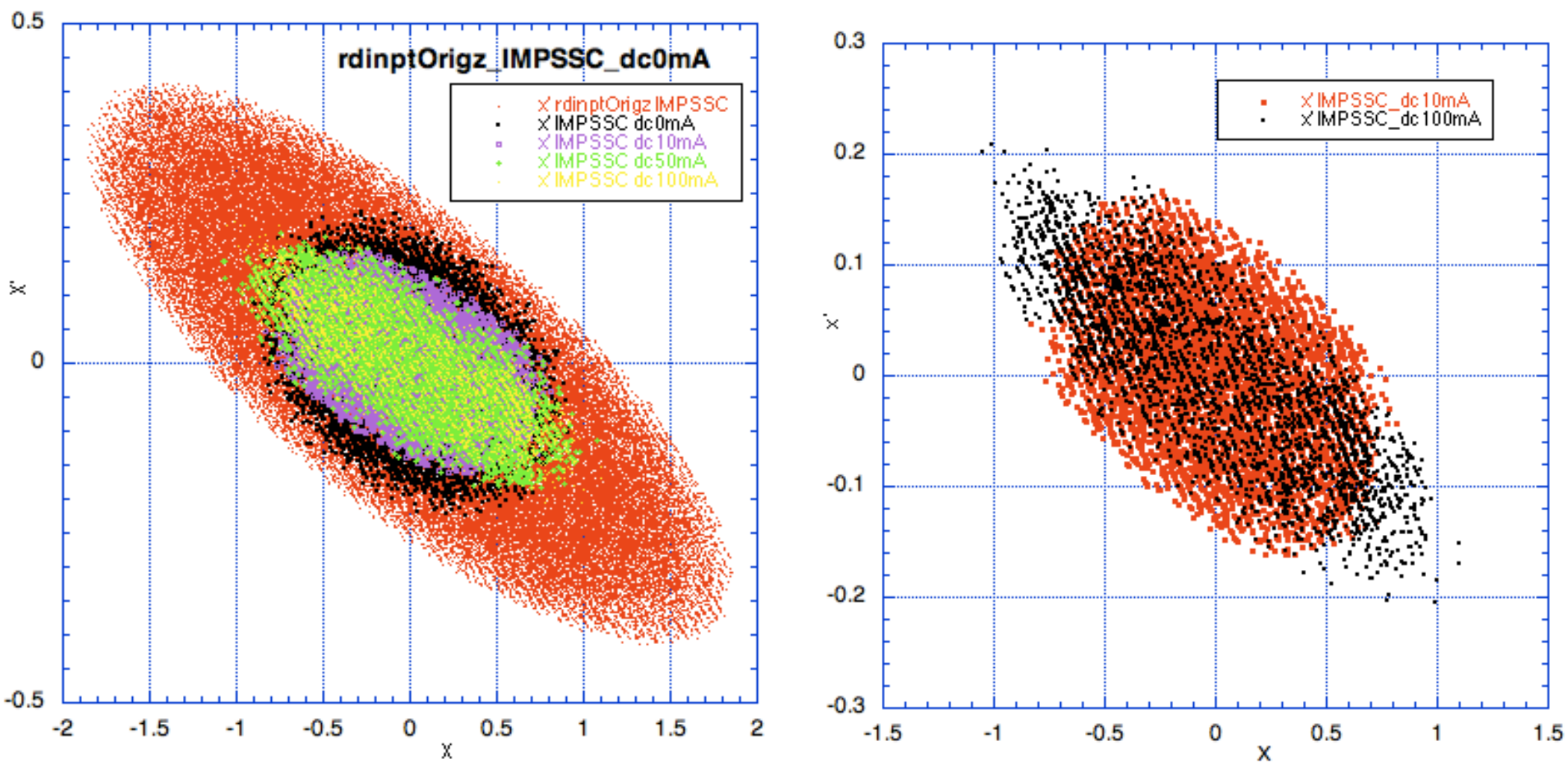

Fig. 19.32a. Red – the input dc distribution. Other distributions are the accepted particles for 0, 10, 50 and 100mA input particle currents.
Fig. 19.32b. The accepted particles for10mA and 100mA input particle currents.

The design particle current for this particular RFQ is 0.0147 mA ≈ 0 mA.

At 0 mA, the accepted particles lie in an ellipse. For this RFQ at 10 mA, the distribution has shape and orientation still more similar to the zero-current ellipse, but is no longer a true ellipse – it is somewhat "squared off" in x at ~x=±0.75. But at higher currents, the accepted particles lie in a longer, narrower area with a more elliptical shape.

This suggests a matching method.

*19.4.5.1.5 Input Matching Using Dynamic Aperture*

An example RFQ with design input beam current = 130mA and input total real transverse emittance of 0.01491 cm.rad is taken, for which the transmission grid match (Fig. 19.33) shows best transmission for alpha = 2.2, beta = 9.8:

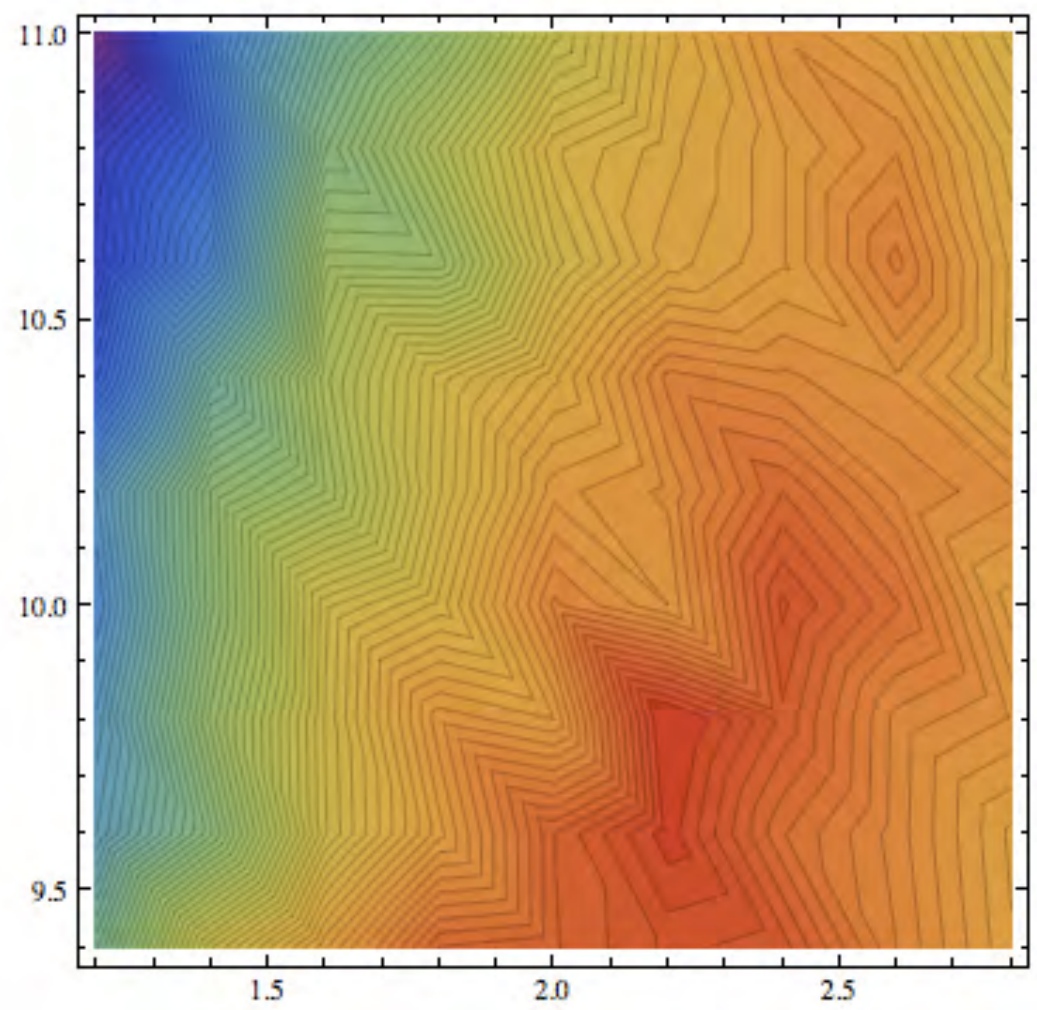

Fig. 19.33 Transmission grid contour map – match point is at alpha=2.2, beta=9.8.

Again specify a large enough input ellipse, here with the best transmission alpha and beta and an area of 0.3 cm.rad, and look at the dynamic aperture as function of beam current, Fig. 19.34.

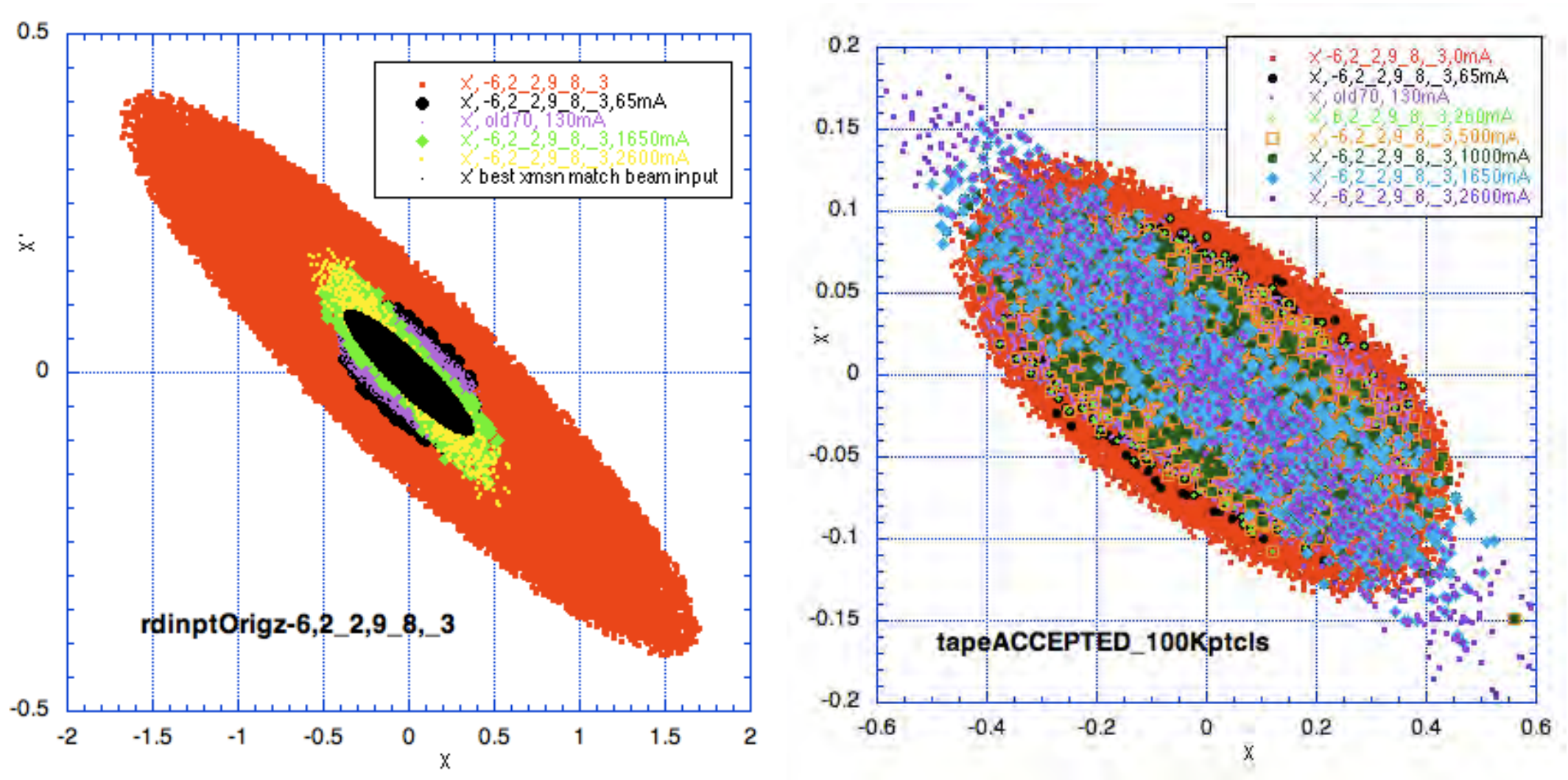

Fig. 19.34 Dynamic aperture vs. beam current.

```
input -6 -50000 2.2 9.8 .3  2.2 9.8 .3
   180. 0.0 0.0
 0.0 0.0 0.0 0.0 0.0 0.0 1. 2.0145 130.0 0.095
```

TRANSMISSION ACCEPTANCE X-XP:

| Current | alphax | betax | exrmsreal | extotalreal |
|---|---|---|---|---|
| 0 mA | 0.7632 | 4.4156 | 0.0092 | 0.0370 |
| 65 mA | 0.7425 | 4.4099 | 0.0066 | 0.0264 |
| 130 mA | 0.8259 | 4.7397 | 0.0065 | 0.0260 |
| 260 mA | 0.8755 | 4.7776 | 0.0066 | 0.0264 |
| 500 mA | 1.0555 | 5.3425 | 0.0065 | 0.0261 |
| 1000 mA | 1.5117 | 6.5810 | 0.0064 | 0.0256 |
| 1650 mA | 2.2570 | 8.4582 | 0.0063 | 0.0250 |
| 2600 mA | 3.6791 | 11.5998 | 0.0063 | 0.0252 |

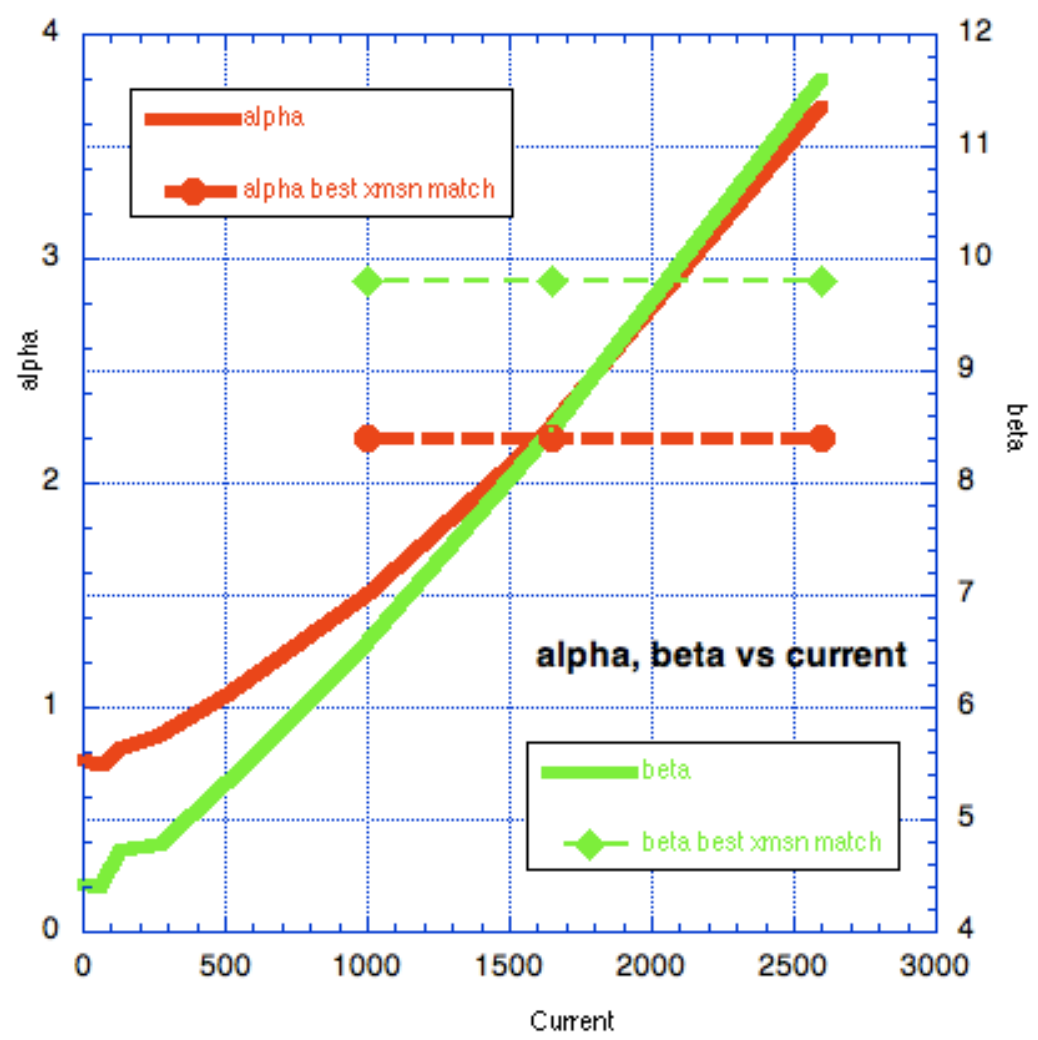

Fig. 19.35 Variation of dynamically accepted ellipse alpha and beta with input beam current.

The behavior is similar to the previous RFQ in Fig. 19.32. Space charge changes the input beam, and over the range of ~0.5-20 times the design current, a quite constant input transverse extotalreal area of ~0.025-0.026 cm.rad is accepted dynamically.

It is seen in Fig. 19.35 that at 1560mA, around ten times the design current, alpha is close to best transmission match alpha=2.2, while beta=8.4 is smaller than the best transmission match beta=9.8. Some scaling factor on the current for the larger input ellipse is to be expected – a direct scaling would be 130*(.3/.01491)=~2600mA, but this appears to be too high.

We next explore the feasibility of an RFQ input matching method using a rule-of-thumb measurement of the dynamic aperture, with an input ellipse area 20x the design area, and an input current of ~10x the design input current.

A family of 16 RFQs is chosen in which only one design parameter is changed systematically – the aperture at the end of the shaper, equal to $\beta\lambda$/(aperfac), where $\beta\lambda/2$ is the cell length. Transmission maps were made.

Dynamic aperture runs were made for an input ellipse area of 0.3 cm.rad and currents ranging across 1-20 times the design current {130,630,1130,1630,2130,2630} mA, recording the ellipse alpha, beta and dynamic aperture (Fig. 19.36).

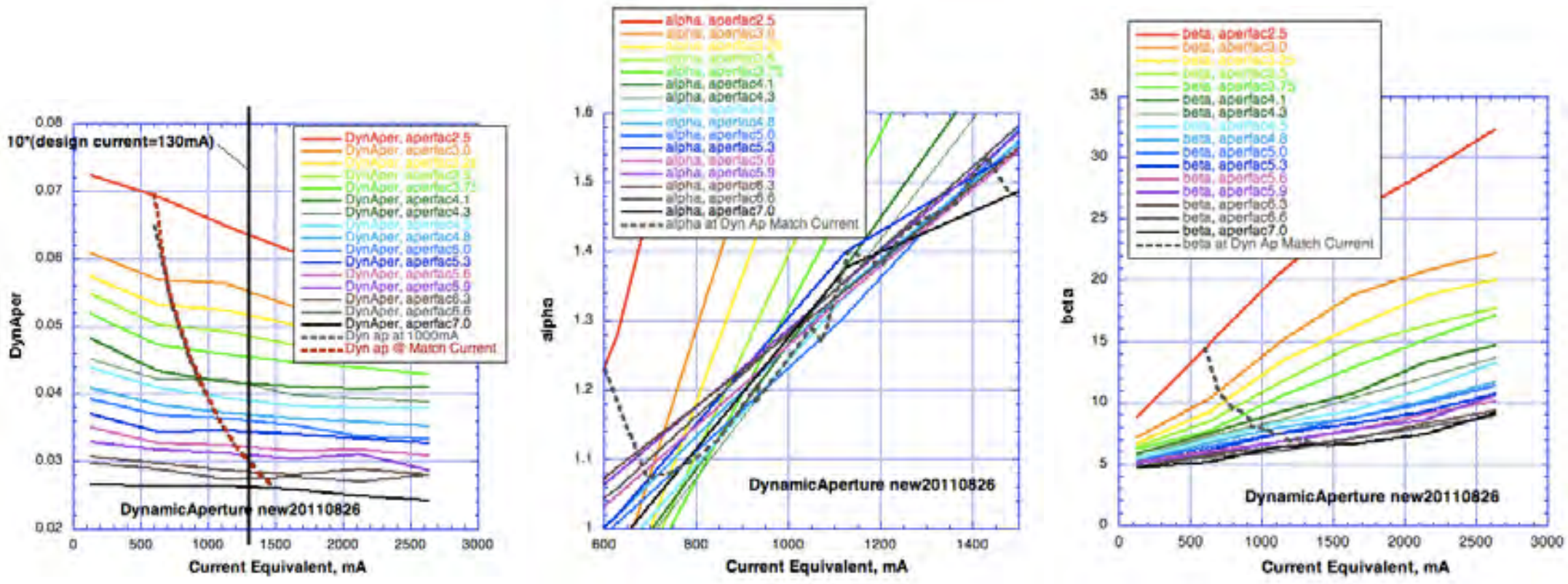

Fig. 19.36. a,b,c. Emittance (dynamic aperture), alpha and beta of the ellipse fitted to the rms dynamic aperture, vs. beam current, for the 16 RFQs set 'new_20110826'.

The dynamic aperture at ~10 times the design current (1130mA) was used to scale a "Match Current = 130mA(0.3/(Dynamic Aperture at 1130mA))" . Then at this match current, the "alpha at Dyn Ap Match Current from 1130mA" and "beta at Dyn Ap Match Current from 1130mA" (Fig. 19.36 dashed lines) were chosen as the input match parameters. These are plotted against the aperfac in Fig. 19.37a and 19.37b. Fig. 8c shows the resulting transmission and accelerated fraction, compared to those from the transmission match.

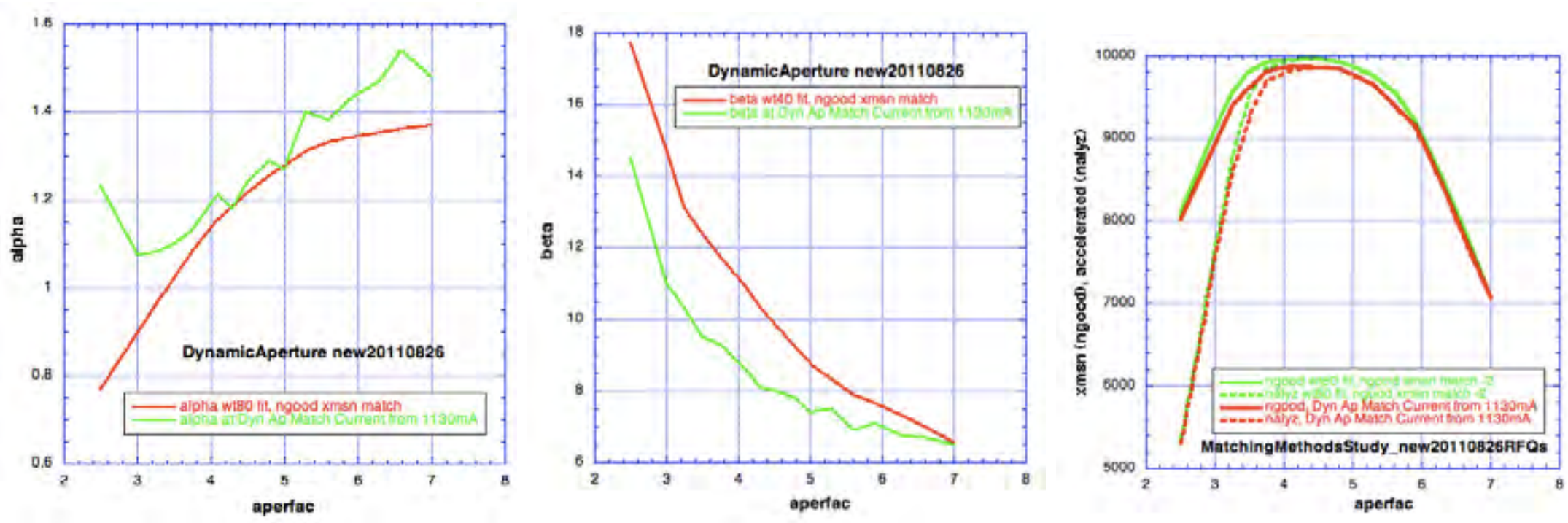

Fig. 19.37 Comparison of match alphas, betas and resulting transmission and accelerated fraction, at the 130mA design current, between the transmission match and the match determined from the dynamic apertures.

For the smaller aperture RFQs (aperfac >= ~3.5) , the alphas agree quite well. The betas agree less well. However, the transmission and accelerated fractions agree very well – they are not sensitive to this range of alpha and beta..

The dynamic aperture matching procedure should give a quite reasonable match, which could serve as the starting point for making an alpha/beta transmission map.

*19.4.5.1.6  Conclusion – Dynamic Aperture Matching*

***The Dynamic Aperture Matching Method yields a good estimate of the input match, but does not satisfy the matching method criteria 2) Should require use of only the early part of the channel – not a simulation through the whole linac, and 5) Should be relatively easy to use, without requirement for additional interim input, and 6) Should require few channel computations***

*19.4.5.2  Longitudinal Dynamic Aperture*

The RFQ is primarily a transport device, so a broad range of energy can be transported. Fig. 19.38 shows a characteristic longitudinal dynamic aperture.

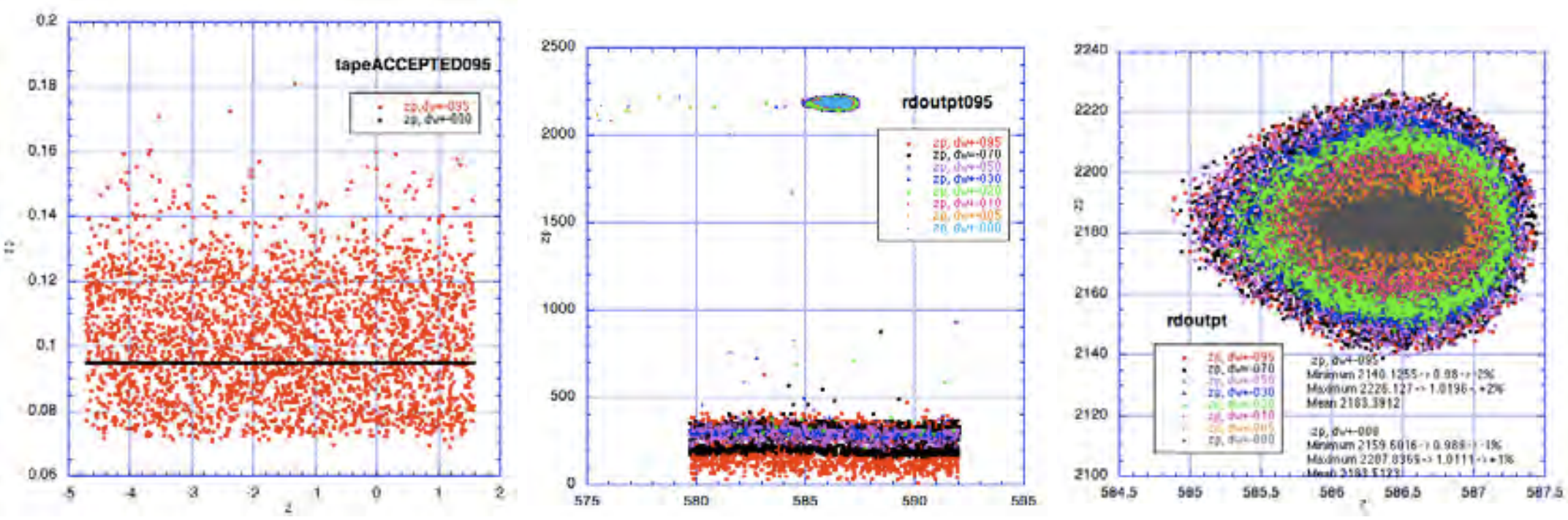

Fig. 19.38a. Red – Input z, z' coordinates of particles transmitted from input with ±0.095 MeV (±100%) energy spread. Black – usual dc input distribution.

Fig. 19.38b. Zz' coordinates of particles transmitted at the RFQ output, vs energy spread of the input distribution. The accelerated particles in the bucket lie above. Below are the low energy particles lost from the bucket but still transported to the end of the RFQ.

Fig. 19.38c. Zz' coordinates of the accelerated bunch at the RFQ output, vs energy spread of the input distribution. Usual dc input distribution results in ±1% energy spread, but ±2% can be accepted.

## 19.4.6  Phase Wobble Matching Method

We want to investigate a phase wobble matching method, as outlined above, at the end of the RMS, but with forward-running.

The derivation of the RFQ transverse dynamics smooth approximation is given in [157] as the solution to the Mathieu equation for the beam size:

$$x(\tau)=C_0\, e^{j\sigma^t \tau}\,(1 + C \sin 2\pi\, \tau) \qquad (1)$$

where $\sigma^t$ is the transverse phase advance with beam current per focusing period, and $C \sim B/(4\pi^2)$ is the fast "wiggle" which oscillates once per focusing period.

The "flutter factor" psi = (1+B/4pi^2)/(1-B/4pi^2) is also sometimes used; it is the ratio of the maximum emittance ellipse beta to the minimum ellipse beta [158].

157  M Weiss, "Radiofrequency quadrupole", Proc. CAS (Aarhus, Denmark, 1986); CERN 87-10 (1987)

158  G. Parisi, "Investigations on Particle Dynamics in a High Intensity Heavy Ion Linac for Inertial Fusion", Dissertation, Inst. Angew. Physik, Johann WolfgangGoethe Universität, Frankfurt. 1999.

A matching method was set up to compute the x and y beam sizes of a 2° slice of beam located just after the end of the RMS, at each step as the beam steps through. The tracked beam is two cells, 360° long. By computing forward, space charge effects (and image effects if using full Poisson simulation) are included. Fig. 19.39 shows the excellent agreement with the design beam at the best transmission match.

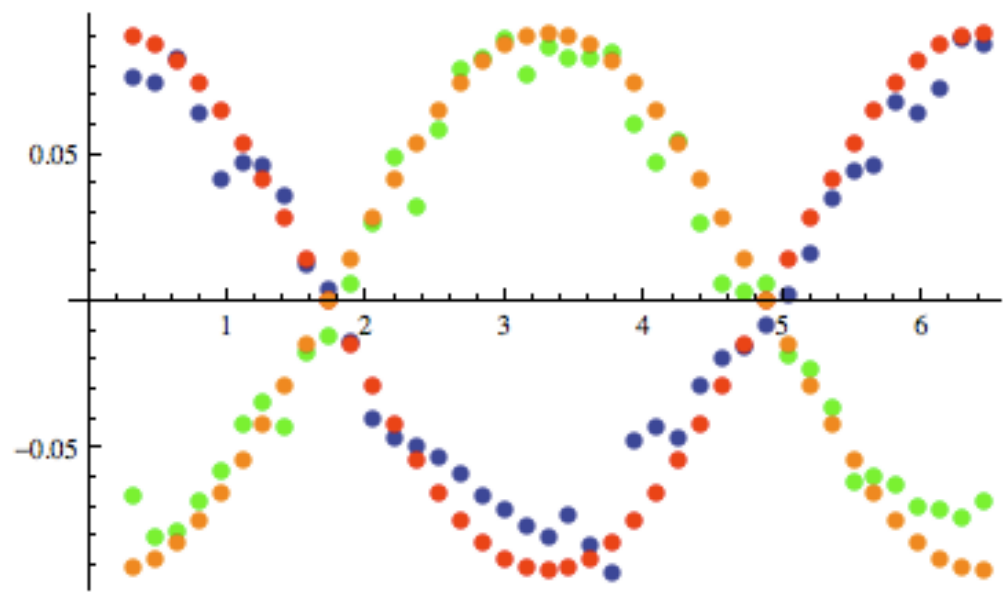

Fig. 19.39 X (blue) and y (green) are the variation of beam size from the average design value, of beam slices located in the 2°, $\beta\lambda/180$ region just after the end of the RMS, for 20 steps per cell. The input beam $\alpha$ and $\beta$ are for the best transmission match. Red and yellow are the x and y beam sizes from Eq.(1) (which does not include space charge effect) evaluated at the end of the RMS.

Varying the input match produces variations such as shown in Fig. 19.40. It would appear that optimization would be possible, by minimizing the sum over the steps of the squared differences between the simulated and model x and y beam sizes.

However, disappointment was in store. Fig. 19.41 shows a contour map of the objective function. There is a long trough toward larger $\alpha$ and $\beta$, with no bounded minimum.

*19.4.6.1 Conclusion – Phase Wobble Matching Method to End of Radial Matching Section*

***Therefore, the forward-running phase wobble method is not suitable for finding the best input matching conditions, at least not with comparison to design information at only one place.***

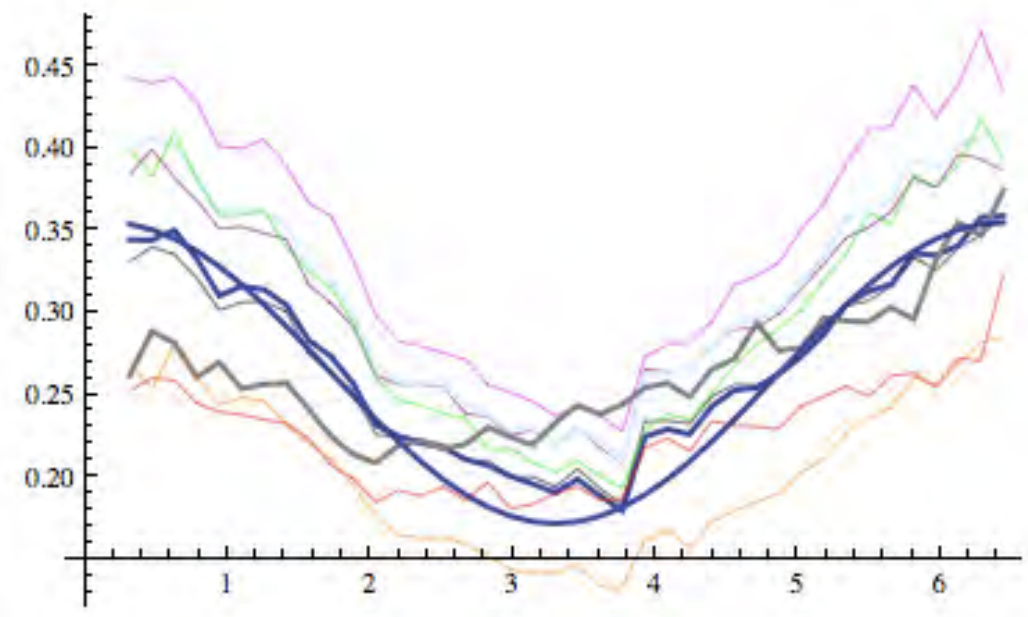

Fig. 19.40 Typical variations in the x-beam-size as the input ellipse a and b are varied. The smooth curve is the desired variation according to the RFQ design.

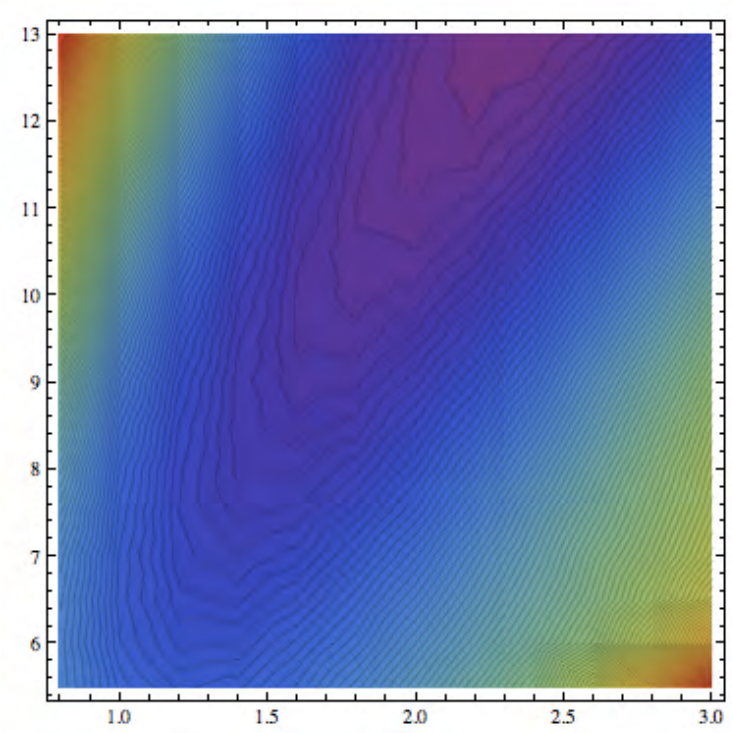

Fig. 19.41 Contour map of the phase wobble objective function as function of the input match α (abscissa) and β (ordinate). Red is high, violet is small.

### 19.4.7 Design Envelope Matching Method

We investigate a rigorous beam observation procedure based on knowledge of the design external field and space charge physics.

*19.4.7.1 Design/Simulation Code Correspondence*

*LINACSrfqDES* controls all RFQ parameters, including the space charge physics, so that whatever the design rules are, the beam itself is controlled in the design by requiring that the rms beam envelope is matched transversely and longitudinally at each cell. Thus the rms beam envelope is known at each cell. *LINACSrfqDES* is unique in using a good approximation of the external field (8-term multipole), applied in the design at the rms beam sizes, so that the design and *LINACSrfqSIM* are fully connected. Thus, if the design is working, the computed beam envelopes should agree with the design envelope.

*19.4.7.2 Comparison of Design and Simulated Beam Sizes*

A waterbag distribution is injected into the RFQ, so the (design total beam size) = (sqrt(6)*(design rms beam size).

The x and y rms beam sizes are computed at the center of each cell, where the beam is essentially round.

Fig. 19.42 shows, for the aperfac7 RFQ we have been studying, the total design and simulated beam sizes, with the simulated rms beam sizes multiplied by sqrt(6) and sqrt(7). The beam was injected with match α and β found from a transmission map. There is still some initial oscillation, which finally damps out after ~cell 100.

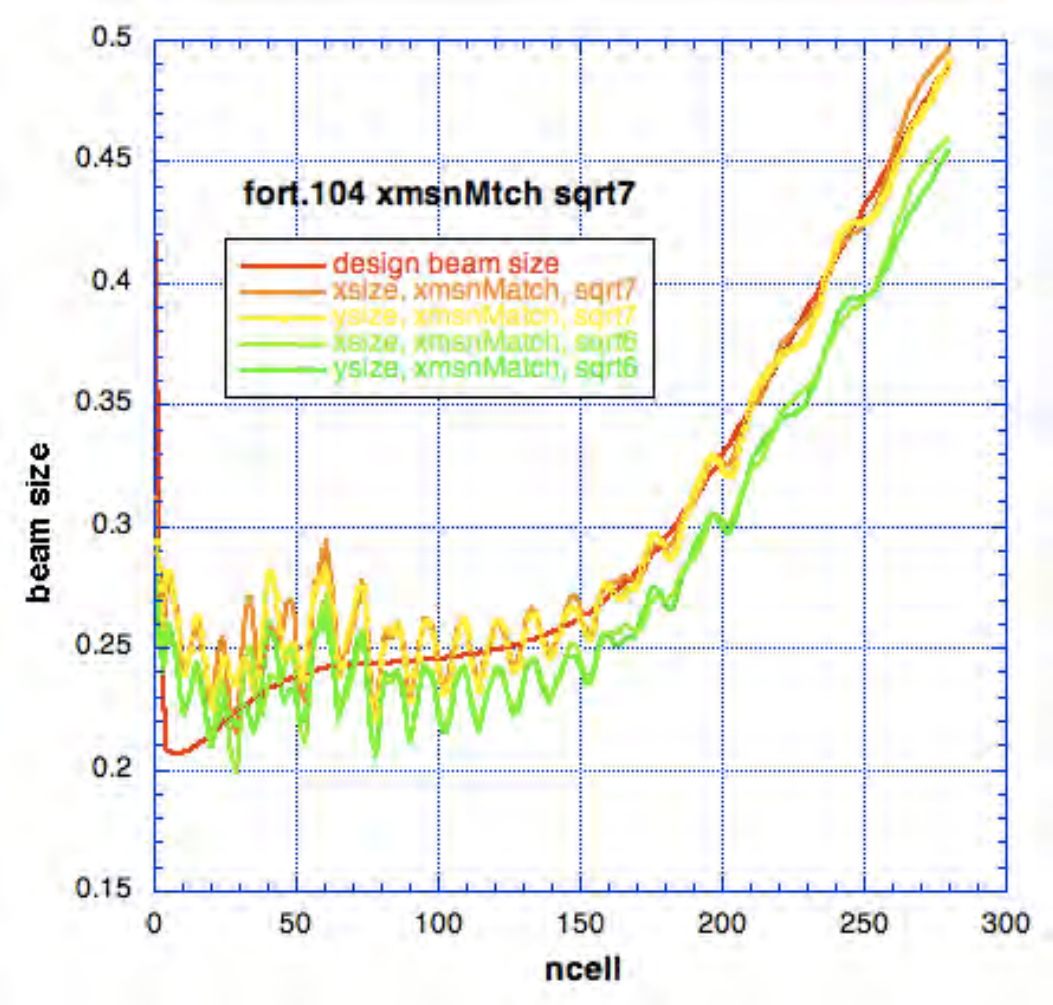

Fig. 19.42 Total design and simulated beam sizes vs cell number[159], with the simulated rms beam sizes multiplied by sqrt(6) and sqrt(7).

The design synchronous phase and modulation for this RFQ are shown in Fig. 19.43; the end of the "porch", for which phis=-90°, is at cell 41; the end of the "shaper" (EOS) (where the beam is brought to equipartition (EP)) is at cell 66. There is a 4-cell radial matching section, and the modulation begins to rise from 1 in cell 6.

---

159 Aperfac70 new20110826 RFQ. See 'Match Design Envelope' folder.

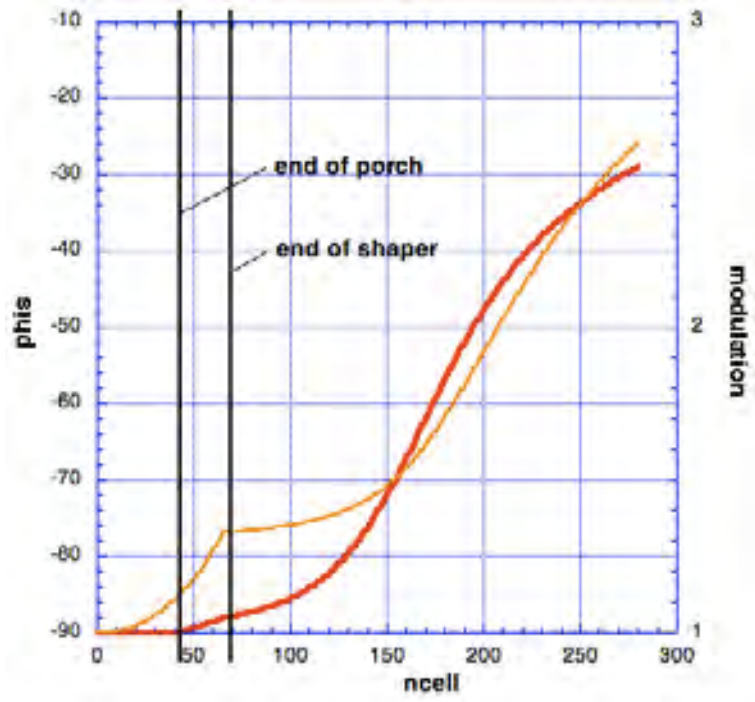

Fig. 19.43 Design synchronous phase and modulation for aperfac70_new20110826 RFQ.

When developing an RFQ shaper that would bring the beam to equipartition (EP) as soon as possible, it was observed that this could be accomplished with only a very small increase of the synchronous phase, to ~-88°. The longitudinal focusing force starts to rise from zero at cell 6, and at phis=-88°, the bucket is still very long and has little height. The design model is based on the rms envelope equations for an ellipsoidal beam, so even though the beam in the shaper is more cylindrical than elliptical, the rms description starts to work for this long, thin bucket.

The beam flies in far from the design envelope until about halfway through the porch, where it is clearly seen that the "ellipsoidal beam" begins to work, and then it is centered on the design envelope beam size = sqrt(6)*(rms beam size) from there through the end of the shaper (EOS).

Then is seen very clearly that the beam form factor evolves to (total beam size) = sqrt(7)*(rms beam size). After the EOS, the synchronous phase starts to rise faster, and the beam rapidly re-equilibrates from sqrt(6)*(rms beam size) to sqrt(7)*(rms beam size). This feature has long been observed[160], and is typical for an RFQ. This is the reason that a final design step that corrects the form factor is important and usually gives improved emittance growth (and equipartitioning performance if the design is an equipartitioned one).

### *19.4.7.3 Development of an Input Matching Procedure*

It is desired to find out whether a good RFQ input match can be obtained by comparing the simulated beam size to the design beam size. Fig. 19.44 indicates that the simulated beam size averages fairly well to the design beam size in the region between the end of the porch and the EOS. The length of this region is controlled in the design program by specifying the porch length as a percentage, usually 60%, of the total shaper length. It seems that comparing over this region could be long enough, and is easy to find from the cell characteristics.

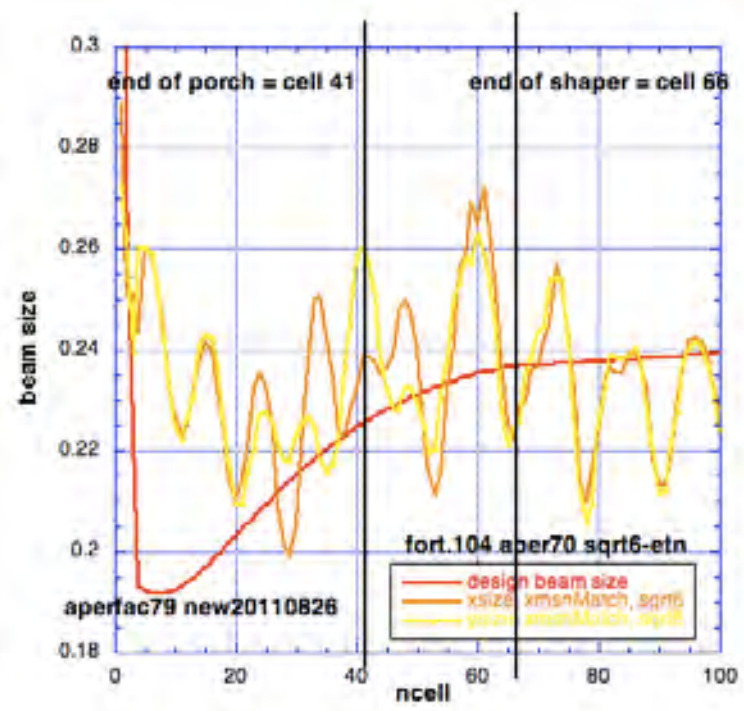

Fig. 19.44. Fig. 19.42 with expanded scale.

160 *LINACSrfqSIM* records all the relevant space-charge physics beam characteristics at each cell.

The objective function for the matching procedure is programmed in outmatch.f90, and is

```
objfunc=objfunc+(desgnsize(i)-xsize(i))**2+(desgnsize(i)-ysize(i))**2
objfunc=objfunc**2
```

The lowest objfunc is sought by varying the RFQ input alpha and beta. The result is compared to the match found from a full RFQ transmission {alpha,beta} map.

*19.4.7.4 Design Envelope Objective Function Map*

It is useful for development to make an objective function map over the same {alpha,beta} matrix as was used for transmission Then the routine can be given to the NPSOL constrained nonlinear optimization routine in *LINACSrfqSIM*, to see if convergence to the match point is obtained.

Fig. 19.45a is the transmission map, made with alpha intervals of 0.05 and beta intervals of 0.5; Fig. 19.45b is the objective function map, made with alpha intervals of 0.1, and using cells 41-66.259

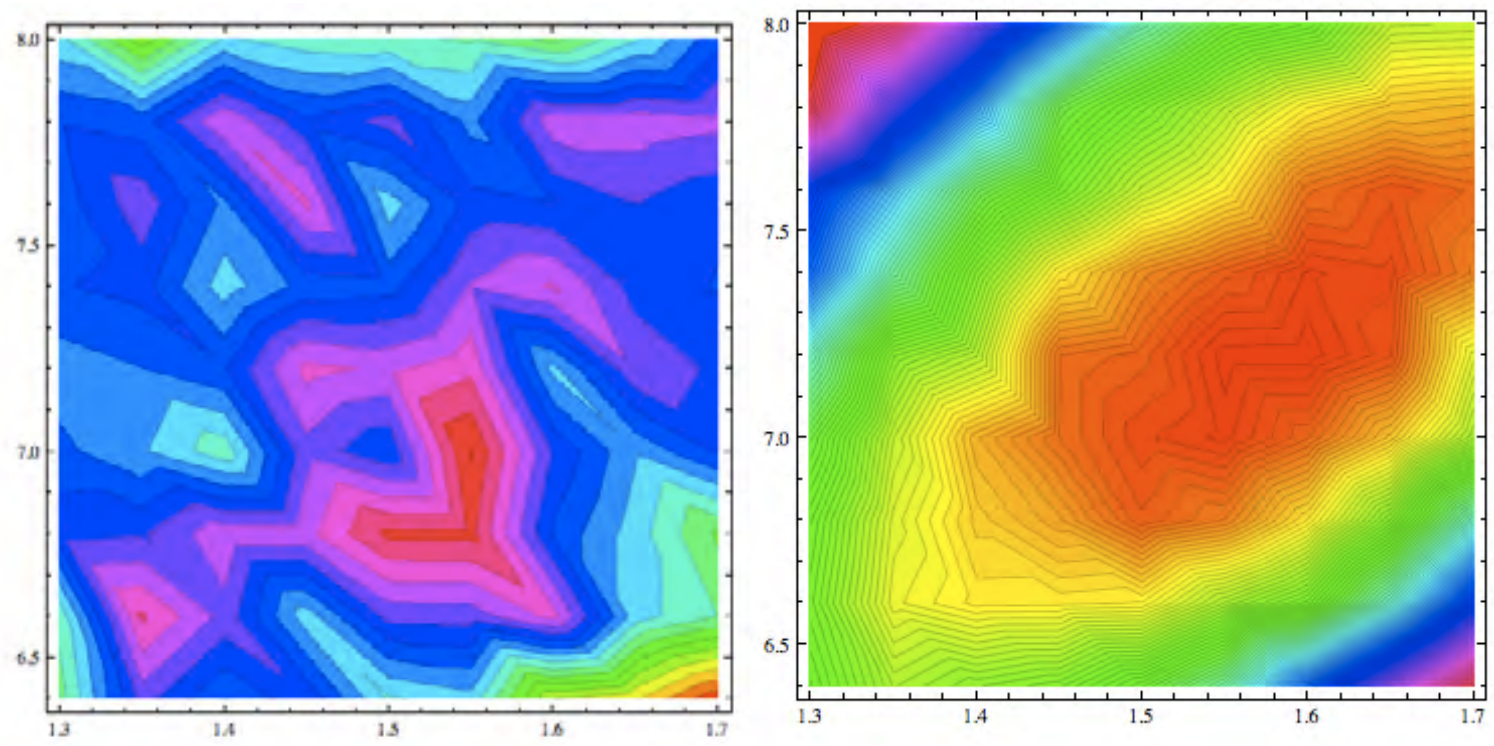

Fig. 19.45 a. Transmission map (red is high). The optimum transmission is obtained at alpha = 1.55, beta-7. Contour range is 69.5%-71.5% transmission b. Objective function map. The minimum is at alpha = 1.56 beta = 7.2. Objective function contour range is equivalent to 87.7%-91.7% transmission.

The objective function contour map is smoother and has a stronger variation than the transmission map, and the minimum is at the maximum transmission!

Fig. 19.46 shows that the agreement is very good for the whole aperfac RFQ family (new_20110826).

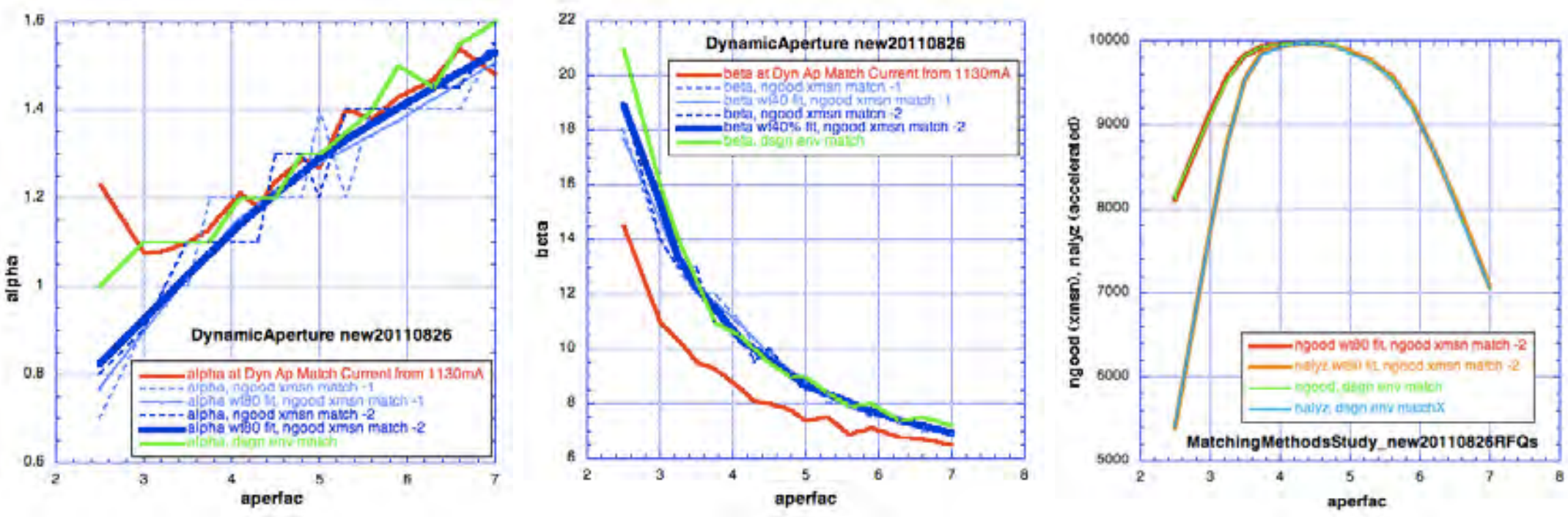

Fig. 19.46. Comparison of match alphas, betas and resulting transmission and accelerated fraction, at the 130mA design current, between the transmission match and the match determined from the design beam envelope.

*19.4.7.5 Design Envelope Optimum Search*

Minimization of the objective function using the nonlinear constrained optimization program NPSOL is successful, also for starting points relatively far from the optimum, but the search time was slow (average ~15 minutes), so not desirable.

*19.4.7.6 Conclusion – Design Envelope Matching Method*

***The design Envelope Matching Method satisfies the criteria except for 6) Should be inexpensive. It requires making a map, or search by an optimization program – much faster and essentially as good as the full transmission method, but still taking considerable time.***

## 19.4.8 Backward Simulation Matching Method

### *19.4.8.1 LIDOS Method*

As stated in Ch.18.1.3.2, use of the *LIDOS* code showed very good input matches. The cooperation with the *LIDOS* team was always very helpful, and the re-check below explains that they use a backward simulation starting just after the radial matching section:

From: "Alexandr Durkin" <durkinap@mail.ru>
To: <jameson@riken.jp>
Subject: Re: injected beam
Date: Thu, 17 Feb 2011 14:35:13 +0300

Dear Bob,

Procedure of input beam parameters calculation is as follows. We divide the first time period after radial matching section on N part (for example N=10). For every time we find initial data for periodical solution of envelope equation so we find parameters of 10 matched ellipses. We solve envelope equation for every ellipse back to the begin of channel. In the case of ideal shape of vanes transverse field is linear and envelope equation is valid, in the case of real shape of vanes transverse field is nonlinear so we used only its linear part. For ideal shape of vane begin of channel is the begin of vanes, for real ones we add gap and outer wall.

So at the begin of channel we have 10 input ellipses for different values of time. We find parameters of a new optimize ellipse which have minimal mismatching with all 10 ellipses - maximal mismatching among these 10 ellipses has minimal value.

You can see results of this procedure in the part Advisor Simulation. You see vertical lines corresponding mismatching. At first we consider ellipse for synchronous phase as input ellipse, further code ask "do you want to optimize?" If "Yes" then you see the result and parameters of optimize ellipse.

Best regards, Sasha

Email from Stas on same date – uses different words:
Stanislav Vinogradov, 2/17/11 12:29 AM +0300, Re: injected beam 1
Date: Thu, 17 Feb 2011 00:29:10 +0300
….

The matched slices are ellipses, not particles. They are computed back from *the beginning of regular part of the linac after input section* (italics added). It is Sasha's code, probably he'll describe it in more details.

….

My attempts as above to find a match at or near the end of the RMS were not very successful however. Then, at the RuPAC 2012, a stroke of luck – happening to sit in on a paper by a LIDOS user, heard a sentence that "input matching is done by backward simulation "from an appropriate point inside the RFQ". Discussed with Sasha and Stas, who said they did not have a firm idea of the appropriate point, or whether another point might give an even better match than the above, which is quite good and fast.

*<u>But then realized, as very clearly shown in Figs. 19.42 & 19.44, that near the end of the RMS is not an appropriate point, but that the firm physical basis of the recommended default shaper which ends with beam at EP would be a very appropriate and consistent point.</u>*

### *19.4.8.2 End of Shaper (EOS) Method*

Figs. 19.42, 44 & 45 provide essential information concerning the possible success of a backward simulation matching method. At the end of a shaper (EOS) which brings the beam to equipartition (EP), the beam is close to the design beam size with the input ratio of (total emittance)/(rms emittance) = 6. At earlier positions, such as at the end of the radial matching section, the beam size is different from the design value, so will not correspond to a model obtained from the design, from which to compute backwards.

A beam distribution at the end of the shaper must be modeled as input for a backward simulation. The end of an even-numbered EOS cell (or the previous even-numbered cell) is chosen as the starting point, where the forward simulation shows that the ellipse alphas are zero, the x-size small, and the y-size large.

The design beam size (x) is found from the exact solution of the two envelope equations and the EP equation. In these equations the wiggle factor is averaged out, and the beam is represented as an upright, $\alpha$=0, ellipse, with $\varepsilon$ = (ellipse area)/$\pi$ = (beam size)*(beam divergence)/$\pi$ = (x)(xp)/$\pi$, where $\varepsilon$ = total real emittance which is adjusted from the RFQ input emittance by the velocity $\beta\gamma$ at the EOS. (For non-zero alpha, the divergence (xp) at the maximum beam size extent (x = sqrt($\beta\varepsilon$) is -$\alpha$ sqrt($\varepsilon$/$\beta$), which reduces to $\alpha = \pi$.) The model beam sizes are found by application of the "wiggle factor" B/$4\pi^2$, with reversal of the x- and y-sizes so that the x-size is large and the y-size is small.

The synchronous phase at the default *LINACSrfq* EOS is ~-88°, so the bucket is still long, and the phase spread of the model input for the backward simulation is set at ±180°. An energy spread equal to the bucket height at this synchronous phase is specified. From this information, a 6D ellipsoidal distribution is generated.

Bunching has occurred, and an additional cos(kz) effect is modeled as coszz= ( cos($\pi$( $z - z_{sync}$ )/( $z_{period}$ )) )$^{(0.25)}$ and applied to the x, x', y, y' coordinates, and without the (0.25) exponent to the z' coordinate, where $z_{sync}$ is the position of the synchronous particle, and $z_{period}$ is the focusing period length (two cell lengths),

The synchronous particle is initially positioned at the end of the starting cell, and the initial time is set to 0.75*(time period).

For the backward simulation the signs of the divergence impulses are reversed.

Fig. 19.47 shows typical comparison of the model distribution to the distribution obtained by transporting the best transmission input match distribution forward to the middle of the EOS cell.

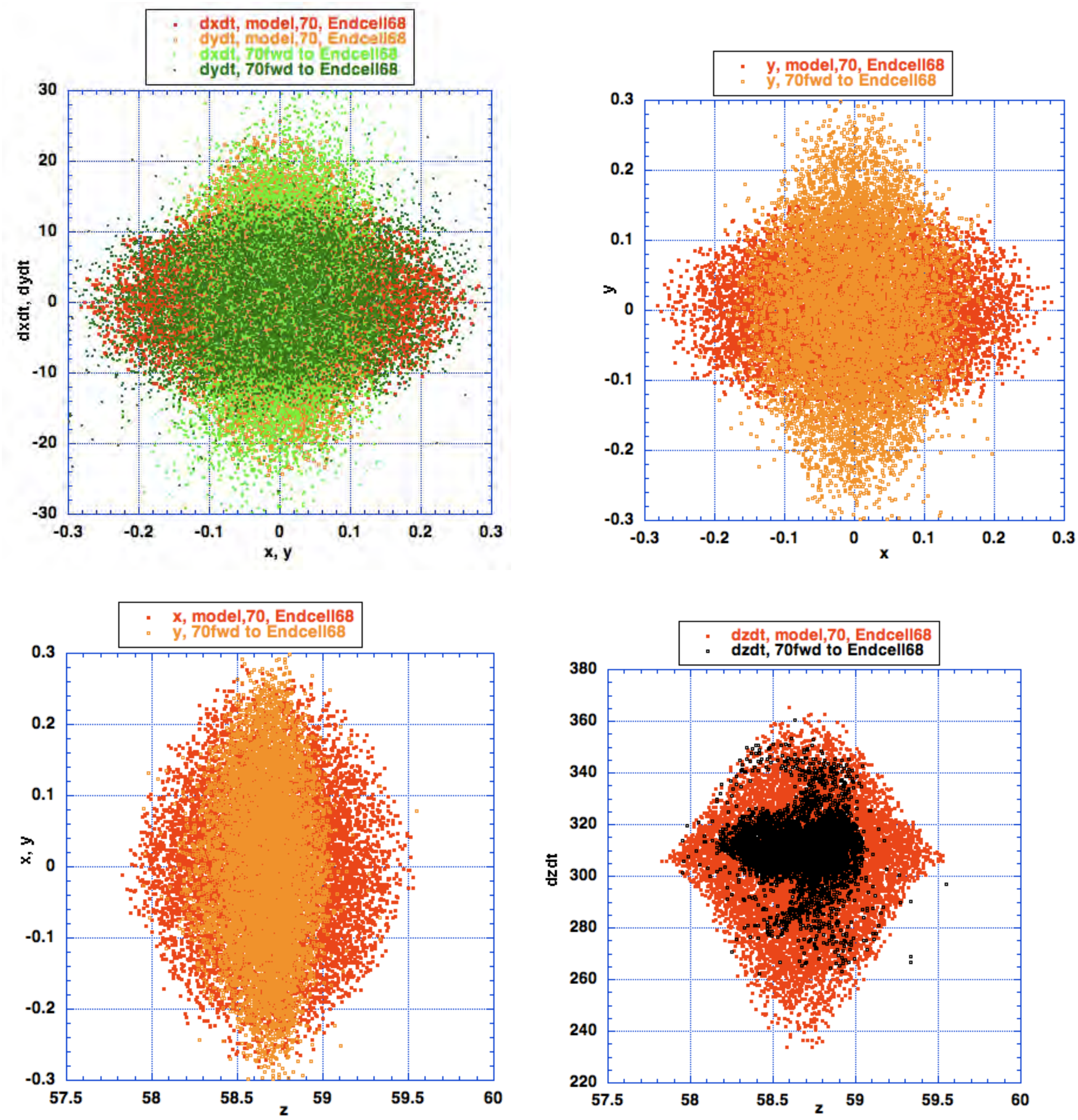

Fig. 19.47 Typical comparison of the model distribution to the distribution obtained by transporting the best transmission input match distribution forward to the end of the EOS cell.

The model distribution is run backwards (2-term mode) to place the synchronous particle at minus one cell length. The input match parameters are then applied to a forward run through the whole RFQ. The backwards procedure is fully automatic and very fast.

Fig. 19.48 compares the input match ellipse alphas and betas, and the transmission and accelerated beam fractions of the best transmission input match and the run backward input match for a family of RFQs. The agreement is good for these designs.

The use of 2-term dynamics is approximate, avoiding the issues of saving the Poisson fields to use with inverted sign, or to recalculate the Poisson fields for the model distribution. Later work with other designs shows that this backward match is often rather far from the best transmission match point.

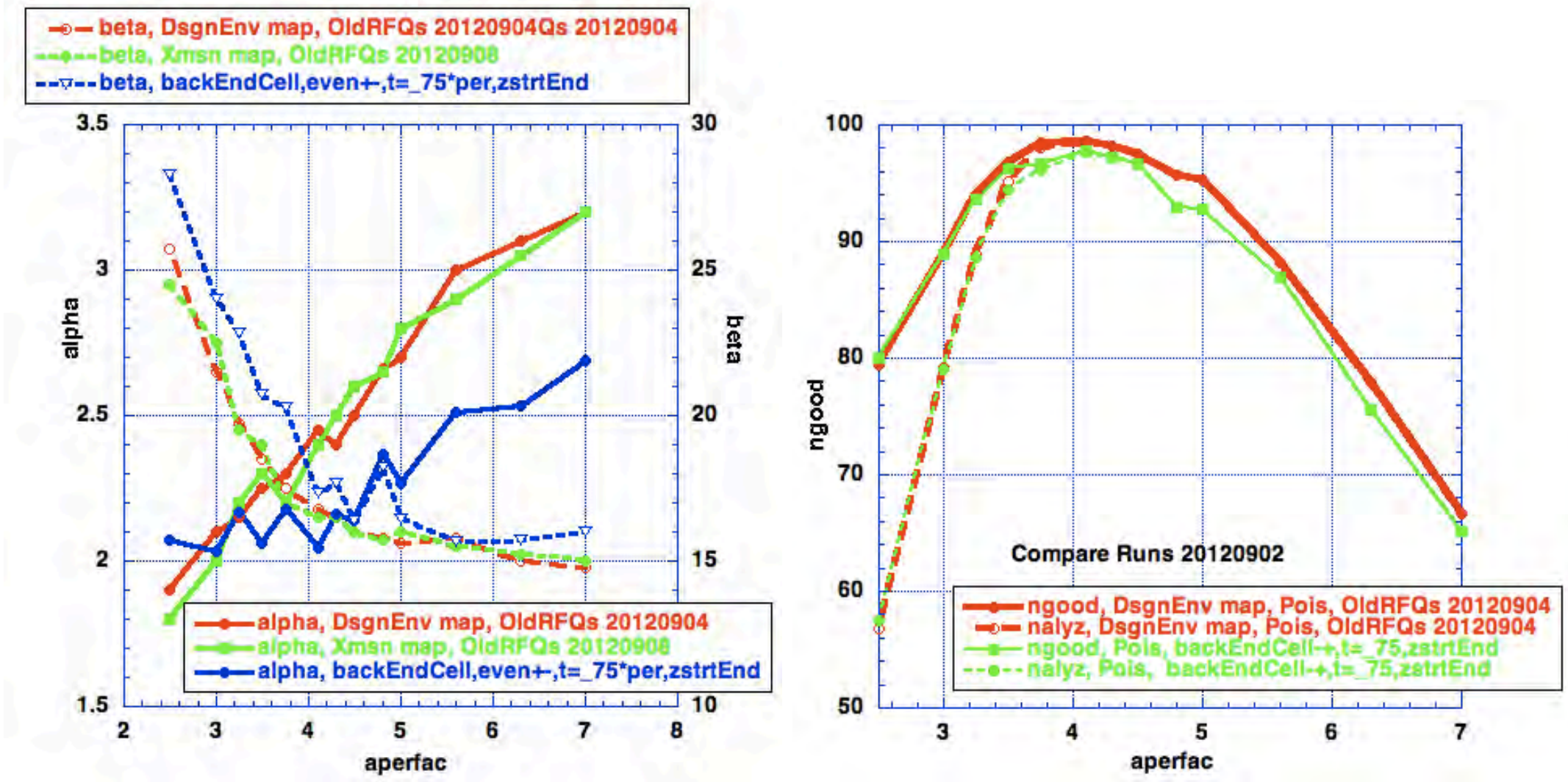

Fig. 19.48 Comparison of best input transmission match, design envelope match, and backward simulation input match. Left – input match ellipse alphas and betas. Right - full Poisson simulation with trnsmission matches for RFQ family, with a 2-cell input section consisting of an entrance wall with hole and then a large aperture section, before the start of the vanes. ngood = transmitted particles, nalyz = accelerated articles, from a 10,000 particle input distribution.

For these RFQs, the design strictly controls the space-charge physics, and the EOS, where the beam is brought to equipartition, is an ideal place from which to compute backwards. Computing backward matches for these RFQs from 40 cells beyond the EOS gave essentially the same results for aperfac ≤ 5, but lower for aperfac > 5 where radial losses play a more important role – indicating some problem with the model distribution there. Perhaps there are improvements that could be made to the model. It may be necessary, and more difficult, to generate an adequate model for RFQs for which the design does not fully control the beam physics.

*19.4.8.1 Conclusion – Backward Simulation Matching Method*

***By starting from a point far enough into the RFQ that the actual beam is close to the design beam, the backward simulation produces an input match that: is for some designs is essentially as good as the best transmission match found from a transmission or design envelope map, but for other designs is not as good, probably because of the poor fidelity of 2-term dynamics used for the backward run. The backwards match satisfies all of the matching method criteria, and is adopted as a quick, preliminary design method in LINACSrfqSIM, but must be checked by the*** Design Envelope Objective Function Map ***or the*** Transmission Map.

### 19.4.9 Conclusion – RFQ Input Matching Method

Pursuing a number of aspects concerning finding a good match at an RFQ input, information is obtained which affords an input matching method that satisfies all the required criteria:

1) Should work for any linac (not only an RFQ, although this is perhaps the hardest because it is highly non-periodic.
2) Should require use of only the early part of the channel – not a simulation through the whole linac.
3) Should move the beam forward or backward, from the input over this early part of the channel.
4) Should be available in the simulation program at all levels (e.g., in RFQ simulation at full Poisson level (slow running time, RFQ entrance and vanes sections are simulated with accurate geometrical, space-charge and image effects), or approximate but fast 2-term level), but accuracy using fast version should be good compared to slow version.
5) Should be relatively easy to use; programmed as a standard program result or on call as a separate run, without requirement for additional interim input.
6) should be inexpensive – fast running time, minimum number of channel computations.

Many matching schemes do not satisfy one or more of these criteria. In particular, schemes which suppose periodicity do not work well.

The Design Envelope Optimum Search minimization of the objective function using the nonlinear constrained optimization program NPSOL is successful, also for starting points relatively far from the optimum, but the search time averaged ~10-15 minutes). *LINACS* includes this method as a simple to use option, taking less time as the full transmission map with nearly the same result.

A backward simulation method works well, is very fast, and is a simple *LINACS* option.

The crucial point is that the characteristics of the actual beam do not approach those of the design beam until some distance beyond the end of the radial matching section. In *LINACSrfqDES* and *LINACSrfqSIM*, the standard shaper brings the beam to the equipartitioned condition at the end of a shaper section, and at this point, the actual beam is behaving like the design beam, with total-to-rms emittance ratio as specified for the usual input beam (waterbag transverse with total-to-rms emittance ratio = 6, dc longitudinal). This is apparently an ideal point from which to simulate backwards. As acceleration occurs, the total-to-rms ratio quickly re-equilibrates to ~7 for designs which are maintaining EP through the entire RFQ, but the match performance is not as good for smaller aperture where radial loss is the dominant factor (instead of first longitudinal loss and then radial loss).

### <u>19.4.10 Re-visit September 2020</u>

In 2020 came a welcome invitation from Masahiro Okamura to join in a project to make a constant r0 RFQ with 2-term vane modulation. The cavity structure has an unusually long entrance section, the input matching procedures were re-visited:

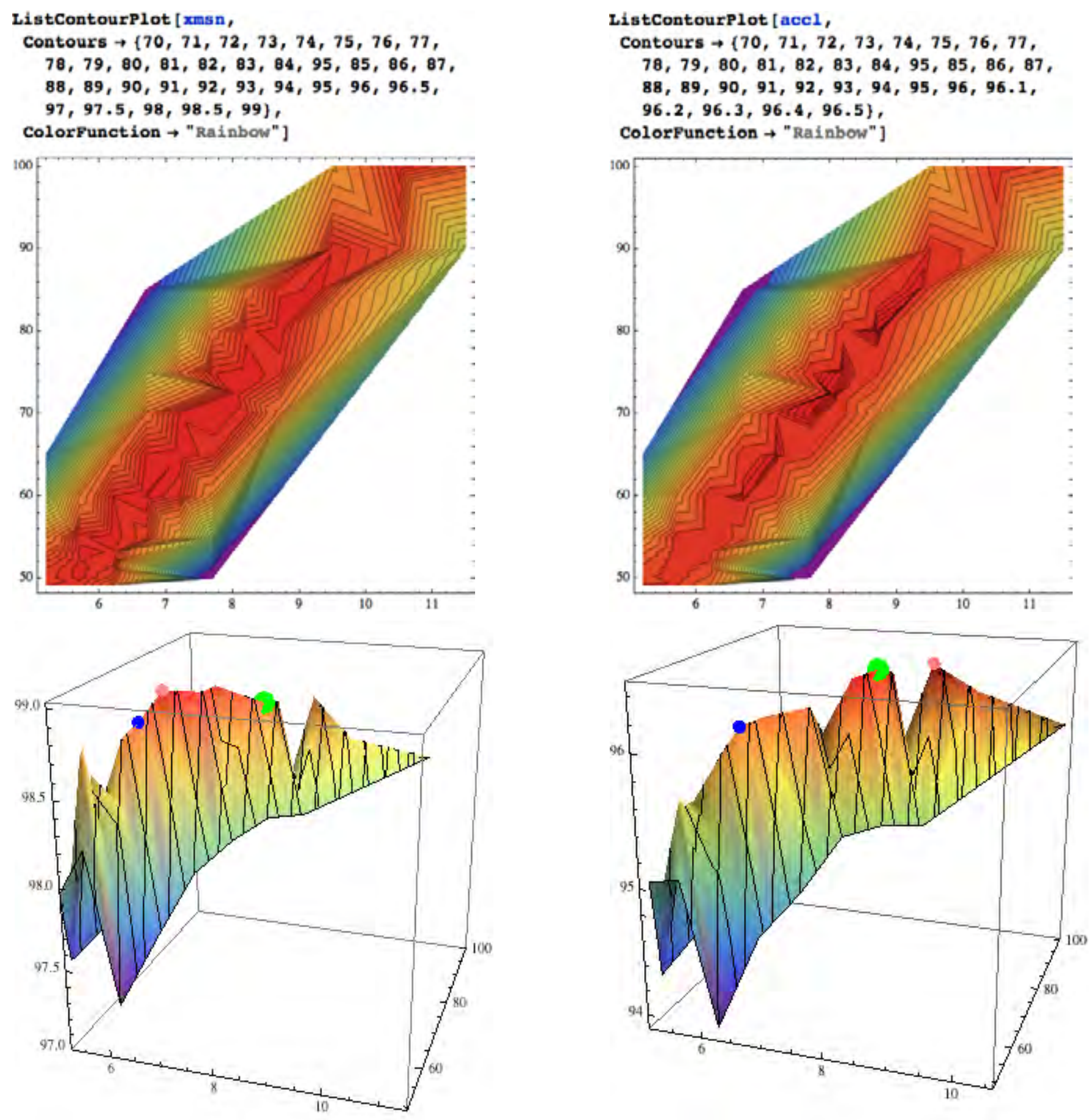

Fig. 19.49 Transmission Accelerated beam fraction

Advanced input matching strategies are a unique feature of *LINACS*. As a demonstration background, arrays of transmission (xmsn) and accelerated beam fraction (accl) vs. {alpha,beta}were made, requiring many full Poisson runs (very tedious). Because of the long entrance wall, the area of good matches is an unusual long, narrow peak. Fig. 19.49 The highest transmission point of this rather coarse matrix search is {alpha 6.7, beta 60., xmsn 99.02, accl 96.23}, shown by the pink dot in Fig. 2.

An optimum input match is also found with the Design Envelope Optimum Search, yielding alpha = 8.025, beta = 72.5 with transmission of 98.88%, and accelerated beam fraction of 96.45%, shown in Fig. 19.49 by the green dot. Different starting points for the optimizer always resulted in a point on the ridge peak.

Another advance match search method projects a model beam backwards from the end of the shaper (EOS) to the cavity input, with {alpha 6.415, beta 56.79, xmsn 98.88, accl 96.14} and shown by the blue dot in Fig. 19.49 This method is very fast to use; equivalent for this design, but not always.

## 19.5 The RFQ Output Condition

It is preferred to remain in time coordinates after the RFQ, so that space charge computation is accurate in following transport sections matching sections, subsequent linac stages. If it were to be desired to change from from canonical (time) coordinates to position (z) coordinates, the conversion method was studied in terms of a time sliced arrival at the RFQ end for each particle, and it was determined that a direct coordinate conversion to dphi, dW is fairly accurate, but recommended not to convert before ~8MeV for protons, where the relativistic effects become small enough, as long as small enough steps are used with the position coordinates.

## 19.6 Estimation of RF Power Requirement

Memorandum: RAJ-12-July-2002-RIKEN

### RFQ SHUNT IMPEDANCE

Various formulas are used (see my Mathematica document RFQ Practical:RFQ Shunt Impedance::RFQ_Power/m.nb) to estimate the rf power requirements for an RFQ design. They tend to be more optimistic, that is to give a higher RFQ shunt impedance, Rs, that the values measured on operating RFQs. This is the case even including the "standard" assumption that measured Q will be 0.6-0.7 that of the SUPERFISH calculated Q. These formulas can be summarized as having the form Rs = ~10^5/(frequency, MHz)^(~1.5), but tending to give shunt impedances up to several times higher than Rs=10^5/(f,MHz)^(3/2). Until now, I have used these formulas in my *LINACS* design code for RFQs.

Shunt impedances for operating RFQs have been collected in the literature and from his experience with his own RFQs by Alwin Schempp, and updated by me from the PAC, EPAC, and LINAC conference proceedings for 2000-2002, as shown in Fig. 19.49 and Table 1.

The relation Rs=10^5/(f,MHz)^(3/2) is compared to the individual fits for 4-rod type RFQs ( Rs=10^4.09496/f^0.98306 ) and 4-vane-type RFQs ( Rs=10^4.45071/f^1.241 ). It appears that 4-rod and 4-vane RFQs should be fitted separately. It is also clear that the experimental values should be used to help estimate the rf power that will be required in a new design. I will use these formulas in my *LINACS* code from now on.

There is a fairly large spread in the data, probably due to variations in construction technique that affected the achieved Q value for the structure. For example, the power required for the VE BERLIN 4-rod structure is almost twice the amount predicted for by the 4-rod-type fitted formula.

(rf power per meter, MW) = 1000(vane voltage, MV)^2/(Rs, kΩm)

For an 80 MHz, 4 meter long RFQ operating with 160 kV vane voltage, the 4-rod fitted formula gives 611 kW, the relation Rs=10^5/(f,MHz)^(3/2) gives 733 kW. Keeping in mind the spread in the data points, a conservative estimate of more than 700 kW, just for the copper power, may have to be made. For C4+, we might achieve 3 MeV in 4 meters; at 100 mA, another 300 kW is required for beam power.

Therefore, it appears that an rf power system capable of about 1 MW would be required for these parameters.

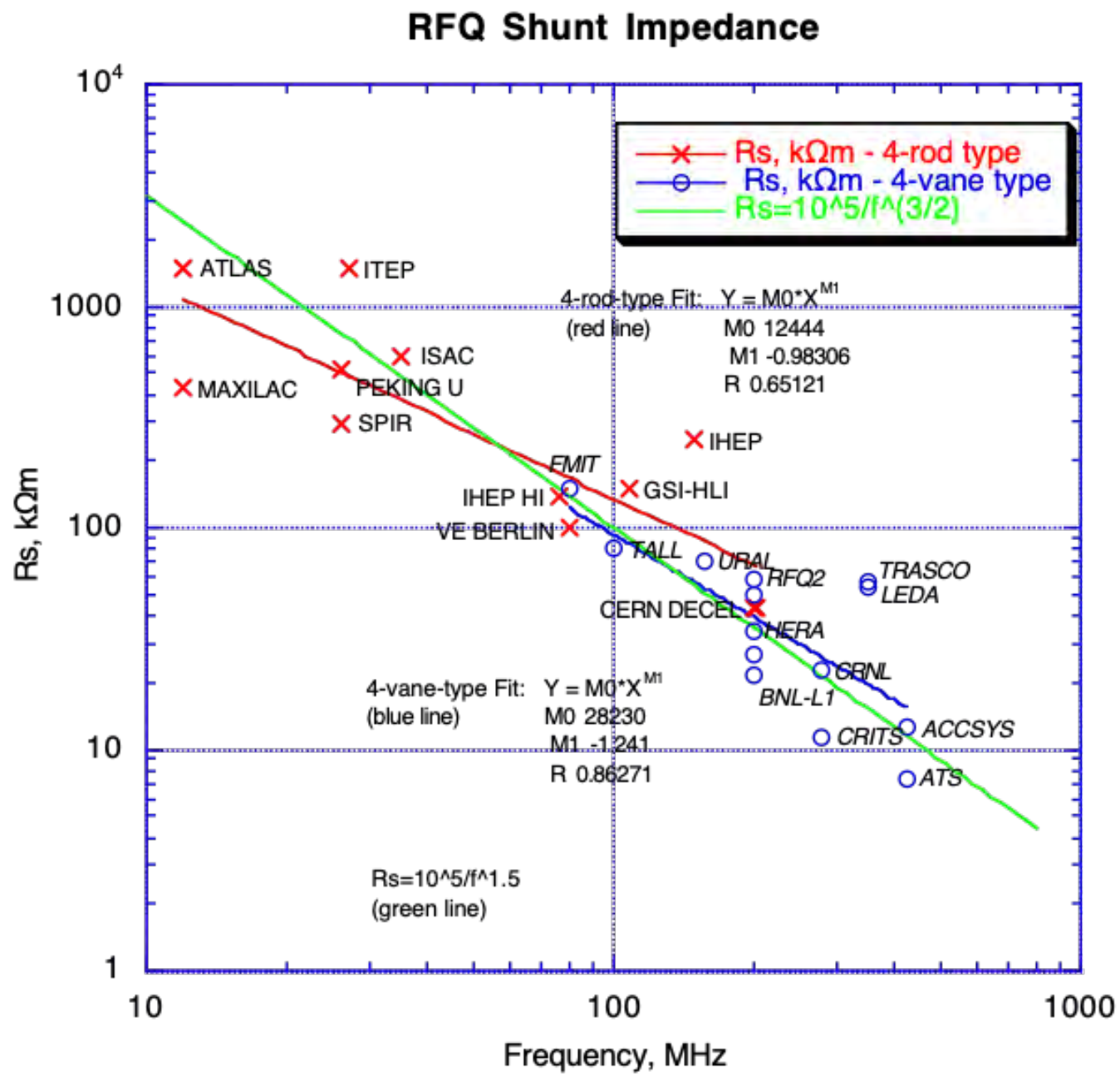

Fig. 19.49 Chart of various RFQ actual shunt impedance vs. frequency
Table 1. Characteristics of various RFQs

| *Name* | *Type* | *Ion* | *Freq, MHz* | *I, mA* | *Duty Factor* | *Win, MeV* | *Wout, MeV* | *# Cells* | *Copper Power, MW* | *Avg Voltage, kV* | *Avge Rs, kΩm* | *Lgth, m* | *Reported In* |
|---|---|---|---|---|---|---|---|---|---|---|---|---|---|
| ATS | | H+ | 425 | 110 | | 0.1 | 2 | 358 | | | 7.4 | | |
| APT | 4vane | H+ | 350 | 110 | | 0.075 | 6.69 | 427 | | | | | |
| BTA | 4vane | H+ | 200 | 100 | | 0.1 | 2 | 181 | | | | | |
| CERN RFQ2 | 4vane | H+ | 202.5 | 225 | | 0.09 | 0.75 | 127 | | | 59 | | |
| CRITS | 4vane | H+ | 268 | 75 | cw | 0.05 | 1.25 | | 1.2 | ~73 | 11.2 | 2.54 | Linac'98 |
| ESNIT | 4vane | D+ | 175 | 75 | | 0.075 | 2 | 353 | | | | | |
| IFMIF | 4vane | D+ | 175 | 140 | | 0.1 | 8 | 427 | | | | | |
| GTA | 4vane | H- | 425 | 45 | | 0.035 | 2.5 | 247 | | | | | |
| JHP | 4vane | H- | 432 | 20 | | 0.05 | 3 | 329 | | | | | |
| SSC | 4vane | H+ | 427.6 | 70 | | 0.035 | 2.5 | 239 | | | | | |
| ATLAS | Split coax | HI | 12 | | | | | | .017 | 102 | 1486 | | PAC'99 |
| IPHI | 4vane | H+ | 352 | 100 | cw | | | | | | | | |
| LEDA | 4vane | H+ | 350 | 100 | cw | | | | 1.5 | ~100 | 53.3 | 8.0 | |
| CERN Decel | 4rod float dc | H anti | 202.5 | | | 5.31 | .01 | 34 | 1.1 | 167 | 13.8 | 3.4 | PAC'01 |
| GSI | IH | Dark currents | | | | | | | | | | | |
| IHEP | 2H | H+ | 148.5 Qth 15800 | 120 Qm 13600 | 5 Hz reached 3.4 Kp | 0.1 bore 7.59 mm | 1.8 | 91 | 0.2 | 140 | 251 | 2.563 | Linac'00 |
| ISIS | 4rod IAP | H- | 202.5 Qth 7500 | Qm 2K+-200 @ lo pwr | 10% | .035 Qm 2.7K+-200 @ hi pwr | .665 | | .23 (consistent w/ Qm hi pwr) | 90 | 42.3 | 1.2 | EPAC'00 |
| | | | | | | | | | .175 | 80 | 43.9 | | |
| | | | | | | | | | Actual Pin< 160kW | | | | |
| ITEP | Coupled window | HI | 27 | | | | | | 2.2 Kp | | 1490 | | PAC'97 |
| JDS Rev | Fig.1 | I vs df | | | | | | | | | | | |
| PekU | 4rod | O+ | 26 | | 1/6 | | 1 | | .024 | 70 | 522 | 2.6 | EPAC'00 |
| SNS | 4vane | H- | 402.5 | | 6% | .065 | 2.5 | | | | | 3.7 | PAC'01 |
| TRASCO | 4vane | H+ | 352 | | cw | | 5 | Bore .29-.32cm | .58=SF x1.2; Q=8261=SF/1.2 | 68 | | 7.13 | |
| TRIUMF ISAC | 4rod split ring | | 35.3 | Rs th 470 kΩm | Cw Rs m 279 kΩm | Q th 14816 | 5 Qm 8400 | | .075 | 75 | 600 | 8 | |

The graphs need to be updated; since 2002, better fabrication techniques may be consistently better; for example, RFQs recently built and tested at IMP, Lanzhou, China, indicated substantially higher shunt impedance and correspondingly lower rf power requirement.

Several *LINACS* design studies have used different estimates, based on company experience, etc. As far as it concerns design practice, the main concern is just to have an estimate which is traded off against other requirements, with a better estimate then obtained for the final design from the actual cavity design.

## 19.7 *LINACSrfqSIM* Results for the *aperfac* Family

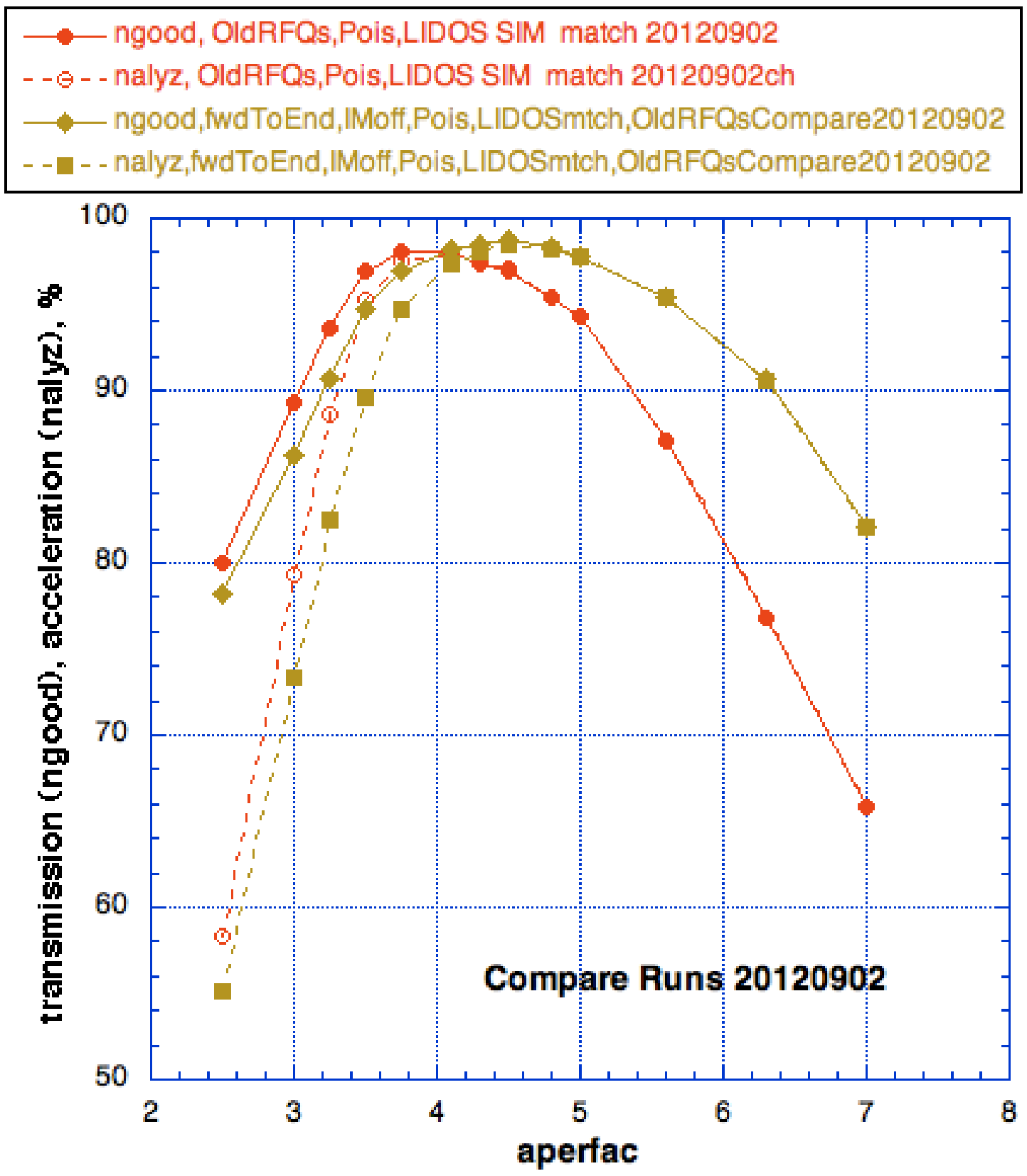

Fig. 19.50 Final result 2012 for the *aperfac* family, **with full Poisson vane geometry for both external and space charge fields (red)**, and with full Poisson vane geometry for external field **but a cylinder boundary for the space charge field (ochre)**. The shift in optimum EOS aperture is apparent.

## 19.8 Compatibility of Accurate and Fast Solutions

Optimization! The optimization program will require many runs of the model, so would like fast run time for the simulation. The problem is to be sure that the fast and slow simulation modes are closely enough the same with regard to the optimum. *LINACSrfq* does this by having the exact Poisson solution (slow mode) related to the design mode using the 8-term potential at the beam rms sizes, and the 2-term potential for the fast mode.

The correct placement of the *LINACSrfq* design program with respect to the *LINACSrfq* simulation program and to an optimization program – is in the middle. The design program is a good approximation of simulation program at the rms beam sizes. Poisson simulation solution is more accurate, slow. 2-term simulation is an approximation of the 8-term design, fast.

“2-term” is 2term potential plus scBATYGIN rz Fourier space charge WITH CYLINDRICAL BOUNDARY AT THE APERTURE as default.

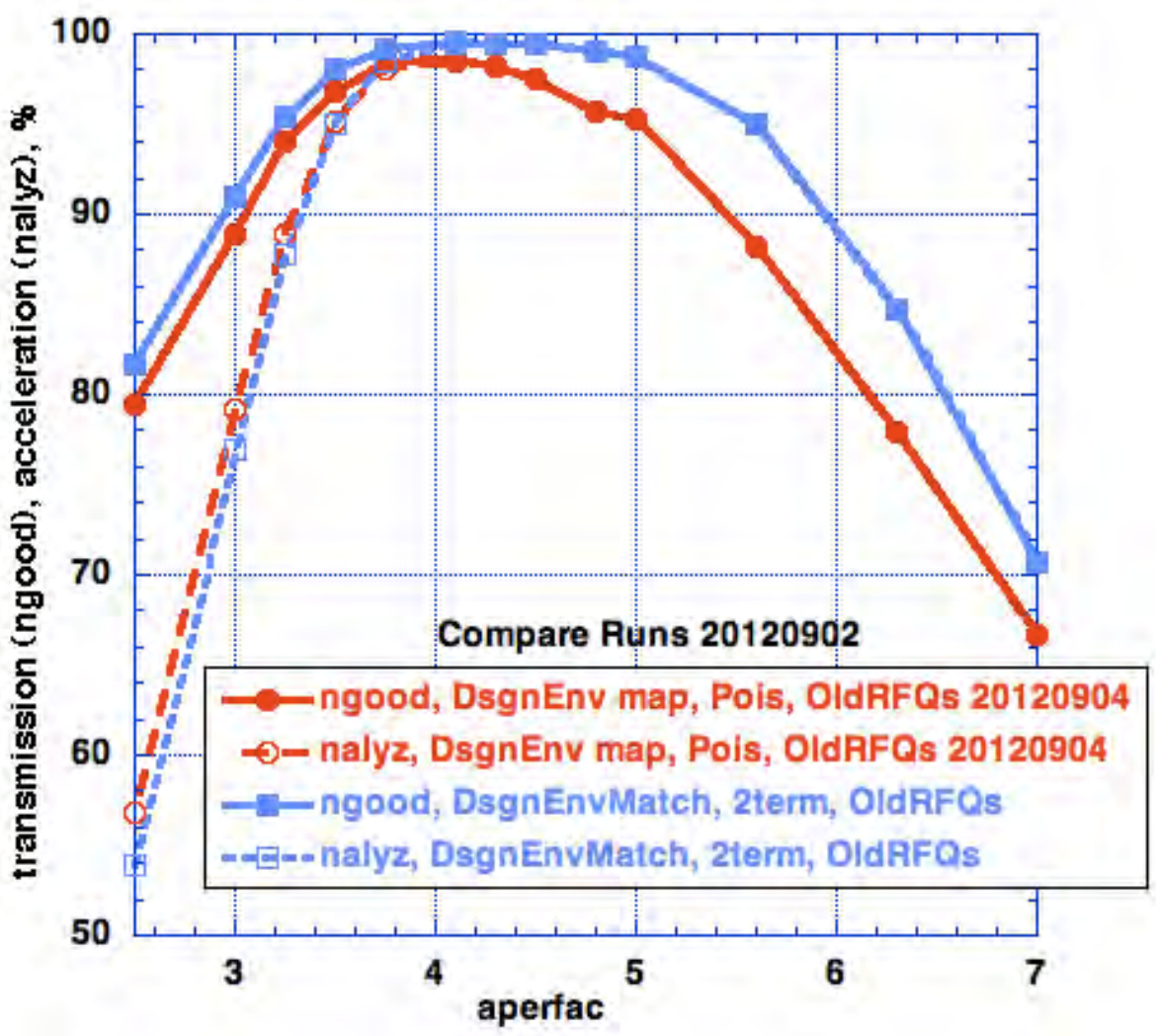

Fig. 19.51 A representative example for the *aperfac* family:

Fig. 19.51 indicates that the Poisson and 2-term performances are similar (for this family, in this design space), so that an optimization program could do a search, and final design then iterated from that point.

## 19.9 Connect to vane machining

Customers have preferred to write their own milling machine coding; such is not included in *LINACS*.

## 19.10 *LINACS* Availability

As of September 2023, the full design and simulation code *LINACS* is available at https://ln5.sync.com/dl/01ce57220/v23cfj82-gsx74hws-runykxcf-btqmfe2a . (See p.2 above.)

# *LINACS*

***Linear Accelerator Design and Simulation***

Author R. A. Jameson

**A <u>Framework</u>:**

- Comprehensive coordination of design and simulation, from physics, engineering and computer science points of view
- Control of all parameters, including space charge physics.
- Possibility for "inside->out" design – specify desired beam behavior, create appropriate fields to realize it.
- Full mode with best physics, approximate modes for speed (affording advanced optimization methods), correlated.
- Experimental methods devised for testing and verifying physics and programming of codes, over wide parameter range.

**<u>Design approach</u>**

Via smooth approximation, equivalent rms property, local matching, equipartitioning if desired, control of space charge physics.

Any type of linac:

- RFQ – very fully developed.
- APF – new method developed for designing long sequences.
- DTL, different types

Framework and tools for all types functional; the following require addition of detailed programming.

- CCL, SCL – all types of separated function linacs

**<u>Simulation:</u>**

•Subroutines in overall framework,
- RFQ very fully developed.
  - Full 3D Poisson mode (with J. Maus); correct quadupolar symmetry
  - Fast 2-term mode, with careful correlation to full Poisson result.
- APF – initial method fully programmed, additions straight-forward
- Other linacs – setup of simulation subroutines straight-forward

- Framework tools available for any linac:
  - Nonlinear, constrained optimization routines
  - Graphics – using modern python/matplotlib
  - GUI – using python/Kivy
  - multiple runs, etc., etc

**<u>Availability:</u>** As of September 2023, the full design and simulation code *LINACS* is available at https://ln5.sync.com/dl/01ce57220/v23cfj82-gsx74hws-runykxcf-btqmfe2a . (See p.2 above.)

• Full source code is essential for trust in any real project, and for future development and collaboration. Valuable for students.

[eltoc]

# Chapter 20. *LINACSrfqSIM* and *LIDOS* Comparison

A final comparison of the ellipse parameters and transmission/accelerated result for the *aperfac* family is shown in Fig. 20.1. It is seen that the default *LIDOS* mesh (red curve) is too coarse and produces optimistic transmission results. Increasing the number of mesh points to the available limit (dashed red curve) indicates the sensitivity. The *LINACS* default Poisson curves with somewhat different input matching (orange, green blue) fall within the *LIDOS* band. All of the comparison work and this result was discussed with Durkin and Vinogradov at MRTI in Moscow in 2012, and they agreed with all aspects.

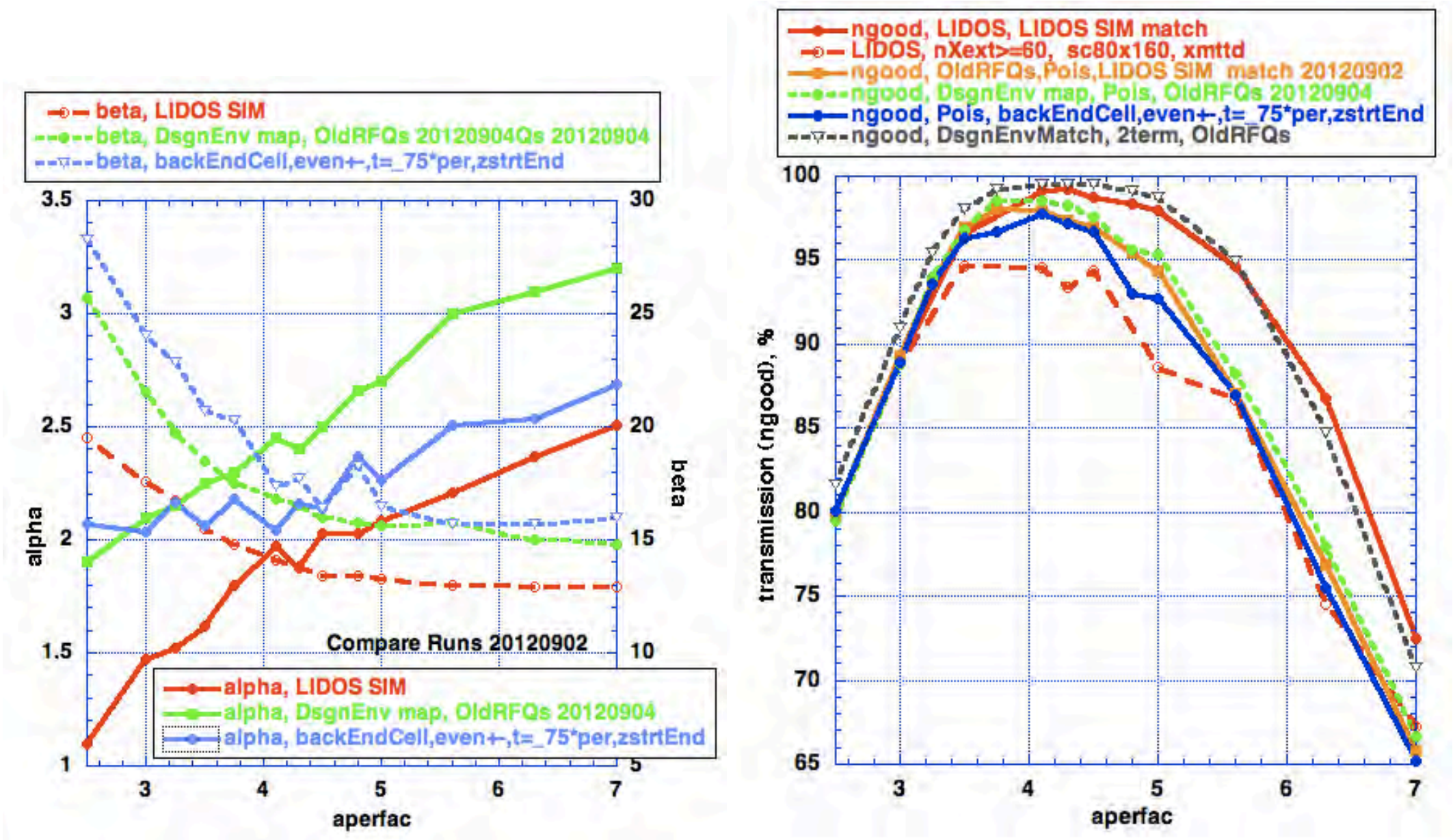

Fig. 20.1 Final comparison of the ellipse parameters and transmission/accelerated result for the *aperfac* family.

## <u>20.1 Run Times</u>

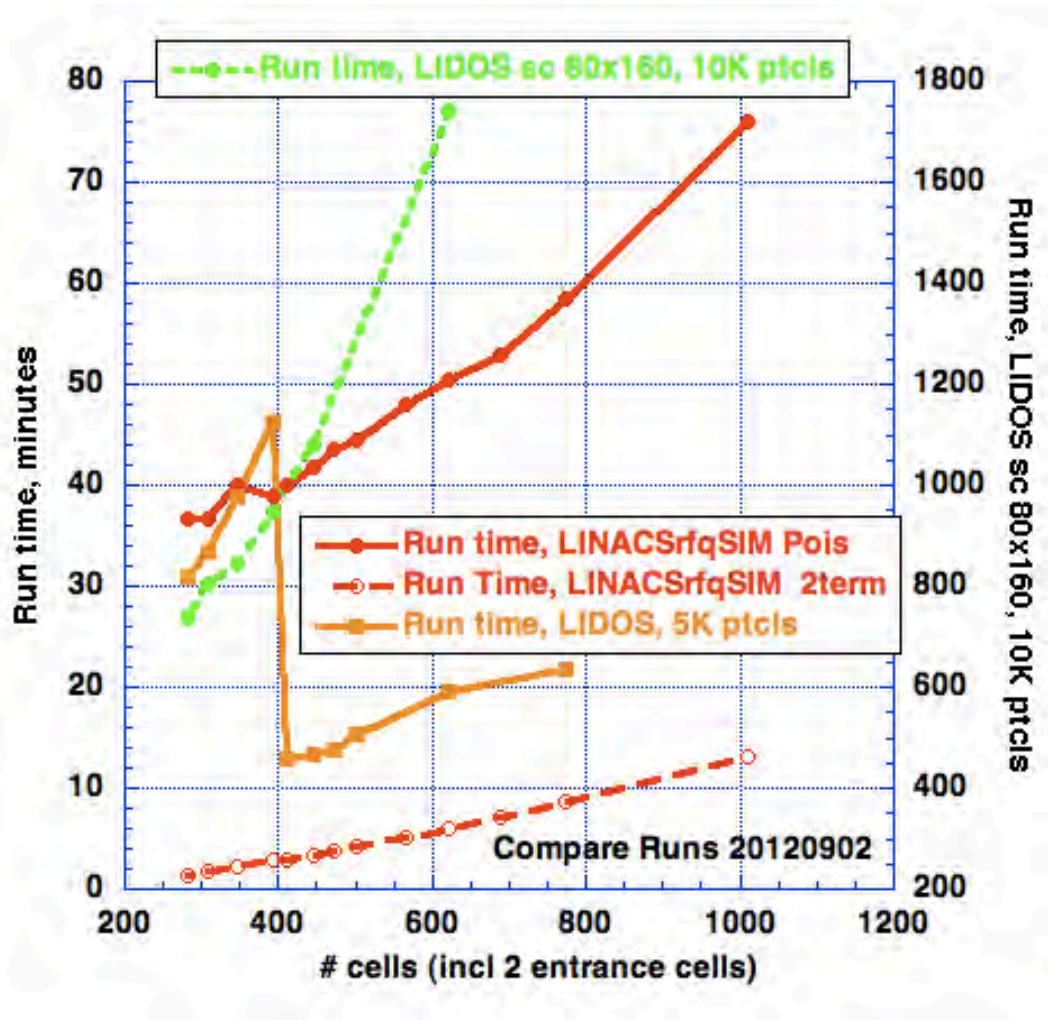

Fig. 20.2 Run time (minutes) comparison between *LIDOS* (4 quadrants) and *LINACSrfqSIM* (one quadrant, 10K particles). Orange *LIDOS* is for default 30x30 space charge grid.

The reason for the jump around 400 cells is unknown – all mesh setting were the same – only evidence is a garbage word or two after the line Section input parameters in the rfq3D.out file of the RFQs with more than 400 cells, e.g.:∞ ∏®Ú10000 0 0 0 0 1 1 f

# PART 4 — DETAILED OBSERVATIONS

[eltoc]

## Chapter 21 – Single Particle Phase Advance Notes[161]

R.A. Jameson
November-December 2012

### 21.1 Representative Linac Example

An example 15mA, H+ RFQ, named "IMP_ADS", will be used as an example of a certain design technique which strives for minimum length for the specified energy gain. The simulations are with 2-term external fields and Fourier space charge fields in cylindrical pipe with boundary at cell aperture. Fig. 21.1 shows the rms features of this RFQ. The length is made short by utilizing as much longitudinal focusing as possible, by allowing sig0l to approach, but should not cross, sig0t. Then the transverse focusing is raised again, by allowing the longitudinal focusing to decrease as the cells become longer with acceleration. At the point where sig0l~sig0t, at the center of the most dangerous sigt=sigl (1:1) resonance (~ cell 110), the beam is equipartitioned, coincidentally, because the design was done with sig0t and sig0l, not with sigt and sigl. It happens in this example that sigl is approximately equal to sigt at this point. This type of design leads to short length, but crossing of resonances on Hofmann Chart, and entrance and exit from the main 1:1 resonance zone, with milder effect because of short instance of equipartitioning there. The transmission is 99.8%, with final

161 Notes and extra material at end.

accelerated fraction of 97.8%. The one-quarter transverse plasma period is ~5 cells, and longitudinal ~2 cells, over which rapid redistribution within the beam can be expected to occur. The behaviors of the total beam sizes, which are expected to show evidence of single particle behavior beyond the rms behavior, are of interest, Fig. 21.1.g&h.

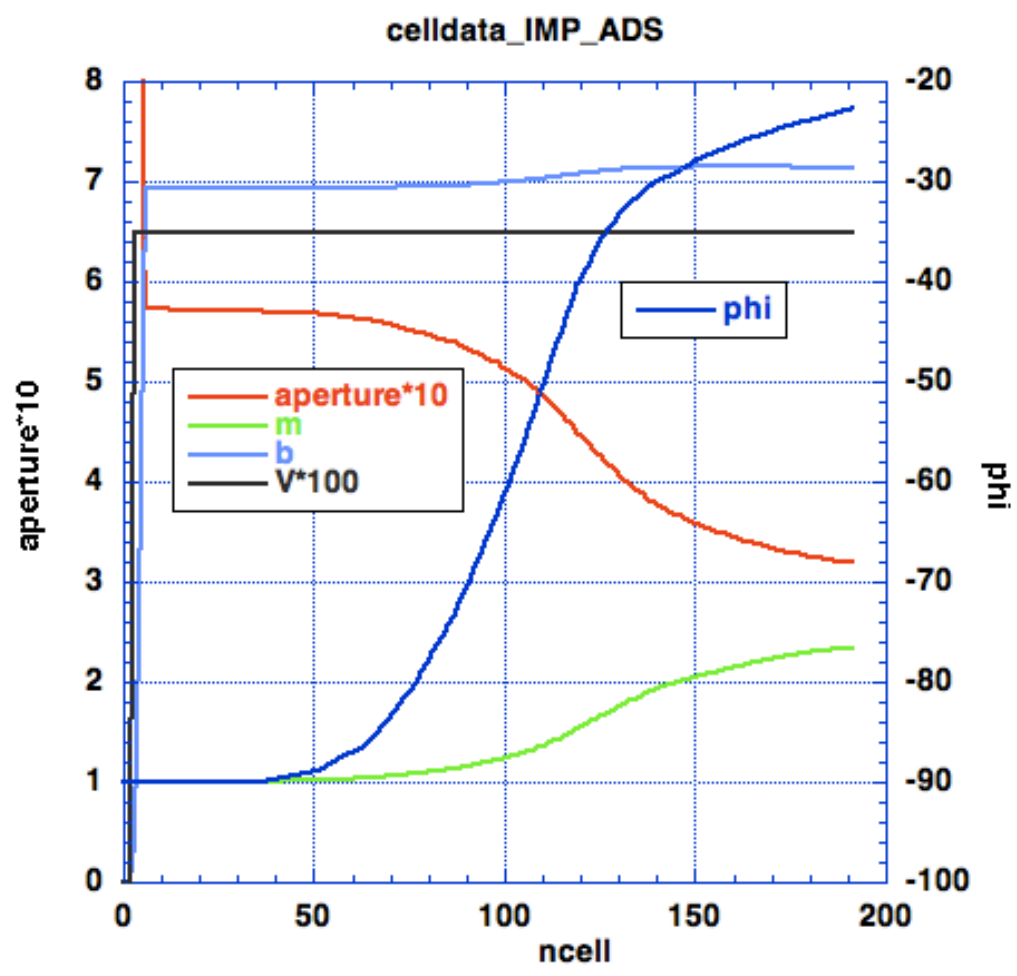

a. Cell parameters

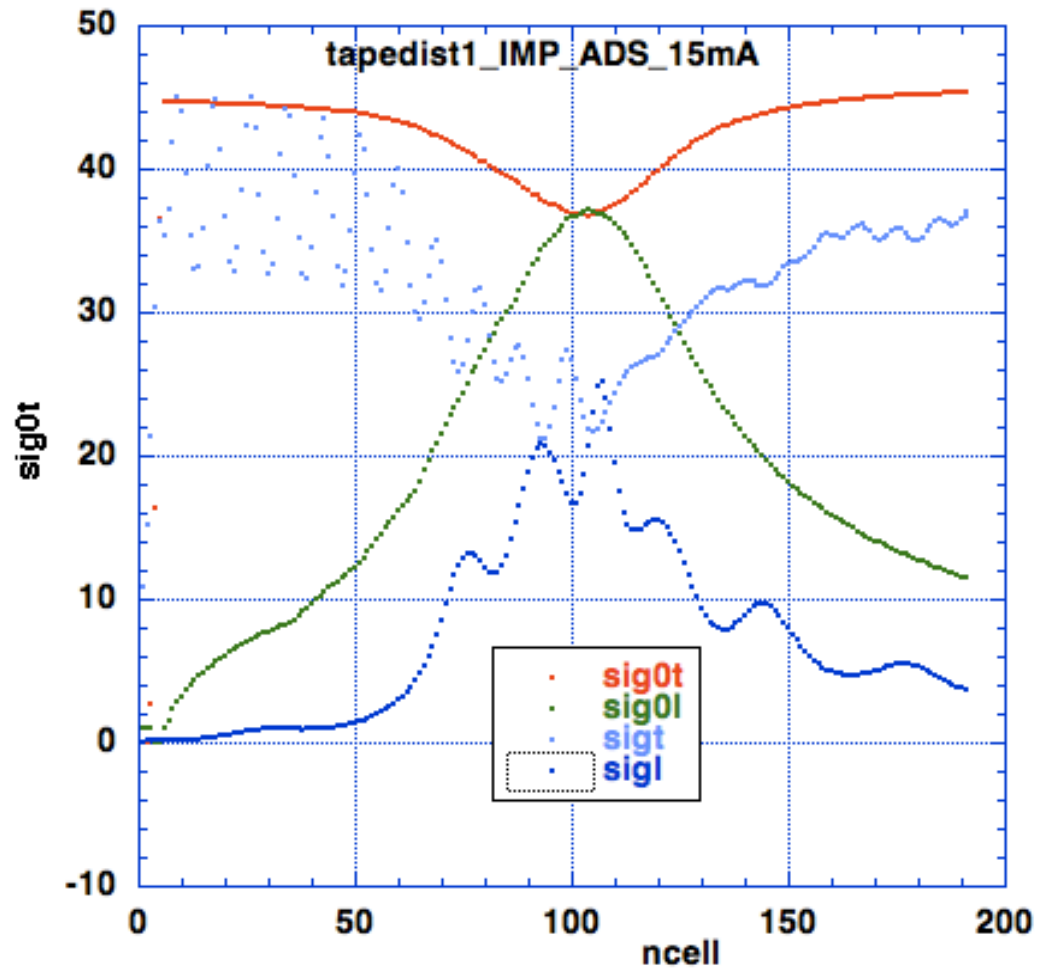

b. rms phase advances

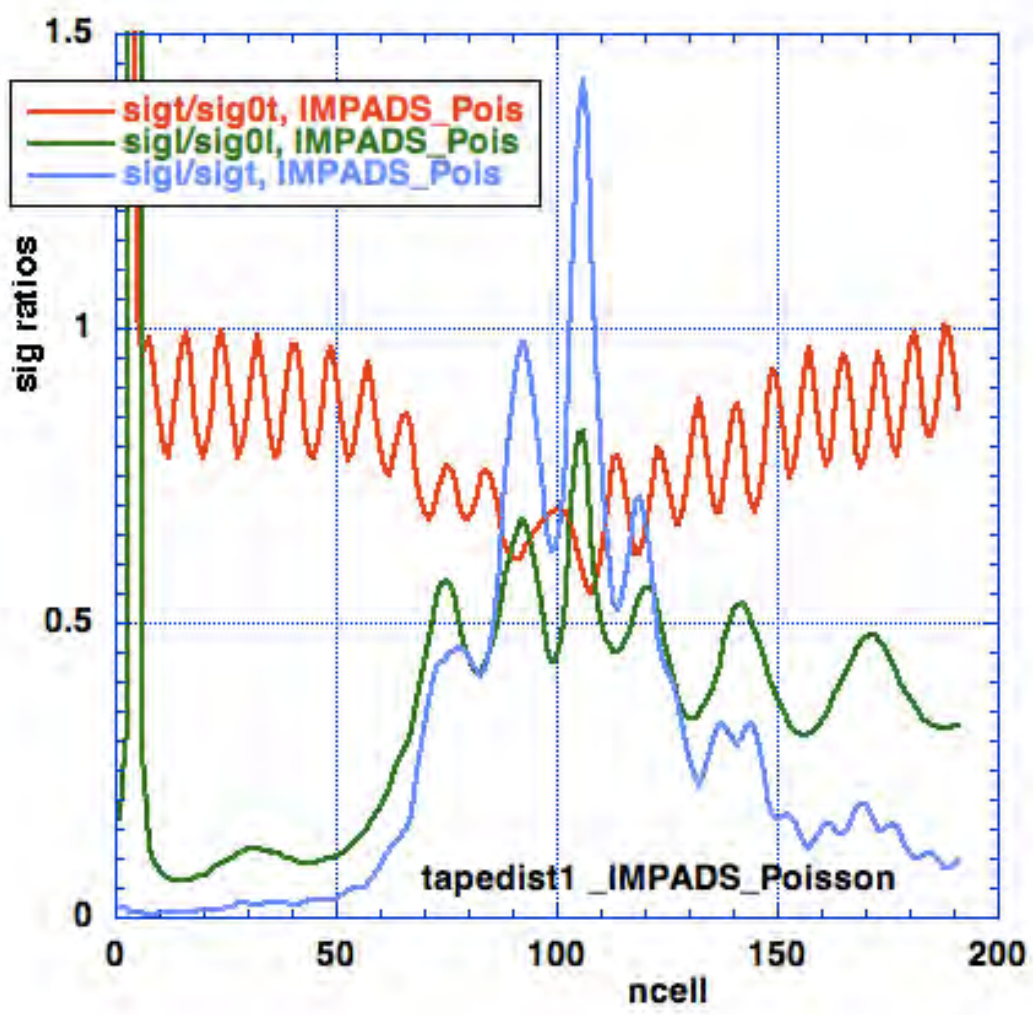

c. Tune depressions and sigl/sigt ratio

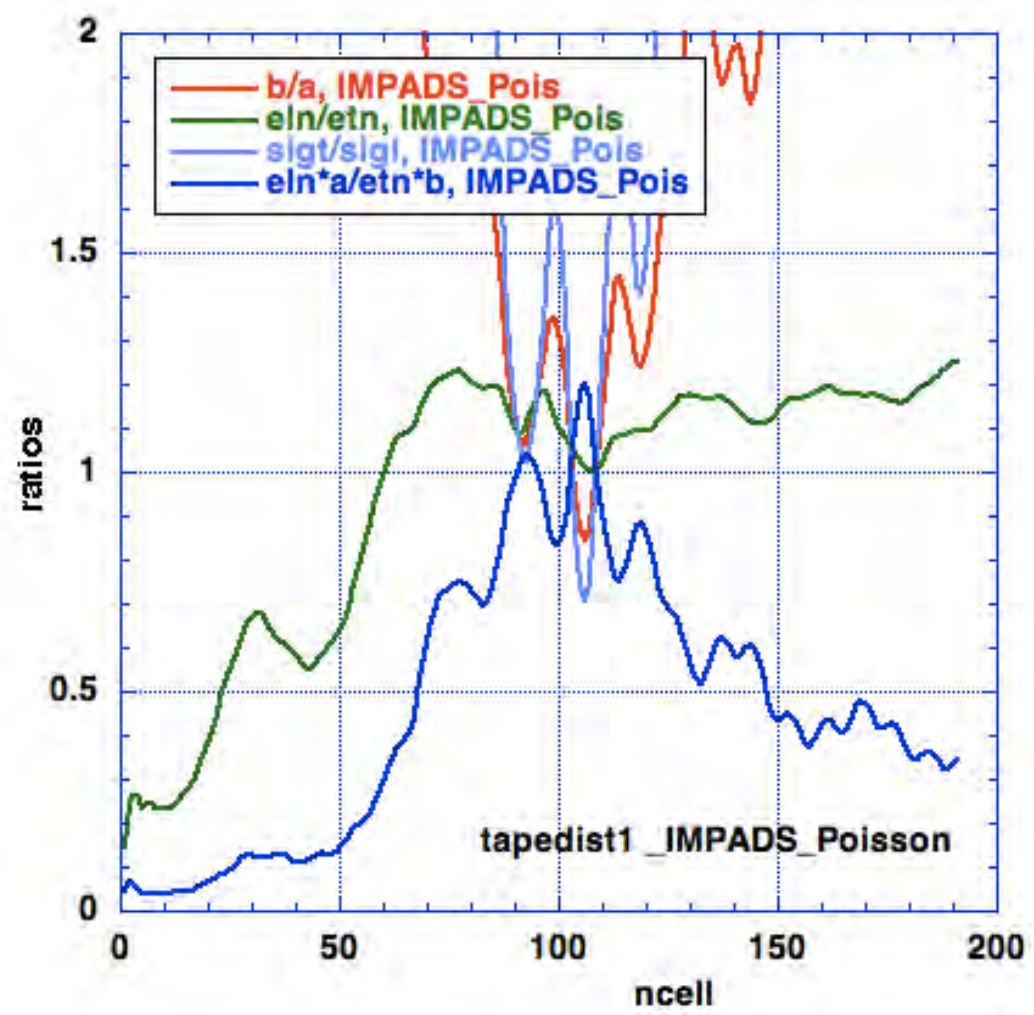

d. Equipartitioning ratios

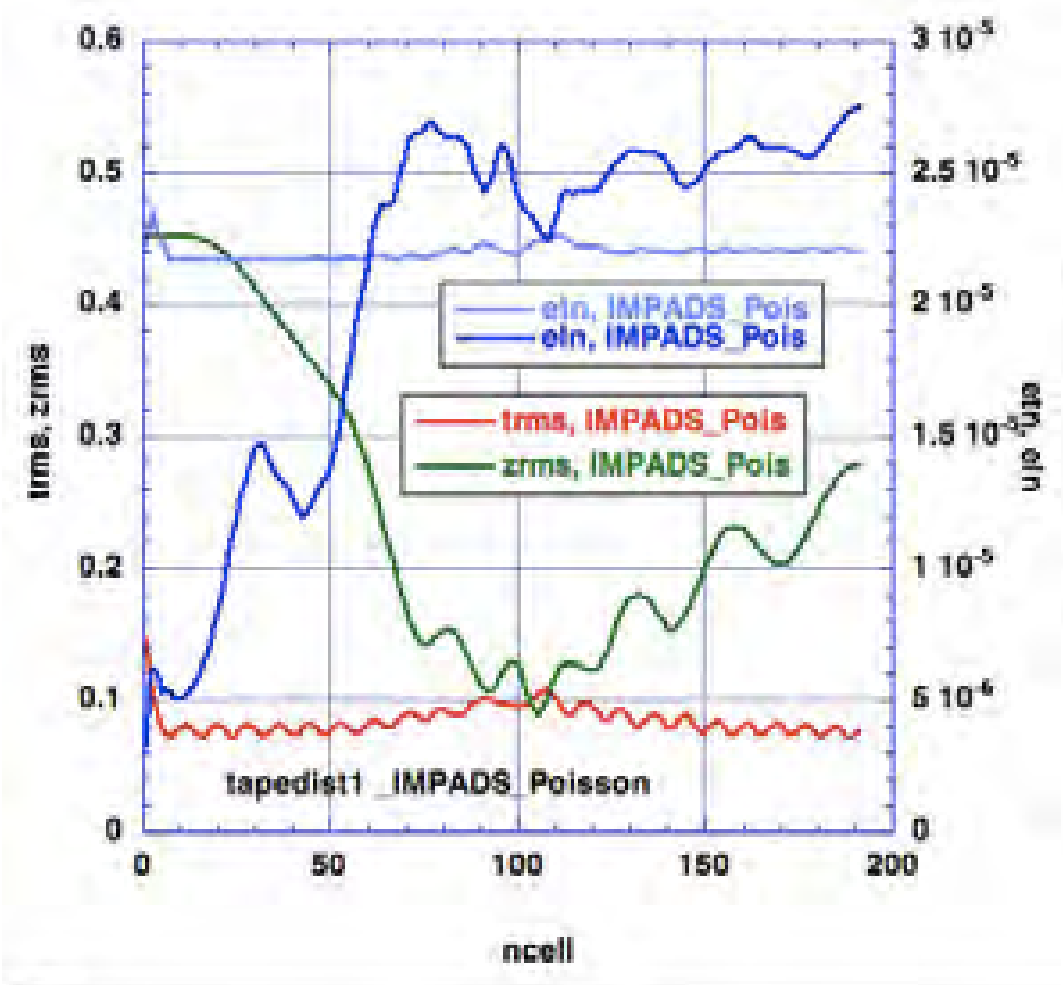

e. rms emittances, beam sizes

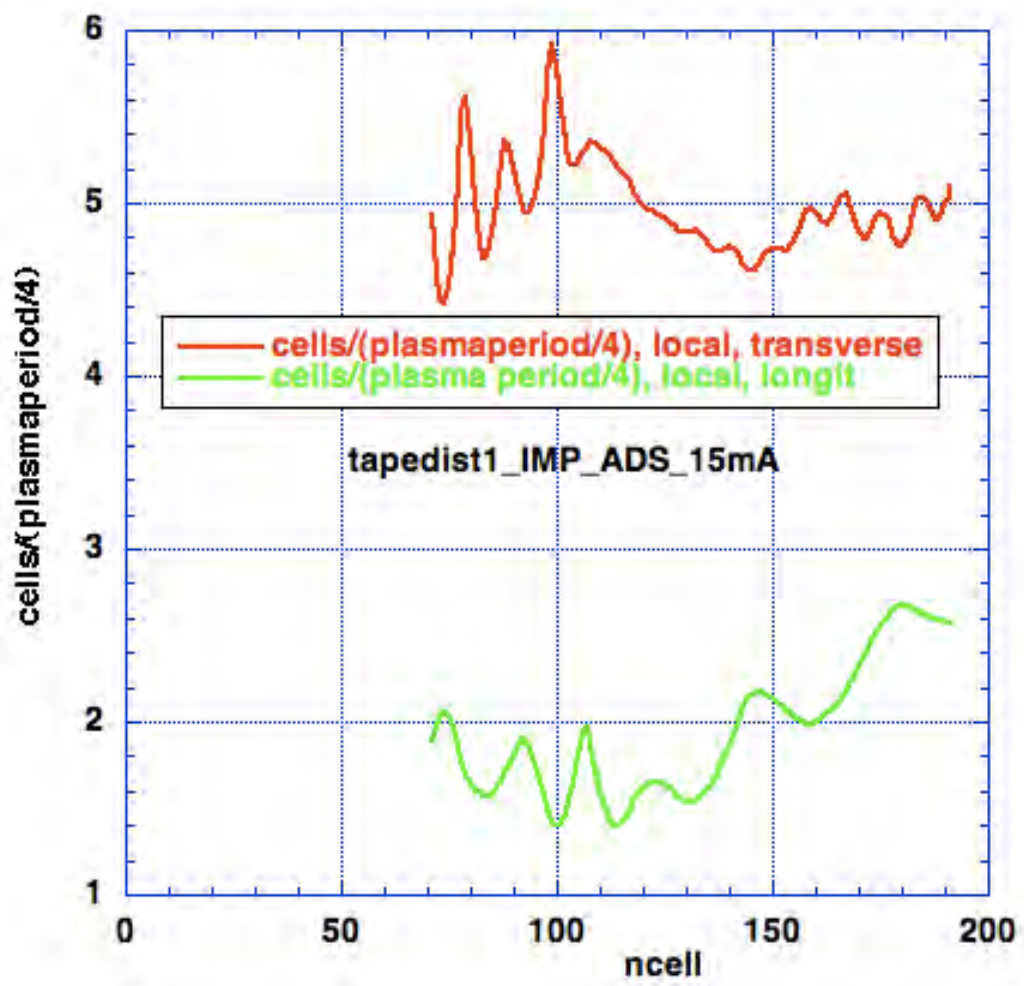

f. Plasma periods

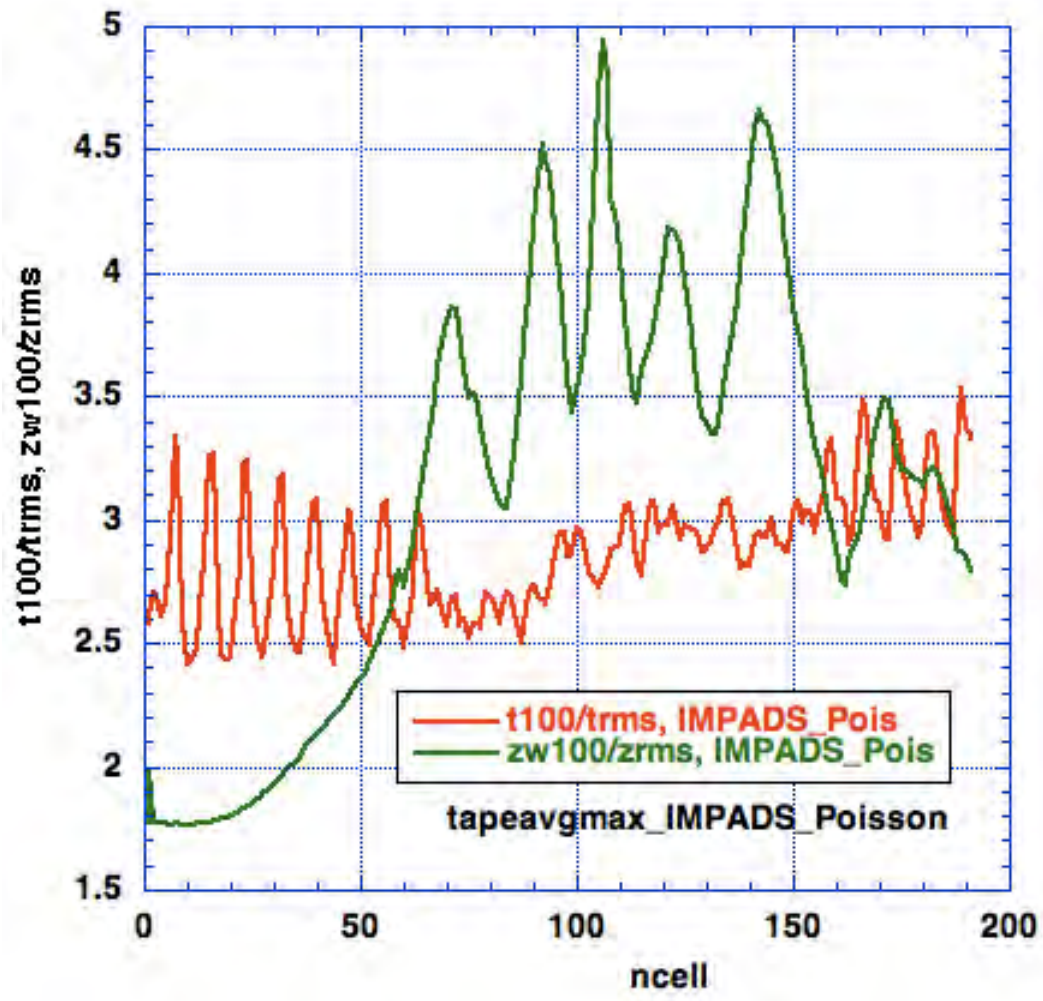

g. Ratios of total to rms beam radii

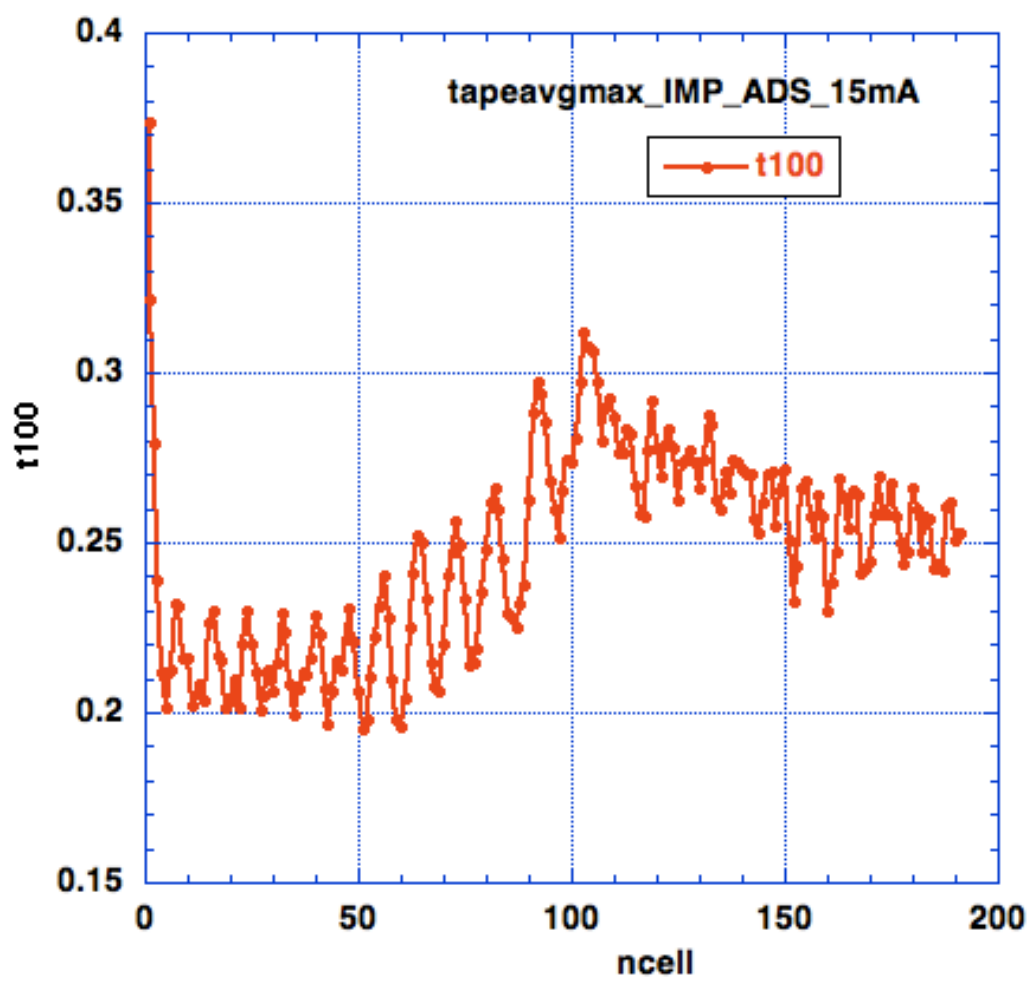

h. Total transverse beam radius

Fig. 21.1. Design rms and simulated beam features.

The transverse and longitudinal rms trajectories are plotted on a Hofmann Chart in Fig. 21.2. For full understanding, both need to be plotted and considered. The transverse and longitudinal performance is completely coupled in an RFQ; they will be more or less coupled in other types of linacs. On the Hofmann Chart, an interaction with a resonance is evidenced by a kink in the trajectory if there is no capture by the resonance, and by a loop is there is actual capture by the resonance. Note that the trajectories experience rms capture in both planes in the region sigl/sigt=0.5 (~cell 80-90) and rms interactions in the region of sigl/sigt=1 (`cells 90=110), and at sigl/sigt=0.33 (longitudinal, ~cells 115-125),. The 1:1 resonance strength is reduced as the beam is transiently equipartitioned very briefly when traveling in this region.

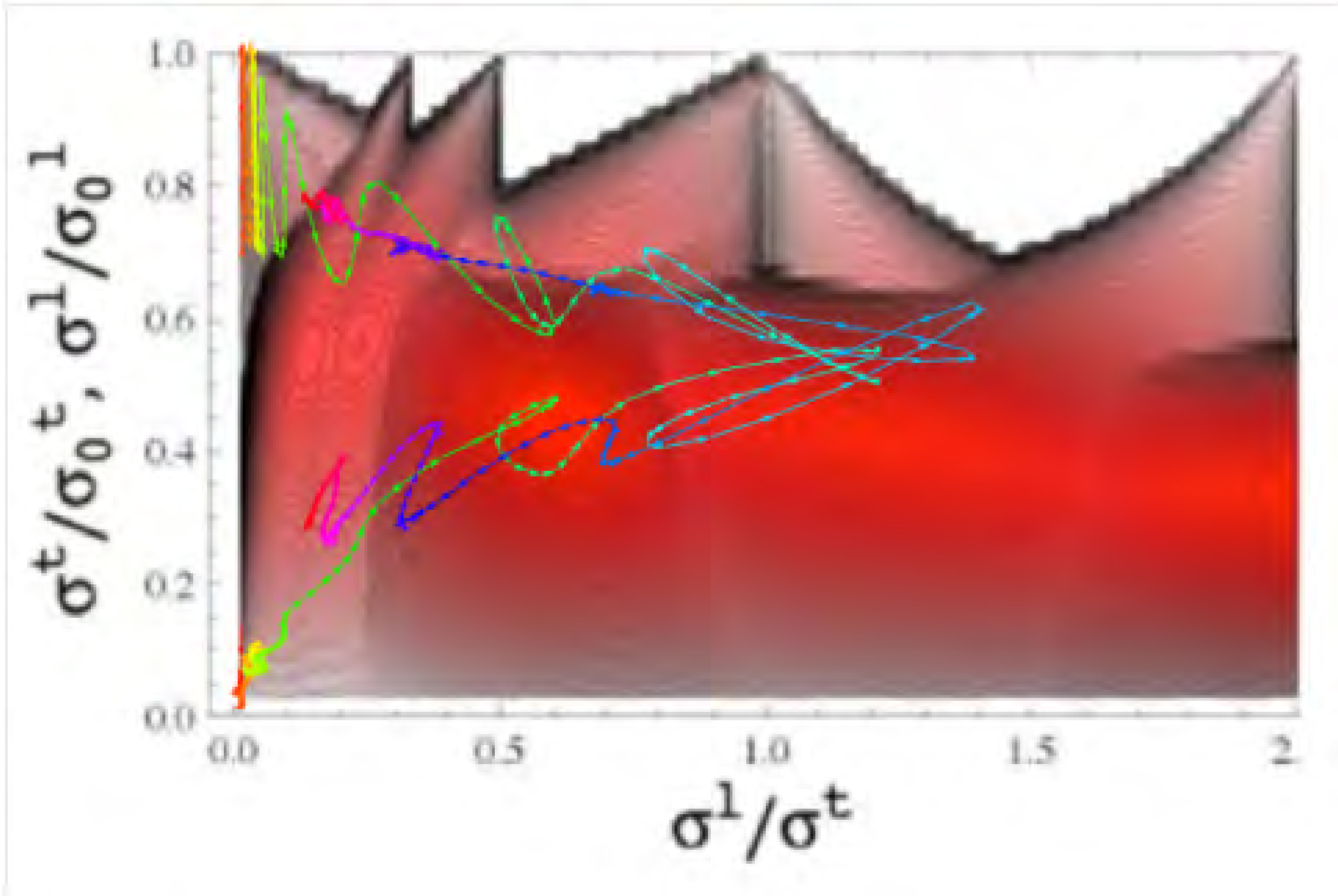

Figure 21.2: Rms trajectories through the RFQ, on Hofmann chart drawn for EP ratios eln/etn = b/a = sigt/sigl = 10. Stronger growth rate of the instability is indicated by deeper red. The trajectories progress from red at the RFQ input through orange, yellow, green, blue to magenta at the RFQ output. The chart shows stop-bands for resonances at $\sigma^l/\sigma^t$ near 0 and at 0.33, 0.5, 1.0 and 2.0. Note that the influence of the sigl/sigt=1 resonance is very broad and strong, with tails extending far away from 1:1. The lack of interaction to the right of the $\sigma^l/\sigma^t$ =0.33 and 0.5 resonances is not completely explained, but commented on below; simulation shows there is interaction on both sides.

## 21.2 Single Particle Phase Advance - Expectations

• The single particle phase advance method must allow for possible zero or negative rotation, because space charge forces can cause a particle to stop momentarily, or be repulsed backward. The procedure therefore uses all four quadrants, ±180°.

• As a practical accelerator must maintain a positive overall balance of transverse and longitudinal external focusing fields relative to the space charge fields (1 > (sigt/sig0t and sigl/sig0l) > 0), no operation (or design) at the space charge limit!), in general over a long enough period, the single particle phase advance will be positive.

• The high nonlinearity of the longitudinal dynamics will strongly influence the single particle phase advances (also at zero beam current). Particles in the accelerating bucket undergo synchrotron oscillation, with widely varying rate depending on the position in the bucket – fast near the head, and extremely slowly if located in the bucket tail. There is strong coupling (in an RFQ) between longitudinal and transverse, so this will be reflected in the transverse phase advances.

• In order to understand the rms dynamics along the channel, it is necessary to analyze only the particles that are successfully accelerated to the end of the channel.

• Investigation of the single particle dynamics is different – here both the local performance of all particles that are still being transmitted, and those that are transmitted successfully to the end, probably should also be observed. Single particle interactions with resonances may result in later loss, either radially or longitudinally from the accelerating bucket.

• Particle losses complicate understanding.

• Constantly changing parameters and space charge and longitudinal dynamics mixing will prevent a particle from staying near a particular resonance very long.

## 21.3 Extension of tools and procedures from extensive continuous beam studies in mid-1990's [162] to 3D.

Two procedures are investigated:
1. Phase advance as difference in angle in phase space.
2. Phase advance in angle of transfer function written in framework of rms phase space ellipses, as extended to include both input and output ellipses, per Handbook of Accelerator Physics and Engineering.

### 21.3.1 Phase Advance in 6D phase space

The phase difference between each particle's xx', yy' and zz' phase space vectors from the origin, from a starting cell to two cells later are computed. Representative phase space distributions are shown in Fig. 21.3.

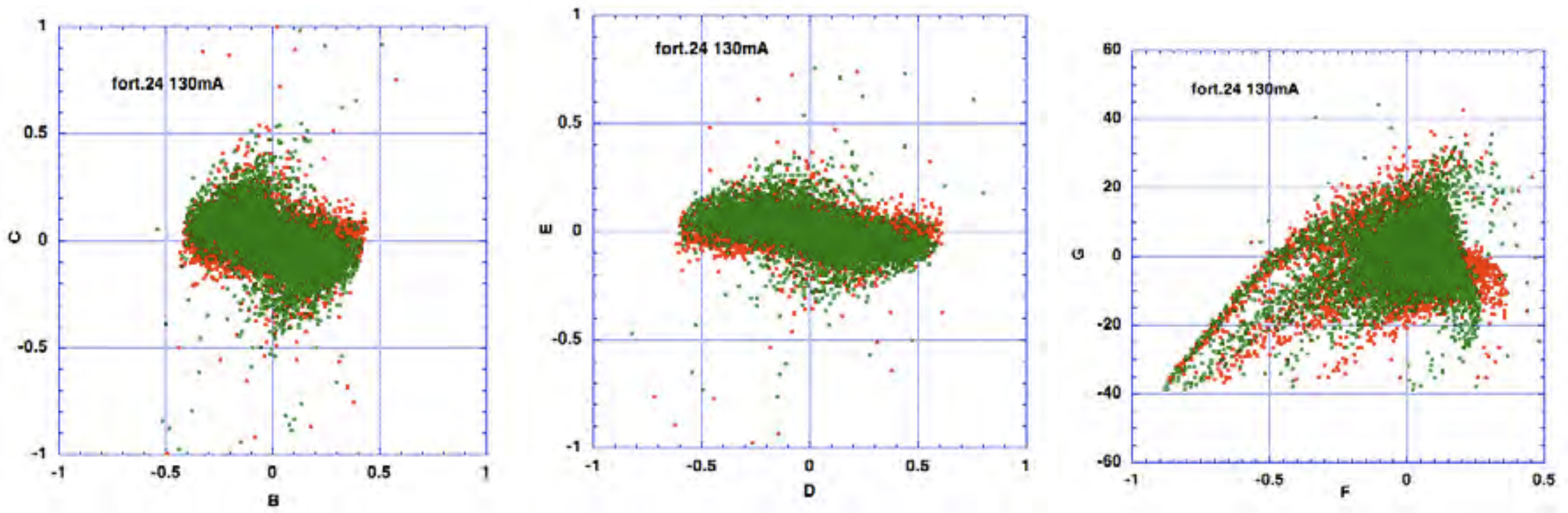

---

162 R.A. Jameson, ""Beam-Halo From Collective Core/Single-Particle Interactions", LA-UR-93-1209, Los Alamos National Laboratory, 31 March 1993.

R.A. Jameson, "Design for Low Beam Loss in Accelerators for Intense Neutron Source Applications - The Physics of Beam Halos", (Invited Plenary Session paper), 1993 Particle Accelerator Conference, Washington, D.C., 17-20 May 1993, IEEE Conference Proceedings, IEEE Cat. No. 93CH3279-7, 88-647453, ISBN 0-7803-1203-1. Los Alamos National Laboratory Report LA-UR-93-1816, 12 May 1993.

R.A. Jameson, "Self-Consistent Beam Halo Studies & Halo Diagnostic Development in a Continuous Linear Focusing Channel", LA-UR-94-3753, Los Alamos National Laboratory, 9 November 1994. AIP Proceedings of the 1994 Joint US-CERN-Japan International School on Frontiers of Accelerator Technology, , Maui, Hawaii, USA, 3-9 November 1994, World Scientific, ISBN 981-02-2537-7, pp.530-560.

C. Chen, R.C. Davidson, Q. Qian, R.A. Jameson, "Resonant and Chaotic Phenomena in a Periodically Focused Intense Charged-Particle Beam" (Invited paper for C. Chen), Proc. 10th Intl. Conf on High Power Particle Beams, NTIS, Springfield, VA 22151,(1994), 120-127.

C. Chen & R.A. Jameson, "Self-Consistent Simulation Studies of Periodically Focused Intense Charged-Particle Beams, Physical Review E, April 1995, PFC/JA-95-9 MIT Plasma Fusion Center

R.A. Jameson, "Beam Losses and Beam Halos in Accelerators for New Energy Sources", (Invited), Proc. 7th Intl. Symp. on Heavy Ion Fusion, Princeton, NJ, 6-10 Sept. 1996, Princeton, NJ, Fusion Engineering and Design 32-33 (1996) 149-157; Los Alamos National Laboratory Report LA-UR-96-175, 22 January 1996.

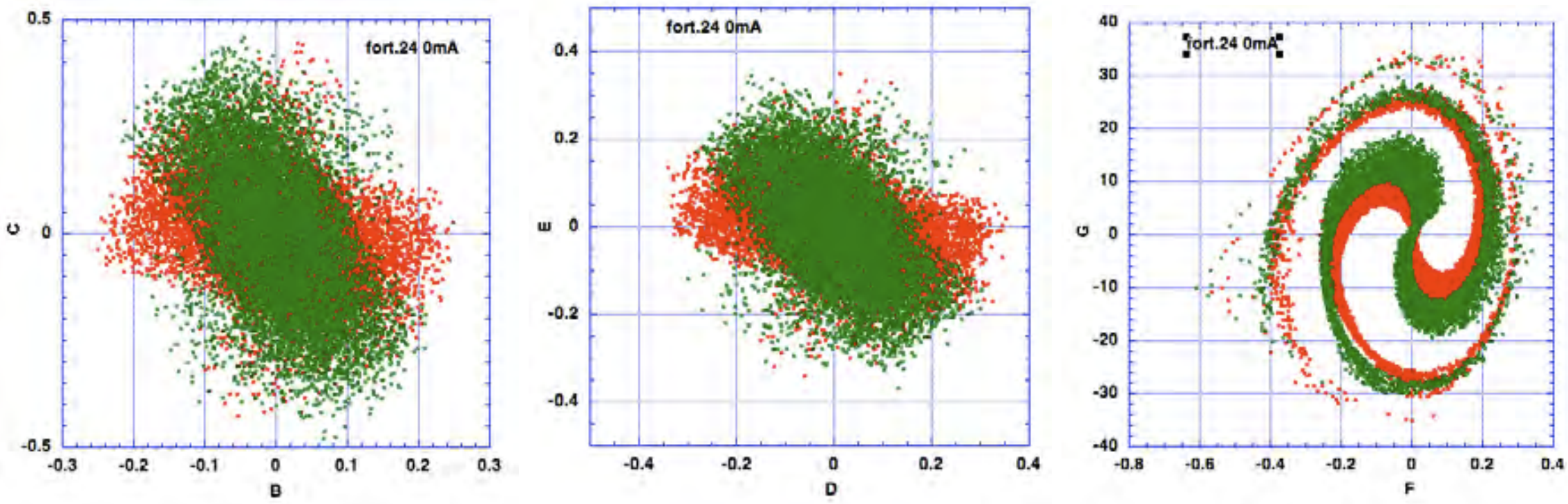

Fig. 21.3. Particle distributions of particles successfully accelerated to end of RFQ, at entrance of an initial cell (red), and at entrance of the initial cell plus two cells (green); xx’, yy’, zz’.. Upper – 130mA beam current. Lower zero beam current.

The rotations of the xx’, yy’, zz’ vectors are measured for each particle across a 2-cell period. The results for a particular particle are shown in Fig. 21.4, showing the dependence on the inverse tangent method used.

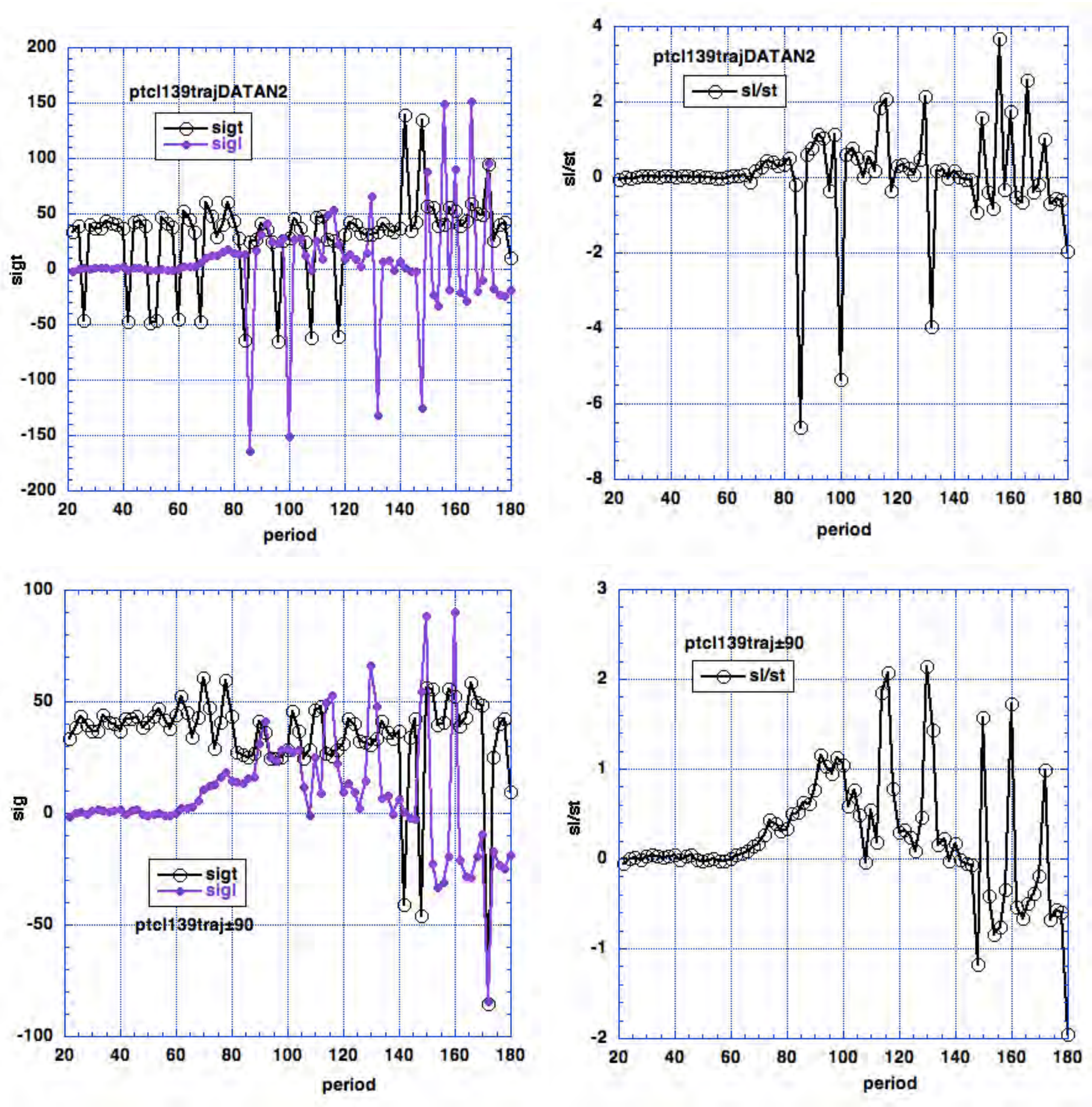

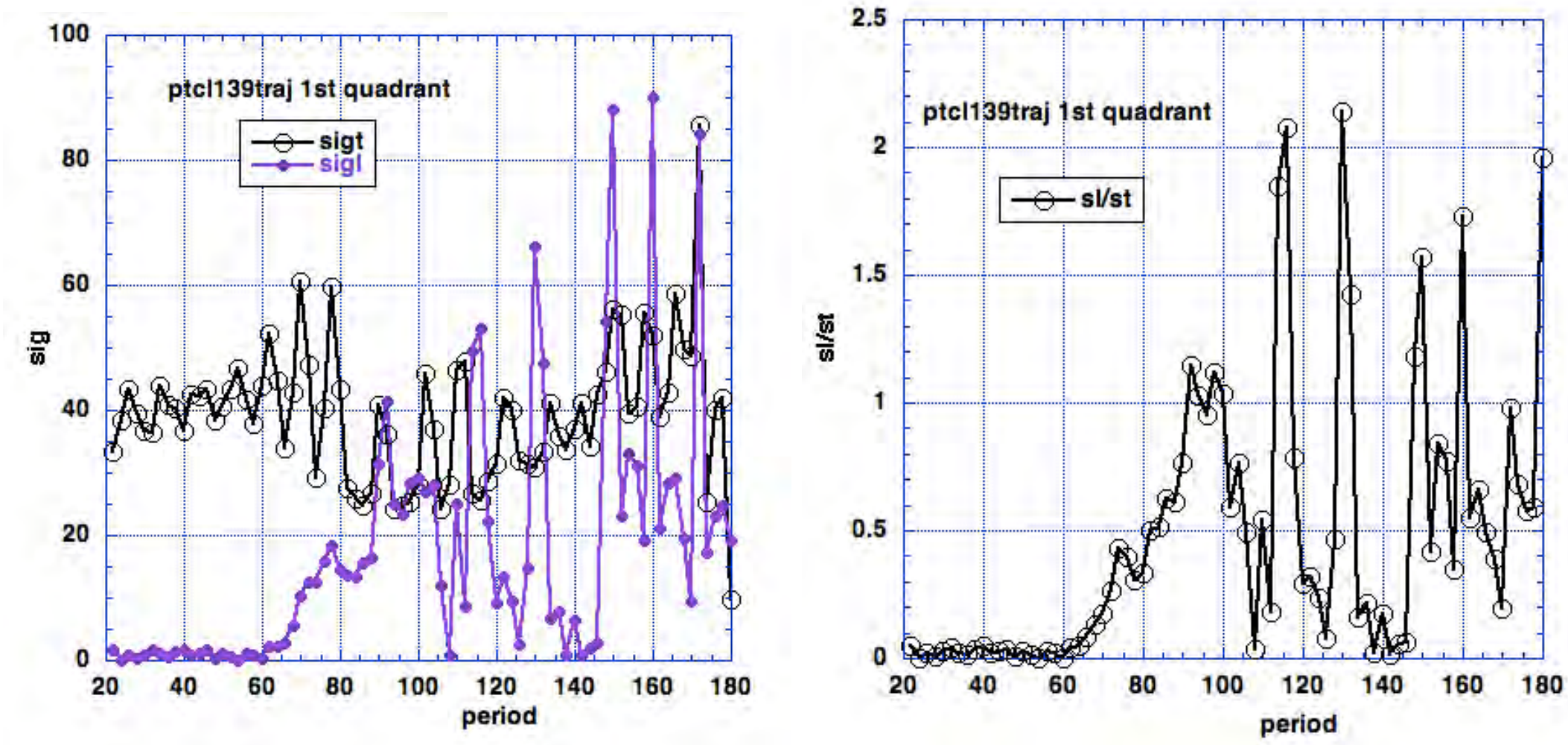

Fig. 21.4. Transverse phase advance (average of x and y) and longitudinal phase advance of a particular particle over two cells in 6D phase space (degrees). Top row – angle found with datan2, which returns results over ±180°. Middle row - angle found with datan, which returns results over ±90°. Bottom row - angle found as abs(datan), which returns results in the first quadrant.

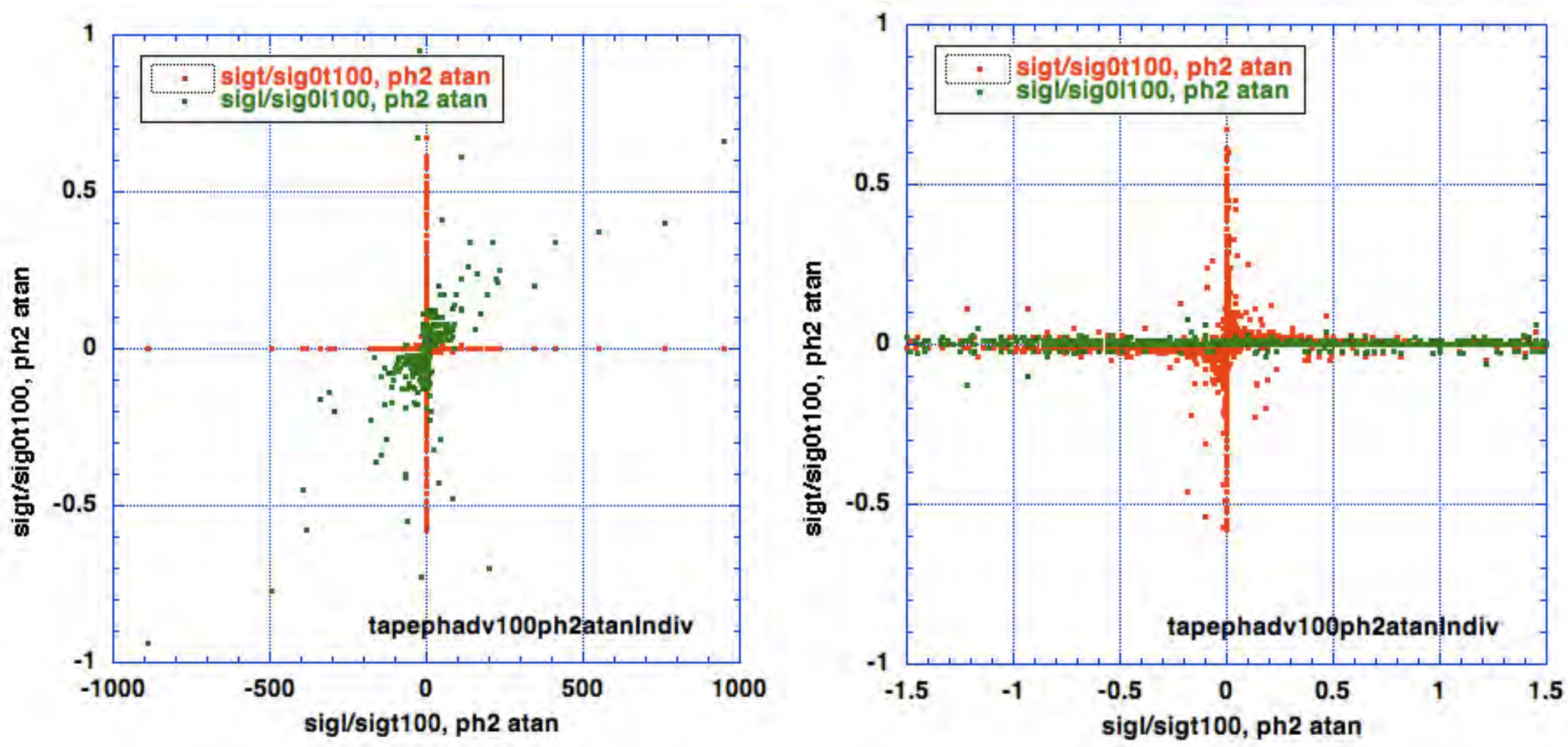

Fig. 21.5. Hofmann Chart for single particle 2-cell transverse phase advances in 6D phase space. Right is expanded horizontal scale.

The atan result on the Hofmann Chart across cells 100 and 101 is shown in Fig. 21.5. The expected "bowtie" qualitative shape, familiar in ring tune charts, is seen [163], but otherwise there seem to be no distinguishing features. The result is similar for all the 2-cell periods, and is clearly not a useful method.

### 21.3.2 Phase advance in framework of rms ellipses

The phase advance between two points in the elliptical framework of the rms betatron and synchrotron oscillations can be expressed in terms of the rms ellipse parameters, as shown in Fig. 21.6.

163 I, Hofmann, "A NewApproach to Linac Resonances and Equipartition ? – arXiv:1210.7991v1 [physics.acc-ph] 30 Oct 2012

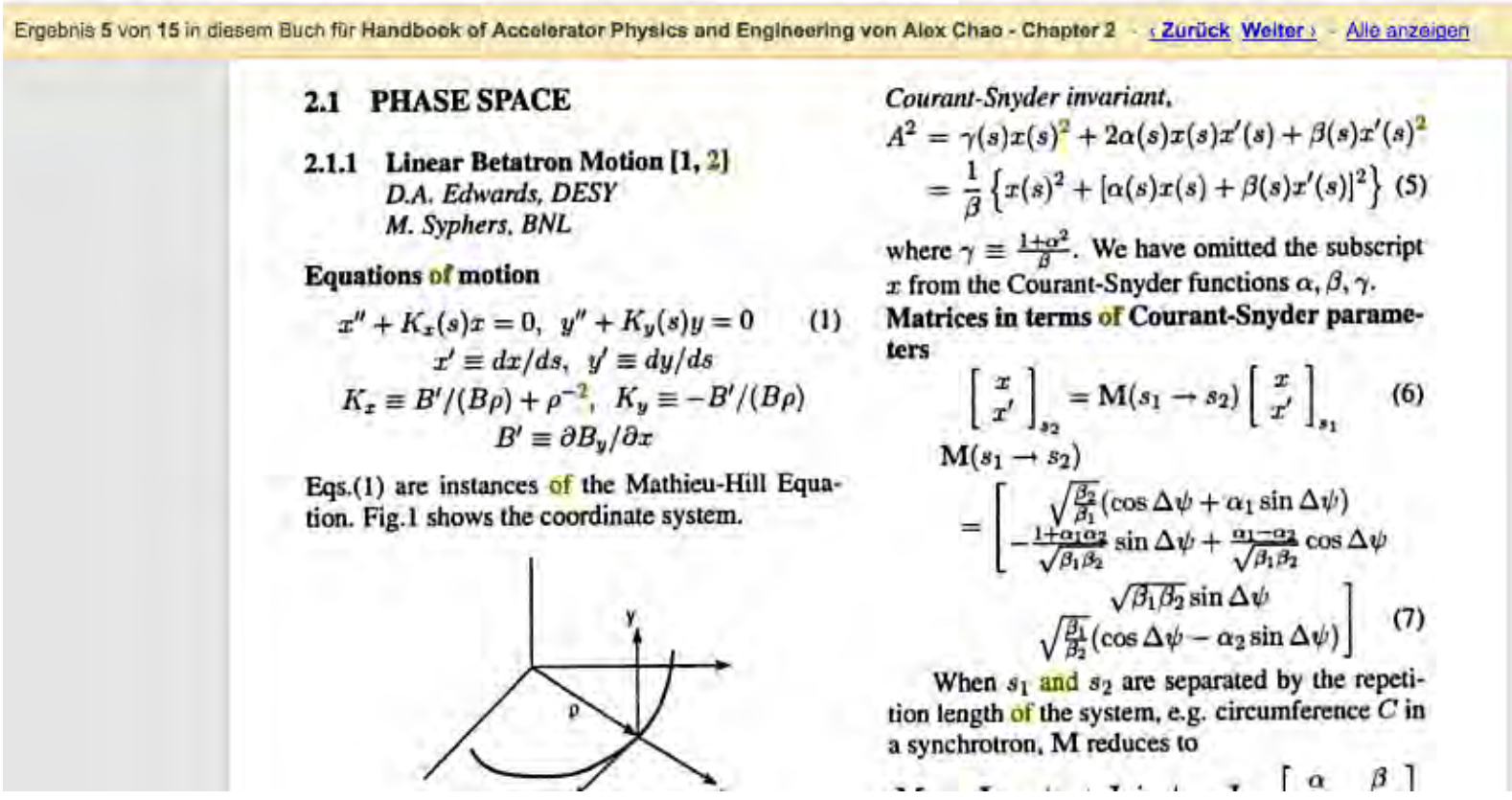

Ergebnis 5 von 15 in diesem Buch für Handbook of Accelerator Physics and Engineering von Alex Chao - Chapter 2 - ‹ Zurück Weiter › - Alle anzeigen

**2.1 PHASE SPACE**

**2.1.1 Linear Betatron Motion [1, 2]**
*D.A. Edwards, DESY*
*M. Syphers, BNL*

**Equations of motion**

$$x'' + K_x(s)x = 0,\ \ y'' + K_y(s)y = 0 \quad (1)$$
$$x' \equiv dx/ds,\ \ y' \equiv dy/ds$$
$$K_x \equiv B'/(B\rho) + \rho^{-2},\ \ K_y \equiv -B'/(B\rho)$$
$$B' \equiv \partial B_y/\partial x$$

Eqs.(1) are instances of the Mathieu-Hill Equation. Fig.1 shows the coordinate system.

*Courant-Snyder invariant,*

$$A^2 = \gamma(s)x(s)^2 + 2\alpha(s)x(s)x'(s) + \beta(s)x'(s)^2 = \frac{1}{\beta}\left\{x(s)^2 + [\alpha(s)x(s) + \beta(s)x'(s)]^2\right\} \quad (5)$$

where $\gamma \equiv \frac{1+\alpha^2}{\beta}$. We have omitted the subscript $x$ from the Courant-Snyder functions $\alpha, \beta, \gamma$.

**Matrices in terms of Courant-Snyder parameters**

$$\begin{bmatrix} x \\ x' \end{bmatrix}_{s_2} = \mathrm{M}(s_1 \to s_2) \begin{bmatrix} x \\ x' \end{bmatrix}_{s_1} \quad (6)$$

$$\mathrm{M}(s_1 \to s_2) = \begin{bmatrix} \sqrt{\frac{\beta_2}{\beta_1}}(\cos\Delta\psi + \alpha_1 \sin\Delta\psi) & \sqrt{\beta_1\beta_2}\sin\Delta\psi \\ -\frac{1+\alpha_1\alpha_2}{\sqrt{\beta_1\beta_2}}\sin\Delta\psi + \frac{\alpha_1-\alpha_2}{\sqrt{\beta_1\beta_2}}\cos\Delta\psi & \sqrt{\frac{\beta_1}{\beta_2}}(\cos\Delta\psi - \alpha_2\sin\Delta\psi) \end{bmatrix} \quad (7)$$

When $s_1$ and $s_2$ are separated by the repetition length of the system, e.g. circumference $C$ in a synchrotron, M reduces to

Fig. 21.6 Excerpt from Handbook of Accelerator Physics and Engineering [164]; definition of transfer function between two points in terms of Courant-Snyder ellipse parameters.

The rms phase advances come simply from local, instantaneous, evaluation of the rms envelope matching equations, assuming a constant periodic channel. For single particle phase advances, the ellipse parameters change across the section, so the initial and final ellipse parameters are used, Fig. 21.6 Eq. (6). The Hofmann Chart result across cells 100-101, using atan, is shown in Fig. 21.7.

The patterns exhibit the strong features of the longitudinal dphi-dW phase space, shown in Fig. 21.8. Single particle tunes range over all the major resonances. Detailed investigation of this pattern (with 100K particles) across the RFQ showed movement of the cluster following the rms tune, but no evidence that the eye could perceive of particle clustering near the main $\sigma^l/\sigma^t = 0, 0.33, 0.5, 1.0$ or 2.0 resonances. It means that redistribution within the beam and the effect on total beam size is essentially an rms effect; the rms trajectory may remain in the vicinity of a resonance long enough for redistribution to occur in accordance with the plasma periods. The single particles will be influenced collectively by the resonances, but in the case of a typical focusing and accelerating channel they will not be captured and held by the parametric resonances.

Thus some other method of display is sought to further investigate the single particle effects.

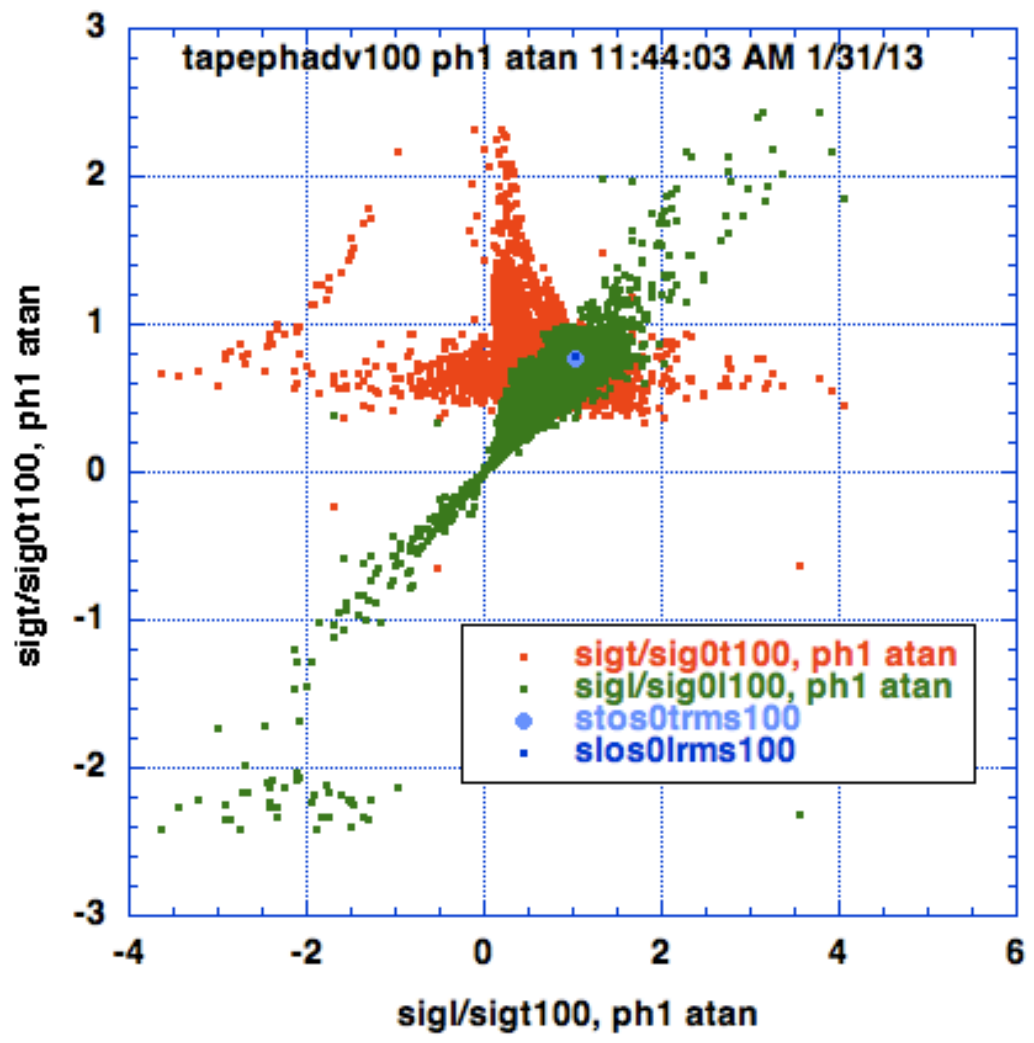

Fig. 21.7. Hofmann Chart for single particle phase 2-cell advances in framework of initial (cell 100 entrance) and final (cell 102 entrance) rms ellipses. The blue dots are the rms tunes at the end of cell 102.

164 "Handbook Of Accelerator Physics And Engineering" (3rd Printing), edited by: Alexander Wu Chao and Maury Tigner.

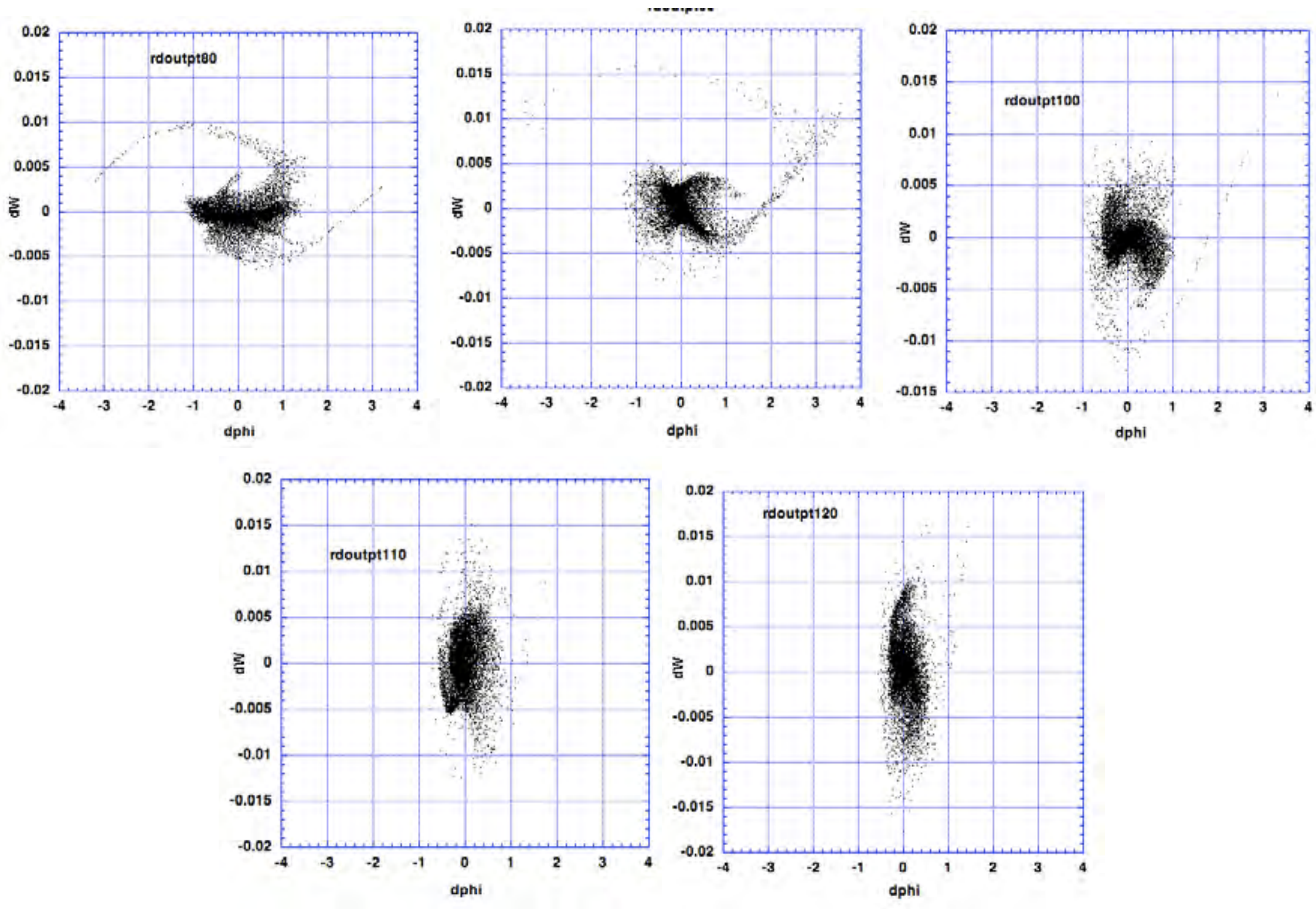

Fig. 21.8. dphi-dW phase space of individual particles at end of cells 80, 90, 100, 110, 120.

## 21.4 Single Particle Behavior When Near Main Resonances

The number and radii of particles within a ±0.03 band of $\sigma^l/\sigma^t$ = 0.33, 0.5, 1.0 and 2.0 are investigated in Fig. 21.9 and 21.10.

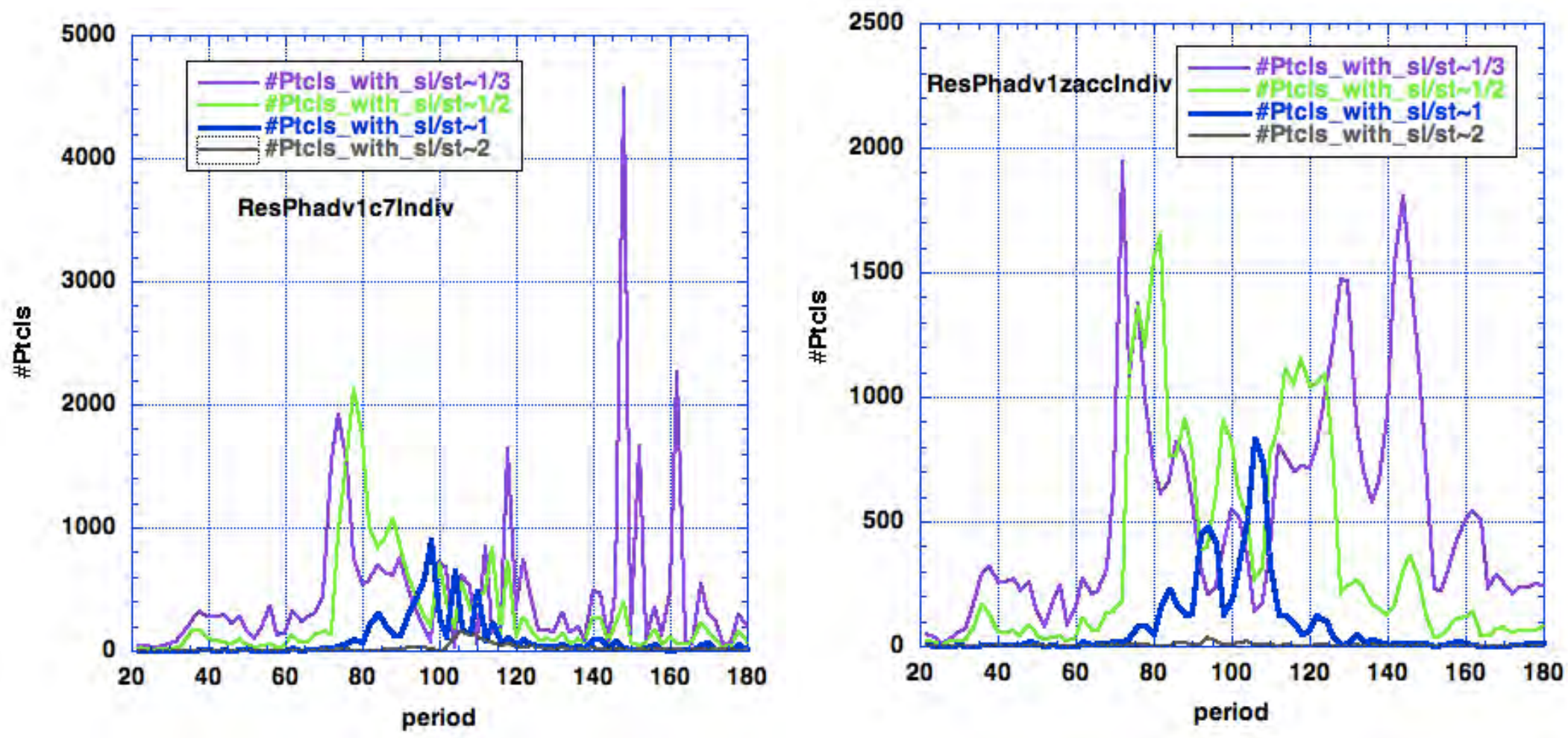

Fig 21.9. Number of particles within ±0.03 band of $\sigma^l/\sigma^t$ = 0.33, 0.5, 1.0 and 2.0 resonances. Left – particles being transmitted at each period. Right – particles accelerated successfully to the end of the RFQ.

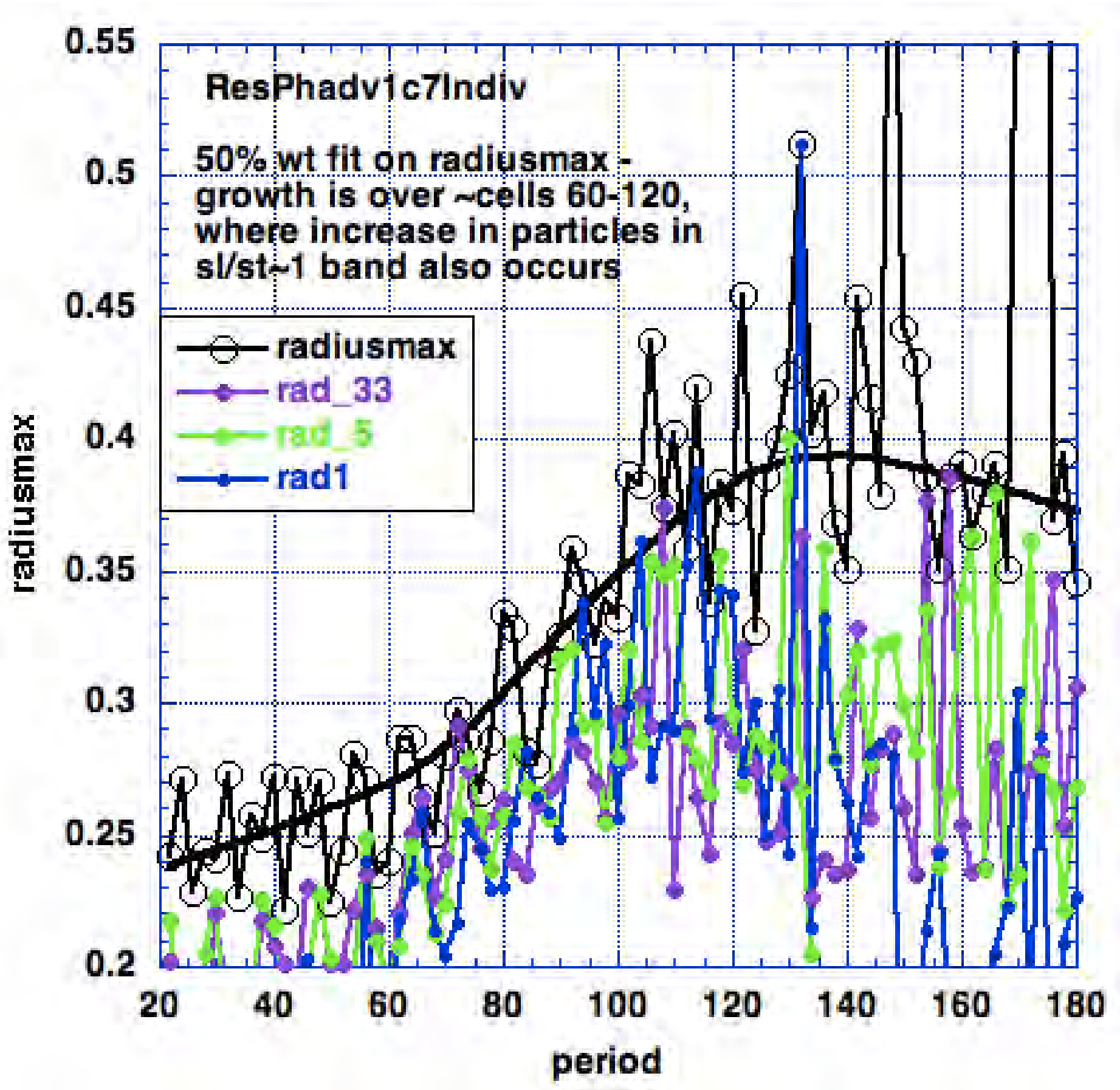

Fig. 21.10.a. Radii of particles within ±0.03 band of $\sigma^l/\sigma^t$ = 0.33, 0.5, and 1.0 resonances for particles being transmitted at each period. Radiusmax is the maximum radius of all particles at that period, solid line is 50% weighted fit. A particle in band near $\sigma^l/\sigma^t$ = 0.33 is at maximum radius at period 66, 72, 108 and 158; near $\sigma^l/\sigma^t$ = 0.5 at cell 166, and near $\sigma^l/\sigma^t$ = 1.0 at cell 132.

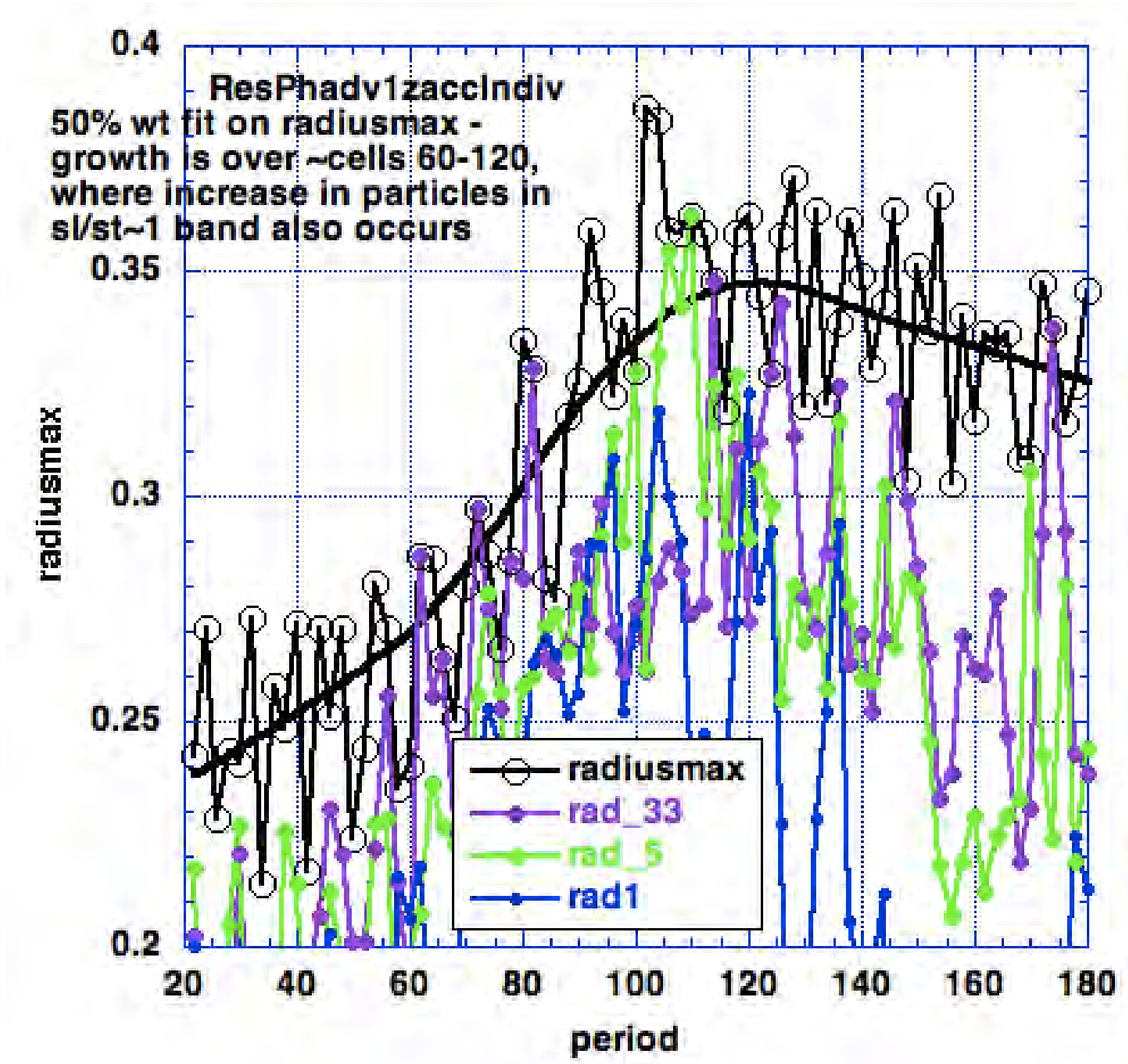

Fig. 21.10.b. Radii of particles within ±0.03 band of $\sigma^l/\sigma^t$ = 0.33, 0.5, 1.0 and 2.0 resonances for particles accelerated successfully to the end of the RFQ. Radiusmax is the maximum radius of all particles at that period, solid line is 50% weighted fit. A particle in band near $\sigma^l/\sigma^t$ = 0.33 is at maximum radius at period 61, 66, 72, 78, 82, 114, 124 and 174; near $\sigma^l/\sigma^t$ = 0.5 at cell 100, 110 and 170, but none near $\sigma^l/\sigma^t$ = 1.0.

Fig. 21.9 indicates that particles near the main resonances do concentrate when the rms tunes are in those regions, as expected. Fig. 21.10 indicates that a particle in these bands may indeed achieve the maximum radius in the distribution, even more so for the successfully accelerated particles. Although particles near $\sigma^l/\sigma^t$ = 1 are not so often at maximum radius, the maximum radius fit follows the integral of particles in this region, indicating an rms effect. Although there is a dip in the number of particles near $\sigma^l/\sigma^t$ = 0.33 and 0.5 while the number near $\sigma^l/\sigma^t$ = 1 peaks, maximum radii are occurring near $\sigma^l/\sigma^t$ = 0.33 and 0.5.

Fig 21.11 shows the number of particles, that are eventually accelerated successfully to the end of the RFQ, locally within ±0.03 band of $\sigma^l/\sigma^t$ = 0.33, 0.5, 1.0 and 2.0 resonances, as function of input beam current. As the amount and rate of mixing increases with increasing beam current, fewer particles are found locally near the resonances, again indicating that the action is essentially rms.

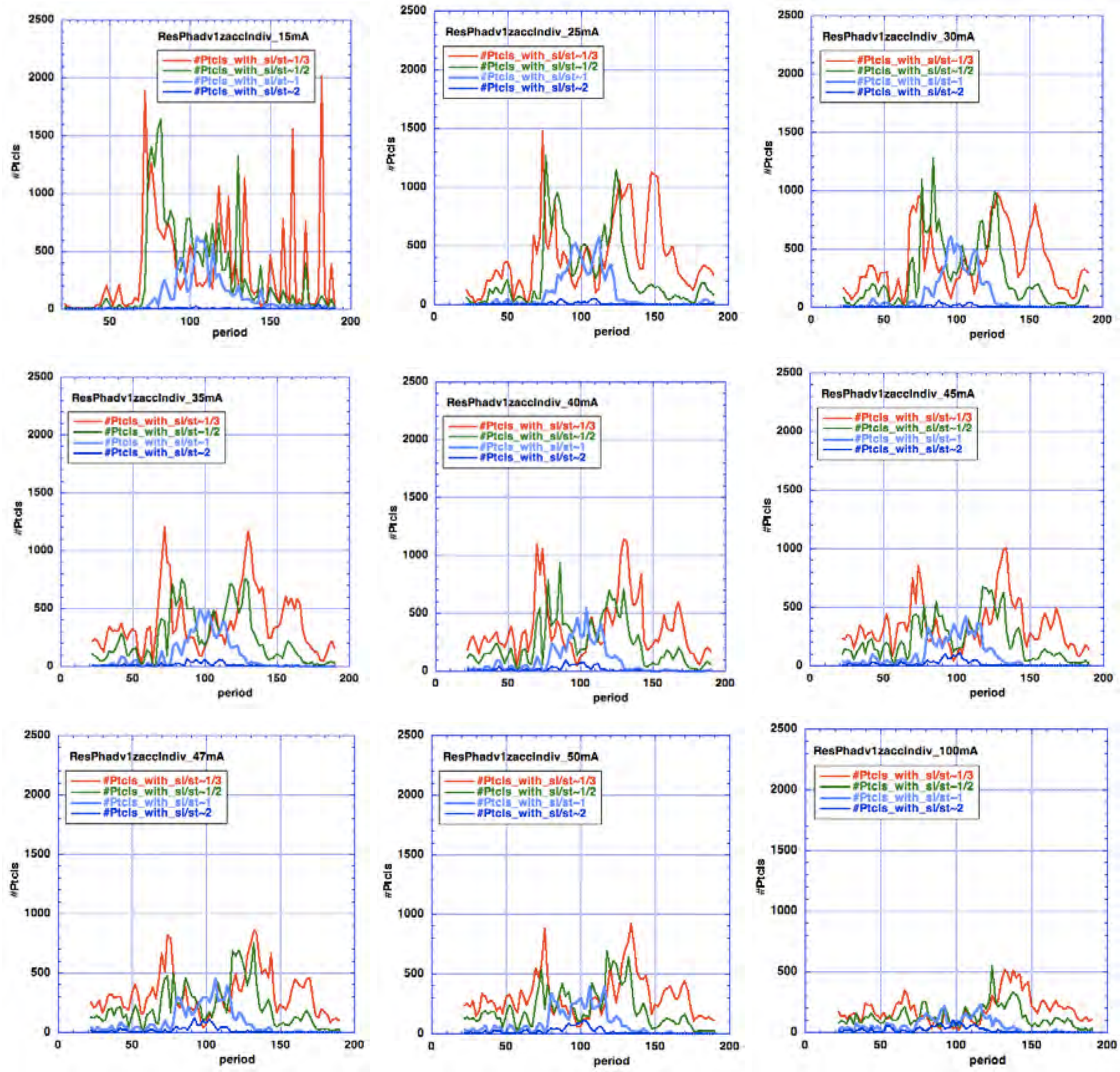

Fig 21.11. Number of particles particles accelerated successfully to the end of the RFQ, locally within ±0.03 band of $\sigma^l/\sigma^t$ = 0, 0.33, 0.5, 1.0 and 2.0 resonances, as function of input beam current.

The aperture and maximum particle radius and the number of transmitted particles at each RFQ cell are shown in Fig. 21.12. Losses begin around the cell where $\sigma^l = \sigma^t$.

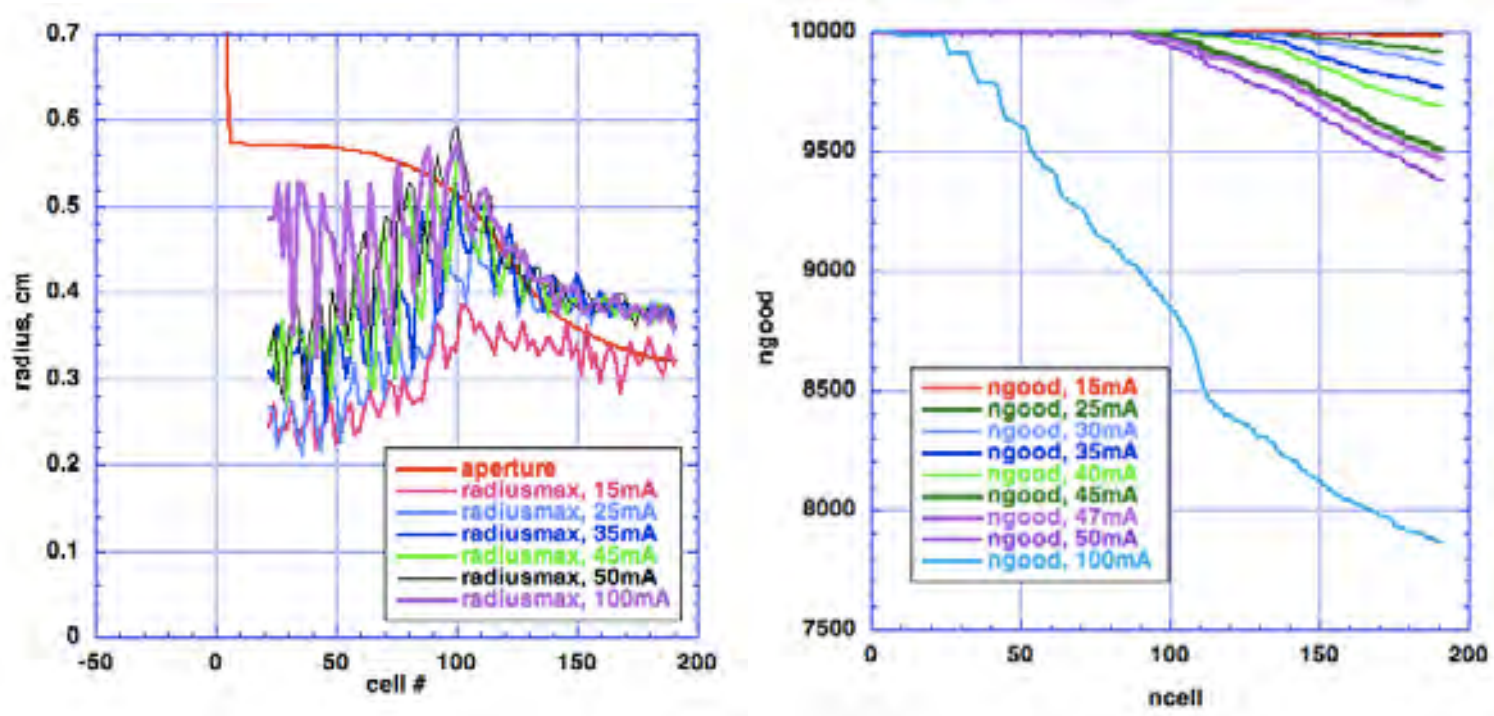

Fig. 21.12. Left: RFQ aperture and maximum particle radius as function of input beam current. Right: transmitted beam as function of input beam current.

## 21.5 Conclusions for Single Particle Behavior

- Single particle tunes inevitably have a large ± $\sigma^l/\sigma^t$ range.
- Entrance into the main $\sigma^l/\sigma^t = 1$ resonance will strongly influence maximum particle radii, even though the beam may briefly reach equipartition during the transit. The actual mechanism for a single particle to reach the maximum radius may be an enhancement of the action of other resonances from the excited tails of the $\sigma^l/\sigma^t = 1$ resonance.
- Observation of single particle tunes, in terms of the number and radii of particles near the main resonances, gives added utility in evaluating the channel performance. It helps understanding that the main effects seem to be from the rms behavior, and also of the influence of neighboring resonances tails at the local value of $\sigma^l/\sigma^{t.}$
- It appears that observation of the single particle tunes has no particular value for a study of total beam size and particle redistribution, but that maximum radii and emittance profile effects would be sufficient.

## 21.6 Question of formal appearance of Hofmann Chart:

It would seem that the space charge spreading would be to both sides of the bare resonance as the tune depression gets deeper? Why, for example, does the chart for eln/etn=0.1 show no spreading to the right side for kz/kx=0.33, 0.5 and 2.0, but does show spreading to the right for kz/kx=1.0?? Corresponding, the chart for eln/etn=10 shows no spreading to the left except for kz/kx=1.0??

From Prof. Hofmann: "You raise a good question. That everything is mirrored is not surprising, because the band at kz/kx =0.33 for ratio 0.1 is the same as the band at kz/kx=3 for ratio 10. But why does the kz/kx=1 have both shoulders? The question is more complicated, and I checked back our PRSTAB-paper of 2006 (attached). It appears that following Fig. 5 and 6 the broader stop-band at tune ratio 1 comes from two stop-bands of even and odd modes touching each other (remember that in second order the even modes are symmetric spatial deformations in x-z, the odd modes are tilted in x-z as if there was a skew quadrupole, similar in higher order). For all other resonances I notice that the even and odd mode stop-bands are separate, therefore sharp edge in one direction. That is only a formal explanation – sorry I don't have a good physical picture for it. Somehow it has to do with symmetry of modes."

In Fig. 21.13, the radii of particles to the left or right of the main resonances was observed, indicating no difference.

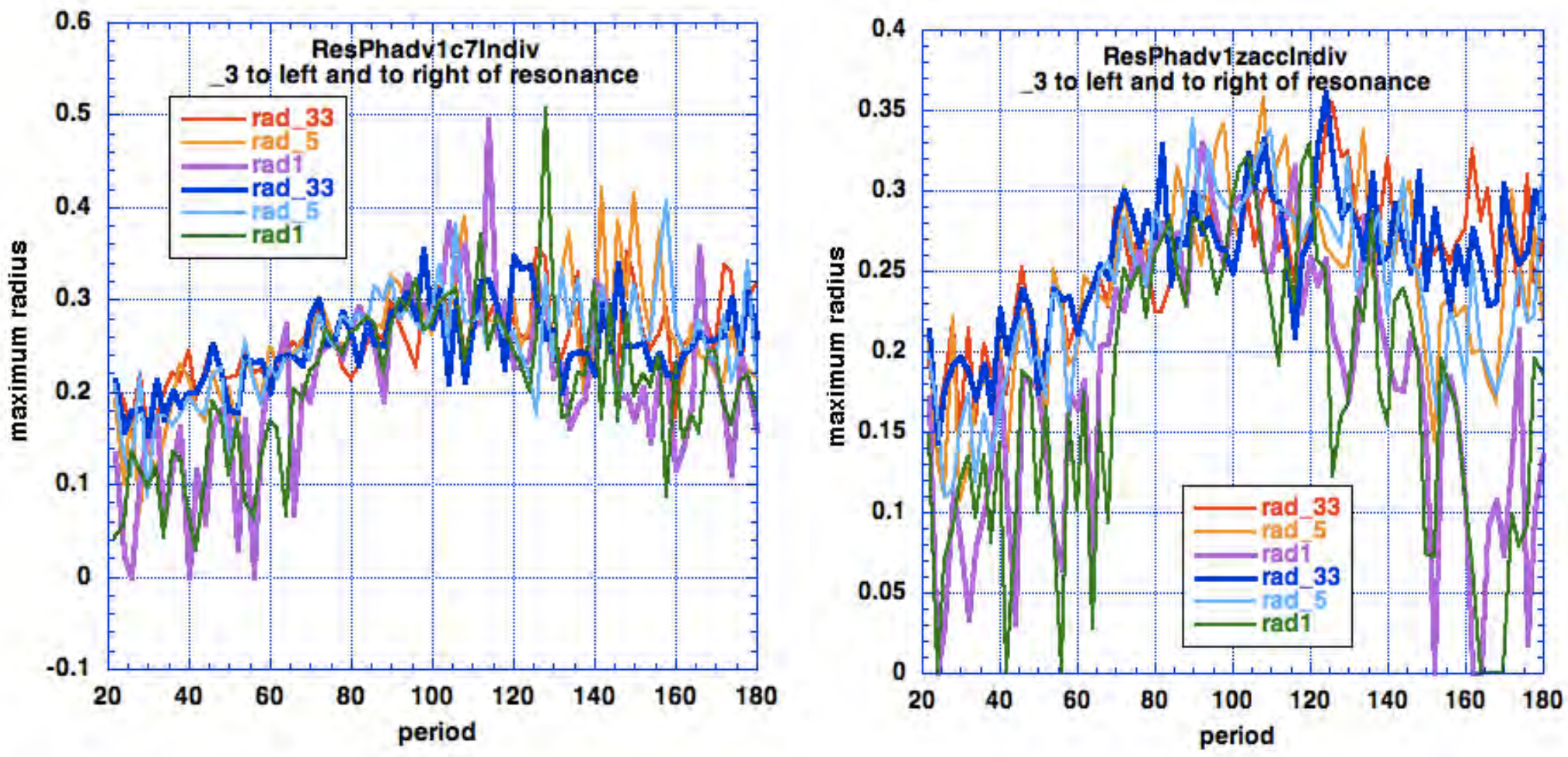

Fig. 21.13. Single particle evidence is that there is no difference between to left and to right of the bare resonance for bands of ±0.03 or ±0.1. Left – particles being transmitted in each period. Right – particles transmitted successfully to end of RFQ. Red, orange, purple – to left of resonance; dark blue, light blue, green – to right of resonance

## 21.7 Continuation of Total Beam Size and Redistribution Study

Being pursued in tedious detail in “Redistribution in Focusing and Accelerating Channels.docx”, Chapter 22.

## Notes, Extra Material:

See notes, 30 November 2012 and following. Complicated procedure.
Folder ‘Single Particle Phase Advance’

Synch particle or zcentroid enters new cell ~ when ntime=5. Only approximate because of oscillation of “synch” particle. So a sure test is on ntime, and can trigger in desired cell by checking if cord(5,npointp) or zcentroid is within 0.5*cell(4,ncell) of desired ncell.

Phase advance across one transverse focusing period (2 cells) from beginning of cell ncell at ntime=5, to beginning of ncell+2 at ntime=5 is good. Some early tests with other ntimes seemed not so good, but were maybe too early.

“2cell rotation” folder – early pictures showing rotation are good but no axis labels and not sure what conditions were. Later 12/19/2012 runs are not very clear – rotation is small…

In raw phase space, rotation over two cell period – then angle is divided by 2, or by 2*(# 2-cell periods) for summed. Summed is nearly same as for individual periods, so use individual periods. Have very wide distribution of points – not useful

In ellipse frame, period is one rotation around the ellipse, so do not divide by two. Summed is nearly same as for individual periods, so use individual periods. Get cluster of points with strong zz’ features on H chart – useful.

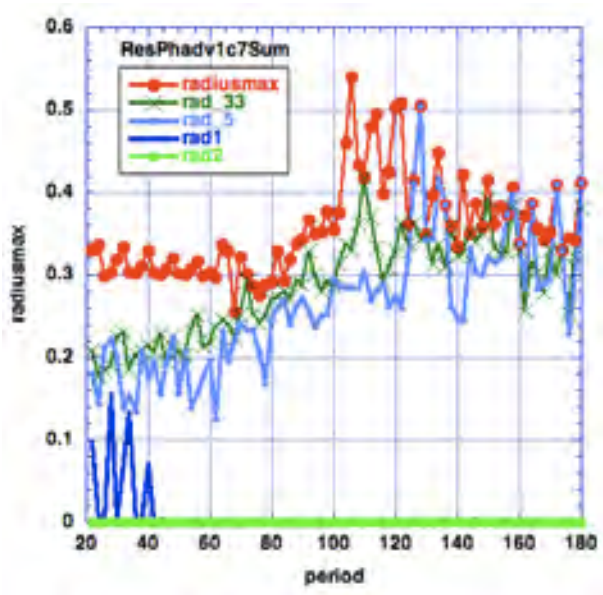

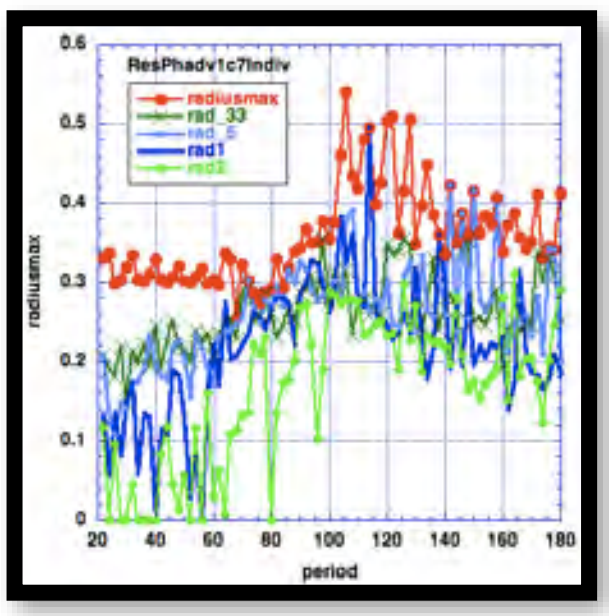

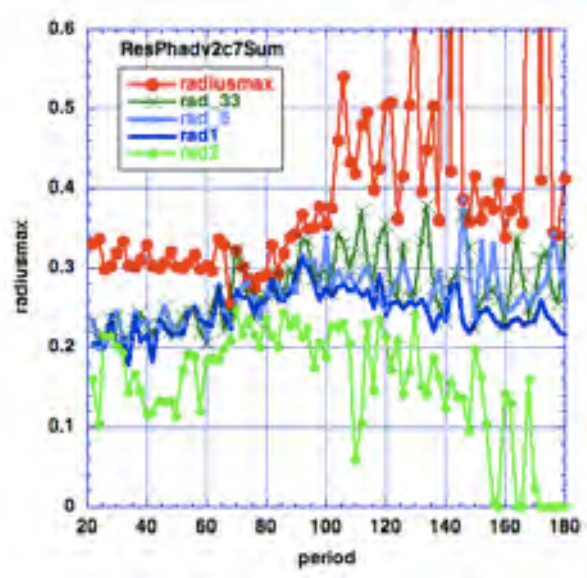

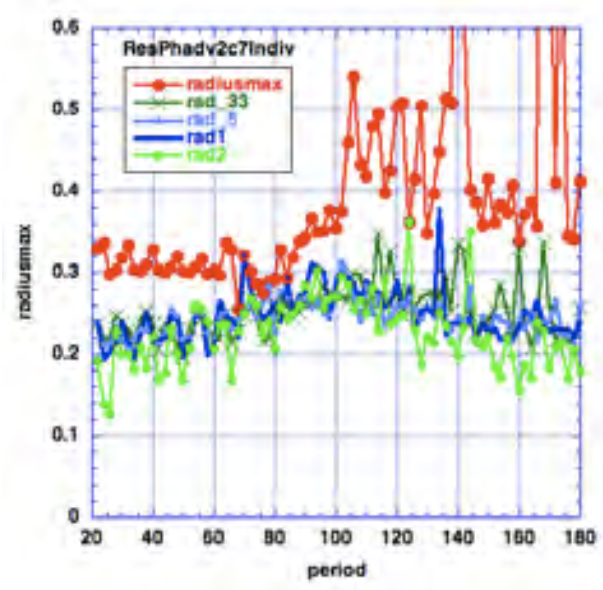

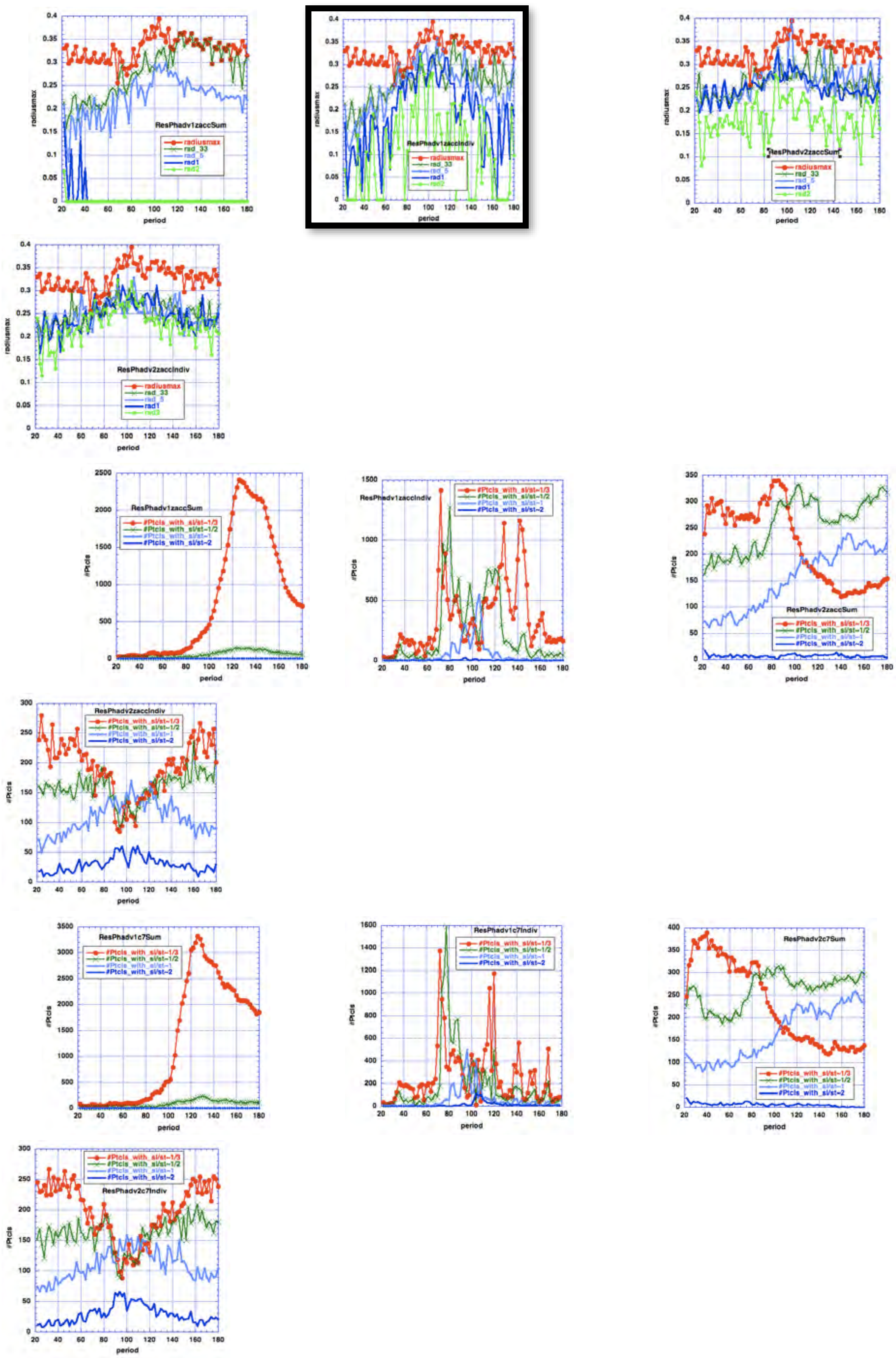

Fig.21.14 Maximum radius and count of particles involved in sigl/sigt = 1:3, 1:2, 1:1 and 2:1 resonances:

• phadv1 – in framework of rms ellipses.
•• computed using cord(7 or zlost -> zacc as filter
••• computed for each individual 2-cell period, or
••• running sum from beginning 2-cell period, divided by number of 2-cell periods

Conclusion: Best information is from phadv1, in frame of rms ellipses, for individual periods, maximum radius of particles involved in sigl/sigt = 1:3, 1:2, 1:1 and 2:1 resonance.

There may be further questions that could be investigated:

- Chao Li methods without smooth approximation.
- Where are there actually missing odd and/or even modes?
- Lapostolle showed there is actually a tilt in (average x&y = t)-z plane.
- The x and y tunes are generally different along z – we used the average, interpreting it as symmetric with respect to the slow motion, or points where the beam is xy round. But everywhere else, in the beta functions of the lattice, x and y are not symmetric.

# Chapter 22 – Redistribution in Focusing and Accelerating Channels

R. A. Jameson
October 2012, updated September 2013

The author has long observed that often in RFQs, the injected transverse waterbag transverse total-to-rms beam size ratio = sqrt(β etn/6) changes to ~sqrt((β etn/8), where etn, the normalized transverse rms emittance, is sought to be preserved by the design method, which includes control of the space charge physics. (β is the rms ellipse beta.). A change in the rms transverse beam size with respect to the design size is also sometimes observed. It is of interest to investigate redistribution in focusing and accelerating channels further [165,166,167].

The linac design method [168] uses the rms envelope equations with the smooth approximation phase advances, and is therefore general. It is convenient to use the RFQ for investigations.

Sec. I gives a brief background for discussion of the rms and total beam size and emittance behavior in linac channels. A focusing section with no bunching or acceleration, followed by smooth transition into a bunching and accelerating channel is investigated in Sec. II. Sec. III gives a long section with bunching but no acceleration, followed by a bunching and accelerating channel. In Sec. IV, three cases are studied to build a basis for understanding the behavior of a focusing and accelerating channel with very small beam loss, and comparing with a channel with continuous radial beam loss, and a channel with continuous beam loss from the accelerating bucket. Sec. V shows a non-equipartitioned RFQ, designed using a method that allows the longitudinal zero-current phase advance to approach, but not cross, the transverse zero-current phase advance midway in the main part of the RFQ.

The RFQ simulation code *LINACSrfqSIM* [169] uses the exact RFQ vane surfaces and finds the external and space charge fields using the multigrid Poisson method, and with many other improvements to insure accurate modeling. Meticulous comparison to the other full Poisson RFQ

---

165 Oscar A. Anderson, "Internal Dynamics and Emittance Growth in Space-Charge-Dominated Beams", Particle Accelerators, 1987, Vol. 21, pp. 197-226;

"Emittance Growth in Intense Mismatched Beams", 1987 Particle Accelerator Conference, Washington, DC, IEEE Cat. No. 87CH2387-9, p. 1043.

These papers clearly elucidated for the first time the important behavior of beams with mismatch of the particle distribution and rms characteristics, showing the physics of equilibration that occurs during the first one-quarter plasma period, and other results.

166 R.A. Jameson, ""Beam-Halo From Collective Core/Single-Particle Interactions", LA-UR-93-1209, Los Alamos National Laboratory, 31 March 1993.

R.A. Jameson, "Design for Low Beam Loss in Accelerators for Intense Neutron Source Applications - The Physics of Beam Halos", (Invited Plenary Session paper), 1993 Particle Accelerator Conference, Washington, D.C., 17-20 May 1993, IEEE Conference Proceedings, IEEE Cat. No. 93CH3279-7, 88-647453, ISBN 0-7803-1203-1. Los Alamos National Laboratory Report LA-UR-93-1816, 12 May 1993.

R.A. Jameson, "Self-Consistent Beam Halo Studies & Halo Diagnostic Development in a Continuous Linear Focusing Channel", LA-UR-94-3753, Los Alamos National Laboratory, 9 November 1994. AIP Proceedings of the 1994 Joint US-CERN-Japan International School on Frontiers of Accelerator Technology, , Maui, Hawaii, USA, 3-9 November 1994, World Scientific, ISBN 981-02-2537-7, pp.530-560.

C. Chen, R.C. Davidson, Q. Qian, R.A. Jameson, "Resonant and Chaotic Phenomena in a Periodically Focused Intense Charged-Particle Beam" (Invited paper for C. Chen), Proc. 10th Intl. Conf on High Power Particle Beams, NTIS, Springfield, VA 22151,(1994), 120-127.

C. Chen & R.A. Jameson, "Self-Consistent Simulation Studies of Periodically Focused Intense Charged-Particle Beams, Physical Review E, April 1995, PFC/JA-95-9 MIT Plasma Fusion Center

167 A. V. Fedotov, R. L. Gluckstern, S. S. Kurennoy, R. D. Ryne, "Halo formation in three-dimensional bunches with various phase space distributions", PRSTAB, Vol. 2, 014201 (1999)

168 "RFQ Designs and Beam-Loss Distributions for IFMIF", R.A. Jameson, Oak Ridge National Laboratory Report ORNL/TM-2007/001, January 2007.

169 "Framework for Linear Accelerator Design and Simulation – Development for RFQ", R. A. Jameson, with J. Maus, in preparation, October 2012.

simulation code *LIDOS* [170] shows that these two codes are in agreement, with full understanding of differences, as verified by direct discussions with the *LIDOS* authors in October 2012. Sec. 22.8 updates the simulation results of fifteen RFQs designed to various criteria, some of which were reported previously in [171]. The examples are then grouped in terms of their correspondence to the three cases of Sec. 22.5. From these examples, some conclusions are drawn in Sec. 22.9.

## 22.1 Background

The framework for linac focusing and accelerating channel design is given in [172], involving consideration of the rms, single particle, and plasma related behavior of a particle distribution. It is essential to understand that it is the transient behavior – the local behavior – that is of interest, and not a steady-state condition.

The evolution of the rms behavior is understood via the external field evolution, its periodicity or lack thereof, its adiabaticity or lack thereof, and through the mechanism of interaction with, or indeed capture by, coupling resonances between the degrees of freedom, which one explores using the Hofmann chart.

An important concept is "transport" in phase space. How do particles get their starting conditions, and how do they move in phase space. In the mid-1990's, there was almost no literature on transport. The "road map" of intersections of orbits of particles with arbitrary initial conditions is informative but not sufficient. An important way to get information about the rates of evolution is the "exit time" concept. The dynamics is that of the now classical nonlinear system, in which ion linacs operate below the stochastic limit, resonances are the main mechanism, there are only local chaotic areas, and the exit times are long.

A second interaction, first explained by O.A. Anderson is that redistribution inside the beam can occur in about (plasma period)/4. This is also a collective *non-resonant* effect, but with much faster change that the resonance interactions.

**This mechanism – space-charge mixing – is always present and is the basic underlying mechanism of all redistributions of the beam, also when interacting with resonances.**

As it is also defined by the rms quantities, it would be observed in changes of the rms emittance and beam size. Changes in the total emittance or beam size may accompany – this needs more investigation. Typical (plasma period)/4 for a tune shift of 0.4 or below is only a few linac cells. The internal distribution can change quickly to follow the parameters, while the rms properties may take longer.

---

170 "Code Package for RFQ Designing", B. Bondarev, A. Durkin, Y. Ivanov, I. Shumakov, S. Vinogradov, Proc. 2nd Asian Particle Accelerator Conf.., Beijing, China, 2001. "LIDOS.RFQ.Designer", User's Guide, publ. by AccelSoft, Inc.

171 "A Discussion Of RFQ Linac Simulation Jameson, R. A. Jameson, LA-UR-07-0876, Los Alamos National Laboratory, 2/8/07 (re-publish of LA-CP-97-54, September 1997).

172 R.A. Jameson, "Framework for Linac Focusing and Accelerating Channel Design", November 2012, in progress.

This type of redistribution is clearly seen in simulations of the first few cells of machines with space charge when waterbag or other nonequilibrium distributions [173] are injected, the fast reaction of the distribution when the channel trajectory enters or leaves a resonance zone, at transitions, etc.

Single particles within the beam bunch experience a tune spread; their behavior can also be related to any of the rational fraction resonances of the Hofmann Chart. The *single particle phase advance* must be realistically defined, as above, to allow negative values, as a particle may stop temporarily or move backward. Techniques have been developed for detailed analysis [Ch.21].

## 22.2 Focusing Channel Followed by Accelerating Channel

We wish to look first at redistribution that occurs when beams with various initial distributions are injected into a focusing channel with no acceleration, which is followed by an accelerating and focusing channel.

An RFQ was designed with a long initial quadrupolar focusing channel with no acceleration or bunching (up to cell 391) and rms transverse zero current phase advance s0t = 42°. A dc beam is injected; the tracked 2-cell beam slice has rms longitudinal beam size = (cell length)/sqrt(3) = 0.5 cm. The transverse rms beam size is determined via the transverse rms matching equation by the external field and beam current.

The focusing section is followed by a standard accelerating channel which brings the beam to EP at the shaper end (EOS->end-of-shaper) (cell 434) and maintains EP to the end. The design required the EP ratio eln*a/etn*b = 1 and the ratios b/a=eln/etn=1.6, using the space charge form factor ff = a/3b. An optimum aperture at the shaper end is used, equal to βλ/aperfac, where aperfac=4.1. The design was made taking into account the multipole effects of circular vane tips (radius = Rho) with constant ratio Rho/r0 = 0.75, where r0 is the average aperture = (minimum aperture)(1+(vane modulation)).

The design goal was to have high transmission and accelerated beam fractions; >95% is achieved (Fig. 22.1.a.). Particles are declared lost only if they intercept the vane surface. Particles are declared accelerated if their velocity is within 10% of the design velocity. Only the particles successfully accelerated to the end of the RFQ are then analyzed to get the rms and total beam quantities given in the figures, at mid-cell, where the emittance ellipse alphas are ~zero; in this way (discarding radially lost particles) the performance compared to the design makes sense.

DC beams with zero energy spread and with transverse waterbag, conical, parabolic, and Gaussian truncated at 3s, distributions having the same rms beam size were injected [174, 175]. The initial t100/trms ratios are:

| | |
|---|---|
| Waterbag | sqrt(6) = 2.45 |
| Parabolic | sqrt(8) = 2.83 |
| Conical | sqrt(8.451) = 2.91 |
| Gaussian | ~sqrt(10) = ~3.16 |

### 22.2.1 Full Poisson Simulation

With full Poisson solutions for the external and space charge fields in *LINACSrfqSIM*, the non-ideal vane shape and image effects are accurately modeled. The rms beam sizes and rms emittances are shown in Fig. 22.1.b-f. The rms size is essentially independent of the initial distribution, with only a small effect on the rms emittances, as expected. In the focusing section, there is a longitudinal rms emittance heating and transverse rms emittance cooling, indicating an equipartitioning process that is tending to stabilize.

---

173 G. Parisi, "Investigations on Particle Dynamics in a High Intensity Heavy on Linac for Inertial Fusion", PhD. Dissertation, Goethe Uni Frankfurt, 28 September 1999.

174 G. Parisi, "Investigations on Particle Dynamics in a High Intensity Heavy on Linac for Inertial Fusion", PhD. Dissertation, Goethe uni Frankfurt, 28 September 1999.

175 F. J. Sacherer, "Rms Envelope Equations With Space Charge", PAC1971, pp. 1105-1107.

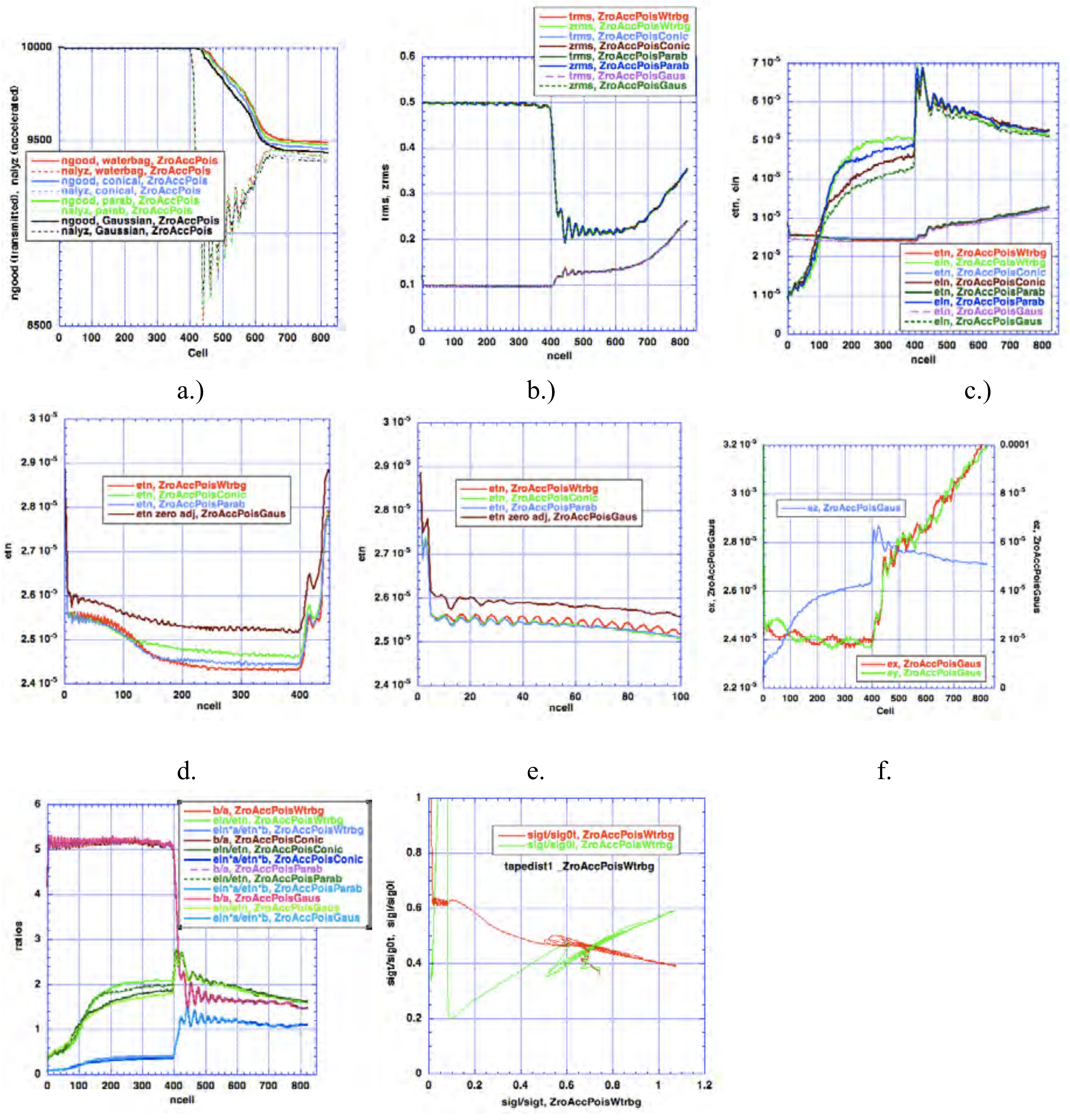

Fig. 22.1. Rms behavior for focusing channel followed by accelerating channel.
a.) transmitted and accelerated beam, b.) rms beam sizes, and c.) rms emittances,
d.) transverse rms emittances to end of transport section.
e.) transverse rms emittances expanded scale, showing immediate redistribution in ~(plasma period)/4.
(A real transverse emittance of ten times the rms emittance was used for the Gaussian distribution; the figure shows that this ratio is not exactly correct.)
f.) ex, ey and ez emittances for the Gaussian distribution.
g.) EP ratios
h.) Hofmann chart

The input beams are not exactly matched to the preferred distribution of the focusing channel. The transverse plasma phase advance = sqrt((s0t^2-st^2)/2/ff) is ~44°/cell, so the plasma period is ~ 9 cells; Fig. 22.1.e. indicates an immediate redistribution of the transverse rms emittance within about ¼ plasma period as predicted. Fig. 22.1.f. shows that the beam is well matched and that there is no x,y emittance splitting [176].

The focusing section was not carefully matched to the following accelerating section; the shaper brings the beam to a roughly equipartitioned state. In the accelerating channel, the distribution moves toward more exact equipartitioning, with transverse heating and longitudinal cooling via interaction with the 1:1 resonance at sigl/sigt=1.

The behavior of the ratios of 100% transverse beam size to the rms transverse beam size (t100/trms) and corresponding longitudinal ratio (zw100/zrms) are shown in Fig. 22.2. The curves show fast variations because the fast terms in the dynamics are also being sampled – the variations at the frequency of one focusing period, characterized as the "flutter factor". These are averaged out in the smooth approximation, which uses rms quantities.

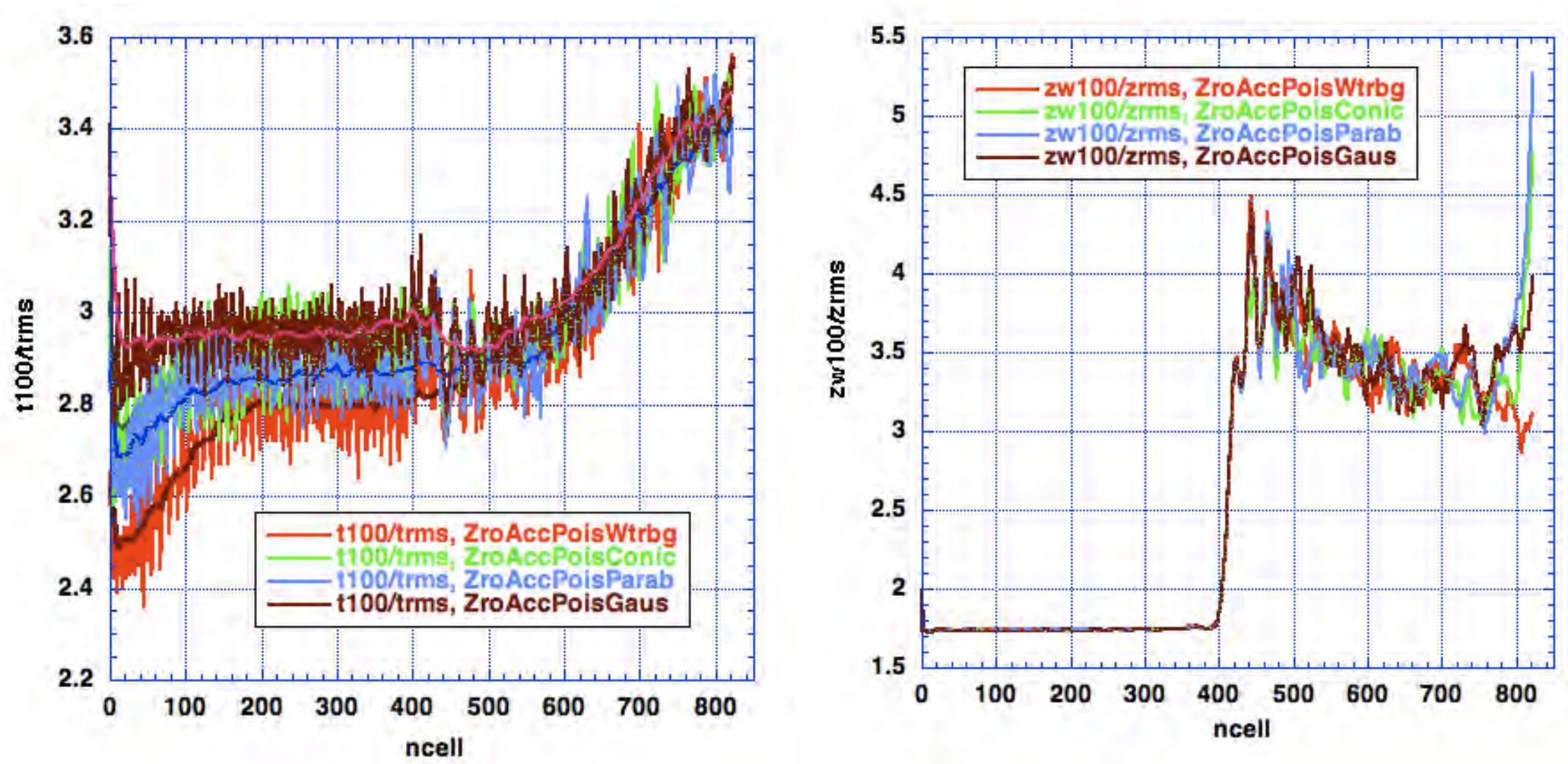

d. e.

Fig. 22.2. a.) Ratio of 100% transverse beam size to the transverse rms beam size (t100/trms) b.) corresponding longitudinal ratio (zw100/zrms).

In the transport section up to cell 391, the transverse total beam size redistributes more slowly that the initial ¼ plasma period redistribution, to an equilibrium with a more parabolic t100/trms ratio between 2.8 and 3.0, but depending on the form of the input distribution. This redistribution occurs on the same time scale as that of the rms emittances (Fig. 22.1.d.); the Hofmann chart (Fig. 22.1.h.) indicates that there will be interaction with many low order resonances in the transport section.

In the shaper section from cell 392-434, the synchronous phase is raised linearly from -90° to -88°. There is a rapid redistribution of all of the input distributions to a nearly common, approximately parabolic, distribution. (See Reiser 2nd Ed., 5.4.8 Longitudinal Motion in rf Fields and the Parabolic Bunch Model, using Neuffer's work which shows the naturalness of the parabolic distribution.)

Then the synchronous phase increases with acceleration according to a rule that regulates the longitudinal bunch length with respect to the accelerating bucket length. t100/trms grows in the same manner as the acceleration profile, reaching ~3.4 at the end (total-to-rms transverse emittance ratio ~ 12). zw100/zrms reaches an approximately Gaussian distribution as the beam bunches and accelerates.

### 22.2.2 2-term External Field and Cylindrically Symmetric Space Charge Field Simulation

176 I. Hofmann, "A New Approach to Linac Resonances and Equipartition ?", arXiv:1210.7991v1 [physics.acc-ph] 30 Oct 2012

*LINACSrfqSIM*'s "2-term" mode uses the 2-term external field (with transverse focusing multipole a01=1, but same acceleration multipole a10), and an r-z cylindrically symmetric space charge model with the radial boundary condition set to a closed boundary pipe with radius at the minimum aperture of the RFQ cell. The multipole effects from the non-ideal vane tip profile are thus absent, and the transverse image effect is not quadrupolar. These differences have some effect, as shown in Fig. 22.3.

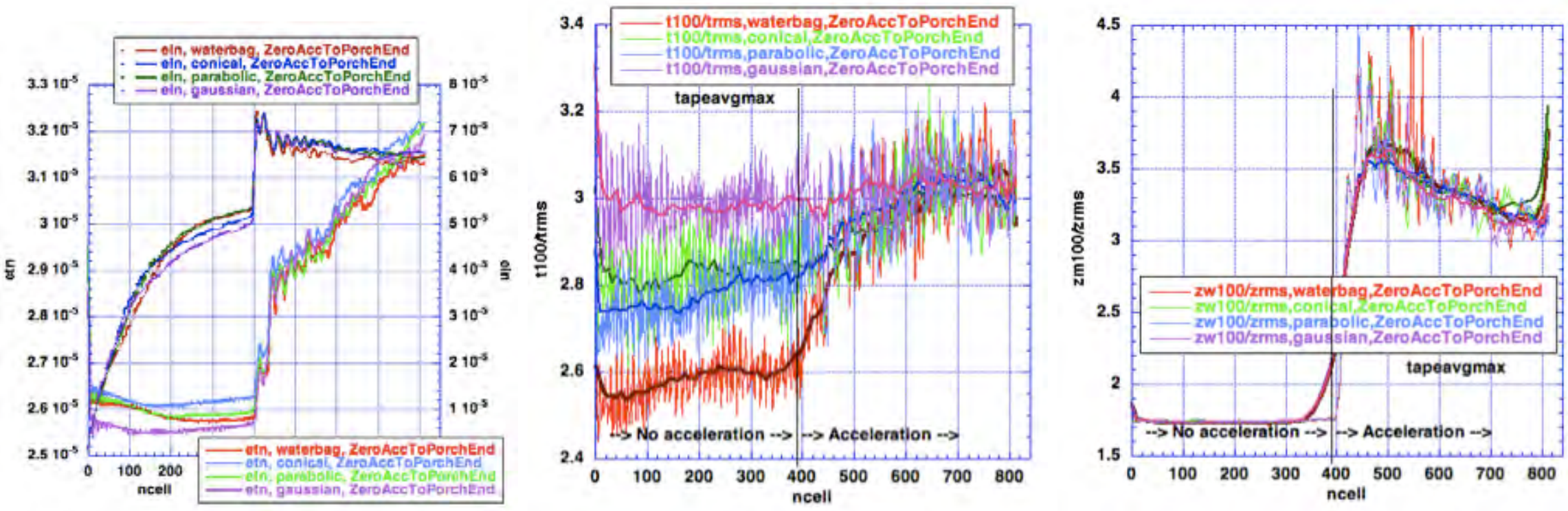

Fig. 22.3. a.) as Fig. 22.1.c., b.) as Fig. 22.2.a. and c.) as Fig. 22.2.b., with "2-term mode" simulation.

In the transport section, the rms emittance behavior in Fig. 22.3.a. and Fig. 22.1.c. is similar, although with more difference in transverse and less in longitudinal. The t100/trms ratios tend to remain more constant and separate. In the acceleration section, t100/trms redistributes to an ~3σ Gaussian distribution, but more slowly – the total distributions are farther apart at the start of the accelerating section. The multipole and quadrupolar image field nonlinearities are absent, resulting in significantly smaller effect on the total size compared to the Poisson simulation.

## 22.3 The Natural Matched x-y Distribution in an RFQ

It was noted above that the input waterbag, conical, parabolic beams are not exactly matched to the preferred distribution of the RFQ focusing channel, but quickly redistribute. Simulations that include the quadrupolar symmetry with actual vane tip shapes for both external and space charge fields, e.g. *LIDOS* and *LINACS* with full Poisson, show that the {x,y}, {x',y'} distributions in an RFQ quickly form a diamond shaped pattern with the tips at the vane tips. Fig. 22.4.a shows the typical pattern with full Poisson *LINACS* simulation. Fig. 22.4.b. shows the distribution resulting from 2-term external field and a cylindrically symmetric space charge simulation; as expected, the shape remains essentially round; particles are seen in the regions between the vanes.

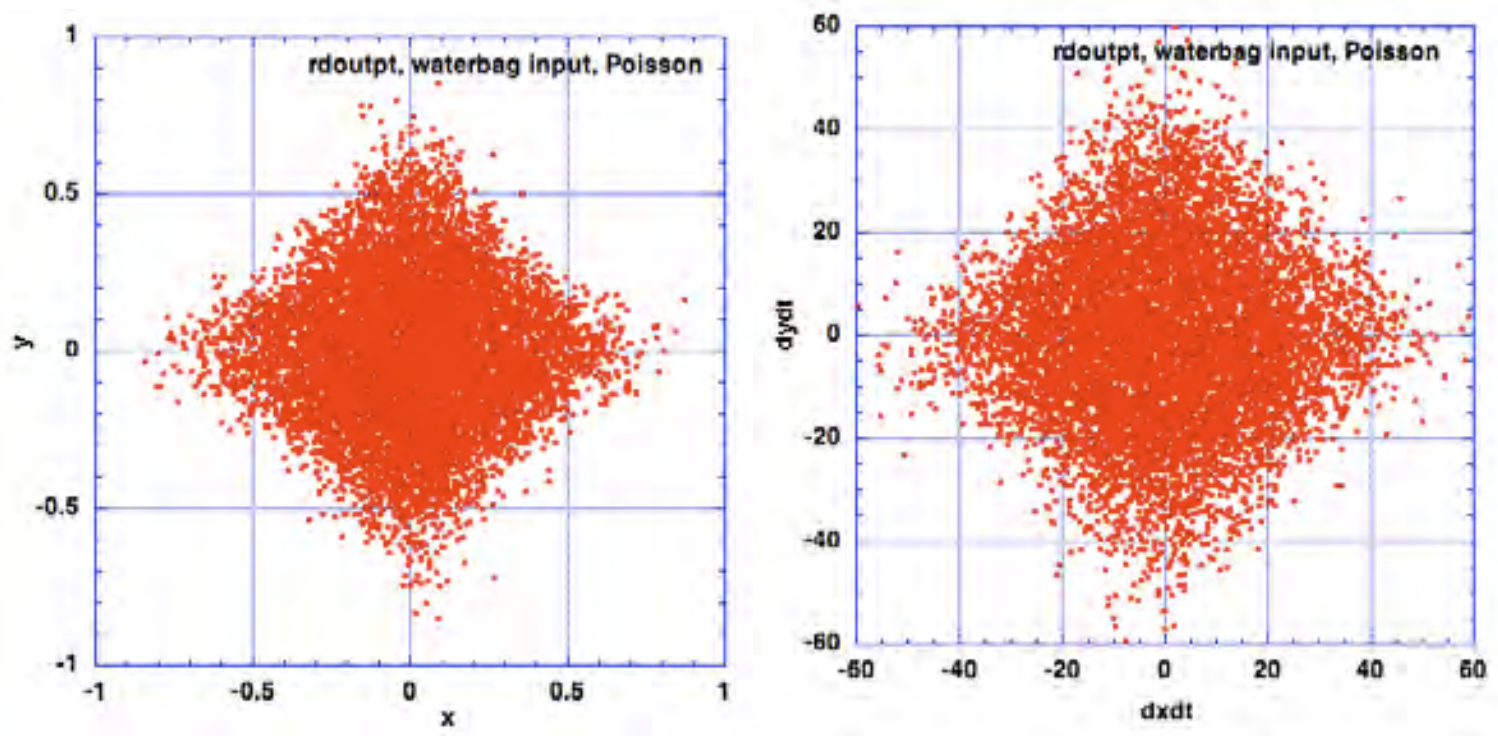

Fig. 22.4.a. Typical phase space distributions in an RFQ with full Poisson fields.

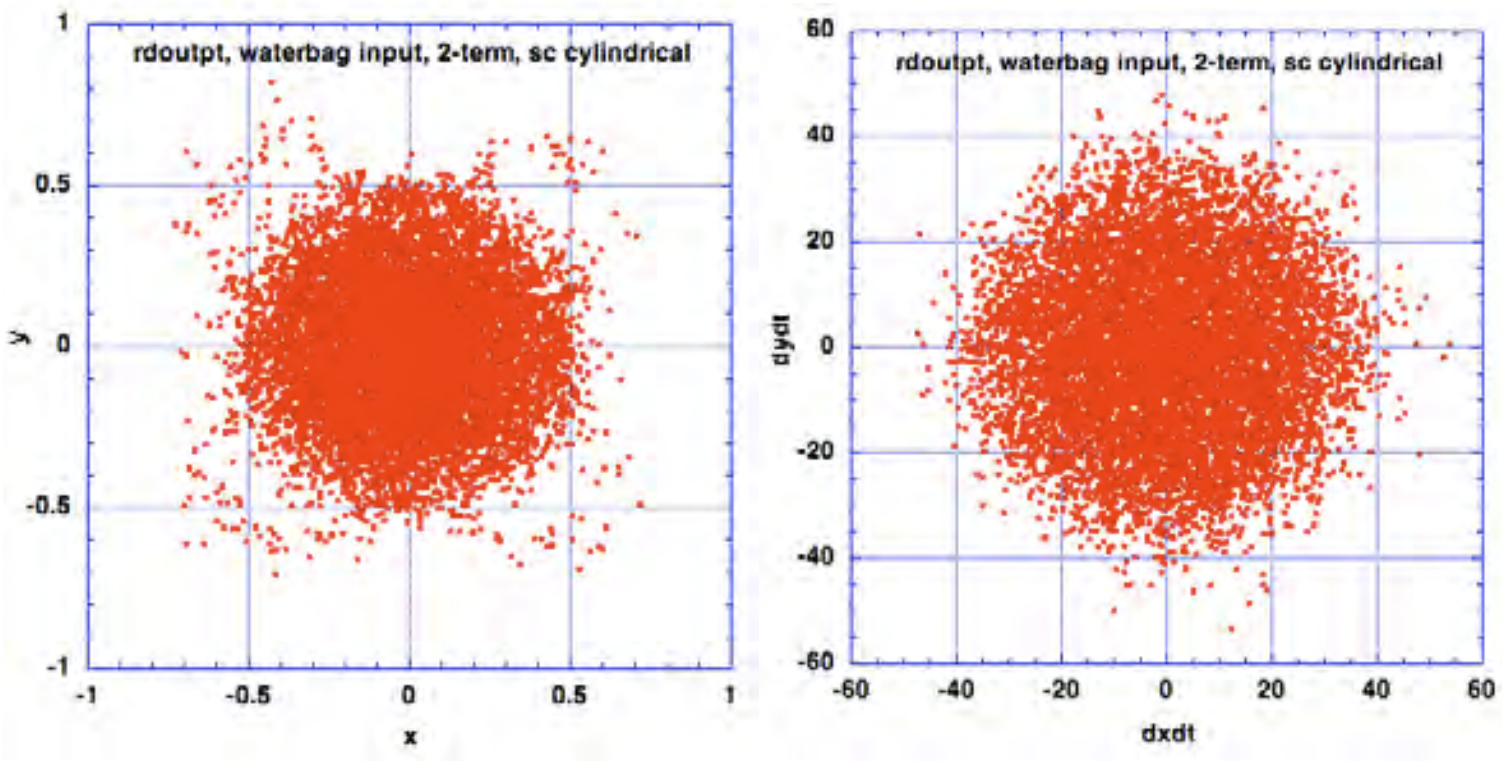

Fig. 22.4.b. Typical phase space distributions in an RFQ with 2-term external and cylindrically symmetric space charge fields.

### 22.3.1 Effect of a03 Sextupole Term in RFQ

Batygin has argued for higher order nonlinear terms in the focusing potential to counteract natural higher order space charge field terms in typical beam distributions, and has related this to the quadrupole channel with higher order terms [177]. This is tested with *LINACSrfqSIM* using the 8-term multipole field method, for which the a03 term can be adjusted.

Batygin's example in Sec. 3, "Matching of the Beam into the Four-Conductor Quadrupole Line with Duodecapole Component", argues for ratio of quadrupole/duodecapole field = 50/0.014 = 35.7. Crandall's tables [178] for RFQ with rho/r0 = 0.75 and sinusoidal modulation give almost this ratio for a01/a03 ~ 1/0.035 = 29. Many RFQs have been built with this configuration. Here a03 refers to the $r^{2m}$ component of the transverse potential, with m=3.[179]

Regarding Batygin's discussion, in the typical RFQ with a circular vane tip profile and sinusoidal longitudinal modulation, the sign of the a03 component with respect to sign of a01 quadrupole component external field appears to be as required to counteract beam space charge field.

A beam with significant space charge injected into the RFQ with an elliptical waterbag transverse distribution is naturally quickly redistributed in the RFQ to a diamond shaped distribution (45° tilted square in Batygin's words). Therefore, it is not necessary to inject with this shape, in order to test the effect of the a03 component on transmission, accelerated beam fraction (emittance, etc.).

The effect of varying the a03 component is tested for two RFQs, aperfac70 and aperfac43. These RFQs were designed with multipole effect included in the design. Aperfac70 is a poor RFQ, with generally too small aperture, resulting in large radial losses. Aperfac43 is a good RFQ, with ~ optimum transmission. The simulation was made with the 8-term multipole approximation of the external fields, allowing the coefficients, including a03, to be varied. Fig. 22.5 shows the result. It appears that the very typical RFQ, with vane-tip curvature ratio rho/r0 = 0.75 and sinusoidal modulation, is an optimum shape. Further work later gave further insight.

---

177 Y.Batygin, A.Goto and Y.Yano, "Suppression of Space Charge Induced Beam Emittance Growth in Transport Line", EPAC 1996, MOP096L

178 K.R. Crandall, "Effects of Vane-Tip Geometry on the Electric Fields in Radio-Frequency Quadrupole Linacs", Los Alamos National Laboratory LA-9695-MS, April 1983.

179 Crandall and Batygin use different formulas for the potential expansion. Batygin uses potential, which is result of averaging of particle motion (effective potential), while Crandall uses actual potential (before averaging). Their potentials before averaging are the same. The field in a pure quadrupole structure contains only multipole terms with m = 2, 6, 10, and the discussion here concerns the m = 6 term.

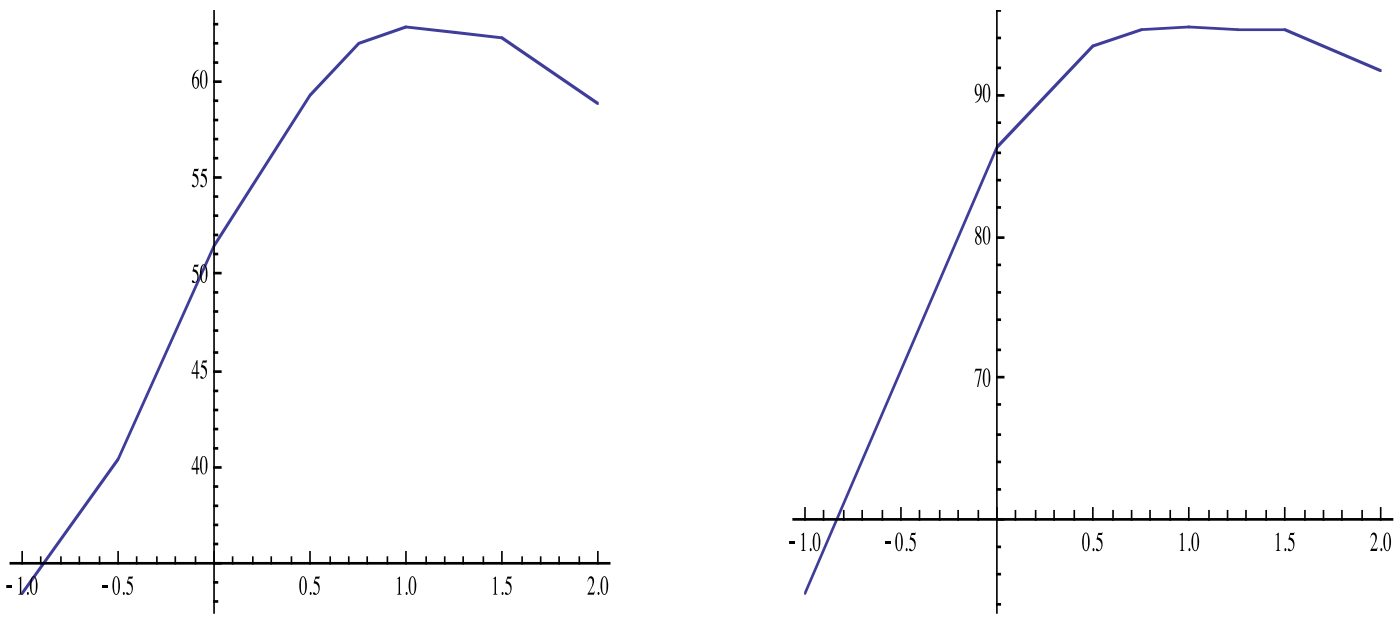

Fig. 22.5. (Left) Accelerated beam fraction (%) vs. fraction of standard a03 RFQ external field component, for aperfac70 RFQ. (Right) Accelerated beam fraction vs. fraction of standard a03 RFQ external field component, for aperfac43 RFQ.

*22.3.1.1 Additional Study – Redistribution Effects*

Dear Yuri,

I am with Li Chao at IMP and working further on the duodecapole effect. Tomorrow, May 7 2014, is my last day, so if you would have time to comment today, May 6, your time, we would have one more chance to debate face to face – but it is not urgent.

As you probably suspected, my hypothesis, that the natural a03 component in the RFQ vane tip external field when machined with a circular profile provides a space charge compensation as you propose [180], is more complicated. There seem to be two effects, as outlined below. We would appreciate your comment and help with the wording.

**Results for long constant FDFD channel, 400 periods, with and without constant G6 added over the whole channel, with G6/G2 = ~0.01. With beam current, tune depression is ~0.6 :**

Li Chao has results as follows: Noted xy shapes are at end of focusing period

| 1. Zero current: | Injected shape | Final shape | Redistrib Speed |
|---|---|---|---|
| a. G2 only: | round | round | |
| b. | diamond | diamond | |
| c. G2+G6 | round | hexagon | |
| d. | diamond | diamond | |

| 2. With current: | Injected shape | Final shape | |
|---|---|---|---|
| a. G2 only: | round | round | ¼ $\sigma_p$ |
| b. | diamond | round | ¼ $\sigma_p$ |
| c. G2+G6 | round | diamond | ¼ $\sigma_p$+slow |
| d. | diamond | diamond | |

With beam current, there are two different effects:

- *Immediate redistribution of internal space charge mismatch via the ¼ plasma period mechanism.*

Cases 2.a, b exhibit this effect. In a few periods, the xy shape changes to round and there is an abrupt emittance increase as the internal distribution equilibrates through the ¼ plasma period mechanism. *It is important to note that an injected diamond shape with space charge into a pure G2 channel changes to the round shape in a few periods.*

---

180 "Adiabatic Matching of a nonuniform intense charged-particle beam into the focusing channel", Phys. Rev. E, Vol. 54, No. 5, November 1996.

Case 2.c has this immediate emittance increase, but the shape still remains round and only becomes diamond shaped on a slower time scale.

In Case 2.d, the immediate redistribution is minimized, because the diamond shape is better matched to a channel with G6 component, as you point out. However, it happens naturally and quickly.

• *The effect of adding a continuous G6 component to the external field.*

Case 2 shows that a final diamond xy shape will only happen if the channel has an external field G6 component.

**Results for typical RFQ, with vane tip a circular radius Rho and average vane tip radius r0, with ratio Rho/r0 = 0.75. This ratio is suggested by Crandall to minimize peak surface field. The G6/G2 ratio for the whole RFQ is ~0.03 :**

Numbers from your formulae:

It is clear in your development that the amount of G6 in the beam space charge depends like 1/R^4, so completely on the particular case.
At injection with parabolic distribution: etrms = 0.000025 cm.rad, trms - 0.2 cm, Icurr = 0.130 A, $\beta$ = 0.0101. G6/G2 (Eq.(24) = ~1. The beam is rms matched, but has a strong internal mismatch.

***The result in the RFQ is the same as for Cases 2.c and 2.d.***

The diamond shaped injected beam was simply prepared by passing the round injected beam through a diamond shaped aperture. The remaining number of particles are 84.4% of the number in the round beam, so the current should be raised to ~150 mA. The rms size ratio is 82%, so the injection emittance should be adjusted and the match might be somewhat different; these were kept the same as for the round beam. The resulting normalized transverse rms emittance and transmission are shown in Fig. 22.6.

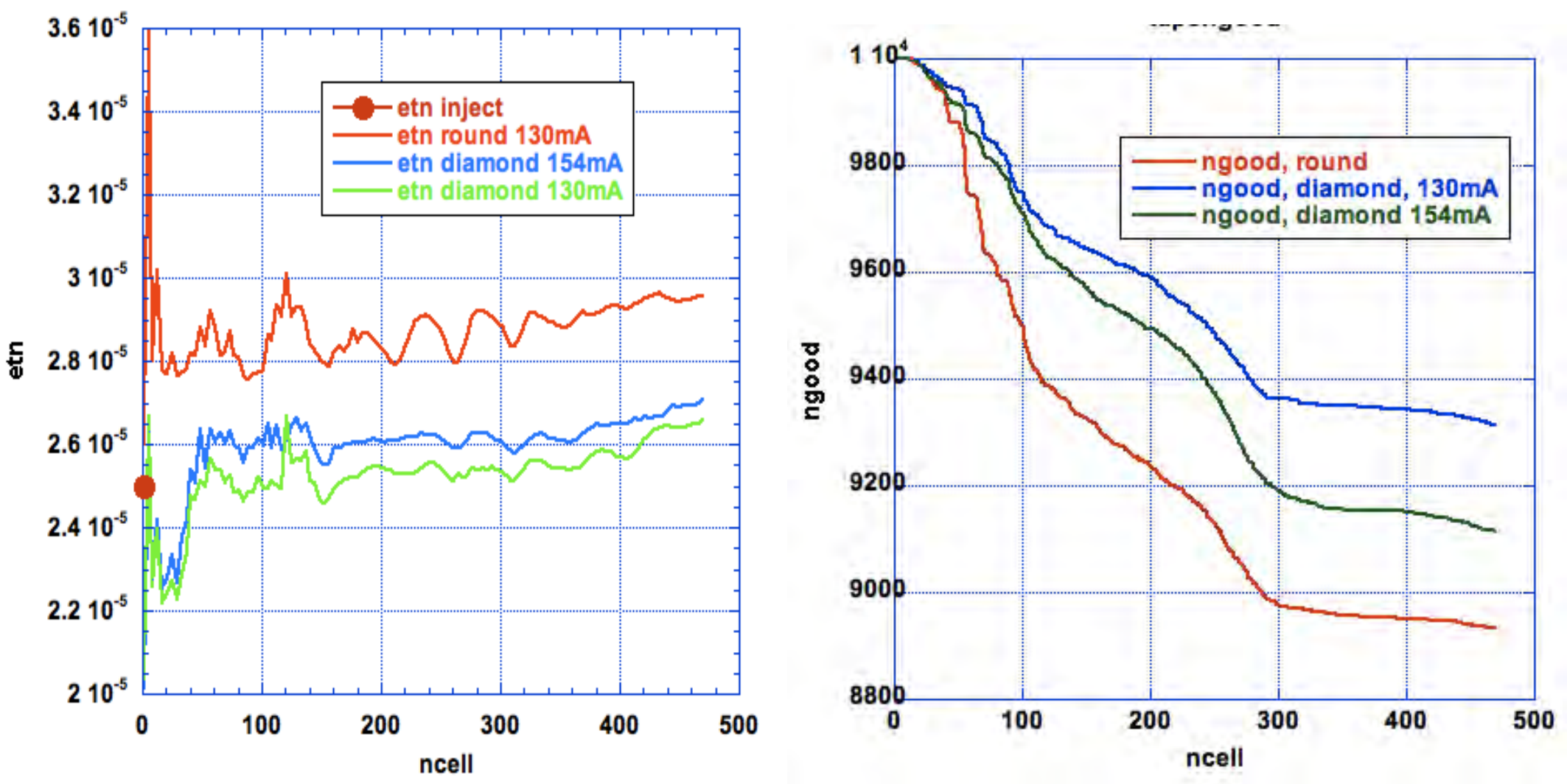

Fig. 22.6. Normalized transverse rms emittance and transmission for round or diamond-shaped input beams injected into an RFQ.

*22.3.1.2 Conclusions*

The injected beam (usually assumed waterbag, close to parabolic) into an RFQ redistributes quickly in a few periods. This beam then adapts to the external field, and in the RFQ with ~constant G6 (a03) component, the xy shape becomes diamond shaped.

The RFQ beam is still very long at cell 100, where the synchronous phase $\phi_s$ is -88°, and even at cell 300 where $\phi_s$ is -65°, reaching -25° at the end. Although the g6/g2 ratio is ~constant at ~0.03, the diamond shaped becomes more distinct as the RFQ vane modulation increases, Fig. 22.7, indicating a 3D effect.

It is true that varying the magnitude of external field a03 term in simulation of a given RFQ is a strong effect. It is partly because my RFQ designs take the multipoles into account directly, and are optimized accordingly. If the a03 is varied in the simulation, the transmission typically falls; with a03 = 0 by ~10%.

However, it is seen that the a03 component is responsible for the redistribution to the typical diamond shape at higher vane modulations, and that higher transmission could be achieved with a diamond-shaper input distribution. The input distribution effect is occurring at low vane modulations – indicating that in this region, the duodecapole external field component appears to be compensating for the corresponding component in the space charge field distribution of the beam.

Practically speaking, it appears that the initial transverse rms emittance growth usually seen in the RFQ could be eliminated by passing the injected beam through a diamond shaped aperture at the injection point. The price would be more current required from the ion source – which price might be too high in comparison with the emittance increase.

When the beam leaves the RFQ with its diamond shape and enters the next linac, it will quickly redistribute again in same way. Typically no beam loss is seen in simulation of the linac sections after the RFQ (excepting intrabeam scattering for H-).

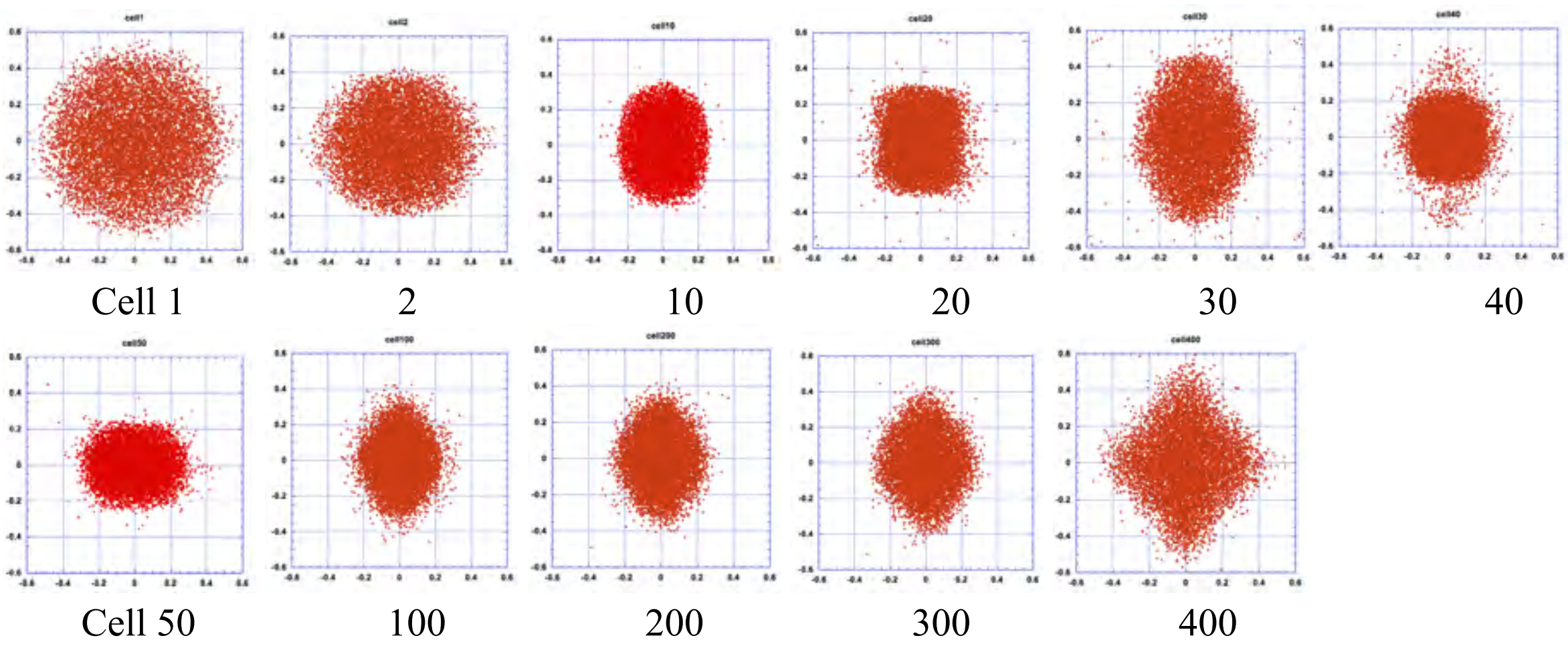

Fig. 22.a. With round-shaped input distribution, at cell ends:

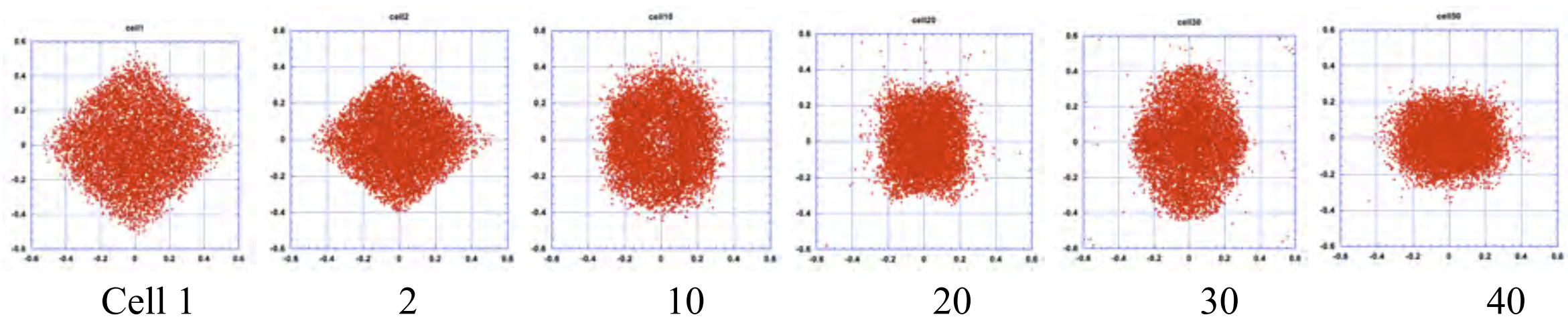

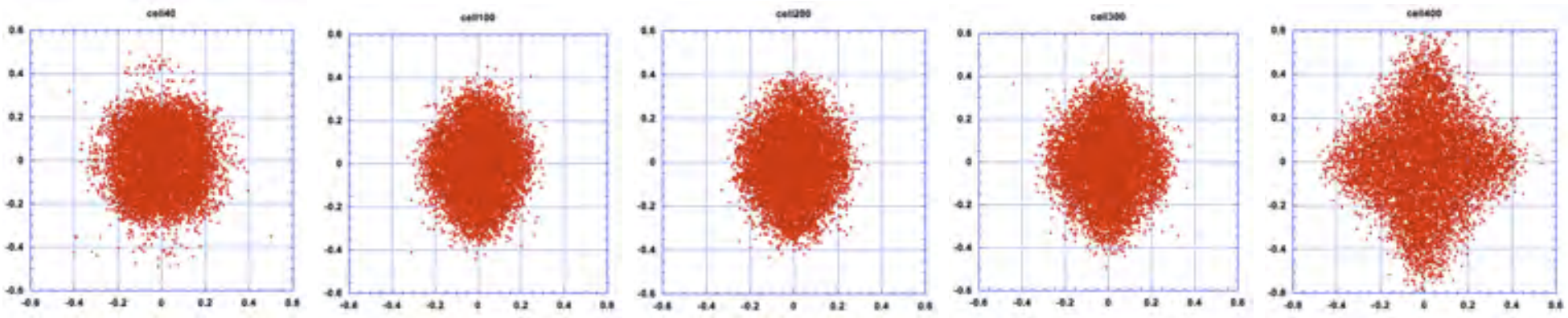

Cell 50 100 200 300 400

Fig. 22.8.b. With diamond-shaped input distribution, at cell ends:

**Subsequent Papers** [181, 182, 183]

### 22.3.2 Summary – Quadrupolar Focusing Channel

The principle that the rms behavior of different particle distributions will be very nearly the same is demonstrated. A bunching and equipartitioning process occurs with heating of the initial dc zero-energy-spread longitudinal distribution and cooling of the transverse rms emittance. The transverse ratio of 100%-to-rms beam size evolves to a more parabolic/semiGaussian form with t100/trms~sqrt(8 to 10). The longitudinal distribution tends to a Gaussian. The “2-term” simulation shows significantly less effect on the total transverse beam size than the accurate Poisson simulation. An beam injected a given rms transverse emittance, but not matched in detail over the distribution and containing an a03 component, will evolve to the naturally matched {x,y} distribution, which has a diamond shape. Part of this effect appears to be a compensation for the $r^6$ term in the beam space charge field, as predicted by Batygin.

## 22.4 Long Focusing and Bunching Channel Followed by Accelerating Channel

A standard RFQ with extra long shaper section has synchronous phase of -90° to cell 391, then linear rise to -88° at cell 434, but with rising vane modulation over the whole shaper to meet the required modulation for EP at the end of the shaper. The full-Poisson simulation behavior is shown in Figs. 22.6 & 22.7. The total transverse beam sizes of the four distributions again converge over the -90° section of the long focusing and bunching channel, similar to the focusing channel without bunching, but with some overshoot. When the synchronous phase is raised to -88°, there is again the rapid convergence to the ~parabolic distribution.

181 Y. Batygin, et. al., “Nonlinear Optics for Supression of Halo Formation in Space Charge Dominated Beams”, IPAC2014.

182 Li Chao, et.al., “Particle Orbits in Quadrupole-Duodecapole Halo Suppressor “, NIMA, Volume 770, p. 169-176.

183 Y. Batygin, et.al., “Suppression of Space Charge Induced Beam Halo in Nonlinear Focusing Channel”, NIMA 816 (2016) 78 - 86

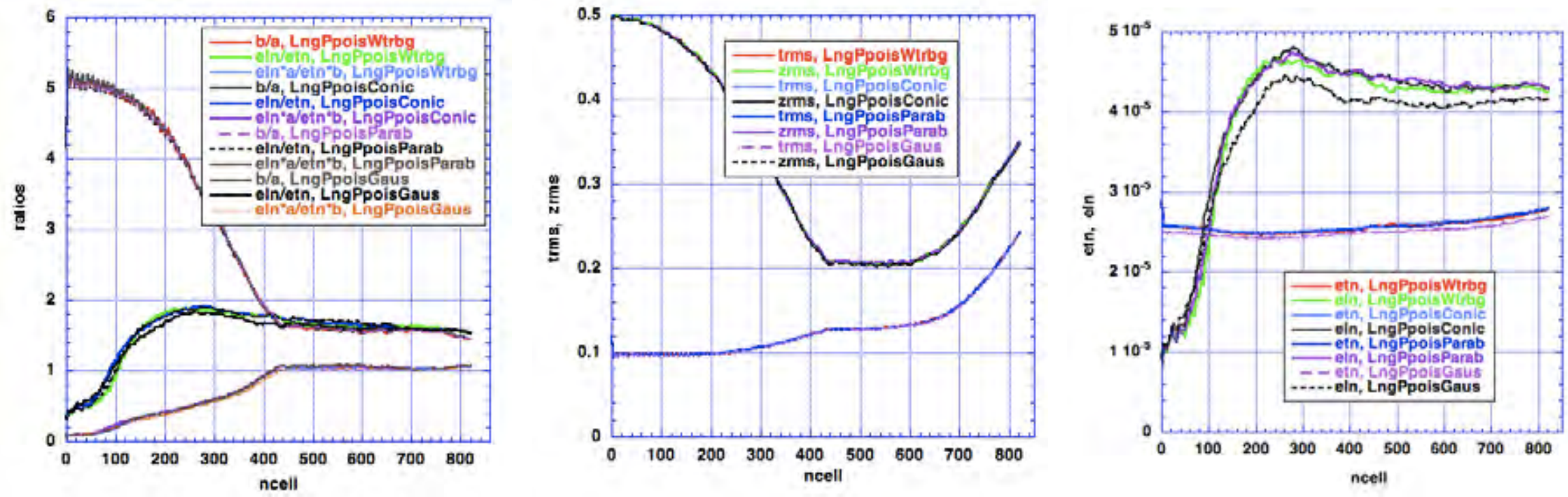

Fig. 22.6. For Long Focusing Channel With Bunching Followed by Accelerating Channel: The a.) EP ratios, b.) rms beam sizes, and c.) rms emittances for injected beams with transverse waterbag, conical, parabolic, and Gaussian truncated at 3σ distributions having the same rms beam size.

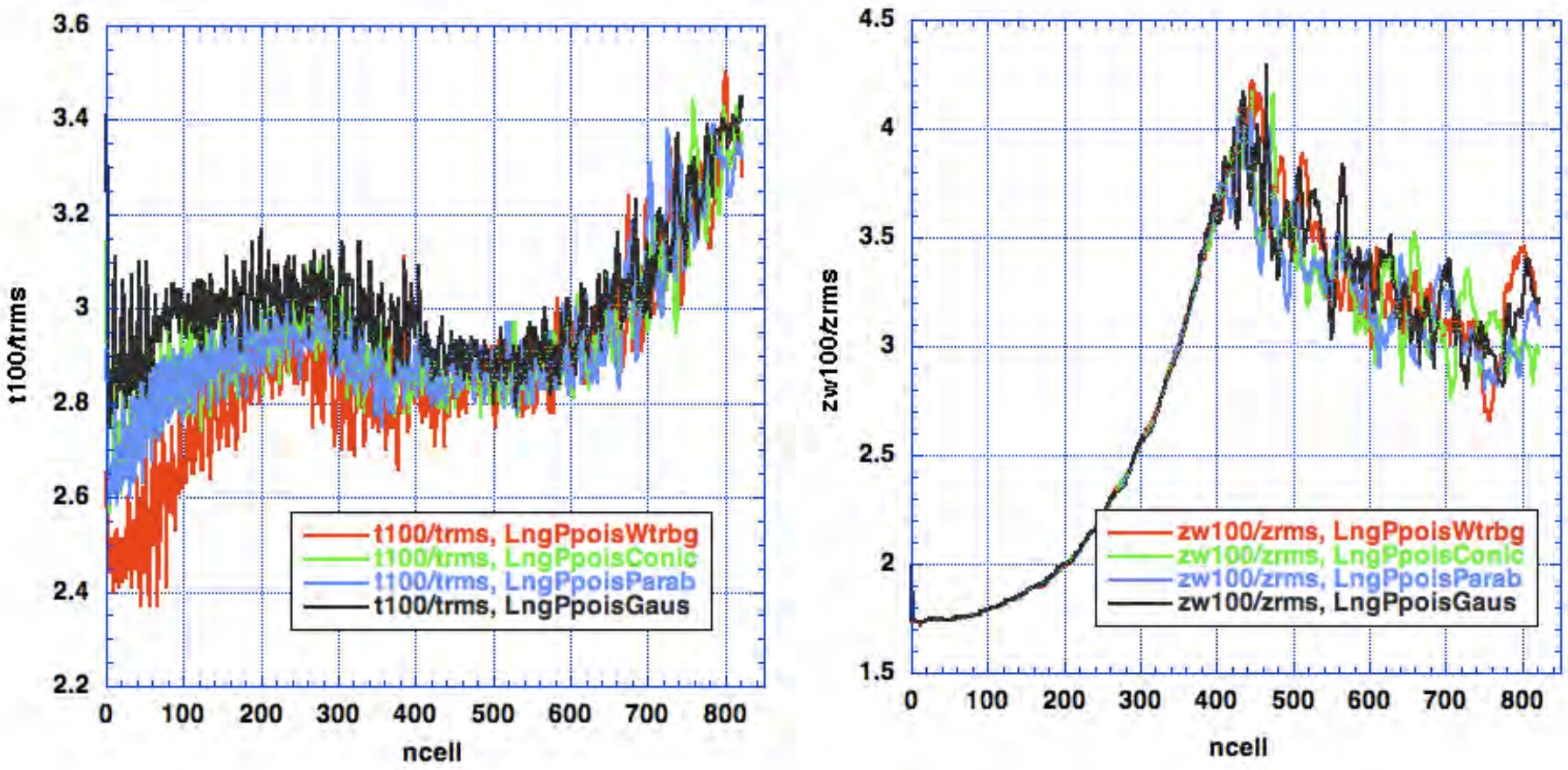

Fig. 22.7. For Long Focusing Channel With Bunching Followed by Accelerating Channel: The ratios of a.) 100% transverse beam size to the transverse beam size (t100/trms) and b.) corresponding longitudinal ratio (zw100/zrms).

The corresponding "2-term" mode simulation is shown in Fig. 22.9. Comparing Figs. 22.4.b. and 22.9.b., it appears that bunching does play a redistribution role in bringing an initial distribution closer to an approximately parabolic one.

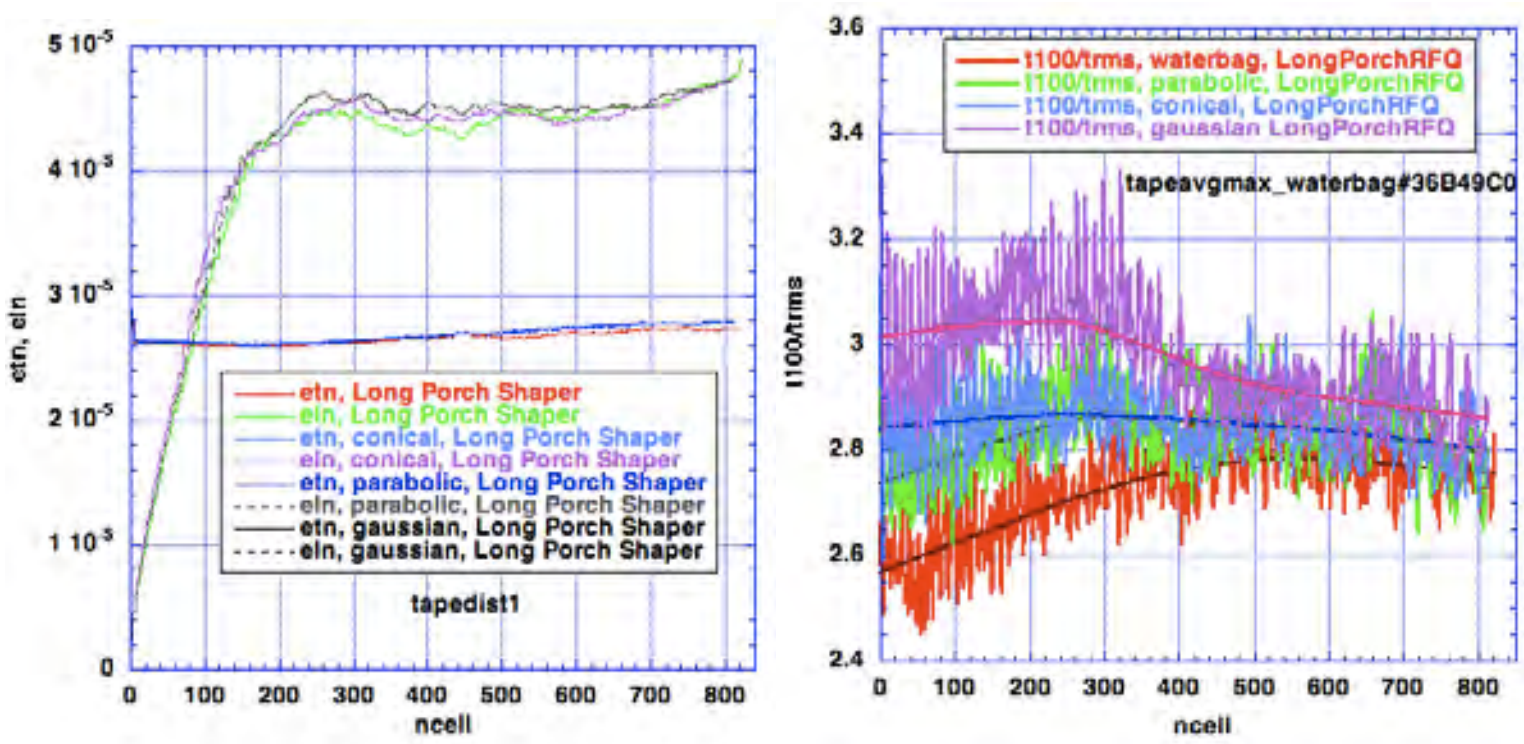

Fig. 22.8. "2-term" mode simulation of long focusing and bunching channel, followed by acceleration channel, corresponding to Fig. 22.6.c. and Fig. 22.7.a.

### 22.4.1 Summary – Quadrupolar Focusing Channel With Continuous bunching

Slow bunching in a focusing channel does not change the total beam size behavior significantly from that of the pure focusing channel.

## 22.5 Three RFQ Case Studies[184]

Next we investigate three RFQs with the same design rules except for a different aperture required at the end of the shaper, which has a normal, shorter length. The beam is brought to EP at the end of the shaper, and the design seeks to keep the beam equipartitioned to the end of the RFQ, succeeding more or less depending on the end-of-shaper aperture. The RFQ section from shaper end to the final energy is then representative of a general focusing channel with acceleration, into which a self-consistent, equilibrium beam is injected.

Three 130mA, $D^+$ cases accelerating from 0.095 – 5 MeV will be studied, all designed for EP ratios eln/etn=b/a=σt/σl=1.6:

Case 1(43). Equipartitioned (EP) channel with optimum aperture. (No final optimization for form factor or emittance growth.
Case 2(70). Channel with too-small end-of-shaper aperture, resulting in continuous radial loss.
Case 3(25). Channel with too-large end-of-shaper aperture, resulting in continuous loss from bucket as well as radial loss.

The shaper end is at cell 108 for Case 1(43), cell 68 for Case 2(70), cell 191 for Case 3(25). Fig. 22.9.a shows the transverse focusing factor B and longitudinal focusing factor A; the vane voltage is defined by the same Kilpatrick criterion for all three cases. Thus there is an optimum aperture, Case 1(43). Case 2(70) with smaller aperture has strong focusing but the aperture is too small. Case 3(25) has larger aperture but the focusing is too weak.

Simulations were done with *LINACSrfqSIM* with Poisson solutions for the external and space charge fields. Fig. 22.9.b shows the transmission and accelerated beam fractions for the three cases.

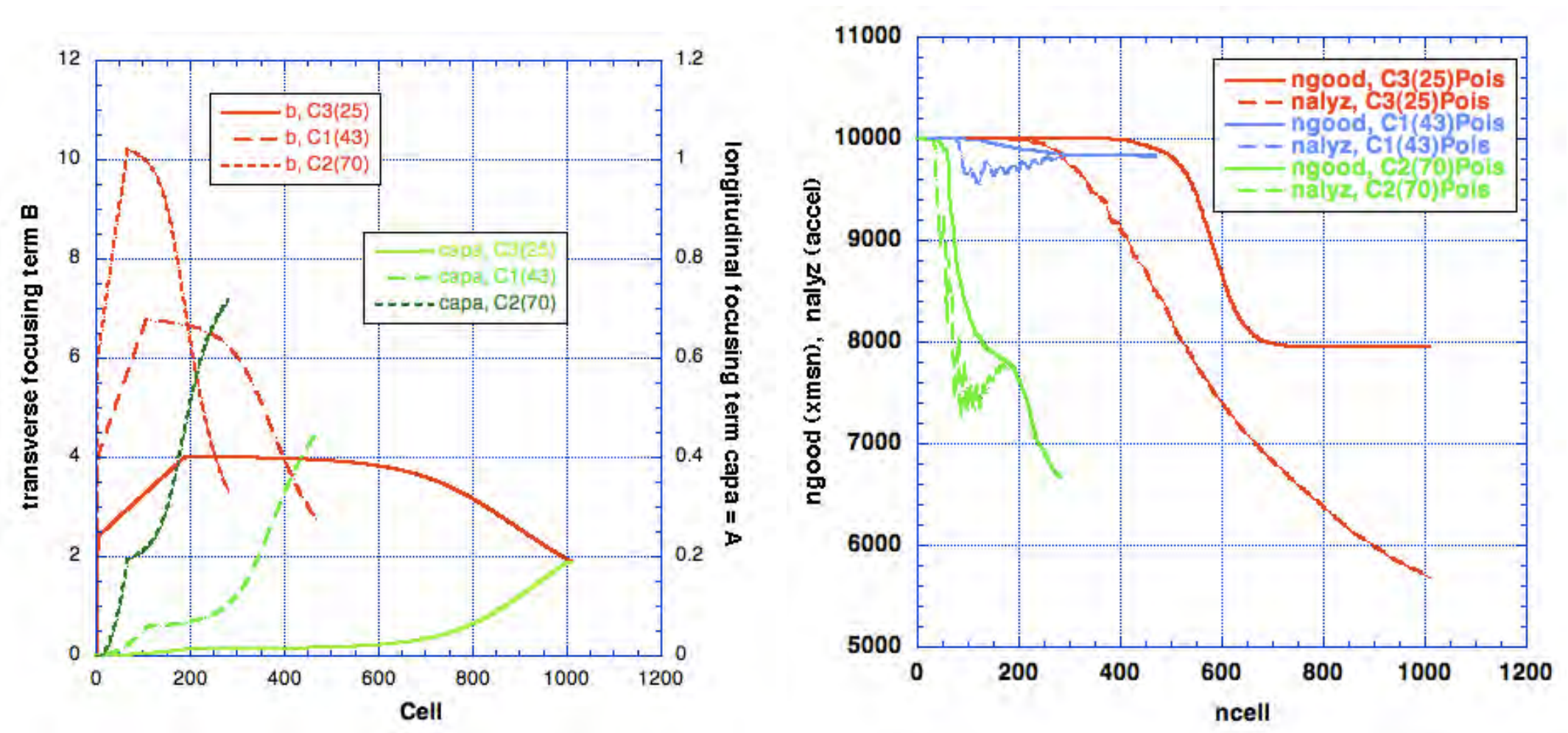

Fig. 22.9. a. Focusing factors for the three cases. Maximum s0t = 23° for C3(25), 42° for C1(43), 59° for C2(70). b. Transmission (ngood) and accelerated (nalyz) particles from 10000 particle waterbag input distribution.

The success of the design to have EP in the main part of the RFQ is shown in Fig. 22.10. (The RFQ beam is analyzed using only particles that are successfully transmitted to the end of the RFQ.)

184 From the standard test family of RFQs used to experimentally test RFQ simulation programs – aperfac = 2.5, 4.3 and 7.0. The RFQs are designed by varying only one parameter – the aperture at the end of the shaper = βλ/aperfac. The design rules, including a requirement for an equipartitioned beam in the main part of the RFQ after the shaper, then determine the RFQ cell parameters.

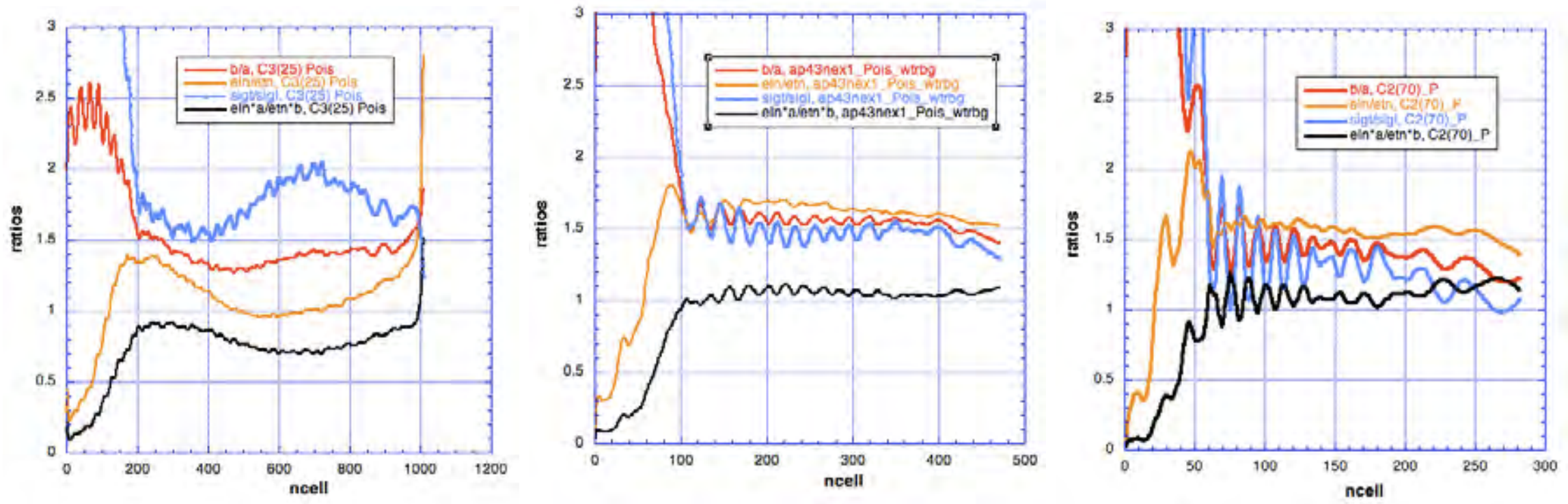

Fig. 22.10. Equipartitioning ratios. a. Case 3(25), b. Case 1(43), c. Case 2(70)

The Hofmann Charts in Fig. 22.11 show that there is more tune depression initially for increasing aperture, and also somewhat more in the main part, although there the tune depressions are very similar at ~0.4.

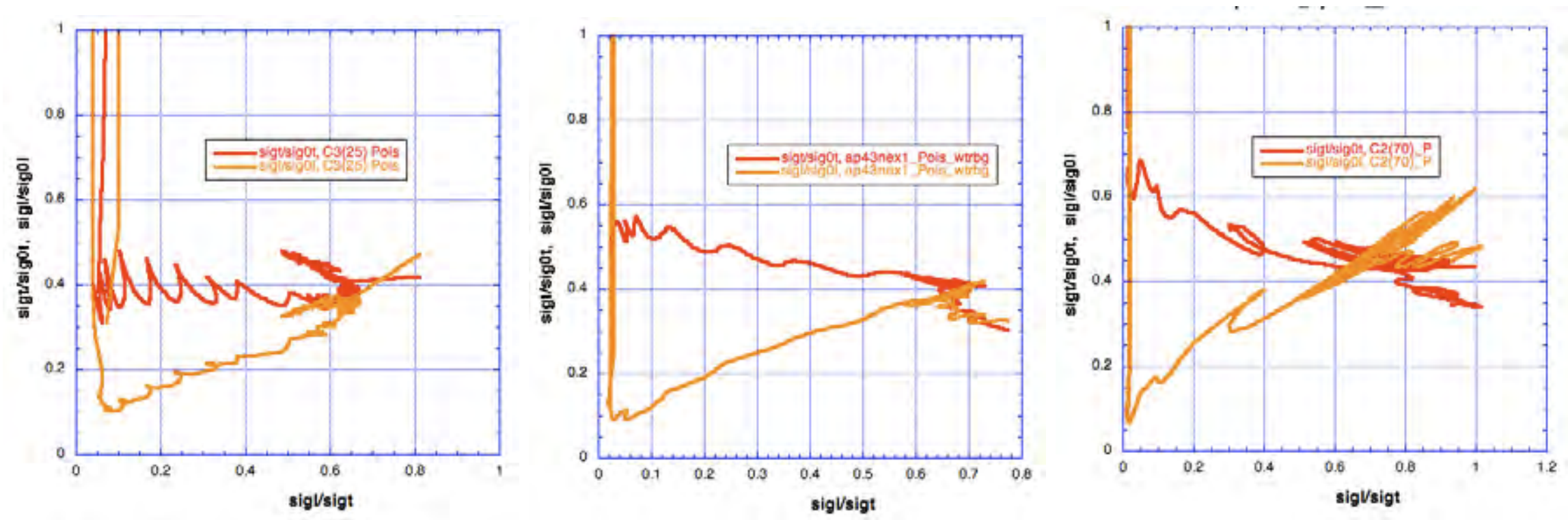

Fig. 22.11. Hofmann Charts for the three cases.

The performance of the beam sizes and the total-to-rms beam sizes for each case is now investigated in more detail. Three simulations are compared:

- Full Poisson simulation of both external and space charge fields, with the vane surface as boundary, closed by a circle with sufficient radius that the fields in the central region and out beyond the vane tips are not influenced.

- External field Poisson solution, with the effect of the vanes on space charge turned off, by replacing the transverse vane surface boundary with a cylindrical pipe with radius twice the maximum vane radius of the local cell (2*modulation*aperture) [185]

- "2-term" mode uses the 2-term external field (with transverse focusing multipole a01=1, but same acceleration multipole a10), and an r-z cylindrically symmetric space charge model with the radial boundary condition set to a closed boundary pipe with radius at the minimum aperture of the RFQ cell. The multipole effects from the non-ideal vane tip profile is thus absent, and the transverse image effect is not quadrupolar.

## 22.5.1 RFQ with small losses – Case 1(43)

### *22.5.1.1 Transmission and Accelerated Fractions – Case 1(43)*

Fig. 22.12 compares transmission and accelerated fraction with full Poisson, Poisson with space charge transverse vane image effect turned off, and "2-term" simulation. In each case, radial losses are determined from the actual vane surface. The external field multipole and transverse space charge image effects each account for 1% lower transmission and acceleration.

[185] The ability to turn the vane image effect on and off is an essential feature of an RFQ Poisson simulation code, both as a check during code development, and in order to study the effect. There have been claims that this is impossible in a Poisson code, but it is very easy, essential, and indeed it would be rather impossible to verify correct performance of the Poisson solvers without it.

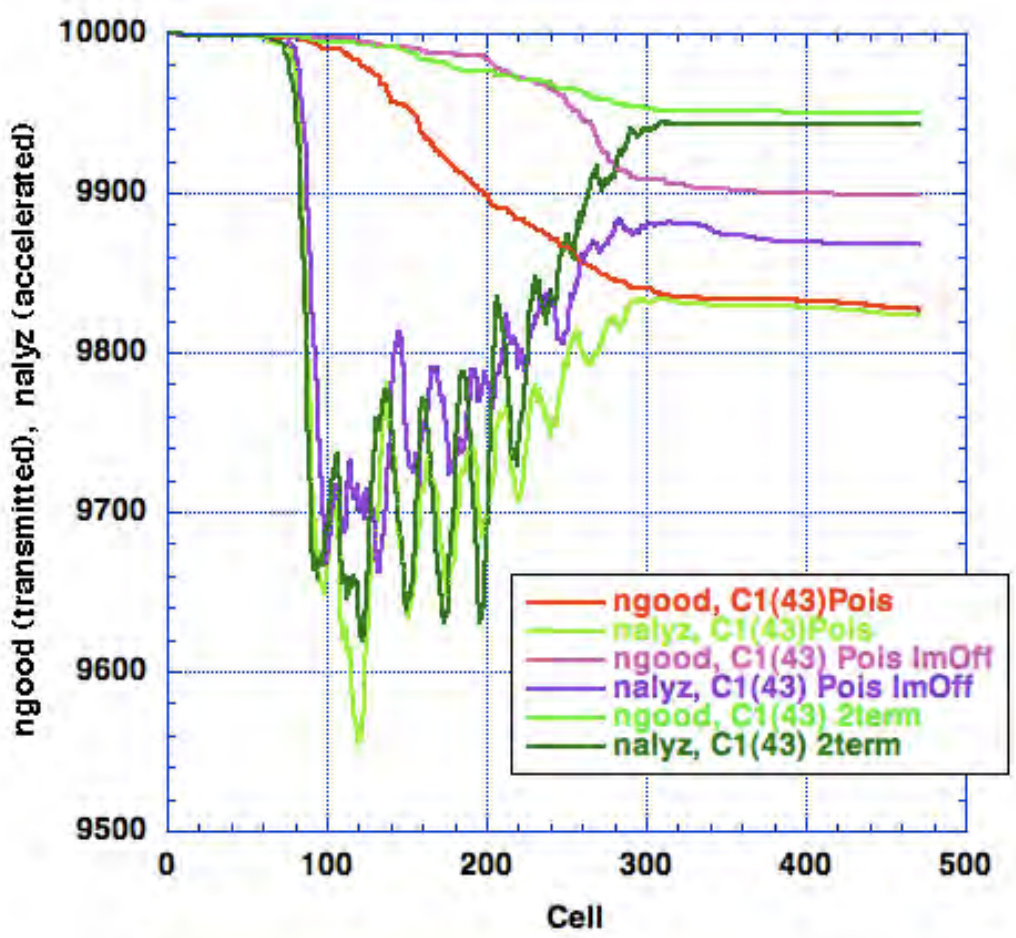

Fig. 22.12. Comparison of transmission and accelerated fractions, Case 1(43).

*22.5.1.2. Full Poisson simulation – Case 1(43)*

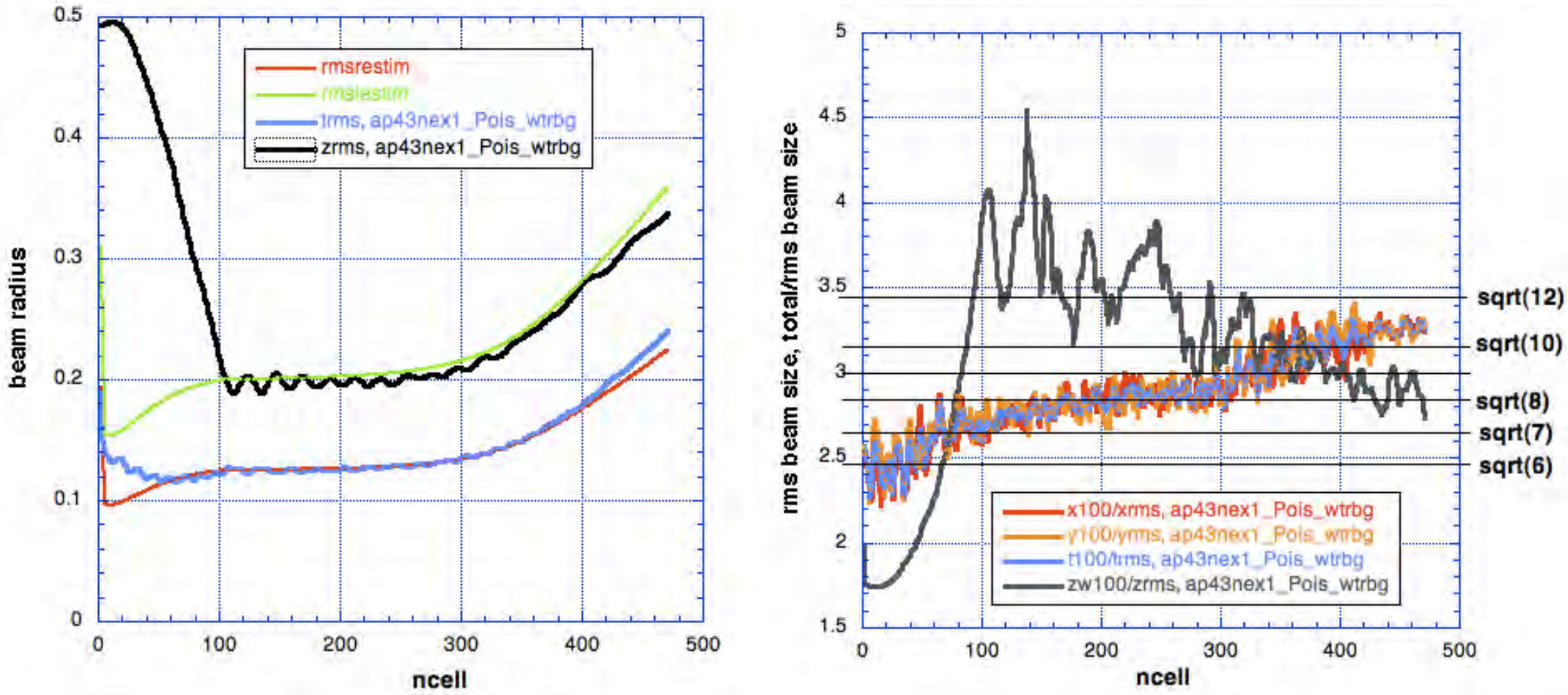

Fig. 22.13. Full Poisson simulation, Case 1(43). a. Beam size and beam size ratios. Rmsrestim and rmslestim are the design transverse and longitudinal rms beam sizes, valid from EOS (cell 108) on. Trms and zrms are the simulation result for the transverse and longitudinal rms beam sizes. b. Ratios of transverse (t) and longitudinal (zw) 100% beam sizes to the rms beam sizes.

Case 1(43) beam size and beam size ratios are shown in Fig. 22.13 for full Poisson simulation. The rms beam sizes and equipartitioning ratios follow the design closely. However, the transverse t100/trms ratio changes steadily from the injected waterbag distribution with t100/trms = sqrt(6) to a more Gaussian distribution with t100/trms ~sqrt(10 to 12). As bunching occurs, the longitudinal z100/zrms ratio moves toward a parabolic, tighter than Gaussian, distribution.

*22.5.1.3. Full Poisson simulation – Transverse Space Charge Image Off – Case 1(43)*

Fig. 22.14 presents the results with the space charge image effect turned off, by replacing the transverse vane surface boundary with a cylindrical pipe with radius twice the maximum vane radius of the local cell (2*modulation*aperture). The longitudinal rms beam size increases slightly. The 100% beam sizes are not affected – the transverse space charge image effect on these quantities is not significant in this case, but does have ~1% effect on transmission (Fig. 22.14).

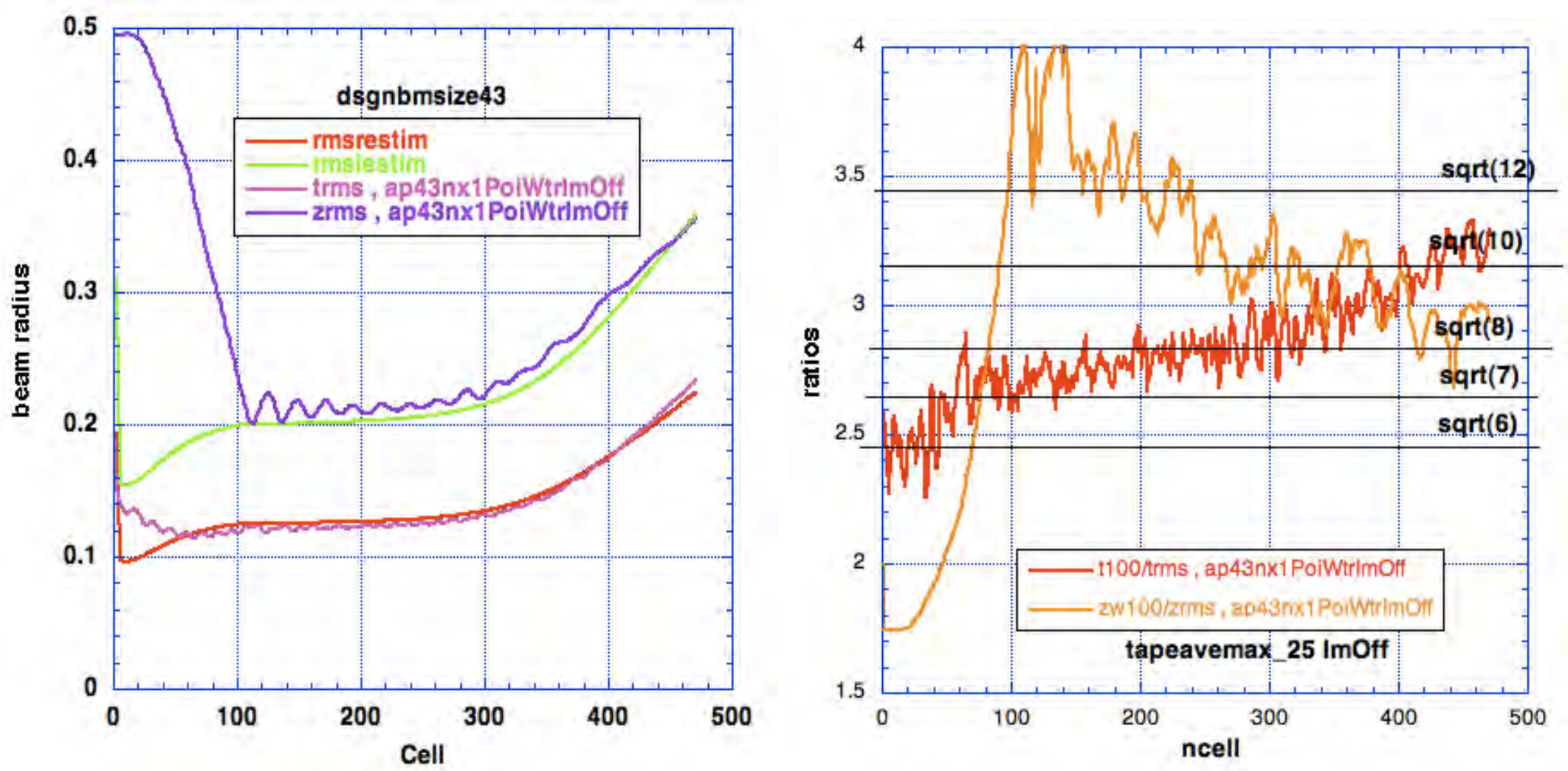

Fig. 22.14. Case1(43) Full Poisson simulation results but with space charge transverse boundary a cylindrical pipe with radius twice the maximum vane radius of the local cell, corresponding to Fig. 12.

*22.5.1.4. 2-term External Field and Cylindrically Symmetric Space Charge Field Simulation – Case 1(43)*

Fig. 22.15 shows the result of "2-term" simulation. The rms design is followed even more closely. t100/trms remains constant in the EP part after EOS at Cell108, revealing clearly the influence of multipole external field components and the quadrupolar geometry. This is reflected in the misleading higher transmission and accelerated fractions (Fig. 22.12).

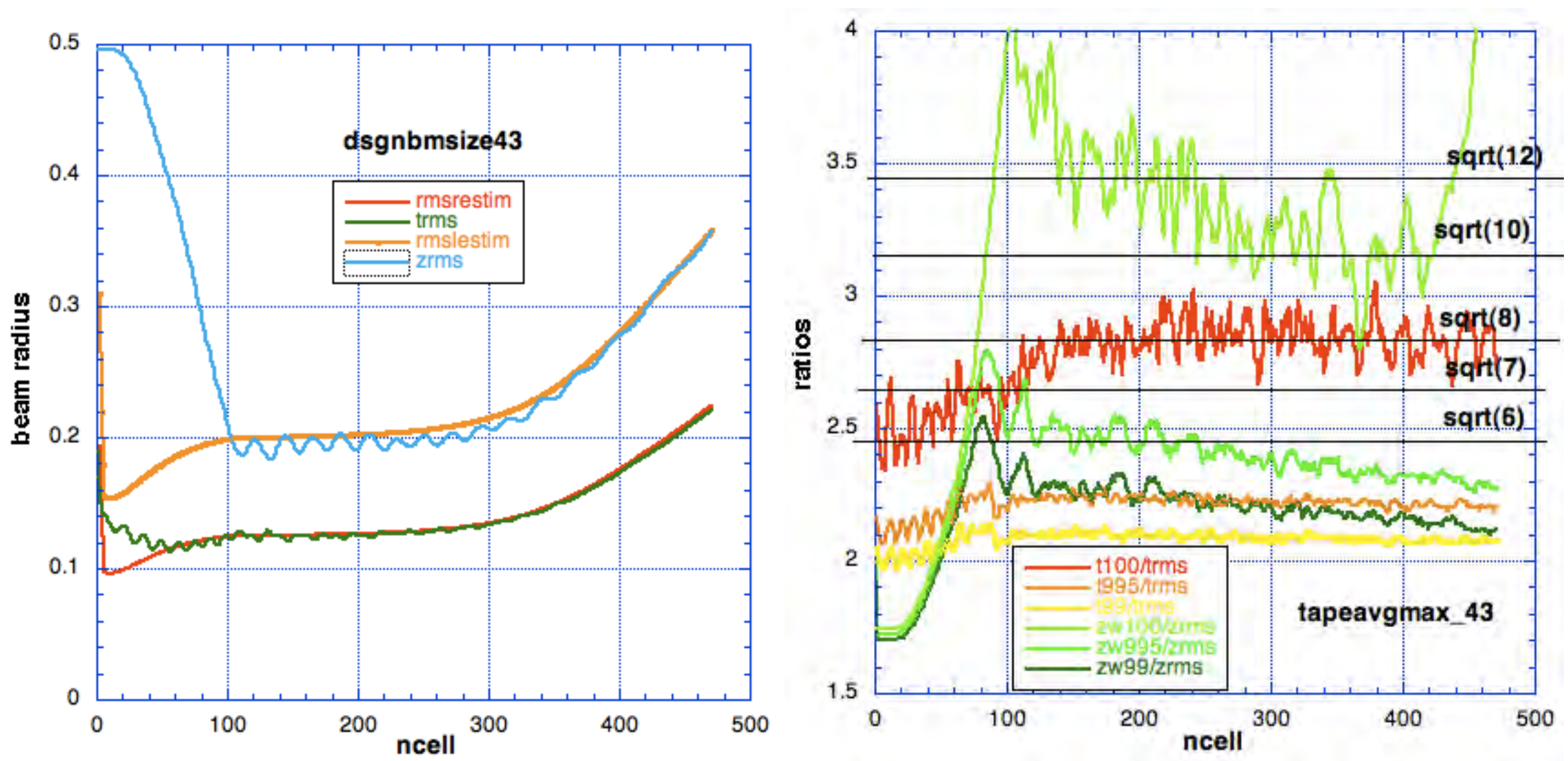

Fig. 22.15. Case 1(43) "2-term" simulation, results as Fig. 12.

## <u>22.5.2 RFQ with continual radial losses – Case 2(70)</u>

*22.5.2.1. Transmission and Accelerated Fractions – Case 2(70)*

In Fig. 22.16, transmission and accelerated fraction are compared for full Poisson, Poisson with space charge transverse vane image effect turned off, and "2-term" simulations. In each case, radial losses are determined from the actual vane surface. Comparison of full Poisson with vane surface to the essentially open boundary condition indicates the large influence of the actual space charge geometry in this case. The "2-term" simulation with closed space charge boundary at the minimum aperture is a reasonably accurate approximation.

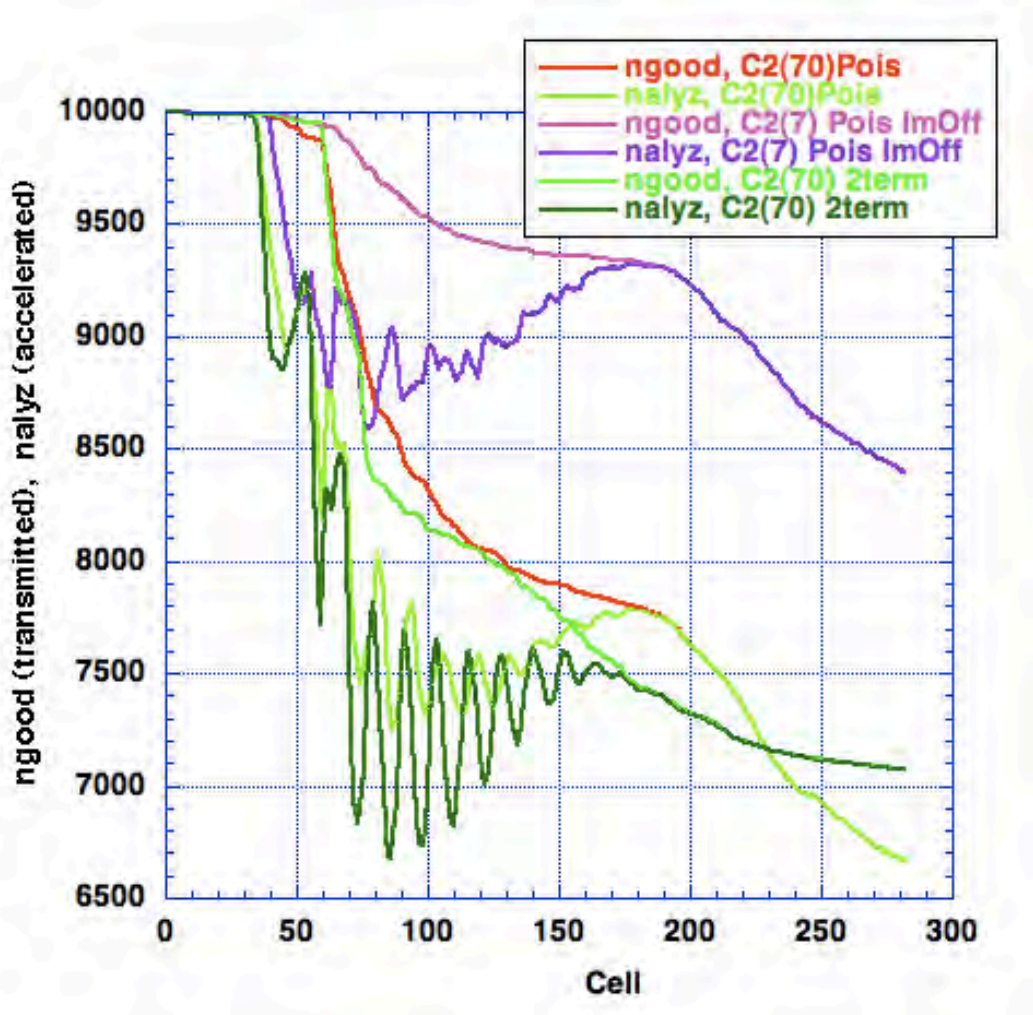

Fig. 22.16. Comparison of transmission and accelerated fractions, Case 2(70).

*22.5.2.2 Full Poisson simulation – Case 2(70)*

Case 2(70) ratios are shown in Fig. 22.17. The rms beam sizes and equipartitioning ratios of the successfully transmitted particles follow the design reasonably well, but particles are continually being lost radially. The transverse rms size (trms) redistributes by EOS (Cell68) from the injected waterbag distribution transverse size characterized by t100/trms=sqrt(6), to a more parabolic distribution with sqrt((ellipse beta)* (total real transverse emittance)/7). This ratio remains until about Cell125; Fig. 22.17.c. indicates that radial losses are concentrated at the vane tip from EOS to ~Cell125, but after that, an increasing number of successful particles are traveling along the sides of the vane tip. After ~Cell175, trms moves slowly back to the waterbag form, and the difference between t100 and the aperture decreases. Up to ~Cell 175, the acceleration rate is increasing (e.g., the curvature of the synchronous phase is positive), but then inflects and the curvature becomes negative. The longitudinal rms beam size assumes a Gaussian form, which flattens from the inflection point.

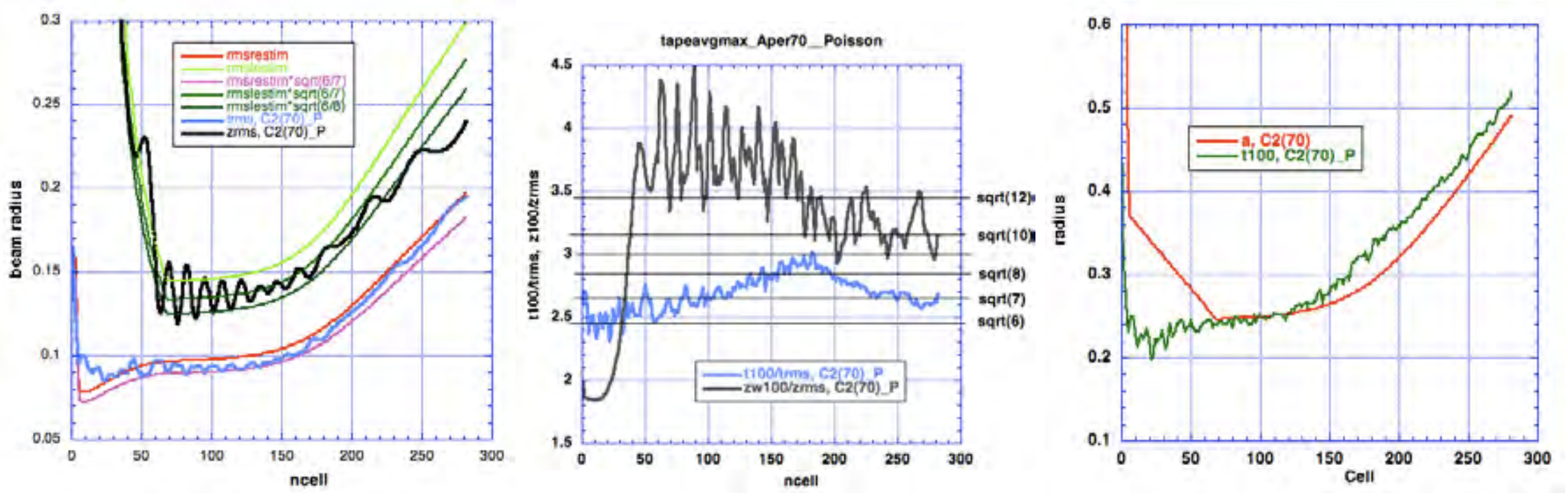

Fig. 22.17. Case 2(70), full Poisson simulation. a. rms beam sizes, design waterbag beam sizes and adjusted design beam sizes. b. 100% to rms beam size ratios. c. 100% beam size (t100) compared to aperture.

The t100/trms and 100/zrms ratios move at first with the same form as the lossless channel in spite of the continuous heavy losses, but then have a different form after ~Cell175. Particles are identified as accelerated if their velocity is within 10% of the design velocity; Fig. 22.16 shows that a number of particles are outside this band but still being transmitted up to ~Cell175, where they have all been lost. Heavy radial loss continues from ~Cell175 on.

*22.5.2.3. Full Poisson simulation – Transverse Space Charge Image Off – Case 2(70)*

Fig. 22.18 presents the results with the transverse space charge image effect turned off by replacing the transverse vane surface boundary with a cylindrical pipe with radius twice the maximum vane radius of the local cell (2*modulation*aperture). The rms beam sizes now remain near the design values. The t100/trms ratio rises somewhat less than the vane boundary case, but has the same qualitative shape, indicating that the losses play the major role for the total transverse size. The z100/zrms does not level out after the inflection point as it did for the vane boundary case.

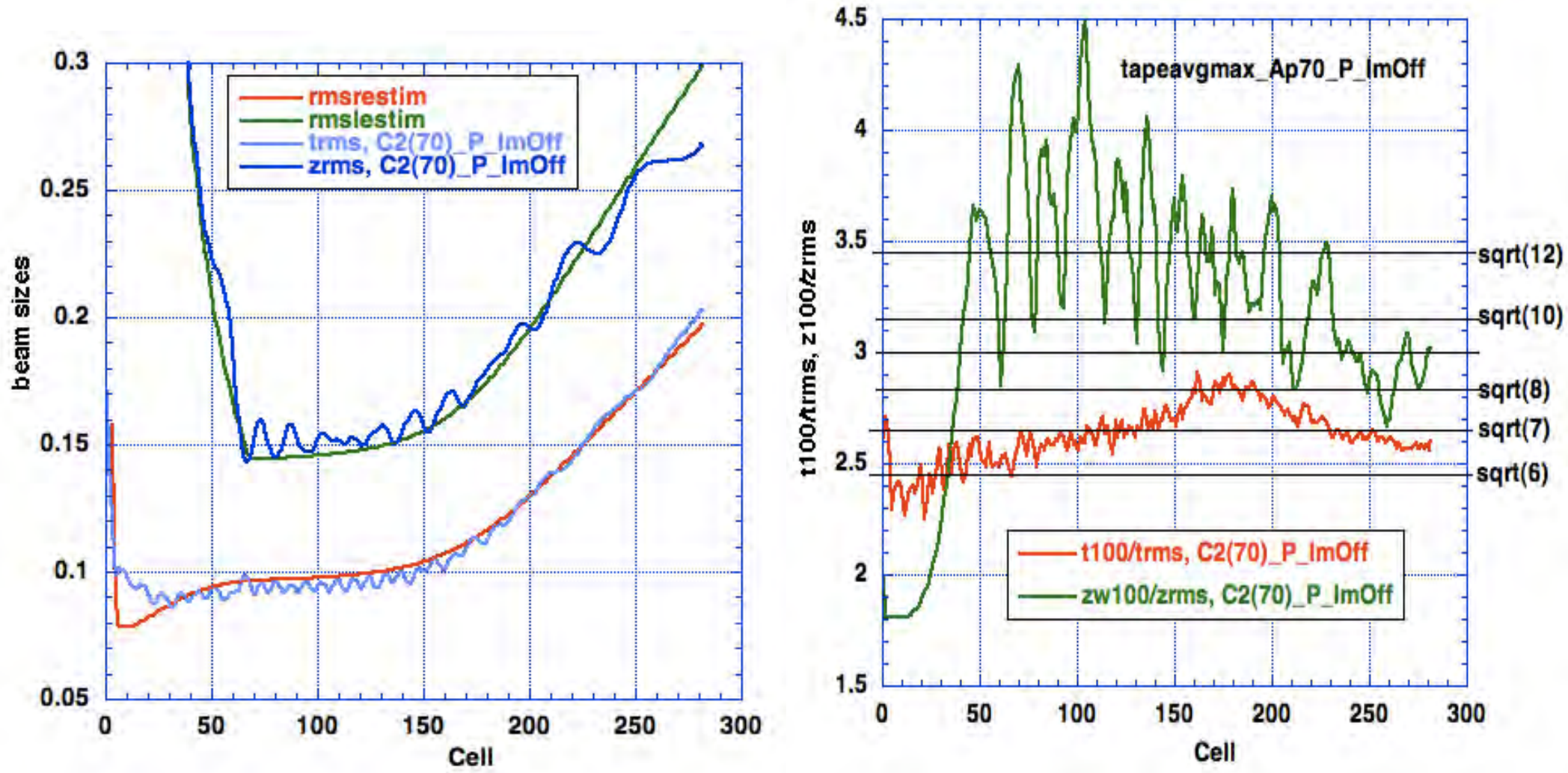

Fig. 22.18. Case 2(70) Full Poisson simulation results but with space charge transverse boundary a cylindrical pipe with radius twice the maximum vane radius of the local cell, corresponding to Fig. 16.

### *22.5.2.4. 2-term External Field and Cylindrically Symmetric Space Charge Field Simulation – Case 2(70)*

The "2-term" simulation results for rms and total beam sizes are dramatically different from the full Poisson results for this heavy loss case, as shown in Fig. 22.19. The particle loss criterion at the aperture radius prevents particles from being transmitted along the sides of the vane tip. The transverse rms size (trms) redistributes very quickly, when the acceleration begins strongly at the end of the shaper, from the injected waterbag distribution transverse size characteristic t100/trms=sqrt(6) to a more parabolic distribution with t100/trms=sqrt(7). The longitudinal rms beam size assumes a parabolic form more quickly that the full Poisson result. The transverse t100/trms ratio remains constant at the injected waterbag distribution with t100/trms = sqrt(6). The longitudinal zw100/zrms ratio quickly assumes a nearly constant value of sqrt(12) – a Gaussian form with >3σ tails – without the reduction characteristic of the Poisson result. This is clear evidence that accurate simulation of multipole and image effects is important and that "2-term" simulation could be misleading.

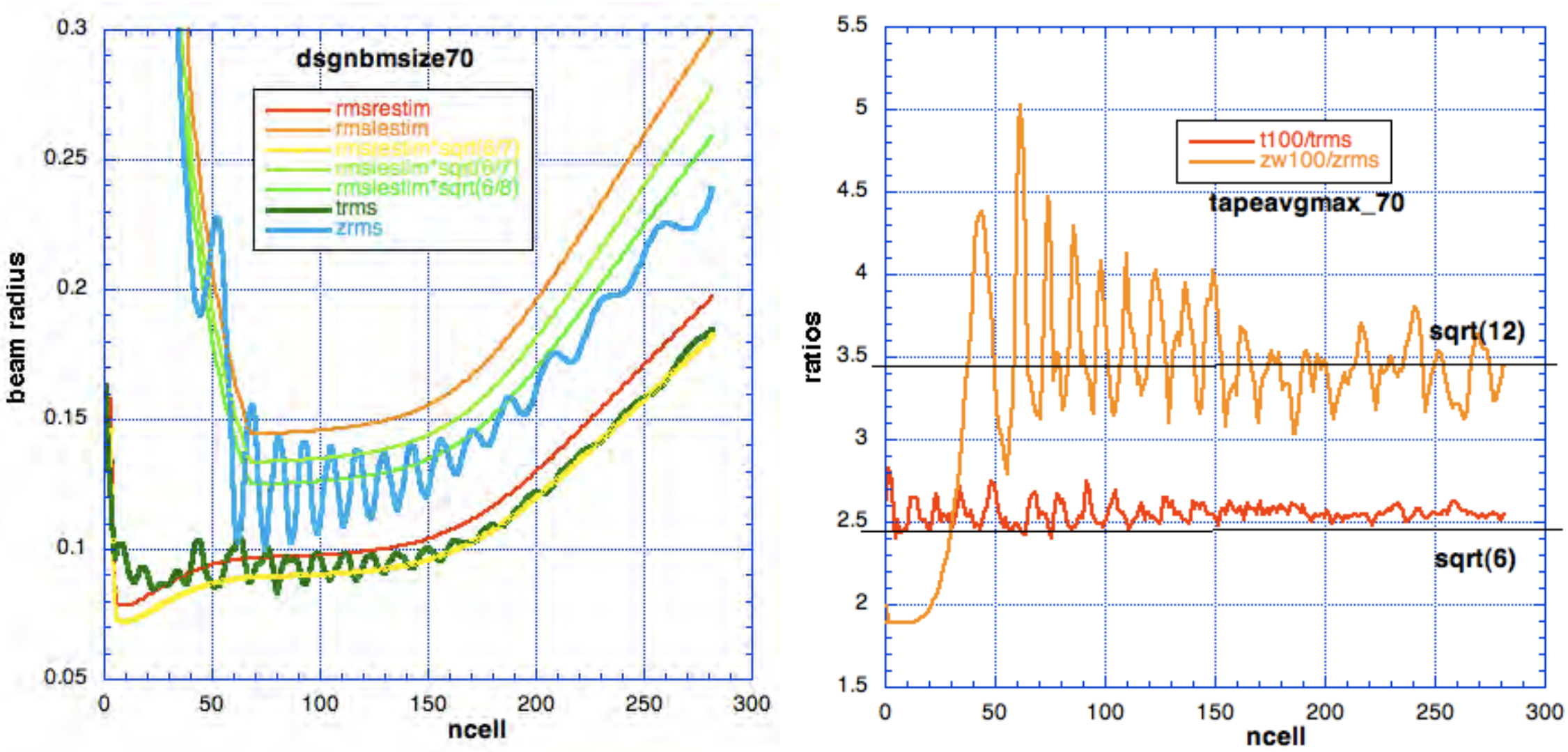

Fig. 22.19. Case 2(70) "2-term" simulation results, corresponding to Fig. 22.16.

## 22.5.3 RFQ with continuous loss from bucket as well as radial loss. – Case 3(25)

*22.5.3.1. Transmission and Accelerated Fractions – Case 3(25)*

Fig. 22.20 shows transmission and accelerated fraction compared for full Poisson, Poisson with space charge transverse vane image effect turned off by replacing the vane boundary with a cylindrical pipe at twice the maximum RFQ aperture = (2*aperture*modulation), and 2-term external field with cylindrically symmetric space charge in a cylindrical pipe at the RFQ aperture. In each case, radial losses are determined from the actual vane surface. Particles are continually being lost from the bucket. Radial losses are small from the shaper end (cell 191) to ~ cell 500, increase strongly from ~ cell 500 – 700, and then are again small from ~ cell 700 – 1009.

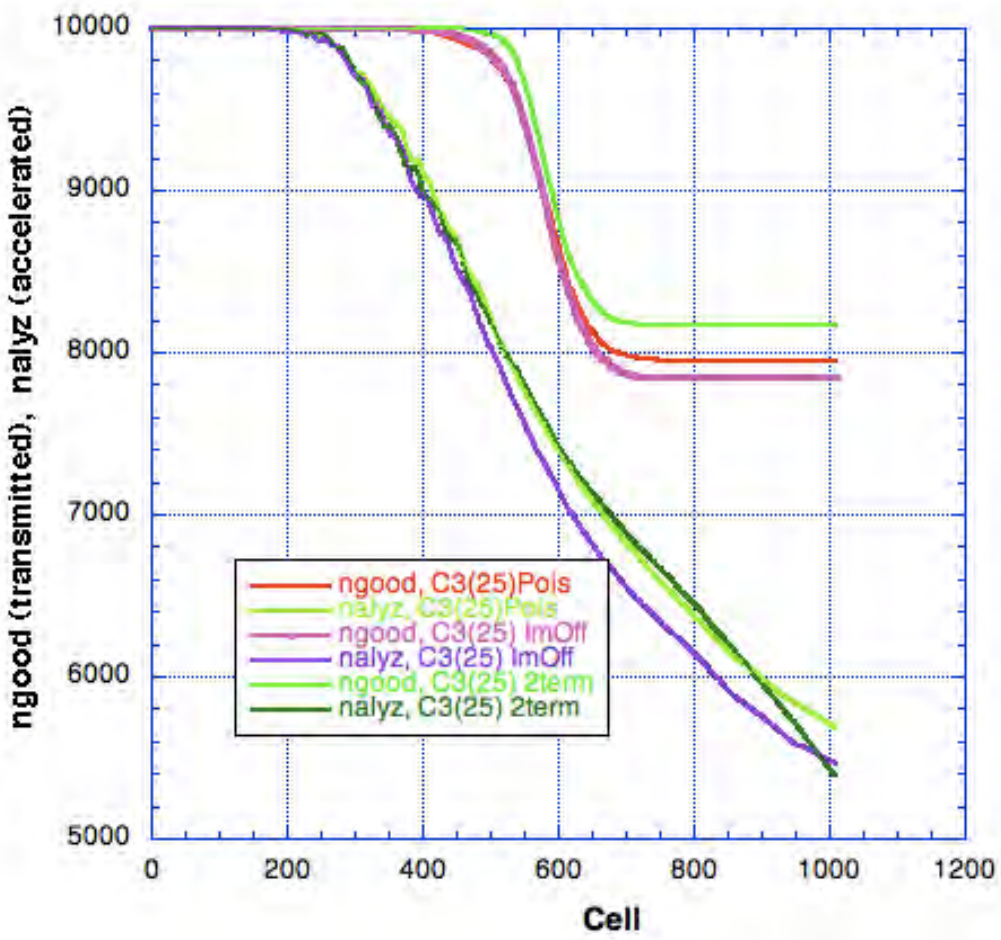

Fig. 22.20. Comparison of transmission and accelerated fractions, Case 3(25).

*22.5.3.2. Full Poisson simulation – Case 3(25)*

Case 3(25) ratios are shown in Fig. 22.21. The equipartitioning ratios do not follow the design. The particles remaining after radial loss are well within the aperture. The transverse rms size (trms) stays near the injected waterbag distribution transverse size as long as the radial losses are small (as in Case 1(43)). As radial losses occur, redistribution to a more parabolic distribution with sqrt((ellipse beta)*etn/8) occurs. When the radial loss ceases, again the rms distribution remains the same. The longitudinal beam size assumes a more compact form before finally increasing again near the end.

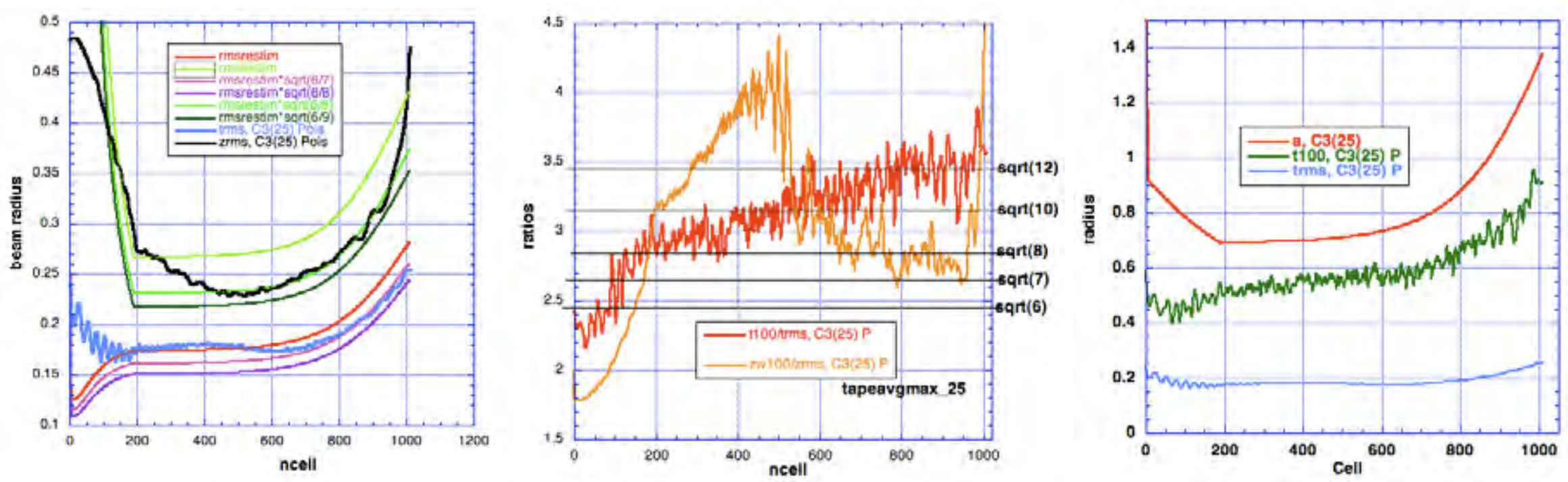

Fig. 22.21. Case 3(25) full Poisson simulation. a. rms beam sizes, design waterbag beam sizes and adjusted design beam sizes. b. 100% to rms beam size ratios. c. t100 and trms size beam sizes compared to aperture.

The particle loss from the bucket causes the remaining particles to assume both smaller zrms and z100. This apparently results in more longitudinal-to-transverse space charge coupling and more growth in t100 (than for Case 1(43)).

*22.5.3.3. Full Poisson simulation – Transverse Space Charge Image Off – Case 3(25)*

Fig. 22.22 presents the results with the transverse space charge image effect turned off by replacing the transverse vane surface boundary with a cylindrical pipe with radius twice the maximum vane radius of the local cell (2*modulation*aperture). With the heavy loss from the bucket, the space charge transverse boundary has only a small effect on these features.

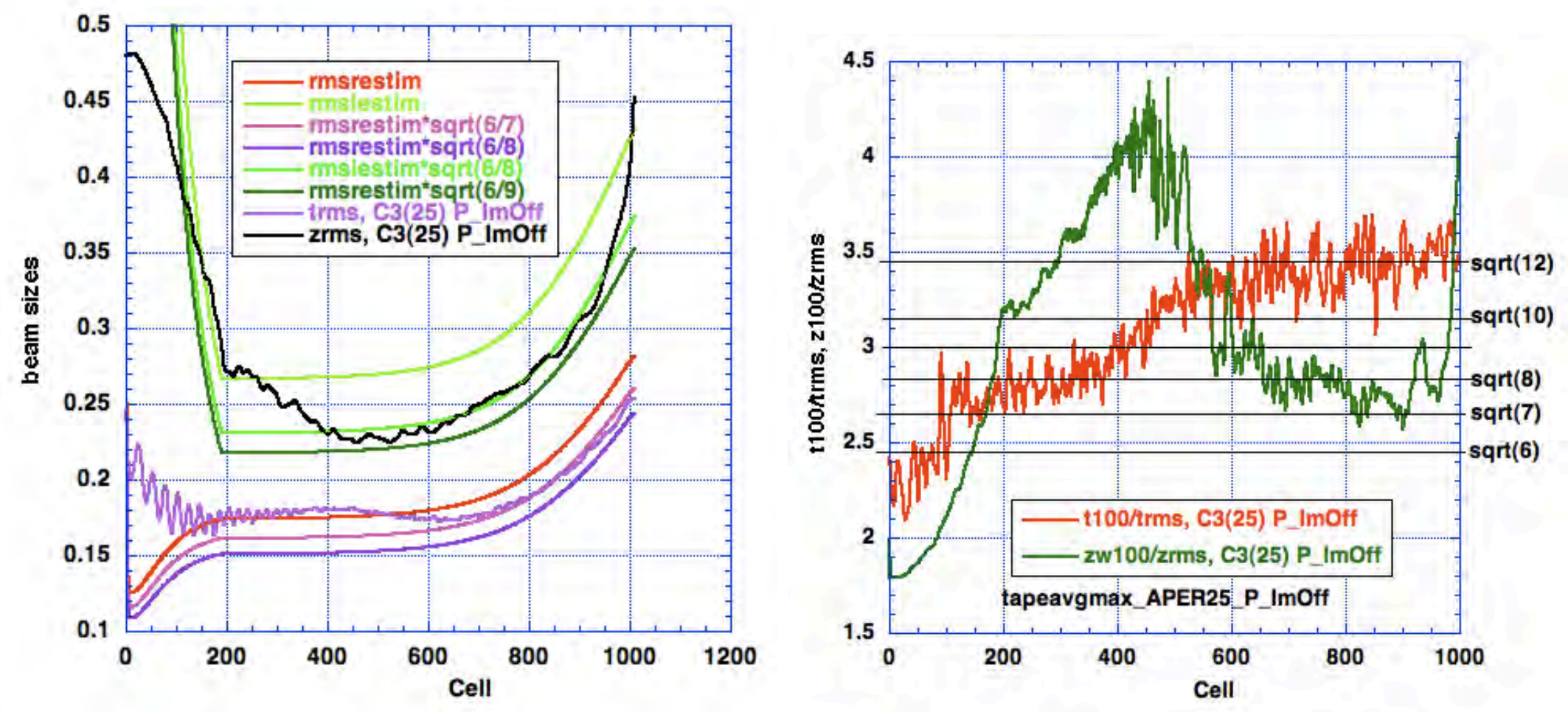

Fig. 22.22. Case 2(70) Full Poisson simulation results but with space charge transverse boundary a cylindrical pipe with radius twice the maximum vane radius of the local cell, corresponding to Fig. 20.

*22.5,3.4. 2-term External Field and Cylindrically Symmetric Space Charge Field Simulation – Case 3(25)*

There is little difference from the Poisson simulation in this large aperture case with heavy loss from the bucket..

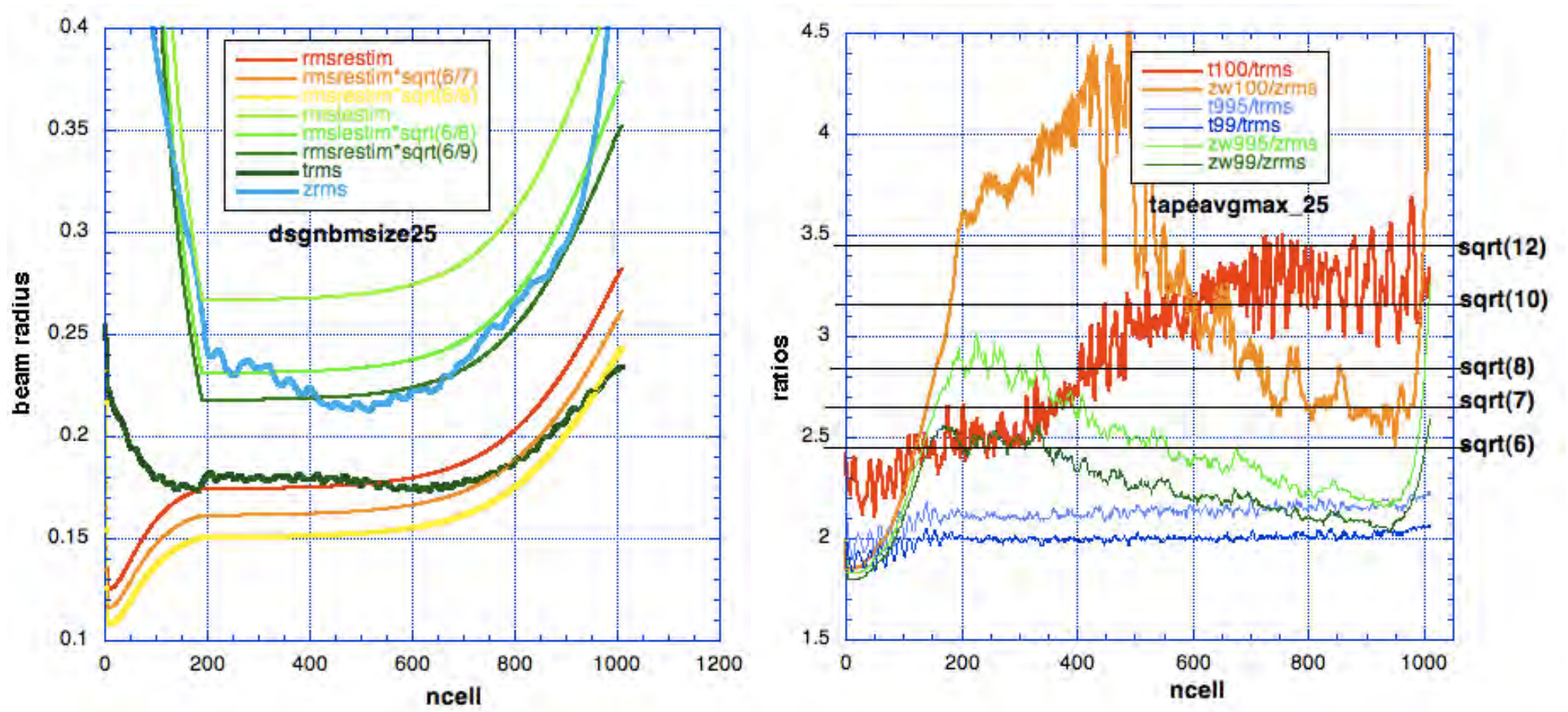

Fig. 22.23. Case 3(25) "2-term" simulation results, corresponding to Fig. 17.

## 22.6. Space Charge Form Factor

Fig. 22.24 shows the space charge form factor (ff=a/3b) in the envelope equations for the three cases. There is little difference between the Poisson and "2-term" results. Case 1(43) keeps the design form factor (1/(1.6*3)=0.208), while the longitudinal characteristics of Cases 2(70) and 3(25) seem to be responsible for a redistribution to a form factor of ~(1/(1.33*3). The form factor, which depends on rms quantities, is amenable to manipulation in the design.

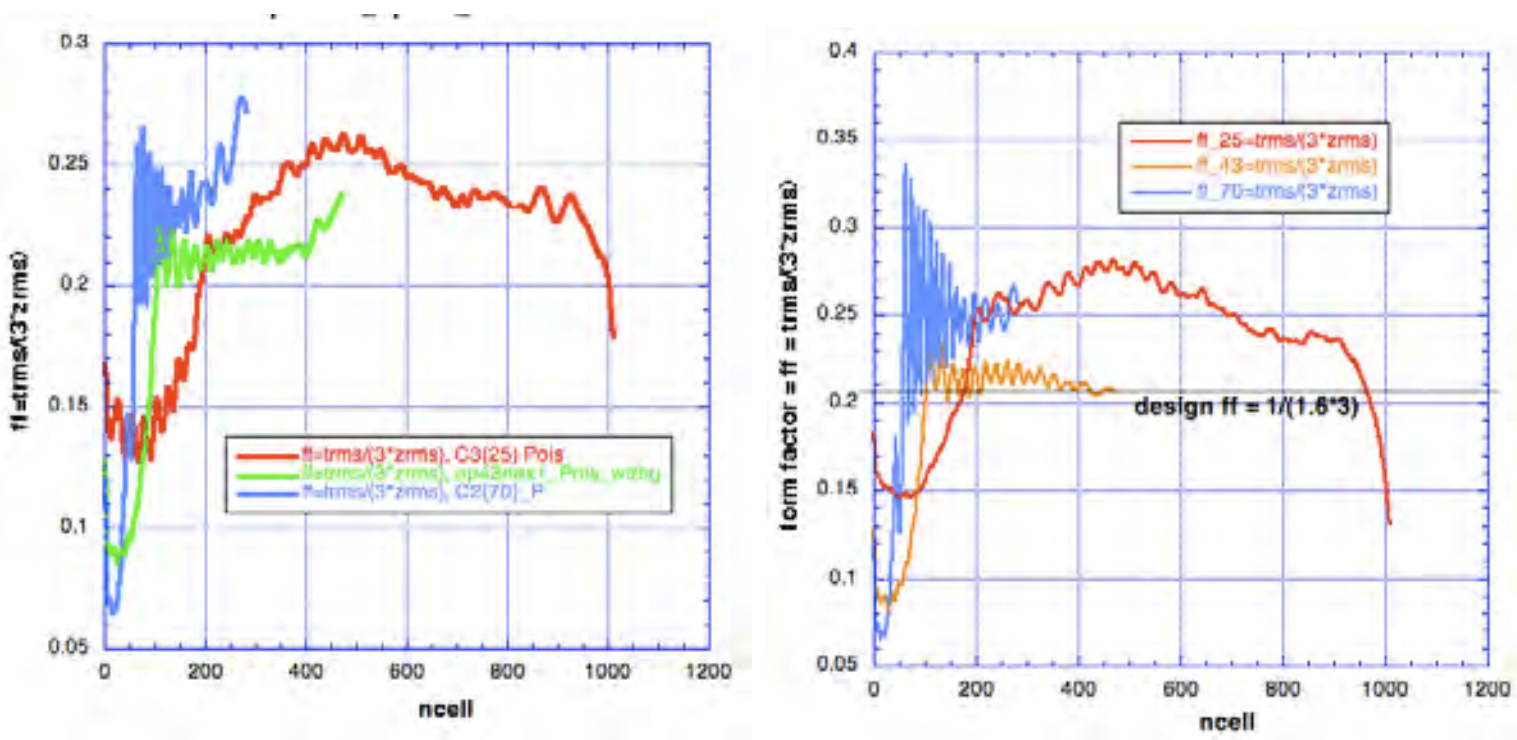

Fig. 22.22. Space charge form factor (trms/(3*zmrs). a. full Poisson b. "2-term"

## 22.7 Non-Equipartitioned RFQ, $\sigma_0^l \sim\leq \sigma_0^t$

Redistribution is investigated here in a non-equipartitioned RFQ, designed using a method that allows the longitudinal zero-current phase advance ($\sigma_0^l$) to approach, but not cross, the transverse zero-current phase advance ($\sigma_0^t$) midway in the main part of the RFQ. **(NOTE!! – THIS RFQ IS *NOT* EP AT EOS!!)** This method, sometimes with less approach of $\sigma_0^l$ to $\sigma_0^t$, has been used for several projects recently, because it can make a short RFQ with good transmission. This RFQ has ~100% transmission, and ~98% accelerated fraction. The strong 1:1 resonance is reached briefly from Cells ~90-110, but the beam is also close to equipartition in this region, and there is little effect on the rms emittances. The transverse 100%-to-rms beam size stays small at first because $\sigma_0^t$ is kept high. Then it shows an effect from being near the resonance; t100 continues to grow in spite of the quick departure from the resonance and the transverse focusing force becoming stronger again. This is evidence that the ¼ plasma period redistribution inside the bunch occurs faster than the rms effect from a resonance interaction.

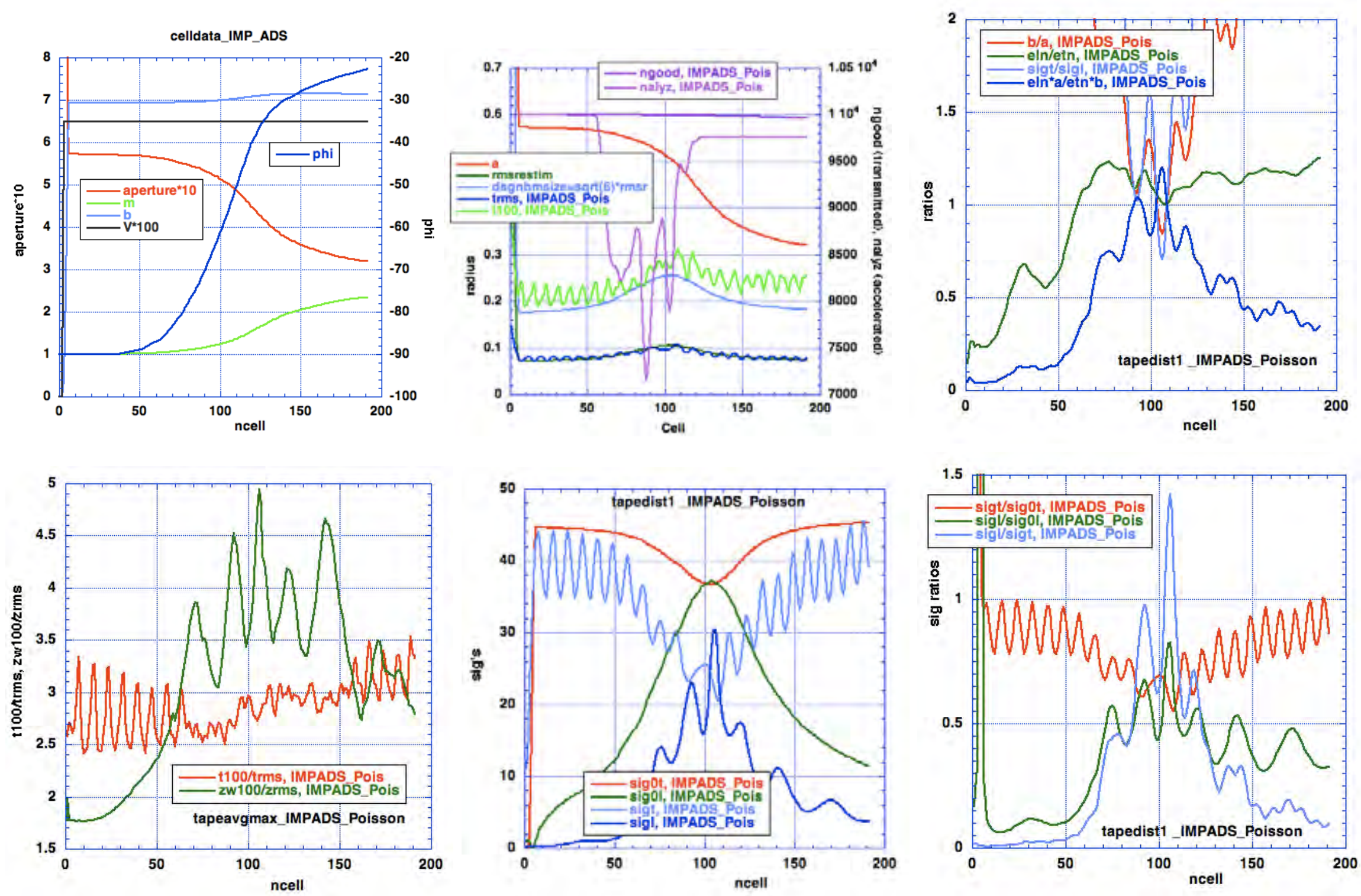

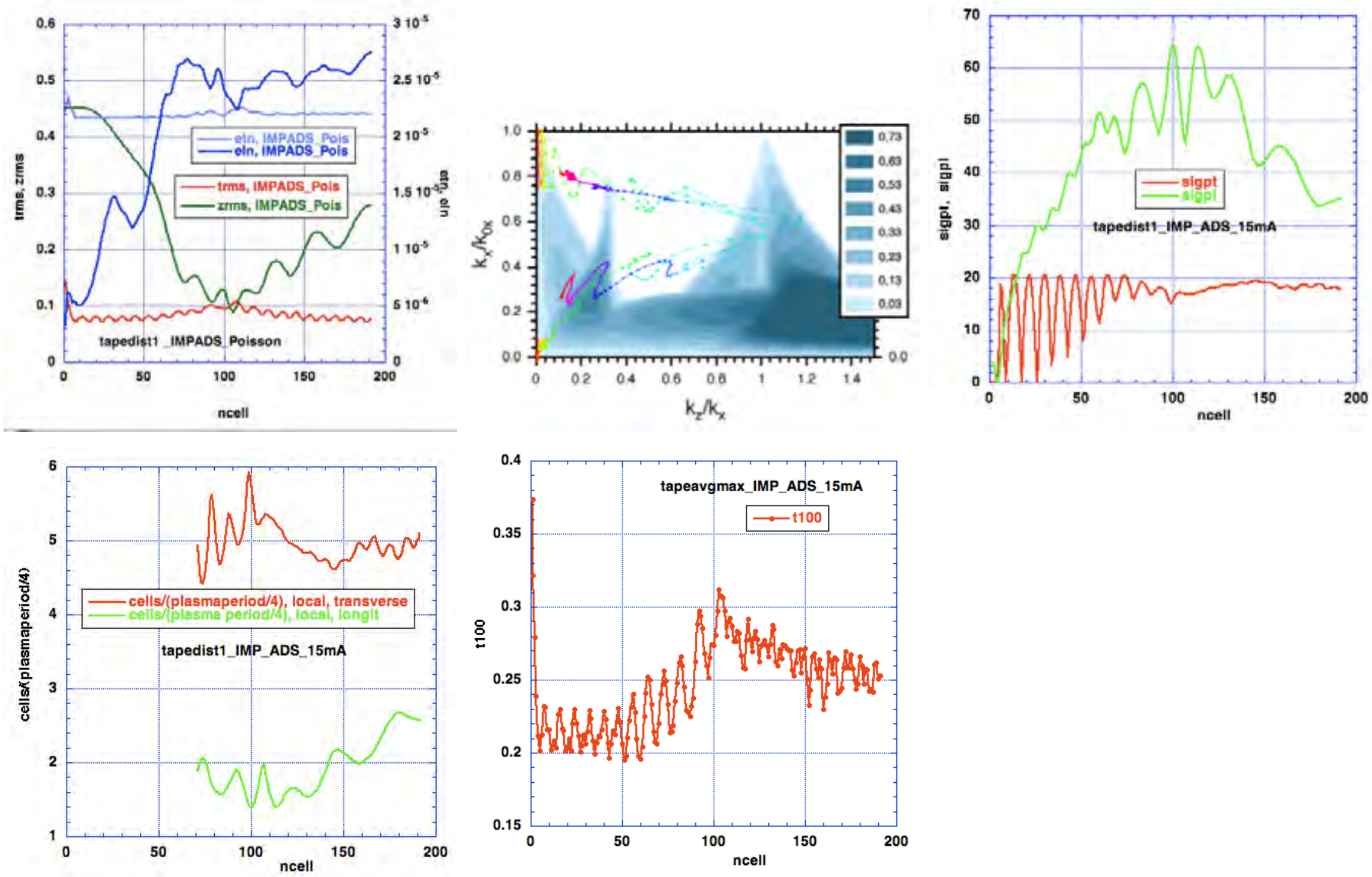

Shaded HChart – loop at 0.4 on way in, but no loop inside 1:1 zone. Then influence of 0.33 on t100 around cell 150,not on trms, growth on t100 stays. See C1, same kind of design, except sig's stay quite equal after meeting to end.

1:1 resonance ~cell80-120, although briefly reaches EP. Not enough time – ¼ plasma period = 5 cells transverse, 2 cells longitudinal? Sigl/sigt > 0.8 ~cells91-95, 103-110. No trms redistribution.

Fig. 22.25. A non-equipartitioned RFQ, designed using a method that allows the longitudinal zero-current phase advance ($\sigma_0^l$) to approach, but not cross, the transverse zero-current phase advance ($\sigma_0^t$) midway in the main part of the RFQ.

## 22.8 More RFQs

References not given to protect the innocent.

### 22.8.1 AltCDR

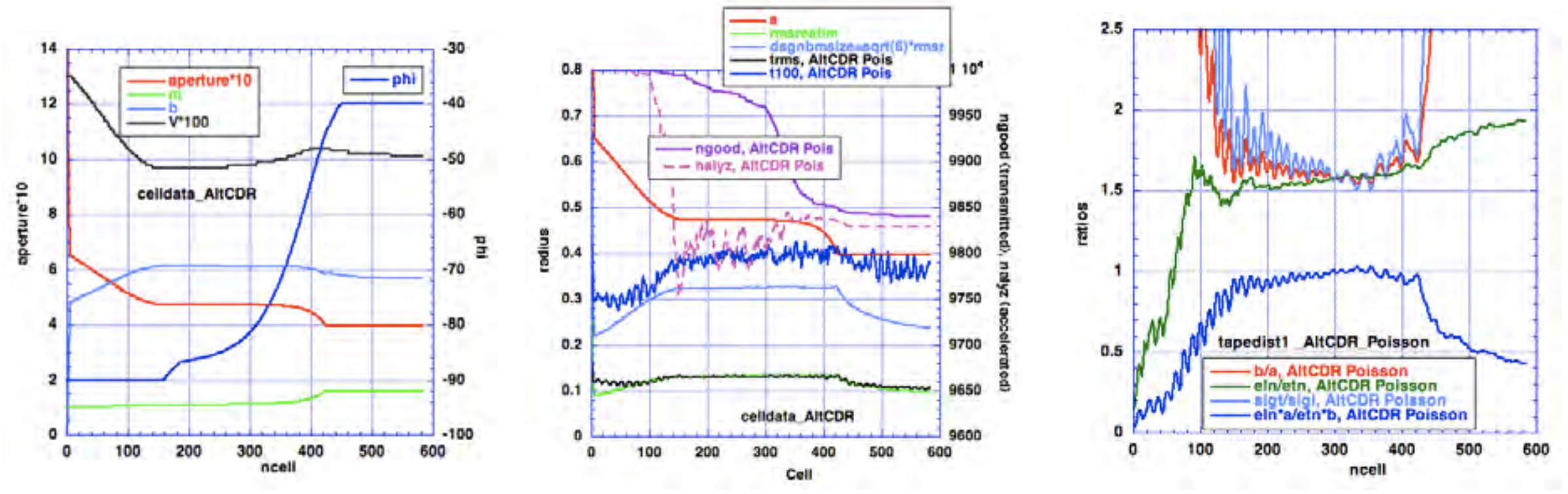

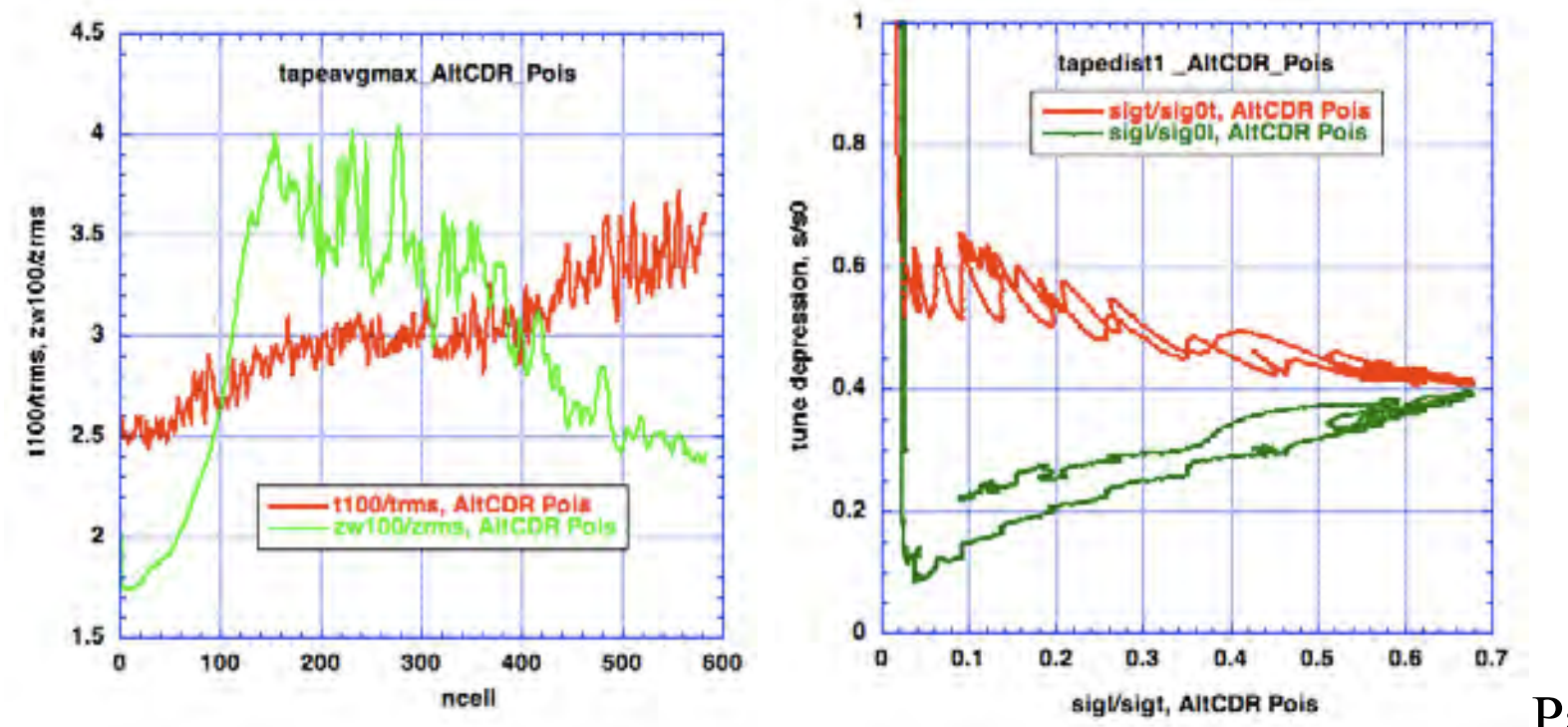

Parameters flat at end. Low losses, well EP at first, ***then saturated, have growth in t100 – good evidence of "conventional design"*** -> Case 1(43). No trms redistribution. 2term result clearly shows smaller trms and redistribution.

## 22.8.2 APT

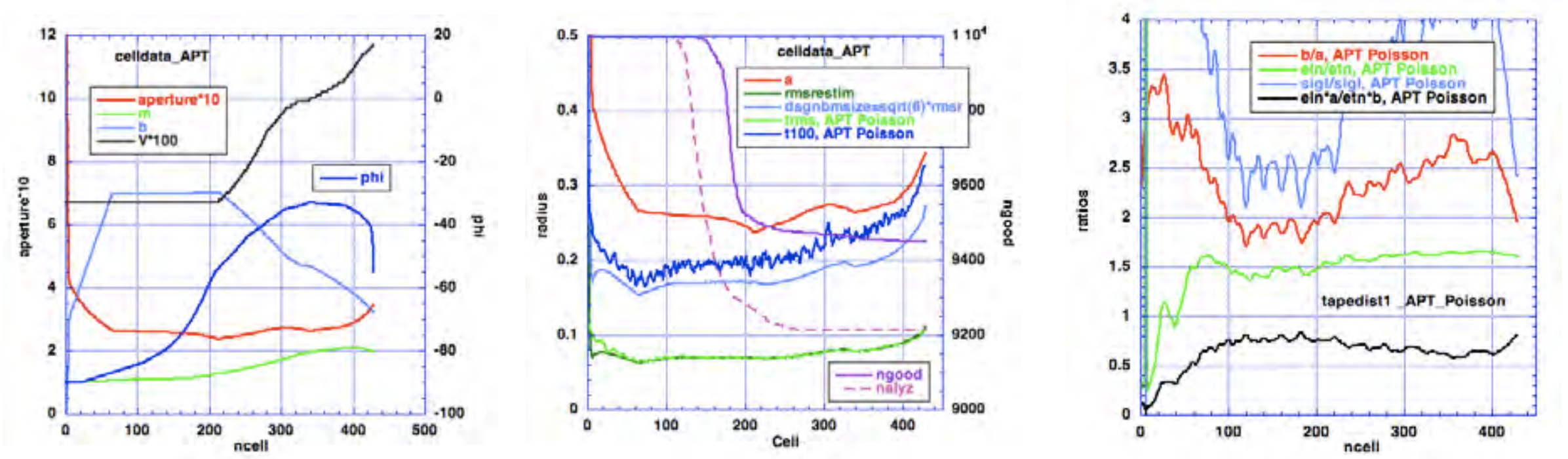

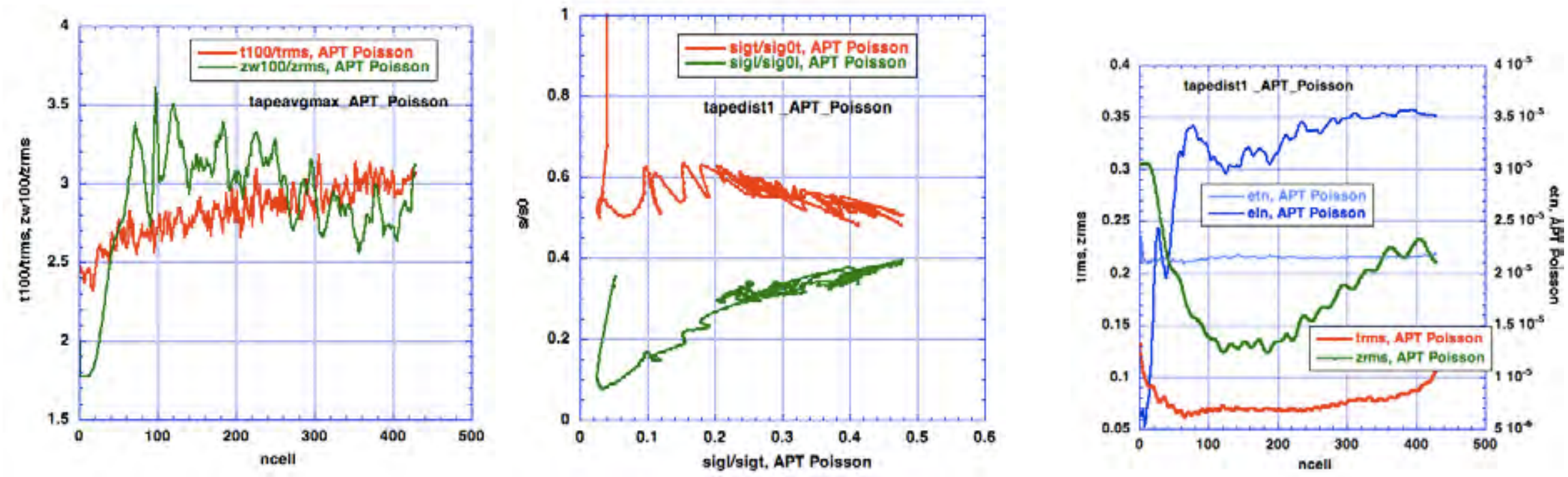

Parameters do not saturate, strange variations. Significant longitudinal losses, -> Case 3(25). No trms redistribution. 2term result clearly shows smaller trms and redistribution.

## 22.8.3 ATS

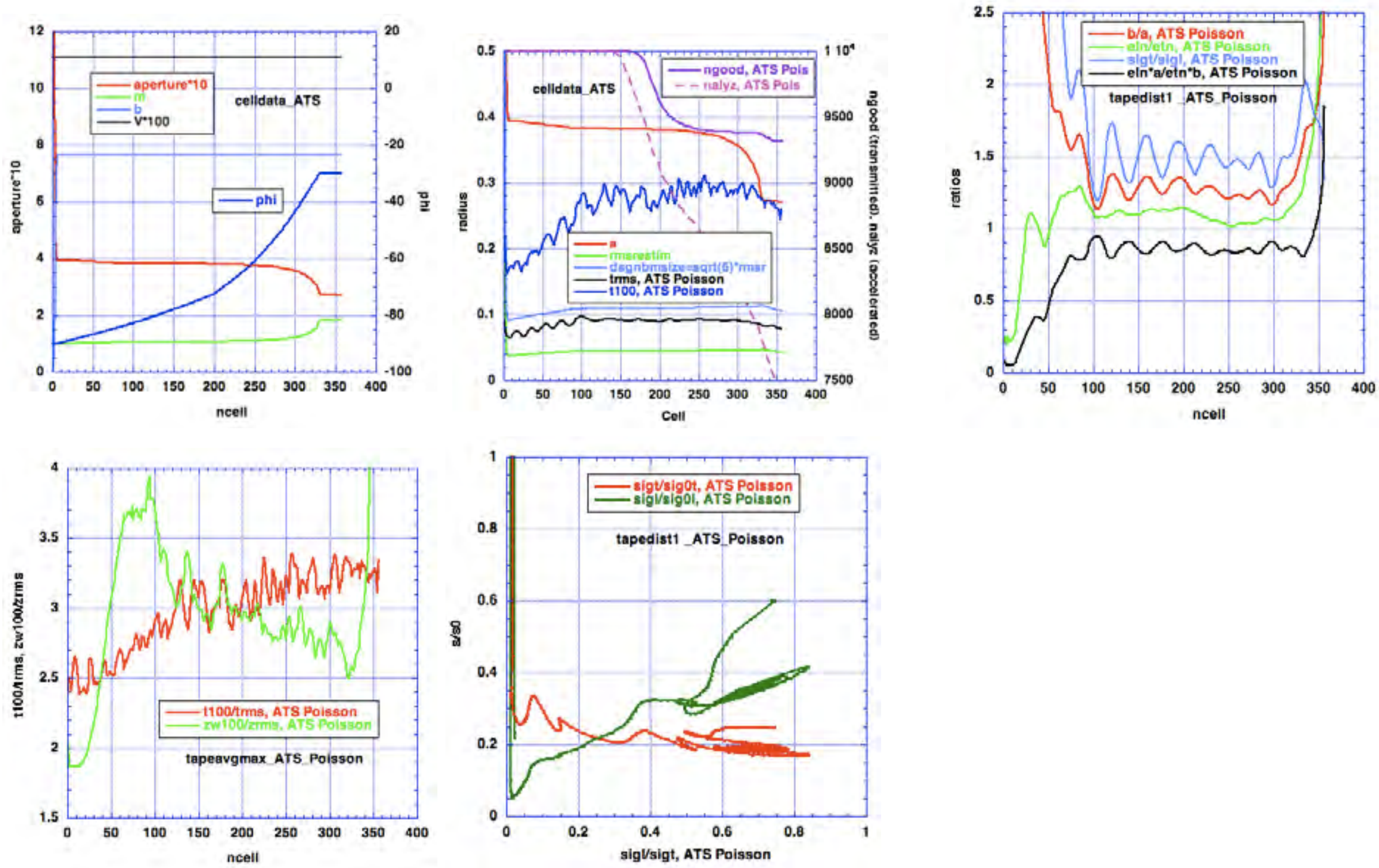

Constant parameters at end. Very large longitudinal losses, -> Case 3(25). Cannot say trms redistribution – was no design goal. 2term is much worse…

## 22.8.4 CERN

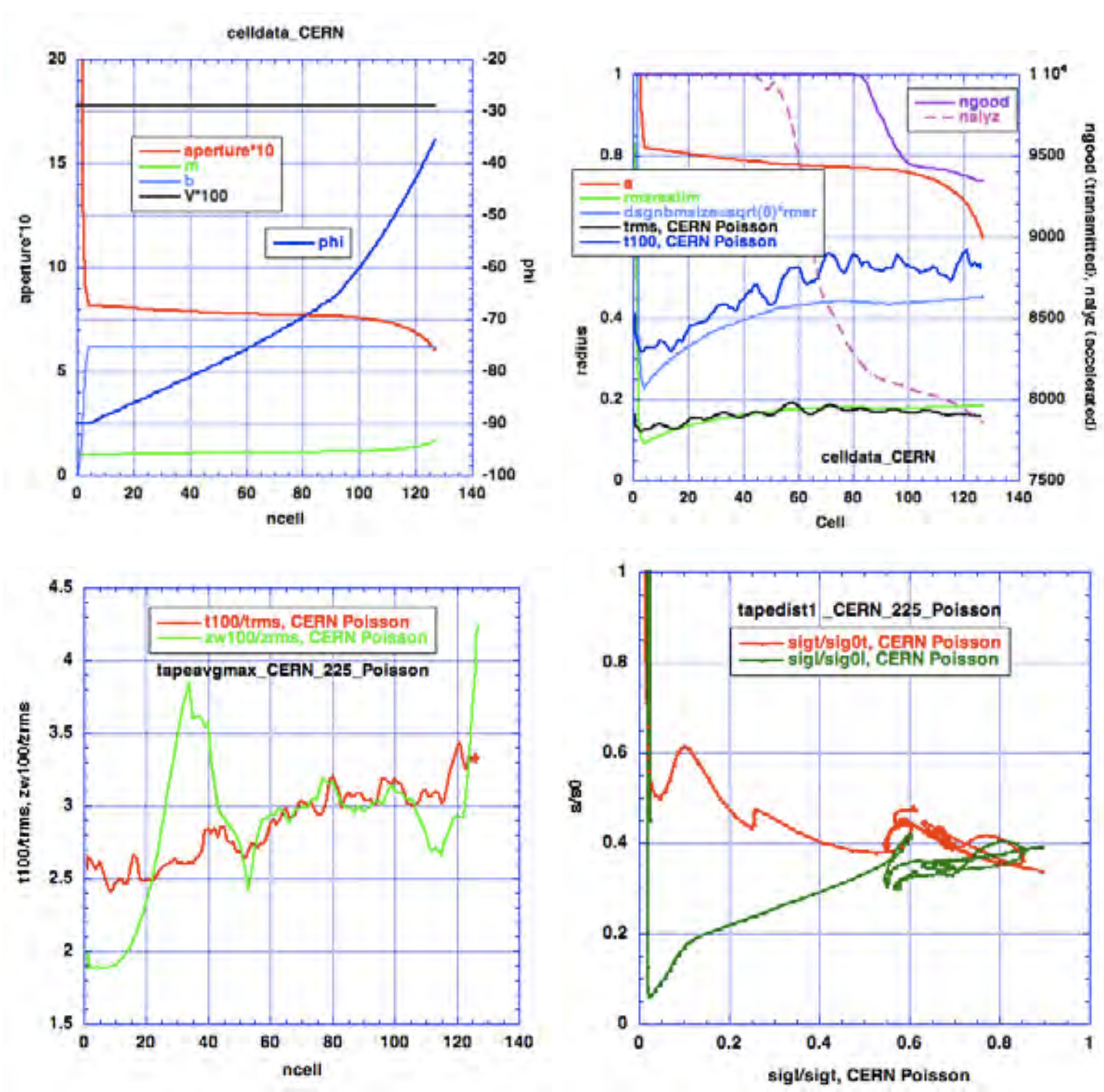

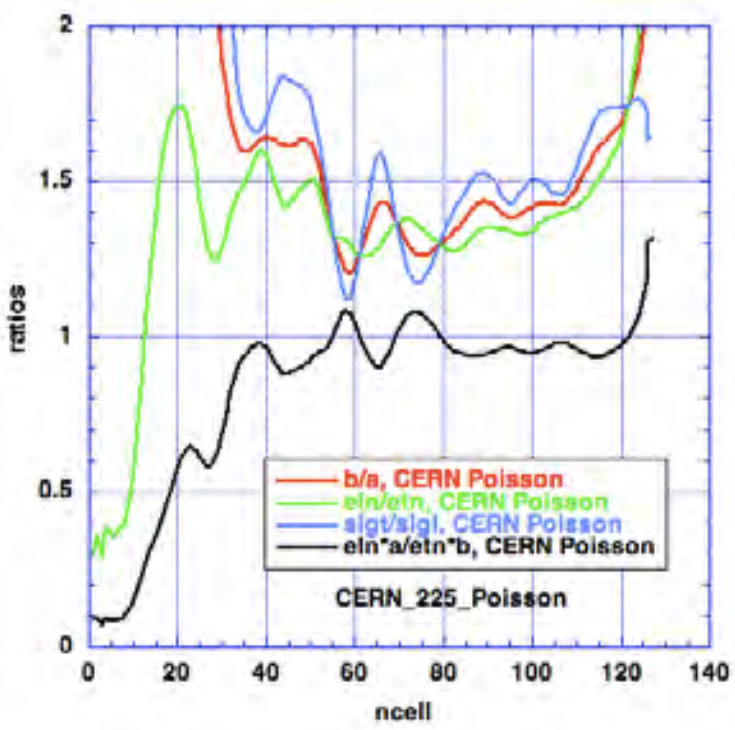

Const V,B, only shaper. EP (was not known then). Trms well controlled. Large aperture, probably thinking because of very large current; Large longitudinal losses to ~cell80 -> Case 3(25), followed by both losses, but not flat 100/rms ratios as Case 3(25) – very high current, 225mA… There is some trms redistribution – 2-term is worse….

## 22.8.5 C1

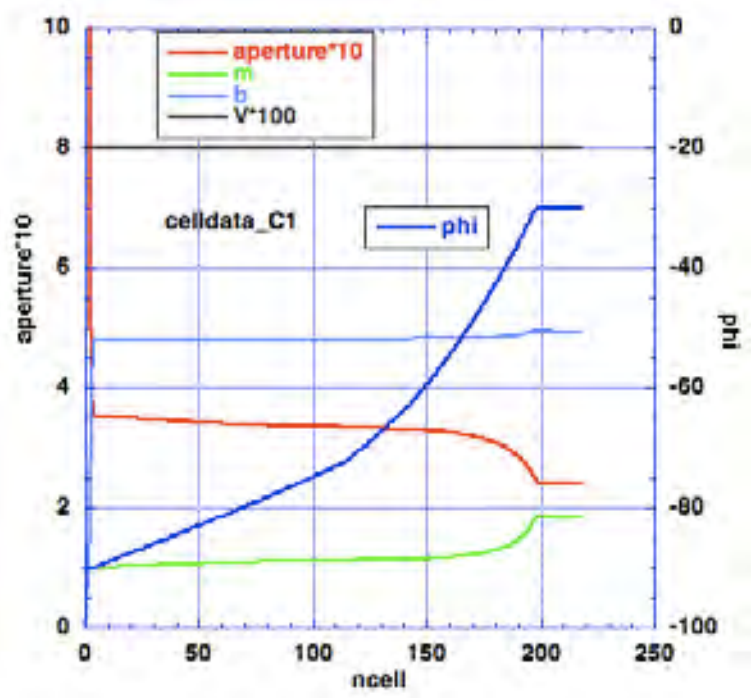

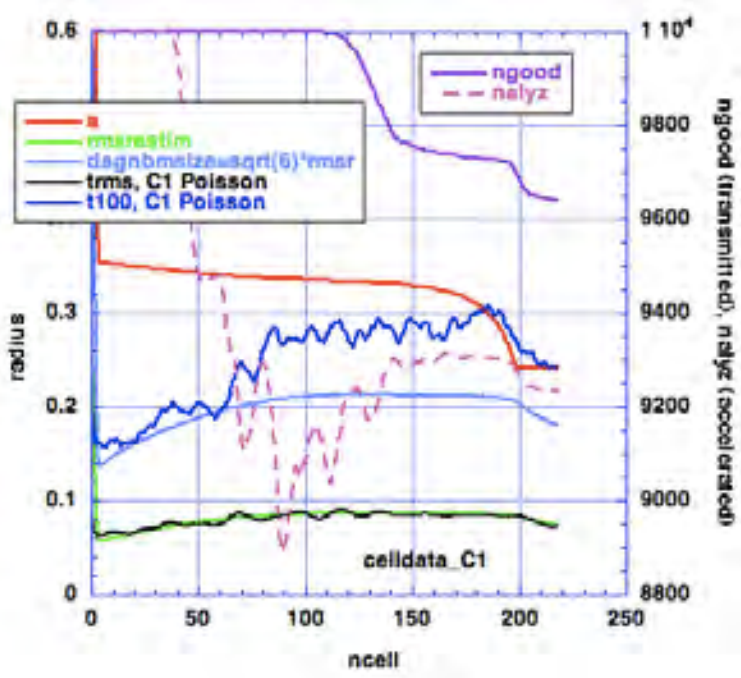

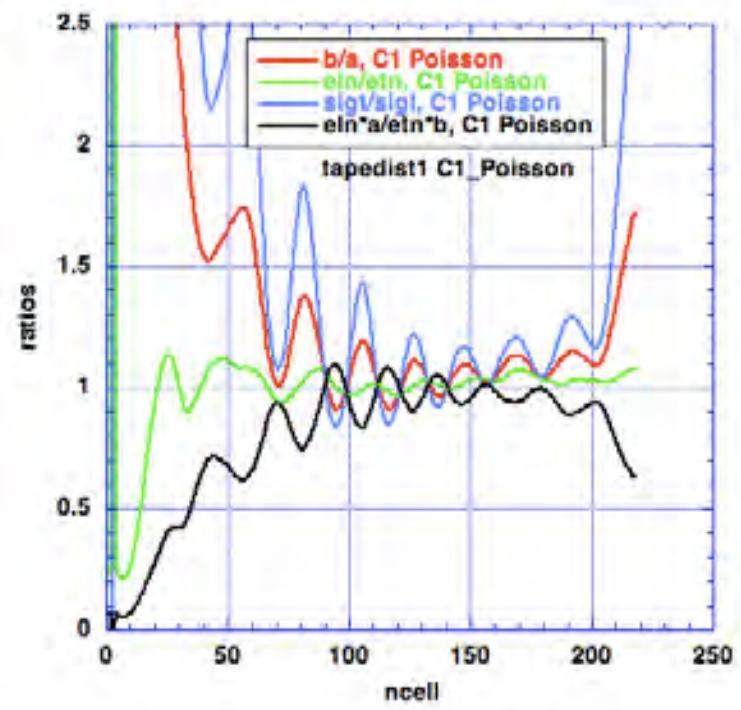

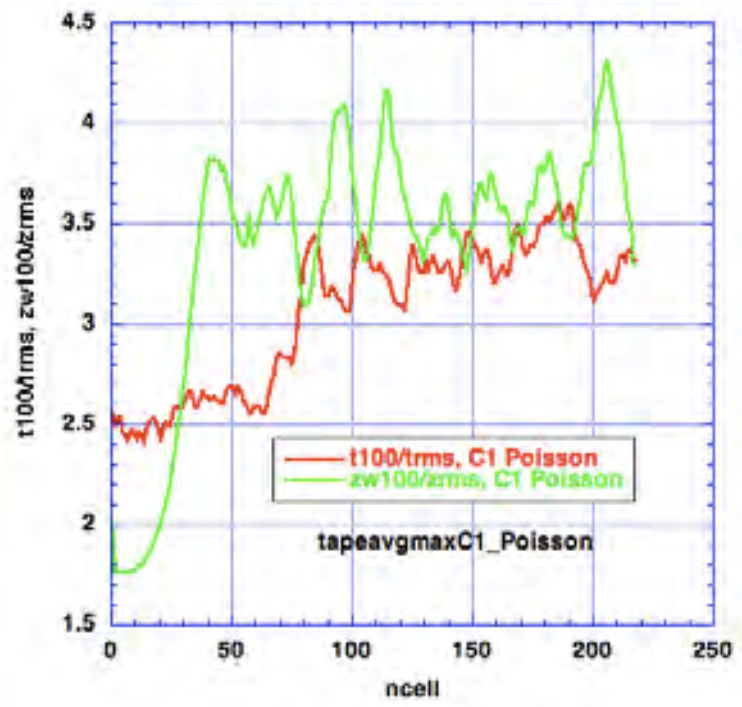

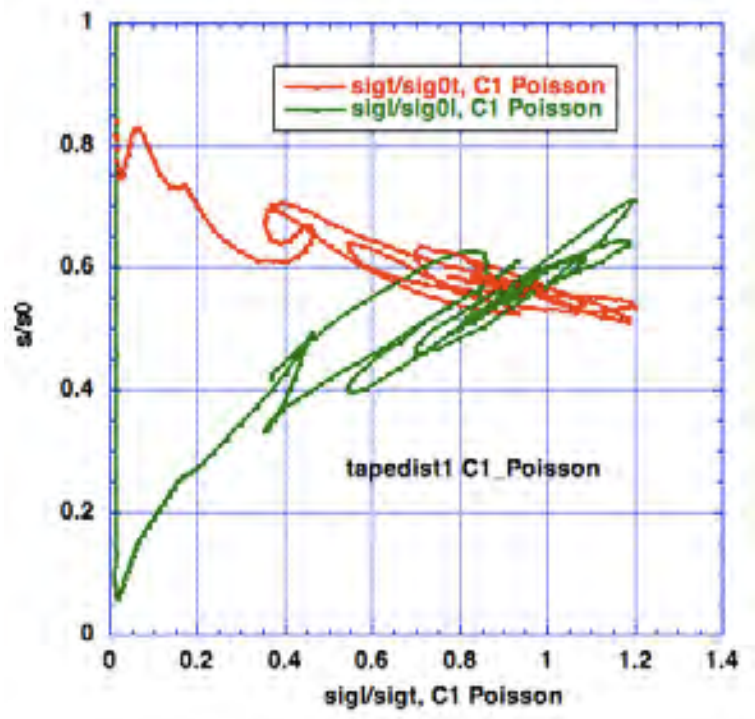

THESE ARE FOR POISSON *IMAGE OFF*!! Constant parameters at end. EP with ratio 1, but clear evidence of 1:1 resonance on approach tp EP. Then EP ~ maintained (sig's stay ~ equal) and growth in t100 stays, same as IMP_ADS, where sig's are reduced after kiss and Xmsn, accel reasonable No trms redistribution.

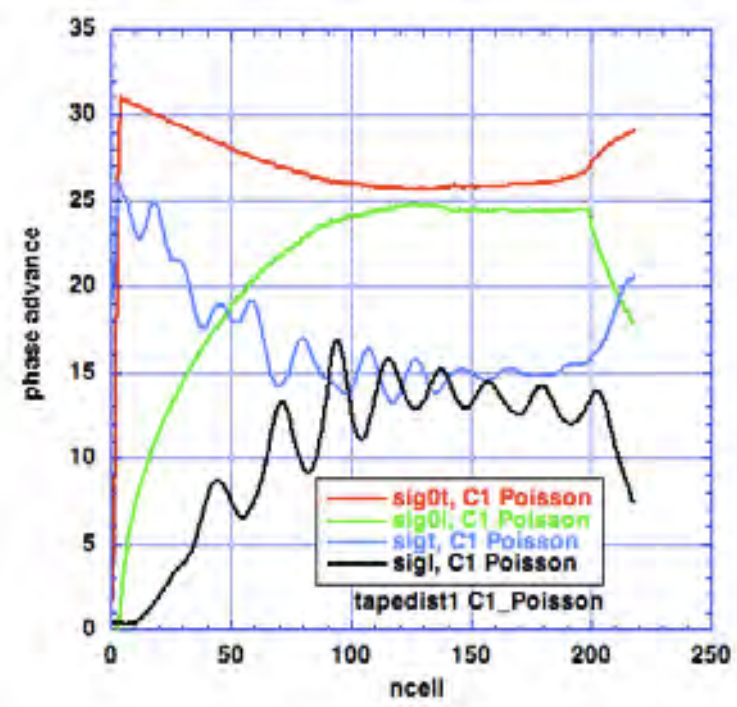

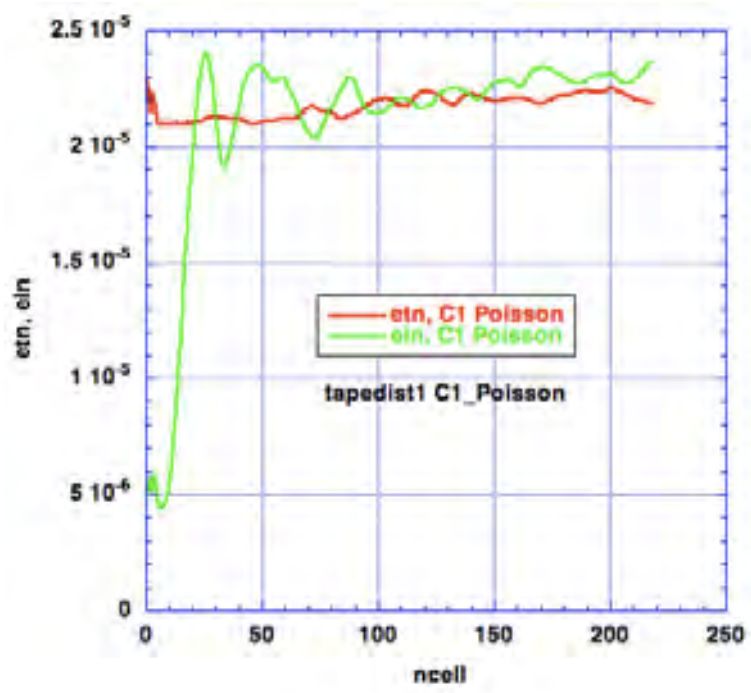

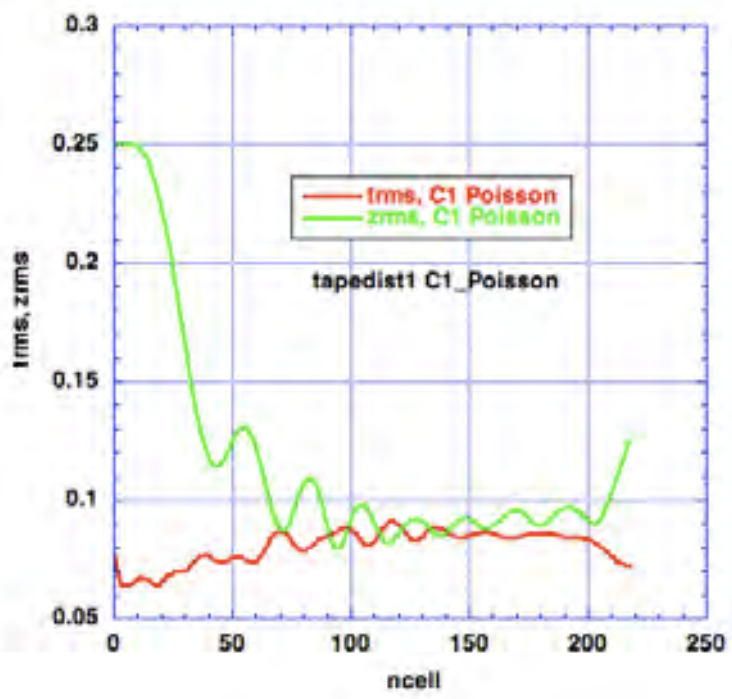

### 22.8.6 C2

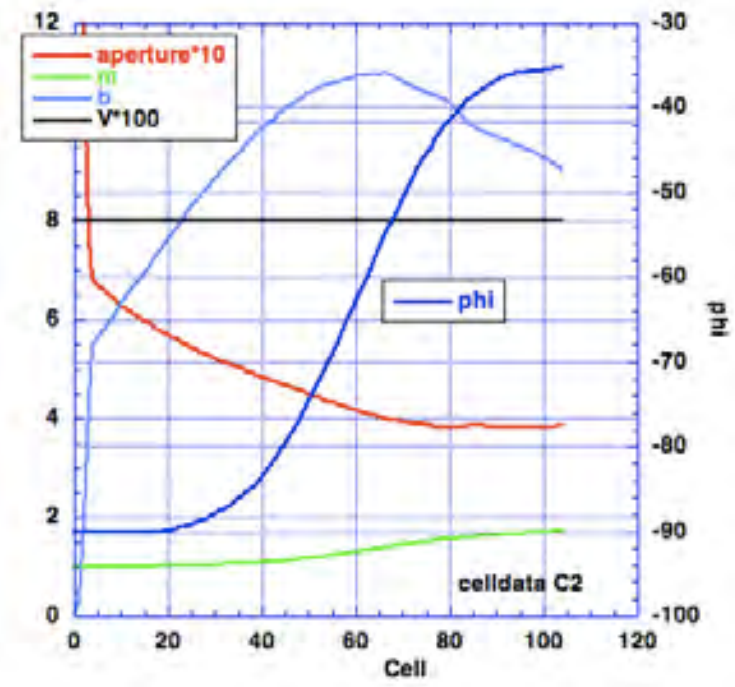

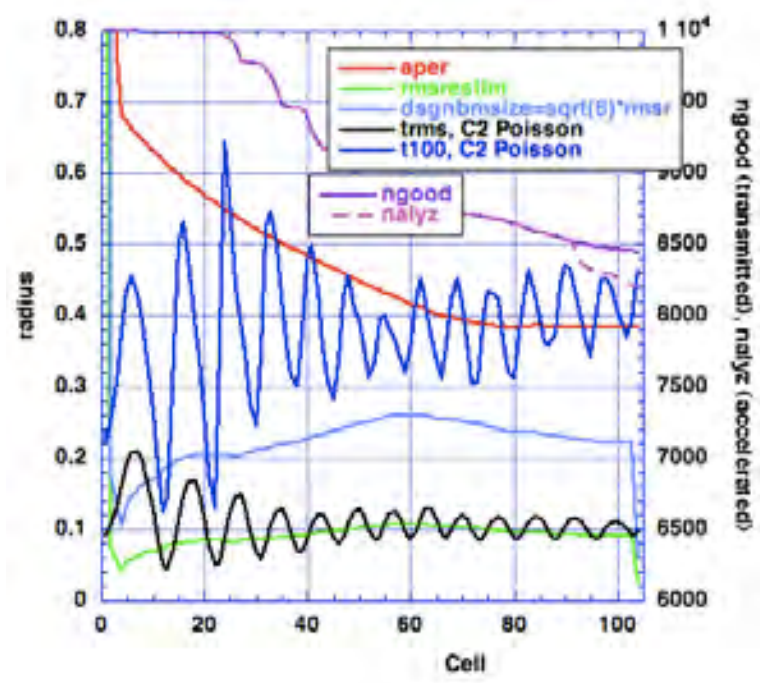

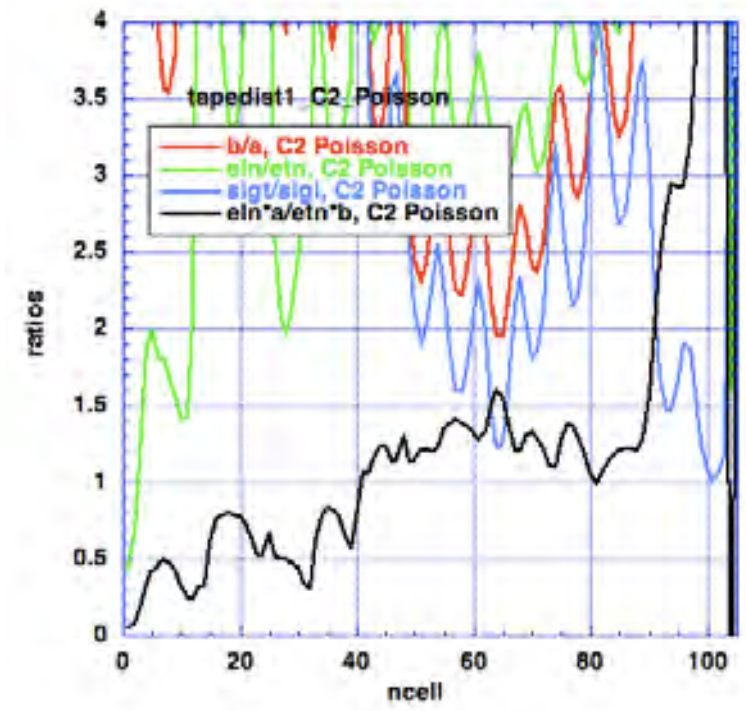

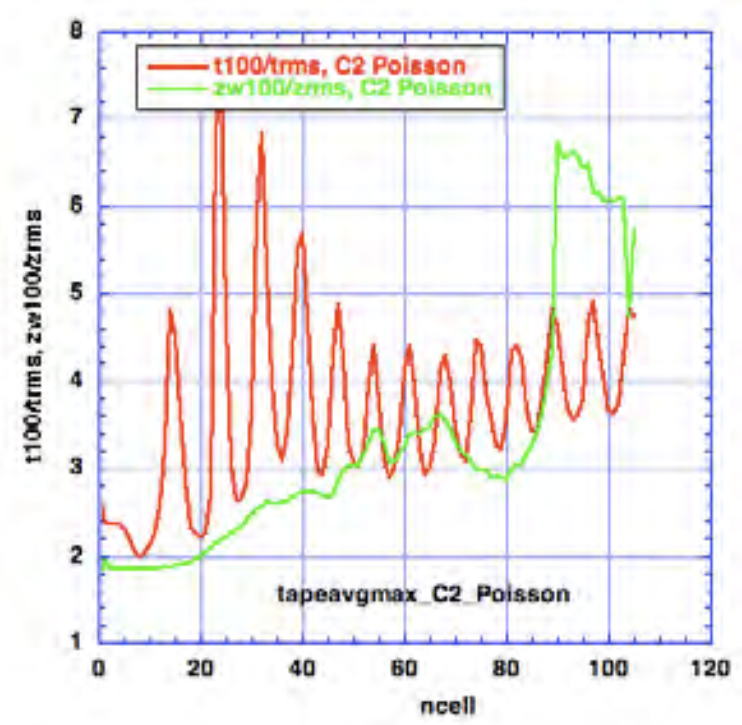

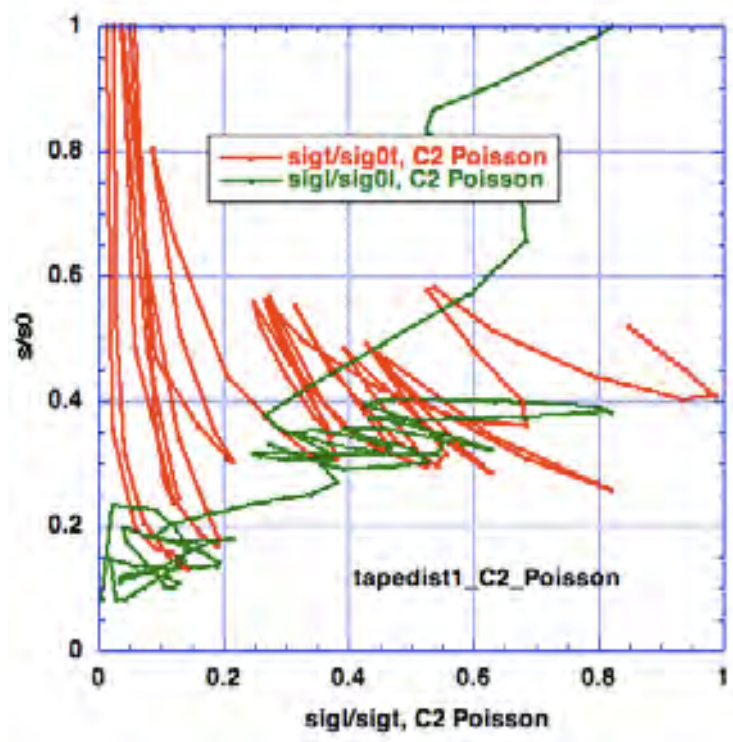

Not EP. ~constant parameters at end. Continuous both losses -> Case 3(25). No trms redistribution. Drastically short design for high current "to save space".

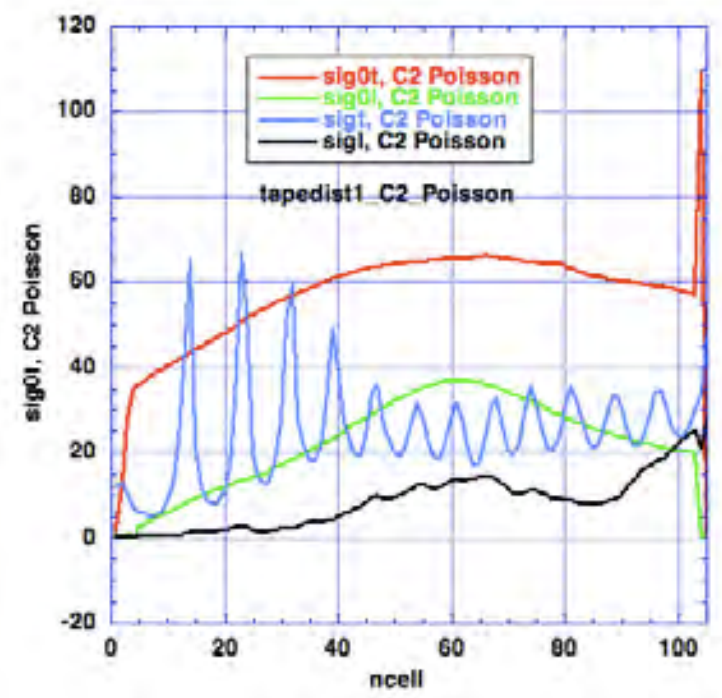

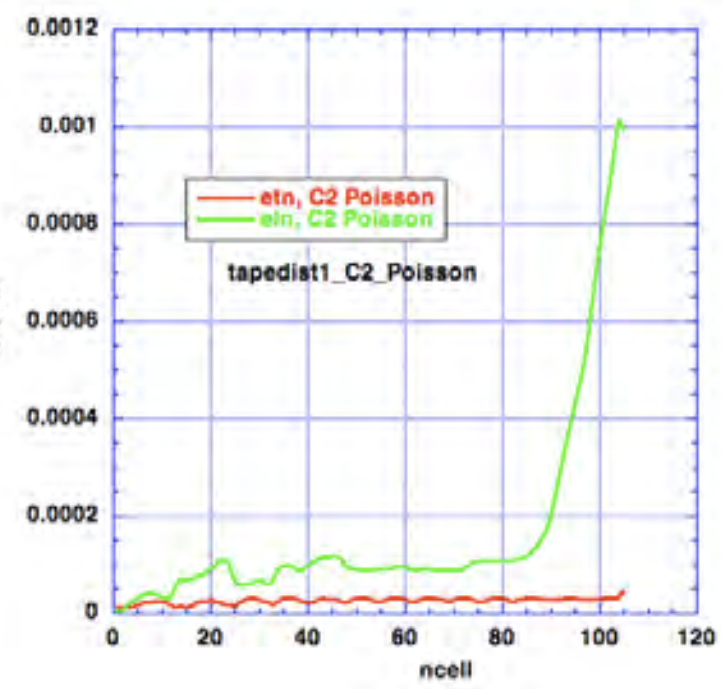

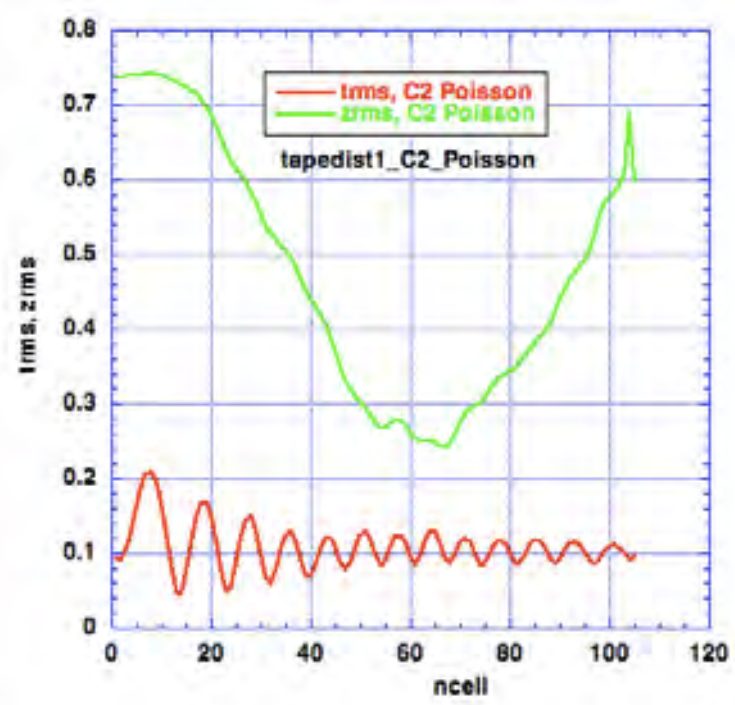

## 22.8.7 C3

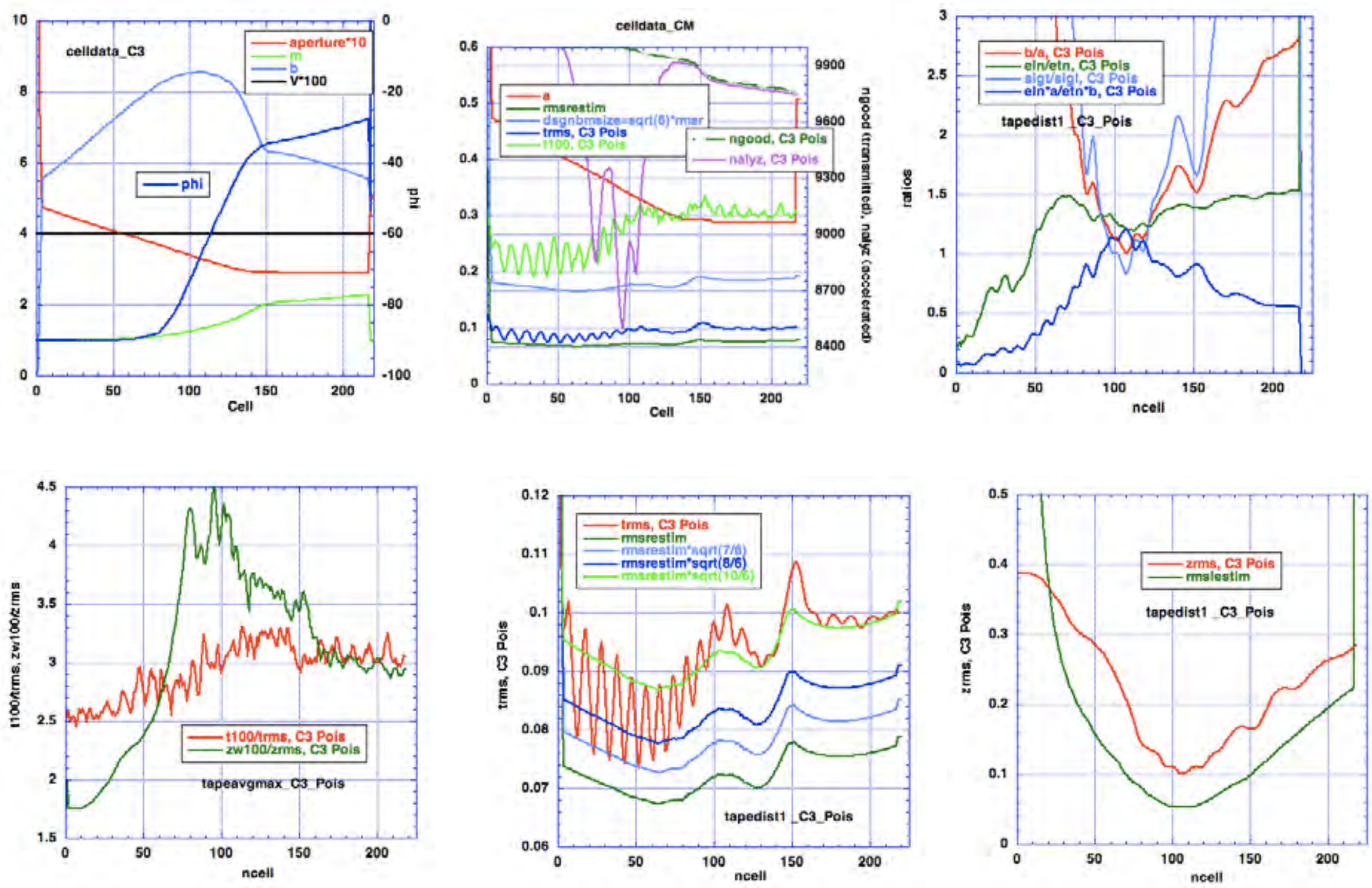

**(NOTE!! – THIS RFQ IS *NOT* EP AT EOS!!)** Constant aperture at end. Not EP, except briefly at cells ~100-120, but also evidence of 1:1 resonance cell 80-100, where scraping starts, then both losses -> Case 2(70). Trms redistribution – immediately to ~parabolic, then to Gaussian; no redistribution at low current. Like C1 and IMP_ADS.

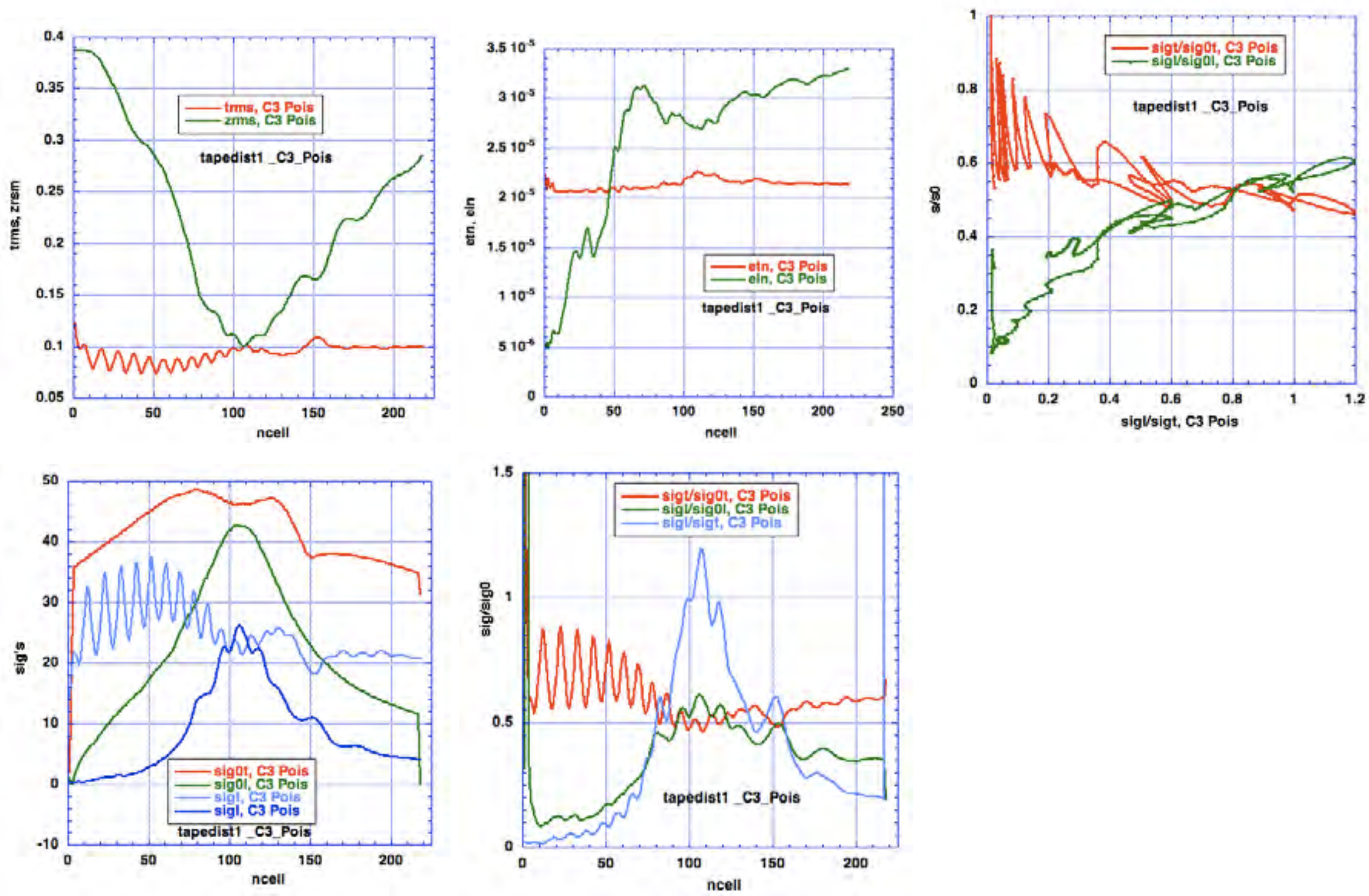

### 22.8.8 CRNL

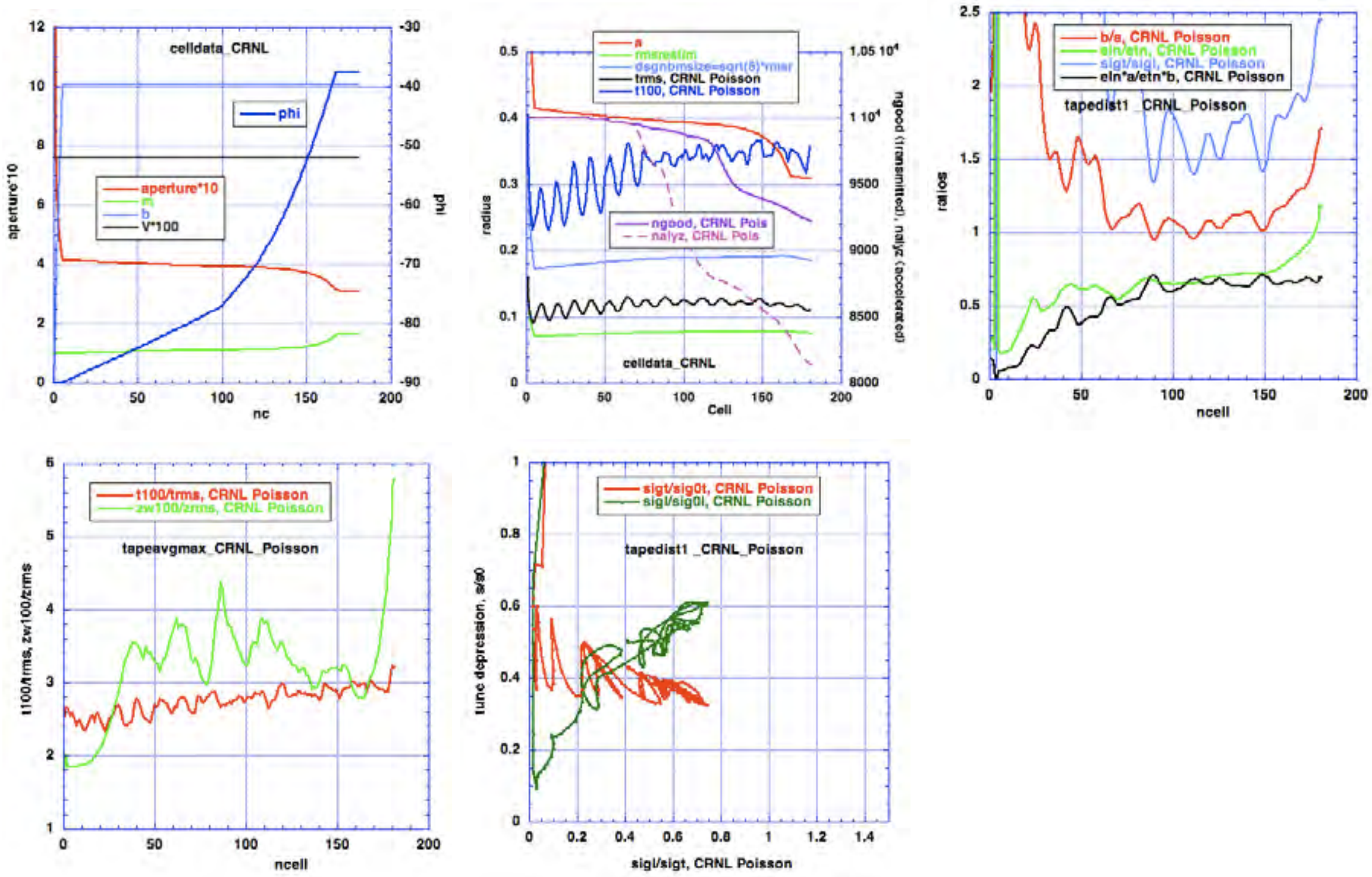

Very non-EP. Strong longitudinal loss while no radial loss, BUT NOT LIKE 3(25)?? There is some immediate trms redistribution – but no design value known. Parameters flat at end.

### 22.8.9 IFMIF CDR

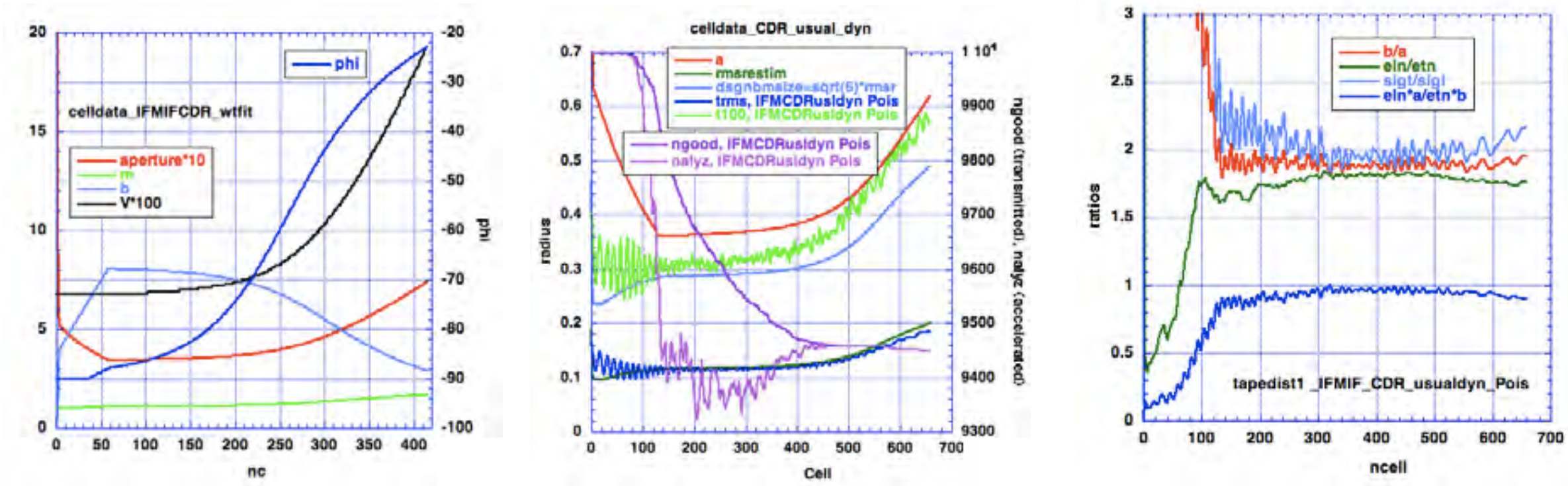

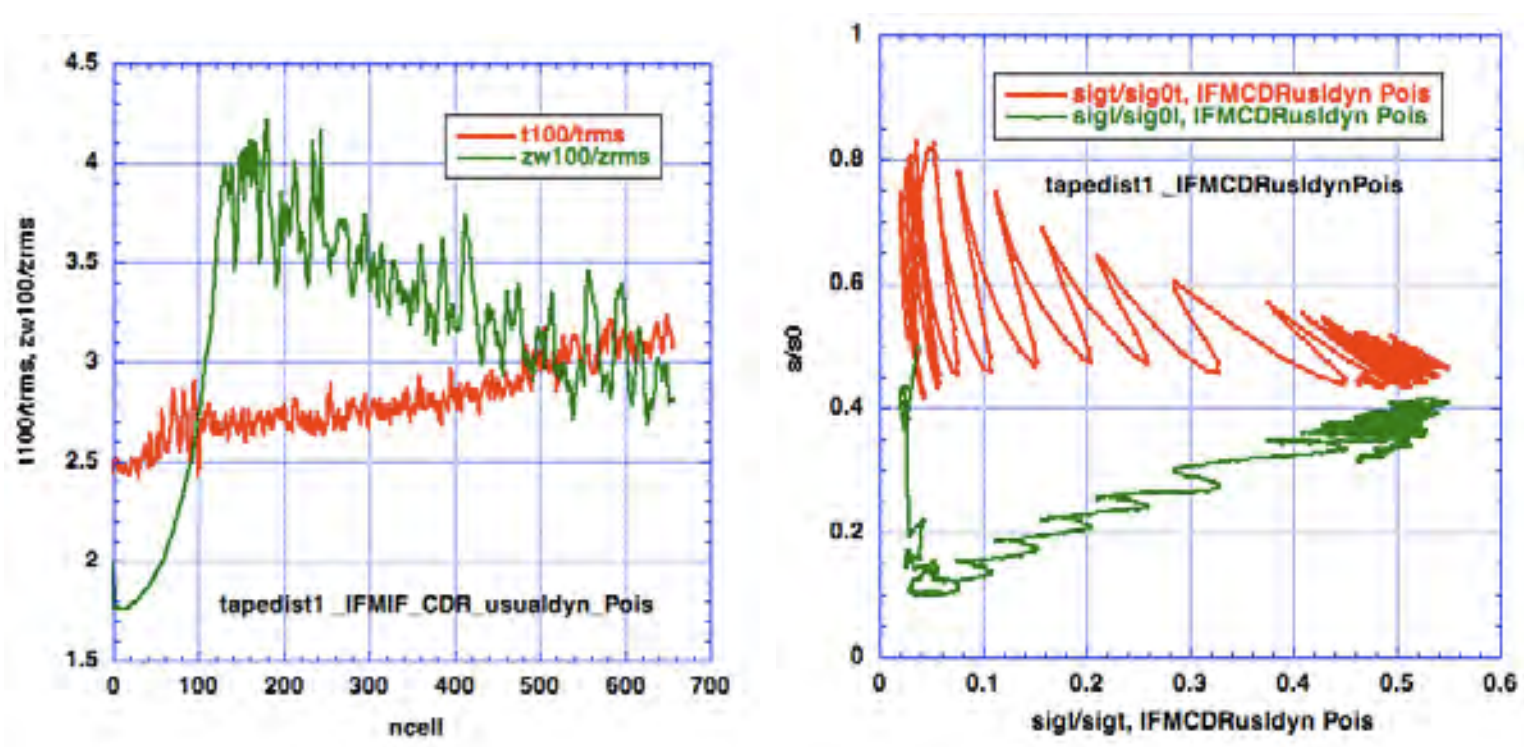

EP, small losses -> Case 1(43). No trms redistribution. 2term less t100 steady growth – good evidence.

## 22.8.10 IFMIF E

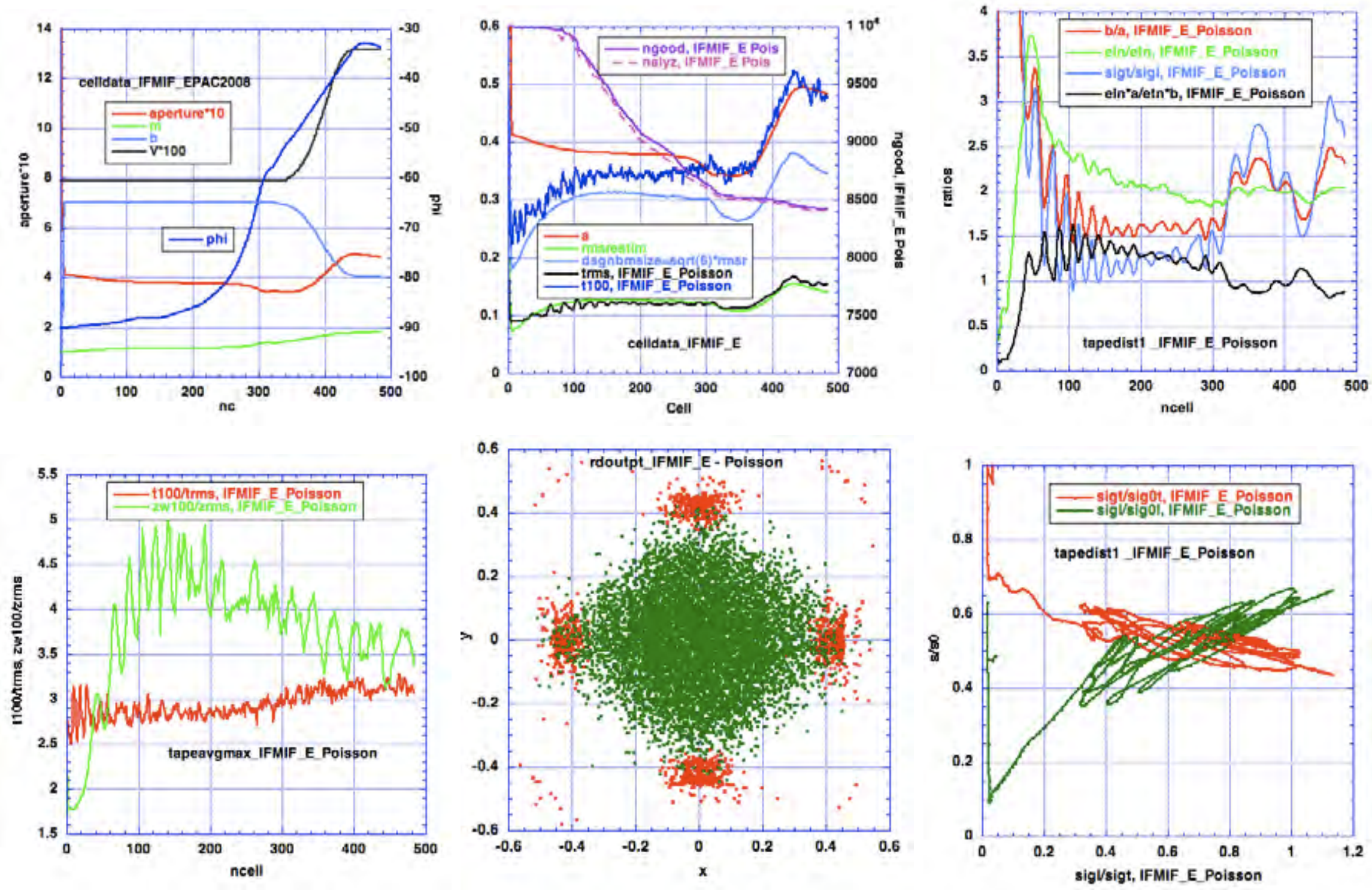

Not EP, continuous both losses - Case 3(25). No trms redistribution. Particles riding beside vane tip. Parameters ~flat at end.

## 22.8.11 IFMIFpostCDR

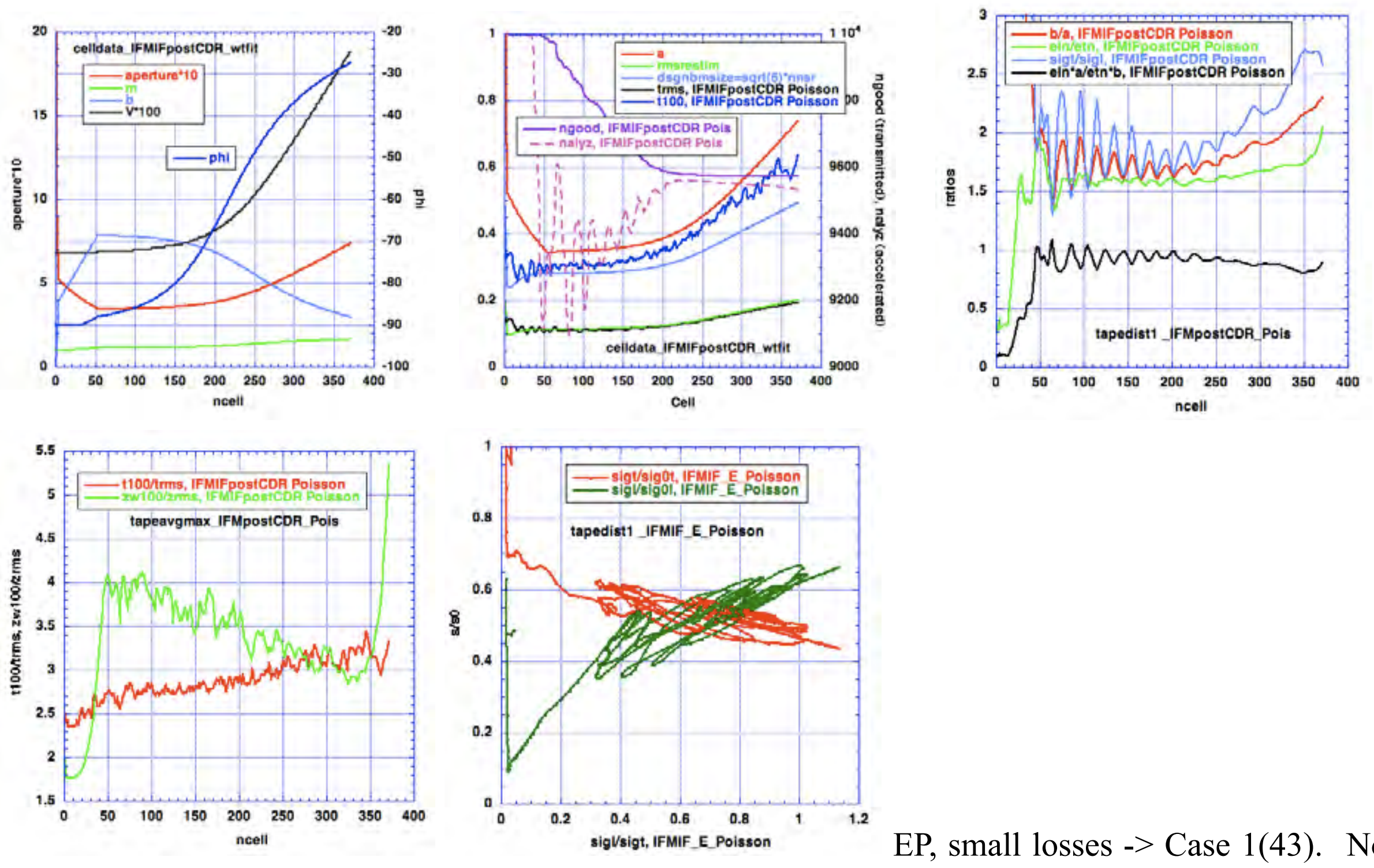

EP, small losses -> Case 1(43). No trms redistribution. Departure from EP as EP ratio changes – Poisson effect stronger than design (which includes MPs at rms) – rms no redistribution, follows well ??

## 22.8.12 IMPSSC

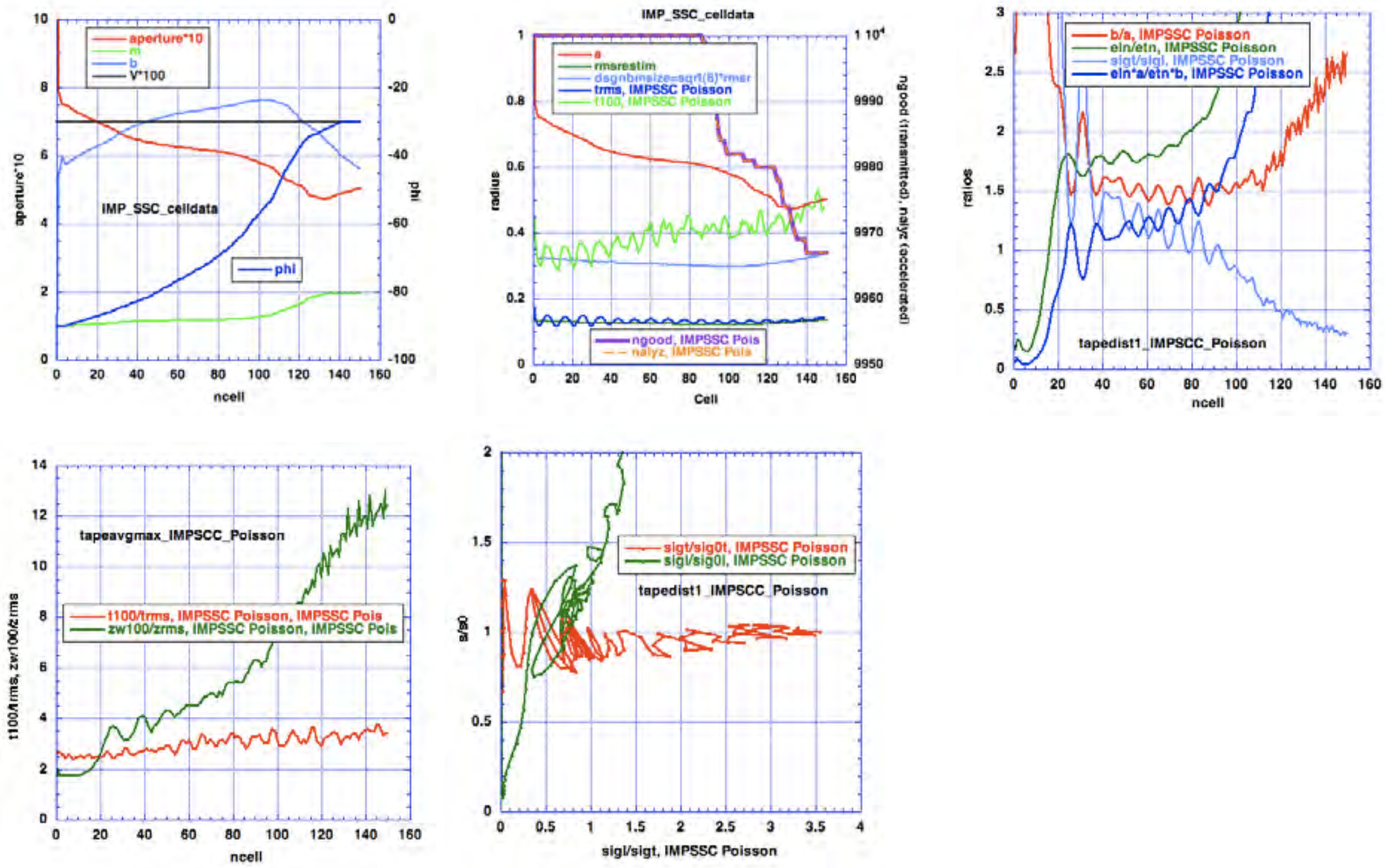

Zero current. No trms redistribution, longitudinal running away (elimit 1 1).

## 22.8.13 J-Parc orig RFQ

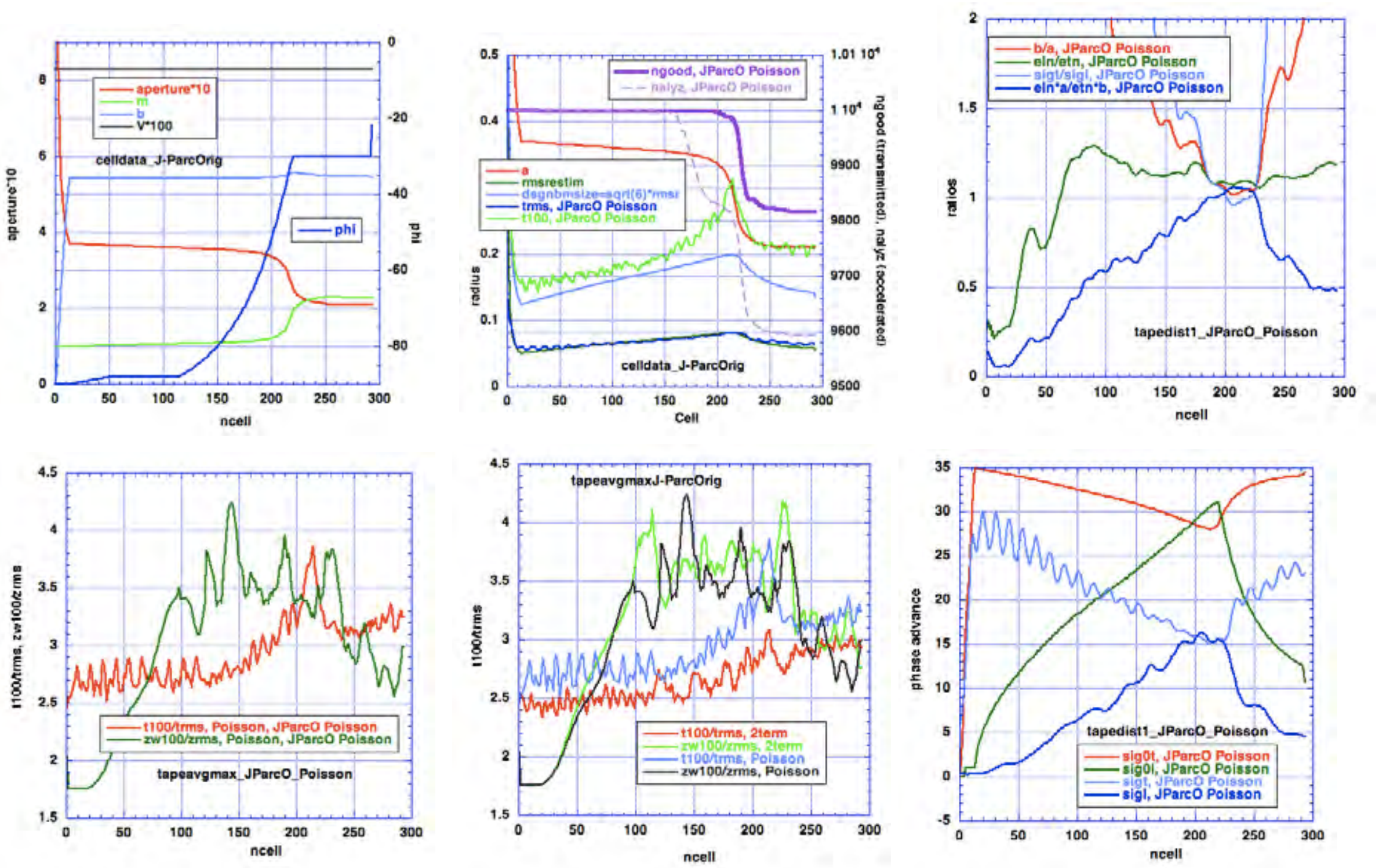

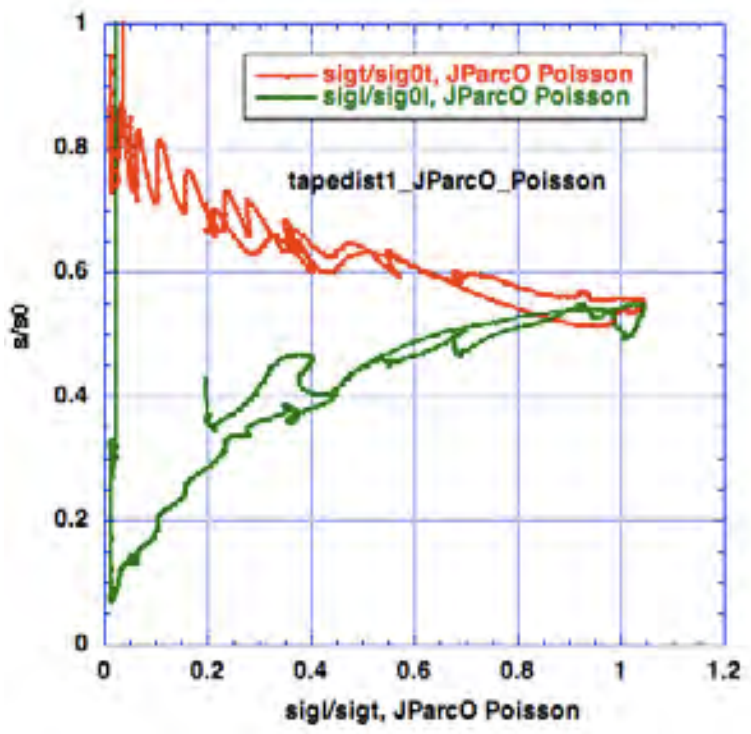

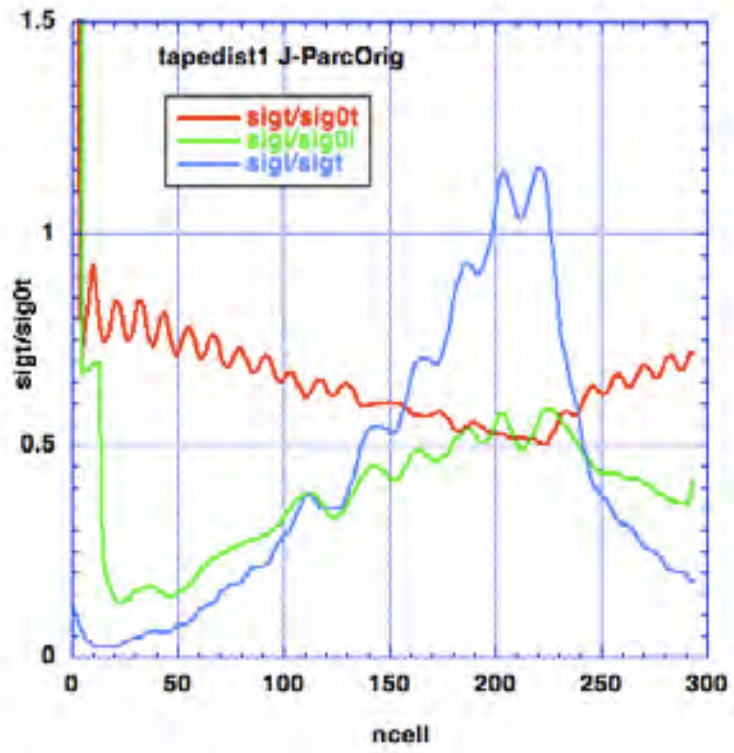

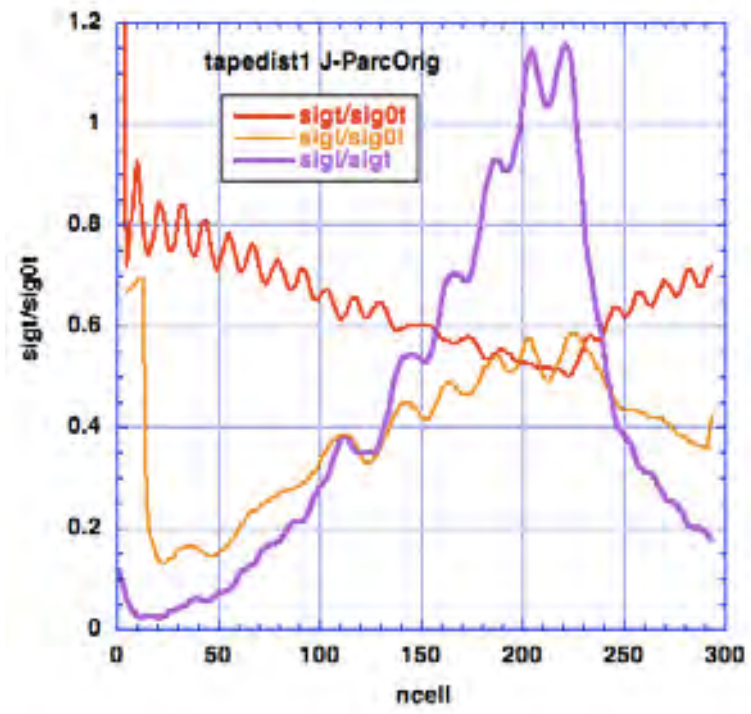

Reached EP near end scraping section, very clear evidence of 1:1 as approaching. Sig's show the gradual approach, and short period with sig0l>sig0t – BEST EVIDENCE. Then immediate intentional (but not realized) scraping, constant V, m, aper declining. No trms redistribution.

## 22.8.14 SNS

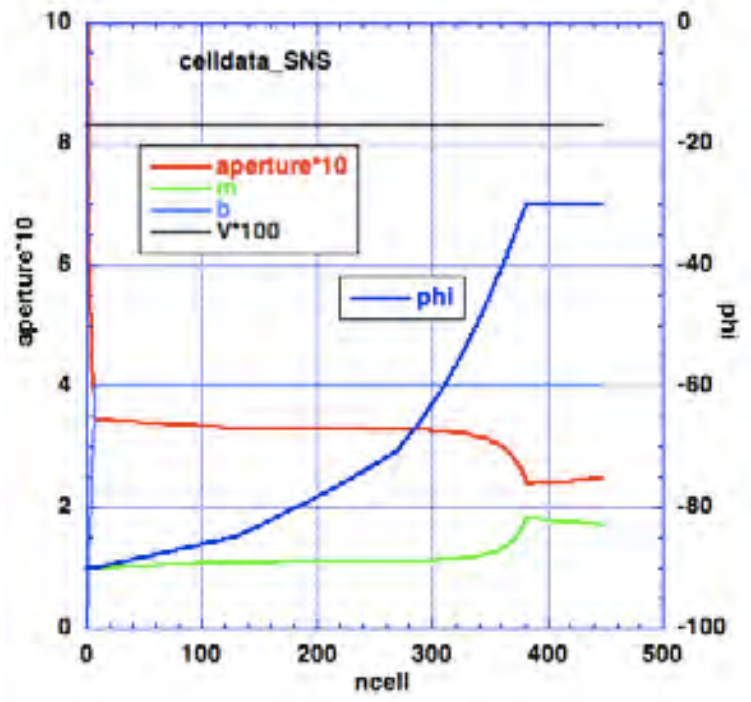

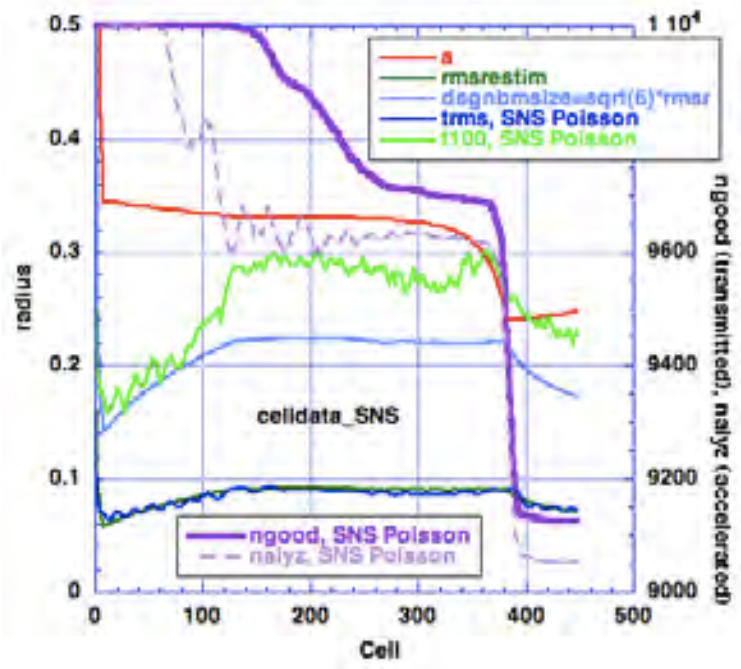

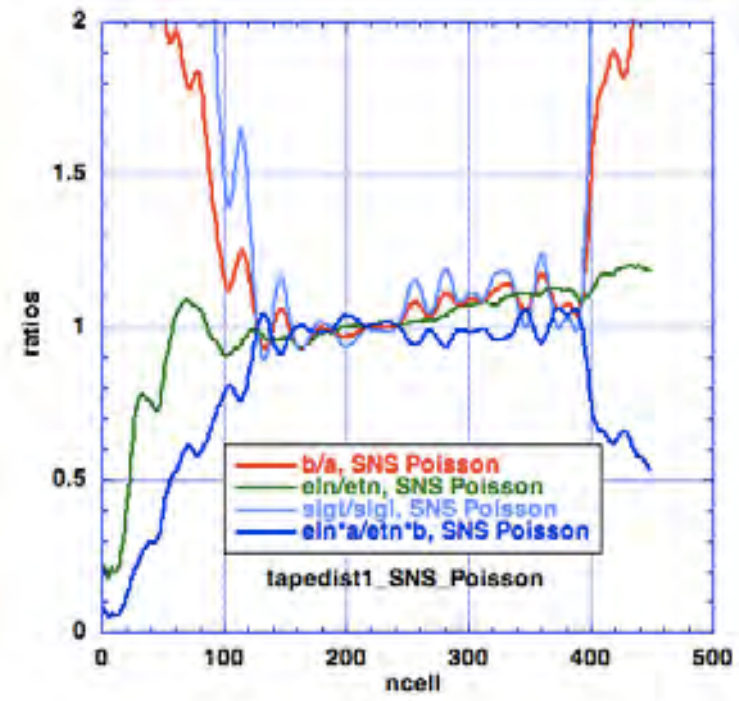

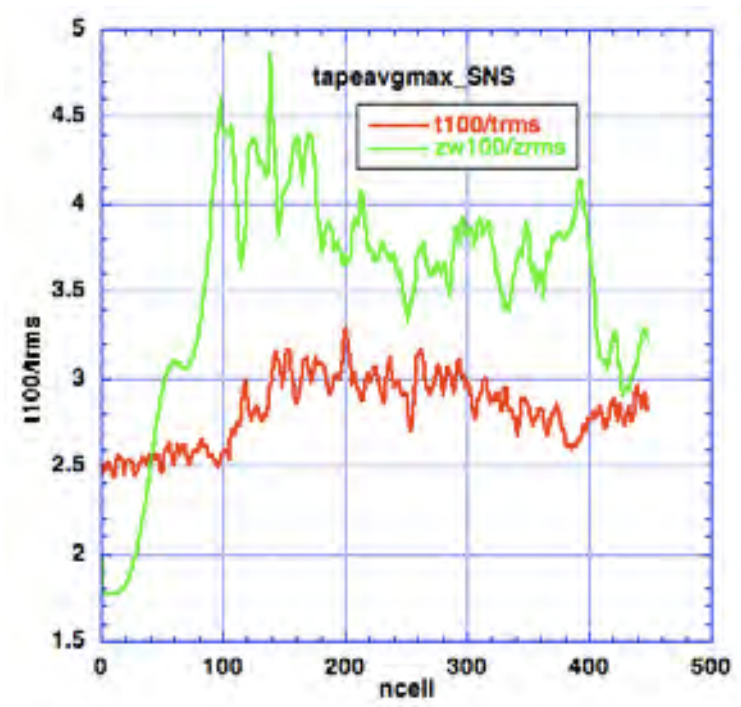

Looks like resonance ~cell100-150. Not EP at cell100, but EP at cell150. Probably continual longit loss up to ~cell 400. Scraping from ~cell200-380, especially at cell 380. On purpose?- flat parameters V and m, slight opening of aper. Old CURLI LBL design. No trms redistribution.

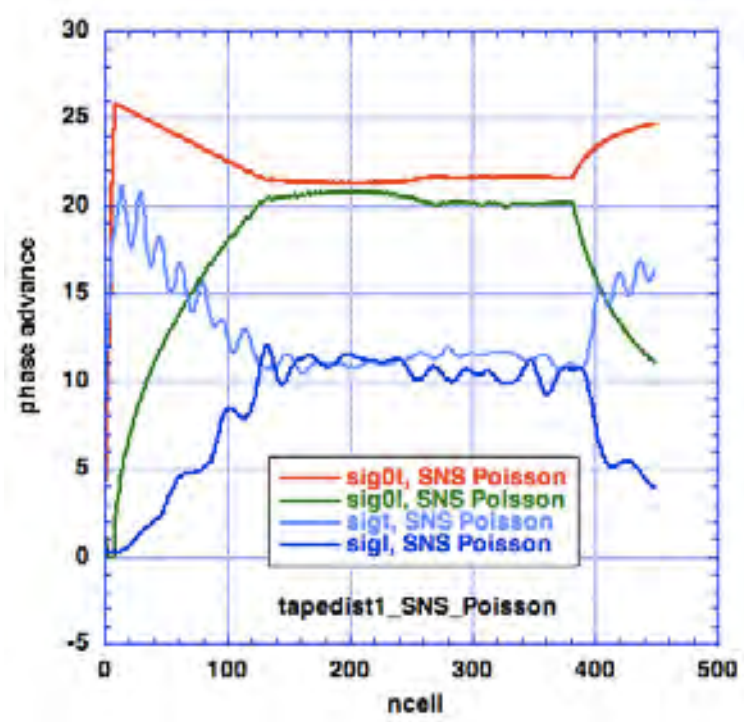

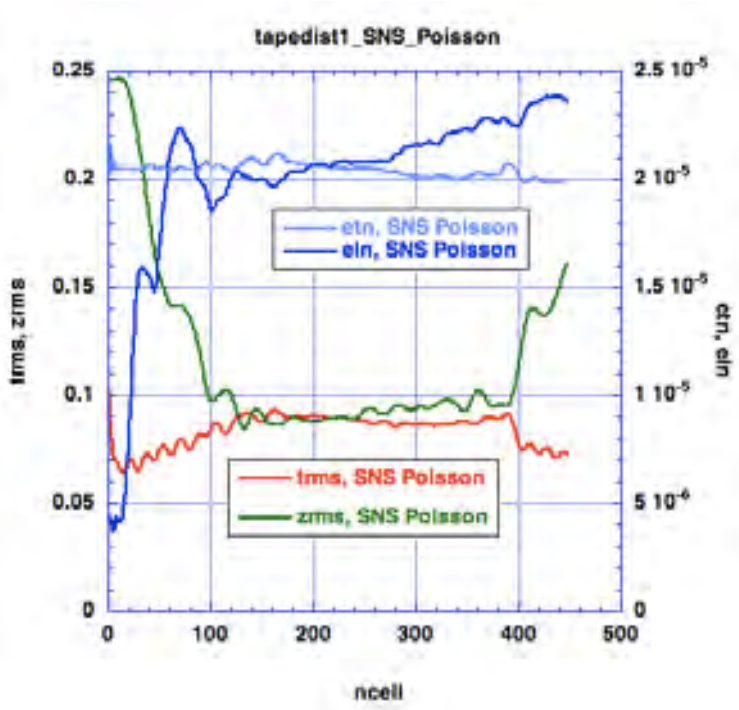

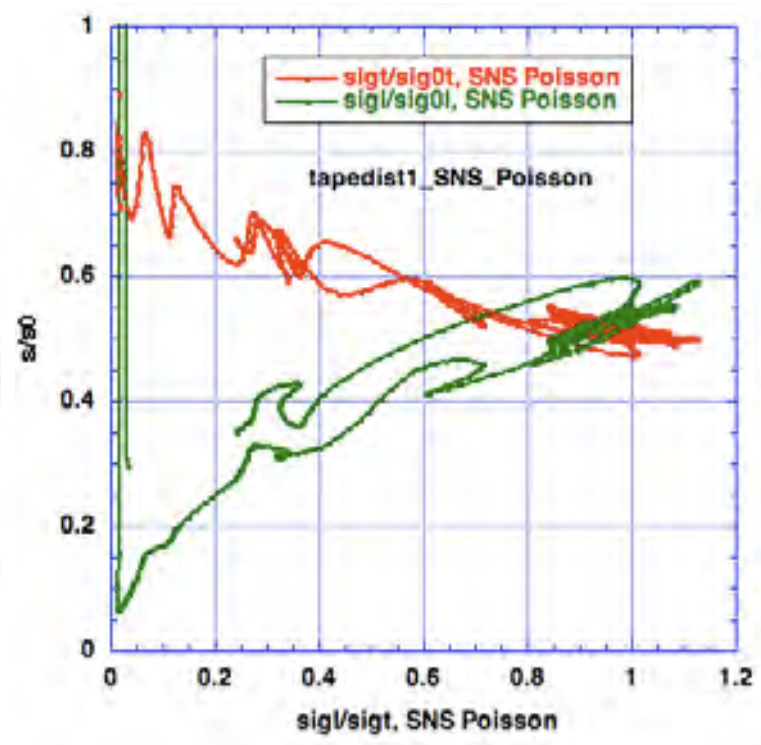

```
sigpl = Sqrt[(sig0l^2 - sigl^2)/ff];
sigpt = Sqrt[(sig0t^2 - sigt^2)/(2/(1 - ff))];
```

| | | | | | "2-term", % | | Poisson, % | | Poisson Img Off, % | |
|---|---|---|---|---|---|---|---|---|---|---|
| **Name** | **Freq** | | **Win/Wout** | **I,mA** | **trm** | **acc** | **trm** | **acc** | **trm** | **acc** |
| C1(43) | 175 | D | .095/5.0 | 130 | 99.5 | 99.4 | 98.3 | 98.2 | 99.0 | 98.7 |
| C2(70) | 175 | D | .095/5.0 | 130 | 70.7 | 70.7 | 66.7 | 66.7 | 84.0 | 84.0 |
| C3(25) | 175 | D | .095/5.0 | 130 | 81.7 | 53.9 | 79.5 | 56.9 | 78.4 | 54.6 |
| AltCDR | 175 | D | .095/5.0 | 130 | 99.0 | 98.7 | 98.4 | 98.3 | 97.3 | 96.7 |
| APT | 350 | H- | .075/6.7 | 110 | 95.5 | 93.6 | 94.5 | 92.1 | 92.1 | 88.4 |
| ATS | 425 | H | .1/2. | 100 | 99.4 | 49.7 | 93.2 | 74.1 | 92.5 | 70.0 |
| CERN | 203 | H | .09/.75 | 225 | 94.6 | 76.3 | 93.4 | 78.6 | 93.1 | 76.5 |
| C1 | 352 | H | .05/1.0 | 50 | 96.7 | 91.5 | 96.4 | 92.3 | 96.3 | 92.0 |
| C2 | 175 | H | .12/1.0 | 200 | 92.8 | 89.2 | 84.5 | 82.0 | 82.4 | 77.0 |
| C3 | 176 | H | .030/1.5 | 30 | 96.3 | 96.1 | 97.6 | 97.5 | 98.6 | 98.4 |
| CRNL | 269 | H | .05/.6 | 82.5 | 98.8 | 72.9 | 94.1 | 82.9 | 92.4 | 80.2 |
| IFMIF CDR | 175 | D | .095/5. | 130 | 96.0 | 96.0 | 94.5 | 94.5 | | |
| IFMIF E | 175 | D | .1/5. | 130 | 80.8 | 80.6 | 84.3 | 84.1 | 91.2 | 90.9 |
| IFMIF Post CDR | 175 | D | .095/5. | 130 | 95.7 | 95.5 | 95.7 | 95.3 | 97.4 | 95.4 |
| IMP ADS | 163 | H | .035/2.1 | 15 | 99.8 | 97.8 | 99.6 | 97.6 | 99.7 | 97.4 |
| IMPSSC | 54 | U | .89/34. | 0.5 | 85.0 | 25.8 | 99.7 | 99.7 | 99.7 | 98.7 |
| J-P Orig | 324 | H- | .05/3 | 36 | 98.8 | 96.2 | 98.2 | 96.0 | 98.1 | 96.0 |
| SNS | 403 | H- | .065/2.5 | 60 | 96.5 | 95.5 | 91.3 | 90.5 | 91.3 | 90.5 |

## 22.9 Conclusions

There are many "definitions" of what constitutes a "halo" in focusing and accelerating channels. A definition that seems correct is that a halo consists of particles that occupy phase space other than the expected phase space.

In a matched and equipartitioned quadrupolar transverse focusing channel with bunching and acceleration, and with very high transmission and accelerated beam fractions,

- the longitudinal 100% to rms beam size ratio tends to redistribute to an approximately Gaussian distribution.
- the transverse 100% to rms beam size ratio tends to redistribute to an approximately parabolic distribution.

This is then the expected, and desired, behavior. Losses and resonance interactions can cause deviations from the expectations.

If the channel has continuous radial beam losses, the transverse 100% to rms beam size ratio may remain nearly constant at the injected ratio, but the transverse rms beam size may quickly re-equilibrate to a smaller value, e.g. Fig. 22.6, where the transverse rms beam size reduces by factor sqrt(6/7). However, this seems not to be the case in general?????

Longitudinal loss from the bucket, if radial loss is not occurring at same time, appears to cause redistribution to a larger t100/trms ratio.

The form factor, which depends on rms quantities, is amenable to manipulation in the design. The design 100%-to-rms beam size ratios are also amenable to *a priori* design manipulations – little investigation so far.

[eltoc]

# Chapter 23 – RFQ Design & Simulation With Arbitrary Vane Tip Shapes

R.A. Jameson, Y. Iwashita [186]
December 2015

## 23.1. Introduction and Method Chosen

For many years, most RFQs have been built with circular transverse vane tip cross section, and sinusoidal longitudinal modulation, as developed by K. R. Crandall (KRC) [187,188]. RFQ simulation approximated the RFQ potential in a cell by an 8-term multipole expansion based on the use of symmetries, and KRC provided tables of the eight coefficients of these terms over a grid parametrized by Rho = (radius of vane tip curvature), Ls = (cell length)/(average vane tip radius from the axis = r0 = r0rfq), and the vane modulation = emrfq, for interpolation locally. Using such tables afforded fast computation.

Lately, it has become popular to consider trapezoidal vane modulation, because it gives a somewhat higher acceleration rate and therefore a possibly shorter RFQ. Trapezoidal modulation was actually

[186] Y. Iwashita contributed actively and was extremely helpful in this investigation, and generated the final coefficient tables for use in *LINACSrfqDES*. The files for this work are very extensive.

187 K. R. Crandall, "Effects of Vane-Tip Geometry on the Electric Fields in Radio-Frequency Quadruple Linacs", Los Alamos National Laboratory Report LA-9695-MS, UC-28, Issued: April 1983.

188 K. R. Crandall, "PARMTEQ With Image Charges", Los Alamos National Laboratory Report, AT-1-92-213, October 1991.

one of the earliest methods used, because it is easy to machine [189]. Many other possible modulation methods have also been considered. [190,191,192,193,194,195]

A multipole expansion for the fields is very useful, in a very specific and limited way, for *design* purposes, but should *not* be used for simulation. A set of multipole tables must be generated for each specific modulation shape.

Ch. 29 relates the discovery of a new important use, in a different regime - to control bunching and longitudinal emittance in the RFQ - of the concept of controlling the longitudinal field in detail within the focusing period

### 23.1.1 Simulation

An "**element of linear accelerators**" (a basic principle) is that the physics of the actual linac should be represented as accurately as possible, given computer limitations, in the simulation program. *And also in design program*s. The reason is that design parameter searches will find optima based on the design program conditions. Thus another "**element**" is that the correlation of a linac design program to its simulation program must be close, *in order to find correct optima*.

*It is emphasized* that the *LINACSrfqSIM* (simulation) program accomplishes this by using an accurate Poisson solution for both the full external and space-charge fields, and is equipped to include sinusoidal, trapezoidal and 2-term vanes with circular vane tip cross-section. No multipole coefficients are used in the simulation. Incorporation of other vane shapes would be straightforward.

---

189 P. Junior, et. al., "An RFQ Concept Using Circular Rods", LINAC1981.

190 "IHEP Experience on Creation and Operation of RFQs', O.K. Belyaev, O.V. Ershov, I.G. Maltsev, V.B. Stepanov, S.A. Strekalovskikh, V.A. Teplyakov, A.V.Zherebtsov, IHEP, Protvino, Russia, LINAC2000

191 "Full 3D Modeling of a Radio-Frequency Quadrupole", LINAC2010,B. Mustapha #, A. A. Kolomiets and P. N. Ostroumov Used Trapezoidal profile for vanes with higher modulation at end of RFQ. Fields from 3D model: "Once validated, the 3D approach could be safely used for the trapezoidal design option for which an analytical potential formula is yet to be developed.

192 "The New Option of Front End of Ion Linac", A. Kovalenko, JINR, Dubna, Russia. A. Kolomiets, ITEP, Moscow, Russia, LINAC2012 ...the sinusoidal modulation of the RFQ vanes is used only at the gentle buncher part whereas the trapezoidal one is applied at the main acceleration part with constant modulation coefficient. The trapezoidal modulation provides 15-20% higher RFQ accelerating efficiency at the same modulation...

193 "RFQ With Improved Energy Gain", A.S. Plastun, NRNU (MEPhI), Moscow, Russia, A.A. Kolomiets, ITEP, Moscow, Russia, LINAC2012 Improved(compared to trapezoidal) profile vanes

194 "Full Three-Dimensional Approach to the Design and Simulation of a Radio-Frequency Quadrupole", B. Mustapha, A.A. Kolomiets, P.N.Ostrumov, PRSTAB 16, 120101 (2013)

195 "Preliminary Design and Simulation of a 162.5 MHz High-Intensity Proton RFQ for an Accelerator Driven system", Xiao Chen et. al., Chinese Physics C, CPC(HEP & NP), 2011, 35(11): 1053-1058, Vol. 35, No. 11, Nov., 2011

The 8-term approximation is inadequate for modern simulation. This was one of the problems with the older codes *PARI/PARMTEQM*, which nonetheless were used for many constructed RFQs. There the philosophy was to design with the 2-term potential description, to simulate with an 8-term potential description, and to make software fitting adjustments of the aperture and/or modulation to "correct" the average transverse and/or longitudinal focusing strength to the 2-term values. The optimum found by a "one-parameter-search" was different for the 2-term and 8-term simulations either uncorrected or corrected. An additional problem was that it has been known for a long time that the 8-term potential description, while quite accurate on-axis or out to the rms aperture and even out toward the vane tip, is highly inaccurate in the region of highest interest - near the vane tips, where particles are actually lost. *As-built RFQs have performed well, but not as predicted by the simulations.* They probably did agree with the way they were built – but trying to compare as-built to an incorrect model will fail.

Some extant programs are using a Poisson solution for the external fields, but with a space charge method that is fundamentally limited to cylindrical symmetry. This cannot result in an accurate simulation.

*23.1.1.1 Trapezoidal vs. Sinusoidal vane modulation*

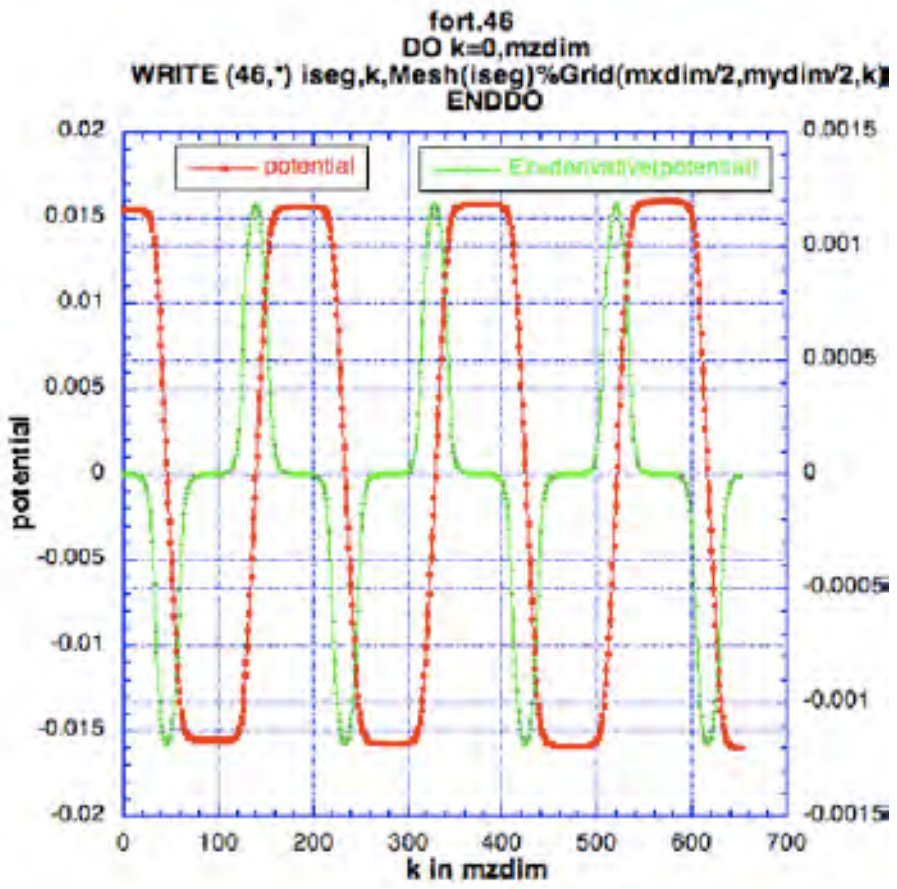

Fig. 23.1. Longitudinal potential and electric field on-axis for trapezoidal cells, from *LINACSrfqSIM*.

*LINACSrfq* characterizes the trapezoidal modulation by the percent of the cell length that is occupied by the linear slope between the minimum aperture (a) and the maximum aperture a*modulation = am. Investigation indicates that trapezoidal has advantage when slopepercent < ~50%.

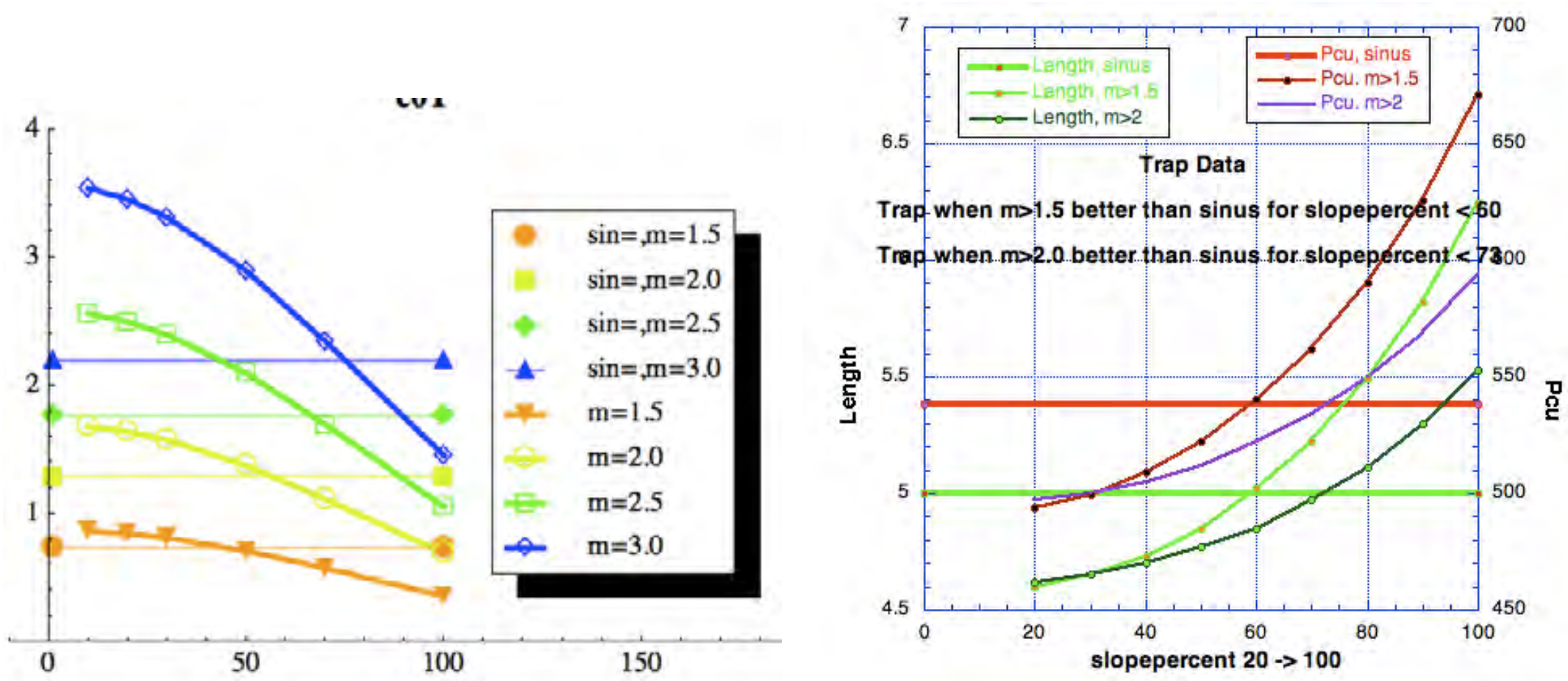

Fig. 23.2. e0t (left), length and copper power (right) vs. slopepercent, for different sinusoidal and trapezoidal vane types and modulation levels.

As the slope percent becomes smaller, the slope becomes steeper and the transit time factor T = t[rfq] becomes higher. The average accelerating term e0 remains the same, but acceleration rate e0T becomes higher, reducing the length and the rf power Pcu requirement. Pcu is computed using a shunt impedance typical for sinusoidal modulation, but the shunt impedance for the trapezoidal shape should also be higher, so Pcu should be even less. tfac and lfac are somewhat higher than for the sinusoidal case. ***It is necessary to simulate with full Poisson to determine what the performance actually is.***

### 23.1.2 Design

A driving principle for development of the *LINACSrfqDES* design code has been to include every parameter of the linac, and also the relevant space charge physics – this makes *LINACS* unique. Preliminary design for any linac can be done using the smooth approximation of the dynamics, and this method can be made very fast, so that optimization searches are practical. The *LINACSrfqDES* design program uses tables of the average cell voltage (named e0rfq) and transit-time factor (called trfq) obtained from a Poisson solution for the cell, to closely approximate the actual RFQ longitudinal fields in the actual device. The average transverse parameter for the smooth approximation is obtained from the standard multipole expansion, but with a crucial difference in application. Repeating from above, the multipole expansion is not suitable for simulation because it is inaccurate near the vane, but the multiple expansion is quite accurate within the aperture and near the axis. For design, all the quantities in the equations are rms; the design beam rms size should be well within the aperture, so applying the multipole expansion at the rms beam size will be accurate.

Then there is a further advantage – bringing the design closer to the actual device (and Poisson simulation) then allows a backward approximation to the 2-term description, and fast 2-term simulation can give useful approximate preliminary design results without a shift to a different optimum. The design procedure with 8-term potential description represents a "middle ground" - it is an less accurate approximation of the Poisson solution (which is the best possible (can be described as "exact")) - and the 2-term mode is a further approximation the 8-term mode. Thus the design code "binds together" the slow Poisson and the fast 2-term simulation modes.

Any actual vane shape can be set up in the Poisson simulation program. Design can be done precisely by computing the Poisson solution for a cell at each local step as the design proceeds – but then the design program takes a long time to run [196], and is not useful for preliminary design and searching for optima. Therefore, a fast method for design with arbitrary vane shapes is needed.

The smooth approximation RFQ design requires three quantities: the average voltage e0 = e0rfq across the cell, the transit time factor T = trfq, or the actual product e0T, and the quantity B which defines the average transverse focusing available with zero vane modulation. In the program, the rf defocusing and the space charge defocusing are subtracted from the B term, and the zero-current phase advance and the phase advance with space charge can be calculated.

e0, T, e0T are found by integrating Ez on the axis. It is not necessary to compute the longitudinal multipoles using KRC Eqns(18-20) – i.e., for design, it is not necessary to worry about how many and which multipoles may be necessary to represent the smooth approximation longitudinal quantities.

It appears that there is no way to avoid having to make a Poisson solution to get the B term. A long search for an alternative direct method failed, because, unwittingly, underlying all the attempts was implicitly the condition that the charge density on the vane surface was constant. Although the potential is constant on the conductor surface, the charge density is not, so a Poisson solution is required.

---

[196] Especially when all the considerations outlined below are taken into account, such as using the middle cell of a longer repeating cell sequence, and finer transverse mesh resolution.

The only method known for finding the smooth approximation quantity B is to use the multipole expansion. The terminology for B is = V*xx*(l/r0rfq)^2, where xx=(sum of transverse a0m multipole terms). The definition will be xx = (a01+a03) - that is, use only the lowest frequency components, in keeping with the idea of the smooth approximation. The a01 and a03 terms (and possibly more a0n terms) then have to be found from the Poisson potential mesh.

For fast execution of the design program , it has been decided to again use the KRC table method with interpolation, for the desired vane shape, i.e., at present, for sinusoidal and trapezoidal longitudinal modulation and circular transverse vane tip cross-section.

Three parameters were needed to describe sinusoidal modulation - the ratio Rho/r0 (Rho is the transverse vane tip assuming circular machining, r0 is the local average aperture across the cell), ls = cell length/r0, and the local value of the cell modulation. For trapezoidal vanes, a fourth parameter would be needed - the fraction of the cell that is slanted, called the "gap". However, for complicated shapes, ~30 coefficients are required for precision of ~1%. Although this is a large number, [197] indicates that even more coefficients can be useful, to include aspects like errors. Another fitting method for trapezoidal vanes is given in [198]. For more complicated vane shapes, piecewise expansions would be necessary. One of the problems for this study was to determine an appropriate number of coefficients for describing trapezoidal vane modulation.

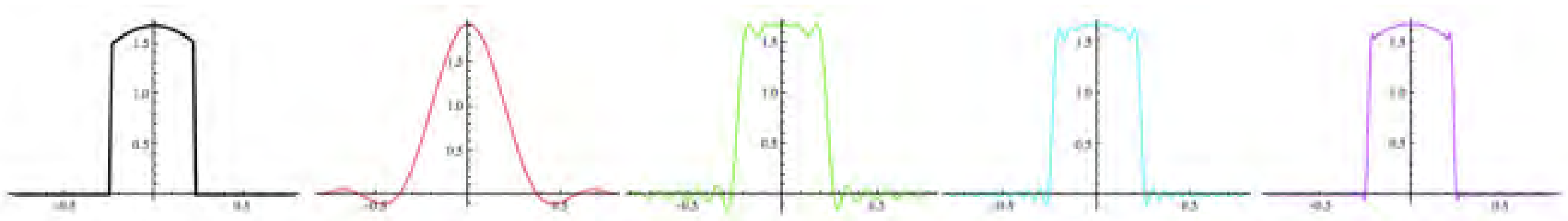

Fig. 23.3. Ez*Cos[kz] in trapezoidal cell, Fit with 5, 30, 49, 101 coefficients.

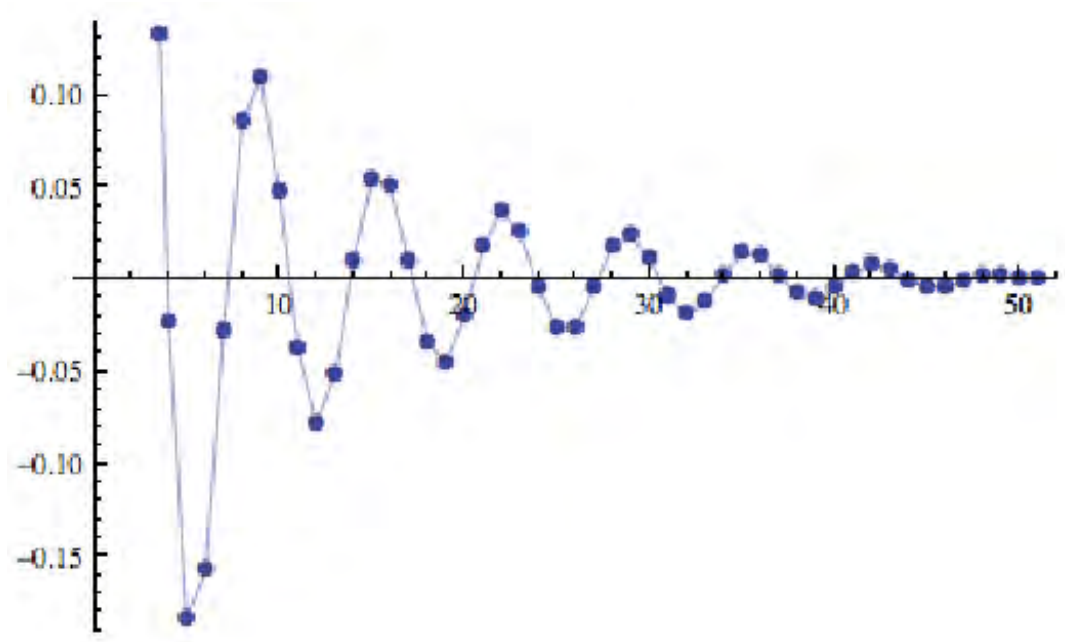

Fig. 23.4. Magnitude of coefficients for 50 coefficient fit.

This is close to the design method used in *LINACSrfqDES* (design) for a long time. Now it can be extended clearly to arbitrary vane shapes.

Three methods for computing the coefficients were compared, and a *Tosca/Mathematica* method chosen for both technical and practical reasons. Design tables have been made for aarfq, trfq, a01 and a03, for sinusoidal modulation, and for trapezoidal modulation with "gaps" of 30%, 40% and 50% of

197 "RFQ BEAM DYNAMICS MODEL DEVELOPMENT", A. Todd & M. Reusch, Proceedings of the Linear Accelerator Conference 1990, Albuquerque, New Mexico, USA. (Unfortunately, source code is lost (A. Todd - private communication.)

198 "PARAMETRIZATION FOR TRAPEZOIDAL VANES", Z. Ahmed, V.T. Nimje, R.C. Sethi, Nucl. Phys. Div., B.A.R.C., Bombay-400 085, India, 1985. Courtesy of V.T. Nimje.

[eltoc]

the cell length having the sloped part of the trapezoid. The tables are accurate, and the design procedure is very fast.

Sec. 23.2 describes the Kapchinsky-Teplyakov-KRC multipole expansion. Development of three methods for finding the coefficients for RFQ cells is outlined in Sec. 23.3, and testing for circular vane tip transverse cross-section and sinusoidal longitudinal modulation in Sec. 23.4. Sec. 23.5 gives conclusions.

## 23.2 Kapchinsky-Teplyakov-KRC Multipole Coefficients

The potential in an RFQ cell is expanded, and multipole coefficients for terms appropriate to the quadrupolar symmetry expressed as follows, using the notation of KRC Eqns. (18-20):

$$U(r,\theta,z) = \frac{V}{2}\left(\sum_{m=1}^{\infty} A_{om} r^{2m} \cos 2m\theta + \sum_{m=0}^{\infty}\sum_{n=1}^{\infty} A_{nm} I_{2m}(nkr) \cos 2m\theta \cos nkz\right) . \qquad (18)$$

$$A_{om} = \frac{16}{\pi^2 V \rho^{2m}} \int_0^{\pi/2}\int_0^{\pi/2} U(\rho,\theta,kz) \cos 2m\theta \, d\theta \, d(kz) , \qquad (19)$$

$$A_{nm} = \frac{32}{\pi^2 V \, I_{2m}(nk\rho)} \int_0^{\pi/2}\int_0^{\pi/2} U(\rho,\theta,kz) \cos 2m\theta \cos nkz \, d\theta \, d(kz) . \qquad (20)$$

where ρ is a radius somewhat less that the minimum aperture (arfq) of the RFQ cell.

The coefficients are extracted using the properties of trigonometric orthogonalities. Multiply both sides of (18) by the cosine function of the term to be separated and integrate over an appropriate range; integration of mixed cosine terms will be zero, and the desired coefficient will be isolated. Including (vrfq/2) so that factor 2 will be included, KRC Eqs.(19-20) will have the form, for coefficients a01, a03, a10, a12, a21, a23, a30, a32:
:

(vrfq/2)*Cos[2*m*theta]*Cos[n*kz]

```
(* a01,a03,a10,a12,qa21,a23,a30,a32 *)
Integrate[(vrfq / 2) * {a01, a03, a10, a12, a21, a23, a30, a32} *
  (Cos[{2, 6, 0, 4, 2, 6, 0, 4} * theta] * Cos[{0, 0, 1, 1, 2, 2, 3, 3} * kz]) ^2,
 {kz, 0, Pi / 2}, {theta, 0, Pi / 2}]
```

$$\left\{\frac{1}{16} a01\,\pi^2\,vrfq, \frac{1}{16} a03\,\pi^2\,vrfq, \frac{1}{16} a10\,\pi^2\,vrfq, \frac{1}{32} a12\,\pi^2\,vrfq, \frac{1}{32} a21\,\pi^2\,vrfq, \frac{1}{32} a23\,\pi^2\,vrfq, \frac{1}{16} a30\,\pi^2\,vrfq, \frac{1}{32} a32\,\pi^2\,vrfq\right\}$$

The coefficient 32 for the an0 terms in (20) is wrong, but later tables used 16.

KRC normalizes by r0, see Eqns.{4-9) and the note between Eqns. (4) and (5), so the bare r in (18) and the bare ρ in (19) are written as r/r0 and ρ/r0, while r and ρ in the Bessel functions and in U(ρ,θ,kz) are not divided by r0.

The rho/r0 parameters in the standard KRC tables are:

```
data RhooR0s/    0.6d0,0.75d0,0.89d0,1.d0,1.15d0,1.3d0/
```

The Ls parameters = (cell length)/r0 in the standard KRC tables are:

data xLs/0.75d0,1.d0,1.3d0,1.8d0,2.4d0,3.1d0,4.1d0,5.4d0,6.9d0,8.7d0,10.9d0,13.5d0,16.5d0,20.d0/

The vane modulation parameters in the standard KRC tables are:

data ems/1.00000000d0,1.05d0,1.08d0,1.12d0,1.16d0,1.21d0,1.26d0,1.32d0,1.38d0,1.45d0,
1.52d0,1.6d0,1.68d0,1.77d0,1.86d0,1.96d0,2.06d0,2.17d0,2.28d0,2.40d0,2.50d0,2.65d0,2.75d0,
2.92d0,3.d0,3.21d0/

## 23.3 Methods for Determining RFQ Cell Multipole Coefficients

### 23.3.1 KRC Program

KRC used his program CHARGE3D to calculate the charge distribution on the surface of a vane cell parametrized onto a 2D grid, assuming a bicubic spline function of the grid variables with known spline knot locations, and finding the charge density values at the knots with a least-squares procedure. The details (grid resolution, starting conditions for the least-squares solution, source code, etc.) are not available; it is stated that the procedure appears extremely well conditioned and gives very satisfactory results. The charge density is then integrated to find the potential distribution, and further integration over the radius ρ cylinder.

The KRC tables are left unfilled for the shortest cells with higher modulations, which would have practical problems of higher peak surface field and more difficult machining. Also, however, as discussed below, the integrals become very difficult in this region.

### 23.3.2 *Tosca/Mathematica* Program

Prof. Y. Iwashita at Institute for Chemical Research, Kyoto University is developing ideas for alternative vane shapes, and uses the *Tosca* program. *Tosca* has a MODELER, which generates the mesh. MODELLER cannot handle tabular functions; only trigonometric and exponential related functions are implemented. Therefore the functions must be approximately synthesized from the Fourier series. This however brings the problem full circle, as the selection and number of terms must be determined. Sinusoidal modulation is straightforward, but other longitudinally smooth modulations must be handled in this way. Trapezoidal modulations can be synthesized by connecting three vane parts of flat-slope-flat.

*Tosca* finds the potential distribution using the Finite Element Method (FEM), with linear and/or quadratic potential shape functions. (The primary variable in *Tosca* is the potential.) While BEM (Boundary Element Method) needs to solve a large dense matrix, which has a size of the square of the number of variables (surface mesh points), FEM needs to solve a sparse matrix, which requires less memory space. When the mesh size is reduced to half, the number of variables becomes 4 times and the matrix (memory) size inflates 16 times for BEM. Although the number of variables becomes 8 times for such a case in FEM, the matrix to be solved is a sparse matrix, and the required memory size increases just proportional to the number of variables (8 times). KRC seems to have used a BEM like method.

A fine grid resolution was required to get good behavior and accuracy of the integrals in the bore. The base mesh size is set at 0.7r0 for Lc>5r0, 0.5r0 for Lc>1.5r0 and 0.35r0 for the rest of the cases. The mesh size on the vane is reduced to 1/3 of the base mesh size and that of bore is reduced to 1/4. The mesh distance on the axis is further reduced to 1/7 of the base mesh size. The largest file size of the result is about 120MB for Lc=20r0 case, while the smallest one is about 21MB for 0.75r0 case.

Since the calculation took a very long time, trials were conducted to optimize the mesh densities:

- Just prism elements everywhere instead of tetrahedra gave a very bad a03, even though it was quadratic and the estimated %error by the solver was not bad. The a03 value was worse than those given by linear elements( 00LM and 00L). Since this prism-only configuration is close to rectangular grids, such a configuration may be a reason for the small a03. Since the prism only mesh often failed to be generated, only one example could be tested.

• The quadratic mesh is better than linear ones in accuracy and cpu time.(as expected)

• The MOSAIC meshing (quadratic prisms for axial cylinder region and tetrahedra for elsewhere) gives shorter cpu time and memory size with comparable a03 value.

• A round outer boundary seems better than block boundary for shorter cpu time and memory size with comparable a03 value. (see the attached png file)

The final runs for sinusoidal and trapezoidal modulations with 30%, 40%, 50% gap factors ('Iwashita Final Harmonics Multipole Coeffs' Folder', .xls files) were performed with:

• The numerical integrations were performed with precision and accuracy set at 1e-6 to limit the slwcon (slow convergence) messages.

• The inter-vane capacitances are the total charge on the surface area listed on the next next column.

• The charge is the surface integral of D on the area.

• Columns for aarfq, trfq, A01, A03, A10, A12, A21, A23, A30, A32, A41, A43, A50, A52, vane capacitance, epk enhancement factor (not useful for the trapezoidal modulations).

(The inter-vane capacitance C[Farad] is found by using Q=CV formula, where C=Q for the V=1 case. The C is not divided by Lc in the table to get the scaled [Farad/m].)

• Column headings for settings:

nodes: the number of nodes of elements in *Tosca*
edges: the number of edges of elements in *Tosca*
q.tetra: the number of quadratic tetrahedrons
q.prism: the number of quadratic prisms
q.pyramids: the number of quadratic pyramids
elements: the number of elements
equs: the number of elements in *Tosca*
non-zeros: the number of non zero elements in the sparse matrix
RMS: Estimated error in D Field RMS value in %
wRMS: Estimated error in D Field weight RMS value in %
cpu: Total time in second

• One full set of the modulation series out of the four types took about 63GB of disc space.

• Since the mesh sizes for long cells are set coarse for the less disc space and the shorter computing time, the errors became somewhat large. If the modeller had an option to generate long elements in z-direction, it could be mitigated. Maximum wRMS (Sin 1.2%, T50 4.8%, T40 5.4%, T30 6.3%).

The *Tosca* post processor POST generates the field map on the 0.8aperture = 0.8r0*2/(1+emrfq) cylinder, and the Ez field on the axis. The Integrated D (=epsilon0 E) and maximum electric field Emax on a vane surface is also generated by POST. The Integrated D corresponds to the total charge Q on a vane.

A cylinder radius of 80% of the aperture was chosen to avoid mesh granularity, especially in the vicinity of the boundary. A cylinder with radius close to the aperture radius would hit some of the peripheral elements and might exhibit less accuracy, although the quadratic elements should return better results than linear elements.

The potential on the cylinder is directly interpolated from the potential values on the vertices.

Mathematica reads these files and makes the integrations for harmonic analysis and longitudinal quantities. Mathematica also reads the D and Emax files and summarizes them in the table file.

The results for a test at rho/r0=0.75, for cylinders at radius 0.80, 0.90 and 0.95 times aperture, gave essentially the same result, indicating that the *Tosca* procedure has good smoothing at the lower harmonics. The test at 0.95*aperture radius did, however, take a very long integration time for the Mathematica harmonic analysis, with many complaints on slow convergence (for 1e-6 convergence level); indicating the above problem with jaggy or less accurate data or points in elements with some corners inside the vane. 90% analysis did not show any complaint from Mathematica, while 95% did.

Runs for a full set of coefficients for a given longitudinal modulation required ~ 15 hours.

### 23.3.3 ***LINACSrfq Program***

*LINACSrfq* solves the Poisson equation for the potential with a finite difference multigrid method on a 3D rectangular grid, which may be anisotropic, with adjusted grid points at the vane boundary to assure exact boundary conditions.

The parameters that must be set are: the extent of the transverse grid, the number of cells covered by the longitudinal grid, the resolution (number of grid points) of the transverse and longitudinal grids, the radius of the cylinder for the multipole computation, the number of theta steps for the cylinder integration, and in the Poisson multigrid solver the number of W cycles and the number of smoothing levels in the cycle.

*Normal LINACSrfq full Poisson solution (uses no multipole method)*

The settings have been carefully checked for normal RFQ simulations, where it was found that the transmission and accelerated beam fraction did not change significantly with more grid resolution than a relatively coarse grid, and for more than five multigrid cycles (W=5) and four smoothing levels.

*Use of the LINACSrfq routine for computing multipole coefficients*

However, if the *LINACSrfq* routine were to be used for computing multipole coefficients for some new design idea, the parameter selection for generating the design tables was not straight-forward, especially for small Ls<1.5. The primary reason is that the integrations on the cylinder in Eqns. (19, 20) are oscillatory – a notoriously difficult class of integrals.

The mesh end conditions of the RFQ cell mesh are periodic, but there is a tradeoff between noise from the Poisson solution at the ends of the Poisson mesh, the mesh resolution, and the running time. This tradeoff has been carefully evaluated (Sec. 19.3); it requires that a middle cell of a longer cell sequence be used. For usual simulation, groups of 15 cell are sequentially generated and the middle 13 cells are used. For generating the multipole tables, a mesh of at least 7 cells is necessary, from which the middle cell is used. This condition was actually necessary only for the shortest cells, xLs<~1.5. Longitudinal grid resolutions from (cell length)/80 to (cell length)/200 were tested.

The transverse mesh extent was studied from 2*r0 to 3*r0, where r0 is the average cell aperture. With zero modulation, the vane tip is at r0; for modulation (emrfq) = 3, the vane tip aperture (arfq) = 2r0/(emrfq+1)=r0/2. Tests were made with transverse vane grid resolution from r0/40 to r0/100. The coefficients were close for Ls>1. For an 80x80x80 grid, the results were well-conditioned, but blowup appeared for Ls<1 with very fine (e.g., 200x200x200) transverse grid resolution. Further tests showed that the Ls<1 problem could be resolved with more multigrid W cycles, as discussed below. A final generation of tables with this method should use a varying transverse mesh extent of ~1.1*(arfq*emrfq).

Integration over the radius and length on the cylinder, with radius (smallrho) "slightly less than the minimum cell aperture", was tested. The number of steps along the radius (0 -> π /2) was chosen as 90; fewer steps gave poor results for a03 and other higher order multipoles; little change was noted at 180 steps. The number of longitudinal steps over kz, (0 –> π/2), was chosen as the number of longitudinal Poisson mesh steps.

An 80x80x80 grid and cylinder radii from 0.80-0.95*aperture and gave nearly same results at for the seven coefficients except for a03 . The a03 variation indicated that both a finer mesh and more smoothing of the higher frequency components of the potential were necessary. A 200x200x200 mesh and increasing the number of W-cycles to W=10 essentially eliminated the a03 variation with cylinder radius.

Runs for a full set of coefficients required 17 hours for an 80x80x80 cell mesh and 5 W-cycles. Runs for a 200x200x200 mesh become practically too long.

## 23.4. Comparison of Methods – Sinusoidal Modulation

Iwash refers to *Tosca*/*Mathematica* method implemented by Iwashita.

RAJ denotes *LINACS* method implemented by Jameson

RAJbigmesh parameters:
XDIMDSGN=R0RFQ*3.0D0, 200X200X200, NUMTHETASTEPS=90, SMALLRHO=0.95D0*ARFQ, W=5

RAJ80 parameters:
XDIMDSGN=R0RFQ*2.0D0, 80X80X80, NUMTHETASTEPS=90, SMALLRHO=0.80D0*ARFQ, W-5

**a01**
**diff betw Iwashita, RAJ <±1%**
percent diffIwashRAJbigmesh -> {-0.66%, -0.38%}
percent diffIwashRAJ80 -> {-0.90%, 0.3%}
**diff from KRC ~ -2%, 1.2%**
percent diffIwashKRC -> {-1.5%, 1.2%}
percent diffKRCRAJ80 -> {-2%, 1.2%}
percent diffKRCRAJbigmesh -> {-1.7%, 1%}

**a03**
magnitude over ~useful range of Ls and m: 0.0 -> 0.1
diffIwashRAJbigmesh ® {-0.002, 0.005} **-> ~2-5%**
diffIwashKRC ® {-0.002, 0.005} **-> ~ -2% to 5%**

**a10**
magnitude over ~useful range of Ls and m: 0.0 -> 0.9
diffIwashRAJbigmesh, ® {-.0, .015} **-> ~2%**
diffIwashKRC -> {-.0, .006} **-> ~ 1.5%**

**a12**
magnitude over ~useful range of Ls and m : 0.0 -> 30000
diffIwashRAJbigmesh -> {-50, 800} **-> ~ 3%**
diffIwashKRC -> {-5000, 100} **-> ~ -17%**

**a21**
magnitude over ~useful range of Ls and m : 0.0 -> -6 (minus 6)
KRC and Iwashita qualitatively same, but RAJs different from those
diffIwashRAJbigmesh -> {-0.0, 3} **-> ~50%**
diffIwashKRC -> {-0.65, 0.1} **-> ~ -11% to 2%**

**a23**
magnitude over ~useful range of Ls and m : 0.0 -> --5*10^6
diffIwashRAJbigmesh -> {-30000, 65000} **-> ~1%**
diffIwashKRC -> {-40000, 60000} **-> ~ 1%**

**a30**
magnitude over ~useful range of Ls and m : 0.0 -> --0.4
diffIwashRAJbigmesh -> {-.0, .017} **-> ~4%**
diffIwashKRC -> {-.001, .0} **-> ~ 0%**

**a32**
magnitude over ~useful range of Ls and m : 0.0 -> - 400
diffIwashRAJbigmesh -> {-1, 8} **-> ~2%**
diffIwashKRC -> {-1, 40} **-> ~10%**

## 23.5. Conclusions on Coefficient Generation

The results with 2015-era laptop computers compared to earlier mid-1990's results are qualitatively the same.

The *Tosca/Mathematica* method is chosen for final generation of new tables with circular vane tip profile and both sinusoidal modulation and trapezoidal modulation with 30%, 40%, and 50% gap factors.

The a01 difference between the two new methods is <1%, and ~2% between the new and old methods. The a03 coefficient is important because it is added directly to the a01 component for design work, and indicated differences of up to ~5% overall between the old and new methods; computation of a03 was the most difficult of all coefficients, requiring much testing of code parameters, and finally very fine meshing. This is a characteristic of the notoriously difficult oscillatory integrals. The other coefficients show some variation; they would probably not be used for design, and the effect of these higher multipoles is less important even if used for simulation so the differences are probably negligible.

Based on the comparison studies, the old method appears to have had less grid resolution and probably also less smoothing that the Tosca method or the *LINACS* method with very fine grid and many multigrid smoothing cycles.

## 23.6. Test of a Design Model

The RFQ design is based on the envelope equations (Part 2.a.). A synchronous particle is launched at the input of the simulation, and tracked through the RFQ to observe its behavior with respect to the design synchronous phase and energy at each cell. In a proper time-based code, there is no true synchronous particle, so the diagnostic particle may depart from synchronism depending on space charge, etc.

With sinusoidal vane modulation, the diagnostic particle remains synchronized with the sinusoidal cell geometry, and both its phase and energy remain near the design values.

The actual energy gain e0Tcos(phi) (of each particle) is the effective criterion. With trapezoidal or other vanes, a particle's phase will depart strongly from the design synchronous phase (which still applies to the rf), but its energy gain will remain close to the design value. This is then the test of whether the design model is accurate with respect to the full Poisson simulation. (See Ch. 23.7 for 2-term vane modulation.)

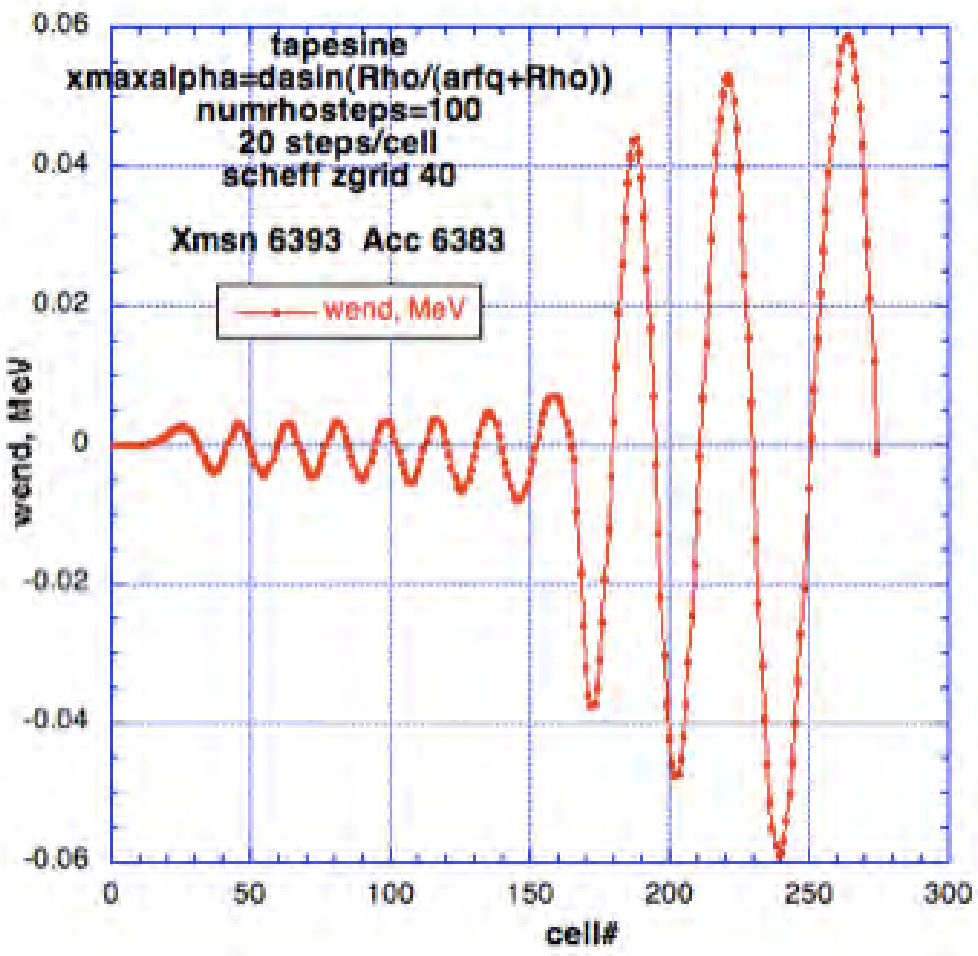

Fig. 23.5. Test of design model accuracy with respect to the full Poisson simulation.

The RFQ has sinusoidal vanes up to modulation = 1.5, then trapezoidal with slopepercent = 50.

wend = (diagnostic particle energy - design energy), MeV. Final energy is 3.2 Mev. The deviation from design synchronous energy is shown in Fig. 23.5: with the maximum deviation of ~0.06 MeV, (0.06/3.2)*100 % = ~2 %, so it is seen that the design model is accurate.

The accuracy for 2-term simulation is not as accurate - wend reaches -2% for this example, but the transmission and accelerated beam fractions are still representative (Fig.23.6):

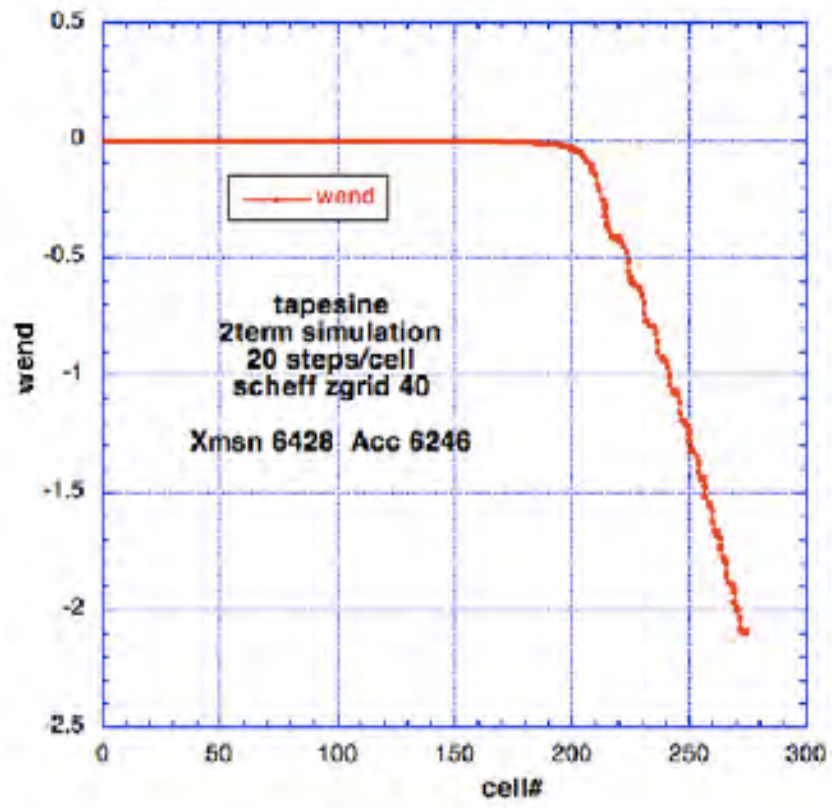

Fig. 23.6. Test of 2-term simulation accuracy.

## 23.7. 2-term Longitudinal Modulation

In Spring 2020, another RFQ project came along with M. Okamura. The average vane radius, r0, and vane voltage, V, were to be held constant – advantageous for cavity tuning, disadvantageous for performance but it was possible to meet the specification. Based on a stated, and published, JAEA/J-Parc preference for 2-term longitudinal vane modulation, it was decided to proceed with the 2-term profile, which was then added to the *LINACS* options.

The vanes part of this RFQ was "typical", in terms of its layout and parameters, a "typical" shaper length and other features, and therefore could be included in a "typical" class including RFQs for widely varying applications such as the IFMIF CDR, heavy ion RFQs designed for multiple ion species, many smaller RFQs for small neutron sources, etc.

The upgraded multipole tables discussed above were used in the design, with the circular transverse vane tip profile. Overall for this short ~160 cell RFQ, the sinusoidal profile has slightly higher acceleration rate – 160 cells – vs 162 cells for the 2-term profile which has weaker initial acceleration. Important to note is that the 4 permutations of 2-term or sinusoidal design method or simulation method (set by simply changing code switches) gave nearly the same transmission, accelerated beam fraction and other characteristics.

In 2021, work on the JAEA ADS project RFQ design with Bruce Yee-Rendon pushed the work on 2-term vane modulation further.

As usual, the next project presents a specification that has new features. The JAEA-ADS spec builds on J-Parc RFQ experience; for high availability, a very low KP limit ≤ 1.2 is required; the vane voltage is to be constant at this KP limit; and equipartitioned (EP) dynamics should be used in the acceleration section. The output longitudinal emittance at 2.5 MeV is to be very small, for smoother (or equal) phase advance into the following superconducting (SC) linac, so the requested EP ratios =1.125. The longitudinal vane modulation would be the stated JAEA/J-Parc preference for 2-term longitudinal vane modulation.

Sinusoidal longitudinal vane modulation with circular vane tips has been the standard method since Crandall's original report "Effects of Vane-Tip Geometry on the Electric Fields in Radio-Frequency Quadruple Linacs", K.R. Crandall, LA-9695-MS, April 1983, so our preliminary design studies carried both sinusoidal and 2-term longitudinal vane modulation in parallel.

Crandall mentions that pure 2-term vane geometry has two undesirable effects (1) the peak surface field

> is drastically increased, and (2) machining of the vane tip is more difficult. To remedy this situation, one could consider keeping the transverse radius of curvature constant throughout a cell at the expense of introducing higher order terms in the potential function. A benefit is that the intervane capacitance is almost constant along the entire vane length. In addition, one could use a sinusoidal modulation on the vane tip rather than the modulation specified by the two-term potential function, which has higher acceleration rate but also higher multipole components.

Crandall warned us, but the tools for real checking were not available until the full (and accurate) Poisson simulation method was in place, ~ early 2012. Until somewhat coincidentally in 2020, everyone (me included) stayed in the sinusoidal modulation box, because "higher acceleration rate" and "easier to machine", and because good designs were found *with Poisson simulation*. Essentially everyone is still using inaccurate (8-term multipole) simulation – which not only gives inaccurate fields, which was then avoided by Poisson simulation – but also significantly smoothed fields [199]. We can now use the Poisson simulation to carefully compare the multipole and image effects of 2-term vs. sinusoidal modulated vanes. Anticipated, is that there should not be much fundamental difference, but perhaps practical preferences.

As the design should be EP, the synchronous phase phis and the modulation must not saturate. Operating beam current was specified as 20mA, so the design was as usual done for somewhat more , 25mA, as this procedure usually is better than design at the desired operating current. Although 20 mA is typically not large, the low KP factor design indicated tune depression ~0.6.

Figs. 23.7 & 23.8 show the preliminary candidate designs for 2-term modulation or sinusoidal vane modulation. All parameters and rules except the type of vane modulation are exactly the same. The results are very nearly the same, with a minor length advantage for sine vane modulation.

Space charge mixing will play a major role in washing out multipole and image effects.

**Comparing 2term and sine vane modulation, as was concluded also in the 2020 study, any difference in performance, or that might be due to multipole or image content, is small, not a basis for the choice of 2term modulation.**

**It is no longer considered more difficult to machine essentially any profile, so that is not a consideration for the type of vane modulation. The acceleration rate is less for 2term modulation but this is a minor disadvantage, and numerous slight design adjustments are available that have stronger effect on the RFQ length.**

## Conclusion: staying with sinusoidal vane modulation is recommended.

[199] It would seem unnecessary to point out, but it is necessary to note that a 2-term potential theoretical representation in a simulation is a very drastic approximation.

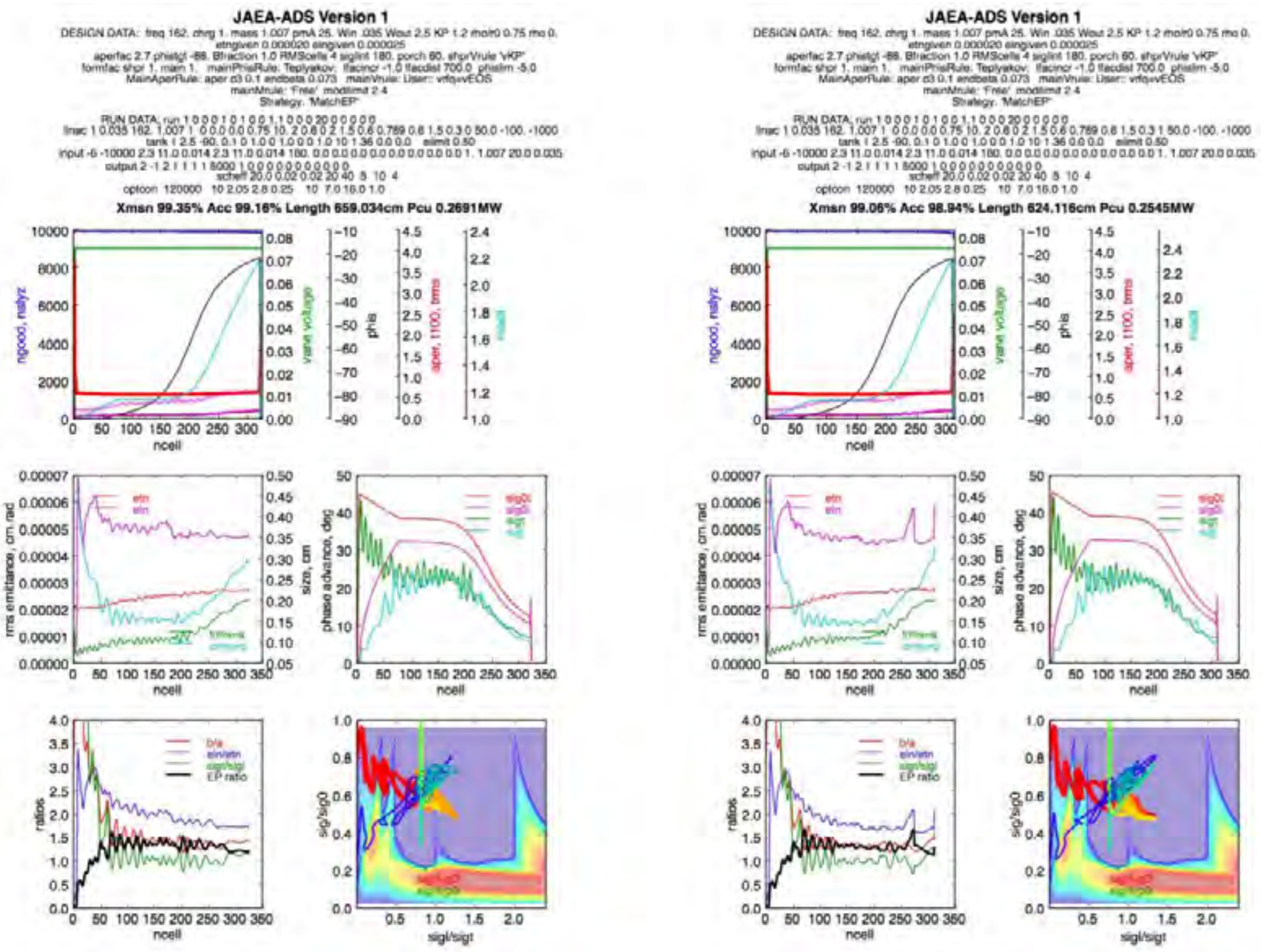

Fig. 23.7. The *LINACS* run graphic shows the important beam performance aspects. 10000 particles simulated. Left – 2term vane modulation, Right - Sine vane modulation. Dark blue line in upper left graph illustrates the transmission and accelerated beam. The longitudinal emittance growth, and corresponding transverse emittance growth are seen in each middle left graph, and tune depressions at lower right.

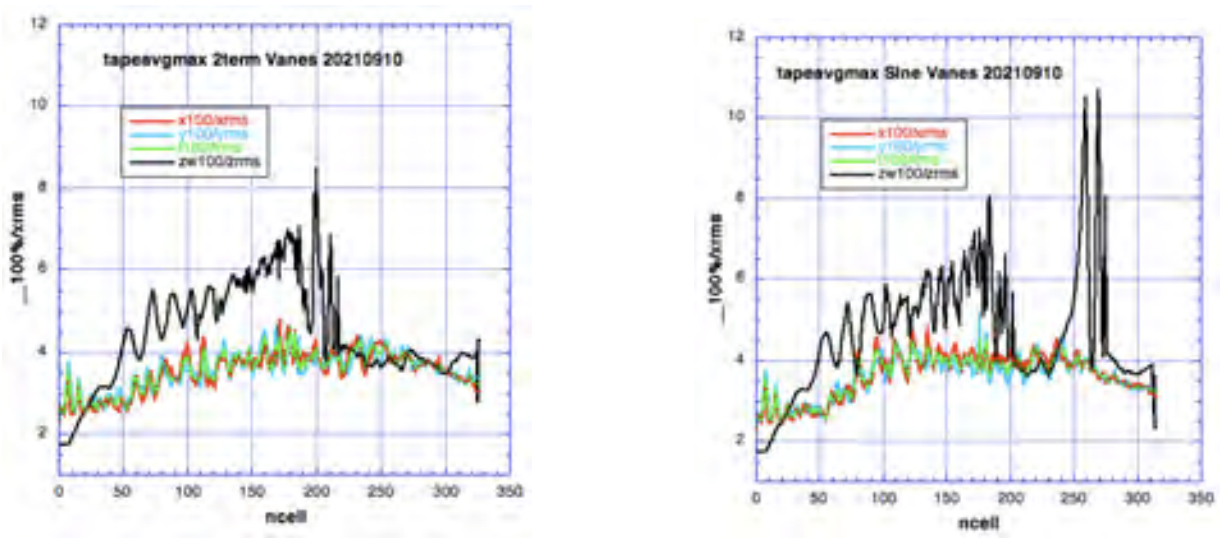

Fig.23.8 Ratios of 100%/rms emittances, which would reveal the effect of multipoles. Left for 2term vane modulation, right for sin vane modulation.

In this particular case, neither candidate meets the EP requirements; both have large longitudinal rms emittance growth starting in the first few cells, with subsequent transverse emittance growth indicating an EP process.

The effort to reduce the longitudinal emittance and meet the requirement for EP ratios=1.25 consumed very much time, because there were several conflicting factors, and no further conventional parameter search improvement upon this preliminary base design has been found. Fortunately, the following SC linac was checked and found insensitive to larger longitudinal emittance.

Further improvement in this case, and others, requires more physics to be invoked in the design, as mentioned first in Sec. 1.4.5 and covered for the severe deficiency found in this project in Ch. 28.

Also, later an unconventional shaper design was discovered that gave improved longitudinal emittance and EP conditions – see Ch.28.

The work highlighted several interacting complicated aspects, which are now reviewed.

### 23.7.1 2-term Vane Modulation Analysis

#### *23.7.1.1 Strength of Acceleration*

Crandall's original proposition that sinusoidal vane modulation has a higher acceleration rate has been widely promulgated, which is true but require some clarification.

Fig. 23.9 shows the strength of the acceleration multipoles A10 for 2-term and aarfqSin for sinusoidal modulations:

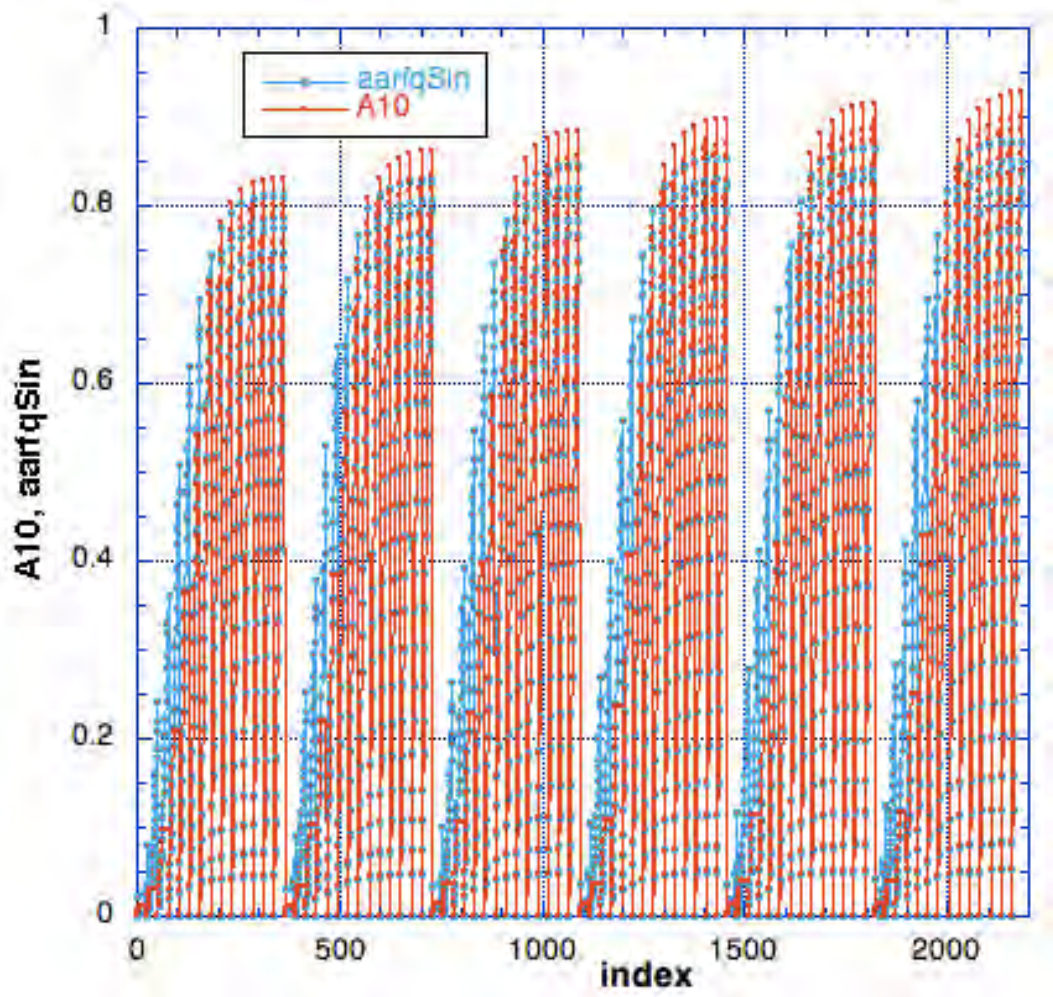

Fig. 23.9. A10 is the multipole acceleration coefficient for 2-term vane modulation with circular vane tip cross-section. aarfqSin is the acceleration coefficient, used in calculating the average voltage gradient across an RFQ cell e0rfq=2.*aarfq*vrfq/(betalam), for sinusoidal vane modulation and circular vane tip cross-section. Both were computed by Y. Iwashita in very high precision using TOSCA. The figure shows the multipole tables – the six groups are for different ratios of rho/r0={0.6, 0.75, 0.89, 1.0, 1.15, 1.3}. Each of these groups has 13 values of (cell length)/r0 from left to right {0.75,1.0, 1.3, 1.8, 2.4, 3.1, 4.1, 5.4, 6.9, 8.7, 10.9, 13.5, 16.5, 20.}. Each of those has 26 values of modulation from bottom to top {1.0, 1.05, 1.08, 1.12, 1.16, 1.21, 1.26, 1.32, 1.38, 1.45, 1.52, 1.6, 1.68, 1.77, 1.86, 1.96, 2.06, 2.17, 2.28, 2.40, 2.50, 2.65, 2.75, 2.92, 3.0, 3.21}. At larger (cell length)/r0 and larger modulations, A10 is actually somewhat higher than aarfqSin, which does not agree with the usual statement that Sin has higher aceleration rate (or with earlier less accurately computed tables). See also Sec. 23.7.1.6.1.

***But important is that aarfqSin is larger for (cell length)/r0 <~6.***

Fig. 23.10, for a different, "typical" 0.035-2.5 MeV RFQ, indicates that "typical" RFQs might have (cell length)/r0 ratios up to ~5, and compares various acceleration term coefficients. "aarfq a10" and "aarfq a10+a30" are for 2-term longitudinal vane modulation; "bigA" is the theoretical A for the ideal 2-term potential vane description. "aarfq, full sin run" is the acceleration term for sinusoidal longitudinal vane modulation.

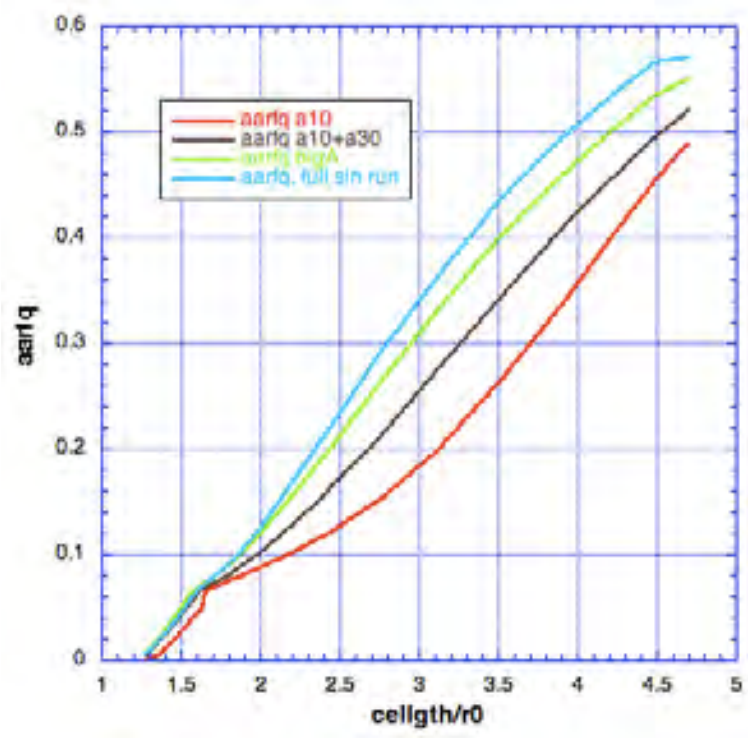

Fig. 23.10. Comparison of various acceleration term coefficients vs. (cell length)/r0 ratio through the RFQ, for a "typical" 0.035-2.5 MeV RFQ

Sinusoidal longitudinal vane modulation can give some RFQ length advantage and therefore less rf power requirement, both saving cost. Trapezoidal modulation can give some further advantage. But the important points are:

• Striving for the shortest RFQ physical length has always been a vastly over-rated goal. It has usually been pursued without much (mostly without any) consideration of tradeoffs, in terms of the four pillars: beam physics, engineering, cost, and RAMI (reliability, availability, maintenance, inspectability).

• The length advantage of sinusoidal or trapezoidal modulation is not so large. A fully integrated design procedure with control of all external field and rms beam physics parameters provides a number of equally strong or stronger methods for influencing the length.

• The choice of longitudinal modulation should not be made solely on the basis of higher accelerating gradient, but also considering the side effects which include higher peak surface fields and higher multipole content, machining complication, etc.

*23.7.1.2 Tracking the Beam Bunch in the simulation*

*23.7.1.2.1 Tracking the phase of a quasi-synchronous particle through the 2-term vanes*

As stated in earlier Chapters, there is no true synchronous particle in the time code, but a "quasi-synchronous" particle, that remains on-axis transversely, receives only the accelerating force, and is not affected by space charge, was added for the indexing of simulation run ending and as an important diagnostic since the earliest stages of *LINACS* development. The tracking compared the phase of this particle to the design synchronous phase, and tracks closely for sinusoidal and trapezoidal vane modulation, but goes berserk for 2-term vane modulation, as shown in Fig. 23.11; crossing a cell boundary invokes a 180° phase shift, and it is seen that this is occurring many times (this was first seen in the 2020 design study):

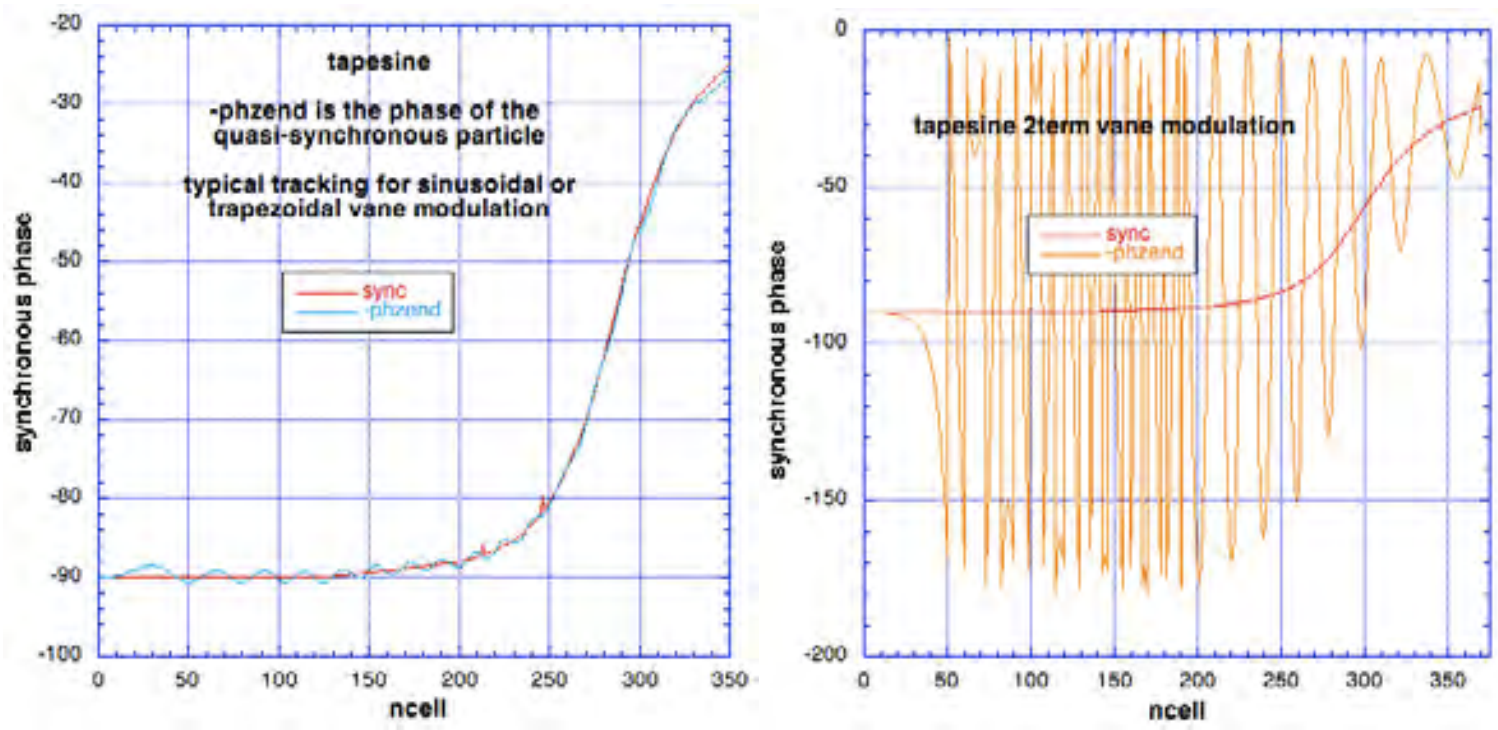

Fig. 23.11. Tracking the phase of the "quasi-synchronous" particle against the design synchronous phase. Left – sinusoidal or trapezoidal vane modulation. Right – 2-term vane modulation. Oscillation is crossing 0° or 180° in region ~cell 50-200.

Tracking using the "quasi-synchronous" particle has been used extensively in *LINACS*, so investigation was necessary.

*23.7.1.2.2 Alternative tracking check*

Another tracking method was added, to use only the time steps to determine a cell boundary (ncelltrack), i.e., only the design progression. Fig. 23.12 compares the positions of the quasi-synchronous particle (znptp) and of the bunch centroid, referenced to the design position (zphis), in units of cell length.

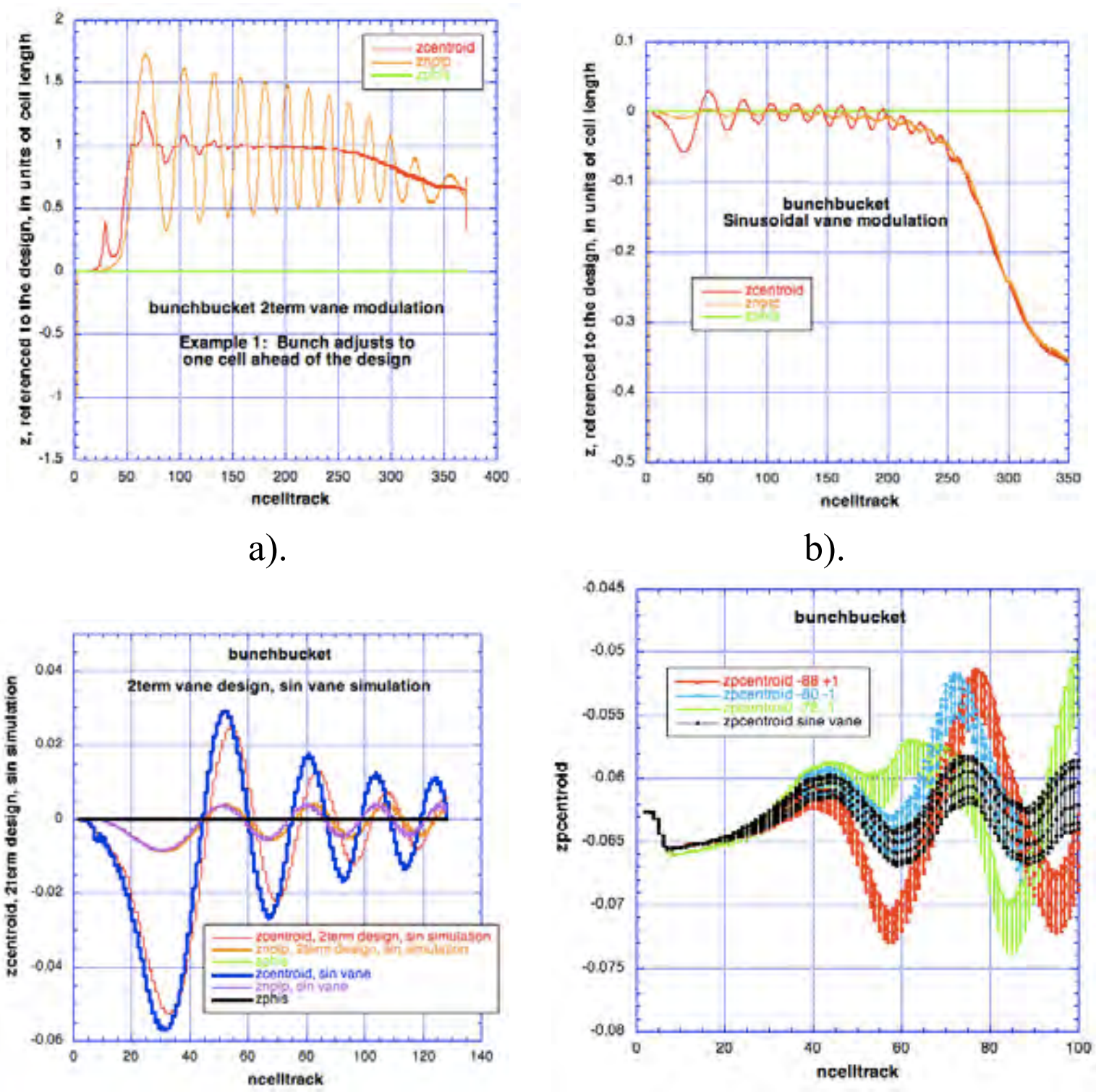

a). b).

c). d).

Fig. 23.12. Tracking referenced to the design z-coordinate of the quasi-synchronous particle. a). for a 2-term vane modulation RFQ. b). for a sinusoidal vane modulation RFQ, that differs from a). only in the choice of modulation for the design and simulation. c). (upper 3) the 2-term vane design simulated with sine modulation; (lower 3) sine vane design and simulation as in b). d). the momentum centroid.

With 2-term vane modulation and phis at end-of-shaper (phisEOS) {-88°, -80° and -78°}, -80° is a critical point: from phis values closely <-80°, the bunch centroid position settled, and then remains, one cell ahead of the design; from values closely >-80° the bunch centroid position settled, and then remains, one cell behind the design. The particles oscillate around the centroid; the quasi-synchronous particle oscillation covers more than one cell length in the region ~cell50-200.

For the sinusoidal vane modulation case, (shown is -88° phisEOS), the bunch centroid position remains around the design position. The momentum centroid is slightly lower that the 2-term phisEOS -80° case near the critical "deciding point ncelltrack~50", but remains centered around the design momentum. At this "deciding point ncelltrack=50", the design modulation for the four cases is 1.043-1.048, very close to the same (the designs use average and rms quantities, and are very nearly the same in the early cells).

If the RFQ of Fig. 23.12, which was designed using 2-term vane modulation, is simulated using sine modulation, the tracking is as shown by c).

Fig. 23.12.d tracks the momentum centroid of the bunch relative to the design momentum, for the four RFQs; the only difference between the designs is the type of vane modulation (and the choice of phis at the end of the shaper (phisEOS)). There is more oscillation for the 2-term vane modulation, but all remain on track. This indicates that the actual *average* strength e0Tcos(phis) of the acceleration field in the Poisson simulation agrees with the design, and the overall picture is exactly as expected – see Ch.23.6.

This diagnostic gives insight into the different characteristic of 2-term vane modulation.

The dc beam entering the RFQ, simulated by a canonical time-coordinate code, has no knowledge of any "cell", synchronous phase" or such longitudinal definition – only when enough modulation introduces a longitudinal structure can the beam synchronize itself to that structure. Why the 2-term vs. the sinusoidal or trapezoidal vane results in a different synchronization requires further consideration:

*23.7.1.3  Analysis of the vane modulation profile*

The sinusoidal vane modulation profile is a simple function, described by one sine Fourier trigonometric coefficient, symmetric across a 2-cell period (the sign of the acceleration term changes from one cell to the next), and symmetric about its average radius (r0).

The trapezoidal vane modulation profile requires very many Fourier trigonometric coefficients; it is applied symmetrically across a 2-cell period with the ramp center at the period center. Also its aperture (a) and maximum aperture (ma) are applied symmetrically about the sinusoidal modulation r0= a(m+1)/2.  Therefore only cosine coefficients are required, and the main a10 coefficient is sufficient to keep the design and simulated performance tracking closely.

A characteristic 2-term vane profile is shown (2 cells) in Fig. 23.14, with its first 10 cosine and sine Fourier trigonometric series coefficients.  First, it requires both cosine and sine Fourier terms to describe, due to its lack of symmetry around its r0 radius and around the cell center, and its complicated shape.  Second, the coefficients have declined to only about half at 10th term, so it is a very non-pure sinusoid and will have higher order effects.  It requires a full Poisson geometrical mesh to simulate – an 8-term potential function, not to mention a 2-term or rms function, could not simulate it.

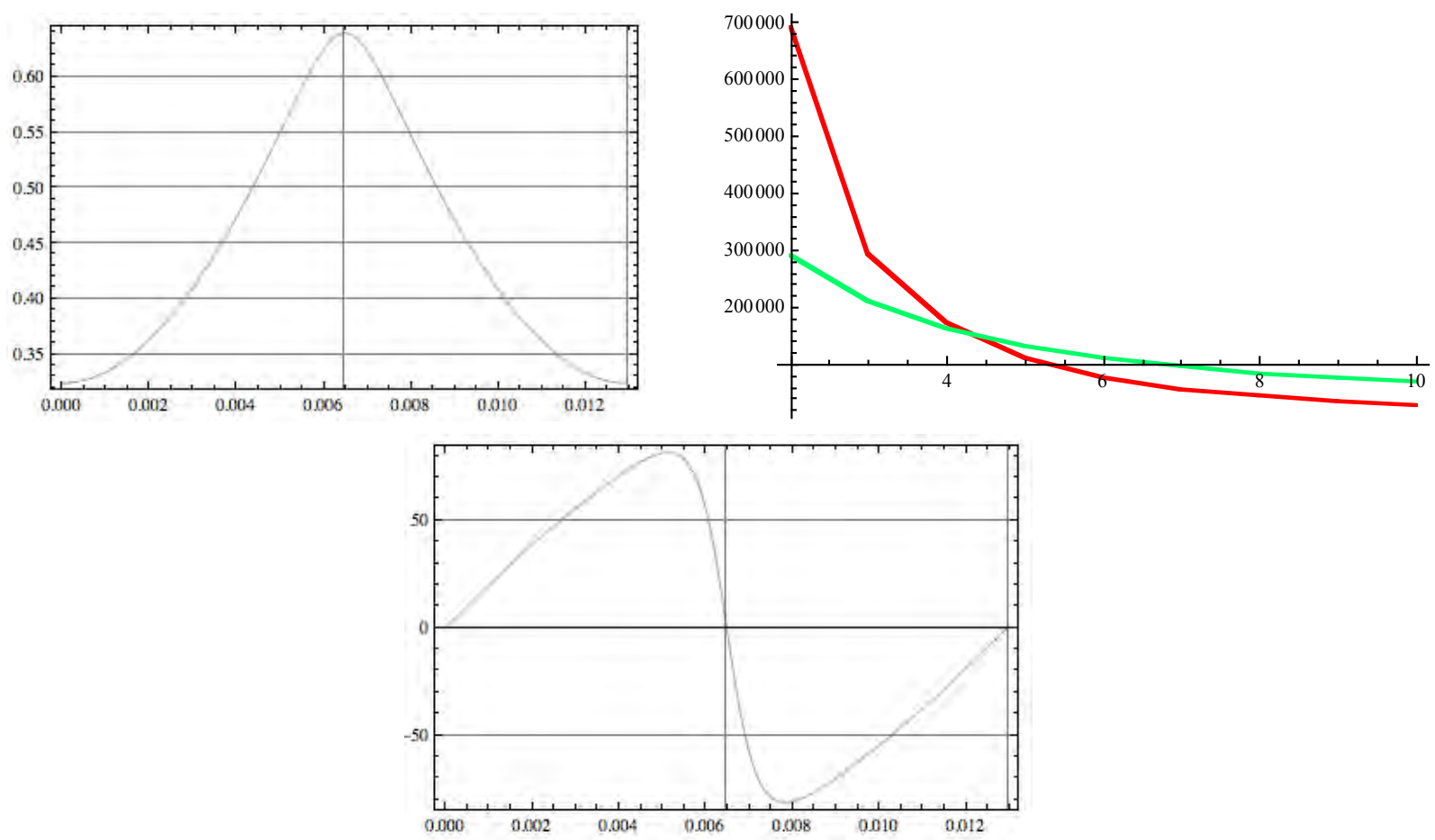

Fig. 23.14.  Characteristic 2-term vane modulation profile;  Fourier cosine coefficients (red), sine coefficients (green);  and longitudinal electric field (Ez) profile (derivative of the modulation profile)

It means that the 2-term vane modulation *design*, even though including the main multipole terms and rms quantities, is not quite as accurate a detailed representation of what will happen in the manufactured RFQ or in an accurate Poisson simulation.  The coupling between design and simulation is not as tight.  A false ordering of the importance of theory, simulation and experiment (the actual importance is in the order experimental, simulation, theory) is a common misconception – a simulation based on a theoretical 2-term potential function is not a simulation based on an actual modulation.

*Recall also from previous sections that e0Tcos(phi) is the effective energy gain criterion, with important difference in the transit-time-factor T, adaptation of phi (the phase of the particle on the rf wave).*

These observations should have made it completely obvious long ago that design, simulation, and vane machining using the simple 2-term formulas is very ill-advised.

*23.7.1.4  Practical decisions*

*23.7.1.4.1 Additional diagnostics are added.*

*23.7.1.4.2 Run tracking registration*

The code has been thoroughly revised and reviewed to use the beam position centroid for tracking.

An uncertainty of the beam centroid tracking at one cell for both sine and 2term vane modulation may occur occasionally and about equally often for either modulation type, and, but even less often, can result in run abort.

*23.7.1.4.3 Run ending registration*

The simulation run is ended when the beam centroid reaches the end of the RFQ tank. The output emittance parameters are needed here. Other measures are also necessary to help insure robust ending.

*23.7.1.4.4 Implications for the design model*

- The simulation transverse cross-section of the vane tip is circular for all three longitudinal vane modulation models.
- The simulation is accurate for all the rms design input parameters presented to it, regardless of their source.
- Therefore, no change was needed in the design model.

*23.7.1.4.5 Particle loss check*

Because of the use of canonical time coordinates, checking for particle loss on the vane surface must be done for each particle individually. This becomes much more time-consuming for 2-term vane modulation. As *LINACS* is very careful to have an exact match between the external field and space charge field physics, using exactly the same Poisson mesh for both, it is possible to use a pointer method for very quickly accessing the vane geometry when checking for particle loss. This method was checked for sine modulation against the direct analytical method for computing the vane surface, and the results agree exactly.

*23.7.1.4.6 "Adjustments" to "make a 2-term Design RFQ"*

An early, but still extant, code affords only very limited *design*, using constant vane voltage and r0, and constant modulation and synchronous phase during the acceleration section after a bunching section. The *simulation* strategy uses the (not accurate enough) 8 multipole coefficients to model the potential, and sinusoidal vane modulation. Limiting the number of coefficients has severe effects as discussed herein, *and also has a smoothing effect, with loss of information about higher order effects.* Then "adjustments" are provided for making "fits" to the modulation and aperture to return the simulated RFQ to theoretical 2-term, both transverse and longitudinal. As indicated by Figs. 23.9&10 and corresponding main quadrupole transverse focusing terms, such adjustments would be large, of order 20% or more. Many have felt that such adjustment was unsatisfactory because it results in loss of oversight of the design. It also has an unrealistic goal.

*LINACS* gives direct control of all parameters and rules, affording many possible design variations, in an accurate Poisson fields environment.

## 23.7.2 Work in the Context of the JAEA-ADS RFQ Specification

*23.7.2.1 Simulation Input Matching*

See Ch. 19 for exhaustive discussion of input matching.

Unusual cases can occur. We explored a different case than that of Fig. 23.7, with an unusually long shaper in an attempt to influence the longitudinal emittance.

It is recommended that the time consuming full simulations to construct an $\alpha,\beta$ matrix be undertaken as soon as preliminary design work begins to indicate a possible candidate design. Each of the other methods will indicate different and/or several possible match points. Experience with more "typical" RFQs (more typical shapers) indicates relatively small areas of points. It is recommended to first use the fastest §19.4.8 Backward Simulation Matching Method, then a fast 2term simulation transmission matrix, and then to occasionally check as design proceeds using the (much) slower §19.4.7 Design Envelope Matching Method of the slowest or the very slow transmission match method §19.4.2.1.

Fig. 23.15 shows the α,β matrix obtained. At least four reasonably good match areas, at widely separated {α,β} pairs are seen. This variation in good match areas is very large, "non-typical". The lower {α,β} match points are in the region indicated by the EOS backward {α=0.536, β=4.438} and pari design matches; they indicated strong and fast oscillations, especially in the shaper.[200] The best performance match is far different at {α=2.3, β=11}.

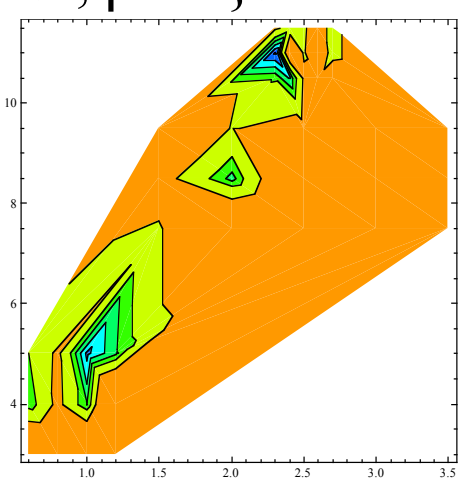

Fig. 23.15. Accelerated beam fraction transverse {α,β} transmission matrix for a very long shaper, sinusoidal vane modulation RFQ. Full Poisson simulation. Contours are higher in the direction orange to blue, with highest peak red.

*23.7.2.3 Multipole and Image Effects*

Both 2-term and sinusoidal longitudinal modulations use circular vane tips. **The Poisson simulation accounts accurately for *all* multipoles *and also image effects (including images on images)*.** 8-term multipole tables are available for both modulation types, and parmteqM contains a clever approximation for image effects. Another look at relative effects, Fig. 23.16, shows plots (from the K.R. Crandall original report) of the field enhancement factor and the a01, a03, a12, a10 and a30 multipoles for 2term modulation (Tabl VI) and sinusoidal modulation (Tabl VIII).

Space charge mixing practically washes out these differences.

field enhancement:

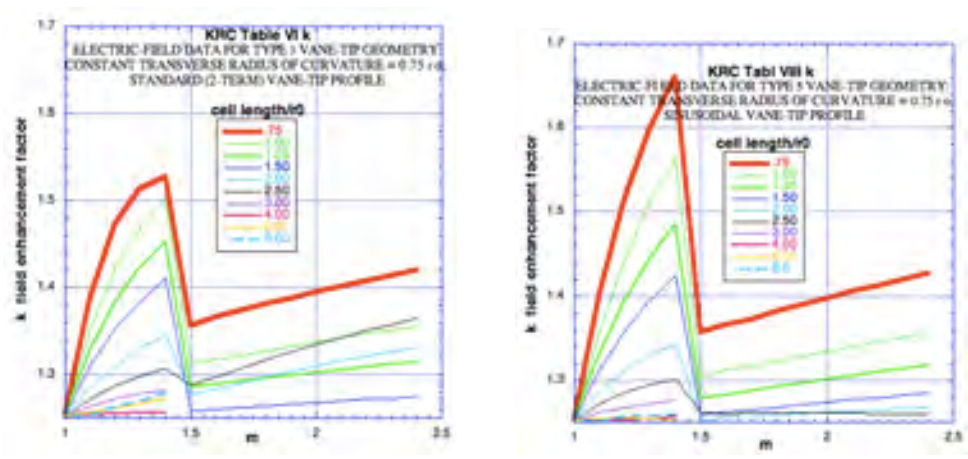

a01:

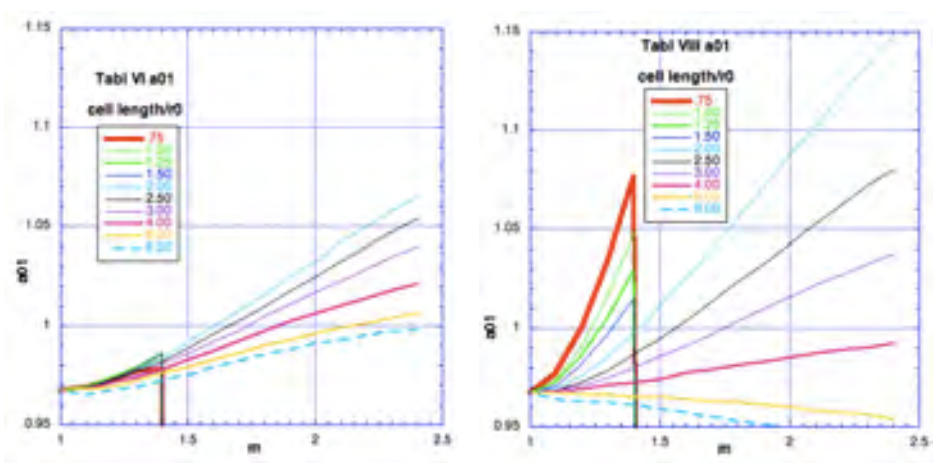

[200] ( Note: A preliminary design from pari with sinusoidal vane geometry and simulation input real transverse emittance 0.01160067 shows an input match (RFQgenOUT) {a =1.2999115, b=6.7945695} (remarkable precision!!), and 97% transmission using parmteqM ("RFQgen"); *LINACS* Poisson simulation gave transmission 90.9%, accelerated beam fraction 90.5%.)

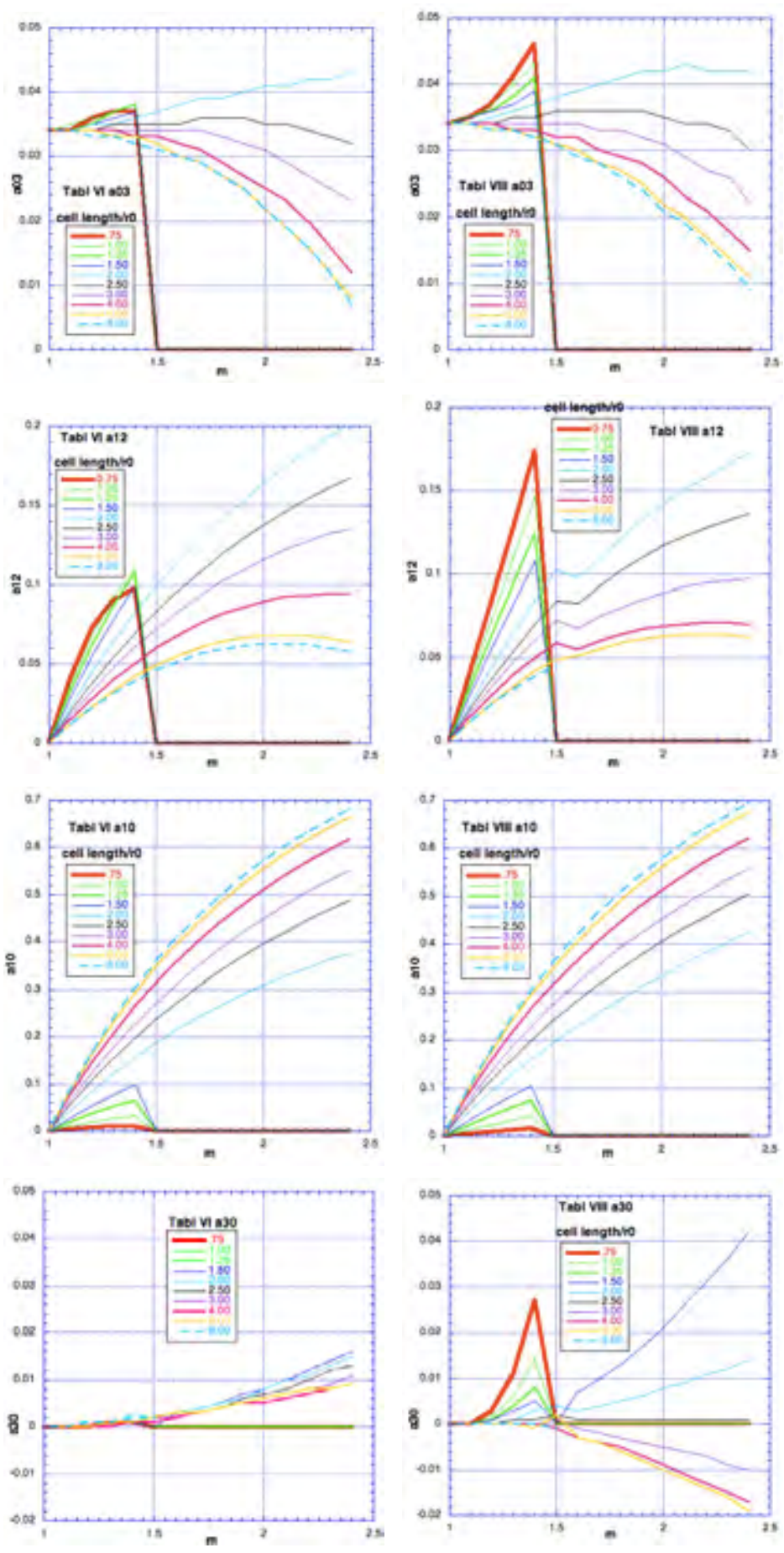

a03:
a12:
a10:
a30:
2-term vane modulation      sinusoidal vane modulation

Fig. 23.16. Plots for field enhancement factor and multipoles from original K.R. Crandall tables for 2-term (standard) vs. sinusoidal vane tip modulation, to give a visual feeling for some of the differences.

*23.7.2.4 Effect of vane modulation on RFQ Cavity Tuning*

Extensive further work is necessary to design the whole RFQ cavity, and it is of interest to know if the choice of longitudinal vane modulation influences this design. Based on the dimensions involved, small variation in vane modulations in comparison with the large cavity dimensions would indicate little effect.

ANL, in pursuit of their interest on trapezoidal vane modulation, has done perhaps more work than anyone else on very detailed CST simulations; "Full three-dimensional approach to the design and simulation of a radio-frequency quadrupole", B. Mustapha, A.A. Kolomiets, and P.N. Ostroumov, PRAB 16, 120101 (2013) contains a thorough analysis, at the accuracy limitations of 2013 CST.

There are fine grain effects, in many other ANL papers and also commented by Crandall in LA-9695-MS 1983, but we were interested here in whether there would be any effect on nearest cavity modes separations.

The nearest mode separation for a standard full vane 60 Mhz RFQ is reported as -1.5 MHz. The total effect on main mode resonant frequency from adding vane modulations was 500 kHz. CST errors in that simulation are estimated at 100-200 kHz.

To find the frequencies of neighboring modes, paraphrasing: "we disable the symmetry option in the calculation, which significantly reduces the effective number of mesh cells in the simulation and hence the accuracy of the results. Since we are interested mainly in the frequency separation of these modes, we may simplify the problem by not including the vane modulation. We know the frequency shift from the vane modulation (500 kHz), which should not affect the relative mode frequency separation."

This is taken as confirmation that the full cavity is not affected by the vane modulation details, and attention can be concentrated here on beam dynamics effects.

### 23.7.3 Conclusions

• *Sinusoidal vane modulation is recommended over 2term*. Overall, aided by the effects of space charge, there is essentially little difference beyond the average acceleration rate. A minor consideration is that 2term modulation is somewhat more complicated to machine. Further evidence for this conclusion is given in Sec. 28.4.2.

• Model/simulation correlation is good for simply described (sinusoidal), and for symmetric (sinusoidal, trapezoidal) vane longitudinal modulations, *but somewhat weakened for non-symmetric modulations like 2-term*.

• *Non-predictable effects* Slippage, (or even complete loss from the simulation frame) have been noted to sometimes occur in the full Poisson simulation, particularly for non-optimal designs. (This has also been noted in *BEAMPATH* and *LIDOS* – Sec. 14.2.). Important to note: This is an effect from the simulation code book-keeping; actual particles in an operating machine have no knowledge of such book-keeping or theory.

# Chapter 24 – Beam Halo from time-varying beam density, Halo Diagnostics

This paper remains the most complete treatment of halo formation and phase-space-transport in the linear accelerator literature, and opened the way to investigations by others. Simplified models could identify the location of fixed points.. Eventually 3D studies were made, which confirmed that the phenomena presented here are changed only in details.

## Self-Consistent Beam Halo Studies and Halo Diagnostic Development in a Continuous linear Focusing Channel

R. A. JAMESON

ABSTRACT

Beam halos are formed via self-consistent motion of the beam particles. Interactions of single particles with time-varying density distributions of other particles are a major source of halo. Aspects of these interactions are studied for an initially equilibrium distribution in a radial, linear, continuous focusing system. When there is a mismatch, it is shown that in this self-consistent system, there is a threshold in space-charge and mismatch, above which a halo is formed that extends to ~1.5 times the initial maximum mismatch radius. Tools are sought for characterizing the halo dynamics. Testing the particles against the width of the mismatch driving resonance is useful for finding a conservative estimate of the threshold. The exit, entering and transition times, and the time evolution of the halo, are also explored using this technique. Extension to higher dimensions is briefly discussed.

## 24.1 Introduction

A strict design condition for intense linacs operating as beam factories is that the whole beam is transported through the machine with very few particles lost on the walls. This allows "hands-on" maintenance during the facility lifetime without the necessity for remote manipulators. The beam loss requirement translates into less than about 1 nA per meter per GeV of beam energy. The fractional loss/m for a 100-mA, 1-GeV beam must therefore be controlled to less than one part in $10^8$. Similar challenges apply to various circular accelerators and storage rings, where beam control over as many as $10^9$ turns is required. Aspects of the space-charge mechanisms can also apply to these machines, and computational challenges are equivalent.

Operating experience [SC Workshop, 1978] with high-intensity linear ion accelerators and multiparticle simulations of these accelerators show that, under certain conditions, a diffuse "halo" may form around the dense "core" of the beam. These halo particles would be scraped off by the bore walls if their radius becomes large enough.

Linac designers have known that halos are reduced if the machine design keeps field nonlinearities as low as possible, parameter changes as smooth as possible, external focusing as strong as possible, and the rms parameters of the beam matched as well as possible to the acceptance parameters of the channel. Earlier studies also showed that it is desirable to keep the energy balance between the degrees of freedom of the system (e.g., radial and longitudinal) within certain limits; otherwise a dynamic equilibrating process will occur in which energy flows from the warmer to the colder state. This is called "equipartitioning" [Jameson 1981, Hofmann 1981]. Linac designs that maintain this condition along the machine have been proposed [Jameson 1981, Jameson 1992, Jameson 1993, Pabst 1994], and realized since then for the RFQ and other linac types in the main lattice. It may not be completely possible in practice, because of transition regions or because of various constraints or errors (e.g., facility length, which translates to cost, or errors in the rms matching). In the main lattice, however, finding an EP design should serve as the point of departure.

Following these precepts means that sources of free energy in the system are minimized. Free energy may be converted into beam emittance and growth in the outer radius.

The actual mechanics of halo formation were unknown until recently. [Carlsten 1993] showed that halo formation from point-to-point Coulomb interactions is not of concern in the intense rf linac beams under consideration. It had been observed in simulations that particles once in the core could later be found at the outer edge, indicating that a large effect had occurred, strong enough to push the particle out of the core potential well. Other probable mechanisms suggested slower, more diffusive effects. We now concentrate on the former, and will return briefly to other mechanisms later.

## 24.2 The Core/Single-Particle Mechanism for Beam Halo Production

It was realized [Jameson 1993, O'Connell 1993] that a single particle, passing through time-varying fields set up collectively by the other particles, could gain or lose large amounts of energy. The total energy of the system is of course conserved. A direct analogy is that of a spacecraft interacting with a large body in space; the spacecraft may be slowed down or accelerated (or even stopped) depending on its orbit around the large body, except that in the beam the reactions are from space-charge and thus repulsive, rather than gravitational and attractive. If a particle passes through the beam core when the core density is increasing, the particle will gain energy, and conversely, when the core density is decreasing, the particle will lose energy. Indicated schematically in Fig. 24.1, this is the basic energy transfer mechanism, and also a source of randomness (because there is a finite number of particles in a real beam) that adds significant chaotic aspects to the dynamics of the system.

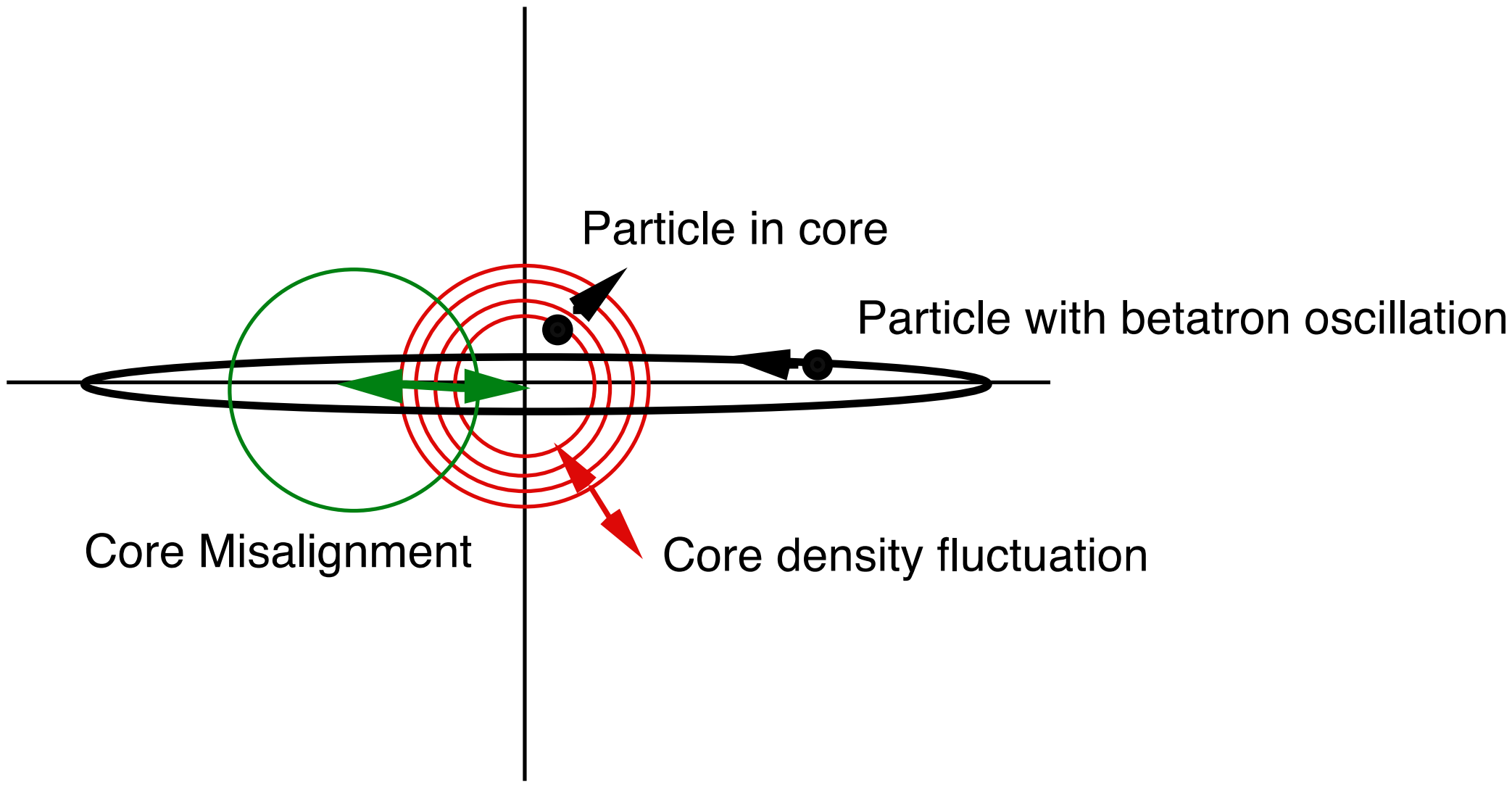

Fig. 24.1. Schematic of the collective core/single-particle halo-producing mechanism.

The second aspect is that the accelerator system is quasi-periodic, so the particle may pass through the core repeatedly. If the particle falls into resonance with a driven core oscillation frequency (called a “parametric” resonance), repeated energy transfers can occur, in total enough to move the particle from inside the core to far outside into a halo.

Third, as particles move out, the restoring force increases, and the orbit frequency (tune) changes. This causes the outward movement to be self-limiting; the particle falls out of phase and may move back in again, to be replaced in the halo by other particles. This motion may repeat many times. The self-limiting feature is very important in terms of practical definition of adequate linac aperture.
Finally, the nonlinear resonance dynamics can exhibit chaotic features, well-known in other fields.

**24.2.1 Simple One-Degree-of-Freedom Model**

A simple model illustrates these features of halo production. We study the motion of a continuous cylindrical beam in a uniform longitudinal magnetic field. Following the notation of [Bondarev 1993], the equations of motion (with zero angular momentum) are

$$x'' + 2\Lambda y' - a(Q(r)/r^2)x = 0 \,, \qquad (1)$$

$$y'' + 2\Lambda x' - a(Q(r)/r^2)y = 0 \,, \qquad (2)$$

$$\Lambda = (e\,B_z\,L)/(2\,m_o\,c\,\beta\,\gamma) \,, \qquad (3)$$

$$\alpha = (2\,I\,L)/(I_o\,E\,(\beta\gamma)^2) \,, \qquad (4)$$

where e, $m_o$ are particle charge and rest mass, $\beta$ is ratio of particle longitudinal velocity to light velocity c, $\gamma = (1 - \beta^2)^{-1/2}$, $B_z$ is magnetic field, E is beam emittance, L is focusing period length, I is beam current, $I_o = 3.13*10^{-7}$ A is Alfven current for protons, Q(r) is the proportion of particle charges inside the circle with radius r, $(Q(\infty) = 1)$, $r^2 = x^2 + y^2$.

The equations of motion inside and outside the core are not the same:

$$r'' + \Lambda^2 r - \alpha/r = 0 \; ; \; r > rk \, . \tag{5}$$

$$r'' + (\Lambda^2 r - \alpha/r_k^{\,2})\cdot r = 0 \; ; \; r \leq r_k \, , \tag{6}$$

where $r_k$ is the core boundary. Bondarev gives a proof showing that a particle will gain (lose) energy if it traverses the core when the core density is rising (falling).

### 24.2.2 Demonstration with Cold Beam

The continuous, linear, radial external focusing system above was programmed by K.R. Crandall circa 1970 as a particle-in-cell code, in x,x',y,y' coordinates to avoid difficulties at the origin. A ring-model is used for the space-charge computations. A zero-emittance beam (at the space-charge limit) distributed uniformly from the origin to the matched radius will propagate without change.
If the uniform distribution is filled to a mismatched radius, there will be a linear oscillation but the distribution will remain a straight line without distortion. However, as shown in Fig. 24.2, if a mismatched, Gaussian, zero-emittance space-charge-limited beam is injected, the particles will rotate at different frequencies and mixing will occur. [r-r' phase-space is shown. $r' = (xx' + yy')/r$. "rr' radius" = $(r^2 + r'^2)^{1/2}$. When x or y change sign, the sign of r is reversed, and r' adjusted, to aid the eye.]

At z/wp = 3.75 plasma periods, some particles are still at rest at the origin. Then the outer tail sweeps through the origin, causing a local density anomaly there that repels nearby particles, and also slows or speeds up particles in the tail. This can be observed at z/wp = 0.5 and subsequently. Repeated interactions of this type result in folding of segments onto a beam core (z/wp = 3.875), from which new tail segments begin to emerge. By z/wp = 9.750, the new tail extends almost as far as the original tail. Tracking of the particles initially at the origin shows they are far out in the new tails. The result of an abrupt scraping of the halo at z/wp = 50 can now be anticipated by considering the mechanism as described above – a strong central density oscillation is still present, so halo continues to form.

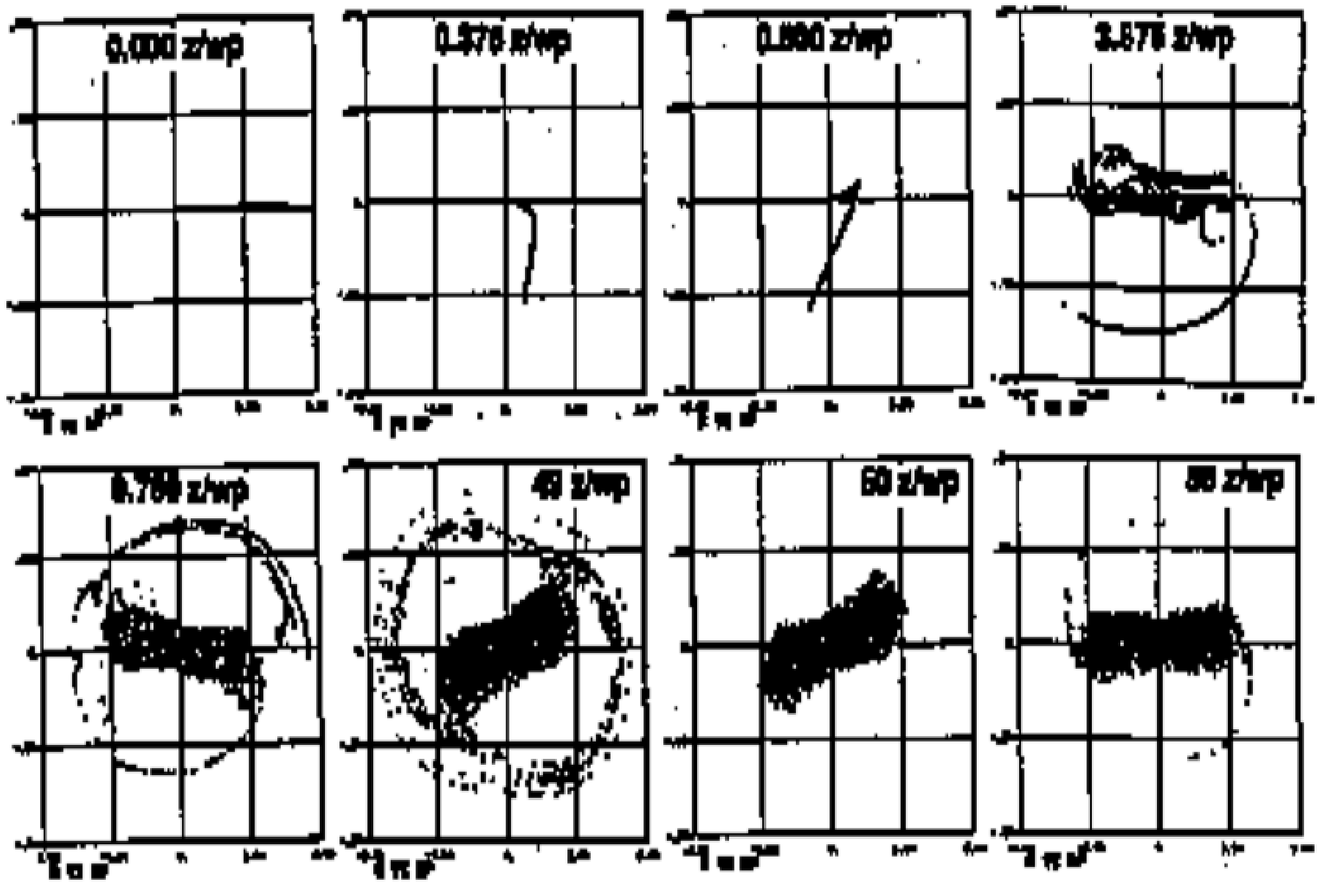

Fig. 24. 2. r vs. r' phase-space evolution of initially zero-emittance, mismatched beam in the radial channel; scraped at z/wp = 50. (z/wp is distance in plasma wavelengths.)

### 24.2.3 Warm Beam Demonstration

Figure 24.3 shows the core/single-particle interaction. The simulation was made with a finite-emittance, rms tune depression [Jameson 1981] = 0.4, mismatched, Gaussian initial distribution. The density oscillation in a disc about the origin is shown. The trajectories shown are the rr' radii of six particles (out of 10K) that had the largest rr' radii at z/wp = 10.

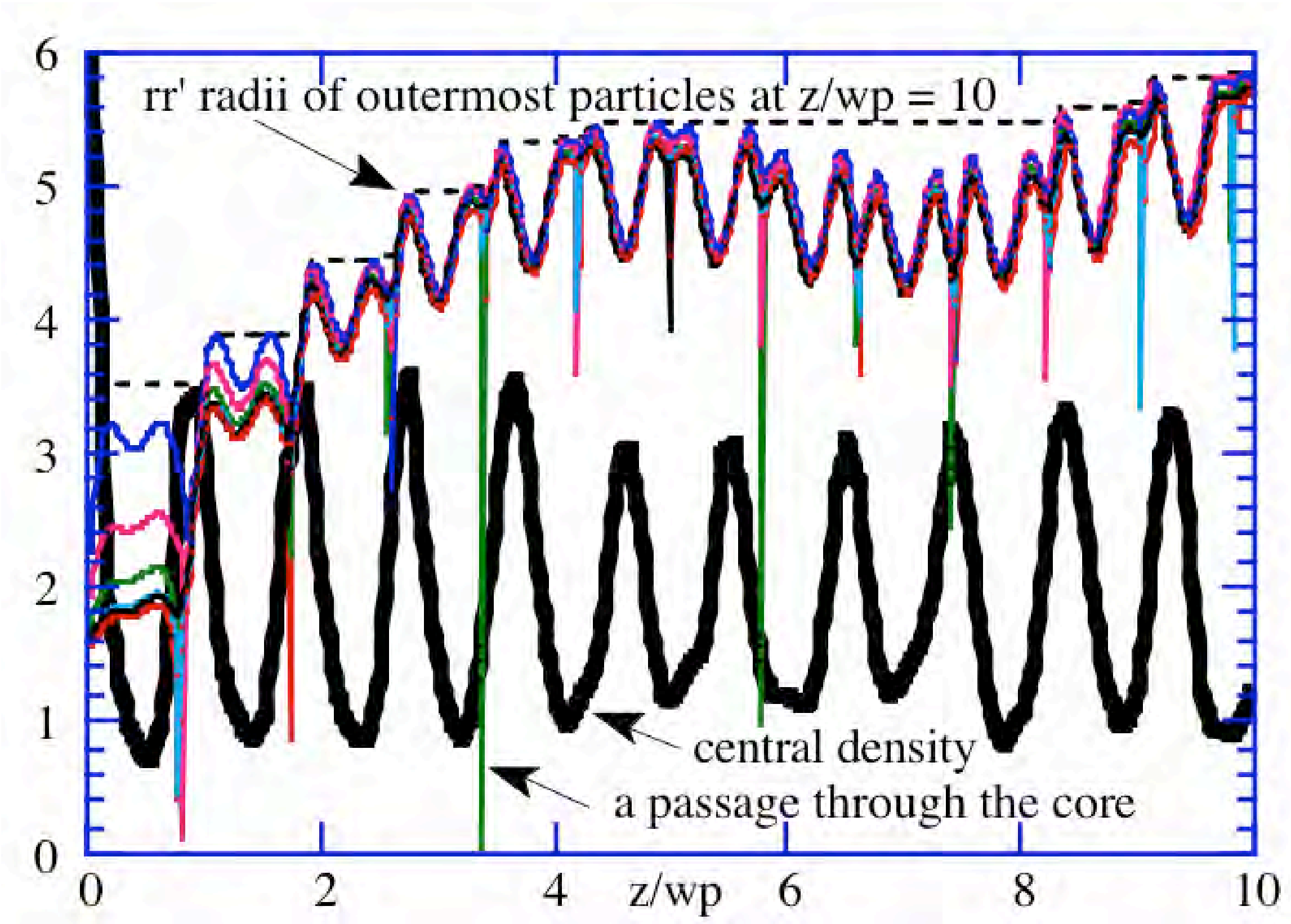

Fig. 24.3. The core/single-particle interaction.

The downward spike in these orbits indicates a passage through the core. If the core density is rising when this passage occurs, it is seen that rr' increases after the passage. These increases resonantly accumulate, but the particles' tune changes and they begin to traverse the core when the core density is falling, which takes energy from the particles, and they move back in.

## 24.3 Definition of Terms

A number of interpretations of the terms "core", "tail" and "halo" are now used. As indicated in the equations above from Bondarev, one fundamental indication is that the "core" (or "kern" in Russian) is within the radius where the space-charge force peaks, with the outer part of the distribution having a ~1/r dependence at greater distances. This outer part, which may be a natural part of the initial distribution, is called the "tail". Practical beams with mismatch or misalignment errors will of course have particles at the radii corresponding to these errors. The term "halo" is then reserved for particles outside the natural extent of the tail or the expected radii from errors. This definition is appropriate to the resonance "width" concept developed below.

[Lapostolle 1970] and [Sacherer 1971] showed that the rms properties of a transported beam depend little on the detailed shape of monotonically-decreasing distribution functions; thus the "equivalent rms uniform beam" can be used. Other analyses [Struckmeier 1984, Wangler 1985, Anderson 1987, Hofmann 1987, Reiser 1991] have shown that the uniform component of a distribution's field energy can be represented by the equivalent uniform beam, and the remaining nonlinear field energy can be related to asymptotic rms growth (from errors of energy imbalance, misalignment, rms mismatch, and input distribution mismatch) in time-independent transport systems that are considered infinitely long. This is not easy to apply to an accelerator, and also deals with a macro-effect on the beam bunch that does not reveal what really happens to particles that may form a halo. However, there are some definitions of core, tail and halo that use an equivalent uniform beam as the basis.

## 24.4 Initial Particle Distributions

In a completely linear system with linear external forces and a uniform beam distribution, there is no free energy and no emittance growth occurs. The equations describing the motion of this distribution were found by [Kapchinsky and Vladimirsky 1959]. The KV distribution has particles distributed on a shell in phase-space. It is convenient because it can be analyzed analytically, but it is not realizable physically and has unphysical instabilities because of the sharp shell edge, so it must also be used with great caution in simulations.

The Gaussian distribution (which may be truncated) has a tail; this mathematical function is not linearly matched to accelerator channels and is also not a good shape for use in fitting observed beams (even by combining several) [Boicourt 1980, 1983]. Thus it will initially undergo charge redistribution during the first few plasma periods as it tries to equilibrate to the channel, and the initial tail will continue to evolve as indicated in Fig. 24.2. These confusions limit its usefulness in studies of halo phenomena.

The thermal (Maxwell-Boltzmann) distribution may be an equilibrium state at the space charge limit, or in circular accelerators or storage rings where the lifetime is long enough for the beam to equilibrate via Coulomb collisions. In linacs and other high-power beam devices, the shorter length and rich arrays of resonances indicate that other dynamic effects will dominate, although there can be a component of free energy thermalization via emittance growth and via equipartitioning if the energy balances are bad enough. Work has recently been done [Brown, 1994] to find the extent of the radial thermal tail for bunched beams, as a best-case guide for channels with time-independent external focusing.

### 24.4.1 The Hamiltonian Equilibrium Distribution

[Gluckstern 1970] has shown that an infinite class of equilibrium distributions exist for a simple time-independent, continuous, linear, radial focusing channel in which the particle coordinates are chosen based on the Hamiltonian (including space-charge) of the system. These distributions extend from the zeroth-order KV distribution, with no tail, to the thermal distribution at infinite order. The intermediate orders have monotonically decreasing density distributions and sharp cutoff radii. Thus they are well-suited for study of phenomena in the linear continuous radial focusing channel, and the first-order distribution was selected for further studies to be described below. The beam radius and focusing strength are chosen, giving the tune depression. In rr' phase space, the distribution has a squarish shape that sharpens with more tune depression (Fig. 24.4.a).

Figure 24.4.b shows an example, in which an initial distribution with rms tune shift $\sigma/\sigma_o = 0.4$ was scraped with an elliptical rr' filter after 5 plasma periods. This excited a central density oscillation, and the maximum beam radius grew back larger than the initial value (Fig. 24.4.c). As expected, the largest radius particles were driven as described above. With a linear external restoring force, simulation of a misaligned initial distribution over a large range of misalignment and tune shift did not exhibit halo formation.* In a channel with nonlinear restoring forces, both halo formation and centroid damping would be expected.

Energy equilibration via equipartitioning was demonstrated by injecting an unbalanced beam using the xx' distribution from a $\sigma/\sigma_o = 0.1$, and yy' from a $\sigma/\sigma_o = 0.83$ Hamiltonian distribution (Fig. 24.5). In 10 z/wp, the rms emittance growth was damping out at ~7-8%, but the maximum radius growth was about 25% and still growing almost linearly. A strong central density oscillation was excited, with an increase in maximum radius on every rise in central density.

* An error in the simulation reported by [Jameson 1993] was corrected.

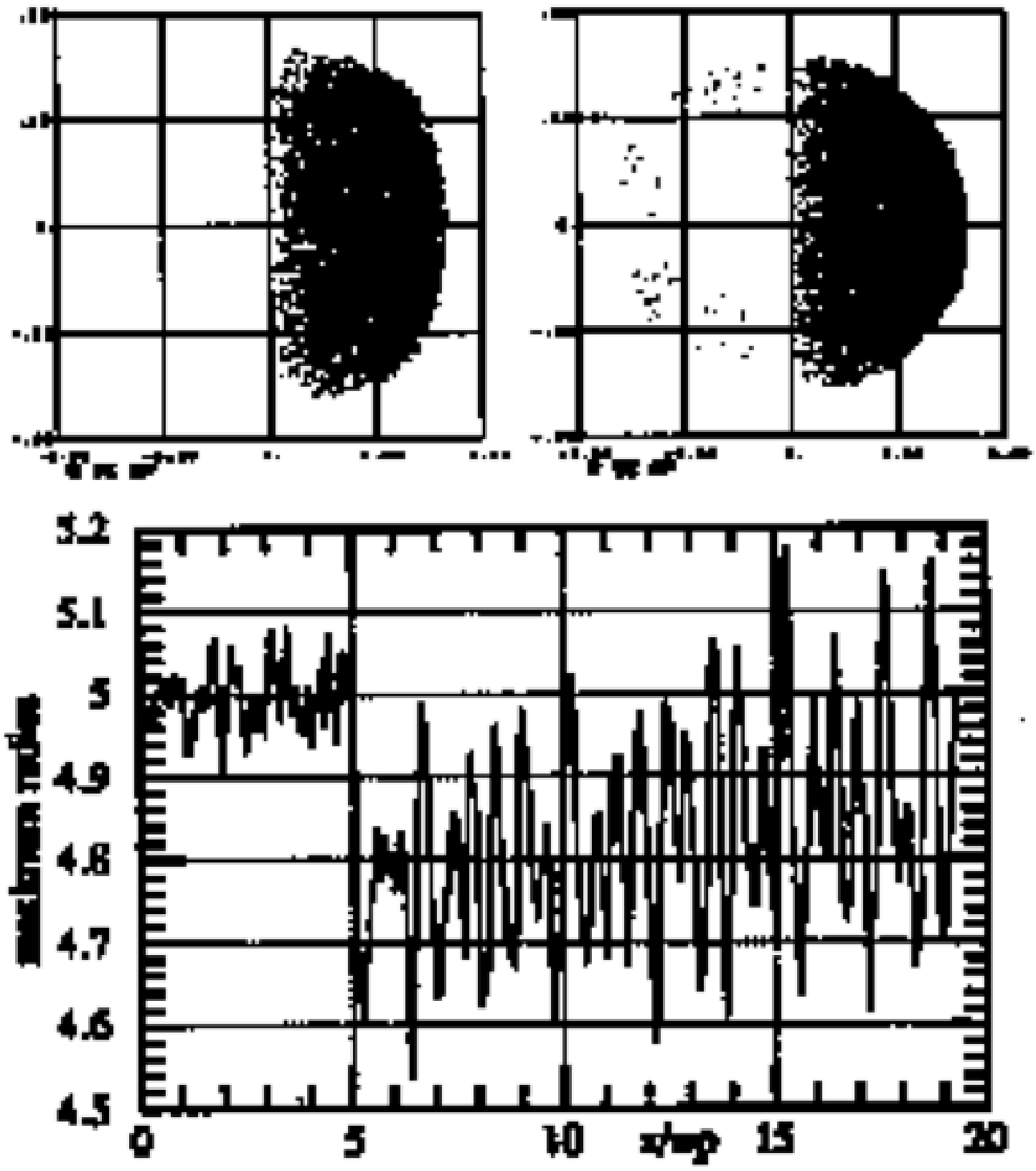

Fig. 24. 4. (a.) Hamiltonian s/so = 0.4 distribution, rr' phase-space.
(b.) Scraped at z/wp = 5. (c.) Regrowth of maximum radius.

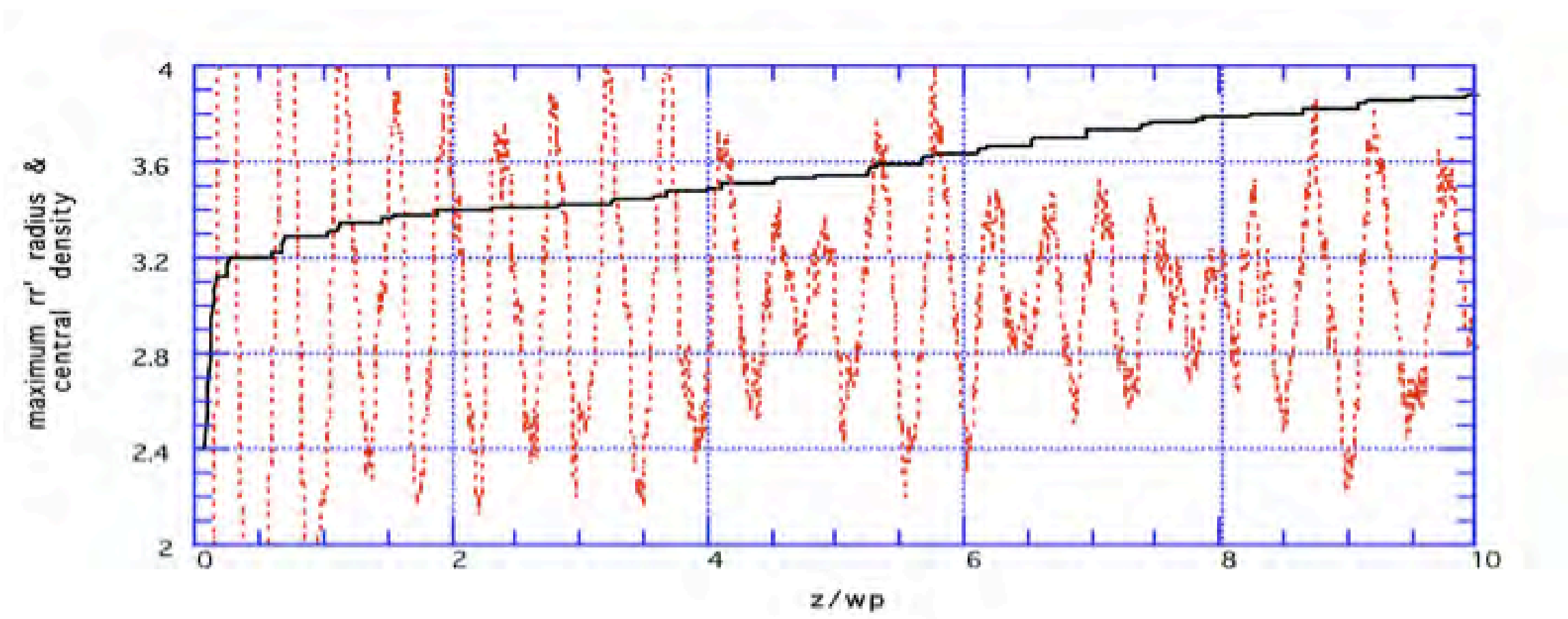

Fig. 24.5. Maximum rr' radius and central density, non-equipartitioned distribution.

## 24.5 Resonances

We have identified time-dependent density distribution of the beam core and resonance with a single-particle trajectory as a major halo producing mechanism. We now proceed to examine this in more detail. Here the reader is referred to the literature of chaos in Hamiltonian and near-Hamiltonian systems, which essentially describes everything that we will see in the simulations to follow. This literature is too rich to describe in this note, and we only point out aspects of it. In nonlinear systems, resonance strengths depend on the parameters, and when resonances overlap, particles may be transported from one resonance zone to another and exhibit chaotic properties.

The behavior of one-degree-of-freedom (two-dimensional phase-space) systems has been extensively studied, and we will pursue the case of the radial focusing channel further below. In this case, for example, we are assured that the self-limiting property of the core/single-particle interaction is strictly satisfied for reasonable values of mismatch and tune depression because there will be an outer "KAM curve" that is impenetrable. However, in systems of more than two dimensions, the phase-space cannot be strictly divided, and a stochastic web may exist along which particles can move to more distant regions of phase-space. Beyond this, little is known about the mechanics or even how to describe higher-dimensional transport. An accelerator has three degrees-of-freedom (six-dimensional phase-space) plus time variation and thus is a formidable and forefront problem. In this paper, some degree of definition is attempted, and some considerations of possible tools for computational analysis of the full system are outlined.

Resonances are introduced in accelerator systems from many sources; each, and their combinations, require detailed future study.

Some basic construction techniques introduce density fluctuation and resonances, such as alternating-gradient focusing. This is the fastest period in most linac or transport lattices, and is often averaged out for rms purposes; however, it needs to be studied in more detail in terms of halo production. Careful work [Struckmeier 1994, Struckmeier & Hofmann 1992] indicates that there is a small contribution to emittance growth, because an energy (not temperature) balance can never be reached in a quadrupole focusing system where the energy (not temperature) imbalance is periodically restored by the focusing forces. A particle with phase advance ≥ the quad lattice phase advance would experience a strong "envelope" resonance, so the zero-current phase advance ($\sigma_o$) must be below 90˚. There are integer fractional resonances below the main envelope resonance (72˚, 60˚, 45˚, ...) that might also have to be avoided in a long transport line without acceleration.

A second construction technique is the practice of making all the cells the same length in high-beta tanks. This was done to reduce production costs, and results in no true synchronous phase — the particles slip in phase as they traverse the tank. They enter the next tank and slip again. If the periodicity is not carefully considered, serious resonances can occur that would affect even the rms behavior (SSC linac). The effect on halo production should be studied. However, since modern numerical machining can produce graded-beta cells essentially without additional cost, this effect can be eliminated and this basic way to avoid one halo-production mechanism now forms the baseline approach for the ESS project [ESS 1994].

Acceleration and bunching processes, e.g. in the RFQ [Jameson 1993], introduce time-variation into the system. This has a fundamental effect of increasing the degrees of freedom. Smooth acceleration could also have the benefit of minimizing the time spent near a resonance, but on the other hand, longitudinal resonances will couple with the transverse ones.

A wide variety of errors can introduce time variations. The strongest are errors in the linear energy mismatch, such as rms mismatch at transitions in a linac. Errors include misalignment, quadrupole rotation or other errors that introduce unwanted coupling between the degrees of freedom, or higher-order aberrations, field pattern errors, time-dependent errors that mean that a single rms-matching procedure is only good on average, etc.

## 24.6 Parametric Resonance

[Gluckstern 1994] has written the action integral for the mismatched radial focusing system with the addition of a cubic restoring force that increases with distance from the origin, as shown in Fig. 24.6.

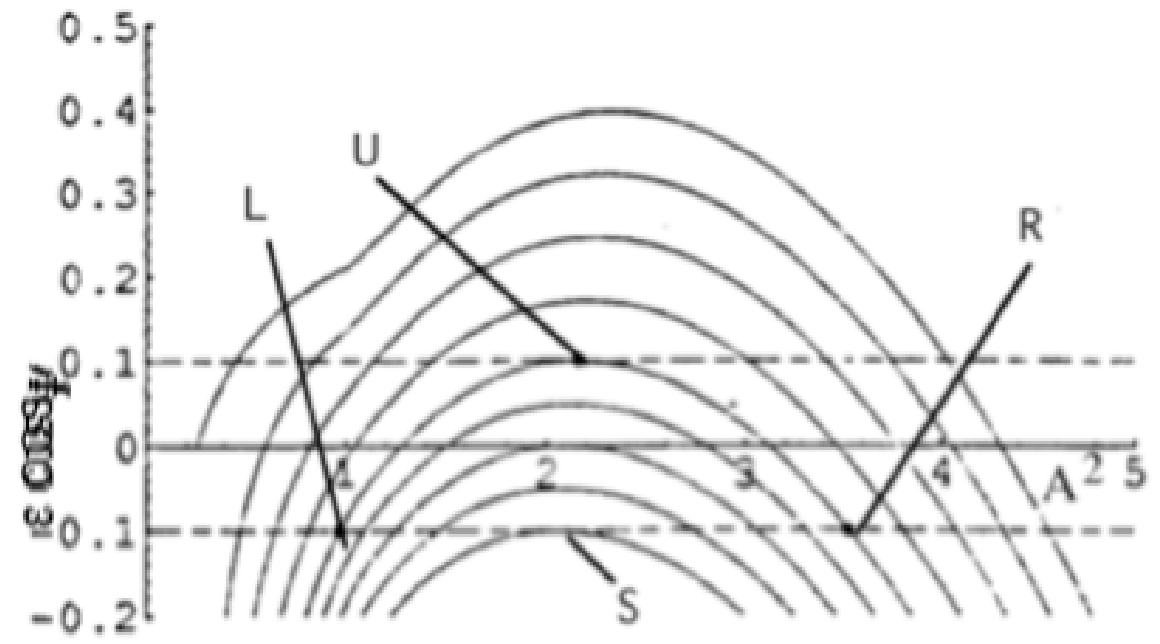

For ε = 0.1, particles to the left of point L oscillate in the core, but if a particle is near (at) point L, it is swept far out to point R. Point U is an unstable fixed point and point S is a stable fixed point. The phase-space plots of Fig. 24.6.b and 6.c show these features and the separatrix. This model is very illuminating qualitatively. However, it is oversimplified for quantitative use even in the radial channel system, where the space-charge force nonlinearity is more complex and beam emittance changes during the evolution [Chen 1994], nor does it address how a particle might achieve a given initial condition.

Fig. 24.6.a. Plot of ε cos Ψ action integral vs. square of (particle radius / core boundary radius) = $A^2$, for a tune depression of 0.4 and zero angular momentum. (From Gluckstern

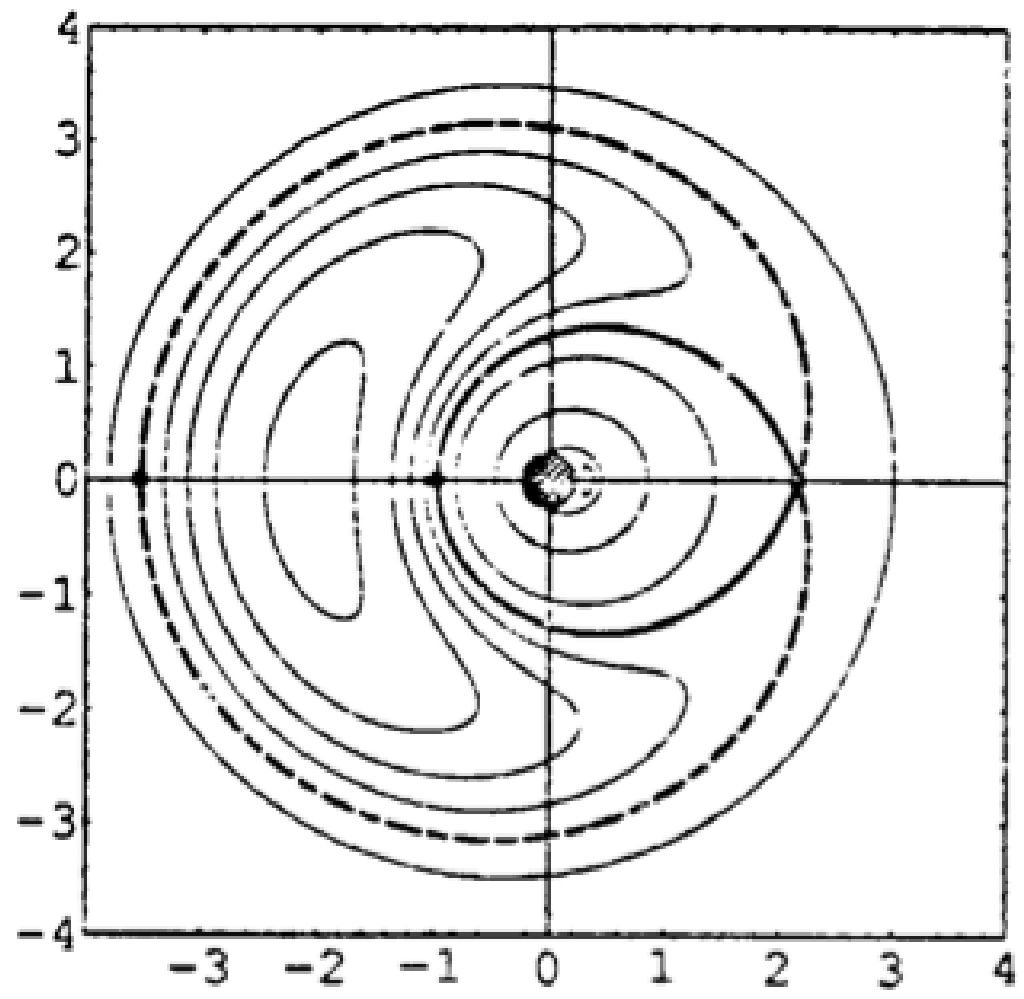

## 24.7 The Test Particle Method

The literature of nonlinear dynamics and chaos in Hamiltonian (or near-Hamiltonian) systems with 2-dimensional phase-space has made tremendous progress over the past few years in finding ways to describe and measure the system dynamics. The systems amenable to study have the properties of wist maps. Resonances are typically very rich in structure (being possible at any rational number relationship with the driving terms). Particle transport through "turnstiles" occurs when thedisturbance parameter becomes strong enough that resonances overlap. Chains of "islands" are formed near separatrixes, corresponding to the rational number resonances, and in which particles are trapped for at least one full circuit of all the islands before they have a chance to turnstile out into another area of phase-space.

Fig. 24.6.b. Polar plot of w vs. Ψ for trajectories corresponding to ε = 0.4. (From Gluckstern)

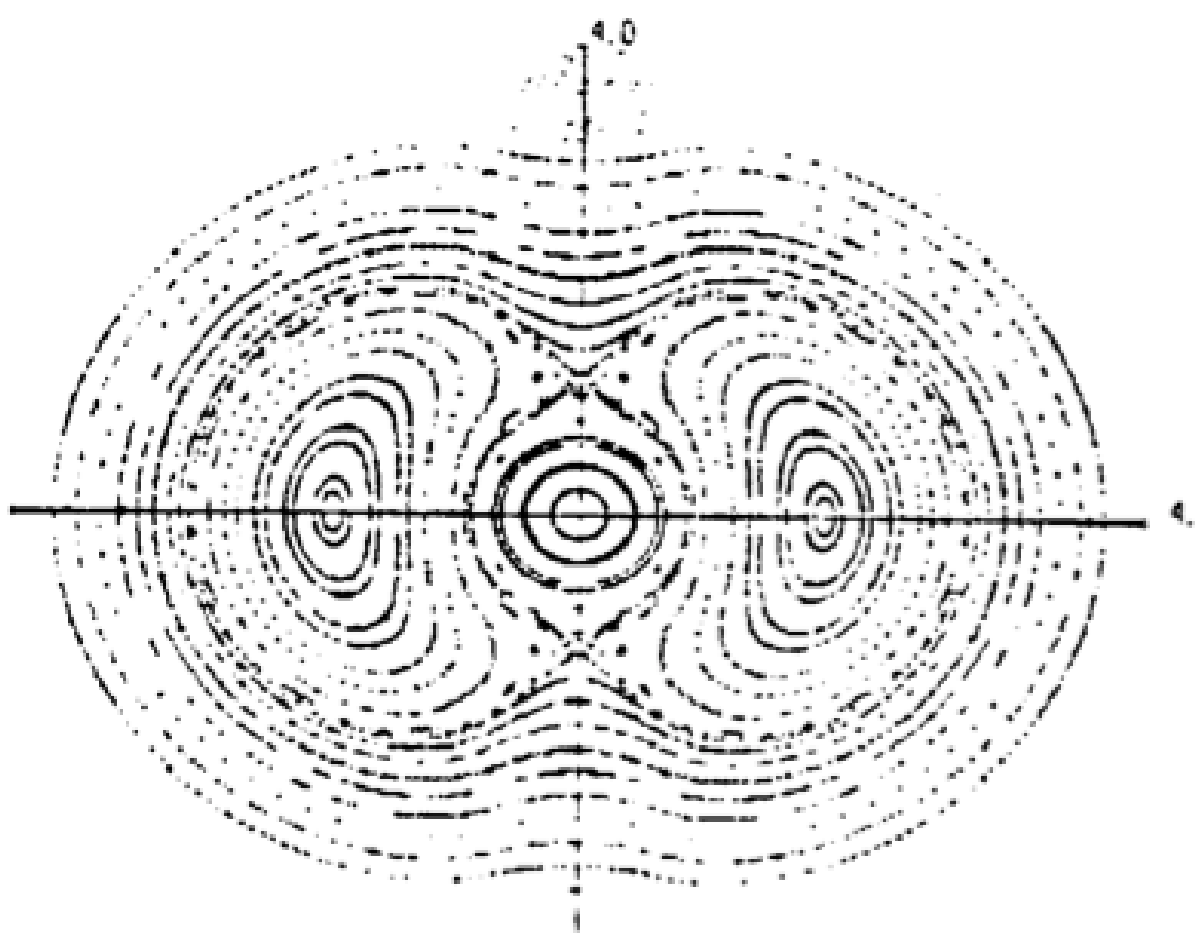

These features are typically explored by "test particle" methods. There are several versions. In each, single particles are placed at many initial conditions and their motion computed in the field of a separate ensemble, i.e., the test particles do not contribute self-consistently to the fields, but are only affected by them.

Fig. 24.6.c. Stroboscopic plot of x,x' trajectories, corresponding to Fig. 24.6.b. (From Gluckstern)

One version is to solve an analytic model for the core fields. These models do not account for emittance growth. A second version is to use a particle-in-cell code to solve for the fields self-consistently, and add extra test particles. The test particle method is a valid procedure when the effect of a test particle on the system would be small, as in the case of one particle out of ~$10^9$ in a real beam, and in the case of strong external fields [Bruhwiler 1990].

The trajectories of the test particles map out the phase-space characteristics. Plotting the trajectory intersections with the phase-space once per period of the disturbing term gives a "Poincaré Map", as shown in Fig. 24.7 for the radial channel system [Ryne 1994]. It is seen that the general features of the Gluckstern development for the mismatch oscillation are present, along with the full richness of the resonant structure. Chaotic features are observed, particularly near the unstable fixed points of the separatrixes, depending on the strength of the disturbance.

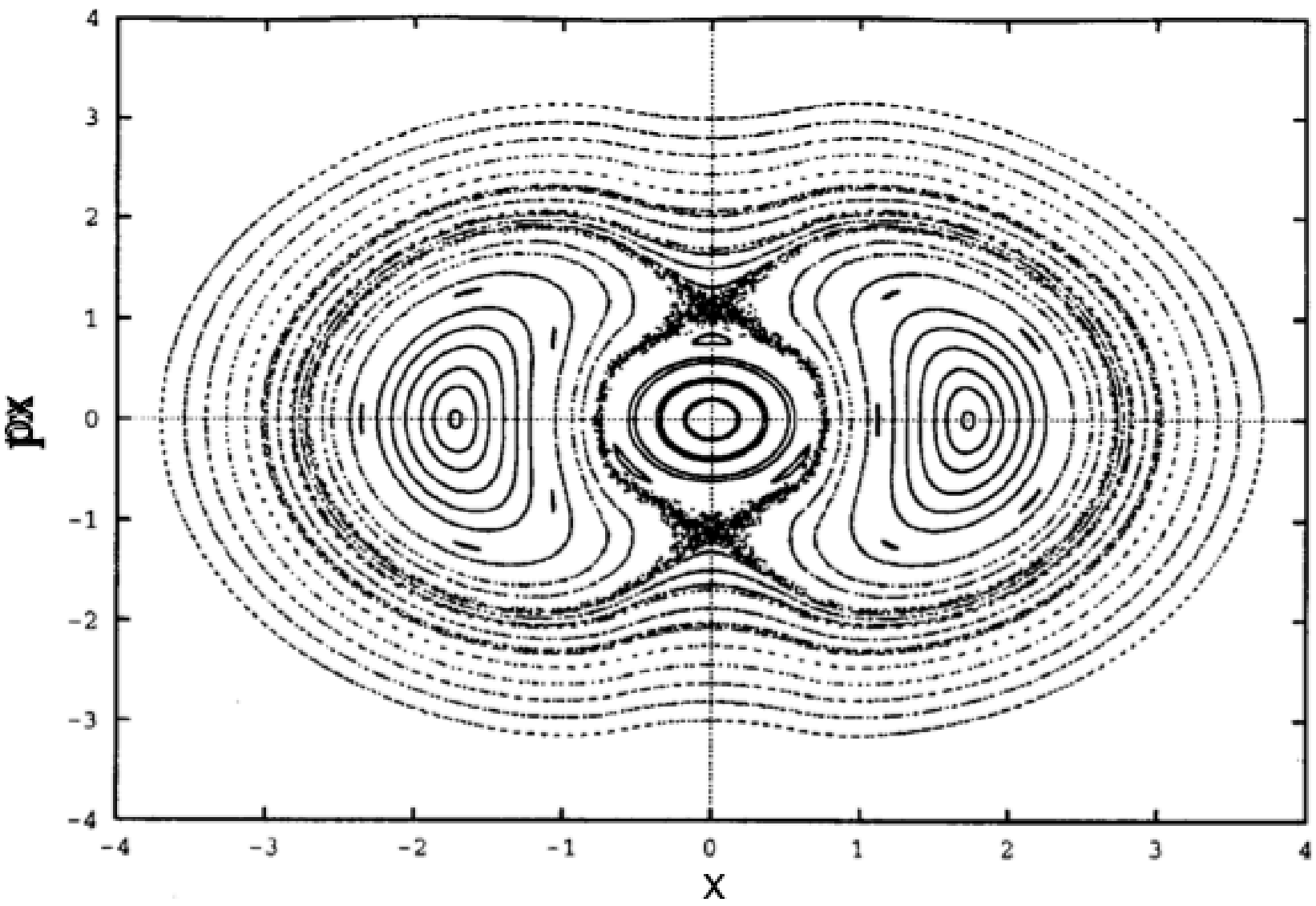

Fig. 24.7. Stroboscopic phase-space plot based on the analytic (perfectly periodic mismatch) core-halo model (uniform density core, tune depression = 0.4, mismatch = 1.5). (From Ryne)

Test particle methods are very useful in preparing a "road-map" of the whole phase-space area under consideration. <u>They are not solutions of the core/single-particle interaction outlined above</u>. To understand this interaction, we have to observe the single-particle dynamics <u>self-consistently</u>. This corresponds to finding a specific route on the roadmap, and is the real problem we want to study.

## 24.8 Self-Consistent Core/Single-Particle Interactions

For an initially Hamiltonian distribution with $\mu = (1 - \sigma^2/\sigma_o^2) = 0.54$ and mismatch = 1.30, Fig. 24.8 shows the particle-in-cell simulation-code self-consistent Poincaré plot of the particle that achieved the largest radius during the course of the simulation. It is completely different from Fig. 24.6 or 24.7. First, the forces for the whole distribution, including the particle of Fig. 24.8, are evolved self-consistently so that emittance growth, for example, is taken into account. And also, the particle plotted traces out only a portion of phase-space; in this case, it spent part of the time in the core and part in the halo.

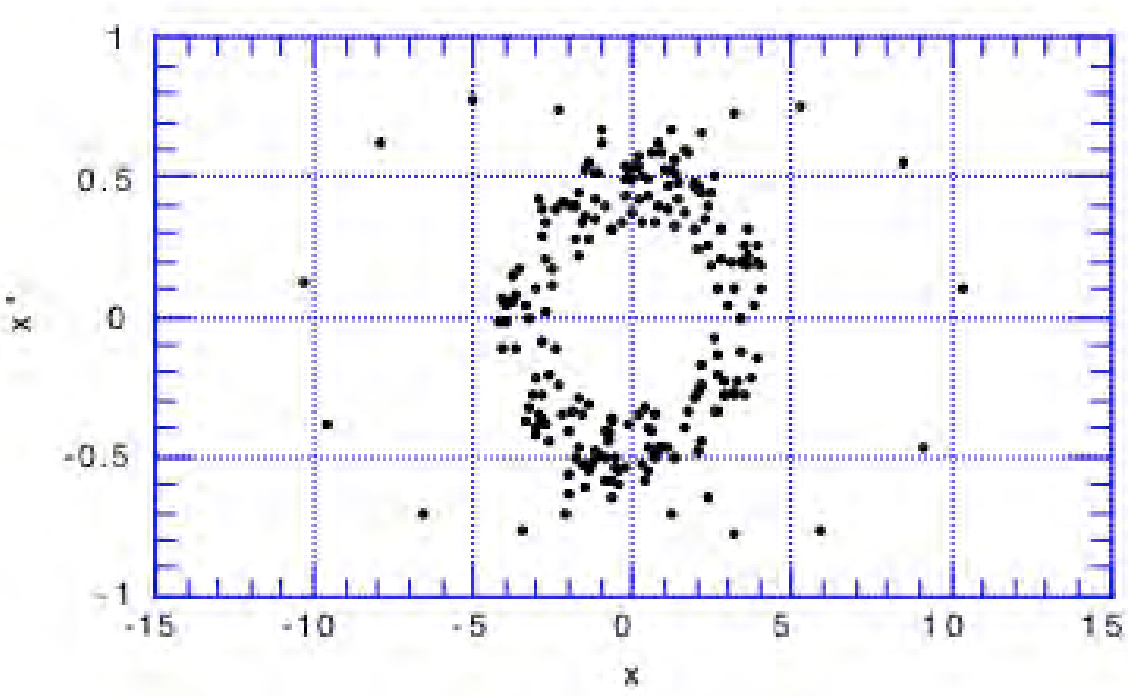

Fig. 24.8. Self-consistent Poincaré plot for a particle caught by the mismatch parametric resonance. $\mu$ = 0.54, mismatch = 1.30

Figure 24.9 shows the detailed radial trajectory of this particle. The $r_{max}$ curve for the whole distribution shows that other particles were moving out and in during the run; this particle succeeded, near step 9500, in reaching the largest radius. The behavior is seen to be quite rich. Initially, the particle is oscillating in the core but is immediately involved in a resonance with the breathing mode

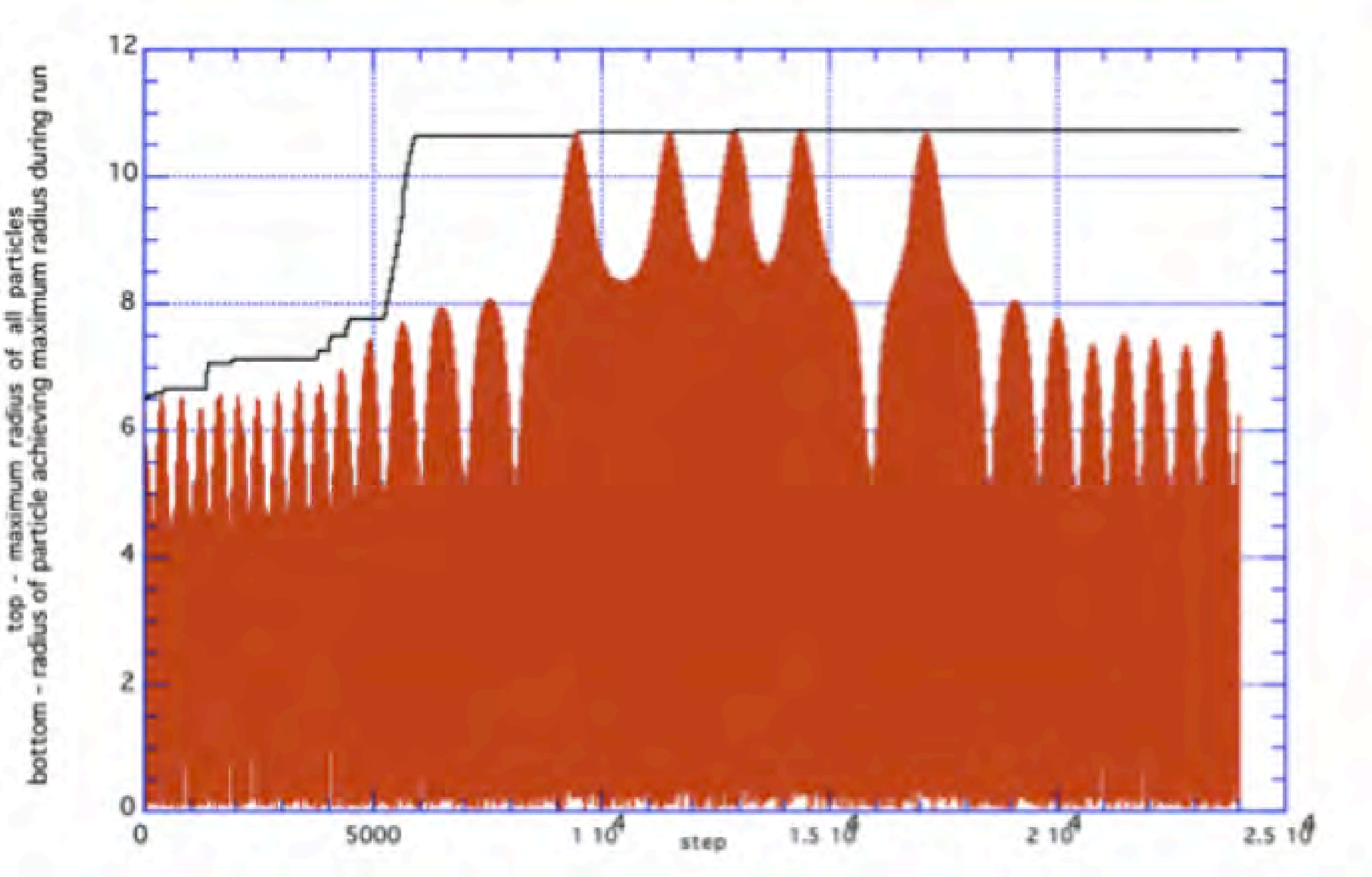

Fig. 24. 9. Mismatched Hamiltonian initial distribution, with mu = 0.54 and mismatch = 1.30. Maximum radius of all particles during the run, and radius of the particle that achieved the maximum radius during the run.

having a rational number relationship, in this case with a cycle time around the island chain of about 8 or 9 core oscillation periods. The uncertainties of the single-particle interactions with the other particles as the single particle traverses the core eventually causes it to transport out of this chain, and into another with a longer period – closer to the main separatrix of the breathing mode. This repeats, with some growth in the maximum radius, in each case with an entrainment toward the main separatrix, until the island chain is 20 or more periods long. Travel on the main separatrix itself would take an infinitely long time, but the chaotic aspects of the sweeping separatrix and the core interactions result in a jump across the main separatrix to the other side. In this process, the particle is swept far out in radius. The process continues with some overshoot to lower number resonances outside the separatrix, but entrainment back toward it, and in this case, transition in, out, and back in to the core to the end of the run. The entrainment toward the main separatrix is the result of resonant core/single-particle interaction and self-limiting changes in the particle tune as the particle interacts with the main resonance. At higher mismatch and space-charge factor, Fig. 24.10, the action is much quicker and the particle goes farther beyond the separatrix, without return, and interacts with a higher-order mode for the second half of the run. In this example, emittance growth is strong, evidenced by a decline in the central density (Fig. 24.10.b) and change in the central core oscillation wavelength (Fig. 24.10.c)

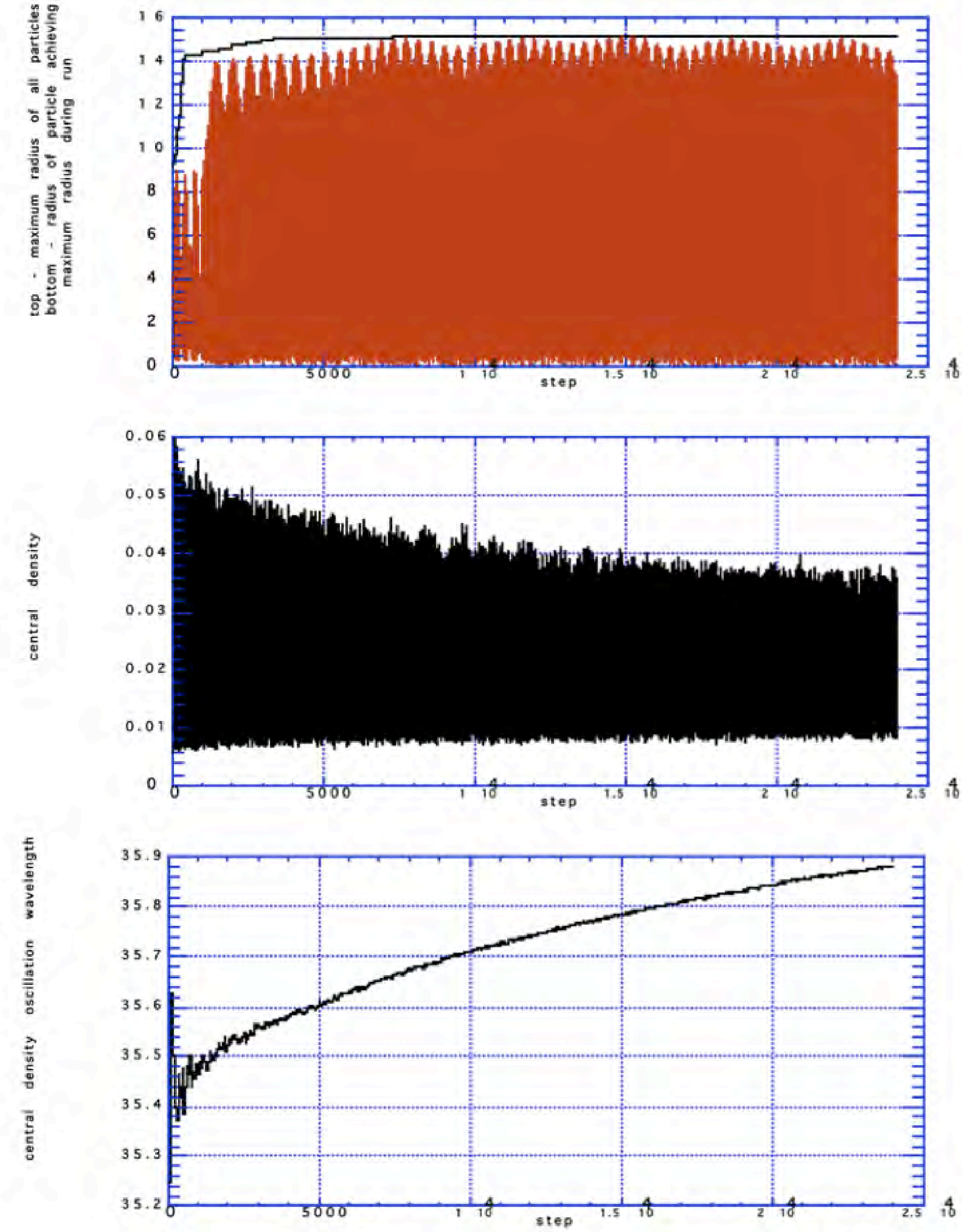

Fig. 24.10. Mismatched Hamiltonian initial distribution with mu=0.84 and mismatch = 1.85.

## 24.9 Particle Tune

The space-charge forces vs. radius for various values of the rms space-charge parameter $\mu = 1 - (\sigma/\sigma_o)^2$ for our Hamiltonian distribution are shown in Fig. 24.11.

Fig. 24.11. Space charge force vs. radius for the Hamiltonian particle distribution. The distribution outer radius is 5; (2*r**2 = 50); test particles were added out to r = 10 to show force outside the core.

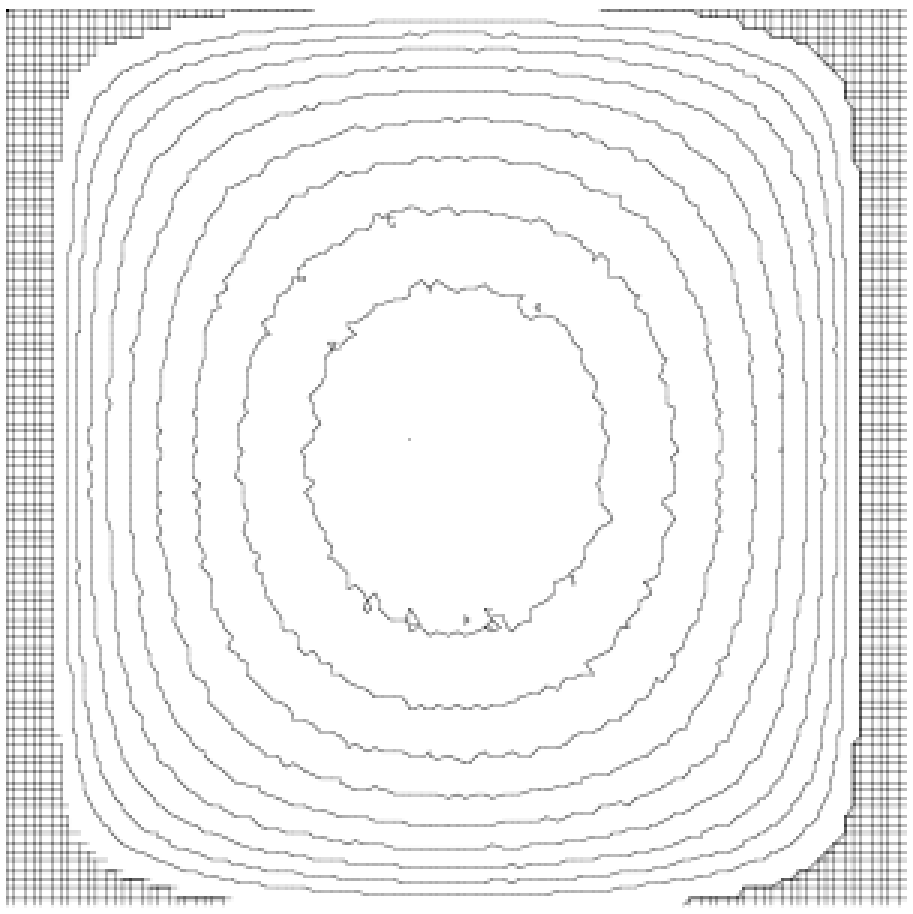

The contours of constant density do not have the same shape from center to edge (Fig. 24.12).

Fig. 24.12. Contours of constant density for Hamiltonian distribution with $\mu = 0.98$.

To help quantize the particle motion, we need a definition of its phase advance, or tune. It is useful to define the instantaneous local tune for each particle from the net restoring force minus the space-charge force acting on it, expressed as an equivalent phase-advance over the external focusing period length. The result is a cleanly defined line vs. radius, as shown in Fig. 24.13 for the Hamiltonian initial distribution at $\mu = 0.84$. Lack of smoothness at smaller radiii is due to statistics.

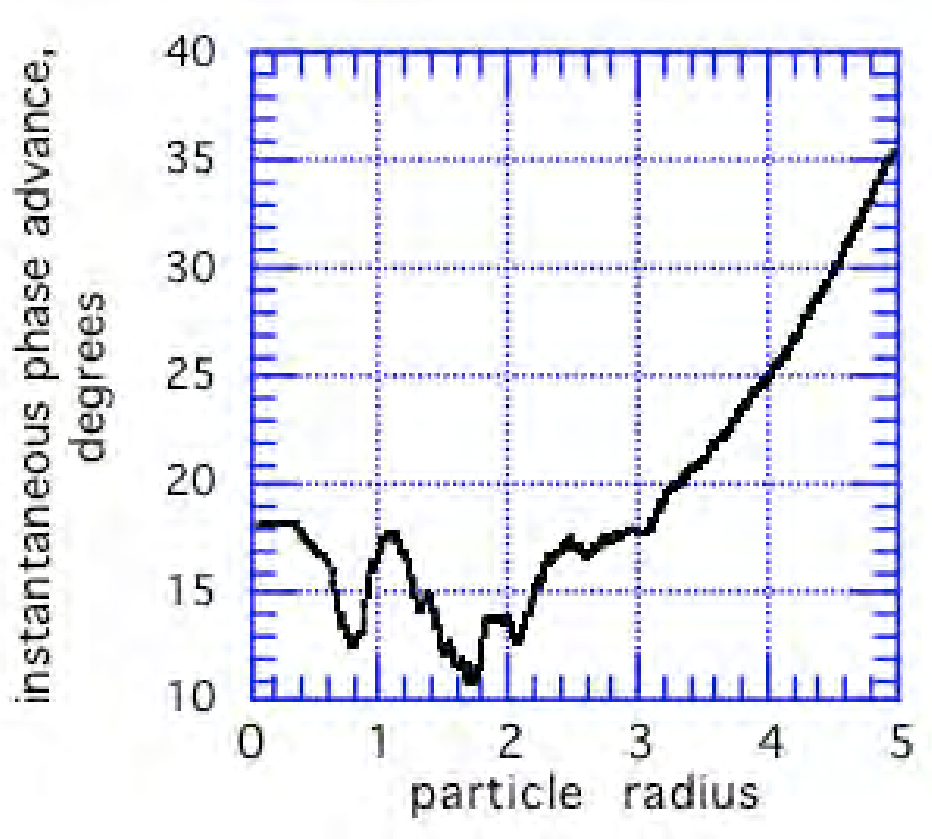

Fig. 24.13. Initial phase advance of all particles in 10000 particle Hamiltonian,

Figure 24.14 shows how the distribution tune shift varies during the mismatch cycle, becoming very strongly negative as the central density peaks. This enormous, time-varying tune spread shows that particles have a chance at each resonance. However, since few actually do get caught in the main mismatch resonance under study here (discussed further below), the time-varying separatrix clearly does not maintain the resonance opportunity for very long.

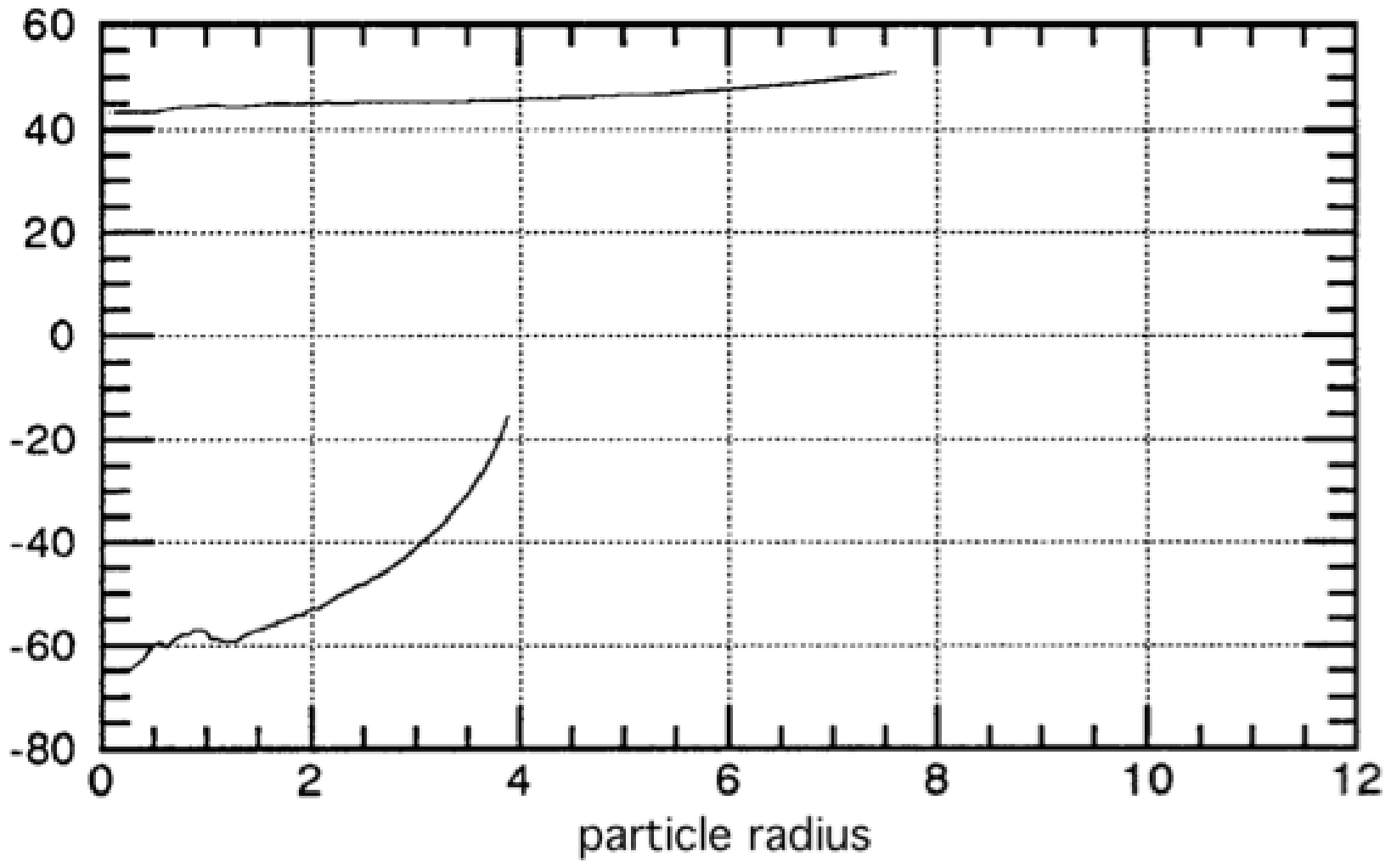

Fig. 24.14.a. Run with initial Hamiltonian distribution with maximum radius = 5, $\mu = 0.84$. A mismatch of 1.5 increases the maximum radius to 7.5. Instantaneous phase advance vs. radius for all particles; upper curve is at a central density minimum near the beginning of the run, lower curve is at an adjacent central density peak.

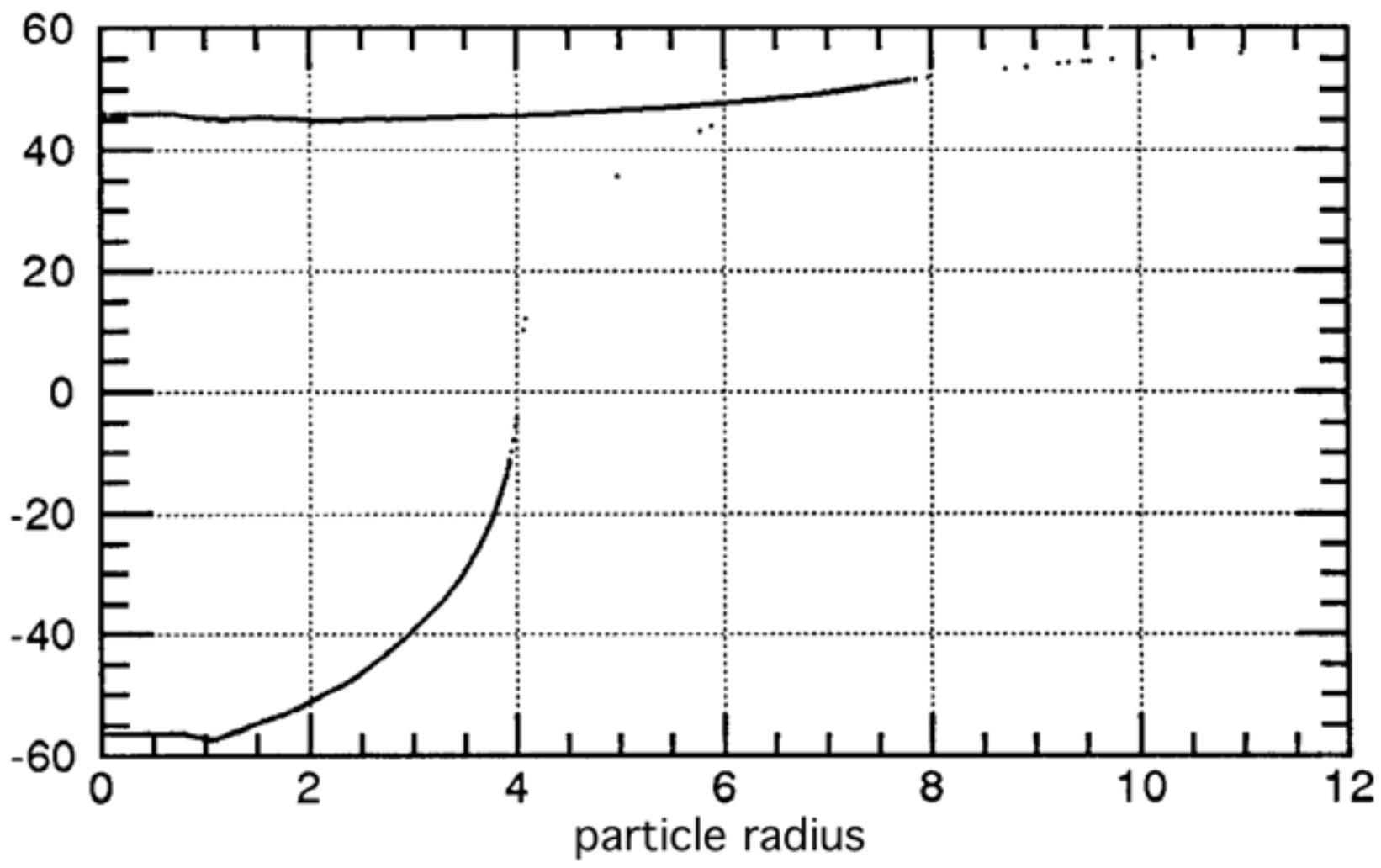

Fig. 24.14.b. As in Fig. 24.14.a.; upper curve is at a central density minimum near the end of the run after the halo has formed, lower curve is at an adjacent central density peak.

## 24. 10 Integration of Results Over Distance

Runs were made over a broad grid of mismatch and space-charge factor, for two initial conditions before the mismatch was applied – constant beam radius, and constant rms emittance. Each run was for 125 betatron periods. The maximum particle radius and maximum and minimum particle tunes achieved during the entire run were recorded. These runs were intended to explore a number of issues.

One was to see if the integrated results would indicate some of the resonance characteristics seen in the full trajectories, such as Fig. 24.9. Figure 24.15.a shows the maximum radius vs. the space-charge factor for a mismatch of 1.30 applied to an initial radius of 5. (Each point is a separate run.) The onset of the rational number resonances occurs before the main resonance is excited. Slope adjustments made on lines joining sets of peaks in different areas can identify which rational number resonances contribute most strongly in a region.

The onset of different resonances also appears to be correlated with the minimum tune, as indicated by the wavelength ratio plot of Fig. 24.15.b. A change in the active resonances appears to occur whenever the ratio is decreased another half integer. Above zero, the shape of the ratio curve is parabolic, while below zero, where the space charge force is larger than the external force, the ratio is linear. It is seen that the integrated results do carry features of the detailed particle orbits. Further work may be able to show the explanation and utility of these observations.

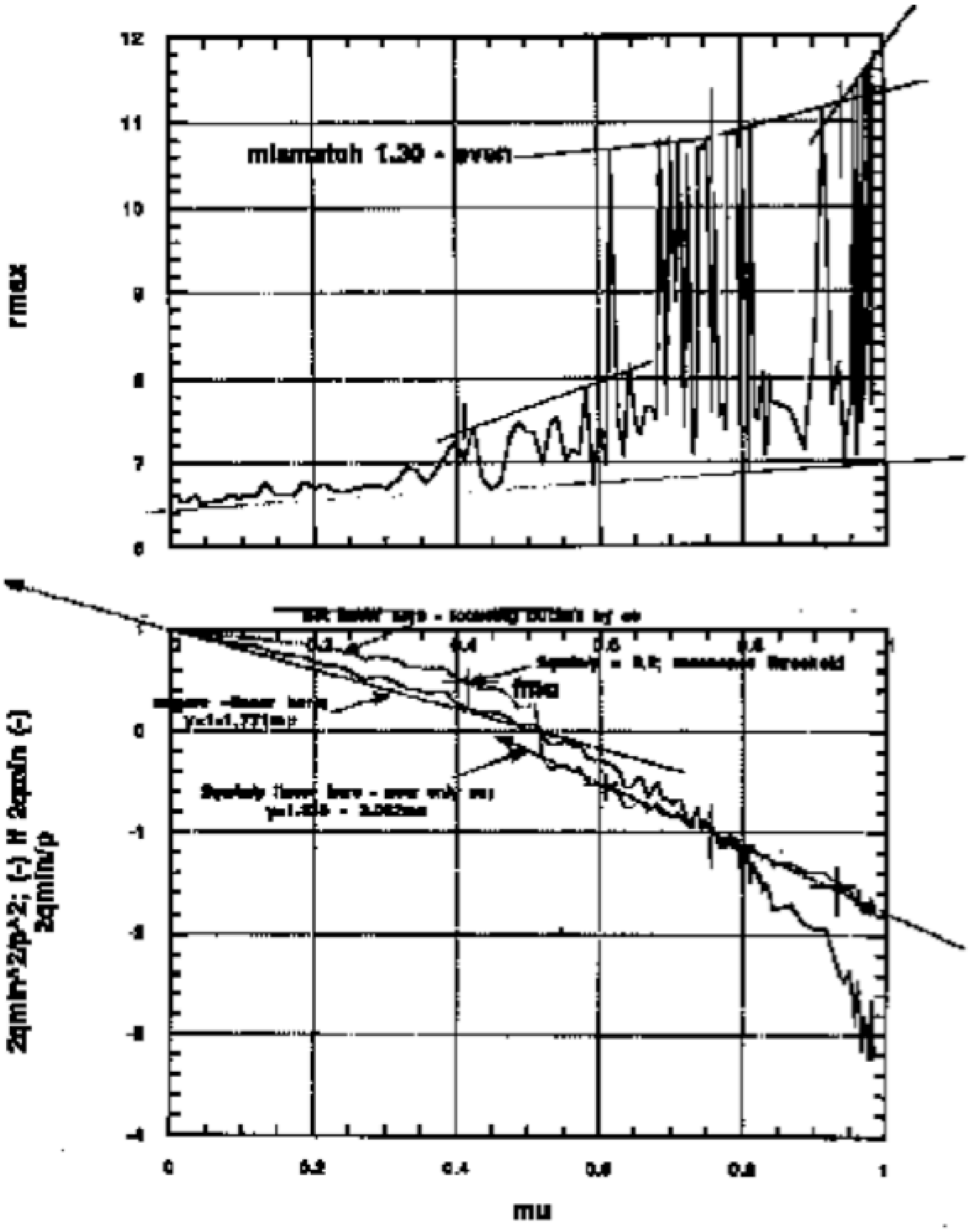

Fig. 24.15. Mismatch = 1.30. (a.) Maximum radius achieved at each value of μ. (b.) Ratio of twice the minimum tune wavelength (2qmin) to the central density oscillation wavelength (p), and the square of this ratio.

The statistics for the resonance onset are not very good, to an extent beyond even what might be expected for the 10,000 particles per run to 125 betatron periods. Figure 24.16 shows the same 1.3 mismatch with a much finer grid of space-charge factor points. Because not many particles actually hit the resonance, the statistical error in any one run is high, and we may or may not see the onset for space-charge factors near the threshold. But more importantly, the fundamental resonant nature of this Hamiltonian system causes uncertainty. We return to this point below. However, by making a large number of runs on a grid, the threshold for halo generation by mismatch can be seen clearly, as shown in Fig. 24.17. This plot keeps the initial rms emittance (before the mismatch is applied) constant over the entire grid.

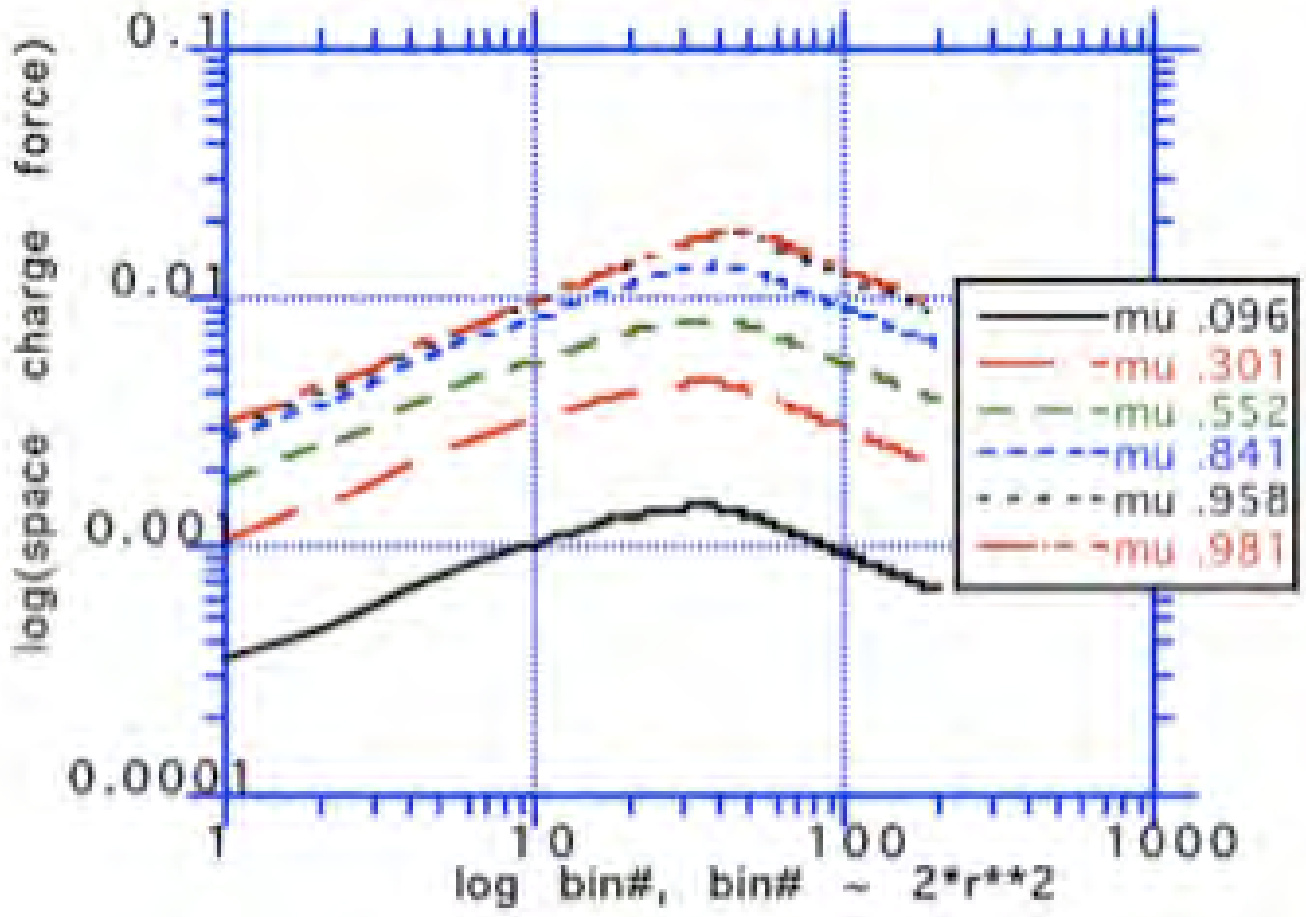

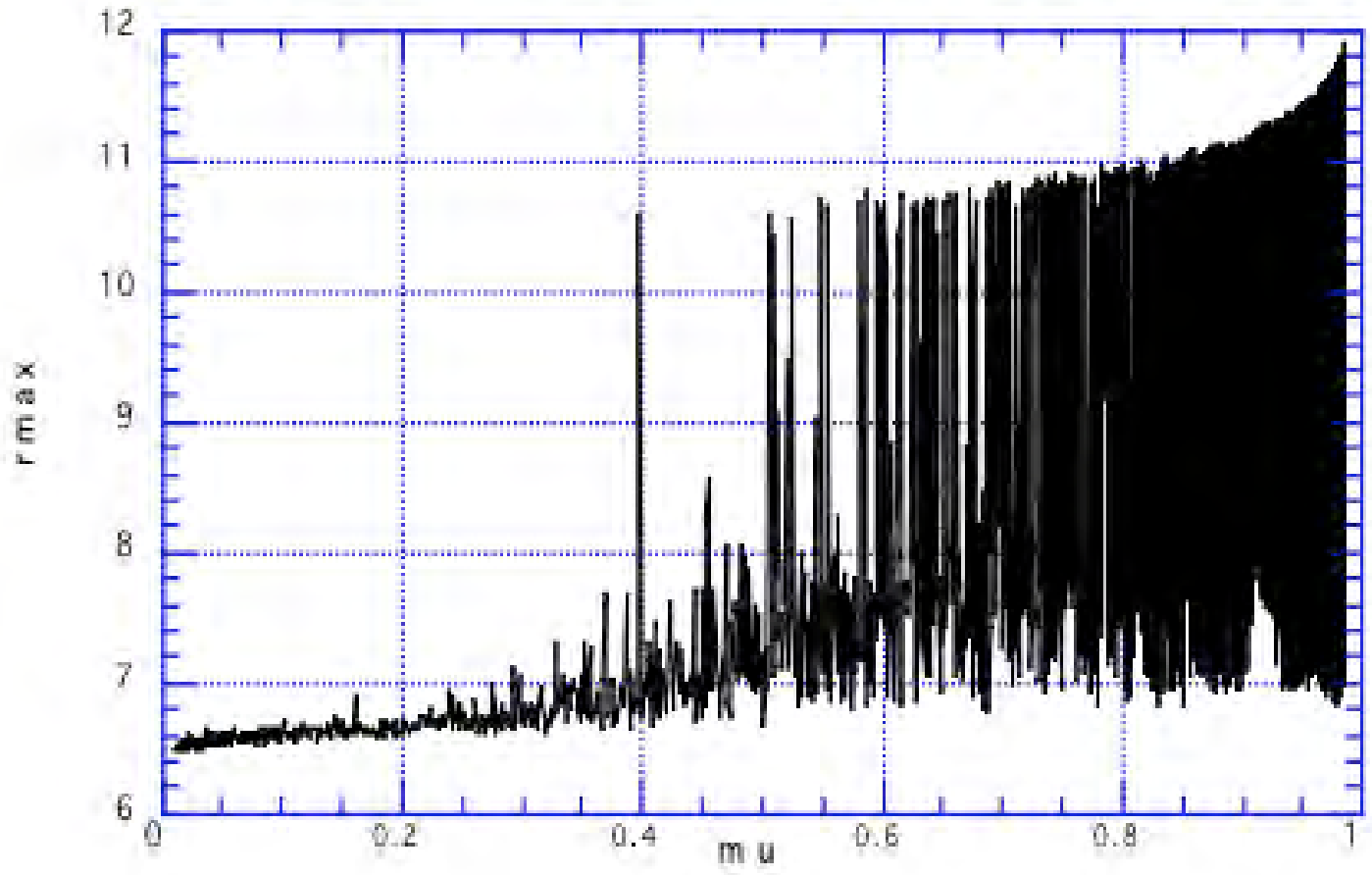

Fig. 24.16. Repeat of Fig. 24.15.a. with finer grid in μ.

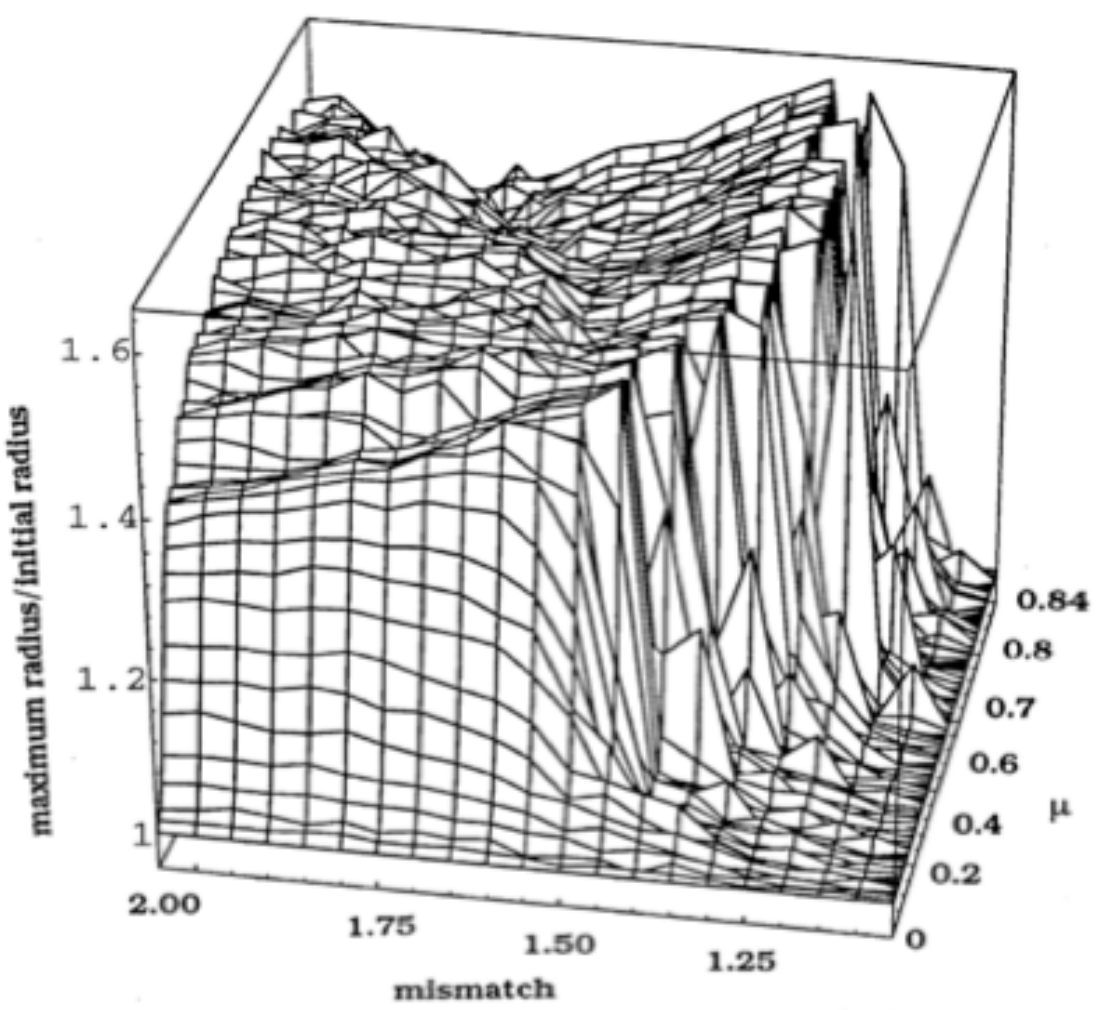

Fig. 24.17.a. Maximum halo radius achieved as a function of space-charge μ and initial mismatch. Constant rms emittance was maintained at each point before application of the initial mismatch.

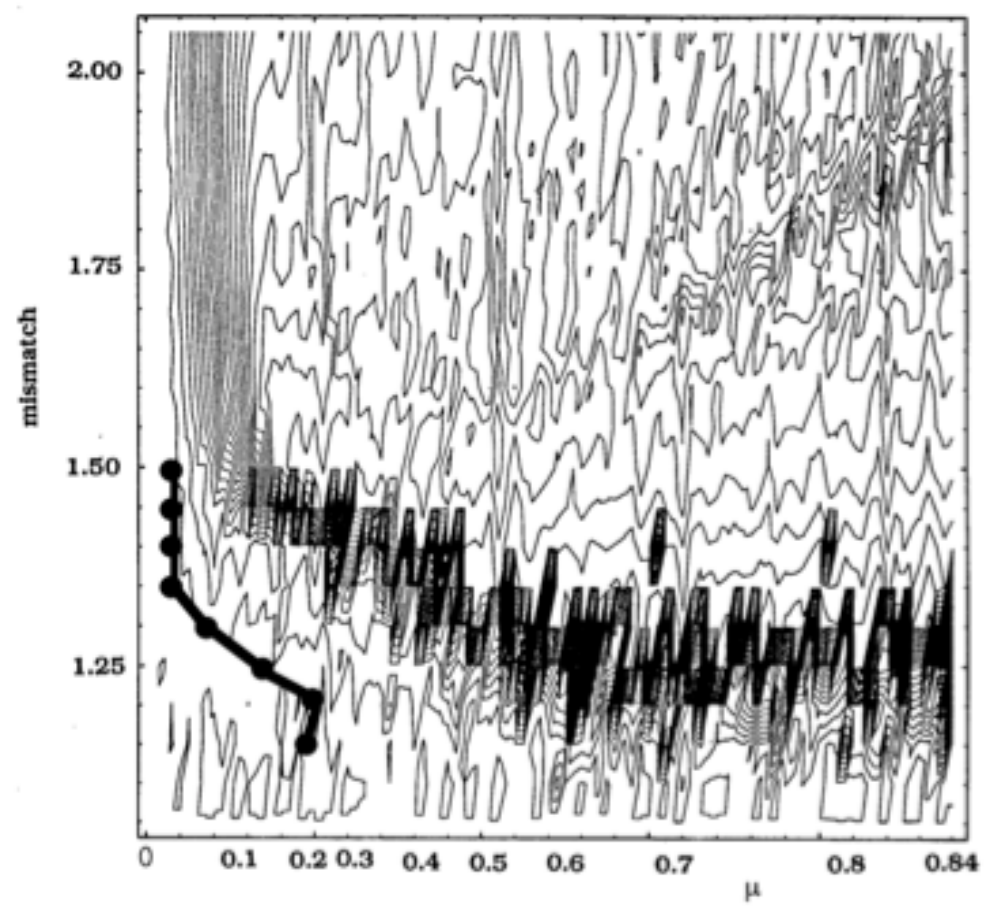

Fig. 24.17.b. Contour plot of maximum halo radius, per Fig. 24.17.a. Dots connected by lines are threshold points determined by the resonance width method.

## 24.11 Threshold of Mismatch Halo Generation

At zero space-charge, the threshold must go to infinity. The shape of the threshold curve has so far not yielded to analysis. However, the results indicate the wisdom of an old practical operating rule to keep mismatch factors below 1.2.

## 24.12 Extent of Halo From Mismatch in the Radial Focusing System

Normalizing to the initial radius plus the expected added radius from mismatch shows that the halo in this particular system (continuous, linear, radial external focusing system with initially equilibrium Hamiltonian particle distribution) is limited to ~1.5-1.6 times the expected radius (including the initial mismatch) over a wide range of mismatch and space-charge factor. This limiting is more strict than predicted by Gluckstern's formulas or test-particle methods of the first type, which do not allow for emittance growth and other factors.

## 24.13 Central Density Oscillation Wavelength

Figure 24.18 shows the central density oscillation wavelength scaled to (betatron wavelength) /sqrt(4 - 2μ) from the simple breathing mode formula for the KV beam. It is seen that the wavelength does not deviate too far from this scaling, although it does depend also on the mismatch factor.

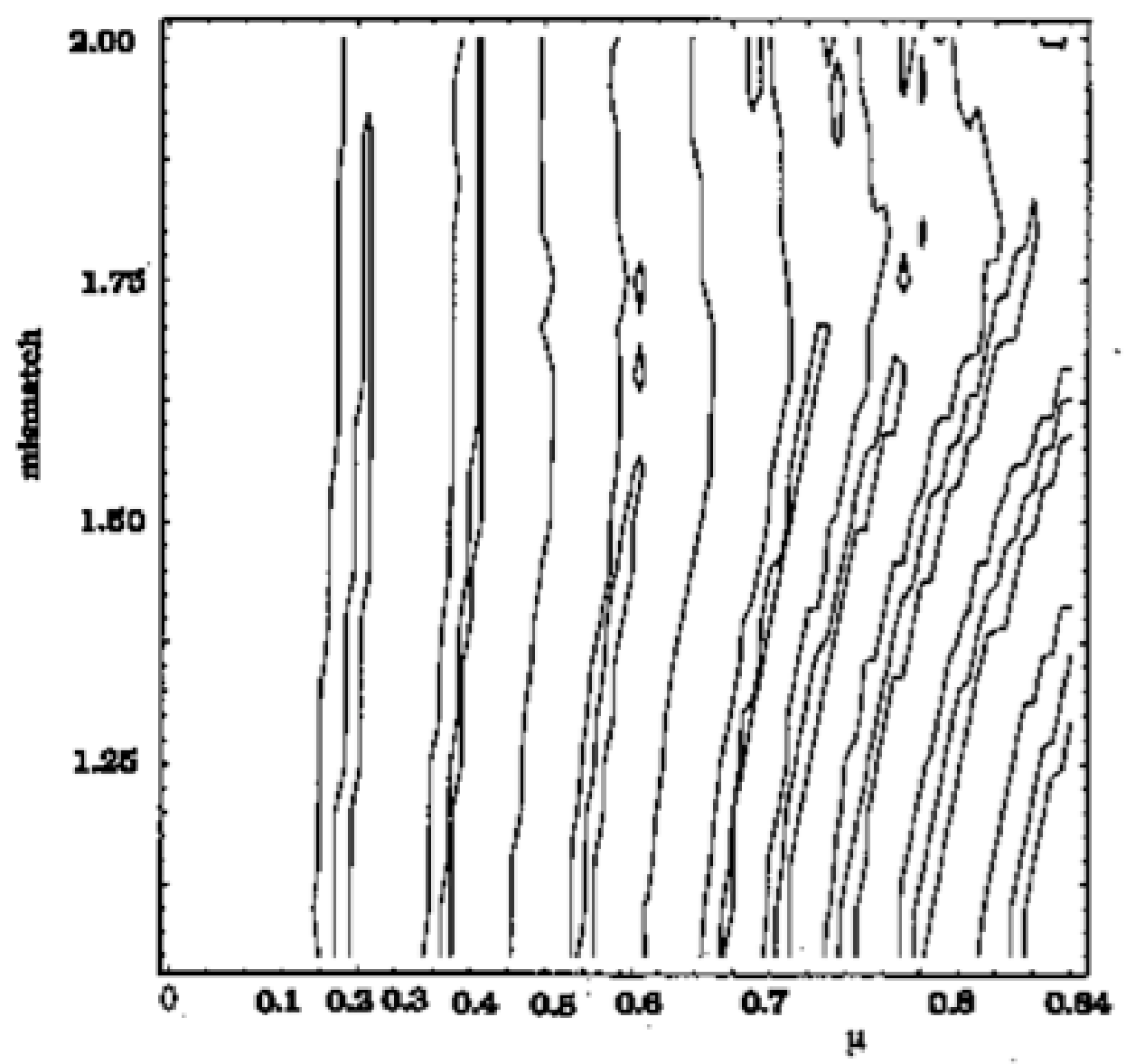

Fig. 24.18. Central density oscillation wavelength scaled to (betatron wavelength)/sqrt(4 - 2μ).

## 24.14 Resonance Width

Figure 24.19 shows the ratio of the doubled instantaneous particle phase advance to the central density oscillation phase advance for the particle of Fig. 24.9. It is sensible to surmise that if the integral of the instantaneous phase advance over a core oscillation period is ≥ 180˚, the particle will have beem caught by the main resonance and will spend some time with a phase advance of > 180˚ per core oscillation period until it falls out of resonance and back into the core. This technique may afford a major diagnostic, with suitable adaptations for multiple resonances. While arrived at independently, this notion of a particle advancing more or less than once around the phase-space wrapped into a cylinder representing the wavelength of interest has been used recently in general chaos studies, and is known as the "width" of a resonance [Easton 1993, Benkadda 1994]. The time needed before the width is exceeded is the "exit time". The "transit time" is the time from when the particle's width exceeds one, until it again becomes less than one.

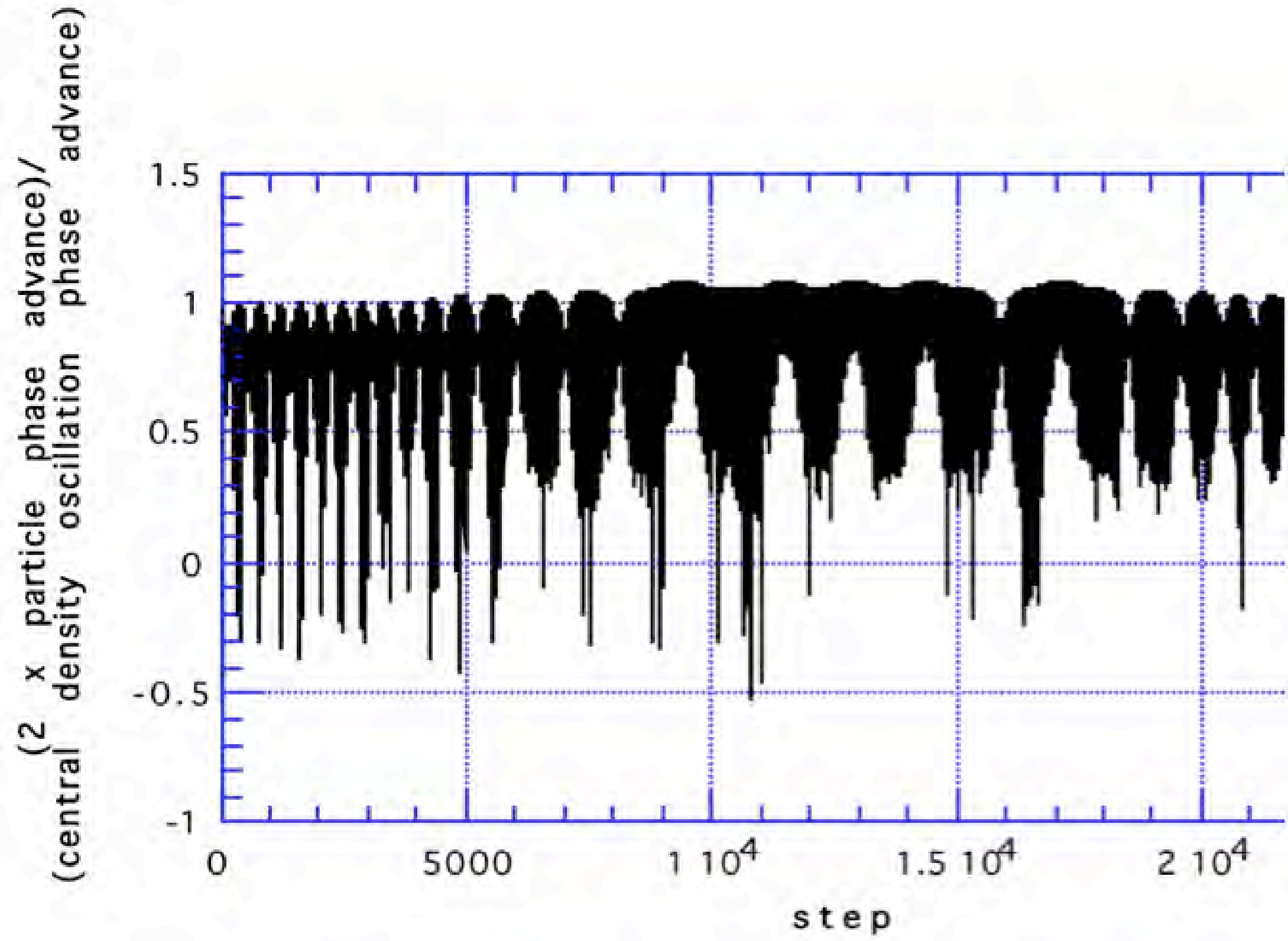

Fig. 24.19. Ratio of doubled instantaneous phase advance to control density oscillation phase advance for the particle of Fig. 24. 9.

Figures 24.20 and 24.21 show exit time computations on the width of the central density oscillation wavelength for two sets of conditions, showing the number of particles exiting and entering at each step, and the transit time for reentering particles. Because the mismatch oscillation persists, the halo is being generated throughout the run. Particles may go in and out of the halo many times; more with increased space-charge, but they stay out longer when there is less space-charge. Figures 24.20.c and 24.21.c show a somewhat erratic equilibrium may be achieved late in the run. The distinct oscillation is noteworthy and may be explained by Langmuir waves [Lagniel 1994].

Figure 24.22 shows the statistical distribution of the first exit of the run, for a set of runs at different mismatches, all at the same space-charge factor μ = 0.84. The runs were each repeated 12 times with different random numbers. Two families of runs were made. The figure shows the statistics when the central density oscillation wavelength was found during the run. A second family was made in which the central density oscillation wavelength was estimated from (betatron wavelength)/sqrt(4 - 2μ), since Fig. 24.19 showed that this is a reasonable approximation. The statistics are similar and the averages are very close.

The scatter in the exit times, especially at the low mismatches, is very typical of stochastic processes in near-Hamiltonian systems like this one, as is the slow equilibration in Figs. 24.20.c and 24.21.c. The bounding circles and island chains are very sticky, and while there might be eventual ergodicity, the time to reach it would be extremely long [Meiss 1994] and well beyond our interest. The dynamics in the present problem perhaps cannot be described by any diffusion process that invokes Brownian motion, an eventual thermal distribution, or uncorrelated transport probabilities. (There are other cases where diffusive-like equations can accurately describe chaotic dynamics, e.g., [Struckmeier 1994].)

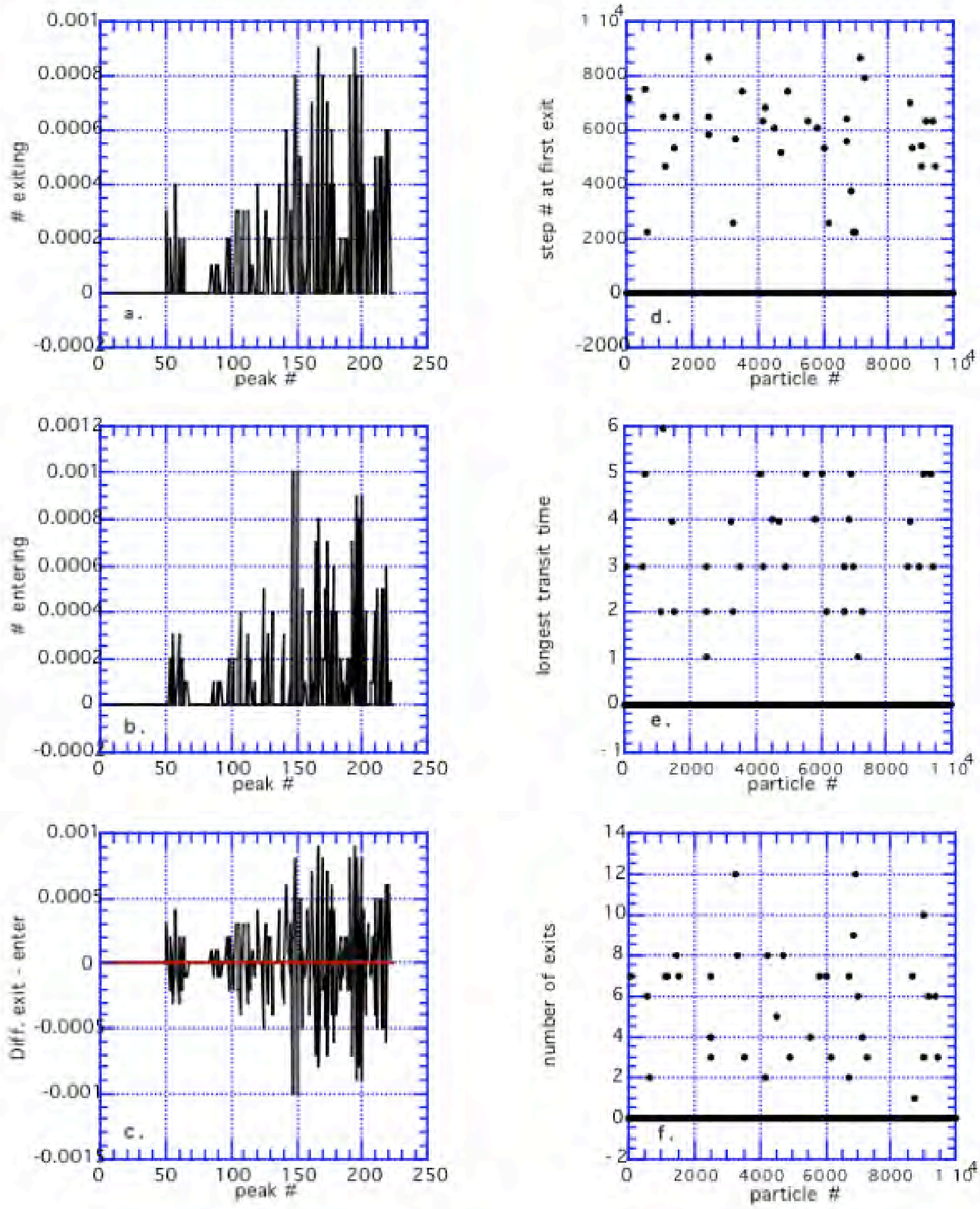

Fig. 24.20. Mismatch = 1.30, cay = 0.50, mu = 0.552. a.) Fractional number of particles exiting the core during each central density oscillation. b.) Fractional number of particles re-entering the core during each central density oscillation. c.) Exiting minus entering particles. d.) Step number at first exit. Most of the particles never exit. (Betatron period is 80 steps, breathing mode wavelength is ~46

steps. e.) Longest transit time, in number of core density oscillations, of particles that exited into the halo. f.) Number of times that each particle exited into the halo.

The width method using the estimated central density oscillation wavelength was used over the matrix of Fig. 24.17. The threshold is found to be at the lower edge of the major increase in $r_{max}$ as indicated on Fig. 24.17.b. Other features are being analyzed.

Definition of "the halo" as particles whose phase advance exceeds 180° over a period of the central density oscillation is precise for this system, which has an easily measured disturbance wavelength. The situation will be more difficult with weaker disturbances and multiple driving frequencies, but modified width or spectral concepts should have utility.

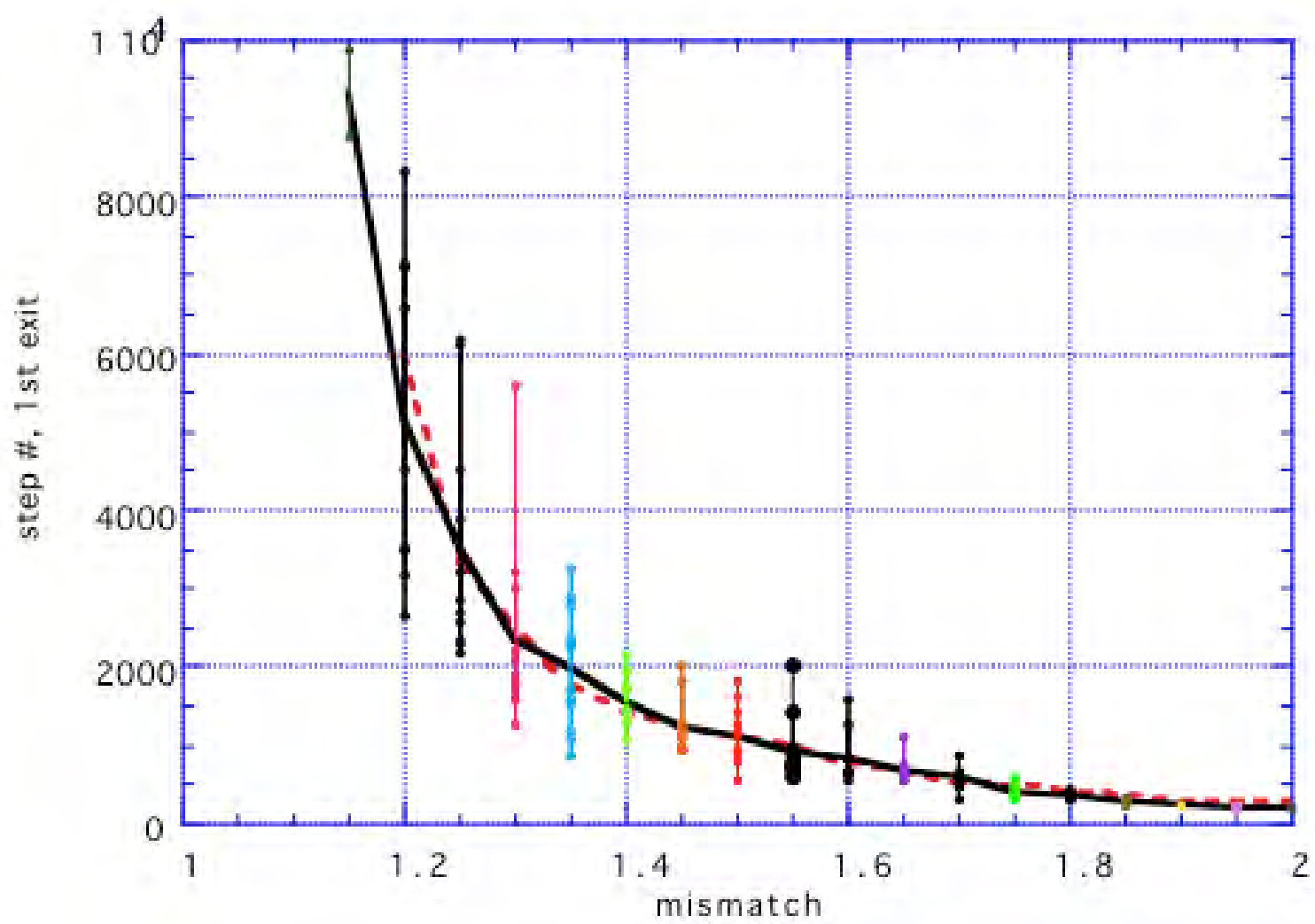

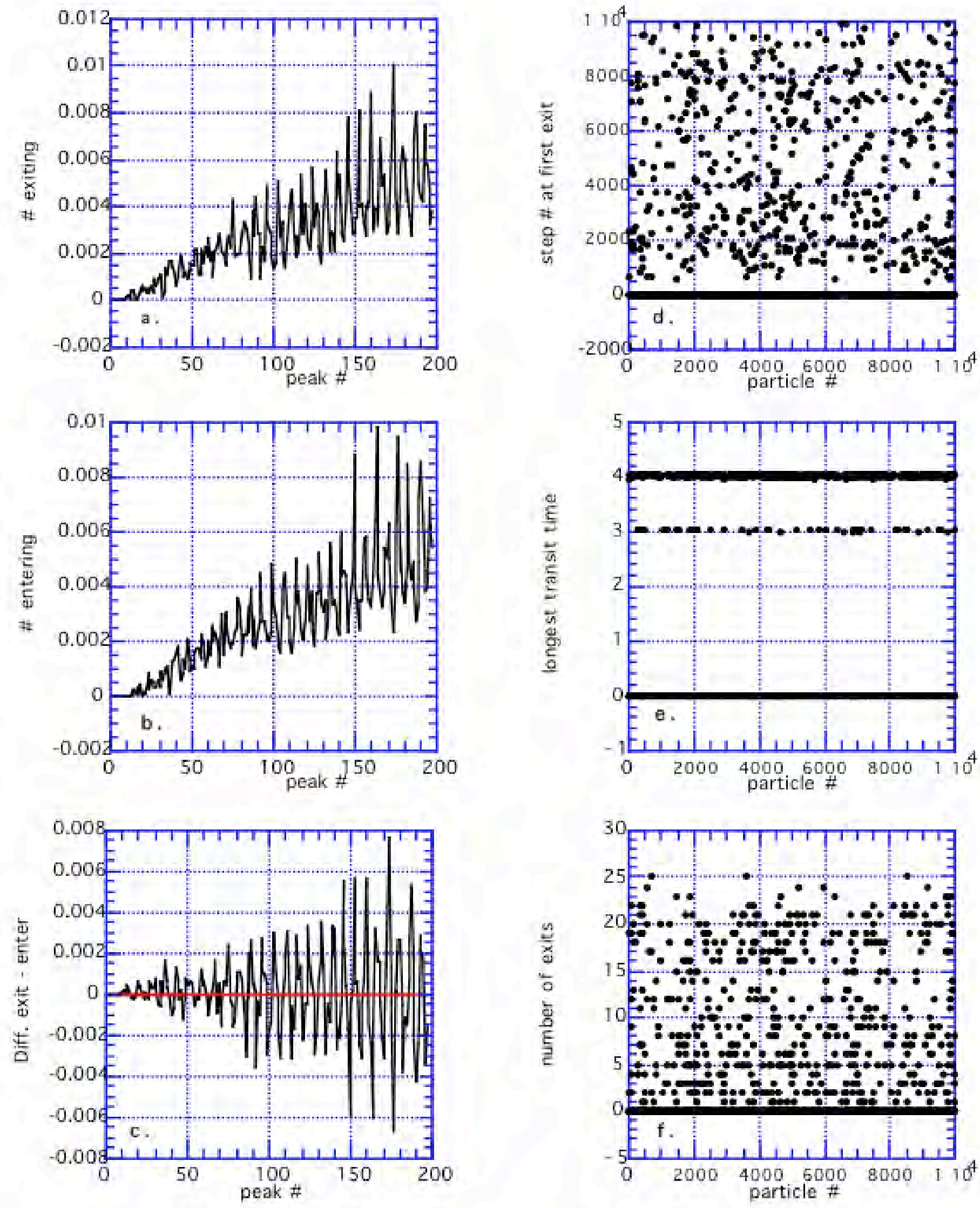

Fig. 24.21. Mismatch = 1.70, cay = 1.0, mu = 0.84. Same as Fig. 24.20. Breathing mode wavelength = ~51 steps.

Fig. 24.22. Statistics on exit times for mismatched Hamiltonian distribution with $\mu = 0.84$. Runs made with 10000 particles, repeated 12 times each with different random numbers. Averages are shown – solid line is with central density oscillation wavelength found during the run; dotted line is with central density wavelength approximated by (betatron wavelength)/sqrt(4-2$\mu$). Exponential fit to averages for mismatch $\geq$ 1.30 gives y = 35100. * e^(-3.8465) with confidence factor R = 0.998.

## 24.15 Separatrix Crossing

In this mismatch problem, the main parametric resonance separatrix is moving rapidly, sweeping over a large area of phase space. [Bruhwiler 1994] and others have developed a separatrix crossing theory that shows how the adiabatic invariant is invalid at the time an orbit crosses the separatrix, and how trajectories will eventually fill the whole area swept by the separatrix. In the mismatch problem, this is equivalent to the statement that trajectories will be found everywhere in the expected phase-space area of the mismatch. (In the rf linac longitudinal phase-space, it means that filamentation will eventually fill the entire bucket.) Bruhwiler also found that some particles leaked outside the separatrix in his examples, and indicated this meant that correlations between the conditions on either side of the separatrix are important. His methods might be able to shed more light on the leakage particles. It is possible that changes in the separatrix from the emittance growth allowed in the self-consistent model are instrumental in the leakage mechanism.

## 24.16 Higher Dimensions

Little is formally known about transport in systems with two or more degrees of freedom (or four-dimensional phase space) [Meiss 1994]. It is known that there will be a stochastic web, and the possibility that particles could escape, for example, beyond the self-limiting boundary of the mismatch problem studied above. It is clear that the main features of halo generation from time-varying density fluctuations will persist, in particular the rich resonant structure and the tendency toward self-limiting.

[Lagniel 1994] has explored the features of the quadrupole channel, including the addition of an additional driving term from the longitudinal coupling in a linac, using the test particle method. He shows, Fig. 24.23, that the rich structure of resonances and stochastic layers is present, and that Arnol'd diffusion is present when two degrees of freedom are coupled. This work must be extended to self-consistent studies. It has not been shown yet, for example, how far particles go beyond the expected boundaries, or how the trajectories are modified by acceleration. Development of tools such as the width method, spectral and statistical techniques will be paramount.

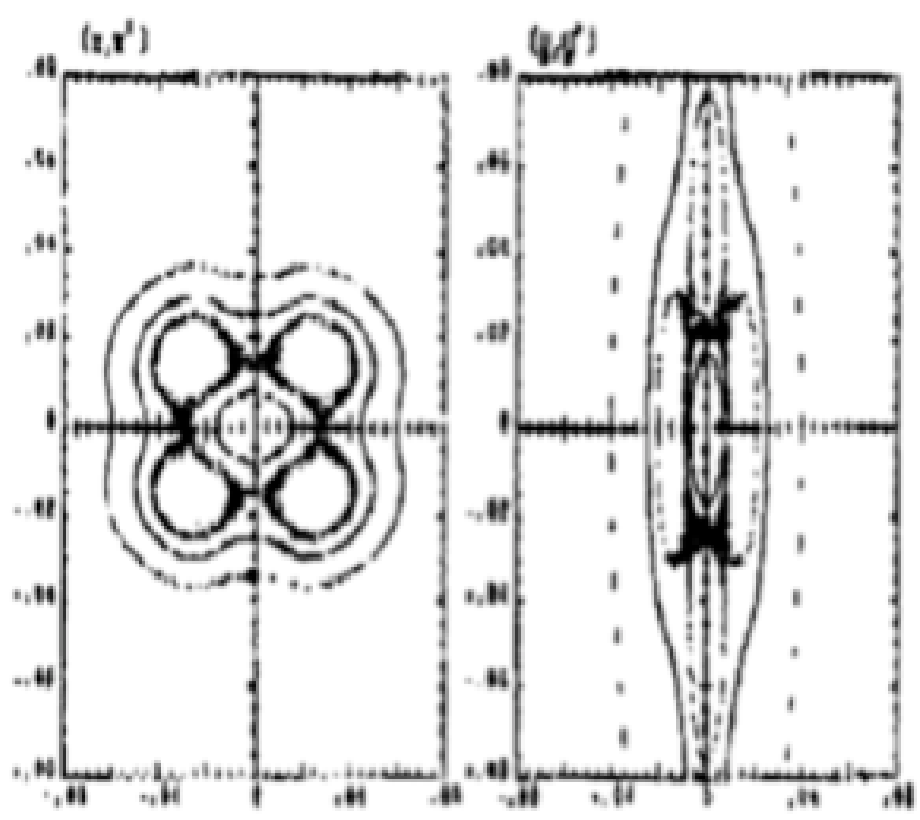

Fig. 24.23.a. (x-x') and (y-y') Poincaré surfaces of section for $\sigma_o^t = 100°$ and $\sigma^t = 70°$ (test particles without (x,x')-(y-y') coupling) (From Lagniel.)

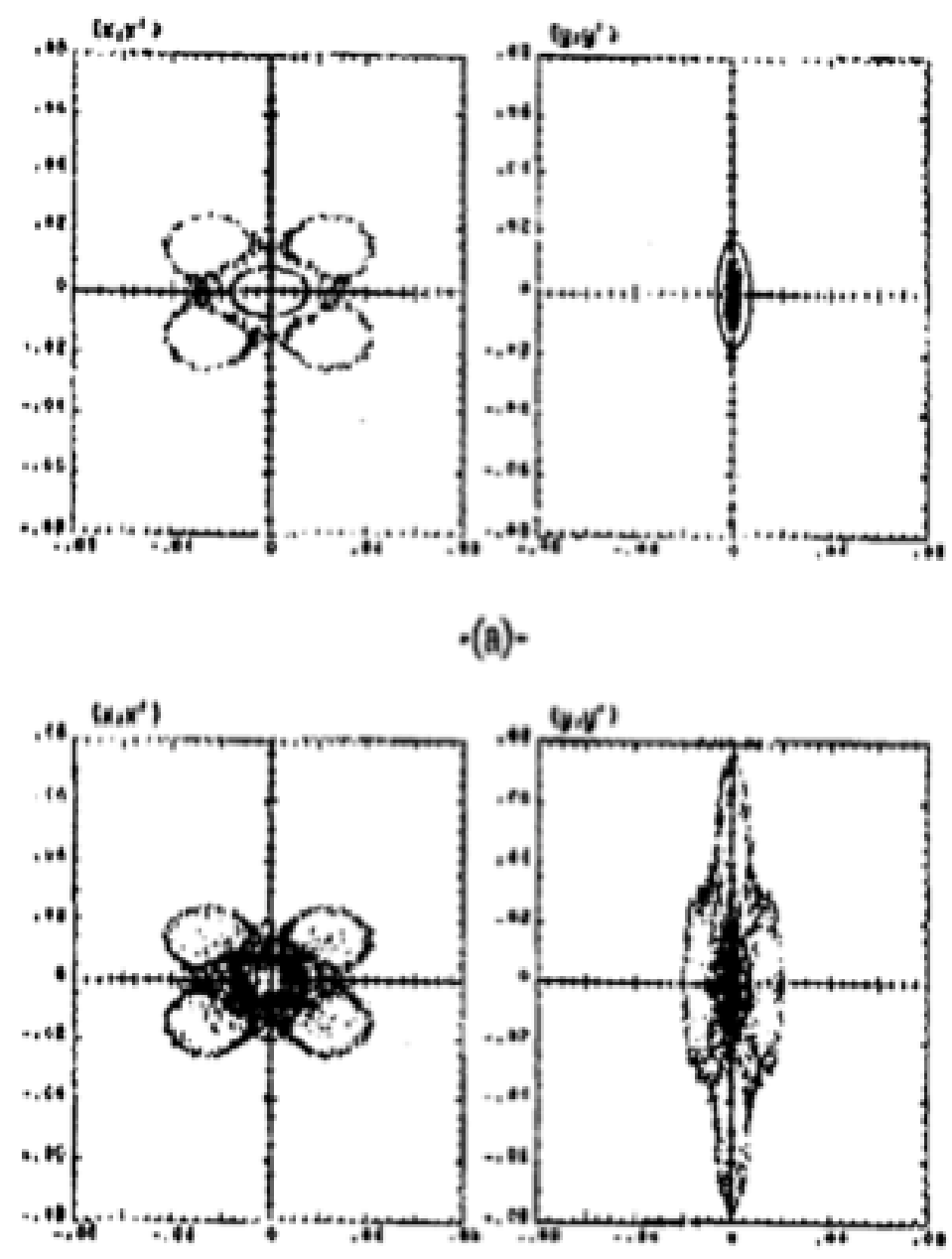

Fig. 24.23.b. Arnol'd diffusion for $\sigma_0^t = 100°$ and $\sigma^t = 70°$. One particle injected at x = 0.028, x' = 0.0, y = 0.001, y' = 0.0 is followed during 1000 (upper) and 2000 (lower) periods. (From Lagniel.)

The 60-cell bunching section of a high-current RFQ was studied [Jameson 1993] as an initial foray into the multidimensional problem. In this section, the beam is at injection energy and encounters a steadily rising bunching voltage. The forming bunch makes a time-dependent density distribution. A few (order 0.2%) particles were anomalously repelled longitudinally far from the bunch point, in some cases into the next bucket. Using knowledge of the core/single-particle self-consistent dynamics and statistical procedures, it was found that these extraordinary orbits were strongly correlated with very close encounters with the transverse xx'yy' origin. (Fig. 24.24).

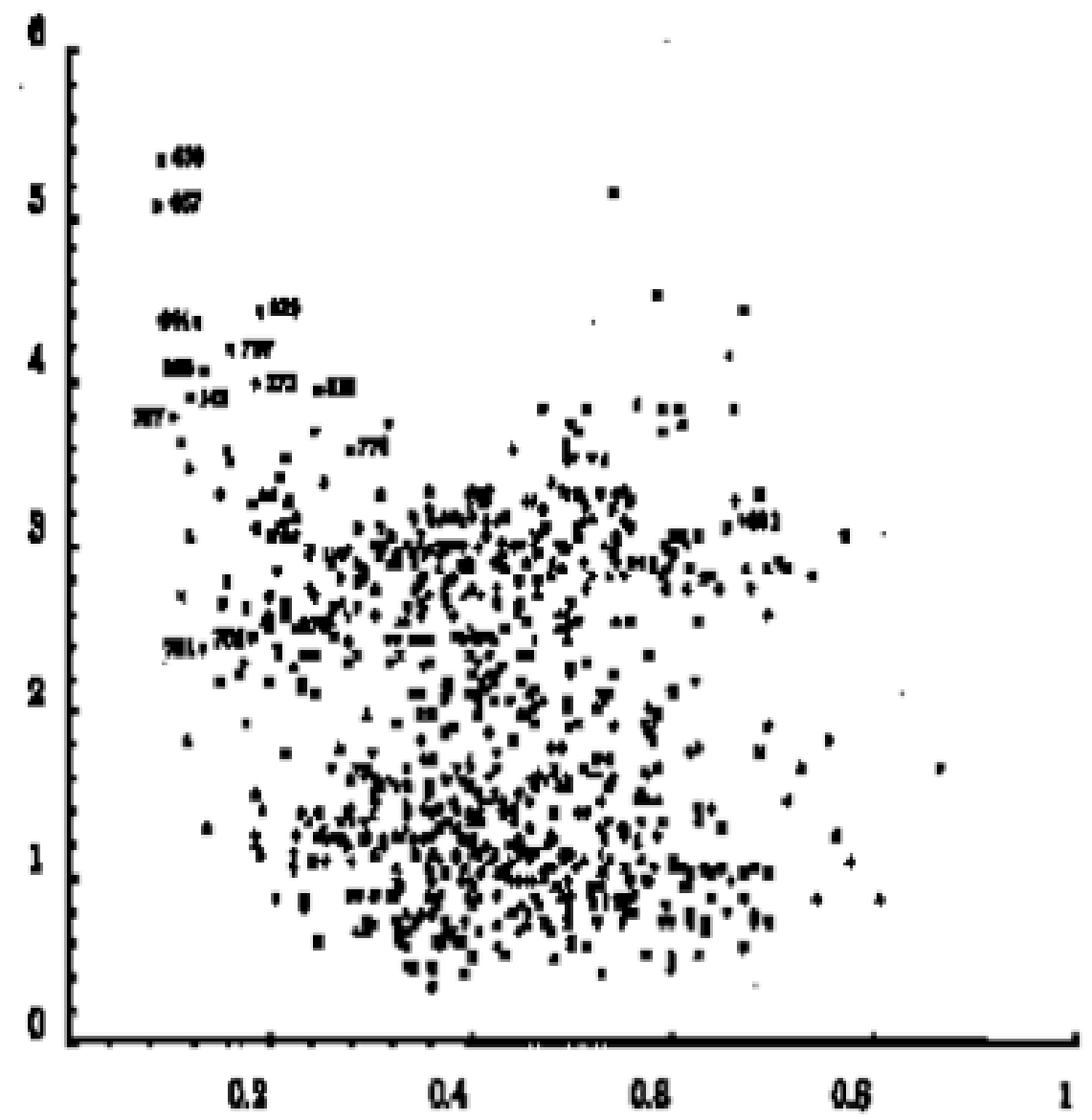

Fig. 24.24. Correlation plot for particle traversing the first 60 cells of a 125 mA deuteron RFQ. The particles were selected if they experienced a minimum in xx'yy' radius (abscissa) before they experienced their maximum zz' radius (ordinate). The particles among the special group being tracked are labeled.

## 24.17 Conclusions

Investigation of the basic single-particle dynamics is important to guide development of theory and diagnostics.

Self-consistent methods are essential to obtain practical results for the accelerator problem [201].
Time-varying core/single-particle dynamics is a major halo-producing mechanism. In the simple linear, continuous-focusing channel, taking emittance growth into account, halo from mismatch is approximately limited to 1.5-1.6 times the expected beam radius.

Diagnostic development is important, and will become more so when higher dimensions are considered. The resonance width method is powerful and easy to compute.

Similar preliminary investigation and diagnostic development in the full-dimensional high-intensity accelerator case should lead to productive investigation by massively parallel computing.

---

201 In spite of access to all of this information, an egregious statement has been made that “halo particles originate from particles already in the outer part of the phase space distribution (clearly shown not true)” “and there are relatively very few of these” “and therefore it is not necessary to use self-consistent methods”. (Could this be “an assumption by a theoretician”, or a NIH evasion? It certainly reveals lack of knowledge of space charge mixing and phase-space transport…)

[eltoc]

# PART 5 — SOME LINAC INVESTIGATIONS

## Constructed RFQs and J-Parc Linac

RFQs have been constructed and operated from LINACS design and simulation for a number of projects (Laser ion source injection, ADS, J-Parc).

Most of these were for project specifications other than very low beam loss, with severe length requirements, sometimes based on existing real estate, ion sources, cost, etc., resulting in the use of non-EP designs. In these cases, the amount of testing specifically to test the RFQ performance was limited by project schedules, manpower, instrumentation, etc., but in each case the transmission was close to the simulated result. Also, full project simulations and analysis were able to closely elucidate experimental results.

An EP designed RFQ was used in the first Chinese beam test of the direct plasma injection scheme [202], but again without specific testing of the RFQ.

The first dedicated experiment to design, simulate, build and test a fully equipartioned EP 60mA H-RFQ is reported in [203]. The experimental and simulated results are very close for transmission and output transverse phase space, verifying the elements of the design and simulation. The very helpful reviewers comments are significant: "*The RFQ designed with the fully beam-oriented is a world's first one, and the verification was conducted through the beam commissioning. This paper is important to the field of the high intensity RFQs because the design scheme can be standard.*" – "*This paper deals with the design and experiment of an RFQ based on equipartition principle. This is a rare paper that covers the issue from the design of the RFQ to the beam experiment. This is an excellent work.*"

---

202 "Beam Acceleration of DPIS RFQ at IMP", Zhouli Zhang, …, R.A. Jameson, et. al., MOPC028 Proceedings of IPAC2011, San Sebastián, Spain

203 "Development of a radio frequency quadrupole linac implemented with the equipartitioning beam dynamics scheme", Yasuhiro Kondo, Takatoshi Morishita, Robert A. Jameson, 2019, https://doi.org/10.1103/PhysRevAccelBeams.22.120101
(DOI: 10.1103/PhysRevAccelBeams.22.120101)

The DTL, SDTL and superconducting sections for the Neutron Science Project at JAERI were designed to be EP [204], and the design of the eventually constructed entire J-Parc linac is EP. During initial commissioning studies, unanticipated intrabeam scattering causing beam loss was discovered and retuning off-EP has dominated further investigations. Chapter 27 explores this linac and compensation for the intrabeam scattering. Results are unfortunately not conclusive; there is indication that the underlying EP design makes the compensation less problematic, but as with all user-facility machines, the scientific urge to understand has given way to daily operation.

[eltoc]

# Chapter 25 – Initial *LINACSrfq* Attack on RIKEN 400mA $D^+$ RFQ

R. A. Jameson – April 2017

## 25.1 Introduction, Comments on IFMIF/EVEDA Design

14 March 2017 - Decided to work on 500mA RFQ RIKEN project, to prepare for visit of Xingguang Liu in May.

20 March 2017 - Think maybe on increasing acceptance in area of shaper end – reviewed what Comunian did for EVEDA… Big differences – *2term machining* "to avoid multipoles" (maybe also transverse cross section ?), and limited voltage pattern for practical tuning reasons, r0 smooth and follows voltage law…

21 March 2017 - Comunian Vrule formula in paper seems wrong, but characteristic can be set up. 'IFMIF RFQ EPAC2008.nb'

EPAC2008: 98.9% waterbag parmteq "which uses field expansion close to axis", 98.8% Tou "which includes MPs and images" {and needs many particles to compare to *PARMTEQM* due to large number of 3D cells to populate), both run with 1 million particles. HB2010 shows 94% for Gaussian, 99% for WB. Fig 5 shows involvement with $\sigma_l/\sigma_t$ 0.33 resonance.

Comparison to this other code used by Comunian cannot be accepted here as particularly meaningful. This code was at one point (~2007-2010) very extensively compared to several other codes. Similarity to parmteq is also telling.

Codes should be compared over a broad parameter range, rather than non-informative anecdotal case comparisons. Such a test can be set up with *LINACSrfqDES*, because all parameters, including space charge physics, are under control, allowing a family of RFQs to be generated with only one parameter varying – one of the characteristics of a good experiment. A multiple-RFQ family with varying end-of-shaper (EOS) apertures was generated; there is a clear optimum end-of-shaper aperture that gives the maximum transmission and accelerated beam fraction for the RFQ family. The family is then simulated by the various codes that one wants to compare.

The main interest is comparison of the aperture giving optimum transmission, rather than on the absolute value of the transmission – because it is known in engineering practice that the "apparent

204 "Beam Dynamic Study of High Intensity Linac for the Neutron Science Project at JAERI", K. Haswgawa, H. Ogiru, Y. Honda, H. Ino, M. Mizumoto, R.A. Jameson, APAC 1998, http://citeseerx.ist.psu.edu/viewdoc/download?doi=10.1.1.613.3462&rep=rep1&type=pdf

optimum" can be influenced very much by the coding practices, and we want to be confident that the real physical optimum is found. The absolute value of the transmission also depends on the coding. For example, *PARMTEQM* multipole field expansion gives accurate field only near the axis, using position rather than time as the independent variable is another reason that results are optimistic, etc., etc. But more seriously, the *PARMTEQM* transmission curve for the multiple-RFQ family shows a considerably different optimum EOS aperture.

This other code showed a different transmission curve from *PARMTEQM* and several other codes. Investigations clearly indicated several possible problems, but attempts to discuss produced only anger that there could possibly be any questions about this "perfect" code, that agreed *to four decimal places* (on an anecdotal case) *with another code known to have accuracy problems,* and "agrees well" *with the known-less-than-accurate PARMTEQM.* It is possible that two quite different curves might coincidentally cross at some point… So the matter was dropped.

22 March 2017 - Worked and read – programming of strategy employing Comunian limiting of voltage swing working, preliminary problem is that voltage easily >KP for some aperture rule.

23 March 2017 - Got first version of an outer big driving loop working to explore design parameter variations – see "!bigloop" comments in source code..

## 25.2 *LINACSrfqSIM* Simulation of Xingguang Design

27 March 2017 - Xingguang sent *PARI/PARMTEQM* files

27 March 2017 - *PARI* file -> 'celltable_from_pari_file.nb' -> celltables.txt (with MPs) gave good 95% 2term, but fell to 85% with Poisson simulation (Fig. 25.1).
This is adequate comparison to *PARMTEQM* result, which is known to be quite optimistic.
The Poisson result indicates further design search should be made.

Very sensitive – transplanted digitized curves to rfqtable.txt -> J. Maus model with no MPs gave much different lower result. Also very sensitive to match, pari match vs tapemap input ellipse alpha-beta search.

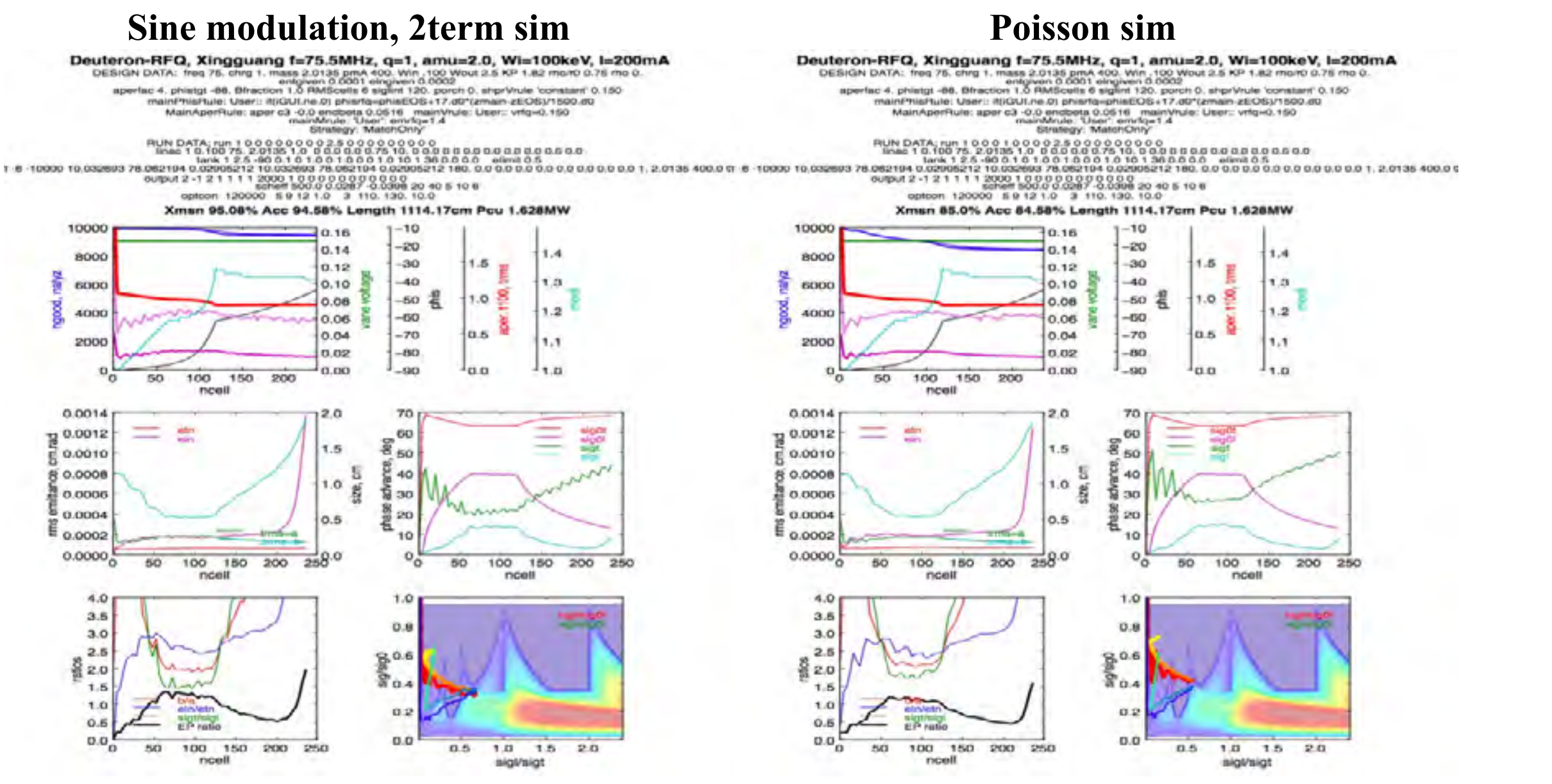

Fig. 25.1. *LINACSrfqSIM* of Xingguang Design. Sinusoidal vane modulation. Left - 2term simulation, right - Poisson simulation. (Title showing I=200mA is a title only – DESIGN DATA and RUNDATA show designed and run at 400 pmA.)

## 25.3 Best Case for *LINACSrfqDES* Initial Design Search

30 March 2017 - Programmed for different voltage and aperture rules. No improvement.

1-4 April 2017 - beta rule on amplitude. Shorter shaper – 90°. Have ~93% case. Ran many cases.

5 April 2017 - Sorted through cases, picked "best", ran with 400mA and get 96.97% transmission and 96.77% accelerated beam with Poisson simulation, and 97.33% and 96.77% for 2term simulation!! So think can declare finished, write up.

*'rajameson$ /Users/rajameson/Documents/Folders/Japan/Japan\ 2016\ November/Liu\ Xingguang\ RIKEN/75MHz/Cases/Best\ Case\ 20170405'*

Best case turned out to be a quite "standard" design according to my experience. The very high current with large tune depression results in very strong space charge mixing, and thus easier control of the space charge physics by the design formulae.

The shaper length does not affect the beam loss to the EOS, so a shorter (90° length) than usual (180° length) shaper is chosen to reduce the overall length and rf power.

The main problem is radial loss during the initial stage of acceleration where the curvature of the synchronous phase is high (the grounds for how this and other curvatures are handled in the simultaneous solution of the design beam equations and by the Poisson simulation are given in Sec. 16.1.3). Maximum focusing is provided by keeping the vane voltage at the chosen KP limit. Many variations were explored trying to reduce this beam loss without requiring much more length and rf power; the parameter choices for this case give a satisfactory compromise for meeting the spec. Probably less beam loss might be obtained with a stretched-out acceleration profile – subject for further design study, but maybe not economical.

Fig. 25.2. shows the result with design and simulated 500 pmA beam current. The main RFQ aperture is held constant after the EOS.

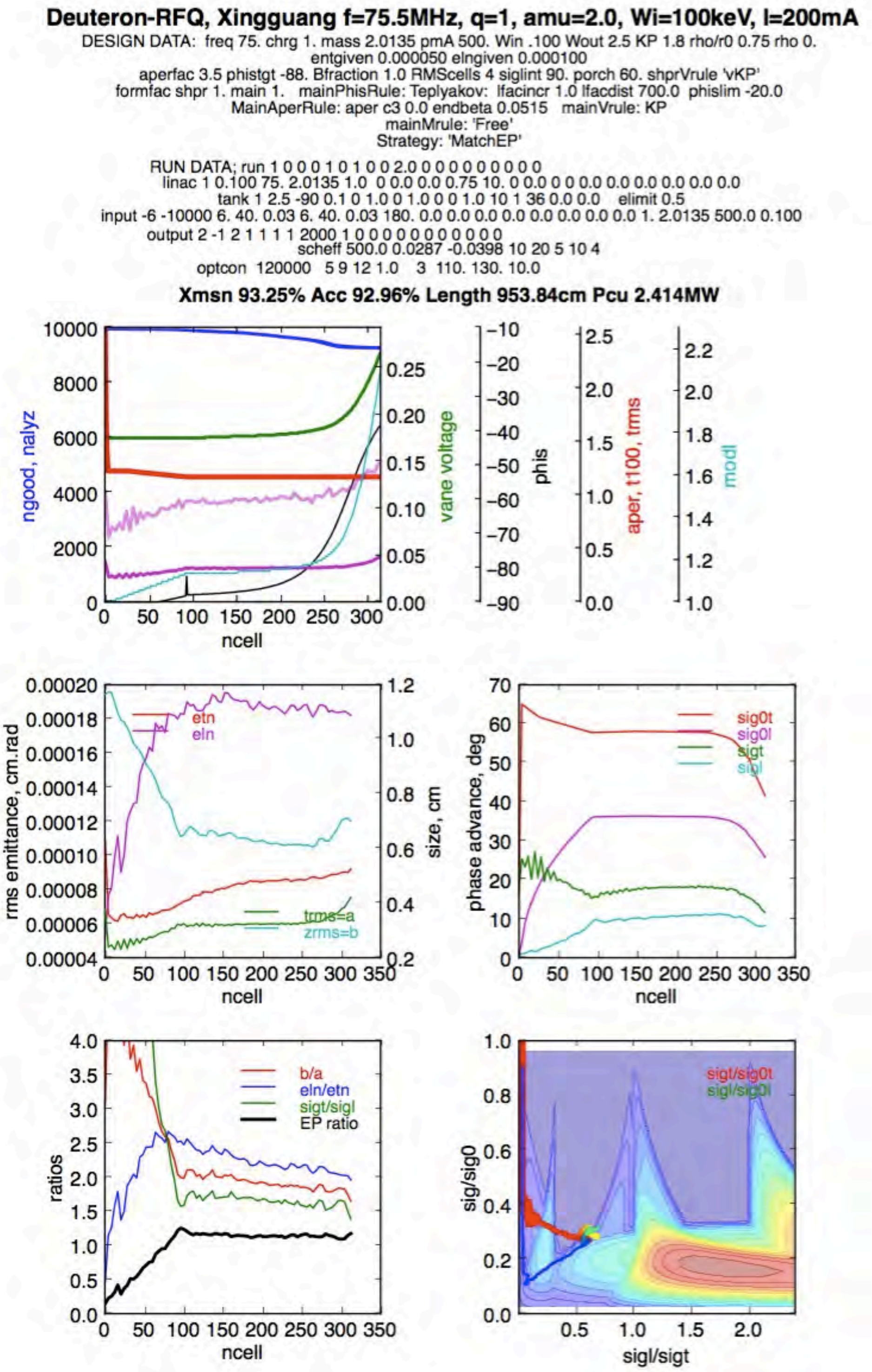

Fig. 25.2. 500mA design and Poisson run: '*LINACSsimRFQ* plot em1 EP2 T1 a0 zmain/150 9325 9296 954.jpg' (Title showing I=200mA is a title only – DESIGN DATA and RUNDATA show designed and run at 500 pmA.)

Fig. 25.3. shows the 500 pmA design and Poisson simulation result with input beam current of 400 pmA. Usually design to a higher current that the operational current will give a better result at the operational current than a design to the operational current.

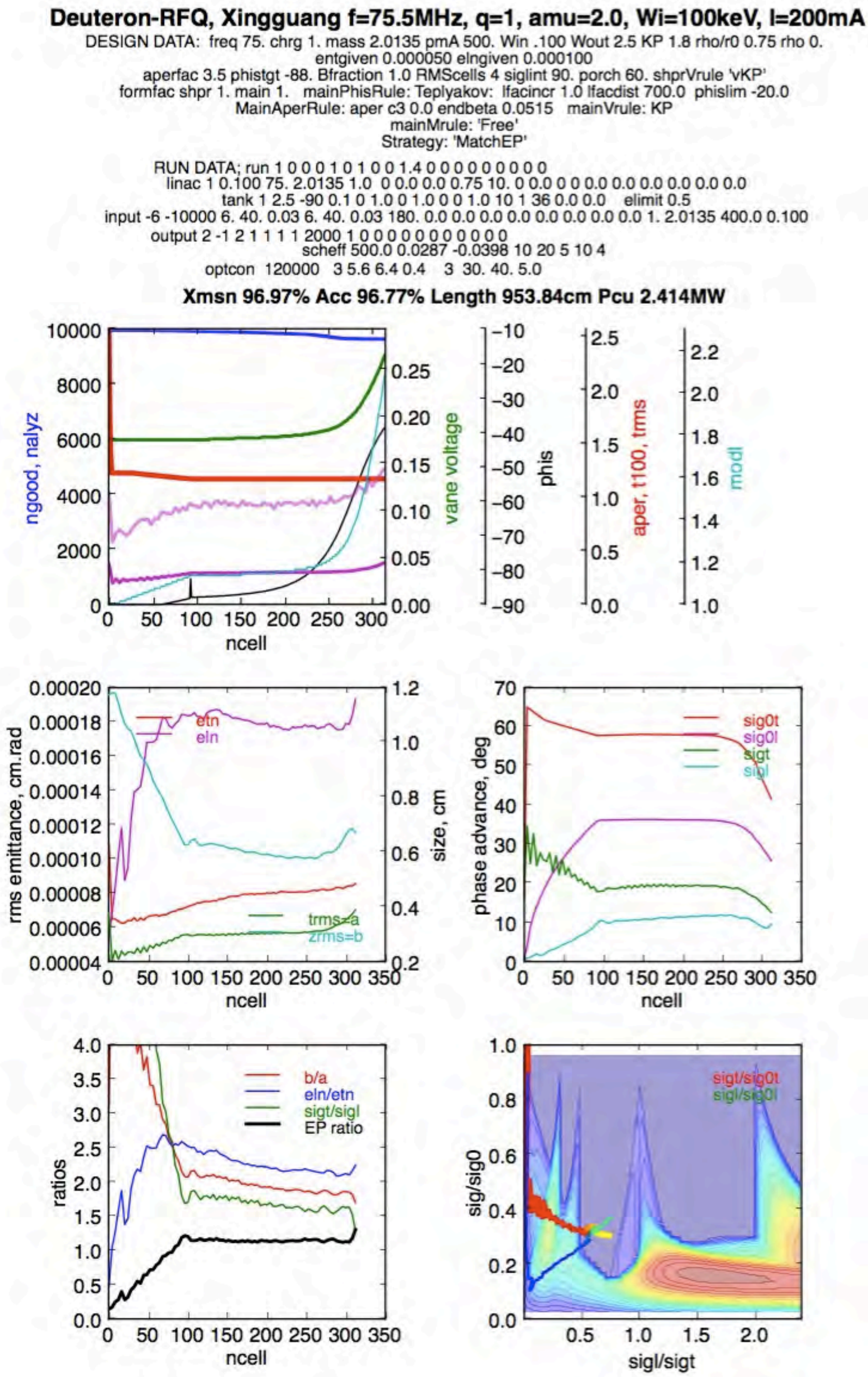

Fig. 25.3. The 500mA design result with 400 pmA input current, Poisson simulation. (Title showing I=200mA is a title only – DESIGN DATA and RUNDATA show designed with 500 pmA and run at 400 pmA.)

Fig. 25.4. shows the 2-term simulation result with input beam current of 400 pmA.

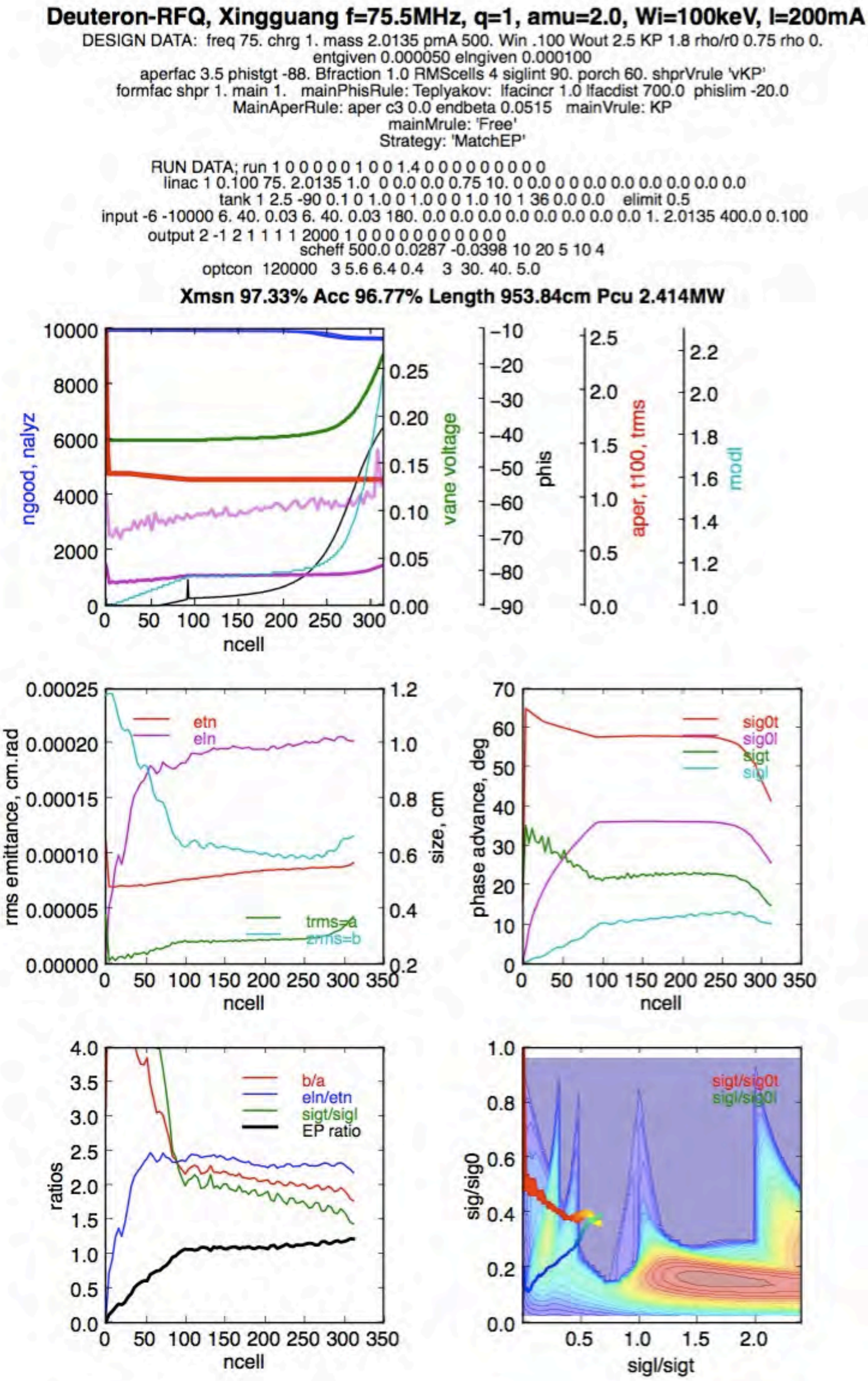

Fig. 25.4. The 500 pmA design result with 400 pAm input current, 2-term simulation.

### 25.3.1 Loss Pattern:

Losses are all radial, and mostly below 0.3 MeV. (From end-of-shaper (EOS) to RFQ end, the Teplyakov longitudinal compression factor 1.0 over distance 700 cm gives longitudinal (bucket length)/(rms beam length) "lgapfac" from 2.7 -> 3.8, and (aperture)/(rms beam radius) "tfac" from 1.7->1.35.

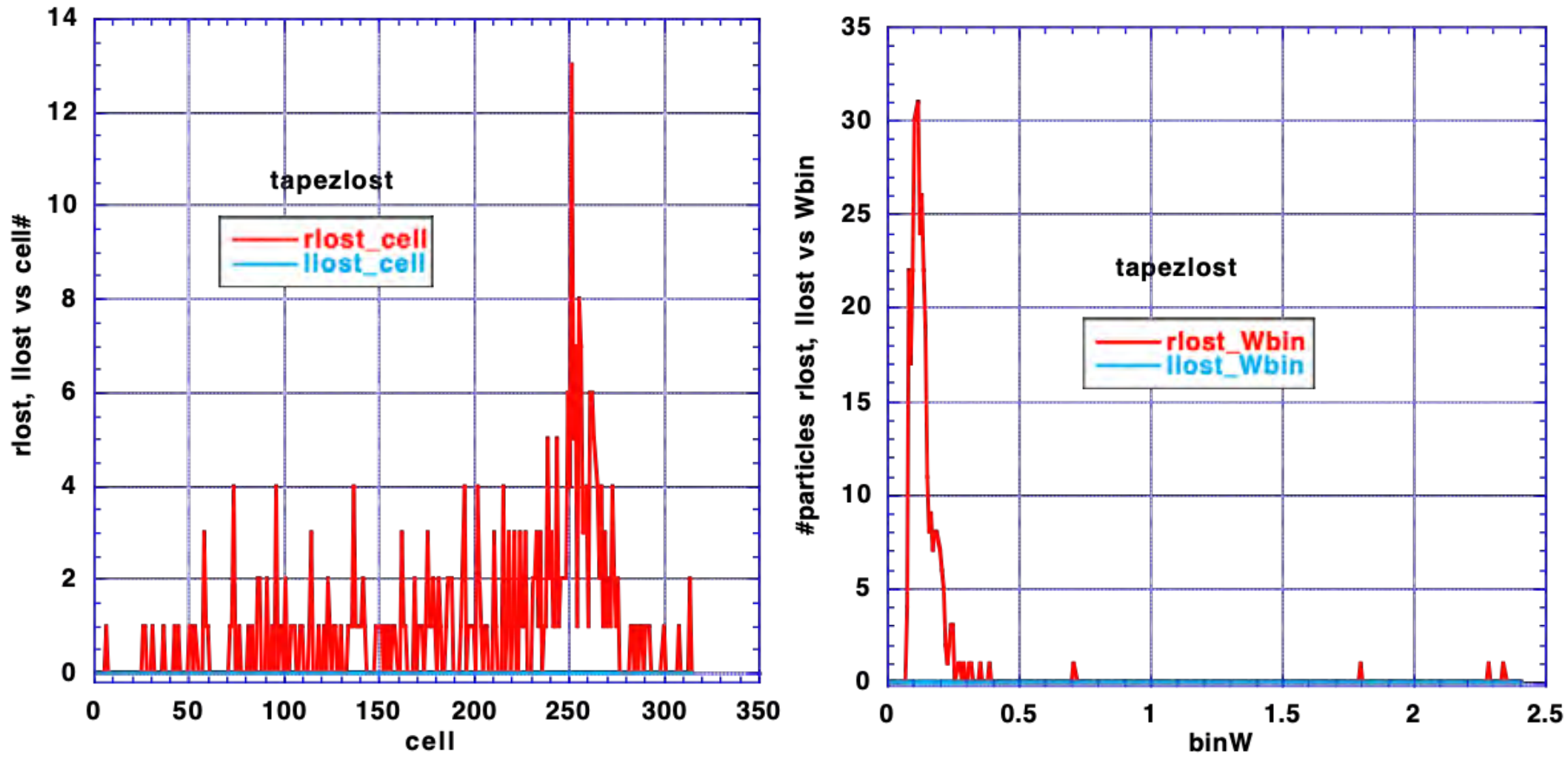

Fig. 25.5. Beam loss patterns. Left: Number of particles (out of 10000 input) lost vs. cell number. Right: vs. energy. All losses are radial.

### 25.3.2 100%/rms characteristics:

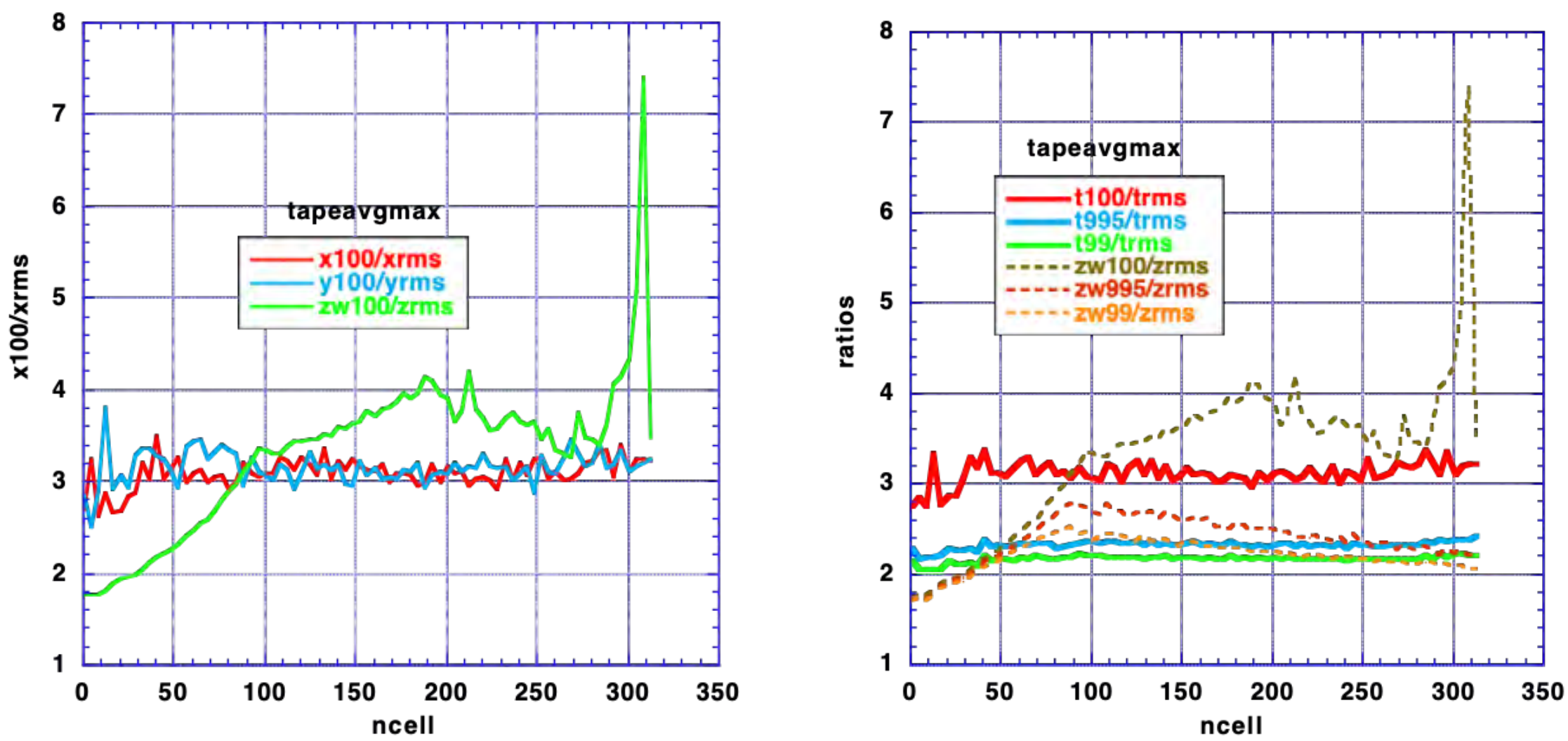

Fig. 25.6. Ratios of various fractions of beam size to rms beam size.

Equipartitioned design controlling the rms beam internal energy also controls the maximum emittances - keeps the beam distribution "tight" – this fact is often not understood.

## 25.3.4 Design summary ‘Terminal Saved Output 400mA’:

IRUNOPT 1 0 0 0 1 0 1 0 0 1 0 0 0 0 0 0 0 0 0 0 0 0 0 0 0 0 0 0 0 0

Deuteron-RFQ, Xingguang f=75.5MHz, q=1, amu=2.0, Wi=100keV, I=500mA

vane modulation sinusoidal

in rfqdesignGUI

FREQ 75.0000000000000 Charge 1.00000000000000 Mass 2.01350000000000 Win 0.100000000000000 Wout 2.50000000000000 CURRENT 0.500000000000000

powcuFacToshiba 0.000000000000000

shprtgt,blinj,EOSaperfac,apertgt 4.12752252415669 3.50000000000000 1.17929214975905

rfqdesignGUI 3.50000000000000 1.17929214975905 -88.0000000000000 1.00000000000000 1.00000000000000 4.00000000000000 90.0000000000000 60.0000000000000

in shprtgt, jshprpass= 1

shprtgt back from EP, emrfq 1.13803511577706

in shprtgt, jshprpass= 2

shprtgt back from EP, emrfq 1.13802088025679

End of shaper: a,m,b,v,rmsr,rmsl,sig0t,sigt,sigt/sig0t,sig0l,sigl,sigl/sig0l,gamma*rmsl/rmsr,elnrms/etnrms,sigt/sigl,energy,etnrms,elnrms

1.179182 1.138010 9.798801 0.175501 0.134663 330.249674 57.756494 12.027674 0.208248 36.229485 6.012930 0.165968 2452.549708 2.000013 0.499925 0.100000 0.000050 0.000100

Beginning of shaper: a,m,b,v,rmsr,rmsl,sig0t,sigt, sigt/sig0t, workingKP

1.265222 1.000010 9.798801 0.175501 0.552408 1.384413 63.183070 57.591350 0.911500 0.016528

RFQ entrance b,phi,m,v,a =

0.010000 -90.000000 1.000010 0.175501 38.666206

Radial matching section cell# z, b, phis, m, v, a

| | | | | | | |
|---|---|---|---|---|---|---|
| 1 | 2.0638 | 2.4572 | -90 | 1.0 | 0.175501 | 2.466170 |
| 2 | 4.1275 | 4.9044 | -90 | 1.0 | 0.175501 | 1.745800 |
| 3 | 6.1913 | 7.3516 | -90 | 1.0 | 0.175501 | 1.425980 |
| 4 | 8.2550 | 9.7988 | -90 | 1.0 | 0.175501 | 1.235160 |

Generate shaper

| jjj | znew | bemstart | phisstart | emstart | vstart | energystart | astart | Pcu | aarfq | e0rfq | trfq | e0t |
|---|---|---|---|---|---|---|---|---|---|---|---|---|
| 10 | 12.48826 | 10.02891 | -90.00000 | 1.00930 | 0.17550 | 0.10001 | 1.23522 | 0.02519 | 0.00343 | 0.00029 | 0.78540 | 0.00023 |
| 20 | 33.12705 | 9.88543 | -90.00000 | 1.02467 | 0.17550 | 0.10001 | 1.23522 | 0.06648 | 0.00900 | 0.00077 | 0.78540 | 0.00060 |
| 30 | 53.55946 | 9.79880 | -90.00000 | 1.03988 | 0.17550 | 0.10001 | 1.23192 | 0.10735 | 0.01445 | 0.00123 | 0.78540 | 0.00096 |
| 40 | 74.19825 | 9.79880 | -90.00000 | 1.05525 | 0.17550 | 0.10001 | 1.22326 | 0.14864 | 0.01980 | 0.00168 | 0.78541 | 0.00132 |
| 50 | 94.83704 | 9.79880 | -90.00000 | 1.07062 | 0.17550 | 0.10001 | 1.21482 | 0.18993 | 0.02475 | 0.00210 | 0.78544 | 0.00165 |
| 60 | 115.47588 | 9.79887 | -89.88474 | 1.08599 | 0.17550 | 0.10002 | 1.20646 | 0.23122 | 0.02970 | 0.00253 | 0.78546 | 0.00198 |
| 70 | 136.12712 | 9.79887 | -89.32762 | 1.10136 | 0.17550 | 0.10033 | 1.19816 | 0.27253 | 0.03472 | 0.00295 | 0.78547 | 0.00232 |
| 80 | 157.04442 | 9.79880 | -88.76332 | 1.11694 | 0.17550 | 0.10120 | 1.18986 | 0.31438 | 0.04000 | 0.00338 | 0.78547 | 0.00266 |
| 90 | 177.87757 | 9.79880 | -88.20129 | 1.13245 | 0.17550 | 0.10276 | 1.18197 | 0.35606 | 0.04549 | 0.00382 | 0.78544 | 0.00300 |

End of shaper EOS: jjj 93 nshprcells 89 celldivdsgn 10 energy 0.10351172E+00

paramoutlist = {z,brfq,phirfq,emrfq,vrfq,energy,arfq,r0rfq,rmsr,rmsl,tfac,lgapfac,epk,Pcu}

| | | | | | | | | | | | | | | |
|---|---|---|---|---|---|---|---|---|---|---|---|---|---|---|
| 100 | 198.828104 | 9.8162 | -87.6573 | 1.1394 | 0.1760 | 0.1055 | 1.1792 | 1.2614 | 0.3081 | 0.6167 | 1.7117 | 2.7176 | 1.1906 | 0.3982 |
| 110 | 220.169081 | 9.8233 | -87.5502 | 1.1403 | 0.1763 | 0.1084 | 1.1792 | 1.2619 | 0.3080 | 0.6161 | 1.7121 | 2.7481 | 1.1907 | 0.4412 |
| 120 | 241.803892 | 9.8321 | -87.4294 | 1.1412 | 0.1766 | 0.1115 | 1.1792 | 1.2624 | 0.3078 | 0.6158 | 1.7132 | 2.7790 | 1.1907 | 0.4850 |
| 130 | 263.752013 | 9.8411 | -87.2861 | 1.1421 | 0.1770 | 0.1148 | 1.1792 | 1.2630 | 0.3076 | 0.6154 | 1.7142 | 2.8103 | 1.1907 | 0.5296 |
| 140 | 286.037470 | 9.8515 | -87.0930 | 1.1431 | 0.1774 | 0.1185 | 1.1792 | 1.2636 | 0.3074 | 0.6151 | 1.7156 | 2.8421 | 1.1908 | 0.5750 |
| 150 | 308.690833 | 9.8587 | -86.8539 | 1.1445 | 0.1777 | 0.1226 | 1.1792 | 1.2644 | 0.3073 | 0.6147 | 1.7163 | 2.8745 | 1.1908 | 0.6214 |
| 160 | 331.748602 | 9.8550 | -86.5879 | 1.1467 | 0.1780 | 0.1272 | 1.1792 | 1.2657 | 0.3073 | 0.6150 | 1.7160 | 2.9074 | 1.1909 | 0.6688 |
| 170 | 355.253605 | 9.8505 | -86.2455 | 1.1492 | 0.1784 | 0.1324 | 1.1792 | 1.2671 | 0.3074 | 0.6150 | 1.7153 | 2.9410 | 1.1909 | 0.7173 |
| 180 | 379.259618 | 9.8463 | -85.8121 | 1.1519 | 0.1787 | 0.1384 | 1.1792 | 1.2688 | 0.3075 | 0.6152 | 1.7148 | 2.9753 | 1.1909 | 0.7670 |
| 190 | 403.836613 | 9.8419 | -85.2608 | 1.1551 | 0.1792 | 0.1455 | 1.1792 | 1.2706 | 0.3076 | 0.6155 | 1.7145 | 3.0104 | 1.1909 | 0.8181 |
| 200 | 429.077836 | 9.8362 | -84.5053 | 1.1591 | 0.1797 | 0.1541 | 1.1792 | 1.2730 | 0.3078 | 0.6154 | 1.7134 | 3.0464 | 1.1910 | 0.8709 |
| 210 | 455.110220 | 9.8322 | -83.5271 | 1.1637 | 0.1804 | 0.1647 | 1.1792 | 1.2757 | 0.3077 | 0.6160 | 1.7136 | 3.0836 | 1.1911 | 0.9257 |
| 220 | 482.109961 | 9.8256 | -82.1593 | 1.1699 | 0.1813 | 0.1784 | 1.1792 | 1.2794 | 0.3079 | 0.6162 | 1.7128 | 3.1222 | 1.1911 | 0.9831 |
| 230 | 510.329555 | 9.8179 | -80.2928 | 1.1782 | 0.1826 | 0.1968 | 1.1792 | 1.2842 | 0.3081 | 0.6165 | 1.7118 | 3.1625 | 1.1912 | 1.0438 |
| 240 | 540.140178 | 9.8084 | -77.7391 | 1.1895 | 0.1843 | 0.2225 | 1.1792 | 1.2909 | 0.3083 | 0.6167 | 1.7106 | 3.2050 | 1.1914 | 1.1089 |
| 250 | 572.100426 | 9.7482 | -74.3678 | 1.2092 | 0.1864 | 0.2603 | 1.1792 | 1.3026 | 0.3093 | 0.6203 | 1.7051 | 3.2506 | 1.1918 | 1.1802 |
| 260 | 607.028605 | 9.6512 | -69.8280 | 1.2412 | 0.1899 | 0.3180 | 1.1792 | 1.3214 | 0.3116 | 0.6232 | 1.6925 | 3.3005 | 1.1922 | 1.2605 |
| 270 | 646.174002 | 9.5236 | -64.0296 | 1.2902 | 0.1957 | 0.4116 | 1.1792 | 1.3503 | 0.3143 | 0.6286 | 1.6777 | 3.3563 | 1.1930 | 1.3548 |
| 280 | 691.479027 | 9.2129 | -57.4684 | 1.3777 | 0.2041 | 0.5701 | 1.1792 | 1.4019 | 0.3212 | 0.6426 | 1.6419 | 3.4210 | 1.1948 | 1.4723 |
| 290 | 745.668108 | 8.7265 | -50.5384 | 1.5159 | 0.2170 | 0.8404 | 1.1792 | 1.4833 | 0.3341 | 0.6625 | 1.5785 | 3.4982 | 1.1965 | 1.6279 |
| 300 | 812.228915 | 8.0450 | -44.7568 | 1.7117 | 0.2336 | 1.2885 | 1.1792 | 1.5988 | 0.3510 | 0.7023 | 1.5022 | 3.5931 | 1.1996 | 1.8475 |
| 310 | 893.818376 | 7.3272 | -40.1534 | 1.9469 | 0.2542 | 1.9668 | 1.1792 | 1.7375 | 0.3711 | 0.7553 | 1.4209 | 3.7094 | 1.2007 | 2.1636 |
| 315 | 951.116189 | 6.8186 | -38.0570 | 2.1187 | 0.2677 | 2.5006 | 1.1792 | 1.8388 | 0.3916 | 0.7813 | 1.3468 | 3.7911 | 1.2005 | 2.4178 |

sigoutlist = {z,sig0t,sigt,sigt/sig0t,sig0l,sigl,sigl/sig0l,sig0t/sig0l,gamma*rmsl/rmsr,elnrms/etnrms,sigt/sigl,etnrms,elnrms,wkgKP}

| | | | | | | | | | | | | | | |
|---|---|---|---|---|---|---|---|---|---|---|---|---|---|---|
| 100 | 198.828104 | 57.8648 | 12.0601 | 0.2084 | 36.2757 | 6.0213 | 0.1660 | 1.5951 | 2.0019 | 2.0012 | 2.0029 | 0.000050 | 0.000100 | 0.999984 |
| 110 | 220.169081 | 57.9016 | 12.0679 | 0.2084 | 36.3178 | 6.0319 | 0.1661 | 1.5943 | 2.0002 | 2.0000 | 2.0007 | 0.000050 | 0.000100 | 0.999981 |
| 120 | 241.803892 | 57.9553 | 12.0835 | 0.2085 | 36.3451 | 6.0376 | 0.1661 | 1.5946 | 2.0008 | 2.0004 | 2.0014 | 0.000050 | 0.000100 | 0.999982 |
| 130 | 263.752013 | 58.0074 | 12.0974 | 0.2085 | 36.3805 | 6.0456 | 0.1662 | 1.5945 | 2.0005 | 2.0003 | 2.0010 | 0.000050 | 0.000100 | 0.999976 |
| 140 | 286.037470 | 58.0711 | 12.1159 | 0.2086 | 36.4111 | 6.0515 | 0.1662 | 1.5949 | 2.0013 | 2.0008 | 2.0021 | 0.000050 | 0.000100 | 0.999979 |
| 150 | 308.690833 | 58.1119 | 12.1264 | 0.2087 | 36.4426 | 6.0589 | 0.1663 | 1.5946 | 2.0008 | 2.0004 | 2.0014 | 0.000050 | 0.000100 | 0.999984 |
| 160 | 331.748602 | 58.0913 | 12.1213 | 0.2087 | 36.4251 | 6.0546 | 0.1662 | 1.5948 | 2.0012 | 2.0007 | 2.0020 | 0.000050 | 0.000100 | 0.999987 |
| 170 | 355.253605 | 58.0626 | 12.1125 | 0.2086 | 36.4155 | 6.0532 | 0.1662 | 1.5944 | 2.0005 | 2.0002 | 2.0010 | 0.000050 | 0.000100 | 0.999982 |
| 180 | 379.259618 | 58.0384 | 12.1061 | 0.2086 | 36.3983 | 6.0492 | 0.1662 | 1.5945 | 2.0007 | 2.0003 | 2.0013 | 0.000050 | 0.000100 | 0.999982 |
| 190 | 403.836613 | 58.0139 | 12.1001 | 0.2086 | 36.3783 | 6.0443 | 0.1662 | 1.5947 | 2.0011 | 2.0006 | 2.0019 | 0.000050 | 0.000100 | 0.999973 |
| 200 | 429.077836 | 57.9752 | 12.0851 | 0.2085 | 36.3725 | 6.0434 | 0.1662 | 1.5939 | 1.9995 | 1.9996 | 1.9997 | 0.000050 | 0.000100 | 0.999956 |
| 210 | 455.110220 | 57.9592 | 12.0860 | 0.2085 | 36.3338 | 6.0335 | 0.1661 | 1.5952 | 2.0020 | 2.0012 | 2.0032 | 0.000050 | 0.000100 | 0.999968 |
| 220 | 482.109961 | 57.9195 | 12.0746 | 0.2085 | 36.3133 | 6.0292 | 0.1660 | 1.5950 | 2.0016 | 2.0009 | 2.0027 | 0.000050 | 0.000100 | 0.999963 |
| 230 | 510.329555 | 57.8730 | 12.0614 | 0.2084 | 36.2882 | 6.0239 | 0.1660 | 1.5948 | 2.0013 | 2.0007 | 2.0023 | 0.000050 | 0.000100 | 0.999941 |
| 240 | 540.140178 | 57.8156 | 12.0456 | 0.2083 | 36.2577 | 6.0180 | 0.1660 | 1.5946 | 2.0008 | 2.0005 | 2.0016 | 0.000050 | 0.000100 | 0.999845 |
| 250 | 572.100426 | 57.4748 | 11.9578 | 0.2081 | 35.9895 | 5.9537 | 0.1654 | 1.5970 | 2.0057 | 2.0036 | 2.0085 | 0.000050 | 0.000100 | 0.999933 |
| 260 | 607.028605 | 56.8869 | 11.7909 | 0.2073 | 35.6827 | 5.8912 | 0.1651 | 1.5942 | 2.0004 | 2.0000 | 2.0014 | 0.000050 | 0.000100 | 0.999720 |
| 270 | 646.174002 | 56.1332 | 11.5866 | 0.2064 | 35.2153 | 5.7902 | 0.1644 | 1.5940 | 2.0001 | 2.0000 | 2.0011 | 0.000050 | 0.000100 | 0.999508 |
| 280 | 691.479027 | 54.3035 | 11.0914 | 0.2042 | 34.0609 | 5.5365 | 0.1625 | 1.5943 | 2.0013 | 2.0004 | 2.0033 | 0.000050 | 0.000100 | 0.999529 |
| 290 | 745.668108 | 51.3930 | 10.6392 | 0.2070 | 32.4020 | 5.3722 | 0.1658 | 1.5861 | 1.9840 | 1.9893 | 1.9804 | 0.000052 | 0.000103 | 0.998697 |
| 300 | 812.228915 | 47.4168 | 9.2856 | 0.1958 | 29.7532 | 4.6278 | 0.1555 | 1.5937 | 2.0020 | 2.0002 | 2.0065 | 0.000050 | 0.000100 | 0.998993 |
| 310 | 893.818376 | 43.2608 | 8.2792 | 0.1914 | 26.8589 | 3.9905 | 0.1486 | 1.6107 | 2.0371 | 2.0043 | 2.0747 | 0.000050 | 0.000100 | 1.001922 |
| 315 | 951.116189 | 40.1749 | 7.4460 | 0.1853 | 25.2599 | 3.7193 | 0.1472 | 1.5905 | 1.9979 | 1.9991 | 2.0020 | 0.000050 | 0.000100 | 0.998452 |

logout

[Process completed]

### <u>25.3.5 Parameter discussion:</u>

*Specification:*

- Beam accelerated fraction ≥ 95%
- Shortest length (also gives lowest rf power requirement) for lowest cost
- Consideration of construction, tuning.

*Frequency:*

75 MHz seems ~highest frequency affording a good result.

*Design current:*

particle mA = 500 mA. Operational current is considered to be 400 mA. Design for higher current usually gives better performance at reduced current than design at that reduced current.

Ion source development is separate (big) problem – see input emittance below.

*Injection energy:*

100 keV. As IFMIF/EVEDA ion source.

*Output energy:*

2.5 MeV

*KPfac:*

1.8

*Input transverse normalized rms emittance:*

IFMIF/EVEDA ~150mA source -> ~ 0.3 mm.mrad. Ion source output emittance vs energy may scale with ~ current^1.5 -> Estimate at 500 mA source = (500/150)^(3/2)*0.3 = 6*0.3 = 1.8 mm.mrad.

Design investigated with 0.5 mm.mrad – may be *harder* problem (more space charge – but larger beam). ***Further design study should include larger input emittances, which may require different handling.***

This example design requests that the initial emittance (at EOS) be kept the same through the main RFQ. The result however shows at 500 mA a slow increase – from an equipartitioning exchange with the longitudinal emittance, which is decreasing. At 400 mA, both etnrms and elnrms are slowly increasing. ***Further design should explore introducing such emittance variations a priori, as experience shows that this could improve the performance.***

***D+ ion source development for the injection current, energy, and emittance desired is a bigger problem than the RFQ, which is shown by this study to be feasible.***

***Multichannel RFQs have been envisioned since early days. Lower current ion injectors could then be used. (Subject for further study.)***

*Input longitudinal normalized rms emittance:*

_*LINACSrfq* design procedure is unique in that it includes the ability to design for desired space charge physics behavior – "inside-out" design – from what the designer wants the beam to do, out to finding the external fields that provide the desired beam performance.

elnrms at EOS is specified, for this design at 1.0 mm.mrad, which would give equipartitiong ratios = 2, and requesting to stay constant from EOS to RFQ end. As seen in the figures and stated above, at 500 mA the longitudinal emittance along the main RFQ slowly decreases, via an equipartitioning exchange with the transverse emittances. ***There are some procedures that could be explored in further design refinement to counteract this (as noted also above).***

Many cases also tried *a priori* emittance variations along the main RFQ, as done for the IFMIF CDR design (much lower current), but were not immediately better. ***Further refinement of this "best case" example should further explore this point, because probably some adjustment of the EP ratio from 2 towards 1.8, 1.6, 1.4 can probably be tolerated without more beam loss but giving shorter length and rf power requirement.***

*Aperfac*

_= betalambda/(RFQ aperture) at the end of the shaper (EOS). Aperfac thus has a physical meaning, whereas specification of an aperture in cm has no meaning. The cell length is betalam/2, so an aperfac of 3.5 means an aperture at EOS somewhat larger than half of the cell length. An aperfac = 4 would be a default value. Aperfac is the strongest variational parameter and the first to be investigated.

*Phistgt:*

The shaper strategy is to bring the beam to an equipartitioned (EP) condition at least at one point in the RFQ, at the end of the shaper. This provides a robust condition, after which some strategy for the main part of the RFQ is used. The input synchronous phase $\phi_s$ is -90° to provide bunching; at some point $\phi_s$ is raised. The design equations assume an ellipsoidally shaped beam, but will practically work even for a long beam, and practice has established that using the shaper to bring the beam to a formal EP condition at an only slightly increased $\phi_s$ gives the desired robust shaper. From this point, the main RFQ uses parameter variation rules and space charge physics strategies to provide the acceleration profile. Phistgt variations to > -88° are usually not productive in terms of beam loss. But can be investigated, for example to set up the strategy of constant σt in the shaper until this forces an avalanche closure of the aperture, limiting the final shaper $\phi_s$ (the pari strategy). Such investigation requires changing the (pass.f90) source program.

*Bfraction:*

RFQ injection required a beam converging in x and y (usually with equal ellipse parameters, but x and y behavior is not always symmetric – ***further design refinement should investigate different input x,y, matching*** ). It was found practically that the maximum possible focusing strength is often not needed in the radial matching and shaper sections, and that if the input focusing is lowered and then ramped up to the desired EOS value, the input matching becomes practically easier, because less input convergence (lower ellipse alpha) is required. This design investigation with very high current worked better with full focusing from injection.

*RMS cells:*

4 works well. 6 sometimes used, but often result is not as good.

*Siglint:*

The length of the required shaper section is not known initially, but controlled using a value in degrees that indicates the zero-current longitudinal phase advance (as would be seen by a particle very near $\phi_s$). This is a strong parameter on the shaper length. 180° is a default value. For this very high current design, shapers from very short to very long were investigated, with 90° giving better performance for this reference design.

*Porch:*

Practice has shown that keeping the initial $\phi_s$ constant at -90° to provide some initial pure bunching is advantageous. As the length of the shaper is not initially known, the initial $\phi_s$ = 90° "porch is specified as a percentage of the eventual shaper length.

*Shpr Vrule:* Usually it is desired to keep the voltage as high as allowed by the Kilpatrick criterion and the KP factor. The EOS voltage is presently hardwired in shprtgt.f90 to be the KP voltage at the end of the shaper; the choice "constant" is overruled.

*Form factor adjustments:*

An EP design requirement along the main RFQ is usually not met completely, as shown by the EP condition and the separation of the EP ratios in the *LINACSsimRFQ*plot. It has been found that an *a priori* variation of the form factor can often be introduced which can bring the EP ratios closer together and the EP condition closer to = 1. As stated above, initial investigation did not give improvement, but this should be investigated further. (See Ch. 28)

*Rules for parameters not used for solving equations:*

For the main part of the RFQ, the acceleration profile must be provided, while keeping beam losses low. There are two, or three, equations to satisfy: the transverse and longitudinal matching equations should always be satisfied along the entire trajectory, and the equipartitioning condition may be invoked if desired. There are six variables: transverse and longitudinal rms beam size, which are found to satisfy the two matching equations, leaving no further equations, or just the EP equation. The other four variables are $\phi_s$, vane voltage, vane aperture, and vane modulation. Therefore, rules must be given for all four if EP is not invoked, or one variable can be used to satisfy the EP equation.

*Mainrfq**phisrule**:*

The main reason that an RFQ works is Teplyalov's insight that the beam longitudinal charge density should be kept constant. It is highly desirable to invoke this law as the energy is increased. The synchronous phase will then follow automatically, giving the designer one less thing to think about, rather than some arbitrary user (designer) rule. However, the Teplyakov condition of constant charge density does not have to hold – as with all the quantities in the three equations, all can vary *a priori*. By allowing the beam density to increase or decrease along the main RFQ, a strong handle is obtained on both the longitudinal beam loss, and on the RFQ length. The (bucket length)/(rms beam length) ratio is seen in the "lgapfac" column of the terminal output. "Tep lfacincr" is the amount that "lgapfac" should be adjusted over the length "Tep lfac dist". An example:

**Teplyakov lgapfac**

| Tepl lfac incr | lfac at end | tfac at end | #cells | length, cm | Pcu, MW | phis at end, ° | modl at end |
|---|---|---|---|---|---|---|---|
| 0 | 4.9 | 1.42 | 338 | 681 | 0.279 | -25 | 2.12 |
| +1 | 5.9 | 1.47 | 362 | 729 | 0.299 | -29 | 2.04 |
| -1 | 4.0 | 1.36 | 320 | 644 | 0.264 | -21.2 | 2.19 |

Teplyakov's insight was to control the bunch charge density – dependent on the beam bunching – and placing that as higher priority than the acceleration. The bunching can be expressed as the ratio of the (rf bucket width = phis)/(beam bunch length gb) = (longitudinal gap factor lgapfac). Then look at what happens to the linac design as lgapfac is varied:

lfac incr = 0 keeps the beam density constant, per Teplyakov. But it can be varied "*a priori*".

lfac incr = +1 requires that the density increase; the beam gets shorter w.r.t. the bucket. More longitudinal bunching is needed, so phis is more negative, requiring less modulation – which makes the length longer and need more rf power.

lfac incr = -1 requires that the density decrease; the beam gets longer w.r.t. the bucket. Less longitudinal bunching is needed, so phis can be more positive, requiring more modulation – which make the length shorter and need less rf power. If lgapfac gets too small, particles will be lost from the bucket…

It is not so intuitive when considered from the charge density point of view (probably why no one could imagine it before Teplyakov!). The RFQ transverse and longitudinal are completely coupled. Various forms of separated-function linacs are more or less coupled, and can be controlled in the same way.

*Mainrfq**aperrule**:*

The aperture variation along the main RFQ is critical, and has physical meaning to the designer; therefore is defined by a rule, starting with the EOS value. The default rule is a function of velocity beta variation from EOS to the RFQ end. An alternative rule based on z gives essentially the same result.

(If desired, the aperture could be used as a free variable. For example, as tried in various ways in this study, where it was of interest to specify the behavior of the average vane radius r0, while still using the modulation also as a free variable, and the source code was so modified. This interest was coupled with desire to have a different vane voltage profile (a la IFMIF/EVEDA). In the case of this study, it appears better to always utilize the maximum total focusing afforded by the KP voltage; for this and other reasons involving finding self-consistent sets of rules, this line of investigation was abandoned.).

This design keeps the aperture constant at the EOS value. This is a major factor in keeping the length short.

*Mainrfq**vrule**:*

Reason for preferred use of KP voltage as stated. Could be used as a free variable, but because the voltage has clear physical meanings, better to fix as a rule.

*Mainrfq**emrule**:*

The vane modulation has no direct physical connotation or requirement, other than a maximum desired value, so is the logical choice to leave free for satisfying an equation.

*Mainrfq**strategy**:*

The "strategies" are the way the free variables and the rules are used to design the RFQ cells. "MatchOnly" satisfies transverse and longitudinal rms matching by adjusting the beam sizes, and the other parameters must be set by rules. "MatchEP' requires one variable to be free, usually the modulation. Other strategies can be investigated.

**Directions Indicated by a Subsequent Study:**

Another directly following study[205] investigated an H+ RFQ design with very difficult specification. Beam loss control in the accelerating section required a strong and carefully patterned increase in the transverse zero-current phase advance sig0t. The specification (on length) was so tight that EP was not required, and even formal requirement of transverse and longitudinal matching in this section was abandoned, but the sig0t rule worked.

***This strategy should be explored for the D+ RFQ.*** It could be that the H+ strategy (no EP or matching requirement) would suffice, but EP and matching are important ingredients in this preliminary design and probably should be kept. Then with fixed aperture, voltage and synchronous phase rules, the modulation is involved in two requirements – sig0t and EP, which might be handled by a weighted optimization function. Or, sig0t and EP might be handled by solving for aperture and modulation.

[205] ICR D+ H+ RFQs RAJ-ICR-3_201705.pdf

### 25.3.6 Case Files, updated Source and Support Files

‘Best Case 20170405’ folder – contains ‘Best 400mA Case files’, ‘subroutines 20170406 since 20160929’, ‘Support files 20170406’.

[eltoc]

# Chapter 26 – Preliminary Design Study for ICR D+ and H+ RFQs

**ICR Accelerator Lab Tech Memo RAJ-ICR-3_20170503**
R. A. Jameson – 3 May 2017

## 26.1 D+ RFQ 30 keV Input Energy

The study continues from the work done in November 2015, where an equipartitioned (EP) design was found; the Poisson simulation demonstrated good EP performance and high transmission and accelerated beam fraction, but at cost of a higher than desired injection energy of 75 keV and long length ~4.5m. A lower injection energy ~30 keV and short length are desired, and the goal of this study.

Design was pursued for 0.030 MeV injection, 1.5 MeV output energies and operational beam current of 30mA. Performance is better if the design current is higher than the operational current, so the design current = 50mA.

50% trapezoidal vane modulation was used for the whole RFQ.

### 26.1.1 Long Shaper

Characteristics of this design are shown in Fig. 26.1. The design requested that the beam be brought from dc to EP at the end of the shaper (EOS), where the synchronous phase is defined to be -88°. A large EOS aperture is needed, but limited to betalambda/3. **One of the distinctive features of the design search is that with such low injection energy and significant beam current, the initial shaping must be done very carefully.** The transverse focusing term B must be high enough to prevent loss from occurring already in the shaper. A long shaper with significant porch is required - the shaper has 143 cells. The common idea that the RFQ will be shorter with lower injection energy is not necessarily true.

### 26.1.2 Short Main RFQ

From EOS to the end of the RFQ, the aperture is held constant at its EOS value. The aperture rule used for the main RFQ is a major factor in making the remainder of the RFQ short.

The input (and EOS) design rms normalized emittances were set at etn = 0.000025 m.rad, eln = 0.000050 m.rad, and shorter main RFQ length forced by requiring the design transverse rms emittance to decrease *a priori* to 0.000010 m.rad from EOS to the end, using a velocity beta rule.

The KP factor = 1.8, and the vane voltage was allowed to stay at the KP voltage as the modulation increases. This also results in a shorter RFQ, but is more complicated to construct and tune.

A neutral Teplyakov synchronous phase ($\phi_S$) rule keeps the bunch charge density constant. The charge density can be decreased or increased *a priori* in the main RFQ, giving respectively shorter or longer length, and this can play a helpful role in the design. Here the result was better if this capability was not used to influence the RFQ length, but rather to reach the final fs = -30° limit just briefly at the RFQ end.

The design strategy then required EP to be maintained in the main RFQ, with the vane modulation used to satisfy this requirement.

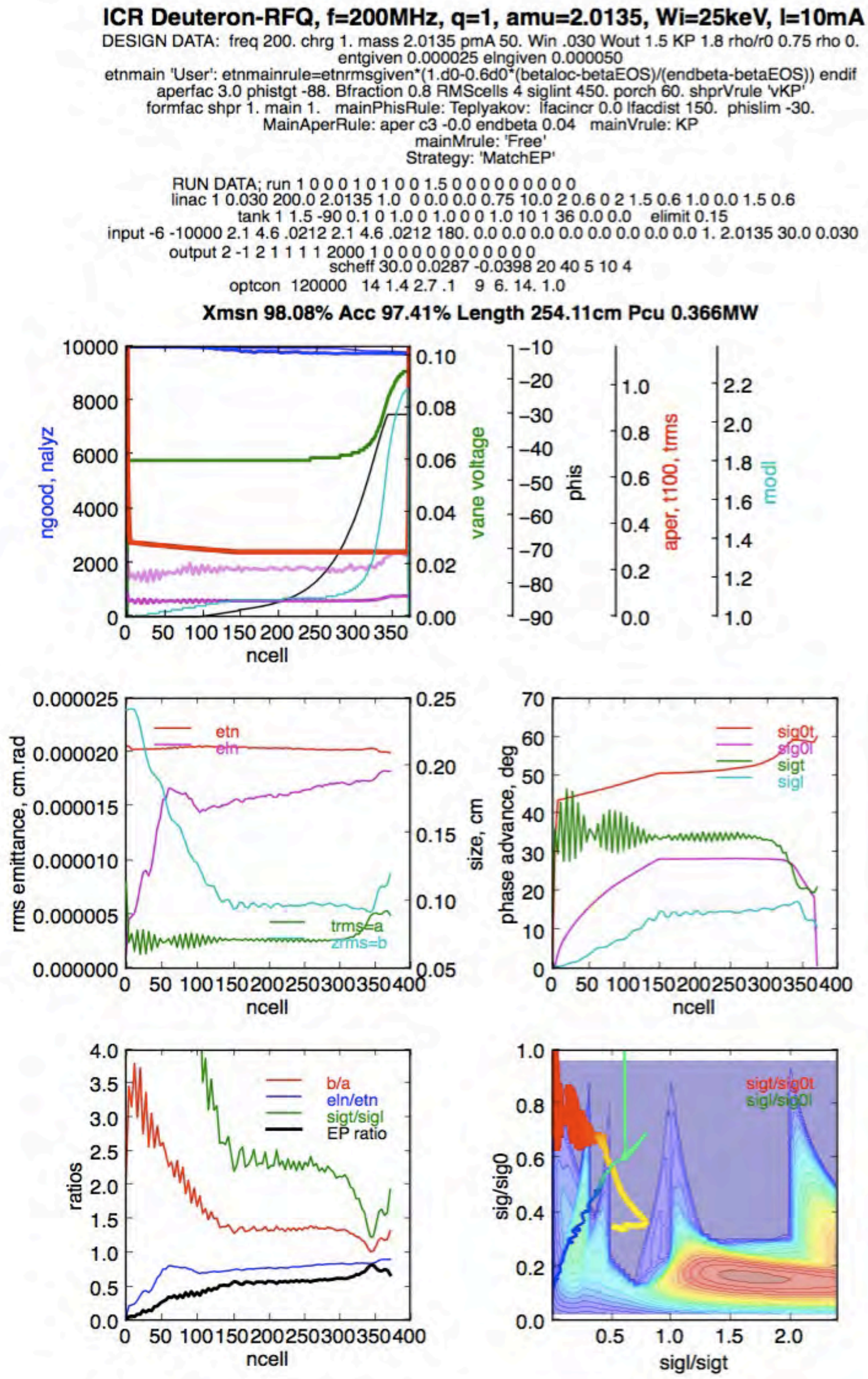

Fig. 26.1. D+ design 'Iwashita etn25->10,arfq=aEOS,dsgn50mA,run30mA'.

### 26.1.3 D+ Design Result – NOT EP, but short and good performance.

The resulting length is 2.54m, including input and output sections. The Poisson simulation indicates 98% transmission and 97.4% accelerated beam.

The design tables are given in Appendix 1, and show that the design equations were satisfied.

However, the simulation result does not at all achieve the design conditions (EP, transverse emittance decrease, etc.), but has very good beam performance.

This is an example of the power of the *LINACS* design methods. Under ideal conditions, the design conditions will be met by the simulation. But in the case of extreme, even seemingly unrealistic specification requests, the simulation does not meet the requested design conditions, as here. The simulation is of course showing the actual performance for the design cell parameters.

So as here, with the hard specs of low injection energy and desired short length, the designer would still like the design procedure to be driven by a self-consistent design program. Some of the rules, such as for aperture, vane voltage and synchronous phase, have physical requirements, but other parameters, like emittance, EP, are free to manipulate. The advantage of equipartitioning in keeping the total as well as the rms emittances controlled is not to be given up lightly, so in the present case, the design rules and strategy (maintaining transverse and longitudinal match, EP) were used as forcing functions to achieve good performance in the main RFQ by observing the simulation results until a successful set was found. Other rules and strategies could be programmed.

EP was not achieved, matching is maintained. The accelerating section was very critical to avoid beam loss, and considerable search and manipulation was necessary. **The key point was that the transverse zero-current phase advance, sig0t, needed to rise significantly in the accelerating section**, to counteract the tendency of the transverse phase advance with current, sigt, to fall.

How the successful cell parameters are achieved, even by seemingly strange rules, is in the end not important – the simulation result is studied to suggest what changes could reduce beam loss, knowing how various parameters influence the design.

## 26.2 H+ RFQ 30 keV Input Energy

This year, an additional design is considered:

> 4/9/2017: This year I would like to fabricate an RFQ using a tank, which was a test DTL cavity for prototype. The DTL are removed and RFQ electrodes will be installed.
> 4/25/2017: The length is about 1100 mm and the diameter is 400 mm. The assumed frequency would be 200 MHz.
> 4/26/2017: Particle: Since this tank is so short, proton would be a target.
> Current: First target is 40 mA.
> Injection energy: Is 25 kV feasible?
> Output Energy: 750 keV, for example.

### 26.2.1 Case 1

Again, low injection energy, short (and restricted) length. The design study was done with 30 keV injection energy; vane modulation sinusoidal until modulation = 1.5, then 50% trapezoidal. After considerable manipulation, along the lines described above and including an *a priori* final correction to the space charge form factor, a good design was found, shown in Fig. 26.2 and Appendix 2.

With the same design rules, just change of particle type, D+ performance in this design is good up to ~ 0.135 MeV energy, but then falls off as the synchronous phase saturates. Further design work might also be able to find a good D+ result for this specification.

### 26.2.2 Case 2

With the knowledge that the key point in the main RFQ is an increased sig0t, the design was approached from another perspective, with the requirement that the input emittances be preserved, that sig0t increase following some rule, transverse and longitudinal rms match and EP not required. Again, the search for a good design was tedious; B at EOS must be high enough, and the form of the sig0t rule is important. Fig. 26.3 and Appendix 3 show a design using the modulation to find the solution for the sig0t rule:

```
ds0t=0.45d0
xs0=ds0t*(zmain-zEOS)/(100.d0)
sig0tgoal=sig0tEOS*(1.d0+2.176d0*xs0-1.17d0*xs0**2-0.00783d0*xs0**3)
```

The input match can probably be improved, and an *a priori* final correction to the space charge form factor added, both should decrease the loss somewhat further.

## 26.3 Conclusion

*LINACSrfqDES* was able to find good designs even for the hard specs asked for here. It is an interesting and good further example of the power of the design program to assist the designer in meeting a specification set, which is always somewhat different for each request, and even usually requesting almost the impossible. RFQ design is quite flexible; but for long linac design, controlling beam loss may not allow wished for length restrictions, etc.

The design experience suggests that addition of other perhaps desirable rules to make construction easier, like constant vane voltage, constant r0, are unlikely to succeed.

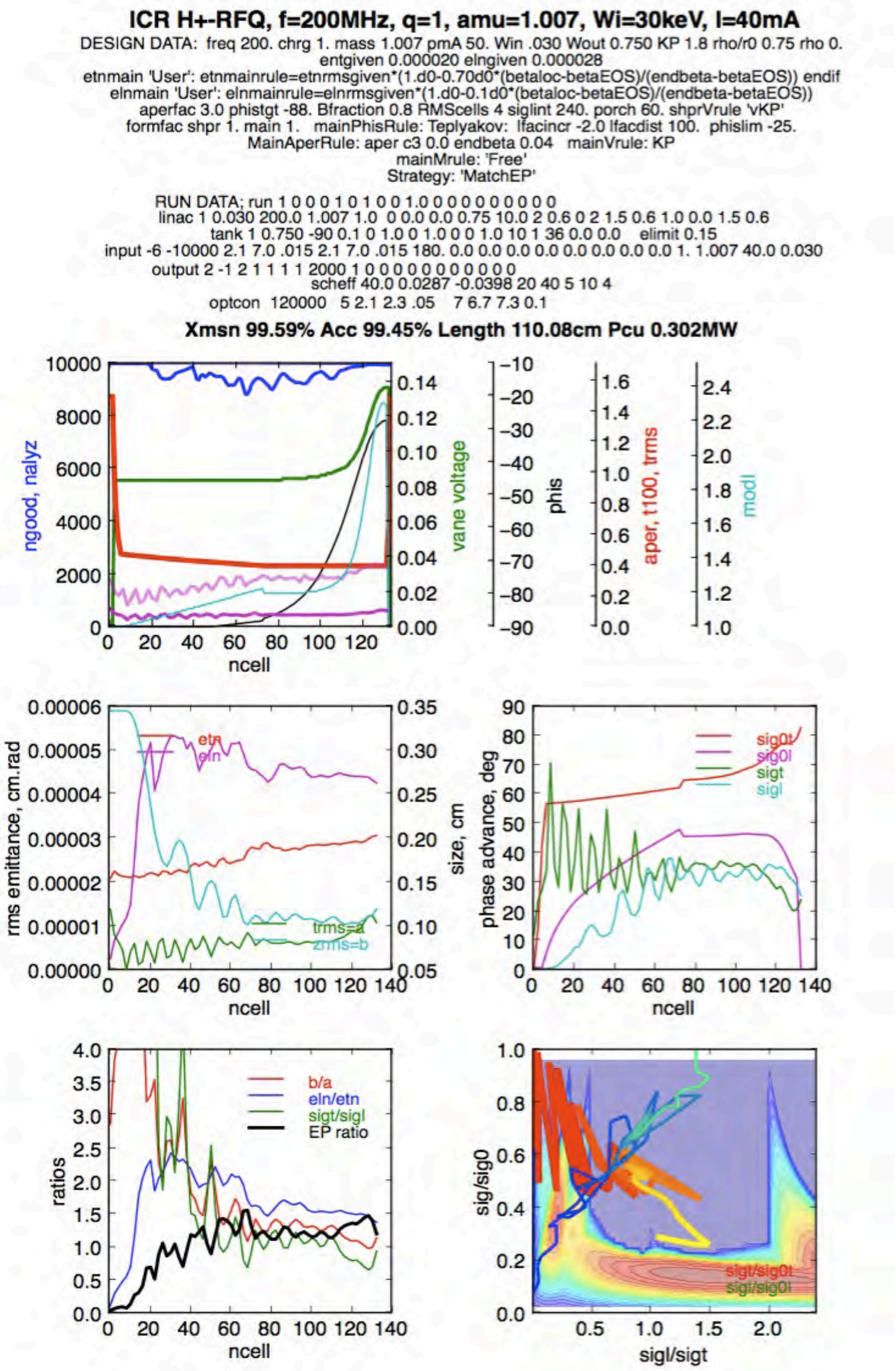

Fig. 26.2. Case 1 H+ design 'Iwashita 30keVinj H+ RFQ\ ffadj_8 length110'

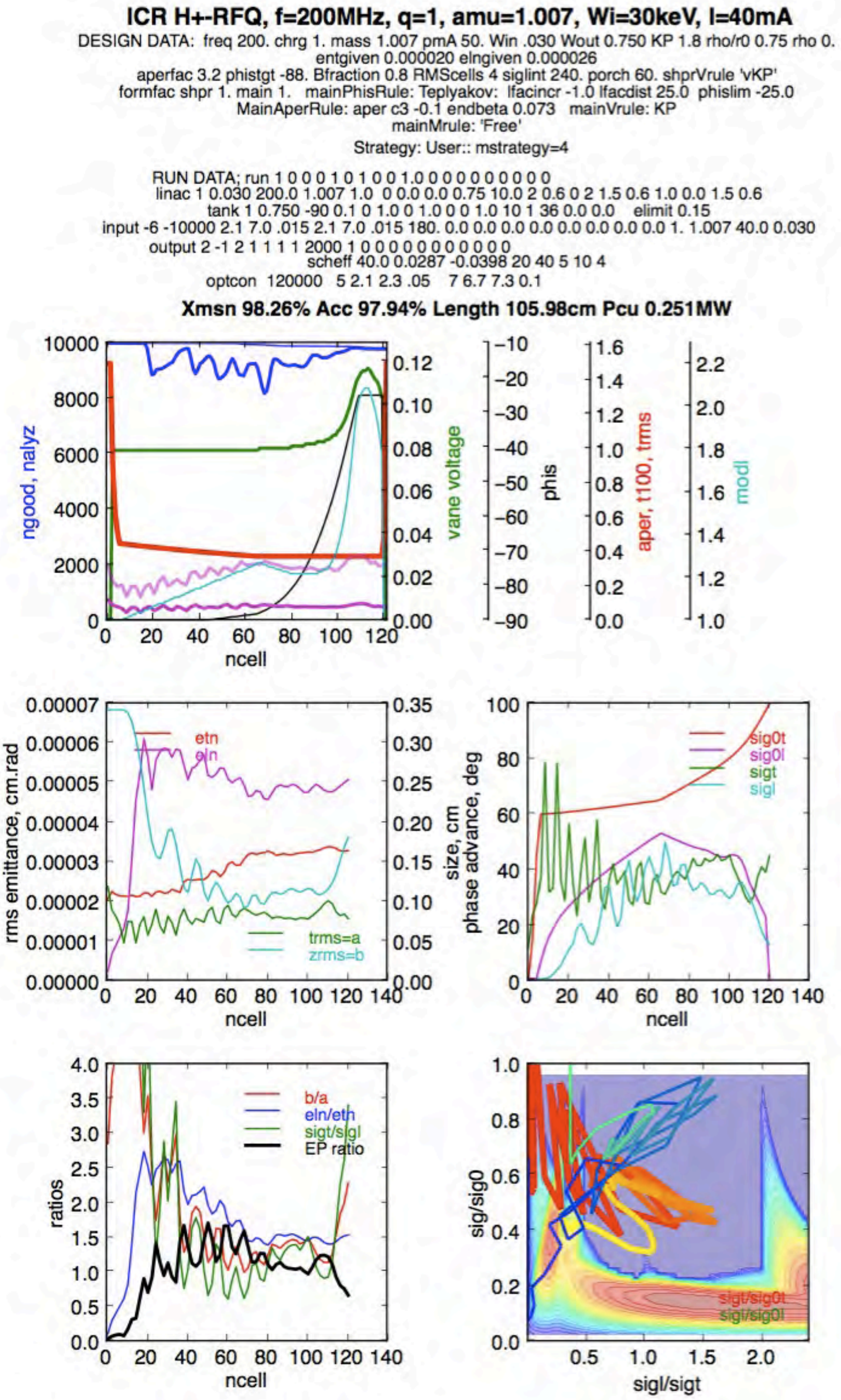

Fig. 26.3. Case 2 H+ design 'Iwashita 30keVinj H+ RFQ\ s0t rule'

# Appendix 1. D+ Design

```
FREQ  200.000000000000  Charge  1.00000000000000  Mass  2.01350000000000  Win  3.000000000000000E-002  Wout  1.50000000000000  CURRENT  5.000000000000000E-002
 powcuFacToshiba  0.000000000000000
 shprtgt,blinj,EOSaperfac,apertgt  0.847800177201162  3.00000000000000  0.282600059067054
 rfqdesignGUI  3.00000000000000  0.282600059067054 -88.0000000000000  0.800000000000000  1.00000000000000  4.00000000000000  450.000000000000  60.0000000000000
 in shprtgt, jshprpass=   1
 shprtgt back from EP, emrfq  1.09268458292027
 in shprtgt, jshprpass=   2
 shprtgt back from EP, emrfq  1.09236675140711
 End of shaper: a,m,b,v,rmsr,rmsl,sig0t,sigt,sigt/sig0t,sig0l,sigl,sigl/sig0l,gamma*rmsl/rmsr,elnrms/etnrms,sigt/sigl,energy,etnrms,elnrms
   0.282715    1.092356    8.450972    0.059696    0.070183   53.845396   50.626046   37.623196    0.743159   28.509801   18.810118    0.659777  767.230221    1.999968    0.499961    0.030000    0.000025
0.000050
 Beginning of shaper: a,m,b,v,rmsr,rmsl,sig0t,sigt, sigt/sig0t, workingKP
   0.325236    1.000010    6.760777    0.059696    0.145450    0.284361   43.593768   50.587299    1.160425    0.014827
 RFQ entrance b,phi,m,v,a =
   0.010000  -90.000000    1.000010    0.059696    8.456619
 Radial matching section   cell#  z, b, phis, m, v, a
```

| | | | | | | |
|---|---|---|---|---|---|---|
| 1 | 0.4239 | 1.6977 | -90 | 1.0 | 0.059696 | 0.648850 |
| 2 | 0.8478 | 3.3854 | -90 | 1.0 | 0.059696 | 0.459550 |
| 3 | 1.2717 | 5.0731 | -90 | 1.0 | 0.059696 | 0.375430 |
| 4 | 1.6956 | 6.7608 | -90 | 1.0 | 0.059696 | 0.325220 |

Generate shaper

| jjj | znew | bemstart | phisstart | emstart | vstart | energystart | astart | Pcu | aarfq | e0rfq | trfq | e0t |
|---|---|---|---|---|---|---|---|---|---|---|---|---|
| 10 | 2.56505 | 6.83203 | -90.00000 | 1.00389 | 0.05970 | 0.03000 | 0.32281 | 0.00261 | 0.00099 | 0.00014 | 0.78620 | 0.00011 |
| 20 | 6.80420 | 6.94990 | -90.00000 | 1.01033 | 0.05970 | 0.03000 | 0.31937 | 0.00688 | 0.00266 | 0.00038 | 0.78625 | 0.00029 |
| 30 | 11.00096 | 7.06636 | -90.00000 | 1.01670 | 0.05970 | 0.03000 | 0.31604 | 0.01111 | 0.00436 | 0.00061 | 0.78630 | 0.00048 |
| 40 | 15.24011 | 7.18424 | -90.00000 | 1.02313 | 0.05970 | 0.03000 | 0.31277 | 0.01538 | 0.00611 | 0.00086 | 0.78635 | 0.00068 |
| 50 | 19.47926 | 7.30200 | -90.00000 | 1.02957 | 0.05970 | 0.03000 | 0.30959 | 0.01966 | 0.00790 | 0.00111 | 0.78640 | 0.00087 |
| 60 | 23.71841 | 7.41976 | -90.00000 | 1.03600 | 0.05970 | 0.03000 | 0.30649 | 0.02393 | 0.00973 | 0.00137 | 0.78645 | 0.00108 |
| 70 | 25.95756 | 7.53738 | -90.00000 | 1.04244 | 0.05970 | 0.03000 | 0.30347 | 0.02820 | 0.01159 | 0.00163 | 0.78650 | 0.00128 |
| 80 | 32.23910 | 7.65631 | -90.00000 | 1.04893 | 0.05970 | 0.03000 | 0.30049 | 0.03252 | 0.01352 | 0.00190 | 0.78655 | 0.00150 |
| 90 | 36.47825 | 7.77421 | -90.00000 | 1.05537 | 0.05970 | 0.03000 | 0.29765 | 0.03679 | 0.01517 | 0.00214 | 0.78658 | 0.00168 |
| 100 | 40.71791 | 7.89184 | -89.65405 | 1.06180 | 0.05970 | 0.03002 | 0.29489 | 0.04106 | 0.01682 | 0.00237 | 0.78661 | 0.00186 |
| 110 | 44.96096 | 8.00970 | -89.30538 | 1.06824 | 0.05970 | 0.03010 | 0.29220 | 0.04534 | 0.01854 | 0.00261 | 0.78665 | 0.00205 |
| 120 | 49.21147 | 8.12777 | -88.95610 | 1.07470 | 0.05970 | 0.03024 | 0.28957 | 0.04962 | 0.02035 | 0.00285 | 0.78670 | 0.00225 |
| 130 | 53.47425 | 8.24619 | -88.60581 | 1.08117 | 0.05970 | 0.03045 | 0.28700 | 0.05392 | 0.02227 | 0.00311 | 0.78676 | 0.00245 |
| 140 | 57.75489 | 8.36509 | -88.25405 | 1.08766 | 0.05970 | 0.03075 | 0.28448 | 0.05824 | 0.02435 | 0.00339 | 0.78686 | 0.00267 |

End of shaper EOS:   jjj  147  nshprcells  143  celldivdsgn  10  energy  0.31027284E-01

| | paramoutlist = {z, | brfq, | phirfq, | emrfq, | vrfq, | energy, | arfq, | r0rfq, | rmsr, | rmsl, | tfac, | lgapfac, | epk, | Pcu} |
|---|---|---|---|---|---|---|---|---|---|---|---|---|---|---|
| 150 | 62.038528 | 8.4623 | -87.6721 | 1.0930 | 0.0598 | 0.0312 | 0.2827 | 0.2959 | 0.0755 | 0.1510 | 1.6753 | 2.2634 | 1.1895 | 0.0626 |
| 160 | 66.377332 | 8.4695 | -87.3699 | 1.0932 | 0.0598 | 0.0317 | 0.2827 | 0.2959 | 0.0754 | 0.1510 | 1.6768 | 2.2634 | 1.1895 | 0.0670 |
| 170 | 70.755765 | 8.4777 | -87.0109 | 1.0934 | 0.0599 | 0.0323 | 0.2827 | 0.2959 | 0.0753 | 0.1510 | 1.6785 | 2.2634 | 1.1895 | 0.0714 |
| 180 | 75.179725 | 8.4871 | -86.5810 | 1.0938 | 0.0599 | 0.0330 | 0.2827 | 0.2960 | 0.0752 | 0.1510 | 1.6804 | 2.2634 | 1.1895 | 0.0759 |
| 190 | 79.656405 | 8.4981 | -86.0622 | 1.0942 | 0.0600 | 0.0339 | 0.2827 | 0.2960 | 0.0751 | 0.1510 | 1.6827 | 2.2634 | 1.1895 | 0.0804 |
| 200 | 84.194671 | 8.5109 | -85.4316 | 1.0947 | 0.0600 | 0.0349 | 0.2827 | 0.2961 | 0.0750 | 0.1510 | 1.6854 | 2.2634 | 1.1896 | 0.0851 |
| 210 | 88.805577 | 8.5260 | -84.6589 | 1.0953 | 0.0601 | 0.0361 | 0.2827 | 0.2962 | 0.0749 | 0.1510 | 1.6885 | 2.2634 | 1.1896 | 0.0898 |
| 220 | 93.503063 | 8.5442 | -83.7052 | 1.0961 | 0.0602 | 0.0376 | 0.2827 | 0.2963 | 0.0747 | 0.1510 | 1.6923 | 2.2634 | 1.1896 | 0.0946 |
| 230 | 98.304947 | 8.5665 | -82.5199 | 1.0972 | 0.0603 | 0.0395 | 0.2827 | 0.2965 | 0.0745 | 0.1510 | 1.6968 | 2.2634 | 1.1897 | 0.0995 |
| 240 | 103.234299 | 8.5938 | -81.0371 | 1.0988 | 0.0605 | 0.0418 | 0.2827 | 0.2967 | 0.0743 | 0.1509 | 1.7027 | 2.2634 | 1.1898 | 0.1046 |
| 250 | 108.321700 | 8.6288 | -79.1707 | 1.1008 | 0.0607 | 0.0448 | 0.2827 | 0.2970 | 0.0739 | 0.1509 | 1.7099 | 2.2634 | 1.1899 | 0.1099 |
| 260 | 113.608149 | 8.6693 | -76.8297 | 1.1040 | 0.0609 | 0.0488 | 0.2827 | 0.2974 | 0.0736 | 0.1510 | 1.7185 | 2.2634 | 1.1900 | 0.1154 |
| 270 | 119.148976 | 8.6986 | -73.9902 | 1.1099 | 0.0611 | 0.0541 | 0.2827 | 0.2983 | 0.0732 | 0.1512 | 1.7275 | 2.2634 | 1.1902 | 0.1212 |
| 280 | 125.018325 | 8.7412 | -70.4653 | 1.1181 | 0.0615 | 0.0615 | 0.2827 | 0.2994 | 0.0727 | 0.1516 | 1.7396 | 2.2634 | 1.1905 | 0.1275 |
| 290 | 131.320454 | 8.8081 | -66.1054 | 1.1303 | 0.0620 | 0.0720 | 0.2827 | 0.3011 | 0.0720 | 0.1519 | 1.7569 | 2.2634 | 1.1907 | 0.1343 |
| 300 | 138.135905 | 8.9195 | -60.8022 | 1.1495 | 0.0628 | 0.0877 | 0.2827 | 0.3038 | 0.0709 | 0.1522 | 1.7837 | 2.2634 | 1.1912 | 0.1418 |
| 310 | 145.834671 | 9.0356 | -54.5769 | 1.1874 | 0.0640 | 0.1128 | 0.2827 | 0.3092 | 0.0696 | 0.1531 | 1.8174 | 2.2634 | 1.1922 | 0.1505 |
| 320 | 154.715087 | 9.2628 | -47.2631 | 1.2573 | 0.0664 | 0.1557 | 0.2827 | 0.3191 | 0.0674 | 0.1540 | 1.8753 | 2.2634 | 1.1943 | 0.1612 |
| 330 | 165.404519 | 9.5232 | -39.3509 | 1.4077 | 0.0712 | 0.2372 | 0.2827 | 0.3404 | 0.0645 | 0.1568 | 1.9598 | 2.2634 | 1.1991 | 0.1754 |
| 340 | 178.958960 | 9.6705 | -32.0238 | 1.6907 | 0.0798 | 0.4004 | 0.2827 | 0.3804 | 0.0610 | 0.1648 | 2.0743 | 2.2634 | 1.2083 | 0.1972 |
| 350 | 196.776982 | 9.5271 | -30.0000 | 1.9679 | 0.0879 | 0.6874 | 0.2827 | 0.4195 | 0.0572 | 0.1821 | 2.2118 | 2.2634 | 1.2096 | 0.2333 |
| 360 | 219.506978 | 9.4059 | -30.0000 | 2.1223 | 0.0926 | 1.0610 | 0.2827 | 0.4414 | 0.0528 | 0.2058 | 2.3960 | 2.2634 | 1.2056 | 0.2860 |
| 370 | 247.020049 | 9.3416 | -30.0000 | 2.1809 | 0.0942 | 1.4847 | 0.2827 | 0.4496 | 0.0481 | 0.2365 | 2.6295 | 2.2634 | 1.2014 | 0.3543 |
| 370 | 248.214832 | 9.3416 | -30.0000 | 2.1813 | 0.0942 | 1.5022 | 0.2827 | 0.4497 | 0.0479 | 0.2379 | 2.6410 | 2.2634 | 1.2012 | 0.3573 |

| | sigoutlist = {z, | sig0t, | sigt, | sigt/sig0t, | sig0l, | sigl, | sigl/sig0l, | sig0t/sig0l, | gamma*rmsl/rmsr, | elnrms/etnrms, | sigt/sigl, | etnrms, | elnrms, | wkgKP} |
|---|---|---|---|---|---|---|---|---|---|---|---|---|---|---|
| 150 | 62.038528 | 50.6957 | 37.6873 | 0.7434 | 28.5418 | 18.8382 | 0.6600 | 1.7762 | 2.0005 | 2.0004 | 2.0006 | 0.000025 | 0.000050 | 0.999994 |
| 160 | 66.377332 | 50.7434 | 37.7215 | 0.7434 | 28.5497 | 18.8390 | 0.6599 | 1.7774 | 2.0022 | 2.0022 | 2.0023 | 0.000025 | 0.000050 | 0.999994 |
| 170 | 70.755765 | 50.7978 | 37.7604 | 0.7433 | 28.5587 | 18.8397 | 0.6597 | 1.7787 | 2.0042 | 2.0042 | 2.0043 | 0.000025 | 0.000050 | 0.999993 |
| 180 | 75.179725 | 50.8603 | 37.8052 | 0.7433 | 28.5689 | 18.8404 | 0.6595 | 1.7803 | 2.0065 | 2.0065 | 2.0066 | 0.000025 | 0.000050 | 0.999991 |
| 190 | 79.656405 | 50.9329 | 37.8570 | 0.7433 | 28.5806 | 18.8411 | 0.6592 | 1.7821 | 2.0092 | 2.0091 | 2.0093 | 0.000025 | 0.000050 | 0.999990 |
| 200 | 84.194671 | 51.0180 | 37.9178 | 0.7432 | 28.5941 | 18.8417 | 0.6589 | 1.7842 | 2.0123 | 2.0123 | 2.0124 | 0.000025 | 0.000050 | 0.999988 |
| 210 | 88.805577 | 51.1189 | 37.9906 | 0.7432 | 28.6095 | 18.8420 | 0.6586 | 1.7868 | 2.0160 | 2.0159 | 2.0163 | 0.000025 | 0.000050 | 0.999986 |
| 220 | 93.503063 | 51.2397 | 38.0771 | 0.7431 | 28.6282 | 18.8423 | 0.6582 | 1.7898 | 2.0205 | 2.0203 | 2.0208 | 0.000025 | 0.000050 | 0.999983 |
| 230 | 98.304947 | 51.3888 | 38.1880 | 0.7431 | 28.6472 | 18.8383 | 0.6576 | 1.7939 | 2.0258 | 2.0246 | 2.0271 | 0.000025 | 0.000050 | 0.999979 |
| 240 | 103.234299 | 51.5687 | 38.3106 | 0.7429 | 28.6800 | 18.8443 | 0.6571 | 1.7981 | 2.0329 | 2.0328 | 2.0330 | 0.000025 | 0.000050 | 0.999972 |
| 250 | 108.321700 | 51.8020 | 38.4806 | 0.7428 | 28.7128 | 18.8411 | 0.6562 | 1.8041 | 2.0413 | 2.0404 | 2.0424 | 0.000024 | 0.000050 | 0.999965 |
| 260 | 113.608149 | 52.0781 | 38.6815 | 0.7428 | 28.7301 | 18.8094 | 0.6547 | 1.8127 | 2.0520 | 2.0477 | 2.0565 | 0.000024 | 0.000050 | 0.999976 |
| 270 | 119.148976 | 52.2855 | 38.8029 | 0.7421 | 28.7110 | 18.7595 | 0.6534 | 1.8211 | 2.0665 | 2.0646 | 2.0684 | 0.000024 | 0.000050 | 0.999967 |
| 280 | 125.018325 | 52.5886 | 38.9960 | 0.7415 | 28.6796 | 18.6764 | 0.6512 | 1.8337 | 2.0853 | 2.0827 | 2.0880 | 0.000024 | 0.000050 | 0.999954 |
| 290 | 131.320454 | 53.0542 | 39.3039 | 0.7408 | 28.6682 | 18.5881 | 0.6484 | 1.8506 | 2.1107 | 2.1071 | 2.1145 | 0.000024 | 0.000050 | 0.999932 |
| 300 | 138.135905 | 53.8060 | 39.8062 | 0.7398 | 28.7294 | 18.5378 | 0.6453 | 1.8729 | 2.1470 | 2.1469 | 2.1473 | 0.000023 | 0.000050 | 0.999824 |
| 310 | 145.834671 | 54.6275 | 40.3189 | 0.7381 | 28.6468 | 18.3190 | 0.6395 | 1.9069 | 2.2005 | 2.2004 | 2.2009 | 0.000023 | 0.000050 | 0.999746 |
| 320 | 154.715087 | 56.1823 | 41.3460 | 0.7359 | 28.6645 | 18.0920 | 0.6312 | 1.9600 | 2.2847 | 2.2846 | 2.2853 | 0.000022 | 0.000050 | 0.999479 |
| 330 | 165.404519 | 58.0448 | 42.4400 | 0.7312 | 28.3365 | 17.4522 | 0.6159 | 2.0484 | 2.4309 | 2.4306 | 2.4318 | 0.000021 | 0.000050 | 0.999091 |
| 340 | 178.958960 | 59.3690 | 42.7394 | 0.7199 | 26.9643 | 15.7967 | 0.5858 | 2.2018 | 2.7039 | 2.7033 | 2.7056 | 0.000018 | 0.000050 | 0.998635 |
| 350 | 196.776982 | 59.0092 | 41.2183 | 0.6985 | 24.1540 | 12.9182 | 0.5348 | 2.4430 | 3.1872 | 3.1861 | 3.1907 | 0.000016 | 0.000050 | 0.999302 |
| 360 | 219.506978 | 58.7429 | 39.5015 | 0.6724 | 21.3379 | 10.1068 | 0.4737 | 2.7530 | 3.9018 | 3.8996 | 3.9084 | 0.000013 | 0.000050 | 0.999669 |
| 370 | 247.020049 | 58.7585 | 37.8737 | 0.6446 | 18.7454 | 7.6296 | 0.4070 | 3.1345 | 4.9224 | 4.8888 | 4.9640 | 0.000010 | 0.000050 | 0.999981 |
| 370 | 248.214832 | 58.7736 | 37.8029 | 0.6432 | 18.6496 | 7.5362 | 0.4041 | 3.1515 | 4.9729 | 4.9379 | 5.0162 | 0.000010 | 0.000050 | 0.999984 |

# Appendix 2. H+ Design – Case 1

```
FREQ   200.000000000000  Charge   1.00000000000000  Mass   1.00700000000000  Win   3.000000000000000E-002  Wout   0.750000000000000  CURRENT   5.000000000000000E-002
 powcuFacToshiba  0.000000000000000
 shprtgt,blinj,EOSaperfac,apertgt  1.19880729624088  3.00000000000000  0.399602432080295
 rfqdesignGUI  3.00000000000000  0.399602432080295 -88.0000000000000  0.800000000000000  1.00000000000000  4.00000000000000  240.000000000000  60.0000000000000
 in shprtgt, jshprpass=   1
 shprtgt back from EP, emrfq  1.22600851652930
 in shprtgt, jshprpass=   2
 shprtgt back from EP, emrfq  1.22599064940523
 End of shaper: a,m,b,v,rmsr,rmsl,sig0t,sigt,sigt/sig0t,sig0l,sigl,sigl/sig0l,gamma*rmsl/rmsr,elnrms/etnrms,sigt/sigl,energy,etnrms,elnrms
    0.399935    1.225978   11.018770    0.083694    0.054976   65.266642   62.334263   30.022021    0.481630   48.217595   21.447190    0.444800 1187.219829    1.400501    0.714382    0.030000
0.000020    0.000028
 Beginning of shaper: a,m,b,v,rmsr,rmsl,sig0t,sigt, sigt/sig0t, workingKP
   0.476890    1.000010    8.815016    0.083694    0.213272    0.402092   56.839583   62.236020    1.094942    0.014176
```

```
RFQ entrance b,phi,m,v,a =
  0.010000  -90.000000   1.000010   0.083694   14.158920
Radial matching section   cell#  z, b, phis, m, v, a
    1     0.5994     2.2113   -90   1.0   0.083694   0.951950
    2     1.1988     4.4125   -90   1.0   0.083694   0.673970
    3     1.7982     6.6138   -90   1.0   0.083694   0.550520
    4     2.3976     8.8150   -90   1.0   0.083694   0.476870
Generate shaper
  jjj     znew      bemstart   phisstart     emstart    vstart    energystart   astart      Pcu       aarfq      e0rfq      trfq       e0t
   10   3.62744    9.01365   -90.00000    1.02037   0.08369   0.03001   0.46782   0.00724   0.00469   0.00066   0.78601   0.00052
   20   9.62209    9.34203   -90.00000    1.05403   0.08369   0.03001   0.45434   0.01912   0.01284   0.00179   0.78608   0.00141
   30  15.55680    9.66688   -90.00000    1.08735   0.08369   0.03001   0.44216   0.03088   0.01986   0.00277   0.78608   0.00218
   40  21.55145    9.99514   -90.00000    1.12101   0.08369   0.03001   0.43054   0.04276   0.02730   0.00381   0.78626   0.00300
   50  27.54690   10.32344   -89.57760    1.15468   0.08369   0.03005   0.41980   0.05463   0.03490   0.00487   0.78637   0.00383
   60  33.56040   10.65273   -88.83048    1.18844   0.08369   0.03041   0.40999   0.06655   0.04301   0.00596   0.78649   0.00469
   70  39.63207   10.98521   -88.07614    1.22254   0.08369   0.03126   0.40082   0.07858   0.05253   0.00718   0.78670   0.00565
 End of shaper EOS:   jjj  71  nshprcells  67  celldivdsgn  10  energy  0.31378866E-01


paramoutlist = {z,       brfq,     phirfq,    emrfq,    vrfq,    energy,   arfq,    r0rfq,    rmsr,    rmsl,     tfac,    lgapfac,   epk,     Pcu}
   80  45.832847  11.2945  -85.2015   1.1987   0.0845   0.0332   0.3999   0.4397   0.0743   0.1045   2.4060   4.4854   1.1900   0.0910
   90  52.316015  11.4143  -80.6961   1.2040   0.0852   0.0376   0.3999   0.4407   0.0735   0.1045   2.4324   4.3558   1.1907   0.1042
  100  59.397665  11.6683  -71.9983   1.2227   0.0869   0.0480   0.3999   0.4445   0.0718   0.1044   2.4898   4.2143   1.1924   0.1190
  110  67.872431  12.0696  -56.8187   1.3128   0.0914   0.0789   0.3999   0.4625   0.0686   0.1057   2.6062   4.0453   1.1963   0.1380
  120  80.087669  12.7841  -36.7191   1.7683   0.1114   0.2116   0.3999   0.5536   0.0610   0.1146   2.9300   3.8022   1.2232   0.1727
  130 102.165563  12.8344  -27.2699   2.2928   0.1369   0.6946   0.3999   0.6584   0.0469   0.1759   3.8171   3.3632   1.2238   0.2779
  130 104.805484  12.9544  -27.3475   2.2567   0.1359   0.7501   0.3999   0.6512   0.0449   0.1867   3.9848   3.3106   1.2199   0.2918


sigoutlist = {z,       sig0t,     sigt,    sigt/sig0t,   sig0l,     sigl,     sigl/sig0l,  sig0t/sig0l, gamma*rmsl/rmsr,
                                                                                   elnrms/etnrms, sigt/sigl,  etnrms,     elnrms,     wkgKP}
   80  45.832847  65.2029   30.9206   0.4742   45.8776   21.9847   0.4792   1.4212   1.4063   1.4063   1.4065  0.000020  0.000028  0.999945
   90  52.316015  65.9880   31.2322   0.4733   46.0980   21.9826   0.4769   1.4315   1.4206   1.4205   1.4208  0.000020  0.000028  0.999877
  100  59.397665  67.6646   31.8782   0.4711   46.5234   21.9369   0.4715   1.4544   1.4530   1.4529   1.4532  0.000019  0.000028  0.999666
  110  67.872431  70.5178   32.6600   0.4631   46.5612   21.2026   0.4554   1.5145   1.5400   1.5399   1.5404  0.000018  0.000028  0.999128
  120  80.087669  76.2860   32.9772   0.4323   44.1719   17.5559   0.3974   1.7270   1.8772   1.8767   1.8784  0.000014  0.000027  0.994695
  130 102.165563  79.8198   26.3749   0.3304   30.9017    7.0046   0.2267   2.5830   3.7571   3.7544   3.7654  0.000007  0.000025  1.000557
  130 104.805484  80.8625   25.8310   0.3194   29.6176    6.1674   0.2082   2.7302   4.1632   4.1449   4.1883  0.000006  0.000025  1.002181
```

# Appendix 3.  H+ Design – Case 2

```
FREQ  200.000000000000  Charge  1.00000000000000  Mass  1.00700000000000  Win  3.000000000000000E-002  Wout  0.750000000000000  CURRENT  5.000000000000000E-002
 powcuFacToshiba  0.000000000000000
 shprtgt,blinj,EOSaperfac,apertgt  1.19880729624088  3.20000000000000  0.374627280075276
 rfqdesignGUI  3.20000000000000  0.374627280075276 -88.0000000000000  0.800000000000000  1.00000000000000  4.00000000000000  240.000000000000  60.0000000000000
 in shprtgt, jshprpass=   1
 shprtgt back from EP, emrfq  1.26781708896965
 in shprtgt, jshprpass=   2
 shprtgt back from EP, emrfq  1.26779498866489
 End of shaper: a,m,b,v,rmsr,rmsl,sig0t,sigt,sigt/sig0t,sig0l,sigl,sigl/sig0l,gamma*rmsl/rmsr,elnrms/etnrms,sigt/sigl,energy,etnrms,elnrms
    0.375006    1.267782   11.671086    0.079091    0.053417   68.082817   65.163552   30.977992    0.475388   53.237625   23.826120    0.447543 1274.592139    1.299996    0.769131    0.030000
0.000020    0.000026
 Beginning of shaper: a,m,b,v,rmsr,rmsl,sig0t,sigt, sigt/sig0t, workingKP
   0.450450    1.000010    9.336869    0.079091    0.201447    0.402092   60.204509   65.040893    1.080333    0.014191
 RFQ entrance b,phi,m,v,a =
   0.010000  -90.000000    1.000010    0.079091   13.764080
 Radial matching section   cell#  z, b, phis, m, v, a
     1     0.5994     2.3417   -90   1.0   0.079091   0.899270
     2     1.1988     4.6734   -90   1.0   0.079091   0.636630
     3     1.7982     7.0052   -90   1.0   0.079091   0.520010
     4     2.3976     9.3369   -90   1.0   0.079091   0.450430
 Generate shaper
   jjj     znew      bemstart   phisstart     emstart    vstart    energystart   astart      Pcu       aarfq      e0rfq      trfq       e0t
    10   3.62759    9.57061   -90.00000    1.02681   0.07909   0.03001   0.44052   0.00647   0.00711   0.00094   0.78637   0.00074
    20   9.62238    9.95687   -90.00000    1.07113   0.07909   0.03001   0.42595   0.01708   0.01810   0.00239   0.78640   0.00188
    30  15.55722   10.33928   -90.00000    1.11500   0.07909   0.03001   0.41270   0.02758   0.02846   0.00375   0.78656   0.00295
    40  21.55200   10.72554   -90.00000    1.15931   0.07909   0.03001   0.40041   0.03818   0.03917   0.00517   0.78670   0.00407
    50  27.55268   11.11219   -89.19718    1.20367   0.07909   0.03020   0.38939   0.04880   0.05010   0.00659   0.78680   0.00518
    60  33.59816   11.50172   -88.36279    1.24835   0.07909   0.03095   0.37917   0.05950   0.06307   0.00819   0.78708   0.00645
  End of shaper EOS:   jjj  64  nshprcells  60  celldivdsgn  10  energy  0.31508556E-01


paramoutlist = {z,       brfq,     phirfq,    emrfq,    vrfq,    energy,   arfq,    r0rfq,    rmsr,    rmsl,     tfac,    lgapfac,   epk,     Pcu}
   70  39.729704  11.8993  -86.0083   1.2495   0.0798   0.0328   0.3750   0.4217   0.0745   0.0971   2.2527   4.8889   1.1908   0.0704
   80  46.260357  12.3068  -79.7867   1.2195   0.0807   0.0379   0.3750   0.4157   0.0745   0.0971   2.2527   4.6278   1.1916   0.0823
   90  53.477614  12.8039  -67.3728   1.2236   0.0826   0.0518   0.3750   0.4155   0.0745   0.0971   2.2527   4.3396   1.1933   0.0959
  100  62.548362  13.5220  -46.8224   1.3861   0.0893   0.0980   0.3750   0.4432   0.0745   0.0971   2.2527   3.9779   1.2004   0.1144
  110  77.587346  14.3016  -25.0000   2.0819   0.1156   0.3654   0.3750   0.5605   0.0745   0.0971   2.2527   3.3804   1.2251   0.1616
  118  99.795658  15.5983  -25.0000   1.7064   0.1014   0.7519   0.3750   0.4827   0.0745   0.0971   2.2527   2.4957   1.1978   0.2403


 sigoutlist = {z,       sig0t,     sigt,    sigt/sig0t,   sig0l,     sigl,     sigl/sig0l,  sig0t/sig0l, gamma*rmsl/rmsr,
                                                                                   elnrms/etnrms, sigt/sigl,  etnrms,     elnrms,     wkgKP}
   70  39.729704  67.4238   30.9780   0.4754   53.2376   23.8261   0.4475   1.2665   1.3036   1.3000   1.3002  0.000020  0.000026  1.000000
   80  46.260357  71.4561   30.9780   0.4754   53.2376   23.8261   0.4475   1.3422   1.3036   1.3000   1.3002  0.000020  0.000026  1.000000
   90  53.477614  75.7554   30.9780   0.4754   53.2376   23.8261   0.4475   1.4230   1.3036   1.3000   1.3002  0.000020  0.000026  1.000000
  100  62.548362  80.9225   30.9780   0.4754   53.2376   23.8261   0.4475   1.5200   1.3037   1.3000   1.3002  0.000020  0.000026  1.000000
  110  77.587346  88.9048   30.9780   0.4754   53.2376   23.8261   0.4475   1.6700   1.3041   1.3000   1.3002  0.000020  0.000026  1.000000
  118  99.795658  99.4511   30.9780   0.4754   53.2376   23.8261   0.4475   1.8681   1.3046   1.3000   1.3002  0.000020  0.000026  1.000000
```

# Attached Files

Folder:  'D+ H+ Files to Iwashita'

[eltoc]

# Chapter 27 – On Compensation of an Existing Linac for H- Intrabeam Scattering and Residual Beam Loss

R. A. Jameson
October 2017
KEK Report 2-17-4 December 2017, A (also ref# 1724004)
With editing thereafter

## Abstract

Intrabeam stripping of the H- beam causing beam loss in the SNS and J-Parc linacs was discovered after they were built and brought into operation. A rough compensation, via reducing quad strengths below design resulting in larger beam size, was found. However, problems with the lowered focusing strength remain, because residual losses apparently from other sources remain that are of concern for future beam power upgrades. Detailed analysis has not yet been performed, which could help optimize a compensation scheme. Here are offered some observations, analysis, discussion and conclusions, that formulate a framework for such study, which is hoped may be of use.

## 27.1 History – Technical and Human Factors

The LAMPF 800 MeV, 1 mA average H+ current linac was a jump of ~x10 in energy and x1000 in average current – completely unknown territory when the project started in ~1962. The beam dynamics design strung components together and the first simulation programs were written and gave some information about performance. After LAMPF came online, strong emittance growth at low energy was observed – reason completely unknown, same result at sister linacs then built at BNL, ANL, CERN. Sacherer at CERN showed how the transverse and longitudinal envelope equations could be used with equivalent rms beam distributions. CERN designed the "New Linac" keeping the beam "matched" to the envelope equations throughout, and expected no emittance growth. We tested the operation in 1977 [206] – the same strong emittance growth was observed, and it was learned how misalignment and rms mismatching were strong contributors. However, ~30% of the observed growth remained even after very careful matching and alignment. In 1981, it was shown that this component was lack of energy balance within the beam bunch – the beam was not equipartitioned, but tended to move toward EP by emittance exchange [207,208].

---

206 R. A. Jameson, R. S. Mills, O. R. Sander, "Report on Foreign Travel - Switzerland", LASL Office Memo AT-DO-351(U)MP-9, Dec. 28, 1978; and "Emittance Data from the New CERN Linac", R.A. Jameson, Letter to G. Plass, CERN, AT-DO-262(U), January 8, 1979, and R. A. Jameson, ""Emittance Growth in the New CERN Linac - Transverse Plane Comparison between Experimental Results and Computer Simulation", LASL Office Memo AT-DO-377(U), Jan. 15, 1979; and R. A. Jameson, "CERN Linac Tests", LASL Office Memo MP-9/AT-DO-(U), Mar. 1, 1979; and R. A. Jameson, "CERN Linac Tests" LASL Office Memo AT-DO-514(U), Apr. 26, 1979. In 1992, the first three of these were consolidated in LA-UR-92-3033, "Emittance Growth in the New CERN Linac - Studies in 1978", R.A. Jameson, R.S. Mills, O.R. Sander.

207 "Beam-Intensity Limitations in Linear Accelerators", R.A. Jameson, (Invited), Proc. 1981 Particle Accelerator Conf., Washington, DC, March 11-13, 1981, lEEE Trans. Nucl. Sci. 28, p. 2408, June 1981; Los Alamos National Laboratory Report LA-UR-81-765, 9 March 1981. (Correction, Jameson, RA; IEEE TRANSACTIONS ON NUCLEAR SCIENCE; 1981; v.28, no.4, p.3665-3665)

208 "Equipartitioning in Linear Accelerators", R.A. Jameson, Proc. of the 1981 Linear Accelerator Conf., Santa Fe, NM, October l9-23, 1981, Los Alamos National Laboratory Report LA-9234-C, p. 125, February 1982; Los Alamos National Laboratory Report LA-UR-81-3073, 19 October 1981

When longitudinal focusing is falling, as in room-temperature linacs, maintaining internal beam equilibrium (equipartitioning – EP) requires transverse focusing to also fall. It is interesting that studies of the LAMPF linac and later extensive, but still preliminary, studies of ATW (Accelerator Transmutation of Waste) linacs [209] were dismissed by LANL[210] because resulting final total beam emittances at the high energy output of the linac (~1 GeV) were ~ the same in EP and existing non-EP designs, and the EP rms output beam size was larger – ignoring the fact that the EP design had remained tighter (controlled 100% to rms emittance ratio), while the non-EP design had large total beam size and emittance growth already at low energy and throughout the linac, so would have more distributed low level loss.

It is ironic that now "it has been discovered" that intra H- beam stripping requires a larger beam size at higher energy, and that this was discovered after the SNS was built. The SNS beam dynamics[211] was designed only by "best practice" – so-called "state-of-the-art". But not best practice. Although design "from inside out" (that is, to require properties of the beam dynamics, including the space charge physics, and designing external field to give those properties) had been known and demonstrated by then for many years, it was ignored. The "front end" was salamied off to a lab without RFQ experience, which used the limited *PARI/PARMTEQM* design method for the beam dynamics, with predictable performance results. The actual "choke-point" – the limiting dynamic aperture – occurred in the drift-tube linac (DTL) – an unheard of place. The limiting dynamic aperture should occur in the RFQ, with no losses observed in the DTL. The DTL was built with permanent magnet quadrupoles; the reason was "to be the first to do that", eliminating future possibility to retune as performance needed to be improved. Investigation of possible resonance interactions was not performed, except finally after direct command from the Advisory Committee, the DTL was checked. Fortunately the DTL trajectory lay mostly in the safe region – as long as the longitudinal emittance was not larger than design – and there is some resonance encounter. The rest of the room-temperature and superconducting (SC) linacs (adopted only after the DOE unilaterally declared SC to be the official approach) were not checked, even for the main resonances, and will certainly have the rf gap effects investigated below. There was a precedence regarding the H- stripping – Star Wars. It had been shown that stripping off the extra H- electron was easy. For Star Wars, high brightness was important, loss in the linac was not important because the war would be over in a few seconds, but for SNS the implication was not considered.

J-Parc had learned the principles of beam equilibrium and how to apply them to design, and the J-Parc linac was designed and built to be completely equipartitioned. They had no precedence so the problem of H- intrabeam stripping also became known only in operation. However, the entire linac is built with electromagnetic quads, so there are full possibilities for possible adjustment.

The observation that beam loss could be reduced by reducing quad strengths (increasing transverse beam size) was reported in early 2009. Intrabeam stripping was later proven to account for the major portion of the loss. Machine tunes with lowered quads became standard at SNS and J-Parc. But small losses are still observed, while not troublesome at present beam power levels, potentially too high for intended upgrades.

Some measurements and simulations have been made at J-Parc, observing for example that an equipartitioned beam is more stable against errors than a beam which is battling resonances – as should be expected, and that basic tunes in regions free of the main resonances have less emittance consequences, as expected. The transverse focusing reductions desirable for enough loss reduction in the existing linac seem to be successful but still problematic. However, the analyses could be extended and further detailed actions may be possible.

---

209 "On Scaling & Optimization of High Intensity, Low-Beam-Loss RF Linacs for Neutron Source Drivers", R.A. Jameson, in Advanced Accelerator Concepts, AIP Conf. Proc. 279, ISBN 1-56396-191-1, DOE Conf-9206193 (1992) 969-998; Proc. Third Workshop on Advanced Accelerator Concepts, 14-20 June 1992, Port Jefferson, Long Island, NY, LA-UR-92-2474, Los Alamos National Laboratory, 28 July 1992.

210 The dismissal was actually much more basic than technical – human characteristics, NIH, etc..

211 This section refers to beam dynamics, not other aspects such as engineering, controls, etc., which on the whole were executed very well.

## 27.2 Analysis Method

Deep analysis using established techniques is possible and is of interest here. The method is very simple and has been in use since LAMPF in the 1970's.

***The actual accelerator, and the design and analysis of it, is simply a time-varying system – not a steady-state one.***

***The observation method*** was and ***is simpl***e – the rms beam sizes (termed herein as "a" or "rmsr" for transverse and "b" or "rmsl" for longitudinal) and emittances could be measured even continuously in both experiment, given such instrumentation, and in simulation – at every step. Then theoretical constructs such as the local phase advances and local satisfaction of the equipartitioning equation can be checked. The "Hofmann Chart" is used to study resonance locations and effects.

The breadth and depth of the lack of understanding that people seem to have about these preceding two paragraphs is simply astounding. Vehement argument has been presented in recent conferences, and by journal reviewers, that the observation method "is very suspect", that the equipartitioning equation is derived essentially from thermodynamic principles and thus cannot be used locally but must wait until thermodynamic equilibrium is reached – which never occurs even in a very intense linac, nor in any typical nonlinear system with long settling times. Or that the envelope equations only apply to infinitely long continuous unchanging focusing channels, and not to bunched beams being accelerated. Or that in any case, there have been no experimental checks until the recent GSI experiments. Etc.

Such arguments are totally beside the point, and only reveal lack of thought. The observation method is obvious, works for analysis, works for design with acceleration, and is verified by simulation and running machines. Every running machine has been an experiment, and has been analyzed and re-simulated.

From years of observation, perhaps as a trained engineer in a design environment dominated by persons trained in physics, I believe a basic problem is in the essentially steady-state training of physics, vs. the continual emphasis on time-varying phenomena in engineering training. There is also the ability of engineering to take a *concept* and *make something work*, without needing to re-derive the concept – for example, the crucial concept of applying approximate theory successfully to elucidate the local, instantaneous state. Another problem is the classical one of getting trapped in specific examples and being unable to get out of that box.

Another aspect concerns the placement of theory in the scheme of things [212]. A key **element** that helps the understanding of the big picture is that there is an "actual (experiment/simulation/theory) hierarchy", in that order. It was noted above that "the (LAMPF) design strung components together and the first simulation programs were written and gave some information about performance". A linac or ring beam has no knowledge of "theory". With knowledge of accurate beam-in to beam-out component transfer functions, a machine is built, and simulated, with an appropriate sequence of components. The components can include many real details like errors, simplified or more accurate descriptions, etc. that are not possible to include in a theoretical model. The real machine, or this "map", is the reality that is being dealt with. The input/output characteristics of the beam distribution at each component - that is, the local, instantaneous characteristics - constitute the performance – it is

212 Please see the various writings of Etienne Forest, an exceptionally insightful beam dynamicist, who wonderfully expresses these same issues. E.g., "Beam Dynamics: A New Attitude and Framework / Edition 1, ISBN-10: 9057025744 ISBN-13: 9789057025747 Pub. Date: 07/07/1998 Publisher: Taylor & Francis. "A Contemporary Guide to Beam Dynamics", Etienne Forest· and Kohji Hirata. LBL-32793, ESG-210, UC-410,KEK Report 92-12. See footnote, Sec. 16.1.3.1.3.

not practical to measure all of them, but they are available in simulation. The beam itself has no knowledge of "Hamiltonian", "envelope equation", internal energy balance", "rms", etc. A "simulation", more or less accurate, can then reveal more information that cannot be practically measured in the real device. "Theory" then comes in as a way to reveal other information about the local, instantaneous state. The "theory" for beams with space charge is all approximate (mathematically limited to KV distribution as replaced by "equivalent rms" distribution, infinitesimal perturbations, use of steady-state theory locally (instantaneously), etc., etc., etc.). It is not the reality - but it has been shown in actual linacs and rings and by comparison to accurate simulation results that theory can be used in this way to understand and evaluate performance.

Even more important, this derived information from the local beam distribution is also very useful for design, as I do and J-Parc did, although the prevalent confusion is even still almost completely clouding this most important aspect. Unfortunately, the number and size of large high intensity linac projects has not been as great as for large rings, with correspondingly less development of simulation, theory and design practice for very low beam loss.

The method is used again in the following, and it is hoped that observed features will be found believable and useful.

The reader is already aware that the acceptable, usual, totally dry and disconnected presentation is abandoned here, and actual very real issues are mentioned. The following analysis is also given as a story, following a learning curve.

In Chapter 27.3, we consider, as an introduction to the many facets and usefulness of the analysis technique, a generic drift-tube linac, on which we will reduce the quad strengths in the latter half, investigating the performance and considering compensation. Surprising results are found, indicating intriguing and possibly useful new design technique. Understanding the results was possible with the full set of analysis concepts and tools.

In Chapter 27.4, we apply the techniques to the J-Parc linac. Chapter 27.5 gives discussion and conclusions. The conclusions with respect to preparing for a major upgrade (x2 in beam power) are demanding. Chapter 27.6 notes new perspectives and prospects for research on future high intensity linacs.

## 27.3 Generic Equipartitioned Drift-Tube Linac

As preparation with simple conditions to get an understanding of the situation and illustrate the analysis method, a fully equipartitioned 324 MHz, 3–400 MeV 50mA DTL linac is set up, with the *LINACS* design code, as the generic test bed, using a simple quad-drift-gap-drift-quad transfer matrix sequence. The DTL has the strongest transverse focusing of all separated-function linac types, in which the transverse and longitudinal focusing are provided by separate components. The longitudinal focusing is considered fixed and unchangeable. The transverse focusing will be manipulated to explore the strategy of H- intrabeam stripping compensation.

The design is done with the program *LINACS*, unique in emphasizing "inside out" *design* with control of all accelerator parameters including the space charge physics.

### *27.3.1. dtl2: 324 MHz, 3–400 MeV, 50mA EP design, adequate aperture*

Design results are shown in Fig. 27.1:

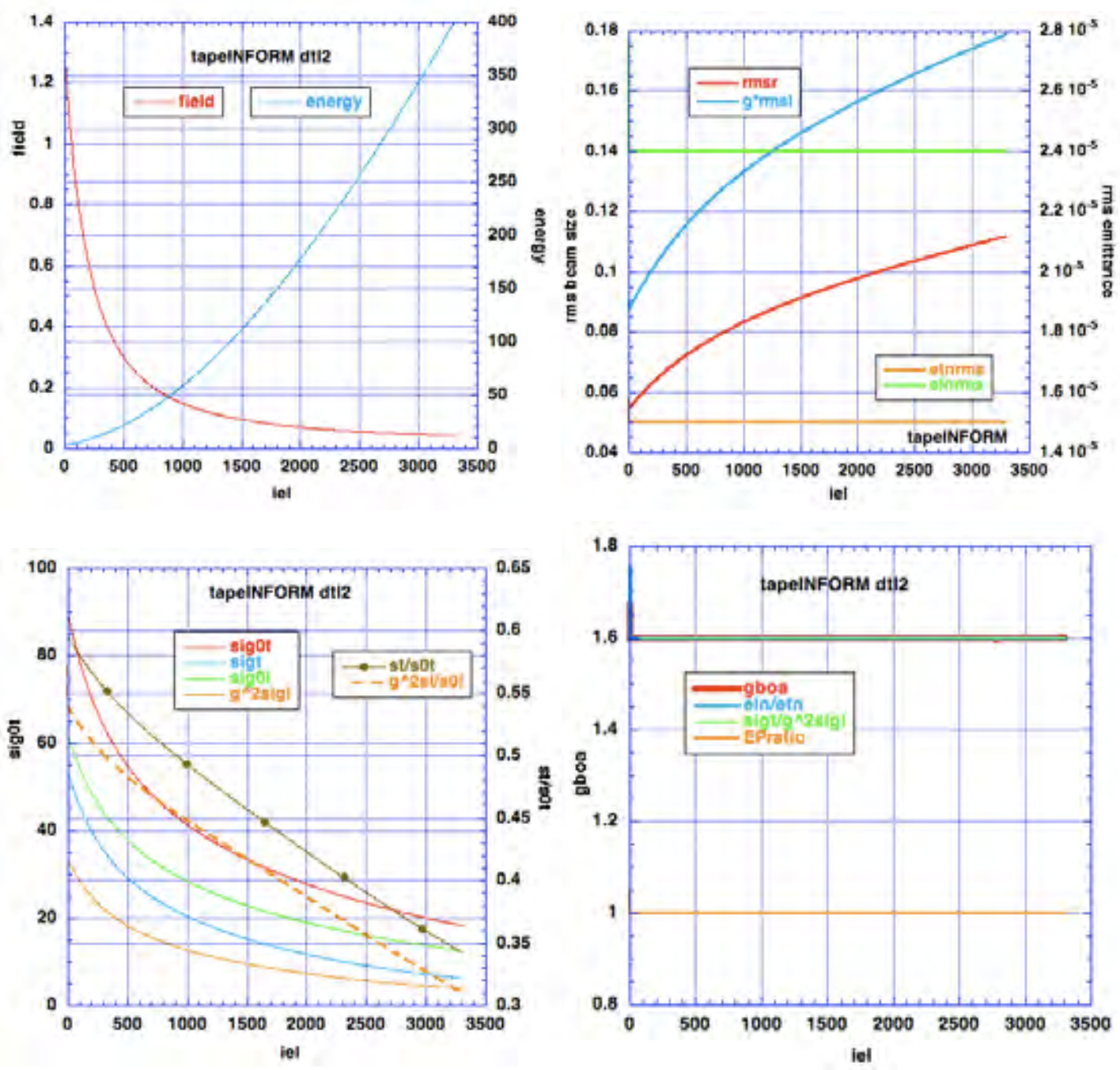

Fig. 27.1. Design performance of dtl2. (field is quadrupole fields, g = gamma, rmsr = a = trms = transverse rms beam radius, g*rmsl = gbrms = γ*longitudinal rms beam size, sig = s = phase advance (s0 with zero current, s alone with beam current), t is transverse, l is longitudinal, n stands for normalized). The necessary growth in transverse beam size is seen at top right. No emittance growth. The EP condition = 1, with EP ratios = 1.6.

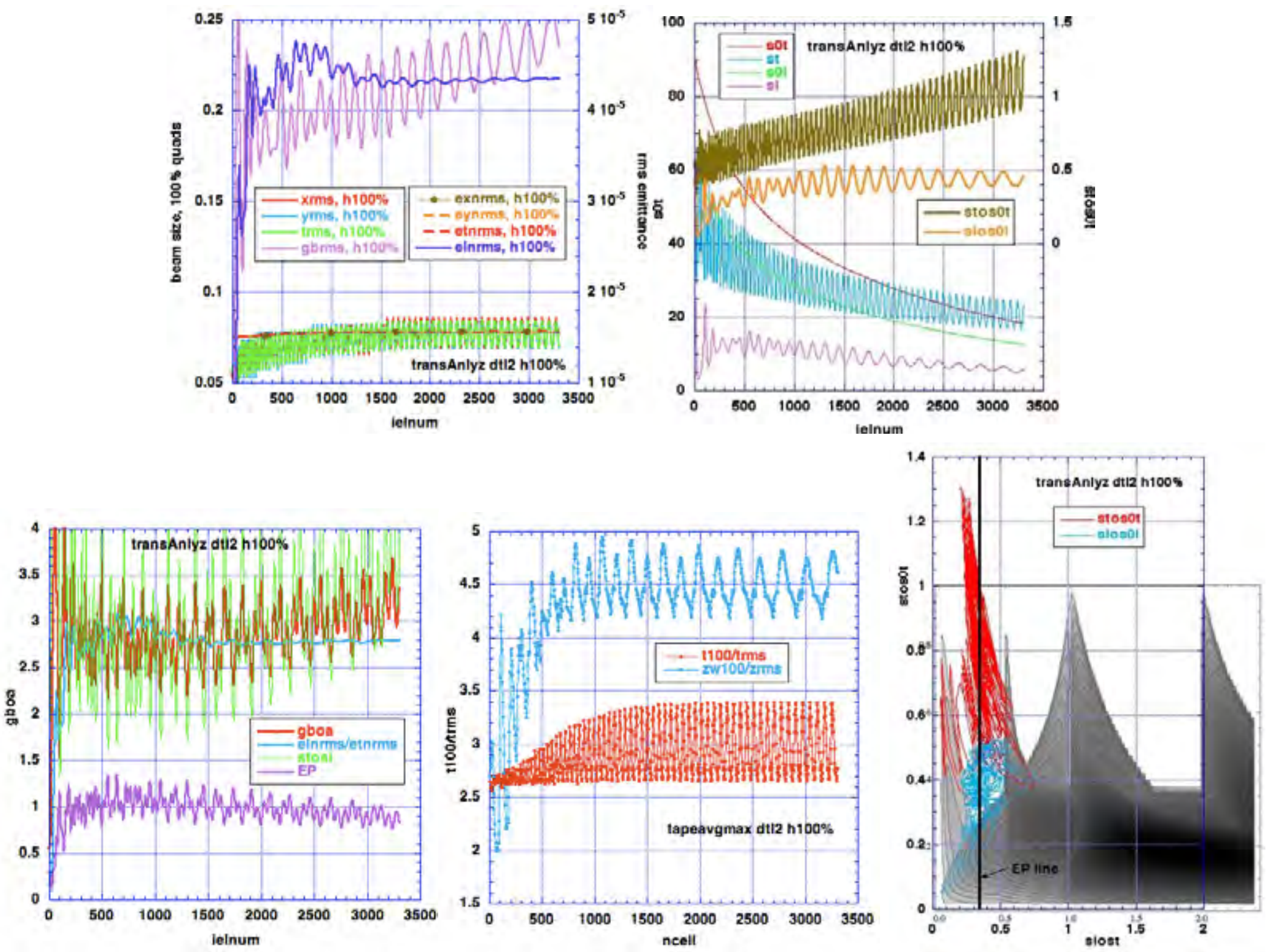

Fig. 27.2. Simulation performance of dtl2. stos0t = st/s0t etc. Hofmann chart for eln/etn emittance ratio = 3. For this run, the longitudinal input emittance was mistakenly set too small, but EP design is specified, the longitudinal emittance quickly adjusts to an ~constant value.

Simulation results are shown in Fig. 27.2. All 10000 injected particles reached full design energy. Injection at center of quad with empirically determined input matching conditions.

The transverse x,y averaged rms beam size (trms) does not follow the design too well - the tune depressions become very deep and the trajectory gets into the sl/st = 0.5 resonance; so the EP ratios are not exactly equal. This is explored a little further below, but this design served the initial purpose. Beam transverse rms x and y sizes are same with small variation. Ratios of 100% to rms beam sizes are controlled.

### *27.3.2. Change quad values to 50% at ~ component 1000:*

J-Parc papers show an H Chart with points representing equipartitioning conditions[213] = (ex*b)/(ez*a) = (ex*sx)/(ez*sz) altered by changing the external field transverse focusing. There is no indication of what is expected will actually change in practice within the local equipartitioning condition when sx decreases or transverse rms beam size (a) increases with the reduction in s0x; the usage on the unchanging emittance ratio chart indicates the assumption that ez, sz, b, and ex remain constant and the change is only in sx and a. Other presented results show that that this assumption is not strictly right, so this terminology is only a kind of visual aid.

The quad strength is lowered from ~ component 1000 uniformly for the rest of the linac:

```
110 h=component(2,n)      ! = quadrupole strength
! Abrupt change makes x and y different ... this makes amplitudes nearly same (and also in-phase):
if(n.eq.998) h=0.85d0*h
if(n.eq.1002) h=0.70d0*h
if(n.ge.1006) h=0.50d0*h
```

#### <u>27.3.2.1. Beam rms size, emittance, accelerated fraction with 50% quads</u>

Fig. 27.3 shows that the average rms beam size has been increased from ~0.08 to ~0.17 cm in the reduced quad strength region. There is a strong oscillation, which could be expected as no particular matching has been applied at the transition. The x and y oscillations are in-phase.

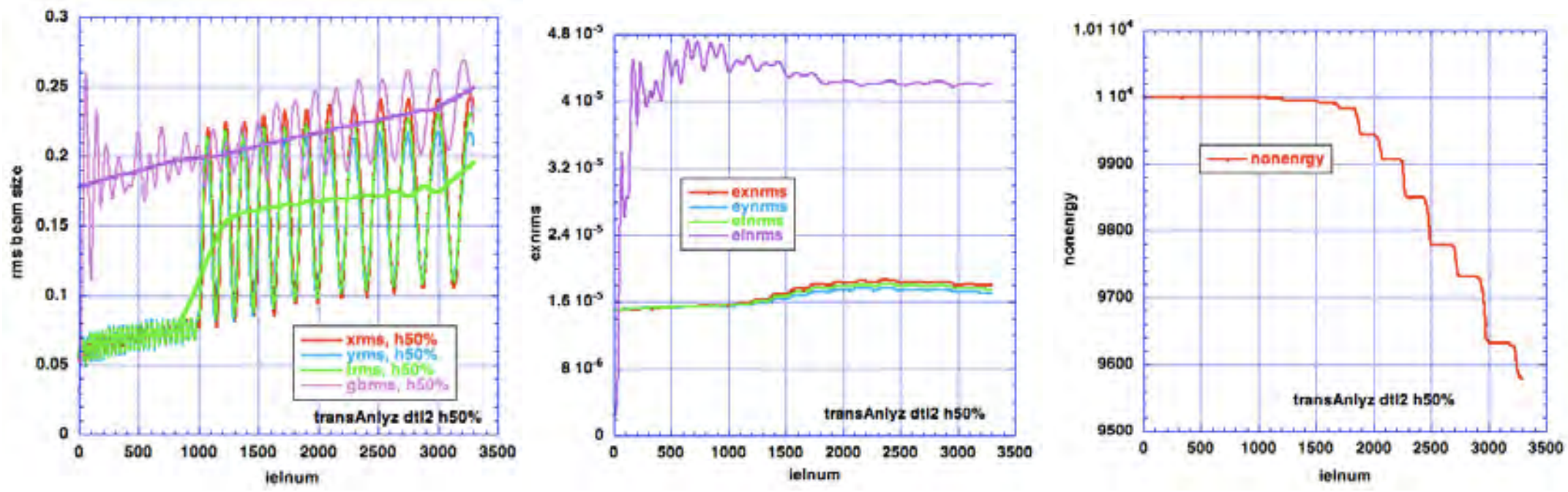

Fig. 27.3. Rms beam size, rms emittance and transmission behavior with quad strength reduction. 20% weighted fit is shown for trms and γbrms.

There is a small transverse rms emittance growth, and corresponding reduction in the rms longitundinal emittance via an equipartitioning reaction.

The maximum x and y transverse beam size remains well within the aperture. There is a small loss in particles reaching full design energy ± 15%.

***This simple local evaluation of beam size and emittance already reveals some very interesting features.***

[213] "Temperature" is an incorrect usage here. This has been discussed with the authors.

• Although the reduction in s0t is drastic, there is very little emittance change. This is also observed in J-Parc simulations and experimental measurements. It would seem that this might not be "expected", and so further investigation is required. *Also, this result and its magnitude is similar to the J-Parc IMPACT simulation result*, made with more accurate beamline component modeling, and thus this simple dtl is a good surrogate for study and should be investigated closely. It is noted that this similarity required some fine tuning and was sensitive to details – this is probably also true for the actual J-Parc simulation and experimental setups; the reason(s) will become clear below.

• xrms and yrms are in-phase and reach the minimums of the oscillation at the same positions, indicating a mismatch mode with minimum transverse area – not good for reduced intrabeam stripping.

• The step-like reductions in accelerated beam fraction are indicative of some kind of systematic behavior.

**27.3.2.2. Phase advance and coherent space charge modes interactions**

To proceed further, we invoke the rms envelope equations relating rms emittances and beam sizes through the betatron and synchrotron phase advances per transverse focusing period, and the equipartitioning condition with its related equal rms emittance, beam size and inverse phase advance ratios. [214]

Fig. 27.4 indicates that the tune depressions are deep – ~0.4 in the longitudinal plane, and really deep ~0.2 – 0.3 in the transverse plane with reduced quads. The sl/st (written as slost) ratio is used on the Hofmann Chart to investigate interaction along the dtl trajectory with the coherent space charge instability modes, the most important of which are at slost ~0.33, 0.5, 1.0, 2.0, 3.0, and higher modes with slost below ~0.25. It is seen that slost oscillates over a very wide range spanning these modes, and that the beam is far from equipartitioned.

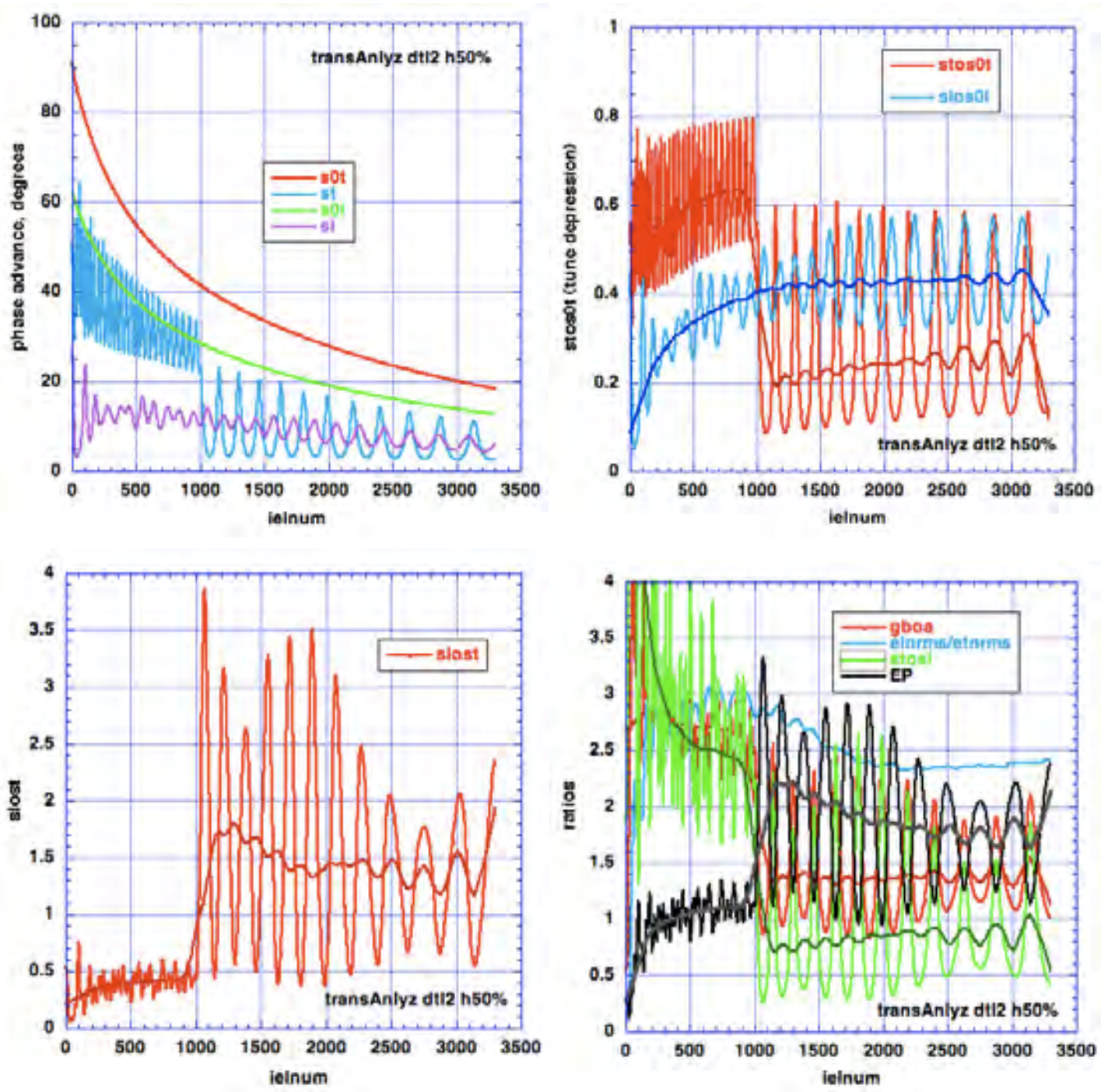

Fig. 27.4. Phase advances, tune depressions, sl/st ratio and EP conditions (with weighted fits).

214 "RFQ Designs and Beam-Loss Distributions for IFMIF", R.A. Jameson, Oak Ridge National Laboratory Report ORNL/TM-2007/001, January 2007.

The direct Hofmann Chart presentation is messy (Fig. 27.5). A trajectory loop around a resonance indicates capture by the resonance (more remote approach without capture is indicated by a kink in the trajectory); there appears to be significant capture by slost=1&2, with perhaps very short involvement with slost=0.5&3:

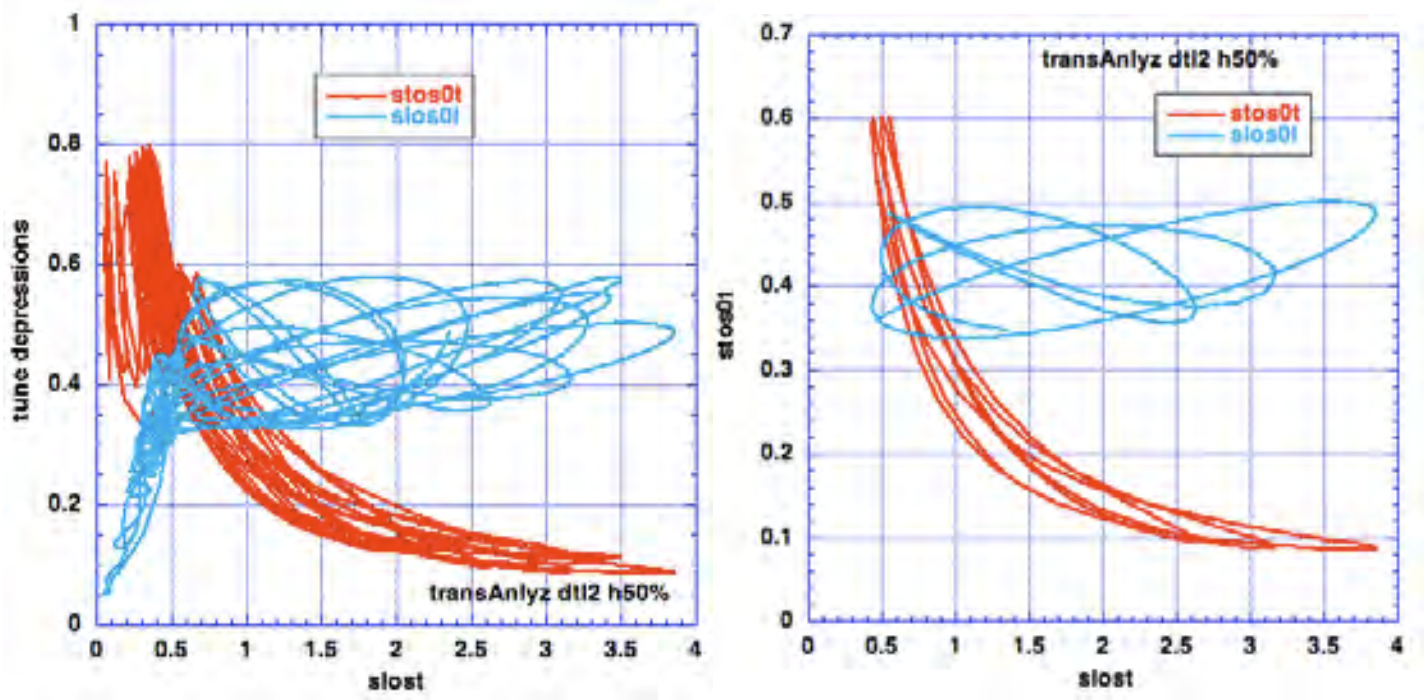

Fig. 27.5. Hofmann Chart, for full trajectory (left),
and for a portion of the reduced quad trajectory (right).

This highly oscillating case is unusual, and a different presentation is needed to show the mode interactions better (Fig. 27.6). exrms and eyrms demonstrate the systematic behavior. Plotting a portion of the etnrms trajectory against slost shows the characteristic resonance capture looping pattern. Then plotting etnrms and slost together vs. components along the trajectory (Fig. 27.6 right) clearly shows a direct correlation of emittance growth with the tune ratio oscillation.

Fig. 27.7 shows the resonance interactions in detail, with shading indicating when the trajectory is in the slost=1 region (light purple), and in the slost=2 region (light red). The slost=2 mode causes etnrms to increase when entering as slost rises and the increase stops when the mode is exited. The slost=1 mode appears to cause etnrms to fall.

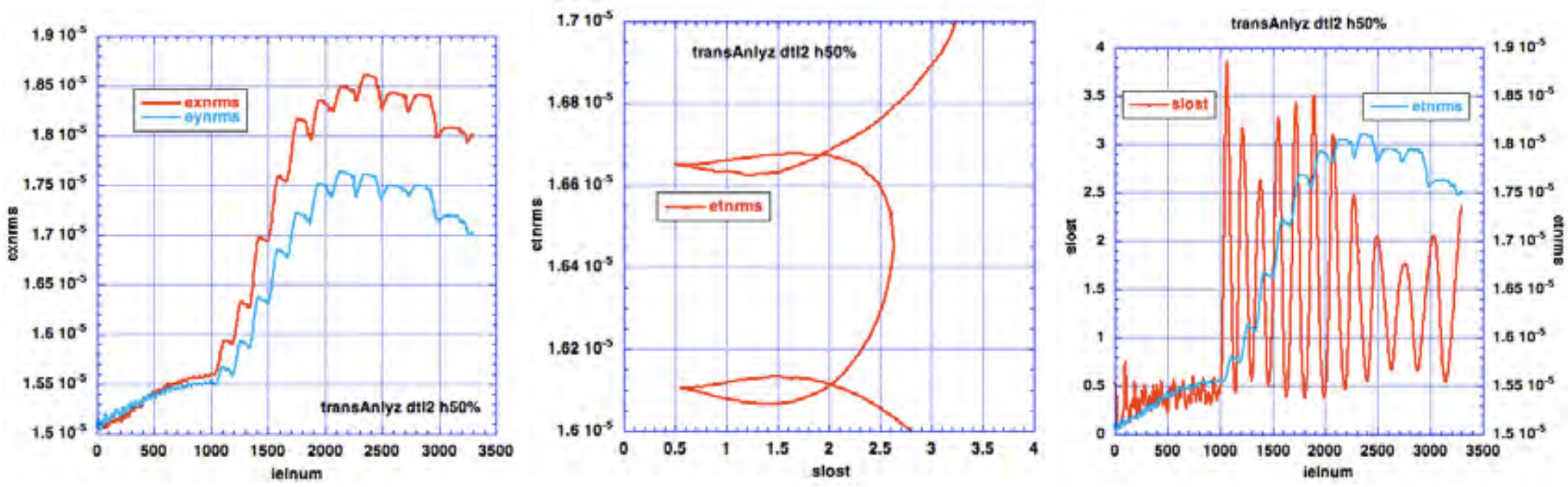

Fig. 27.6. exnrms and eynrms with expanded vertical scale, a portion of the x,y averaged etnrms trajectory vs. slost, and slost compared to etnrms vs. trajectory component number.

***The form of etnrms growth shows direct correlation to space charge resonance modes, from the full step-by-step analysis; whereas weighted curve fits (or a Poincaré plot at the end of each transverse focusing period) are non-informative.***

There are ~400 2βλ periods, each 8 ielnum steps long. At these tune depressions, ¼ plasma period varies from ~2-8 βλ, or 16-64 ielnum steps:

sigpl = Sqrt[(sig0l^2 - sigl^2)/ff];

sigpt = Sqrt[(sig0t^2 - sigt^2)/(2/(1 - ff))];

thus the reaction upon entering a coherent mode stopband is very fast (Fig. 27.8).

stos0t is ~0.2 – 0.3, slos0l is ~0.4;  So the transverse emittance growth is stronger and moves toward equipartition with the longitudinal emittance. Correlation of slost with the longitudinal elnrms is also apparent.  (Fig. 27.9):

The effect of the reduced quads is stronger on the total beam sizes than on the rms sizes, as shown in Fig. 27.10.   t100/trms with 50% quads becomes very large, and eventually the longitudinal zw100/zrms  also increases.

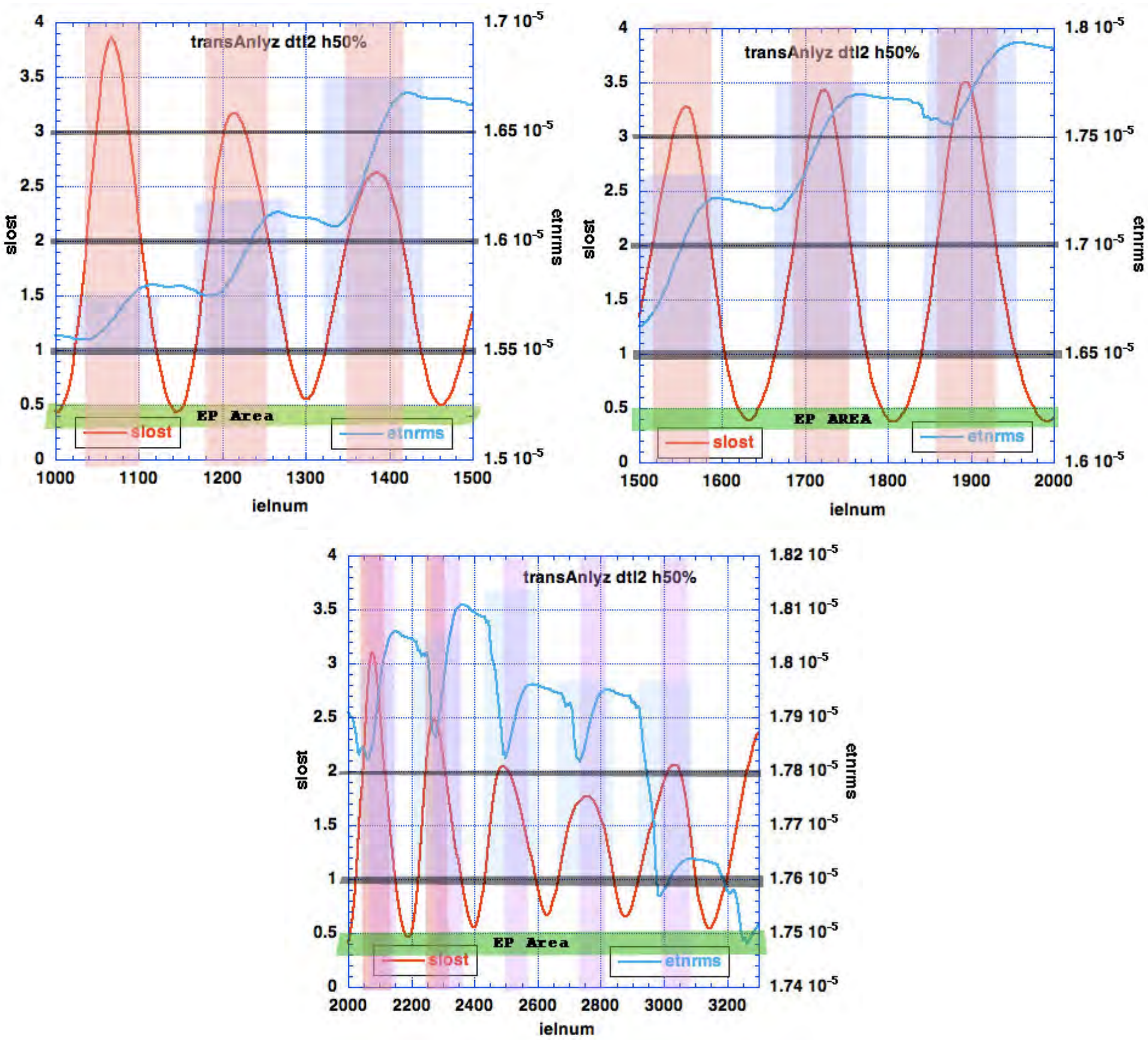

Fig. 27.7.  Detailed evidence of correlation between transverse rms emittance growth and coherent modes.

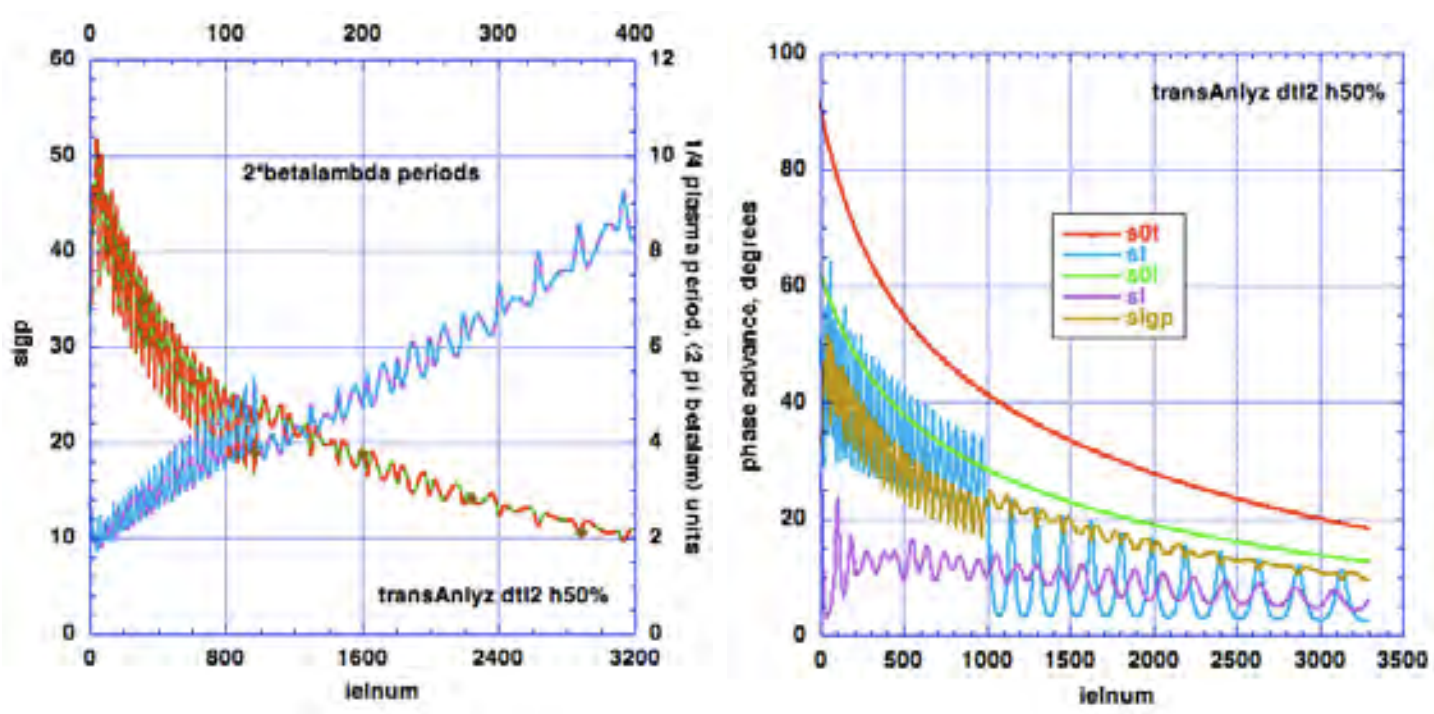

Fig. 27.8.  Plasma oscillation characteristics.

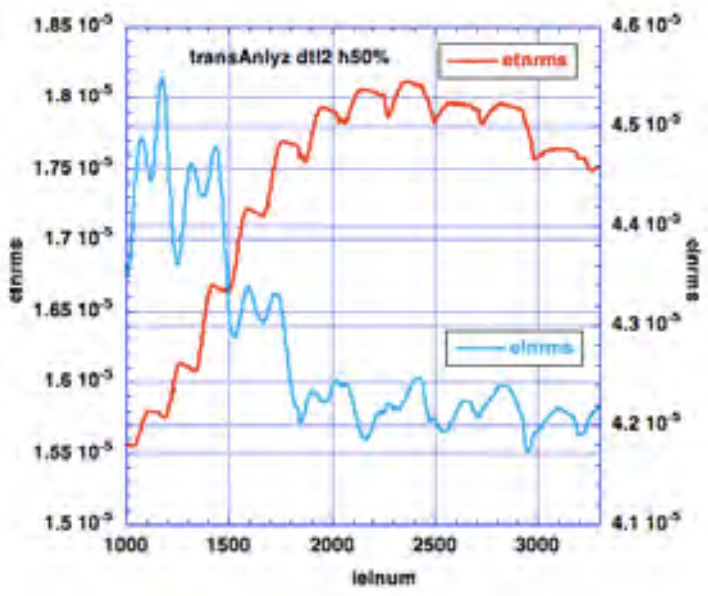

Fig. 27.9.  Emittance transfer

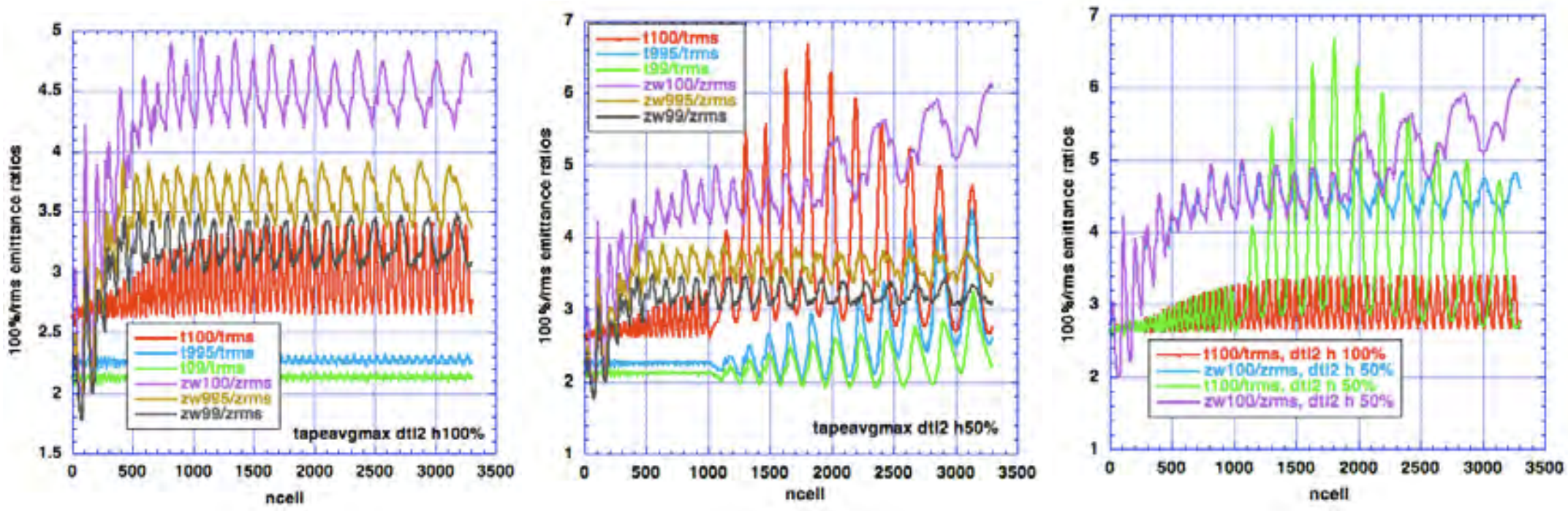

Fig. 27.10.  Total/rms beam size ratios – 100% quads (left), – 50% quads (middle), – compare 100%/rms ratios (right).  t100/trms with 50% quads becomes very large, and eventually the longitudinal zw100/zrms  also increases.

## *27.3.3  Results of attempts to compress x,y beam size oscillation*

Another equipartitioned linac (“dtl4”) with a different (somewhat larger) aperture rule was designed, and then simulated with the better matched inputs, Fig. 27.11:

```
input    -2    -10000  0.0 35.  0.0015 0.0 12.  0.0015
                        22. 0.03154 0.0
 0.0 0.0 0.0 0.0 0.0 0.0 1. 1.008373582   50.0   3.000
```

***Note:  With this dtl, all 10000 particles are accelerated, both with the original design, and with the quad strengths reduced to 50%.***

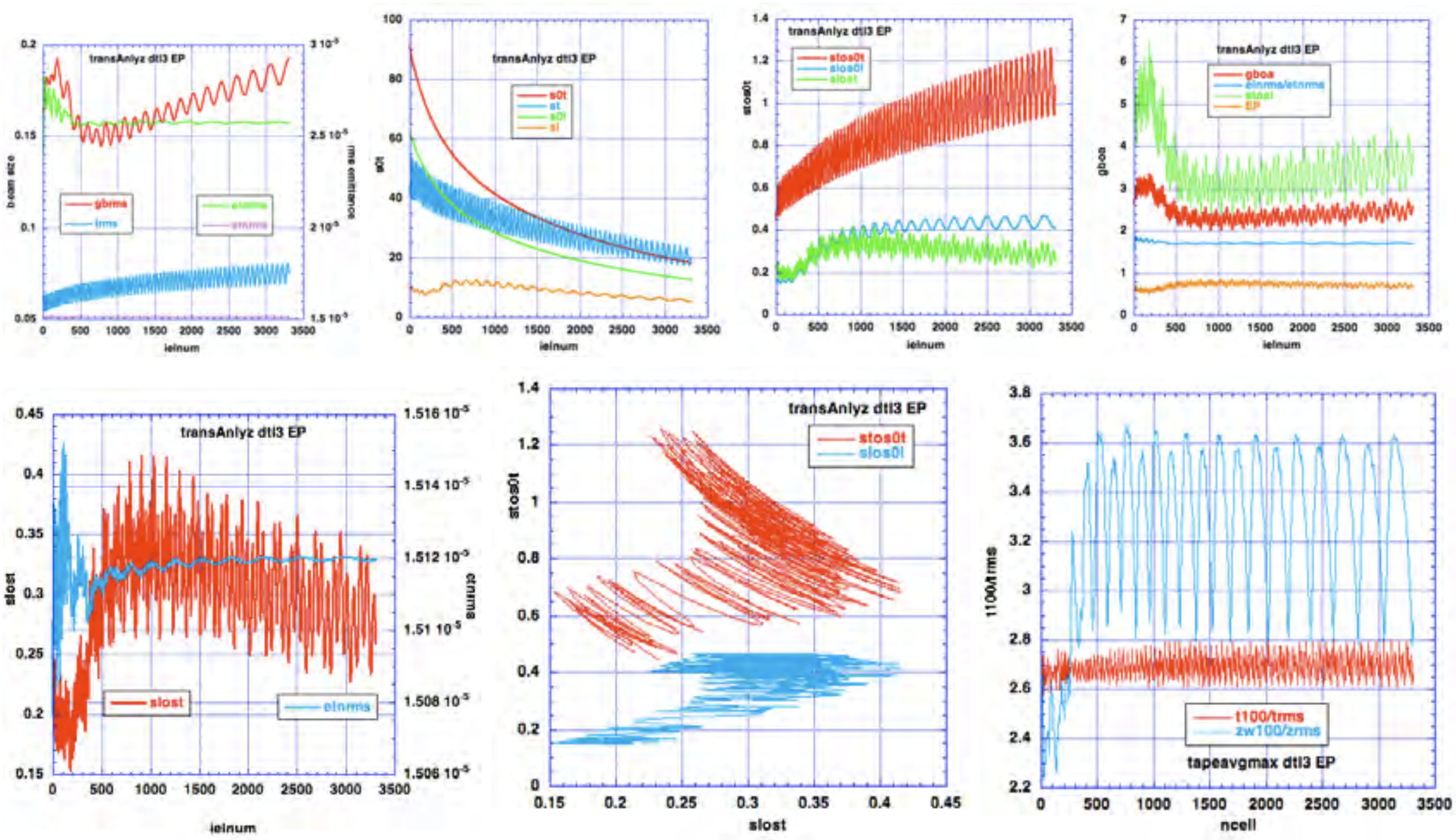

Fig. 27.11.  dtl4 characteristics.

EP ratios are not equal and EP is not exact.  This could be improved in the design, especially after cell ~500, with a form factor adjustment procedure, but the beam is stable, even though sitting on the slost 0.33 resonance.  t100/trms stays tight at sqrt(6).

Main quad reduction to 50% is again done with 3 quads 998, 1002, 1005 to smooth somewhat, resulting in similar behavior (Fig. 27.12):
if(n.eq.998) h=0.85d0*h
if(n.eq.1002) h=0.70d0*h
if(n.ge.1006) h=0.5d0*h    (All quads with component number ≥ 1006 are reduced by 50%.)
Result is strong transverse rms beam size oscillation.

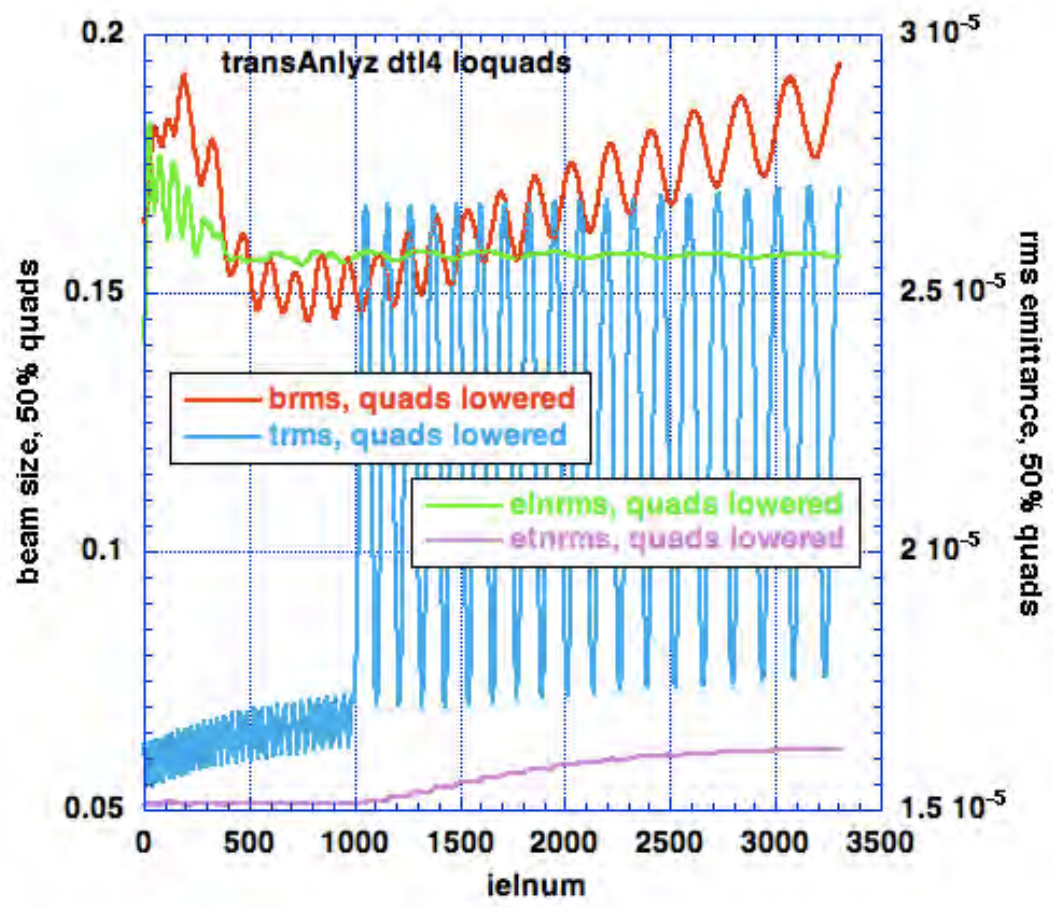

Fig. 27.12.  dtl4 with quad reduction to 50%.

### 27.3.3.1.  Match after the 50% reduction.  Try a little common sense.

We now want to explore what happens if attempts are made to reduce the extent of the transverse oscillation.  The beam size preceding the quad change is ~0.07 cm (Fig. 27.13). A larger transverse beam size is desired to reduce the effect of intrabeam stripping – try for 0.15 cm.

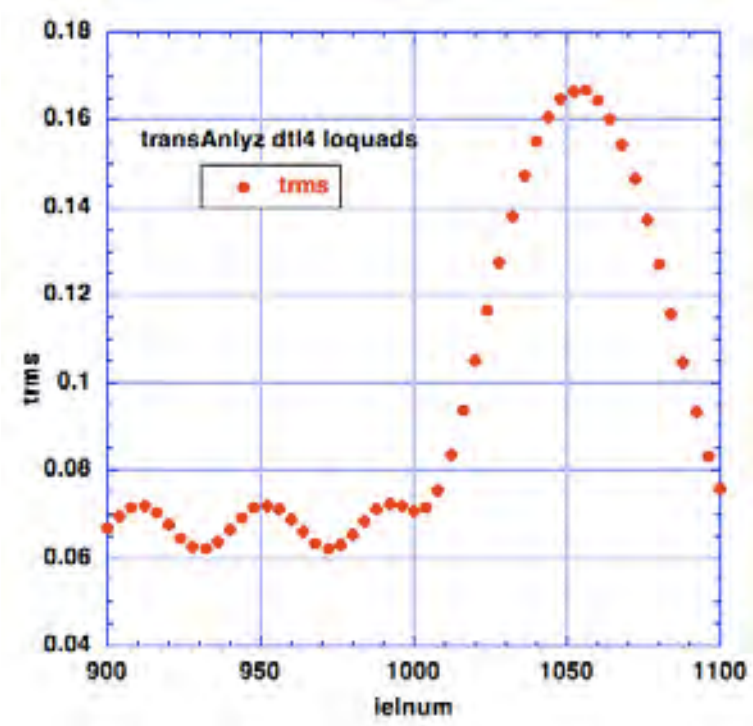

Fig. 27.13.  Detailed transition of the x,y averaged transverse rms beam size (trms) at the quad strength reduction point.

This would require stronger quad when size reaches 0.15 to flatten out.

Trial quickly resulted in x1.5 increase at quad 1034. All quads with component number ≥ 1034 are multiplied by 1.5.

Then a trial and error sequential correction was done on the next 3 quads 1038 -> 0.85, 1042 ->0.80, 1046 -> 0.85, at each step multiplying that quad strength and all those downstream. This produced a fairly flat result and returned the subsequent strengths to ~0.5*(original value): 1.0*0.5*1.5*.85*.8*.85 = 0.4335 (Fig 15):

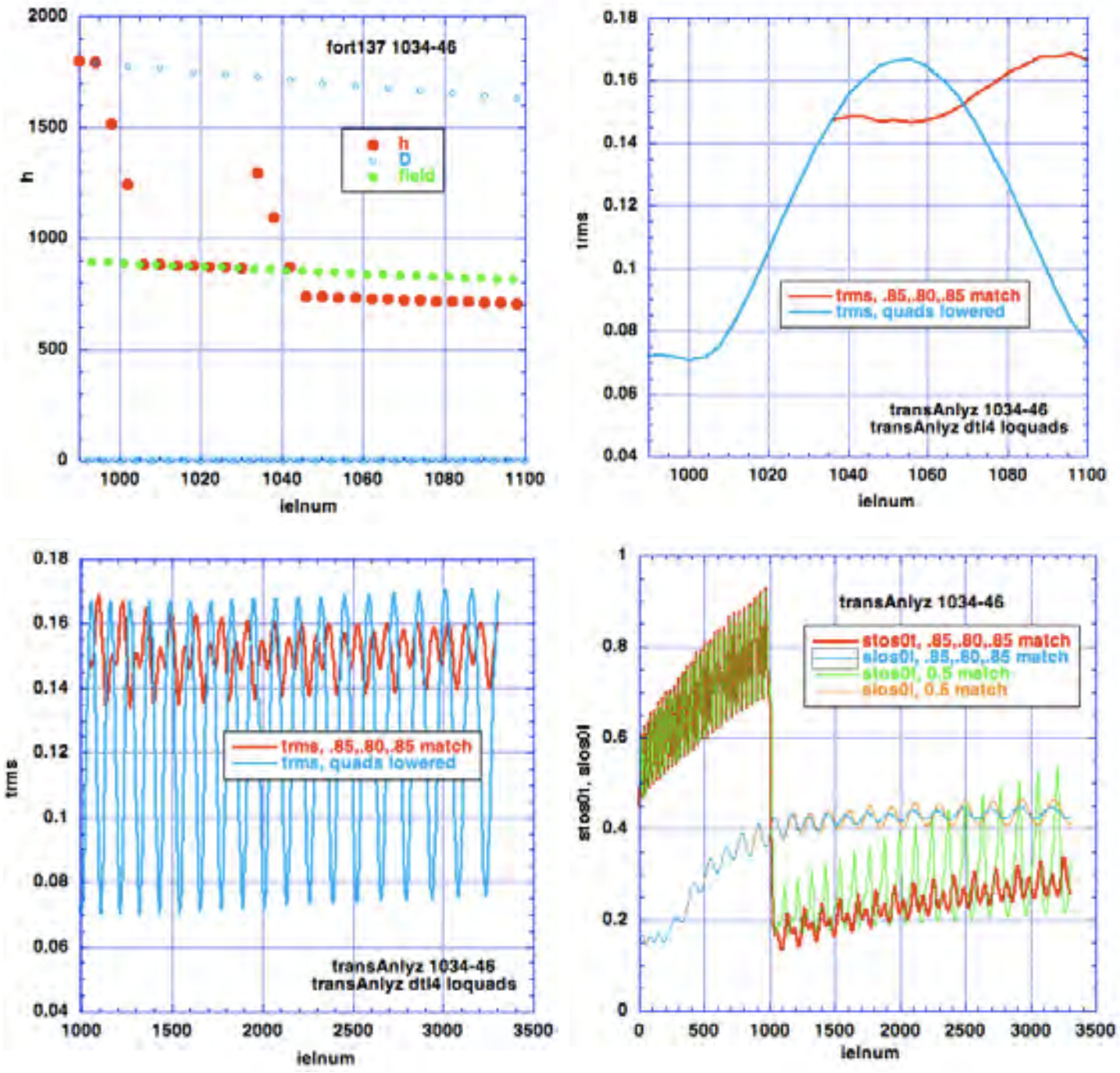

Fig. 27.14. Flattening of trms by adjustment of quads 1034, 1038, 1042 and 1046.

Refinement of the return to 0.5*(original value) requires the factor 0.87358 three times: 1*0.5*1.5*( (0.5/0.75)^(1/3) = 0.87358); steps down from 0.75 -> 0.655185 -> 0.572357 -> 0.5 (Fig. 27.15):

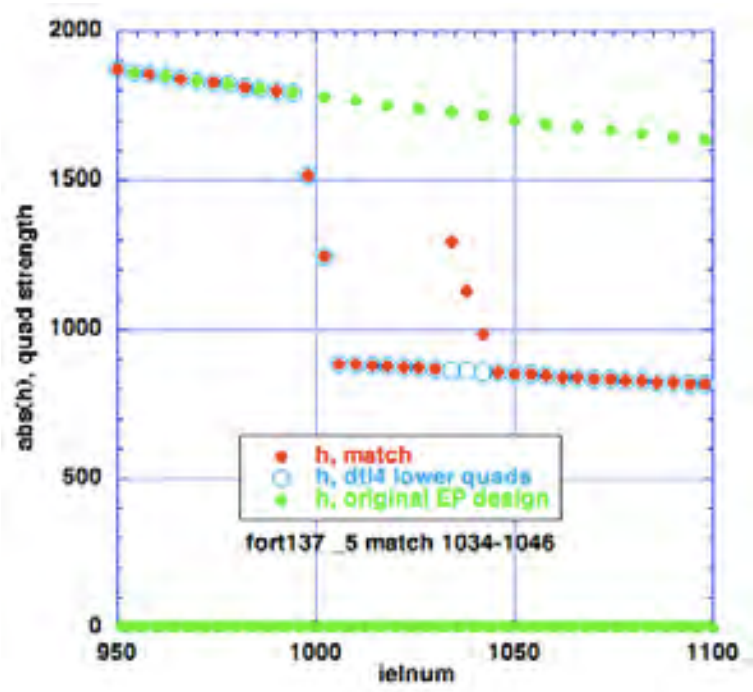

Fig. 27.15. Flatten at quad 1034 and step back to 50% quad strength reduction.

Cool idea, but the hand match above (1.0*0.5*1.5*.85*.8*.85 = 0.4335) gave tighter result and higher average rms beam size ~0.145 (Fig. 27.16):

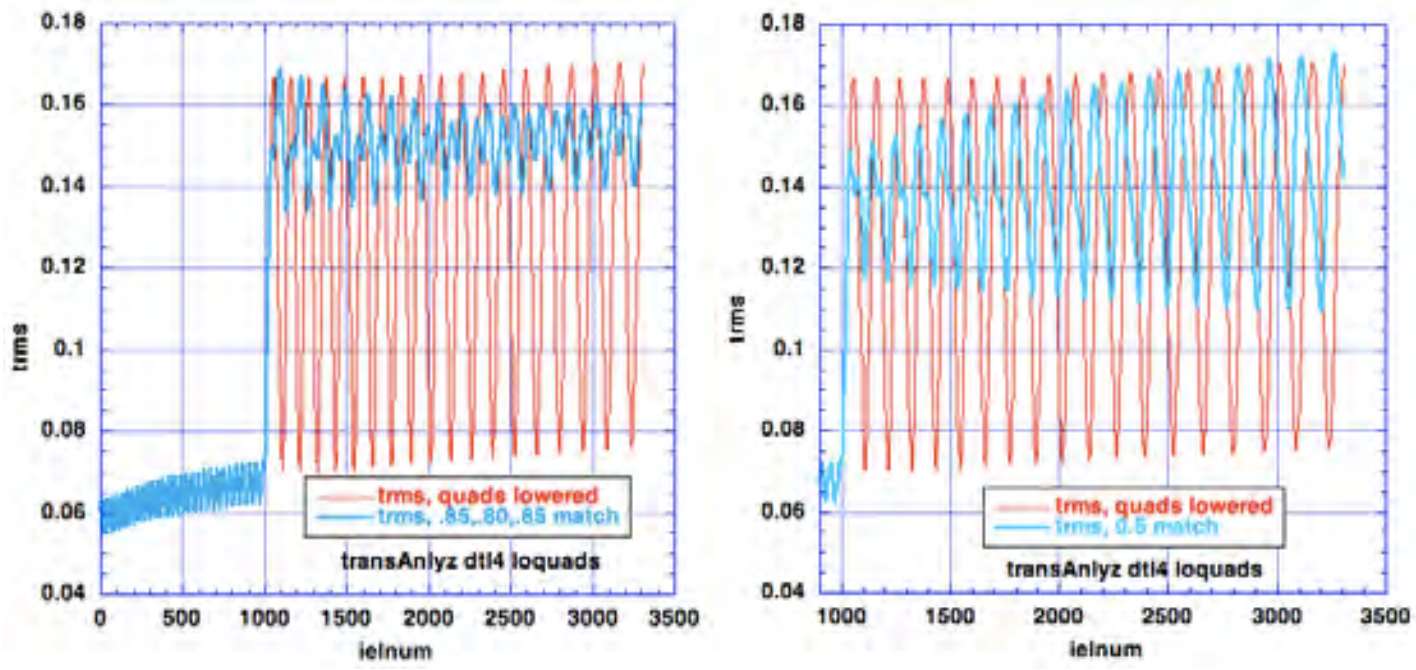

Fig. 27.16. Comparison of Fig. 27.14 and Fig. 27.15 strategies on trms.

Additional results of the 0.85,0.80,0.85 strategy are shown in Fig. 27.17:

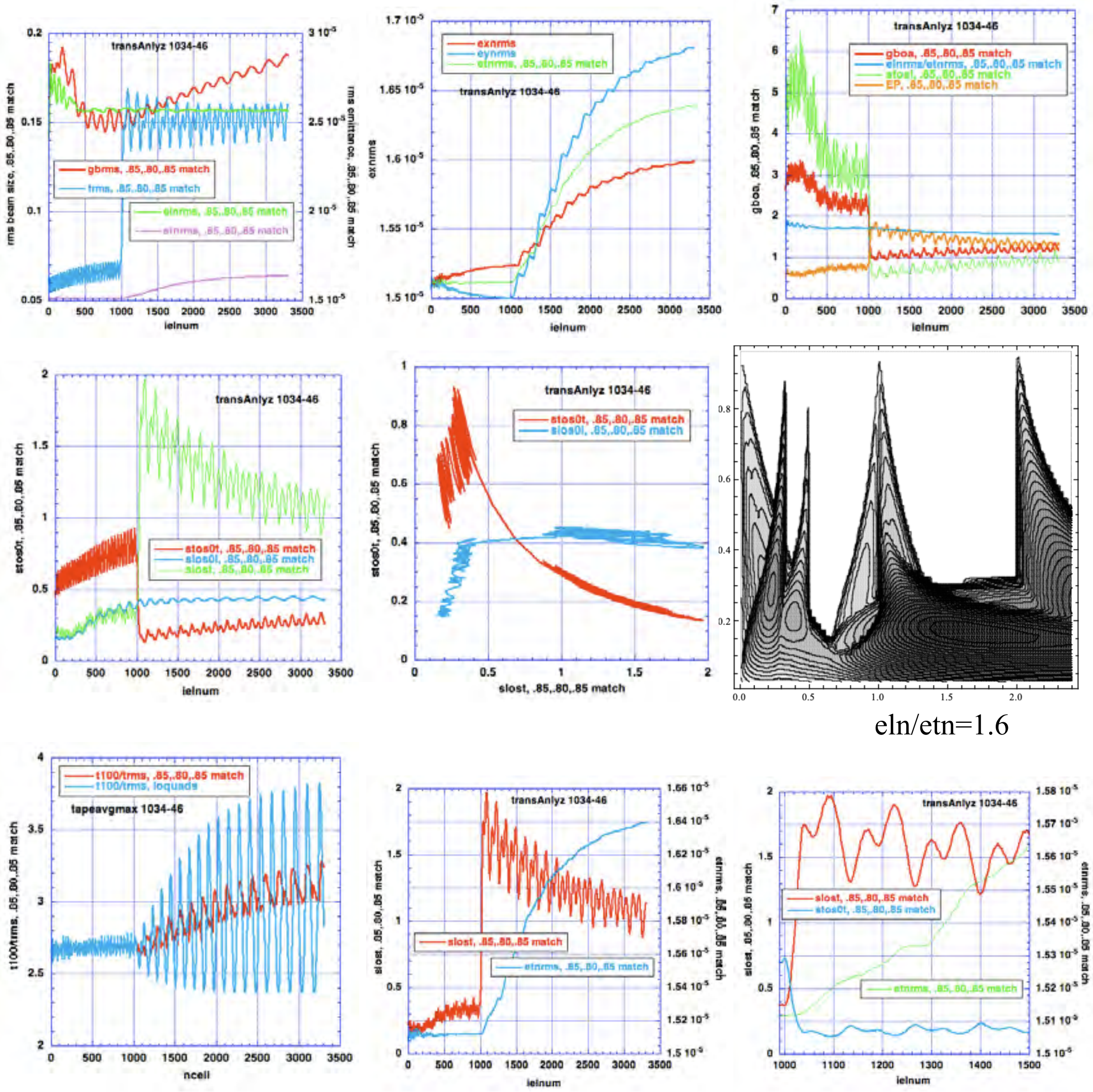

Fig. 27.17. Other results of 0.85,0.80,0.85 strategy.

The result seems a good start; *is it really an improvement?*

The notation "t" is used for averaged (x+y)/2 quantities. Looking at the average beam size trms, there is reduced oscillation around the desired value, similar etnrms growth but with reduced t100/trms growth, no emittance exchange. Oscillations of slost on the Hofmann Chart now lie between 1&2; in an area free of major resonances for longitudinal, but transverse stos0t is very low, ~0.2, in the strong resonance area. There seems to be no correlation of transverse rms emittance growth (etnrms) with the brief excursions near the slost=1 resonance, so it seems to come from deep tune depression into the overlapping modes. The oscillations are not around any local, instantaneous EP condition.

*However, we are in a box, and had better get out of it!*

The box is that with a conventional well-tuned linac, averaging x&y works well. But this linac is now severely compromised, at least formally, with quads at 50% of design, and we had better look at all three planes…

**27.3.3.2 x, y, z performance**

Fig. 27.18 is very interesting! The x and y oscillations with the (.85,.80,.85) match are larger than with the plain 50% quad strength reduction. But, the x and y oscillations are in-phase with the plain 50% reduction, but out-of-phase with the (.85,.80,.85) match. This might have been anticipated because the matching was made on trms, which is the average of xrms and yrms.

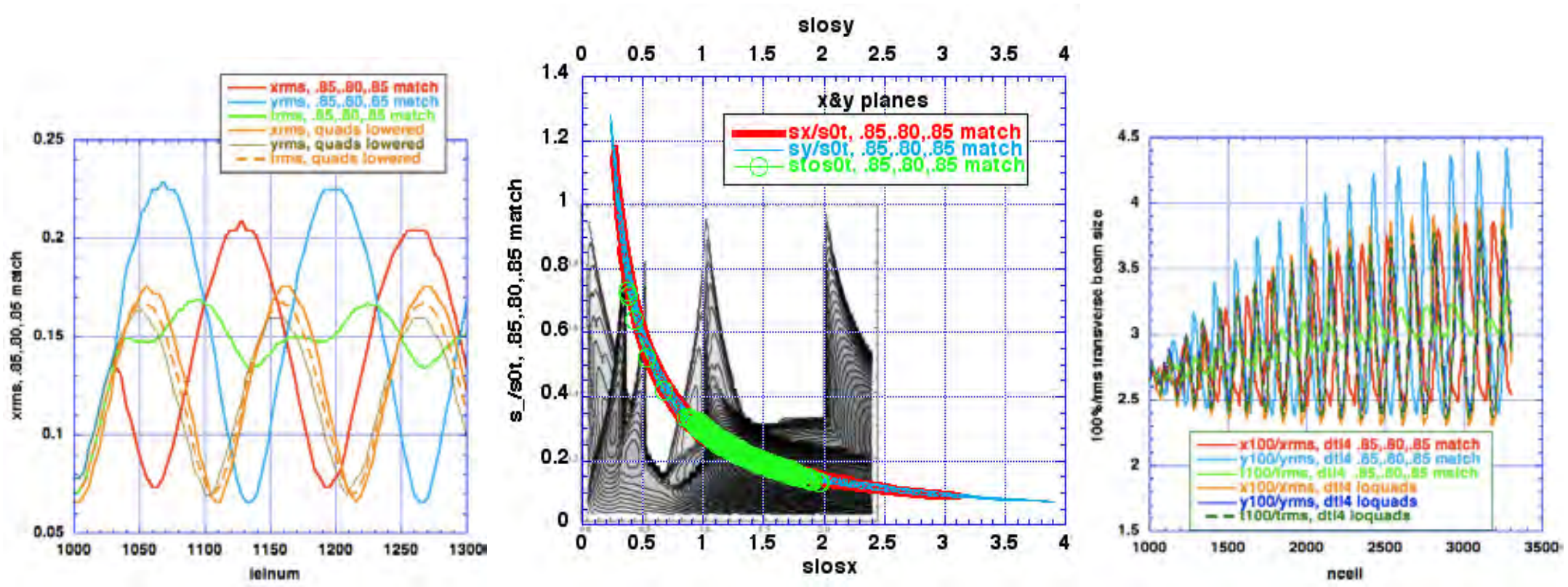

Fig. 27.18. Comparison of transverse beam sizes before and after detailed matching at quads 1034-1046 (left), Hofmann Chart for components (1000->end) with detailed matching at quads 1034-1046 (middle), ratios of 100%/rms transverse beam size.

The x and y rms emittance growth is seen to be correlated with their tune ratios in the same way, Fig. 27.19:

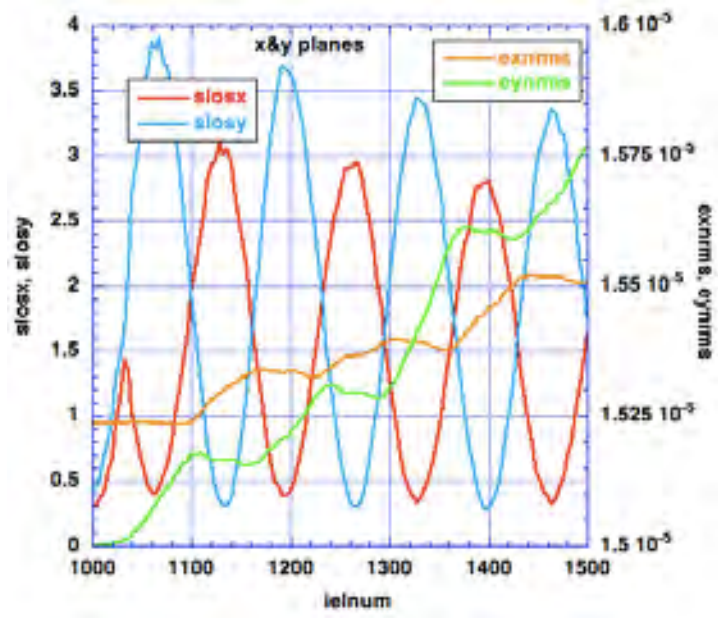

Fig. 27.19. After the (.85,.80,.85) match , rms x and y oscillations are out-of-phase, and the rms emittance growth is correlated with the respective tune ratios.

## *27.3.4. Conclusions about the "Match after the 50% reduction"*

The out-of-phase x,y oscillations do result in a larger average beam size, also larger envelopes, which should be better for reduction of intrabeam stripping.

The rms emittance growth (small in this case), with or without the match after the 50% reduction, comes from local, fast, interactions with the space charge coherent modes, and is almost the same even though the x,y oscillations are larger with the match.

Such interactions, affecting the total beam sizes and total emittance, could contribute to the small beam losses still seen in J-Parc and SNS with reduced quads. Runs with the detailed models for these accelerators, using many particles, must be made to confirm the effectiveness of this strategy.

## *27.3.5 Relation to "Elements of Linear Accelerators"*

My statements about linac design with respect to equipartitioning and resonance have long been reduced, for utmost simplicity, to two "**elements**":

1. The existence of a possible local and sustained equipartitioned design has obvious merit, and should always be investigated as a possible design, and adopted if its price is not too high. EP is a sufficient condition for lower beam loss (controlled rms behavior and also tighter total beam), but not necessarily the optimum over all design specifications. For example, low intensity designs do not need the best space charge strategy, and often can be made cheaper (shorter) using the second "**element**":

2. If EP is not used, try to have the design trajectory lie in the relatively broad $\sigma_l/\ \sigma_t$ band around the EP condition, where the EP condition is killing a low order resonance, or in a band between low order resonances. Or, if there is some merit in traversing resonance areas, then have the design trajectory cross the resonances as quickly as possible, in order to minimize both rms and total emittance growth and subsequent beam loss.

With regard to the 50% quad reduction in an existing DTL linac as explored here, **element** (1) has been abandoned, but **element** (2) appears to be working extremely well. The large x,y "mismatched" oscillations result in short, fast interactions with the resonance regions, with only small emittance growth, and all (10000) particles still accelerated. With the out-of-phase x,y oscillations, further advantages are obtained. The low order mode resonant interactions are clearly correlated with rms and total emittance growth [215]. In addition, there are an infinite number of higher order resonances in the tune areas of the Hofmann Chart that are free of the major resonances. Emittance accumulation has been observed on trajectories in these areas. Here the big oscillations make the traversal time of the resonant areas very short, but a small effect could remain.

That the quad strength reductions did not cause more emittance and beam size growth, with larger beam losses, was already to some extent surprising and suggested closer investigation, but could also be dismissed by assuming that the apertures were adequately large (especially in a superconducting section). At first, it might seem that lowering the transverse focusing strength would give a deeper but still concentrated transverse tune depression and more effect from the instabilities. But with analysis, it is seen that the quad strength reduction results in very large, fast oscillations. That the performance remains so good with such large, fast oscillations is surprising, and a new insight.

#### 27.3.5.1. Attempts to reduce x, y oscillations

A number of attempts were made to reduce the size of the x,y oscillations around a desired rms beam size, by hand, and by using the powerful capabilities of the nonlinear constrained optimization procedures in the *LINACS* code with various strategies. For a few periods, the xrms and yrms could be held relatively constant, but beam loss began immediately and the sequence could not be prolonged; all attempts failed. As expected. With a non-equilibrium beam and very deep tune depressions, strong space charge instabilities are present, and if the linac trajectory were concentrated in such region, it would be extremely sensitive. So even an automated optimization procedure would have to be very carefully defined. With the model used here, such procedure has not been as yet found.

#### 27.3.5.2. EP Design with larger transverse beam size

---

215 "Self-Consistent Beam Halo Studies & Halo Diagnostic Development in a Continuous Linear Focusing Channel", R.A. Jameson, LA-UR-94-3753, Los Alamos National Laboratory, 9 November 1994. AIP Proceedings of the 1994 Joint US-CERN-Japan International School on Frontiers of Accelerator Technology, Maui, Hawaii, USA, 3-9 November 1994, World Scientific, ISBN 981-02-2537-7, pp.530-560.

A new high intensity H- linac can be designed for larger beam size at increased energies, under various scenarios but always maintaining EP throughout.

### *27.3.6 A Conclusion re Compensation of an Existing DTL Linac for H- Intrabeam Stripping and Residual Beam Loss*

Nature has been kind in this instance. On first thought, it might seem that reducing the transverse focusing would indeed reduce intrabeam stripping, but would on the other hand increase other losses, including those from space charge resonance interactions perhaps because the transverse tune depression would be much greater. But that does not happen – intrabeam stripping losses were reduced, satisfactory accelerated beam fraction was maintained, and the residual beam loss is not a problem at the 1 MW intensity initial design goal (although it needs careful attention for an upgrade or new more powerful high intensity linac).

The quad reduction in this example worked, without significantly increasing focusing or space charge resonance related losses, *because* the large oscillations are produced, and *because* they are relatively fast. The large, fast oscillations appear to move the trajectories quickly enough across resonances that there is little time to react, and are apparently relatively insensitive to details, whereas design with concentrated trajectories in deep tune depression areas might be expected to be quite sensitive. The oscillations should have out-of-phase x,y, and placement of the average (trms) oscillation with respect to the instability regions may be adjustable and play a role.

It is hoped that the situation has been further illuminated here; we proceed to apply the techniques to J-Parc in Sec. 27.4&5, and very interesting implications for future linac development in Sec. 27.6.

## 27.4. Analysis of Compensation of the J-Parc Linac for H- Intrabeam Stripping and Residual Beam Loss [216]

The $H^-$ J-Parc linac is unique in that it is designed to be equipartitioned throughout. This should afford an outstanding opportunity to test various theoretical and practical aspects, some of which have been done, for example to test the effects of random errors [217], a very important study. It shows the critical importance of investigating not only the rms behavior, but also the 100% emittance behavior - see Sec. 27.5.

The linac consists of a sequence of several structure types: a 3MeV Radio Frequency Quadrupole (RFQ), a 50 MeV Drift Tube Linac (DTL), a 191 MeV Separated-type DTL (SDTL), and a 400 MeV Annular Coupled Structure linac (ACS), with plans for an eventual 600 MeV Superconducting Cavity Linac (SCL). Operation frequencies are 324 MHz for the RFQ, DTL and SDTL and 972 MHz for the ACS and SCL. Intrabeam stripping losses were reduced by lowering the design quad strengths in the ACS linac – this section is studied here for quads set at 100%, 50%, and 30% of the design values.

The ACS linac operates at high beta's (~0.5-0.7), so can have a separated-function lattice with 17 accelerating cells between each pair of the 47 sets of FFDD transverse focusing quadrupoles. This produces quite different phase advance characteristics compared to the strongly focused DTL. Fig. 27.20 indicates the layout:

---

216 ACS Analysis.nb

217 "Lattice and Error Studies for J-Parc Linac Upgrade to 50 mA/400 MeV", Y.Liu, M.Ikegami, IPAC2013, THPWO027.

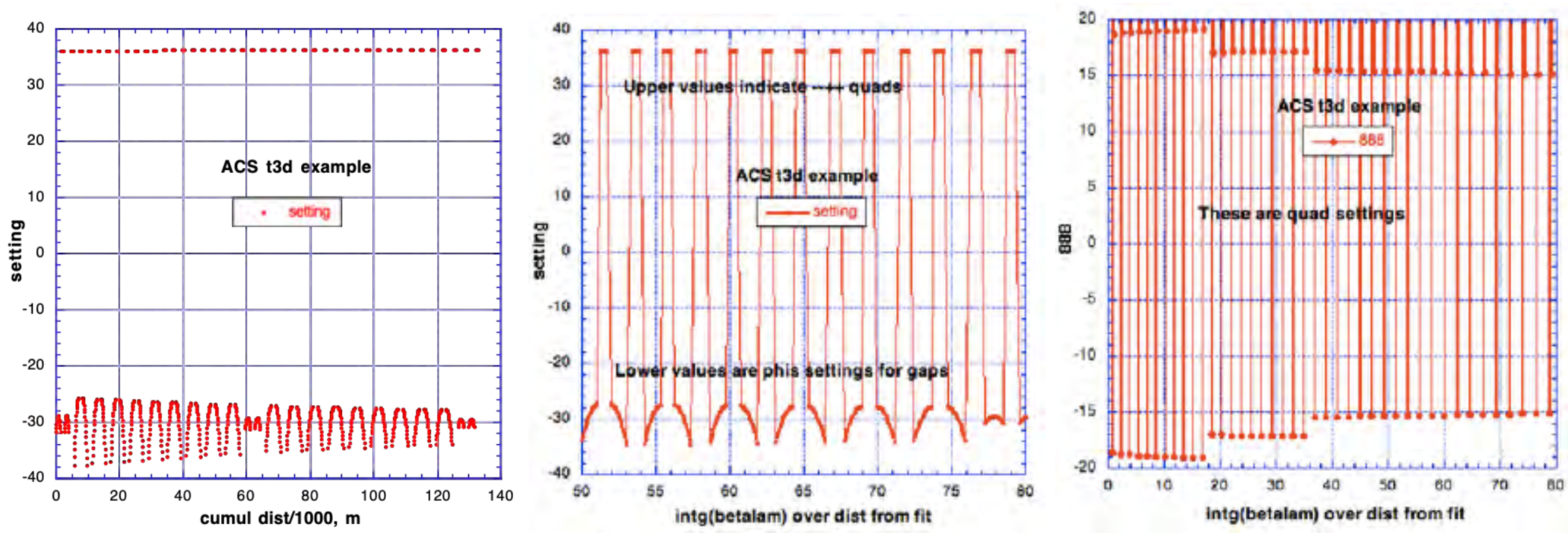

Fig. 27.20. J-Parc ACS layout. Upper dots indicate the placing of FFDD quads. The lower pattern indicates the synchronous phase setting of the 17 rf accelerating cells between each set of quads. Right graph shows the quad settings.

### *27.4.1. Phase Advance in a Separated Function Lattice*

We did not discuss the construct and derivation of “phase advance” above, because of the discussion on more general **elements**, because the situation was “conventional”, and because it was assumed that the reader was experienced, at least to that extent. Now it will be necessary to elaborate [218].

The theoretical analysis tools rely on the Courant-Snyder model showing that the rms emittances rotate in phase space and indicate the betatron transverse and synchrotron longitudinal oscillations. The angular amount that these oscillations rotate over a designated distance are called the phase advances. The periods change when space charge is present; knowledge of these and other possible driving force periods are necessary for design and analysis.

The rms emittances, rms beam sizes, and rms phase advance are related, indicating that they are computed directly at each step of a simulation run:

> Transverse rms emittance = (transverse rms beam size)$^2$*γ*((transverse rms phase advance)/unit length)
>
> Longitudinal rms emittance = (γ *longitudinal rms beam size)$^2$* γ *((longitudinal rms phase advance)/unit length)

---

[218] This seems to be hard to grasp. A reader of this book found the explanation here not so clear; I tried other wording in email reply June 2018 (please excuse running together in one paragraph here – the reply emphasized each sentence with separate paragraph): “the envelope equations are emittance = (beam size)^2*(phase advance, degrees or radians)/(unit length). The derivation of the envelope equation is rooted in Courant-Snyder about periodicity of focusing system, so that period is built into the derivation. **But an even more fundamental <u>element</u>** is to realize that it is the local, instantaneous, state of the beam distribution that is important, and that is where the theory can be used as (unit length); integrating along the trajectory accumulates a phase advance over a length of interest. Especially useful when in matching section or frequency transition or such, when one wants to keep the phase advance as constant as possible; (phase advance per unit length) = emittance)/(beam size)^2 and the emittance or beam size should not change too quickly. Can convert to any other unit length just by multiplication, of course /meter loses the connection to the betatron or synchrotron period but has the above, or can use the period of <u>any</u> resonance if trying to see if particles are getting influenced by that, which you can see I used to think about single particle effects, etc. But if trying to use the Hchart, which is all in terms of ratios of tunes and emittances, seems clear that the unit lengths should be consistent!? It seems that physics trains people to think in terms of steady-state and they are not comfortable with the transient state, which is second nature to engineers. Unfortunately much confusion has been spread because of this... PRAB reviewers don't get this at all! Claims that "phase advance is dimensionless", etc., etc., like "single particle phase advances lie between the rms phase advances with and without current" - I still cannot figure out where that notion comes from... it seems completely illogical, but probably some theory might suggest that?

Typically, the phase advance is desired over the transverse lattice period, in order to understand the stability properties in terms of the Courant-Snyder model. An equation of motion for the beam "envelope" is found by averaging the detailed motion over the transverse focusing period, *neglecting the detailed motion inside, which has a faster time structure*, resulting in a "continuous" "smoothed" lattice. The longitudinal motion is treated similarly. The averaged (smooth approximation) betatron and synchrotron wavelengths can be found theoretically for any lattice.

Possible resonance interactions related to the structure periods have then been studied, and indicate tune areas (values of the ratio of the phase advance in two planes) that should be avoided (e.g. the Hofmann Chart). The cases studied have been restricted, however.

Continuous focusing channels are the basis for the Hofmann Chart [219]. Continuous transverse alternating-focusing or solenoidal channels typically have focusing periods of n $\beta\lambda$, and reduce to the continuous channel result if the smooth approximation envelope equations are used. The approximate resonant areas from these simple models have been shown to agree closely (since 1981) with the x,y,z performance of lattices with acceleration and slow (essentially adiabatic) parameter changes, and are therefore a useful analysis tool.

The transverse and longitudinal dynamics in the continuous RFQ channel are tightly coupled, with transverse period typically $\beta\lambda$ and longitudinal period $\beta\lambda/2$. The DTL periods are the same except the components are separated and there is much less direct coupling between transverse and longitudinal planes. The action within $\beta\lambda$ is ignored, and the phase advance unit length is chosen as $\beta\lambda$ for all planes (x,y,z, or with $t=(x+y)/2$) to have the necessary common period for relation to resonances. These lattices are very short, and in practice precisely built, so it is unlikely that there would be significant perturbations in the construction of the lattice.

However, they, and the continuous transverse alternating-focusing or solenoidal channels, are still time-varying and with opposing motions in the transverse planes. Studies of the latter lattices with resonance interactions have observed emittance splitting, usually dismissed lightly as "typical of a system with noise". A more precise analysis might indicate that within the lattice particles experience different focusing forces at different times, at different radii, and therefore could have different phase plane projections and emittance splitting.

There are many more possibilities. Linacs with separate transverse and longitudinal components are generally called "separated-function linacs", at least by me, because there are very many different kinds of lattices used, over all energy ranges – different kinds of RFQs, DTLs, IH, CH, CCL, ACS, SC (superconducting) [220]. The J-Parc ACS example here has a transverse FFDD lattice many $\beta\lambda$ long, because transverse focusing is required less often at higher particle velocity. Between each FFDD pair, there are 17 rf accelerating gaps in an rf cavity driven by one amplifier at the center. Now there can be considerable action within the transverse focusing period. For economic reasons, the cells within one rf cavity are machined to the same length, resulting in a synchronous phase slip from one end to the other with the correct synchronous phase at the middle. The LAMPF 805 MHZ CCL linac is also built this way, with up to 80 rf cells per transverse focusing period. Superconducting linacs also, although usually with a smaller number of rf cells (e.g., 6 to 8) driven by small rf amplifiers; here the rf cells usually also have the same length ("constant beta") and corresponding

---

219 It has been taught that plotting the tune trajectory in one plane is enough, but especially because the coupling between transverse and longitudinal planes is widely different for different types of linacs, and because their parameters may be different, I find it useful to plot the tune depression of all planes. For $n\neq1$ separated-function cases, if is even more clearly necessary to plot all planes.

220 Certain lattices have been claimed to be unique and studied with unique names and supposedly unique dynamical methods, although they are only examples of separated-function lattices and in fact the same as the old LAMPF lattice … phase slip. This slippage, *within the transverse focusing period*, produces new driving forces that are not present in the usual theoretically studied lattices. The usual "smooth approximation", which averages out all motion that occurs within the transverse focusing period, is no longer appropriate.

In the ACS case, it is readily seen that the driving pattern of the rf gaps is quite regular but nevertheless would take many harmonics to fit with a series of sinusoids. It will be seen below that the varying synchronous phase pattern excites new resonances beyond the resonances of a smooth channel indicated on the "Hofmann Chart".

I have stated since LAMPF that this is something to watch out for, but I have not looked at it in any detail until now, nor has anyone else. Many other possibilities for driving forces within the transverse period can be considered. The effect of periodic patterns overlaid on a smoothly focused lattice were studied during my first sabbatical at KEK in 1988, indicating that significant Alternating-Phase-Focusing (APF) effects could be achieved, in particular to increase or decrease the longitudinal acceptance. Several kinds of random errors, construction errors, alignment errors, matching errors, could be present.

Any resonance can be tracked in a particle simulation code by looking at particle trajectories at homologous points on the resonance sinusoid – a Poincaré plot. Here it is clear that there is interest in possible resonance effects that could arise from driving forces within the transverse focusing period, but that the "smoothing approximation" usually used for the envelope equations would lose this information.

Therefore, the unit length is chosen initially to be simply meters. For later analysis, other lengths can be applied, for example multiplication by the transverse period length (distance between FFDD quad sets) to get the phase advances relevant to stability analysis. [221]

### *27.4.2. The Data Set*

The desired data set was that of [222], using the EP design and three cases with reduced quadrupole settings. However, although explicitly requested, this data set was not released.

The picture is confused because the case finally obtained and presented here is not at linac design conditions, where injection etn ~ 0.0000015, eln ~ 0.000024, but with larger emittances observed on the actual linac of etn ~ 0.000034, eln ~ 0.000041. Reason why larger: "here just used the typical measured emittance at MEBT2 (just before ACS), which is of order 0.35 due to emittance growth. Continuous efforts are needed to improve the beam emittance..". [223]

---

221 A "reviewer" for a primary journal for the field (one can guess the name (of the journal)) very arrogantly and insultingly opined that the unit length for phase advance can only be the transverse period or other such length if the beam current is zero, and that it has to be meters if there is space charge. I am still curious what the basis for this could be, and would appreciate if someone could help me out of my ignorance. Beyond that, condescending, insulting review comments are particularly devastating to students, and in any case out of line.

222 "Measurement of Resonant Space Charge Effects in the J-Parc Linac", C. Plostinar, et.al. THPO087, IPAC2013.

223 October 2022: At start of 2022, a Japanese University Professor was asked to revisit this topic, with emphasis on why a "200% transverse emittance growth was observed in the DTL". He also requested the design data set, but no one seemed to have that and finally he also was only given "an operational data set". Also, no one seemed to know when this "200%" was first observed. As of this date, no further analysis has been made. See also next footnote on original design and EP condition. It might be helpful to know that some change in the longitudinal emittance input to the DTL could result in DTL transverse emittance growth.

It is also not a complete data set (particle distribution data or 100% emittances data were not included), and in view of all the difficulties discussed below, it is not even clear if a consistent data set was provided.

However, analysis proceeded, initially with the given output data for design quad strengths (100%). The EP condition is not equal to 1, but there is no *rms* emittance growth beyond a fast initial redistribution, indicating that the beam is in equilibrium. We would prefer a dataset at design conditions, to check if the 100% quads case is equipartitioned. But a simple scaling procedure could change the EP condition to ~ 1, so it was decided to proceed with the provided dataset. [224]

Another problem is present in the 50% reduced quad dataset, but the analysis indicates important useful aspects of the procedure.

All runs were made with 80,000 particles and none were lost. The code provides a table of maximum particle radii, and probably particle coordinates; they should also be analyzed, but were not provided and therefore not analyzed in this article.

Very considerable hand manipulation of the data sets is necessary.

### *27.4.3. Contending With Non-Cooperative Simulation Programs*

The main simulation program used for the J-PARC linac is IMPACT. IMPACT was not written to study linacs - it was written to explore the fascinations of parallel computer programming. Much later, some more accurate physics was introduced, such as field maps. The available output data, as with many linac simulation codes, is not convenient, incomplete – not including necessary space charge physics quantities such as phase advances, inconvenient units, inconvenient form.

**x,y plane conversions: fort__.24 and fort__.25**

c3 = rmsx
c5 = xprms
c6 = ellipse alpha
c7 = exnrms100,0(m-rad) NOT "when alpha=0"; source code shows usual computation of area, including cross-term.

Emails that "xrms*x'rms=erms happened when alpha=0" gave, at least, wrong impression. After wasting much time using this assumption, worked through IMPACT source code 4/21/2017 and found that reported emittance and alpha, Col.7 and Col.6, are computed correctly, and that reported x, xp are the projections of actual ellipse, not from an ellipse that was rotated to circular frame.
c10 = beta
c11 = gamma

---

[224] IMPORTANT: During this study 2015-2017, no one at J-Parc could recall or find information on the original JAERI/J-Parc linac beam dynamics design and especially on the procedure used for the "EP design". I had forgotten details and did not remember a publication, but during editing in October 2022, chanced again on the paper that had been published at APAC 1998, "Beam Dynamics Study of High Intensity Linac for the Neutron Science Project at JAERI", K. Hasegawa, H. Oguri, Y. Honda, H. Ino, M. Mizumoto, R.A. Jameson, First Asian Particle Accelerator Conference, March 23-27, 1998, KEK, Tsukuba, Japan - that shows that the contraction mapping formulae plus the EP equation had been used, so at least the starting point for the original design was correct. From that to the actual design is unknown.

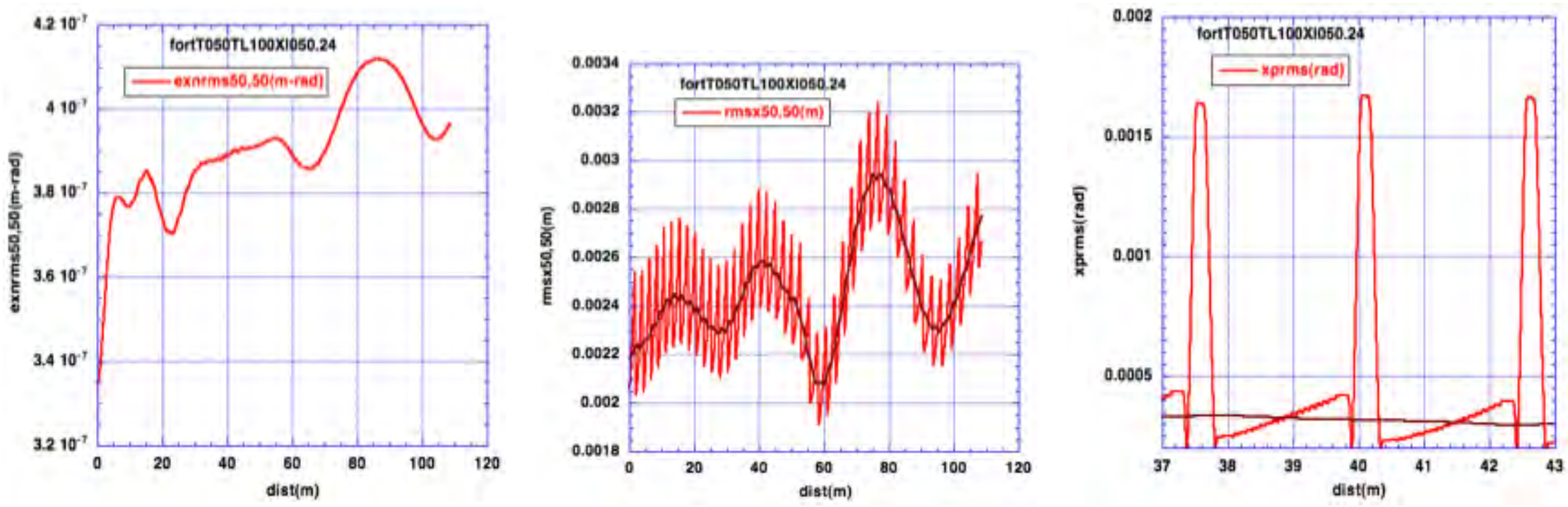

Fig. 27.21. Plots of IMPACT output data.

*Conclusions for x,y planes:*

- reported normalized emittance, alpha are ok, x, xp are the usual projections.
- **compute extra information needed at every step (do not sample at "homologous points"** (As noted often enough before... (e.g. SNS analysis))
- DO NOT USE WEIGHTING ON ORIGINAL DATA - USE ONLY ON FINAL GRAPHS IF DESIRED .

The above suggests that careful check of the source code should be done, to confirm what is actually being computed.

**<u>z plane conversions: fort__.26</u>**

*Longitudinal emittance conversion – confusing*

Degrees are with reference to base frequency 324 MHz, not ACS frequency of 972 MHZ.

This infers that distance is one cycle – lambda. But particle travels beta*lambda in one cycle… Confusing – because have to use lambda working with rmsz to compare emittance with Column 7, but to get $\gamma b/a$ ratio ~ constant, agreeing with eln/etn, have to normalize rmsz using beta*lambda/360°.

This seems to introduce an inconsistency, but length factors, etc., cancel in taking ratios…

el = (bmlen*gamma)^2*gamma*siglr /(nblt*(100*betalam)) ;
(* real longitudinal total beam emittance, cm-rad, for matched longitudinal beam -- longitudinal matching equation; assumed always satisfied. *)
eln = beta*gamma^3*el ;(* normalized longitudinal emittance, cm-rad *)
elnmd = eln*erest*him*360./(wavel*100.) ;(* eln, MeV-deg *)

*Conclusions for z plane: Same as for transverse.*

## *27.4.4. Design Quad Settings – 100% Quads*

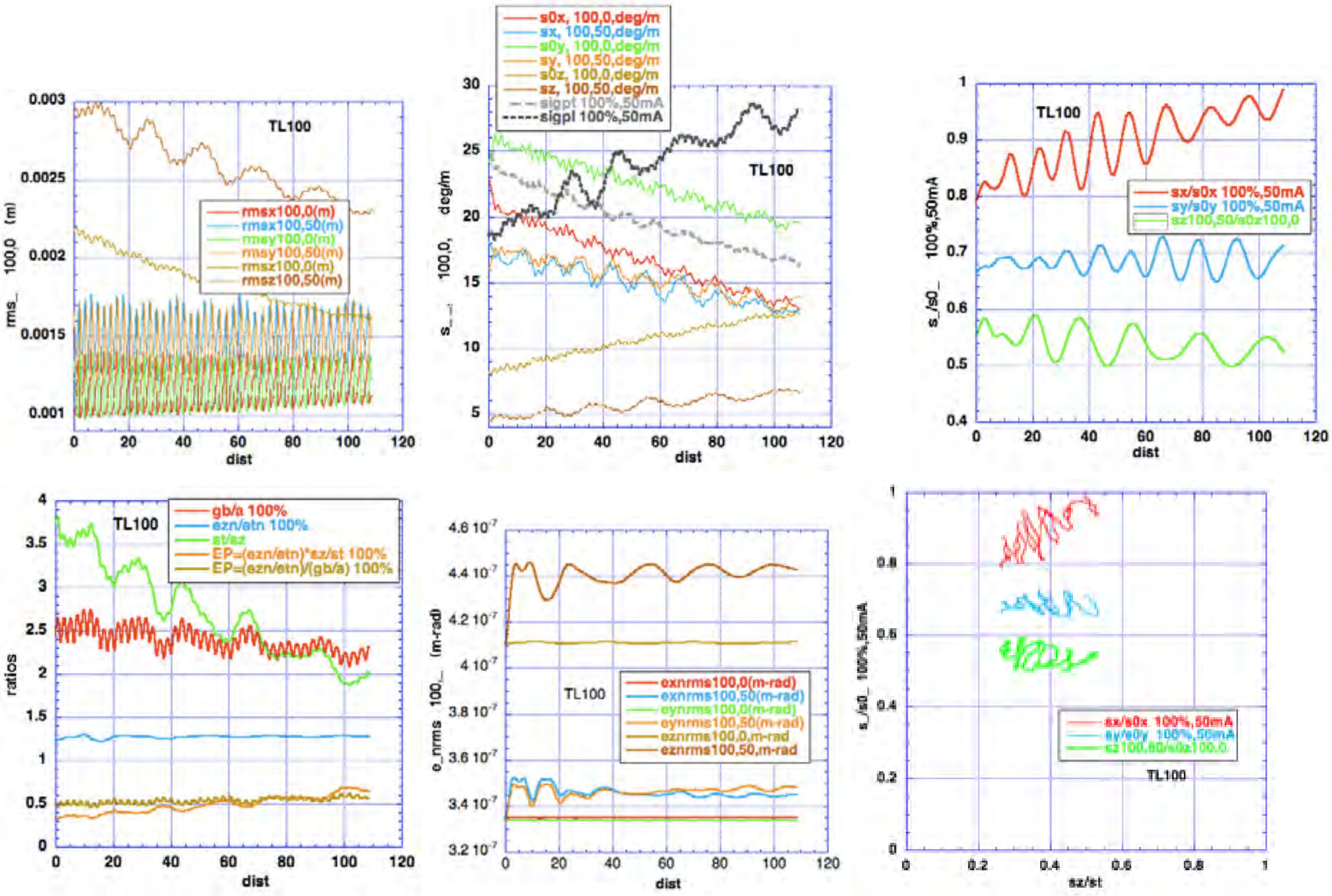

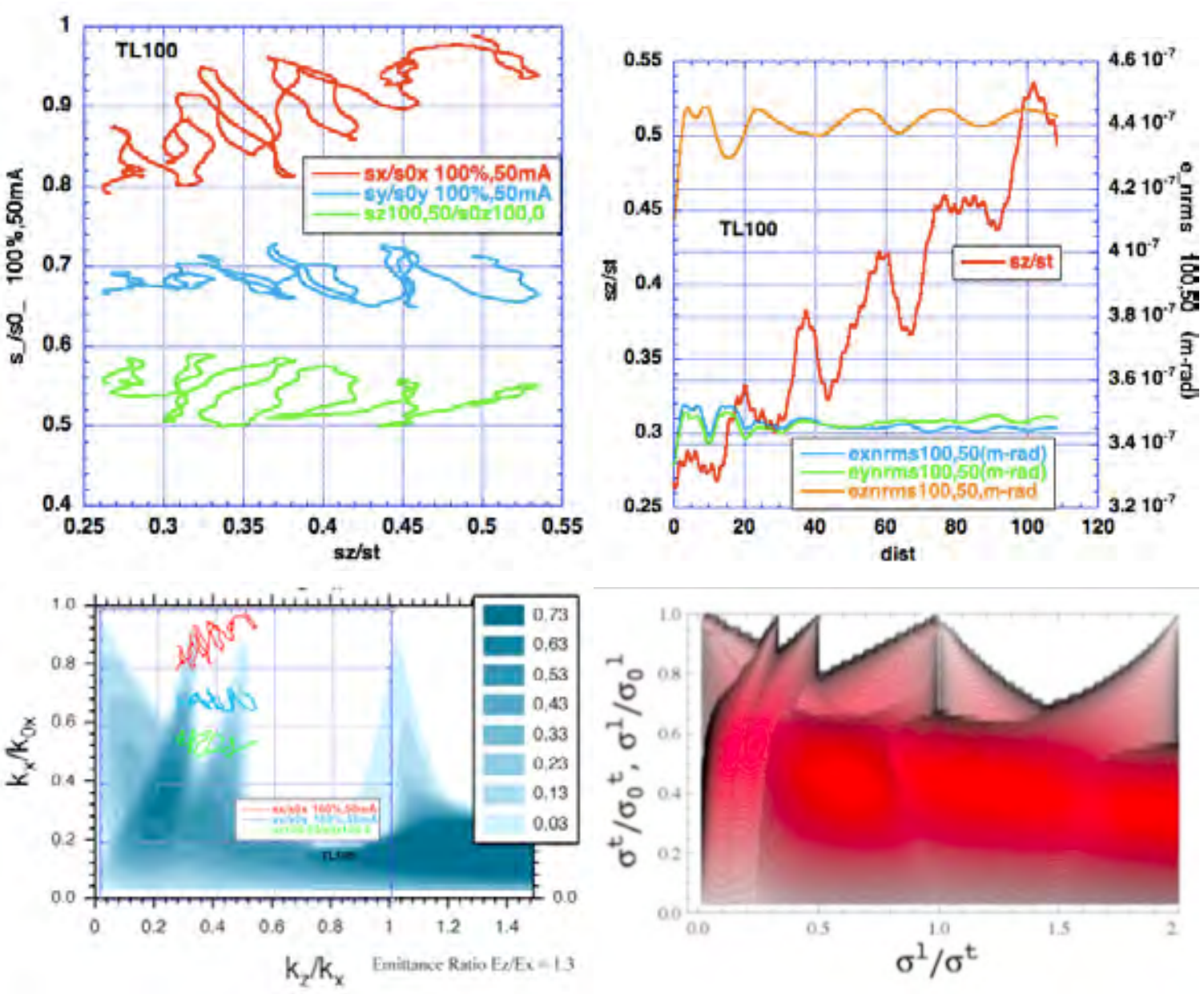

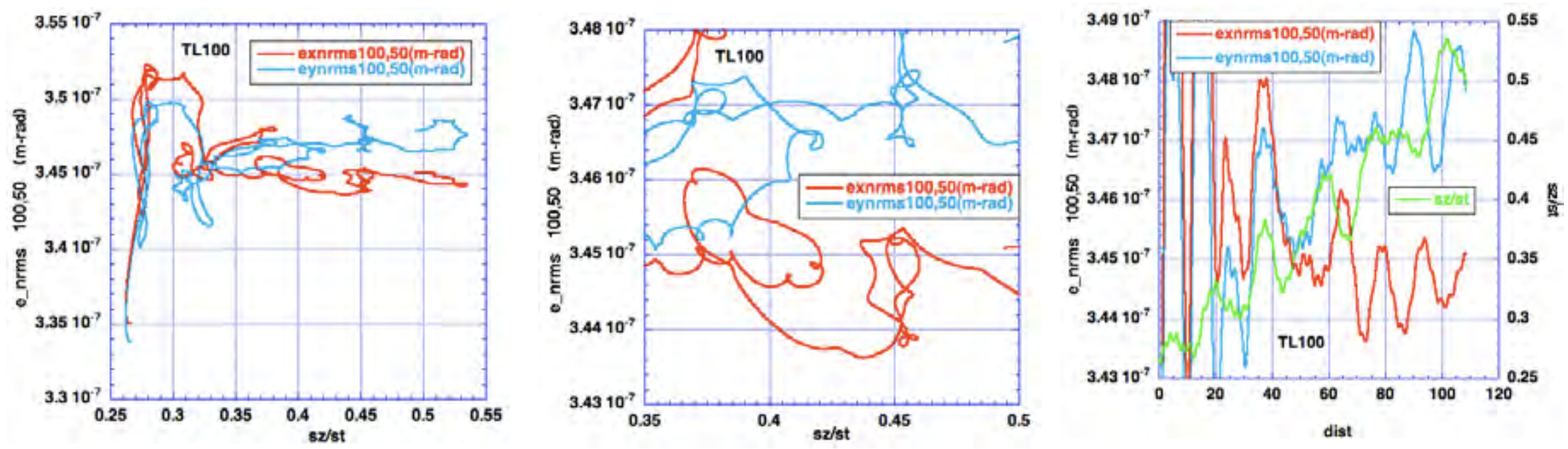

Fig. 27.22. Characteristics with design (100%) quads.

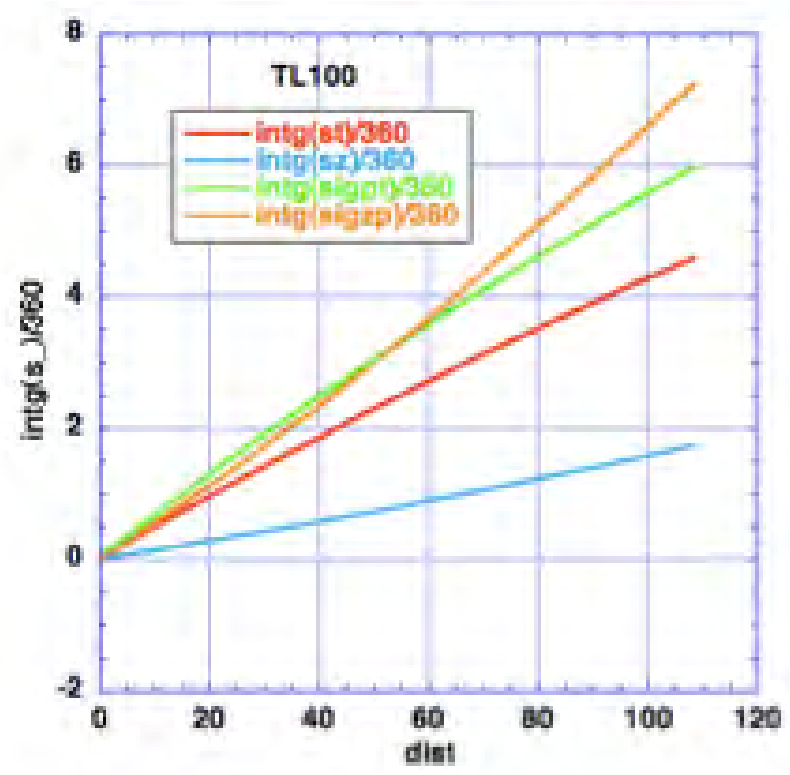

Fig. 27.23. Number of cycles of various phase advances over ACS length: s0t : 5.926, st : 4.5756, sz : 1.7276, sigpt : 5.9607, sigpl : 7.2213, where sigp_ are the plasma oscillation phase advances.

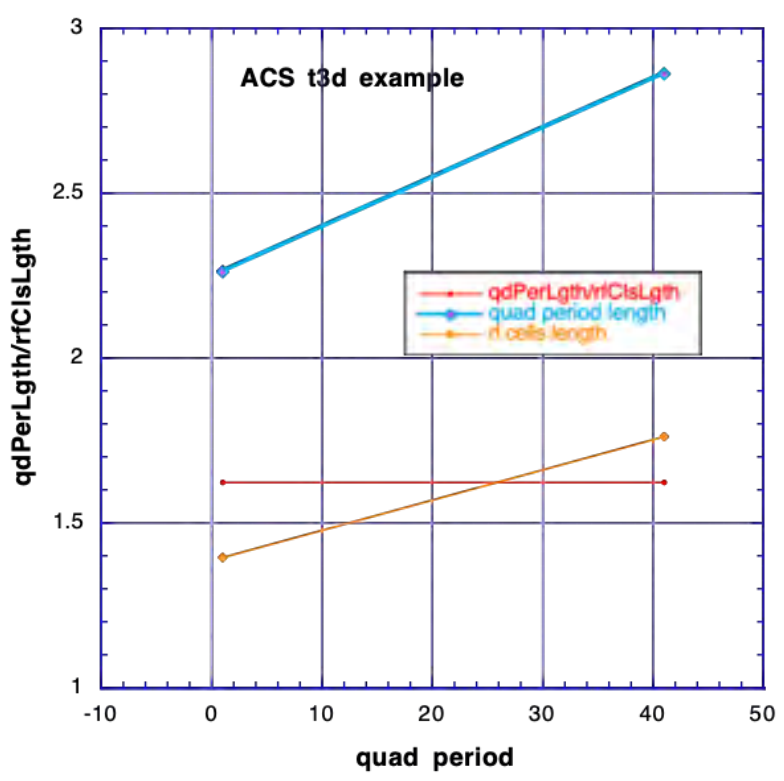

Fig. 27.24. Total length of quad periods and lengths occupied by rf sections and quad with associated drifts. Average transverse focusing period length ~ 2.5 meters. 3 meters is used in graphs below for representative values of (phase advance/(transverse focusing period length)

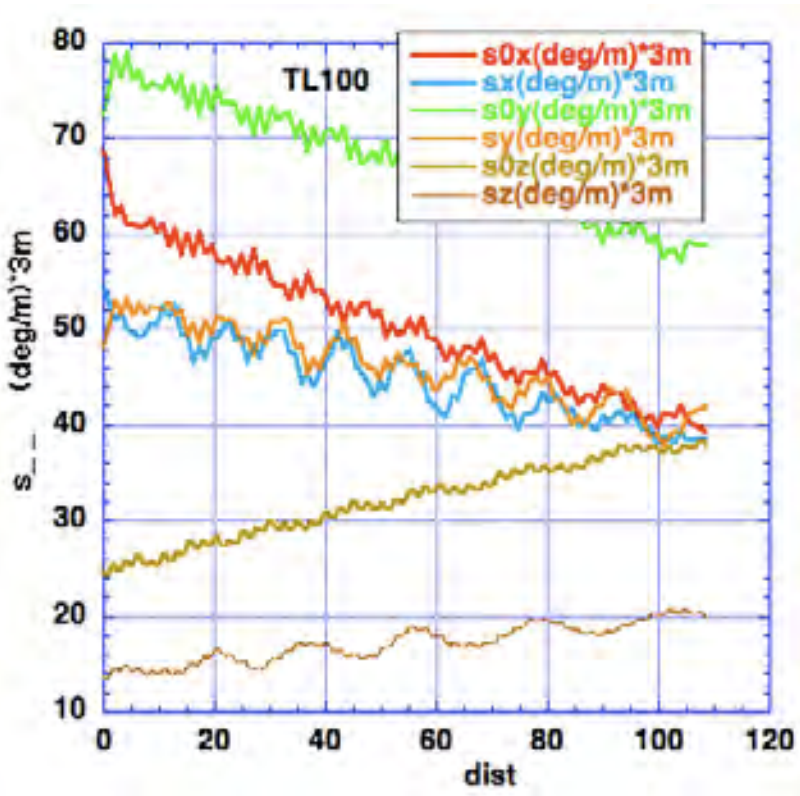

Fig. 27.25. ~ (Phase advances)/(transverse focusing period length)[225].

---

225 Longitudinal phase advances are low; indicating that it is probably not possible to change setting to get longitudinal zero-current phase advance from 90° to 120°, as desired by Hofmann to check a theoretical result. This might be possible in a superconducting linac with many separate cavities each with its own amplifier.

sx/s0x is high ~0.9, sy/s0y high ~0.7; exn and eyn start to separate; resonances crossed but interaction is small and quick; splitting seems to be caused by interaction with mode at sz/st = 0.33.

A trajectory loop indicates capture by a resonance; a kink indicates a resonance influence.

Resonances at sz/st ~ {0.28, 0.31 (small), 0.35, 0.4, 0.45, 0.52}. Resonances at 0.33 and 0.5 would be expected without any rf gaps. J-Parc ACS has 17 accelerating cells between each pair of the 47 sets of FFDD transverse focusing quadrupoles – therefore it is logical that rational fractions m/n with n=17 might be excited, and these seem to fit quite well: {5., 6., 7., 8., 9.}/17. = {0.29, 0.35, 0.41, 0.47, 0.53}.

A lot of small resonances are being encountered. This is known from experience to lead to an accumulation, and may be the cause of the slight rising trend of the longitudinal rms emittance in Fig. 27.22.

Also note that here and for the reduced quad cases below, the EP conditions (eln/etn)(sz/st) and (eln/etn)/(gb/a) differ, also indicating additional influences.

## *27.4.5. 50% Quad Settings*

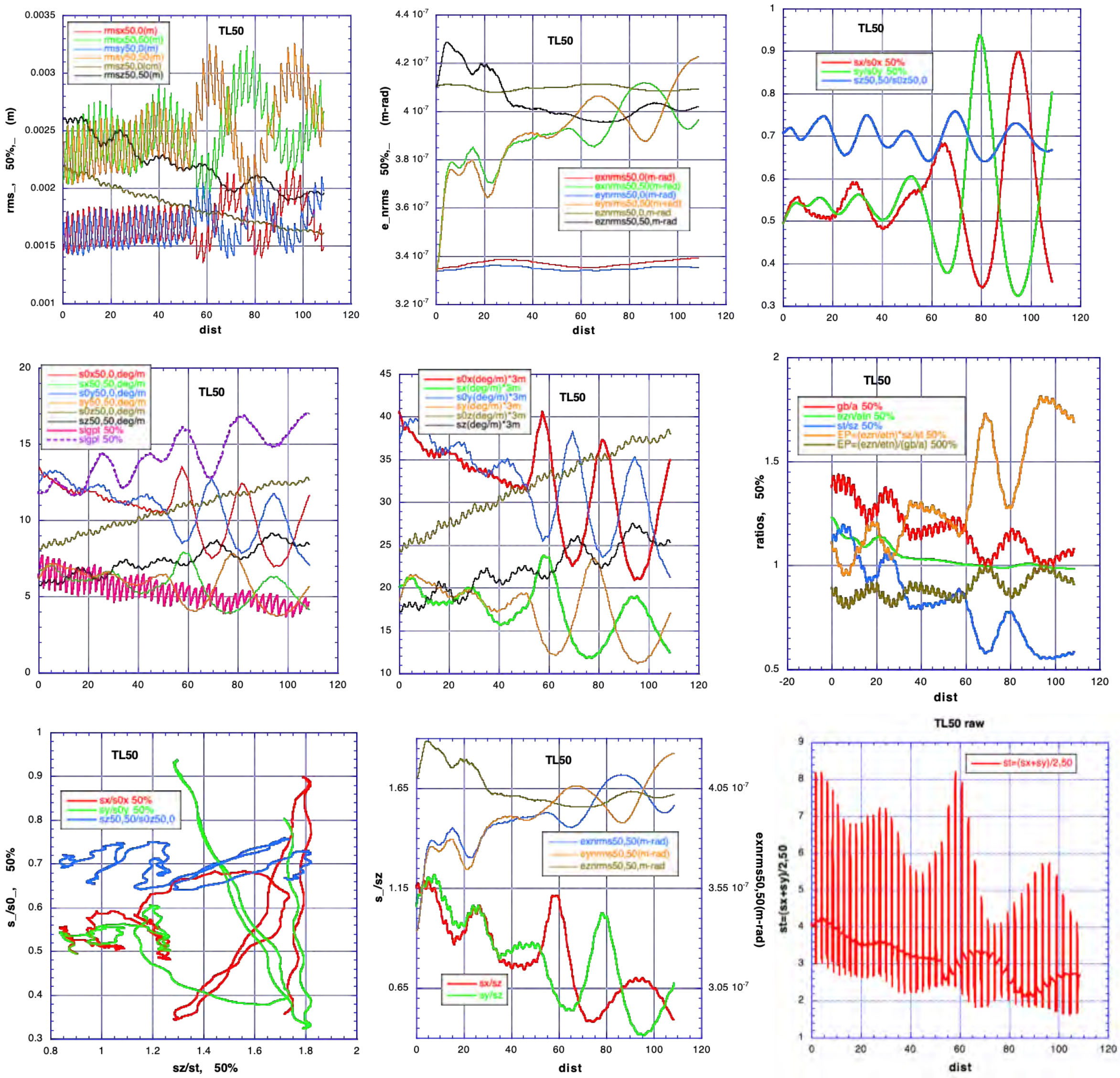

Fig. 27.26. Characteristics with 50% quads.

Beam is not equipartitioned, emittance exchange begins immediately via interaction with mode at sz/st = 1. Mode excitement due to phis pattern driving are apparent.

There is a sudden disruption at distance = ~53m. It is better not to use any weighted fit here - to see that abrupt change occurs at one quad set - probably a setting error. The trace3D files exactly corresponding to the three 100%, 50% and 30% runs were not provided, so it is not possible to know for sure.

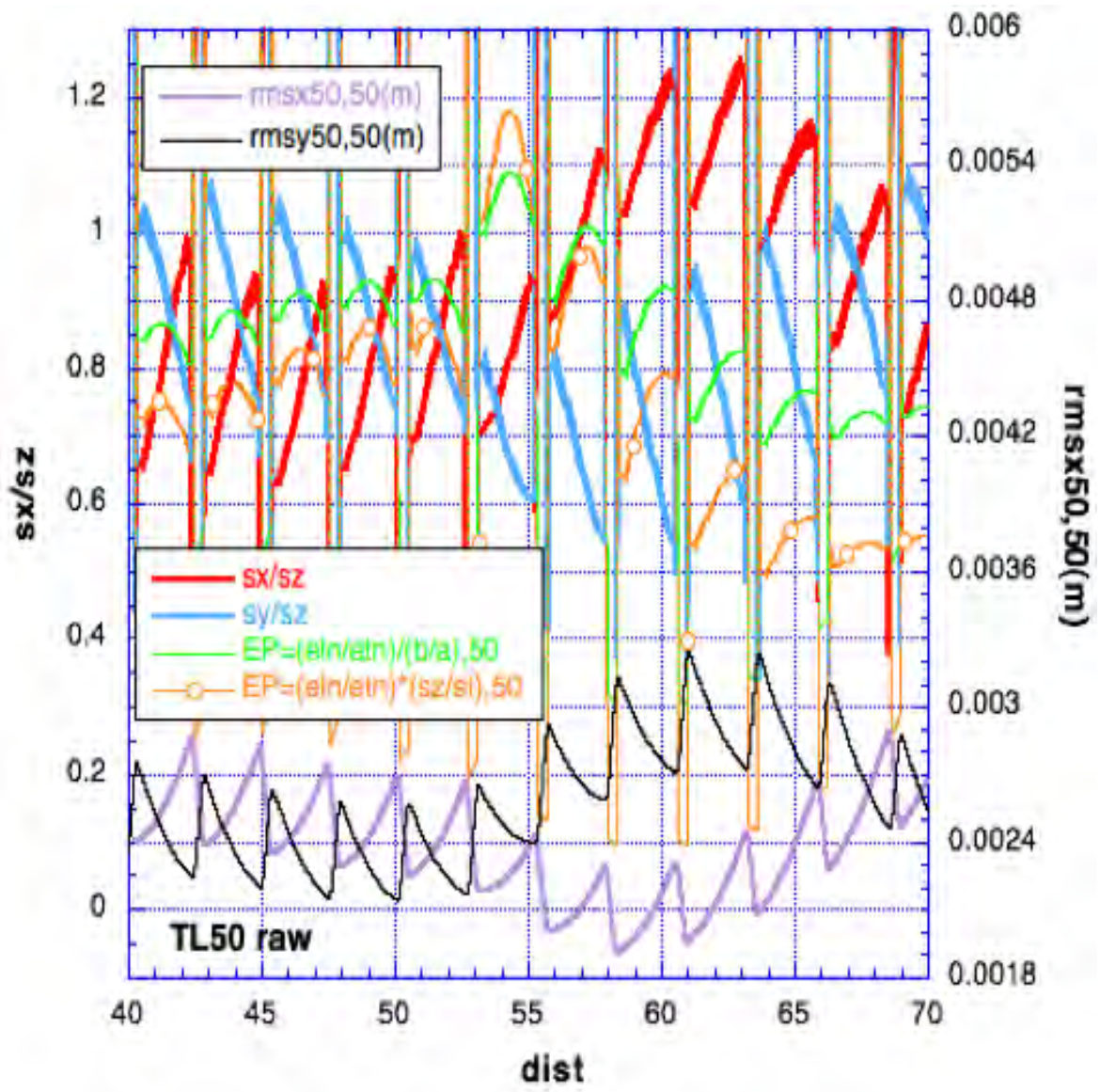

Fig. 27.27. Plot shows that an abrupt change occurs at distance ~53 meters.

## 27.4.6. 30% Quad Settings

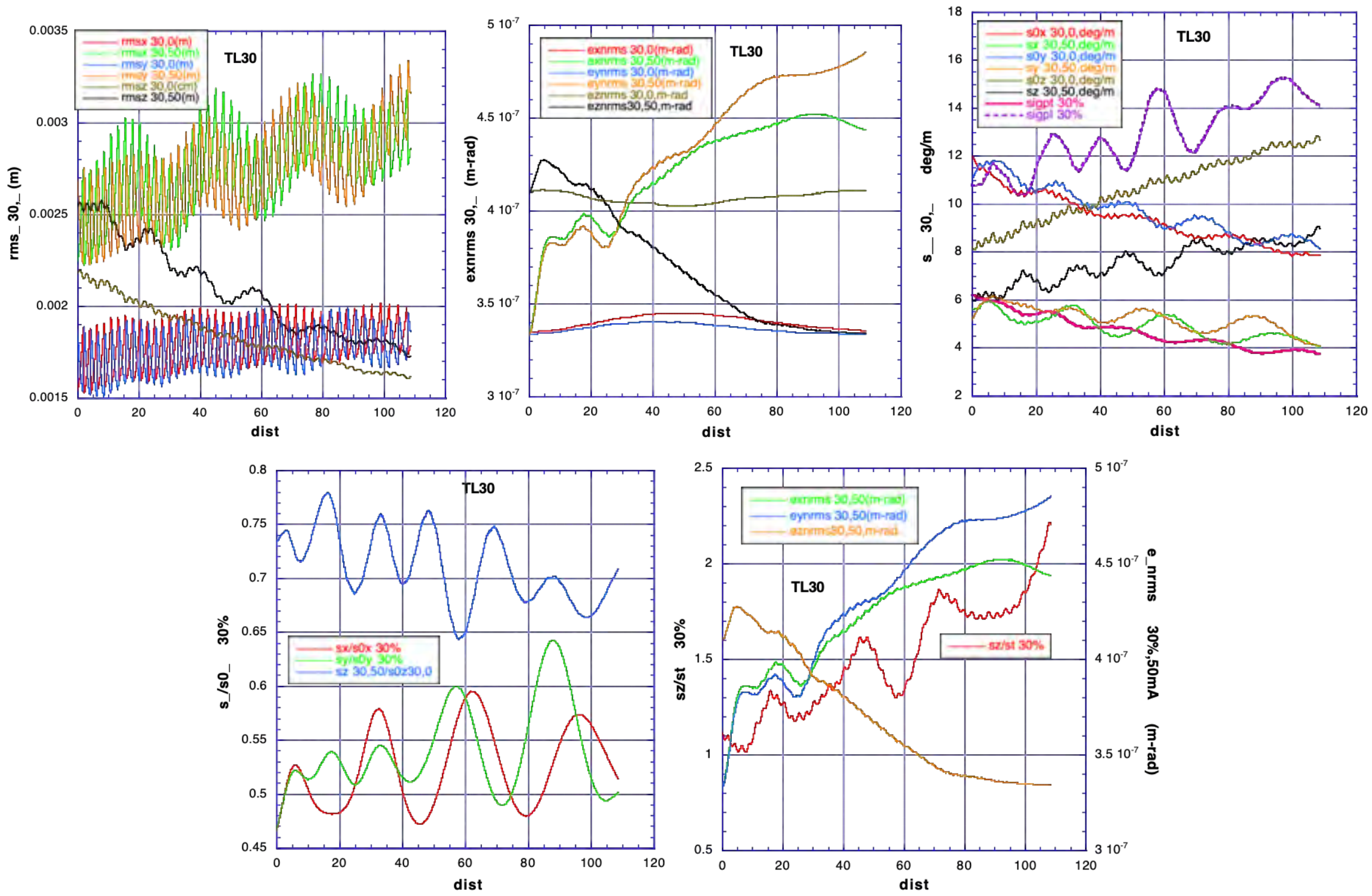

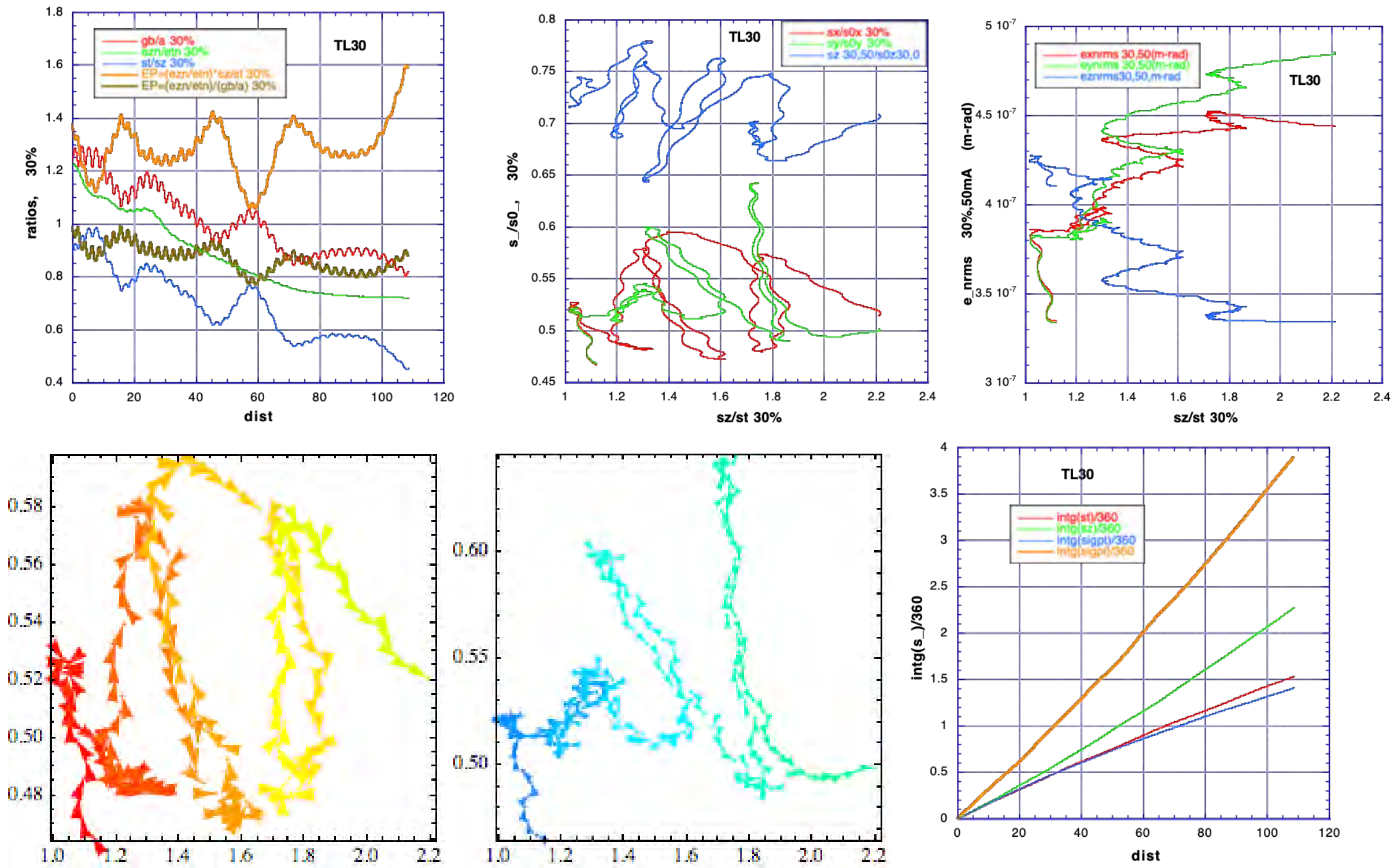

Fig. 27.28. Characteristics with 30% quads.

Emittance exchange begins immediately - seems to be from sz/st = 1 resonance; growth seems to change character at dist~70, continues slower... The transverse rms emittance growth is correlated with sz/st.

Resonances at sz/st ~ {1.05, 1.25, 1.45 to 1.5, 1.7}. J-Parc ACS has 17 accelerating cells between each pair of the 47 sets of FFDD transverse focusing quadrupoles – therefore it is logical that rational fractions m/n with n=17 might be excited, and these seem to fit quite well: {18, 21, 25, 29}}/17. = {1.06, 1.24, 1.47, 1.71}.

## 27.5. Discussion, Conclusions

Based on the observation of the 100% quad setting case, where no trending rms emittance growth is seen, apparently the rf gap driven resonances are not dangerous, but conclusions cannot be drawn about possible contribution to very low losses; more insight might be available from the 100% beam sizes and emittances, and also from error studies.

It is fortunate that the J-Parc linac was designed with strong transverse focusing, giving small tune depression >~0.7; then the quad strength reduction strategy increases the tune depression to a manageable ~0.5 (x), ~0.7 (y) for 50% and 30% quads.

The analysis above could be expanded in detail, including total beam sizes and emittances, errors, applied to other cases, etc., but the main intent here is on the larger aspects, to outline a method and future steps for consideration.

It is necessary to state clearly, however, that Sec. 27.5 and 27.7 indicate serious questions at the rms level, largely from not having a precisely defined and consistent data set. First the rms behavior must be clear, then the aspects in the preceding paragraph can be investigated with confidence.

### *27.5.1 Robustness of Equipartitioned Design Against Errors*

The point above (27.4, 27.5) about the necessity of investigating the total beam sizes and emittances is emphasized, for example by the important reference [217]. The Abstract is excerpted here:

> "Equipartitioning (EP) settings are applied as the baseline designs both for present 15mA/181MeV operation and coming upgrade to 50mA/400MeV. On the other hand, the J-PARC Linac offers considerable flexibility to search for the overall optimum. A preliminary trial was made to mitigate the intra-beam stripping (IBSt) with a lattice with constant-envelop and off-EP setting (Tx/Tz~0.3) at the 3-fold frequency jump from SDTL to ACS. With simulations without error, no significant emittance growth and halo formation were found for the off-EP setting. But when the errors at generic level are added in the simulation, emittance growth becomes by far not acceptable. It is found that being off-EP could make the lattice less robust against errors and EP condition seems more important in a world with imperfections.

### *27.5.2 The Root Problem, and Approach*

**The root problem being addressed in this work is that intrabeam stripping of the H- beam causing beam loss in the SNS and J-Parc linacs was discovered after they were built and brought into operation. A rough compensation, via reducing quad strengths below design resulting in larger beam size, was found to be successful for operation to intial design levels. However, problems with the lowered focusing strength remain, because residual, small, losses, apparently from other sources, remain that are of concern for future beam power upgrades.**

Experimental measurements of beam losses along the J-Parc linac are available. Thus in this case, this experimental data is the primary model. It is accepted as correct, and complete, although data is obviously restricted by practical constraints.

To try to explain the losses, the next level tool is simulation, from which many properties can be computed. The simulation is over a map of components which the beam is transported through – in itself, there is no knowledge of any other "theory". A great deal is required from this simulation, discussed below.

The next level tool is approximate theory, which yields useful quantities such as information integrated over various characteristics, e.g., various oscillation wavelengths. At present, theory useful for the envelope of the rms beam distribution is available and gives very useful insight as demonstrated above, but only at the rms level and for a smoothed channel. Extended theory which removes the smooth approximation is discussed below in Sec. 25.7.3 & 26.6.2.

### *27.5.3 Simulation*

As stated in Sec. 27.4.2, the main simulation program used for the J-PARC linac is IMPACT. IMPACT was not written to study linacs - it was written to explore the fascinations of parallel computer programming. Much later, some more accurate physics was introduced, such as field maps.

**The IMPACT platform might be good for linac simulation, in that:**

- **it is designed for parallel computing and capable of handling a large number of particles, even up to the actual number of particles in a typical bunch (say of order 5x10$^9$), so that very low losses could be explored.**
- **collaboration is encouraged, meaning source code is available.**

**In its present state and use however, as seen above, it is not concerned with linac study. With a dedicated and knowledgeable effort, it might be brought into line.**

**All runs were made with 80,000 particles and none were lost.** ***Thus no information has been directly obtained on the root problem of observed low losses, and no conclusions can be made.***

The available output data even at the rms level, as with many linac simulation codes, is not convenient, incomplete – not including necessary quantities such as phase advances and total beam information, inconvenient units, inconvenient form. It should not be difficult to remedy the output data situation.

**The rms information indicates that phenomena other than intrabeam stripping are indeed occurring. The DTL example reveals an unusual and intriguing situation that suggests new design possibilities. The J-Parc high-beta lattice is much different, with a small number of rms oscillation wavelengths and thus not too much time for resonance loss phenomena to develop, but also with additional driving forces, from both the reduced quad settings and new resonances involving the rf gaps, that give the possibility for small beam losses.**

Analysis is needed at every step, or particle coordinates for later analysis. Total beam sizes, total emittance, emittance distribution, particles identified by various methods as halo, can be computed various ways. Even here with 80,000 particles, such information would be usefully indicative for comparison of different schemes and settings. Apparently, more particles should be simulated.

***However (a very big however): if the model is not complete and accurate to the level required for accurate low beam loss study, more particles are GIGO.***

Every corner of the source code must be questioned and developed, under a proper framework and in terms of all parts working together.

***What is needed is a very good code addressed to the problem of low losses in linacs – period. I know of no report or reference which addresses how any separated-focusing-function, higher velocity beta, linac simulation code deals with the approach to, and solutions chosen for, this problem*** [226]**. I would be pleased to participate in discussion of a project team.**

Unless a code has been developed and documented to this *LINACS* level, then "comparison" of codes[227] is basically a waste of time, except for rough check for gross errors. If another code could present equivalent documentation, then comprehensive comparisons could be useful for collaboration and bringing the codes together. In the past, most "comparisons" have been only for *anecdotal examples* and not over the whole design space, for the most part just superficial views of output graphs, and between codes *with known deficiencies.* A code with "approximations" is useless for the present need for low-loss simulation at the highest level of present overall knowledge and computer capability. Only when a full picture has been (tediously) obtained, long computer runs not withstanding, should approximations (like for faster runs) be justified and made.

I am familiar with the history of IMPACT, and general observations by people that "it is very difficult to use", but am not familiar with the internals, although do know that it may also be very difficult to check, for example the dynamics, because of loss of the physical meaning due to a penchant for repeated "re-normalization" of the original Hamiltonian.

Needed features include:

- Open source, so that complete checking and collaborations are possible.
- Strict use of only canonical variables (time as independent variable) – otherwise computation of space charge will not be accurate and questionable. This presents no difficulty to implement.
- Ability to use full field maps and exact physical modeling for components, with inclusion of errors, recommended for the low-loss problem.

---

[226] The *LINACSrfq* code for the low-beta, continuous focusing function APF RFQ has dealt with these issues.

[227] And especially comparison with codes that are known to be approximate or deficient in critical aspects, such as choice of independent variable, approximate potentials, non-Poisson so without correct treatment of image charges, or debatable Poisson technique, etc.

• Inclusion of an intrabeam stripping model, to allow distinction between this and other loss mechanisms.

• All details of the space charge method. E.g., full Poisson computation of both external and internal fields on the same grid, for accuracy of the boundary conditions and correct consideration of image charges.

• Integration methods – consistency, accuracy, symplecticity, and also stability.

• Handling of out-of-mesh particles.

• Definition and handling of "lost" particles (only on metal surfaces).

• Use of a tracking particle starting at the "synchronous" position. (There is no formal synchronous particle in a time-based code.)

• All parameter choices – number of mesh points in grids, number of particles per mesh cell, Poisson solver parameters, boundary condition parameters, etc.

• Computation method and presentation of all quantities needed for analysis, and to eliminate problems such as shown in Fig. 27.27.

• Multiple methods for halo analysis, single particle analysis; e.g. new method developed by Chao Li using 4D action-angle phase space.

• Details of the initial particle distribution.

• Method for initial injection of particles under space charge conditions.

• Matching, at input and over matching sections.

• Analysis is needed at every step, or particle coordinates for later analysis [228]. Full rms beam size, emittance, and phase advance information, total beam sizes, total emittance, emittance distribution, particles identified by various methods as halo, can be computed various ways. If there is particle loss, the analysis must be adjusted accordingly.

• Testing of the code over a wide parameter range. A powerful "experimental" approach for this is used for the RFQ simulation code *LINACSrfq*, and should be adapted for higher beta linacs.

• Use of full computer capability – runs will be long, but then approximations can be justified and implemented for special purposes.

• Careful checks of computer usage, e.g. parallelization.

• Etc., etc.

In addition to the crucial technical features of the model (the "map) and its use, it would be desirable:

• to have a code that is :straight-forward" and in "conventional" terminology, (which will have to be expanded for the low-loss, single-particle, "halo" aspects and thus "non-conventional", but simple, direct, and well-justified and explained.

• to develop new presentations of the results especially directed at the low-loss problem.

• to include design. It is especially important that design procedures be closely correlated to the simulation procedures.

• to include design optimization, and optimization of matching and other possible adjustment schemes.

• Suitable for design and study of all types of linacs, separated function, Alternating-Phase-Focusing (APF), inclusion of nonlinear components (sextupoles, octupoles) for suppression of disturbing modes, strongly oscillating designs, etc.

### *27.5.4 Theoretical Support*

As stated above, approximate theory yields useful quantities such as information integrated over various characteristics, e.g., various oscillation wavelengths, phase advances, and the location of dangerous resonances on a tune chart.

At present, we have concepts for the envelope of the rms beam distribution, which gives very useful insight as demonstrated above, *but only at the rms level and for a smoothed channel.* This is quite adequate for many purposes, including design, as I have shown for many RFQ variations, preliminary design and analysis of separated function linacs, and for the new practical design method for Alternating Phase Focusing (APF) linacs [229].

---

[228] The available output data, even at the rms level, from many linac simulation codes, is not convenient, incomplete – not including necessary quantities such as phase advances and total beam information, inconvenient units, inconvenient form.

All previous studies of this subject and corresponding development of analysis and design tools have used the “smooth approximation” of the transverse focusing period. This approximation removes all detail inside the transverse focusing period. However, nearly all actual ring and linear accelerator lattices contain details inside the transverse focusing period that we want to understand, such as, the separated functionality of the actual transport, inclusion of driving functions other than just transverse focusing components, etc., as illustrated above for the KEK/J-Parc linac.

Chao Li and I are exploring and clarifying these effects, including the full equations of motion without the “smooth approximation”, theoretically and with simulation [230]. This is a significant advance from the present state. To establish a basis, in this article the full Hamiltonian is invoked for a pure transverse focusing lattice, revealing additional modes beyond those of the smooth approximation.

This approach will open the possibility for theoretical guidance on complex, non-continuous transverse focusing periods, whether pure alternating quadrupole or interrupted solenoid, or a linac channel with just ~2 rf gaps per transverse focusing period, but also those that contain a large number of rf gaps that may not be completely synchronous, or ring with other components ( such as sextupoles, octupoles, rf cavities, superperiods).

The subject of useful theory that helps with the problem of very low beam losses is essentially wide open.

## 27.6. New Perspectives and Prospects for Research on Future High Intensity Linacs

This Chapter contains new work and touches on new concepts and ideas for high intensity linacs, that suggest new research directions: oscillatory design, with connection to Alternating Phase Focusing (APF) linacs; and theoretical support for ring or linac channels without the limitations of the smooth approximation over the focusing period.

### *27.6.1 Advantage of Oscillatory Design – interesting new concept*

A very general theoretical comment on the possibility for an added fast oscillation to restabilize an unstable smoothed channel is given by Paul Channell [231] in his comprehensive treatment of a “Systematic solution of the Vlasov–Poisson equations for charged particle beams”. This is one of the best articles on this subject that I am aware of, and the only one by an accelerator physicist that mentions this phenomena, although published outside the accelerator field.

---

229 “Practical design of alternating-phase-focused linacs”, R. A. Jameson, 2013, arXiv identifier 1404.5176, available at:http://arxiv.org/abs/1404.5176;
“Acceleration of Cold Muons”, R.A. Jameson, KEK Report 2013-2, April 2013. (Includes method for practical design of alternating-phase-focused linacs).

230 “Interaction between high intensity beam and structure resonances”, Chao Li & R.A. Jameson, arXiv 1905.10008 v1, [physics.acc-ph] 24 May 2019.
“Structure resonances due to space charge in periodic focusing channels”, Chao Li & R.A. Jameson, Phys. Rev. Accel. Beams 21, 024204 (2018)

231 Paul J. Channell, “Systematic solution of the Vlasov–Poisson equations for charged particle beams”, Physics of Plasmas, Vol. 6, Num. 3, March 1999.

***Design with such oscillations may be a new possibility for very high intensity linac design…*** The idea of high-intensity design taking advantage of large oscillations is really intriguing, and a really interesting area for future work.

In parallel, the new practical design method for Alternating-Phase-Focused linacs [232] also uses large oscillations of the synchronous phase, and provides strong focusing (with large oscillations) in both transverse and longitudinal planes even with high energy gains.

Another noteworthy example is that in the study of space charge resonances modes, the largest growth rates are found for the non-oscillatory instabilities with Re[ω] = 0, while the oscillatory instabilities with Re[ω] > 0 growth rates are found to be generally smaller.

An initial quick survey again reveals that this phenomenon is well known and a subject of current research in a number of other fields. Refs.[233, 234] are very interesting.

After discovering the effect found in this study, and these references, *once again* I find that it may be possible to take a *concept* from the relatively enormous body of work outside the accelerator field, and neither being or not being an expert in this other field, nor re-deriving for an article the whole background of the concept, make the concept *work* practically in an actual *practical* accelerator design (not just make a theory). Concepts from outside the usual accelerator field in engineering (time-varying systems, automatic control), plasma physics (plasma period, space charge mixing), and nonlinear systems have greatly helped my accelerator design and simulation tools [235].

## *27.6.2 Theoretical support without the limitations of the smooth approximation over the focusing period.*

Ref. [230] reports the initial feat of obtaining the eigenvalues and eigenphases for ring or linac channels without the limitations of the smooth approximation over the focusing period, and presents extended results for a pure transverse periodic focusing channel.

This approach will open the possibility for theoretical guidance on complex, non-continuous transverse focusing periods that are not continuous, whether pure alternating quadrupole or interrupted solenoid, or a linac channel with just ~2 rf gaps per transverse focusing period, but also those that contain other driving terms, such as a large number of rf gaps that may not be completely synchronous, or ring with other components ( such as sextupoles, octupoles, rf cavities, superperiods).

---

232 “Practical design of alternating-phase-focused linacs”, R. A. Jameson, 2013, arXiv identifier 1404.5176, available at:http://arxiv.org/abs/1404.5176. “Acceleration of Cold Muons”, R.A. Jameson, KEK Report 2013-2, April 2013. (Includes method for practical design of alternating-phase-focused linacs).

233 “Self-Oscillation”, Alejandro Jenkins, arXiv:1109.6640v4 [physics.class-ph] 11 Dec 2012. “In general, the possibility of self-oscillation can be diagnosed as an instability of the linearized equation of motion for perturbations about an equilibrium. But the linear equations then yield an oscillation whose amplitude grows exponentially with time. It is therefore necessary to take into account nonlinearities in order to determine the form of the limiting cycle attained by the self-oscillator. The study of self-oscillation is therefore, to a large extent, an application of the theory of nonlinear vibration, a subject which has been much more developed in mathematics and in engineering than in theoretical physics.”

234 “Restoration of rhythmicity in diffusively coupled dynamical networks”, W. Zou, et.al., Nature Communications 6, Article number: 7709 (2015), doi:10.1038/ncomms8709, 15 July 2015.

235 But if they are included in a write-up, they are condemned as vague and unsubstantiated by narrowly trained “reviewers” - and unfortunately maybe also by readers who could benefit from the big picture.

Of course, the low-frequency components of the addition of other driving terms will also show up in the smooth approximation, as seen in the J-PArc linac analysis.

### *27.6.3 Local, instantaneous state*

This report stresses the concept that the real machine and simulation concentrate on the actual local, instantaneous, time-varying behavior, and how this is then used for analysis. It is also the essential design concept.

**It is hoped that this note will be useful in preparing for and implementing significant upgrades to the J-Parc linac and rings.**

## 27.7 Reverse Engineering, Phase 2

***Two Steps Back, One Step Forward***

R.A. Jameson
2nd Investigation October 2018 – January 2019

September 2019

**Preamble**

Investigating high-intensity linacs has been my research interest since starting to work in accelerator technology in 1963 on the Los Alamos Meson Physics Facility. In 2014, after several years of collaboration on an accelerator for cold muons, the collaboration members changed, and the emphasis turned to considerations of the J-Parc linac present operation and steps necessary for a significant upgrade in beam current and beam power. I was involved in the design of the J-Parc linac, which is the only linac designed and constructed to be equipartitioned (EP) throughout, affording an equilibrium beam dynamics that have a tighter beam that would be more tolerant of various error effects and consequently should have less losses along the linac. The constructed linac showed unanticipated losses, of which ~90% can be attributed to intrabeam scattering, with 10% remaining losses from various sources [236]. Lowering the design transverse focusing to give larger beam size gave successful operation to full design beam power with losses not exceeding maintenance requirements. For the upgrade, the remaining losses become a maintenance concern and further investigation is required.

The 2015 visit mainly concerned matching at MEBT1 between the RFQ and the DTL [237]. The 15-page visit report [238] outline is:

1. Seminar - *LINACS* Design and Simulation Framework
2. RFQ design with arbitrary vane profiles, e.g., trapezoidal longitudinal modulation.
3. RFQ output transverse output ellipse orientations and MEBT
4. Recommendation for experimental verification

---

236 As determined at SNS by inserting a foil stripper after the ion source, to inject H+ instead of H- into the linac.

237 The full MEBT1 specifications were not made apparent until the end of the study, but could be incorporated into the report.

238 “KEK/J-Parc Tech Memo RAJ-20150507”, R.A. Jameson

5. MEBT PIC Simulation to Include Emittance Growth and Beam Loss using the *LINACS* Code
6. Matching with MEBT PIC Simulation Using the *LINACS* Code
7. Discussion of Present MEBT1
8. Conclusions on MEBT Study
9. Brainstorming Ideas for Future

Trace3D files from MEBT1 through the entire linac, including the ACS, were obtained for this study, and are a key source for the present report.

The collaboration had lost key staff and became dormant, but my interest in exploring very low beam losses remained, and in 2016, I requested the data files for the results reported in [239]. These results were for *the design case EP linac*, with design transverse focusing and two cases of reduced transverse focusing. I considered it important to have the design case, because one could start from the EP design and perform more detailed analysis than had yet been done.

But this request was not fulfilled. After a long time and many requests, finally some incomplete simulation data was afforded, for a "100% quads" case, and for "50% and 30% reduced quads" "of an operational case". The crucially needed input data files were not afforded, also not information on the particle coordinates internal to, and exiting from, the linac, which are crucial to any beam loss investigation, as the rms picture is incomplete. There were significant problems in interpreting the output data and it required adjustment. The first analysis with 100% quads showed that the beam was not equipartitioned.

At this point I decided to proceed anyway, thinking that some performance aspects that I suspected would probably be revealed, even if compromised. In such cases, the method is called "reverse engineering".

**Reverse engineering**

https://en.wikipedia.org/wiki/Reverse_engineering:

> "**Reverse engineering**, also called **back engineering**, is the process by which a man-made object is deconstructed to reveal its designs, architecture, or to extract knowledge from the object; similar to scientific research.
>
> The **scientific method** is an empirical method of knowledge acquisition which has characterized the development of science since at least the 17th century. It involves careful observation, ***which includes rigorous skepticism about what is observed***, given that cognitive assumptions about how the world works influence how one interprets a percept. It involves formulating hypotheses, via induction, based on such observations; experimental and measurement-based testing of deductions drawn from the hypotheses; and refinement (or elimination) of the hypotheses based on the experimental findings. These are *principles* of the scientific method, as opposed to a definitive series of steps applicable to all scientific enterprises.
>
> Though there are diverse models for the scientific method available, in general there is a continuous process that includes observations about the natural world. People are naturally inquisitive, so they often come up with questions about things they see or hear, and they often develop ideas or hypotheses about why things are the way they are. The best hypotheses lead to predictions that can be tested in various ways. The most conclusive testing of hypotheses comes from reasoning based on carefully controlled experimental data. Depending on how well additional tests match the predictions, the original hypothesis may require refinement, alteration, expansion or even rejection. If a particular hypothesis becomes very well supported, a general theory may be developed."

In simpler words, starting with incomplete information, one tries to figure out what is going on. There are many aspects that could influence the observed results, and one must be rigorously skeptical about what is observed. *One must also be always and very carefully rigorously skeptical about one's own test, comparisons, observations.* One must be brave enough to offer conclusions.

---

239 "Space charge Resonances in Linacs", Ciprian Plostinar, Y. Liu, T. Maruta, M. Ikegami, A. Miura, HB2016, Malmö, 07/07/2016 (slides)

In even simpler words, with incomplete information and no one to discuss and collaborate with, reverse engineering is primarily an exercise in frustration. It is very tedious, takes a long time, false starts, take two steps back, taking the long way and then later finding a new reference or simpler way, restarts, iterations, in the end may be (probably is) not completely correct, and will require more digging. Patience, perseverance…

**2017 Reverse Engineering Result**

The first result, concerning the incomplete simulation data set, was written up by August 2017 as a 29-page report "On Compensation of an Existing Linac for H- Intrabeam Scattering and Residual Beam Loss", R. A. Jameson, August 2017.

An opportunity for a collaborative visit arose, again on the subject of investigations concerning the machine upgrade with very low beam losses. I was very concerned about whether or not to accept, having a strong apprehension that somehow my efforts were regarded as some sort of threat or competition. I asked two competent reviewers for an opinion of the report, but received no answer, and asked senior management if extended discussion was desired, again with no answer. Finally I decided to go, and immediately stated that I was worried about possibly offending, and asked if a scientific approach and discussion was possible. The report was presented in depth, to the working level and the next three levels of management. It was agreed that the report could be published [240], but there were no questions, comments or discussion, at the presentation[241], for the rest of the visit, and, although having clearly indicated that more analysis could be performed with a complete data set, to the present.

The analysis of the as-built linac showed that other modes are present.

It was very difficult to deal with the output information of the source code, and no losses were seen, meaning that no conclusions regarding very low beam loss were possible. If the simulation model does not reproduce the experimental results (where residual losses are seen), the simulation has to be changed. Significant investigation into the source code of the simulation program and reprogramming of its output to present actual quantities needed for detailed analysis was recommended. Many questions were left open.

The "50%" quads case of the data set indicated an anomaly suggesting a mistyping of a quad strength, although the value lay in a consistent progression with its neighbors. It is interesting to notice the remark in a slide:

Experimental Evidence: J-PARC
•2012 campaign conclusions
– Experimental observation of emittance exchange in a linac driven by the kz/kt=2 resonance.
– First emittance exchange measurement in a linac with emittance ratios close to 1
– Cases 1.0 and 0.9 consistent with simulation
• Weak exchange for 0.9
*–Unexpected exchange for 0.7* ***(Probably sitting on big resonance?)***
• Transverse mismatch at DTL-SDTL transition?
–Unexpected transverse halo

**Continuing Reverse Engineering Exploration 2018-2019**

However, the scientific curiosity remained, particularly as to what effect the additional modes resulting from multiple gaps inside the transverse focusing period, which introduce additional driving terms, might have on beam loss. The theoretical support has all been based on the "smooth approximation", which washed out all effects inside the transverse focusing period, including the fast

240 "On Compensation of an Existing Linac for H- Intrabeam Scattering and Residual Beam Loss", R. A. Jameson, KEK Report 2017-4, December 2017 A

241 (except for a young post-doc, Bruce Yee-Rendon, who asked several good questions. So caught him afterward, to meet and have short discussion, and Poldi took his photo.)

oscillation of the alternating-gradient focusing, as well as any other components such as accelerating cells.

Available were trace.3d files and the incomplete simulation results, with question as to whether these were consistent. It was decided to set up the trace.3d lattice in the transport (trans2) section of the *LINACS* simulation code and to try to reproduce the provided simulation results.

It would have been easier, possibly more productive, to set up a lattice from scratch. But as I know to be very skeptical of my own powers, and have great respect for the careful work of others, I always like at first to compare to actual experimental results, or the "experiments" consisting of simulation results of others.

After a long period of quite intensive flailing around, the following comments can be offered.

*LINACS Subroutines*

partran.f90
Removed paraxial approximation.

scheff.f90
- There were bad bugs in older versions
  - using wsync instead of cord(6,np),
  - all particles outside z-mesh…

Careful checking found incorrect meshing – final result – do not use mesh.f90, set meshing by hand in tapeinput scheff line. Include check and print of particles inside mesh.

- Thorough checks of scheff again.
  - Use Parmila version.
  - Can set relativistic or non-relativistic with comments.
  - dr and dz set initially and throughout by sce(2) and sce(3)
  - Better diagnostics to watch for particles outside mesh
  - Using mesh.f90 again to adjust mesh sce(2) and sce(3) as necessary
- Added analysis of good particles to make 'transAnlyzgoods.txt'
- Lowering quads made oscillations of Hchart smaller (no apertures; repeat?)

*Impact Data, comparison to LINACS simulation:*

Want just the ACS, so starting at element 1478 in middle of first rf gap sequence

Coalesced drifts to check if drifts are systematic "adjust drifts.nb' – do seem to be systematic:

*J-PARC IMPACT TL100 USES 5% WEIGHTED FITS FOR PLOTTING THE HCHARTS: 5% and 10% - SHOWS ACTION CENTERED AROUND slost ~1.5, BUT FINE ACTION IN SLOST 1.0 and 2.0 REGION IS LOST:*
Stos0t: action moves from 1.5 area to 1.0 slost area
Slos0l: action moves from 1.5 area to 1.0 slost area

***Matching very tedious, by hand. Near front matching not good – got mismatch further on, growing to end. Had to adjust match looking at whole linac.***

Much work on matching, trying to understand problem with 50% quads:
- At 100%quads, I = 0: Almost perfect match, almost perfect equal size at buncher 8. At 50mA, match is not good:
- At 50% quads, match is not good, **mismatch at (now understood) 2$^{nd}$ quad change** is very evident:

• At 30% quads, match is not good, but no evidence of quad mismatch. Was not afforded a complete data set, will never know if the same input files were used, or what the input conditions were for each run…

• Make highs in quads equal. This would not remove a basic mismatch, but showed problem near element 2360 (seems no problem in input file??); quad length change is at 2394.

***Declared in KEK Report that xp, yp behavior could not be real, but had not checked it. Trans2 is similar, so that statement in Report was wrong.***

• Very detailed observation of x, y, xp, yp through lattice
  - Which arctan to use.
  - How does this fit with Chao's definition of phase rotation over period for doublet lattice? For FDFD, alpha will symmetrically rotate thru 180°, then for 2nd 180° have to use correct tangent. For doublet period, there is repeat over one period, but not rotation by 360°…

*Matching makes sense. And may have interesting connotations – beam is breathing between big equal sizes at quads and small equal sizes at bunchers – exactly like a mismatch mode – maybe source of emittance growth inside transverse period (space charge mixing is fast enough)??!! Quite sure no one else has quite thought of this before, but this the reason for the difference between Chao's new chart and the smooth approximation chart. Have been considering effect of n buncher cavities driving extra modes, but now in conjunction with mismatch mode with period of transverse focusing period. Which effect is larger?*

**Mismatch at (now understood) 2nd quad change:**

*KEK Report – shows same excursions to large rms sizes at magnets. 50% quads case show jump at start of buncher group 22, counting from first quad set (1502-1502), so at quad 2351 (-15.402035 80.092 = 1233.57978722 ). At the next quad group 2394 (-13.880145 87.567 = 1215.442657215*
*2436 -13.835410 87.567 = 1211.52534747*
*2479 -13.802357 87.567 = 1208.630995419*
*2521 -13.770860 87.567 = 1205.87289762*
*2309 -15.425013 80.092 = 1235.420141196 ) SO THERE IS A NON-SMOOTH TRANSVERSE FOCUSING CHANGE – MISMATCH AT QUAD2394, WHERE DIFFERENT MAGNET LENGTH STARTS. THIS WAS PRESENT IN THE J-PARC ANALYSIS BUT NOT NOTED IN THIS DETAIL – IT SEEMED NOT TO AFFECT THE 100% AND 30% CASES, BUT SHOWED CLEARLY IN THE 50% CASE. EXPERIENCE NOW INDICATES ALSO THAT IT SHOWS UP MORE OR LESS DEPENDING ON WHAT IS ANALYZED… HAVE TO TRY SOME DETAILED MATCHING HERE…*

Tried matching methods with optimizer. Criteria is to minimize difference between xrms and yrms at buncher 8 of each buncher sequence, and/or minimize difference between xrms and yrms at minima in buncher sequence. Nice little puzzle to set up. Tried with 8 variables – first four quads, input alphas and beta, but never found improvement

Tried optimizing on 4 variables, the quads or the inputs, but never found improvement. (Tried different options.dat)

• First run on 12 buncher groups to element 2012, optimizing on xyloweq using input alphas and first two quad sets at 1498-1499 and 1502-1503.

• Run on whole, 41 buncher groups, runtime12:58-14:24 = 86 minutes.
Optimize using all 8 variables on both, equal weight:
objfunc=1.d0*sumbun8 + 1.d0*xyloweq

• Run on whole, 41 buncher groups, optimizing on b8xyeq using input alphas and first two quad sets at 1498-1499 and 1502-1503.
objfunc=1.d0*sumbun8 + 0.d0*xyloweq

• Run on whole, 41 buncher groups, optimizing on both b8xyeq and xyloweq, equal weights, using input alphas and first two quad sets at 1498-1499 and 1502-1503.

• Also other strategies

• November 9, 2018 -

*BUT THAT MAY BE THE LEAST OF IT!! INCREDIBLE MISTAKE STARTING IN 2016. NOW HAVE THE REASON WHY 2015 RUNS WORKED, WHY NOTHING HAS WORKED SINCE, WHY*

*BEAM SIZE HAS SEEMED TOO BIG, WHY LENGTHS WERE SUSPECTED BUT TOOK NO ACTION – COLOSSAL SNAFU !@#$%^&*()_+*

-'partran.f90 June 2, 2015' was correct – at start of element loop, changes were made to convert units from trans2 input IN T3D UNITS to trans2 units - lengths for drift and quad from mm->cm->/10, quad strengths from Tesla/m to gauss/cm -> *100. Buncher phases changed to degrees in buncher section was ok.

So that is why these runs worked.

- 'partran.f90, Sept. 4, 2016' – the unit changes were removed !@#$%^&*()

partranMatch.f90 has never been right. !@#$%^&*()_+

- partran and partranMatch fixed. Could use ~ input parameters from 2015, back to same situation – need soft-edge quads, set up PMQ.nb. Basic beam size now agrees with Impact.

• Question on hard- or soft-edge quads as used in Impact. Set up soft-edge quads on partran. T3d PMQ formulas work in M.

• **Check if type of input distribution makes difference – Gaussian may be softer??**

DOES MAKE BIG DIFFERENCE! MISMATCH PATTERN SAME, BUT MAGNITUDE DIFFERENT, ESPECIALLY WITH GAUSSIANS: IMPACT MAYBE USING DISTRIBUTION FROM BEGINNING OF LINAC, OR FROM ION SOURCE – NO INFORMATION…

LONGITUDINAL WOULD INFLUENCE SIGMA'S, b/a, EP, ETC…

TRANS2 TRANSVERSE XRMS, YRMS TYPES DON'T DIFFER SO MUCH UNTIL GAUSSIAN. COMPARING TO IMPACT, COULD CHOOSE TYPE 2, WHICH AGREES WITH TL100 LONGITUDINAL WHEN SCALED BY RMSZ*BETA*GAMMA*SQRT(6) (REFERENCE CONFUSION WITH IMPACT DATA COLUMNS, 'ACS Analysis.nb'; the βγ scaling "works",,,).

SO USE TYPE 2… and work on transverse matching at 0mA further…

• November 22, 2018

***DECISION:*** trans2 is working ok. PMQ magnets are some different from Impact. PMQ lengths for trans2 are longer that for Impact – Impact seems to be using harder PMQs – no fringe fields??

• 27 December 2018 -

Probably TL100, etc. results are for field map with "length" including fringe field, as stated here:

- .t3d file uses Type 4 consistent with PMQ and its inputs.

Therefore: 82% quads work -> effective strength of 82% hard-edge quad equivalent to field map with fringe field of "100%" TL100. Ikegami quad: Using input specification and a modified trans2 PMQ routine to approximate Ikegami-type quad field map would have been very close to same as exisiting hard-edge quad *0.82 input specification. See 'PMQ Ikegami.nb"

*But TL100 result not equipartitioned. Gap effect – smoothed rfdata profile, 3rd harmonic??*

Using a smoothed rf gap – use "rfdata", "IMPACT gap map conversion.nb".

Trace3D files indicate harmonic=3, and energy gain

So Impact seems not using "3rd harmonic"??

Can get good detailed xrms agreement between Impact and partran by lowering 1st 2 quad sets much lower and also adjust ellipse alphas, beta – **CLEARLY THE TL100 DATA SETS WERE FOR A QUITE DIFFERENT DESIGN – NEVER WAS AFFORDED THE COMPLETE DATA SET, NEVER HAD THE CORRESPONDING INPUT FILE. HAS COST AN ENORMOUS AMOUNT OF SCREWING AROUND !!!!!!!@#$%^&*()_+{}[];':",./<>?**

**Run with smoothed quads, same input parameters as with original quads. NO DIFFERENCE – PROBLEM IS MISMATCH AT QUAD CHANGES.**

There are bumps in the buncher voltage settings at ~1480 and at ~2400 and at end – full up-down phases buncher sections – WHY?? – but Ecos(phi) is smooth.

*Forgotten!! - dr and dz set initially and throughout by sce(2) and sce(3)*

LOWER ALL QUADS TO 82%, SOME REMATCHING AT INPUT – GIVES CLOSE TO IMPACT RESULT, qualitatively same:

• 7-8 December - ‘Plot resonances as angle.nb’.

13 December 2018 -
Rematch 82%quads 50mA: Much stronger emittance growth than TL100 …
*CANNOT USE *GAMMA^2 OR *GAMMA^3 TO LOCATE MODES. EVEN WEIGHTED FIT IS SUSPECT?*

21-23 December 2018 - Redid 82%, new figures in 12/20. **Return to 100% quads.**

23-24 December 2018 -
Compare to TL100: trans2 harm=1
**Compare 100% and 82% quads – *little difference!***
**Harmonic harm=3 Gaps:**
82% quads – **MUCH BETTER EQUIPARTITIONED! (but still not good, ramped…) NO CORRELATION BETWEEN EMITTANCE AND GAP RESONANCES… EMITTANCES MOVING COUNTER eln growing, etn shrinking !!!**

• 7 January 2018 -
**Investigate other non-EP lattices:**
1. J-PARC EP LATTICE. NEAR START, s0t IS EP. RAMPING s0t BY UP TO 50% STAYS NEAR EP.
Smoothed the quad profile over whole length, reduced to 75%, same match as75% quads standard lattice below; original gaps: y match stays good, x match gets worse…
REALLY EP, BETTER THAN 82% NON-SMOOTH QUAD

**Different buncher rules:**
5-10 January 2019 – and very early in 3.) Struggles
1.Orig phi’s E’s
2.Orig phi’s,aveEcosphi
3.phis’s=-30, orig E’s;
4.Modify E for same Ecosphi with phi=-30
2. if long RT tank fields could be retuned.
2.-4. If independent SC cavities. Wobble buncher amplitudes +-5%, Average Ecosphi settings
- *Variations of Ecosphi strategies made no big differences (repeat?)*
- 5 January 2019 –

For 75% quads, run buncher variations: have to rematch
*Nothing particular noticed…*

**Investigate all gaps raised:**
11 January 2019 -
MISMATCH 75% original quads, harm3 gaps

13 January 2019 -
Another anomaly with TL100:

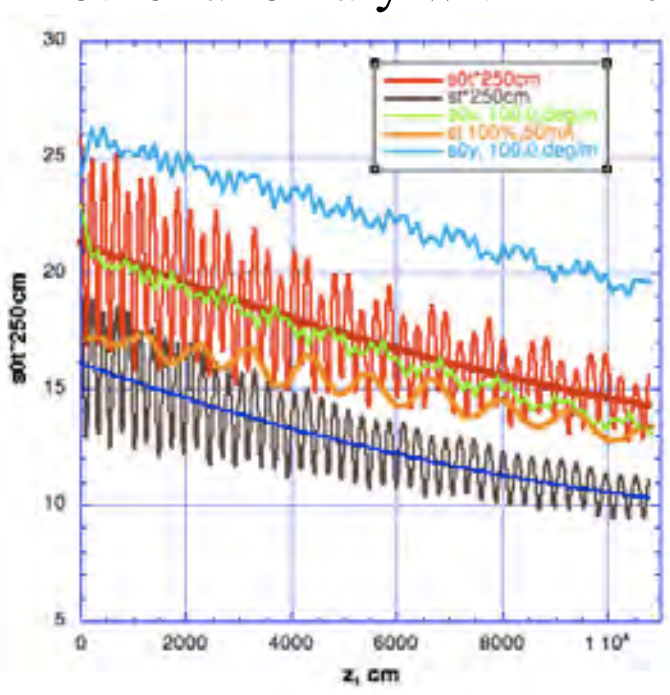

s0t*250 agrees well with s0x of TL100, but not with TL100 s0t.
Behavior of depressed tunes quite different… ???

CANNOT BE SURE ABOUT PRODECURE THAT WAS APPLIED TO TL100 – GAMMA'S – MAYBE PHASE ADVANCES INCORRECT (CERTAINLY SO IN LONGITUDINAL) – BUT AT LEAST MAGNITUDE FOR TRANS2 DATA SEEMS OK.

**Conclusions?**

I very much wanted to collaborate, not compete, on such a study…

*The EP Linac*

Wanted to study the design EP linac, and asked for the data used in Ref. [222] in order to have connection to published data. But was never afforded this data, and only after repeated requests and clear reluctance, obtained an incomplete data set. Extensive reverse engineering of this set pointed out procedures necessary for analysis and some features worthy of further investigation. But mainly, it shows that this data set is not for the design linac, but is for some undefined ad-hoc "operational" tune and has so many anomalies that it is not possible to use it for any conclusive purposes. Some of the reasons that analysis and comparison fail are probably still due to modeling differences, but trying to work with an "operational" tune that is no longer tied to any consistent, coherent overall physics model is probably pre-doomed, as a solid basis for reverse engineering is missing. If only the data set from [222] could have been obtained, even working alone, so much valuable time could have been spared for more important things.

It is a shame that this unique linac, designed to be EP, has not been thoroughly investigated to determine its operation in comparison to the design. The later discovery of a severe performance complication due to intrabeam scattering with H- operation does severely complicate the issue, but is no excuse.

Although the SNS linac was not designed taking the space charge physics into account, a key benchmark was obtained rather early in the operation of this facility after the intrabeam scattering effect was known ~2009 – a foil stripper was installed at the front end, and H+ ion run through with the same settings. The result showed that the H+ beam losses were ~10% of the total losses observed with H-, and also the location of the H+ losses.

***This benchmark has still not been obtained at J-Parc.***

*Extra Modes from Longitudinal Layout With Multiple Gaps In Transverse Focusing Period*

The results so far, with only rms analysis of a suspect data set, indicate that at least one extra resonance mode does exist from extra longitudinal focusing elements distribute inside the transverse focusing periods. At the rms level and with the present level of modeling, it appears they are not serious, as no beam loss is observed. However, analysis of the 100% beam sizes and emittances is lacking, ***and perhaps very important, the effect of errors at the expected levels (generic – Sec.27.5.1), and sensitivity to parameters such as input emittances and beam current should be investigated.***

*Simulation vs. Experiment*

The operational linac shows beam losses not predicted by extant simulations. In this situation, there is only one answer – the simulation must be improved. Correct data presentation and analysis methods must be brought to bear. At present, no conclusions about beam loss can be made, at current beam power level, or for foreseen upgrades.

[eltoc]

# Chapter 28 – Time Dependence, A Priori Adjustment

Hopefully, this book has remained intriguing to the end, with many **elemental** and detailed insights, while wondering about the many questions that still remain.

It is late (January - February, year 2022); regardless of decades of my using and publicizing, only a few details of the **elemental** view and its possibilities for any linac and other such systems have become used to some extent, and then often not in full, or even correct, context.

A key example is the insight, Ch.28.2, into the key equations given or hinted at by Sacherer. I consider the understanding and application achieved on this subject as a primary accomplishment, on par with making equipartitioning and alternating-phase-focusing linacs feasible, development of linacs design and simulation tools, and the deep analysis of mismatched beam dynamics. However, although the bait has been presented over and over, no one has ever asked about it or expressed curiosity or interest in using it. In view of its **elemental**, fundamental importance, that is reason for emphasizing and reviewing it as a final Chapter, with further details.

Then all of the elements come together in the discovery of something entirely new.

## 28.1. Time scale of beam distribution changes in typical linacs

See also Sec. 16.1.3.

### 28.1.1 The transverse and longitudinal rms envelope and matching equations

The rms envelope equations are derived from the time-dependent Hamiltonian. i.e., they are functions of time. They are approximated by using the "smooth approximation" – the slower part of the Hamiltonian – over the transverse and longitudinal focusing periods or some other appropriate length. The smoothed approximation transverse and longitudinal beam envelope equations can be derived for any linac type. These are called the rms matching equations, which must be satisfied at all times (instantaneously, locally), everywhere along the linac. For matched design, they are solved simultaneously – either alone, or with 3rd equation (§ 28.3) that additionally provides stability. These equations relate the rms beam emittances to the rms beam sizes and the rms smooth approximation phase advance per unit length:

$$\varepsilon_{tn} = \frac{a^2 \gamma \sigma_t}{L} \qquad (1)$$

$$\varepsilon_{ln} = \frac{(\gamma b)^2 \gamma \sigma_l}{L} \qquad (2)$$

where $l$ indicates transverse and t indicates longitudinal; $\varepsilon_{ln}$ and $\varepsilon_{tn}$ are the normalized rms emittances (cm.rad), $a$ and $b$ the rms beam sizes (cm), $\beta$,$\boldsymbol{\gamma}$ the relativistic beta, gamma. $\sigma^t/L$ and $\sigma^l/L$ are the phase advances in the smooth betatron or synchrotron oscillation over a suitable unit length (here $n\boldsymbol{\beta}\lambda$), $\gamma$ is the relativistic gamma. (The division by a unit length L has been widely overlooked, which is still causing confusion.)

Expanding the phase advance (or "tune") terms;

$$\sigma^{t2} = \sigma_0^{t2} - \frac{I\lambda^3(1 - ff)}{a^2(\gamma b)\gamma^2} k \qquad (3)$$

$$\sigma_l^{\,2} = \sigma_0^{l\,2} - \frac{2I\lambda^3 ff}{a^2(\gamma b)\gamma^2} k \qquad (4)$$

where $\sigma_0$ is the zero current phase advance defined by the external field, $I$ is the beam current, $a$ and $b$ are the transverse and longitudinal rms beam radii, $ff$ is the geometry (form) factor ≈ $a/3\gamma b$, $\gamma$ and $\beta$ are the relativistic gamma and beta, and

$$k = \frac{3}{8\pi} \frac{z_o q 10^{-6}}{mc^2}$$

Averaging the x and y motions is usually sufficient for analysis and design with linac structure types. However, both the transverse and longitudinal should always be observed, and all three dimensions are also sometimes necessary, because of different degrees of transverse-longitudinal coupling in various types of linacs (RFQ is completely coupled, superconducting linac hardly coupled). This is very often overlooked because of a non-linac teaching bias.

*28.1.1.1 The missing derivatives*

The exposition in Ch.16.1.3 is expanded here.

Usually the rms envelope equations are presented as the "matched" equations, with $d_^2/dt^2 = 0$, interpreted as over *smoothed* the transverse focusing period (same unit length for all planes). The "betatron" (transverse) or "synchrotron" (longitudinal) periods are included in the rms envelope equations in terms of the "phase advance" of the oscillation per unit length.

They are also presented with d_/dt = 0, but these are just omitted by requiring that changes with time be "adiabatic" [242]. The approximation of adiabaticity has been questioned for many decades. Originally, probably for a case at hand, and now, probably indeed for many cases a negligible approximation. So people got in that box and never question, just grind the handle. But for deuteron beam loss, for example, it should be thoroughly understood.

I know of no published reference that includes the first derivative terms, with justification and explanation, as they would appear when used for design or design simulation, but know of two cases where their influence has been studied to some extent. First is my own derivation, which was stimulated when linac (e.g. RFQ) designs were seen with very fast parameter changes (Ch.16.1.3). Various derivative assumptions have been added on top of the *LINACS* Poisson simulation, just to observe the effects; they show up to a few percent decrease in accelerated beam fraction. They are not in the present code. Second is the work of Li Chao, who extended the theory and simulation of non-stable modes in periodic lattices by discarding the smooth approximation.[243]

Chao Li notes (paraphrased) "that when the damping terms (the 1st derivatives) are added to the envelope equations, *the equations are then for a* ***non-Hamiltonian*** *system.* The mode analysis (Ingo 1993,1998, our paper 2018) are only (formally) valid in the Hamiltonian system" – [but Sacherer in 1970 stated also valid for 3D elliptical distributions (Ch.1.4.3), and I showed 1981 it is also valid with acceleration].

Take first derivative before the adiabatic approximation – of external field parameters w.r.t. particle coordinate – and do not do any derivative of emittance, as that is declared impossible to handle by Sacherer. So questionable. And getting the Poisson coordinates, and then subsequently apply the derivative terms to the divergence coordinates – has no rigor and is only to see what happens.

*28.1.1.2 Relation of the rms envelope equations to simulation*

The rms envelope equations should be used for design, and for aid in analysis. Their theoretical peculiarities, assumptions, simplifications, limitations, are not relevant to a physically accurate, self-consistent Poisson simulation.

In such simulation, testing of the simulation settings and properties (e.g., mesh density, step size) are used to check aspects such as the rate of design parameter changes. It has been clear for decades that the results always depend to some extent on the number of steps computed per lattice period or cell, with more steps resulting in more beam loss. A design with faster than adiabatic parameter changes will be seen to have even more beam loss. This is expected, in regard to the smooth approximation, as parameter changes can occur within the smooth approximation period. The effect of the number of steps must always be checked, and decisions made.

---

242 https://en.wikipedia.org/wiki/Adiabatic_theorem

243 "Structure resonances due to space charge in periodic focusing channels", Chao Li & R.A. Jameson, Phys. Rev. Accel. Beams 21, 024204 (2018)

### 28.1.2 The fastest time interaction in the rms envelope equations

Simulation of typical linac designs show that the beam dimensions and area typically change significantly within the betatron and synchrotron period, and even on the time-scale of the focusing period.

When there is space-charge, the fastest and ever-present influence on the beam is characterized by the plasma period. The basic interaction, termed *space charge mixing*, is that particle redistribution inside the beam can occur in about (plasma period)/4. Space charge mixing *is continuously present* when there is space charge. This was first explained by O.A. Anderson[244], in terms of the time it takes for initially parallel trajectories to bend and first cross. Definitions of the plasma periods are:

$$\sigma_p^l = \sqrt{\left(\sigma_0^{l^2} - \sigma^{l^2}\right)/ff}$$
$$\sigma_p^t = \sqrt{\left(\sigma_0^{t^2} - \sigma^{t^2}\right)/(2/(1-ff)}\qquad(5)$$

where ff = ~ a/3b is the space charge form factor.

This is a collective effect – not a resonant effect - with much faster change than those caused by other driving forces like resonance interactions or mismatch, etc. In terms of its rms definition, it will be observed in changes of the rms emittance and beam size, but changes in the detailed particle distribution, total emittances and beam sizes, internal energy balance, detailed match to the local features of the channel, etc., are particularly important. Typical (plasma period)/4 for a tune shift of 0.4 or below is only a few cells of a linac.

## 28.2. The *time-dependent* Equipartition (EP) Equation

The third time-dependent equation describes the internal energy balance, called equipartitioning (EP), between the degrees of freedom of the beam. Balanced means in equilibrium and that the beam will be immune from certain disturbances, such as a resonance corresponding to the EP parameters. The beam internal energy is balanced between the transverse (the average of x and y is often justified) and longitudinal degrees of freedom if

$$\frac{\varepsilon_{ln}\sigma^l}{\varepsilon_{tn}\sigma^t} = 1 \qquad (= \gamma^*(db/dt)/(da/dt)) \qquad (6)$$

which also implies

$$\frac{\varepsilon_{ln}}{\varepsilon_{tn}} = \frac{\gamma b}{a} = \frac{\sigma^t}{\sigma^l} \qquad (7)$$

Eq.(6) is called the equipartitioning condition, and Eq.(7) the equipartitioning ratios.

The simple and practical form of the equipartitioning Eqn. (6) was the breakthrough needed to realize that this equation can be *exactly* [245] solved *simultaneously* with the matching equations, giving the possibility to design linac channels that maintain the beam in equilibrium *locally*, thus minimizing the potential for beam size and/or emittance growth that could lead to undesired beam loss causing radioactivity buildup and maintenance problems.

The use of this equation in design is optional; if used, it also should be locally satisfied *exactly* (not in some approximation, such as, heaven forbid, at the space charge limit!!).

---

244 Oscar A. Anderson, "Internal Dynamics and Emittance Growth in Space-Charge-Dominated Beams", Particle Accelerators, 1987, Vol. 21, pp. 197-226; "Emittance Growth in Intense Mismatched Beams", 1987 Particle Accelerator Conference, Washington, DC, IEEE Cat. No. 87CH2387-9, p. 1043.

245 Not, for example, heaven forbid, severely approximated at the space charge limit !! Linacs with tune depressions ≤~0.6 are in the region of maximum nonlinearity.

These equations deal with effective rms quantities – not Liouvillian. The concept of energy is also definable only at the rms level, and not at higher order, so the set of equations is consistent. The beam internal energy is expressed in terms of local rms quantities, and not as a thermal quantity, although of course is also correct in the thermodynamic limit.

Practical use of the equipartition equation and corresponding ratios simply uses quantities from the rms envelope equations, with their corresponding **time dependences**.

It should thus be obvious that the EP condition, and corresponding ratios are **time dependent**, will evolve in the same way as the included variables, and describe the **local beam state**.[246] For example, the beam may be non-EP, then approach, pass through, and leave the EP condition.

*(This paragraph could be a footnote, but describes a very unfortunate and quite astonishing misinterpretation that has wasted enormous amounts of time and energy and has created a lasting problem in accelerator practice among those who do not have basic understanding of modern nonlinear systems, or do not think about it) :*

Unfortunately, this obvious time dependence and its practicality has not been clear to some, and loud and very misleading claims have been made that the EP equation is derived from thermodynamics and describes a purely thermodynamic equilibrium, and is therefore not appropriate for an accelerator system, in which (as typical for such nonlinear systems), the settling time is of the order of the settling time of the universe.

Please think about it, and dismiss this clearly false thermodynamic proposition.

## 28.3. Sacherer 's "*a priori*" clarification[247] on the rms envelope equations

Theorists seek underlying explanations by simplifying the real world, usually drastically, for example by assuming a linear model, as Kapchinsky did ~1958 for his "KV envelope equations[248].

By 1970, Sacherer showed that the KV assumption is not necessary if distributions have equivalent rms values. Beyond that, but hardly noticed, crucially seminal is that he shows that the time dependence of the emittances in the rms envelope equations is practically (mathematically) limited to expressing the time dependence "*a priori*" – initially. In particular, he notes that expressing the beam emittance as a function of time or equivalent (e.g., z or cell) is mathematically allowed.[249] Expression of *a priori* evolution of the emittances can have useful effects on the control of space charge, on the length of the linac, on the EP condition and ratios.[250]

**We point out here that, although Sacherer does not state it specifically, *a priori* expression can be applied to *any* of the variables in the equations. And that this will have interesting, useful, consequences on a design.**

---

246 This was shown in the first practical use of the EP equation in simulataneous solution with the rms envelope equations in the original paper, R. A. Jameson, "Equipartitioning in Linear Accelerators", Proc. of the 1981 Linear Accelerator Conf., Santa Fe, NM, October l9-23, 1981.

247 "RMS ENVELOPE EQUATIONS WITH SPACE CHARGE", Frank J. Sacherer, CERN/SI/Int. DJ_J/70-12 18.11.1970

248 I. M. Kapchinsky and V. V. Vladimirsky, Proc. 1959 Int. Conf. on High Energy Accelerators and Instrumentation, CERN, Geneva, p. 274.

249 (Sacherer shows that the KV distribution assumption is absolutely not necessary; nor is it recommended for practical use; if people had paid attention, an enormous amount of time, publications, could have been avoided and the time spent on productive problems.)

250 "RFQ Designs and Beam-Loss Distributions for IFMIF", R.A. Jameson, Oak Ridge National Laboratory Report ORNL/TM-2007/001, January 2007

***It is then also necessary to understand that a simulation of a design is independent of the design method*** – it just simulates (hopefully with the best possible fidelity) the design that is presented to it. **So a trusted simulation can be used to check the fidelity between design and simulation, and between simulation and a very well executed experiment (any working linac).**

## 28.4. Other possible *a priori* manipulations

### *28.4.1 The form factor*

During the IFMIF studies, it was found that EP designs for bunched beams with significant space charge, with or without *a priori* manipulation of the emittances, were achieved, and verified by simulation, showing that the EP and ratio conditions were satisfied to within ~2-5%. (The EP condition = 1 is fully equipartitioned.) It was desired to investigate this remaining discrepancy, realizing the conditions of using the equivalent rms quantities, and anticipating that higher order effects could be present. Attempts to insert further plasma conditions, such as a Debye length, were unsuccessful. Knowing that the higher order effects would be accurately simulated using full Poisson simulation *with correct boundary conditions and symmetries*, it was decided to make an *a priori* adjustment of the space charge form factor, using information from simulation of the bare design (i.e., with theoretical form factor). $(b/a)_{experimental}$ was computed for the run, and fitted with a polynomial depending on z or velocity $\beta$. [251] The design was modified by multiplying ff in the equation above by $((b/a)_{experimental})$, giving $((b/a)_{experimental}))/(3*(b/a)_{design})$. In the case of the IFMIF design, as stated above, *EP and the ratios were already nearly correct*. Simulation of this design with the *a priori* form factor adjustment showed the EP condition and ratios satisfied to ~0.5%.

#### <u>28.4.1.1 Base designs which deviate further from the requested design</u>

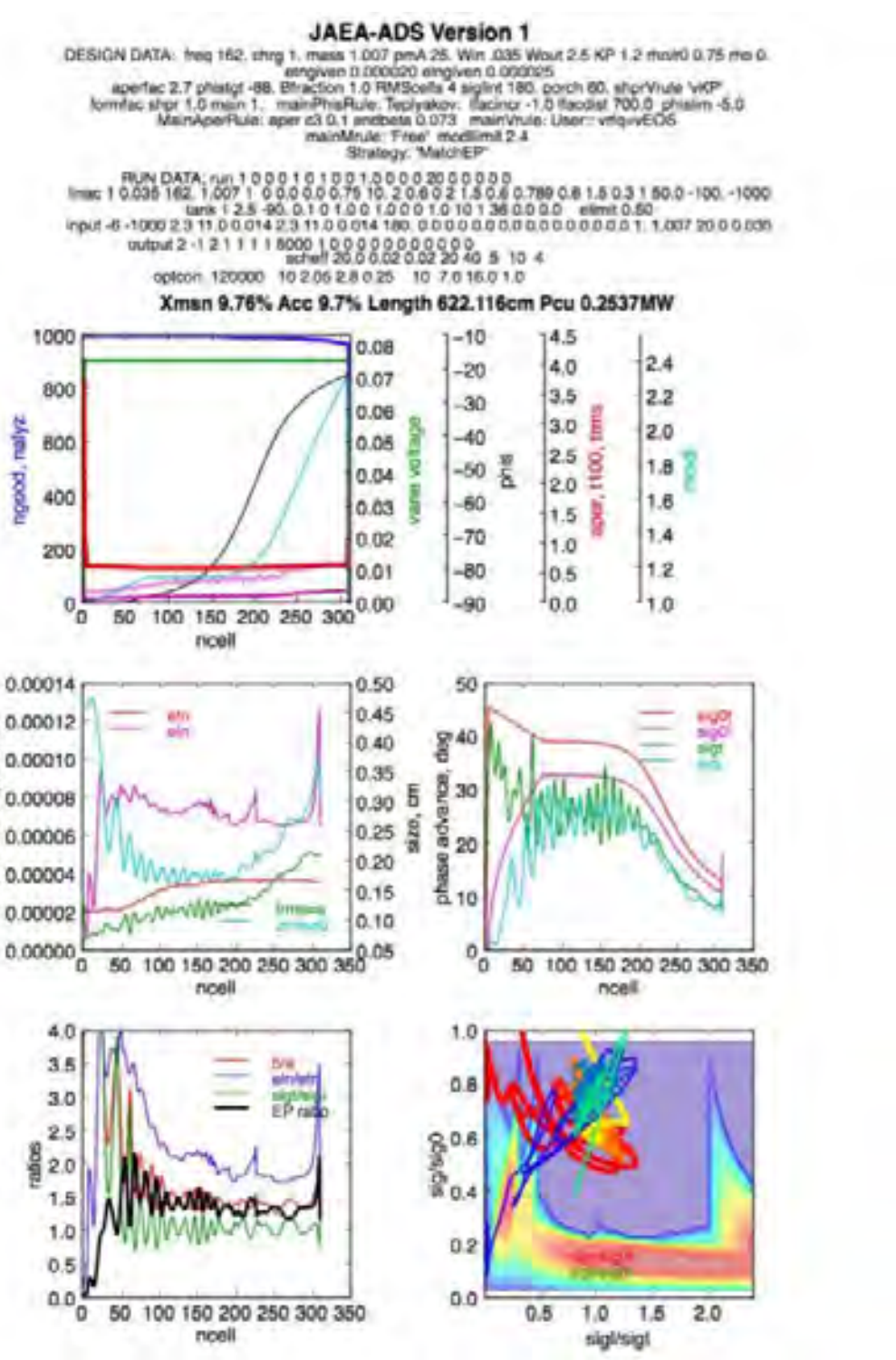

The case used for this exposition is the JAEA-ADS RFQ design study of Ch. 23.7. It is desired to achieve EP in the whole accelerating section of the RFQ after a shaper, and also to achieve as low longitudinal emittance as possible, so the design EP ratios = 1.25. The specification is difficult, and simulation of preliminary design studies, Figs. 28.1 shows undesired emittance growth, particularly in the longitudinal plane, which then causes an equipartitioning transfer with transverse emittance growth and longitudinal emittance decrease, and the requested EP conditions are not well satisfied.

A priori adjustment of the form factor will be used to improve the design.

The form factor ff in the code is: ff = ffadj*(1.d0/(3.d0*p))

where p = gammaloc*X(3)/X(2), where X(3) and X(2) are the values for b and a found by the simultaneous solution of the design equations.

The form factor at the end of the shaper can also be adjusted using “formfac shpr” (ffadjshpr).

Fig. 28.1. Base bare design (2term vanes). There is no saturation of either the modulation or synchronous phase. A few particles are lost from acceleration at ~cell150; radial losses start ~cell 250.

---

[251] In general, I refuse to do fitting unless the fitting has a firm physical basis. The literature is full of pseudo-fittings to some function that have no basis to the problem at hand, for example, naively fitting Gaussians to beam distributions. Non-relativistic beam behavior is not thermodynamic and should not be approximated as such (i.e. at the space charge limit).

*28.4.1.1.1 Form factor ff hard-wired to spherical bunch, ff=1/3*

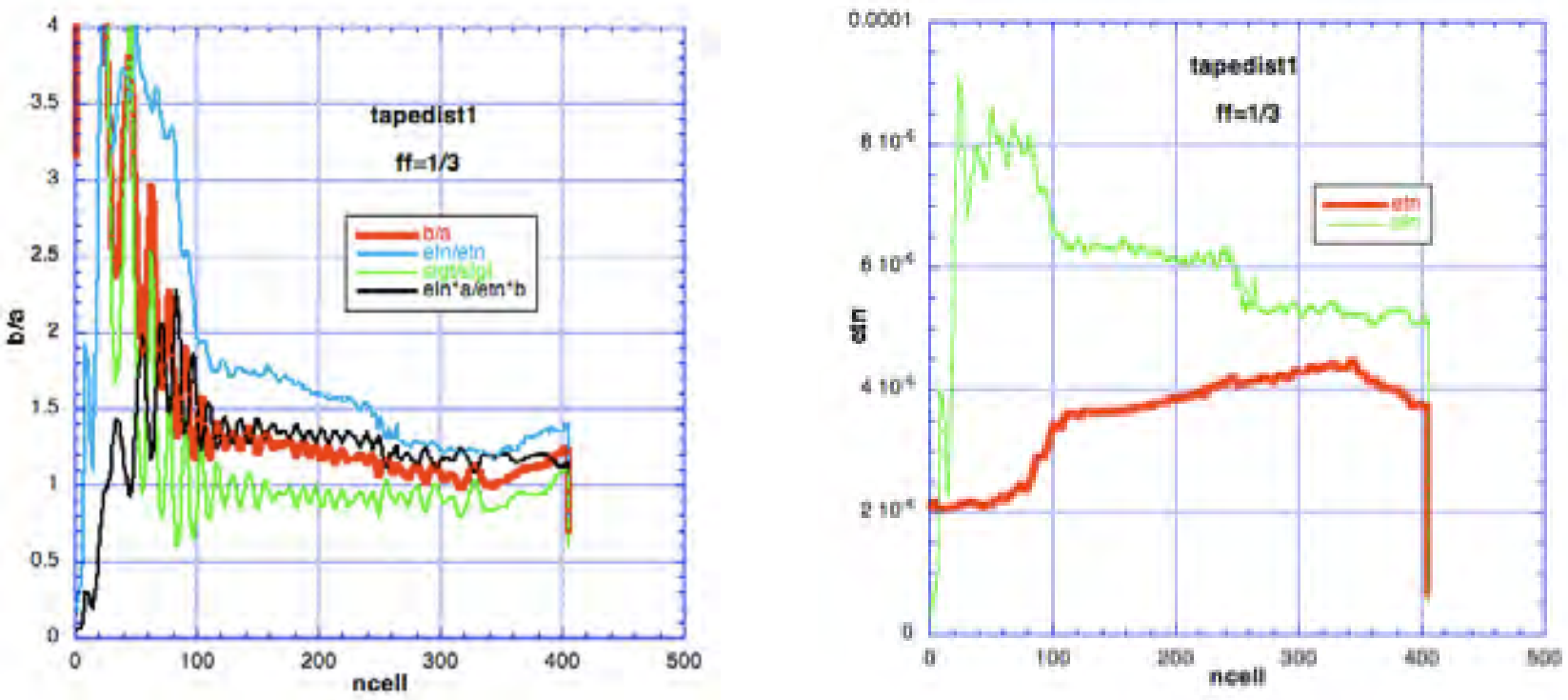

Fig 28.2. 2term vanes. Form factor set to 1/3; EP conditions and rms emittances. There is no saturation of either the modulation or synchronous phase. The total length in increased to 735 cm.

Transmission and accelerated beam fraction at the end, cell 405, are ngood= 94.0%, nalyz 94.0% .[252]

*28.4.1.1.2 Form factor ff hard-wired to desired bunch, ff=1/(3*1.25)*

Very nearly same as the base design; total length, emittances, EP conditions, because it is the same design except only the hard-wired ff ratio, with all other a and b design conditions solved as the base design, where ff participates freely.

*28.4.1.1.3 Form factor adjustment using polynomial fit to (b/a) from base design simulation*

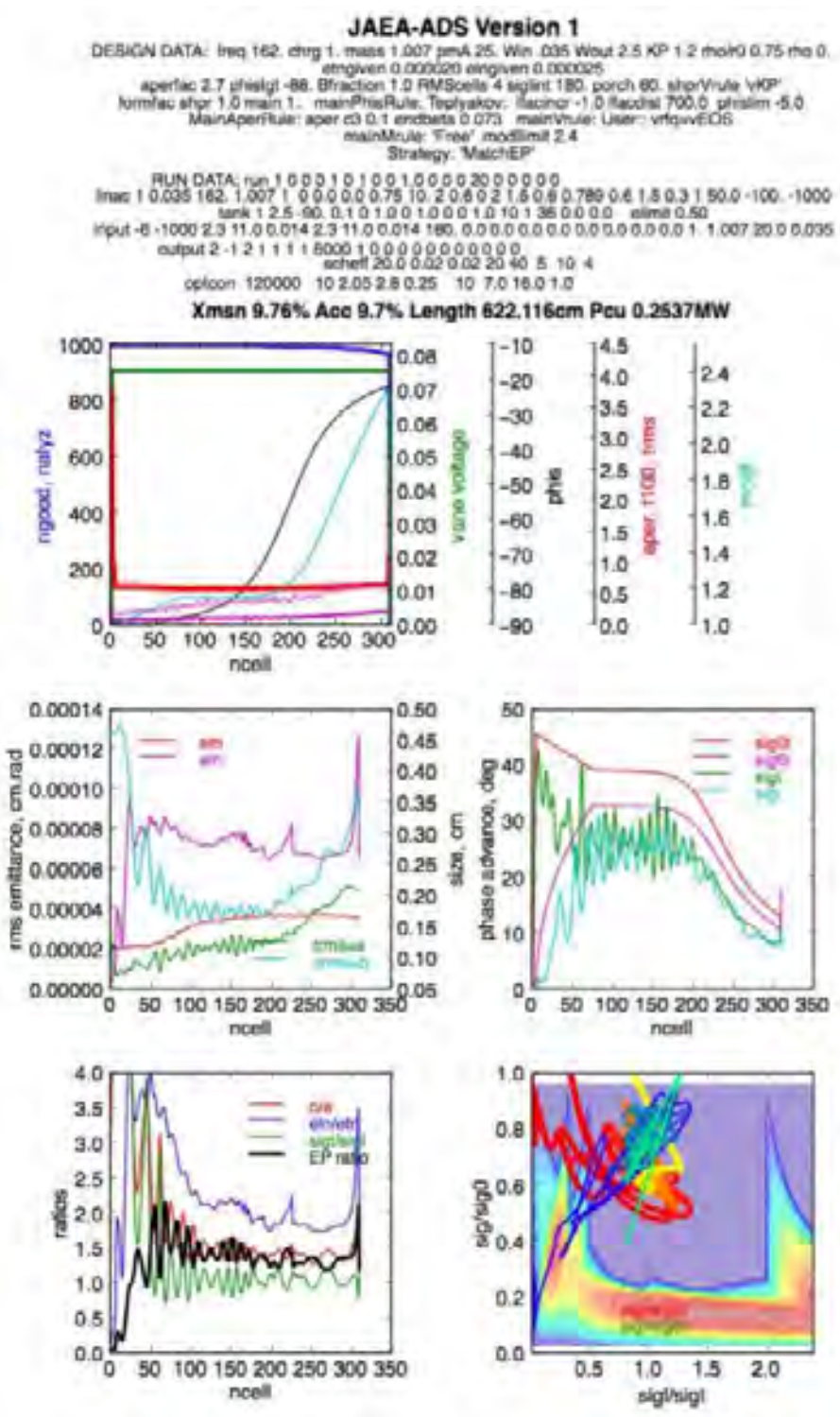

**A. *SINE* VANES**[253] **-** Fig. 3 shows the base design simulation, with linear and 4th-order polynomial fits to the simulation (b/a) result:

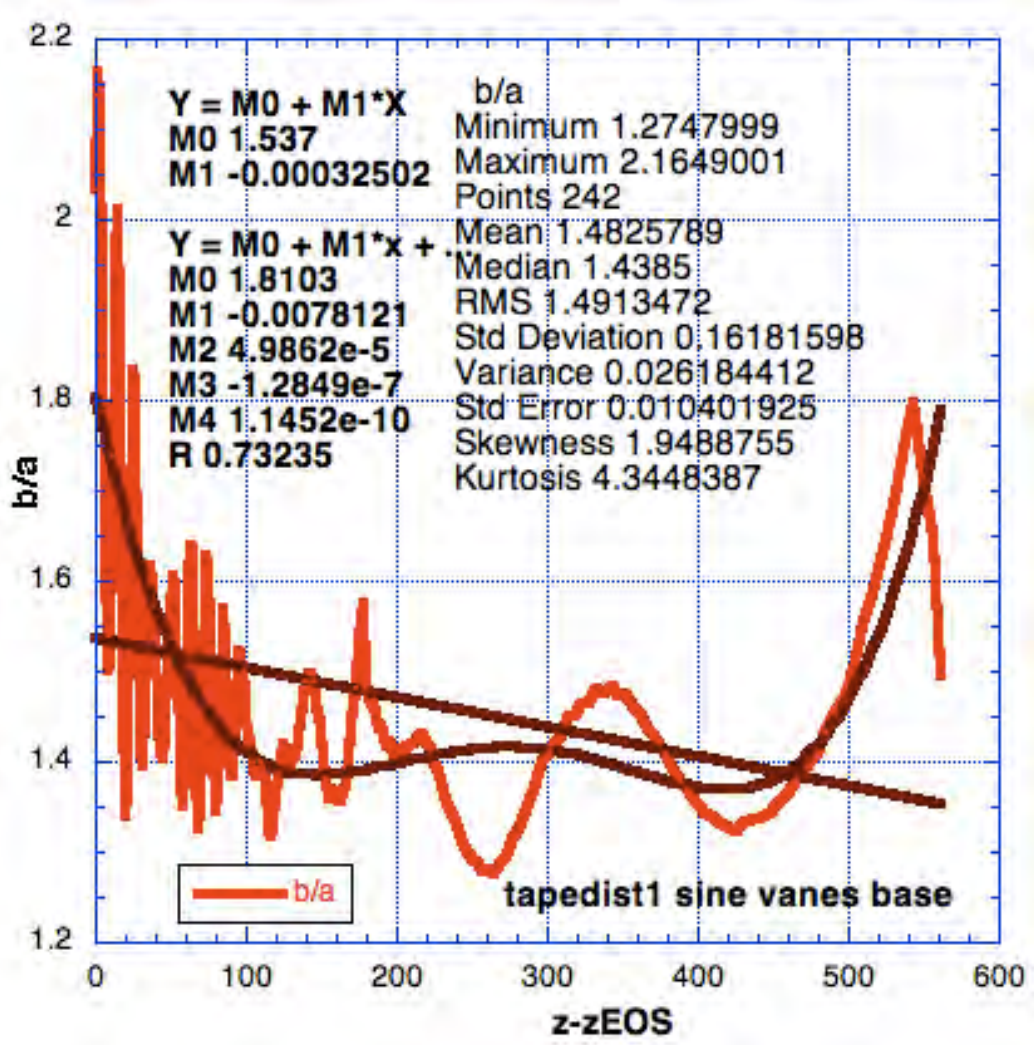

Fig. 28.3. Left - **Sine vanes** base design. Length 622 cm. Right – $(b/a)_{experimental}$ linear and 4th order fits and statistics.

252 Problem at end of the run. See below …

253 (There was some ending trouble with 2term vanes – fixed later, see below. Ch. 23.7 shows almost no difference between 2term or sinusoidal vane modulation, so proceeded for awhile with sine vanes. Please excuse – just look at the improvements achieved.)

The mean of the 4th order fit is 1.483 ~= 1.5 ~= ~ same as 0 intercept of linear fit (at EOS) = 1.537. (coincidental)

### A.1. Move the $(b/a)_{experimental)})$ average down to 1.25, (0.8 = 1/1.25)

ffadj = 0.80*(4th order (b/a) fit))

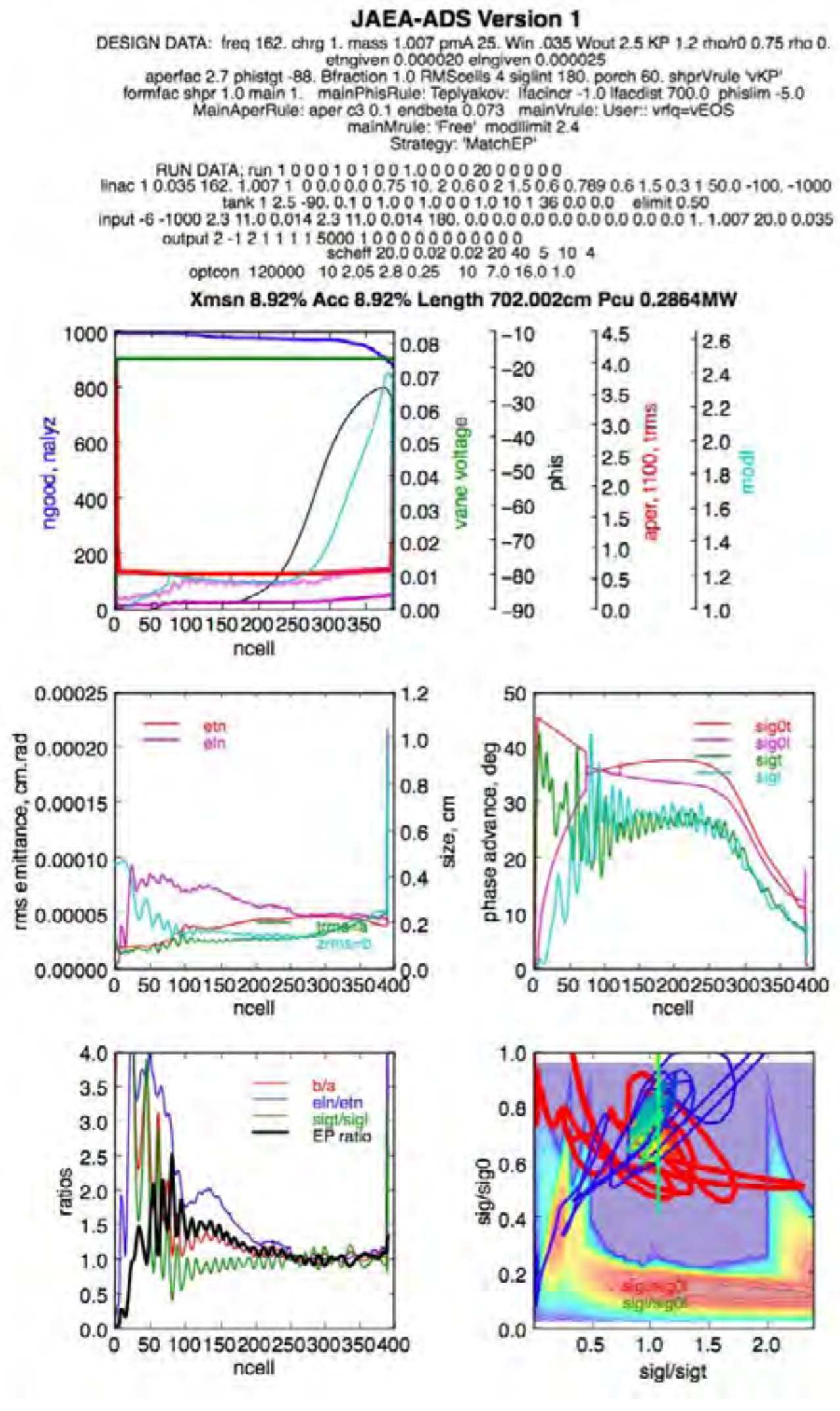

**Then the resulting ff is** $\sim(b/a)_{experimental)})/(b/a)_{design} = (b/a)_{experimental)})/1.25 =\sim 1$ -> spherical bunch

The result is a very good spherical and equipartitioned bunch.

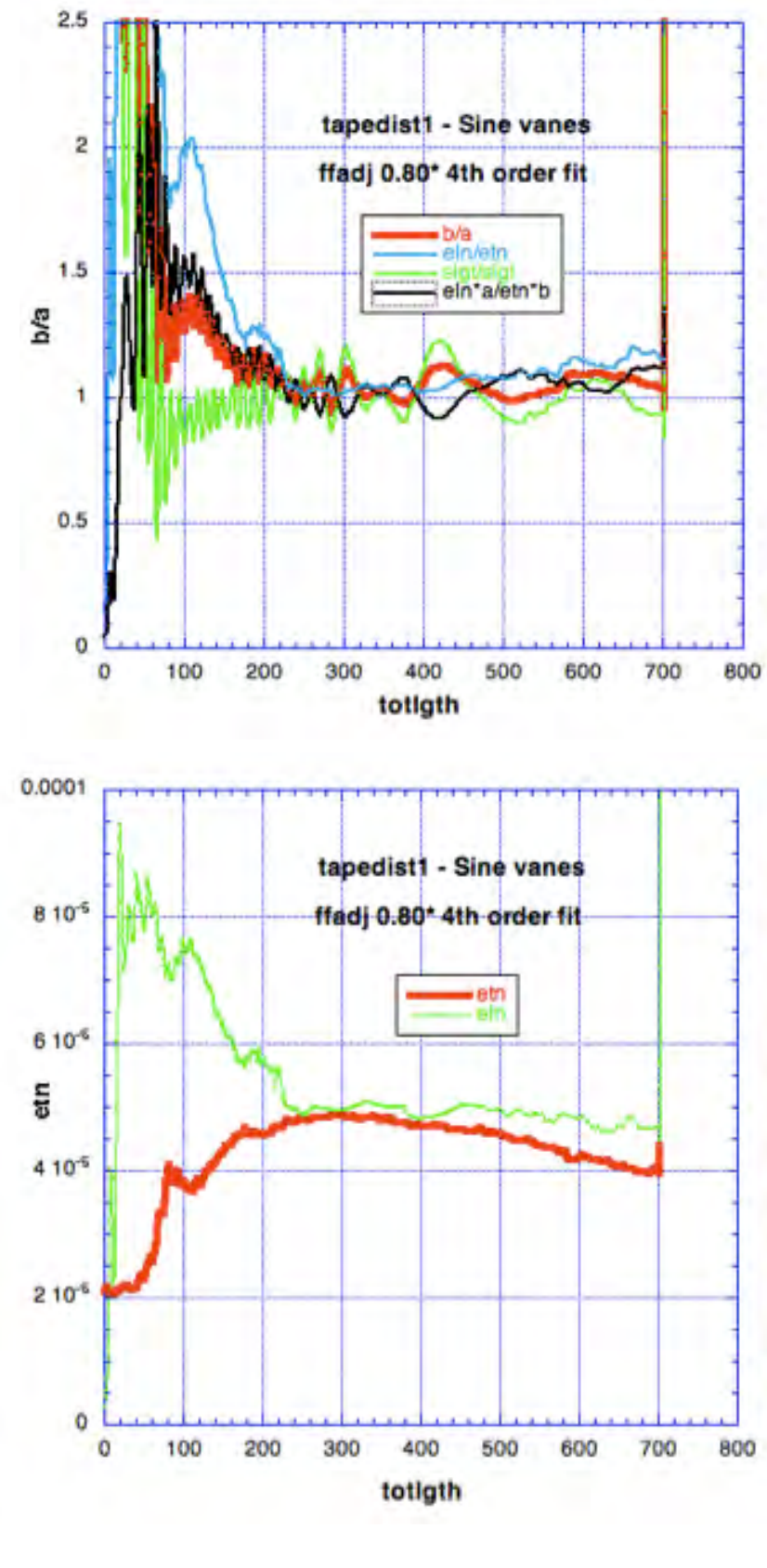

Fig. 28.4. **Sine vanes.** ffadj = 0.80*(4th order polynomial fit) $\sim(b/a)_{experimental)})/(b/a)_{design} = (b/a)_{experimental)})/1.25 =\sim 1$ -> spherical bunch.

The shaper end is cell 73, then there is an equipartitioning transient ((ffadjshpr=1.0; should also be adjusted but here was not.), the EP condition and EP ratios reach 1 at ~cell 200 and stay there to the end cell 391. Total length including end gap and wall 702 cm. At the end of the transition cell, ngood is 89.2%, nalyz 89.2%. eln/etn ~4.7/4.0 = 1.175. The length and accelerated beam fraction are subject to further design variations, with the form factor adjustment performed. Tedious at present, but could be programmed using two runs.

**A.2. ffadjshpr = 1/1.537 = 0.6506; this is 0 intercept of linear fit (at EOS). Move whole ffadj down also by this amount, use 4th order fit: ffadj = 0.6506**d0*(1.8103d0 - 0.0078121d0*(zmain-zEOS) + 4.98628d-5*(zmain-zEOS)**2 - 1.2849d-7*(zmain-zEOS)**3 + 1.1452d-10*(zmain-zEOS)**4 )

Then the resulting ffadj is ~$(b/a)_{experimental)})$/1.25 =~ 1/1.25 - the desired design EP ratio.

Also coincidentally in this case, 1./(mean of $(b/a)_{experimental)}$) = 1./1.4826 = 0.6745, almost same as 0.6506

eln/etn ratio ~1.5, eln 6, etn 4. Ngood 97.3%, nalyz 96.8% at trance.

***NOTABLE THAT FFADJ FOR ~ 1/1.25 GIVES LENGTH ~SAME AS BASE DESIGN.***

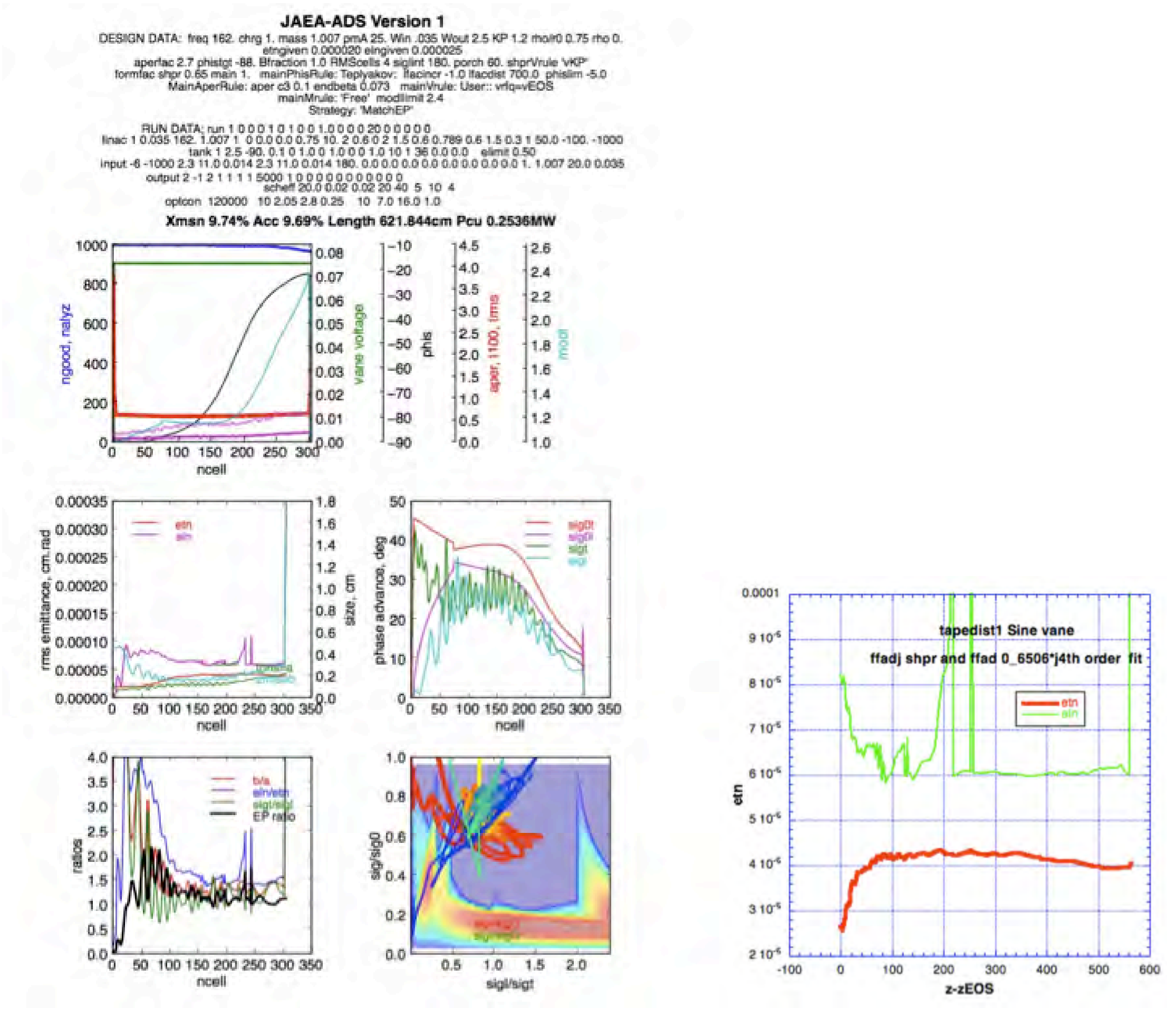

Fig. 28.5. **Sine vanes.** ffadj = 0.6506*(4th order polynomial fit) ~$(b/a)_{experimental)})/(b/a)_{design}$ = $(b/a)_{experimental)})$/1.25 =~ 1 -> adjusted ~ design bunch. Length 621 cm – same as base design…

## B. 2term Vanes Case continued

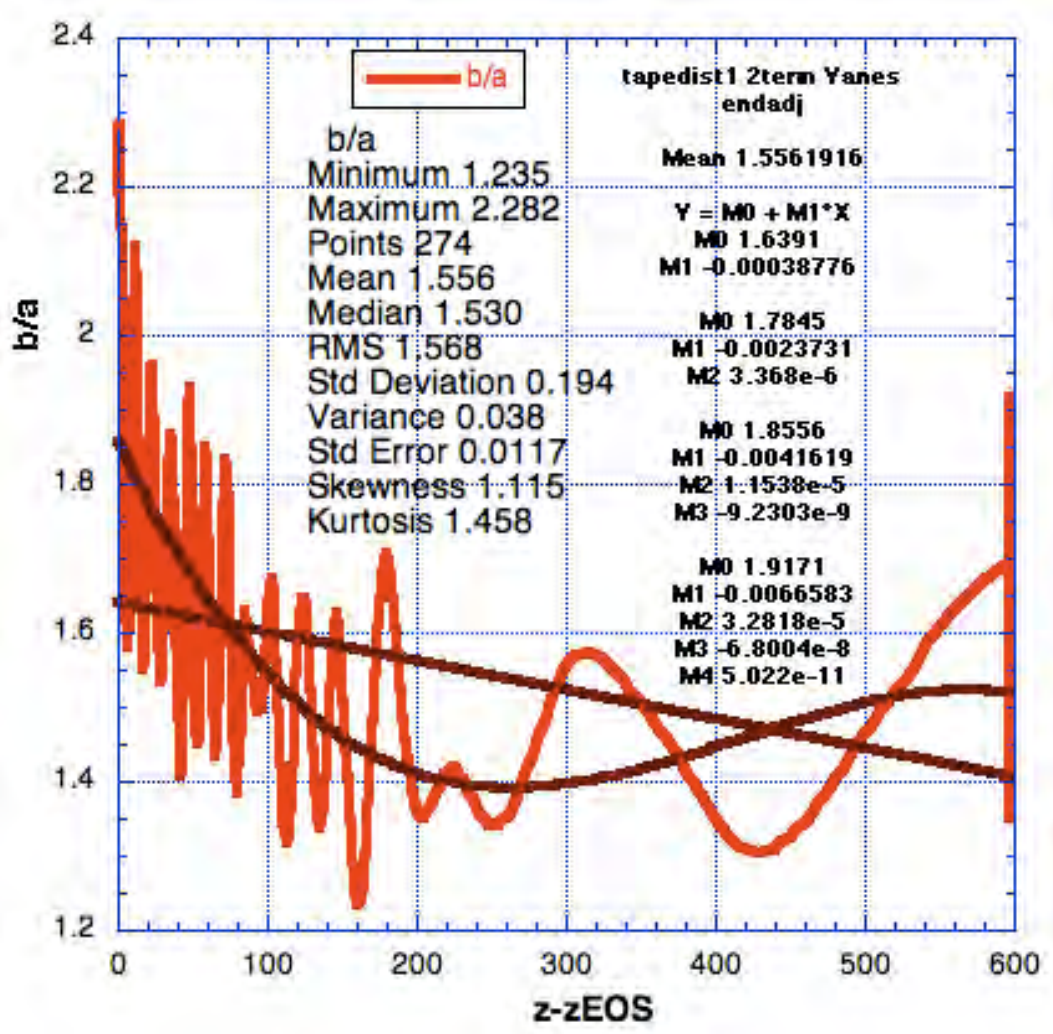

Fig. 28.6. $(b/a)_{experimental}$ from the 2term vanes base design simulation. Statistics, and polynomial fits from 0-4th order on distance along the RFQ from the end of the shaper. Linear and 3rd order polynomial are plotted.

$(b/a)_{experimental)}$) indicates an adjustment in the shaper, based on the linear fit, of ffadjshpr = 1.25/(1.65 to ~1.8); the runs in Table use ffadjshpr = 0.7.

**B.1. The multiplier 0.8 moves the ratio of averages 1.245 down to ~1 = 1.0, based on assuming that $(b/a)_{experimental)}$)/(b/a) =~ 1. MAKES A SPHERICAL BUNCH :**

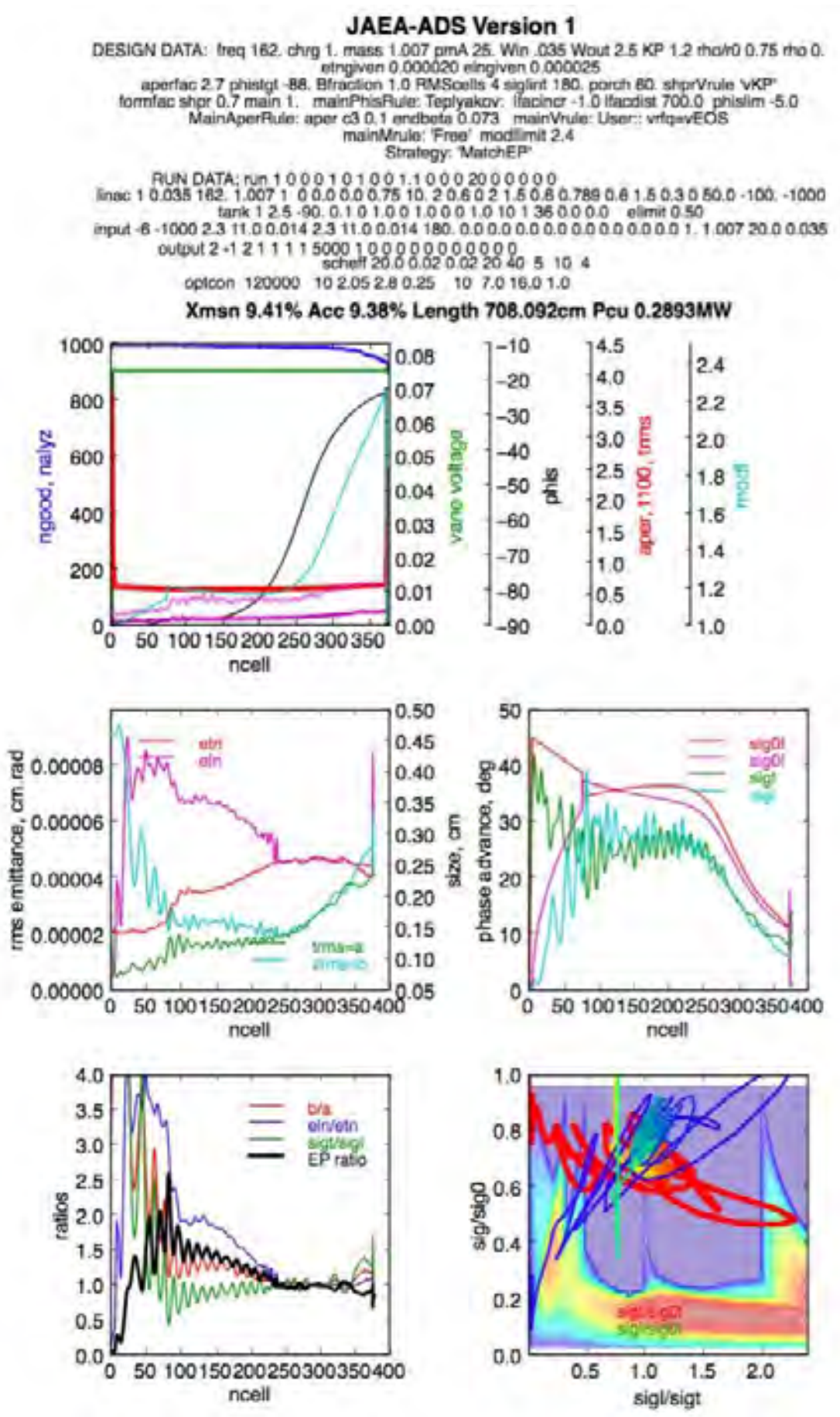

Fig. 28.7. 2term Vanes. ffadj to make spherical bunch. Length 708 cm.

The jump at cell ~250, where radial losses start, is a real change in solution, also a tracking shift at one cell and then back again at next cell – see Figs. 1&2 where it is not as abrupt ….

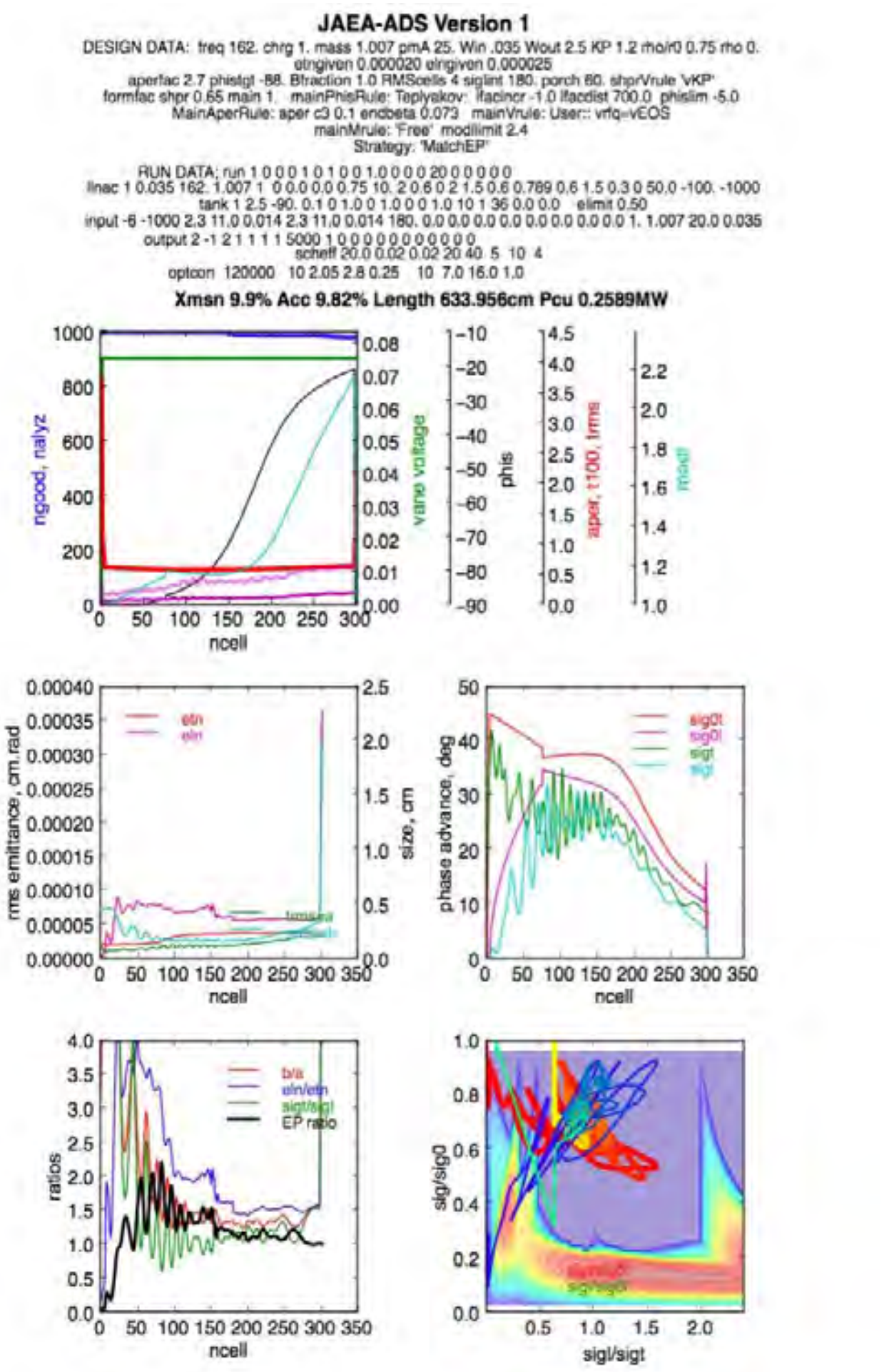

**B.2. The multiplier 0.64 = 0.8*0.8** moves (b/a)experimental))/(b/a) =~ 1/1.25.

Fig. 28.8. 2term Vanes. ffadj to make design condition. Length 634 cm.

Table 1. Some (b/a)experimental fits. Length is total RFQ length, including end cells. There is emittance growth. End vanes emittance data is used, because rapid debunching occurs in tank end gap and wall. The multiplier 0.6426 was used to move the average 1.5562 down to 1.0, to give (b/a)experimental))/(1.25) =~ 1/1.25. The result is similar to Fig. 2, and no improvement over the base design.

| <u>**(b/a)exp**</u> | **Length, cm** | **ngood, %** | **nalyz, %** | **eln/etn at end vanes** | **10^-5 eln** | **10^-5 etn** |
|---|---|---|---|---|---|---|
| Base design | 659.0 | 98.5 | 98.3 | 2 | 6.5 | 3.5 |
| 1st order 0_6426*b/a fit | 654.4 | 98.0 | 97.6 | 1.8 | 6.5 | 3.5 |
| 2nd order 0_6426*b/a fit | 661.2 | 98.7 | 98.3 | 1.6 | 6.0 | 3.8 |
| 3rd order 0_6426*b/a fit | 645.5 | 97.9 | 97.4 | 1.6 | 6.2 | 3.8 |
| 4th order 0_6426*b/a fit | | 98.7 | 98.3 | 1.5 | 5.7 | 3.8 |
| | | | | | | |
| 4th order 0_8*b/a fit | 708.1 | 94.1 | 93.8 | 1.0 | 4.5 | 4.5 |

### *28.4.2 Other possibilities*

Beyond the refined concept of varying parameters in an equation, it is apparent that any applied rule is "*a priori*" – with close correlation of design, simulation and construction. There are certainly more possibilities – possibly using more sophisticated overall optimization techniques (including various forms of AI), possibly using techniques from other fields ([254,255], which also include the element of outside-in design). The main point is that anything is fair game in the design, or design plus simulation, as long as the resulting simulation shows an improvement and that the improvement is robust, not just for a single example, but over a broader design space.

So the **elements** guidance is again – be out of the box!

*Review and additional points:*

• The rms envelope equations can be *a priori* adjusted.

- Of course, the external field quantities in the zero-current phase advances are manipulated in the outside->in design. One of (vane voltage, aperture, synchronous phase or modulation) is used to solve the EP equation if used; the default is modulation, with reasons for this choice given. (Other combinations of fixed and free variables could be chosen – such have not been explored.)

- Sacherer suggested the emittances, which is how the EP ratio (eln/etn) has then also been adjusted.

- Here the form factor has been usefully manipulated. The rms envelope equations are usually expanded in terms of a & b only, but the isolation of the form factor in terms of the a/b ratio is especially useful. It shows how the emittance ellipse form can easily be manipulated, independently from the ellipse area (a*b).

- The remaining a & b in the equations could also be expressed in terms of *a priori* rules, as long as the complete set is consistent.

• The procedure can be iterated, and this can give further improvements.

• A boundary value correction 256 is made for the starting value of the synchronous phase in the main part of the RFQ at the end of the shaper (EOS), where the sunchronous phase goal, phiEOS, is specified. But the synchronous phase in the main part is governed by a rule, usually the Teplyakov condition controlling the charge density, which depends on the overall design definition. A boundary value correction is therefore made for the starting value for the synchronous phase in the main part of the RFQ.

• Beam redistribution plays a large role in an RFQ, influencing both the rms and total emittances and beam sizes, and the ratios of total/rms. Thus the remaining influence of the beam distribution, beyond the equivalent rms, as shown by Sacherer, is present. *LINACS* computes these quantities257, and the ratios can also be *a priori* adjusted. Further research is needed here.

So far, the valuable influence of *a priori* adjustment of the emittances and the form factor has been demonstrated and carried through to constructed RFQs. The other possibilities are wide open. They might afford additional improvement, although probably smaller.

- - - - - - - - - - - - - - - - - - - - - - - - - - - - -

(Note. The run ending sensing has been enhanced to be more robust.

---

254 Method for computationally efficient design of dielectric laser accelerator structures, T. Hughes, G. Veronis, K.P. Wooton, R.J England, Shanhui F., Vol. 25, No. 13 | 26 Jun 2017 | OPTICS EXPRESS 15414

255 Microchip accelerators, R.J. England, P. Hommelhoff, and R.L. Byer, Physics Today 74, 8, 42 (2021); doi: 10.1063/PT.3.4815

256 !! WARNING !! - special dsqrt(bhatEratio) total/rms ratio used in rfqrules.f90 **TO CORRECT PHIS AT EOS** 5.00

257!! NOTE !! rfqdesignGUI Eratios for possible re-distribution: xvseEtotalorms, xlongEtotalorms 5.00 5.00

The vanes should end with a transition cell, and then can have an exitm0 zero modulation section, and an exit flair-off. These latter two and the exit gap and end wall are in cm, and adjusted to also correspond to a "cell".

The default setting for a Poisson cell block is 15 cells, of which the middle 13 are used. The ending is more robust if the transition + following cell + end gap and end wall are positioned near the beginning of a full Poisson cell block, and this is now programmed that way.

Ending is mostly robust for sine or trapezoidal vanes, but faulty ending or fault is still sometimes observed for 2term vanes, where the connection between design and simulation is weakened and the beam centroid may track the design cells less exactly… )

[eltoc]

# Chapter 29 – Discovery of a method to control a longitudinal-emittance-dominated beam, leading to a better shaper design – the "truncated vane shaper section" [258]

A modification to the longitudinal vane modulation profile in an RFQ has been discovered to appreciably improve bunching and longitudinal emittance control and reduced RFQ longitudinal output emittance. [259] This procedure occurs in the individual cells, is not dependent on the overall design, and therefore is general, affording an extra parameter for beam manipulation. It can be applied in addition to the usual goals of vane modulation variation, e.g., to achieve higher acceleration efficiency. ***It is applicable to any RFQ***, with implications for any linac. Examples of the cumulative effects on overall design are provided, to indicate further exploration avenues for the designer.

*Highlights:*

- A longitudinal vane modulation profile that improves bunching and longitudinal emittance control.
- Vane modulation manipulation to eliminate deleterious accelerating field inside the focusing period.
- Analysis in the full time-domain.

It is a very good concluding example and summary of how the **elements** come together, and can be used more generally – an intriguing future research subject.

The Radio Frequency Quadrupole (RFQ) plays a critical role in the beam quality of linear accelerators and precise control of beam properties is required under a variety of specifications. Achieving a small longitudinal rms output emittance has been difficult, as the initial bunching of the injected dc beam (initially completely emittance dominated) tends to fill the evolving bucket; the effective longitudinal rms emittance becomes large and cannot be reduced without an irreversible technique with beam loss.

Working with a specification calling for minimization of RFQ rms longitudinal output emittance, to achieve equipartitioning with ratios close to 1, the usual vane modulation procedures were unable to meet the spec; a successful new approach is presented.

---

[258] Ch. 23 "RFQ Design & Simulation With Arbitrary Vane Tip Shapes" covers more traditional concepts of controlling the longitudinal field in detail within the focusing period, such as trapezoidal or 2term modulation.

259 "Improved bunching and longitudinal emittance control in an RFQ", R.A. Jameson, Bruce Yee-Rendon., arXiv 2203.04632, 3/10/2022 – in revision.

The specification, and the preliminary design example presented here, is for an updated RFQ design being developed by JAEA for an accelerator-driven subcritical system (JAEA-ADS) [260] with a 30-MW superconducting linac, that will be injected by a constant vane voltage 0.035-2.5 MeV 20mA RFQ. The RFQ must have very high availability, so the specified KP limit = 1.2. For optimum injection into the SC section starting at 2.5 MeV, the RFQ longitudinal rms output emittance specification is 0.25 mm.mrad, with transverse rms input emittance specified = 0.2 mm.mrad, Fig. 29.1.

An EP design is adopted because there is beam-based control over the emittances ensures low emittances, tight beams, and low loss. The design can access all "inside-out" measures to meet the spec requirements without any disadvantage because of EP.

The robust standard, default, shaper design uses simple piece-wise linear rules for modulation and synchronous phase (phis) to bring the injected beam into the equipartitioned (EP) condition as soon as possible, at (phis) = ~88°, very near the pure bunching phis = 90°. This provides a smooth transition into the following acceleration section of the RFQ, avoiding problems of other shaper designs. [261].

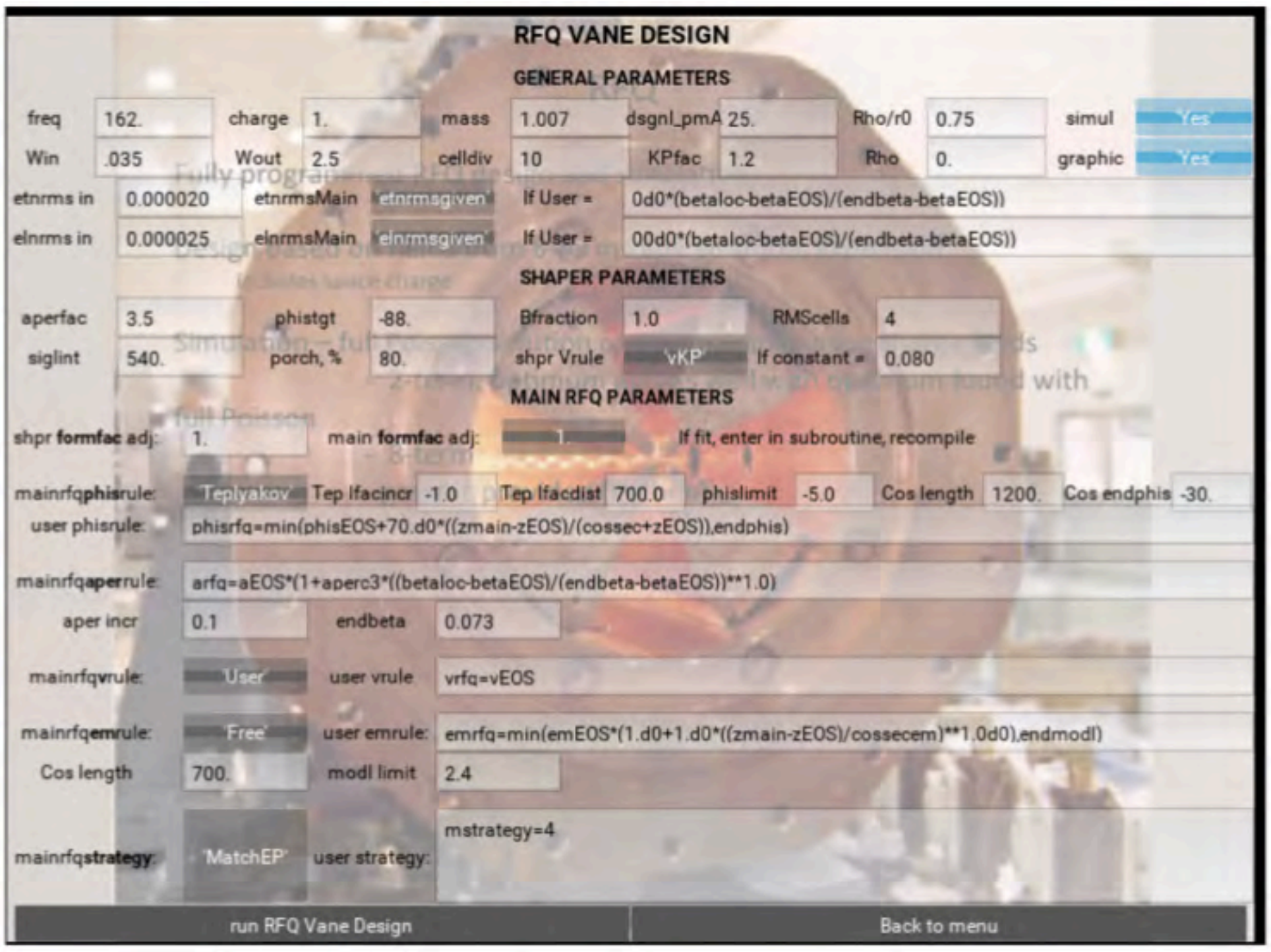

Fig. 29.1 The case run, here with aperfac=3.5. (See also Ch.23.7.2) The shaper is unusually long.

## 29.1 Discovery case

The design settings for the discovery case and the primary example presented here are shown in Fig.29.1.

The longitudinal vane modulation in the example here is 2-term, but the choice is arbitrary. After the shaper, the aperture is allowed to increase slightly; the inner profile (out to radius 0.6cm) is shown in Fig.29.2-green. Initial designs use a very long, gradual shaper section (ending at z = ~129cm (cell 161)), which helped reduce the output longitudinal rms emittance, but could not meet the spec, as

260 "Design and beam dynamic studies of a 30-MW superconducting linac for an accelerator-driven subcritical system" Bruce Yee-Rendon, et.al. PRAB 24, 120101 (2021)

261 See Ch. 4 introduction re original Los Alamos shaper, which had undesirable properties.

shown in Fig. 29.3-left. The output longitudinal rms emittance and EP condition requirements are too severe and not achievable using the standard shaper[262].

A fortuitous *a priori* NPSOL boundary setting error, during the 2021 JAEA-ADS work, serendipitously showed how to make a better RFQ shaper and initial main RFQ section to control the longitudinal emittance and EP conditions, as desired by the JAEA-ADS spec. And, after using the high priority **elements** of getting out of the "rms/smooth approximation/steady-state" boxes and into the time-domain box, it is shown how a "truncated vane shaper section" works - and that it is a major new insight and can be used generally for any RFQ as an advance of the state-of-the-art .

Previous JAEA-ADS study runs with aperfac = ~2.7 (a minimum aperture = 0.585 cm) gave a correct detailed vane-tip profile from the Poisson mesh with lower bound setting = 0.5 cm for the NPSOL solver [263]. But then a check to reproduce the results and vane profile for the 2020 constant-r0 RFQ would not reproduce. The cause was traced to the lower bound setting given for the NPSOL solution, with severe truncation of the result – after lowering the lower bound, the reproduction was correct.

Subsequent JAEA-ADS design optimization studies had resulted in a larger aperfac = ~3.5 (smaller bore), which has a minimum inner radii = 0.46, and the lower bound should have been adjusted. But with the lower bound still = 0.5, the vane profile was being truncated at rh = rv = 0.5 cm inner radius, as shown in Fig. 29.2.

However, this truncation did not result in a disaster, but allowed the runs to finish, and the results of the optimization work seemed reasonable. In fact, the results seemed advantageous. So this case, with design settings as in Fig. 29.1, is now presented and discussed.

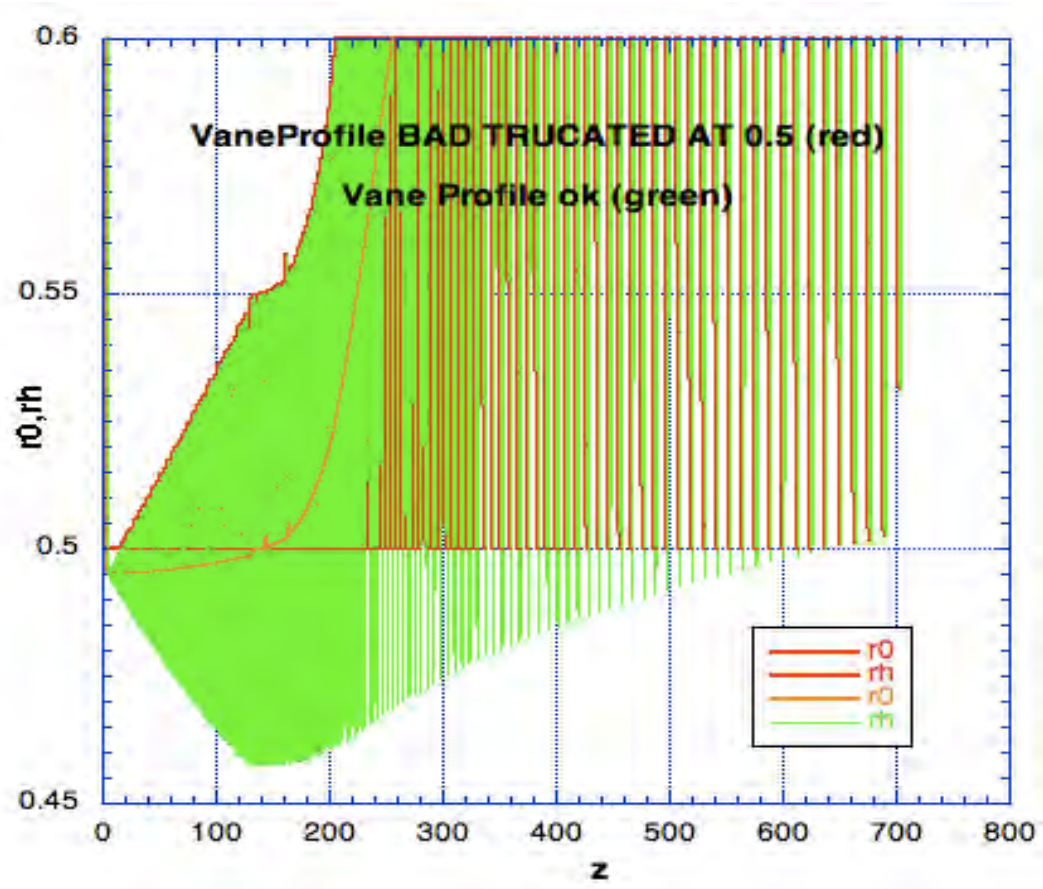

Fig. 29.2 The 2-term vane profiles. Green - Standard design. The minimum radius = 0.46 cm at the end of the standard shaper, then increases. Red – vane radii below 0.5 cm cut off (truncated).

The overall graphics[264] are shown in Fig. 29.3:

[262] The shaper and following acceleration section designs are manipulated separately.

[263] JAEA-ADS is using 2-term vanes; subroutines get2trmvane.f90 bounds, objfn2trmvane.f90 solve for the detailed vane tip profile.

[264] Reference the graphics axis being ncell vs z; remember that shaper is to phis=-88°, very little change in velocity β, so the influence of β in ncell vs. z is small over the shaper.

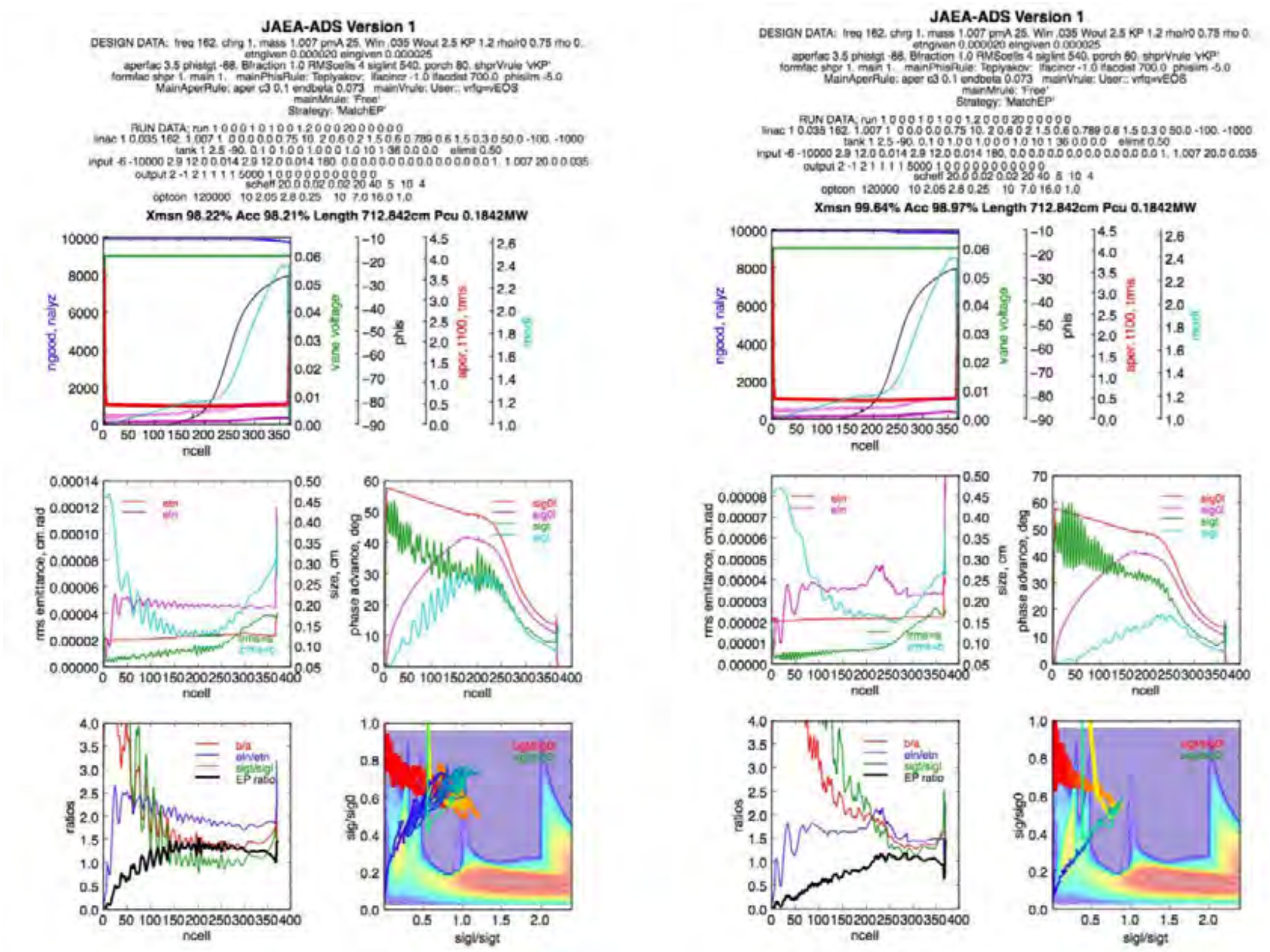

Fig. 29.3 Left – conventional 2-term vane profile, no truncation. Right – 2-term vane profile truncated at 0.5

Instead of using this NPSOL boundary truncation, it is better and easy to specifically apply vane profile rules in the Poisson external and space charge mesh generation process.

The key requirements for this JAEA-ADS RFQ are EP performance in the main part of the RFQ after the shaper, and the lowest possible output longitudinal rms emittance, therefore the design specs of etnrmsgiven = 0.000020cm.rad and elnrmsgiven = 0.000025cm.rad. The longitudinal vane modulation is 2-term[265]; after the shaper, the aperture is allowed to increase slightly; the inner profile is shown in Fig. 29.2 - green.

Initial results use a very long, gradual shaper section which helped reduce the output longitudinal rms emittance, but could not meet the spec, as shown in Fig. 29.3-Left. The output longitudinal rms emittance and EP condition requirements are too severe and not achievable using the standard shaper[266].

With the truncated vanes, Fig. 29.3-right, the longitudinal emittance (eln), rms longitudinal beam size (zrms), and resulting phase advance (sigl = $\sigma^l$) are considerably smaller for the truncated vanes, resulting in considerably better EP and EP ratio performance, Fig. 29.4. The transmission and accelerated beam fractions are also much improved.

---

265 Later the longitudinal vane modulation will probably be modified to sinusoidal or trapezoidal in the high vane modulation region, or a mixture of these, to obtain shorter RFQ length. In any case, the vane modulation form is irrelevant here – the method can work for any RFQ.

266 The EP condition and ratios can be improved in the acceleration section after the shaper.

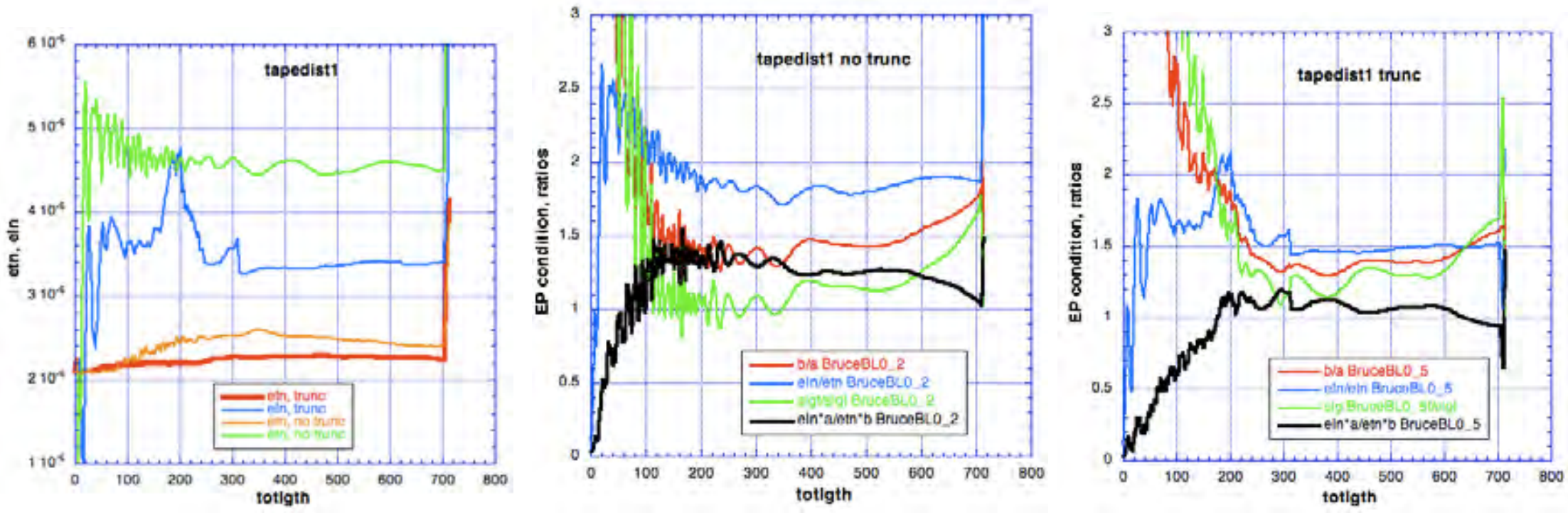

Fig. 29.4 2term vanes. Expanded views of longitudinal emittance, EP condition and EP ratios. (A priori form factor adjustment may give further improvement for both non-truncated and truncated cases).

## 29.2 Test over wider design space

### 29.2.1 Single parameter changes to the JAEA-ADS design, without further optimization.

The single change from 2-term to sinusoidal or trapezoidal longitudinal modulation resulted in the better acceleration rate and shorter length, but with beam losses that would need to be designed away, particularly for trapezoidal vanes. The effect with truncated vanes is clear.

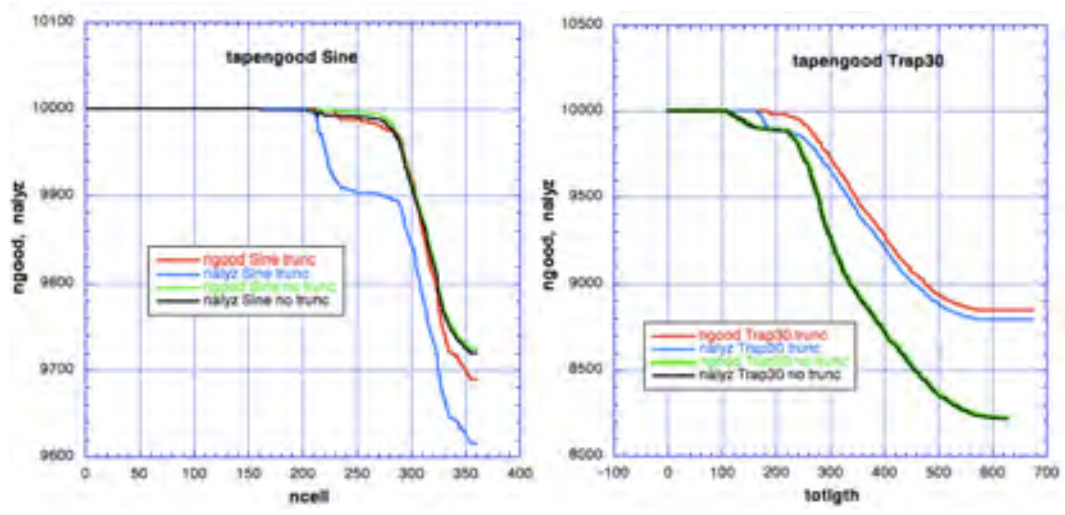

Fig. 29.9 Transmission and accelerated fraction for sinusoidal and trapezoidal vanes.

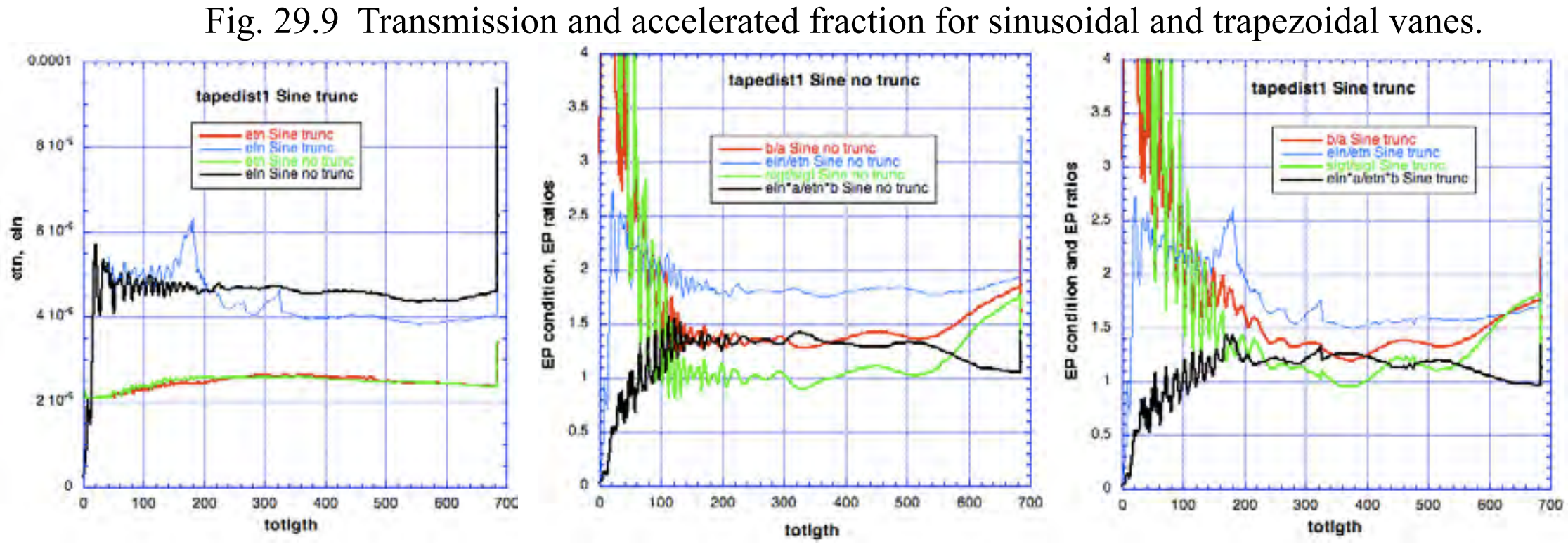

Fig. 29.10 Sine vanes. Expanded views of longitudinal emittance, EP condition and EP ratios. A priori form factor adjustment may give further improvement.

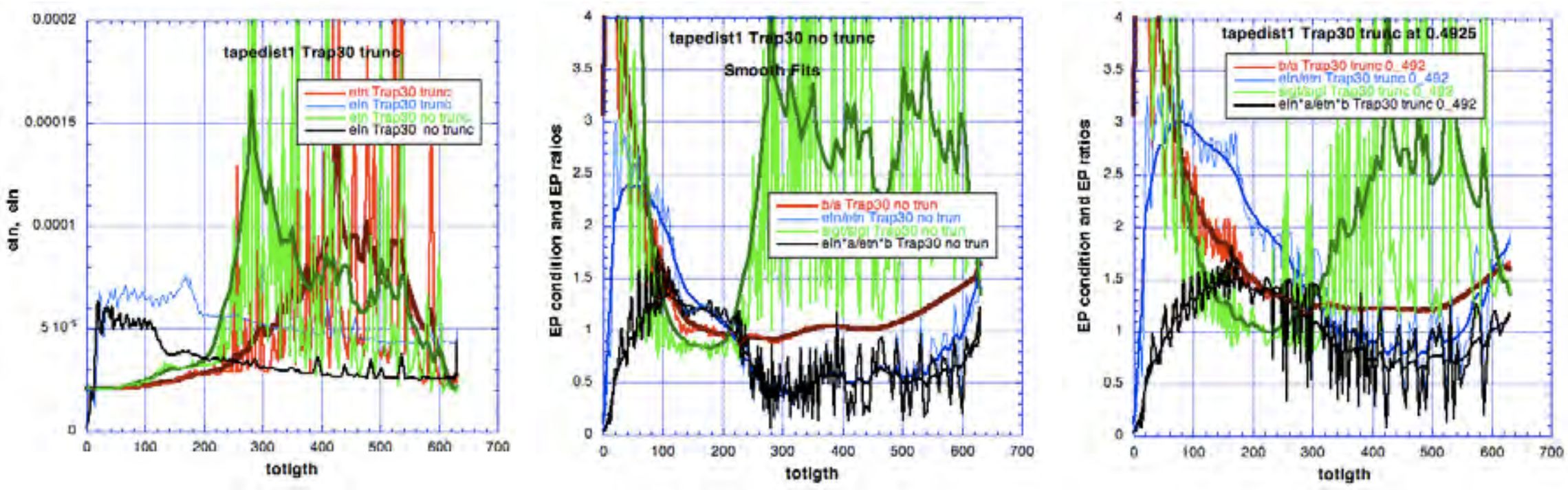

Fig. 29.11 Trapezoidal 30% vanes, longitudinal emittance, EP condition and EP ratios.

Table 29.1  Design characteristics for different vane modulation types

| Vane type | # cell | Length, cm | Pcu, kW | epk |
|---|---|---|---|---|
| 2term | 372 | 712.8 | 181 | 1.196 |
| Sine | 359 | 685.3 | 174 | 1.194 |
| Trap 30% | 323 | 611.8 | 156 | 1.22 |

peak field elk found in design using multipole elk table;
    vKilpatrick=xKPlimit*xKPfac*(r0rfq/100.d0)/epk
 in pass & passmain:
    vKP = vKilpatrick
    wkgfield=vrfq*epk/r0rfq
    wkgKP=vrfq/vKP          -> this is printed in Terminal output paramoutlist as "epk"
peak field on mesh can be found using optional code in poissonEdesign.f90, line ~119

### 29.2.2 Earlier JAEA-ADS preliminary design with short shaper, larger aperture.

2term Vanes with aperfac 2.7, siglint=180°, truncation applied without further optimization.

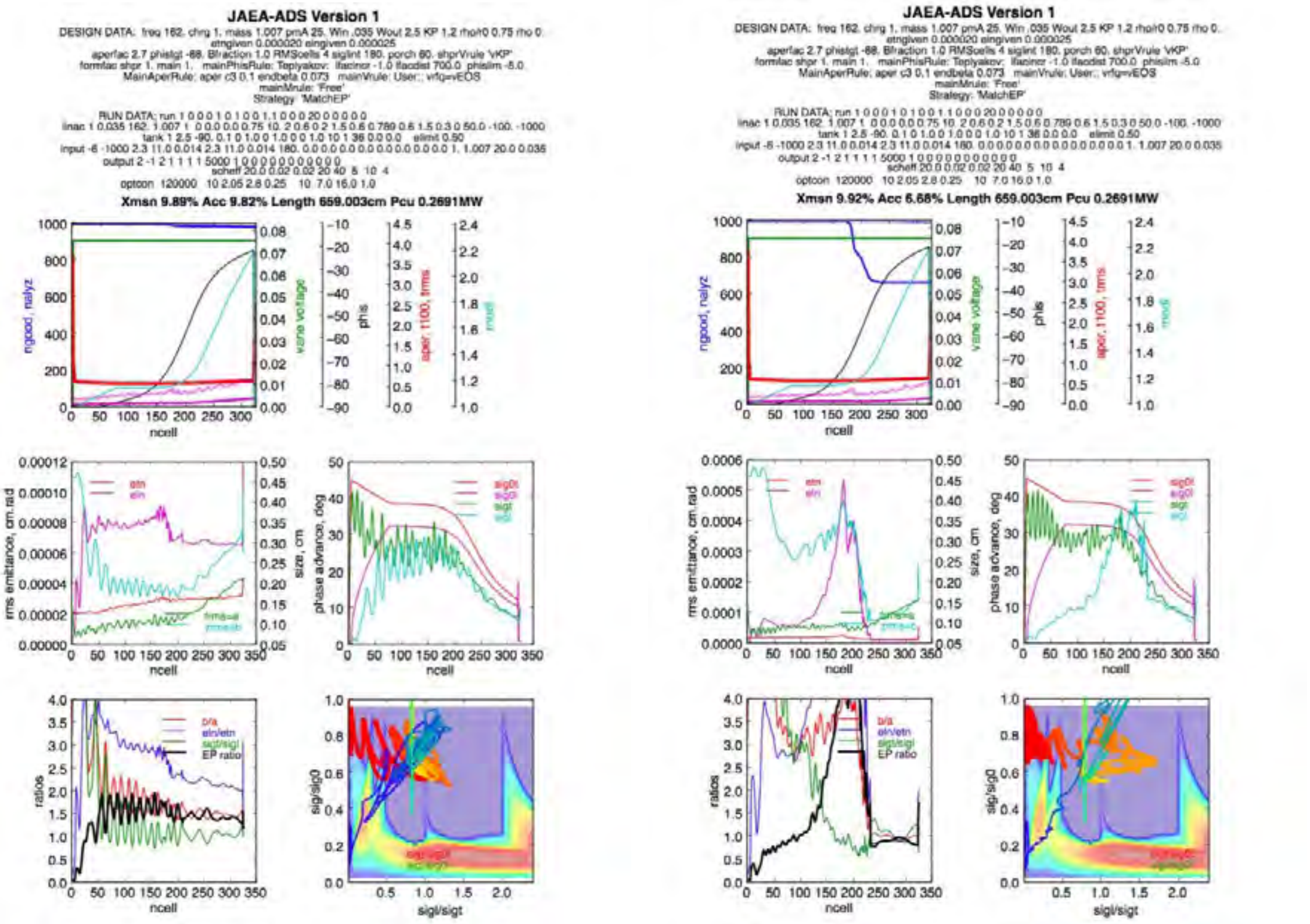

Fig.29.12 Short shaper, larger aperture design case.  Left – no truncation.  Right – with truncation - (1000 particles simulated, so multiply Xmsn and Acc by10.)

### 29.2.3 IFMIF CDR RFQ

Truncation applied without further optimization.

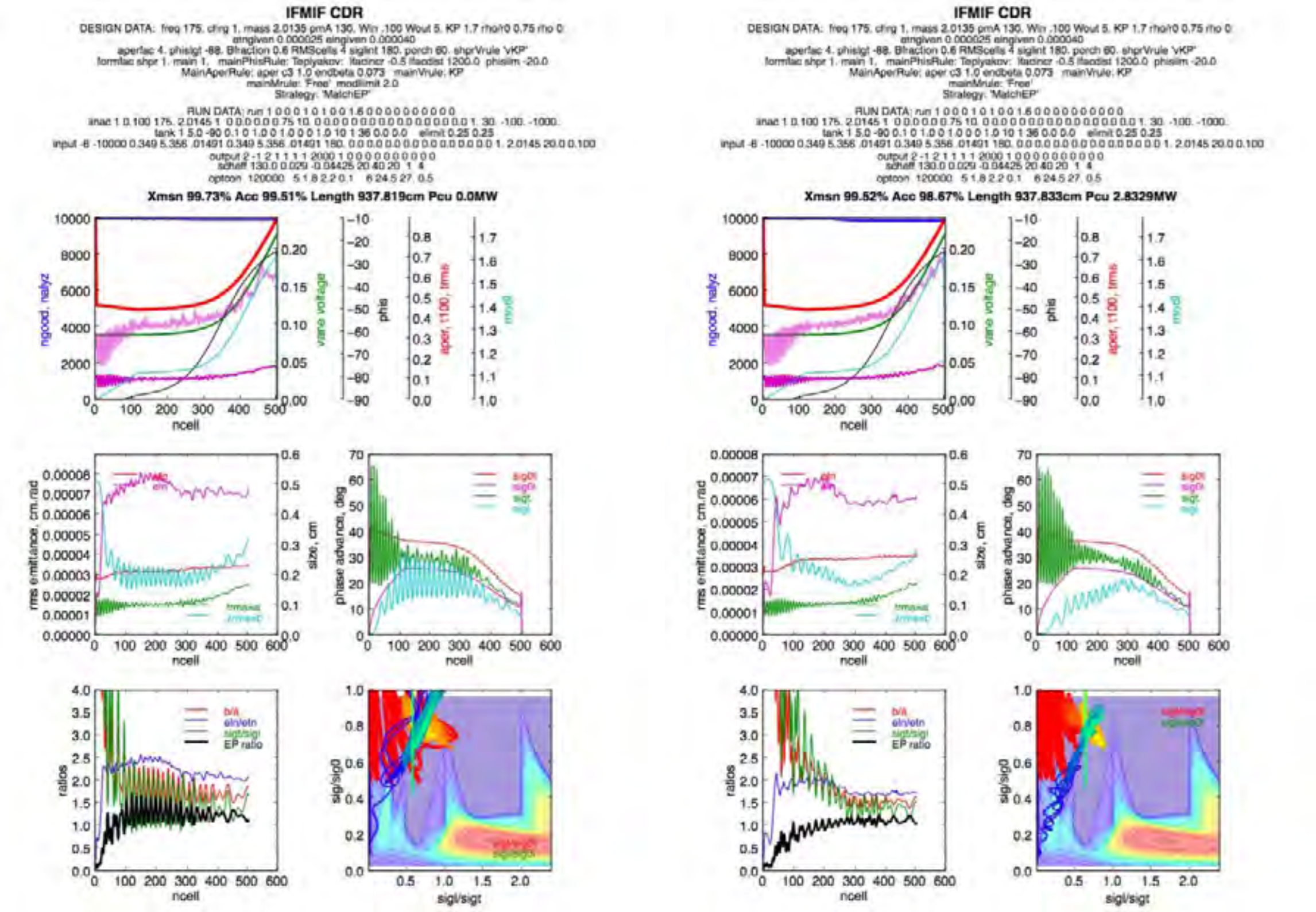

Fig. 29.13 IFMIF CDR RFQ. Left – no truncation. Right – truncation at 0.47cm.

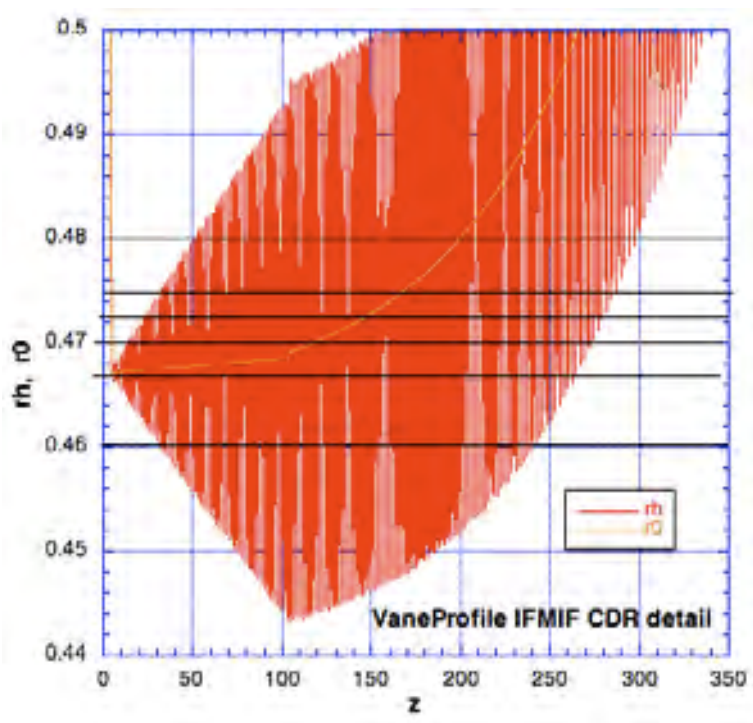

Fig. 29.14 IFMIF CDR RFQ sinusoidal modulation vane profile. The shaper ends at cell 116, z=103. The vane voltage is kept at the Kilpatrick level; the aperture increases rapidly after the shaper. Six truncations were tested; best truncation = 0.47, just above begin-shaper aperture.

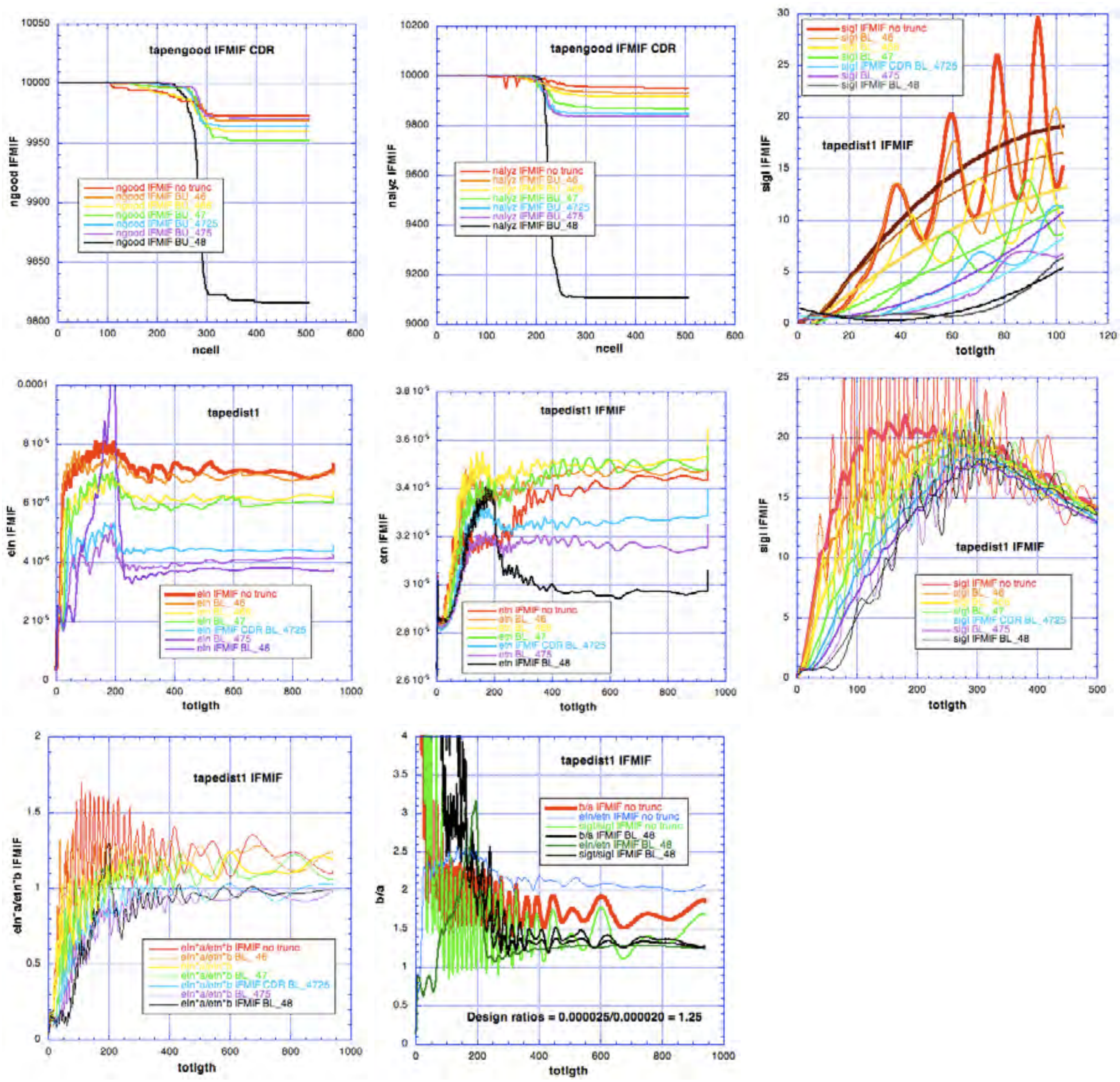

Fig.29.15 Characteristics for no truncation and the six truncations tested. First row, right, shows that truncation at 0.47cm produced linear longitudinal phase advance (sigl).

Fig. 29.16 is another view comparing truncated and non-truncated vanes for the JAEA-ADS EP and IFMIF CDR EP RFQs. Table 29.2 summarizes the transmitted (Xmsn) and accelerated (Accel) particles, and the normalized rms emittances for the two RFQ models with and without truncated vanes.

**Table 29.2** The percentage of transmitted particles (Xmsn), accelerated particles (Accel), output $\varepsilon_{long,\,rms}$, and the output $\varepsilon_{trans,\,rms}$ for the JAEA-ADS EP and IFMIF CDR RFQ models.

| Parameter | JAEA-ADS EP RFQ | | IFMIF CDR | |
|---|---|---|---|---|
| | Trunc | No-Trunc | Trunc | No-Trunc |
| Xmsn (%) | 99.71 | 97.71 | 99.52 | 99.73 |
| Accel (%) | 98.87 | 97.70 | 98.67 | 99.51 |
| $\varepsilon_{long,\,rms}(\pi\, mm\, mrad)$ | 0.32 | 0.47 | 0.61 | 0.7 |
| $\varepsilon_{trans,\,rms}(\pi\, mm\, mrad)$ | 0.22 | 0.24 | 0.35 | 0.35 |

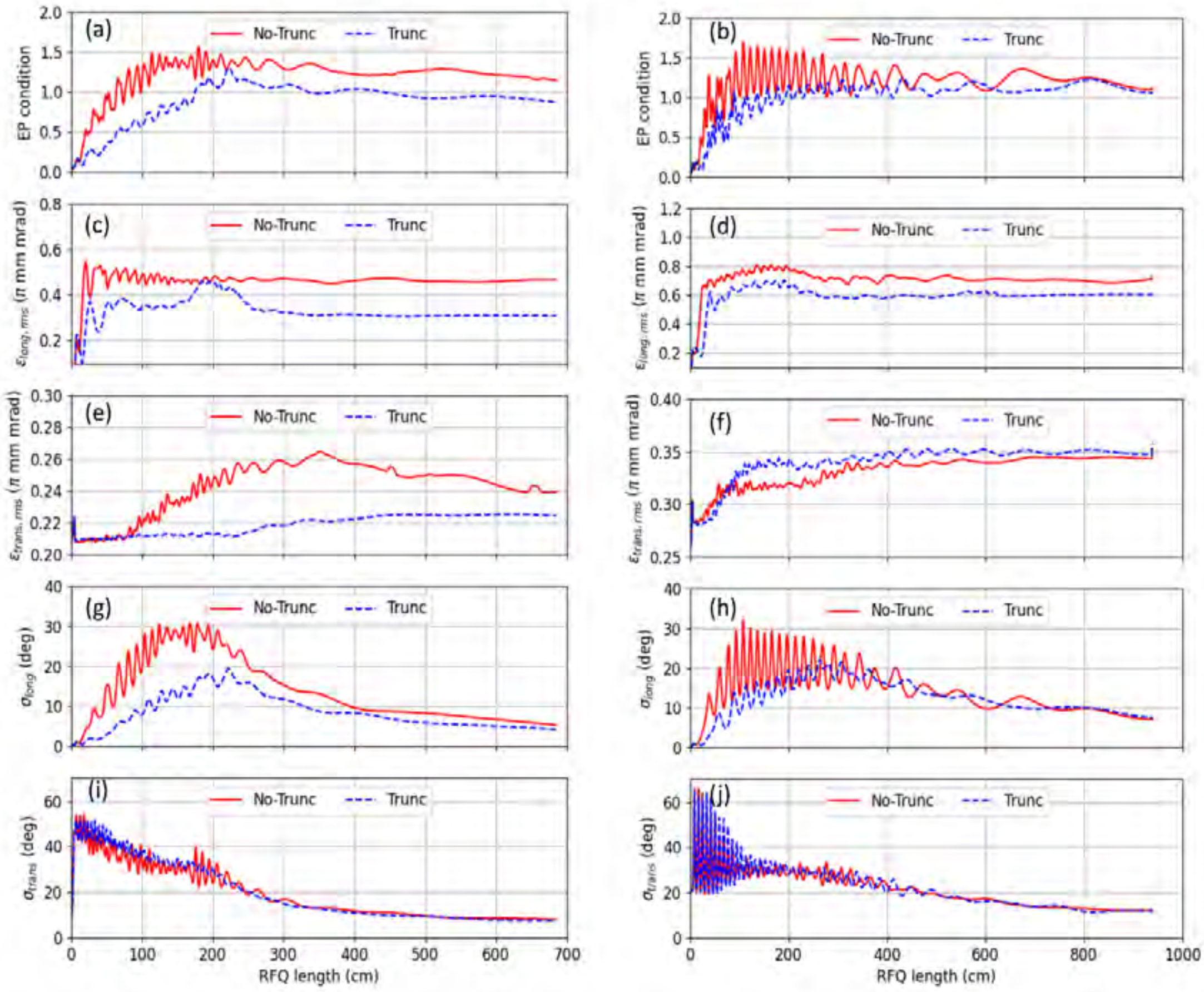

Fig. 29.16  Design comparison between truncated (Trunc) vane and No-truncated (No-Trunc) vanes for the JAEA-ADS EP RFQ (left) and the IFMIF CDR EP RFQ (right). (a) and (b) plot show the EP condition, (c) and (d) are the longitudinal normalized rms emittance, (e) and (f) are the transverse normalized rms emittance, (g) and (h) are the longitudinal phase advance, and (i) and (j) are the transverse phase advance.

## 29.2.4 A Small Neutron Source RFQ

Ref. Ch23.7.  0.035-2.5MeV proton constant-r0, constant vane voltage, 2-term vane modulation RFQ; design etn = 0.000050, eln = 0.000100.  Shaper ends at cell 63, z=42cm.  Truncation applied without further optimization.

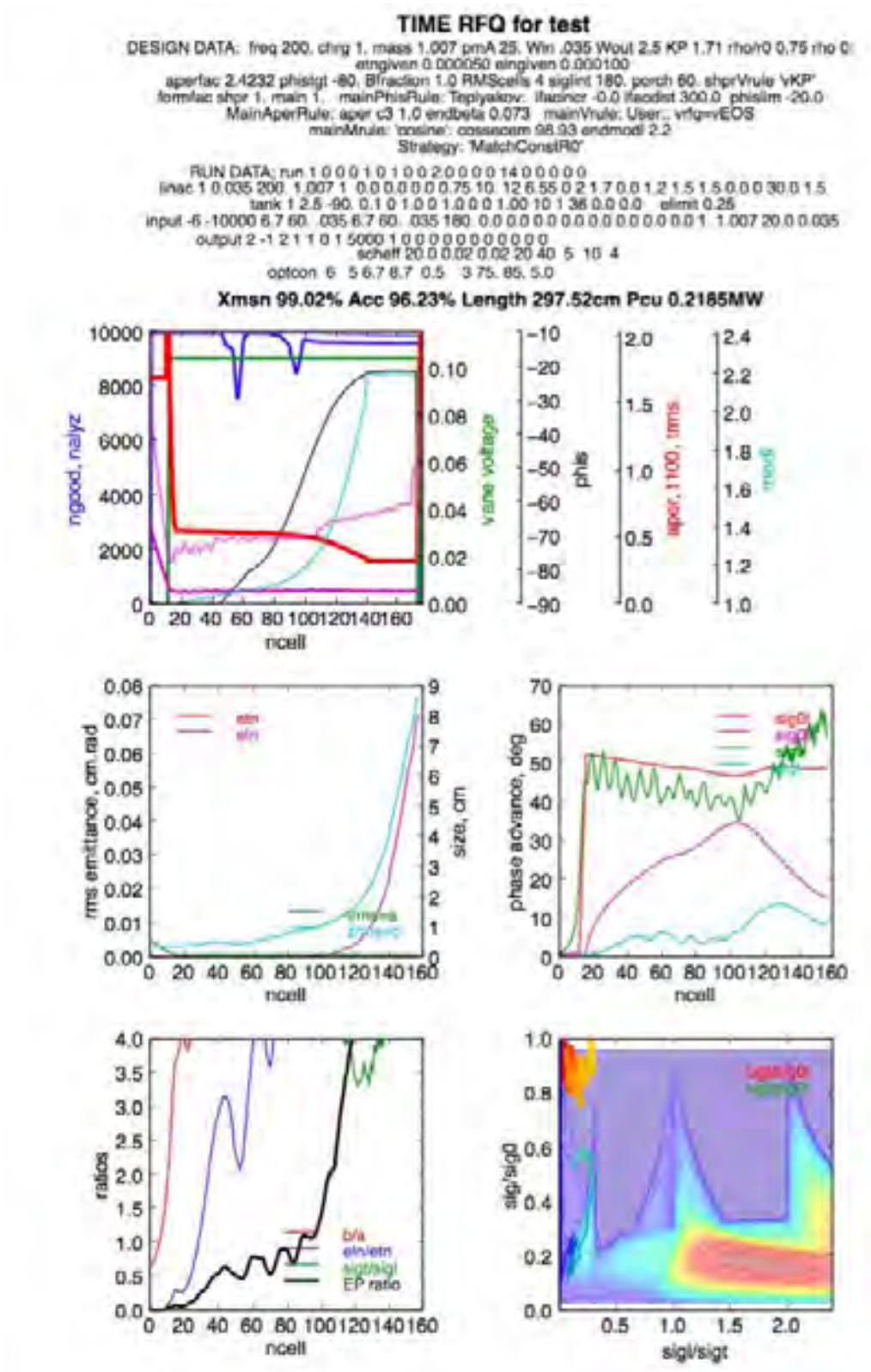

No truncation

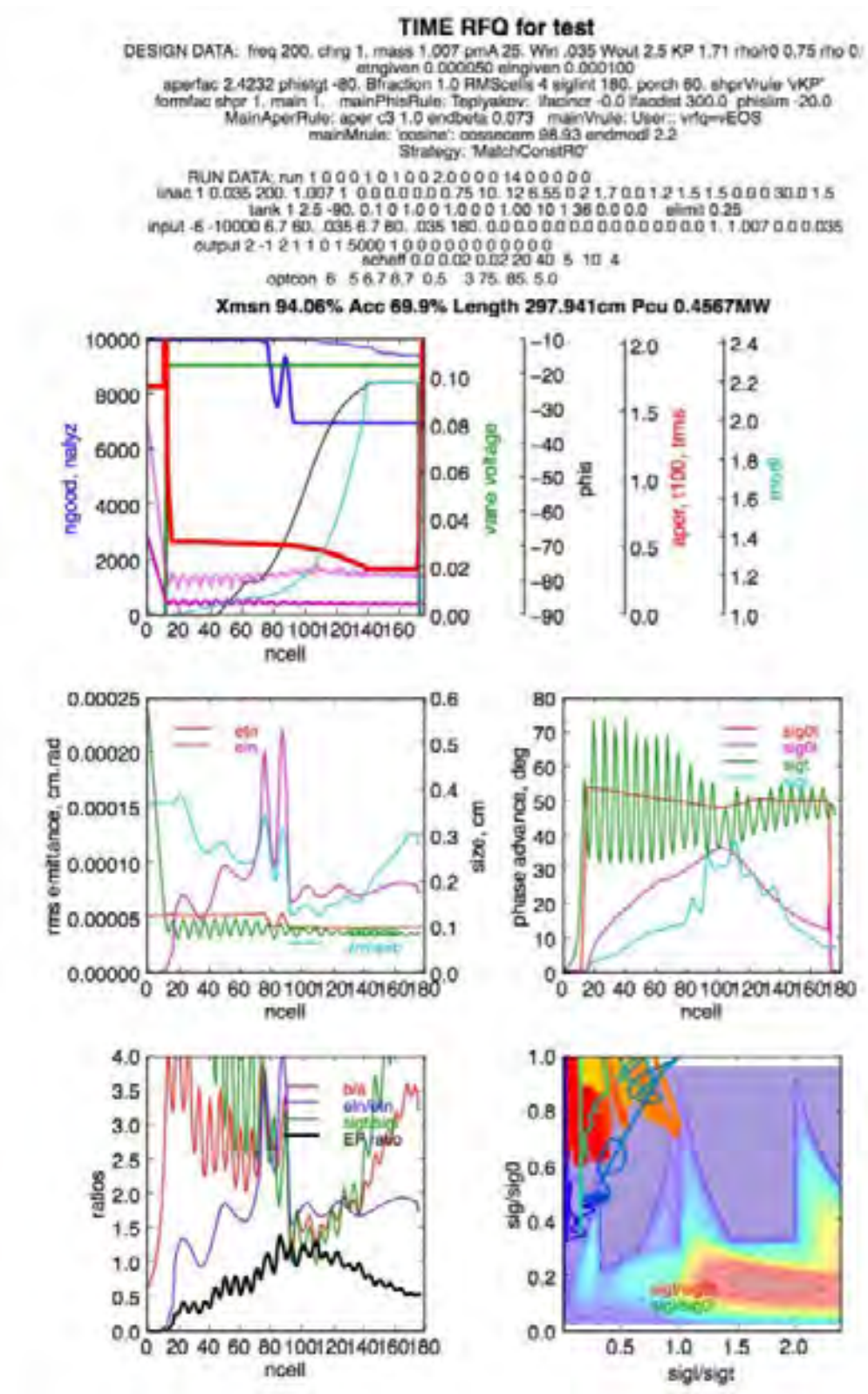

truncation at 0.56 cm

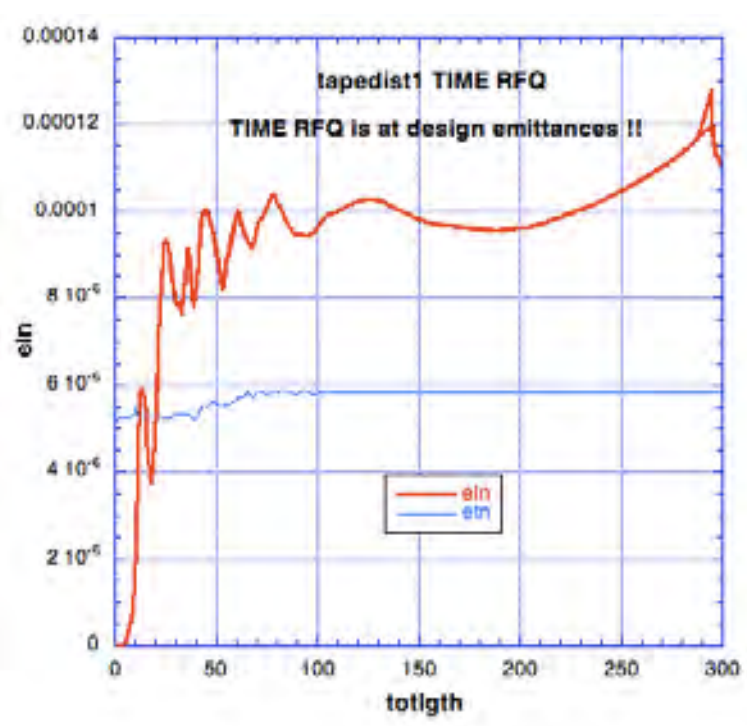

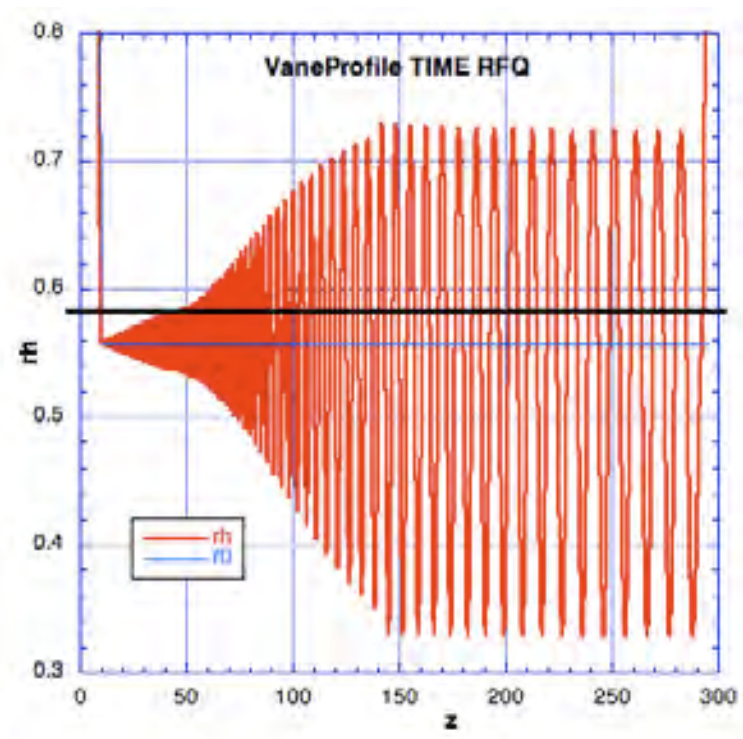

Is at design emittances

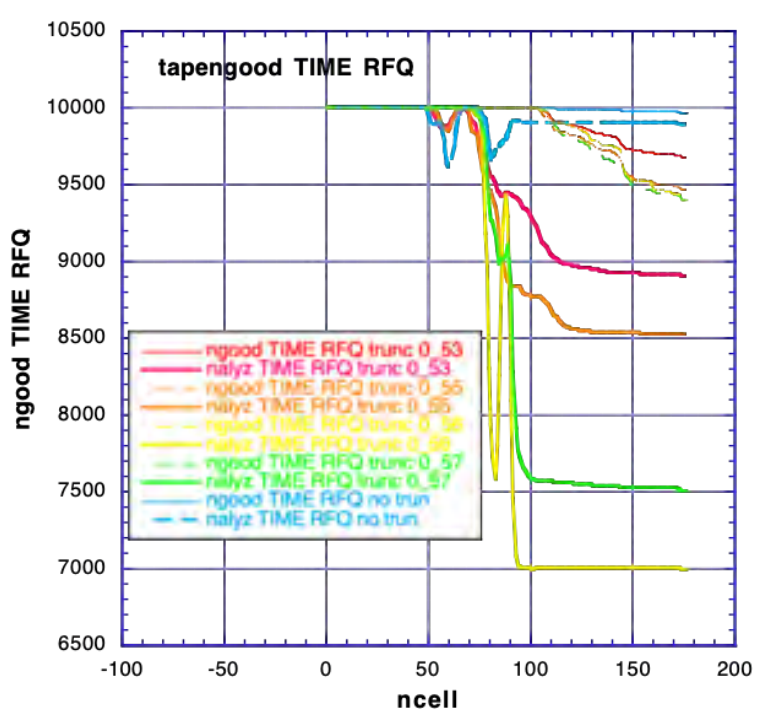

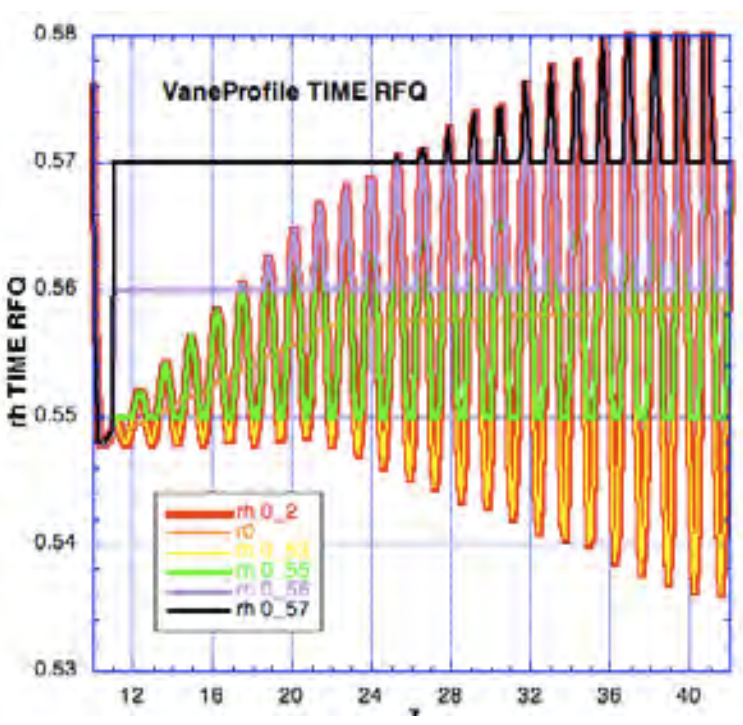

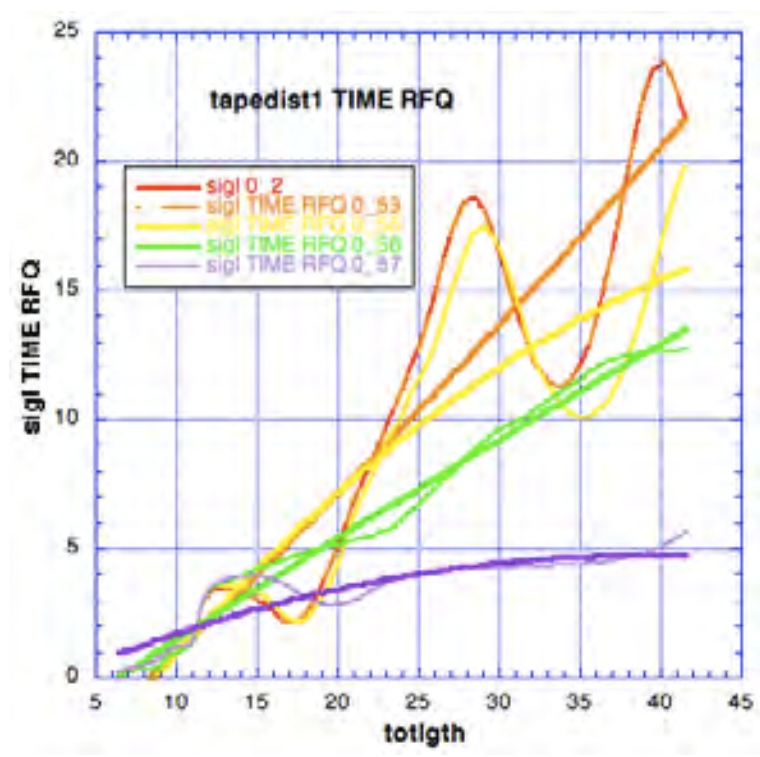

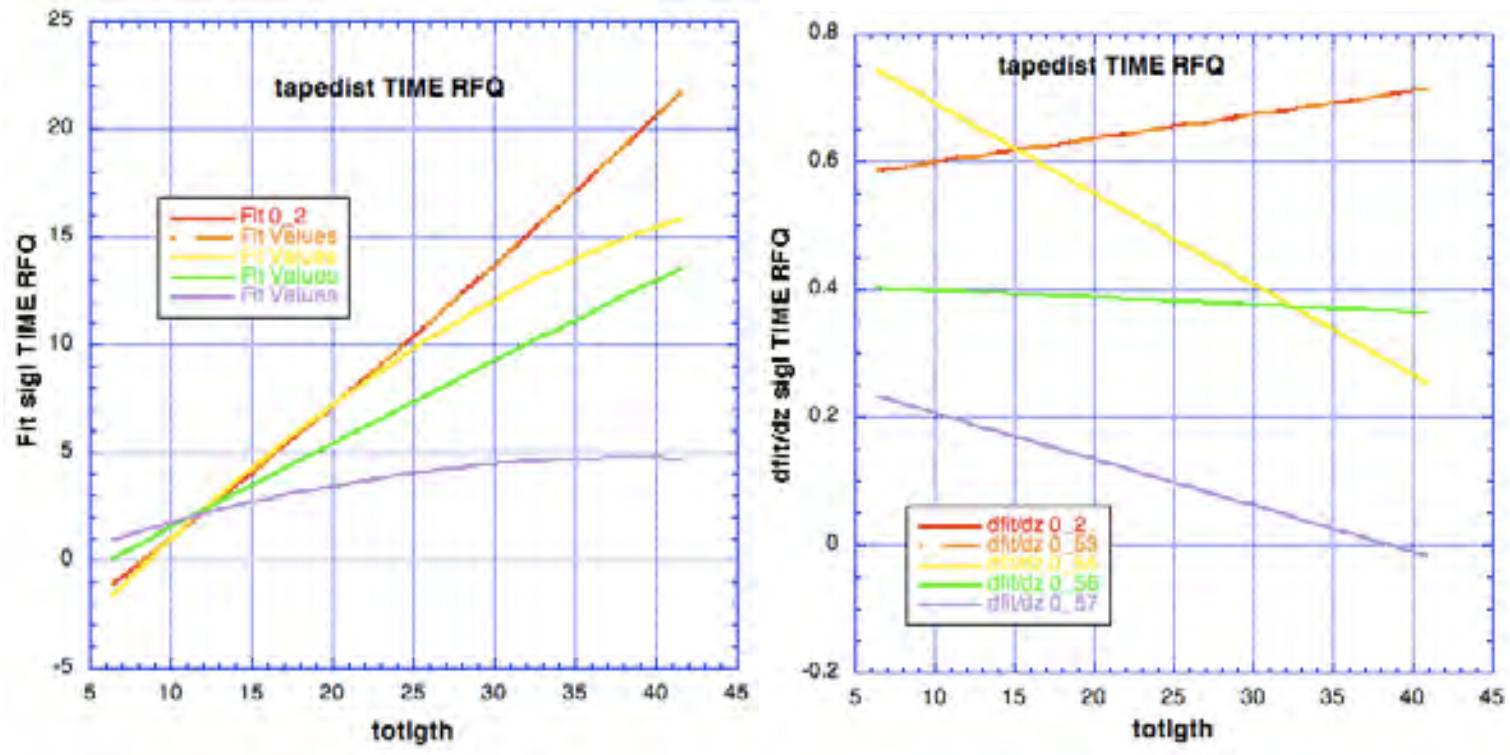

Fig. 29.17 Best truncation = 0.56 – at or somewhat above begin-shaper aperture. Truncation was done in the shaper, then tapered down over 18 cm to z=60cm to the untruncated profile.

## 29.3 How the Truncated Vane Shaper Section Works - viewed from the "rms/smooth-approximation/steady-state" box

Here with reference to the JAEA-ADS discovery case, which is optimized. (For the other cases, the truncation was just tested to compare with the no-truncation design, without further optimization.

An important point is that with truncation, the design modulation depth and synchronous phase are not changed – only the actual vane profile is later truncated.

The design, Fig. 29.18 left is unchanged – the truncation is applied to the vane profile in the Poisson simulation.

The design modulation is preserved at radii beyond the truncation; *a key point is that the depth (~ modulation∗aperture) is preserved.*

*With no truncation – radial losses are trimming transmission and accelerated beam fraction equally. With the truncation, performance is improved.*

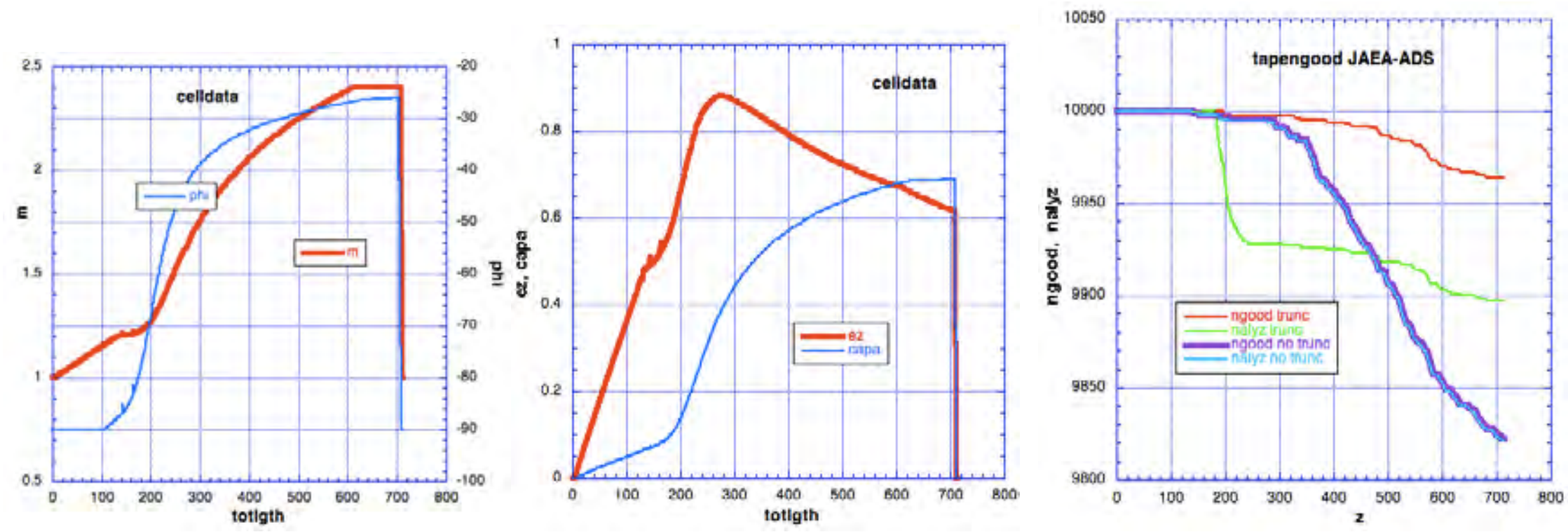

Fig. 29.18 Design modulation, synchronous phase and acceleration rates. For orientation, the shaper end (EOS) is at cell 161. Right, the transmission and accelerated fractions.

The truncated shapes will have a different transit time factor for the cell (Ch.23). It is important to note that there is slightly more acceleration in the truncated cells, *even in the shaper sectio*n, and the total energy gain to 2.5 MeV is closely identical to the fully 2-term modulation non-truncated case, Fig.29.19:

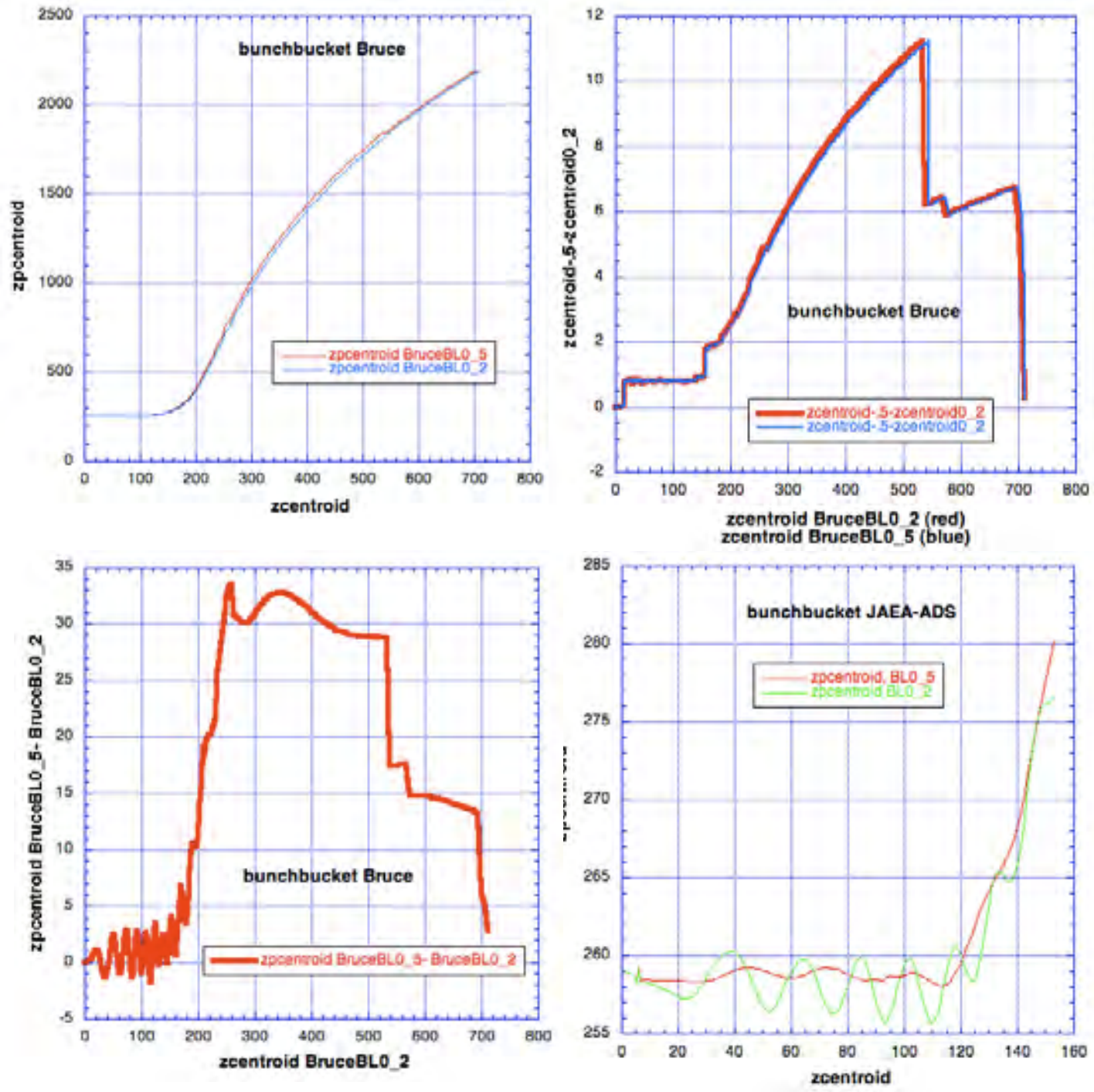

Fig. 29.19 There is a slight increase in acceleration in the truncated part but ending with nearly identical total length and momentum.

The modulation is >1.5 in the range z > ~250cm, where a trapezoidal modulation transit-time-factor and accelerating strength are even greater than that of corresponding sinusoidal vane modulation (Fig. 23.2), which is stronger than that of a 2-term vane modulation (Fig. 23.9).

The energy gain e0Tcos(phis) adjusts to the design parameters.

It is clear that a large portion of the improved performance is occurring in the shaper, which ends (EOS) at z = ~129cm (cell 161). However, the truncation beyond the shaper also plays an important role in forming and maintaining the improvements in longitudinal emittance and EP performance, as was found by comparison to a case with truncation only to the shaper end (EOS).

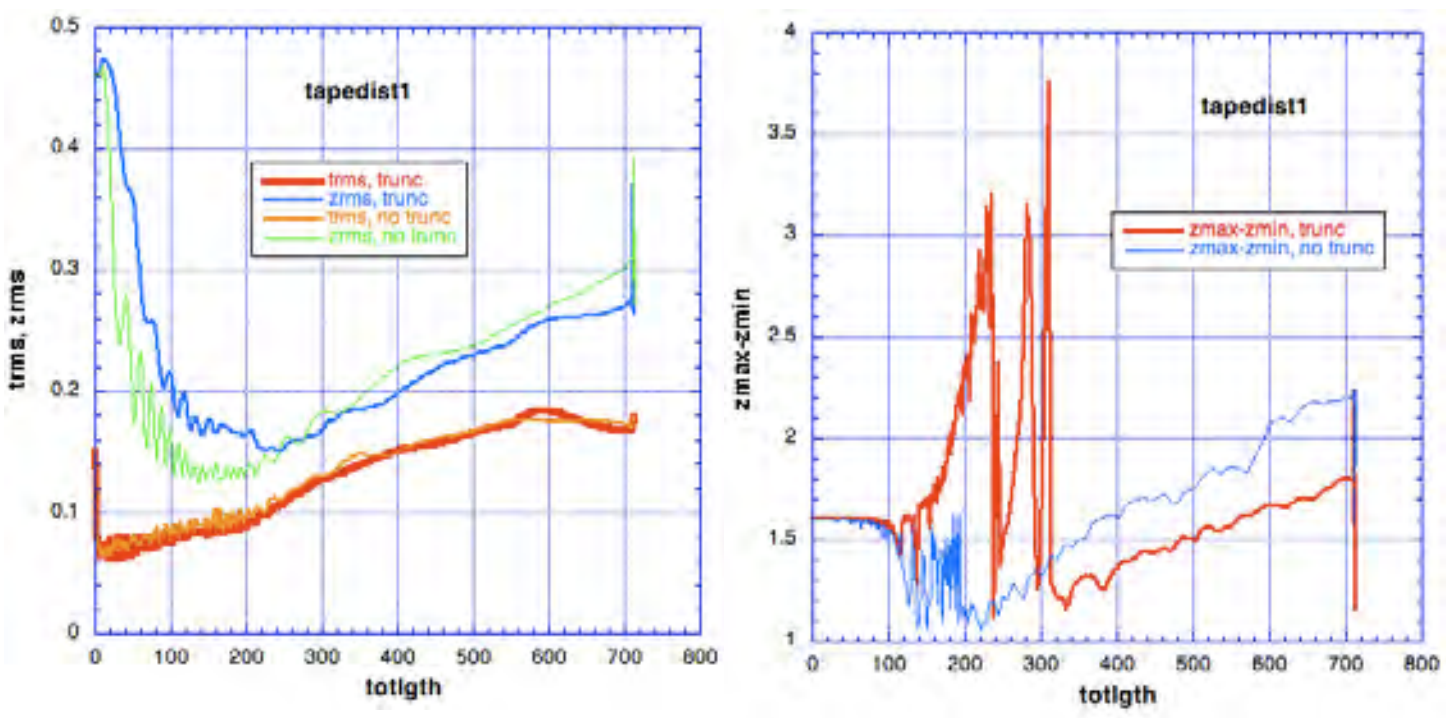

*The linear longitudinal phase advance (sigl) in the shaper is desired* – and important to note that *it was inherent in the result of the parameter optimization study* for this example, *with the truncation level slightly above the design aperture at the beginning of the shaper (end of RMS).*

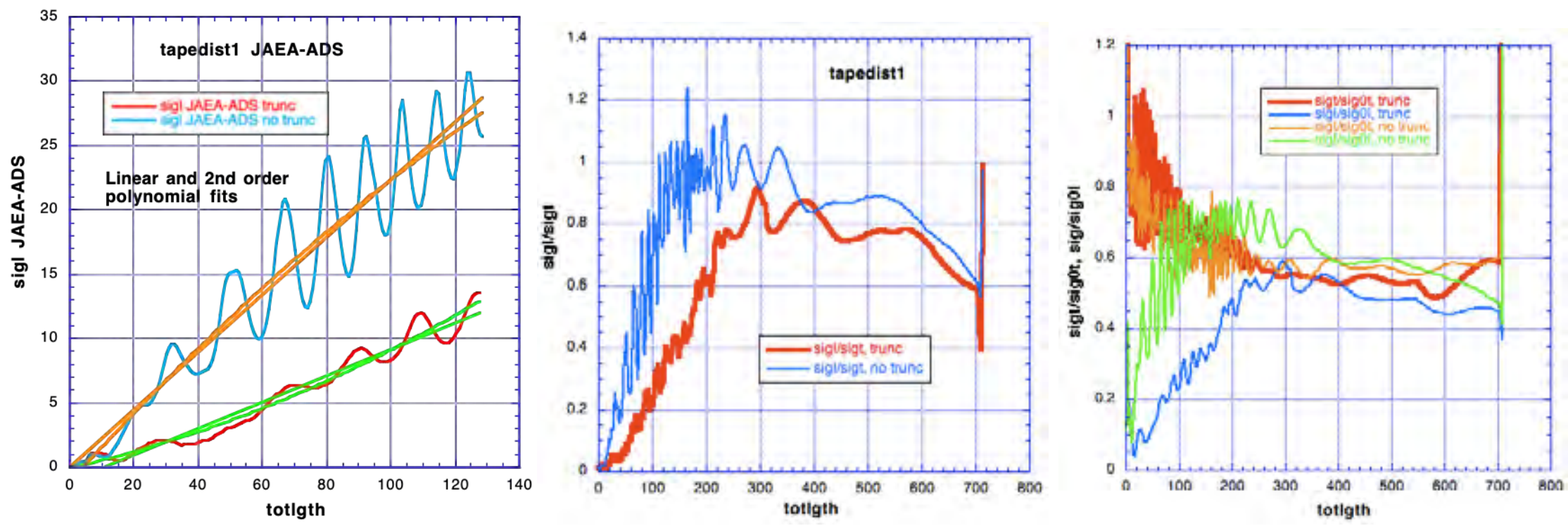

Fig. 29.20 Rms beam sizes, longitudinal total beam size, longitudinal phase advance sigl, tune ratios, and tune depressions. There is slightly more space charge effect with the truncated vanes.

The envelope equations show that

*sigl = ( (eln=constant or with apriori rule)*(unit length)/(γ*(beamlength)*$^2$ = linear

The vane modulation and the aperture in the shaper might be used to try to adjust the beam length to make this happen. Both must arrive at the EP condition goals at EOS, and maintain EP beyond if desired. An NPSOL attempt to achieve these goals was not successful.

Each design study will search for optimization of the whole specification, so the full set of conditions must be present, including the truncation if desired, and use the full Poisson simulation.

The improvements to longitudinal emittance and the EP conditions required something drastic. Each design will have to be individually explored. It is interesting that the truncation was present during all of the optimization searches of the JAEA-ADS case, and resulted in the linear sigl, but the other examples indicate that such success would be rare.

It was lucky that the spec in this case drew attention to the longitudinal emittance and EP conditions. Discovering that there was unintended truncation, an error, would usually produce the reaction to correct the error and redesign, even though there had been improvement in the performance. It was "out of the box" to realize that an opportunity was presented, that should be explored.

The explanation was not clear until getting out of the "rms/smooth-approximation/steady-state" box, and into the broader view of the "time-domain" box!

## 29.4 How the Truncated Vane Shaper Section Works - viewed from the time domain

It was noted at the end of Ch.19.1 that "It is actually harder to design for low space charge than for significant space charge, because the emittance dominates for the former, and for the latter, the space charge mixing affords a counteraction to the emittance." There is little longitudinal space charge until the beam bunch is formed to some extent. Why the possibility now elucidated below was not realized years ago, at least ~2015 when the trapezoidal vane profile was incorporated into *LINACS*, is one of those things, including having too narrow a point of view.

Referring to Fig. 29.2, the profile is truncated, very significantly in the shaper and initial acceleration parts, and the result is a quasi-sinusoidal/trapezoidal vane modulation. Fig. 29.21.

The 2-term vane *design* is unchanged – the truncation is applied to the vane profile in the Poisson simulation. Preservation of the design modulation beyond the truncation point to the full depth (~modulation*aperture) is a key point.

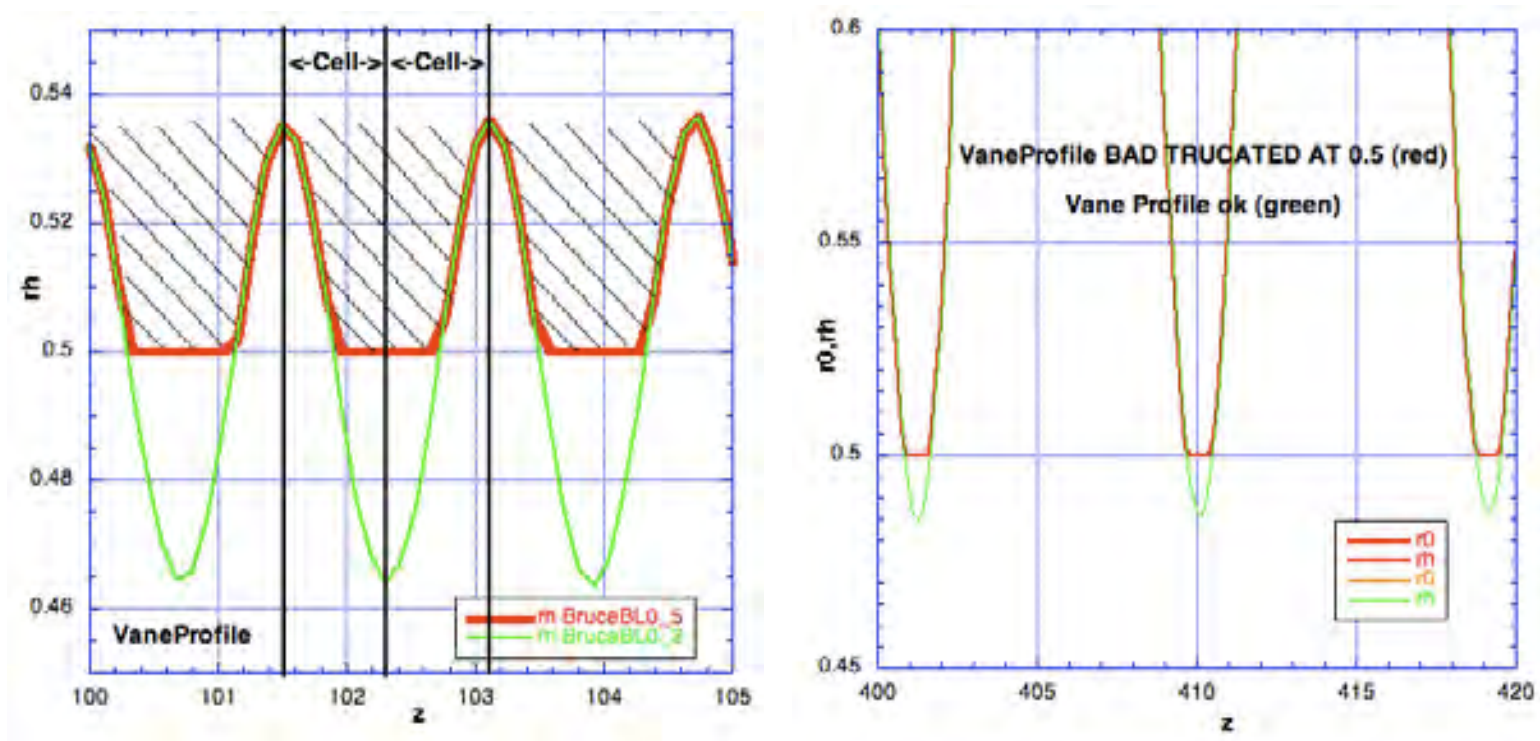

Fig. 29.21 shows the typical truncation, for z 100-105 (left) and z 400-420 (right).

The explanation is immediately seen from an expansion of Fig.29.21 to Fig. 29.22, showing the truncated vane profile and the form of the potential and longitudinal electric field resulting (from, and near, that vane only) that will interact with the beam in the time domain. [267, 268] [269] The total field of both horizontal and vertical vanes on-axis is shown at the center – there is a change, but what is going on in this very low modulation region is not so apparent from this. The location of Fig. 29.22 is about 75% through the shaper and the modulation is only 1.15; at the end of the shaper, the modulation is typically ~1.2.

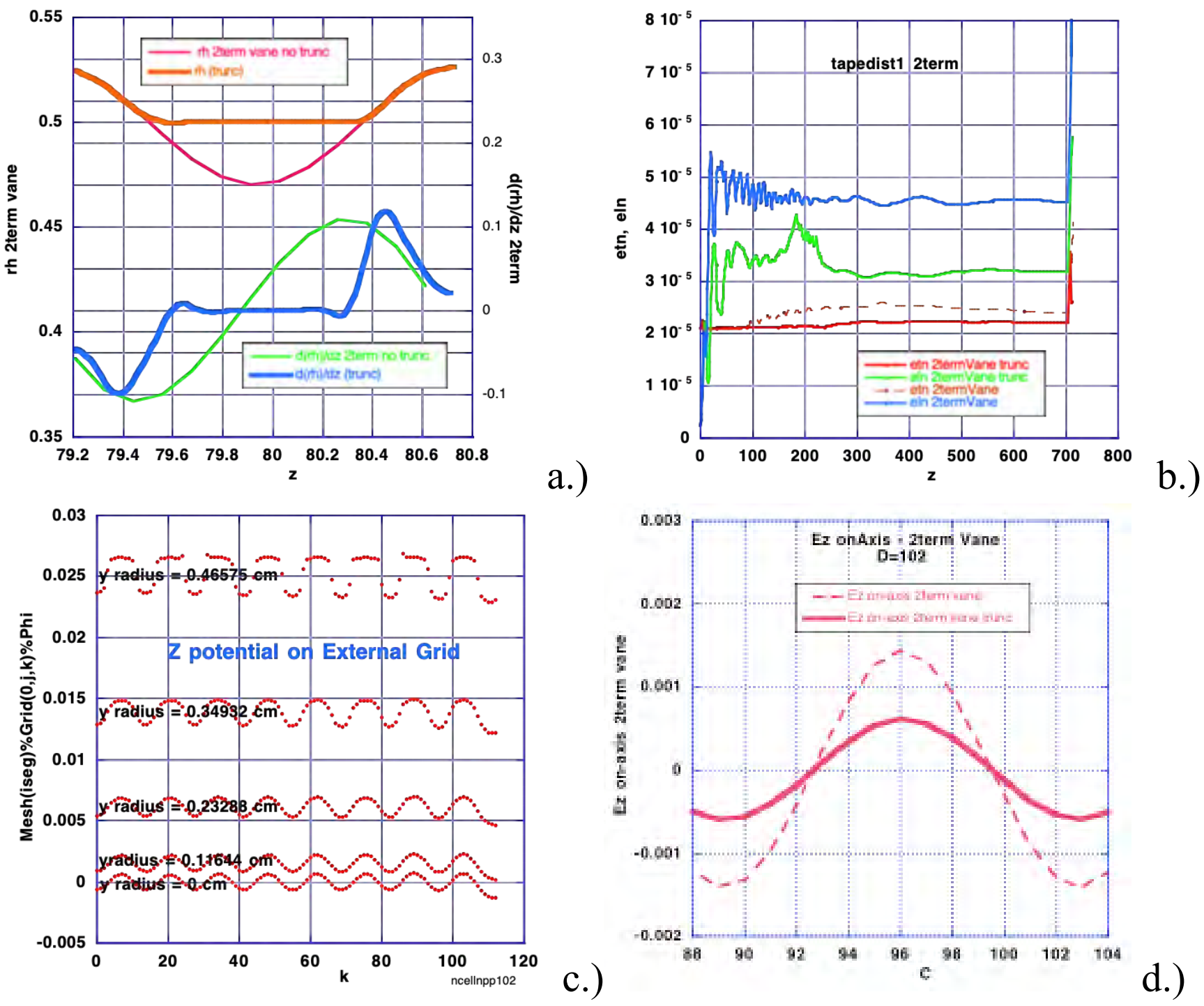

Fig. 29.22 a.) Two cells (one transverse period) of the truncated vane profile (same shape as potential) and the shape of the resulting electric field for that vane, compared to a non-truncated vane.
b.) The rms emittances through the RFQ.
c.) External field potential at different vertical (y) radii; the field near the vane has the truncated shape. At this location, about 75% through the shaper the modulation is only 1.15.
d.) The resulting Ez field on-axis for the full configuration of both horizontal and vertical vanes.

Here, the main interest is to influence the bunching such that the longitudinal emittance is kept as low as possible. Therefore, a region of zero longitudinal field is introduced by truncation of the inner vane radius, Fig. 29.21, leaving the modulation in the remaining part of the cell unchanged, at the full depth (here of a 2term vane modulation; sinusoidal would be equivalent, and it is shown below that trapezoidal is equivalent).

---

267 "RF Linear Accelerators", T. P. Wangler, 2nd edition, ISBN: 978-3-527-40680-7

268 B. Mustapha,* A. A. Kolomiets,† and P. N. Ostroumov, "Full three-dimensional approach to the design and simulation of a radio-frequency quadrupole", PRST-AB 16, 120101 (2013)
V. Bencini, H. W. Pommerenke, A. Grudiev, and A. M. Lombardi, 750 MHz radio frequency quadrupole with trapezoidal vanes for carbon ion therapy, Phys. Rev. ST Accel. Beams 23 (2020) 122003.
A.S. Plastum and A. A. Kolomiets, RFQ with improved energy gain, in: Proceedings of the 26th International Linear Accelerator Conference, Tel-Aviv, Israel, Aug. 21-26, 2012, pp. 41-43.

269 The basics of RF acceleration, longitudinal particle dynamics, and application to the RFQ are well covered in [267], and extension to vane profiles with flat sections (trapezoidal) in [268]; this background is assumed. The usual emphasis is on the vane tip profile. The resulting on-axis longitudinal field is found by the Poisson solver.

There is longitudinal external field and acceleration from a vane *only when the vane profile is changing*; in this case driving the bunching around $\varphi_s$ that is approximately –90° in the shaper. Here the outer part of the profile is not truncated, and bunching is driven, with a larger transit time factor over the cell. The inner radii are truncated, and there is no longitudinal action over the truncated region; the ± swing is set to zero over that region.[270] In the truncated zone, the beam will continue to bunch with the velocities it has picked up outside the truncated region. Thus, the whole bucket does not get filled.

In this example, the underlying vane design is unchanged, the truncation is applied to the vane profile in the Poisson simulation. The design modulation beyond the truncation point to the full depth, approximately modulation times aperture, is preserved.

Here at the beginning, the truncation was nearly 40% of the cell length and was somewhat shifted because of the non-symmetry of the underlying 2-term longitudinal vane modulation. (There is no advantage to 2-term modulation, compared to the length advantage of sinusoidal or trapezoidal base modulation.) Some of the deleterious fields that cause an energy spread are eliminated. But there is still enough longitudinal field that a "gap" still exists and the higher transit time factor allows the energy gain to remain at design. If there is too much truncation (e.g. at 0.525 cm in this case), not enough Ez remains and the simulation fails.

The outer part of the modulation could also be manipulated. The fields could be analyzed more exactly to determine where they are most deleterious in causing energy spread during bunching. A different vane modulation shape might give only the desired fields and at higher strength for faster bunching with lower energy spread. Here there is a tapering off of the truncation and the truncation continues beyond the shaper end, both could be controlled. Other such manipulations within the focusing period used with the smooth approximation rms envelope equations could offer the designer similar advantages.

Figs. 29.23-25 show z-z' phase-space plots, , showing the bunching and acceleration processes of a distribution of $1 \times 10^5$ macroparticles.. The transmitted particles (xmsn) and accelerated particles (nalyz) (of 10K particles injected) show: that with no truncation, there are more radial losses and xmsn ≈ nalyz, i.e. particles outside the specified energy band (here ±50% of synchronous energy) are later radially lost). With truncation there is less transmission loss and particles that are not in the selected energy range are transmitted to the RFQ end, and the bunching (including space charge effect) is tighter, with less energy spread.

---

270 This has nothing to do with the linearity or non-linearity of field that is set to zero. It is just set to zero, eliminating any external field action in this region.

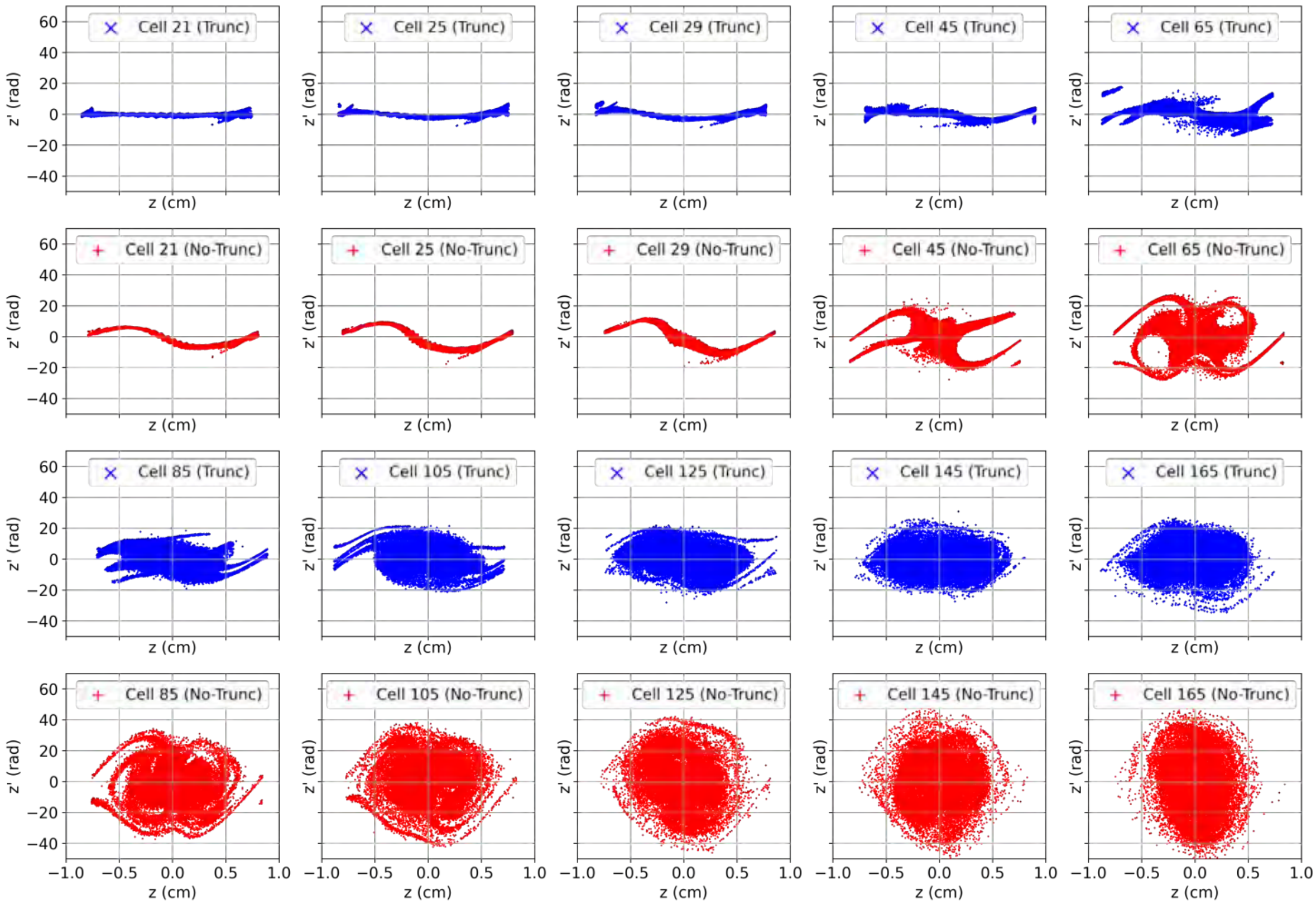

Fig. 29.23 Longitudinal phase distribution at different locations along the shaper section for Truncated vane (blue x) and No-truncated (red plus).

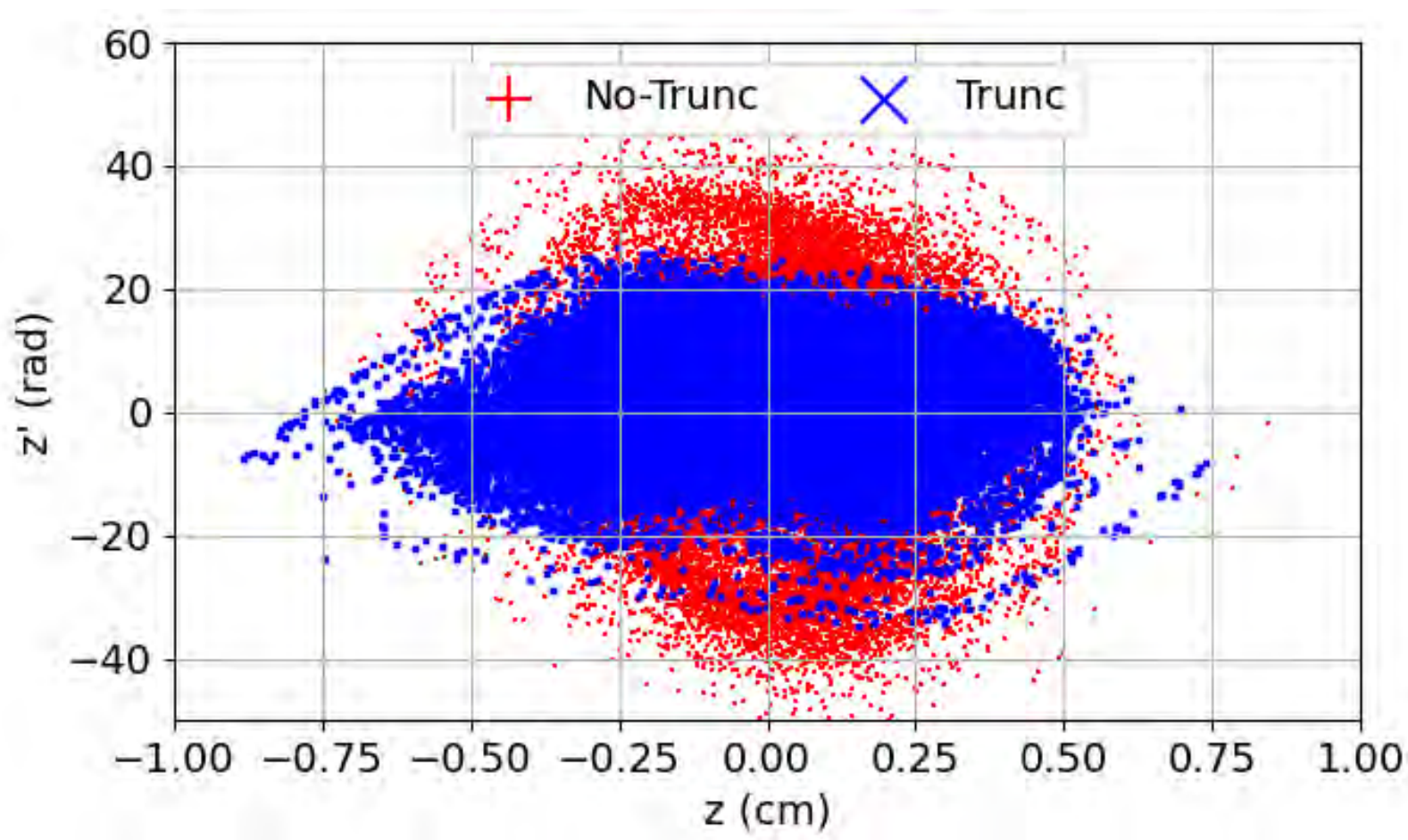

Fig. 29.24 Longitudinal phase space comparison between truncated vane (blue x) and No-truncated (red plus) at the end of the shaper, cell 165.

Fig. 29-25 and Table 29.2 provide information on the particle distribution evolution. Xmsn and Accel particles show that with no truncation, there are more radial losses, and Xmsn was almost the same as Accel, where particles outside the specified energy band (which in this case was ± 50% of synchronous energy) were omitted. With truncation, there was less transmission loss, particles that were not in the selected energy range are transmitted to the RFQ end, and the bunching with space charge effect was tighter, with less energy spread.

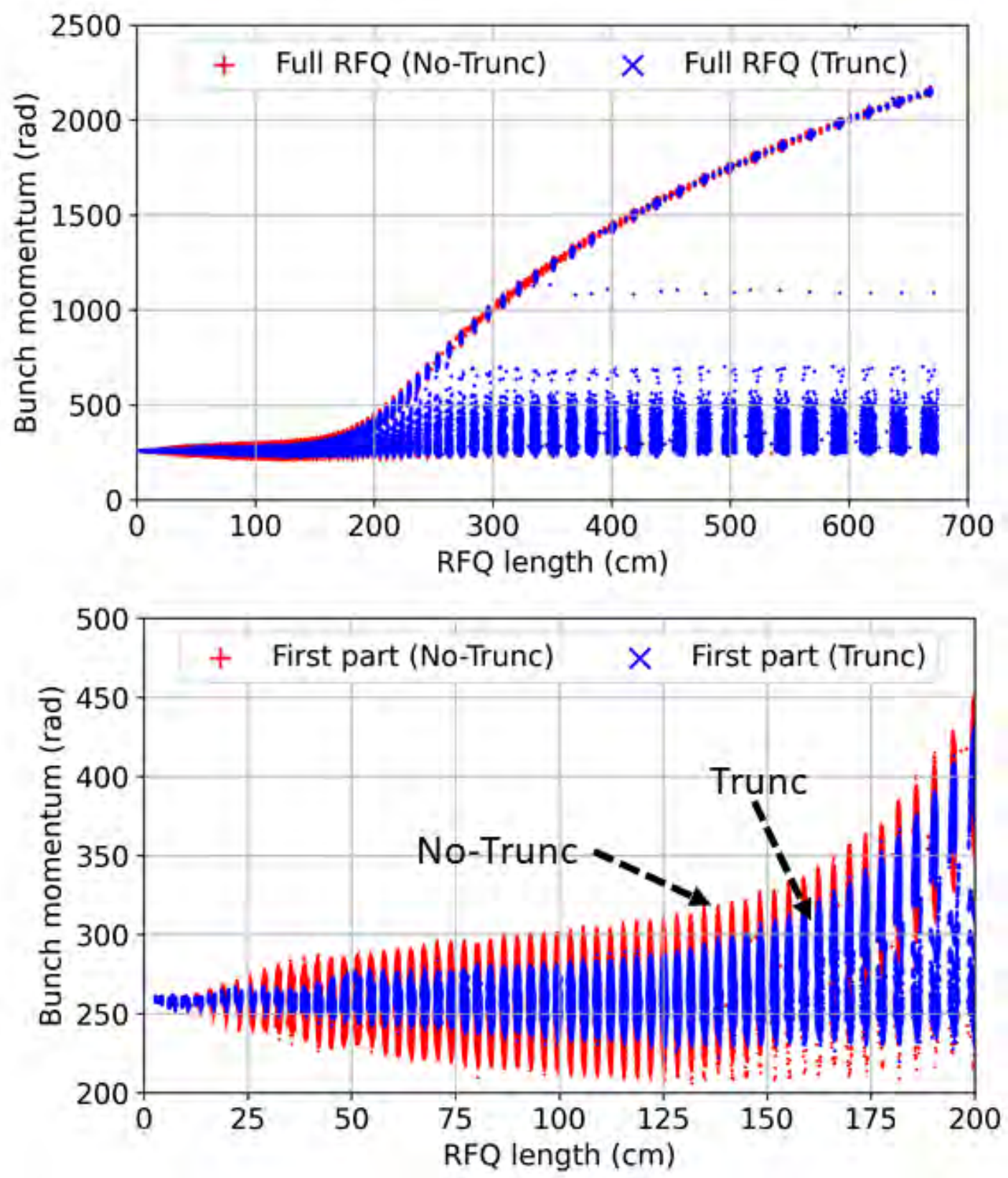

Fig. 29.25 Longitudinal phase space comparison between truncated vane (blue x) and No-truncated (red plus). The top plot shows the full RFQ; bottom plot expanded for the shaper section and somewhat beyond to 200 cm.

## 29.5 Extension to trapezoidal case – truncation of both inner and outer radii

A different strategy using zero-longitudinal-field regions inside the cell - trapezoidal vane modulation - has long been used to enhance the RFQ acceleration efficiency at higher modulations ≥~1.5-2, depending on the percent of the cell where the vane is sloped. This is a completely different regime than the shaper, but it is informative to look at the shaper of this case with the single change of the underlying vane modulation from 2term to trapezoidal.

Fig. 29.22 (left) compares trapezoidal (40%/cell slope regions) vane profiles without and with truncation. (Right) compares the longitudinal emittance behavior – indicating similar behavior in the shaper section as for the 2term vanes. Fig. 29.22 (right) indicates that additional redesign should be done in the main acceleration section, which would also open up the parameters to realize the RFQ length advantage.

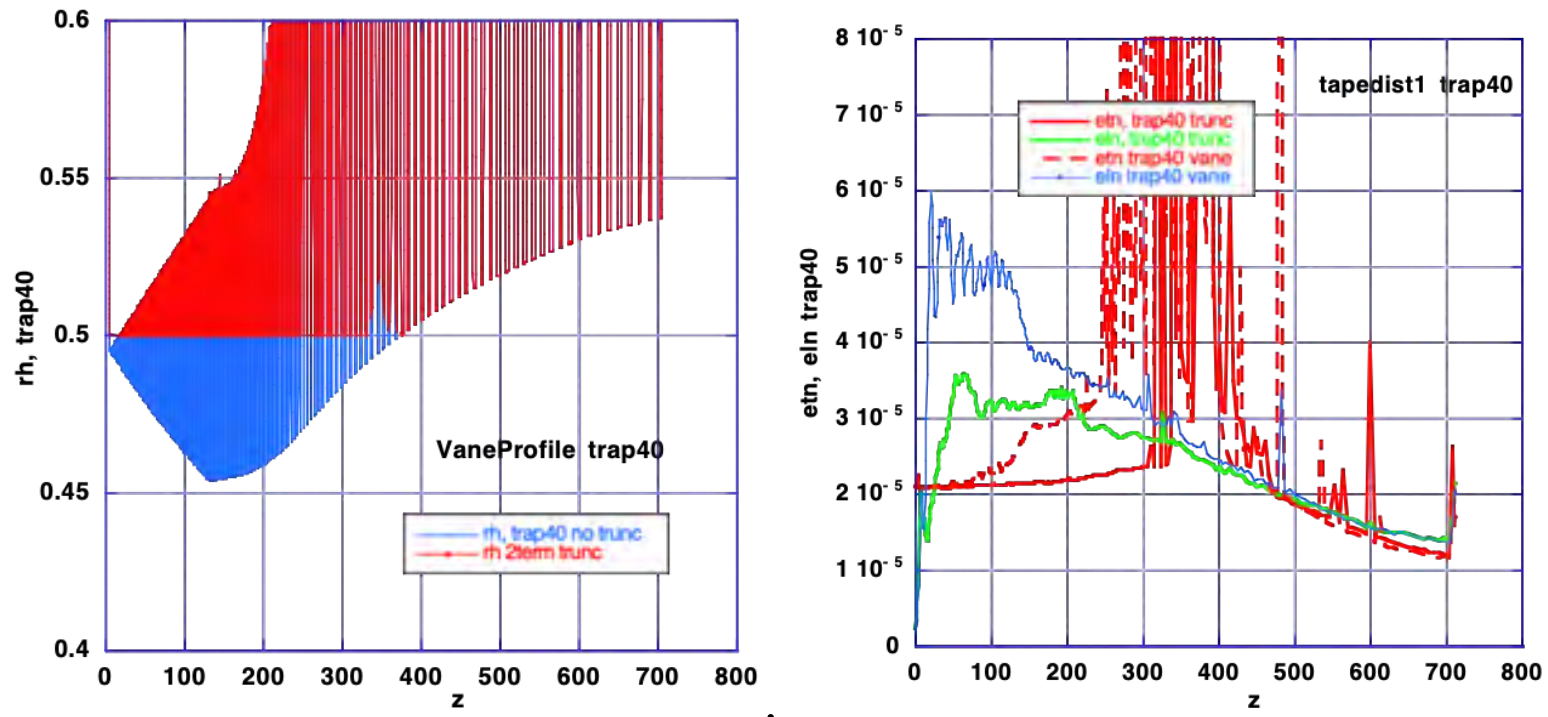

Fig. 29.22 Characteristics of trapezoidal vane modulation. There are heavy beam losses from z=~350.

## 29.6 On the R&D of advanced processes using *LINACS*

A "design RFQ" is established, *using a choice for the vane transverse and longitudinal geometry (presently sinusoidal, trapezoidal, or 2-term longitudinal modulation)*, circular vane tip, and tapered or straight vane body; plus rules, defined in terms of β or z, for the aperture, vane voltage, modulation, and synchronous phase along the RFQ as the design proceeds. Or, one or more of these parameters are left free to solve, as the design proceeds, the transverse and longitudinal matching equations for the beam rms radii and one or more equations involving some function(s) of the accessible rms quantities, such as requiring equipartition.

A previously computed 8-term multipole coefficient table (obtained from accurate Poisson simulation of the detailed vane geometry) and rms parameters is then accessed. The table is parametrized by Rho = (radius of vane tip curvature), Ls = (cell length)/(average vane tip radius from the axis = r0 = r0rfq), and the vane modulation = emrfq. The table has a tabular 8-term[271] multipole approximation of the potential, transverse and longitudinal focusing strengths, transit-time factor, for interpolation locally - to produce the generated design cell table tabulating all of the cell parameters (including the derived cell length, cell end energy) at the end of each cell. The 8-term approximation is not appropriate for simulation, but appropriately applied in the design procedure, and it enables good agreement between design and simulation.

This "design RFQ" cell-end table is then used in the simulation program for generating the external and space charge Poisson meshes, with the design detailed vane geometry, cell lengths, aperture, and modulation. *The Poisson simulation results provide an accurate description of the physical process that occurs.* If beam-based, "inside-out" requirements have been imposed on the design, the simulated results may or may not agree well with the design, depending on the actual physical practicality of the design. Usually, it is desired that design and simulation agree well, so if disagreement occurs, design variations that result in agreement will be sought. If an appropriate rms smooth-approximation design is found, experience has indicated that simulation and experimental measurements on the constructed RFQ will also agree well.

*The simulation is free to introduce off-design condition*s; in this case, the results can be expected to deviate from the design condition results.

Here we have developed a control for bunching and longitudinal emittance, quantities inaccessible to the design equations. Imposing a truncation on an inherently smooth vane modulation (2-term or sinusoidal) fortunately did not interfere with the underlying design modulation, produced desirable simulation results, and allowed optimization of the truncation. But if the underlying modulation had already used truncation (like trapezoidal modulation) to control another aspect, such as the acceleration efficiency, then adding an additional truncation in the simulation to influence an extra aspect could be expected to be conflicting and require perhaps a compromise, which might require a complicated optimization process to find. And, if good agreement between the new vane geometry results and design results is desired on a built-in basis, new Poisson-based underlying design tables would have to be generated.

[271] Or 16-term, as shown that the use of 16 terms is also more accurate for simulation. However, Poisson simulation is preferred.

*The point is that for good agreement between the design and simulation, the underlying design tables must be derived separately for any particular form of vane modulation, i.e., for any specific form of intra-period manipulation.*

This, and subsequent investigations into manipulation of intra-period details, then require open-source, correlated, design and simulation tools capable of accurate physics, adequate detail, reasonable ease of modification, use and running time, such as *LINACS* provides. *LINACS* can also compute new multipole tables for a new variation, although this is very time-consuming and precision may require another program.[272]

## 29.7 Conclusion - A new shaper section for longitudinal emittance control

An improved vane longitudinal modulation profile technique for longitudinal emittance control is presented. This advanced approach using manipulation of the RFQ vane modulation profiles can be used to minimize longitudinal rms emittance; it is implemented at the cell level, and therefore general; without change to the overall rms design, e.g., while maintaining the requirements to reach equipartition at the shaper end and maintain equipartition in the acceleration section to the RFQ end. The technique can be adopted by any RFQ, of course also with attention to the particulars of the case at hand. It also has implications for any linac.

***Thus this is a first, and compelling, discovery that better rfq vane modulation profiles are possible for minimizing longitudinal rms rmittance, while maintaining the requirements to reach equipartition at the shaper end and maintain equipartition in the acceleration section to the RFQ end.***

The key is to avoid most of the deleterious longitudinal field variations, particularly in the bunching process, by truncating the vane profile so that no longitudinal electric field is produced in that region.

This work also points out that the limited "rms/smooth-approximation/steady-state" approach gives useful analysis insights. However, it has to be expanded to the full time-domain, instantaneous-state view, where clearly much is going on within the period. All the rms observations are also apparent in the time domain view, but smoothing all that away, i.e., averaging over a distance of interest and discarding detailed motion within that distance, lost the ability to explain what was happening. Here, the phase-space details provide helpful information.

This recent example strongly shows that getting "out of the box(es)" and being open to surprises is always necessary – there is always still a lot to discover when working with nonlinear systems, an "inside out" design framework, and unconventional insights and ideas may be the way to progress.

- - - - - - - - - - - - - - - - - - - - - - - - - - -

- - - - - - - - - - - - - - - - - - - - - - - - - - -

[eltoc]

272 *LINACS* tables for 2-term, sinusoidal and trapezoidal longitudinal vane modulation are actually computed to high precision using TOSCA.

# Chapter 30 – On Space Charge Influenced Resonance Modes and Beam Halo (action-angle)

## 30.1 Once upon a time…

As usual, in such case, neither the names of the innocent nor the guilty are given.

A Russian Academician, I.M. Kapchinsky, first investigated (1959) space charge influenced resonance modes in an approximated form of a focusing channel, realizing that solutions to the Vlasov equation could be found mathematically if a special, unrealistic, beam distribution was used. This distribution is now called the KV distribution – the beam particles lie on a delta-function shell in 4-D phase space. Sacherer did his PhD under Lloyd Smith at LBL, and in 1970 published his seminal paper, showing, among other things, that use of the problematical KV distribution is not necessary. Outstanding researchers at LBL extended the investigations.

They were joined late in the decade by a young researcher who wanted to follow in these footsteps.

He was given full access and support, but Sacherer's work was apparently lost.

~1981, from the context of very useful workshops on heavy-ion fusion, I found that his then-crude instability thresholds, for KV distribution stopbands in simple focusing channel with the "smooth approximation", gave the first-ever clear correlation with emittance growths seen in simulations of practical linacs, as analyzed cell-by-cell. In other words, although the theoretical model was for a non-realistic beam distribution, the simplest form of transverse focusing channel only, and smoothed out all detail inside the transverse focusing period, the resonance thresholds were closely correct when applied to the instantaneous state analysis of an elliptical {r,z} beam bunch with equivalent rms emittances and with acceleration and changing parameters (as actually predicted by Sacherer). I immediately used this as the basis for studying emittance growth in linacs, and for linac designs that avoided the resonance areas.

In the 1990's he made a very useful parameterization to represent, on a "tune chart", for particular rms emittance ratios between two planes of a distribution, the areas of instability and the growth rates of the three strongest resonance modes. The charts also show the strong cancellation effect of an equipartitioned (EP) distribution on resonances, and it was clear how designs could use resonance-free areas and/or areas where EP cancelled resonance effects.

At that point, the practical usefulness of this approximate theoretical result was clear and the story could be declared finished. That is still the case in 2022, although I remain the only one to extensively use it for analysis and design.

Recall Ch.1.2 – the actual hierarchy of (experimental)/simulation/theory) and the local, instantaneous beam state. If an approximate theory is accurate enough to help clarify the real situation of simulated or measured behavior, then do not be concerned with further details of the theory, use it and get on with practical accelerator problems.

The theoretical problem, however, remained very interesting for the theorists, and has produced a massive literature since the 1980's. Following every publication closely, it was observed that more and more problematic and confusing practices and statements were being promulgated. Other people began to be interested in the theory (not in the practice), the number of publications grew and grew, with much confusion, even leading to heated and irrational debates at workshops and conferences.

I took no interest, as such details of the theory were clearly only academic, but wondered

1) why the actual (experimental/simulation/theory) hierarchy is never acknowledged, to stimulate work on practical aspects.

2) how so many could not see that injecting a beam distribution directly into a resonance was very clearly not related to linac practice (or ring practice, with certain unusual exceptions).

3) Or why examples guaranteed to be instable receive again and again so much publication space, once simulations had clearly been shown that the stopband limits for the very restricted models agreed with the theory, and also that the stopband limits for extended models also agreed rather closely.

4) Or why the trajectory of only one plane was plotted on the tune charts, when clearly realistic channels often have coupled transverse and longitudinal motions.

5) Or why steady-state results and explanations are preferred, when clearly there is no steady-state and everything is transient. This relates to my wondering why the more powerful Laplace Transform method already applied by Landau in 1946 and used by most authors outside the accelerator field for this type of problem, plus the powerful Nyquist and Root Locus methods for instability analysis, are not used.

6) Or, in other words, why viewing the accelerator system as a local or instantaneous state, so central to understanding phase space transport, analysis and design, seems so hard to grasp.

7) Or why the underlying phenomena have not been ordered in terms of their time structure, to identify clearly what is faster, what is slower, which greatly clarifies the situation.

8) Or why work with realistic accelerator and ring channels was not pursued. Why in fact projects resisted first investigating a good EP design in good spec context, and continued to defend "like always" non-EP designs - perhaps with tune-chart check that just observed that the trajectory was mostly "in the clear areas" but without any analysis, e.g. of 100%/rms or error tolerance behavior.

9) Or how statements could be made about using the theoretical observations for practical design – that clearly did not relate to reality. Up until this started to happen, the academic theoretical investigations could be ignored in practice, but when the practice began to be impacted, especially when so grievously and clearly for personal non-scientific motivations (Sec. 1.7.4), there was cause for concern.

10) Etc.

There have been many good papers outside of what can be considered "the mainstream" that have correctly interpreted and enhanced the knowledge of this subject. Also, many details of the "mainstream" papers are useful. However, all of this academic activity has had negative effects. It has not succeeded in reaching clarity in defining a nomenclature, or for clearly indicating practical use. It also has by now produced a rather large population of partially informed "experts" who passionately defend their view of the elephant.

Also during this time, the publication culture of the accelerator field has developed, largely for the worst. In the previous century, the accelerator field published mainly at conferences and by internal laboratory reports. One was free to publish his views, new things, developing ideas. This has evolved in the new century to "dedicated journals", and "publish or perish" for young workers. And a "review" system has evolved, that in the accelerator beam dynamics field is a very big problem, particularly on this subject. The "experts" and "hobby reviewers" very clearly show little actual knowledge of the whole picture, are even unable to discern when the scientific procedure is not used or even abused, and clearly use half-baked notions about the technical aspects and about how a paper can be presented – to shoot down papers, whereas certain obviously flawed or even manipulative papers are published. Why? Very clearly, according to present review criteria, very few of the mass of papers published up to ~2000 should have been approved for publication. And what is the actual quality of many published papers?

In the 2010's, a young researcher again developed great respect for this field, and determined to fill in all the missing steps in the published derivations[273] . However, this time, upon inquiry, he was informed that he "could fill in the steps himself".

---

273 Usually preceded by egotistical statements like "it is easy to show", "clearly". "it follows directly" , "obviously" – which "reviewers" should summarily require to be excised, but never seem to…?

In fact he did, and in the process broke out of a severe approximation used in all previous work. He has copies of, and has closely studied, literally all publications on the subject[274]. He has attempted to restate the definitions, theory, and simulation more clearly in line with practical aspects as numerated above. But his publication attempts in the "mainstream" accelerator field vehicles have been shot down, with nonsense reasons and also some very suspect aspects. See Appendix 5 of "Calm in the Resonances, and Other Tales".

Fortunately, some initial publication in broader forums has recently (2019) been achieved, with good review understanding and support. Subsequently however there is again blockage. Many dilemmas confront presenting a new view. Short papers are too short – uneducated reviewers have to be educated. A series of short papers doesn't work, because reviewers do not read references and cannot follow the series. Long papers are judged to be too long, and there are no educated reviewers. The time lost in battling with this problem is at the expense of further work. Our current thinking is that probably a "monograph" [275] on this particular subject should be attempted, and published on an open platform such as arXiv.

So my considerable effort, to teach the real elements of the space charge physics in focusing and accelerating channels (it has to be considered in the transient state, the tying together of the phenomena by the fundamental space charge mixing via the non-neutral plasma interaction, etc.), trying to help the younger generation to proceed with clear understanding, has born only some fruit.

## 30.2 Some expansion, reiteration

This book addresses in detail the items outlined above, so only some short reiteration and expansion:

1) of course is clear.

2) This is really an amazing phenomenon. Real machines avoid resonances or deal with them quickly and specifically. The study of beam injection directly into a resonance is an easy, publishable, academic exercise.

3) Again amazing, academic, not practical.

4) Originally because of only vaguely known thresholds, especially as the tune ratio crossed EP. But it has continued until the present, in spite of continual urging and detailed examples, that both {r,z} or all {x,y,z} trajectories are important in application to real linacs with separated-function lattices, various degrees of coupling by additional components, acceleration and changing parameters. (See item 9).

5),6),7) are dealt with extensively in this book. They are the **elements** of the problem.

8) Theory does not live in the real world and must make approximations, often severe approximations. It is indeed very hard, perhaps impossible, to formally extend the theory for focusing channel resonances beyond that afforded by the non-realistic KV distribution through which the boundary conditions can be matched to 4D; this approach cannot be extended to 6D. It is hard to extend the theory to more complicated channels. It is hard to be a theorist who really concerns himself with being useful to the practitioners. However, the business of a theorist is to find approximations that are useful. It was/is known that the concept of equivalent rms emittances completely removes the necessity to involve the KV distribution in actual simulations. It would seem that an approximate theory, based on rms distributions and realistic approximations of the space charge fields "inside and outside" the beam and matched at that boundary, should be possible, to break out of the formal KV box and give useful guidance about the resonant mode structure for realistic lattices. Or at least correct teaching – that KV is a theory, and how to use it.

---

274 In marked contrast to "reviewers" who clearly are hardly or erroneously informed, often suggesting some of the weakest references.

275 A quote from "https://openscience.com/the-forever-decline-academias-monograph-crisis/":
"The monograph provides researchers with the finest of stages for sustained and comprehensive—sometimes exhaustive and definitive—acts of scholarly inquiry. A monograph is what it means to work out an argument in full, to marshal all the relevant evidence, to provide a complete account of consequences and implications, as well as counter-arguments and criticisms. It might well seem—to risk a little hyperbole—that if the current academic climate fails to encourage scholars and researchers to turn to this particular device for thinking through a subject in full, it reduces the extent and coherence of what we know of the world. But then I am not the first to raise such concerns about scholarly responsibilities for the scale of thought."

Concerning item 9) (and 4)), a startling realization came to me in 2017-2018. There was suddenly a breathless revelation from a much-published theorist that in a standard drift-tube linac with alternating focusing there are two rf gaps in a transverse focusing period!!!! I was really flabbergasted – had not realized for all these years that the theorist had never had any idea about how actual linacs are laid out.

Suddenly, the reason for many questionable statements became clear, and it was clear that the theorist had never been trained or understood my emails and urgings about actual linacs, for which I had been using the local state for analysis and design for decades. There is now ensuing confusion about how to define the "phase advance" of the longitudinal plane, in relation to that of the transverse plane, and the question of whether now the system could still be stable up to 120° longitudinal phase advance…

There is the quite horrible misguidance that "equipartioning is not necessary, because there is ample space in the tune space area around equipartioning, in which a stable trajectory can be placed". This very unfortunate oxymoron is discussed in Ch.1.7.4. There would be no stable space around equipartitioning, in which there is no free energy in the beam bunch that could drive a resonance, if equipartitioning does not exist. There is very little realization that equipartitioning is also a local phenomenon, of how the local system state is utilized for analysis and design, of the flexibility of sophisticated EP design against a specification.

## 30.3 Extension, prospects

All the theoretical development until recently has assumed the severe approximation of a "smooth approximation" channel, where the phase advance has been integrated over a transverse focusing period, which discards all detailed information within the transverse focusing period, and thus is a severe approximation of commonly used lattices. E.g., at higher energies, linacs require less transverse focusing, and many rf gaps are included within a transverse focusing period. The young researcher succeeded in keeping the mathematics of the full Hamiltonian, and has shown that the resulting tune charts for simple channels are different than those for the corresponding smooth approximation.

He has developed a synthesis of the nomenclature and identification of the various types of resonance.

Given that the exercise of injecting a beam into a resonance is academic, it still can be useful as an educational tool, if time-domain analyzed as phase space transport. But the existing academic literature, with the over-emphasis of purposely exciting a resonance, does not clarify the phase space transport resulting from interaction with a resonance, the relationship between higher and lower order modes, the relationship between resonance driving and damping from nonlinear density, and the situation at tune depressions near the space charge limit. It fails to identify the always present, underlying and fastest mechanism for beam redistribution, caused by the driving mechanism of collective action of space charge, is space charge (or phase space) induced mixing at the plasma frequency of the beam. This appears to be little known or understood by "reviewers", who have claimed to have never heard of space charge or phase space mixing or the plasma frequency, and use statements like "the resonance saturates".

He has not yet extended the theory or simulation to make charts for practical separated-function linacs or rings with additional elements, or finished extended ideas for more exact identification of beam halo beyond the largely "by-eye" commonly used descriptions.

Very unfortunately, his attempts to publish in this area in the directly related literature have been increasingly thwarted – see Appendix 5 of “Calm in the Resonances and Other Tales”, as have also been his wishes to collaborate, with demoralizing consequences – costing the subsequent loss of this talent to the ion linac field.

Those defending their territory in the “KV box and injecting into resonances” are welcome to stay there; *they are not the actual designers in any case*. But they should not get “too big for their britches” and purport to be gurus outside that box. It is important that the **elements** and their relation to actual design are understood. Also, the KV box breakout and the potential to achieve extended theoretical tools for actually used lattices are, and could be further, useful in practice. It is good that publication in broader journals is sometimes possible, although very difficult for the same reasons and because the development is very specific to accelerators - but such sources are rarely read and it is not clear how such sources can be brought into direct accelerator field education and practice.

## 30.4 Breakout from the KV Box and Smooth Approximation

Collaboration on the theoretical approach above quickly led to my wondering why the more powerful Laplace Transform method already applied by Landau in 1946 and used by most authors outside the accelerator field for this type of problem, plus the powerful Nyquist and Root Locus methods for instability analysis, are not used.

Some work, found in publications outside the accelerator field, earlier than Li’s recent application to pure FODO and solenoid channels, indicate possible approaches but without direct application to the mode analysis of practical channels. [276] sets up a computational scheme to integrate the Poisson–Vlasov equations in a drift-tube linac type periodic focusing system including rf gaps in 2D and 3D with a numerically generated initial beam distribution. The approach could probably be extended to identify resonance modes and their approximate growth rates, and to also give an explanation of the phase space mixing (space charge mixing). [277] and [278] use the more advanced Laplace Transform method already applied by Landau 1946 and contain a clear explanation of phase mixing, but are very highly formal and would require very extensive work, as Li has done, to apply in detail to practical channels. It is clear that the KV distribution is useful in the theory to match the fields at the beam boundary, but even that is perhaps not necessary. With an understanding of space charge mixing and Sacherer’s results, *it has never been necessary to use the KV distribution* *as the input to a simulation*, and much confusion could have been avoided.

Practically, given the direction toward using simulation, and the severe difficulty and lack of flexibility for quickly evaluating many complex lattices, it seems clear that the time is long overdue to put the theoretical approach to rest.

It is high time to put this theoretical approach in its proper place as an approximate aid. It is high time to stop claiming that dedicated experiments are necessary “to verify the theory”. The theory is less accurate than the experiments. It is high time to stop claiming that accelerator simulations are suspect – at least some are well known to be very accurate at least to the rms level and this has been verified by comparison to the living experiments of many actual machines.

Even the usual direct PIC simulation is accurate, and shows resonant structure directly. It is sufficient to use the local analysis method and the smooth approximation tune charts. A simulation can be easily modified for study of different configurations.

---

276 “3D solutions of the Poisson–Vlasov equations for a charged plasma and particle-core model in a line of FODO cells”. G. Turchetti, et. al., Eur. Phys. J. C 30, 279–290 (2003)

277 “Linear response theory for hydrodynamic and kinetic equations with long-range interactions”, P-H Chavanis, arXiv:1209.5987v1 [cond-mat.stat-mech] 26 Sep 2012

278 “General linear response formula for non integrable systems obeying the Vlasov equation”, A. Patelli, S. Ruffo, arXiv:1403.5460v3 [cond-mat.stat-mech] 19 Nov 2014

[eltoc]

# Chapter 31 – On Future Work Toward Very Low Beam Loss

## 31.1 Nonlinear lattices, intrinsic nonlinearity

It must be pointed out that future accelerators for very low beam loss might best be designed by departing completely from the types of essentially linear lattices used to date. Significant work has been done by Batygin, Bruhwiler [279,280] and others on nonlinear lattices.

Nonlinearity introduced intrinsically that results in amplitude dependent particle phase advances also reduces the effect of resonances and the tendency to produce halo. The nonlinearity of the linac's longitudinal phase space has this effect, and also the non-KV particle distribution. Perhaps there are other ways to introduce intrinsic nonlinearity to achieve very low beam loss? (e.g. building on Ch.28.4, APF, highly oscillatory trajectory?)

## 31.2 Work Plan for very low beam loss extended investigation of design, simulation, analysis, optimization of linac lattices

The following work plan was proposed (2018) as initial steps [281] to get a quick start on eventually developing a simulation code that is primarily and directly single-focused on studying the very low beam loss problem in linacs. In contrast to extant codes that are multi-purpose and not suited.

It is based on getting into low-loss investigation and accompanying optimization as quickly as possible, by proposing the use of *LINACS* and the RFQ as the starting platform – because this already exists, was built considering the basic elements, contains needed tools, and the software shows how the physics, tools, etc. are integrated, which serves as examples and guidance for extensions.

**Purpose :**

- *Simulations* using existing codes, models, are not accurate enough to give accurate results for very low beam loss studies. Must make significant step in simulation code.
- *Advanced analysis methods* can provide more insight:
  - Detailed analysis of simulation run data, step-by-step (local state)
  - Support from advanced theoretical tools (Li Chao)
- *Optimization* drivers desired.

**Outlook:**

- Necessary steps are known, tools are available.
- Could start immediately.
- Could expect significant new simulation results even in ~1 year. (maybe even sooner, or maybe optimistic), certainly in <2 years.
- Advanced design methods can also be coupled later (as for RFQ)

**Impact:**

---

279 "Effects of Nonlinear Decoherence on Halo Formation", S. D. Webb, D. L. Bruhwiler, et. al., arXiv:1205.7083v2 [physics.acc-ph] 7 Aug 2013

280 "Suppressing Transverse Beam Halo with Nonlinear Magnetic Fields", S. D. Webb, D. L. Bruhwiler, et. al., arXiv:1205.7083v1 [physics.acc-ph] 31 May 2012

281 For a person with primarily AI interest. Whereas first primary interest should be improvement of basic technique capable of showing very low beam loss.

• It is essential that such steps be taken. Better simulation, better analysis, and optimization will give new insight for both design and operation.

**Base Code Required**

• one integrated code – not patching results of different codes together

• Single-focus on studying the very low beam loss problem in linacs.
  – we are most probably still at the level of trying to understand and use quite simple physics and engineering – not small higher order effects.

• SOURCE CODE MUST BE COMPLETELY OPEN
  – no black boxes
  – fosters collaboration
• Could compare different codes as desired for essential features, e.g.:
  – independent variable – must be time, canonical coordinates
  – easy to stay with physical meaning – no tricky renormalizations of Hamiltonian, strange units, etc.
  – physics correct to required level - 6D, Poisson solution
  – external and space charge fields found under same conditions, same boundary conditions. (Poisson solution)
  – code language, packaging should be simple, maintainable (e.g. Fortran, natural accelerator components, rather than hard to maintain C++, object oriented, etc.
  – analysis methods incorporated, required data output, correct units
  – optimization methods incorporated in base code
  – options provided, (for simulation, analysis, design)
  – error and parameter sensitivity studies supported (input emittances, beam current, mismatch, mis-steering, etc.)
  – benchmark, experimental test of code over wide parameter space
  – how much time, manpower needed to bring desired capability

The *LINACS* code has unique DESIGN + SIMULATION + OPTIMIZATION Framework
The *LINACS* code is already very close to desired capability.

**Model Development:**

• For study of very low beam loss, must have best physics representation of fields
  – no approximations
  – no non-matching boundary conditions
• *LINACSrfq* is full Poisson, external and space charge, same boundary conditions, image forces are correct
• DTL, CCL, SC, even beam pipe, that is – ***every*** component – needs full Poisson solution of fields with exact boundary conditions - (as in dancing, maybe transitions between components is the key)
  for example: - to get correct image fields
  – must be in base code (not imported from outside, e.g. CST), because will want to make many runs with changing parameters
  – make geometrical models (for SC HWR cavity, magnets, beam pipes)
  – incorporation into Poisson solvers (especially matching of boundary conditions, optimization of run time)

RESULT: Unique, first time code with consistent physics everywhere

**Tools**

• Review analysis subroutines, begin enhancements.
• Begin development of specific halo measurement subroutines

**Optimization Engines**

• Powerful linear/nonlinear constrained/unconstrained optimization engine NPSOL is already built in to *LINACS* code and used extensively.
• RFQ capability includes fully integrated design and simulation.
• Optimization of variations within a design strategy are already programmed.
E.g. find input matching, find EP condition, vary vane voltage, input current, input emittance, input current.
• Addition of outer driving loop to drive different designs plus simulation, through parameter range with optimization of desired objective function, is already sketched in *LINACS*, and can be quickly implemented fully.
• ("Optimization" encompasses very many methods, e.g., AI, neural networks, steepest descent, etc., etc. …)

**A Work Plan**

*First goal: learn LINACS*

1.a. Port *LINACS* to desired computer – has been done many times.

1.b. Set up an RFQ design problem using *LINACS* GUI and do preliminary design, e.g.: Deuteron-RFQ, f=162.5MHz, q=1, amu=2.0135532, Wi=60keV, Wout=3 MeV,
operating current I=50mA, design current I = 70mA
- Explore other parameters, rules, strategy as available on GUI interface:
- For promising design, do Poisson simulation (irunopt(5)≠0 = 1 for further design iteration.
- If no beam loss is seen, consider what is missing, plan to incorporate in parallel with interest in optimization.

1.c. It is very important to learn the source files and program flow. Read all execution options and subroutines, and make a flow chart of the program

*Second Goal: to set up overall driver loop needed to do full linac optimization studies, and do an initial study.*

2. Learn about *LINACS* optimization tool NPSOL. Task is finished when understanding is achieved.
Study examples in code, where used, method of use.
getEP.f90, getam.f90, getEPmain, getEPdtlccl
Look at NPSOL manual.
Explore options.dat settings for RFQ design example above.

2.a. Program the overall optimization loop "bigloop" and do first demonstration with NPSOL optimization tool already built into *LINACS*, using RFQ example.
Preliminary "hooks" are already outlined in *LINACS*, showing where the bigloop driver would be inserted and interfaces needed to various subroutines.
Program the needed optimization subroutines.
This is new development work, time to implement cannot be stated in advance, but hopefully within say 1-2 months.

2.b. Add AI optimization alternative and do first demonstration using RFQ example.

*Third goal: to build up superconducting linac design, initial simulation and optimization in the LINACS framework, using the existing already programmed methods as templates.*

3. Outline design strategy rules for superconducting linacs.

Types of transverse focusing to be studied, number of rf gaps between magnets, actual geometries of rf cavities, magnets, drift pipes, cryostats, etc.

3.a Program matrix type dynamics routine for simple, quick simulation; needed especially for checking basic code layout.

can also use for some preliminary optimization exploration

*Fourth Goal: Do simulation, optimization studies on superconducting linac using the matrix dynamics model.*

*Fifth Goal: Implement full Poisson modeling for the superconducting linac*

5. Program external field grid, metal surface boundaries, open boundaries, as required for Poisson simulation of superconducting rf cavities, magnets, beam pipes.

5.a Adapt poissonE_SC.f90 for superconducting linac

5.b Integrate superconducting Poisson subroutines into *LINACS*.

*Sixth Goal: Prepare LINACS for very-low-beam-loss studies on supercomputer*

6.a Many particles, parallelization

6.b Extended analysis methods, multiple methods of halo identification (Li Chao method?), etc.

[eltoc]

# Chapter 32 – Expert System for RFQ

## 32.1 The Concept of Optimization, Application to RFQ

The question of what a "best" RFQ is, and how a "best" RFQ could be obtained, has been a constant companion and motivation for decades, the development of the vast panoply of ideas and tools was followed in a broad literature, and tried on occasion (Ch. 11 footnote).

The languages and tools have evolved, been applied, improved, superceded. In general, the **element** can be described as "optimization" – how to make the best or most effective use of something.

The term "expert system" was popular in the 1980's, intending to capture the ephemeral talent of an "expert" that goes beyond a straight-forward procedure to produce an extraordinary result. Modern language includes "neural network", "machine learning", "deep learning". Each individual method (and often language) has been developed to address a particular problem area. But they are all optimization. They all require human judgment, problem definition, many decisions, very significant programming (also when assisted by "high-level languages, toolkits" which eliminate programming all the underlying details), debugging – and thus are prone to all the usual problems.

The following references and excerpts are chosen to give an introduction:

https://en.wikipedia.org/wiki/Expert_system

> "In artificial intelligence, an **expert system** is a computer system emulating the decision-making ability of a human expert.[1] Expert systems are designed to solve complex problems by reasoning through bodies of knowledge, represented mainly as if–then rules rather than through conventional procedural code.[2] The first expert systems were created in the 1970s and then proliferated in the 1980s.[3] Expert systems were among the first truly successful forms of artificial intelligence (AI) software.[4][5][6][7][8] An expert system is divided into two subsystems: the inference engine and the knowledge base. The knowledge base represents facts and rules. The inference engine applies the rules to the known facts to deduce new facts. Inference engines can also include explanation and debugging abilities."
>
> "In the 1990s and beyond, the term *expert system* and the idea of a standalone AI system mostly dropped from the IT lexicon. There are two interpretations of this. One is that "expert systems failed": the IT world moved on because expert systems did not deliver on their over-hyped promise.[31][32] The other is the mirror opposite, that expert systems were simply victims of their success: as IT professionals grasped concepts such as rule engines, such tools migrated from being standalone tools for developing special purpose *expert* systems, to being one of many standard tools."

A "modern approach" (fad problem driven) often uses the methods of "neural networks (NNs)", "deep learning". This very clear tutorial reveals that the neural network is optimization in the **element**al sense. It includes some deep comments, such as that NNs are not artificial intelligence (AI), but that one could suppose that in a few decades, they might lead to AI,

> Neural Networks Tutorials/Neural networks and deep learning.html :
>
> Chapter 1 of Michael A. Nielsen, "Neural Networks and Deep Learning", Determination Press, 2015

NNs are applied in many directions. Application to RFQ design can be separated into two approaches – supervised or unsupervised learning.

> https://www.tutorialspoint.com/artificial_neural_network/index.htm
>
> **https://en.wikipedia.org/wiki/Unsupervised_learning**
>
> **Unsupervised learning** (UL) is a type of algorithm that learns patterns from untagged data. The hope is that through mimicry, the machine is forced to build a compact internal representation of its world. In contrast to Supervised Learning (SL) where data is tagged by a human, eg. as "car" or "fish" etc, UL exhibits self-organization that captures patterns as neuronal predelections or probability densities.[1] The other levels in the supervision spectrum are Reinforcement Learning where the machine is given only a numerical performance score as its guidance, and Semi-supervised learning where a smaller portion of the data is tagged. Two broad methods in UL are Neural Networks and Probabilistic Methods.

The interest of Peiyong Jiang in achieving an optimum RFQ design via unsupervised learning has motivated this Chapter in 2021. He envisions an approach, using NNs, without specifying any physics conditions or example database.

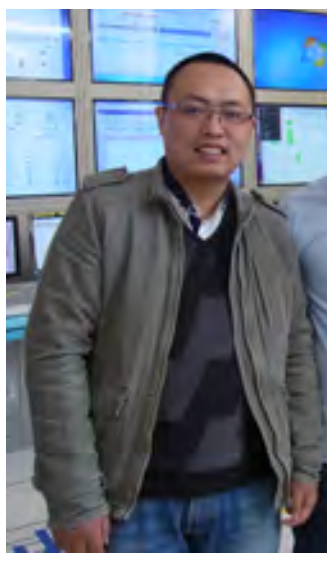

Fig. 32.1 Peiyong Jiang - was my visit coordinator in 2018, resulting in his obtaining a Post-Doc at GSI, 20 km south of Frankfurt Germany, from 2019-2022, to work on an advanced AI method for analyzing experimental physics data. Although limited by the corona pandemic, we got together often, Peiyong riding his bike up and back to visit us, cooking for us, and always teaching us many very interesting Chinese details. He saw my aquarium figures of two old men fishing, and declared that the unglazed one was the original Jiang, who was smart not to wish to become a king but was highly valued by the courts for his wisdom. On retirement, he often sat by a pond and "fished" - with no hook on his line. So my figure was immediately modified accordingly. Peiyong is a documented ~350th descendant.

My approach is that an RFQ design is based on a complete physical model of the beam dynamics, which is the basis of "supervised learning" of how to adjust design parameters.

There are also firm engineering considerations, which would use "supervised learning".

And also more nebulous interacting, often conflicting, considerations that are decided by "experts" based on "the state-of-the-art", "experience", "best practice" (which are often actually not up-to date or using the best knowledge of real experts, due to things like the often very long time delays and costs involved in proliferating, and accepting, new techniques). It is of interest to attempt to capture some of these aspects, perhaps with "unsupervised" methods.

Optimization techniques designed to use existing databases for training or mining are not appropriate for RFQ design. There are not that many constructed RFQs, most of those were designed with incomplete and inaccurate physics (even with the same inaccurate software), most were not tested very much against their design details, most did not even achieve their design transmission (much less accelerated beam fraction or beam loss pattern) so closely (within say 2-3%).

It does not make sense to use incomplete, inaccurate models.

Therefore, tools for optimization should be chosen for the RFQ which incorporate these facts, and not because they are in the latest hot research area or actually intended for far different problems, such as social engineering (manipulation), for which the large dangers of having wrong answers are clearly obvious seeing the multitudinous definition and programming details described in even just the above references.

The optimization tool NPSOL has been used for decades for a wide variety of equation solutions and problem solving in *LINACS*:

USER'S GUIDE FOR NPSOL 5.0: A FORTRAN PACKAGE FOR NONLINEAR PROGRAMMING

Philip E. GILL (UC San Diego), Walter MURRAY & Michael A. SAUNDERS (Stanford U.), Margaret H. Wright, Bell Laboratories, Technical Report SOL 86-6_. Revised June 4, 2001

Abstract

NPSOL is a set of Fortran subroutines for minimizing a smooth function subject to constraints, which may include simple bounds on the variables, linear constraints and smooth nonlinear constraints. (NPSOL may also be used for unconstrained, bound-constrained and linearly constrained optimization.) The user provides subroutines to define the objective and constraint functions and (optionally) their first derivatives. NPSOL is not intended for large sparse problems, but there is no fixed limit on problem size.

NPSOL uses a sequential quadratic programming (SQP) algorithm, in which each search direction is the solution of a QP subproblem. Bounds, linear constraints and nonlinear constraints are treated separately. NPSOL requires relatively few evaluations of the problem functions. Hence it is especially effective if the objective or constraint functions are expensive to evaluate.

It clearly contains all the required optimization elements to investigate the Expert System level. It is basically, **element**arily, similar to NNs as described by Nielsen above.

It is important to point out that the Abstract can be extended to add a key point on the unsupervised training aspects: *the bounds, linear constraints and nonlinear constraints, and weights (rewards) in the objective function(s) can also be optimization variables.* (like the follow-on sentence in Sacherer's article on the rms envelope equations!)

The "optimization" will be explored in two separate steps, as actually when designing for a project.

<u>First step is to do the rms physics design:</u>

a. Consideration of the spec. Based on particle type, current, output energy, experience is used to decide on frequency, input energy and transverse emittance, Kilpatrick factor, vane modulation type (depends on cavity design expertise available), i.e., deciding on basic form of RFQ to be used.

There could be outer loop of optimization on these also...

b. Initially specify all the other physics parameters (the GUI, rfqfileGUI).

c. Then the model is run, and tables made of various things.

d. The optimizer variables would be simpler things like a length limitation, copper power limitation, aspects of modulation and aperture, voltage, (assume synchronous phase is hardwired to best physical practice Teplyalov rule, but also has adjustable parameters).

At this stage still "simple" only. I eyeball the table, and check that the design reaches final energy; modulation and synchronous phase required to be monotonic and not saturated too long; phase advances within reasonable bounds; total emittances = 5*erms some factor less than the transverse acceptance, bucket dimensions.

See Ch.29 for new method using truncation in the shaper and following region to improve longitudinal emittance and EP characteristics, while maintaining reaching EP at the end of the shaper, and beyond if desired. **This standard robust EP goal shaper will be used.**

**Variables that could be optimized:**

KPfac
(etnrms in – usually given from ion source experience) , elrms in, variables in User rules
aperture factor – *primary variable*
phistgt
Bfraction
RMS cells
Siglint
Porch,%
(Keep shpr Vrule at vKP)
ffadj
(keep mainrfqphis rule Teplyakov, but “Tep lfac dist” and phislimit are variables
mainrfqaperrule “aper incr”
mainrfqvrule variables
mainrfqemrule variables depending on strategy
<u>*the bounds, linear constraints and nonlinear constraints, and weights (rewards) in the objective function(s)*</u>

**Observations that are made from the tables produced after design run on main RFQ:**

- shaper design successful, #rms & #shaper cells, phis, V, m, B obeying given rules
- aperture at EOS – optimizable variable
- Initial and final B, phis, m,a, tfac, lfac.
- Final length, Pcu
- Sig0t<90°. Initial and final sigt/sig0t, sigl/sig0l
- Monotonic increase of m, phis
- sig0l<sig0t, sigl<sigt
- Depending on strategy, rules for B,phis, em, V, arfq, r0rfq being followed
- Tfac, lfac beam stay-clear from aperture, bucket adequate.
- Pcu reasonable
- If EP used, EP ratios correct

<u>Second step</u>: - after doing the above "optimization" based on rms model with no beam, simulate full beam and gather info. Optimizing on some of that is obvious - transmission, accelerated beam fraction, beam loss. Much more sophisticated would be beam loss pattern in energy and location.

Much much more sophisticated, when it isn't working right, usually the case for quite a long time of at least some days, use experience, *a priori* adjustments, intuition, imagination to find ways to change the physics parameters and rules, and try to meet the spec.

*The bounds, linear constraints and nonlinear constraints, and weights (rewards) in the objective function(s) can be optimization variables.*

I know how I basically proceed, but it is really complicated, will be daunting to program, will take lengthy runs, will be difficult to make converge, etc.

There are examples in this book of how I had to do this for various cases... They are put there for exactly this reason - that getting the desired final result is not obvious.

Need to make a specific outline of how to attempt this, what observations, what decisions, in what order, etc. - *before diving into programming and running*.

# Finale

To those who even only browsed to here, I hope you will be interested to study further. To those who have read carefully to here, congratulations on meeting the challenge given in Italian at the beginning. And to those who study, please excuse the probable pedagogical deficiencies and incompleteness. My hope is that you will have found some pearls [282], will go beyond me, extend these interests of mine, and succeed in bringing them into "the state-of-the-art" – generate some pearls of your own….

[eltoc] [gotoCalm]

[282] "When we could be diving for pearls", Editorial, Charles Day, Physics Today, March 2022

# CALM IN THE RESONANCES AND OTHER TALES

## R.A. JAMESON

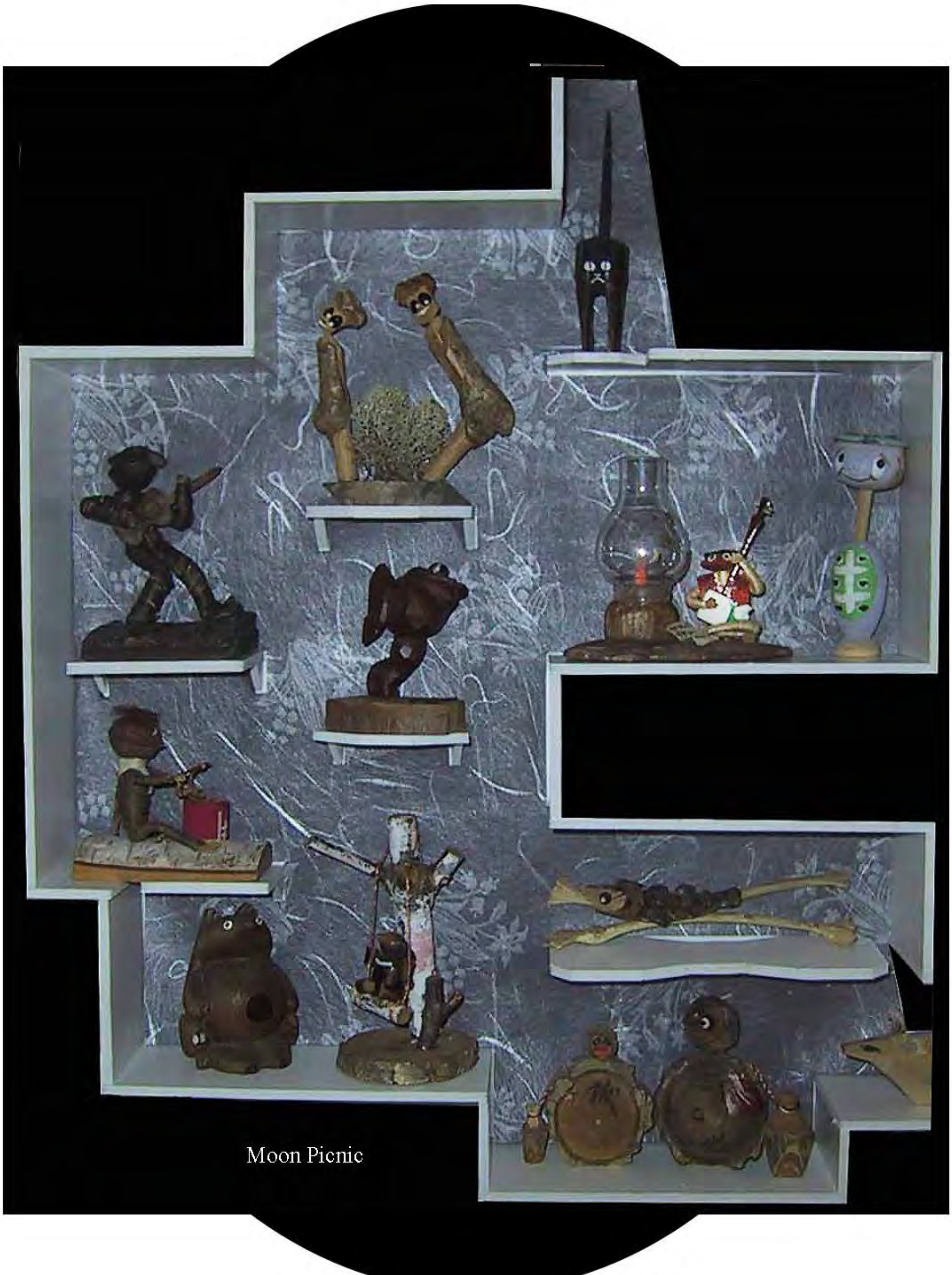

# Calm in the Resonances and other Tales

R.A. Jameson

Contains eyewitness accounts of ~60 years of participation in accelerator technology. "Calm in the Resonances" of course alludes to equipartitioning and the buffeting of life – keeping an EP state is a good life philosophy as well as a good linac philosophy. Stay calm, enjoy what you're doing; if you have to deal with resonances, do it fast.

Table of Contents [TOC]

Many stories – of historical interest in themselves as inside view, fodder for many conversations, urged to write down by many, including Mahlon Wilson (1/12/2007), Nikolay Lazarev’s urging that “otherwise no one will even remember that there was such a thing as emittance growth and no idea why…” [1] and encouraged by his example of so much writing effort after his retirement.

A lot of the history is nothing new in general – the victories were great. Some great successes are related, and the names of heroes are given in full, and enhanced where possible by photographs, or sketches made by Carol Coppersmith [2]. Fate was not – characterized as always by problems with

---

1 Email 3/6/2007. Dear Bob, I hope in time to read much more interesting for me thing - remembering of your life. I know a lot of fields where you left significant achievements beginning of RF Control of LAMPF, Feed-forward scheme, PIGMI, FMIT, and other High-Current CW Projects, Transmutation Problem, etc. It can be if not you, nobody notice and will know of emittance blow-up and this effect quite not exists? Seriously, I would be glad if you share with me your memoirs on the eve of our close very circular dates. My the best wishes, your old friend Nikolay.

2 Carol Coppersmith (1925-2010) was my greatly valued secretary from 1978-1982 as Alternate AT-Division Leader.

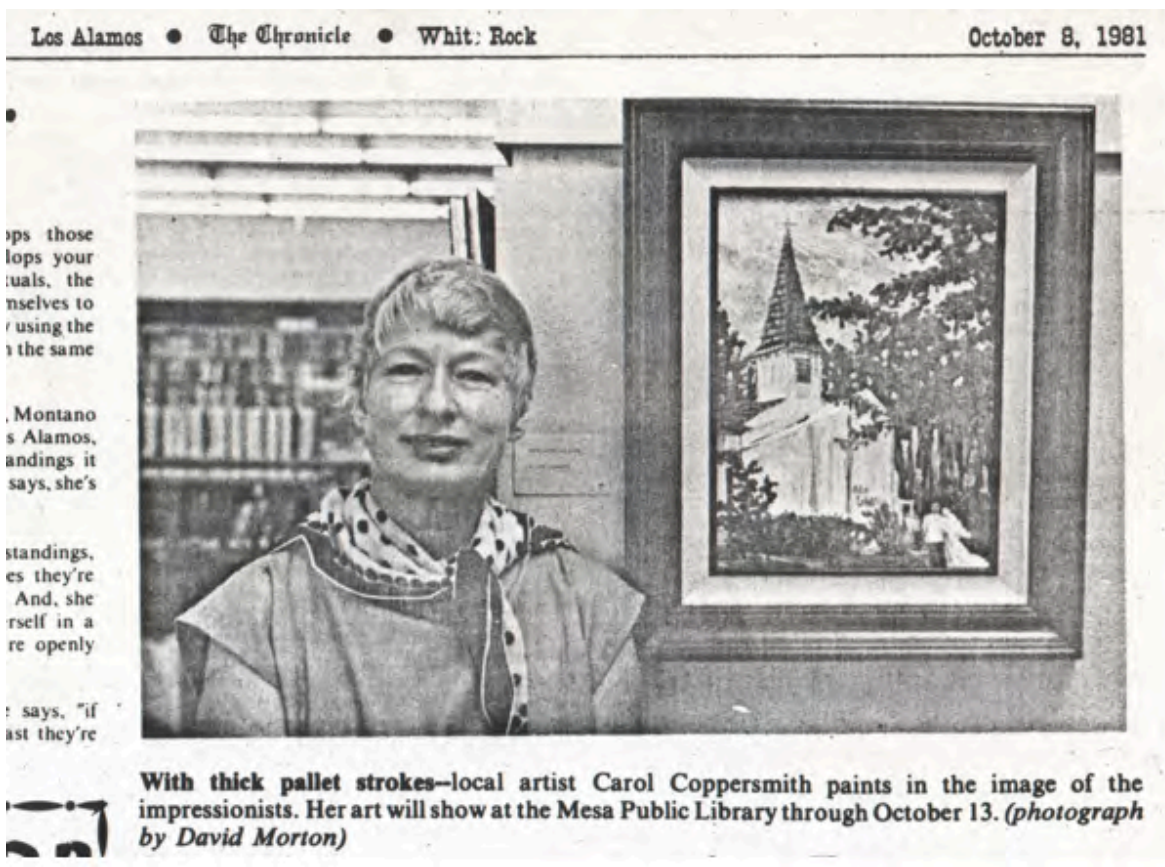

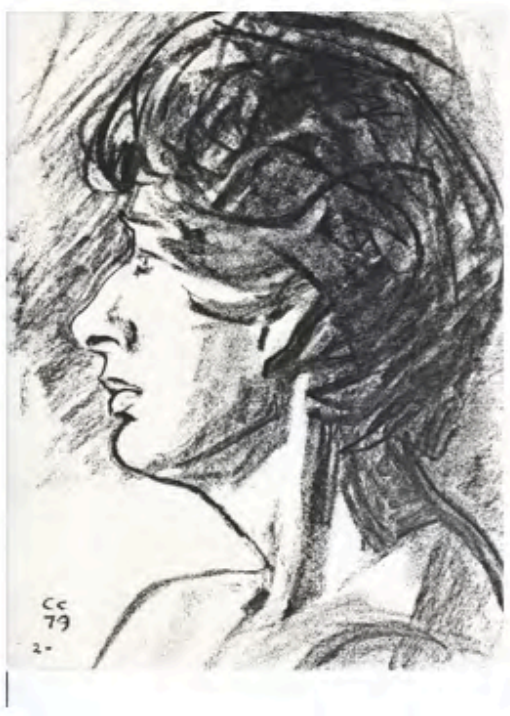
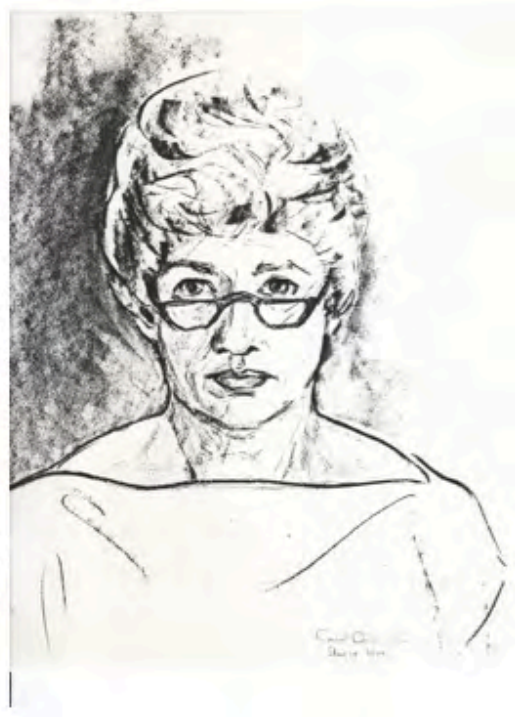
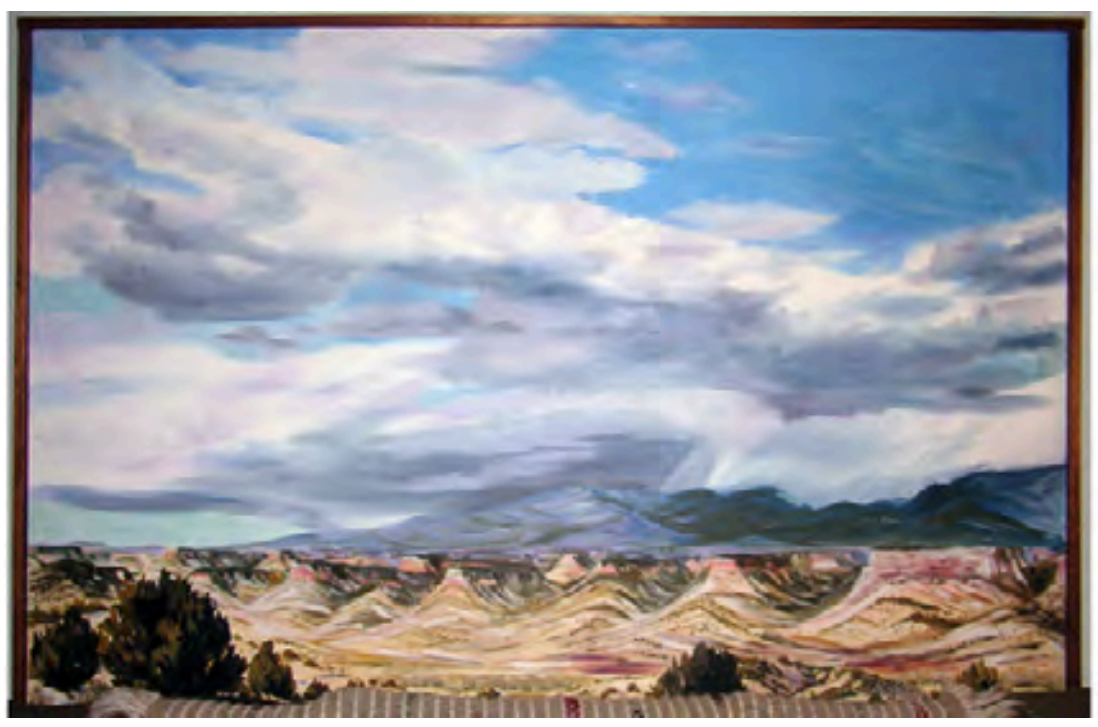
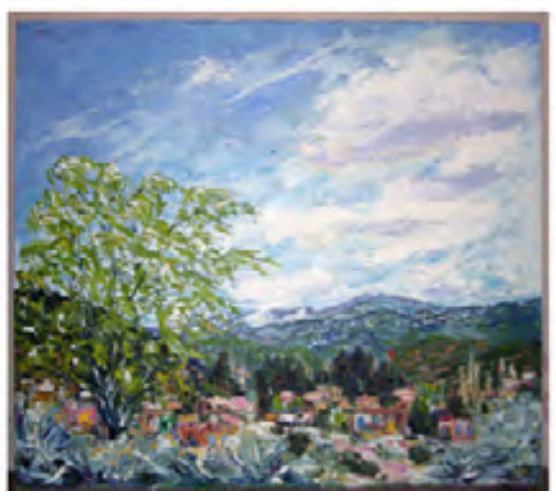
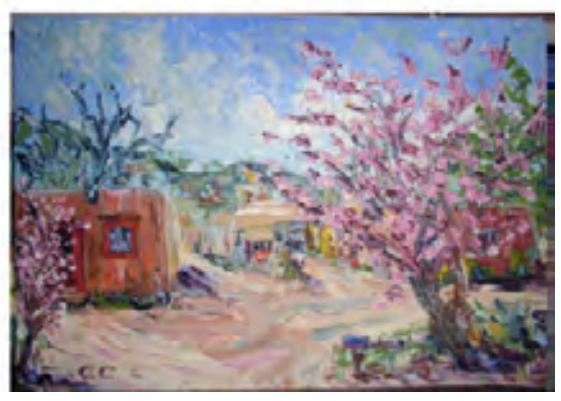

She was an avid artist, especially in New Mexico landscapes in oil, and in portrait sketches, which she would do during lunch hours. Most of her paintings were on small canvases; once I said a great artist should paint a big canvas, and bought her a ~6’x4’ one along with plenty of paints. The result is a landscape looking back toward Los Alamos and the Jemez mountains from some point east of the San Ildefonso Pueblo on the road to Pojoaque. I still have this and a number of her other works.

persons motivated by power and/or fame rather than hoping to make some lasting contribution. [3]. One wonders about the interest we all have in failures and why conversations always migrate to these – perhaps because they occur so often, are so often related to problem people, and have such long lasting effects. The stories are about events we had to live through. No names, honor or fascination are due to villains or power or fame hungry, and they appear here only in terms of their damage, or sometimes by the acronyms or nicknames or phrases that are used, even as a "zero" when that, rounded up from negative, was the case. For example, "regime" to characterize LANL directors after the decline and then free fall which started as Harold Agnew left (and warned us what was coming). Herein the "regime" numbering differs by one from the director succession – it starts after WWII from Norris Bradbury as the first, with Harold Agnew as the second. The derogatory use of this term starts with the third, and particularly with the disastrous fourth – "the $4^{th}$ regime".

I hope the reader will mostly remember the successes.

Many notes, diaries and records were kept over the years from which accounts could be directly extracted, so it has not been necessary to work only from memory. Bound logbooks cover the early LASL, LAMPF, and some of the LANL days up to 1988 – some of the pages with the painful happenings under the $4^{th}$ regime are clamped shut with spring paper clamps. All foreign travel was extensively journaled, handwritten up to late 1998, scanned to computer, and then continuously by computer from then on.

The first story is about the fascinating discovery and development of the Radio Frequency Quadrupole (RFQ), which revolutionized ion linacs.

The following chronicles are arranged in roughly chronological order from ~1963 to the present.

---

3 "Weep for Isabelle, A Rhapsody in a Minor Key", Mel Month, *Paperback*, *692 Pages*, *Published 2003 by Authorhouse ISBN-13: 978-1-4107-3253-8, ISBN: 1-4107-3253-3;*
"Dreams and Shadows: An Inside Story of Science", Mel Month, *Paperback, 432 Pages, Published in 2009, Paperback Edition*. Overview: Leon joins forces with Doug and Jay - Doug who is lording it over Bill, and Jay who has the ear of President Reagan. A conspiracy is set in motion. The science community rumbles, with many crushed in the quaking earth. And the tale fate has writ plays out to its inevitable end.

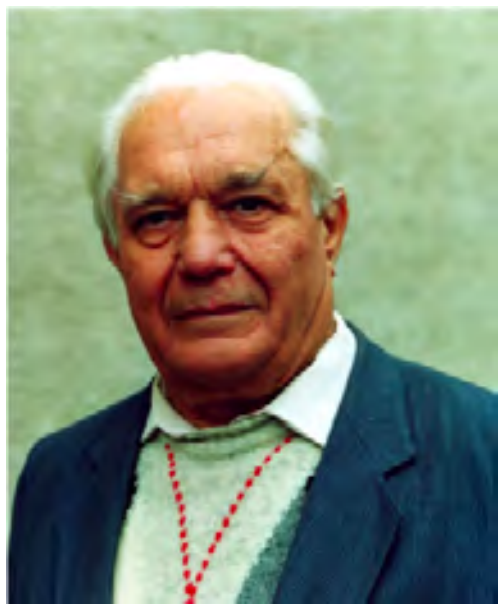

# The RFQ

[toc]

Vladimir Alexandrovitch Teplyakov (11/6/1925 – 12/10/2009)

## Teplyakov's Story

(From notes written during the night after having dinner with Teplyakov on October 19, 1996.) (Teplyakov -> T, Kapchinsky -> K)

Conversations about the RFQ with various people during the summer of 1996; seeing Teplyakov at Protvino in July; and then the great chance to work with him and the IFMIF CDA team for two weeks at Frascati in October - led to long conversation on Friday evening, 10/18, with a couple of younger guys, in which they were talking about the history of the RFQ, wanting to know how I remembered it, and wishing they were brave enough to ask Teplyakov directly about it. I told them there was no reason at all why they shouldn't, that he would probably be glad to tell them about it. And realized later that I am always too diffident in this respect too, and also had never asked him myself. So resolved to try to connect with him on Saturday, and fortunately did...

Dinner with Vladimir Alexandrovitch Teplyakov at Cantina Bacca, Frascati, October 19th, 1996. It was his habit to eat very little in the evening and he only shared a little of what I ordered. Asked him to tell his story of the RFQ...

~1953 - working under Simonov at Institute for Chemistry. Needed thorium fission cycle - no uranium had been discovered yet in Russia.
T invented a 2-gap buncher, with realization that he could get both transverse and longitudinal focusing. No one wanted to listen - some that he tried to talk to said what he was saying was impossible according to Earnshaw's Theorem.
(Regret - said it was a short step to alternating gradient focusing, but he never got to more than two gaps and so lost his chance to invent AG focusing.)

~late 1950's - Uranium had been discovered. Had to do something else, so started work on electronuclear breeding. By this time, T had already invented the idea of finger-type focusing.
Worked ~15 years trying to make this type of focusing practical (~1953-1968).

~1966 - moved to the Institute for High Energy Physics (IHEP), Protvino to join Kapchinsky's group.

Before 1968-1969, T had a working experiment using a radiofrequency quadrupole (RFQ). (T said this later in the conversation - earlier I thought he meant an experiment using finger-type focusing, but he was emphatic later that it was a working RFQ.) In earlier days, K had not wanted to listen to T's ideas about RFQ-type focusing.

By 1968, K had moved to the Institute for Theoretical and Experimental Physics in Moscow. There was continued interaction between the IHEP and ITEP groups.

~1968 - K came in one day and asked "Why can't we put together the transverse focusing and longitudinal acceleration?" This question caused T to make the final leap to a really practical, simple and elegant solution for the RFQ. Then he and K worked day and night for about two days. K made the theory complete. (T made practical theory based on simple formulas, but K was the theoretician.)

1968-1969 - K wrote papers, ~ three with T; including the one at the 1969 Kharkov Conference (think this was an Intl. HEP Conference).
T said he did not like to write papers; said he would have never written a paper without K.

Ron Martin was at the Kharkov Conference, and told T later that he had heard their paper and did not understand it (this took guts - Martin is one of those men that T describes as ‘a mighty man’). Livdahl was also there.

Later, T would not participate in any more of K’s papers - said he didn’t like to publish repetitively, and nothing new had transpired.

T told that there were two times when K came in and asked crucial questions, after which there was a spurt of development. The second was when K came in and asked about filing for invention, saying he (K) wanted to include Vladimirsky. So the filing was in their three names. V was not on any of the publications.

(I think my collection has a paper ~1975 on experimental tests - may have been first publication on tests going back to 1968 - may have references, I remember not being able to find any evidence of experimental work for quite awhile after 1968-1969...)

Using a napkin to sketch on, T told how he had ‘made a silly mistake’ with his first drift-tube type finger-focused structure. It had severe pulse breakup. T recognized immediately what it was - a mechanical longitudinal resonance of the drift tubes that was changing the gaps - but it took quite a long time to get it fixed and try again.

Then I told my story...

T said he had had a request from Manca for a preprint (1969 paper) and had sent him one via regular post - never knew if it was received. But the paper Manca had at Los Alamos was in English translated from German, and we never knew who had translated it from Russian to German; so suggests Manca never got T’s post.

T liked very much when I said the RFQ had lain in the literature 10 years before we (West) discovered it in 1977 via Manca. T got in some trouble from his political bosses -’How did West find out about the RFQ?’ He told them “They have a requirement for cw machines and thus they would have to invent such a structure”.

T says Horst Klein was the first foreigner to give a paper on the RFQ in Russia (this was after we at LASL had told HK about the RFQ and were working on it together, ~1979).

Asked T if he would write this history. Other Russians have apparently also asked him to. Then he said he had taught at MEPI for two years, and developed (invented) an explanation for linear accelerators based purely on the idea of symmetry, and this pleased him very much. So should write that up too. I said I’d help him get his writing translated to many other foreign languages.

---

Toward the end of the IFMIF CDA meeting, there was not much for T to do. He spent several days working steadily on page after page of equations. He has thought of a new way to express the RFQ and linacs, as indicated above, but the few sentences he explained were not clear to me - need another opportunity to meet...

---

So my version of the RFQ history had been quite accurate, with respect to the roles of T and K, timing, etc. But great to have heard it directly from him. (There have been some kind of derogatory versions, relayed to me by some younger guys, that helped make me determined to ask T directly - one was that T had had very little to do with it, just make the hardware; another, that K had only been the boss and had not really participated...)

Need to have HK tell me his version. (Was able to do this during the summer of 1997 while in Frankfurt. Horst had been of the opinion that T had mainly made hardware, and that K had had the idea as well as doing the theory. Horst had arranged for the JWG Uni-Frankfurt to present K with an honorary doctorate (date? - around 1992). T later asked Horst why he did not also get an honorary doctorate; Horst replied that K had done many things besides the RFQ theory, such as the KV distribution, etc; T said ok. Until this summer, Horst was unaware of the real story...)

[toc]

## Russian View

The question of priority was discussed thoroughly in Russia also, and renewed at the time of the EPAC 2006 Prize. It is generally as above, couched in terms also of great respect for Kapchinsky. Following from Nikolay Lazarev, who was very close to I.M. Kapchinsky and often represented him at conferences: (nikolay.v.lazarev@itep.ru)

"Dear colleague, on November 5 at IHEP (Institute High Energy Physics) it will be the 80th anniversary of Professor V.A. Teplyakov, the inventor of principle and the first RFQ constructions." (email 10/20/2005)

" As concern of true priority: Kapchinsky's view on this problem was written in his report "History of RFQ Development" (LINAC-84, p.43). It is true, that all complex experimental works (were) carried out by Teplyakov with his (Teplyakov's) command. Finding of the best configuration of cavity, questions of RF supply, struggle against multipactor and other discharges, high-vacuum problems etc. All this were weight of cares on shoulders of V.A and overcome by his genius.

Three authors of Discovery (#350, registered 1991, with priority earlier 1970) ˆ V.V.Vladimirsky, I.M.Kapchinsky and V.A. Teplyakov and in Certificate of Invention (#265312 registered 1970) are written there in order of Russian alphabet. The first of them (V.V.) early proposed drift tubes with horned electrodes, the 2nd (I.M.K) "provided theoretical support, both they always participated in useful discussions and supported of V.A. by all kind of help: scientific, administrative and friendly support. They had very fruitful discussions. I.M. was about 5 years in that times Head of Injector Division at IHEP including Teplyakov's Group and V.V. ˆ main author of Project of 76-GeV Synchrotron had very wide range of talents and duties, he was Vice-Director of ITEP headed neutron-physical and reactors researches and had interests in elementary particle physics. I.M. and V.A. worked always with very tight contact." (email from N. Lazarev, 10/25/2005)

[toc]

## Kapchinsky's Review

"History of RFQ Development", I.M. Kapchinsky, 1984 Linac conference, p.43-48.

Independently from V.V. Vladimirskiy, V.A. Teplyakov with colleagues in 1962 proposed to implement a spatially-periodic RFQ-focusing by means of a system with drift tubes having rectangular aperture holes[4]. The orientation of the rectangular aperture holes is periodically changed by $90^{\circ}$ (Fig. 2). The drift tubes with rectangular holes were also considered by the group of P.M. Lapostolle[5].

….

The idea of spatially-homogenous quadrupole focusing (SHQF) has become a further step in the development of RFQ focusing. USSR patent on SHQF structure was received by V.V.Vladimirskiy (ITEP), I.M.Kapchinskiy (ITEP) and V.A.Teplyakov (IHEP) with priority from October 25, 1968[9]. First publications appeared in 1969[10,11].

….

[toc]

## Teplyakov's EPAC 2006 Prize Paper

"The First CW Accelerator in USSR and A Birth of Accelerator Field Focusing", Prof. Dr. V.A. Teplyakov, EPAC 2006, THPPA03, pp. 2755-2758. Extracts:

"Forty five years ago I with Gennady Anisimov [3] put forward the idea of FAF - focusing by the accelerating field. At first this idea was published in our book [2]." (D.V. Karetnikov et al, Linear Accelerators of Ions, Gosatomizdat, 1962.)"

"In 1963 my article [5] appeared. I showed there that focusing by high-frequency quadrupoles may be quite effective. (V.A. Teplyakov, PTE, vol. 6, 1964)

"To distribute potentials at the accelerator electrodes properly, I, together with Victor Stepanov had to invent a new type of cavity - the H-cavity [6]."

"When Boris Shembel reported our invention of this new type of accelerator to one of the heads of our Institute he was advised "not to stick out": let others judge us. Nevertheless we managed to build a small proton accelerator with FAF effect and tested it successfully. The PT-500 cw accelerator was launched into operation. It accelerated protons of 150 mA current from 70 up 500 keV. In 1966 the ideas of thermonuclear synthesis went out of favour. It was decided to disband Boris Shembel's group. I moved to Protvino near Moscow, to the Institute for High Energy Physics."

"…November 1967, by the launch time of the U-70 ring accelerator. Having provided reliable operation of I-100 we had an opportunity to continue the development of the accelerator with focusing by the accelerating field. …. For quite some time I did not manage to realize how to organize the work of a buncher with several drift tubes. And suddenly (after fifteen years of searches) it came to me: in the Initial Part of Accelerator (IPA) it is necessary to form clots of particles with constant density of the charge! By this time I already knew that it was possible and I knew how to do it. An elementary theory for the particle dynamics was ready by the next day. Ilya Kapchinsky very actively contributed to the development of a more comprehensive theory of dynamics of particles in IPA. As a theorist, he was able to work quickly and energetically. On my part I still could not for a long time find the necessary manufacturing know-how for the electrodes and resonator. It was clear to me that a four chamber resonator (like in magnetrons) suited IPA. Ilya Kapchinsky liked this idea very much. So, in two evenings in Protvino he sketched the theory of the four chamber resonator. But to me it was still not clear how to make it."

"At the VII Yerevan Conference on Accelerators (1969), Kapchinsky, Maltsev and myself reported on the development of the 30 MeV proton energy accelerator with quadrupole focusing [9]. However, very few people took this project seriously."

[toc]

## 2008 Paper

In 2008, Teplyakov and Maltsev published an expanded history [4]. Excerpts from a roughly edited machine translation:

*"I must say that IM Kapchinsky first did not take seriously the high-frequency quadrupole space-uniform focusing. No, he did not deny it, but doubted its efficiency. Anyway* when *I asked him in 1966 to be opponent in the defense of my PhD thesis, he initially agreed, and we discussed various issues, but then he refused to under the pretext that I was under the protection of Alexei A. Naumov, another of the opponents. … There were many doubters. Gradually the attitude of Kapchinsky to (space-uniform focusing (SUF)) began to change for the better. Thus, for example, he did not in 1968 oppose the deployment in the Department of the injector works on SUF. Somehow one day Kapchinsky walked into our room, where VB Stepanov was working on the 4 MeV H-resonator model. The container has been removed, and the resonator loaded electrodes were all in sight. It was the first run of the beam resonator. Fingers had been fastened onto rings, and the entire channel looked entirely made up of horns* ***(Fig. 13).*** *Kapchinsky came and looked in surprise, asked: "Where is the drift tube?" And then I mused: "It turns out that you can do without them*" *... I think it was at that moment I came to the idea of focusing and acceleration in the channel* u*sing a four-wire line."  By happy degree circumstances, Vladimir Alexandrovich Teplyakov also at that time had the idea of an incompressible bunch. The combination of these two basic ideas gave rise to the radiofrequency focusing quadrupole (RFQ).*

*Further events unfolded rapidly. Operating system and basic proposition of the theory were completed in one or two months and published in the famous papers [15]. Working-out theory was presented in two preprints (Separate preprint issued IM Kapchinsky [20] separately and in collaboration - AP Maltsev, VB Stepanov and V.A.Teplyakov [21]). At the same time, parameters were calculated for a full-scale accelerator ... .. This project was reported at the International Conference on Accelerators (Armenia) in 1969. [16] Running with the RFQ ... took place in 1972 [18]. Despite the fact that all the ideas were published by us, and we were visited by experts from various institutes who have seen our models and working accelerators, as before, many remained skeptical. The RFQ accelerator at IHEP was operated for 8 years, but no one dared to do further develop RFQ development. Even in ITEP, the practical realization of this idea only began in 1977, although the principles of the formation of the bunch in the RFQ, which* are *now considered classic and look clear and obvious, were first developed and demonstrated much earlier at IHEP.*

(A technical note:
*" ... programs for numerical calculation components of the electric field in the accelerating channel with SUF. Developed a method based on a combination of the Fourier method with the Monte Carlo method, which has the advantage that the calculation uses the potential of the needed points on the border of the aperture as required to solve the problem in all areas. Method has been so successful that we gave users for decades (and still use)."*)

[toc]

## The Los Alamos RFQ Story

### The Challenge of the Fusion Materials Irradiation Test (FMIT) Program, and Joe Manca

Another section relates how the embryonic Accelerator Technology Division of the Los Alamos Scientific Laboratory was faced with this program. Extending the conventional approach to a cw

4 V.A. Teplyakov & A.P. Maltsev, "Ion linear accelerators with high-frequency quadrupole focusing in IHEP", Russian title: "ИФВЭ Magasine "Новости и проблемы фундаментальной физики" №2, 2008, Тепляков.", Journal of the Institute of High Energy Physics, ISSN: 19992858, Key name: Novosti i problemy fundamental'noj fiziki (Print), Abbreviated key title: Nov. probl. fundam. fiz. (Print), EAN13: 9771999285501, ISSN: 19992866, publisher IHEP, not available in English.
Vladimir Teplyakov - Wikipedia, the free encyclopedia, http://en.wikipedia.org/wiki/Vladimir_Teplyakov

multi-100meV Cockcroft-Walton injector into a drift-tube linac had to form the original plan presented to the DOE. But it was so formidable that it surely opened us to a new approach.

Joe Manca approached us at the Proton *Linear Accelerator Conference*, 14 - 17 Sep *1976*, Chalk River, Canada, seeking work. and was welcomed to a two-year appointment. However, after coming, he was very reticent. I would check with him about once a week to see if he and his work were ok, but he would only say that he was studying “a new structure for low-velocity acceleration that had no transverse emittance growth and 100% transmission”, and demurred any further comment. That sounded like an impossibility, or at least a miracle. I was very curious. Politeness prevented forcing the issue, so I played a kind of ruse on Joe. I had a continuing interest since LAMPF in accelerating higher beam currents and organized a workshop on space charge in linacs [5]. (It is interesting to note, when now they so ubiquitous, that this was one of the first workshops in the field and permission had to be obtained from the DOE, in order to facilitate travel approval for the participants!) I asked Joe to present a paper, to see if he would make the seemingly miraculous claim in public. He did, with little detail, halting English, and little attention from the audience, but he did. Then immediately after the session, I said that now he had made in public a very important statement and to please show me the details and his source material. It turned out that he had only one paper in German translated from Russian, and the magic words “essentially no emittance growth” and “almost 100% transmission” were indeed there. A copy of the paper in Russian was found via the LANL Library, and L. Cernicek confirmed the translations. Later I listened to a talk by the famous physicist George Gamov about how they had to work in those days. He said they had to be so careful that they always kept secret their recent work and kept a backlog of about two years, only revealing work up to that point. Joe had been trained and had lived under that system, which most probably explained his initial reticence [6].

We quickly decided that the risk of developing this exciting approach was less than that of trying to build a cw, very high voltage Cockcroft-Walton injector. Then it was my job to convince the DOE to fund a prototype development program for a brand new accelerator structure, totally unknown in the West, invented in the Cold War enemy USSR!

[toc]

## Los Alamos Proof-of-Principle RFQ

During the next two years, we had an ideal R&D program – a very good team, enough resources, left alone to do our work. A proof-of-principle (POP) RFQ was demonstrated at Los Alamos in 1980. [7]

To make it very clear that the Los Alamos RFQ development was totally a team effort and to be sure that credit for development of the RFQ outside Russia was properly given, I arranged for and ghost-wrote an article in the Los Alamos Scientific Laboratory magazine “Atom”, July/August 1980. As this document is hard to find and is the best contemporary account, it is included in full here (RFQ-Atom1980_7_8.pdf).

---

5 R. A. Jameson, Ed., “Space Charge in Linear Accelerators Workshop", Los Alamos Scientific Laboratory Report LA-7265-C, Conference Proceedings, May 1978.

6 J. Manca, “Some New Accelerating Structures for High Current Intensity Accelerators”, LASL Report LA-7157-MS, UC-28, Issued: March 1978.

7 R. W. Hamm, K. R. Crandall, L. D. Hansborough, J. M. Potter, G. W. Rodenz, R. H. Stokes, J. E. Stovall, D. A. Swenson, T. P. Wangler, C. W. Fuller, M. D. Machalek, R. A. Jameson, E. A. Knapp, and S. W. Williams, “The RF Quadrupole Linac: A New Low-Energy Accelerator", Proc. 2nd Int. Conf. on Low Energy Ion Beams, Bath, England, April 14-17, 1980, Inst. of Phys. Conf. Ser. No. 54: Chapter 2, p. 54, April 1980; Los Alamos Scientific Laboratory Report LA-UR-80-1091, 9 April 1980.

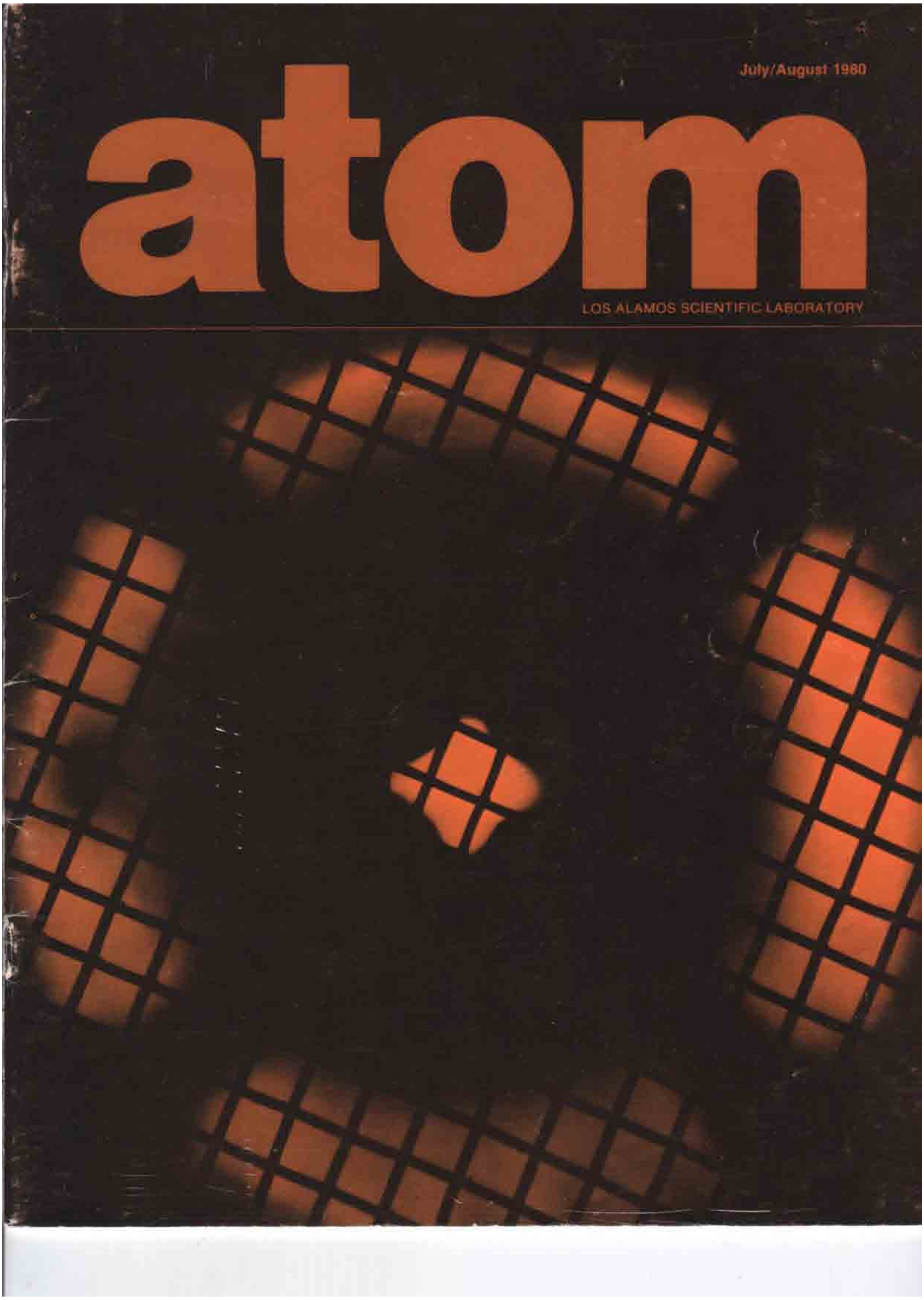
July/August 1980
atom
LOS ALAMOS SCIENTIFIC LABORATORY

# atom

contents

*ON THE COVER: Sparks illuminate the vanes during the successful test of the RFQ device. Photo by Henry Ortega. AT LEFT: Launching track for vehicles leaving the moon would use less energy to attain escape velocity than would a vertical rocket takeoff. Illustration by Krafft Ehricke. (See story on p. 28.)*

western union

Telegram

# RFQ IS ALIVE AND WELL...

**The combined requirements of continuous duty, high current and high reliability posed a unique challenge to accelerator researchers.**

By JOHN T. AHEARNE

One of the telegrams sent during the jubilant hours after the extraordinarily successful Valentine's Day inaugural test of the radio-frequency quadrupole (RFQ) was to I.M. Kapchinskii at the U.S.S.R.'s Institute for Theoretical and Experimental Physics in Moscow. It simply stated, "The RFQ is alive and well at the Los Alamos Scientific Laboratory."

Eight years earlier, Kapchinskii had outlined a theory which described a linear accelerator system capable of focusing and accelerating low velocity subatomic particles with virtually no loss of particles from the beam. The concept was so revolutionary that, when it resurfaced at a LASL workshop in 1977, it was met with considerable skepticism.

The theory found its way to Los Alamos via J.J. Manca, a Russian-educated Czech who years earlier had emigrated to Canada and, in 1977, was a two-year visiting staff member working here on means to accelerate particles at low velocities.

In October of that year, Bob Jameson (now with the Accelerator Technology Division) organized a workshop to address the questions of beam quality—particle loss and the unwanted spreading of the beam in accelerators.

Jameson asked Manca to present a paper at the workshop concerning his persistent notion that nearly complete particle capture at low velocities was possible—an idea that many felt was a misinterpretation perhaps caused by translation difficulties.

Manca repeated the same "astounding" statements in his presentation, writing, "The injection energy can be as low as 50 keV and the capture coefficient remains close to 100 percent." (The injection energy required to initiate the beam at LAMPF is 750 keV.)

"Because the concept was just what we were looking for, and because Manca was so persistent, we initiated a literature search and the Kapchinskii paper was the first analytical description we found," Jameson said.

The LASL researchers discovered the paper in German, translated from Russian, and Jameson asked A.D. Cernicek to alleviate any possible language misinterpretations. Cernicek, an ISD-4 translator, teaches German and Russian at Los Alamos High School.

The translator's efforts showed that Manca was correct in his interpretation of the literature. The formulas in the Kapchinskii paper were quickly analyzed, and excitement for the new concept began to grow.

A team of LASL experts on accelerator structure development began to consider ways to turn the theory into reality. The group included Ed Knapp, now AT Division Leader, and AT-1 Group Leader Don Swenson. (AT Division was formed in January of 1978.) The pair studied the theory, drew pencil sketches of their mental pictures of the structures, and

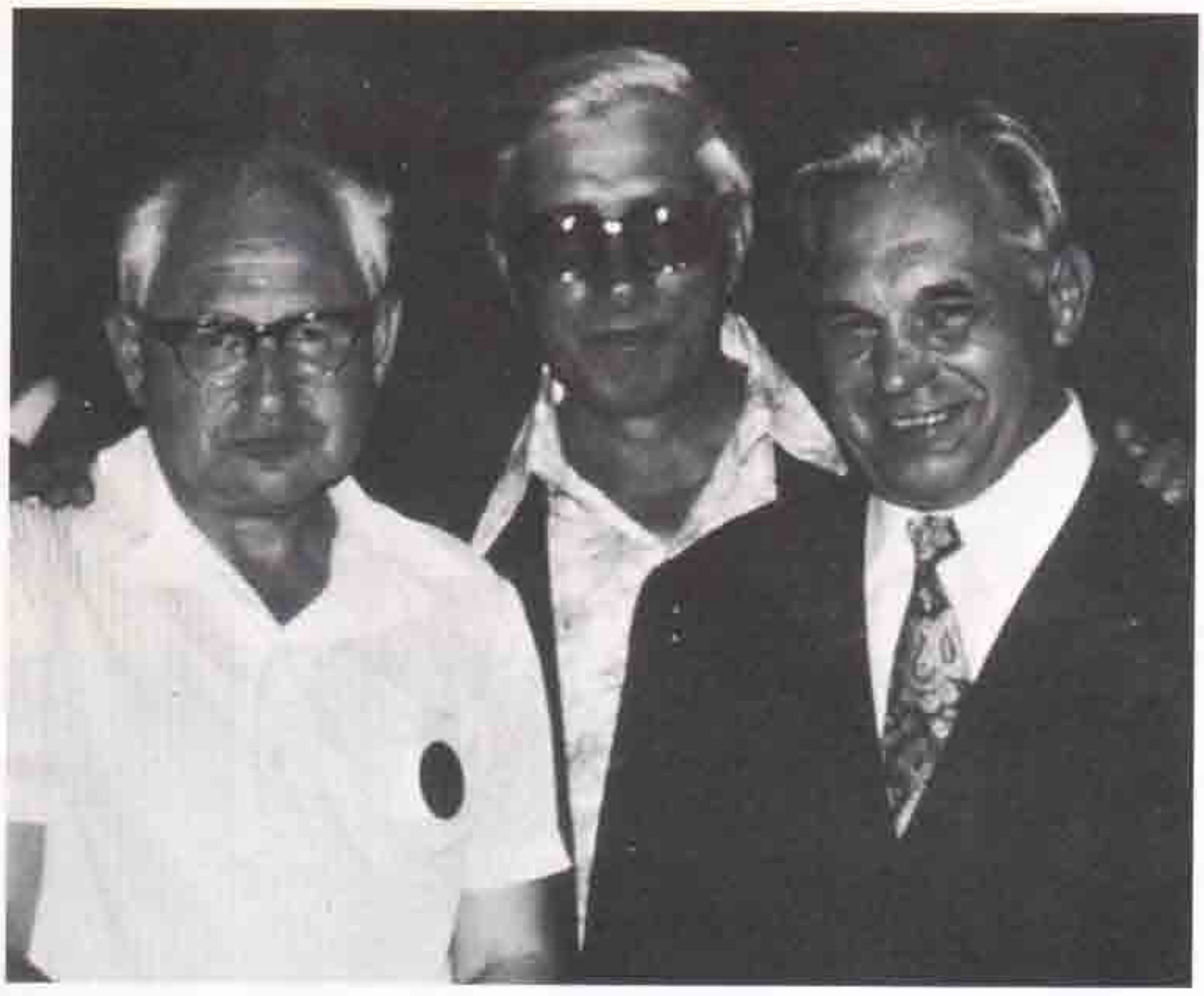

*Eight years ago, I.M. Kapchinskii (left) outlined a revolutionary theory for a linear accelerator. Later, V.A. Teplyakov (right), a co-inventor with Kapchinskii, reported on an experimental device. (Colleague Andreev in center)*

eventually came up with a plastic model they could use as a visual aid in explaining the structure to others.

Then in February of 1978, LASL was asked, by the Office of Fusion Research, Office of Energy Research at DOE, to design and build an accelerator, for Hanford's Fusion Materials Irradiation Test (FMIT) facility, capable of producing an intense and high quality beam. The particle accelerator necessary for the FMIT "neutron factory" would require significant advances beyond the state-of-the-art technology of the time. The accelerator would have a 10- to 20-year useful operating life. Its output would be 0.1 amperes, which is a billion-billion particles a second.

Because most of today's high-current accelerators operate only in a pulsed mode and produce 10 to 100 times less current, the FMIT's combined requirements of continuous duty, high current, and high reliability posed a unique challenge to LASL's Accelerator Technology Division. Totally new accelerator techniques would have to be developed for the FMIT. There at the right place and the right time was the RFQ. The highly reliable RFQ would meet these requirements economically and efficiently (see accompanying story).

Two factors were primarily responsible for the FMIT sponsor's eventual acceptance of the RFQ theory as the avenue toward meeting the requirements at Hanford. First was the realization that meeting the FMIT needs in the old ways would present immense technical and engineering difficulties. Second was the dwindling of scientific skepticism in the accelerator community after the AT researchers held a spring 1978 workshop to explain the promise and potential of the new structure. The go-ahead was given to design and build the RFQ for Hanford.

Three general steps would have to be accomplished before the first test. First, a complete understanding of the theory by the team was necessary. Swenson and Knapp had developed a general understanding of the theory—they were the ones who had a feel for the structure. With their physical descriptions and models, they began to help others understand, and moved the team toward the second necessary step to fruition—to develop computer codes to simulate how particles would move through such a structure.

Dick Stokes, who had worked on accelerator structures at the Van de Graaff, came from P-Division to guide the overall program and develop the theory in more detail. Beam dynamicist Ken Crandall, and later Bruce Chidley from Canada's Chalk River National Laboratory, began creating "some pretty fancy codes" to describe the future RFQ. Tom Wangler came from Argonne to help with theory and design.

Gary Rodenz of AT-4 bridged the gap between the second and third steps along the experimental path by using the computer codes in the actual designing of the first RFQ. Jim Potter, an expert on radio-frequency cavities and accelerator structures, was to be the man who "put the metal together."

The hardware part of the experiment had to be completely done from scratch. AT researchers made repeated requests to visit V.A. Teplyakov at the Serpukov Institute for High Energy Physics, U.S.S.R., who was a co-inventor with Kapchinskii and had reported in the literature on an experimental device. They hoped to see what sort of hardware approach the inventors had taken. (Ed Knapp did have the opportunity to discuss the theory with Kapchinskii.) Despite two years of effort, permission for the trip was never received from the Soviets, and it is not known for sure whether an RFQ structure was ever built in the U.S.S.R.

Potter, aided by Fred Humphry, made a number of sheet-metal models and tuned them in his laboratory, developing procedures to achieve the required field distributions in the structure, and working out a method to provide radio-frequency power. In parallel with Potter's activities, Hanford's Steve Williams made a high power test section to investigate how much voltage could be applied to the vanes before sparking between them occurred. Potter and Williams also began to develop more computer codes for the most vital step in producing the first proof-of-principle RFQ—the delicate and precise job of cutting the vanes in LASL's Shops Department (SD).

AT scientists agree that the RFQ development would have been impossible without SD, headed by Joseph F.B. Szoo, and its one-of-a-kind numerical milling machine. The machine, programmed and monitored by SD experts, is capable of milling metal to precise tolerances as described by mathematical formulas. At least three dozen machinists and technicians were involved in the RFQ work; the main job of cutting the vanes was done in Shop 13 of SD-1, headed by Ed Riggs with supervisor Leroy Wampler. Working closely with Potter and Williams on the milling machine programs was Otto Maier of SD-DO. Given that the waves and scallops on the vanes of the RFQ must conform to the movement of particles through electrical fields alternating at 425 million cycles a second, the accomplishment of such precision is, in itself, a remarkable machining feat.

Long-time mechanical engineer Chuck Fuller, AT-4, began the assembly for the first RFQ test. For the experiment with the real beam, sheet metal assemblies had to be replaced with a carefully engineered structure. Fuller, whose accelerator engineering experience began at PHERMEX, had to insure that the final precision required was achieved after all the parts were assembled.

Final assembly began in the fall of 1979 in Jim Stovall's PIGMI laboratory. Under overall proof-of-principle coordinator Milt Machalek, an AT-1/AT-4 team readied the hardware for the final

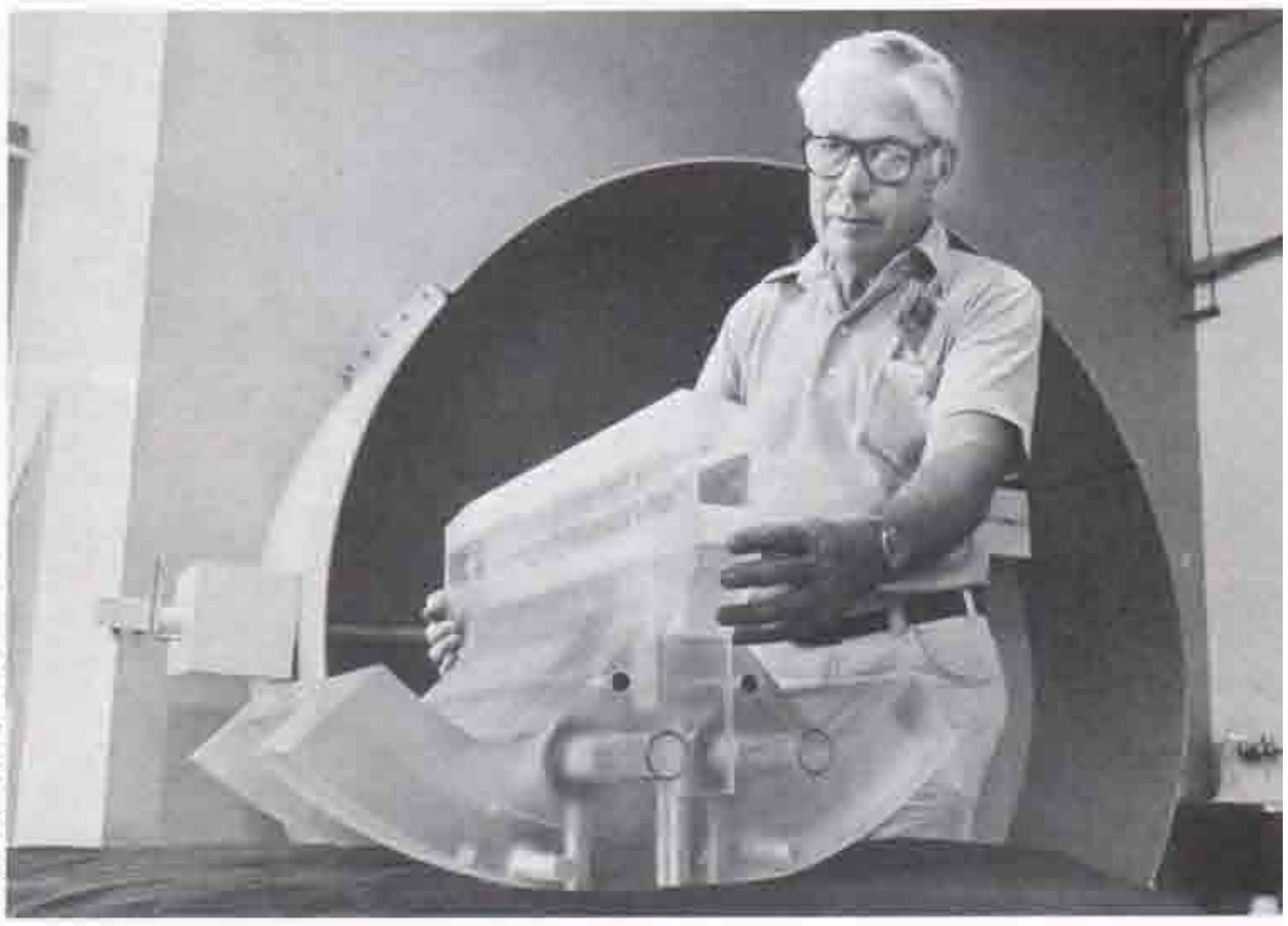

Photo by LeRoy N. Sanchez

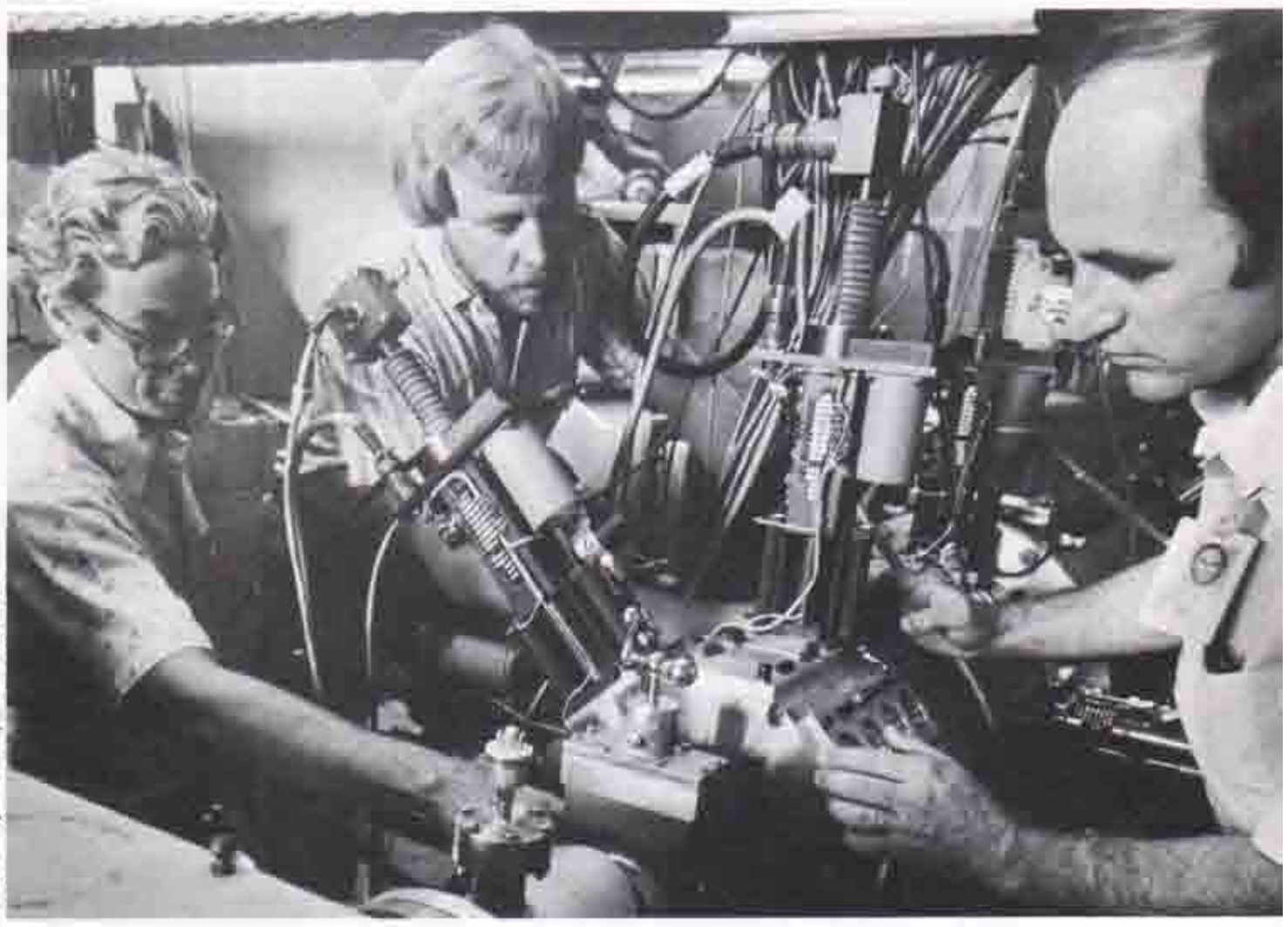

Photo by LeRoy N. Sanchez

*TOP: Chuck Fuller displays an engineering design model. The plastic model is a full-scale quarter section model of the RFQ for Hanford's Fusion Materials Irradiation Test facility. The circular structure behind him is part of the drift tube accelerator. BOTTOM: (Left to right) Jim Stovall, Steve Williams and Milt Machalek examine the experimental RFQ in the PIGMI lab.*

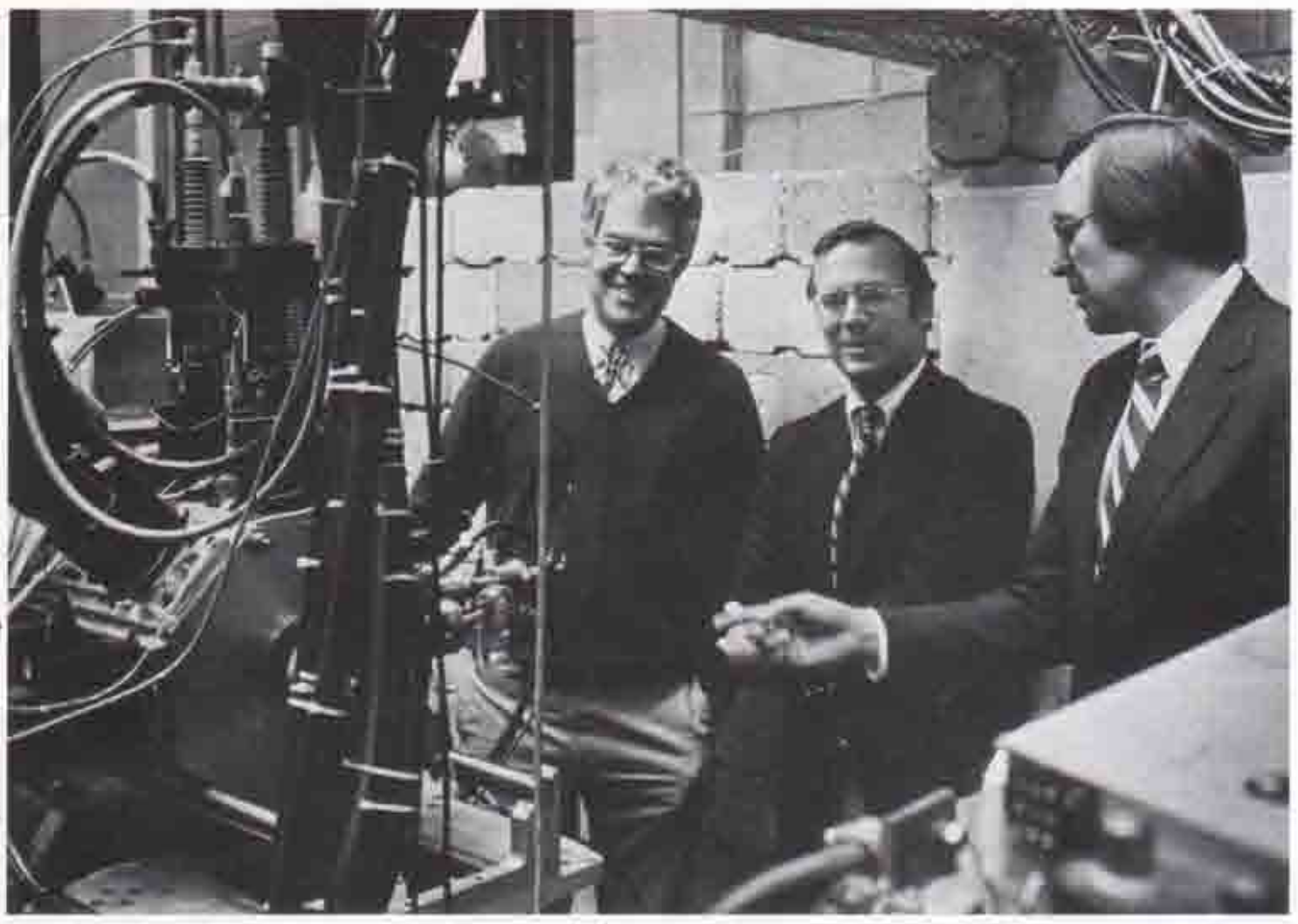

Photo by Bill Jack Rodgers

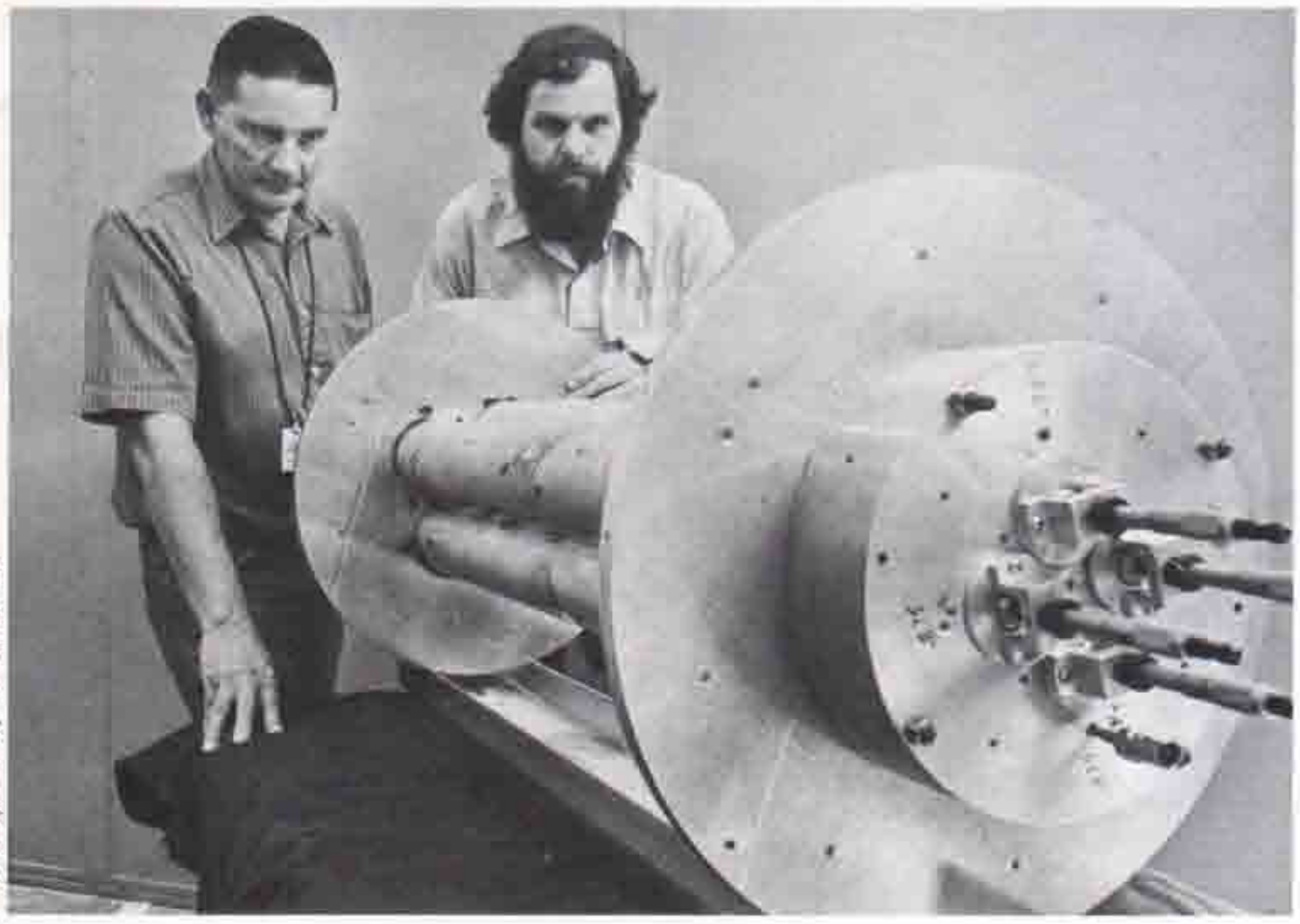

Photo by LeRoy N. Sanchez

*TOP: AT Division Leader Ed Knapp and Laboratory Director Don Kerr proudly show off the RFQ to the President of Westinghouse Hanford Co., John Yasinsky. BOTTOM: AT researchers, Arlo Thomas (left) and Jim Potter, check out a cloverleaf model of an RFQ whose outer jacket has been removed.*

## The missing link in linear accelerator development is progressing quickly from theory, to experimental device, to application.

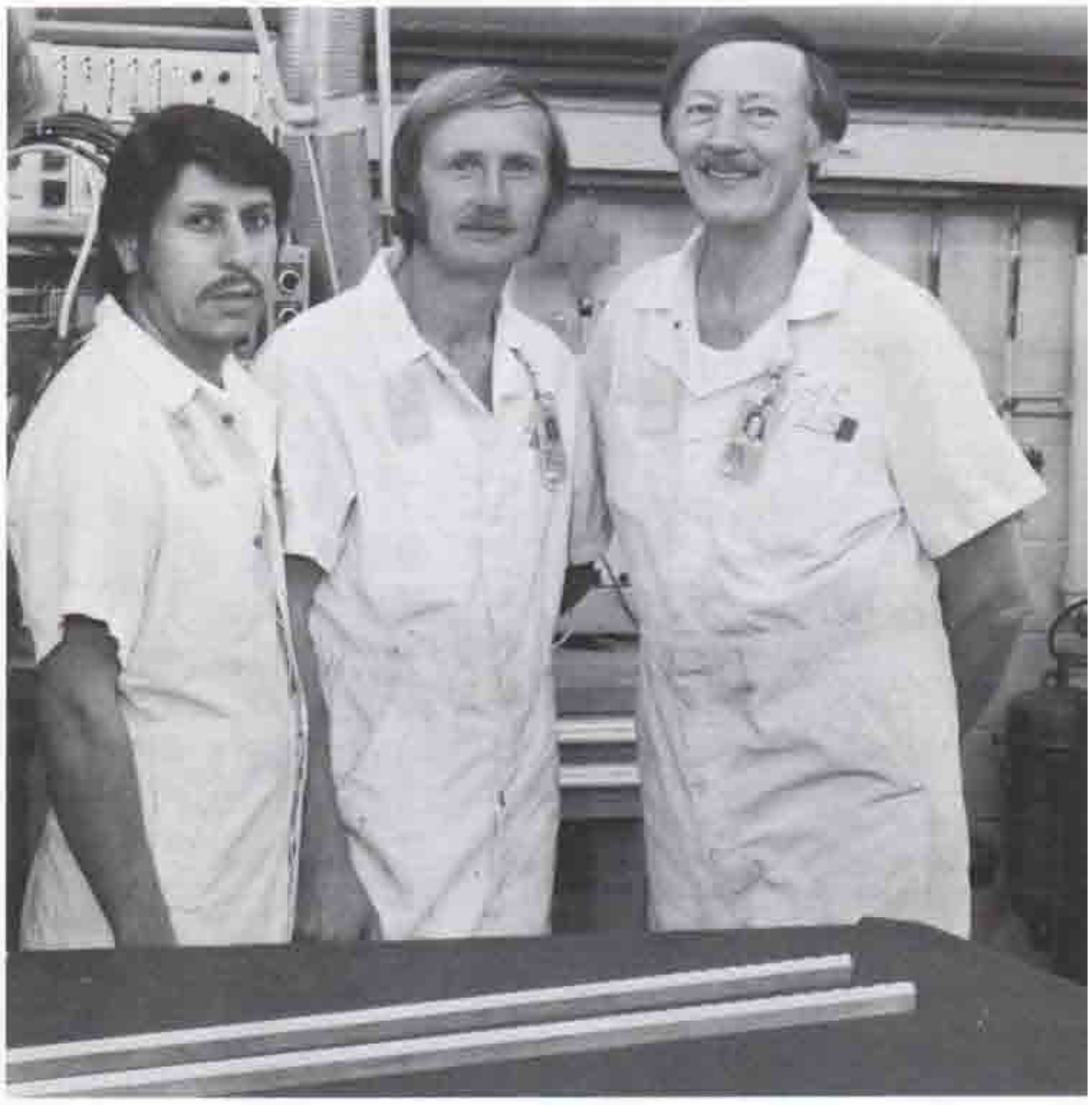

*Two of the RFQ vanes are displayed by Shop 13 machinists (left to right) Alex Lopez, George Zakar and Ed Andolsek. Photo by Bob Peña*

phase of the experiment. Bob Hamm reconfigured the injector systems he had helped build for PIGMI to provide the proper input beam for the RFQ. Hamm, Crandall and Kim Melson laid out the input and output beam transfer systems. Jim Stovall, Dave Chamberlin, Brian Smith and their crews readied the beam transport lines, vacuum and beam diagnostic systems necessary to verify the RFQ performance. An SD team working with P Division, including Ray Squires and Donald Marien, machined most of the remaining RFQ parts, guided assembly, and made a tuning jig to precisely adjust the equipment.

In mid-February, 1980, everything was ready. The structure itself provided the drama for the initial run. As the radio-frequency current was turned on low, the device began sparking across the vanes—an expected and undesirable occurrence. When sparking occurs, researchers must wait until the structure

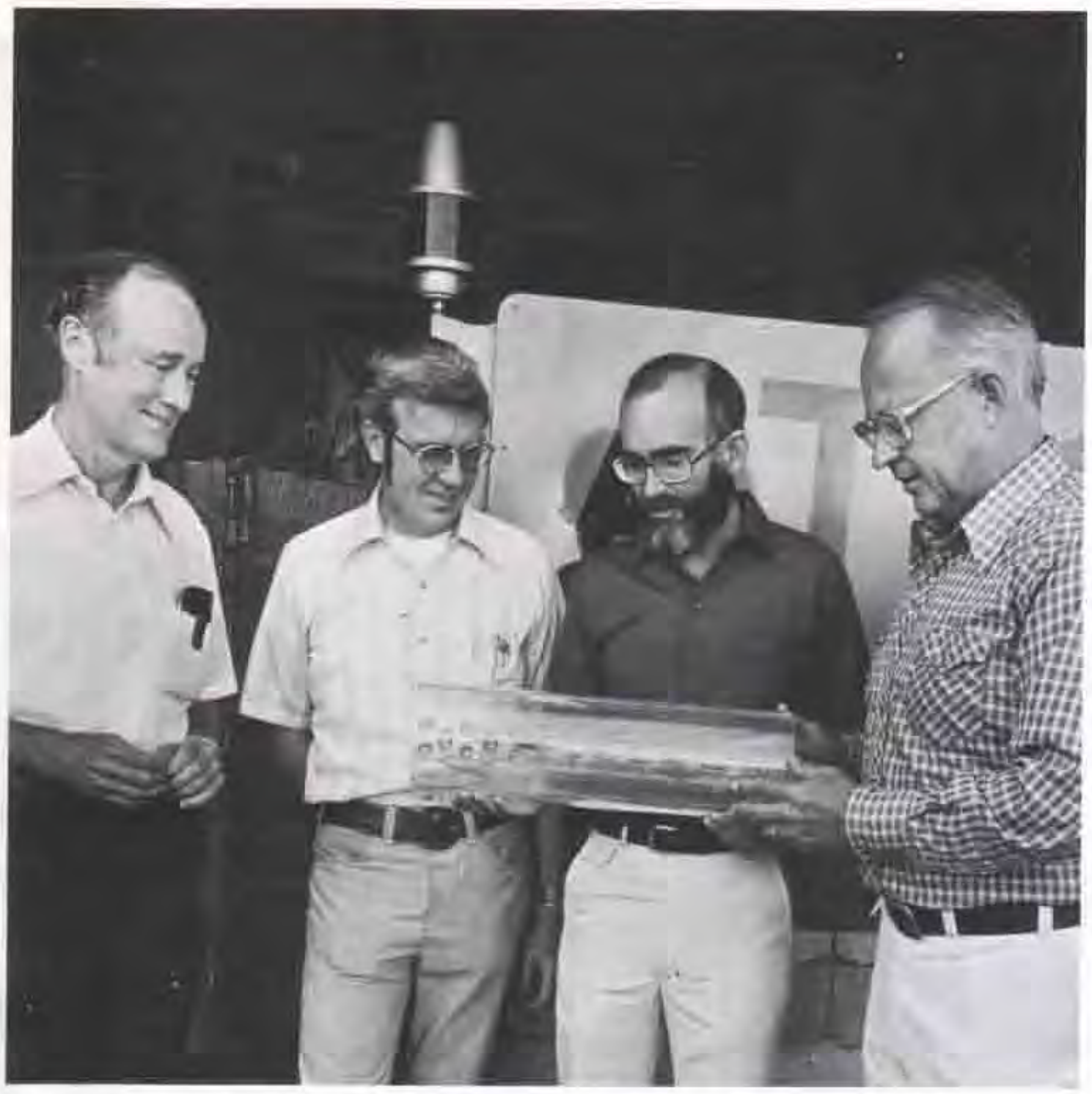

*Dick Stokes, Bob Jameson, Tom Wangler and Don Swenson reminisce about the days when the RFQ was just a preliminary conceptual model.*
*Photo by LeRoy N. Sanchez*

becomes "accustomed" to the current and stops sparking. Then they add more power until it sparks again. The question was whether the desired power could be reached with no sparking. Power was reached. They turned on the beam. The device, representing more than two years of work by dozens of scientists, engineers, technicians and machinists, operated immediately and perfectly.

With the success, and while testing continues, LASL scientists can now speculate more on the potential of the RFQ. Because of its simplicity and economy, the RFQ promises to be an important development in the evolution of particle accelerators for medical applications (to produce neutrons for cancer therapy or pi-mesons for pion therapy in a hospital environment), in the generation of research accelerators, and in the practical application of accelerator technology to fields such as heavy ion driven fusion.

# What's an RFQ?

Don Swenson called the radio-frequency quadrupole (RFQ) "revolutionary" and the "missing link" in linear accelerator (linac) technology.

The initial tests on the RFQ represented a giant step forward in this technology because of its ability to accelerate subatomic particles and to focus them into a coherent beam using a powerful radio-frequency electrical field.

Linacs, such as LAMPF, are primarily research devices that are used as an artificial source of well-defined beams of subatomic particles for studies in nuclear science. In most linacs today, the focusing of the beam—keeping it in a tight, compact line—is done with magnetic fields. However, magnetic focusing is "velocity dependent." That is, the slower the particles in the beam are traveling, the stronger the magnetic field is required to counteract the disruptive effects of the electrical fields. At the low particle velocities required at the beginning of practical linacs, it is not possible to build strong enough magnets in the available space.

Therefore, in machines to date, the particles have to be raised to a higher velocity where magnetic focusing is possible. This is done using large power supplies, such as the 750 keV required at LAMPF, which are complex and costly. The RFQ, using only electrical fields, can focus the beam without regard to particle velocity, thereby having the ability to capture lower energy (slower) particles, and essentially eliminating the need for large and costly power supplies.

The RFQ also has an inherent capability of performing another vital operation, called "bunching," as the beam is being focused and prepared for entry into the main section of the linac. (This main section, which uses a different kind of accelerator structure and could be hundreds of meters long, is where particles are accelerated to the final velocities necessary for experiments or applications.)

The linac accelerates the particles by taking advantage of the principle of opposite charges attracting and like charges repelling. The electrical field that accelerates the particles is cyclical—that is, it changes from a positive to a negative orientation many times a second. The charged particles in the beam, therefore, must be ordered and arranged into sequential ball-shaped "bunches" so they enter the linac during the half cycle when the electrical field is charged to accept them. With present systems, the bunches are not well formed; they tend to have long tails. The particles in the tails tend not to be accelerated properly in the main section of the linac, often resulting in particle loss and degradation of beam quality. The RFQ, whose sophisticated design allows bunching with no tail formation, little degradation of beam quality and little particle loss, will provide a highly efficient and reliable link between the particle sources and the acceleration section of future linacs.

"Quadrupole" in RFQ means that the device has four poles (metal bars) inside a tube and running the length of the tube. The four poles are placed 90 degrees from one another and converge toward the center of the tube like the spokes of a wheel. The electrical field between the tips of the poles alternates its positive and negative orientation as the particles move down the tube, thus causing a net acceleration and focusing. The tips of the poles, which are nearly smooth on the entry end of the structure, are machined so that they slowly scallop in larger and larger waves toward the exit end. This scalloping is responsible for the inherent bunching and acceleration capability of the structure.

*Alex Lopez of the Shops Department oversees the machining of one of the RFQ vanes.*

*Photo by LeRoy N. Sanchez*

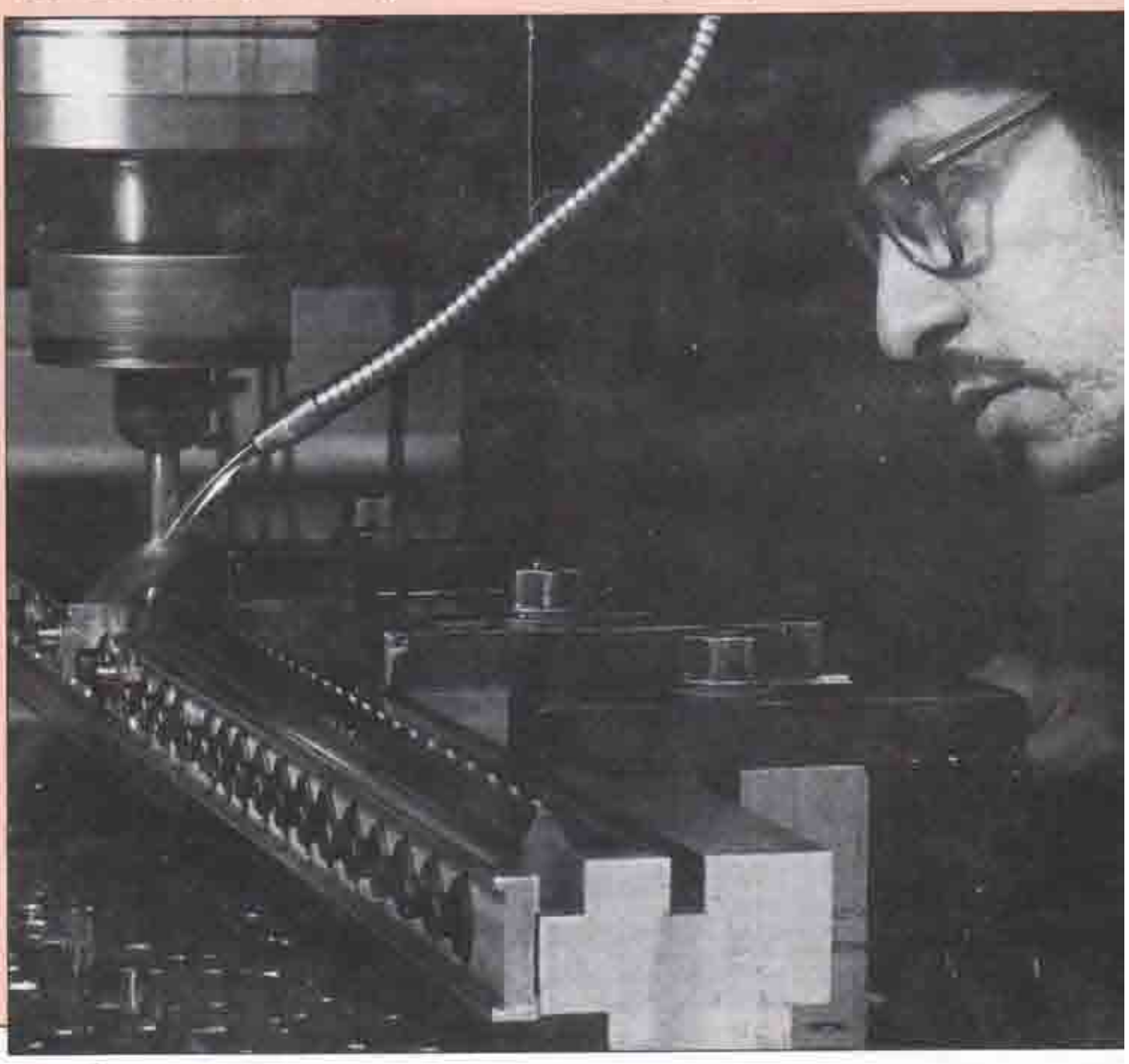

# Putting the Pieces Together at HEDL

The FMIT facility will produce large quantities of neutrons—uncharged nuclear particles—to test candidate materials for the world's first fusion reactors. This "neutron factory," a joint effort between the Hanford Engineering Development Laboratory (HEDL) and LASL will be located at HEDL in Richland, Washington. LASL, which has an FMIT group (AT-4) under Ed Kemp, will spend $35 million to $40 million to design and develop the particle accelerator necessary to drive the neutron-producing facility.

Until now, there has been no appropriate neutron-producing facility where samples of materials for fusion reactor walls could be evaluated. The FMIT will fill this need. It will produce a steady stream of neutrons at the same energy but at even greater intensity than those expected from a fusion reactor. The neutrons will be used to test various materials, such as metal alloys, that will be needed in the reactors. By placing samples in the facility's neutron stream, the effects of fusion neutrons on the candidate materials can be determined and the best materials for the reactor walls can be selected.

Neutrons will be produced in the factory using a form of hydrogen. In nature, about one hydrogen atom in 6,400 has a neutron attached to its single, positively charged proton nucleus. This proton-neutron nucleus is a form of hydrogen called heavy hydrogen, or deuterium. To produce a neutron stream, deuterium is injected into the FMIT and accelerated about 35,000 miles a second, or about one-fifth the speed of light. These high speed deuterium nuclei, called deuterons, pouring out of the accelerator at a rate of a billion-billion a second, each with an energy of 35 million electron volts, are directed onto a target of flowing liquid lithium metal. As they pass through the target, the lithium atoms strip off the deuterons' positively charged protons and let the uncharged neutrons pass through the target toward test samples.

Test samples of candidate materials only a fraction of an inch in size will be placed in the intense neutron stream. Some samples will remain in the stream for only a few minutes, others may be tested for as long as a year. In each case, scientists will look for test samples showing the most resistance to neutron-induced damage and thereby will choose the best reactor inner-wall material.

Scientists researching fusion energy believe that it will be the ultimate energy source for our planet. Hydrogen fuel, obtained from water, is abundant, and the amount of radioactivity produced by fusion will be less than the amount produced by fission reactor plants. Scientists continue to work diligently to achieve controlled thermonuclear fusion—which is perhaps the most difficult scientific and engineering problem they have ever faced. Researchers think that they are close to achieving control of fusion; however, beyond this stage much work will be needed to build a power-producing reactor.

The FMIT will make a significant contribution to the first fusion reactor design by providing information for the selection of reactor materials. These contributions will hasten the time when the world will benefit from abundant electrical power produced by fusion.

**atom**, July/August 1980 9

We invited visitors during the POP phase, and many came. Especially interested were Horst Klein and Alwin Schempp of the Institute for Applied Physics at the Goethe University in Frankfurt-am-Main, Germany. Alwin Schempp became the world's leading RFQ innovator, with over 50 RFQs to his credit over his long career, for R&D, and installed as operating machines in many laboratories all over the world.

We had to develop every aspect from scratch – the practical theory needed for our prototype, simulation codes, engineering. We had to arrange for import of the 5-axis numerical milling machine – the first in the US. The visitors were welcome, but one day a request came to establish "consultation hours" for questions, with other hours to be undisturbed for getting on with the work.

All but one of the interactions were good – the exception was when a publication was made without a single reference to the extensive time spent at Los Alamos. This incensed our team. For some time, it appeared there was nothing that could be done about it, but one day a top-level official from that country's agency visited our Director H. Agnew, and Ed Knapp got the idea to see if we could get 15 minutes of his time. The guy was sharp and immediately said he would take care of it. The leader of the visiting contingent was commanded to travel the considerable distance to headquarters once a week for six months to report on his progress.

[toc]

### Subsequent RFQs

Los Alamos continued to work on challenging RFQ applications – the 100mA, cw FMIT protoype, the high-intensity, high-brightness Star Wars investigations that culminated with the Beam Experiment Aboard Rocket (BEAR) RFQ, the Low Energy Demonstration Accelerator (LEDA) RFQ. The challenges of cw and high brightness afforded some learning by doing, but after making the corrections usual in advanced R&D, all the projects demonstrated their goals. There were some envious statements (from midland and the west coast) that Los Alamos seemed to have only problems. But when these labs finally also built RFQs, neither the design method nor the engineering were advanced beyond the Los Alamos methods.

As of 2022, an RFQ had not yet been installed in the LAMPF linac, although, after a very long time for getting approvals and arranging procurement, a 4-rod RFQ with only the old original beam dynamics design was delivered in first half of 2016 but has not yet been installed.

[toc]

## The RFQ Prizes

### R&D100

Product Name: Radio Frequency Quadrupole Linac Structure
Developer Name: Edward A. Knapp
Organization Name: Los Alamos National Laboratory (U.S. Dept. of Energy)
Award Year: 1981
http://www.rdmag.com/rd-award-winners-archive?field_organization_name_value=&field_award_year_value_1[value][year]=&title=quadrupole&field_developer_name_value=&items_per_page=20

Ed and I discussed who should accept the Award at the ceremony – Ed kept demurring and finally I went – the only time I ever rented a tuxedo…

### USPAS Prize 1988

Presented in Snowmass, Colorado on July 13, 1988: **I.M. Kapchinskii** and **V.A. Teplyakov,** For the Invention and Early Development of the Radio Frequency Quadrupole ;

I was asked by the USPAS Prize Committee to recommend who in the US should receive the prize. I outlined the "Atom" article, and insisted that the prize belonged to Teplyakov; they got the order mixed up.

**EPAC2006 Prize**

THPPA03 (Teplyakov EPAC2006 Prize Paper).pdf, THPPA03_TALK (Teplyakov EPAC2006 slides)

Again there was some discussion, in which I emphasized the USPAS decision and reasons for it; the decision was to award the prize to Teplyakov (Prof. Kapchinsky died in 1993). Teplyakov planned to go to Edinburgh to accept it, which was very pleasing as he had already for some years been unable to travel. I helped edit his talk. Then suddenly, a flurry of emails caught up with me to say that indeed he could not attend, and would I "give a Laudatio". I had stopped attending conferences for some time, but had already made an exception to be there on the day the prize would be awarded. I did not want the forum to be a Laudatio, but agreed to simply read Teplyakov's paper, with a few extra slides. It was an honor; the session attendance was large.

[toc]

## Teplyakov Tales

**Frascati Toothache, etc.**

(IFMIF Workshop 10/12-11/3 1996 (1986Oct-1998Dec.pdf))

24 October 1996. "… Then had 1½ hours to kill before supper, so wandered streets with T, who wanted to buy a bag. We found one in a toy shop – black vinyl leather small knapsack. Dropped it at T's room, no good to sit there, so out again. Then T said that "his tooth was very sick" and he had wanted to treat it with vodka, but had not been able to find any – only grappa, and that only made it worse. So deflected only a few doors later into a bar, inspected the shelves and saw a bottle of Finlandia vodka. T didn't know the bottles were for serving – guess he though would have to buy the whole thing. So he bought a "mezzo" – maybe something more than a jigger – for each of us. He held each sip carefully on a back tooth and swished it around awhile. Then he bought a bag of potato chips that we shared. He clearly felt better, and said that the vodka had helped after only a few minutes. Also that it would give us a good appetite."

17 October 1996. … Tom Shannon had taken Teplykov fishing at the ORNL IFMIF meeting that I missed – T had caught a fish, spoken to it at some length in Russian, and let it go again, telling Tom that now there was a fish in the Tennessee waters that speaks Russian. T reminded Tom of this, and Tom said "No, now there are many fishes that do, because the first one taught them".

**Santa Fe Host**

IFMIF ACCELERATOR TEAM MEETING, 11-13 September 1995, Santa Fe, New Mexico. Visited LANL Thursday, 14 September 1995 (Teplyakov photo with the BEAR RFQ).

Drove him back to his Santa Fe hotel in evening – he did not want to go out for supper, insisted on inviting to his room for "a small supper". It was unforgettable – he treated with full Russian hospitality, with supplies that he and Zavyalsky had bought at a market – simple bread, cheese, pickles, a white wine, etc. – felt like being in a king's court.

**Moscow HIF 2002, May 2002..**

From notes:
"… at noon as planned to meet Teplyakov, who had come up by ITEP car 150 km from Protvino just to meet me - big honor.

He and Lazarev waiting in the sun - L had not gone to his dacha. T looked good, stout, looked far younger that 75 or 76. But will not travel far any more. He didn't want to sit, so walked another hour to west where walked yesterday, but not as far as the park. Back at one, invited to Lazarev's for "lunch". Very nice - full banquet again. Very good soup, main course of pork, potatoes, zuccini squash, tomato sauce to ladle onto the pork.

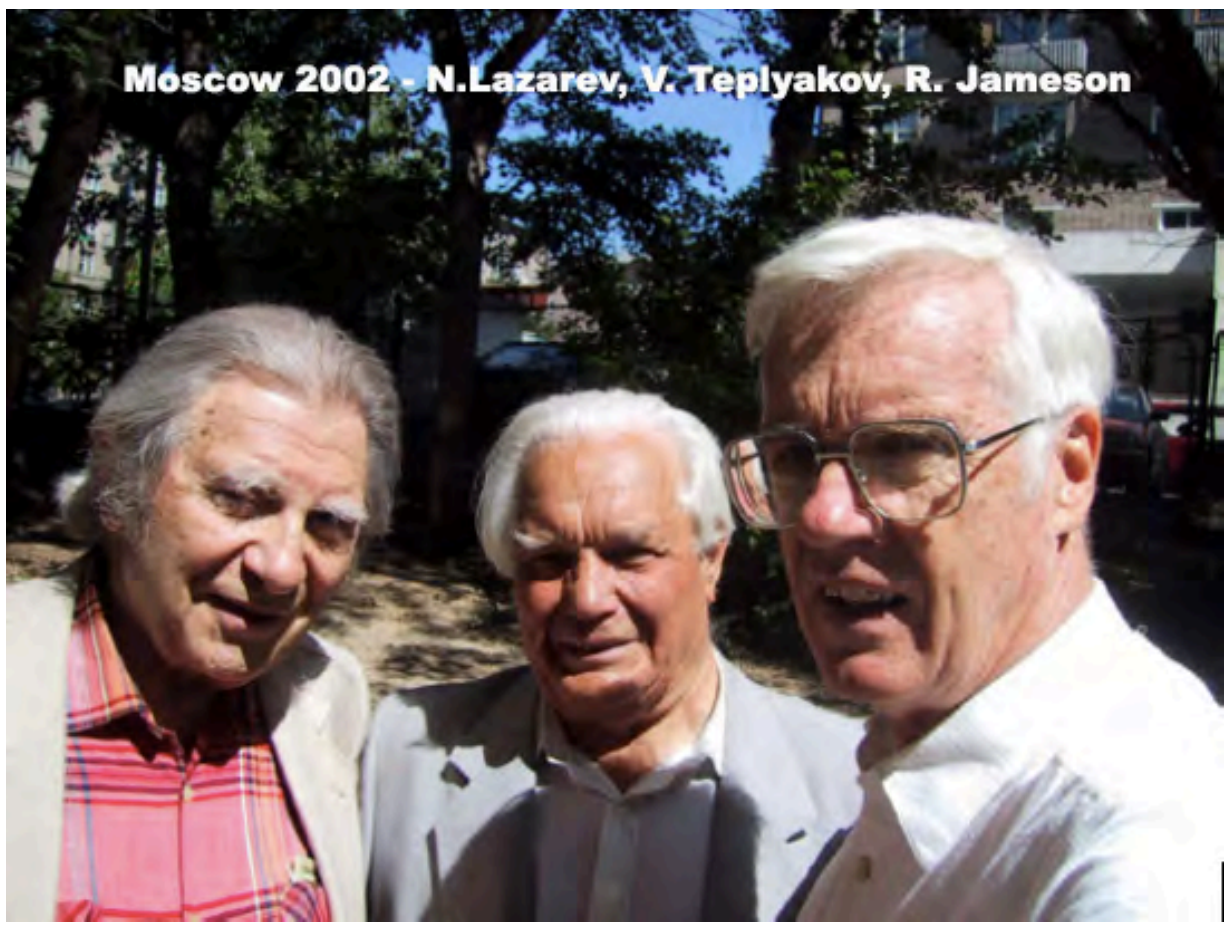

T told about Georgian toasts. Only men at the table. Once he escorted a woman Minister - of course they had to fete her - the men privately complained that the table was ruined. They make two formal toasts with hard liquor, then with wine. Only one kind of wine is served the whole evening. After many toasts already, a heavy contest - chug-a-lug a liter of wine, pass the big glass on…

T recalled the toothache story. T recited in German very nicely from "Die Lorelei". Said he needs a copy of Longfellow's "Song of Hiawatha" - loves the sound when recited in English - must send. Invited us to Protvino, but no time left now."

[toc]

## RFQ for Generic High Intensity Linac Research

My further use of the RFQ as a generic linac example was to some extent a matter of good luck. In 1994, when Horst Klein invited me to spend a year at IAP, I thought to extend the research on equipartitioning and nonlinear systems using already bunched beams. However, a Chinese graduate student, Li Deshan, sought me out and wanted help with RFQ application for his thesis. The RFQ is difficult, as the beam changes from dc to bunched, so at first demurred, but overnight realized that it would be very good to have some help, that the RFQ represents a generic linac, and that the hardest practical problem was as always the most fertile ground for me (with Ingo Hofmann pursuing the theoretical background). The problem closest related to the real application is best – not simplification of a chicken to an egg (as we characterized the approach of one rather famous, or infamous, theorist). So we agreed that the original part of his thesis would be to investigate whether we could succeed to practically bring the dc injected beam to an equipartitioned state at least at one point within an RFQ, and we did succeed [8].

First it was necessary to understand the rms behavior of the beam inside the RFQ. This had not been done before, and the strange notion that had been propagated, that the innards of the RFQ were too hard to understand so one could only look at the output, had to be demolished. Is the RFQ really different from any other linac? The **framework** needed, that included all parameters, especially the rms space charge physics, had been developed, so it seemed natural to think that rms parameters along RFQ would be the right quantities to investigate - and *only for successful particles* – the envelope and EP equations have no knowledge of beam loss. Simulation quickly showed that these rms quantities corresponded to design values.

---

8 D. Li, R.A. Jameson, H. Deitinghoff, H. Klein, "Particle Dynamics Design Aspects for an IFMIF D+ RFQ", EPAC'96 (MOP057L.PDF); Los Alamos National Laboratory Report LA-UR-96-3205, 11 September 1996.
R. H. Stokes, K. R. CrandalI, R. W. Hamm, F. J. Humphry, R. A. Jameson, E. A. Knapp, J. M. Potter, G. W. Rodenz, J. E. Stovall, D. A. Swenson, and T. P. Wangler, "The Radio-Frequency Quadrupole: General Properties and Specific Applications" , Proc.11th Conf. on High-Energy Accelerators, CERN, Geneva, July 7-11, 1980, Experimentia: Supplement 40, p. 399, 1981; Los Alamos National Laboratory Report 80-1855, 30 June 1980.

Thereafter it was my good fortune to be able to work with other students, including Johannes Maus (IAP) who developed the first Poisson solver for an RFQ as outlined in this book, Zhang Zhouli (IMP) who helped develop a fair way to compare EP and non-EP designs, Li Chao (IMP) who worked to extend the underlying space charge physics theory and application, Zhang Zhouli and Yasuhiro Kondo (JAEA) who were responsible for all steps in building an RFQ, from beam dynamics design, cavity design, construction commissioning and operation, and Bruce Yee-Rendon on a new design study for a JAEA-ADS RFQ.

[toc]

## The Clinton P. Anderson Los Alamos Meson Physics Facility — LAMPF

The story of the Los Alamos Scientific Laboratory's Los Alamos Meson physics facility proton linear accelerator from 1963 to ~1977 is a technical story, but even more a story of people. The technical people were magnificent.

[toc]

### Early Days – RF field control, computer control system

A leap of 10,000 times! That seemed like a worthwhile challenge in 1962 to a graduate student in electrical engineering at the University of Colorado. As it did to the four men who explained it, who had decided to set the stakes so high. A particle accelerator with a thousand time more average particle beam current and ten times more energy than the most advanced proton accelerator operating at that time - the University of Rochester 70 MeV, 1 μA average, very low duty factor proton accelerator (Robert P. Featherstone). The new specifications were to accelerate protons to 800 MeV with 1 mA average current. Darragh Nagle, who would be the Group Leader and concentrate on the characteristics of the coupled rf cavities that provide radio-frequency (rf) energy to the proton beam, Edward Knapp, who would develop the accelerator cavities, Donald Hagerman, who would be responsible for the rf power amplifier system, Austin (Mac) McGuire, who would lay out the civil construction.

I was not hired as a graduate student with low salary as would be the case now, but as a full-time staff member with responsibility to do something which had never been done before and which would also be the basis for my PhD thesis. This was to invent a system which would control the radiofrequency fields within the new accelerator to very close tolerances. It involved radiofrequency technology, what was then already called Modern Controls, and an intimate knowledge of the entire accelerator technical system.

One of the first jobs was to simulate on the new and quite powerful computer systems at Los Alamos the characteristics of radiofrequency waves driving a resonant cavity. Nagle and Knapp developed a theory of chains of coupled resonators using a lumped circuit model, to include mutual coupling to nearest neighbors, next-nearest, etc. This was converted into a computer model for long tanks with many cells by Henry Hoyt. Another model using coupled-mode analysis was developed by Nishikawa at BNL. My approach was directed to the development of a control system, using Laplace Transforms and carrier removal. Los Alamos had one of the world's most advanced computer centers, and a big IBM Computer called Stretch. Thomas Springer was one of the first technical persons that I met at Los Alamos during the first week, and he was very helpful [9]. He had a program which could run Laplace transform problems. The first job was to set up the Laplace transformation characterization of the resonant cavity problem on punched cards. It was possible to make one computer run per day, so one thought rather carefully before dragging the heavy card boxes over to the computer center and submitting the run. One time I made a mistake and the computer spurted out more than 50 boxes of output paper. When I went to the computer center to pick up the run, I didn't see my output, although there were stacks of boxes standing by the output shelves. So I went back and resubmitted it. This time I got a call from the computer center. I finished my thesis and graduated in 1965. [10]

---

[9] In those days, and until the characterization "regime" began to be used along with the curse of project and matrix management, people were generous with their help, and it was not necessary to ask that first question "We could talk, but to which program code should I charge my time?"

The book-length thesis required development of a **framework** and **elements**. The framework had to be the integration of the whole accelerator – cavities, rf system, cooling, electronics, instrumentation, computer control. It was completely original work, covering theory (including Laplace transforms and advanced application of Laplace transform jump functions for the coupled cavity representation for computer simulation, the simulations themselves on the fledgling, but most advanced, digital computers of the day, designing and building hardware that was commissioned and has operated successfully for 50 years, and proposed extensions for further development by using the real and imaginary parts of phase measurements.

The funding for the new accelerator did not come immediately. We were in competition with Yale University - both a technical competition and a political one

Finally in 1967 we won the competition and the big accelerator would be built in Los Alamos. We who were working in the group were aware of the political competition, and knew that the Los Alamos Scientific Laboratory Director Norris Bradbury and our powerful New Mexico Senator, Clinton P Anderson, had been very active for our cause. We were proud that it was acknowledged that our technical case was outstanding, although it was clear that we had won for political reasons, to give the New Mexico economy a boost.

At this point, a new division of the Los Alamos National Laboratory, the Meson Physics Division, was created to build the new accelerator. It was a very big surprise to me, and still is, that the new division's leader was not Darragh Nagle, our group leader who had been involved every day during these years. I think that then, as well as ever afterward, I was interested and informed about my surroundings, and especially in the people, but the new division leader was a person I had never seen or heard of. I remember being completely perplexed, and asking who this person was. People told me that he was a nuclear physicist who had been interested in the idea of such a new and powerful accelerator. However he had no technical experience with any of the technologies involved. This was to have very long lasting effects.

Nagle became the deputy division leader of the new division and was involved mainly in the meson physics for which the machine was to be built. Later he was made was made a Senior Laboratory Fellow, an ample reward and a nice road to his retirement years later.

The new accelerator would be of the linear type - that is, it would be laid out in a straight line. It would be about 1 kilometer long. This required clearing of a Mesa top along the East Jemez road leading from the central Los Alamos laboratory down to the Pajarito plateau, and we hiked down the first bulldozer track. The facility was named the Clinton P. Anderson Los Alamos Meson Physics Facility. The division leader objected to this name, did hesitate to propose his real preference and then advanced the "Los Alamos Meson Physics Facility". Of course that is this did not sit completely well. The longer name remained the official one, and the shorter name became the acronym LAMPF.

One of the central issues for LAMPF was the possibility of controlling the whole facility with a computer-based control system. This was the dawn of the computer age, and there was little experience at all with such a control system. A final cost study for the project cost was contracted to EG&G. It was to include a detailed comparison of a "conventional, "bat-wing" lever, switch and meter" system typical of reactors and totally without a computer, to a fully computerized central computer control. As the due date approached, EG&G announced they had used up all the funds, and had not yet worked on scoping the computer-based control system and comparing it, so they would just abandon that. I was young, inexperienced, naïve, and incensed. I really got upset, and told Nagle and the other three that "we have a contract - they cannot do that!!" Etc. Nagle said "Pack your bag, we'll go see them." So I went on one of my first official travels. Told EG&G my same naïve tale. And they agreed to complete the study. It turned out, as later I learned all good studies do, that the costs would be the same for either control system, and that the computer approach looked feasible. So we were able to choose the central computer control system - without which it would have been extremely difficult, if not almost impossible, to turn LAMPF on and operate it [11].

---

10 R. A. Jameson, "Analysis of A Proton Linear Accelerator RF System and Application to RF Phase Control", Ph.D. Thesis, Los Alamos Scientific Laboratory, LA-3372, UC-28, Particle Accelerators and High-Voltage Machines, TID-4500 (45th Ed.), Nov. 1965.

[toc]

## Megawatt RF Systems

After getting a prototype rf field control system working, it was necessary to work on individual components of the control loop to insure that they were controllable and realized in reliable hardware [12].

The error analyses in these references show why seemingly ordinarily commercially available microwave components, such as directional couplers, detectors, and slotted lines, are not adequate for obtaining automatic control data or measurements with high-power rf systems, and required custom components to be developed to achieve the LAMPF automatic rf field phase and amplitude control specifications. [Ref. 12.b]

The central rf reference frequency source and distribution system was the first of its kind, and required extensive analysis and care for its execution and very many hours of patient work to prove its coherence and stability.[ref 12.a]. It would be interesting to know if newer applications have considered these issues. [13]

The major elements were the rf accelerating structure and the megawatt class rf amplifiers that drove them. Both required completely new development, and a framework and elements were set out. LAMPF was to be a pulsed machine, with pulses 0.5-1.0 ms long and 6-12% duty factor. No one had ever run any type of megawatt amplifier below fully saturated output before, and LAMPF required precise field control to tolerances of ±1% in amplitude and ±1° in phase. No instrumentation was commercially available for pulsed rf. It was very interesting to work out instrumentation and techniques and measure the characteristics of the high power RCA triode, Varian klystrons and the crossed-field amplitron (very tricky because of mode jumps within the control range) in terms of their controlled output into an rf load of arbitrary impedance (Rieke diagram). The 201.25 MHz triode was

---

11 T. M. Putnam, R. A. Jameson, T. M. Schultheis, "Application of a Digital Computer to the Control and Monitoring of A Proton Linear Accelerator", IEEE Trans. Nucl. Sc., Vol. NS-12, No 3, p. 21, June 1965.

12 R. A. Jameson, and W.J. Hoffert, "Fast Automatic Phase and Amplitude Control 0f High Power RF Systems", IEEE Trans. Nucl. Sci., Vol. NS-14, No. 3, p. 205, June 1967.

R. A. Jameson, W. J. Hoffert, and D. I. Morris, "Microwave Instrumentation for Accelerator RF Systems", IEEE Trans. Nucl. Sci. Vol. NS-16, No. 3, pp. 367-371, June 1969.

R. A. Jameson, J. D. Wallace, R. L. Cady, D.J. Liska, J. B.Sharp, and G. R. Swain, "Full Power Operation of The LAMPF 805-MHz System", Proc. 1970 Proton Linear Accelerator Conf., p.483, National Accelerator Laboratory, Batavia, Illinois, October 1970

R. A. Jameson and J. D. Wallace, "Feedforward Control of Accelerator RF Fields", IEEE Trans. on Nucl. Sci., NS-18 (3), p.598, June 1971).

13 Bill Hoffert showed up sort of unannounced one day and inquired if he could be of help - just retired from the FCC in California after a fascinating career, which included figuring out that a massive transmission disturbance was because a local iron bridge was acting as an antenna. Very fortunately he joined me and was instrumental in developing the rf field and amplitude system components. The rf reference source is installed in a mezzanine between the 201.25 MHz and 805 MHz sections of LAMPF. Bill designed and built the frequency multiplication and amplifiers for the environmentally controlled coaxial distribution system. We spent very many hours checking the coherency and stability of the source. There has never been a complaint about this reference system. Decades later, on meeting Catherine and Scott Richmond, Scott related that he had worked at LAMPF for awhile, and had viewed the reference system as a "crown jewel" and fascinating.

used in the first four sections of the accelerator, and the 805 MHz klystron in the following 44 sections.

Characterization of the amplifiers continued over several years, to even after the LAMPF turn-on. A screen room with many oscilloscopes and instruments became my operating method and hermitage for many hours. In those days, Polaroid camera photographs were the data collection method. The amount of data required for one Rieke diagram was large. I was afforded an assistant – Mrs. Rene Mills, who returned to work after raising a family and significant public service [14]. Somehow we heard of a big and no longer used projection machine that would project a photograph on a large screen, over which horizontal and vertical cross hairs could be positioned and the coordinates of the intersection punched onto a computer card. We digitized our data, with one run requiring about a yard of cards. The signature of the Rieke diagrams became clear and we derived the method for installing the amplifier and the connecting waveguide to the accelerating structure so that load impedances that could result in amplifier oscillation were avoided. One can never be careful enough – when trying to understand some apparent deviations, I suddenly realized in the middle of a late afternoon meeting that I had not installed low-band frequency filters in the directional coupler outputs and that maybe higher harmonics were present. I rushed out of the meeting, made a full run and the next day we saw that the deviations had disappeared and the results were fully understandable [15]. The LA-5649 report is also book-length.
[toc]

## The Control Philosophy Committee

As the new division leader had not the slightest knowledge of any of the technology that would be needed for the new accelerator, he basically ignored the technical team responsible for the accelerator. We said very often during the design and construction years that this was very lucky for us, as in other areas such as the civil construction or the experimental area, his attempts at involvement resulted in chaos. We did have many occasions when responsible coordination from above would have helped us very much through some difficult times. As there was no effective management above the group level, it was sometimes difficult to resolve issues which spanned several groups. Finally from within the groups, it was suggested to have a central committee whose members would be the group leaders and several others who were playing a major role. This committee was named the Control Philosophy Committee. The name was chosen to reflect the fact that this accelerator would be the first ever designed with the central computer system that would be required to start up and operate the machine. The central control issues did span all the groups, and the committee afforded effective interaction and decision. It soon became, however, the management apparatus for the whole accelerator part of the project. As is well known, management by committee is sometimes difficult. However there seemed to be no other way. But we said we were lucky to be left alone and not bothered by whom we termed " 3M ". In those days the 3M Company was a powerful, profitable and well-run company. Bob Frank was a tall, crusty, well-respected computer expert in the LASL Computer Division, whose help I valued very much. He gave his excellent help generously when he felt the recipient was not wasting his time - so one brought him well-formulated questions. His wife was a 3M heiress - they

---

14 Rene was certainly an exception to Mark Twain's remark about school boards… She started as a Data Analyst, but learned computer programming and eventually became a staff member and a key member of the team works described here. One of the hard things about taking on Division Leader responsibilities was having to give up the daily association with her and others, such as Ken Crandall, and the gradual estrangement because of being perceived as now being some part of the quite detested general "upper management".

15 T. J. Boyd and R. A. Jameson, "Optimum Generator Characteristics of RF Amplifiers for Heavily Beam-Loaded Accelerators", IEEE Trans. Nucl. Sci., Vol. NS-14, No. 3, p. 213, June 1967.

R. A. Jameson, R. S. Mills, and R. L. Cady, "Performance of Pulsed 805-MHz, 1.25-MW Klystrons into Mismatched Loads", LA-5649, Los Alamos Scientific Laboratory, September 1974

My logbooks.

R.A. Jameson, "Measured dynamic performance of 1.25-MW, 805-MHz klystrons", 02/1970; DOI:10.1109/IEDM.1970.188316 In proceeding of: Electron Devices Meeting, 1970 International, Volume: 16 lived in the finest estate in Pajarito Acres. We did not mean 3M that way but with irony the opposite, we meant " most meager manager".

[toc]

## Side-Coupled Accelerator Structure

A full scale model of an 805 MHZ module was built, with a klystron rf system and a side-coupled accelerator structure scaled for electrons, named the Electron Prototype Accelerator (EPA). A quick test using a bending magnet in the transport line to a target indicated that the energy gain was correct. This essentially concluded the experiment, but it was a testbed for an entire closed loop rf field control system, so I took it over, built up a large screen room again, and spent many days running the system to work out details of the control system.

In parallel, the central computer control system was being developed, and a few data channels were connected to it. I made the first system experiments using a computer control system to collect and analyze data from a linac.

A pleasant diversion was the closed-loop analysis of the accelerator structure water cooling system temperature control, which became successful with the addition of a lead-lag component, realized by one resistor and one capacitor. [16]

I would come in the morning, turn on the EPA, and pursue my experiments (with no beam). The system would sometimes run without any problem all day, but often, after an hour, or several hours, at random, the oscilloscope traces would suddenly drift, the system would run away and be shut off by the protection circuits. Clearly the cause had to be found. Trouble-shooting was done on each component, and with great reluctance, it became necessary to acknowledge that there seemed to be something concerning accelerator structure itself. With great reluctance, because the side-coupled linac was a clever in-house invention, also requiring very significant engineering development, and heroes had been made. I knew that revealing any purported problem with it would likely result in the death of the messenger, and documented the situation with extreme care. The standing-wave accelerator structure was fed rf power through a port at its center, with power propagating through the coupling cells to the ends. With no beam, the accelerator structure vault could be entered, and it was found that in the runaway situation, one end of the structure was cold and the other end very hot, indicating that the rf field was no longer uniformly distributed as required. Supporting evidence documenting the same observation a year earlier was found in a filed and ignored memo from mechanical engineer Seth Rislove. At the same time, the production of the side-coupled tanks for the complete linac was in full swing and several 4-tank modules had already been installed in the accelerator tunnel.

The situation was presented to the management. Instant death was avoided, but perhaps not a lingering one over the ensuing long and hard campaign to understand and fix the problem. A small

16 G. R. Swain, R. A. Gore, and R. A. Jameson, "Temperature Control for Maintaining Resonance of Linac Tanks", Proc. of the Proton Linear Accelerator Conf., Brookhaven National Laboratory, BNL-50120 Pt. 1, p. 159, December 1968.

team was put together, with Rene Mills, George Swain [17] (very tall, very quiet, very thorough, later commissioned to document the whole adventure [18]), Bob Patton (a very talented technician), John Sharp (rf engineer), Don Liska (one of the key mechanical engineers of LAMPF [19], along with Spike Worstell and Ed Bush), and Jerry Wallace (rf engineer, later Group Leader for LAMPF maintenance) [20].

The first step was to get the first full 4-tank module (#5) removed from the tunnel and installed in a test building and to provide it with a full rf power and control system. Testing from low to full duty factor revealed that sampled signals from the structure changed randomly but also systematically with duty factor, and we spent a lot of time trying methods to investigate the field flatness in the structure at full rf power. We considered using a rifle to shoot a bead through (trajectory would not be flat enough). Making a bead pull using some kind of string provided some diversion, and a short-pulse measurement at about half power actually was achieved.

The side-coupled structure has two cell types – the main accelerating cells, and the off-axis coupling cells. The result is two sets of resonant frequencies, which must match at the operating frequency. The structure can resonate at many different frequencies over a band of frequencies (different "modes"); the operating mode is at the middle frequency, the next mode on either side should have the same frequency spacing – this condition is called "closed stopband". When the frequency spacing is not equal, the difference (stopband) is open, in one of two possible directions. So we knew that the tuning method was critical in this regard, and needed to understand the stopband as tuned and also at full operating rf power.

17 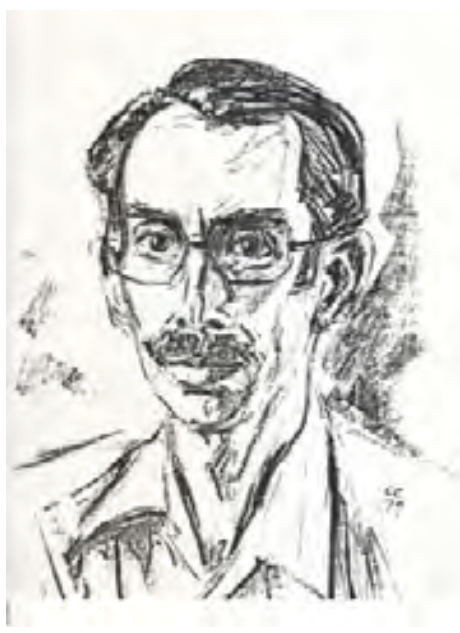 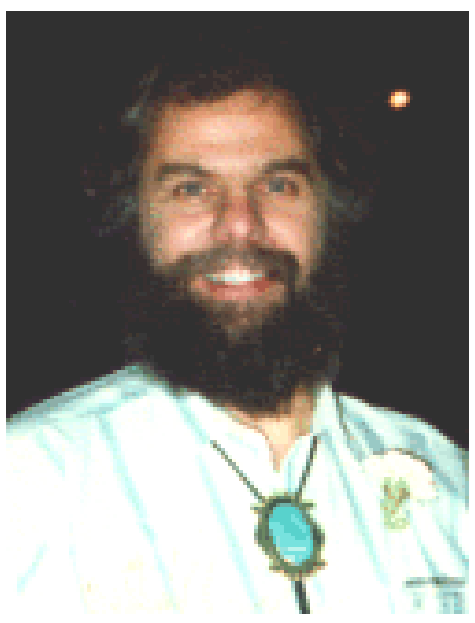 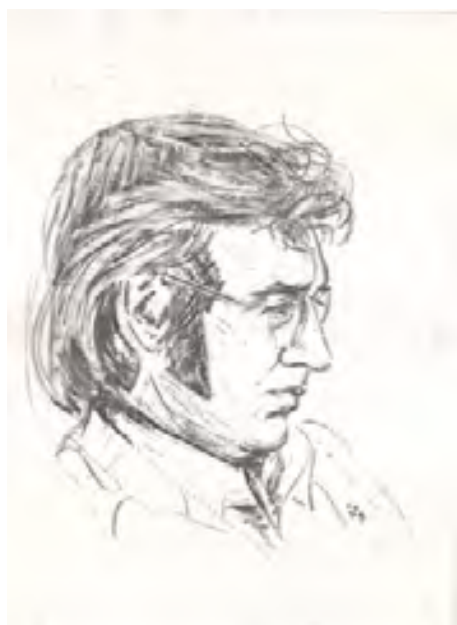

George Swain, 1979 Jim Potter, me by (Carol Coppersmith),

18 "LAMPF 805-MHz Accelerator Structure Tuning and Its Relation to Fabrication and Installation", G.R. Swain, Los Alamos Scientific Laboratory Informal Report LA-7915-MS, UC-28, Issued: July 1979.

19 D. J. Liska and R. A. Jameson, "Particle Accelerator Engineering", University of California Engineering Faculty Meeting, Los Alamos, May 25-27, 1983, Los Alamos National Laboratory document LA-UR-83-1358.

20 "LAMPF Side-Coupled Structure Tuning EAK_Memo_19701021", R.A. Jameson, G. Swain, J. Wallace, D. Liska, J. Sharp, K.Crandall, LAMPF internal memo. 50 Pages. DOI: 10.13140/RG.2.2.30045.90087

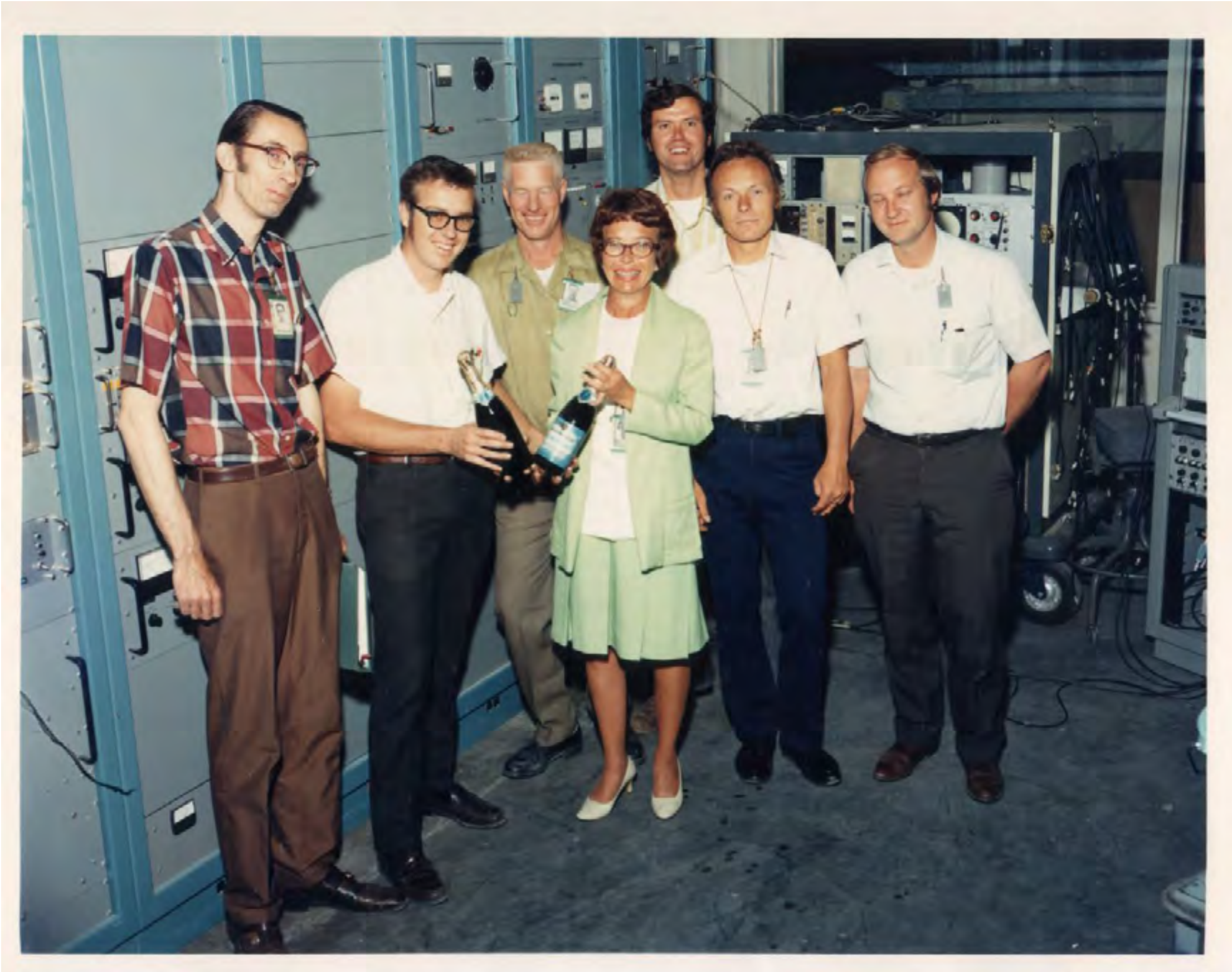

George Swain, me, Bob Patton, Rene Mills, John Sharp, Don Liska, Jerry Wallace

In those days, there was no instrumentation for pulsed systems beyond the oscilloscope. I devised methods to test the structure properties while running at full rf power, for example, by inserting additional diagnostic pulses between the normal pulses, so we were able to do a frequency sweep and dynamically measure the stopband at full power. [21] This led to the realization that that the problem lay with the structure stopband and the tuning method. The high power distribution will remain stable if there is a stopband closure error in one direction, but in the other direction, the error will grow and cause a field tilt, with positive feedback. The error is called a “stopband”. The effect is so powerful that attachment of increased cooling tubes to the structure was insufficient (at this point, they could only be glued on, so heat transfer was minimal.). The cells should be tuned such that a small stopband exists on the stable side, to avoid relaxing to the unstable side during years of operation between zero and full power.

In parallel with this long campaign, modules were continually being installed in the tunnel, and by the time the stopband had been identified as the problem and methods for measurement developed, about half of the one kilometer long accelerator was installed. The manufacturing method was to machine the cells and braze them together in long tanks, with no tuning provision after brazing.

At the start of the Module 5 campaign, an independent structure tuning method was being implemented in the tunnel. There was space between the tanks and a tool was developed, with short tube sections inserted into the bore, the next being screwed onto the former, and so on. A separate rod inside the tube had an expandable section at the far end and a crank on the input end. The expandable section was placed in the drift tube between accelerating cells and expanded with the crank, making it fast, and then the drift tube web could be deformed by a sliding hammer on the rod. The procedure was worrisome to us – a tuning was being done without knowing that the problem lay with the stopband, tuning started at the tank middle and worked out to each end thus requiring ever more web deflection and propagating the correction, and bending the drift tube web work hardened the copper. About half of the accelerating cells were tuned in this manner.

---

21 “Bead Pull Measurements Techniques at High Power”, Liska, DJ; Jameson, RA, Bulletin of the American Physical Society,; 1971; v.16, no.2, p.247
D. J. Liska, R. A. Jameson, J. D. Wallace, and J. B. Sharp, ”Accelerator Field Measurements at High Power", IEEE Trans. on Nucl. Sc., NS-18 (3), p. 601, June 1971.
R. A. Jameson and J. D. Wallace, “Dynamic Measurement of Stopband in
LAMPF 805-MHz Accelerator Structures at High Power, Using Hybrid Computer Techniques,” LA-4593-MS, Los Alamos Scientific Laboratory, January 1971.

We realized that it would be necessary to tune every individual cell – both accelerator cells and coupling cells – after brazing, and after being installed in the tunnel. Don Liska expanded the tool system with insertable sections. First it would be necessary to return the drift-tube webs to their original position, possible with the original tool but with the complication of work hardening. The frequency of an accelerating cell could be measured with the cell isolated by using the insertable tube with special plugs to make short circuits that would isolate the cell to be tuned. The cell frequency was detected by appropriately positioned short-circuiting blocks to isolate individual cells, and transmitting and receiving probes. It might be necessary to raise or lower the frequency, and this must be done for an accelerating cell by shortening or lengthening the drift tube. Shortening was easier – a section containing a blade was placed at a drift tube nose, the knife was raised by turning the crank, locked, and then the nose could be shaved off. This was replaced later by making "dings" with a hammer around the largest circumference of an accelerating cell from the outside. How to lengthen? Liska came up with an ingenious scheme. The insertable end section was fitted with a V-cutter that could be raised just inside the end of the drift-tube nose, and a V-shaped groove made, upsetting the metal toward the gap. The coupling cells had a vacuum port on top, through which a wedge could be driven between the on-axis nose protrusions, or the outer walls could be bent inward with a hammer.

So it would be necessary to completely return the cells to their original position as much as possible, and start a completely new tuning campaign. Again we were very careful to present the case so thoroughly that it was completely clear that it was necessary, thoroughly analyzed and prepared [22]. Not welcomed, but had to be done. The small team reluctantly agreed to take over the supervision of the task, but required signed acknowledgement that it still might or might not be the ultimate cure.

The first day that we took over the tuning was dramatic. Working in the tunnel was like working in a mine – hot, sweaty, and frustrating because it was widely noticed that the tuning method was not understood. The frustration had been so high that serious arguments had occurred among the tunnel crew and there were reports of fist fights. And here came a bunch of people of remote acquaintance who were supposed to have the answer… But we got on with tuning the ~10,000 cells, and the crew was splendid. One of the key technicians was Oliver Rivera, who did so many metal manipulations that he earned the nickname "The Golden Arm". The tuning of the whole 805 MHz linac was completed on April 27, 1972.

---

22 "805 MHz RF-Accelerator System", R.A. Jameson & G. Swain, J. Wallace, D. Liska, J. Sharp, K. Crandall, Office Memorandum to E.A. Knapp, MP-3, October 21, 1970.
G. R. Swain, R. A. Jameson, E. A. Knapp, D. J. Liska, J. M.Potter, and J. D. Wallace, "Tuning and Pre-Beam Checkout of 805-MHz Side-Coupled Proton Linac Structures", IEEE Trans. On Nucl. Sci., NS-18 (3), p. 614, June 1971.
G. R. Swain, R. A. Jameson, R. Kandarian, D. J. Liska, E. R.Martin, and J. M. Potter, "Cavity Tuning for The LAMPF 805-MHz Linac", Proc. 1972 Proton Linear Accelerator Conf., October 10-13, 1972, LA-5115, Los Alamos Scientific Laboratory, p. 242, November 1972.

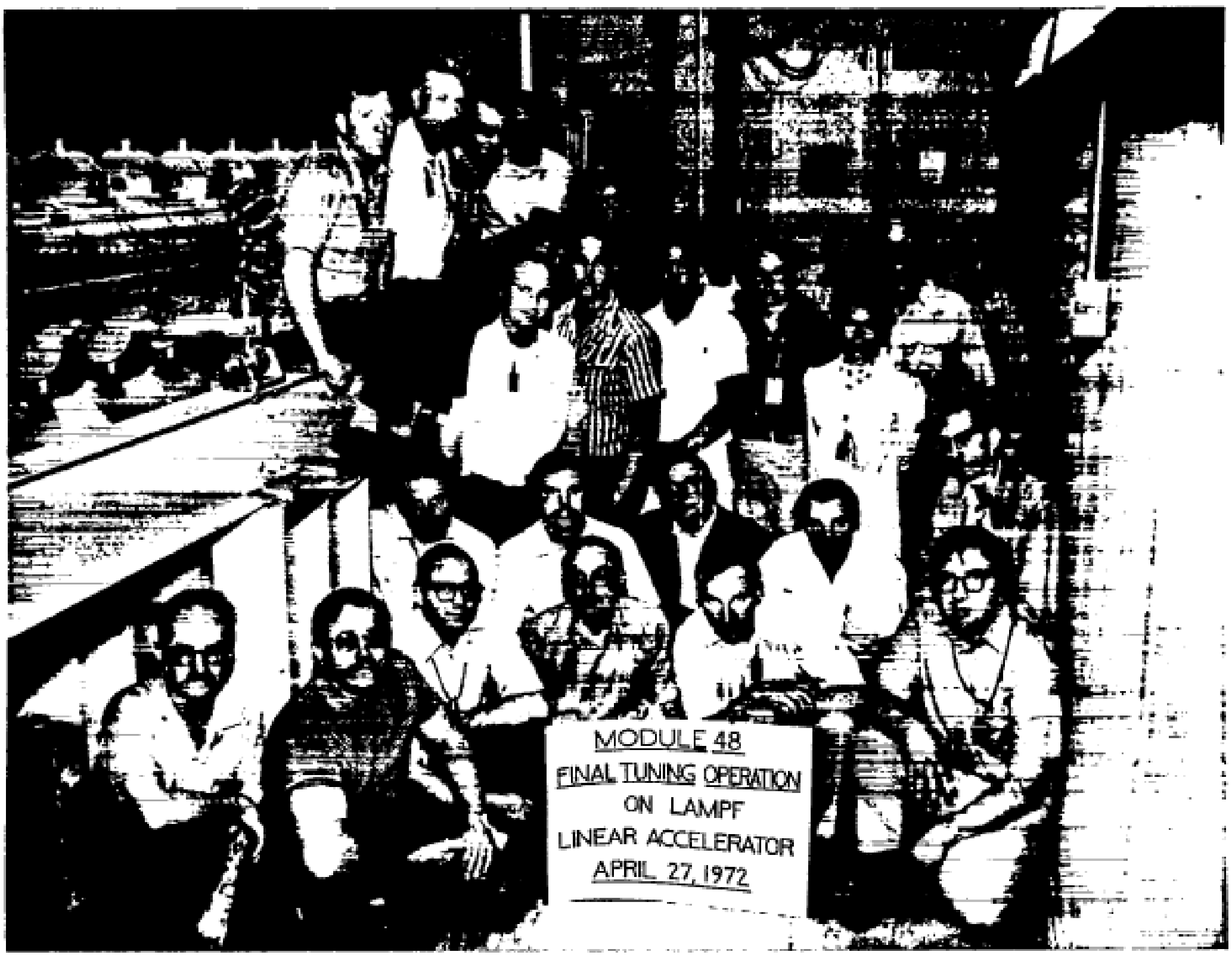

Fig. 71. It's done! Persons involved with structure tuning and some of their contributions: (Front row) Bob Cady - probe calibration, Oliver Rivera - tuning and tool improvements, Dave Jones - tuning, Cecil Stark - tuning, Paul Beauchamp - tuning, Bob Romero - tuning, (2nd row) Don Liska - beadpull equipment design and supervision, Jim Potter - bridge coupler stabilization - beadpull improvements, John Zastrow - beadpull equipment, Charlie Manger - prebraze and section tuning and supervision, George Swain - procedure design and supervision, (on I-beam) Don Holcom - rf testing, Bob Patton - rf testing, Mike Dugan - rf testing, Dennis West - tuning and tool improvements, (back rows) Jerry Wallace - rf test design, Buck Hayes - machining, Bob Harrison - tuning and beadpull data processing, Jess Lee - machining supervision, Ed Knapp - structure design and modifications, Joe Van Dyke - tuning tool design, Bob Jameson - coordination and scheduling and rf system design, Rene Mills - data collection, Bob Kandarian - tuning tool design, and Seth Rislove - cooling system installation supervision.

[toc]

## Completion and Commissioning

The first 800 MeV beam acceleration demonstration had been scheduled [23] for July 4, 1972. By autumn 1971 the accelerator structure tuning was starting to be understood. A PERT project management system – one of the first in the accelerator world – had been set up by Mahlon Wilson. He was sending out alerts that it appeared that the demonstration date could not be met, because PERT said 18 weeks behind and slipping more than day-for day. None of the staff believed in the PERT and would just give off-hand and uncoordinated input. Nothing was being done about it. For some reason, no doubt our professional pride, we (the structure tuning team) decided to look into it. Talking directly to each staff section, a commissioning outline was generated, working up from each subsystem, and the time it would take to commission. The 805 MHZ linac was highly modular, with 44 modules. The outline was then put together on one (big) piece of paper, and discussed. The guys said they actually thought they could do it if left alone to work.

We decided that if a solution for the structure tuning could be finalized by NewYear 1972, the turn-on might be achievable. The key people sacrificed their Christmas and New Year holidays. We wrote up the situation, the tuning conditions that would be followed, the ghosts in the closet that could come out over the next twenty years, and made them all (especially 3M) sign - the scapegoat problem was a very real possibility – and took charge of the whole affair, under the strict condition that we would be strictly left alone to work. (There was an "official" supervisory committee over us, but it was understood that we were not to be "reviewed", second-guessed or bothered with meetings.) The new arrangement was communicated to the DOE.

---

[23] egotisically

It was crystal clear that a **framework** was needed. It had to be very simple, to be very clear to *everyone* "on the mesa", from the janitors up through all the technical ranks whose achievements would make it work, and especially to the "upper levels" that were now excluded from participating directly, but of course would trumpet that they "were not informed" if the turn-on goal was missed. The **framework** was a single large sheet of blueprint paper.

The "big piece of paper" was a real key to the success. The key systems were listed horizontally, and the module numbers from top to bottom. When a module system was commissioned, a red line was extended downward. The graphs were posted at many locations. Rather than an abstract PERT, no one wanted to see his red line advancing slower than everyone else's. Everyone also knew that if his red line was likely to slip, he would receive extra resources to keep up. (Still have some of these posters.)

The 805 MHz rf module commissioning team is shown here at completion of the 805 MHZ module commissioning – all systems ready for beam.

The computer control system was working well, with many firsts (e.g. trackball). A control room display was needed that would show the whole machine status in one glance; display space pixels were few, and when I commandeered the two bottom lines, there were objections, but the color status of 44+ modules would fit, and stood with Paul Elkins as he efficiently programmed it - it remained for decades, maybe still.

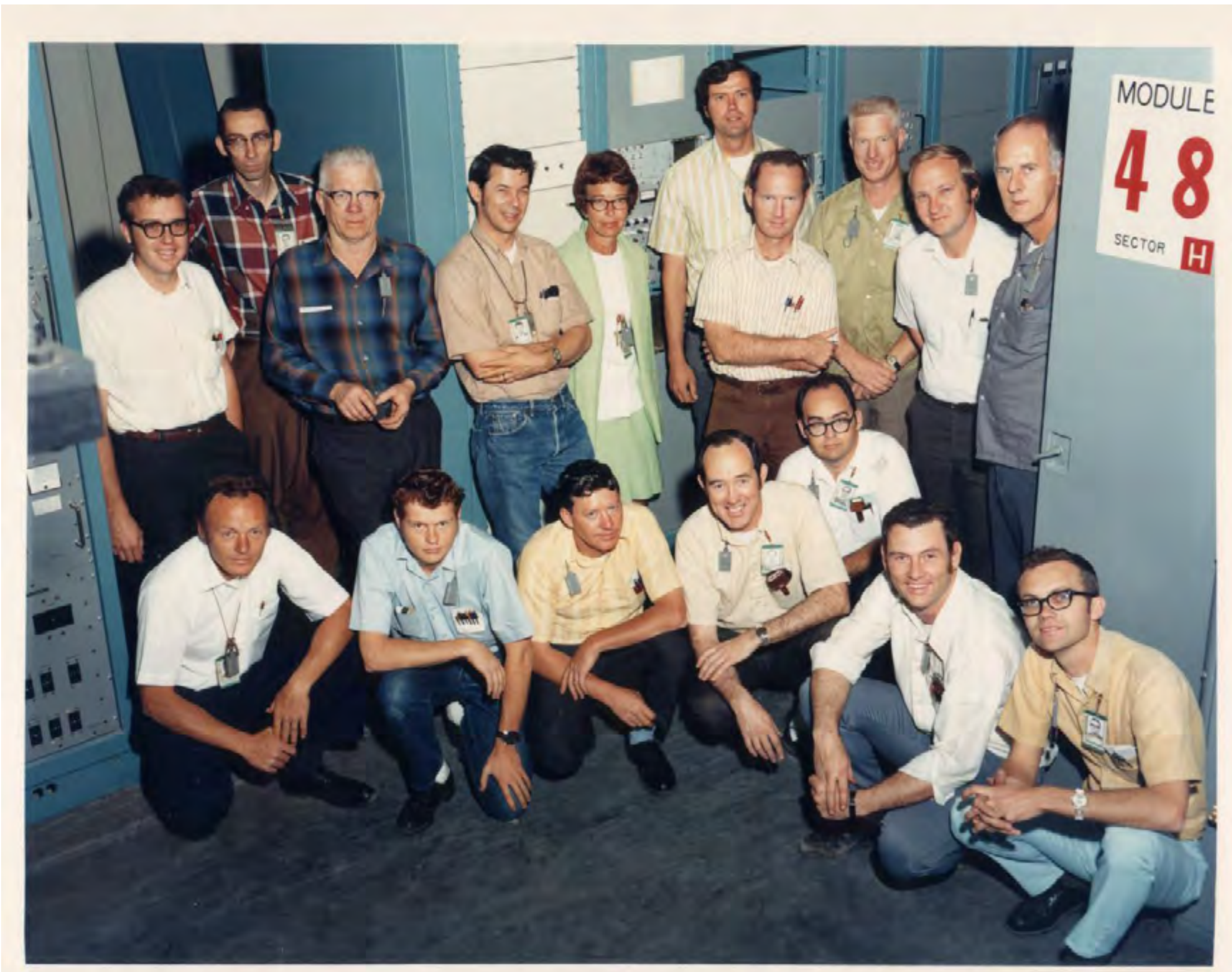

(Don Liska, __, Don Holcomb, Bob Newell, Danny Doss, __, Ray Ridlon; Bob Jameson, George Swain, Bill Hoffert, Duard Morris, Rene Mills, John Sharp, Bill McCabe, Bob Patton, Jerry Wallace, Chuck (Norris?)). Hoffert, Morris, Mills, Patton and Wallace played key roles in the development of hardware components for the low level rf control system

[toc]

## Lighter Moments

1. A new and interesting method for control of some systems arose, called fluidics, or fluidic logic, in which fluids were used for analog or digital operations similar to those performed with electronics. An applications award contest was announced in a trade magazine; I entered an application and we won the Norbert P. No-No Award Humorous Mention Award from the Fluidics College of Practical Knowledge 1972 Honors Program, for a paper on COOLING WATER CONTROL:

*Honors Award competed for:*

Norbert P. No-No Award - for the best paper on an application you've evaluated where fluidics is *definitely not* the best answer at this time.

*Describe the specific application where you think fluidics isn't a good answer, if you're competing for the Norbert P. No-No Award:*

"Wasn't a good answer, at the time (~1967) for control of cooling water to maintain the resonant frequency of high-power microwave standing-wave accelerator cavities for the Los Alamos Meson Physics Facility proton linear accelerator."

*Now: Tell us why.*

"We did consider the possibility, partly because of some radiation environment aspects, but

1. Five years ago, the technology was very young.
2. System was basically hydraulic, but with control by computer, which meant interface to electrical. Nothing was available in direct hydraulic fluidic control, and we didn't think hydraulic-pneumatic-electrical interfacing would be reliable enough.

It would have been beautiful, tho - you see, it's very dry out here, so we have to haul water up from the Rio Grande. The accelerator is 2000 feet above the river on a mesa, so we have this bucket brigade of Indian elephants up to the site. The one at the end of the line dumps into a big tank (capacitor), which is the supply for the African elephants - real big fellows with really tough trunks. They squirt the water through the accelerator cooling tubes - more or less as needed to keep the thing on resonance. When all are at their stations, they (the elephants) are facing north. Well, as you can see, we have two remaining problems - one is to get the buckets back downhill, and the other is at the south end of the elephants. Simplicity in itself - after every fifth squirt into the accelerator, a (fluidic) counter alerts the pumpers to put the next squirt into a different port, which leads to a trench which goes along the south end (of the elephants). The empty buckets are ejected automatically into the same trench. The Pojoaque Green Chile Growers Assoc. is ecstatic over the quality of their recent crops and we have been commended by the EPA for achieving an ultimate in recycling. The system works like a charm - accelerator always on resonance. Only problem we face is keeping the regional export of green chile higher than the import of peanuts.

2. We used bicycles to make transits up and down the half-mile LAMPF tunnel – remarking that if we had a nickel for each trip we would be amply rewarded. One day, I was riding a bike toward the front end when suddenly there was a tremendous "Bang!", like an explosion. Thinking that maybe a capacitor bank had blown up, I stopped and looked around – to see a technician with a horrified look on his face up on top of the racks along the corridor – and a large metal plate on the corridor floor. He had dropped the plate to scare a colleague, not realizing who was on the bike. We had a good laugh. There is also a long narrow metal stairway with railings between the 805 MHz and 201.25 MHz levels – to save time, we used to go down by sliding on hands, with no feet on the stairs.

3. We worked too many hours, often until late at night. One lonely night, I put a battery-powered "Laugh Box" on the PA-system for awhile – it cheered everyone up. Perhaps even more fun was that the next day, a very stern-minded senior staff member objected very strongly to such an inappropriate use of the PA system – it might have been needed just at that moment to announce an emergency…

[toc]

## 1972 – LAMPF Turn-on to Full 800 MeV Energy

We made it on June 9, 1972. Almost a month before the scheduled date.

We were lucky in those days. There were no unwashed "upper bosses", project managers or sponsors that insisted on project review every week or so, with the resulting serious impact on productivity – and schedule – and often then in inappropriate, uninformed, precipitous and impatient orders "directing" what to do next. We went underground (literally), the problem was handled internally, DOE was briefly informed of changes in the supervisory structure, and we fixed it. The fate of one of the next following DOE projects was different. The ISABELLE project at BNL had difficulties with development of the superconducting magnets, public uproar, project cancelled, although soon later the magnet problem was resolved.

A "Thank you" from 3M was never received. Probably it never occurred to him, but think rather he resented having his a__ saved (and his blood, which he had boastfully, ridiculously, promised to donate to the cooling system if the turn-on deadline was not met!). We knew this had to be expected and that we had had to be extremely careful along the way and require signed agreements to avoid possibly being set up as scapegoats [24]. But we often wondered later why we had worked so hard and put in so many extra hours, if the only result were the non-dilution of the cooling water. Of course we knew that we did it because it was very interesting, challenging, our proud accomplishment which no one could ever take away, however unrecognized.

Incredibly, an article appeared in a 2002 USDOE Research News covering a LANL 30 years anniversary of the LAMPF turn-on – (there was some kind of general announcement that attendance was open, but there were certainly were no specific invitations) –- 3M celebrates himself, inserts a copy of the famous turn-on photo at the CCR console *with himself Photoshop highlighted and the rest of us grayed darker!!!!* Expert at self-promotion, he had certainly arranged for that photo, and its cropping, and grandstanded when the observation of beam current at 800 MeV was announced over the CCR PA system – we all knew that at the time. A correctly retouched photo is shown here.

---

[24] "… das menschlich-allzumenschlichen Grundsatz, wonach auf Erden kein Hass so abgrundtief ist wie jener, den ein Notleidender gegenüber seinem Retter empfindet." ("Unbekannter Nachbar Frankreich", Hans O. Staub, AT Verlag Aarau, Stuttgart, 1983

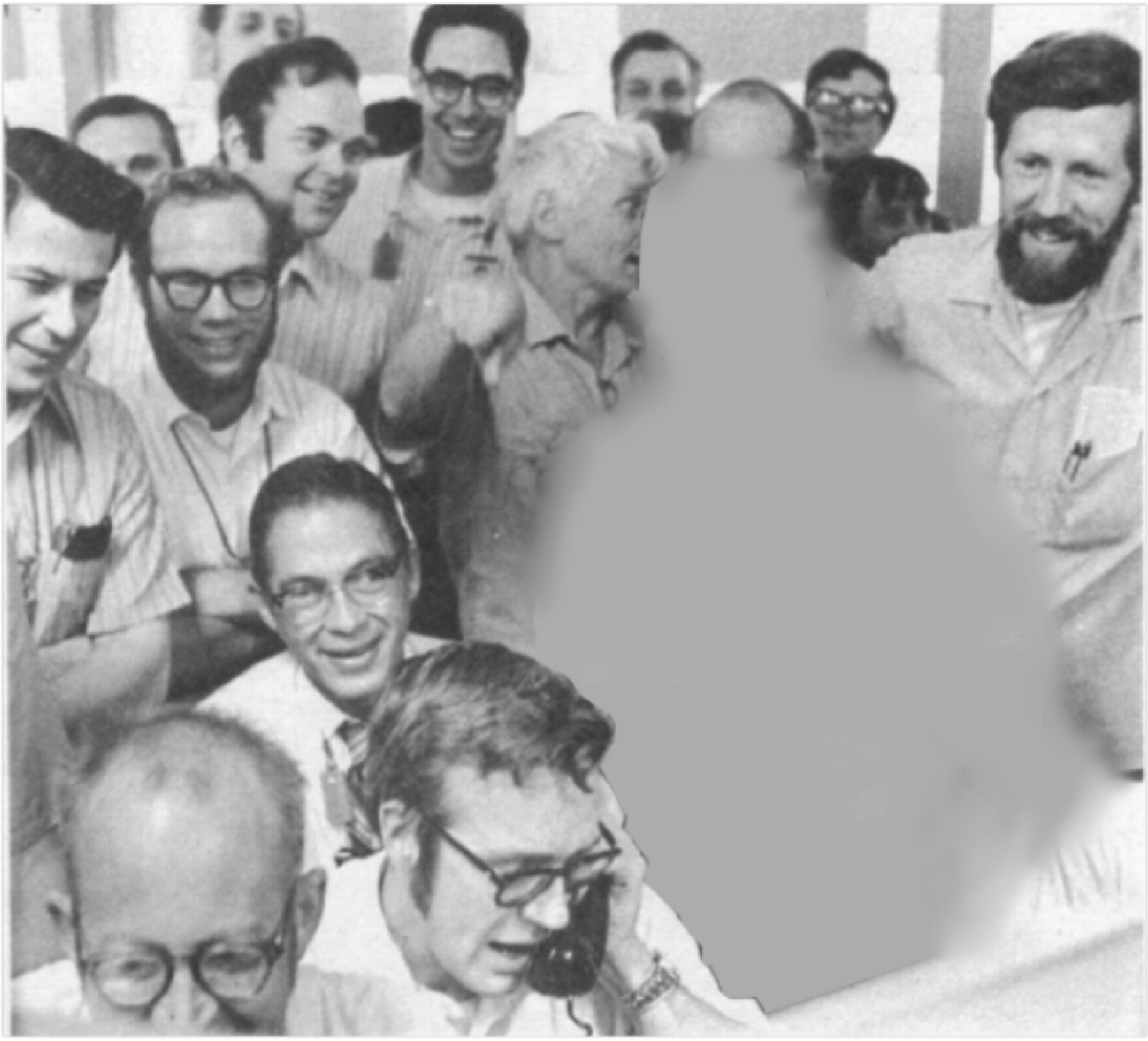

Beam current at 800 MeV is observed at LAMPF, June 8, 1972.
Bottom – Don Swenson, Bob Jameson (turn-on director with phone), behind Tom Putnam. Standing left Duard Morris, Denny Roeder, ___, Ralph Stevens, Darragh Nagle (face partly obscured, John Sharp, far right Jerry Kellog

We felt that such an important scientific event required two proofs, so had worked much longer into the night getting the second proof in the form of delta-rays from the 800 MeV beam.

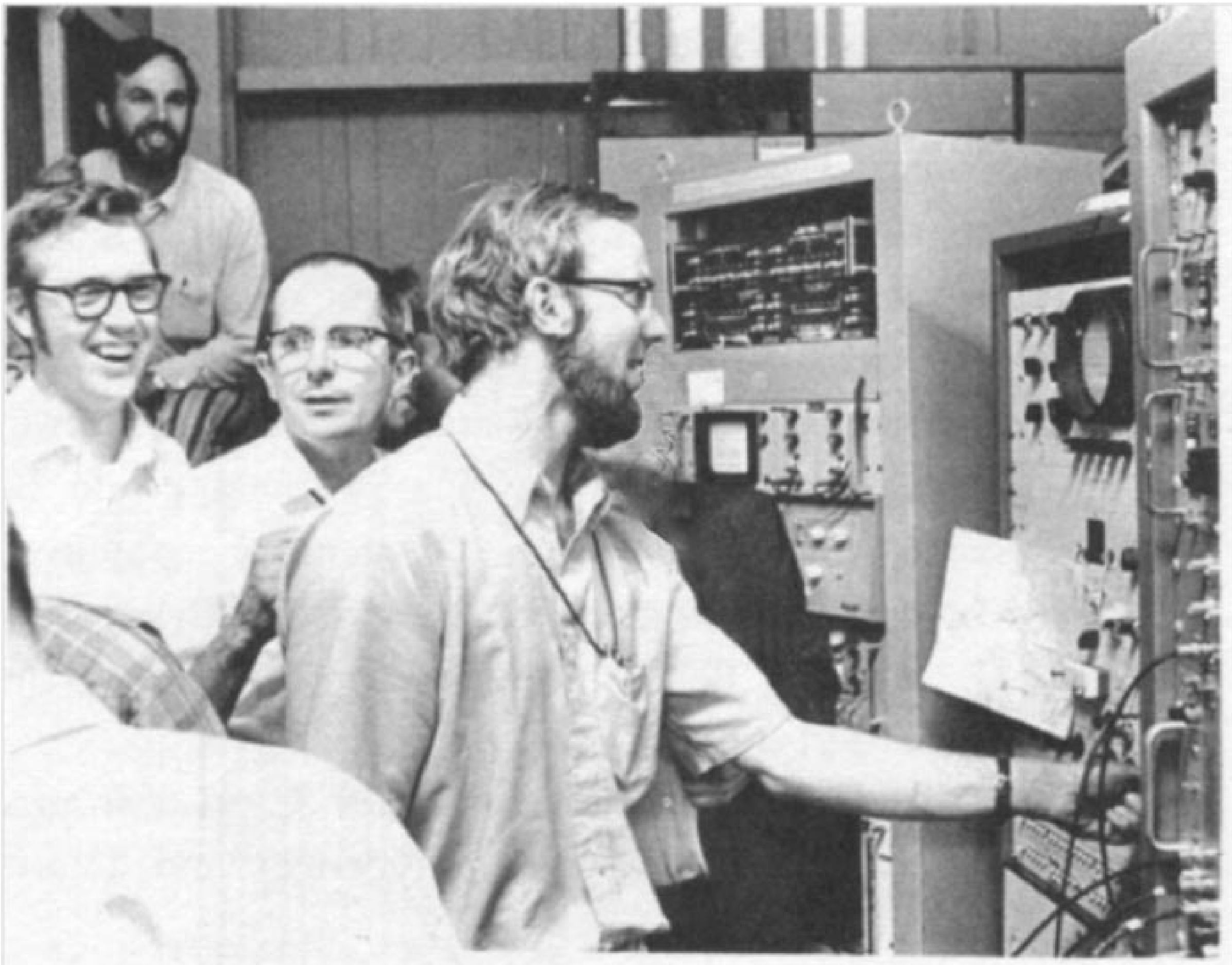

*Peter Gram's Delta Ray Spectrometer confirms that 800 MeV was achieved at 00:08 a.m., June 9, 1972.*

Bob Jameson, Al Criscuolo, Don Hagerman, Peter Gram

[toc]

## Commissioning

The long journey to commission the linac to reach the full 17 mA peak, 1 mA average design current at 6% duty factor for the User's program then began. A maintenance and an operations group were formed. I had looked forward to following the example of Cincinnatus and continued to refine the rf field control system. One evening rather late I was working alone on a 201.25 MHz module when 3M wandered by – surprising because he was never seen along the accelerator. He made some kind of greeting and seemingly that was all. Some days later I was asked what I had decided – but I had no idea about what. It turned out that I had been asked to form a group (MP-9) to lead the commissioning. I was not enthusiastic but agreed; it certainly did not appear to be a reward in any way from the very beginning or ever afterward. Especially when an initial condition had to be that the group had to contain two especially cantankerous characters, who had continually caused strife, both between themselves and widely. (One left the Division rather than submit to that; the other continued to be cantankerous.)

We continued to work long hours. The ability to do effective machine experiments was severely hampered by two factors. First, there was little maintenance support, because of severe problems in getting the experimental area working that drained the budget, so we spent the largest fraction of the scheduled machine development time fixing problems. Second, a low beam current experimental program was started, and an "operations mentality" descended hard – any changes in the operating settings were strongly resisted, and the operations staff was forbidden to participate in any of the machine development experiments. The full linac specification was not “officially” reached for ~10 years; an earlier demonstration was forbidden because the experimenters would then know that the linac was capable and would have increased even more strongly their demand for higher intensity – for which LAMPF was built. A conference article appeared in 2006 stating that it had really taken ten years, and a clarification, via two emails, to the author seemed in order:

*“Dear ___,*

*Read an article of yours, thought probably should correct a misconception. (wrote that it took 10 years to reach design intensity)*

*The LAMPF linac did not require ten years to reach design intensity. We had quite poor top management. The linac started operation on schedule in 1972 because of a heroic effort by a small group of people who took over the management of the linac and saved it. But the hi-intensity targets were not ready, and it took many years before they were. The top management would not allow the linac to be demonstrated at full design capability, because then the users would know that the linac was ready but the targets were not.*

*It was very embarrassing for us, but we could do nothing about it. One night, we just ran the linac up to full power for some minutes, to satisfy ourselves that it really was ok, and of course to get some hi-intensity data.* [25]

*Of course, improvements were needed. The beam loss was acceptable, but we wanted to reduce it. We found a problem with the longitudinal acceptance caused by a systematic error - the cell lengths in the side-coupled structures were too long, because in those days, human machinists always tended toward the long side of the tolerances, to avoid wasting a part. We found a tuning solution to work around this problem, and that resulted in a good reduction of the beam losses. Another large improvement came after extensive work on the (uninstrumented) 200 MHz front end.*

*Anyway, the linac was absolutely not the reason why LAMPF took so long to operate routinely at full intensity - it was the targets.*

*Wishing you a happy and productive New Year 2007,”*

*“As our "management" was against it, we didn't trumpet that we had run the linac up to design performance that night, and I don't remember now if we made a note in the log. I made sure the logs got archived properly when I retired, so could look maybe someday. But it would have been later in 1972. Also I would have to reconstruct from the logs to give an accurate answer to your question (about the commissioning progress). It is rather hard to say, because of the situation with the targets. We ran at reduced intensity and duty factor to around 1975, when we had a "Great Shutdown" that centered on the target area but allowed us to tear into the linac also. There was a big problem in the drift-tube linac - there were bellows on the drift tube stems and they had failed under high duty factor. We also realigned the whole machine then. Sometime later we made the other improvements I noted in the previous email. But for many years, there was hardly any maintenance on the linac because of the money drainage into the target area - we really struggled with linac development all the way up to 1977. Every time we had machine development, we spent most of the time fixing little things - I wrote a "memo to management" once that cataloged our schedule and pointed out that our scheduled machine development time was at best 20-25% effective because there was no maintenance support (or operations support - the operators were forbidden to take data for our experiments, for example!) So getting to "full performance with minimized beam loss" did indeed take a long time. After 1977 there was little development on the LAMPF linac because we had given up on the LAMPF*

---

[25] In April 2020 (Vol.73, #4), an article on the role of particle accelerators, Figure 4., omits a historically important data point. The LAMPF accelerator was capable soon after its initial operation in 1972 to achieve 800 MeV at 1 ma average current = 800 kW beam power, but could not be operated there because of the beam target until later, but then did provide 800 kW for some time. The LANSCE data point shown is because that power level was adequate for the PSR (Proton Storage Ring). The subsequent SNS linac 30 years later is basically a copy of the LAMPF linac, except for superconducting cavities at the high energy end. The LAMPF data point, plus following beam dynamics research including equipartitioning, was important because it allowed vigorous discussions on Accelerator Transmutation of Waste (ATW) to be initiated in the 1990’s with confidence concerning the accelerator requirements.

*management and formed the AT Division. We had learned, at least, that the history of a technical facility depends a lot on the administrative framework."* [26]

The memo about 20% effective machine development time was not popular with the operations chief and it was arranged for me to be fired. The group was incensed, but back to a garden somewhere seemed not so bad. Ken Crandall came over to my house that evening. Next day pointed out the remaining problems and also that the Division did really need the operations chief, who needed big time help. Ok, not this time…, and we kept on.

## The RF Low-Level Control System

My rf field phase and amplitude control system worked well from the initial LAMPF turn-on, and it is interesting to note that the original installation is still in operation in 2022.

Each component was designed very carefully, with possible errors analyzed and against the overall specification of the control loops. The phase reference signals are taken from directional couplers on a temperature-compensated coaxial rf reference line that runs the length of the accelerator. Special phase detectors, phase and amplitude modulators, and waveguide directional couplers were built.

The printed circuit boards were carefully laid out using stripline techniques to protect against rf noise pickup. Also for this reason, the three small potentiometers for adjusting the proportional, integral and derivative (PID) elements of the loop gains were placed on the printed circuit board, rather than bringing the associated wires out to front-panel potentiometers. It turned out that the boards were indeed not affected by rf noise, also when an extender was used to bring the circuit boards outside the NIM bin shielding to make potentiometer adjustment easier. So for over 45 years, walking down the klystron gallery shows the two phase and amplitude modules sticking out on extenders.

The circuits are hybrid analog and digital. The loops are analog, with digital sample-and-hold of the integrator voltage between pulses, which minimizes the pulse turn-on transient.

Nowadays, the low-level rf control has been realized at many places using fully digital techniques. The main advantage could be that central computer control of the PID gains could be realized, digital accuracy, and full decoupling between amplitude and phase; however at significantly higher cost. To make work, replacement of the low-level rf control at LAMPF has been discussed, and has been in progress for a long time, but without success and without learning from the Star Wars and SNS experiences described in other sections.

## Alignment

It was found that the initial alignment was not coordinated among several large segments, resulting in disjoints (one as large as 1 inch), so realignment of the entire complex was necessary during the "Great Shutdown". Again, it fell to my Group. The **framework** and specification that I required for realignment was simply that it would be presented on one piece of paper, with one origin.

## Fixing the 805 MHz Linac Longitudinal Acceptance

Finding and solving the problem with the longitudinal acceptance was a real challenge and victory. Beam loss and other indications pointed to a longitudinal problem, but we seemed unable to find an improved tuning solution. Then someone (have never been able to find out exactly who, but likely Ken Crandall) had an inspiration – in those days, parts were made by human machinists to be within a ± tolerance – but humans would tend to be sure that any error would be on the plus side to avoid throwaways – so in machining the individual cells, there could be a systematic error making assembled tanks too long.  We rushed to the machine, shut it down, and armed with steel tapes, entered the tunnel and measured the tank lengths.  All too long – some by more than a centimeter. Crandall [27] immediately plugged the measurements into his simulation code, and it showed that the longitudinal acceptance was too small and misshapen!

---

[26] The dissatisfaction over the non-existence of any effective, helpful, plan or guidance at the Division level was not at all limited to the MP-9 machine development group. It was expressed from many directions, culminating in the "mutiny" noted in the section "Technical Administration and Documentation" below. A folder titled "A Tree in Los Alamos" in my computer archives contains the memos reassigning the line of command for the 805 MHz structure tuning, installation and commissioning, in 1975 fully detailed and indisputably factual pleas by me to 3M for Division-wide planning, and other environmental frustrations.

It took quite a long time to find a good solution.  The initial least-squares retuning of the whole linac was ineffective.  Crandall worked out a multi-step method which finally yielded new phase and amplitude setpoints, and simulation indicated that the longitudinal acceptance was restored to ~95% of the design value and with a full, normal center region.  The new procedure resulted in significant reduction of beam loss.  [28,29]

**201.25 MHz Linac Tuning**

The second major reduction in beam loss came from a long and very complicated sequence of experiments on the 201.25 MHz linac, A severe complication was that there was almost no instrumentation (that budget had suffered) so we invented beam-based techniques [30]. Our strategy was to first understand the performance of the 201.25 MHz linac with no (very small) beam current), then with full peak current, and then the 805 MHz linac similarly.  The 201.25 MHZ drift-tube linac tank ends were deformable so the field flatness could be adjusted.  This had been done at low rf power, but again there appeared to be a difference at full power.  Operations had already retreated fully to the philosophy of just run it like yesterday, or fiddle knobs, without technical understanding or backup, but against fierce resistance, because the beam loss was indeed of concern, we adjusted the tank ends in steps and measured beam performance (and returned the tank ends precisely)  New tank

27 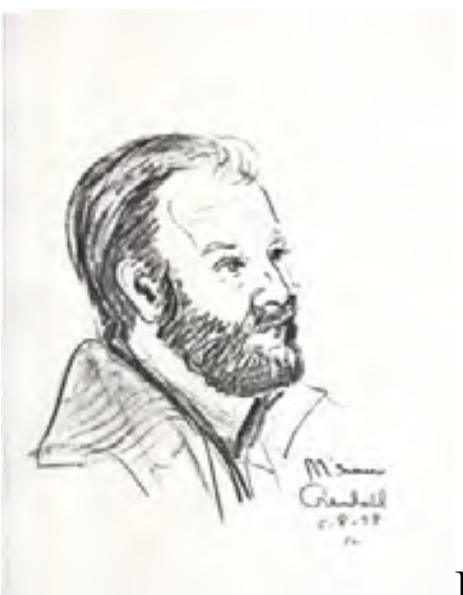 Ken Crandall, 1978 (Carol Coppersmith)

28 K. R. Crandall, "At Revisited, " Los Alamos Scientific Laboratory memorandum, Jan, 3, 1974,
K. R. Crandall, "805-MHz Linac 13csign Modifications Due to Length Errors," Los Alamos Scientific Laboratory memorandum, luly 1, 1974,
K. R. Crandall, "Status of 805 M}{z Linac I.ength Corrections," Los Alamos Scientific Laboratory memorandum, Jun. 22, 1975,
K. R. Crandall. "Summary of 805 MHz Linac Length Corrections," Los Alamos Scientific Laboratory memorandum, March 26, 1975.

29 R.A. Jameson, "Optimization and Nonlinear Solver Experiences in High-Intensity RF Ion Linac Problems, 3rd Intl. Workshop on Beam Dynamics and Optimization, 1-5 July 1996, St. Petersburg, Russia, LANL Report LA-UR-96-3204, CONF-9607158--1

30 We developed many beam-based techniques – a new field in those days – later many reinvented elsewhere and declared as "for the first time"… end settings resulted in significantly less beam loss.[31,32, [33]] The reports were finished in 1978 from AT Division. Using methods like this, we had mapped out a campaign to continue with performance of the 805 MHz linac with low current, then the 201 MHz linac with space charge, and then the 805 MHz linac with space charge, but MP-Division had no further interest in understanding or improving the LAMPF linac.

**The Δt Procedure**

A beam-based, time-of-flight technique, named the Δt procedure, for setting the phase and amplitude setpoints for the 805 MHz linac was developed and could be automatically executed by the computer control system. It produced satisfactory performance with low enough beam loss, but there was a mystery that remained until I finally figured it out during the winter of 1989, after returning from Japan. If the peak beam current was set to only ~1 mA, the procedure gave the correct 800 MeV output energy and an acceptable beam loss pattern when the peak current was then simply raised without further tuning. However, if the procedure was attempted with greater peak current, it diverged unpredictably, with large mean output energy errors, as much as 1-2%, sometimes plus and sometimes minus, and the beam loss pattern was unacceptable. I had developed the Δt hardware and had subjected it to exhaustive tests, but found no grounds for the mystery. By 1989, I had been working in beam dynamics, and decided to look into the Δt software. It was old, but re-engineering it gave a simulation model with LAMPF detailed data, which could be tuned sequentially as on the real machine. The model showed the same mysterious performance! Finally I realized that the problem was that the operator was in the loop. The tuneup sequence is a classic control problem with the possibilities of convergence, a limit cycle, or divergence. The operator or computer adjustment of phase or amplitude in random order, and the characteristics of the system could produce any of these results.

The study was extended to find new set points based on fitting to a new model with more parameters. Simulations indicated that modules with problematic performance were the same as those identified as such when the tank length problem was solved – indicating that our method of setting the stopband had been successful over a period of 17 years without relaxation to the unstable condition. There had been continuing complaint over these years that the 800 MeV beam had a long longitudinal tail which was scraped off in the 90° bend to the Proton Storage ring, with beam loss and a problematic beam injection into the storage ring. Work with the new model indicated that with new phase and amplitude setpoints, this tail could be removed. The sensitivity of the model to data collection errors indicated that a direct-search tuning algorithm, without taking derivatives, would be needed to replace the random setting method. [34]

---

31 R. A. Jameson, W. E. Jule, R.S. Mills, E.D. Bush, Jr., R.L. Gluckstern, "Longitudinal Tuning of the LAMPF 201.25 MHz Linac Without Space Charge", Los Alamos Scientific Laboratory Report LA-6863, March 1978.

32 R. A. Jameson and J. K. Halbig, "LAMPF 201.25 MHz Linac Field Distribution", Los Alamos Scientific Laboratory Report LA-6919, January 1978.

33 Jameson, RA; Jule, WE; "Linear Accelerator Modeling Development and Application", IEEE Transactions on Nuclear Science; June 1977; vol.ns-24, no.3, p.1476-8, Proc of the Part Accel Conf, 7th, Accel Eng and Technol; Mar 16-18 1977; Chicago, IL, USA

34 K. R. Crandall, R. A. Jameson, D. I. Morris, and D. A. Swenson, "The Dt Turn-on Procedure", Proc. 1972 Proton Linear Accelerator Conf., October 10-13, 1972, LA-5115, Los Alamos Scientific Laboratory, p. 122, November 1972 .

R.A. Jameson, "Optimization and Nonlinear Solver Experiences in High-Intensity RF Ion Linac Problems, 3rd Intl. Workshop on Beam Dynamics and Optimization, 1-5 July 1996, St. Petersburg, Russia, LANL Report LA-UR-96-3204, CONF-9607158--1

H.A. Thiessen, et. al.. include R.A. Jameson, "Report of the Committee on a TA-53 Upgrade, March 7, 1994", Los Alamos National Laboratory, -UR-94-1924, 7 June 1994.

The work was presented in detail on 1/25/1991 to MP-Division staff, with offer to collaborate on confirmation and development of a new control system algorithm. The staff complained very much – too much work to get data, machine not in good shape, etc. At this point Ed Knapp was MP_Division Leader and LAMPF Director; he listened patiently, but made no comment, and the subject was not pursued further, to this day. Modern higher energy ion linacs are using short, superconducting rf structures, and absolute phase measurements are possible, so other tuneup procedures are used, but the power of the Δt procedure, and the lessons learned, are probably still very relevant.

**Technical Administration and Documentation**

It had been very interesting to bring LAMPF to the operational stage, for the first time with a full central computer control system, and it was intriguing to think that the computer could also allow a technical overview to be created to help bring the linac to full performance and to reach and maintain high availability for the Users. We worked out an efficient system for getting failure notifications and maintenance records for the machine components, and used a LASL Computer Division developmental "database" GIRLS [35], a computer science research subject in those days, for a maintenance database [36]. Much thought was given to how useful information for improving reliability and availability might be gleaned and presented from the mass of raw data [37].

This concept for technical administration met with fierce resistance however. It was necessary to learn that another point of view was very powerful, as Frederick Taylor learned with his famous time-and-motion studies. Performance data motivates some people, who are pleased that someone notices. Others fear it obsessively, being afraid that they will be forced to work harder. An obscure system is good job security. So our database system was discarded. When the Spallation Neutron source (SNS) was built at ORNL thirty years later, computers were so ubiquitous that this kind of technical administration was unavoidable, and has been very useful in maintaining the availability of the SNS facility at a high level.

After completion of the SLAC 2-mile long electron linac, the staff voluntarily produced a fine, thick Blue Book [38] documenting the knowledge they had accumulated. We volunteered to do the same for LAMPF. But the LAMPF Director refused. The statement was that LAMPF was the best example of a perfectly managed project in the world, it was perfect, and no improvements or upgrades or extensions were required. We did get permission to create a bibliography, which listed all the technical material with copies on microfilm. I have one copy of the microfilms and reports, the other

---

35 W. Draisin and G.M. Connor, guide to girls vol. 1, users manual, internal report, Los Alamos Scientific Laboratory, June 1, 1973.

36 R. A. Jameson, R. S. Mills, M. D. Johnston, "Management Information for LAMPF", LA-5707-MS Informal Report, UC-28, LASL, August l974.

37 R. A. Jameson, "Reliability Engineering for Facility Effectiveness", (Invited), 1975 Particle Accelerator Conf. on Accelerator Engineering and Technology, Washington, D.C., March 12-14, 1975. IEEE Trans. Nucl. Sci., NS-22, No.2, p. 1030, June 1975.

38 SLAC Blue Book: http://www.slac.stanford.edu/library/2MileAccelerator/2mile.htm

was sent to LANL archives when I retired.  MP Division salvaged the microfilm reader. 39 [40]

The continuing budget drain of the experimental area and other management problems caused the capable financial bookkeeper of MP-Division to resign for fear that manipulations he was ordered to perform would leave him hanging.  This spurred a mutiny, and the Group Leaders as a whole went to Agnew concerning 3M. But the latter used an old trick – found the weak link in the GL chain, and offered him the job of financial officer, which was easier than the current position.  With the chain broken, essentially nothing changed. This also broke the motivation of the operations manager as far as a more technical machine development might have been.

[toc]

## Fate

The operations mentality, such a different culture from the technical one, can succeed with clever knob-twiddling.  It was so different and hard to understand that one began to wonder if ideas like RAMI, databases, availability goals, etc. were just pipedreams.  Another disaster was also looming at the time – the catastrophic effect of the Harvard Business School of Management culture on scientific R&D and creativity in general in the US. (See H. Agnew below) In a foresighted (but futile) effort to perhaps postpone by being informed, the lab established a program for experienced managers to attend UNM in Albuquerque for a two-year, compressed, full-content MBA course leading to a Masters degree in Management (MM) – I participated in this program from 1975-1977  [41]  The course work was enjoyable and interesting, with an opportunity to explore subjects missing from the engineering curricula, like statistics, learning methods, operations research, accounting.  The courses in organizational behavior, marketing, etc. were very informative and made it clear that I was not alone in observing the tremendous differences between cultures.  But it was also very clear that the proposals for business models based on five to seven or so precepts on “how to be successful” were rubbish for managing science, and it was crystal clear that none of these “techniques” would find any hint of mention or implementation in my work at LASL.  How different was this to be from 1986 on …

We also observed beam emittance growth in the linac, and in those days this was an unexplained phenomenon.  I determined to learn “accelerator physics”, and started working in beam dynamics and

---

39 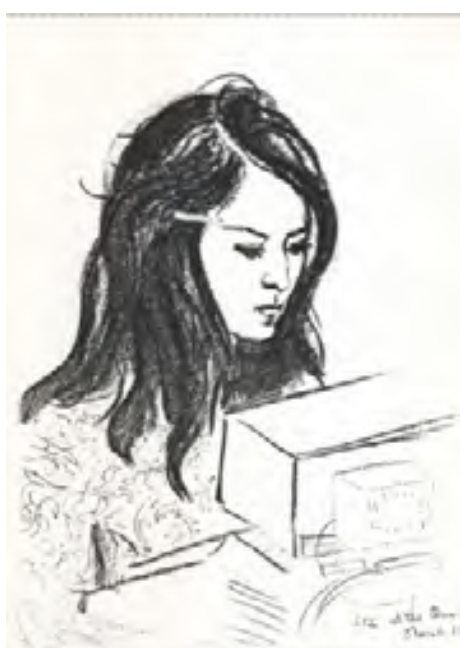 Liz Roybal (Carol Coppersmith)

40 R. A. Jameson and Eliza U. Roybal, “Annotated Bibliography of LAMPF Research and Development", Los Alamos Scientific Laboratory Report LA-7431- MS, July 1978.

R. A. Jameson and E. U. Roybal, “Annotated Bibliography on High-Intensity Linear Accelerators", Los Alamos Scientific Laboratory Report LA-7124-MS, January 1978.

41 R.A. Jameson, "Learning Factors in a Mid-Career Management Training Program", a Practicum submitted for MM Degree, Robert O. Anderson School of Business and Administrative Sciences, University of New Mexico, Albuquerque, New Mexico, 1975-1977 Management Masters Program, August 1977.

simulation [42], aided enormously by computer work executed by Rene Mills. This activity provided fascination and involvement to the present, as outlined in the other half of this book. It helped lead to development of the RFQ, and in 1981, the discovery of how to practically define and use an equilibrium condition (equipartitioned - EP) for a linac beam with space charge. It would have eliminated some decades of confusion in the community over the role of EP if it had been possible to test it on the ideal testbed of LAMPF, but this was never to be.

The LAMPF accelerator has been in operation from 1972-2022 = 50 years. An impressive legacy, and source of livelihood riding along on it "for the Lab" and for the people at LAMPF/LANSCE. Assuming ~100 people in operations division alone, 5000 man years.

Bringing LAMPF to initial operation 1972 and to full operational status by 1977 required talented and enormous effort – not without also personal cost.

"Thank you" was never said, especially by the MP Division Leader/LAMPF Director – whose contributions were negative, whose blood was saved from the cooling system, who finally got his turn-on day photograph retouched highlighting himself. No matter.

We who were involved did a highly professional job, we are proud of it and ourselves. In terms of their interactions with me, I thank all of them.

Essentially no improvements beyond maintenance have been made to the LAMPF linac since ~1980. It continues to run as we left it, like a faithful old television set. LAMPF has continued to be run as an operations facility to the present, and has essentially remained unchanged, however with signs of hope and renewed technical expertise starting mid-2018 as outlined below. The loss of major upgrade possibilities for LAMPF, and eventually the loss of the SNS to ORNL require a separate heading…

[toc]

## Loss of Major Upgrade and Extension Possibilities

**Loss of Upgrade Possibilities**

3M clung to the position of LAMPF Director far, far too long, and his position as stated above was evidenced by lack of further refinement of the LAMPF linac, which could have proceeded stepwise and could have built a strong base for an upgrade. Substantial improvement in the linac understanding and performance could have been achieved even without any increase in budget or staff.

Proposals for facility upgrades to include circular accelerators were indeed prepared, with the involvement of Nagle, Knapp, Thiessen and others. Very much credit is due H. Archer (Arch) Thiessen, a talented and dedicated LAMPF staffer, who prepared a number of thick and thoroughly presented proposals to upgrade the LAMPF linac and add circular accelerator facilities to address various research activities, including spallation neutron research. However, without success. It is not apparent that 3M ever took a strong and leading position to fight for an upgrade. There was no bonding with LANL as a whole – the LANL view was that he considered the title of LAMPF Director as equivalent and equal to that of the LANL Director – a separate entity – and thus as far as they were concerned, LAMPF could hang by itself. Without question, a strong unified push by LAMPF and LANL could have achieved an upgrade program, as has been evidenced by every other large accelerator facility worldwide.

Finally it became very clear that a change was necessary and in 1985 he finally agreed to go, but with a condition – that the successor agree to be a sorcerer's apprentice for one year before receiving the title! This was unbelievably preposterous, but the "upper management" accepted it, and believing that they could "attract candidates at the level of an IBM vice-president", put a job advertisement with this condition on the street. Within days, the whole accelerator community, as surely also all IBM vice-presidents, were laughing and joking that now they could really understand the level to which LANL upper management had sunk. I was on the selection committee, headed by an "upper manager", and went to this manager to plead that the ad be taken off the street, as we were the laughing stock of the

42 R. A. Jameson, "PARMILA Input Subroutine, and Particle Distribution Fitting Techniques", Los Alamos Scientific Laboratory Office Memo, MP-9, June 22, 1978.

whole country. His reply was “my job is to fill the position”. At least one other selection committee member made the same plea. Names of possible candidates for possible solicitation had been sorted into ranks. The first candidate was the very effective director of a foreign accelerator-based laboratory. At the fifth or sixth level, a candidate was found who would agree to the condition. Having just been “relieved” as a Division Leader at ANL, this candidate was actively resisted by the selection committee, but “the job was filled”.

The new LAMPF Director, however, ruined Arch’s further work. His characteristic was supreme arrogance, and his trademark became his arrogant demands to the DOE for “a blank check” to do whatever he wanted at LAMPF. When he would set up a DOE appointment, the DOE people would spread the word “here he comes again – want to bet if he will ask for a blank check?”.

There were other problems as well, and finally the 4th regime got up enough gumption to relieve him. Just before the meeting of the LAMPF Users Group (LAUG), who arrive to find that LAMPF had no Director. Justifiably concerned, they asked for a meeting with the 4th regime to find out what was planned – and were told that the 4th regime had no time to meet with them during their several day meeting. LAUG contained some very experienced people with good connections, and, outraged, arranged that time was found. There being no follow-up plan, the solution, was to reinstate the LAMPF Director that had just been relieved, which LAUG reluctantly accepted. Universal disbelief – and even more when the blank check was again demanded of the 4th regime as a condition! But that was too much, and he was again relieved (and received the respectable Senior Fellow position that he had cannily arranged as a golden parachute when appointed). The LAUG was still in session, and asked that Ed Knapp be asked to be LAMPF Director. Ed had no respect for the 4th regime but a sense of duty, and agreed with the firm statement that he would do it for no longer than two years.

Ed was LAMPF Director from 1990-1991. The LAMPF facility represented a large resource that could have been extended much more economically that building a new spallation neutron research facility from scratch. With his political skill and connections, he arranged for solid and unswerving support from the DOE for the LAMPF linac to be upgraded to a superconducting machine, with higher energy, beam current, and tuning flexibility. This was a shrewd move, to have an upgrade step that would strongly enhance the existing mission, but also result in a linac base that would have been an almost insurmountable advantage for Los Alamos in adding a circular machine and doing spallation neutron research there. The next step was for LANL to put its support behind the proposal. This was refused by the 4th regime.

Ed always got very angry when relating this. The 4th regime, as expanded below, was characterized by smallness. Small science, no understanding of the benefit of flagship projects, scared of its own shadow, no stomach, petty, and also loved sycophancy, which Ed was far above. It is quite possible, even probable, that the pettiness of the 4th regime got revenge by this refusal, the mean smallness of this in comparison to the huge benefit to LANL not withstanding (see below – 1987).

**Loss of the SNS**

US science did not stand still for LANL or LAMPF. The role of “medium-energy physics” came to be seen as essentially fruitless, and other directions were developing, in particular research using spallation neutrons. As LANL could not field a proposal, others moved in.

Alvin Trivelpiece had a very distinguished career, including DOE Director of the Office of Energy Research 1981-1987, where he supported many large projects, and laid groundwork for spallation neutron research, that would become the SNS at ORNL. He served as Director of Oak Ridge National Laboratory (ORNL) from January 1989 - March 2000. The Iron Curtain fell and perestroika came, Adm. James Watson was Secretary of Energy (1989-1993), and asked the labs in clear terms to redefine their missions. One of the things they should do was to say what their central facilities should be. A magnificent opportunity.

Trivelpiece managed a masterwork. The capstone of ORNL’s program was to build an advanced nuclear reactor, but the ever increasing costs caused Congress to cancel the program. Immediately after the election of Clinton, with Al Gore of Tennessee as Vice-President, the reactor project was restored, but it was known that it was only a matter of time before it would be cancelled again. Trivelpiece agreed not to fight this, but stated that “neutrons are our business”.

At this same time, we at LANL had new ideas for closing the fission power cycle by accelerator transmutation of radioactive waste, and were also involved in the fusion materials testing program IFMIF, both of which would have ideally suited Los Alamos, and were trying to convince the LANL 4th regime to respond to the DOE challenge, using the same phrase – “neutrons are our business” – without success, as outlined below.

Trivelpiece arranged with the DOE that ORNL would be named as “the preferred site for the SNS” in exchange for not fighting the reactor cancellation. I was traveling to ORNL at that time in connection with IFMIF, and rode once on a plane with Trivelpiece, so was well informed of this strategy. Trivelpiece even warned LANL. When the inevitable time came, he stood with Hazel O’Leary, who had become Secretary of Energy, when it was announced that the reactor was again cancelled, but the SNS was gained at ORNL. It was said that Trivelpiece was somewhat distant as ORNL Director, but clearly he was in the Agnew mold and concentrated on the large issues appropriate to a national lab director’s job, and not on little things, or agreeing to all forms of silliness.

It was widely told that the 4th regime was stunned by the announcement, and then did an incredible thing. A meeting of all the national lab directors was held shortly thereafter, and there he demanded they “give his legacy back”! The story circulated all over the country, including how the other Directors had laughed afterward.

## The Accelerator Technology AT Division – [toc]

## Founding, the First Round

**Ed Knapp (1932-2009)**

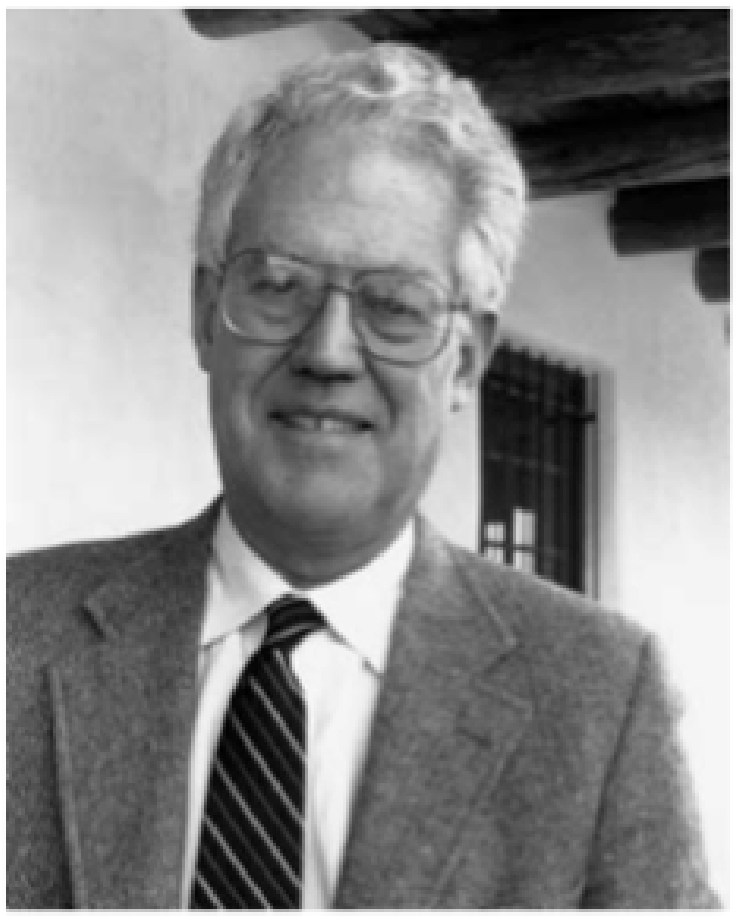

When Ed Knapp left the structure tuning during the LAMPF installation, he received a furlough to CERN, widely remarked by the troops. On return, he did not come back to MP-Division but went to P Division, with a parallel position as Group Leader of an MP group for practical applications. He wanted to develop a more compact and cheaper pion generator for cancer therapy, engineered for routine hospital use. A proposal for PIGMI (Pion Generator for Medical Irradiation) was successfully funded [43]. I drafted most of the writing for the proposal. It was clear that there was no future for accelerator technology in MP-Division; we were thinking of finding other jobs. It seemed clear that

43 E.A. Knapp, R. A. Jameson, D.A. Swenson, J.N. Bradbury, "Program Project Grant Application for Development of a Pion Generator for Medical Applications", Los Alamos Scientific Laboratory Report #p-1046-a, (successful application), February 1979.

there was not enough market to try to do accelerator work from a private company, so our tentative name of The Pojoaque Green Chile & Accelerator Works was only thing that came of that idea. But Ed figured that if we could stay within LASL, we might get federally funded projects, and proposed founding a new LASL Division to H. Agnew in 1977. Agnew was working at that time to expand the activities of LASL outside the weapons area, and so was receptive, with the caveat that he was very busy with many things, so we could try, but it was up to us to sink or swim, as he wouldn't have much time to help. Super! – just fine! And so the Accelerator Technology Division of the Los Alamos Scientific Laboratory was born.

The crucial aspect of AT-Division was its Charter, unique in the US at that time. Accelerator technology had developed only in conjunction with various uncoordinated project demands, and there was no national base except for some underlying support in the DOE Office of High Energy Physics for technology development relevant to that. The Division would be a technical home for development of all aspects of accelerator technology, with application to projects applying the technology to societal needs. Every opportunity for advancing the technology would be seized, and at the same time, project goals of technical specification within budget and schedule constraints would be competently met. The technology advances, and the satisfaction of being able to work on these, would be our "profit". Of course, we fully knew that our funding would depend on getting new federally funded projects, and executing these so well that new projects would follow. This Charter was envied and later copied by other labs, and was worth defending as a national resource, and as appropriate for Los Alamos.

Given history, I did not expect to be involved in the new Division, and was very surprised when Ed asked me to be his Alternate Division Leader. It was my great luck that E. A. Knapp, whose structure development had been refined, was a generous, self-confident and non-vindictive man. Although it was indeed a salvation and tense situation, he did not take our stepping in, to extend the knowledge of structure tuning and to implement it, the same way, and the footnote above did not apply. He was another that the Russians termed "a mighty (or "big") man".

Harold Agnew was a very good Los Alamos Lab Director (as Norris Bradbury had been), and the last good one. He was very direct and decisive, stayed by his decisions and one could count on his support, as indicated in the FMIT story below. He understood how to manage science, and how to motivate scientists. He did not suffer fools or nonsense gladly – had an almost life-size bronze in his office of a man spewing a cloud of baloneys from his mouth, an ever-present reminder to visitors to that office.

As an example, there was no "Deputy" position. I was Ed's Alternate Division Leader – the word choice made a great difference. He was the Division Leader, and I was Alternate DL. We reported directly to the LASL Director H. Agnew. It was a role that I enjoyed very much, and the years 1978-1984 were among my best years. Ed did most of the political and big picture work, and I worked the execution of the projects, in great harmony. Ed was a person that people enjoyed working for.

My role as MP-9 Group Leader was kept until 1980 – we hoped that a bridge from the new Accelerator Technology AT-Division could be maintained. On the AT-Division side, I enjoyed personal research until 1981, aided by Rene Mills, and with new collaborators including Gren Boicourt. After the EP discovery with Paul Channel in 1981, personal research was set aside for the challenge of maintaining and participating in a larger environment for creativity and accomplishment in accelerator technology.

Mahlon Wilson [44] came as Assistant Division Leader for all construction and scheduling activities, and very competently handled these areas.

Ed had a neighbor, Wayne Vanderham, who worked for the Los Alamos National Bank, and asked what I thought of a professional money manager for our financial officer. Experience with retrenched technical personnel as financial officers and the MMM experience indicated that was a great idea, and it indeed proved to be. I gave Wayne his job description: 1) The fiscal years ended with an accounting "closeout". If necessary, I would of course attend, but his performance would be considered good if I did not have to attend – and I never did. 2) The fiscal environment was an "old boys club", and he should become part of it. This was not a light assignment, as he had neither a technical degree nor an MBA (which he later got). Wayne and AT also had very competent financial service from Helen Carpenter and Sharon Brock.

The initial groups were:

- AT-1 Practical Applications, Don Swenson Group Leader (GL). Jim Stovall later became Microtron Project Leader and then GL. PIGMI , the RFQ.
- AT-2 White Horse, JF GL, later Tom Hayward, then Fred Purser. This was the Army-sponsored anti-missile defense program that preceded the Strategic Defense Initiative (SDI). (I showed that high rf frequency was better for high brightness beams – simple, directly from the envelop equations [45], but very surprising to many.)
- AT-3 Proton Storage Ring (PSR)
- AT-4 FMIT Ed Kemp GL, Dale Armstrong, Dave Schneider
- AT-5 RF Systems, Don Reid GL, later Ed Higgins, Mike Fazio
- AT-6 Theory, Dick Cooper GL
- AT-7 FEL Charlie Brau PL and Jerry Watson GL
- Later AT -8, Control Systems, Mike Thuot GL, Dave Hurd, Don Machen, Peter Clout

Harold Agnew was LASL Director from 1970-1979. His farewell address was a serious warning, and unforgettable – here are the key excerpts:

**Harold Agnew Farewell:** From "The Main Gate", Laboratory Retiree Group (LRG) Newsletter Spring, Vol. 12 No. 1 2007: Maestro Harold Agnew, conducting "Those were the days, my friends, we thought they'd never end …"

"The ever-increasing bureaucracy, composed of managers who require more and more detail, justification, and guaranteed schedules will, in the not too distant future, completely eradicate our nation's world position in research and technology.

Bureaucratic regulations and requirements for conformity will stifle basic research.

Bureaucracy will eradicate creative endeavor and innovation in the long run.

Bureaucracy eventually loses sight of the original objective, and becomes only concerned in its own management and control function.

Unless the trend towards centralization is somewhat reversed, I predict that the U.S. will rapidly lose its lead in science and technology. "

By Harold Agnew, Director LASL

---

44 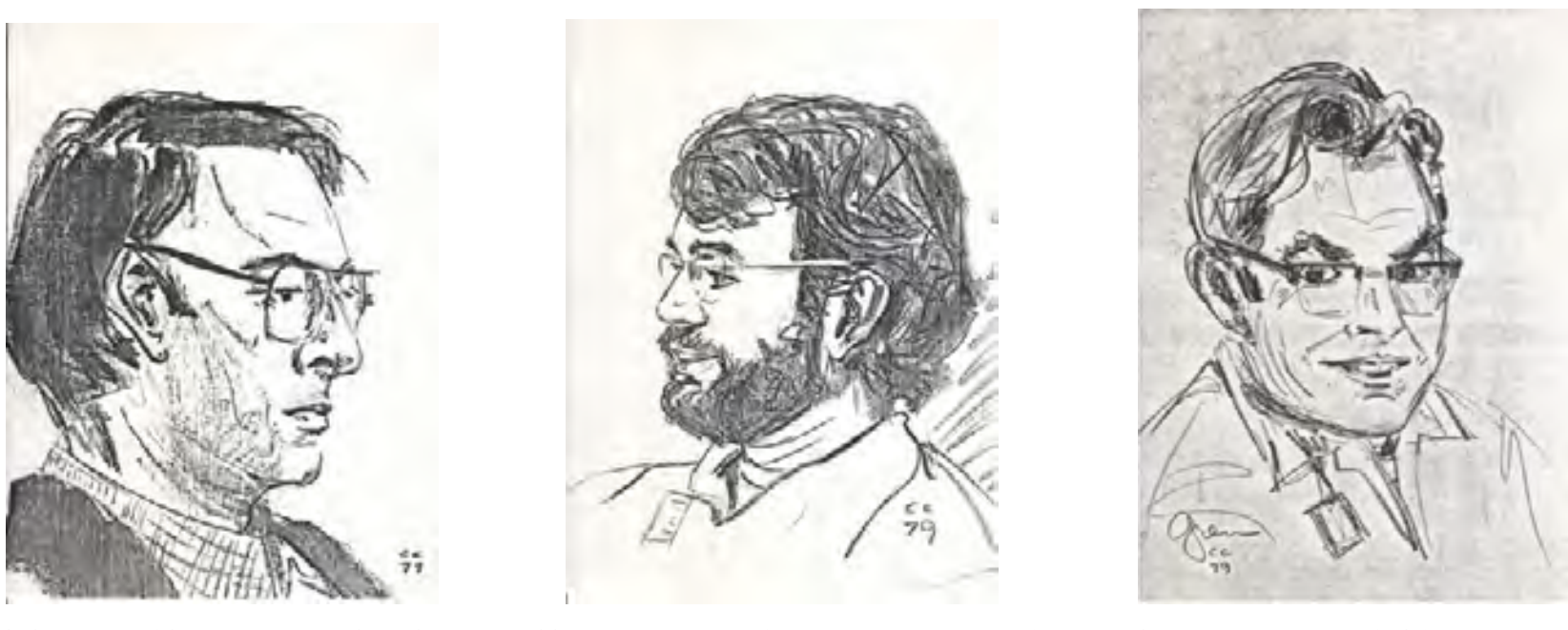

Mahlon Wilson, Paul Channell, Gren Boicourt, 1979 (Carol Coppersmith)

45 R.A. Jameson, R.S. Mills, "Factors Affecting High-Current, Bright Linac Beams", Los Alamos Scientific Laboratory Office Memo, MP-9, April 8, 1977.

1979-1985 Donald Kerr succeeded Agnew as the 3rd regime, and initiated the downfall of LANL. LANL alumnus, went to DOE, had no compunction to come back as director commissioned to destroy the old culture and implement DOE micromanagement. A close source commented that he would sell his own grandmother for a small price. AT-Division did not suffer much under this regime while Ed was there – there were bigger cultures to kill. The typical malfunctions of matrix management began to appear – power games, sycophants; one cohort was sent to prison for abuse of travel. By the second half of 1984, project management had infected the White Horse Program. In 1985, Kerr visited me for the first and only time, sat across my small round table, and declared that I was now forbidden to travel in search of new work, as project managers would take care of us. It was clear that he wanted a reaction for which he could fire me. I knew something of how things lay with him and my instant thought was that I would outlast him, and I just nodded. His scandals were catching up with him, indeed I did outlast him, and did not stop traveling, as it was clear that Star Wars would not last forever and that new projects would be needed then.

The laboratory was renamed Los Alamos National Laboratory in 1981. We did not like that. Said it put the lab in the same category as the Los Alamos National Bank.

Ed took Agnew's warning to heart, and also wanted to try working in Washington. He became Assistant Director for Physics and Mathematics of the National Science Foundation in 1982, and shortly thereafter the seventh director of the National Science Foundation (NSF), serving in that capacity from November 1982 to August 1984. From his NSF biography (NSF Web Site) "Knapp was praised by those who were familiar with him as "a first-class scientist, well-educated, with extensive management experience". At the NSF he briefly served as Assistant Director for the Mathematical and Physical Sciences Directorate before becoming director. During his tenure, the briefest of any director, the NSF budget increased 38 percent. However, he did not enjoy being an administrator, and left the NSF to return to research."

Ed would never discuss this war experience. He had arranged to become a LANL Senior Fellow when he went to Washington, and came back to the lab in that category. He quickly learned that a Senior Fellow, supposed to be a valued advisor, is totally ignored. As related above, he served as Director of LAMPF from 1990-1991. He helped found the Santa Fe Institute and later served as its President. Again, the people there enjoyed working for him.

**<u>PIGMI and Muon Therapy at LAMPF</u>**

Ed Knapp was instrumental in establishing a program for cancer treatment at LAMPF using the muon beam, and AT-Division participated actively in it. Y. Iwashita, from H. Takekoshi's Keage Accelerator Laboratory of Kyoto University, came to work on PIGMI for three years – collaboration with both became of great significance for me later. Don Liska designed and built a variable energy degrader for controlling the penetration depth of the muon stopping point – a mechanical masterpiece. The treatments were designed and controlled by computer, with tomographic data collection and analysis – among the first of such applications. Control treatments were done using x-rays, resulting in a great increase in the effectiveness of the x-ray treatment. The high effectiveness of muon treatment was offset by the high cost of an accelerator system, and, even considering a PIGMI accelerator, the cost-benefit of muons did not become attractive enough.

**<u>The Challenge of the Fusion Materials Irradiation Test (FMIT) Program</u>**

The establishment of the FMIT program in the new Accelerator Technology Division of the Los Alamos Scientific Laboratory is a story from the time of giants, in such sharp contrast to subsequent idiocy that it warrants telling.

Development of neutron-resistant materials was needed by the fusion research program, and the FMIT program had been given construction approval ~1975. Three teams had competed – two were collaborations of a national lab with accelerator experience and an industrial partner, and the third was the Hanford Engineering Development Laboratory (HEDL), where there was no accelerator experience, and an industrial partner. The project was given to HEDL for the usual political geographic reasons. The DOE expected that one of the losing national labs with accelerator experience would join to build and deliver the accelerator, but they refused. The project sat for about two years.

Shortly after H. Agnew agreed to the forming of AT-Divison, he called Ed and phased the question as to whether we would like to build the accelerator for FMIT, which Ed related to me. It was clear that it was not a question but an order, and as the DOE Fusion Director Ed Kintner was coming to Los Alamos either the next or the following day, we needed to very quickly determine our strategy and price. We outlined three points:

- We would have to build a prototype, because of the unprecedented high beam current and average power and the extremely formidable task of trying to extend the known technology.
- As a new Division, we did not have any assigned office or lab space, so FMIT would have to construct new office and lab space for our work.
- We needed some assurance of continuity, so 1.5 years of secure funding must be guaranteed.

We knew that the third point could not be promised, given the fickle nature of US funding, but it could represent a concession if necessary. Agnew agreed. The meeting with Kintner was an amazing and unforgettable insight into the performance of a good Director. It took place in a room with a long, thin, oval table, with Agnew and Kintner sitting across from each other at the top end. I sat at the far end. Kintner was known to be a tough, technical, experienced and effective program manager. He rejected our demands one by one.

"Why did we need a prototype? Ridiculous! Expensive! Just build the machine at HEDL! I (He) was a young member of Admiral Wickover's famous atomic submarine development team – we just built the first one and got inside it and drove it around under water in the Boston bay!" Agnew's answer, very cool – "Ok, then we won't do the project. Second point?"

We (the DOE) have never, and never will, build space facilities for a project – that has to be done by the lab from its operating funds." "Ok, then we won't do the project. Third point?"

"You know that is impossible!" Laughter. Agnew did not budge. The R&D situation was detailed, along with the need for working space. Finally Kintner agreed, and then something really amazing, and that I never again saw even approached, occurred. Agnew stood up and offered his hand to Kintner, who stood and they shook hands on the deal!

Later I was the interface to the DOE and Kintner, and the establishment of this firm and good relationship with Kintner was crucial. As a good program manager, Kintner tried to renege on the agreements. I would just say "No problem – I will go back and we will withdraw from the project". A new 3-floor, narrow office building had been erected on the south side of the LAMPF road for P-Division, which Agnew reassigned to us, to which FMIT built an addition of roughly the same floor space, Ed picked the color scheme as orange, as a motivating color. An auditorium on the side of the new addition is known as "the Orange Box".

Another extreme problem arose: the DOE was already trying to wrestle control away from the "laboratory fiefdoms" and impose micromanagement, and the FMIT project was chosen to be the "pilot program" for their new project management system, with demands for mountains of minutia. I went to Agnew and pointed out the cost and obstruction of work that this would entail – he responded immediately that LASL operated under contract to the University of California, had rules for monthly reporting of costs and progress, and that we would continue to provide this, and only this, information and the DOE and HEDL could do what they wanted with it. So I went back to Washington and told the DOE "No!", which astounded them (and may never have happened again, certainly not after Agnew left). I knew that the LASL Director was behind me, and would not weasel or change his mind and leave us hanging if challenged – a profound difference from the situation during the next, and especially the 4th, regime. The project management apparatus at HEDL had more personnel and cost more than our entire FMIT technical project at Los Alamos.

When it came time to change the approach to the RFQ, it helped that a good relationship with Kintner had been established.

There was another key aspect to the design of the FMIT linac beam dynamics. Drift-tube linacs had been built after LAMPF at FNAL, BNL and CERN, and all exhibited the same unexplained transverse rms emittance growth. Ken Crandall had developed extensive computer simulation codes and I with Rene Mills had made many simulations trying to understand the emittance growth. Sacherer's result concerning equivalent rms emittance had led CERN to develop a new design strategy, in which the rms envelope equations were satisfied at each cell along the linac, had built their "New CERN Linac", and started commissioning the first tank with beam in the summer of 1978. It was predicted that no emittance growth would be seen with the new strategy, but it appeared that there was the same growth as previously. It was crucial to understand this with respect to the FMIT design. CERN was planning to commission the second tank, and agreed to let us participate. The problem was to convince the DOE to allow a 3-week visit by a 3-person team – myself, Rene Mills and Oscar Sander. It seems hard to believe now, but no such collaboration visit had ever been financed by the DOE at that time. I had to go to Washington and outline detailed plans, and finally they agreed. CERN gave us all their source codes, we ran them on LASL computers before leaving, and took our own codes. By the end of the first week, our codes were running on the CERN computers, but the linac was broken and it was clear there would be no further testing during our visit. Panic – I had essentially promised my soul to the DOE for the permission, and figured I would have to return to Mexico instead of New Mexico. But very fortunately, and generously, CERN allowed us to work with their logbooks from the first tank experiments, although they had had no time to analyze them themselves or publish anything about the results. We realized that our codes included the possibility for beam mismatch and misalignment, while the CERN codes did not – this was amazing, because CERN was always the Olympus for accelerator people. With perfect injection, the codes showed less emittance growth than observed, but with mismatched or misaligned injection, the simulations showed as much or more emittance growth. Mismatch and misalignment showed different emittance patterns as percentage of the beam radius. We analyzed each day's experimental running from the logbook data, and were able to say that on some days, the beam had been apparently well matched and well steered, but on another day was well matched but apparently mis-steered, or well steered but apparently mis-matched, and by how much. The simulations with mis-match and mis-steering overlapped the experimental data well within the experimental setting and tolerances, so with great relief I could draw conclusions for FMIT.

There were two follow-ons. First, we learned that at least one senior CERN staff member was greatly angered that we had done such analysis of their results before they had had time for their own analyses and was afraid that we would publish something before them. I quickly contacted the CERN Group Leader Gunther Plass and promised that we would only use the information internally and not publish anything. This placated the situation, and we did not publish outside Los Alamos [46]. The phenomenon of different signatures for mismatch and mis-steering is still of contemporary significance in terms of the search for more detailed understanding of beam halos leading to beam loss.

Second, after the CERN visit, I rented a car and had some vacation in Spain. Arriving in Seville, one immediately confronts the imposing cathedral, and tired from driving, thinks to immediately stop there. But there was no parking space, so drove on, and learned about the unique pattern of Seville traffic. The hotel lay to the right, but there was no way to go to the right – all signs indicated only straight ahead, to the north. Finally study of the map showed tiny arrows indicating that traffic is

---

46 R. A. Jameson, R. S. Mills, O. R. Sander, "Report on Foreign Travel - Switzerland", LASL Office Memo AT-DO-351(U)MP-9, Dec. 28, 1978; and "Emittance Data from the New CERN Linac", R.A. Jameson, Letter to G. Plass, CERN, AT-DO-262(U), January 8, 1979, and R. A. Jameson, ""Emittance Growth in the New CERN Linac - Transverse Plane Comparison between Experimental Results and Computer Simulation", LASL Office Memo AT-DO-377(U), Jan. 15, 1979; and R. A. Jameson, "CERN Linac Tests", LASL Office Memo MP-9/AT-DO-(U), Mar. 1, 1979; and R. A. Jameson, "CERN Linac Tests" LASL Office Memo AT-DO-514(U), Apr. 26, 1979.

In 1992, the first three of these were consolidated in LA-UR-92-3033, "Emittance Growth in the New CERN Linac - Studies in 1978", R.A. Jameson, R.S. Mills, O.R. Sander.

streamed northward through the city center and then back down along the east and west sides, from which turns inward could be made. Finally arriving in the vicinity of the hotel, a parking place was found and the hotel reached after a few minutes walk. As soon as my head entered the door, the proprietor shouted “You did not leave anything in the car?!” Immediately sprinted back, and everything was there. A few days later, Sander arrived and parked at the cathedral – and everything was stolen from his car. I had all the records and logs of our CERN visit, neatly packed in a suitcase, and could have lost it all. At least, one learns that when traveling by car, it is better not to leave everything neatly packed, but to let things lie around separately so thieves can pick and choose, or be intimidated by the general mess in the trunk…

Formation of the group AT-4 constituted an AT-Division fundamental precept. An accelerator project is an even mixture of what is called accelerator physics, and engineering. These are very different cultures. It is essential for success that both disciplines reside in the same organizational structure, and are coordinated at the top level. One of my main jobs was to keep this interface working. The physics design for FMIT was done in AT-1, but at some point, it is necessary to stop fiddling with the design and get on with engineering it to work, and getting it built and commissioned. AT-4 was the engineering group, let by Ed Kemp, a long-time engineer who participated in the early atomic bomb test and the LASL fusion program.

Much was learned from the FMIT prototype program. The prototype RFQ ran at the 50mA limit of the ion source. All final project drawings for the whole accelerator were finished and ready to be placed for procurement, and then the project was abruptly cancelled, as the US government “policy” toward fusion research degraded into chaos – fusion should concentrate only science” problems without regard to possible eventual deployment as an energy producer – hence, should not be concerned with engineering, or the absolutely necessary materials development – therefore, not the multi-disciplinary sensible approach. Actually, very well-known and bitter to the fusion community at the time, the fusion budget had not the faintest to do with fusion research, but was totally decided by budgeteers in Washington, as it was under the rubric of “discretionary spending”. The short-sightedness, actually blindness, of US politicians has resulted in no advanced materials testing and development, and hindered fusion research and development, to this day.

By 1982, the first-round projects had been finished or were nearing completion, and represented outstanding R&D. But a cliff was looming – new projects were not appearing. We were looking all around – Don Reid was especially involved – but the future looked dim.

[toc]

## The Second Round – SDI

LANL immediately had interest in SDI from its announcement in 1983 – big money, big power possibilities. AT-Division needed work, was in a weapon’s lab, and had no problem with the idea of participating in national defense. AT-Division had the two most credible SDI candidates for missile defense – the Neutral Particle Beam (NPB) and the Free Electron Laser (FEL). The unanticipated announcement of Star Wars in 1983 indicated the second wave for AT Div – but we had to get in.

### The Neutral Particle Beam (NPB) SDI Program

I designed the SDI Neutral Particle Beam (NPB) architecture three times in succession. Have always wondered why I bothered with the second two, just as have often wondered if 3M really deserved to be bailed out with so much work.

First phase - SDI came and a power game immediately started at LANL, where, for example, connections through E. Teller existed, among “program managers”, including RB. General Abrahamson was named SDIO Chief, and organized his first meeting at Canaveral, where he was in charge of Space Shuttle launches. Although AT-Division represented the whole LANL competence in the technology, I was not invited to join the LANL delegation, but shortly before the departure day, decided that I was going to be there, called and put myself on the list, to the consternation of the “program managers”, but at that point I had enough clout that I could write my own ticket. It was of course very interesting, and we got to see the Space Shuttle launch, and spent an informal evening in Abrahamson's office. It was crucial to get AT-Div into the game - our first business cycle was running out and only the cliff was in sight. White Horse had been in progress for years, so there was support from the Army. However, the strategy was dominated by powerful laser politics and NPB was a real "dark horse" - so named in the news, and fought viciously by the laser jocks. LANL "program managers" were not making any impact.

<u>Three Bullets</u>
So I returned back to my experience as a 2nd Lt. in the US Air Force with briefing generals. I worked with a flight testing crew on a versatile jet-powered drone, on which we did early tests on radar reflection enhancement or inhibition, infrared enhancement, surveillance cameras, and so on. Generals often came for briefings, and were generally held in dread. I had no particular awe of them, figuring that they were just needing information, or were bored – so needed concise and entertaining briefings. This resulted in my briefing many generals. For General Abrahamson, I developed the "3 Bullets" briefing, gave it to him, and that got us in. For SDI, he had three kinds of bullets. The first bullet has mass (ground-to-air rocket, later "Smart Pebbles", etc.) - lethal if it hits, but very slow, can only be fired in the last seconds before impact and could not do boost phase interception, so a catastrophe if it missed. The third bullet was lasers – although the beam is delivered at the speed of light enabling boost phase interception, no good obviously, requiring huge power and easy to counteract, but with all the political and glamour power. The second bullet was NPB – delivery speed of half the speed of light - not bad, and the beam energy would be fatally deposited inside the missile. Viewing the space exhibit at the Smithsonian would be convincing enough that we could technically put accelerator in space, but no comment on fighting a war that way. That simple; generals like that.

The SDIO organized the Fletcher Panel, in particular to look at countermeasures. I worked with the LANL representative Bob Selden - the study showed that after 5, 10, 20 years of countermeasures development, most everything, especially lasers, looked awful, but NPB still looked very good. Seldon and I developed an ambitious R&D work plan, which he presented and was accepted.

<u>Establishment of AT-8 Controls Group</u>
Computer control of large facilities had become a mature field, but a very rapidly changing one, and its practitioners always wanted to use the latest technology. A project, however, had to decide on a system and build it, and did not have the option to continually upgrade hardware. Therefore, AT-Division policy had been to let each new project develop its own controls, as consistent with our Charter, rather than establishing a distinct controls group, although pressured to do so from the various directions of our very good controls experts located in the various projects. SDI put remote control at the top of the requirements specification, so we knew it had to be focused and that it was time to establish a Controls Group.

At the same time, LANL was losing a large laser program, and I received a phone call directing that the "controls group" from this program would be summarily transferred to AT-Division, to avoid Reduction-In-Force embarrassments to LANL, but without any advance notice or consultation with AT-Division. So one morning, I went to that site to meet "my new group", with no prior acquaintance with any of them, or of them with AT-Division. I began by saying the AT-Division welcomed them, that I would try to explain the situation, in which the first thing they should realize was that they were not a controls group. Consternation, hostility. Continuing, I said they were recognized as being a very good "countdown group", having developed the system for firing the laser, counting down to zero about once per day. But an accelerator system, and especially an SDI, "controls group" had to deal with a continuously running system, with thousands of data channels receiving and sending information, at data rates severely constrained by the technology, and with no failures or errors allowed as a war outcome could depend on it. Their job description in AT-Division would be simply to become the best such "controls group" in the world. The mood changed.

The next problem was to select the Group Leader from three AT-Division project controls leaders and the previous laser Group Leader. I decided for the latter, Mike Thout, because of his broad horizons and technical qualifications, and the fact that he had led a larger staff, and that his people liked working for him. This was a good decision, although any of the candidates might have been good. From this group, AT-8, came the generalized Experimental Physics and Industrial Control System (EPICS), which later has developed into a world-wide standard and applied to many accelerator and industrial applications, and which played a central and crucial role in the successful turn-on and commissioning of the SNS at ORNL.

Matrix
Until the second half of 1984, the LANL SDIO NPB and FEL technical programs were done in AT-Division. We had appointed program managers, but the difference was that there was no matrix management – the program managers reported to the technical leaders. But then the rot set in, when HD, the AT-Division NPB Program Manager, thought he could gain power by moving to the program management side under Bob Selden, then a LANL Associate Director with program side responsibilities [47]. This transfer was a fait accompli without my knowledge while at the 1984 Linac Conference. I had to accept it, but insisted on the condition that the AT-2 GL who was in on the plot would also transfer to Selden. This surprised Selden but he had to accept it. I was not about to tolerate a direct plot, which would certainly continue, between an external program manager and an AT-Division GL. The two continued to plot, and did undermine us once later, but did themselves in.

SDI was only a power game at our level, nothing to do with defending the country [48]. Abrahamson told us at the first meeting that his first job would be to ask for independent procurement authority, as SDI was set up equal to Army, Navy, and Air Force. He said he knew he would lose, but that was the first thing he must try. Of course he lost. It was a power game. So we had SDI bosses and NPB Program Army and Air Force bosses, and Navy bosses in the FEL program. At LANL internally, in 1986 the 4th LANL director regime began, and the lab entered free fall. By the end of the SDI program, I counted that there were at least seven layers of management either officially between AT-Division and the lab director, or assuming they were in between. A stark contrast in getting work done compared to the Agnew days.

Second phase - SDI didn't like the national labs in the programs - too much know-how leads to back-talk, etc. And really stupid management, power hungry but small fry. More on the 4th regime below. RB had been appointed by the 4th regime as overall NPB program manager. His style was to almost never to do anything or to offer any ideas or discussion himself, to have his technical underlings feed info, to leave his technical underlings in the dark, and to blame them for any mishap, such as inability to meet totally unreasonable budgets or schedules. He made a total ass of himself going to Louis C. Marquet, Deputy Director for Technology at SDIO in 1986, and insisting that LANL should run the whole directed energy show - Lou sawed him into small pieces in the hotel room where we were sitting on beds having to listen, and called LANL demanding to have him fired the next day, but nothing happened. Mahlon knows another time there was a direct request to remove RB. And the earlier time when RB finally decided he was going to show everyone he could do something all by himself, produced a glossy brochure without showing it to anyone, and presented the first copies to the Army SDI general when he came for a briefing with Bart Clare as his adjutant. That evening on the Ross airplane to Albuquerque, the general "gingerly raised the corner of RB's brochure between his thumb and forefinger from the briefcase and asked Clare if he should read it". Clare said no - RB had praised the whole LANL program as an Air Force program, never mentioning the godfather Army, who had sponsored the SDI fore-runner White Horse program for many years! The Army asked that RB be fired, but again he was not. It was widely wondered what kind of secret and incriminating evidence RB held on the 4th regime director??

Bart's story is also typical. He used SDIO to feather himself a bed at ANL, and left the Army to run the Continuous Wave Demonstrator (CWD) program – the program to develop the real operating version of the NPB weapon, which would not be a pulsed machine as were the prototypes at LANL, but a machine that could operate for at least 20 minutes, the time estimated that a war would last, that required extensive additional engineering development. It was the "end game" of the NPB program, and we planned to continue with that phase in AT-Division. The CWD was lost to us when HD was independently trying to set himself up as a national program manager and gave away the store on us. I reacted instantly but too late by informing Kerr what it meant. HD had already gone to Seldon's office, was severely chastised and maybe then sent to the spook shop, where he and the previous AT GL did end up. At ANL, Clare was ratted on by someone that he was growing marijuana in his basement big time, and was fired from ANL.

---

47 Selden was an exception to the usual matrix management problem, understood the mutual roles, was open and one could work with him. Too exceptional for the LANL environment, He left in 1993 and had a distinguished later career.

48 "The Strategic Defense Initiative", Edward Reiss, Google Books

Anyway, the time came when industry was clamoring for all the SDIO money, and SDIO wanted the labs out anyway, so LANL was told that our R&D program was over, industry was going to do it, and we were going to train industry. Total panic at LANL. But seemed clear enough - either agree with SDIO or lose everything, time to take our long-term New Mexico Senator Domenici's approach and simplify everything to just jobs. It was totally not my responsibility; I was supposed to be still and "supply the people and the technology". The project management would take care of us, remember? But they were still my people. So I called Doug Pewitt, a savvy guy working in Washington, to find out the lay of the land, and what the latest buzzwords were – I had not been going to many meetings for awhile. Knowing the buzzwords made it easy. I organized a series of meetings in our small "The Nest" meeting room, and led the LANL bunch through a work plan for how we would do a fantastic job of teaching industry and satisfy all the buzzwords, which RB could then present. All was ok, for awhile, maybe a couple of years.

<u>Third phase</u> - Then "Industry has learned everything, thanks, now they are ready to proceed without you and you are out of the program". Total panic. Was it not expected by our LANL "program manager" power-types? It was absolutely in character, and inevitable that it would come. Total panic. I was no longer DL, had no connection to the program, and kept completely out of Division affairs, as wanted on both sides. But then they came to me, "*WHAT SHALL WE DO*??". Said one sentence - "Easy and obvious - just go back to Selden's and my original R&D plan, and double it." They presented that, worked fine…

**<u>The Free Electron Laser SDI Program</u>**

The FEL was a different story. Laser based strategic defense was far-fetched from the beginning, especially from the countermeasures point of view. But if a laser system was desired, then FEL. FEL is half laser and half accelerator, so nearer to the laser power game, and we did not have a clamp on that technology as we did on the accelerator-based NPB. We had succeeded well in early R&D of FEL, and had shown that FEL was the best candidate. But we lost the scale-up program, because the 4th regime was chicken. Scaling up would have brought in more funding and required more resources, so lab support and the 4th regime's approval was required. We went to brief the head of the 4th regime, and watched his stomach quiver. "It's too big! We can't do TWO big programs at once!' (Why not?) "politically incorrect" "won't be able to find the people" (But LLNL already has about three big programs and never hesitates to promote its next one!) "but, but, but… sputter – it's just too big for us…". It was really pathetic. Pointed out that we had the two best programs in SDI - the second bullet, and the best laser alternative for the third bullet, that electron and proton technologies were enough different that we could attract enough people, that people are always interested in joining really good programs. "but, but, but, sputter…". The 4th regime announced that FEL was to be our second priority program after NPB, without first even mentioning that this was going to be their position to any of us or especially to FEL Leader Charlie Brau, a real FEL pioneer, who was totally crushed. He went to Vanderbilt University not long after.

This killed our chance, and LANL lost the FEL scale-up program to Boeing. Boeing was totally incompetent. For example, they supported their soft copper accelerator structure only at the ends, so that it sagged in the middle so much that the direct, line-of-sight aperture was completely blocked. AT-Division was forced to send people up there to bail them out – a bitter pill, but swallowed bravely. Finally the Boeing FEL, with lots of AT-Division TDY, demonstrated spontaneous emission. They lied that it had lased, shut it off and immediately dismantled it to make way for the next experiment, to have ten times the power!! There was no acknowledgement to the AT-Division troops. Bitter. Eventually there was a "shoot-out" competition, with high-level reviews, between FEL and conventional lasers, which the FEL actually won, but too late to save any program, as SDI was ending.

**Further SDI Anecdotes**

R&D treacheries

Power plays also came from technical persons, although seldom. A question often asked about NPB was if we could really produce a beam powerful enough and also with small enough divergence that it would enter a hotel window in Chicago when fired from Los Alamos. A general came for a briefing, the divergence question was answered by saying that was the subject of our R&D, but we had prepared an experiment on the FMIT RFQ prototype to demonstrate that we could indeed produce power. A cube was made with ten ~2'x2' stainless steel sheets arranged in succession and space about an inch apart. The cube was placed in air at the RFQ output. When the beam was turned on, it melted a large hole through all ten plates almost instantly – which convinced the general that we could produce power. Later, LBL said that LANL did not know how to make RFQs, because the hole was large and cross-shaped – knowing full well the hole size that was expected, and that the cross-shape was because several hydrogen isotopes were present in the beam, with different focusing characteristics.

Receiving, and giving, briefings can be hard work.

A briefing was prepared for Director of Central Intelligence William J. Casey (1981-1987). He stood in front of a test stand and was faced by our briefer. A man was shouting into his left ear, and another into his right ear, and another was gesturing that he was wanted on the phone.

In December 1992, the BBC came to LANL wanting to make a film about Star Wars. It was well known that the BBC was antagonistic and would try to make the US look bad, so that was not a briefing that any Program Manager wanted anything to do with, and they ordered that I do it. This was long after I had retired as Division Leader and had stopped working for LANL but was still there. I required that discussion be restrained to only the period after SDIO was established, to only technical subjects, and that our position was that an accelerator could technically be put into space but there would be absolutely no comment about policy would be included in the film. The BBC did not like these conditions, but accepted them, the interview proceeded satisfactorily and at least the LANL part of the film was satisfactory.

Violent budget oscillations.

During my last year as DL, the AT-Division SDIO budget averaged ~$80M – and oscillated between $40M and 120M. SDIO was famous for listening to the "last-in-the-door" and then making instant budget adjustments. Then if our budget was lowered, I would be told, by "program managers" (e.g. RB), to immediately fire enough people, and quickly, because as the time left in the fiscal year became shorter, the more people would have to be fired. The Division strategy for that was to fire no one, and to figure out an SDIO briefing (spoon-fed to the program manager) that would make us the next "last-in-the-door.

A really good example was when the FEL budget got cut, and the program manager ordered work stopped on the development of laser cathode injection to the FEL linac. This injection was absolutely critical, good progress was being made by John Fraser and younger staff member R. Sheffield, and success was in sight. It was the only time during my involvement in the matrix management madness at LANL that I literally told someone who he actually reported to, and who did his salary evaluation and performance appraisal – the Division and not the program manager. I told Sheffield and Fraser to continue without pausing one second, and that it would be great if they succeeded soon, as we needed a new "last-in-the-door". In fact they did soon succeed, SDIO was briefed, and we got a tens of millions immediate increase in our budget.

Of course, this kind of budget increase was claimed to be the work of the Program Manager, who immediately ordered an increase in staff. This was of course ignored, because the oscillation cycle was about three months long, and also because we knew full well that SDI would not last forever.

Overheads never acknowledged in budget

It was not possible to plan realistic budgets taking into account all time-consuming factors. Incessant reviews required the technical staff to spend a lot of time preparing material for the program managers. The classification rules were not made clear for several years after SDI started, but finally the full system was imposed, resulting without any doubt in an extra work overhead of ~30%, but this was never accommodated in the overly ambitious schedules created by the program managers. Of course, a good program manager is never, never at fault, because any problem is always the fault of the other side of the matrix. Also, after classification became fully in effect, it was noticeable that "black" programs were running, with orders on Division resources that the Division Leader "had no need to know", which certainly did not help in meeting the above mentioned schedules!

Publishing

Because for a long time there were no clear classification rules, papers and talks given at conferences were restricted to detailed technical subjects, checked with SDIO but without specific rulings. Finally in 1986, SDIO wanted an overview of the LANL work for their own reasons. By then, classification rules had been established, and permissions had to be onerously obtained. Now the LANL program managers reacted noticeably differently – although this was a chance for more fame, the risk of making an error, either technically (as none knew the subject), but especially with respect to classification, was greater than they wanted to take. OK – push it to the other side of the matrix, and I was elected. I prepared the talk in close cooperation with Army Lt. Col. Warren Higgins, a good SDI program manager, taking great care on the classification issue. Nevertheless, I was anxious because if any question did come up, I knew I would be left hanging. It was the first open meeting talk on SDI, and was given at the 1986 Linac Conference at SLAC [49. Appendix 1 gives the Introduction section of the paper. My talk was in the opening session, we had about 15 or so other SDI related talks, and none of them had officially received classification release. I went to SLAC Director Burt Richter's office early – he was another really good Director and Nobel Prize winner, and I knew he came to his office early. It was clear that the meeting schedule would be drastically disturbed if the papers could not be given. His reaction was instant: "This is *my* meeting – they are not going to screw it up!!". Again, it was a real pleasure to have again, after so long a drought, the support of a good Director. He started making phone calls. He knew that there were two parallel competing branches in the government apparatus, so he called the other side from the office where the classification approvals were given. The other side was delighted to stick the needle in, and orders were given that the approvals (either Yes or No) were to be received by noon, and mine on the spot. Richter's meeting schedule was not disturbed.

[toc]

## Fate

### AT Division

1986

When the 3rd regime was removed, it was announced that for the successor, the DOE wanted "increasing emphasis on small science at the lab". That this was nonsense was pointed out to them and the search committee by many – a large national lab needs large challenges, LANL's main mission is nuclear weapons, with complementary science. Most probably, the DOE motivation was to continue their plan to "kill the old culture (e.g. weapons czars), using their weapon of mass destruction – matrix management – which anyone with any sense knows is another game like church plus state method for obtaining power, and works in no case when upper echelons are only interested in power. To accomplish their control, they would need a puppet director. And that is what they got. Again no one from outside considered being LANL Director a good job, and an inside table-top

49 R. A. Jameson, "Linacs for Esoteric Applications", (Invited - first open conference talk on the SDI neutral-particle-beam program), 1986 Linear Accelerator Conference, SLAC, Stanford, CA, June 2-6, 1986, SLAC Report 303, Conf-860629, Sept. 1986; & Los Alamos National Laboratory document LA-UR-86-1734, 21 May 1986.

researcher was chosen as the 4th regime. His loudly trumpeted qualification was that he had developed “the best performance appraisal system”.

LANL went into free fall from day one, and in less than one year, into an accelerated fall. It was quickly seen that sycophants were desired; they gathered quickly, became “program managers” and spent their time with internal power struggles and “directing”. The performance appraisal system was imposed – as threatening, demotivating, degrading and used for power plays as any before.

Morale plunged, and by the end of 1986, it was on everyone’s lips that “the survival unit is now one”. It was hard, to see how quickly loyalty to the Division and the long-term Charter disappeared, then loyalty to the Groups. This happened because it was made clear that the technical Divisions were relegated to job shops, existing projects were suffering from program managers with no technical knowhow, and that new projects would be found and managed by program managers rather than technically competent leaders. Although the technical Division job shops would still have responsibility for writing performance appraisals and for setting raises, the technical staff realized there would be interference and political pressure, they would have to satisfy many bosses, and therefore survive on an individual basis. Overhead cost skyrocketed.

A 4th regime nickname was everywhere in use – I^n – I to the nth power, of which the mildest I was “incompetent” [50], and the meaning was that incompetence was demonstrated in every dimension of the job that a Director should do. Accelerator terminology is usually in six dimensions, as we keep track of the 3D location of particles and also their divergence in each direction, but it seemed apparent that nD was needed for description here. People started leaving from various parts of LANL – the best are the first to leave.

Support from the 4th regime for the large amount of SDI work was miserable, with no authority over his sycophantic, power motivated and competing program managers. For the overarching question of NPB plus FEL, the 4th regime had no stomach, and visibly flinched when plans requiring big project requirements for work, people, budget were presented. Both the NPB and FEL program managers were non-contributing parasites, making demands on AT-Division, directly pressuring AT-Division Group Leaders, also directly pressuring personnel in the Groups. But they kept having trouble, couldn’t make their own vugrafs, panicked when the big crises came (SDI, later Tritium Production, ATW/ADS), were supposed to be in charge of budget and schedule, but had no real concern for either and did not have accurate accounting systems. (As defense in case needed, Wayne kept our separate set of carefully documented books.)

My job, as defined by the matrix, was to do internal technical management, try to maintain a technical competence, and leave the project people to their jobs, and I had made it clear I was trying to do that and not compete with the program managers. (I was seen as a competitor, having far more program management experience plus technical competence and connections.) I felt that it was crucial to concentrate on trying to win long-term projects for after SDI, which would clearly not last forever, and as it was clear that relying on non-technical program managers to win new projects was not realistic. I delegated all SDI activities to the AT-Division Deputy Division Leader. This however did not work, and although the assignment remained, the pressure returned on me alone.

<u>Bitte Verzeih’ … my biggest mistake</u>

(- - -) -> (0 0 0)

---

[50] The core I also was a reference to the personal pronoun that the 4th regime used in most utterances.

1987

In 1987, I was ordered by JB to re-organize AT-Division, "with RB". JB was then an AD supposedly on the technical side of the matrix with the job to maintain the integrity of that side, but had wanted to be the 4th regime, would patiently wait and eventually be the 5th, and was not about to make any waves. There was absolutely no guidance or offer of discussion with JB as to what reorganization they thought would suit them. I knew, however, that the 4th regime and cronies had decided that they could "direct" the Star Wars projects if they had direct control of the technical staff, subverting the matrix and actually eliminating AT-Division and making it into a pure project support entity for Star Wars, named something like "SDI Project Division".

The biggest problem for AT-Division and its people would have been the total loss of our unique open Charter – to pursue accelerator technology projects that would benefit society – and I refused to give up this Charter. It was crystal clear that the mouthing that PMs would take care for the future of the people was not true.

A second very big problem was that there were two SDI projects plus other smaller projects. Each PM wanted to be the king of the mountain, would demand the best people, and impose completely unrealistic schedules with inadequate budget – without any regard for any other project, and without any capable coordinating management at their level with the technical side of the matrix, or above them. The job of the executive at the top of a matrix-managed organization is to close the matrix, and the 4th regime utterly failed to execute this responsibility. JB's directive to reorganize "with RB" was an impossibility, first because of this competition, and second, because RB knew his project manager business and of course never had any ideas, never would have offered any if he had had, waited for input, twisted to fit, and concentrated on who is king of the mountain today and how to be tomorrow.

Our rule for job assignments was that a technical person could work on at most two projects. This provided challenge while avoiding some stress – even then the person would have three bosses (AT-Division and 2 PMs). It was also necessary keep the total manpower to a reasonable level and not to respond wildly to the extreme and rapid budget fluctuations. Therefore, it was impossible to satisfy everyone, and this of course resulted in power plays and laying all kinds of blame on AT-Division.

For example, the LANL SDI FEL program manager was the worst ever experienced – a sycophant and a serpent in one. The FEL program had started as a project outside of AT-Division, with our support. As it became larger, the project had accumulated a debt of ~$1M, directed by the program sponsor and with the knowledge of LANL. More support and organization was also needed, so arrangements were made to bring the project into AT-Division and AT-8 was formed, with Jerry Watson recruited from ANL as GL. The debt, with fully open Laboratory and sponsor knowledge, was placed under the stewardship of Wayne Vanderham, who smoothly managed its fiscal year-end rollovers. This of course required open negotiations. Part of the 1987 pressure was due to the LANL SDI FEL PM going to the 4th regime and saying that AT-Division was unable to control its budget and should be required to pay off the debt. The 4th regime never checked out the facts.

I had seen at first hand the consequences of what can happen when a job is clung to too long, and could plainly see that the 4rd regime was the end of an environment for technical excellence at LANL. I had some 21 very good years at Los Alamos, with real projects like LAMPF and the first seven years of AT-Division, had served as a Division Leader for nine years, and knew that other opportunities were no problem. It was very tempting to leave, but I was afraid that at the end of 1986 the Charter would be lost and somehow I did not want that written to me, so I spent 1987 trying to find a way to satisfy everyone and retain the technical structure and Charter.

It was a bad year – haggard photos show it. Endless discussions were held inside AT-Division and one-on-one interviews were conducted with all the staff members. People came continuously to complain about near impossibility of the situation, each under thoughtless pressure from project side.

Notes in my personal logbooks during this phase are clipped together with paper clamps, and still are very painful to read. We all knew Star Wars would not last very long. We were ordered not to have any concern about the future, project managers would take care of the personnel – and knew this was ridiculous, as they never had, didn't know how to sustain, didn't know the technology so incompetent to pursue future projects.

I spent considerable time searching for projects after SDI, such as a trying to get support from the NIH for a compact FEL. But LANL overhead costs were skyrocketing and potential sponsors were not interested. Twice there was pressure to fire Jim Stovall – refused, reason that Jim always gave honest answers, and his people liked to work with him. They had not liked the honest answers. The 4th regime was meddling directly in raises at Division and Group levels.

A typical problem with lab "management" was having to deal with a patently unbelievable complaint brought against a Group Leader. I discovered that one lab support office was writing the accuser's case for the accused, to be presented to the lab's legal office. No such support was offered to the Group Leader. This evident abuse was stopped, and the case dismissed when the accuser's self-defense was heard.

The 4th regime believed that management gurus could inform him how to manage LANL. The first of the series of "7 (5, 8, etc) Ways to be a Successful Manager' had been imposed on the whole lab, causing an enormous loss of productivity and enhancing the internal power games. The method would fail within a year, and would be replaced, many times. The last straw was the announcement that any contemplated incoming project ≥$100K had to be evaluated by an hierarchical set of ten committees, after which the 4th regime would then personally decide if the project would be undertaken. Even at that time, $100K was hardly worth the paperwork, to say nothing of such a preposterous approval process. I knew that my next hopeless and time-wasting job as Division Leader would have been to get someone from AT-Division on as many committee levels as possible.

Ed Knapp had returned from NSF and was the Chairman of the AT-Division Advisory Committee. These Division committees were required, were to report to the Division Leader, with information to the Division's upper management. In August 1987, Ed talked to the 4th regime and came on strong, saying that the lab was worse than 6 months before. The 4th regime was too petty to receive criticism constructively and was angry, and directed that Ed should step down as chair of the AT-Div Advisory Committee.

I made a mistake in asking H. Grunder to be the next chair. G was extremely egotistical, sought recognition in high places, forgot who he was supposed to be helping, and was pleased to be entertained by the 4th regime, where he aided the program managers.

At this point, I had decided that no re-organization made any sense – all the best people in AT-Division already in Group leadership roles. My strategy to maintain the Charter would be to leave the Division intact, and hope that the characteristic indecision would keep it and the Charter, perhaps at least until a new regime. There was no confidence in the 4th regime or associated program managers. The future would be wasted in senseless exercises and meetings. Time to go. I resigned in November 1987, announcing that the organization of the Division needed no changes – all the best people were already assigned to the top jobs, were competent, honest, and working very hard – to the consternation of the 4th regime, associate director on the technical side, and the project managers.

[toc]

## Retired

June 1963 – came to LASL – Los Alamos Scientific Laboratory
November 1987 – resigned from AT-DL. "Last day of working for the Lab".
April 1999 – formal retirement from LANL. Took the earned Nambe tray, without engraving. Invisible departure.

***CHECK "A TREE GROWS IN LOS ALAMOS" - E.G. FOR "LOW BLOWS" AND OUTLINES OF LOGBOOK NOTES***

Subsequent

Later I felt that the strategy was at least partially successful. Although some group leaders and staff also soon left, no change was made to the Division organization until 1993. Although it would have been better to do so. But unsuccessful in that there was absolutely no counterweight to the bumbling management, of AT Division, the 4th regime and project management.

There was now no understanding at all of the engineering culture at the upper Division management level. The engineers angrily removed themselves to the general Engineering Divison in the matrix – fatally exposing our technology to matrix management, and disaster for the SNS. The only exception was Dale Schrage, who had cagily remained in AT/AOT-1, knowing that he might be more valued as the only effective remaining interface and could enjoy development with his smaller support staff, rather than going to the very problematic environment of the matrixed LANL engineering division.

AT-1 adopted a strategy of isolation, for example by putting all the accelerator physics software onto specific and uncommon personal computers, stopping open-source distribution, and selling the code to help with updates to the Group's computer resources. Previously, AT-Division had served as a national resource for accelerator physics codes, and even received a small support from DOE High Energy Physics fund for accelerator technology development relevant to HEP for this, as an exception from the policy that national labs would not be funded from this fund. AT-1 was warned that they should inform DOE that they were selling codes supported by DOE finds, they did, and were allowed although clearly they would get no help if trouble developed. The DOE support ended soon after, and big trouble did result later. The tragedy for accelerator technology was that the codes turned into black boxes and collaboration stopped, and although nice computer bells and whistles were added, no further physics development occurred.

In 1993, AT-Division became AOT Division – Accelerator Operations & Technology, to include operation of LAMPF, officially signaling the end. The Lab had been in worse than free fall by now seven years – the first "seven bad years", unfortunately to be followed by another seven and another, an apparently bottomless pit. The strong tree that was Accelerator Technology had died quickly, although LANL has fed on its fruit for a long time. Other chapters relate the loss of the SNS to ORNL, no interest in IFMIF, APT, ATW/ADS, mediocre and worse work for SNS at ORNL, operation of the LAMPF accelerator with no technical knowledge, …

AOT was carefully and purposely sidelined as much as possible in the organization of new accelerator-based projects (APT (LEDA), ATW/ADS), or when it was necessary because of problems (SNS). No physics strength remained and that side of the technology froze. There were advances on engineering side, but outside AOT except for the key contributions of Dale Schrage, which were however resented (see APT). Performance on the SNS was so poor that it required that a separate LANL Division be organized (see SNS).

**LANL from 1986**

Starting in 1986, huge opportunities were lost to Los Alamos because of lack of leadership and incompetent management.

When the Iron Curtain fell and perestroika came, Adm. James Watson was Secretary of Energy (1989-1993), and asked the labs in clear terms to redefine their missions. One of the things they should do was to say what their central facilities should be. A magnificent opportunity. Ed Knapp was LANSCE director (1990-1991) and had the deal wired with at DOE for a superconducting LAMPF upgrade. It was a clever interim step on the way to obtaining a full neutron spallation source at Los Alamos. But a table-top person, with no broader vision and no guts, the 4th regime, would not support EAK. Ed never forgot that, and got very angry when relating it.

When Strategic Stockpile Management (SSM) was a new thing, the same thing happened. The 4th regime and other LANL personnel tried to claim in Washington that they had originated that idea. I knew from contacts in Washington that people were not amused at this impertinence, and angry about it. Again, each lab should have a "signature facility" (buzzword) to support SSM. LLNL, NIF, Sandia (the Z-pinch), ORNL (SNS) all made relevant and ambitious proposals. For a long time no proposal from LANL. Finally after long chiding, LANL said LAMPF (LANSCE) would be their signature machine, but not enthusiastically, and without knowing how to even present a technical case. Again 4th regime, completely blown opportunity, loss of SNS.

In many ways, massively accumulating, productivity was eliminated. A typical example: A friend was part of "the upper management" and had to go to the frequent meetings which cost hours of productivity of the highest paid managers of the lab. He was so frustrated by wasting time over trivialities that occasionally he would vent to me. I had never seen him quite so angry as when he said a whole afternoon had been spent concerning which managers were allowed to bring donuts to a meeting for the attendees, when, from where, to where, …

The 4th regime lasted a long time – very long, far too long. Why? The technical staff wondered, hoping perhaps for a turn-around. There were many speculations, and euphemisms – such as the posture assumed and maintained with the DOE. That in fact was the reason – the 4th regime sailed with the Washington wind, they had their puppet, and no reason or strength was there to counterbalance their micromanagement. The DOE did not care about the idiotic management exercises and procedures imposed within the laboratory, the power games, loss of productivity. The downward productivity spiral never stopped. I would drop by to visit the remaining colleagues, and they would say it had gotten worse again in the last six months, but maybe would get better. I cautioned to be careful, because it would probably get even worse – and it did, as told the next time. I would always ask a test question – how long does it take to order and take delivery on a simple electromagnet. The answer grew from the ~two months that it used to take, to over two years.

On 1/8/93, a report was written by Motorola, commissioned to evaluate lab management, outlining severe mismanagement and with castigating conclusions (but there was no effect…). Other review committee reports from 1989, 1990, 1995 , 1996 also reported serious problem with management at LANL, in particular with regard to their treatment of personnel [51].

It is absolutely clear that the RFQ would never have been developed at LANL from 1986. The negotiations regarding FMIT required decision, support and trustworthy backup from the LANL Director, and from 1986 there was none.

Afterward extensive PR work would attempt to erase the real history. The 4th regime has then criticized the loss of the UC contract privatization of LANL under NNSA. But the 4th regime started LANL down this path and such would not have happened without those years of internal dictatorial and incompetent decline. This is not just my view [52].

---

51 " UCORP Report on UC'S Relations with the DOE Labs, "Report of the University Committee on Research Policy on the University's Relations with the Department of energy Laboratories", January 1996, http://scipp.ucsc.edu/~haber/UC_CORP/doereport.html

52 "The Attitudes and Beliefs of Los Alamos National Laboratory Employees and Northern New Mexicans, A Study of the Interplay of Culture, Ideology, Political Awareness and Public Deliberation", J. Gastil, H. Jenkins-Smith, Univ. of NM, Inst. for Public Policy, Winter 1998, www.la1.psu.edu/cas/jgastil/pdfs/LANL%20Final%20Report%201998.pdf, *especially pp. A37-A38, A40-A42, A55, and again especially A85.*

*"LLNL – The True Story", Monday, Nov. 12, 2012, ,* http://llnlthetruestory.blogspot.de/2012/11/lets-give-parney-alternatives.html

"LANL – The Corporate Story", http://lanl-the-corporate-story.blogspot.de/, see post 1/19/2007 6:34 AM

"LANL – The Real Story", http://www.parrot-farm.net/lanl-the-real-story/2005/04/i-agree-that-we-need-to-put-down-our.html, especially, especially, especially: post 4/01/2005 10:58:00 PM.

The situation has not improved.
1997-2003 ("5th regime") John Browne - waited his turn for 12 years, therefore did nothing. Continued to do nothing, made astonishing choices for deputies, when fired, characterized as "naïve".
2003-2005 ("6th regime") George P. Nanos - over-reacts enormously to supposed security violations, says LANL is 'bunch of cowboys', etc.
2005-2006 ("7th regime") Robert Kuckuck
2006-2011 ("8th regime") Michael Anastacio, previously LLNL Director. LANL has lost UC contract; now a UC Battelle consortium under National Nuclear Security Agency or some such concoction, etc. No end of free-fall yet in sight.
2011- ("9th regime") Charles F. McMillan. Continued inferior performance in areas directly under "top" management. So bad that LANL contract is again put out for bid. Managers are getting enormous salaries (Director >$1M/year), no performance, no accountability, managers and management bonuses continue to be received.

Excerpts from visit notes:

2007 Talked with a capable Group Leader. Another staffer had made many of same comments a couple of days ago. The operation of LANSCE, the group, etc. has been reorganized under the Engineering Division! Who also have no knowledgeable leaders and of course have little interest in LANSCE, or accelerators, or accelerator R&D of all things. The person in charge of LANSCE operations, the group, etc., concentrates on operations and has no technical interest in accelerator technology. The 8th regime is a catastrophe ...

Long chat with staffer about X-Div, which is being "reorganized" and is hit hard by the big reduction in the weapon's budget. Story in the weapons program equally grim. The staffer amuses himself by doing simple numbers on the lab's proposals for windmills, solar energy, etc., including things like what is the efficiency of solar panels covered with dust and bird shit and how much it's going to cost to keep solar panels covering the area of Texas clean, or that a windmill driven by a 10mph wind 24 hours per day provides 80kW, enough for 6 houses, so how many windmills please? None of the lab proposals have any account of anything practical like this. Convinced the lab is now in chaos – uses the term "a new basin of attraction".

Hear here and there that things have worsened with each regime since Agnew. Of course, very few now remember how it was to work in the time of giants, and the implication is that it was better even in the 5th regime or earlier – everything is relative. The only thing they are absolutely convinced about is that the present regime is the worst yet, and that the true adage holds that once reaching a peak, leadership declines faster than linear. Question is only how fast, maybe as square or cube? What they don't know, because they didn't experience the 4th regime, is that the fall can be a cliff. The 4th regime , who was incompetent in every dimension, put the Lab very, very quickly into a new 'basin of attraction'. It is then extremely hard to escape the basin…

Further musing on doing 'simple numbers' to get an idea of the forest instead of getting lost in the trees. Simple numbers on matrix management: Have observed since very small that there are two kinds of people – Those who like to do real things, and those who crave power. Oil, and water or often blood. That implies an equal ratio, and many have observed that there needs to be one to control for each controller. But it is much worse than that. So, along with the political system and churches, the controllers invented matrix management, where all the controlled have two bosses. The mixture is now already 1:2. Adding deputies, and all possible permutations of principle, associate, assistant, acting deputies, CEOs, COOs, chief-of-staffs, executive assistants (used to be secretaries), executive associates – the ratio is easily 1:10. We reported directly to H. Agnew. At the end of 1987, the ratio was at least 1:7!

Story today that although there is a new request for each lab to have a "signature facility", there is no leader at LANL, no proposal beyond a "MaRIE", which is a small solid state lab with a few fancy ($10M) electron microscopes and a much fancier building to put them in. Apparently DOE then has noted that LANSCE again should be LANL's "signature facility" and a proposal is awaited. So the solid-state persons have started mouthing about they would put their electron microscopes out on the mesa, and have started using an aerial view of LANCSE as the background for their PowerPoint

slides!!! Said the regime seems incapable of presenting anything – had a meeting recently where $I^{\wedge n}$ started telling a story, lost his line, asked to be reminded of what the question was, and someone kindly pointed out that there was no question, that he had wanted to tell them a story… Another meeting was on safety in the workplace and security, and orated "Safety and Security are the same thing!" - which disturbed the working guys, as he really hadn't given them any background that day about his lofty discussions in Washington about Iran's atomic weapons program…

2009 Maybe have already noted somewhere about the value of prizes, etc. – not original, as many have noted. (I refused to be even nominated as a LANL Fellow, having observed the dumping ground it was in our area of accelerator work.) Anyway, now it turns out that the DOE has awarded its Enrico Fermi Award, one of the most prestigious awards given to a U.S. scientist, for hot air. It was striking back in the 90's when the $4^{th}$ regime made his first short trip to Japan, and then put two whole pages of absolute BS in the lab paper, about how he now understood all their problems and how to fix, etc. etc. Having already spent well over a couple of years there and feeling Japan as home, I have never claimed to "understand" it or to tell them they had problems and how to fix them. And now $I^{\wedge n}$ is an "international security" expert and awarded!! Perhaps a Japanese analogy: Kyosai, a Japanese artist that overwhelms Picasso and even Dali, had a keen eye for the ridiculous. A long scroll shows a contest, put on by the unproductive aristocracy for their amusement. Productive people probably could not help also finding it somewhat amusing, but as a totally non-productive waste of much money, time and talent, would probably shake their heads and feel a bit sad. The contest was to gather as many servants or court people as possible, and feast them very heavily – on beans. The event began when all were extremely bloated, and the prize was to the best fart. The scroll is indeed very long, many variations of hot air!

The "good old days" are so relative to the present. If one never experienced the age of giants, then Lilliputia seems good:

APS News ~July 2010 - The $4^{th}$ regime is now criticizing the installation of NNSA in 2006 as resulting in "can't get work done"!! 20 years after 1986, when in less than one year, LANL went into an accelerated fall – accelerated to more than free fall by imbecilic management that resulted in not being able to get anything done, and the general feeling that "the survival unit is now one". Such things like people "need a good (that is, threatening, grading, degrading) performance appraisal system", and any contemplated incoming project ≥\$100K has to be evaluated by an hierarchical set of ten committees, after which the $4^{th}$ regime will then personally decide if the project can be undertaken. Etc. Which got the lab into the situation that resulted in the feeling that an NNSA was needed to manage it.

A competent and technically qualified Group Leader had come in from outside LANL to head the small group having expertise in accelerator physics and running some small projects, but ordered to be "hands-off" as far as anything regarding the old linac and its operation or improvements were concerned. But recently a surprise – because of continuing difficulties with the old linac, the GL was asked for ideas. I asked how it was working. He replied that it is difficult, because the DL has the "operations mentality", and they emphatically really do not want to change anything. But the GL had observed a contradiction – that every time anything trips off, within 5 seconds the operator starts twiddling all kinds of knobs!! So he suggested they hold still for at least 30 seconds before changing a knob, and consider what kind of problem they might actually have! Told him he had better be careful, as one can get burned with the politics of such a situation - he said he is interacting very carefully, and anyway is close enough to retirement that he thinks they couldn't touch him...

2010 – Messages from group stopped coming. On checking, found that the GL had retired (and also that availability had considerable improved when the 30 second rule was observed, as the control system has mechanisms for automatically restoring the beam in many cases). His ideas were too much for the operator mentality, and the DL gave him a negative performance report! The ORNL SNS performance had been progressing satisfactorily toward initial design performance, but an upgrade plan calls for roughly doubling, which presents challenges. Later in 2010, this DL was named director of *ORNL's* Research Accelerator Division. Incredible evidence that others can also make a "big mistake".

2011 – email to a competent former colleague: "Too bad, not fair, but quite familiar that your expertise is exploited. There is no excuse for that there, in very rich LANL, with all the daily waste that goes on there". I get some pretty good updates on and off - the general feeling about MaRIE is that the exorbitant cost is on purpose, because then for sure it will never be funded and no one will have to take the risk of actually producing something, But in the meantime, quite comfortable to be sure that it generates design funding. Stories about people proposing completely nonsensical technical things, like electron beam lines with right angle bends, ... Earthquake danger in Los Alamos seems to be a good mechanism for being sure that nothing is ever really funded and being safe by doing nothing.

Speaking of bad "managers", if you are not too much afraid of strong risk of nausea, the July issue of Physics Today contains a disgusting "first-person" article, in which nearly 100 personal pronouns are used in 13 columns of pitiful narcissism. Like when returning from his first few-day trip to Japan, and using two large pages of the LANL rag to exude on everything the Japanese should do to improve themselves... A curious thing though - remember that we so often noted that there was "incompetence in EVERY dimension". Sometimes with effort to grant a little good to everyone, that maybe had done something in own scientific discipline (plutonium). But the article removes the need for that small allowance also. Surely completely unwittingly (admits that the Russians did it better). (Incidentally, I have never thought much about me being "knee-capped", but a fair amount that we (AT-Division) were.)"

2012 - any contemplated incoming project ≥$100K has to be evaluated by a hierarchical set of ten committees, again.

<u>"Peristroika" [53]</u>

The opening chapters of Mikhail Gorbachev's book "Perestroika" evoke an uncanny and hair raising resemblance to the LANL situation. The USSR stagnated during the 1960's – 1970's, as false priorities were exploited by persons and bureaucracies intent on power and with sycophancy and flattery having the upper hand. In the 1980's, at the topmost level, it was realized that the situation had to be changed, resulting in "Peristroika" and "Glasnost". The stagnation at LANL starting in 1986 is amazingly similar, but as of early 2018, there is no evidence of "Peristroika" or "Glasnost".

August 2014 - Tried to find a copy of the BNL NSLS Report via Group, Library, Records Center; finally referred to "Lab Historian" Alan Carr. Was informed had to petition the Freedom of Information Act. Finally commented that this seemed strange for someone with official LANL connection. Then he finally realized I am officially connected to LANL (badge # was in original email to him…) – said could get if have a current badge. But July email to Group secretary requesting badge renewal got no answer and did not push it, so no current badge! And Carr is now requesting a charge code "to "cover the time our staff has spent working on your request. " LANL has something like factor x2-3 overhead on funding, and not even services like library on overhead. Corruption pure. This word was now often used within the Lab to describe the overall situation – and continues to be used.

June 1, 2016 – Guest Agreement cannot be renewed: " … the renewal issue of your Guest Agreement. Since we could not establish any recent activities on your part involving LANL, under current guidance, I am sorry to inform you that we cannot renew the Guest agreement." If the GL and Deputy DL were really that impotent by then was not clear…

The Guest Agreement is a little useful, for the good rental care rate, for library and software access, gate access to the site for conversations with a few oldtimers, a feeling of a small remaining link. No one ever approached me for collaboration on anything there (the usual American "The king is dead – long live the king"). Explained this again to Bob Garnett, then DL, and he got it renewed.

July 28, 2016 – chat with Sergey Kurennoy – "How's it going?". "It's really strange around here. Nothing going on. We try to get used to it…". IAP RFQ has been here long time, water leak – have a couple of years to fix it, test off-line. Think they do not plan to ever install it…". "Is installation plan

53 "Peristroika", Mikhail Gorbachev, extended version 1989.

direct or in parallel?" "Direct". Pointed out they would never have the courage to do that. People thought that when MaRIE got CD0 approval things would happen – now they know better. Reminded about LANL overhead, do-nothing is safe, designing is safe, but actual construction risky, will never happen. Did have a chance to discuss APF.

July 2018 - ***Finally, evidence of a turnaround after a 32 year accelerated fall since 1986, a long generational epoch.*** In 2017, a new Division Leader was appointed, the first scientifically oriented one since the end of 1987. In the succeeding year, he has been able to convince management that new technical staff are necessary because of retirements, etc., knowing full well that it takes a minimum of three years to get a new staff member up to speed assuming the old member can mentor. At this time, a loss of around ten key staff could make it not possible to even operate the LANCE accelerator. Initial funding has been made available to start recruiting. A "Council on Accelerators at LANL" has been formed representing various accelerator interests now apparent across LANL, and meets monthly. The DL is insisting that the linac must be run scientifically and not by "monkey tuning" in the future. The new LANL management contract is now in place, and the Laboratory Director has a firm background in the design, construction and operation of a major accelerator-based facility spanning more than a decade. It is good to see, and wish for the best.

July 2022 - After lip service but two years of LANSCE budget cuts, the Division Leader left earlier this year, having an opportunity for significant technical and technical leadership work. June 9, 2022 passed without a ripple, as far as I know; a lip service article appeared in the Los Alamos Reporter on June 13th. Anyway, congratulations and thanks to all those who made the old television set work, and for so long. The machine is running, but on a razor's edge, facing the real possibility of long outages as there are major components that are experiencing end-of-life failures, with no domestic sources left for spares.

[toc]

## After AT-Division

It was great to work at and for LASL the first 21 years, 1963-1984. Team accomplishments in rf field control system, high-power rf amplifiers, rf accelerator structures at LAMPF, RFQ initial development and harder physics and engineering applications than anywhere else, PIGMI and muon therapy set standards for particle beam cancer therapy, FMIT, FEL development including project for a compact FEL, a legacy.

I was fortunate to experience and participate in the giant age of science in America, but also traveled the bridge to Lilliputia and the demise of America's dominance. The seven Biblical following bad years did not end; a short-lived hope from 2017-2022, but now that is gone again.

Many nice letters came, was asked to remain on committees, and got numerous job offers. It was clear, however, that micromanagement would swamp other places as well, and I was again very fortunate, to be able to be independent since the end of 1987. The first step was to contact Japan, where a year as visiting professor at the High Energy Physics Institute KEK was quickly arranged, with half to be spent at the Keage Accelerator Laboratory of Kyoto University, and with two side activities – consultation to Sumitomo Heavy Industries on their microtron project and other accelerator work, and at the Japan Atomic Energy Research Institute (JAERI) to become informed about their work on accelerator transmutation of nuclear waste. I stayed in LANL until 1999, because of the retirement investment, and because independent funding from JAERI and the IFMIF project afforded freedom. In some countries, the experience of elder people is valued and used, but in most cases not. I stayed strictly aside from any "advice" or interactions inside my direct organizational unit, but was confronted with a lot of astonishingly petty behavior, which could mostly be ignored because other people were also so confronted. However, on two occasions, when important rights concerning invited papers were concerned, upper management was asked for a ruling, and in both cases confirmation was immediate [54]. Involvement on the project side of the matrix was also mutually avoided, but requests came on a few critical occasions. In 1992, LBL asked if I would become Division Leader of their Accelerator and Fusion Research Division; it would have been interesting, but I had done that so declined. It has been rewarding during the following years to pursue the elements of linear accelerators, and also to be involved in a number of large project activities.

**<u>Accelerator Transmutation of Nuclear Waste / Accelerator Driven Systems (ATW/ADS)</u>**

This tale involves my own participation in this very important proposal for closing the nuclear energy fission electric power cycle by changing long-lived radioactive fission waste to shorter-lived waste, which still must be stored for some hundreds of years. In 2021, research to develop ADS is being pursued in Europe and Japan (both very slowly) and aggressively in China, which plans to extensively expand their nuclear power output and realizes that a method for waste handling is imperative; in 2013 ATW was ranked third in priority among their top sixteen projects. The outline of a lecture, invited by the Institute for Modern Physics, Chinese Academy of Science, Lanzhou, China on my involvement and recollections, is followed and expanded here [55].

• ~1968 Transmutation of nuclear waste discussed by W.B. Lewis, Chalk River National Lab, Canada; accelerators not ready.
• Discussions recur ~ every 5 years. More efficient hardware, more beam power, strict beam loss control are needed.
• 1972 LAMPF becomes operational.
• 1970's CERN develops rms approach to linac design.
• 1978 Fusion Material Irradiation Test Facility project begins.
• 1980 First demonstration of RFQ outside Russia. Crucial invention for high intensity with low beam loss.
• 1983 SDI. LANL AT-Division has developed system approach to large linac systems. (I discovered and made practical an equilibrium (equipartitioned) beam in linacs, and found that the rms matching equations indicate that higher frequency is better for high intensity, high brightness beams, which was resisted at first – but the equations are really so simple!))
• Late 1980's Japan JAERI is studying transmutation. 1/14/85 Letter from Shikazono at JAERI re April'85 visit. 1988-1989, RAJ goes to JAERI to learn, informs colleagues at LANL.

• 1989-1990 Linacs are now ready for the ATW task – up to 300 mA H+ feasible. Discussions were held with LANL nuclear physicists, and Dr. Charles Bowman opened the parameter space for ADS and generated much excitement among the LANL technical staff. This was the era of Perestroika and the fall of the USSR, which resulted in calls for the national labs to redefine their roles and propose new missions. The 4th regime was not responding to this challenge, to the great concern of the LANL technical staff, which saw the challenge of closing the nuclear power cycle as a worthy challenge for LANL.

A great volunteer effort began at LANL, involving many people (several hundred eventually) from every technical aspect of the combined accelerator, nuclear physics, subcritical reactor and input/output radiochemical waste processing, that is, from technical groups spanning the whole Laboratory. Such enthusiasm of course was visible. I represented the accelerator side, and also worked the top level coordination. This was tricky, as it was clearly necessary at all costs to avoid the 4th regime's ≥$100K committee system. This was done by clearly informing the committee structure, at each level, that our studies were going on with voluntary support and that the concept would not be presented to them before it was fully formed, but that the committee system would be kept informed.

---

54 ADPLS Letter-12-12-92.docx, letter to LANL Associate Director ADPLS, personal files

55 "Personal history of Participation in the ADS Story", R.A. Jameson, lecture invited by IMP, CAS, Lanzhou, China, 15 November 2010.

This strategy was successful, the $4^{th}$ regime's ≥$100K committee system was completely circumvented, and several of the committee chairmen actually helped support the effort.

Many system-wide concepts were studied.
Options for a thorium nuclear cycle were thoroughly explored.

Finally the $4^{th}$ regime was briefed concerning a fully formed concept for contribution to a pressing societal problem, suitable and of worthy scope for part of a new mission for LANL – named LANTERN – Los Alamos Neutrons - Enterprise for Research Needs.

I coined the acronym LANTERN and described it in a 20 November 1009 memorandum, attached as Appendix 2.

<u>LANTERN</u>

From personal logbooks:
• 12/5/89 - LANTERN Discussion
LANTERN meeting ~12 Jan 1990 w/ J. Jackson, Deputy LANL Director
• 5/7/1990 - LANTERN Strategy Meeting – an abbreviated Senior Management Group (SMG) instructed the LANL director ($4^{th}$ regime) not to yak half-cocked.
• 5/15/1990 - "DIR ($4^{th}$ regime) presently at least neutral on this" !!
• 5/17/1990 - LANTERN presentation to AT-Div Advisory. Board. Grunder got the point but said it was too big for the Adv. Bd.
• 5/18/1900 - "DIR ($4^{th}$ regime) has delegated LANTERN to (an Associate Director (known to be ineffective))" !! Abdication.
• LANTERN presentation to the $4^{th}$ regime and full SMG by me on 6/8/1990 – attached as Appendix 3. (No notes in logbook.)

I prepared the briefing very carefully, and somewhat desperately, knowing full well that the $4^{th}$ regime would not have the belly for it, and that it would be necessary to fully describe his job for him.
• 8/24/1990 –Called Domenici chief aide, known to be frustrated with lack of LANL vision for redefining mission and open to discussion. Asked if possible to have a few minutes with Domenici at Republican Party dinner at the Lodge. Only time ever attended a political gathering, actually bought a ticket. Waited until near end – they came, briefed Domenici, who became very interested and then asked many questions over nearly 30 minutes. He was not officially briefed by LANL on ATW until much later – and after his office had sent messages to LANL wondering when he would be informed about their "light under a basket". Later, Senator Domenici made US energy policy more and more his theme.
• 9/11/1990 - Admiral Watson letter to $4^{th}$ regime agreeing to a briefing on ATW.
• Dry run for Watkins briefing – DIR ($4^{th}$ regime) absent.
• DIR ($4^{th}$ regime) briefed Adm. Watson maybe 30 or 31 October 1990 - only 15 minutes actual audience. Reportedly very ineffective.

The LANTERN sequence illustrates completely the inability of the $4^{th}$ regime to function on the grand scale for benefit even to LANL, not to mention New Mexico or the USA or the world. We had worked very hard together on the 6/8/1990 briefing, discussing every vugraf, rehearsing with dry runs. The briefing was presented matter-of-factly, with full attendance of LANL technical experts ready to answer questions. We knew, from direct contacts, where resistance would be met – in the fossilized DOE Office of Nuclear Energy, which very emphatically refused to even listen to any new ideas and warned us to stay away - and in the nuclear power industry, which had been contributing to a government fund from which the government had promised to handle the nuclear waste problem and wanted no hint of any change in direction. We also knew where support might lie – in the DOE, and in New Mexico Senator Domenici's office, and that this support, with that of strong LANL sponsorship, should be enough to overcome DOE resistance and begin a serious research effort. We proposed a research effort led by LANL – not a full prototype at first, or fully engineered and licensed facility, although the research should bear strongly on those points. We had to try, although knowing that it would very probably be impossible to arouse leadership at LANL's top level, or also from the next tier. But it was - I^n - nothing. Nothing, no matter what the obstacles might have been then, or now, excused the fact that there was no vision, no strength, nothing.

Stockholm 1991

If the research could not be done at LANL, then it was at least our responsibility to inform the rest of the world of our work. Japan had shared their plans generously, and Europe was keenly interested. I decided to organize a meeting. First it was necessary to inform LANL and the DOE and basically have their "permission", obviously somewhat tricky. I obtained support for a technical information exchange meeting from a second-tier LANL manager, and accompanied him to the DOE, where, among other subjects, it was mentioned that LANL wanted to organize such a meeting, and there was no objection.

Then I had a chance, and knew to work fast. The first step was to find a venue. Sweden had voted to stop nuclear energy in their country, but was open to staying informed about developments. On 11/26/1990, I telephoned Curt Mileikowski, founder and owner of the small company Scandatronix that produced accelerator components, that he had founded as a hobby, while serving as a top-level official in Saab and as Sweden's Finance Minister. He was very quick, like Agnew and Richter – after two sentences, just that we had extensive new ideas about handling radioactive waste and closing the nuclear power cycle and were thinking about organizing a meeting, he said "We want to host this meeting in Sweden!" He called back on 12/1/1990 to say that the hosting was approved.

The second step was to invite delegates, as quickly as possible. I wanted two types – technical specialists, but more importantly, persons with active technical qualifications and active in representing science in their governments. It was made clear that only technical information would be presented, and that there would be no attempt to set any kind of policy. Such contacts were known, and in short order, impressive representatives were promised from twelve countries, including Japan. Other Asian countries were informed but could not be invited because of political sensitivities. There were a few interesting wrinkles; for example, a call came from an intermediary that E. Teller had ordered specifically that he wanted an invitation, but this would have also raised many sensitivities. The meeting was organized by the Los Alamos National Laboratory, USA, and would be hosted by the Swedish National Board for Spent Nuclear Fuel, Sweden.

At this point, the predictable stumbling block appeared – LANL "upper management" called me and said that the meeting could not be held, not supported by LANL or the DOE. I just said that the hosting had been arranged and by whom, and that the invitations had been accepted, and by whom. It was clear that cancellation would be a strong loss of face, and, after threatening that only one or two LANL representatives would be allowed – clearly ridiculous - they withdrew.

The SPECIALIST MEETING ON ACCELERATOR-DRIVEN TRANSMUTATION TECHNOLOGY FOR RADWASTE AND OTHER APPLICATIONS was held 24-28 June 1991 in Stockholm, Sweden [56], with full attendance.

I knew from our many discussions and contacts that our work did not answer all questions, that the questions covered many areas requiring serious work and that there were quite different approaches and ideas, sometimes leading to even rather contentious disagreement. For example, Bowman proposed using slow neutrons, while the Japanese and European interest focused on faster neutrons. Concerned that the meeting remain on an even keel, I took an unusual approach. The meeting was organized in typical sessions, with attendees appointed as session chairmen. But I stayed on the speaker's stage during the whole meeting as overall moderator (the role of "Super Chairman" was coined by attendees during the meeting). The meeting progressed very smoothly, and led to the organization of a world-wide collaboration.

---

56 "Specialist Meeting on Accelerator-Driven Transmutation Technology for Radwaste and other Applications", 24-28 June 1991, Saltsjöbaden, Stockholm, Sweden, hosted by Swedish National Board for Spent Nuclear Fuel. LA-12205-C, Conference, Los Alamos National Laboratory Report, also SKN Report No. 54, UC-940, Issued November 1991, Compiled & edited by R.A. Jameson, Conference Chairman.

• 1991 Particle Accelerator Conference, San Francisco, CA. I personally briefed C. Rubbia about ADS, including details of thorium cycle. When I told people from CERN that I planned specifically to do this, they advised against, stating that Rubbia would steal the idea from us. I replied that at least I preferred he steal it directly and not second hand.

• 1992 A long paper describing the Los Alamos work is published [57].

• 1993 I wrote, with C.D. Bowman and F. Venneri, a popular article on request for the British "Physics World" [58].

Rubbia theft. I received a telephone call from London by an editor of "Nature". He said that he was holding my "Physics World" article in one hand, and in the other, a manuscript whose author (Rubbia) demanded publication without any editing and in a particular issue scheduled to appear at a certain date. But beyond noting that such impertinence would be ignored even from a Nobel Prize winner, he said there seemed to be a problem, because Rubbia claimed that he had conceived the whole idea of a thorium nuclear cycle, how to implement it, etc., while the "Physics World" article and its references clearly indicated the correct precedence, and what should be done? The phone call lasted about half an hour, which impressed me as being a long time back then.

I quickly called CERN, and learned that the laboratory was abuzz about an article with such claims that Rubbia had placed in a Geneva French newspaper, and that he was planning a CERN speech, and was making noises about how this would be his second Nobel Prize. We immediately arranged to have the French newspaper article translated to English, and had the translation within an hour. Charlie Bowman, the real originator, prepared his answer – that although the usurper should be spanked for his childish behavior, we welcomed him and all others in any case to the discussion and development of a solution of the nuclear waste problem. This answer was communicated to the press and to "Nature". The CERN speech was called off, and the subsequent "Nature" article was fairly presented.

• 25-29 April 1993 International High-Level Radioactive Waste Management Conference, Las Vegas, NV (United States). The Stockholm initiative resulted in this conference. Criticism came that it was held in playground Las Vegas. There were two reasons – first, it was cheaper to hold a conference there than anywhere else in the US (there were later complaints that physicists do not drop enough cash at their conferences to make them profitable to the local economy). Second, the governor of Nevada was strongly opposed to Yucca Mountain, and interested in options for reducing the nuclear waste problem.

There was a panel discussion, during which Rubbia behaved in astonishing manner, rising from the audience and openly insulting panel members. The community felt that having him involved in the ATW effort was a decided burden.[59]

---

57 "Nuclear energy generation and waste transmutation using an accelerator-driven intense thermal neutron source", C.D. Bowman, E.D. Arthur, et. al., Nucl. Meth. In Phy. Res. A320 (1992) 336-367.

58 F. Venneri, C. Bowman, and R. Jameson, "Accelerator-Driven Transmutation of Waste (ATW), A New Method for Reducing the Long-Term Radioactivity of Commercial Nuclear Waste", Physics World, August 1993, Vol. 6, No. 8, Los Alamos National Laboratory Report LA-UR-93-752, 27 February 1993.

59 R forced an experiment at CERN to demonstrate that a neutron produced by an accelerator could fission another neutron, panned as trivial and not at all welcomed by the CERN beamtime committee, and was active until at least 2013, as far as I know, spreading the misinformation that the thorium fuel concept was his, etc. In 2021, the CERN Courier published an article about progress in accelerator technology for ADS development in China, mentioning the ADS concept was R's. The Courier should know better, so finally too much – a strong complaint and correction request was lodged, and the Editor agreed to publish a clarification in the next (~September 2021) issue.

Russian scientists were at this Conference and visited Yucca Mountain; Batskikh of the Moscow Radiotechnical Institute, and Chuvilo of the Institute for Theoretical and Experimental Physics.

• 1995 I gave the key first paper at the yearly Alexander von Humboldt Stiftung (AVHS) Bamberg Symposium for senior researchers to Germany, and wrote an article [60], again with Bowman and Venneri, for the AVHS "Mitteilungen" publication – both attracted considerable interest outside the physics and nuclear communities.

The talk was scheduled to include 20 minutes of discussion. The then head of GSI, the heavy-ion research institute in Darmstadt, Germany, was the session chairman, who had introduced me and would moderate the discussion. The first questioner had already stood up and raised his hand as I was concluding. It was an elderly man, a Russian. He vehemently, in the heavy older mode of repeating every sentence twice, claimed that everything I had said was wrong, and that he, etc. – and went on for nearly the 20 minutes! The session chairman tried unsuccessfully to interject and have him conclude but to no avail, and became more and more nervous. But I kept smiling more and more, and was able to communicate to the chairman not to worry, that it was no problem and I was prepared to answer. The audience of course quickly lost interest in what the man was saying, but were more and more awaiting what the response would be! When the chance finally came, I warmly said that the remarks were very welcome, having indicated the previous state of things, that the new ideas were now going far beyond, and that it was a super example of the kind of intense discussion that was needed, from all parts of society, not only from nuclear experts. The Humboldt Fellows are from all walks of life, welcomed the answer wholeheartedly, and there was continuous discussion after the session and continuing after the Symposium.

At the end of the session I immediately sought out the Russian, again thanked him for his discussion, and arranged to have breakfast with him and his wife the next morning. We became friends, and his own story was very interesting. He was the responsible engineer for the first installation outside Russia of a Russian nuclear power plant. As usual, there was a Party functionary of equal rank assigned to control him. The startup date had great political significance, and of course the functionary was only interested in that and had no technical expertise at all (remember the 4th regime and subsequent?). As the date approached, the plant was not ready, and the technical situation was very dangerous (the plant was of the Chernobyl design) – the head engineer knew that a disaster could happen if they were too hasty, but also that a personal disaster – he would be sent to Siberia – might ensue if they were behind schedule. He had some delay but the turn-on was successful and he was safe.

• Mid-to-Late 1990's

Charlie blows up Yucca Mountain. Bowman was one of those contentious but very valuable characters with whom the job of a good technical manager is to value, but to moderate. This can be hard work, and is not in the job description of a 4th regime. He had obtained a small internal LANL fund for continuing his research. But then he got the idea that radioactive waste stored in Yucca Mountain might go critical – the kind of natural criticality that was found in Africa. He analyzed it, and showed that under very improbable conditions (numbers given), also depending on the storage density, that there was a possibility. He was urged not to rush this into publication without very thorough discussions but he went ahead. Of course, this was perfect media fodder. Charlie was removed as leader of his project, the usual "review committee" found his findings false or at least that the given probability was so low that it was ignorable, Venneri became the project leader.

- LANL program dies due to internal politics and power plays. Venneri is very disappointed, leaves LANL. The 4th regime does not comprehend potential of ADS and does not step in, LANL effort essentially stops.

---

60 R.A. Jameson, F. Venneri, C.D. Bowman, "Accelerator-Driven Transmutation Technology — An Energy Supply Bridge to the Future, Without Long-Lived Radioactive Wastes", Alexander von Humboldt-Stiftung "Mitteilungen", AvH Magazin Nr. 66, Dezember 1995, (Invited). LA-UR-95-3316. Los A, 19 September 1995 Los.Alamos National Laboratory Report LA-UR-95-31.

There was good support even within LANL and people supporting the future of LANL – Gene McCall [61 ] and Edward Teller [62 ] supported ATW strongly – Teller made it his specific suggestion for a future LANL - in 1993 – during the "$4^{th}$ regime".

- US has no energy program – no funding for ADS
- Japan experiences economic difficulty – JAERI plan for 600 MeV prototype stalls (still eventually planned for J-Parc).
- Europe carries the ball – ADS program
- RAJ insists that IFMIF target team evaluate problems associated with accelerator trip-outs, and highlights this problem also with ADS teams. Worked with Japan to organize the first Workshop to address this problem. IFMIF ok, ATW raises minimizing trip-outs to top level spec and main purpose of EUROTRANS prototype. Also top level spec now in China.
- RAJ highlights at EPAC 1996 that superconducting accelerator technology is now mature, and should be considered for every accelerator project [63].
- Electric power industry, including nuclear power industry, fights hard against any new ideas.
- International interest continues to grow. For example, OECD includes transmutation as part of their work. The Nuclear Science Committee (NSC) of the OECD Nuclear Energy Agency (NEA) organized a Workshop on "Utilization and Reliability of High Power Accelerators," Mito, hosted by Japan Atomic Energy Research Institute (JAERI), from the morning of 13 October to the afternoon of 15 October 1998. This was the first of a continuing series to the present (under various names).

The dragon's tail is in view, but he is not yet confronted.

• 2000-2010 The dragon is finally confronted:
- World conditions change. Peak oil production passed. Global warming. Economic slowdown.
- "Second Generation of Nuclear Power" becomes acceptable – but must have a solution for nuclear waste
- Accelerator technology is ready, but requires large-scale R&D to meet major new challenge to limit beam trip-outs to a few per year.
- Large-scale R&D is needed for the sub-critical reactor system, for waste preparation for burning, and for after-burning residual waste handling.
- No LANL regime is up to the challenge.

• 2011 March Fukushima. 21 lost years later. Efforts of Domenici, others, to no avail. Still no US energy policy. Respect for America, very much money, squandered in self-aggrandizing wars (Cheney & Co.) – one day of Iraq cost ~$1B, no funding for energy R&D…

• Accelerator System Specifications, Priority Ordering
1. LIMIT BEAM TRIPOUTS TO A FEW PER YEAR.
And 1. And 1. And 1. And 1.
This is NOT "business as usual, with good attention to "best practice"!!!
It is a very hard new requirement for an accelerator.
It will take above all the highest ENGINEERING excellence. A very good CHIEF ENGINEER is <u>necessary</u>.

---

61 "Taking on the Future", Los Alamos Science Number 21 1993

62 E. Teller, "The Laboratory of the Atomic Age", Los Alamos Science Number 21 1993

63 R.A. Jameson, "Discussion of Superconducting and Room-Temperature High-Intensity Ion Linacs", (Invited), EPAC'96, 10-14 June 1996, Sitges, Spain; Los Alamos National Laboratory report LA-UR-96-3206, 11 September 1996.

Every single aspect of the accelerator, accelerator design, accelerator facility, operation, maintenance, administration, etc., etc., must constantly concentrate on this specification.

Tools are needed to investigate and quantify this most important spec. e.g., IFMIF did very advanced development of RAMI (Reliability, Availability, Maintenance & Inspection) tools.

Such tools are greatly helpful to the Project Leader.

2. Low Beam Loss – (previously #1)
3. Cost
4. Etc.; e.g. Length (Cost - ~1 M€/m recent estimate)

This ordering of the specifications is by now well accepted and appears in a large literature. It is still not always clear, however, that the mindset of old practice has sufficiently changed… It is a main job of project leaders to make the change.

A Few Comments on Recent Developments

• ESS [64] notes that "SC is now proven for E < 100 MeV". (Actually proven long ago – for heavy-ion post-accelerators but cavities now prototyped for intermediate betas.

- Use de-rating. RF system, superconducting system.

- Use small rf/cavity units, so that if a unit fails, adjoining units can be re-tuned (very quickly) to allow continued beam operation.

• Use redundancy. Parallel elements whose failure would cause unrecoverable beam tripout: ion source, RFQ, any other element before start of individual superconducting units

• Use fewest types of units. Fewest types of rf amplifiers; accelerator structures; control system elements; unified central control system;, common instrumentation; common safety, interlock, fault detection and recovery elements, etc., etc.

• ESS plans to design for up to 150 mA (peak) H+. IFMIF is 125 mA CW deuterons. The space-charge and beam loss physics for high-intensity linacs is well in hand but hardly practiced, and depends also on careful engineering practice in the layout of the actual accelerator. Present ADS research designs are for CW beam currents well below these levels - ~5-10 mA.

• Full control of beam halo mechanisms and halo avoidance are not yet well or generally understood by the linac community as a whole. Developing the understanding is a hard problem, much better design and simulation require much code development, and there are still open questions, particularly with the operation of superconducting systems, in which the beam physics operating regimes are very different from normal-conducting systems.

• Sample Questions:

- FNAL Project-X [65] proposes possible use of 8 GeV linac "to utilize intrinsically redundant and efficient beta=1 linac, leading to lower costs and higher reliability" and having "multiple output proton beams delivering current to multiple reactors". Good topic for discussion, and can be a subject for careful reliability and availability analysis.

- A number of projects propose a number of different kinds of structures between the RFQ and the start of short, individual superconducting cavities. These structures are non-recoverable, non-retunable fault elements. Should they then be in the redundant parallel front ends? RAMI is subject to systematic analysis. Cost factors? Minimum number of different parts?

- Above arguments suggest use of small superconducting units starting after the RFQ. Some projects have already proposed this (Project X, TRASCO [66], SARAF [67], JAEA-ADS).

---

64 Bermejo, ESS-B Workshop, Bilbao, March 16-18, 2009.

65 G. Romanov, "RT options for low energy part of Project X linac", Project X-doc-607-v1, April 14, 2010, projectx-docdb.fnal.gov/cgi-bin/ShowDocument?docid=607, FNAL.

66 P. Pierini, "European Studies for Nuclear Waste Transmutation", High Intensity and High Brightness Hadron Beams, 2005 AIP 0-7354-0258-2/05/.

67 "Status of the SARAF Project", A. Nagler, t. al., Linac2006, MOP054.

IN CLOSING, IT IS CLEAR THAT CHINA WILL SOON BECOME A LEADING PLAYER IN NUCLEAR TECHNOLOGY AND THE CLOSING OF THE NUCLEAR POWER CYCLE WITH THE HELP OF ACCELERATOR-DRIVEN-TRANSMUTATION. I PERSONALLY WELCOME THIS VERY MUCH. IT HAS TAKEN A LONG TIME SINCE THE 1960'S TO REACH THIS POINT, AND IT WILL TAKE PERHAPS ANOTHER DECADE OR EVEN TWO BEFORE IT IS REALIZED, BUT IT IS A WORTHY GOAL AND WILL PLAY AN IMPORTANT ROLE IN PROVIDING ELECTRICAL ENERGY IN THE FUTURE.

**Accelerator Production of Tritium - APT**

Strategic Stockpile Management called for evaluation of how to maintain a necessary stockpile of tritium. The natural way was to use a commercial nuclear reactor, in which tritium is a by-product, but this was contrary to US policy at the time, not to mix civilian and military uses of nuclear energy. Later the policy has changed.

Mahlon recalled the early days of the first round. RB had some connection with DuPont or somewhere, and took charge. They purposefully isolated the project completely out of AOT, determined to avoid the AOT upper management at all costs.

APT was to be a reliability demo, and thus it was ordered that no new, untested, ideas or technology could be used. The APT Low Energy Demonstration Accelerator (LEDA) prototype RFQ was defined to be an engineering-only project using only already demonstrated accelerator physics and engineering, and should demonstrate 100 hours of cw operation. There was no technical leader to show how the physics methods were outdated and strong enough to convince the project to allow improvements that would benefit, so the RFQ beam dynamics design was the old original simple approach, and not even the effect of field multipoles was included. The requirement to use only demonstrated engineering was ridiculous, and the engineers got on with it.

Project approval came down to a very critical JASON review. The afternoon before, they came to my trailer and asked, stuttering, if I would attend the review. At such late notice, they were clearly panicked, as usual with no-content RB and marginal staff on the technical side, when they found out who they were up against - Burt Richter chairman, Dick Briggs, Jim Leiss and others! The project was to be mostly engineering, but there was no rf or mechanical engineer in the delegation. RB never made an original vugraf (except for his famous "Air Force SDI brochure" - another good story), would operate on vugrafs from others until they were no longer useful, and then panic and bully for new ones, and never acknowledge. So I went, and the first day just listened. But at end of the day, the committee had produced a long list of questions, and especially did not understand, having asked but getting only gobbledy-gook, about the rf system and rf field control. Richter looks at me and said I had homework - should be the opening speaker in the morning and go through all the questions!. Knew what the job really was: the audience range for grasping concepts varied from Richter to far slower. R would understand so fast and get bored and insist on cutting short and get on with it, whereas others, especially one, would never understand or only slowly. All the JASON members were needed, so geared for the slowest level, and prepared to hold Burt off. Sure enough, after a little while he got nervous, then fidgety, then couldn't sit still, then said wrap it up. Told him it wouldn't take too long (the rf part took ~20 minutes). Then wrap it up, etc. So told Burt he was just going to have to endure, because it was critical to make the committee understand that we could control the fields and just to endure it. Not many people told R anything like that. Finished, think the slowest pretty much understood. COFFEE BREAK! And R disappears. But I knew where he had gone, so went outside to the sidewalk, where he was puffing away. Told him I was sorry, knew that he needed a smoke and that he would get it too quick; he grinned. The review was very successful and put the APT program over the top. The next day I was dirt at LANL again, of course.

Following the successful outcomes of further JASON panel reviews of Los Alamos and Brookhaven proposals for Accelerator Production of Tritium in January 1992, the DOE decided to fund an 18-month study of the concept. LANL almost loses the project when a LLNL/Jlab team later went to Washington and convinced the funding agency that the LANL design and approach was old-fashioned, and new technology, specifically superconducting structures, should be used for the main part of the accelerator. Washington called LANL and informed that from that point, the main approach would be superconducting. Total panic. But somehow LANL managed to hang on - guess probably because only the Low Energy Demonstration Accelerator (LEDA) RFQ demo was actually funded, with the following linac only on paper.

The engineering effort was good and much was learned from LEDA. Schrage made exceptional contributions and worked very hard – and was fired from the project by program managers on the day that the RFQ became operational, claiming that money had to be saved. Reliability was demonstrated only by a facade – 100 hours accumulated in a multitude of short time segments – instrumentation and rf control didn't work. Reliability depends on the total system, and there was little overall coordination from "program management". As the beam dynamics design physics was outmoded, the RFQ demonstrated poor transverse emittance and a long longitudinal tail, which doomed a follow-on expensive, touted, "beam halo experiment" from the start (as might have been realized…).

The starting authorization for LEDA came at a time when environment was a buzzword, and Congress was ordering that a completed project environment had to be returned to a green grass state, and funds were included for all equipment and buildings used for LEDA to be razed upon completion. And when the time came, the program managers started to do that. No vision, no imagination. Always before, project facilities and equipment were considered the legacy of the project, and belonged to the laboratory. I was visiting, and pointed out this strategy, and also how to state it to Congress, etc. I wrote a one-page letter to our long-term Senator Domenici, telling him that LEDA had been a great success, that warranted an exception to the written Congressional instructions, its facilities should be retained as legacy and of large value to future project (there were buildings and large rf and auxillary systems). They changed course, and the entire legacy is intact to this day.

[toc]

## Work in Collaboration

Not a beat was lost after 1987. Not necessary to participate in stupid meetings, exercises. All activities were carried out in collaboration with very good people, who, with very few exceptions, viewed me as a partner and not as their competitor. Some inside views and experiences follow.

### The ORNL Spallation Neutron Source SNS

One of the many reasons that the Superconducting Super Collider project was cancelled was that they did not utilize the collective expertise of the US accelerator community. So when ORNL won the SNS, they set out to organize it as a collaboration, divided among the large US accelerator laboratories Argonne, Brookhaven, Lawrence Berkeley, Los Alamos, Oak Ridge, and Jefferson. Unfortunately, this was done without top level understanding of how the pieces would fit together – and this situation remained with no knowledgeable top-level technical leader, until the project had been a full construction project for about five years and the design was long frozen. Very fortunately, the resulting sliced-up sausage, although fitting back together somewhat roughly, has worked quite well, at least for the original design specifications. The over-riding important thing was that there was a funded project. Later modifications are not unusual in any case, as lessons are learned from the initial operation.

I was the Accelerator Team Leader for the International fusion Materials Irradiation Facility (IFMIF) project, which was coordinated out of ORNL, so was at ORNL frequently (once sharing an airplane row with Trivelpiece), and often sat with the first SNS team, led by Jose Alonso of LBL, for discussions. Alonso had announced that he would act as SNS Director for only a limited time, and the discussions focused heavily on who should be his successor. Alonso left, but first insisted that I be on the Accelerator Systems Advisory Committee (ASAC), to violent objection of LANL. I served on this committee and also on the side throughout the SNS construction and commissioning from 1998-2006. (The ASAC was the right kind of Advisory Committee – long-lived, giving very important continuity, serious, and listened to by the SNS management.) It was interesting and

challenging – not easy to coordinate six laboratories, each with its own culture. Of course, our twice-yearly meeting, with a few extra, only gave advice, but we worked hard at it. It was later stated that "once we had ASAC buy-in, any dissent on a decision was largely behind us – people bought in." [68].

The project was salamied to fit in the five, later six, national laboratories. For example, it made no sense to give the RFQ to LBL just as LANL finished LEDA, except for the very well-known reputation of LANL for high costs and very bad management, including technical. LBL succeeded in getting a pure R&D program with essentially no deliverables, with a voluptuously padded budget. The submitted LANL budget was far too low in the first rounds, (although AES had done a correct one but they didn't want that answer), and LANL never recovered from that. Several of the partner labs already had advanced designs, the baseline was not optimized, and the project was built at a laboratory on a green-field site without a large accelerator infrastructure or culture [lii].

Need for Strong ORNL Leadership: ASAC Report 1, October 1998 [lii]
"...The SNS is being designed and constructed by a five laboratory collaboration. Oak Ridge has a central role in this collaboration. It has the responsibility of integrating the accelerators delivered by the other laboratories. It will participate in the commissioning, and then it must operate and improve the SNS. This has implications now.

The design and construction will require trade-offs between the accelerators that are the responsibilities of the other laboratories. Some of these trade-offs will be complex and will have to be made with only partial information. This will require strong, experienced leadership at Oak Ridge to deal with the other laboratories and, in addition, to recruit and lead the people who will commission, operate and improve the SNS.

Documentation must be at an unprecedented level for an accelerator project because ORNL will not have the equipment designers and builders in residence to diagnose and repair equipment. Documentation must be exceptional, and, in addition, there could be significant value in having ORNL people involved in the construction and design at various collaborating laboratories..."

Flexibility: ASAC Report 1, October 1998 [lii]
"...There are no reliable beam physics models which can credibly predict losses at this small level, and it is unlikely that such models will be developed before the design is finalized. Therefore, the linac and accumulator ring designs must be conservative by today's understanding, and they must be flexible..."

ORNL had much trouble finding a good leader or leaders, technical and managerial, for SNS. Alonso's replacement did a good job writing interlab agreements, but there were problems with delegation and some very strange job ads that led to his removal. The next Director was a professional project manager, good with cost and schedule management tools, but not technical or strategic (as at LANL, should never be in charge, which LANL never learned (to this day?)) It was clearly necessary to find someone else, and a charter clause that required the various lab directors to meet and decide on very serious problems was invoked. Of course, Lab Directors constitute the absolute poorest kind of committee to pick an important leader, maybe especially a technical one! They pick Monckton. Who comes with the hurricane of Yanglai Cho, Ed Temple and Bob Kustom.

Cho lost the project a year making a re-push (third or more time) for his favorite approach of RCS, but also did a very important good thing by finally getting out the very thin report that the ASAC had urged, pushing superconducting for the main linac, and Jefferson Lab joined the project. LANL had been unable, or didn't want to lose the long room-temperature linac, and/or too timid to propose a superconducting (SC) linac, or write the justification for it. However, LANL was allowed to retain the beam dynamics design responsibility for the SC linac. Cho and Kustom gave so many abrupt, mostly poorly conceived orders that the rest of team, especially LANL, went into silence and obey mode.

---

68 "Leadership at the Spallation Neutron Source", S. Henderson, Bob Siemann Symposium ,July 7, 2009, http://www-conf.slac.stanford.edu/robertsiemann/talks/SHenderson_Symposium%20v1.pdf
I don't know why Henderson chose the ASAC quotes used here, but they are all my words…

Adoption of Superconducting Technology: ASAC Report 2, December 1999 [lii]
"...Superconducting RF has already been accepted for intense, light-ion cw applications. Its technical application to pulsed light-ion applications, such as the SNS linac, awaited confidence that the RF fields and phases could be controlled precisely with Lorentz force detuning and microphonics present. This control has been developed and demonstrated at the TESLA Test Facility, and the TESLA design seems adequate to fulfill the SNS needs...The SNS design study clearly indicates that superconducting RF is economically and schedule competitive. With superconducting technology also applicable to pulsed high-intensity accelerators, it is clear that superconductivity is the technology of choice for many future linacs.... In summary, we give a strong endorsement for the use of superconducting technology."

There were numerous rough edges between the labs, but particularly with LANL. Sometimes the problems were brought to the attention of the ASAC. In the earlier years, before schedule requirements made it too late, the ASAC always recommended that ORNL quickly built its own staff and take over the problem areas themselves. [69]

Flexibility: ASAC Report 3, October 2000 [lii]
"...There is one situation that needs immediate attention and is highlighted here. The Drift Tube Linac (DTL) design has permanent magnet quadrupoles, and this feature is close to being frozen...This planned use of permanent magnet quadrupoles is neither conservative nor does it provide tuneability and flexibility. Space charge is most important at lower energies, and the present plan is to provide no focusing adjustment to 90 MeV. This concern was raised at the last ASAC meeting, and at this meeting it was stated in response to questions that fixed quadrupole gradients are preferable so people couldn't readjust gradients. But this could be necessary and is exactly our concern - study of phase and quadrupole laws in the high-energy end of the linac are to continue, but results of this work could not be applied later to a DTL with permanent magnet quadrupoles."

This concern of the ASAC was ignored. The normal limiting dynamic aperture of a linac of the SNS type would be in the RFQ – but in SNS, it is in the DTL – an unheard of location. Fortunately, at least for the initial specification, the DTL performance is adequate.

However, the phase advances and tolerance for a larger longitudinal emittance are problematical for higher power levels. One could only think that the most important thing was that the SNS project was accomplished, and replacing "small" components like an RFQ or the DTL later could be described as natural parts of an upgrade program if necessary.

Beamloss: ASAC Report 3, October 2000 [lii]
"The SNS has stringent beam loss requirements. The linac design must minimize beam halo generation, potential beam loss along the linac itself, and deterioration in output beam quality to the HEBT and ring. The underlying physics depends on space charge and its nonlinear interactions via the system resonances. The linac dynamics has never been presented to us in a framework of space-charge physics, e.g., in terms of transverse and longitudinal tunes at low and high currents, tune trajectories along the linac, tune spreads, relation to structure resonances and potential unstable modes, and the effects of errors in this framework.... There is brute force running of simulation codes and simplistic evaluation of results by comparing phase-space scatter plots or RMS quantities, without relation to the underlying physics. The best strategy to continue making progress with incomplete understanding of the physics is to 1) make conservative decisions and 2) provide tuneability and flexibility so there will be the ability to respond to either improved understanding from future physics studies or experience gained during commissioning."

Review of LANL Spallation Neutron Source Division, 20-22 December 2000:
The magnitude and frequency of LANL problems led to the separation of SNS work from the AOT Division and formation of a separate SNS Division. A probing review was held in December 2000 at LANL; my contribution is attached as Appendix 4.

---

69 "The Spallation Neutron Source Project", http://science.energy.gov/~/media/opa/pdf/SNS_033110.pdf

Trouble at LANL grew. Roger Pynn, LANSCE director (1997-2001), was asked to come to a meeting, and made his very famous "radar screen" remark, which ricocheted over the whole world – "Well, the SNS (and it's problems) (which was under his purview at LANL) had not ever gotten onto his radar screen yet…" ORNL has built up SNS staff, and wants LANL off the project. The emergency cord of a Lab Directors meeting was again pulled. The Lab Directors convened, and directed that LANL had to form an SNS Division. Again to avoid further contact with AOT Division. The DOE had strongly wanted H. Grunder's program manager Beverly Hartline to be the SNS director, but ORNL had selected Monckton. So Pynn and LANL were ordered to take her as new SNS Division leader. So Beverly is looking up Monckton's skirts, and knew how to do that!

I did not seek any contact with the LANL SNS Division leader, but she asked me to help with a study on beam loss, so did, but it was only a simple evaluation of frozen design.

Monckton finally succeeded in getting rid of Hartline, who migrates to the vaunted regime of the Project Control office of LANL, set up to teach, help, and etc. all projects at LANL over maybe 100K - amazing how many projects never came to LANL at all. Monckton's 2 years at SNS come to an end, and he has to decide what to do. Both he and Temple are so pissed at DOE that both ready to quit; M goes back to his waiting ASD position at ANL, and T also left. But Grunder has moved from Jlab to be director at ANL - and calls Hartline back to be his Principal deputy!! Shortly after, Monckton is forced to leave ANL. Probably one of the biggest big politics stories in accelerator history!

G's end is sorry. He was such an egomaniac that he was harmful to AT Div when I mistakenly asked him to be on the Div. Advisory Committee; got himself elected chairman, and pursued his own interests as usual instead of helping me, taking up the program managers side in 1987 and pleased that he had the chance to meet with the "upper management". Met him at Peking U in 2004 at President's dinner given for him - wondered what to say, but anticipated just greeting him and vice versa and that would be it. Came with young attractive Chinese wife, who helped a lot keeping the table talk going (the President unfortunately had a toothache, managed well for awhile but left early). The exchange of greetings was like that. But something about G seemed strange. Then heard later that he "would not serve" a second term at ANL. He was so arrogant that I could not understand how that could be, but never heard any background until at EPAC2006, sitting with Phil Debenham at the banquet. G had quickly advancing Alzheimer's, they covered for him maybe the last two years.

Finally Norbert Holtkamp comes (2001-2006). Finally a good technical leader. Too late to correct design flaws, but a very good organizer and project leader.

Reliability and User Expectations: ASAC Report 6, February 2002 [lii]
"A high level of reliability is a priority for the neutron scattering community. The ultimate goal for SNS operation is 5000 hours per year at full beam power of 1.4 MW with 95% reliability. In past reports we have stated our opinion i) that reliability did not appear to be an important consideration in the design, and ii) that it was important to develop realistic goals for the initial reliability and for the rate of improvement over the first years of operation."

- John C. Browne LANL Director 5th regime (1997–2003)

LANL continues to screw up. Especially after Holtkamp came (2001-2006), who was competent, able to recognize inferior work, and not reluctant to investigate. But Holtkamp too late to change anything technically. PMQs in DTL - ASAC recommended strongly against. LANL went ahead, tossed engineering over the wall to Eng Div ("AOT" had pissed off the engineers so badly ~1993 that they all went (except Schrage and Wood) to Eng. Div. – which meant there was no longer an integrated accelerator technology capability at LANL. The management of the Eng. Div was not competent and very risk adverse (RB), so no quality control, no visits to factory (AOT physicists complained there was not enough budget for this), so a whole set of drift tubes was overheated by the electron beam welding, which ruined the permanent magnets. Too late to kick LANL out or let them quit - JB actually called me - asked if LANL should pull out, but didn't we have an A team? Told him bluntly that by now it was down to C or worse, and matrix management, that it might be just as well to cut his losses. JB's final epitaph of "naïve" was totally accurate. Domenici also in on it - jobs. They stayed – and formed an ~25 person emergency engineering team, Schrage a key part - to bail them out again. Of course no cost responsibility these days, ORNL had to pay.

RF low level control snafu –
LANL AT-Division had responsibility in the Star Wars NPB and FEL programs for a low-level rf field control system capable of operating in space and without flaw for at least the 20-minute duration of the war. Cost and sense did not count in the Star Wars power game. A system was devised of such enormous complexity that it never functioned (e.g. a 20-layer mother board, and software programming so complex and inscrutable that it did not work and was impossible to maintain). The same system was to be used for the LEDA, but was not available for the tests, and then again put up for SNS. I warned the ASAC that eventually this would result in another crisis. There were continued delays, finally a promise to have a working system at Jlab for the first rf test on a sc cavity. Up to the day before, nothing was said about any problems, then it was announced that LANL had nothing to deliver for the test. Both hardware and software problems, too complicated and top-heavy that it didn't work. (2014 – the same system is now being tried as "LANSCE upgrade"…) Outraged, SNS brought the crisis to the ASAC, and a special review after an ASAC meeting. The job was given to Larry Doolittle at LBL, who had produced a sensible digital system for the RFQ control. Hasid Shoee at LANL, a good software guy, took that over, and the result SNS low-level rf control works well.

New SNS DL Don Rej 2002-2004. LANL share finished early 2004.

SNS_DOE article - biggest problem was mgmt. pdf : *science.energy.gov/.../SNS_033110.... A (biased) story of SNS construction – reading between the lines is the above….*

[toc]

**FMIT, inactivity to 1993, then IFMIF 1994-2006.**

During the inactive years from the cancellation of FMIT until 1993, a small activity remained with monitoring of the need for materials testing in the realistic segment of the world fusion community. From logbook XXXII. 10-17-90 -> 2-1-92, p.50 - 12/12/90 notes from a high-powered review committee: Briggs, Cowan, Winston Little, John Hightower, Lowenstein, Leiss, Harold Lewis, Doug Pewitt, Art Kerman, Roy Culler, J. Jackson (not high-powered) Lewis delegated to review me on IFMIF – used briefing we had just prepared for DOE – Lewis wrote specific statement in report that IFMIF was in good hands and should be left alone.

In 1993, a meeting was held in Moscow to organize the restarting of FMIT under an international collaboration as the International Fusion Material Irradiation Facility (IFMIF) project, with emphasis on high-intensity, low beam loss linac in factory environment. The project was structured with a project leader, Tom Shannon of ORNL, and four teams – accelerator, target, materials, conventional construction. I was asked to lead the accelerator team, and the US also had responsibility for overall system engineering. We stayed in a hotel outside of the city center, and were told that cars would come by in the evening that we could use very cheaply to go downtown, and that if we were lucky, maybe Yeltsin's limousine would come, as the driver was allowed to use it when off duty – we were lucky, only the flags on the front bumper were missing. We walked around too long hunting for a place for a beer (we always did that with Tom Shannon) but couldn't find one, finally I led to the first McDonald's – had beer but not the atmosphere we wanted (not very good beer either!).

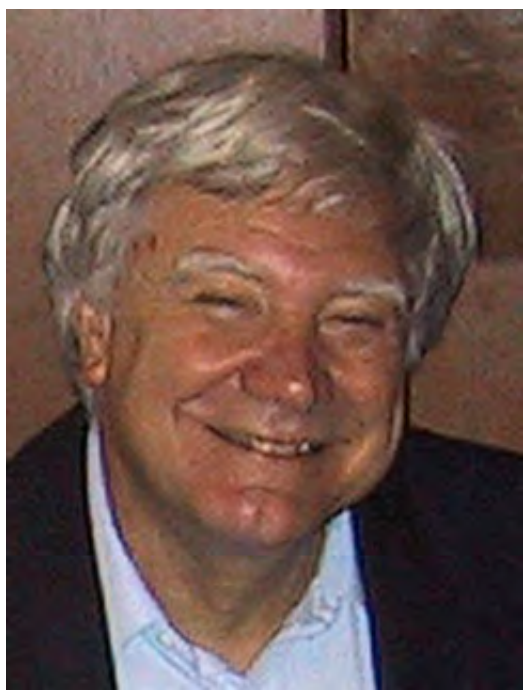 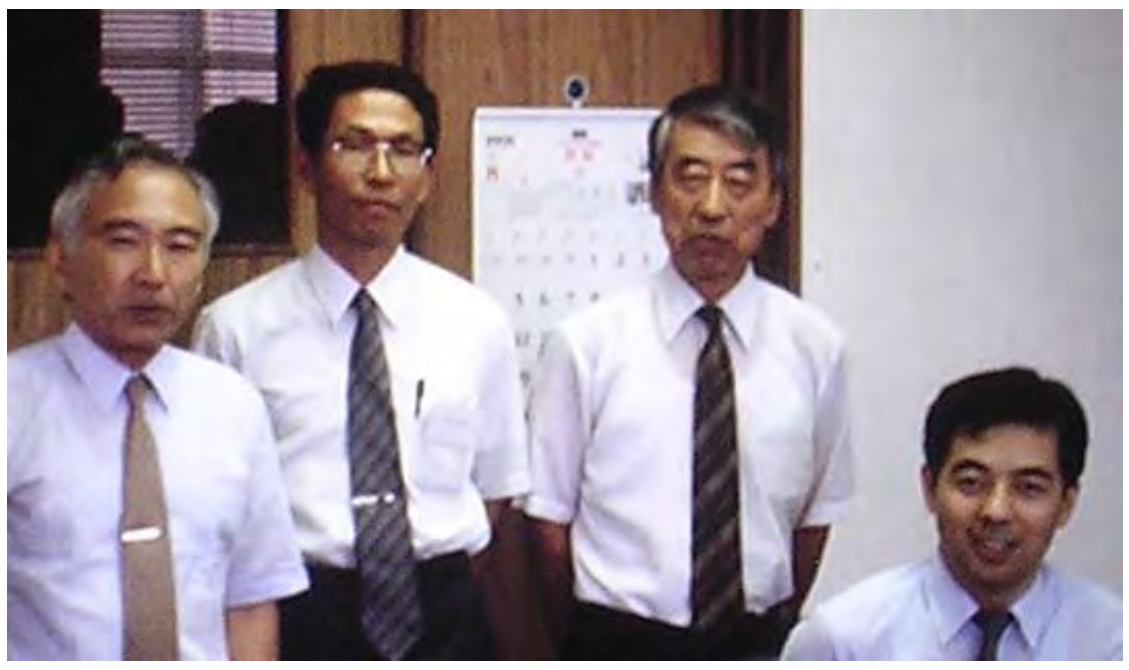

Tom Shannon, M. Odera, Y. Ohno, A. Miyahara, M. Sugimoto – working on IFMIF

The IFMIF project ran to the end of 2006, merging into EVEDA (Engineering Validation and Engineering Design Activity) with funding to build an accelerator prototype in Europe and operate it in Japan. The US funding was adequate initially and we used part of it to establish an industrial team (Northrup-Grumman, later AES), which contributed very much to the project. As there was no fixed project schedule, it was possible to pursue advanced R&D, and during the twelve years of its existence, many advanced and innovative concepts were explored. These included development by Chris Piasczcyk of substantiated methods for using RAMI (Reliability, Availability, Maintainability, Inspectability) techniques for complex accelerator projects. An accelerator systems engineering program ASM (Accelerator Systems Model) was developed by the Northrup-Grumman team that tied the physics design, engineering design, cost estimating, and RAMI model together down to the nuts and bolts level – this was extremely useful for managing the project – we could overnight update a day's discussion and have a new comprehensive model the next morning. My beam dynamics research resulted in methods for achieving exact equipartitioning in the linac design.

The US had entered its "fusion is only science" budgeteering phase, and the last few years provided so little funding that we had to drop the industrial team and the overall system engineering, but overall IFMIF was a very exceptional program, developing and using the above tools to lay out a truly coordinated accelerator project. Unfortunately, but predictably, a lot of this work has not been used further, for example ASM. A wrap-up meeting was held in March 2007 where I reviewed the accomplishments of the IFMIF program and outlined topics where more design work was needed. In particular, I was concerned about the RFQ simulation code situation. IFMIF was not a real project – it was a cooperation, and the various partners used their own codes, usually black-box. A real project should decide on a bench-marked code and have the full software. I was concerned that different codes had always given different results for any given design, for transmission, accelerated beam fraction, and also emittance ellipse parameters and beam loss patterns along the RFQ. Comparisons were cited on specific designs and not in any way for a general comparison over the design parameter space. I devised an "experiment" to test them against each other. I was not so much concerned if the absolute value of transmission varied between the different codes, but was primarily concerned that different codes would show an optimum transmission and accelerated beam fraction at different design parameters. From engineering experience, this is an ever present danger. My RFQ design code has control of all RFQ parameters and also of the space charge physics. Therefore, a set of RFQs could be designed by varying only one parameter – this is a necessary characteristic of a good experiment. The variation of this parameter produced RFQs whose transmission and accelerated beam fraction ranged from low, through an optimum, and again to low. The set of RFQs was simulated with the different codes. Quite different transmission and accelerated beam fraction curves resulted, and the optima occurred for different values of the parameter. I presented the case on the first day of the Workshop, and was instantly and strongly attacked by participants who claimed that codes had been compared and that they agreed – one even claiming to four decimal places (!) I noted there would be a working group meeting the next morning and urged them to come. After the evidence of this preliminary comparison was presented, the attacking stopped. [70]

---

70 Jameson, R.A., "RFQ Design Studies – Investigation (incomplete) of Dependence of Optimization on Codes", Injector/RFQ Working Group, 1st EU-JA Workshop on IFMIF-EVEDA Accelerator, 7-9 March 2007, Paris. (RFQ Workshop Paris.ppt, March 8, 2007)

INFN Padova took over the design and construction of the EVEDA RFQ. I enjoyed working with them on the code comparison during May 2007 in Padova. However, the time they felt necessary vane machining put them under schedule pressure, so they adopted the old outdated standard design approach, but then refined the design to adjust for maximum acceptance. Although equipartitioning was then never mentioned in reports, reverse engineering shows that the EVEDA RFQ is indeed essentially equipartitioned in the main part of the RFQ.

During the ITER negotiation, Japan conceded the site location to France, and received a list of 2nd prize considerations, including IFMIF/EVEDA. They declined this, however, because of insufficient internal technical capability, and it looked like all work toward new materials would stop. However, the Europeans realized that this could not be allowed to happen. The EVEDA project was funded because fortunately, during the IFMIF years, the fusion community finally had to admit two problems – lack of robust material in neutron environments, and the necessity to deal with radioactive waste, although less than from fission systems – and this had placed IFMIF on a parallel track to ITER.

The concern about the simulation codes then occupied me for several years, as detailed in the other half of this book.

We had many very productive and interactive meetings, of the Accelerator Team separately and of the whole team, from a few days to two weeks duration. The two-week workshop in Frascati was where I was fortunate to hear Teplyakov's story of his RFQ. There are many stories in my diaries – another one was at a Mito, Japan, where our Italian M. Martone worried about having to eat a Japanese-style breakfast at the hotel. I knew about Martone's fanaticism about the preparation of coffee – he made every morning for himself and his wife, but making one cup at a time in a one-cup expresso pot, because it was a sacrilege to brew more than that at once. So I had scouted the first evening, and told him that he would go with me in the morning for breakfast – to a very nice and typical coffee shop run by a lady who had been to Europe and decorated the shop in totally European style, with Bach playing in the background, and serving the very excellent Japanese coffee with the "morning set" – a thick toast with marmalade, boiled egg, and cabbage slaw. Martone was very happy. Maybe before, maybe after, he returned the favor by introducing me in Frascati to early morning coffee with the delicious "rumbaba".

IFMIF People.pdf    IFMIF Database CD DVD??

[toc]

## BNL NSLS

1984 - was asked to chair a DOE Review of NSLS BNL. Knew background, and of the heroic efforts of Arie van Steenbergen,

Committee Members: R. A. Jameson, Los Alamos, Chairman; H. Edwards, FNAL; J. E. Paterson, SLAC; T. Godlove, DOE/OHENP; S. Penner, NBS; A Hofmann, SSRL; R. Sah, LBL; R. Martin, ANL; R. Siemann, Cornell, Herman Winick, SLAC.

The critical situation of the NSLS at BNL was well known, and I was in any case strongly aware of the risks of projects being troubled and even cancelled by federal agencies, that in fact should have responsibility. The review was hard work, because I insisted that the funding agency aspect be clearly outlined, and also had to deal with convincing the committee members, one of which was famous for obsessing about details.

Fortunately we could see “light at the end of the tunnel”. Half of the blame was strongly placed directly on the DOE. The words of the Review [71] were “the closest to a diagnosis” of the difficulties, received national notice [72] and helped lead to changes.

NSLS-Part-II.pdf: The National Synchrotron Light Source,
Part II: The Bakeout
Robert P. Crease

This is a remarkably well-done article. It discusses very directly the real personalities involved, the difficulties they faced, the tensions and risks involved in building a large scientific facility in a highly politicized environment: risking a facility’s quality by underfunding it versus asking for more funding and risking not getting it; focusing on meeting time and budget promises that risk compromising machine performance versus focusing on performance and risking cancellation; and the pros and cons of a pragmatic versus an analytic approach to commissioning. Written by a professional historian. (Too bad such has not been written for LAMPF and, competently from a real anthropological viewpoint, not by a sycophant, for LASL/LANL.)

> Page 26:
> Most ominously of all, the perilous situation had attracted the DOE’s attention. The agency, “concerned about the delay in bringing the X-ray ring up to operational level,” announced its intention to review the project. The DOE felt a special urgency given that it was about to fund the Phase II construction project – at $19.7 million to be spent over three years, about the same scale as for building the machine itself – without its Phase I in order. Blume and others feared that the DOE might back away from the NSLS as it had from the recently-cancelled ISABELLE. If the DOE could cancel that big project, one whose technical problems had been fixed, it could cancel anything. It was indeed a perilous situation – a nonworking machine, a management unable to fix it, and a looming DOE review – for the NSLS and for Brookhaven.
> ….
> Blume convinced the DOE to hold off its review until May, leaving him only six months to put the machine in order.
> Page 27:
> The DOE review [27] began at the end of May (1984). To everyone’s relief, it was optimistic.

“[W]hile the NSLS has experienced long commissioning delays,” the final report stated, “the delays do not result from fundamental technical problems that will prevent the facility from fulfilling its promise, but rather from economic and management issues.” It anticipated, by the end of 1984, that the X-ray ring would be operating at 50 mA at 2.5 GeV, with a two-hour lifetime and with six working user lines. This would “constitute a viable operational base from which useful research could proceed while the substantial remaining machine commissioning and further user-access commissioning is completed.” But while “very substantial progress has been made in the last six months, morale is good, and the progress should continue,” the report pointed out, much remains to be done and the project needs additional “key accelerator physics and engineering staff and support” over the next two years. The report also recommended another technical review in 1985, and a followup workshop on the commissioning experience.

What had gone wrong? The report mentioned several factors. One was a set of design decisions about tradeoffs and recycled parts that, while making “the project more palatable to the funding

71 R. A. Jameson, Chairman, "Report of the DOE Ad-Hoc Committee on the Brookhaven National Laboratory National Synchrotron Light Source, 30 May - 1 June 1984”, Los Alamos National Laboratory Document No. AT-DO:84-157(Rev.), June 1984. (Also BNL Historian’s Office, and LANL “Director’s Collection” archives).

72 “X-ray Drought Ending at Brookhaven’s NSLS”, Science, Vol. 229, 2 August 1985, pp. 453-454. http://worldtracker.org/media/library/Science/Science%20Magazine/science%20magazine%201983-1985/Science%201983-1985/root/data/Science_1983-1985/pdf/1985_v229_n4712/p4712_0453.pdf

agencies and the community," also increased "the difficulty, time required, and cost of bringing the facility into full operation." Examples included the omission of glow-discharge clearing electrodes in the vacuum chambers, whose cost "would have been small compared to the cost to provide this capability now." Another factor was the attempt to commission "two separate, state-of-the-art rings into operation simultaneously with available resources." Yet another was the impact of the CBA, which had consumed accelerator resources and lab attention until the previous year.

"Report of the DOE Ad-Hoc Committee on the Brookhaven National Laboratory National Synchrotron Light Source," Doc. # At-Do 84-157 (Rev.), Los Alamos National Laboratory, BNL Historian's Office.

Was later on enough DOE review committees, but never again asked to be Chairman! Herman Winick was on my BNL NSLS committee – led to invitation to be on the Taiwan TRC:

[toc]

**Taiwan Synchrotron Radiation Research Center SRRC Technical Review Committee TRC**

This was the kind of Review Committee that it makes sense to agree to serve on – long-term, so that relationships can be established and a deep understanding of the situation developed, the requestors listened to and used the hard-working advice.

| | |
|---|---|
| *1981 December-* *1982 November* | Formation of the Feasibility Study Group and submission of their report to the National Science Council |
| *1983 July* | Approval of the SRRC Project |
| *1983 October* | First Board of Directors (BOD) Meeting in Taipei |
| | **Chaired by Luke C. Yuan** |
| *1983 November* | Establishment of SRRC |
| *1984 January* | Approval of the budget for the project by the Executive Yuan (1 GeV, 250 MeV Linac Injection System, US$ 44 M) |
| *1984 September* | Formation of the Technical Review Committee (TRC) |
| *1984 December* | First TRC Meeting at BNL, USA; adoption of full energy injection |
| *1986 February* | Site selection made (Hsinchu Science-Based Industrial Park) |
| *1986 August* | SRRC ground breaking |
| *1986 December* | Decision on nominal energy of 1.3 GeV |
| *1987 April* | Adoption of TBA lattice for the storage ring |
| *1988 February* | International Workshop on Constructing 1-2 GeV Synchrotron Radiation Facilities at Taipei |
| *1988 July* | Revision of SRRC Construction Plan (1.3 GeV Storage Ring, 50 MeV Linac, 1.3 GeV Booster Synchrotron)(US$ 83 M) |
| *1988 August* | Injector system contracted out to Scanditronix AB, Sweden |
| *1988 September* | Phase I construction started |
| *1989 April* | SRRC Design Handbook completed |
| *1990 May* | Phase I construction completed; Phase II construction started |
| *1990 June* | Moved to the Hsinchu site |
| | **Director E. Yen appointed** |
| ***1992*** | **Booster commissioning, start Storage Ring installation** |
| ***1993 April*** | **Beam stored. June – Director Y.C. Liu appointed** |

12/27/86 Letter from Olivia Cua, SRRC, including "Thanks for Ojo-de-Dios"
(Old Stuff File Folder Index.doc) Check plaque for end date.
1986Oct-1998Dec.pdf 14 December 1998 notes – must have been ~1983-1997 on TRC, then 1 year on STAC

(Blewett) a member of its Technical Review Committee since 1984. From PAC1993_2546.PDF, Blewett's R.R. Wilson Prize Lecture: Adventures with Accelerators

Taiwan – December 1986 – first time to Taiwan
1990 – first meeting at SRRC Hsinchu LukeYuanLtr19901023.jpg

From: Jnet%"WINICK@SSRL750"
To: Jnet%"JAMESON@LAMPF"
Subj: RE: Draft TRC Report for Oct. 8-9, 1990 Meeting
Bob,
Thanks for the prompt reply. The report is as good as you say largely because of your excellent writeups on I&C and Management. I think I would have trouble writing up the report of a meeting that you missed.
Please don't let that happen.
Regards,
Herman

Received: From SSRL750(WINICK) by LAMPF0 with Jnet id 7186
for JAMESON@LAMPF; Fri, 21 Feb 1992 08:14 MST
Date: Fri, 21 Feb 1992 07:14 PDT
From: <WINICK@SSRL750>
Subject: Good News from Taiwan
To: JAMESON@LAMPF
Subj: beam in the booster
The Scanditronix got 0.5 mA circulating beam in the booster starting last night (actually at 2:00 O'clock in this morning) Champagne was opened this afternoon in the booster this afternoon at 4:00 O'clock.
best regards
Chen-Shiung

Date: Thu, 5 Mar 92 21:02 U
From: CSHSUE%TWNCTU01.BITNET@SSRL01.SLAC.STANFORD.EDU
Subject: 1.3 Gev beam in the srrc booster
This Monday the srrc booster has achieved 1.3 Gev with 2 mA current. Last night (Wednesday) the current is up to 4 mA. This afternoon there was a party celebrating the goal together with the Scanditronix.
best regards
Chen-Shiung

From: Jnet%"WINICK@SSRL750"
To: @TRC
Subj: 80th Birthday Celebration for C.S. Wu and Luke Yuan
Date: Mon, 24 Aug 1992 18:55 PDT
Dear Colleague;
This is to inform you that on the evening of the first day (Sept. 21) of the upcoming TRC meeting there will be a dinner party celebrating the 80th birthdays of C.S. Wu and Luke Yuan. Appropriate remarks by members of the TRC will be welcome.
Herman

Date: Thu, 29 Jan 1998 15:25:09 +0100
cornacchia@ssrl01.slac.stanford.edu, albert@cernvm.cern.ch,
bmkincaid@lbl.gov, avanst@bnl.gov, winick@slac.stanford.edu,
himpsel@comb.physics.wisc.edu, rjameson@lanl.gov
From: Louise Peritore <khadrouche@esrf.fr>
Subject: SRRC - " 21st Meeting of Technical Review Committee"
Cc: ypetroff@esrf.fr
Dear Colleagues,
Please find attached the "21st Meeting of Technical Review Committee" report held on 15

and 16 December 1997.

Members of the 1st Technical Review Committee were H. Winick (SSRL), J. Blewett (BNL), A. van Steenbergen (BNL), A. Hofmann (SSRL), G. Mulhaupt (BESSY), R. A. Jameson (LANL), M. Allen (SLAC), and C. S. Wu (Columbia Univ.).

8th SRRC TRC Meeting, October 2007, for the first time at the newly constructed laboratory and staff building of the SRRC at Hsinchu, Taiwan.

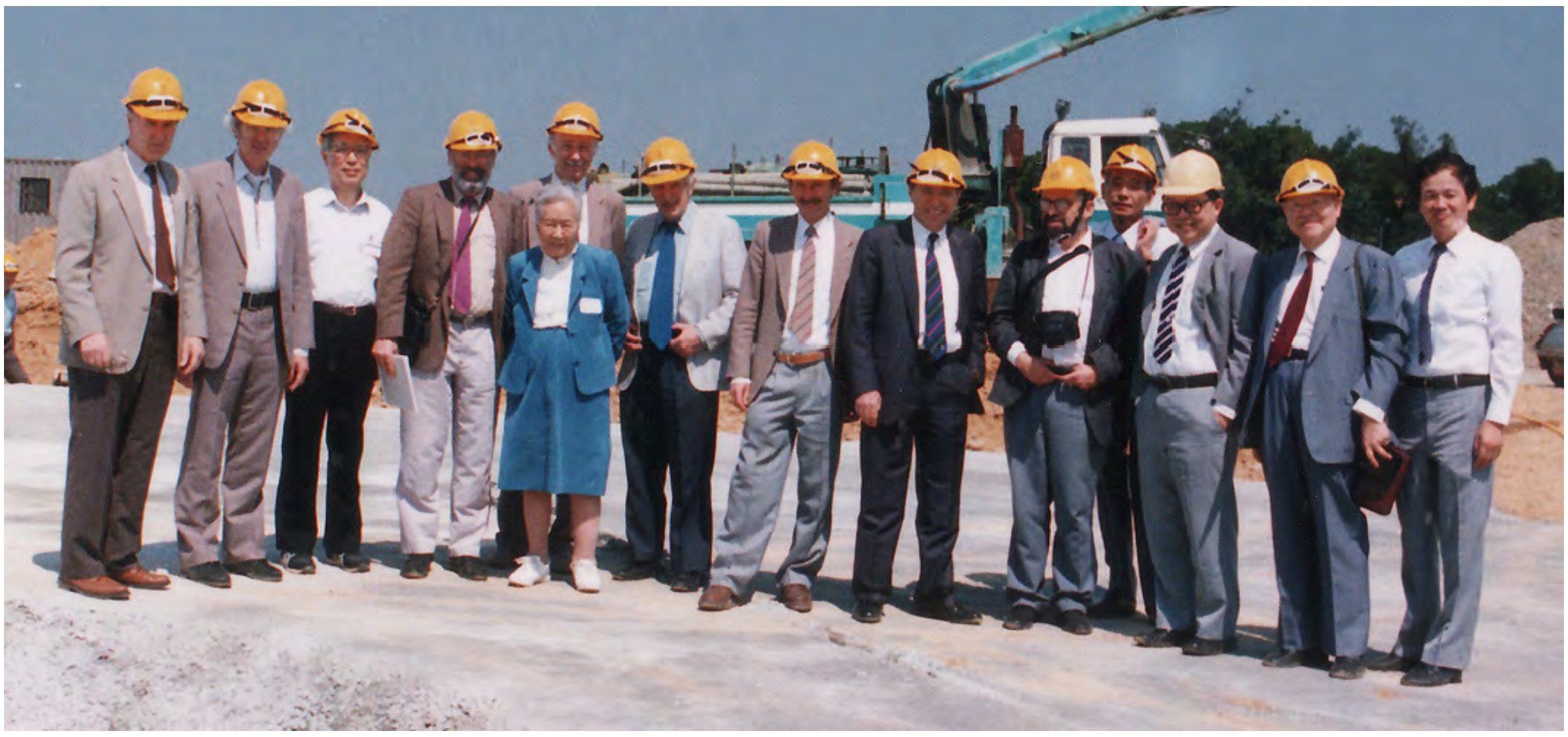

From right: Prof. Y.C. Liu, Dr. Luke C. L. Yuan, Edward Yen, (Prof. P.K. Tseng - behind)

Other adventures in Taiwan included:

- The tailor shop is in the Grand Hotel. Matt Allen came late after lunch, said he had been to "his tailor". We all got interested – at that time, one could get a hand-made fitted suit very cheaply – got two, including the one in the photo above.
- Escaping from the Grand Hotel and exploring the street-food in the maze of small streets behind.
- Taiwan was a center for polished gem and other stones – had some moonstones made to special order, used in making my Scottish Sgian Dubh.
- Hsinchu Cheng Huang (City God) Temple. Wonderful, crazy, completed in 1748. Many street-food stands at the Night Market, and an antique movie projector with carbon-arc light, would smoke like a steam engine, and shake, rattle and roll as hilarious old kung-fu movies were shown during the evening. The Taiwanese kung-fu movies were somewhat different from the Japanese – sometimes very humorous. Zombies – kept calm when a note was pasted on their foreheads, went crazy when the note was removed. Never have forgotten a scene where the hero chased one, which turned itself into a skeleton and climbed a ladder – and p _ _ _ _ _ on the hero when he climbed up after…!
- Prof. Tseng grew up in Hsinchu near the temple, so could tell many interesting stories. Invited me and Prof. Liu to a very old Chinese tea house, and introduced me to the Chinese ceremonial tea drinking ritual. A complicated highly carved huge tree stump table, with places for the utensils and channel for used water to run in river channels, lakes, and finally underneath to a receptacle. Hot water in one pitcher, then the first teapot for quick first brew, then next teapots for subsequent brews, beautiful cups.
- Prof. Liu invited me for a tourist day around Taipei and Keelung with his niece – many very interesting sights. Later also met Prof. Liu in Japan.
- My best friend Prof. Hidekuni Takekoshi was born and grew up in Taiwan, and had many old friends there, including Prof. Tseng. He came to Taiwan twice to meet me in conjunction with a TRC meeting. The first time we stayed in Taipei during the meeting days – visited with Prof. Tseng a great piano bar with a big white grand, and Prof. Tseng requested several beautiful old folk songs; also very interesting meetings and adventures with other of Takekoshi's old friends. After the meeting, he and Prof. Tseng and I traveled to Tainan at the south end of the island, stopping on the way at the Sun

Moon Lake. The second time Takekoshi and I went to the Taroko National Park area, and met an old friend of his (Takekoshi was nearly 70 then). The friend insisted that we hike up a nearby mountain with him – it was hot and we were wearing city suits and street shoes, it was very hard. The friend did it almost every day "to be fit", and was – kept urging us on, saying that we had to meet the return bus. Finally he realized there was no way – there was a flight of stairs straight down from the top, he ran down and got the bus to wait until we finally made it down.

[toc]

## Benedict Nuclear Pharmaceuticals, Inc

Malcolm Benedict was a real character (aren't we all …). He was President of a small firm in Boulder, Colorado that obtained irradiated water containing radioisotopes from various government facilities, made pills, and transported the pills to various hospitals. Some were short-lived and had to be delivered quickly, so he had a helicopter service organized. This was a very lucrative business, and he wanted to expand by having his own accelerator and radiation facility.

Somehow he came in contact with Jim Stovall, and then I met him. His first ebullient observation was that I was "his cousin". More on that later. The second, in the same manner, was that I was required to build his accelerator. That was not so straight-forward, and it took a long time to get him to settle down to discuss possible steps – a long time because he insisted on telling many stories about himself in the meantime. But finally, a plan to approach the DOE was outlined. It was an interesting proposition and something out of the daily routine, so interesting. Finally I did succeed to get a contract negotiated – two linacs would be built – one for him and one for us as a research tool. (Then as now, it was very difficult to get pure research money for linac technology).

But although he was very rich at the time, he still had to get financing. So he invited me as his "consultant" to seek it, and we made two very memorable trips.

We went to visit Curt Mileikowski, founder of Scandatronix and very influential in Sweden, in Stockholm (see Section on ATW). Malcolm wanted to impress, and immediately declared that he wanted to host for dinner, at the very best eatery in Stockholm. Curt calmly said that he would be the host on the first evening – at a fancy place near the drama theater, was fantastic and the most expensive I ever experienced. On the second evening, after another huge lunch, to the Royal Opera house, even more expensive and really gourmet – Malcolm successfully wrestled for the bill. But Malcolm could never settle down and talk business, kept telling stories, although his Veep Dave Allen was very focused and kept trying to channel the discussion, and the Swedish bankers that Curt had invited were not interested.

Afterward, Malcolm took me with him to Scotland, and I learned with great interest as to why I was "his cousin". He said he always said that as a warm-up to people, but in my case it was really true, because his family's original name was Williamson and we were both of the Clan Gunn, in which the Septs (the next level down) have family names ending in -son. He had grown up with his uncle, then the custodian of the Gunn Clan, in the far north of Scotland. His uncle was only custodian because there was no chief – the lineage having been lost during the bad times of "The Clearings" when the Scottish people were driven out by the English who wanted the land for sheep raising. The Gunn chief had retaliated by killing a tax-collector, had to disappear, and the family name was changed to Benedict. Records were lost, and there was a current process to see if a new chief could be established (still not settled as of 2017). Later I learned many more details and had more interest in the family history, visited the Court of the Lord Lyon of Arms in Edinburgh, bought a full kilt with jacket, and made a belt buckle and Sgian Dubh.

We went to a large bagpipe band festival at Stirling Castle. Malcolm had been a Junior Champion bagpipe player when young. Typical Malcolm story – a brewery wagon was there with a span of beautiful and beautifully outfitted Clydesdales. Malcolm asked how much the horses cost, he would buy them. This was always his approach – he was rich then and wanted to show it. Of course the answer was not for sale. Next question immediately – how much for the trappings. Same answer,

Next question – the horses manes and tails had been braided with ribbons – how much for the ribbons. By now the Scots, somewhat known for a certain dourness, were no longer amused.

Then we went to Glasgow, where Malcolm wanted to fete his old bagpipe teacher – at the "best restaurant in Glasgow", which turned out to be the Holiday Inn, which seemed somewhat incredible but was then the case. The teacher and his wife came – very short and small, very much Jack Spratt and his wife. The teacher had been in the charge brigade of Field Marshall Montgomery during WWII and survived. The first course was a very large bowl of mussels – we all had an initial portion, and then the bowl was finished by the other two. Several more courses the same. Finally time for dessert. Malcolm and I declined, but the two enjoyed very much. It was a real pleasure to see them enjoying themselves so much!

Later we went to Japan. Malcolm and Paul stayed at one of Tokyo's fanciest hotels, I stayed at Takekoshi's family house in Tabata, where I had my own key for many years. We met with people from Sumitomo, Japanese banks, but with the same style and result. Although warned about the dives in Shinjuku, they went one night, to second floor, and only Paul's bulk as previous football player saved them as they forced their way back out to the street after being presented with the bill.

Malcolm alternated with being very rich and bankrupt. In a procedure involving corrupt actions of the FDA trying to put him out of business, he won in court, but the court warned him that if any other matter ever came up, he would go straight to jail. They never got an accelerator, he is still active in his company, have not had contact with him since those days…

[toc]

## SLAC 1988

Before leaving for a year in Japan, I worked at SLAC from January to March 1988. Over the years from the initial stages of LAMPF, I had benefitted a lot from interactions with SLAC. One close friend was Martin J. Lee, with whom I had collaborated on intelligent correction of beam line errors. The "experts" gathered data from all instrumentation (such as beam position monitors) at once and attempted to find correcting information to the beam control devices (such as beam steerers) by inverting the data matrix. This resulted in corrections to all of the control devices. In contrast, in the early-1980's the subject of beam-based control was emerging, and new tools were contemplated from new fields such as "artificial intelligence (AI)" and "expert systems". I dispatched S.H. Clearwater from the Los Alamos AT-Division as a Post-Doc to SLAC to work with Martin, and a full system of programs (COMFORT, ABLE) was developed and implemented on the SLC control system, which could efficiently identify where a beam orbit error, or beam instrumentation error, actually occurred and then made local corrections. This work was then in its later stages, I became "portable" with my own Macintosh desktop computer, enjoyed very much being immersed in the SLAC environment, went to the morning debriefing meetings on the SLAC Linear Collider. The experienced reader will wonder how it turned out, and will with understanding nod to learn that the "matrix-inversion" school nevertheless won, and the AI work was forgotten…

The other good interaction was with Burton Richter (see anecdotes about the APT and SDI programs) – my parting discussion with him pointed out that the alignment problem bedeviling them at the time could be addressed with our work, and by closed loop automatic control, recommending that he hire a an automatic control engineer. They did not do that, but T.H. Himel, a talented multidisciplinary guy whose degree happed to be in "physics", implemented automatic control very well.

I learned a great deal extracurricularly from Martin as well. Unforgettable meals in the San Francisco Chinatown, where he would disappear into the kitchen to order directly from the cooks. He was also a master of both East and West – a western PhD in Physics, and an eastern Tai Chi master. Once he took me to his teacher's studio in Chinatown. The master was there and politely conversed, for any demonstration my status was of course too low, but they had arranged for the first disciple to be there, and he performed the whole traditional sequence. He was also a master in two different schools of Chinese art, and had brought two large portfolios of beautiful work. That evening was a real honor.

Later Martin was able to get emigration papers for an old master still living in China, who he brought to live with him, and I was their guinea pig for a few experiments about the real powers of Tai Chi.

[toc]

## Japan

First Visit, Prof. Y. Hirao

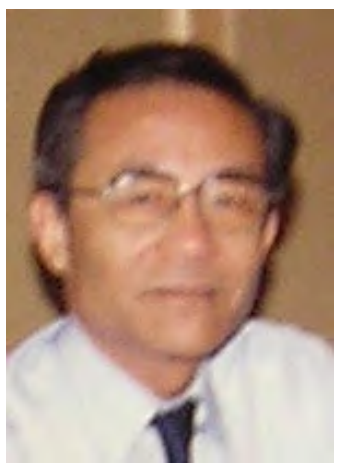

The first Japan visit was in 1980. The Pion Generator for Medical Irradiation (PIGMI) program and the RFQ had attracted wide attention, and a circle of talks at a number of Institutes was arranged.

In those days, the Japanese considered it impossible for a foreigner to navigate the public transportation system from Narita airport to some destination, and would send someone to meet a visitor. I was kindly met by Prof. Y. Hirao, Director of the Institute for Nuclear Science (INS) in Tokyo. He was a passionate automobilist, drove from his home to work and back every day, and had driven to the airport. His first question to any visitor was "How many traffic lights are there between your home and work?" He knew he could trump anyone – he had ~140. But he was still impressed when I told him that I had none, over a distance of about 20 km. We subsequently became close and shared many experiences. Hirao's goal was to build a new big machine, at first for physics. INS finally merged with KEK and the Tokyo site was sold for a lot of money. Hirao became the founding Director of the National Institute for Radiological Sciences (NIRS) in Chiba, and built the large double synchrotron cancer therapy facility, which uses C4+ and developed one of the early operating APF IH structures – later the NIRS group very openly provided much information that helped with the development of the general practical APF design method. We were friends for the rest of his life.

Prof. Hidekuni Takekoshi

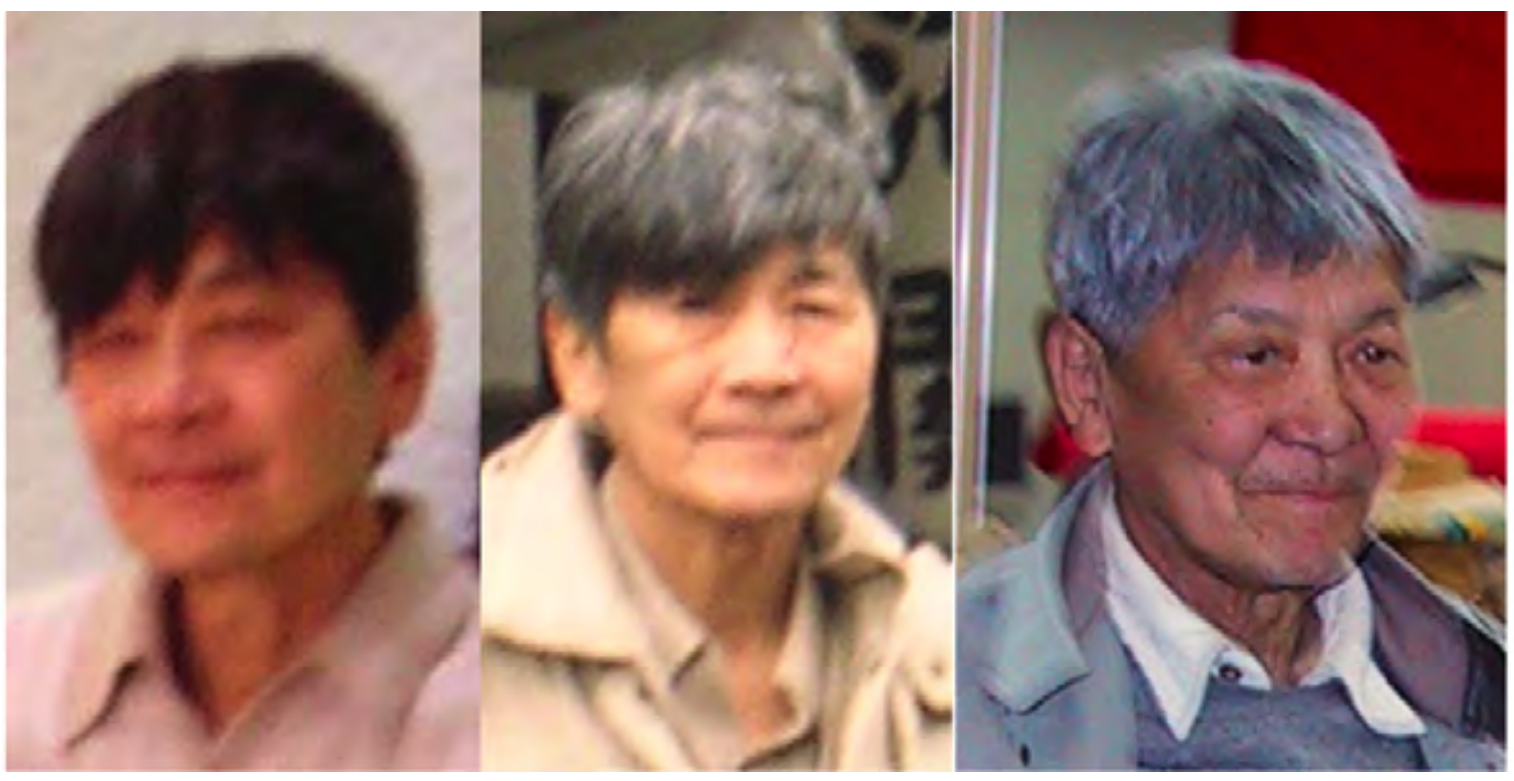

In Kyoto at the Keage Institute of the Institute for Chemical Research, Kyoto University, I first met Prof. Hidekuni Takekoshi. I will abbreviate his name as "T". He was very interested in PIGMI, also enjoyed hosting foreign visitors, and introduced many things even on this first meeting. He visited Los Alamos in 1981, and also sent Y. Iwashita, his student in those days, for three years to work with us in AT Division. It was already clear that we were to be friends, and for my 1981 return visit to Japan, for which I requested that he arrange a Japanese-style accommodation, he replied that I would stay "at his house". Remarkable, as that would be very unusual anywhere, but perhaps especially in Japan. So I traveled without accommodation reservation.

“His house” was two houses – one in Tokyo, one in Kyoto. His father had established a textile factory in Formosa, ruled by Japan for ~50 years to the end of WWII, and had managed to return to Japan at wartime with most of his estate, so was able to built two “rich man’s houses” in a small walled garden in north-central Tokyo. The mother and an unmarried sister still lived in one house, and the other had been for use of the four brothers at various times, but now mostly empty and available to T when he was in Tokyo. Later I had my own key to this house for many years – that was quite something – to be able to go to Tokyo at any time and stay at “my own house”! In Kyoto, T had cleverly invested in a small piece of land in a development planned to be near a new “Kansai Science City” corresponding to the northern Kanto area Tsukuba, where KEK is located. We also started “going around” to beautiful tourist areas of Japan.

We were then, and remained, best friends. His friendship has been a major influence on my life – a true “companion along the way”.

Since 1980, I have averaged one trip per year to Japan, mostly for at least one month, often for several months, once for a year, and explored with T almost every corner of the country. Words were not so many, although his English is quite ok – in Japan it is necessary to be able to learn through feelings. It is necessary to at least learn the two supplementary phonetic tables – katakana and hiragana – they are used for almost all foreign words and it helps very much to know it. The Chinese characters, reading, real Japanese conversation would have required too much time then, with so much to do, computing, etc. So one could not be pestering continuously with requests for translation. And he indicated that very soon, when I asked “What does “chotto-mate” mean?” after hearing it all the time. He replied ”You will figure it out…” So I grasped completely that feeling was the key, and figured out eventually that “chotto-mate” meant “Wait a moment…”

The early 1980’s involved PIGMI – many cities wanted one – and the RFQ. A prominent doctor in Hamamatsu was a very active promoter. He once showed me his very expensive chronometer, said he had a heart condition, and it was very useful to have such a good watch because he could tell if he would have to hurry to meet a train – Japanese trains then and now are punctual almost to the second!

After the collapse of LANL in 1986 and my decision to do something else at the end of 1987, I informed T, without making any requests. He immediately went to work, and although the Japanese fiscal year was already half over, he managed an invitation from KEK for a year’s visit. I informed KEK that I was very glad for their invitation, but that it was necessary for me to spend at least half of the year with T in Kyoto. They had no objection at all. The old Keage Institute had become a very special place for me. It had been built in the Meiji Era, when Japan was opened to the West, as the first hydroelectric power plant in Japan, at the end of a long tunnel/canal from Biwa Lake (Biwako), in the Meiji red-brick style, with very thick walls. After WWII and the infamous incident, when the occupation forces threw the original Kyoto cyclotron and other small betatron accelerators into Tokyo Bay to prevent further nuclear researches, the Keage hydroelectric plant was replaced by a new one next door, and the old building was ideal for the building of a new cyclotron. T was a student and fortunate to be in Kyoto during the war, and helped finish the cyclotron, which became one of the longest running cyclotrons in the world. He then went to the Japan Atomic Energy Research Institute (JAERI) (later JAEA) in northern Japan, of which more later, and extended the electron linac from 20 MeV to 120 MeV. From there he returned as Professor of the Keage Institute. Working in the Keage Building was so peaceful and wonderful, it was the feeling of being a monk in a temple. I was determined to spend a longer period there, and 1988 was the last chance, as a new building on the KU ICR Uji campus outside Kyoto would be ready in 1989. Among many, it was a major reason why, after many years in the saddle holding the reins, knowing what staying too long could mean, it was so easy to give up idiotic demands and meetings at LANL, and I would have set aside time in 1988 in any case and under any circumstance.

The first six months were spent at KEK, working on investigation of APF effects that might be introduced by systematic error patterns in long linacs like LAMPF, and it was found that they could be significant. The Russians were working with their Moscow Meson Factory and were surprised and happy to find that some not understood effects that they had observed could be explained in this way.

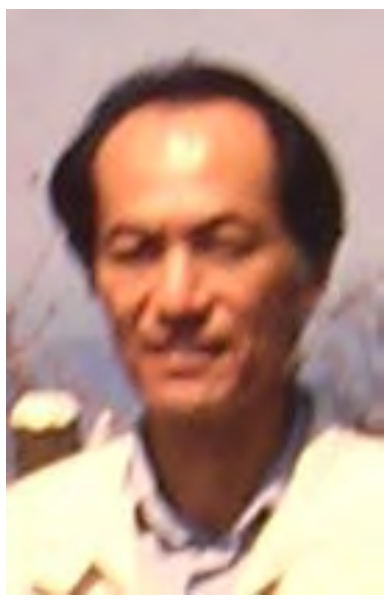

M. Kihara, my host at KEK. The host was required to stand behind my bank account.

During the six months at Kyoto, long linac investigations continued and also work with then-student Hiromi Okamoto on APF schemes, which he expanded and wrote a paper which I and others found very useful, even many years later, as related in the APF chapters.

Explorations of Japan with T then really began, and the many adventures are related in my travel logs and diaries, and documented with many photos. Every visit included, as first priority, time with him. He retired from Kyoto University in March 1990, and went to work in Hiroshima, where I visited him every year, and then he moved back to Kyoto. I was able to visit him almost every year up to April/ May 2019. The many adventures are related in my travel logs and diaries, and documented with many photos. As also noticed by others [73], his driving was remarkable – always with one hand. Driving in reverse was not part of his repertoire. If missing a turn on a major road and wanting to reach an opposite corner, always use small roads and "There will be a way". Once that lead to smaller and smaller roads and finally into a path in an orchard, with nowhere to turn around. I had to drive in reverse, up hill around curves, thru a small group of houses, for about one kilometer - the only casualty was one flower pot. We always made it, and were happy. My friend Hidekuni Takekoshi passed away on 11 January 2020.

> **See in addition: "In Memory of Prof. Hidekuni Takekoshi** 10/26/1926–1/11/2020", Posted to ResearchGate 8/7/2020 (search title); Posted to ArXiv arXiv:2008.02718 http://arxiv.org/abs/2008.02718

Further with Y. Iwashita at the ICR Accelerator Institute

---

73 "In Memory of Prof. Hidekuni Takekoshi 10/26/2020–1/11/2020", Y. Iwashita, M. Mizumoto, T. Igaki, H. Okamoto, R.A. Jameson, ResearchGate, June 2020.

[toc]

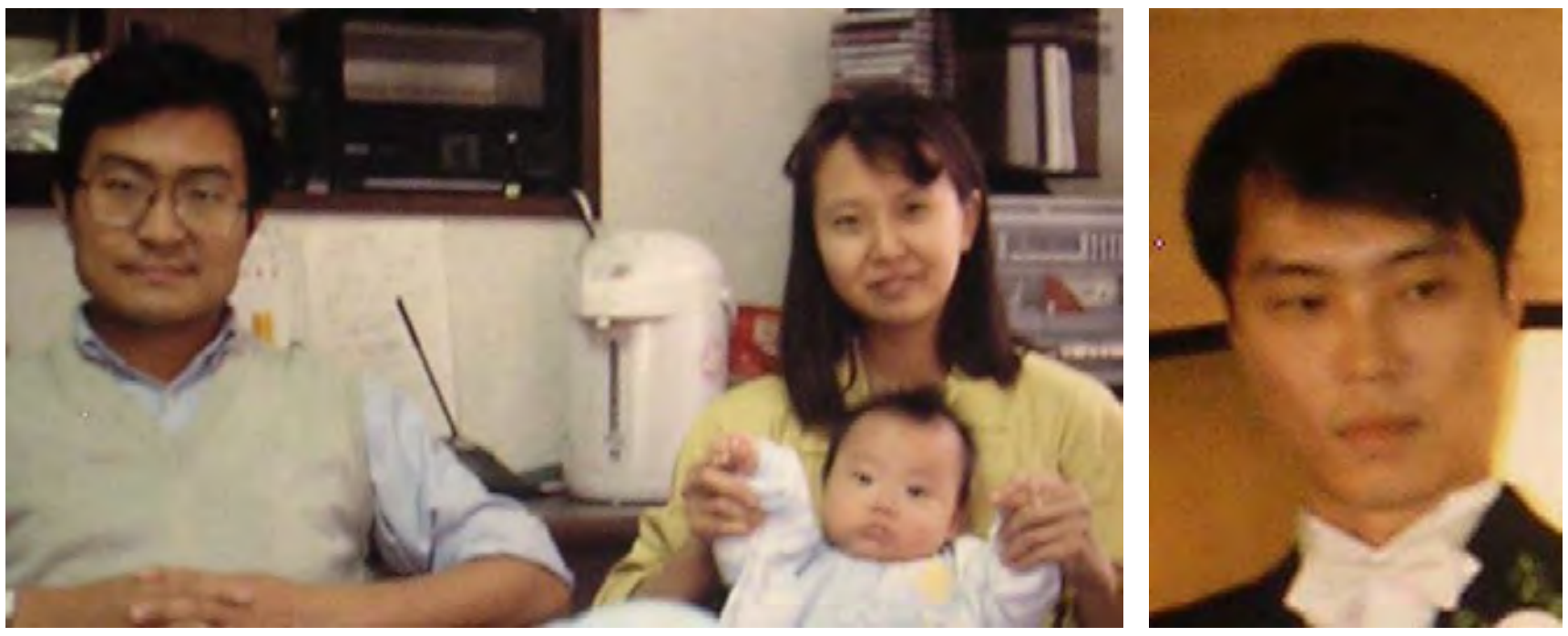

Iwashita-san, wife Junko, and oldest son in 1992 – the parents have hardly aged in appearance since!!, H. Okamoto

Iwashita worked on development of advanced electromagnetic codes (Superfish->Ultrafish, which later turned into collaboration at Los Alamos with Thomas Weiland on Mafia) during his stay at Los Alamos, and since has been an expert with electromagnetic fields, especially permanent magnets, and has developed many clever applications. After T retired, the Accelerator Institute continued to flourish under Prof. M. Inoue, who then went as Director to the KU Reactor Research Institute (KURRI). Subsequently it has become more and more difficult to survive and find students, partly because of the embedment in a chemical organization, and when Iwashita-san retires in 2020, the survival of T's legacy may become problematical. We have stayed in contact through the years, and I did the English editing on many papers. From 2015, we collaborated closely on software development for arbitrary RFQ vane shapes, one of which is Iwashita's invention, and on a proposal for a small neutron production facility at the Institute. The yearly visits to the new building were also very enjoyable, working in the pleasant large room rimmed by the coffee apparatus, frig, sink, etc. The campus cafeteria was quite good, stayed once for a month in Uji, and on or near the campus, Iwashita lives nearby, trains to Kyoto or to T in Shin-Tanabe convenient. The Shin-Tanabe journey from Obaku involves a 10-minute walk to change lines at the mid-point. As T (and I) became older and the "going around" adventures too strenuous, it was a great pleasure to make the journey every day to meet for lunch.

Sumitomo Company Consultant

For the April 1988–March 1989 KEK/Kyoto year, I arranged two further activities, for wider exposure and not to be confined in case one activity didn't pan out. One was as consultant to the Sumitomo company, who had built the first RFQs in Japan, were then building a microtron, which had many turns and was having difficulty controlling the orbit. Their facility in Tokyo had been a munitions factory during WWII, and was across the fence from INS – I learned the secret place where one could climb over the fence to avoid the very long way out to the street and back in.

They were running at night, which afforded travel in Tokyo at times different from the normal work day. One night, a small shop that was always shuttered during the day was open, with a dim light far at the back – a kind of old furniture shop. Entered and found a man restoring an antique kiri chest in the back room – and with essentially no common language except the craft of woodwork, succeeded to be taught how to correctly restore this special wood. Kiri a light colored wood, almost as light as balsa, and breathes; chests are made so precisely that in the damp seasons, the drawers swell and valuable kimonos inside remain dry, in the dry seasons the drawers contract and some air can circulate, keeping the kimonos fresh. When new, no finish at all is applied to insure this feature, and antique kiri tansu from old houses with open firepits become very dark from smoke and exposure. A special abrasive powder and another made from deer antlers can be used to remove the soot and get back to the original color, and then a special very hard white wax, ~ "kiriro", very hard to pronounce, can be lightly applied. Later I found these, through long searches in obscure parts of Tokyo, and causing astonishment that some foreigner (gaijin) would ask for such.

We did orbit adjustment analysis and simulation, but in the end, they adopted the elegant solution – just removed all the steering magnets (there were several for each orbit), and it then transmitted very well…

The Group Leader, Toyoto-san, was a great fan of Ed McBain mysteries. McBain wrote very many, and it was fun to rummage in old book stores in America to find many to bring to Toyota-san.

Japan Atomic Energy Research Institute (JAERI) and J-Parc

The fourth activity for 1988-1989 was to become familiar with the program for atomic waste transmutation at JAERI (later JAEA). The possibility to use a high intensity accelerator to drive a non-critical nuclear reactor, in which radioactive waste could be burned, was an early idea proposed first at the Chalk River laboratory in Canada. About every five years the subject would be re-discussed to determine if accelerator and other technology were ready, and by 1988, we felt that such accelerators were completely feasible. The Japanese had started a development program, and I arranged to meet them and learn about it.

JAERI was the institute that developed the first nuclear power plants in Japan. It is located in Tokai-mura, about 100 km from Tokyo (halfway to the now well-known Fukushima)

In ~1990, during the annual visit to T, he told me that I had a new job. JAERI was founding a new accelerator group for ADT (Accelerator-Driven-Transmutation), the Group Leader Mizumoto had been a member of T's linac group at JAERI, a nuclear physicist using the electron linac but overnight ordered to become an proton accelerator expert. And I was to be his mentor. This was a very good job. JAERI contracted with me through LANL and provided funds that supported me and also were distributed by me in the LANL group to those who helped produce a succession of lengthy reports on all aspects of linac technology, and especially on my work on linac design. The Japanese grasped the advantages of a beam equilibrium, and the resulting linac was the first, and more or less still the only, long linac in the world with a fully equipartitioned design. It also has fully adjustable quadrupoles.

The ADT project was merged with a long-sought project that Hirao had wanted for INS in Tokyo, later merged with KEK, for a spallation neutron source, and the J-Parc project was located at Tokai on the JAERI (JAEA) site. The ADT goal became low priority (and still awaits completion of the linac to ≥600 MeV), but the linac design remained, and it operates well as the J-Parc injector.

The collaboration remained active until October 2017. After a gap of some years while the linac was constructed and commissioned, a visit was proposed in 2007 and I was delighted to learn that the coordinator would be Masanori Ikegami. We had worked together in 1998 when he was still a student, on an advanced and complicated investigation of applying a "Finsler manifold" to accelerator beam dynamics, aimed at determining whether this approach would yield information beyond our usual tools, finally deciding that our tools (not known in the "manifold community") were sufficient. I had lost touch with him, but now came the chance for a long and productive collaboration that lasted until he decided to join the FRIB program at MSU in the US. After this, the collaboration continued with Yong Liu, with whom I worked at IAP in Lanzhou and helped transfer to J-Parc, T. Koseki, who I met first when working at RIKEN, F. Naitoh, who I met his first year after University when he joined KEK in 1988, and K. Hasegawa, who was a student in Mizumoto's original linac group and did the calculations for the EP design.

From ~1990–2019, many weeks were spent at Tokai-mura. The JAERI site is on the east side of the village at the Pacific coast, 4.5 km from the train station, with essentially no convenient connection to this day, so very isolated. In the early 1990's, I was told to stay in a "business hotel" in Katsuka, two stations toward Tokyo from Tokai. "Because there are restaurants around the hotel there" – there were very few in Tokai, and the canteen was not good. So it was very inconvenient, wasting very much time – it took 45 minutes to walk from the station, could use a bicycle if it did not rain, taxi very expensive, etc. I kept asking if there might be a Japanese style accommodation, but apparently that was considered not suitable. Finally in 1998 a German student was also visiting and we talked about accommodations and the wasted time. He asked why I did not stay at a ryokan (traditional Japanese hotel) and I replied that they are usually very expensive. He said it was sometimes possible to negotiate a "long-term stay" rate. I asked where are the ryokans around here, and he replied that just at the south end of the J-Parc site is a famous old Tokugawa-era temple and shrine, with ryokans on the entrance street. I immediately organized an expedition that noon with him, Hasegawa, and a couple of young Japanese to investigate. There are four ryokans. The first would not accommodate foreigners – not at all surprising as this was well known, as we "gaijin" would not know the somewhat complicated rituals involved. The second was willing, was an attractive place with an interesting rate. The third was a surprise – a solidly built lady came out to the street, with a huge smile on her face, and sprudeled Japanese words at an astonishing rate (I always say "10,000 words a minute") – and I could understand her completely! She was saying "welcome, please stay with me!!". I told my companions that there was no need to visit the fourth ryokan, and then looked at the elegantly outfitted Numataya ryokan and discussed the rate, which included breakfast and dinner, with "Midori-san". At that time, her older daughter had left for university, the younger daughter, Kana, was in junior high school, the grandfather was there. Now during 20 years and many, many nights later, I am considered as a family member. I can rummage for as many futons and thin foam mattresses as required, take the only two Western chairs from the dining room up to the room and requisition an extra low table to make a stack of two in order to be able to sit comfortably, arrange the signs at the entrance to the o-furo for privacy (there are two baths, each with a large stainless steel pool), find a yukata (and haori in winter), they pick us up and take us to the train station, etc. The meals were and are amazing – knowing what they would cost in a restaurant, it is still hard to imagine that there could be any profit left. Midori and her husband are dear friends, and still active although we are all much older, and at some point a transition will be necessary. In autumn 2017, Kana was present to help whenever there were crowds (sometimes up to 80 students participating in sports events at a a large sports park nearby), and Midori always had access to neighborhood women to hire for help in the kitchen, all the room service. Still doing well in 2022.

On weekends during those many years, I would go to Tokyo for two days; Takekoshi had given me keys to a second house that his father had built for the children to use, in a beautiful small compound in Tabata – then essentially unused. It was a great honor – I could escape Tokai, where there was little to do, go Scottish dancing on Friday (and occasionally Saturday) evenings, and visit flea and antique markets on Saturday and Sunday, searching for mingei omiyage, as outlined below.

A special connection began in 1988, when Takekoshi drove the side roads during Golden Week (to avoid traffic jams on the expressways) to northern Honshu, and returning, stopped at a small tea shop, named Azuma, in the town of Hitachi-Ota, not far from Tokai-mura. The owner was a small, pert lady named Ryuko – Ryuko-chan (chan -> little) – charming, served coffee, whiskey, cooked up various snacks. It was very pleasant, and I knew it would be nice to go back – but how to find? Noticed that we had used a very unusual pattern of one-way streets, and later was able to take the train there and found it again. After that, visited many times, and got to know a lot of people from all walks of life – a stone shop owner, an owner of a chain of shoe stores (until the depression hit), teachers, town personalities (it is a popular place), sake shop owner, a high steel building skeleton worker, others. Ryuko and Takekoshi conjured up a trip to the Kabuki Theater in Tokyo. Her parents had run a geisha house from their old house still behind the tea shop – where the geishas actually lived, and went evenings to the entertainment restaurant near by. A lot of old things were still in the house, and Ryuko would "clean out" by giving something on many occasions – a wooden geisha pillow, an article of clothing, occasionally a small figure. Once they made a fun karaoke party, which lasted late and I got to sleep in the old house. To 2021, she seemed hardly to age at all.

Yasuhiro Kondo was a student in Mizumoto's group, and I helped him with his PhD work. He has guided four RFQs through beam dynamics, cavity design, construction, and commissioning. In 2020, it was pleasing to be invited as a co-author of a paper on a fully equipartitioned RFQ successfully

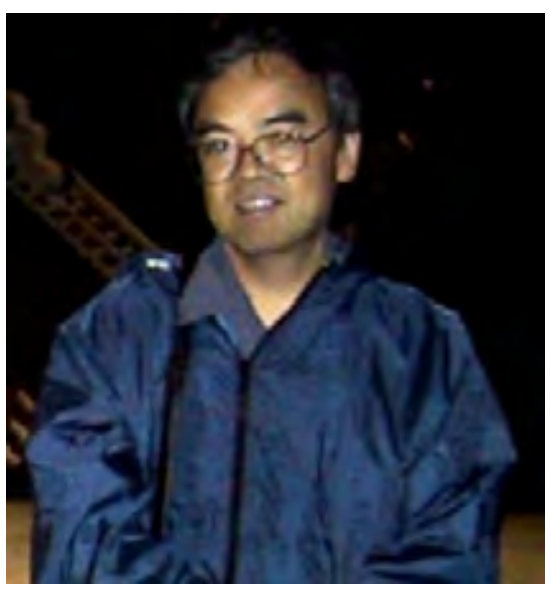
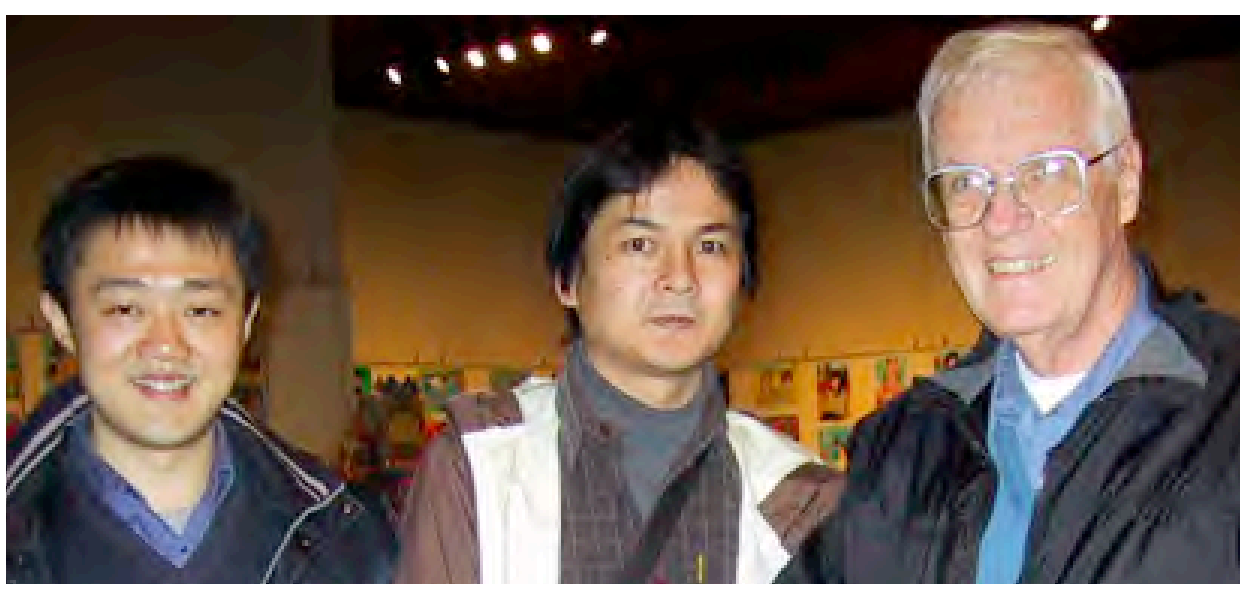
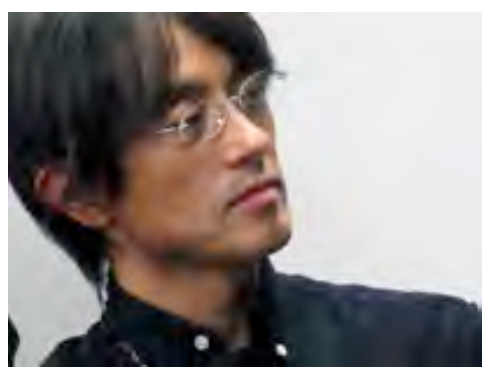

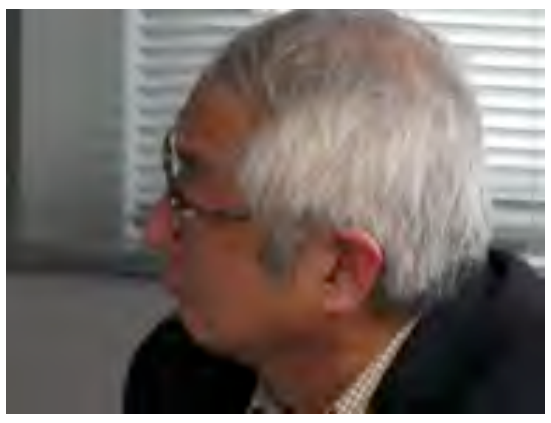
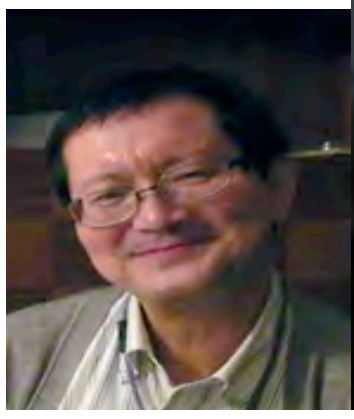
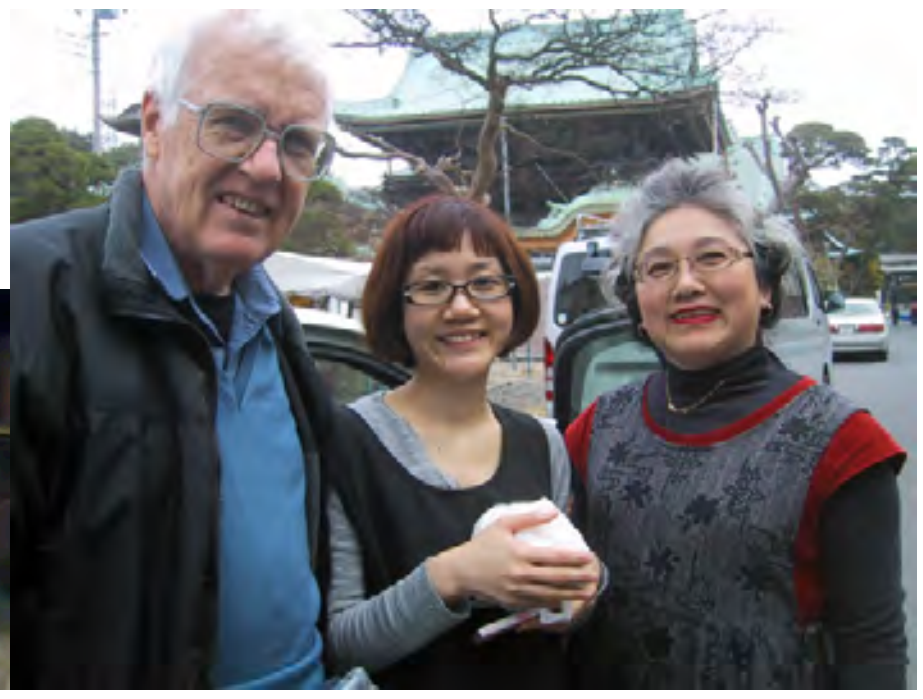
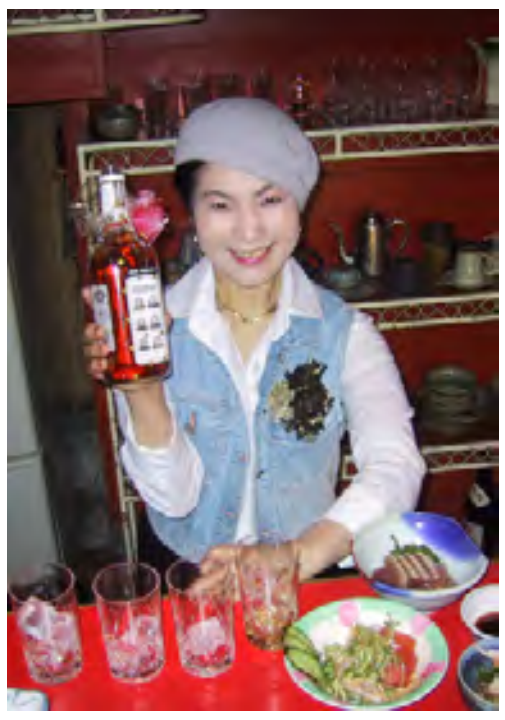

Motoharu Mizumoto, Yong Liu & Masanori Ikegami, T. Koseki, F. Naitoh, K. Hasegawa, Kana and Midori of Numataya, Ryuko-chan

commissioned and performing as designed, and with very helpful and favorable reviews, summarized by comments that are important to include here:

> "The RFQ designed with the fully beam-oriented is a world's first one, and the verification was conducted through the beam commissioning. This paper is important to the field of the high intensity RFQs because the design scheme can be standard."
> -----------
> "This paper deals with the design and experiment of an RFQ based on equipartition principle. This is a rare paper that covers the issue from the design of the RFQ to the beam experiment. This is an excellent work."

In 2021, at the request of Kondo-san, came the pleasure of working with an enthusiastic young colleague Bruce Yee-Rendon[74] on an updated high-availability EP RFQ for the JAEA ADS project. The specification is interesting, with relatively low H+ current of 20-25mA, very low KP factor ~1.0-1.2, and 2-term vane modulation, very low longitudinal rms output emittance, with EP and ratios = 1.25. Meeting the spec was not straight-forward, and resulted in a breakthrough method for a new vane profile to control the bunching and longitudinal emittance. (Ch.29, "Elements …").

Yasuhiro Kondo 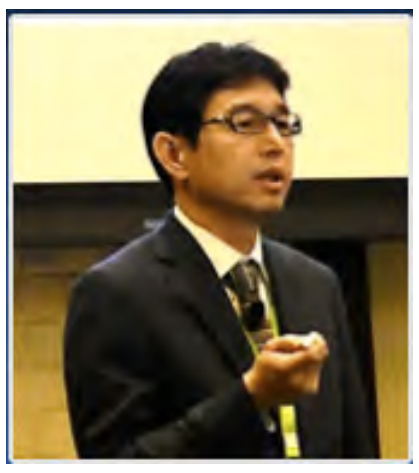 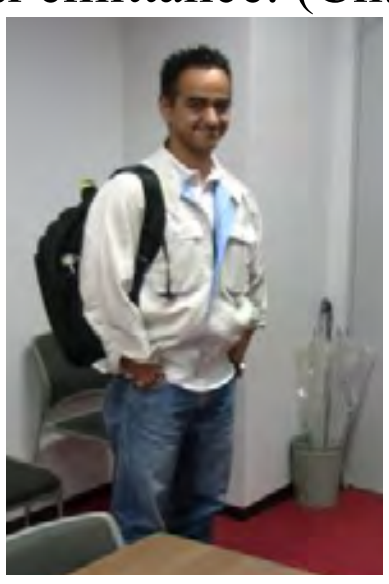 Bruce Yee-Rendon

<u>RIKEN</u>

RIKEN, the Institute of Physical and Chemical Research, was formally founded on March 20, 1917 (by) leaders from various fields of research, modeled on the Kaiser Wilhelm Society, with an Imperial prince as President. Its history is very interesting to read on the web – featuring a very good director after WWI, a connection to industry and commercialization from the beginning, and maintaining to the present a very productive research environment.

Thus it was one stop on the initial lecture tour in 1980, where the original contact was Odera-san, who built a unique variable frequency RFQ, and hosted a first visit to Nikko.

In ~1994, I helped Yuri Batygin to join the RIKEN group of Takeshi Katayama. In 1998, Batygin organized a workshop there, which led to the award of Eminent Scientist, spread out over 2000 – 2001. Katayama was a very good Group Leader, who well utilized the Japanese tradition of group activities beyond the purely work-oriented, with frequent evening activities where an evening meal would be cooked in the group common room, with singing and instrument playing by various members, karaoke parties or eating in a restaurant. His secretary was Akemi Doi, who took very good care and became a lasting friend to the present, and who was of great help in explaining many points about the small mingei omiyage that I collected.

---

74 Bruce has become a real collaborator in LINACS development, after Johannes Maus.

The RIKEN President at this time was Dr. Shun-ichi Kobayashi, also an accomplished cello player, who joined one of Katayama's evening group cooking and played a duet with Katayama, also a dedicated cellist. This lead to an interesting collaboration. Adjoining the RIKEN campus was an empty field with only a tall antenna in the middle. RIKEN, including Katayama's group was working toward an extended accelerator facility, and the site was too small. I asked about this field, and learned that it belonged to the US military. Somehow it became of interest to inquire whether there might be any possibility for RIKEN to acquire it, and I organized an inquiry and expeditions to the Environment, Science & Technology Section of the US Embassy in Tokyo. It turned out that the field had been returned to Japan long ago, still unused, but would be the subject of much competition if the Japanese government would actually propose its release.

Acquaintance was made with Masahiro Okamura, inventor of direct injection into an RFQ from a laser ion source. This required extensive simulation and led to a strong collaboration from 2000 – 2007, when Okamura moved to BNL, with several PhD dissertations and strong development of my RFQ codes to be the first that could handle heavy ions and several inputs with different charge states and currents. We kept contact, and it has been a great pleasure to collaborate with him again in 2020, when he asked me to join in a project to make a constant r0 RFQ with 2-term vane modulation. It was during the corona pandemic – I was stuck in Frankfurt, he had to go back and forth from BNL to Japan and endure quarantines, etc.

Another valued acquaintance was Tamba-san, a technician who worked alone in a small building where a small workshop had been scouted out, for bicycle and souvenirs repairs. He would jog at noon, and invited us to bike alongside to visit a kamakuri park, small violet-like flowers that bloomed for a short time in a special spot. Walking back through a park from one of these excursions, a Japanese singing club was sitting under a blooming cherry (sakura) tree - singing from Schubert's "Winter Reise".

Those were fine years, with long visits to RIKEN and plenty of work in between. The RIKEN campus is very beautiful, with ancient large cherry trees making the spring sakura ohanami viewing very spectacular, and a large pond behind which our favorite lodging in the international visitors building was located.

Mainly through RIKEN came also many interactions with the Tokyo Institute of Technology group of T. Hattori, where Okamura had studied, and his successor N. Hayashizaki.. We were also privileged to join Hattori's social occasions, which included Akemi Doi and Hirao-san.

A special honor is that I am still allowed to be a Guest Scientist at RIKEN, although now with almost the sole purpose being to maintain my email account jameson@riken.jp (to 2021).

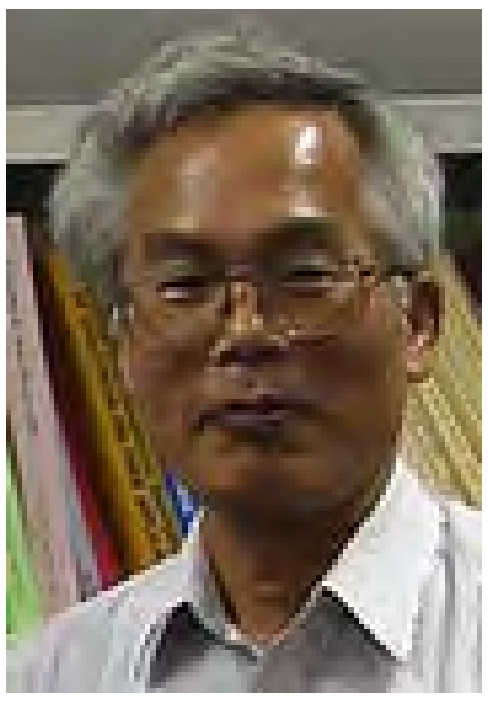 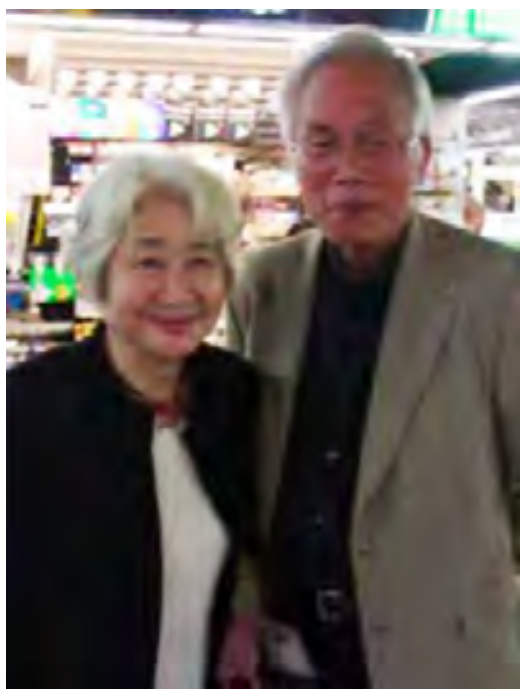 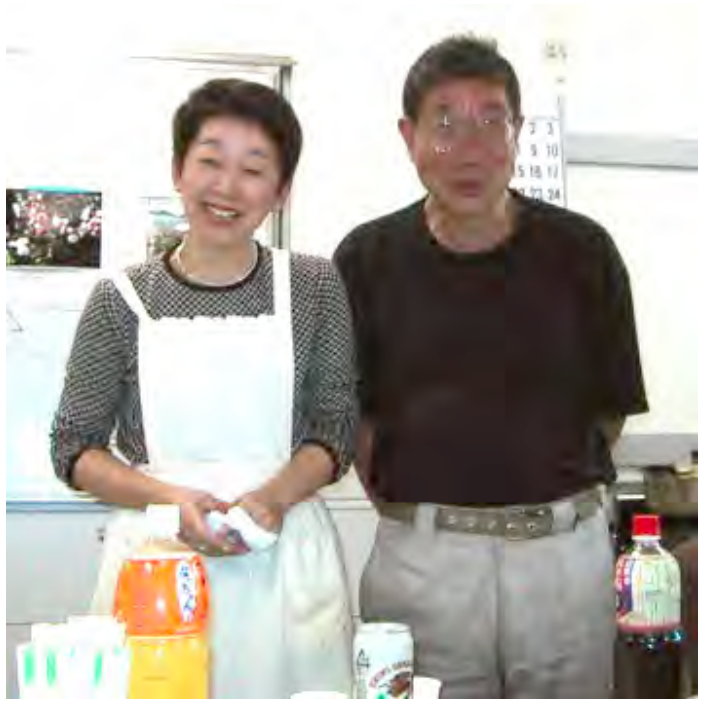

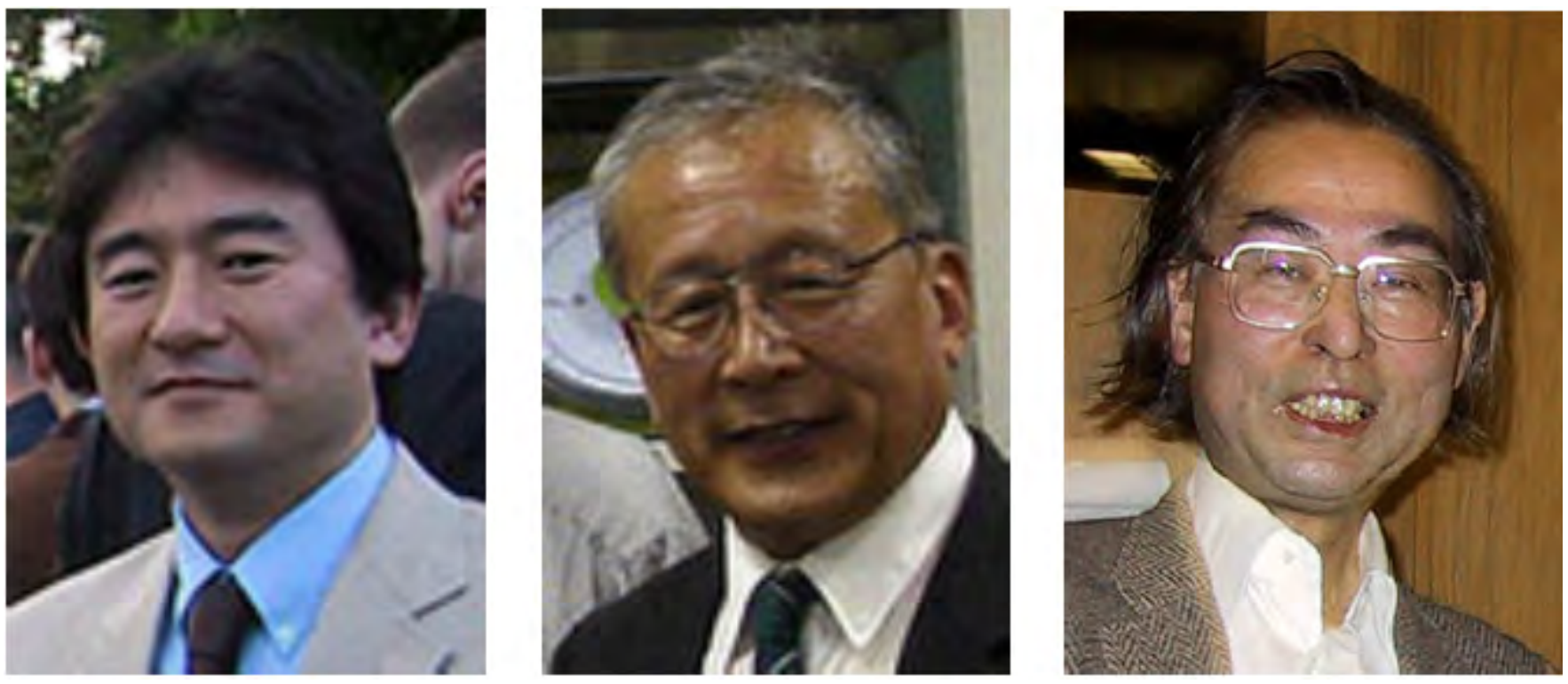

T. Katayama, Takeshi & Hatsume Katayama, Akemi Doi and Tamba-san,
Masahiro Okamura, RIKEN President Kobayashi, T. Hattori

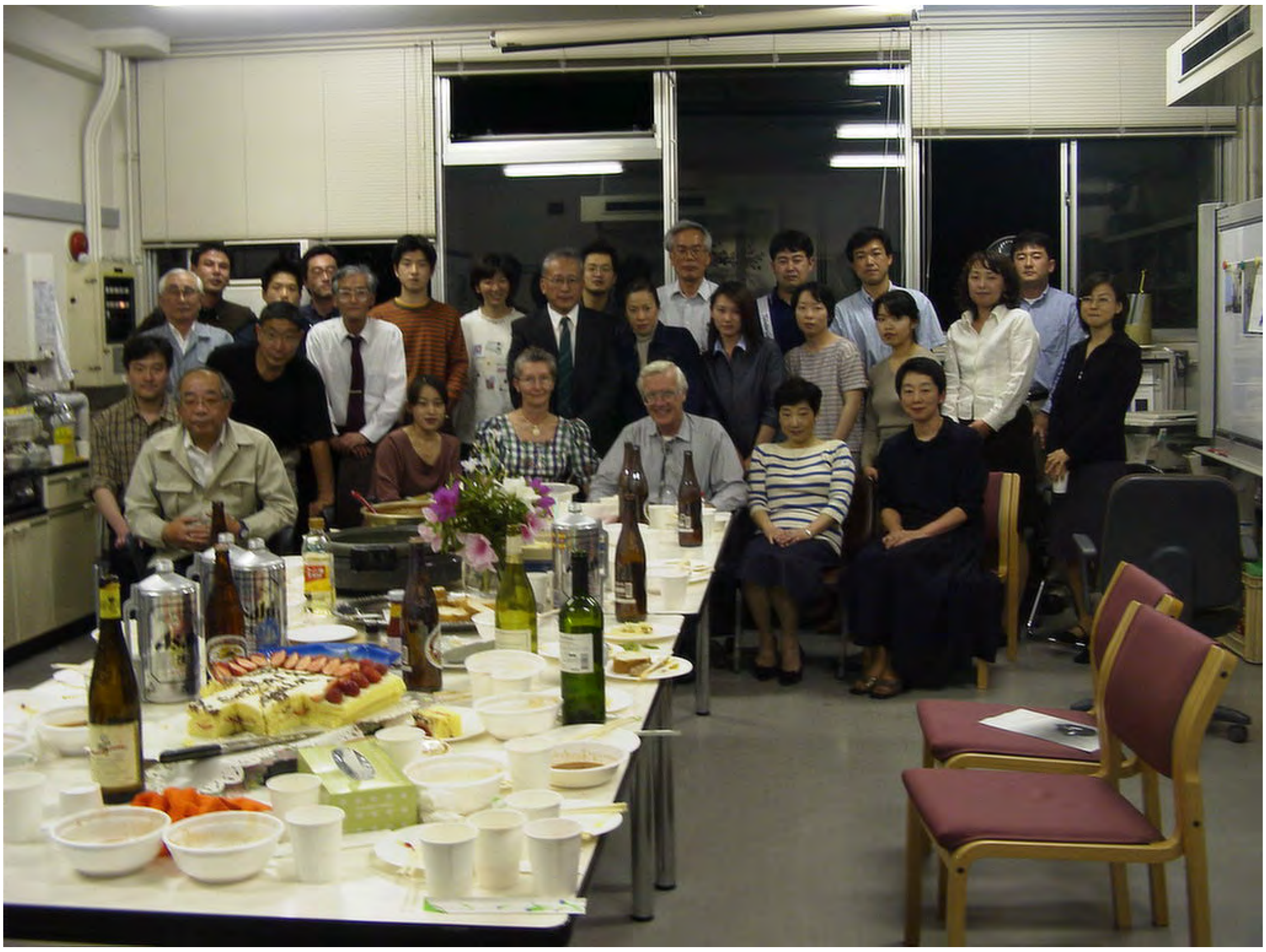

RIKEN Party April 19, 2001; Poldi Heinisch[75] to my right, RIKEN President Kobayashi behind Poldi, Katayama at rear behind me, Akemi Doi front row right, Okamura back right, Tamba left dark shirt. We presented pots of chile con carne and Frankfurt potato salad.

Mingei Omiyage

Japan, lying on the Ring of Fire, has many hot springs, and has developed a sublime bathing culture. On visits to onsen resorts and hotels, people would buy small souvenirs, which in the days before plastics, were handmade, with an enormous amount of artistic feeling, use of many materials, and covering a very wide range of subjects, from the kokeishi doll form to animals, mythological creatures, real creatures, ships, monks, folk figures, and many more. They would show up in flea and antique markets, which I would explore on many weekends for something to do where people were congregated. Starting on the first trip in 1980, I picked up a few, very cheaply, and then one day at a small antique shop in Jiyugaoka, a train intersection station outside Tokyo, I bought a pair of very old figures from Kyoto, for a much higher price – and had to admit that I had become a collector. I coined my name for them – mingei omiyage – folk-art souvenirs – an acceptable name in Japan.

[75] Poldi, my companion on almost every trip starting with this one; requirement from host was daily entry to institute, an extra desk space and a web capable computer.

During the years up to ~2010, the people who had bought before plastics were passing on, and as is usually the case, their children had no interest or room for them, so I discovered and rode a big wave, and assiduously sought and found very many specimens at flea markets, antique markets, and recycle shops. They were essentially being thrown away, 100¥ was the usual price, and sometimes I would get a whole sack full for maybe 300¥. It was also possible to be selective and take only the most interesting and best ones. Only since ~2010 did some dealers at antique markets tell me that they had finally realized that these very figures are also a part of the Japanese culture and should be preserved – and the prices at these markets has soared. The amount of life and feeling that these artists captured is amazing – they have a vibrancy that corresponding European figures (for example Christmas tree ornaments) are lacking.

Often figures would be broken or have missing parts, but would be exceptional, so I would buy them and learned to research and repair, so that the repair would be unnoticeable, or at least in the spirit as close as possible to what seemed to be the original Japanese intent. On a few occasions, I have been inspired to create a figure of my own, again hopefully in the Japanese spirit.

Now there are over 2000 of them, lining the walls of my small study in Frankfurt. Each has been photographed, researched and noted in book form. In 2006, a copy was presented to the RIKEN President, Prof. Noyori, Nobel Prize winner in Chemistry, with foreword and citation, here as updated in 2020:

MINGEI OMIYAGE
Collected by R. A. (Bob) Jameson

*Acknowledgements*

*I am most grateful to Prof. Hidekuni Takekoshi, my best friend over the 40 years of this collection. We traveled together all over Japan, to many markets, through life.*

*Collecting these small works of art over the years has been a pleasure in itself, but even more so that each afforded an opportunity for coffee with a friend, to find out what is written on them, where they are from, and the story behind each. These friends have taught me a lot about Japan, and to see them is the main reason for returning every year. The hospitality of a number of institutes is gratefully acknowledged: Kyoto University Institute of Chemical Research ICR, RIKEN, KEK, JAERI Tokai, Sumitomo Heavy Industries, and NIRS. For seven years, it was an especially great pleasure to spend several weeks each spring at RIKEN in Wakoshi., first in Prof. T. Katayama's group, and after his retirement, in Dr. M. Okamura's group. Many discussions were held with the group members. A very special thanks to Akemi Doi, who explained very many pieces, often found supplementary material on the Internet, and loaned brushes and provided paints and glue for restorations and repairs. Machiko Suenaga and Kazuko Takekoshi also explained many figures. Thanks also to Masako Okada-Naitoh and Naitoh-san - leaders of Scottish Country Dancing in Japan; Kiyoko Amemiya; Motoharu Mizumoto; Ryuko (Azuma), To-chan, Keiko and Masuyama-san in Hitachi-Ota; Harunori Takeda and his father M. Takeda; Chie Takekoshi Kumagai, Prof. M. Odera, and Prof. Y. Hirao. And especially, to Poldi Heinisch, who enjoyed joining in the hunts since 2001, spotting many fine pieces, and joined in all aspects of their care (such as dusting!)*

The book was finalized in November 2020, in two volumes totaling 822 pages, the .pdf downloaded by many; four hard copies printed and bound, in collaboration with Alwin Schempp at IAP.

*Prof. Hidekuni Takekoshi*
My treasured friend Hidekuni Takekoshi passed away in January 2020 (In Memory of Prof. Hidekuni Takekoshi 10/26/1926–1/11/2020, arXiv:2008.02718)

Poldi's long trip days also ended then for health reasons, Iwashita-san retired, so the travel days in Japan closed. They and especially Hidekuni and so many other friends are missed.

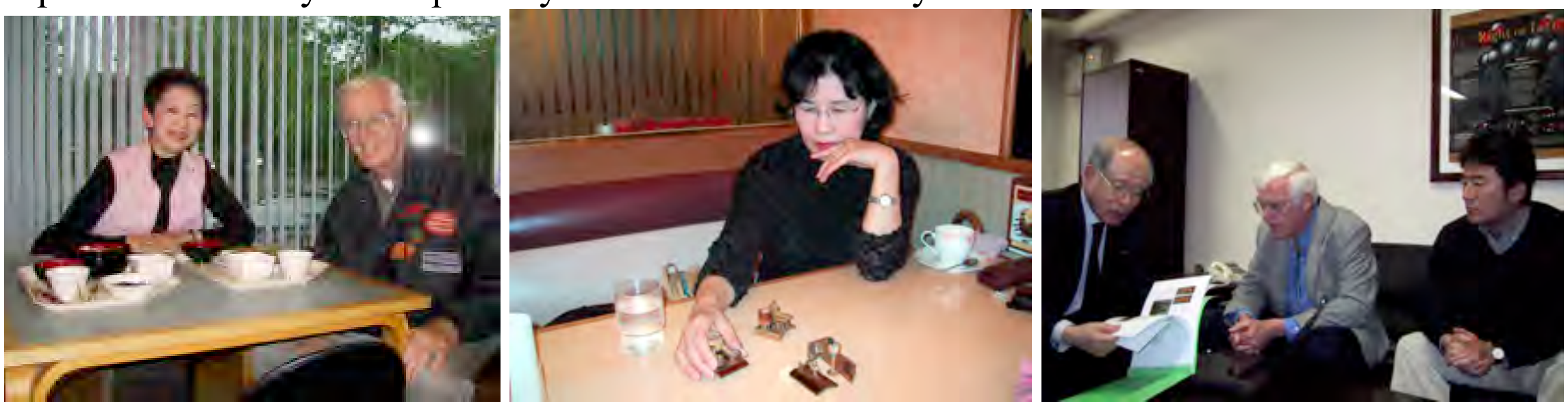

With Akemi Doi May 2005, Machiko Suenaga, with RIKEN President Prof. Noyori & M. Okamura

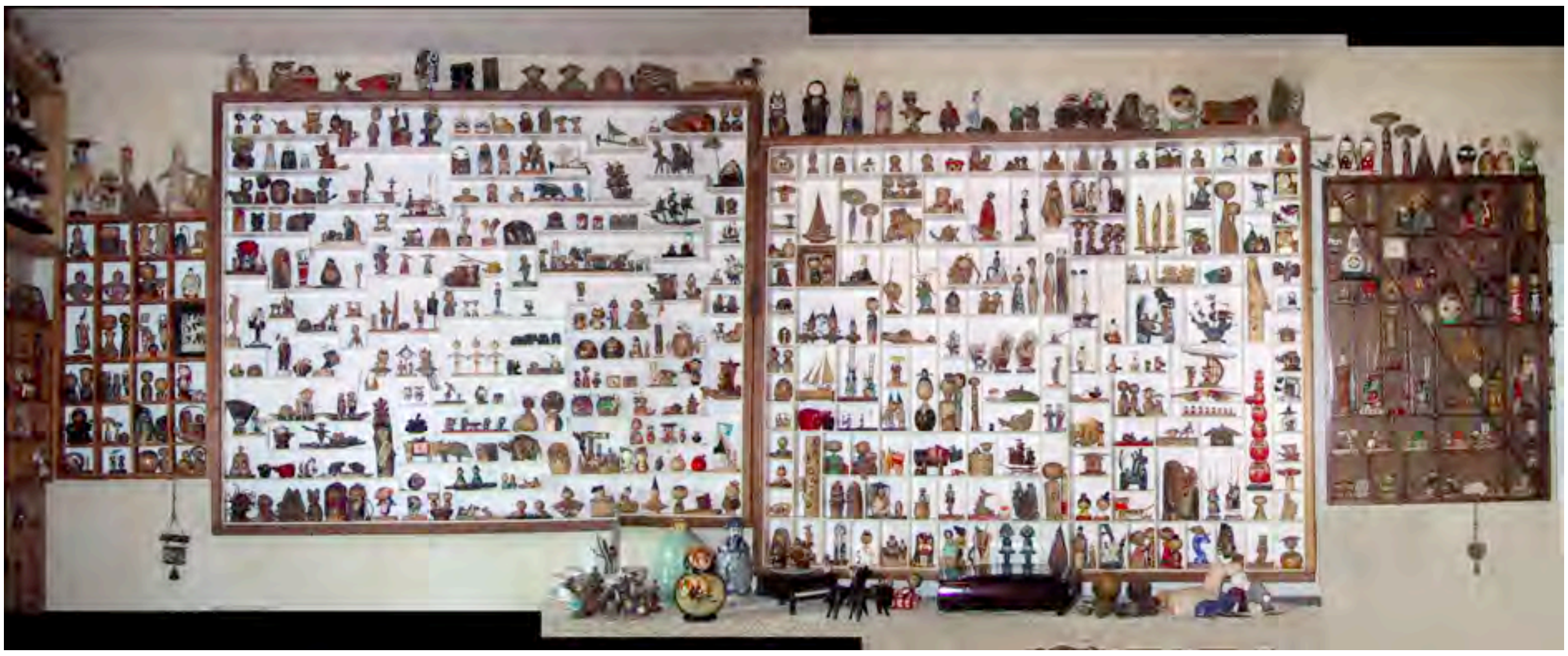

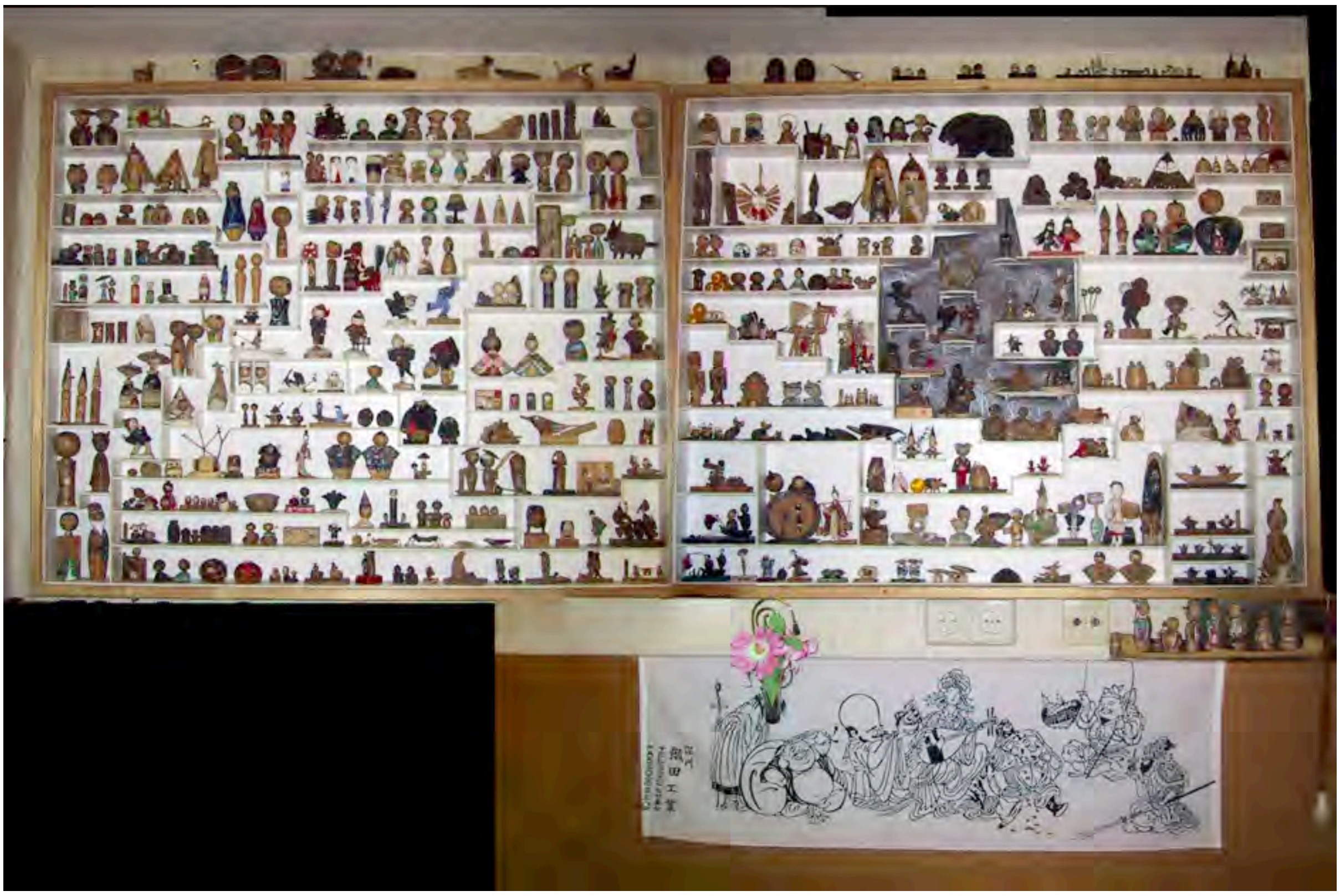

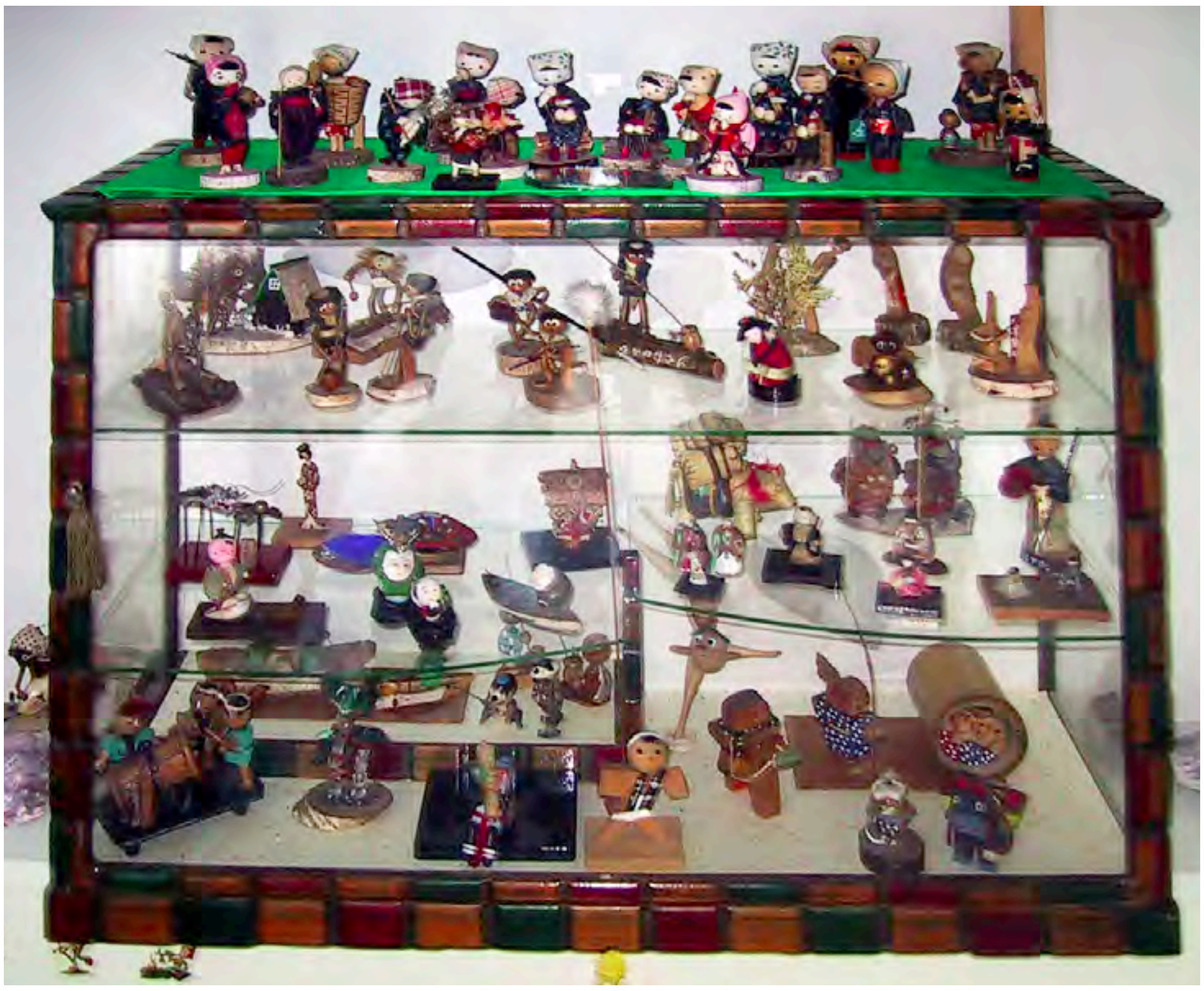

Karaoke

Karaoke with those who cannot find the tone or sing is about the worst there is; with those who can, it can be very pleasant.

Japan is full of tiny bars, where regulars often would buy a bottle of whisky that would be labeled with their name and stored in a shelf, occasionally or often only at year-end settling up. In the old days, the karaoke equipment was just stereo hifi, without the distracting and dumbed down video accompaniment. Takekoshi had two such bars. The first one was in Gion, with a long bar on one side, and tables across the aisle. One night a couple came in, the woman was beautiful. Takekoshi would often make small tricks, somehow he maneuvered that I should dance with her. So we danced the long aisle up and down, moving quickly, and it was very good. Later they told that she was "the best social dancer" in Tokyo. The second was on the inside of the river, the mama-san had a deep, smoky voice and I always wanted her to sing "Save the Last Dance For Me". Her husband had a small eating place nearby and was a superb cook.

Karaoke evenings with Katayama's quite musical group, and with Ryuko-chan's customers and friends were very comfortable. But the best was an unforgettable experience.

A business trip to Hokkaido brought me to Prof. T. Matsumoto at Hokkaido University in Sapporo, who immediately said that in the evening, we would go to "his" Karaoke place. He had heard that I liked Japanese Enka music, which is like American country music in theme and nice melodies! In those days I had played them on guitar and had learned to sing a few. The place was located in the heart of the Sapporo entertainment district, one of the largest in Japan. Tables for ~80 people were arranged around a large grand piano on a low stage in the middle. Only those who could sing were allowed to enter. The owner lady could play anything, from classical to Enka to all the songs in the thick songbooks that included some American songs at the back. Three waitresses were also professional singers and would sometimes do duets with the customers, whose singing was really good. I was of course the only foreigner (gaijin), and of course the etiquette required that I would be asked by the owner to sing. She was a superb accompanist and it must have been ok, as during the course of the evening I sang four times. Then it approached 23:00, when Matsumoto said was closing time, so the pleasant evening was coming to a regrettable but comfortable end. But suddenly, the owner lady appeared in front of me and asked me to sing the last song. That was not comfortable – the other singers were regulars, better than me, and it was their place. I was embarrassed and needed to decline, but that was impossible, so what to do? So tried a diplomatic possibility – I told her that I was very fond of Japanese folk songs and especially those from Hokkaido, which have a kind of pioneering flair reflecting the much later settlement of Hokkaido, and asked if she could have someone sing one of those. I will never forget the intense look that she gave me directly eye to eye and then turned without a word to the big piano. She started to play a beautiful melody, very softly, and then pointed her finger at someone – who knew it was a signal to stand up and sing. She transitioned without pause to a second song, another person, a third – each time the song became more intense and faster. I don't know how many songs were pieced together, but finally they were lusty and powerful. It was really amazing, but then came the finale – she began the melody that I knew well – an old song that is kind of the national anthem of Hokkaido – and motioned that everyone should stand up and sing. All the people had become as involved as I with the progression, they stood and sang from their hearts and the hair on the back of my neck still stands up even as I just recall it. I think she had never done that before, and wonder if she ever did it again. Unfortunately, the next visit to Hokkaido was much later and it appeared that her place had been long closed.

<u>Other</u>

There were many other interactions, collaborations, and friendships, among which was Eji Tanabe, met while student, who became a self-standing researcher and entrepreneur by founding his trading company for advanced hardware, software and services in electromagnetic wave technologies – he has helped me as the distributing agent for the LINACS codes, and we have met over the years at workshops or other venues, where he encourages young people do be adventurous and do their own thing. Another was Y. Mori, a very good-humored guy who was initially an ion source expert, but was asked to become a ring expert, and then became an FFAG enthusiast, along with Shinji Machida, met already in 1988 at INS in Tokyo.

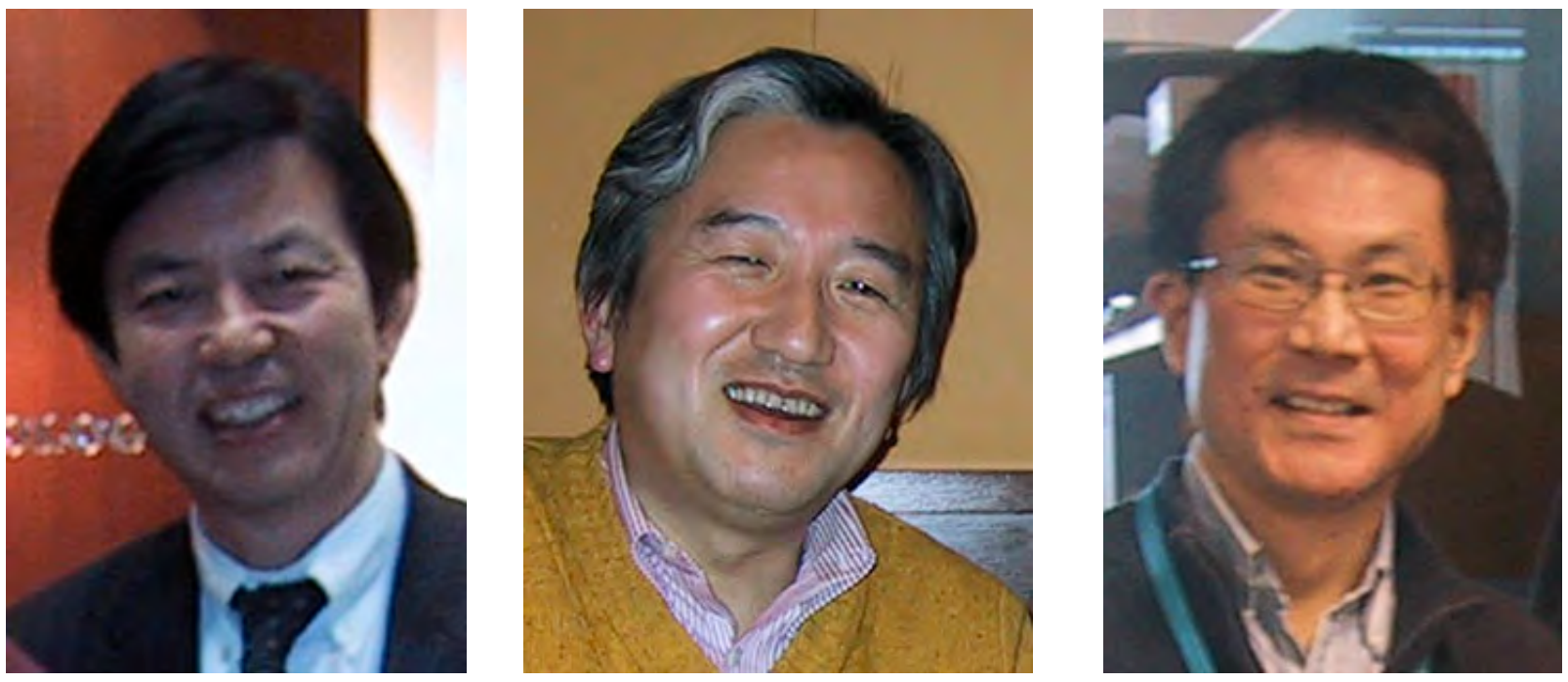

Eji Tanabe, Yoshi Mori, Shinji Machida

[toc]

**Germany**

Institut für Angewandte Physik (IAP), Goethe Universität, Frankfurt-am-Main

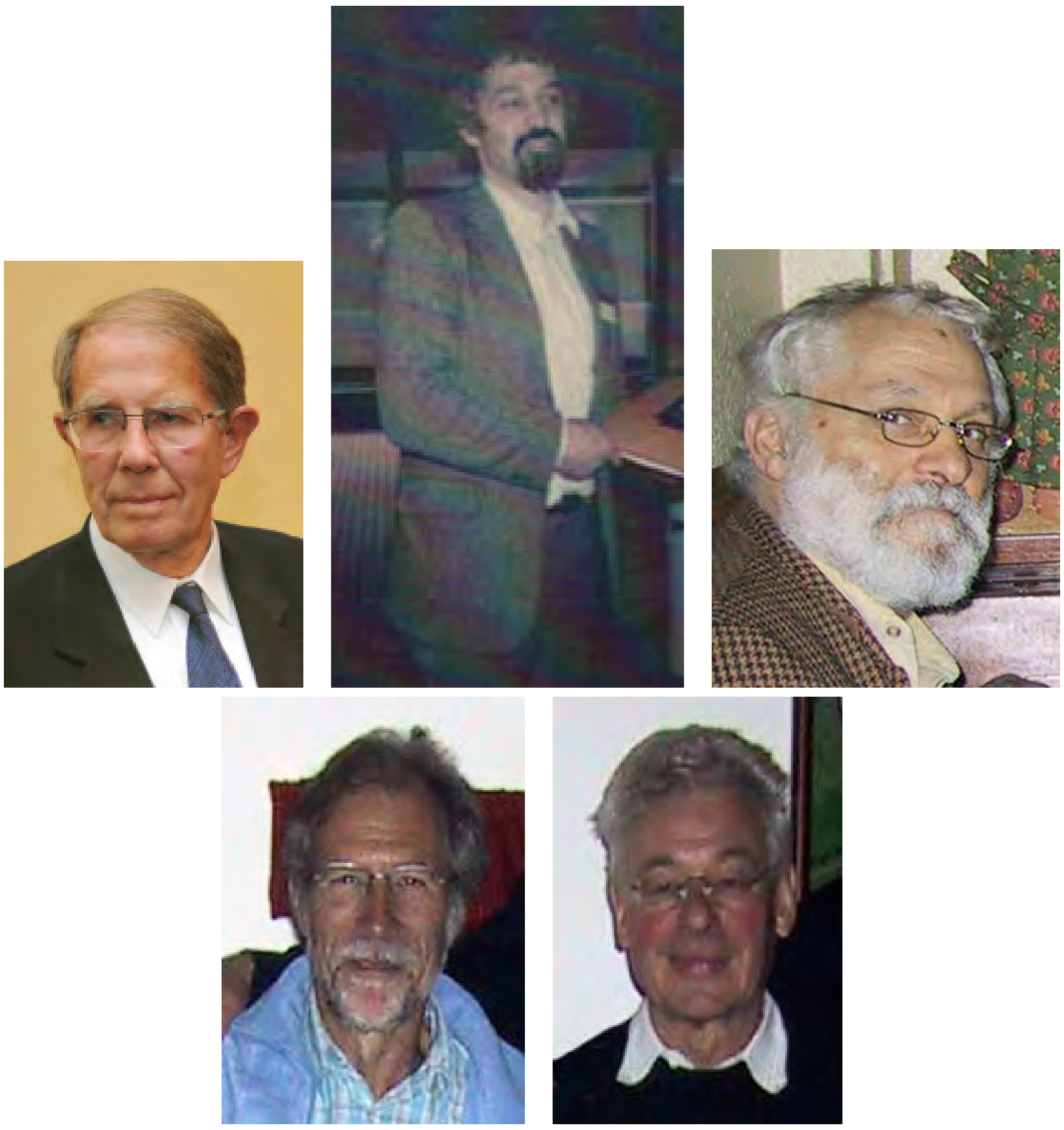

Horst Klein, Alwin Schempp (Damals war's – 19840510) and later,
Michael Kleinod, Reinhard Becker

Technical contacts and friendship with Prof. Dr. Horst Klein and Prof. Dr. Alwin Schempp began very far back during the LAMPF days. As noted in the RFQ history, IAP became one of the earliest, and Schempp eventually the most prolific, Western advocators of Teplyakov's invention. Horst Klein and Alwin Schempp visited us. Alwin by 2020 has built something like 60 RFQs, mostly dissimilar and for various research questions and applications. We had and have lunch together quite often. Many pleasant occasions have been shared through all the years with Michael and Brigitte Kleinod, who warmly welcomed me at the start of the AVHS period.

Progress on beam physics research was substantially extended via collaborations with Deshan Li, a graduate student at IAP, Dr. H. Deitinghoff of IAP, and colleagues at GSI, Darmstadt and in France. Deshan Li's thesis included our first demonstration that the equipartitioned condition could be reached by design at least at one point in an RFQ.

The longest, most productive, and enjoyable collaboration was as informal thesis advisor and co-developer with Johannes Maus, of the (still only) fully correct correlated Poisson solution of both external and space charge fields in the quadrupole symmetric RFQ. Johannes was an out of the ordinary German student, curious, wanting to learn, asking questions, working hard, productive – and his thesis was awarded summa cum laude. We let him graduate before the solvers were fully integrated into the LINACS code, so as not to stand in the way of his intentions to join the NTG company and found a family, both have been equally productive.

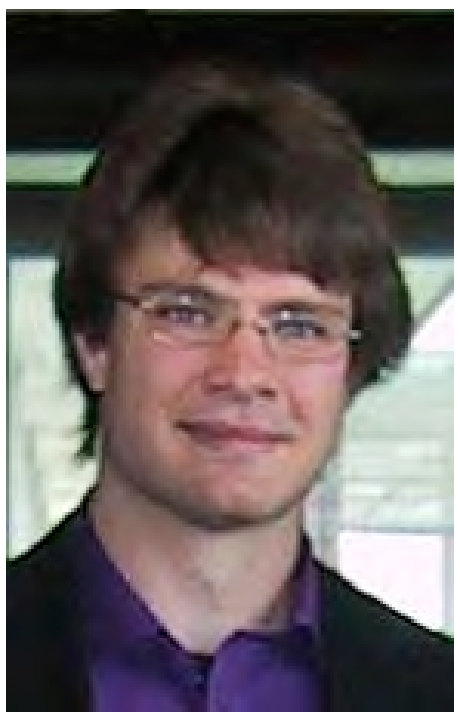
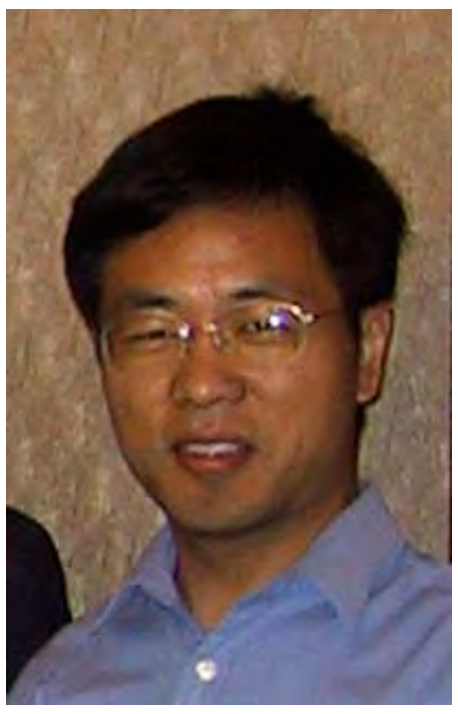

Johannes Maus, Li Deshan

Alexander von Humboldt Stiftung Research Award for Senior U.S. Scientist

On 8/4/1993, Horst Klein called to ask if it would be possible to come to IAP for a year, applying for a Humboldt Research Award for Senior U.S. Scientists, in recognition of research achievements, nominations by eminent German scholars. The application succeeded on 4/15/94 ballot, and the year was split into six segments – September and November 1994, February – April 1995, July – August 1995, June – August 1996, and May – July 1997. [76]

The activities involved continuing work on IFMIF, ADT, interactions concerning ESS and HIF, and the difficult task of presenting an invited paper on superconducting (SC) vs. room temperature (RT) particle accelerator technology at EPAC'96. This was difficult because superconducting technology is outside my expertise, but the competing camps were then so rabid that it was desired to have a non-partisan view. So I avoided a technical discussion completely, put on my big projects hat and announced that SC had reached a par with RT, and each project should carefully consider both and decide – this was very well received.

---

76 The application and subsequent arrangements inside LANL had to be very carefully done, with managers higher than Division level, in particular to avoid any knowledge or otherwise by the Division Leader, as was the practice also by all project managers. There had been two attempts to torpedo the acceptance of direct invitations for papers on my technical work – both of these were immediately slapped down on appeal to higher management (see "ADPLS Letter-12-12-92.docx" (personal files)). Also numerous other petty actions often occurred and were widely discussed, not only against me, but also against others in the Division, such as preventing younger members from attending conferences in favor of Division hierarchy, etc. The AVHS application was handled without electronic communication to avoid Division snooping, and at the moment (2018) I cannot find a copy of it, but will look further in personal files.

It was a great privilege and honor (not to mention hard work!) to be asked to speak on the subject of radioactive waste transmutation and clean electrical power generation as the lead presentation (of 3) at the 1995 Bamberg meeting of the AVHS Awardees, and later to write an article for the December 1995 “Mitteilungen”. It was the question period after this presentation that turned into the suspenseful tale related above in the ATW account.

A second AVHS visit was approved for 15 February – 31 March and 1 May – 15 June 2009, on the basis of work on RFQ simulation with Poisson field solvers with PhD student Johannes Maus. The April interim period between the two official visits was spent with the IFMIF group at CIEMAT in Madrid, collaborating on technical issues for the IFMIF prototype now under construction. Both the Bamberg Symposium and the Jahrestagung in Berlin were attended (at the latter, the pleasure of meeting German President Horst Köhler, who asked “What are accelerators for?”.) (‘Outline of Stay report-2009.docx’)

Interacting with the AVHS was delightful – there was absolutely no bureaucracy involved.

[toc]

**Russia**

The earliest contacts were with the Moscow Meson Facility at the Institute for Nuclear Research in Troitsk, near Moscow, which was ~5 years behind LAMPF. The accelerator design and construction work was carried out by the Moscow Radio Technical Institute (NRTI) in Moscow, led by B.P. Murin and later by G. Batskikh. They were of course very interested in what we were doing, and visited, in delegations led by Sergei Esin. He was also the Party overseer, and was extremely anxious that no one would desert. We jokingly tried to get Elena Shaposhnikova to do that, or at least join us for sightseeing or even a coffee without the other delegation members, but she demurred. She is now at CERN.

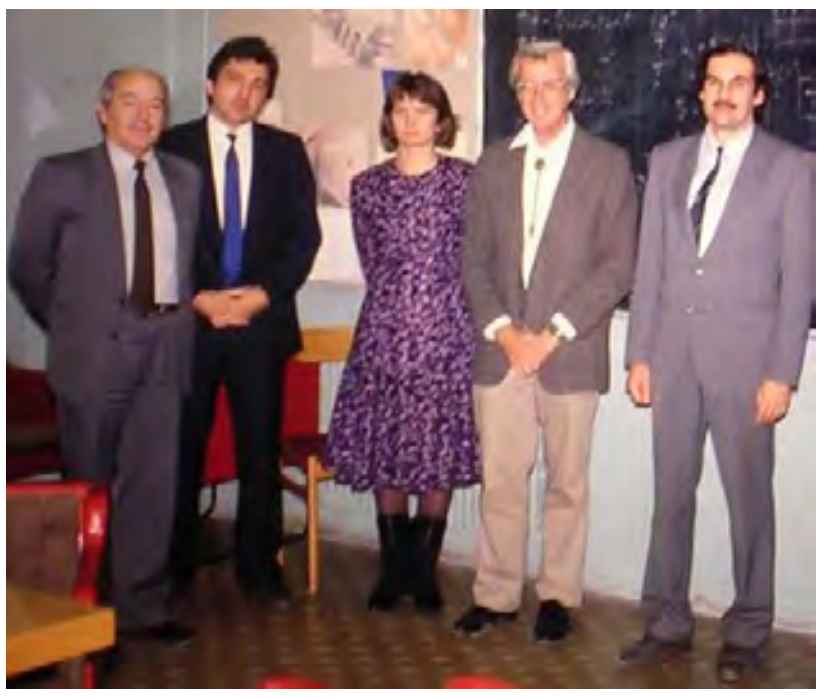

S.K. Esin, Y. Senichev, E. Shapashnikova, L. Kravchuk (MMF)

I was lucky to be able to visit the USSR twice, in 1972 and 1977. Through all the years to the present, I followed the Russian literature as closely as possible, and was impressed by their work on smaller accelerator ideas for industrial application (including Alternating Phase Focusing (APF)), instrumentation and other topics, and on these visits was able to meet many people and discuss and see their work. They were very kind hosts, and lasting friendships were formed. A.P. Fedotov loved Russian art, and toured us through the Moscow museums and discussed the paintings in detail. He was very careful, and would not enter our hotel. On the 1977 visit, his son Peter joined, and worried his father extremely by making deprecating remarks about a prominent person. Gennady Batskikh was a great host and companion at many meetings, including the ATW effort and the attempt to set up a significant collaboration in 1993-1994 described below, and once he hosted a visit to the Golden ring around Moscow.

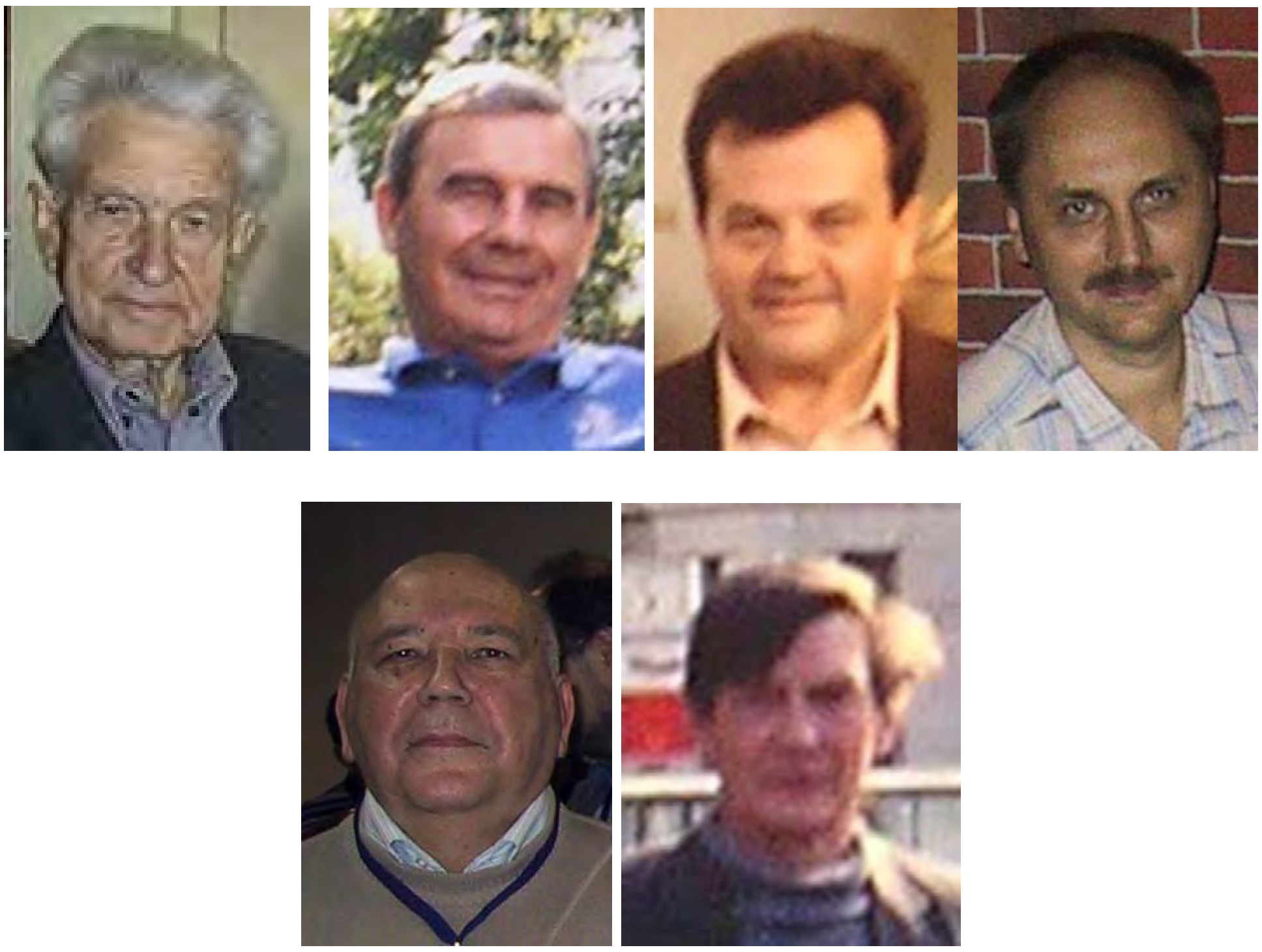

A.P. Fedotov, G. Batskikh, Boris Bondarev, Stas Vinogradov, A. Durkin, V. Boblyev, (MRTI) ;

On the 1977 invitation, I agreed to come on the condition that my whole family could accompany – something that had never been done before. The two sons were 17 and 13. We started in Ireland. In Cork, the younger son came running to request all of his travel money – he had an offer to double it. Of course I went along, to a small coin and stamp shop – the owner smiled at me over my son's head as he rushed in. It was on the up and up – the shop owner would accept foreign coins in payment, although knowing he could not exchange them. When kids would come in, he would offer to give them twice as much in coins for their bills. OK – on the condition that he had to carry them, my son got a whole shoebox full of US coins. The older son was disappointed because there weren't any left. On the sea passages to England, Denmark Sweden, Finland, St. Petersburg, the younger son learned to station himself at the one-armed bandits and be rewarded as good luck when jackpots came, or when coins bounced out and rolled away and were forgotten, and had another shoebox full when entering St. Petersburg. Not only did their presence open many doors, but we also soon learned that the coins could be exchanged as hard currency and made it very easy to get good service, which was otherwise hardly the case.

We enjoyed meeting at conferences in those days, when scientists were well supported in the USSR. Once we met in Canada. They would come with luggage full of good things, and if one was lucky, one would be invited an evening to a hotel room, and instructed to bring a glass along. Then there would be vodka, accompanied by very delicious delicacies, including caviar with butter and bread! Also Georgian brandy, widely known as the best. One evening it was getting late, several bottles of vodka and brandy were gone, and I and others were thinking it was probably time to say thanks and good night, but worried, because when a bottle was opened, it had to be finished. And suddenly another vodka was opened! Diplomatic crisis! I managed to immediately grab the bottle and stood for a final toast: it had been arranged for the Russian delegation to visit Los Alamos after the conference, so I announced that this last bottle would be finished there to our friendship! The ploy was accepted, and we warmly said good night. The next morning I had a splitting headache – The only such I can remember – two rooms shared a bathroom between, and as I reached for the door, it was locked from the other side… We got to the customs at the airport, and the corked vodka bottle was in my handbag; the customs official was bemused, and then amused by the explanation.

From 10/8-27/1989, a workshop had been organized in Yerevan, Armenia, the itinerary was Moscow -> Armenia -> Moscow -> Novosibirsk -> Moscow (then via Narita to Vancouver conference…). The travel notes have many stories of those hard days in Russia. The workshop was almost called off because Armenia was being blockaded by Russia and no coal could be delivered, with winter coming quickly. There was high tension and some conflicts with Azerbaijan. It was planned that the meeting would be held in an observatory isolated from the city and we would be quarantined there, but we immediately escaped and went into the city every day, noting the large gatherings of men in the evening having loud discussions about the Azerbaijan situation. Coming back to the observatory on the first day, we saw an ambulance, well marked with a red cross, arrive at the observatory, and unload carcasses of meat for the kitchen – and we decided to eat in the city. It was harvest time and the food was really good. Later in Moscow we stayed as a group in a large institute apartment, and felt like real Russians. We had to provide our own meals, I was the only one with experience and directed that as each went his way during the day, to keep an eye out and pick up whatever food articles they might spot (the stores were mostly empty), people were selling stuff on the sidewalks – and collectively we had some quite passable meals. Novosibirsk and Budker's institute were far from Moscow, from which they had always benefitted, and it was a privilege to visit there.

When the Peristroika arrived there was Western concern about USSR scientists defecting to other countries with their knowledge, and in the US, the "Freedom Support Act" was passed to set up support projects. This was mostly cosmetic with very small, short-term grants, and without a plan for the obviously needed long-term. So I decided to test it by proposing a real collaborative project, to use the Moscow Meson Facility accelerator as a test bed for ATW Proof-of-Principle Experiments, and had informal preliminary discussions with MMF and MRTI in conjunction with the IFMIF founding meeting in Moscow, 10-23 July 1993. After writing the proposal and getting approval by the various US agencies (approval to try from LANL, but with no active support, as usual), I traveled with Ed Heighway[77], Olin Vandyke, Kevin Jones and Steve Wender for formal talks and preparations from 27 May - 4 June 1994. Good proposal writing and costing were worked out and later submitted. But it was not a cosmetic proposal and did not go further. During these trips I had long discussions with Yuri Senichev, who was agonizing about whether to stay in Russia or emigrate, ostracized within for such thoughts, and facing the unknown outside. He did decide to leave, and I could give some help. The same was true for many others, such as Yuri Batygin, to whom I could provide references and tips through a series of moves – he is now at LANL.

The beam dynamics group led by Boris Bondarev with Sasha Durkin, Stas Vinogradov and Igor Shumakov was especially close, with many interactions over the years. During the very hard years after 1989, when salaries and pensions disappeared, I arranged to contract for some work by this group and for translations of Russian articles by Vadim Bobylev; it necessary to be very careful to make direct fund transfers and not go through their upper channels, otherwise they would not have received a ruble. Vadim was a good friend and I visited his home on an upper floor of a very rundown building many times – although with no income, he installed steel outer doors, each with many locks. He had a small car and always brought the battery, windshield wipers, etc. inside after parking. Gasoline was scarce – once I went with him in the night to trade vodka for a little. His wife stood in lines all day to risk small investments. Shumakov left, knowing that his future in science there was very dim, as did most of the young people – working as a guard at a bank door paid more than being an institute director. At first, a nominal man-year could be contracted for $1000; inflation started immediately and it was soon ten times that.

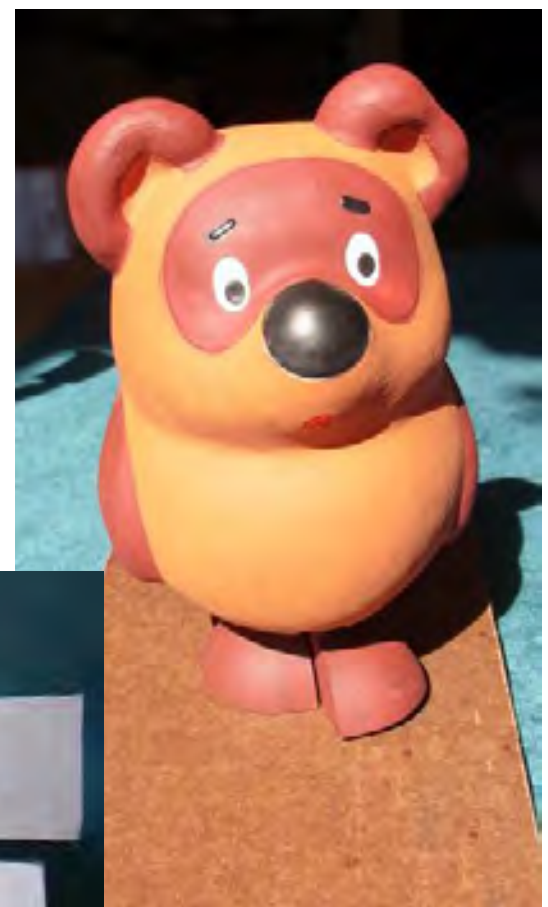

Bondarev was a close friend, a real Russian bear – and once presented me a then fad street toy bear figure, that, when placed on an inclined board and tapped sideways on the shoulder, would waddle sideways while walking down the board – in honor of equipartitioning between transverse and longitudinal planes!

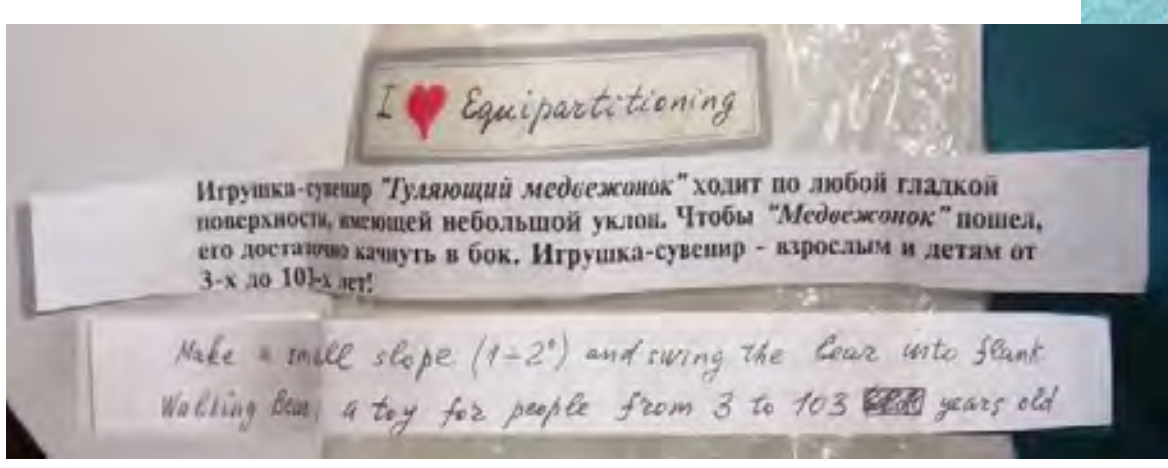

---

77 Who would undoubtedly have been a better choice as my successor as AT Division Leader, but unfortunately I did not know him at the time. Later as the CRNL group dissolved several came to join us at LANL.

He knew that I preferred to go to the circuses above "more cultural" affairs. His wife worked for a large firm, had a big salary of which she was very proud and said it allowed her Boris to play with his physics in peace, was close to Mrs. Gorbachev, and once treated Boris and I to a gourmet dinner.

The group collaborated very heavily and essentially during the IFMIF and LINACS development days, fully open with their excellent code LIDOS.

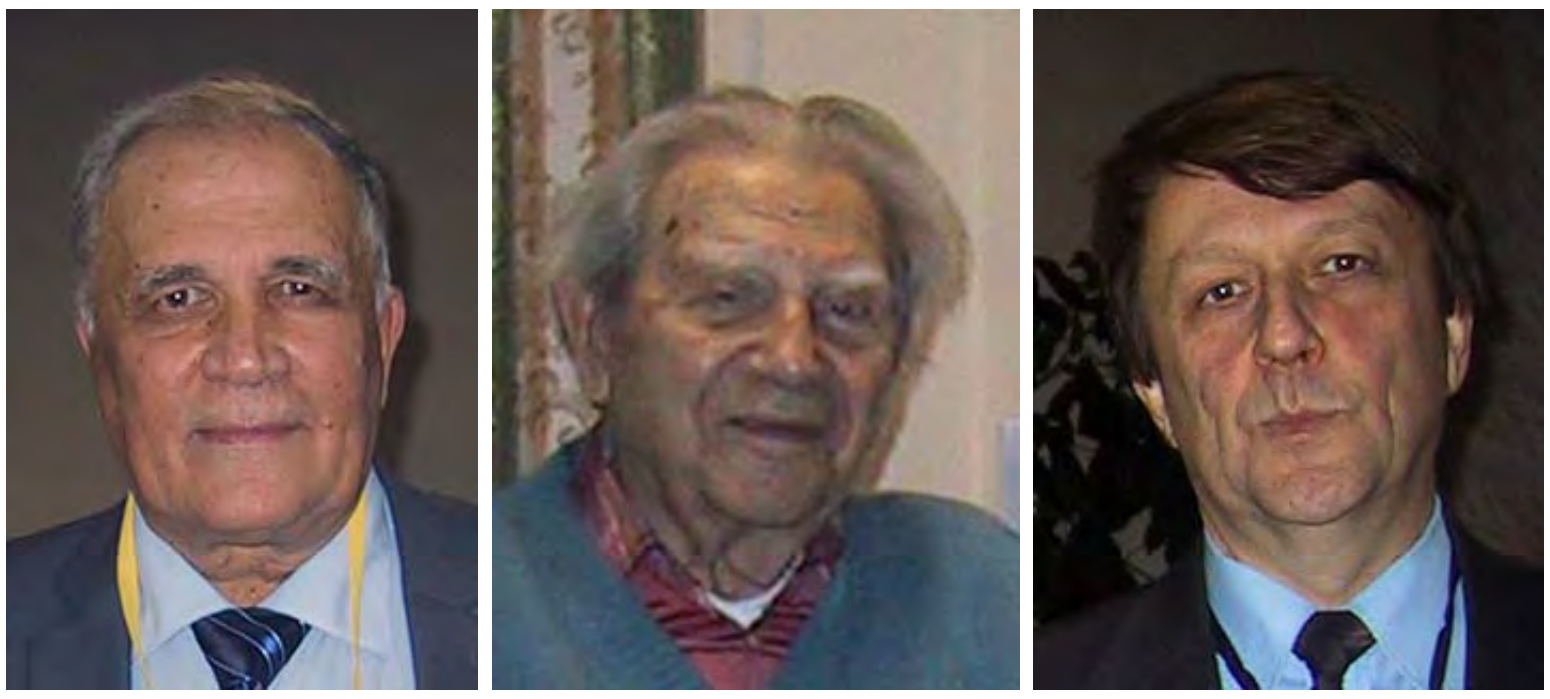

D. Ovsyannikov (St. Petersburg U.), N. Lazarev (ITEP) , M. Chernogubovsky

It was difficult for many groups to learn how to get funding after 1989. The group of Prof. Dmitri Ovsyannikov of St. Petersburg U. decided to organize a workshop on beam dynamics optimization, asked me to be on the organizing committee; I was able to attend the BDO in 1996. After the banquet, I proposed that we stay a little longer for "a business meeting", rather surprising – to discuss a number of other measures that they could take to increase their visibility in the accelerator and Russian science community. I also edited an English version of a book by Ovsyannikov. In 2012 I had made a breakthrough method for practical design of APF accelerators, with heavy reference to earlier Russian work, so attended the Russian Particle Accelerator Conference hosted at his University and dedicated the work to them. This gave a second chance to meet members of his family and again the wonderful opportunity for a personal tour of the Kamera Museum of Peter the Great, where his son-in-law Igor works. After the RuPAC, we visited MRTI once more with Durkin and Vinogradov, and were able to meet Batshikh and Nikolai Lazarev once again.

I was of course familiar with the pioneering work of Prof. I.M. Kapchinsky, and was introduced to him on the 1972 and 1977 visits, but did not work with him directly. [78] Kapchinsky could not travel, it was said because of his health. After discovering how to apply the equipartitioning principle, I wanted very much to discuss it with Kapchinsky, who had been able after 1989 to visit the University of Maryland, which is at sea level – it was not possible to have him at Los Alamos because of the altitude. I had sent him extensive notes and had plane reservations, when he unfortunately suffered his heart attack. He had made notes preparing for our discussion, which I am thankful to have received the originals of.

During this U. Md. Visit, Kapchinsky looked at a book that was in preparation and told the author that it could not be published without a discussion of equipartitioning. The author had not familiarized himself with the subject and asked me for material, which I provided. Then, as a theorist who was never wrong, he changed my material into what is in the book - he makes formulas at the space charge limit, and claims one should design with these formulas, Which is (and was then) clearly totally wrong. Ion accelerators do not operate at the space charge limit. High intensity accelerators operate with less tune depression, actually mostly where the nonlinearity of the equations is the greatest. The three equations must be solved exactly at the operating point - not at the space charge limit. He

78 Fine obituaries and memorials include "Ilya Mikhailovich Kapchinsky – To the 90th birthday, Collection of memories", by his family and others, in Russian. I translated this to English in 2022 and am working on it editing with Vladimir Skachkov, ITEP and his colleagues.

maybe thought this is very hard to do, but his own PhD student published the method - he never grasped the significance. It is very easy to get the solution at every step along the trajectory. [79]

I met Nicolai Lazarev first when he represented I.M. Kapchinsky at conferences. Later Nikolai Lazarev and I have become close friends, sharing good times, with visits where he arranged for staying in an ITEP apartment and living for some days like a real Russian. After he retired he worked very hard at writing his memoirs and about Kapchinsky (and urging me to follow), while listening to his superb hifi, especially Beethoven, which would cause him to interrupt his writing and jump up and down for joy. We are almost exactly ten years apart in age, in 2018 we still correspond frequently.

Michael Chernogubovsky turned up at IFMIF meetings – his wry humor and annual New Year's greetings with very unusual observations about the Oriental year and other matters are very much awaited and appreciated.

It was very nice that V. V. Paramonov came over to J-Parc from KEK in 2013, where he was finishing a collaboration on ACS. He has developed special structure optimization techniques to extend use of TW structures down from usual beta range of ~0.7 – 1.0 all the way down to ~0.2 (1 MeV). "How to teach this to others??" !! It was really fun to chat with him – like with most Russians, with no hesitation to readily get into philosophy. He had mentioned during his questions after talks that everything has its price, how to proceed with new project – I had replied "have to forget everything you knew and rethink for every project". He said he didn't agree – have to start with what we know. I replied that I had had overstated to provoke discussion, of course have to stand on the shoulders of previous work – and had succeeded in provoking! He remarked when the name of a self-proclaimed structure expert came up inadvertently that "he was certainly more of a political type, and although they were nominally in "same field" (structures), "it was always only coupled circuits"!!

Lately, as the contacts in Russia dwindle, new Russian friends have been found in Frankfurt – immigrants – especially Evgeny from Novosibirsk, Gustav (Gush'na) from Lativa. At first, they sold at flea markets, but have now found better opportunities. Evgeny is a whisky connoisseur and has commissioned many duty-free bottles (including the brand of my "rich Irish cousin with the same family name, but never got a penny from, as the Scots and the Irish don't get along well…)

[toc]

---

[79] The basic premise of this "famous" book is wrong with respect to the discussion of ion accelerators. He believes that all of particle beam dynamics is a "thermodynamic" process. The physics of the nonlinear system describing low beta (e.g. ion) accelerators is not thermodynamic. His premise is because he was an electron man, and thought he could just treat ions the same way. I and another of the very few experts tried very hard over years to show him, but he was very egotistical and insisted he was always right. This was before and after he published both editions, without ever asking either of us to review what he submitted for publication. Unfortunately, because the "book" is one of the few published, it has been used everywhere, without much thought. There are exceptions, e.g. in October 2019, a good student expressed frustration when misled by the section on equipartitioning…

**China**

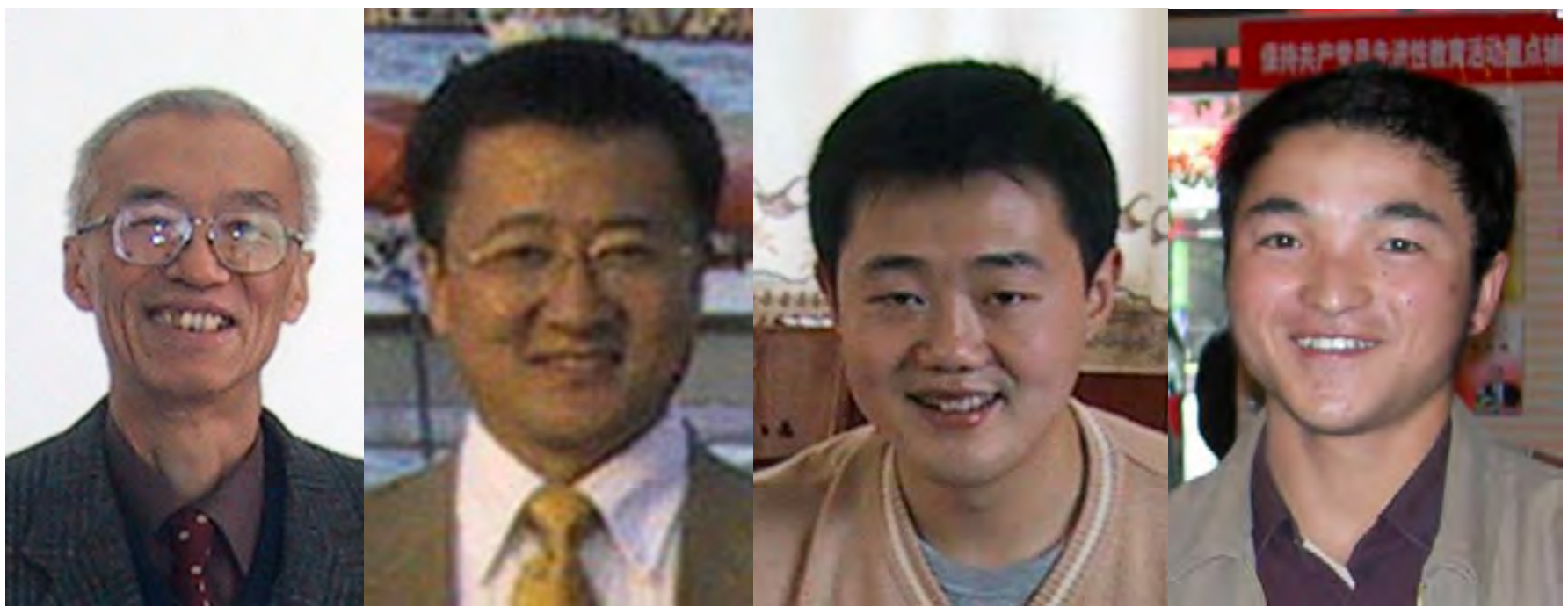

Prof. Chen (2004), Hongwei Zhao, Yong Liu (2004), Xiaojun Li (2005)

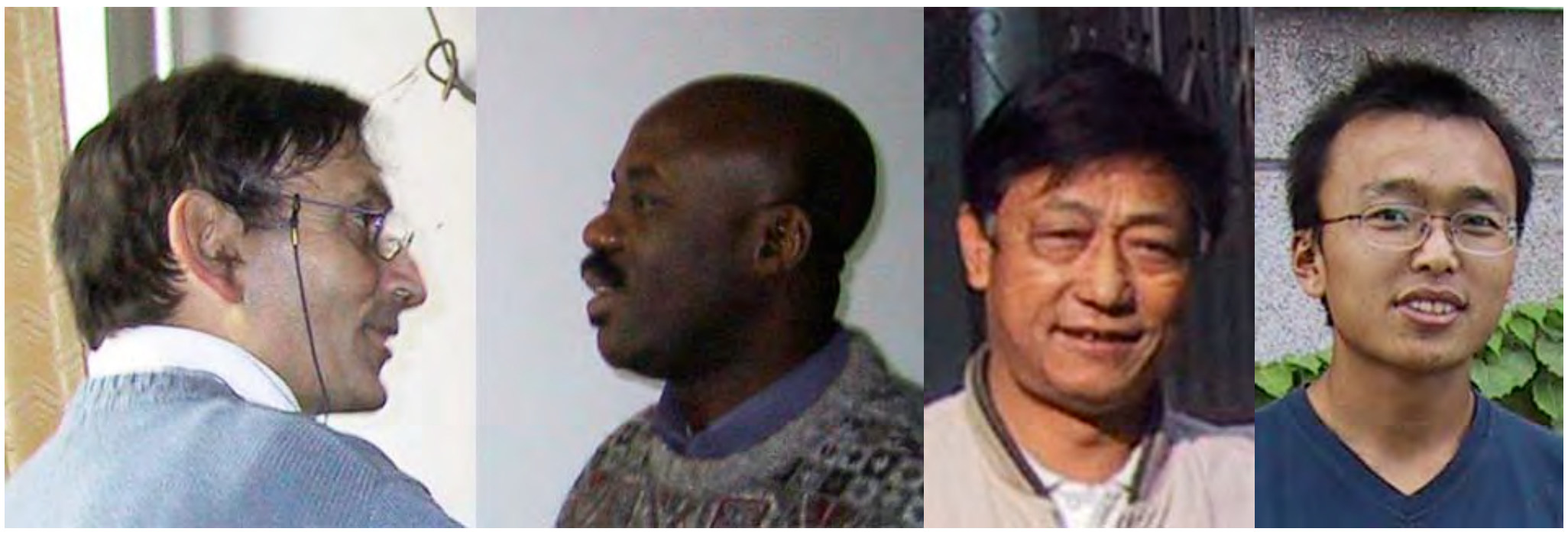

Bernard Bru & Ndontchueng Moyo Maurice (2004), Mr. Liang (2004), Zhouli Zhang (2005)

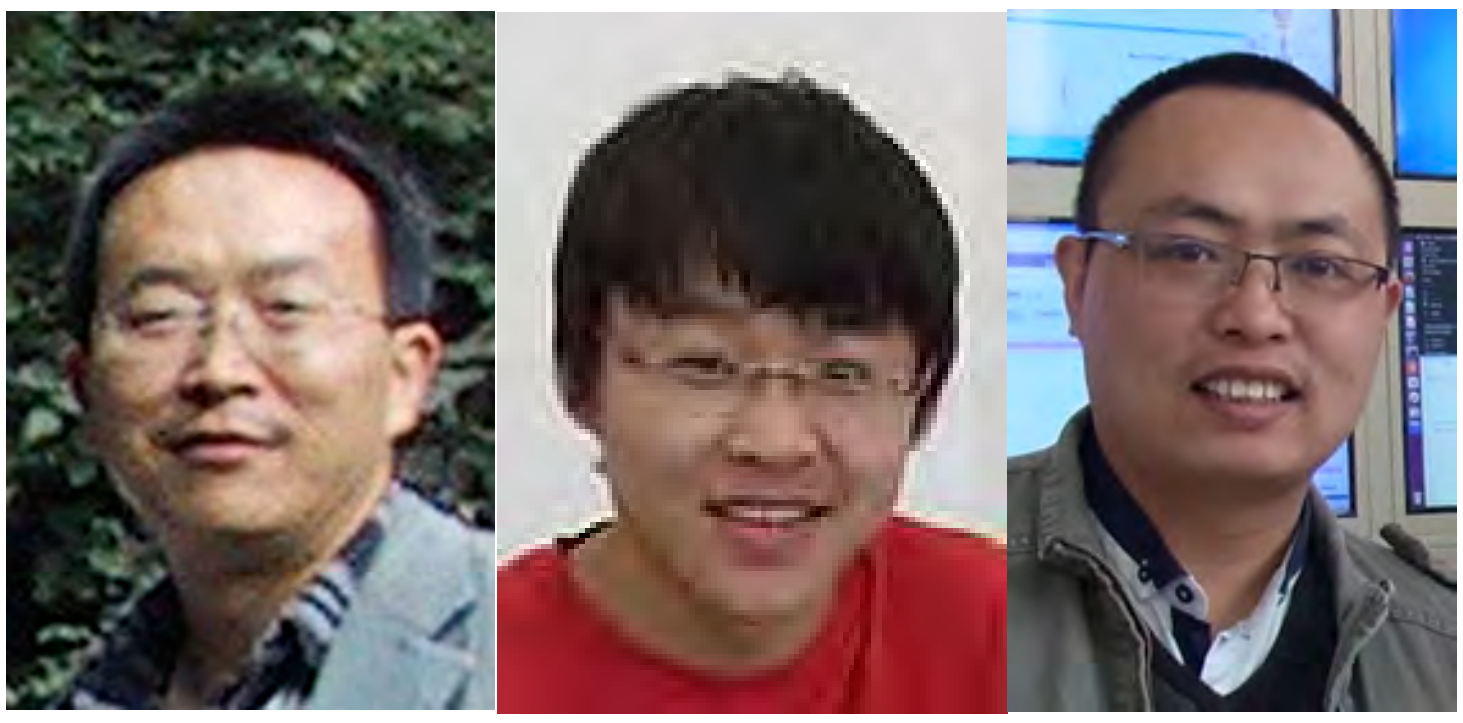

Yuan He (2006), Chao Li (2014), Peiyong Jiang (2018)

In 1986, an invitation came from Prof. Yang Fujia, (then Vice-President (later President) of Fudan University in Shanghai, a nuclear physicist with collaborations at LANL) to "present some lectures on accelerator technology at Fudan and Peking Universities". It was a fantastic trip, hosted at top (ambassador) level, with a young man as interpreter accompanying us. The Shanghai portion included staying at the classic Peace (Cathay) Hotel on the Bund, very good meals and side trips by train to Hangzhou, Suzhou and Xian. At that time there was only one building in Hangzhou – the Friendship 70m joint-venture hotel - with more than ~2 stories, where we stayed. I went for a stroll after dark, there were very few lights, and eventually saw a bright spot, which turned out to be a small night market illuminated by bare incandescent bulbs. I looked at some jackets, and immediately a large crowd of maybe 40 people gathered around that were jovially determined to help me find a jacket. I didn't need a jacket, they were almost all too small and not a style that I liked, but clearly I was expected to buy one, they were not expensive, and finally a quite passable one was found.

Prof. Yang Fujia and his wife were sent, separately, to the country during the Cultural Revolution 1966-1976, after which qualified people were much needed and reached high positions. His nuclear physics institute had up-to-date equipment and he pioneered in analyzing ancient archeological items, including bronze mirrors termed "magic mirrors". The mirror surface showed nothing, but reflecting a strong light onto a wall would show a pattern. He was able to determine that the pattern was produced by carving a pattern on the back side, subjecting the mirror to a special heating technique that would cause grain boundary cracks, then removing that pattern on the back and carving a different pattern, and polishing the front surface. He was able to duplicate the technique and stated making new mirrors. Later, perhaps it was after I was able to finally invite China to participate in ATW activities and invited him to attend Los Alamos organized International Conference on Accelerator-Driven Transmutation Technologies and Applications, to be held in Las Vegas, Nevada, July 25-29, 1994, he visited us in Los Alamos, and asked if I would be the US agent for selling magic mirrors. However, we had some experience with trying to sell old-style Chinese souvenirs in the US, but people did not understand them in New Mexico and would not buy, so I replied that it did not seem to be practical.

Then to Beijing, where we were met by Prof. Chen Jia-erh, who had also suffered the Cultural Revolution, had a Heavy-Ion Physics Institute at Peking University, later President of Peking Universiry, and was a high government official and advisor to the government (which continued long after his retirement).

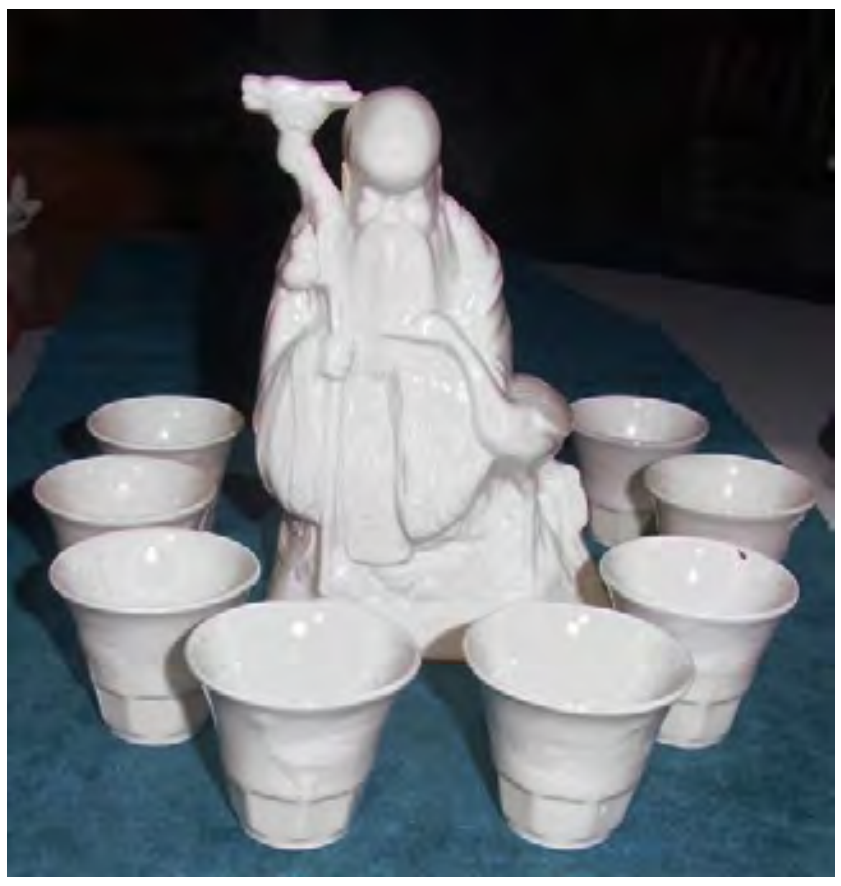

From my notes 22 October 1986, dinner at Zhao Lung Hotel, Beijing, very high level with two Vice-Ministers, Prof. Chen Jia-erh, others at Ministry and Foreign Affairs level – "not exactly clear why they made time for me"... A gift was presented by Mr. Wu Shaozu, Vice-Minister of the Commission on Science, Technogy and Industry for National Defense – a porcelain of "Sanxing, Chinese god of longevity – corresponding to my favorite Japanese god Fukurokuju. The porcelain has a small hole in the bottom that you pour a drink into. Then you turn the figure upright and nothing leaks out. There is a hole at the top in his staff; nothing leaks out there as you fill it from the bottom, but after you turn it right side up, you pour out through the top hole in his staff and it meters out one small cupful at a time – eight cups in all. Eight cups included in the set, very beautiful and could not have been better chosen".

I had prepared "some lectures", but already suspected that the subject of interest might be different and was not surprised when a talk on the Free Electron Laser (FEL) was requested in Shanghai. There was no interest in "lectures", and every day questions focused exclusively on FEL. The group of Charlie Brau, one of the earliest FEL inventors, and Jerry Watson in AT Division had made major advances, such as the invention of the photocathode, and the SDI (Strategic Defense Initiative – "Star Wars") was in full swing. If lasers were a logical candidate at all (easily countermeasured), then the FEL was found to be the best approach (determined after a long battle and finally high-level comparison and decision). After a delay of ~3 years, classification had finally been imposed, and little could be discussed. It was a stroke of luck that just before leaving on the trip, the magazine "Aviation Week" published a long article on the FEL approach, and I had noticed it and bought a copy. This magazine was also known as "Av… Leak", skirting the allowable boundaries, but was legitimate. So I said that the discussion had to be limited to reading the article line-by-line, this was accepted, and repeated every day for some days. (Afterward there was an interview by the spooks (the only time that ever happened); they seemed to be impressed by the strategy and were satisfied.)

Near the end of the stay, an outing to the Peking Opera was announced. A few of those involved in the discussions allowed that they did not like Peking Opera at all, but they had to go anyway. Peking Opera is very non-melodious to Western ears, very loud – but actually very interesting. Attendance was low, we sat in the U-shaped balcony on the right side. The "music" was coming from a room recessed in the right wall and not visible, but clearly from many unfamiliar instruments, including a cymbal with a sharply rising clang. There was no one else in the balcony; I asked if it was ok to go around on the left side to be able to see into the orchestra room.

The answer was "No!". Nonplussed, but noticed some grins – after a bit, they told the reason – "You cannot look in there because that is our SDI!!" And then it was ok. The musicians were all in T-shirts or even bare to the waist – it was hot in there, and full of cigarette smoke. The cymbal could be seen, and I said that I very much wanted to buy one if possible.

So the next day, we went to a department store and the section for musical instruments had the cymbal. Pretending ignorance, I asked for a demonstration about how to strike it. The clerks were reluctant because of the noise it would make, but finally struck one loud "bo –o-i-i-I-I-N-N-G!" So then I tried it and on purpose did it wrong so there was only a "bonk". So asked again that they teach me the technique. Already after the second correct strike, all the customers on the floor were gathering around – in those days, the sight of a foreigner attracted much attention – and soon everyone wanted to try the cymbal, the air was full of "bo–o-i-i-I-I-N-N-G!"'s and everyone, including the clerks, was having a really good time.

On the final evening, Prof. Wang Fujia came again, and asked if we had eaten Peking duck yet. We had not; he said he was very sorry, but then it was a duty that we had to do that on the last evening. We went to the most famous place on the best shopping street near the Forbidden City; on the way and throughout the meal he kept apologizing that we had to eat Peking Duck. We knew already the most famous part – strips of breast with scallion strips and delicious special brown sauce wrapped in a small pancake, but the full course includes every other imaginable part of the duck, which was quite unusual for us. Afterward, he confessed why he had been so apologetic – he said he himself did not like Peking Duck at all, because Shanghai Duck is so much better, and he knew we had indeed enjoyed Shanghai Duck very much!

We later visited Prof. Chen and colleagues at Peking U. several times, collaborating on RFQs, staying at the campus hotel, and enjoying many things, including the weekend dances, buying eyeglasses, going outside the campus among the thousands of bicycles in those days, visiting new huge branches of foreign department stores. The campus is beautiful – always wonder how attractive buildings and environments could be created in past days, when money must have also been tight, in contrast to the modern concrete block mentality such as the Frankfurt U. Riedberg campus.

It was clear that China should have been invited to the Stockholm ATW meeting in 1991, but I did not attempt it. By 1994 it was ok. The same applied for IFMIF, and I raised the question, finally succeeding to have an invitation allowed to have an observer. This turned out to be Prof. Hongwei Zhou, a Deputy Director of the Institute for Modern Physics, Chinese Academy of Science, at Lanzhou, Gansu Province, in northwest China, on the Yellow River, near the Gobi Desert, on the Silk Road. This collaboration has lasted to the present, with many pleasant interactions, visits, and work with a number of graduate students and others.

The first visit to IMP was in 2004, coordinated by Dr. Yong Liu, when Bernard Bru of CERN and Ndontchueng Moyo Maurice from Cameron also visited. We had great times. Maurice needed to have it warm, and kept his room heater on to reach at least 30°C. We visited the Labrang Lamasery; at a restaurant on the return trip (in 2004, it was a 2-day trip, now there is an express road and can be done in one day), ladies in traditional dress entered with trays of Mao-tai and small glasses. We foreigners were approached first, and took a glass, although Bru would rather have declined. We had not been instructed – when the natives' turns came, we could see that the custom was to take three glasses in turn, with a toast made for each. We knew our turn would come again, and knew we should accept. When they came to Bru, he had completely disappeared! Impossible – he was sitting in a corner and there was no escape! He had ducked under the table! That was so unusual and so funny that it was accepted as a good joke and he escaped having to down any more Mao-tai.

We also ate together at noon, and decided to have coffee together afterward, in the CERN tradition. We had noticed "coffee shops" in the town. The first try was a nice empty 2nd floor place with couches at low tables; it took 45 minutes to be served two ordinary cups and one with whipped cream on top. Same at a different place, also empty, the next day. The third day, again the only customers, nothing after 45 minutes, after an hour, so I went to the back – "Please wait just a little longer". After a little longer, we heard a "click-click-click" sound – and knew they must be trying to make whipped cream by hand. Went to the back again, the spray can had run out; said sorry, but we had to leave and get back to work. There were eleven staff back there! We learned that coffee was considered like a cocktail and only really served at night.

There used to be a Foreign Book Store, now disappeared. In 2005, looking for rare TinTin comic books that we might send to Bru, a young guy introduced himself as "Johnson" and offered to help – we chatted a long time, and have stayed in email contact with Xiaojun Li, a chemical engineer, ever since, meeting again in 2013 and 2018.

We enjoyed very much to know Mr. Liang, the officer in charge of foreign visitors, among other things, at IMP. He helped organize side trips, in the early days very well organized and inexpensive, to Dunhuang, Xian, and the Yangtze River cruise, as well as many in the region around Lanzhou.

Yong Liu became a Group Leader at IMP and later in 2011 decided to move to J-Parc in Japan, where our interaction continued.

China was interested in the RFQ, and I became a thesis advisor in 2007 for Zhouli Zhang, including development of a "fair" comparison between equipartitioned and non-equipartitioned designs – keeping all parameters, such as length and rf power requirement, as close as possible to avoid a biased decision. As beginners, Zhouli (and Yasuhiro Kondo at J-Parc) quickly saw the need to use the tools of the LINACS code in preference to older methods. They both have gone on to be in charge of constructing RFQs, so have the full experience of structure design, fabrication and testing. Zhouli received a Post-Doc appointment to the SNS 2016-2019.

In 2011, as customarily sitting at a large table surrounded by young IMP staffers, one young man asked many good questions – it was Chao Li, who wanted to become a theoretician and was attempting to re-derive the theory of Smith & Hofmann from papers where most of the sometimes many steps between formulas were left out (with the annoying statements like "It is easy to show", "Clearly", etc.) It seems that few are interested in theory, not to speak of possibly practical, useful, theory; it was interesting and seemed like possibly a good investment for the future to assist him as much as possible. This had led to a very close apprenticeship and collaboration to the present. He has succeeded to remove the usual smooth approximation and derive resonant mode locations for the full Hamiltonian, opening the way for possible study of transport channels with different kinds of components within the transverse focusing period. Chao finished his PhD, then a Post-Doc at IHEP in Beijing, and then a Post-Doc at Julich/GSI 2016-2018 and at DESY 2021-2023.

Yuan He is now the Accelerator Division Leader at IMP, and has directed my collaborations in recent years. The subject of very low beam loss is now beginning to be of concern in China and was the subject of the 2018 visit.

There are so many more anecdotes from the China visit notes, from visits to Hong Kong, Macao, Guilin, Juizhaigou, Shanghai again, Huangshan, Lushan. Just continuing in Lanzhou for example, it became known that we like to dance, and we went with Yong Liu one night to a dance at the University. Then Yong decided to organize a dance party, with his wife and several other couples, and made an appointment at a night club near the University. That lead to a long adventure – we did not have any appropriate shoes for dancing. So went downtown and searched – to find out that the largest ordinary men's shoe size in China is 43 – and I need 45. But they could order and have the next day. It took about four trips downtown before they really got serious and a size 44 pair came. But upon arriving for the party, they had lost track and were actually closed. That was the closest that I ever observed a Chinese to almost directly show anger – it got arranged that they had another branch, far across town, but they would take us there – so ok. A huge Red Star limousine showed up! All the way across central Lanzhou, somewhere around the large mosque, into narrower and narrower streets, finally into a dark alley ending in a small court with one naked light bulb hanging from a wire – spooky! – go up those stairs – through a ribbon curtain, and there was a very plain room with a CD player, and essentially nothing in the way of soda or snacks. We had a nice time anyway.

Shoe polish ladies used to be allowed to practice their trade on the sidewalks. Observation determined that the going price was 1 Yuan. After completion however, 2 Yuan was requested. Why?! "Your feet are so big that it took more polish" (recall the dancing story above!). A second time: "Actually I put on two coats, first black, then a clear final polish!"

The shoes were not all that durable, so later the young colleagues introduced us to a shoe repair man at the corner above IMP, who did a very good and very cheap repair. Later we took him a pair of ladies dance shoes with plastic soles bought in Lanzhou, and asked him to glue on leather soles, using leather we had brought along. This was something new for him. When that was finished, I said we were not done yet – wanted leather tips glued onto the small heels as well. That really fascinated him. Then still not done- the furnished leather was white, and wanted the edges to be black. Wow – but didn't faze him a bit – went to his bicycle saddle bag and fished out a bottle of black dye. It interested him so much that he did not want to be paid and presented his work as a present! But we did negotiate a price. On later visits, it is a real pleasure to visit "our own shoe repair man in Lanzhou" and ask, in the simplest hand-waving way, how he is, hope healthy, etc.

Similarly, on the street corner of the electronics street to the right of the shoe man, various vendors serve morning snacks, including a tasty fresh made tortilla with a fried egg brushed with a paint brush dipped in a pot of red chili sauce. At the beginning, the couple had a simple barrel oven and fried on top of it, later they got a chic small wagon – also great to see them again after an absence of a year or so.

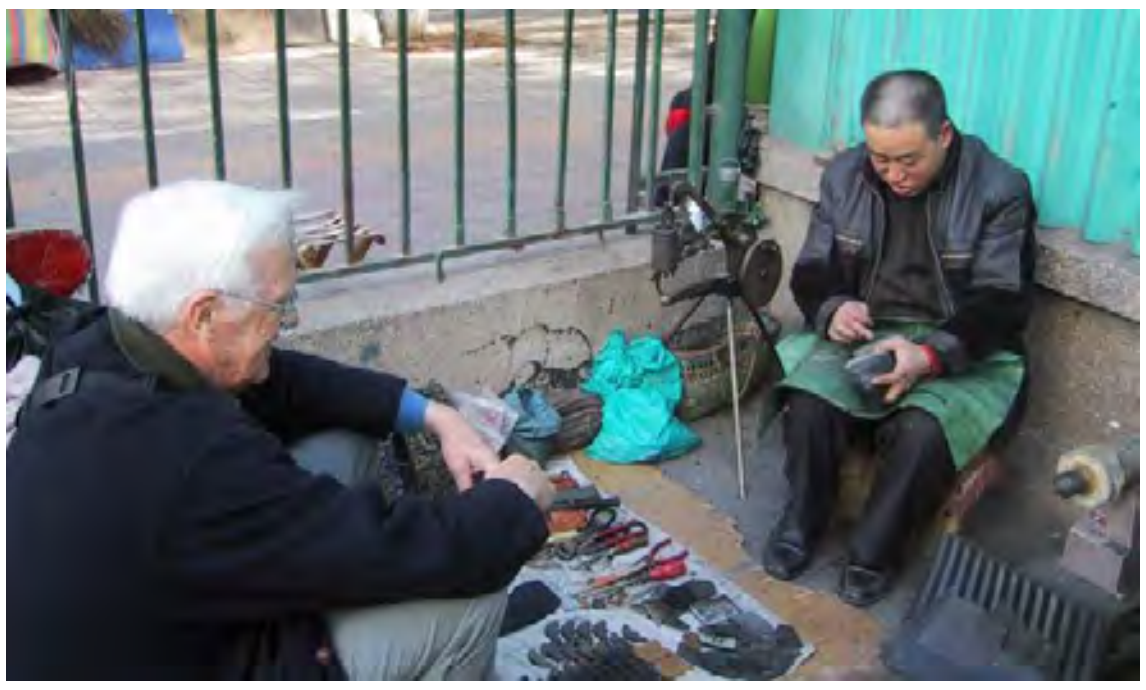

Our shoe repair man

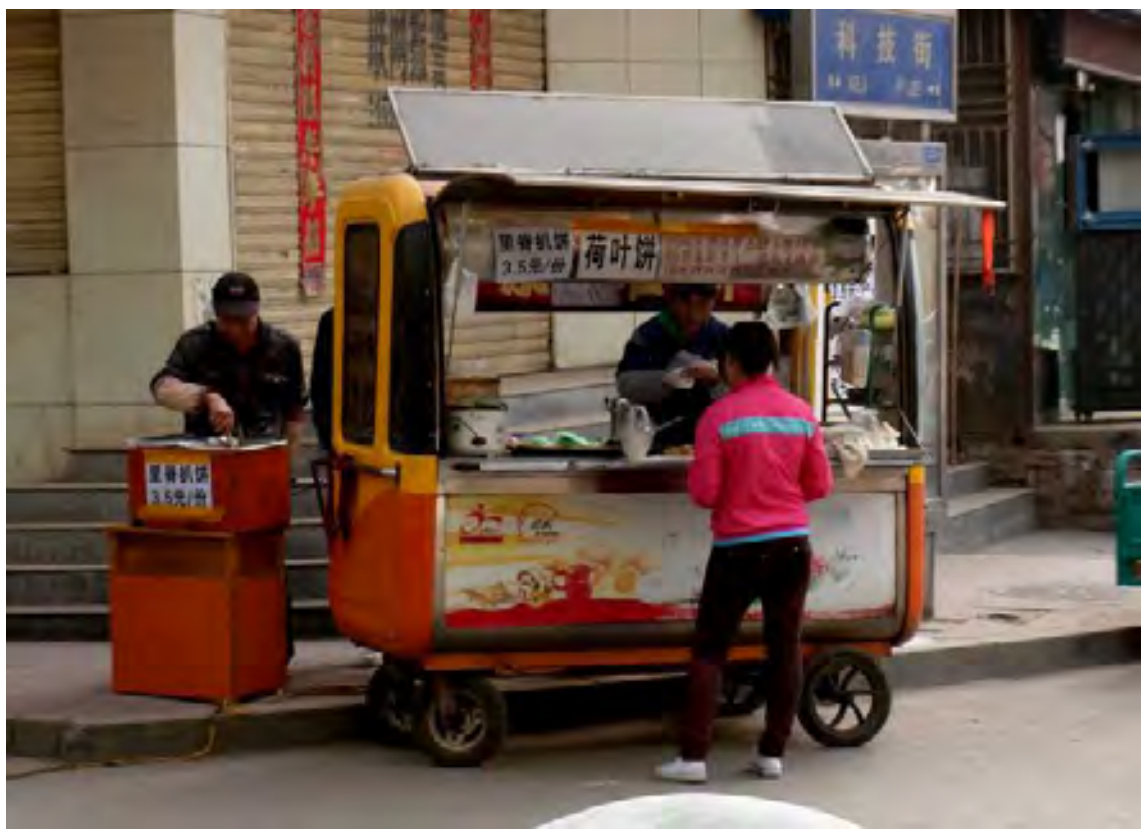

Morning snack

[toc]

## Pioneering

Working on accelerator technology was absorbing, very interesting, intensive, fun. In 1963, the field was still very young. The travel in unexplored regions was fascinating, and I was very fortunate to have had the opportunities to scout landscapes and many details that would sometimes later be rediscovered and developed into camps, towns, cities, re-invention. Or remain still as high challenges, such as the much extended research and simulation that needs to be done on the low beam loss problem, application of APF to a broad range of linacs (superconducting, commercial and therapeutic applications, consideration of oscillating and nonlinear focusing, …), extension of design and optimization techniques. In those areas, much of my exploring was and still remains decades ahead of its time…

Major areas of original work include:

- **RF phase and amplitude control, rf reference source – first analysis of transient response of coupled cavities, first constructed system for feedback/feedforward control of accelerator cavity fields (of the LAMPF linac).**
- Computer control of complete accelerator, **first computer controlled experiments.**
- Instrumentation: - microwave, beam tomography,
- Accelerator Structure Tuning. **Led retuning team.**
- Accelerator Modeling - Correlation with Experiments.
- **RF Generator Characteristics, RF System – first controlled operation of high power rf tubes out of saturation.**

**• Unique experience to be in charge of final installation, tuning, commissioning, turn-on of the major high intensity LAMPF proton accelerator, later commissioning to full operation, and the attempt to aid operation and maintenance using the first database tools. Very good and very rewarding teamwork. Factor of 10 in energy (800 MeV) and 1000 in average current (1mA) was an enormous step; second (unfortunately too similar) accelerator not built until ~30 years later. After almost 50 years, the LAMPF accelerator still operates essentially as we left it (also with my original phase and amplitude control system!!).**

**• Again rewarding teamwork in opportunity to help create and lead the Accelerator Technology Division, seven good years resulted in world renown among the giants, until the Lilliputians chained Gulliver down. (He and others escaped, but the tree was ringed. A technical legacy was left, and a jobs legacy of thousands of man-years…)**

**• Delta-T tuning system. Later explanation of unusual powerful features.**

**• The RFQ – development in the West, extensive investigation.**

**• Equipartitioning - original development of practical expression, first demonstration of simultaneous solution with envelope equations, detailed development for design.**

**• A priori adjustment of the time-dependent design equations - realizing and applying implications of Sacherer's work,**

**• Low-Beam-Loss, High-Brightness Beam Research**

**• Explanation of matching and halo effects in linacs**

**• Linac design – full control of all parameters, design including space charge physics, effect of coding on optimization, optimization**

**• Simulation Code Development, LINACS code, for all forms of linacs, especially developed for RFQ with capability for full quadrupolar-symmetric Poisson fields, heavy ions, multiple beams, advanced shaper design, etc.**

**• Development of close relation between design and simulation; careful calibration of approximations compared to complete beam physics.**

**• Simulation of Laser Ion source injection into RFQ**

**• Alternating Phase Focusing Linacs – breakthrough on practical method for finding and optimizing successful phase sequences. LINACSapf code.**

**• Collaborative research on many topics.**

Publications are listed in Appendix 6; most are available at https://www.researchgate.net/profile/RA_Jameson/contributions and arxiv.org under R. A. Jameson, or from journal publishers.

[toc]

## Closing the circle…

And so the eternal circle can be partially closed.

The technical and philosophical exploration of the framework, elements and details of linear accelerators has been very stimulating and interesting for me, and hopefully these long expositions will contain some useful insights for the younger generations, when linacs of all sizes are needed for research and societal applications.

The advanced and forward-looking technical topics are the most important. The eye-witness historical narrations are the way it was, while knowing that history mostly repeats itself. Hopefully also the good parts of the history, like the now 50-year-long operation of the LAMPF accelerator, the seven good years of invigorating working climate of the early AT-Division, the good non-competitive collaborations with many colleagues and students.

The problems are instructive for the younger generation to be aware of. Most are common in all histories. They persist, and have particular forms, such as the review system problems that concern me for the younger generations, very relevant in 2018-2019 (reference Ch. 30 of "Elements of Linear Accelerators" and Appendix 5), and a driver for a parting observation.

The young researcher, with sinking morale and self-confidence, wonders if the problem is him. I wondered if the total cleft between a scientific and operator approach to an operating linac was only felt by me, and got the MBA in the process of learning that it was not only me. I wonder why the "accelerator physics" community is so resistant to thinking. These are smart people, often very smart, much better with theory and mathematics than me. It cannot be just him or me.

And come to a closing of the circle, and another way of expressing the already long-held thesis that it has to do with the large cleft between the thought processes of engineers and "accelerator physicists" ("accelerator physicists", where it is clear; I do not know so cannot comment on other fields of Physics). "Accelerator physicists" act the way they do technically because they really have no idea. What are known as "accelerator physics" , with its related mathematics and theory, *are not **elements** of a **framework**, but only details*. Such was already noticed in high school and college physics courses – totally dry, not living. I was lucky to be attracted by the lively disciplines of engineering, which makes things work, often when the details of why are worked out later. "Physicists" have always looked down on "accelerator physics", and they all treat "accelerator engineers" with contempt. This deep problem is outside of science and technology – it is partly that they have no real chance of understanding or using a broader view because of a very bounded and narrow education [80], and unfortunately all too often compounded by the base aspects of humanity – low self-confidence leading to arrogance, self-aggrandizing.

And so hopefully can the outlines of a circle be sensed by the newcomers, the younger generations. How to back away from the details and place them in the **elements** of a **framework**. And how much more they will have to learn and navigate concerning the non-technical aspects.

# Stay calm in the resonances, enjoy your work, and contribute.

[toc]

---

[80] One can imagine the reaction I received when suggesting that changes in the physics educational system might be considered along multidisciplinary lines, in work involving the development of PRABST (later PRAB) and USPAS, etc. A modern update is an interesting comment by "Speaker of the APS Council Andrea Liu" in APS News March 2020 (Vol/29, No.3) on her scientific background, which was chemical engineering and materials science before coming to physics – who dares to write publically "I've noticed that physicists are more worried about preserving the purity of their discipline that chemists or engineers. In chemistry, as long as something is interesting they don't care so much whether it deserves to be called "Chemistry". In physics what matters is that problems are interesting, but in engineering it is important that there be a wider impact and it's possible to choose problems that satisfy both criteria."

APS News Sept 2022 Vol 31, No.8 – couple of good articles on education – quantum info science & technology urgently needs overhaul of 70-year old (lectures by Oppenheimer & Dirac) state-of-the art, not getting enough new people to meet demand in this big money field. Hard to teach, very non-intuitive, confusing. In comparison, non-relativistic particle beam dynamics in multidisciplinary framework should not be so hard?? Problems: Small field. New professor will be alone, and it will take on average 5-10 years before producing first PhD graduate. Hope: even only one new professor, who could comprehend the multidisciplinary and advanced state-of-the –art, and young enough to push, could be the originator of a new educational foundation.

# Appendix 1 – SDI Talk at PAC1986

## RF LINACS FOR ESOTERIC APPLICATIONS

*R. A. JAMESON, AT-DO, MS HB/1*
*Los Alamos National Laboratory, Los Alamos, NM 87545*

Summary

Particle accelerators of various types have been considered for many years in terms of their application to national defense. Recently, the Strategic Defense Initiative has focused and emphasized such applications. After appropriate and extensive development, accelerators could fulfill important roles in a defensive-system architecture and could compete effectively with other technologies. A great deal of the required development is engineering. Aspects of the R&D program on rf-linacbased applications are discussed, and potential long-term influences on accelerator technology are outlined.

Introduction

Particle accelerators, in their most prosaic form as viewed by the aficionado, seem to the hoi polloi most mysterious and strange - esoteric. The initiate is excited by the striking, unusual, or imaginative new proposal or the challenge of actually making the idea work. In our world, accelerators are exotic. They become perhaps more so, the more one works and learns; dedication and creativity of high order are apparent in any gathering of enthusiasts. Consider the audacity of confining lightning in a room and using it to perform precise manipulations on invisible particles - as accomplished by Cockcroft and Walton; the extrapolation of a room-size electron linac to a linac two miles long at SLAC; a leap of four orders of magnitude in the intensity of operational proton beams at LAMPF. We are considering the Superconducting Super Collider (SSC), a circular machine 60 miles in circumference, and it was imaginatively postulated that such a machine might be floated on the world's oceans, using the beam to map the earth's core. [1] We seriously discuss colliding two beams of enormous power at a focus of micron or even angstrom dimensions in our efforts to understand the most basic constituents and properties of matter. High-brightness heavy-ion beams may be the most viable possibility for inertial confinement fusion, and high-intensity light-ion beams are the central ingredient of other fusion schemes or materials test facilities.

Likewise, since accelerators were invented, their possibilities as directed-energy systems have been realized and have stimulated research. Because of the enormous power requirements and other technical difficulties, such possibilities have always been elusive and still are. However, in the most recent past, the continuing threat of nuclear-offensive weapons, the rush of progress in science and technology on all frontiers, and (in particular) the remarkable achievements at the frontiers of space came together in President Reagan's 1983 directive to focus more of this research under what is now known as the Strategic Defense Initiative (SDI).

A visit to the space exhibits at the Smithsonian Institute convinces one that accelerators could be made to operate in space. Whether it is reasonable to believe that they could be scaled to serve as defensive weapons devices in any given amount of time, or at a bearable cost, is another question entirely, as is the question of whether an extremely complex network of such systems is manageable, even under ideal conditions. Given the status of small, but powerful, nuclear weapons that can be deployed by cruise missile or suitcase, it is clear that an ICBM defense only is not sufficient. Whether such a system could help prevent humanity from perpetrating another Dark Age (or worse) while it tries to learn how to use similar systems to reach the stars is perhaps the hardest question to answer about SDI. It is part of our responsibility as accelerator scientists to help chart a responsible R&D course and to participate in the national debate on the issues.

President Reagan's challenge was to investigate, through a research program over the next few decades, a defense against nuclear weapons. It was not long before large and costly integrated experiments in the near term were proposed. The relative merits of many ideas have been hotly debated, particularly in terms of scalability and long-term utility. The importance of particle beams, and the relatively advanced state of the technology for producing them, came to be widely recognized - so much so that particle-beam technology is the core of two out of the three Integrated

System Experiments (ISE) now planned for directed-energy defensive-weapons research. These three ISEs will test the space-based neutral-particle beam (NPB), provided by an rf linac; the free-electron laser (FEL), driven by an induction linac (IL) in a ground based version, complemented by an rf-linac-driven FEL, as a backup space or ground-based approach; and a tracking and pointing system.

The initial ISEs do not presume to arrive at full-scale systems capable of defense against ICBMs. Such a system, if ever feasible, is surely a long-term proposition. The SDI will attempt to prove key concepts through sequenced ISEs and the associated R&D and technology-base development that supports each ISE in turn. Within a given program, the ISE and the long-term developments to support the ISEs following in the sequence are carefully scheduled; thus, in any year, activities are spread between near and far term.

A look through the literature and at the agenda for this conference clearly shows that high-brightness accelerators have become the topic of the day in many, or most, areas of accelerator application. Below, we frame some of the central issues and programs relevant to the high-brightness rf-driven linear accelerators needed in the SDI research program.

[toc]

## Appendix 2 – LANTERN – Los Alamos NeuTrons – Enterprise for Research Needs

Los Alamos
Los Alamos National Laboratory
Los Alamos,New Mexico 87545 **memorandum**

---

TO: Distribution
DATE:20 November 1989

FROM: R. A. Jameson
MAIL STOP/TELEPHONE:
H811/5-2275

SYMBOL: AT-DO:89-476

SUBJECT: **LANTERN -- Los Alamos NeuTrons - Enterprise for Research Needs**

Los Alamos has an excellent technical staff and the expertise to do multidisciplinary research.

Los Alamos needs a new Lab-wide focus, something that the whole Lab could relate to, for a long time in the future.

In the past, the focus was the weapons program; in the future, the attention of the world may be more on economic matters. LAMPF, energy programs, laser fusion, isotope separation, and SDI brought large programs, but they did not involve the Laboratory as a whole. In recent years, reactive and maintenance activities have dominated.

A new enterprise is needed -- one that is pertinent to the present and future needs of society, that fits the niches that LANL has carved, that has a long future, that would excite and involve a large part of the Lab and could be seen by the whole Lab and by the nation as a raison d'etre for Los Alamos, raison d'etat.

The usefulness of neutrons could provide such an enterprise.

Needs

*Environment -*
Need to clean up existing wastes (including those at Los Alamos).
Need to prevent future generation of wastes.
Wastes include radioactive waste, air and water pollutants...

Methods are needed that are inherently fail-safe, with proper concern for health and environmental safety.

*Resource Utilization -*
Advanced methods are needed for energy production to conserve scarce energy resources.
Food resources presently involve a great deal of waste.

*Weapons Program -*
Must remain viable.
Requires new methods for testing and verification.
Requires constant supply of tritium.

*Materials Development & Research -*

New materials offer new opportunities in many areas.

Neutron scattering provides increasingly powerful tools for understanding complex material structures, e.g. high-temperature superconductors.

## Research

Los Alamos should do research. Research using neutrons can address all of the need areas outlined above. Many program ideas have been addressed in the past, as semi-isolated projects. The idea for this enterprise is to coalesce as many as possible into an integrated arena at Los Alamos.

*Environment -*

Radioactive waste - Neutrons can be used to process (transmute) radwaste. Research is needed to determine actual cross-sections and to develop the necessary neutron source, target, and process flow technology.

This work should be integrated with research in separation, partitioning and other methods of radwaste processing. The research would show whether production capability should be constructed. (Processing of on-site waste could be a beneficial side activity.)

Other waste treatment problems could be studied using the neutrons directly, or by spinoff programs that would directly use the expertise and technologies involved in the neutron generator.

*Safety -*

It is possible to build intense neutron sources that are fail-safe , avoiding present problems with public acceptance and unreasonable licensing procedures. The capabilities of the Laboratory in safety assessment would be an essential part of not only the construction of an intense neutron facility, but in the research that would be carried out using the facility, toward advanced technology for addressing the other need areas.

*Resource Utilization -*

Intense beams have been considered in the past for energy production, e.g. in heavy-ion inertial confinement fusion, and in electronuclear breeding of nuclear fuels. These approaches might not be addressed directly in the near term, but the R&D results to be expected in generator, target, basic physics, materials development and other aspects of the LANTERN program would benefit future assessments of these approaches.

Food waste can be prevented by radiation processing. Research in this area could be fostered using the technology needed for producing neutrons, and should be correlated with concerns for radiation effects in humans.

*Weapons Program -*

There are many aspects of the weapons program that do or could involve neutrons, which you should fill in.

It is clear that an intense neutron beam would provide an attractive method for generating necessary supplies of tritium. The recent work of a Lab team on this concept received a favorable review by ERAB in October 1989 and has attracted the attention of Laboratory management.

*Materials Development -*

Basic research in materials (including non-metals) is an essential part of the future. Neutrons are often a culprit - materials capable of withstanding bombardment by them are needed. Neutrons are also useful as probes or in other ways for materials development, and the technologies that enable neutron production can easily provide beams of other types of particles.

*Neutron Physics -*

LANL already has a significant neutron scattering research capability in the Manuel Lujan Jr. Neutron Scattering Center, and for fast neutron physics at the WNR facility. The research goals of these Centers would be outlined further in this section, and obviously any new initiative, such as LANTERN, would build and stage upon them.

*Medical & Radiochemistry Research -*

Beams for medical research using neutron or other beam therapy would be natural adjuncts of a neutron production facility, as would be capabilities in radiography research and radioisotope production.

*Associated Technologies and Disciplines -*

A comprehensive enterprise for using neutrons at Los Alamos could involve much of the Lab's expertise, throughout a long time span. Details here should outline the features that the capabilities of each Lab Division would be brought to bear upon. For starters, the contributions of accelerator technology, mechanical engineering, electrical engineering, theory, simulation, system modeling and costing, construction, and operations would be required.

<u>Neutron Source(s)</u>

Intense continuous neutron beams are required at energies of around 14 MeV (for fusion materials studies), and at energies of 1-1.6 GeV for research into processes involving spallation neutrons, such as tritium production, radwaste treatment, and nuclear fuel production. (Delivery in bunches at an rf frequency has been determined to be appropriate.)

A deuteron beam, with several energies selectable up to 35-40 MeV, on a lithium target, would be perhaps the most effective neutron source for materials development.

A proton beam of 20-40 mA, with several energies available up to 1.5-1.6 GeV would be a viable research-level source of spallation neutrons. (Production facilities would require proton beams of 250-300 mA at 1.6 GeV.)

Other beams for neutron research (pulsed or cw) could be provided using storage rings such as PSR, and beams could be provided as needed for medical research and radioisotope production. It is important to note that pulsed neutron beams have many advantages over continuous beams as probes in many areas of neutron physics research. PSR is now the world's most intense pulsed neutron source. Accelerator-based neutron sources can provide either or both pulsed or cw beams.

Advances during the last decade, led by Los Alamos, in the technology of high-intensity particle accelerators that can be operated with low beam losses, affording contact maintenance in a factory environment, have enabled such neutron sources to be built.

<u>Issues</u>

*US Competitiveness -*

The erosion of US capabilities in materials research has been documented in numerous reports.

The Japanese are in an advanced stage of planning and government approval for a new materials testing facility called ESNIT (Energy Selective Neutron Irradiation Test Facility) that will produce a neutron flux using an accelerator to deliver a 35 MeV, 20 mA deuteron beam onto a lithium target, with an advanced hot cell design and small-specimen materials sample techniques.

The Japanese are well engaged in a plan called OMEGA requested as a component of their government's long-term planning strategy for dealing with radioactive wastes. OMEGA will be a research activity into partitioning and transmutation techniques to get answers necessary for subsequent decisions. A research neutron source using a 1.5 GeV, 10 mA cw proton accelerator is planned. The project will probably be approved in FY 1990. The schedule as presently outlined is not aggressive.

Plans of the European Community should be outlined

Certain indicators in the USSR are relevant and should be expanded upon, for example, the surprisingly high ranking that particle beam and pulsed power technologies are given in the national priorities established under peristroika.

US industry would be involved not only as a partner in the construction of a LANTERN facility, but could be expected to participate in the research as is done at synchrotron light sources. Particular enhancements to industry competitiveness should be outlined; for example, as mentioned below regarding accelerator rf power system needs or superconducting machine development.

*Environment -*

Concern for environmental problems, and especially environmental concerns that directly involve LANL and the DOE complex, are very serious issues presently, and we should respond to these in as rapidly, responsibly, and comprehensively as possible.

*LAMPF, PSR, Manuel Lujan Jr. Neutron Scattering Center -*

The future of this capability is obviously a crux issue in the contemplation of a LANTERN initiative, especially with the recent rulings of the Nuclear Science Advisory Committee (NSAC) on the possibilities for hadron physics in the US, and the strong possibility that interim LAMPF budgets may decline steeply in the funding competition with CEBAF, RHIC, the synchrotron light sources, and the SSC. Interesting meson and other particle physics could be done with an accelerator-based neutron source.

*Weapons Program -*

Possible changing emphases in the weapons program, along with the clear need to retain a viable program, and issues such as the possibility that only one weapons lab is needed, are important to LANL's future.

*Neutron Source -*

Along with better resolution of the research mix that would be supported by a neutron initiative, the neutron generator must be defined, to provide an effective, flexible, stageable facility.

Research and continued development of the technology for the neutron generation itself is a major issue, because the answers to fundamental questions of cost/benefit to society for treatment of wastes and other problems depends critically upon how cheaply neutrons can be provided.

Major development issues for neutron production with accelerators include the target systems, as was well-identified in recent reviews of the Accelerator Production of Tritium (APT).

The work on APT and other applications of intense beams has so far not sufficiently emphasized another critical development area -- that of the power supply to the accelerator system, comprised of ac, dc, and rf power, transformations between them, and delivery to the particle beam. In modern high intensity accelerators, 80-85% of the power can be delivered directly to the beam. This has been made possible by extensive work on the accelerator itself. In the meantime, almost no work has been done to improve the power delivery system for the accelerator, and costs are now dominated by this factor. In the ac/dc component of the system, advances have been made by the electrical power industry that have not been utilized yet by accelerator builders - an overall wall-plug efficiency gain of 3-4% might be realized using these techniques. In the rf power component, application of the techniques used to improve the accelerator itself would certainly result in major improvements in overall system efficiency. The term "certainly" is intentional. Improved rf amplifier designs promise efficiency gains of 5-20% over the klystron efficiencies of about 65-68% that are achieved by the best (European) cw klystrons today. Such improvements in the rf amplifiers would have ramifications on the primary power system, stressing the need for integrated studies. A second reason for certainty has to do with the present almost moribund state of the US capability in high-power rf power. (There is a European research and delivery capability that has provided most of the recent high power needs.) Vendors in the US have not applied the physics, engineering and computational techniques used in accelerator technology. They are very interested in participating in R&D to regain their competitiveness in this field. The need for this R&D, directed at the major (80-85%) aspect of the neutron source accelerator, must be made clear and supported as part of our job in developing a strategy such as LANTERN.

A different approach to a lower cost accelerator facility would be to develop a superconducting machine. Our work on preventing emittance growth, and in designing for large aperture-to-beam size ratio lets us avoid the problem that slight beam losses might cause the structure to go normal or induce excessive refrigeration requirements. The amount of rf power that would be needed for the accelerator structure is then small instead of 15-20%, but the large beam power requirement must still be met. The SSC might help lower the cost of superconducting refrigeration business to our advantage. Better yet, LANL is doing some very innovative work on refrigeration (Greg Swift, P-10). We should develop a serious research proposal for cw, superconducting high current linacs. Our current involvement in cryogenic and superconducting accelerators would be expanded upon; the industrial partner arrangements already in place could be beneficial.

*Cost of Power To Operate LANTERN Or For Production Radwaste Treatment-*

The electrical requirements of an accelerator-based neutron source are often cited as a major disadvantage. LANTERN, with a 1.5 GeV, 20 mA beam, raises interesting research questions related to efficiency. A "conventional" copper cw machine at this current level would have low beamloading and maybe 20-25% efficiency. Efficiency for a pulsed machine might achieve 50%, and a superconducting machine would be higher. But we need to demonstrate the high beam-loading feature, and get to work on the rf power issues -- the right research-oriented machine for LANTERN is not at all clear yet, and it will be interesting to think it through as we did for APT. There are other complicated questions such as the desireability of an accelerator/reactor hybrid, which might produce some power. In any case, the benefits of the kind of comprehensive research with neutrons that could be realized with a facility like LANTERN should certainly be worth the power bill.

A production APT facility would require around 1 GW. This issue is presently of real concern, (e.g. for the ERAB APT panel), but may be a red herring, provided the case for national need is made convincingly. For example, a radwaste plant that could treat the waste stream of 10 power reactors, adding an incremental cost of 10% to the cost of the power, may indeed be a worthwhile cost for a society to pay. And the tritium supply is a national concern.

*Radwaste Method -*

The Japanese technical assessment at present favors the use of a sub-critical reactor driven by neutrons generated by a proton beam of 10-20 mA as the most cost-effective approach for transmutation of transuranic radwaste, as opposed to a fission-less approach using spallation neutrons entirely. The latter approach would require about 300 mA of protons. The ramifications of an eventual combined reactor/accelerator research program must be considered.

<u>Initiative</u>

<u>Los Alamos NeuTrons - Enterprise for Research Needs</u>

The purpose of this memo is to propose that we join together to write a prospectus for the LANTERN initiative. The above serves as an initial outline. The most important part is to show and develop how a large number of contemporary and future issues could be addressed in a comprehensive and integrated way, drawing upon the Laboratory as a whole. It is clear that we can provide the necessary major tools, such as PSR and a new research-oriented neutron source, albeit that we would include research on the tools themselves as well.

The LANTERN initiative should be frankly and openly pursued as a Los Alamos objective, with the basic reasons as outlined in the opening sentences above.

The prospectus we write should be by a list of authors; ourselves, plus chapter authors that would be invited later. If we are successful, interests of most or all of the Laboratory technical Divisions would be included and represented.

This approach has been successful in the past, for efforts as diverse as LAMPF, PSR/MLNSC, the drawing together of accelerator technology at the Lab, spinoffs of FEL R&D, the APT initiative, and so on. A broad program of neutron-based research could be a new focal point for Los Alamos.

<u>You are invited to a meeting on Tuesday, 5 December, 1989 at 3:00 pm, in the Nest Conference Room, MPF-6, TA-53 to discuss this idea.</u>

"It is better to light one candle than to curse the Darkness" -- motto of the Christophers.

Distribution: C.D. Bowman, P-3, MS D449
P.W. Lisowski, P-3, MS D449
E.D. Arthur, T-DO, MS B210
S.O. Schriber, AT-DO, MS H811
D. Reid, AT-DO, MS H811
G.P. Lawrence, AT-DO, MS H811

D. Giovanelli, P-DO, MS D434
R. Pynn, P-DO, MS H805
E.A. Knapp, Sr. Fellow, DIR OFF, MS D434
AT-DO

Info: AT-GLs
AT-PDs

[toc]

## Appendix 3 - **LANTERN Presentation to DIR and SMG, 8 June 1990**

Los Alamos
Los Alamos National Laboratory
Los Alamos,New Mexico 87545 **memorandum**

---

TO: AT-DO File
DATE: 11 July 1990

FROM: R.A. Jameson
MAIL STOP/TELEPHONE:
H811/5-2275

SYMBOL: AT-DO:90-

SUBJECT: LANTERN Presentation to DIR and SMG, 8 June 1990

Attached are the transparencies used in the presentation of LANTERN (Los Alamos Neutrons - Enterprise for Research Needs) to DIR and technical AD's on 8 June 1990. Copies of memos setting up the presentation meeting are also attached.

---

# LANTERN

**LOS ALAMOS NEUTRONS - ENTERPRISE FOR RESEARCH NEEDS**

**PRESENTED TO THE**
**DIRECTOR, DEPUTY DIRECTOR, AND ASSOCIATE DIRECTORS**
**LOS ALAMOS NATIONAL LABORATORY**

**8 JUNE 1990**

**R. A. JAMESON**

---

# LANTERN

(Los Alamos Neutrons - Enterprise for Research Needs)

**Accelerator-Driven Spallation Neutron Sources**

**for**

**Research and Technology Demonstration**

## The LANTERN Spallation-Based Technology Demonstration Center

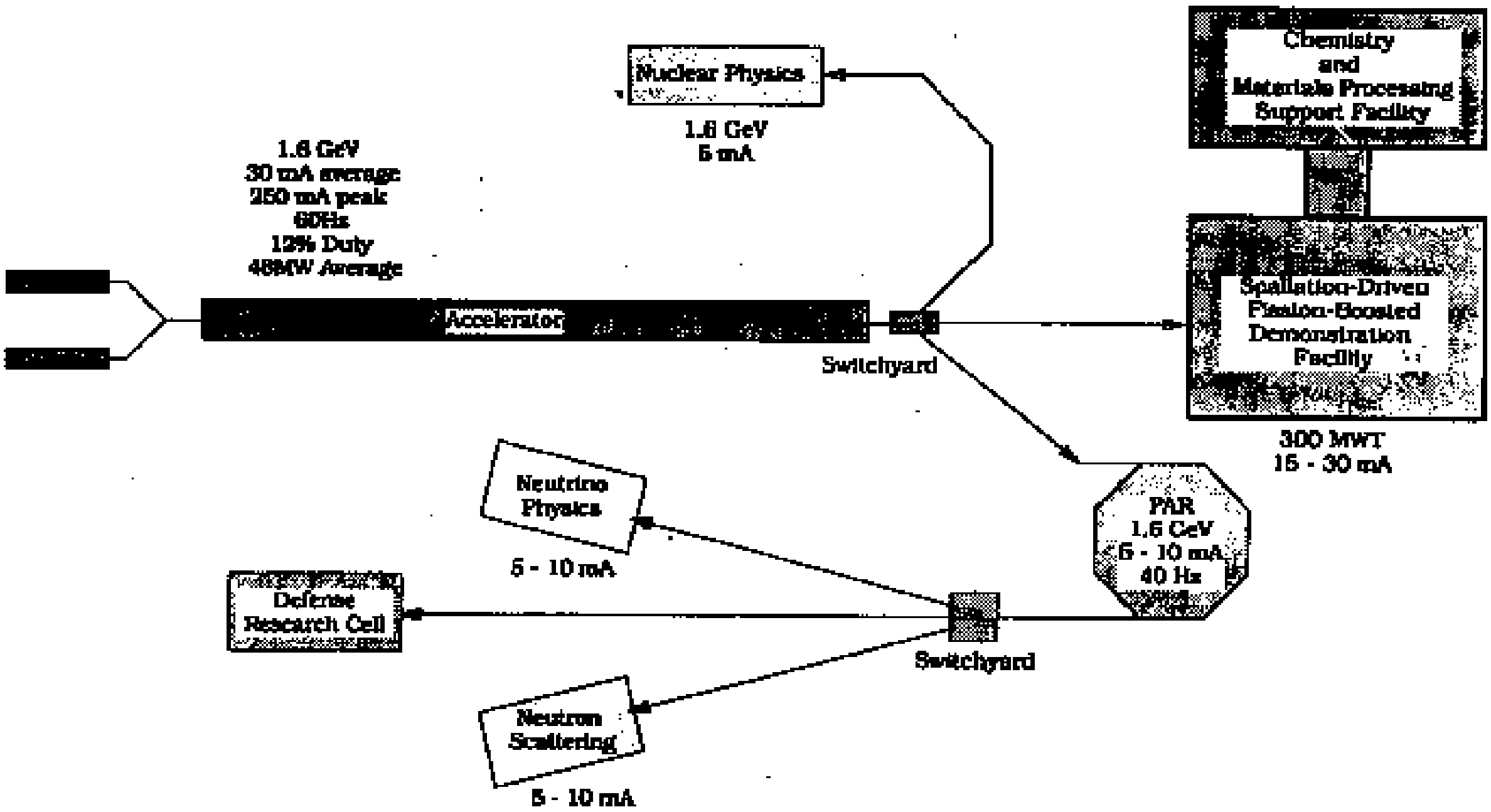

# LANTERN would be a focus on several key national problem and technology areas

- Transmutation of high level nuclear wastes
- Revolutionary concepts for clean nuclear energy production
- Flexible means for meeting stockpile tritium requirements
- Defense applications
- Materials research and neutron scattering
- Industrial applications and technology transfer
- Others

# LANTERN is proposing

New science and technology approaches

for waste transmutation, clean energy production, and tritium production based on recent new discoveries and technology
enablements

## Los Alamos Enabling Technologies

Accelerator

Separated target/blanket concept
for neutron production

High thermal neutron flux
$>10^{16}$ n/cm$^2$ /sec

Fissile daughter production and burn

Burning of threshold fissioners using thermal neutrons
($Np^{237}$, $Am^{241}$, $U^{238}$, $Th^{232}$)

Dilute systems
Small loadings that burn quickly

NEW WASTE BURNING OR POWER PRODUCTION CONCEPTS

## Neutron Economies- Accelerator Sources vs Reactors

- Reactors have a marginal excess neutron production per fission
  - Power reactors : < 0 -> 0.1
  - Material production reactors ~0.5 -> 1.

- Accelerator systems greatly enhance this excess neutron production
  - Variable from 0.2 -> 2.8

- Opens new parameter space for
  - Clean nuclear energy production
  - Transmutation (waste burning or tritium production)

## New Los Alamos Energy Production Concepts

- Technology within our grasp
  ( Accelerator, spallation target, blanket concepts)

- Allows burning of
  - natural uranium and thorium

- depleted uranium
- higher actinides

- Long term energy supply ($10^3$ -> $10^4$ years?)

- Burns its own waste
  (Limited nuclear waste stream managable over human life span)

## Dilute System Blanket Concept

- **A new LANL discovery permitting accelerator-driven systems that burn materials efficiently with orders of magnitude smaller actinide loadings**

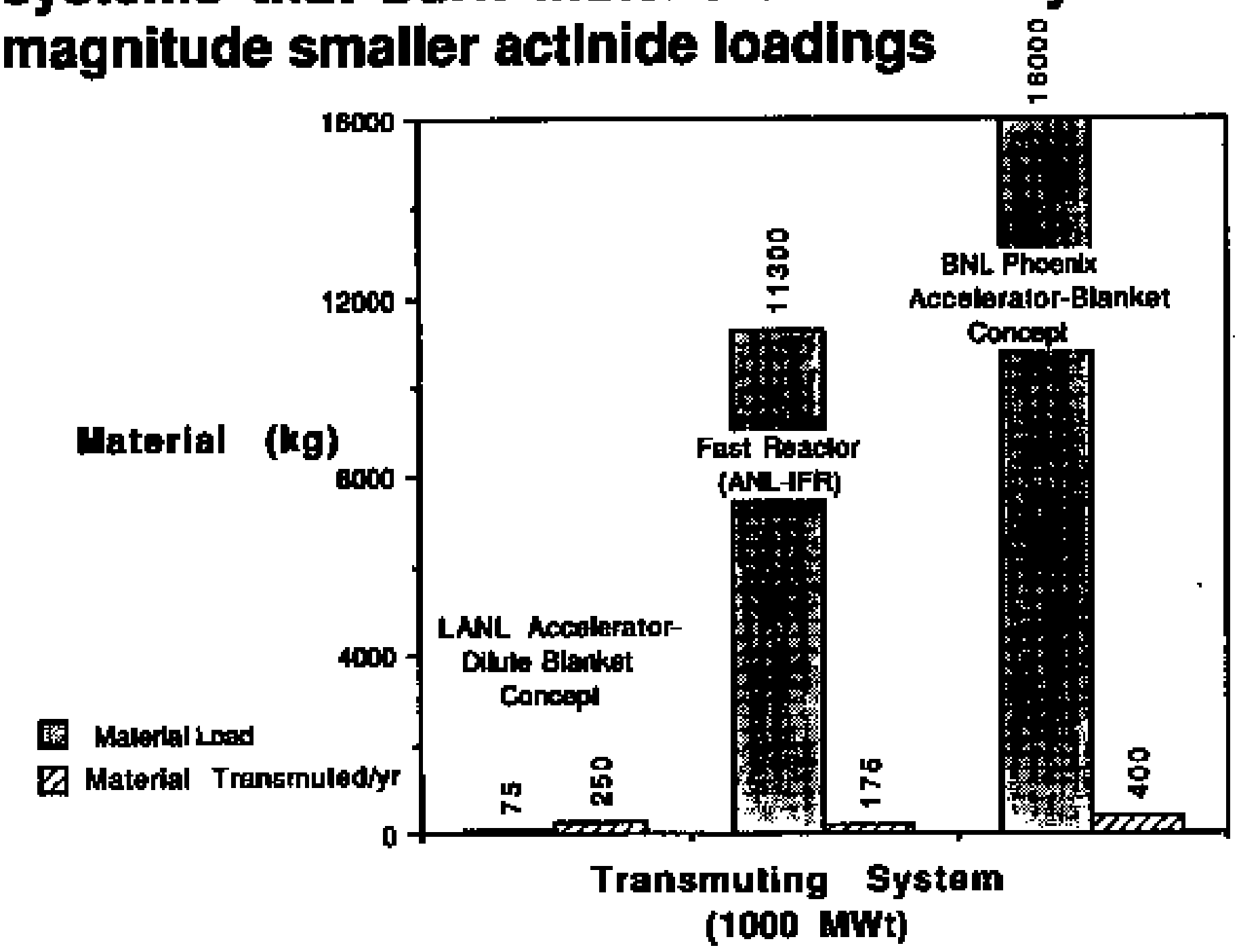

# LANTERN Technology and Research Areas

- Linear accelerator development
- Spallation neutron source technology (nuclear physics, neutron transport, materials science, hydrodynamics, nuclear chemistry)
- Blanket physics technology (Materials, chemistry, neutron transport, nuclear safety)
- Partitioning and processing chemistry (Flouride chemistry, materials separation)
- Materials science using pulsed neutrons (neutron scattering, radiation damage, industrial applications, technology transfer)
- Defense science (high flux neutron systems, effects, dynamic systems)

<u>INTEGRATED</u> R&D for ATW, POWER PRODUCTION, APT

---

# Impact of LANTERN on the Laboratory

For the 90's and beyond, the Laboratory would benefit from a large multi-disciplinary technical program that

- provides a focus and exploits a spectrum of unique Laboratory technology areas
- addresses important national problems and helps ensure national competitiveness
- provides a unique capability for both basic and applied research across many disciplines
- allows us to maintain strengths in key technologies important for national defense

<u>LANTERN provides a vehicle for such a program</u>

---

## Preliminary LANTERN Defense Cell Concepts Provide High Neutron Fluence for Weapons Physics and Effects

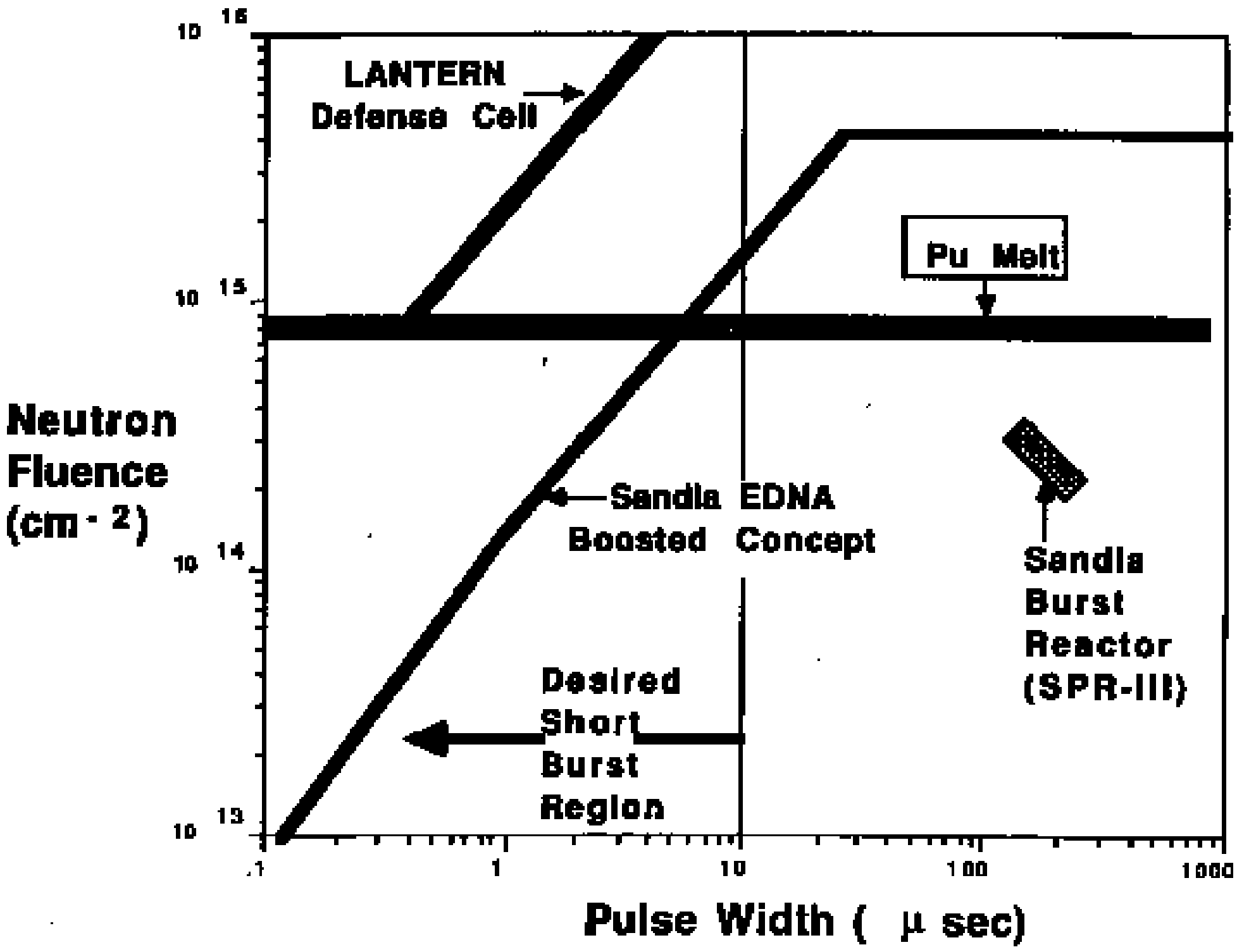

---

# LANTERN Strategy and Vision

- It must be a high priority in the Lab to make it happen here
  - A constituency must be developed at Secretary level in DOE, in Congress, White House, scientific review groups
  - Planning for staged approach to development and full operation is underway
  - Should be made operational quickly (7 - 10 years) to ensure major impact (technological and political) as well as sustained constituency and funding
- Timing and rapid progress are both essential

---

# LANTERN Summary

- Los Alamos has <u>new science</u> and <u>technology</u> pertinent to transmutation and clean nuclear energy

- Potential solutions for critical national needs
- Needed research is uniquely suited to Los Alamos
- Unique world-class research facility for Los Alamos

- **Time urgent - Requires action both in the Laboratory and at national levels**

---

Los Alamos

Los Alamos National Laboratory
Los Alamos, New Mexico 87545

**memorandum**

TO: Distribution
DATE: May 24, 1990
FROM: Fred A. Morse, ADR
MAIL STOP/TELEPHONE: A114 / 7-1600
SYMBOL: ADR:90-265
SUBJECT: LANTERN MEETING

Bob Jameson will notify each of you of the meeting date and your role. You are also welcome to attend the meeting with the ADs and Pete Miller.

FAM:slg

Distribution:

Ed Arthur, T-DO/B210
~~Alex Gancarz~~, INC-DO/J515 71666
Sig Hecker, DIR/A100
Paul Lisowski, P-3/H803
Roger Perkins, DADR/A114
Stan Schriber, AT-DO/H811
~~Bob Wells~~, ENG-DO/M716
ADR File

Charlie Bowman, P-3/H803
Gerry Garvey, LAMPF/H836
Bob Jameson, AT-DO/H811
Pete Miller, ADAL/A104
Roger Pynn, LANSCE/H805
Jerry Stephenson, P-DO/D434
CRM-4/A150

Los Alamos

Los Alamos National Laboratory
Los Alamos, New Mexico 87545

# memorandum

TO: Distribution

DATE: June 4, 1990

FROM: John T. Whetten
Fred A. Morse, ADR

MAIL STOP/TELEPHONE: A107/7-3880

SYMBOL: ADET:90-237

SUBJECT: ACCELERATOR TRANSMUTATION OF WASTE (ATW) REVIEW

Because of a conflict with an ES&H Council meeting on the afternoon of June 8, we have rescheduled the ATW review discussed in the attached memo. The review will now be held in the Turquoise Room (Ad Building, Room C128) from 8:00 a.m. until noon on June 8 (this Friday). We have also combined the ATW review with a separate meeting on the LANTERN neutron center proposal previously scheduled for 11:00 a.m. until 1:00 p.m. also on June 8.

The agenda for the review will be slightly revised from the attached with the LANTERN proposal discussed first and the "Accelerator Technology" section shown on the attached agenda deleted.

Att: A-DO/TJT:90-83

Distribution:
S. Hecker, DIR, MS A100
M. Berger, ERA, MS F643
C. Bowman, P-3, MS H803
J. Bradbury, MP-DO, MS H850
K. Braithwaite, ESD, MS A103
J. Browne, ADDRA, MS A110
D. Burick, DEW, MS F616
P. Cunningham, CN-NM, MS F628
W. Davidson, A-3, MS F607
T. Farish, A-3, MS F607
A. Gancarz, INC-DO, MS J515
D. Giovanelli, P-DO, MS D434
A. Hartford, CLS-DO, MS J563
T. Hirons, N-DO, MS E561
J. Hopkins, ADAL, MS A112
J. Jackson, DD, MS A101
R. Jameson, AT-DO, MS H811
R. Jensen, ADCM, MS A102
G. Lawrence, AT-DO, MS H817
B. Letellier, A-4, MS B299
R. Linford, CTR-DO, MS F640
P. Lyons, ADDRA, MS A110
P. Miller, ADAL, MS A104
C. Myers, EES-DO, MS D446
J. Puckett, HSE-DO, MS K491
D. Sandstrom, MST-DO, MS G756
S. Schriber, AT-DO, MS H811
R. Slansky, T-DO, MS B210
M. Stevenson, ADET, MS A107
TJ Trapp, A-DO, MS F608
G. Wewerka, ADCM, MS A102
CRM-4, MS A150

[toc]

## Appendix 4 – Consulting Report Spallation Neutron Source Division, LANL, 20-22 December 2000

R.A. Jameson

**Consulting Report**
**Spallation Neutron Source Division, LANL**
**20-22 December 2000**

On Wednesday, December 20, 2000, I met briefly with Don Rej, SNS Division Leader and then attended the Open Session of the SNS Division Review Committee presentations. My charge was to listen particularly to the Linac Physics Design, presented by Jim Stovall, and later to help him and Don Rej interpret and respond to items in the October 24, 2000 SNS ASAC Review Report. I began this interaction with a long discussion with Jim Stovall during the afternoon. I also interacted several times with Ed Knapp, Chairman of the SNS Division Review Committee.

On Thursday, December 21, 2000, I spent the morning preparing for a noon meeting with Jim Stovall and his staff, including preparing a list of references on how linacs can be described in terms of space-charge physics and design criteria using this framework, and printing out a copy of a reference on RFQ code investigation and development that is pertinent to the status of simulation codes used in the higher energy sections of intense linacs. Sergei Kurennoy of LANSCE-1 was very helpful in this task.

I met with Jim Stovall, Lloyd Young, Jim Billen and Harunori Takeda from 1200-1430. This discussion was technically specific, addressed to the ASAC Report, and I believe productive.

Later I attended the SNS Division Review Committee closeout, had some discussion with the committee members (two of whom, Grahame Rees and Daniel Boussard are also members of the SNA ASAC, as am I), and closed out discussion with Jim Stovall.

On Friday, December 22, 2000, I spent the morning with Don Rej in detailed discussion, as outlined below, in the framework of his list of ASAC Issues.

### SNS Linac Design

The SNS Linac design at the date of this review should meet most, if not all, of the SNS requirements. The design involves a very complex optimization of many detailed requirements, many of which conflict with each other. The design team has worked very hard in the months since the inception of the LANL SNS Division, and has much to be proud of. Extensive rewriting of the simulation codes was necessary to handle the details of the superconducting linac, as no previous simulation codes were adequate for detailed superconducting linac design. The design team has coped very well with a continuing series of conceptual and scope changes during this time,

The basic design architecture and parameter trade-offs are accomplished using a large pool of experience, a description of the beam using its rms properties, and an assessment of the beam performance in terms of "safety factors" between the beam extent (rms or total) observed in computer simulations. Careful study of possible errors and sensitivity of the design requires many computer runs and thorough analysis.

This design procedure was adequate for the LAMPF/LANSCE linac, which has operated since 1972 at peak power levels 3-4 times lower than the 50 mA planned for SNS. Because the beam loss specification for SNS is very stringent and there are no machines operating at the SNS intensity, the ASAC considers this specification, and the design procedures for meeting it, of primary importance. When asked to review changes to the machine architecture such the change from room-temperature to superconducting for the main part of the linac, and the consequent changes to a longer DTL, the ASAC was uneasy about the depth of analysis and design presented, especially as there were no technical memoranda or internal reports giving details. It is understood that the project is under severe construction schedule and cost constraints, that will not permit further development of a more comprehensive physics basis for the design, or the writing of technical documentation, but the concerns remain and it is felt they should be addressed by a solid effort to be ready for the machine commissioning, in order to be better prepared for surprises at that time.

## ASAC Issues

### 1. Space charge effects and nonlinear interactions

The ASAC report states:

> " The SNS has stringent beam loss requirements. The linac design must minimize beam halo generation, potential beam loss along the linac itself, and deterioration in output beam quality to the HEBT and ring. The underlying physics depends on space charge and its nonlinear interactions via the system resonances. The linac dynamics has never been presented to us in a framework of space-charge physics, e.g., in terms of transverse and longitudinal tunes at low and high currents, tune trajectories along the linac, tune spreads, relation to structure resonances and potential unstable modes, and the effects of errors in this framework. Instead presentations are almost entirely in terms of RMS properties, which must be correct but are not sufficient to predict or confirm the behavior of the beam edges at the required levels of 10 -4 to 10 -8 . There is brute force running of simulation codes and simplistic evaluation of results by comparing phase-space scatter plots or RMS quantities, without relation to the underlying physics. The best strategy to continue making progress with incomplete understanding of the physics is to 1) make conservative decisions and 2) provide tuneability and flexibility so there will be the ability to respond to either improved understanding from future physics studies or experience gained during commissioning.
>
> Linac beam dynamics studies are a general concern that affects the SNS in many ways, and we have more comments and recommendations later in the report."

The ASAC has been concerned since its first meeting that linac dynamics simulation results have not been presented in a context of the relevant underlying physics and in terms of not only the rms properties of the beam, but what is happening at the outer beam edge where losses to the beam pipe occur.

One cannot design a fully detailed linac from theory or analytical formulas. However, it is possible to evaluate a design in terms of the relationship between the external focusing fields and the beam space-charge. A framework for doing this was established by Prof. Dr. Ingo Hofmann of GSI and I around 1981, and has been developed slowly over the intervening years. Various groups (e.g., very recently the CERN Neutrino Factory Study Group, and ESS, CONCERT, JAERI, IAP Frankfurt, Moscow Radiotechnical Institute) have characterized their designs within this framework, and take care to satisfy at least one finding that has been extensively tested by simulation – that there exists a "safe" region of tune-space within which the accelerator tune trajectory should lie to avoid problems with system resonances leading to instabilities and beam growth.

As noted above, the framework is concerned with the system resonances. The strongest of these occur at integer resonances between the transverse and longitudinal beam motions. Weaker, but significant, resonant interactions can occur with the (infinite number of) higher order modes, as excited by various exciting mechanisms such as mismatch, misalignment, free energy, or various errors along the linac.

The framework is directly related to the descriptions and techniques used by ring designers. In recent years, as ring designers must handle considerably higher beam intensities, a growing number of them have found that the knowledge being gained about linac beam dynamics is of direct use to them also.

The framework is directly extendable in its most sophisticated application to the fundamentals and wealth of information and diagnostic technique of modern nonlinear dynamics. The recent shift in explaining nonlinear dynamics geometrically in terms of resonances puts this field directly into familiar language for accelerator and ring designers.

Thus, a program for understanding beam halo and beam losses can be solidly grounded in fundamental theory, and provided with analytical and diagnostic techniques that are not yet in use by the accelerator community at large.

A list of references by the author was provided to Jim Stovall[81]

• It was recommended that the SNS Linac design tune trajectory be traced on a "Hofmann Chart" and checked to see whether it avoids the troublesome lowest-order modes and lies in the "safe region". For the near term, this is practically the main use of the Chart and the framework.

## 2. RMS vs $10^{-8}$ halo studies

A design philosophy for linacs was developed in the 1970's at CERN, based on the fact that rms beam transport properties depend little on the exact shape of the particle distribution as long as the rms properties (of different distributions) are the same. For beams with space charge, two very simple equations defining a transverse and longitudinal matched beam show that the phase advance (with beam current) per unit length should be held constant across a transition in the accelerator structure to avoid a mismatch. The rms beam properties must be thus controlled, and these days almost always are (although by keeping the zero-current phase advance per unit length constant, an approximation).

Slow losses along the linac of $10^{-6}$ to $10^{-8}$ of the beam current per meter do not affect the beam rms value very much. The simulation codes cannot be trusted yet, because of various approximations, to give accurate results at this level, even if the full number of particles in the beam are simulated as is now possible with modern parallel computing.

A long-standing justification that present designs are probably satisfactory with regard to beam loss is to run simulations with the best codes available. After the RFQ, one expects the simulation to show

---

81 R.A. Jameson, "On Scaling & Optimization of High Intensity, Low-Beam-Loss RF Linacs for Neutron Source Drivers", AIP Conf. Proc. 279, ISBN 1-56396-191-1, DOE Conf-9206193 (1992) 969-998, Proc. Third Workshop on Advanced Accelerator Concepts, 14-20 June 1992, Port Jefferson, Long Island, NY, LA-UR-92-2474, Los Alamos National Laboratory.

R.A. Jameson, ""Beam-Halo From Collective Core/Single-Particle Interactions", LA-UR-93-1209, Los Alamos National Laboratory, March 1993.

R.A. Jameson, "Design for Low Beam Loss in Accelerators for Intense Neutron Source Applications - The Physics of Beam Halos", (Invited Plenary Session paper), 1993 Particle Accelerator Conference, Washington, D.C., 17-20 May 1993, IEEE Conference Proceedings, IEEE Cat. No. 93CH3279-7, 88-647453, ISBN 0-7803-1203-1.

R.A. Jameson, "Self-Consistent Beam Halo Studies & Halo Diagnostic Development in a Continuous Linear Focusing Channel", LA-UR-94-3753, Los Alamos National Laboratory, 9 November 1994. AIP Proceedings of the 1994 Joint US-CERN-Japan International School on Frontiers of Accelerator Technology, Maui, Hawaii, USA, 3-9 November 1994, World Scientific, ISBN 981-02-2537-7, pp.530-560.

R.A. Jameson, "Beam Losses and Beam Halos in Accelerators for New Energy Sources", (Invited), HIF Conf., Princeton, NJ, 6-10 Sept. 1995, Princeton, NJ, Fusion Engineering and Design 32-33 (1996) 149-157. , LA-UR-96-175, Los Alamos National Laboratory, September 1995.

R.A. Jameson, "A Discussion of RFQ Linac Simulation", Los Alamos National Laboratory report LA-CP-97-54, September 1997.

R. A. Jameson, "An Approach to Fundamental Study of Beam Loss Minimization", Workshop on Space Charge Dominated Beam Physics for Heavy Ion , 10-12 December 1998, Institute of Physical and Chemical Research (RIKEN), Wako-shi, Japan. LA-UR-99-129.

no losses in a well-designed linac. Then an engineering safety factor is discussed, a ratio between the beam bore size and the beam size. In the past, the rms beam size was used. However, as the beam edge at the $10^{-6}$ to $10^{-8}$ level can move significantly without affecting the rms size, the engineering safety factor should also use the maximum beam size, even though this is always regarded as uncertain.

As low beam loss is a critical machine specification in order to avoid the necessity of remote-manipulator maintenance, and because we cannot yet be completely sure of the beam edges in an actual machine, continued studies of beam halo mechanisms and effects are necessary. A balanced program would combine theory and analysis, as outlined above, and simulation with codes under continuous development to remove approximations and improve the modeling wherever possible.

• It is recommended, and was discussed at length with Ed Knapp, that support be found for halo studies, continuing through the commissioning stage of SNS.

### 3. Understanding why certain errors are important (resonances)

As outlined above, resonances are now known to be the central phenomena underlying the study of nonlinear dynamics. The term "structure resonances" was a new term to the SNS Linac Beam Physics group.

There are resonances intrinsic to the accelerator "structure" or detailed layout – these include combinations of the zero-current transverse (betatron) and longitudinal (synchrotron) tunes, and any other regular periodicities due to tanking, errors, etc. that may persist long enough to allow resonance trapping or instabilities to grow and influence the beam edges.

With beam current, space charge spreads the resonance bands. Considerable work has been done by Ingo Hofmann to identify and characterize the lower-order resonances of this type. Every rational ratio of particle tunes constitutes a resonance. In certain designs, scattering from these higher-order resonances caused longitudinal emittance growth Certain (perhaps unlucky) error patterns can drive the resonances.

As noted above, ring design is characterized by a deep understanding of the lattice resonances. Linacs are less complicated by not having re-entrant resonances, but have complications in the resonance structure arising from high space-charge. It is important to understand the resonances as much as possible, and to work on improving our understanding.

### 4. Beam-based alignment

The term "beam-based alignment" was a new term to the SNS Linac Beam Physics group.

Because the beam is the deliverable quantity from a linac, direct measurements and direct correction of the beam itself are usually the most meaningful, in comparison to indirect quantities such as the power level in an accelerator tank, or the mechanical distance from an alignment monument to a magnet center.

LAMPF was the first linac with really intense beam, and I believe we probably invented beam-based alignment there, with many techniques for beam centering, matching, and longitudinal tuning using the Delta-T method. We had to do it and didn't generalize it under a common name. That came later in the late 80's and 90's, when a number of new machines required it and a lot of labs were working on it.

The SNS linac commissioning design in fact does already involve beam-based technique, in the beam-steering algorithms under development, in matching with wire scanners, and in L. Young's technique for adjusting the rf phase and amplitude setpoints. These need to be understood and explainable, however, in terms of the underlying physics such as the betatron oscillation wavelength (with and without current), both for the designers themselves and for answering questions.

A useful wrinkle, on measuring only the working beam orbit and trying to correct that, is to purposefully disturb the orbit and analyze the difference between the perturbed orbit and the working or desired one. This can make finding the problems (for example, a malfunctioning beam position monitor in addition to a genuine misalignment) easier.

- It is recommended that the SNS Linac Beam Physics team cast their beam commissioning techniques a bit into the language of "beam-based alignment" to be able to communicate better with a large part of the accelerator community, and that they consider whether the "difference orbit" technique could be used to advantage.

**5. <u>"Brute force" code running with "simplistic evaluation of results", "Understanding is limited and supporting calculations supporting decisions have yet to be done".</u>**

These words in the ASAC report were amplified in the Linac Beam Dynamics section. The ASAC is uneasy about the lack of tuning flexibility all the way to the end of the drift-tube linac, about error studies done primarily with envelope codes, non-quantitative descriptions or comparisons of results, lack of correlation of simulation results to an underlying physics foundation, lack of convincing answers to questions about the simulation codes, and no technical documentation.

LANL stated at the Division review that ""understanding halo" is still an open issue and beyond the scope of our assignment", and said that 'if SNS management found out we were doing that with their money…'. The SNS linac is at least three time more intense than the usual LANSCE (LAMPF) operating point, and beam loss is a major issue.

- It is recommended that the team
  - avoid use of non-quantitative terms as much as possible and provide quantitative results and comparisons.
  - Deeply understand the strengths, weaknesses and approximations of all the simulation codes used. It would be well to understand these points as far as possible on codes used by outside groups also, both to seek improvements, and to know how the SNS codes stack up and where support might need to be sought to make improvements.
  - Use the more approximate (usually faster running) codes (e.g. envelope codes) for scoping and initial studies like everyone does, but make final design simulations, and presentations, using full 3D codes. This is now quite possible, the project specifications demand it, and it avoids many questions.

**6. <u>"Significant" disagreement of space charge routines</u>**

Significance is in the eye of the beholder. But terms like "no difference", "little difference", "almost the same", "nominally the same" are not quantitative.

At the rms level, used for example in beam matching, a difference of 10°-15° in an ellipse orientation might be viewed as significant, especially in a machine without any adjustments.

At the beam edge, the beam loss issue is being addressed. While the actual edge may be uncertain for various reasons, quantitative measures, that can be used for example to compare trends, are preferable.

- The recommendations under 5.) apply here also.

**7. <u>PARMILA accuracy & French "actions"</u>**

At the ASAC meeting, it was known by reviewers that PARMILA (and other design codes) contain various approximations that are worth questioning for new linac designs in the SNS class, and that other groups have decided to write their own simulation codes. The reviewers were uncomfortable with answers to questions aimed at getting a feeling about the underlying strengths, weaknesses, approximations and improvement plans for the SNS simulation codes.

Concerns about RFQ simulation codes which use the same beam dynamics and space charge algorithms as PARMILA has resulted in very detailed study, code improvement, and writing of completely new codes, for example at Saclay, in the past few years. It has been pointed out that differences in results found in the RFQ codes, considered significant for predicting matching conditions and beam loss patterns, were present at beam energies that often are handled by the stage following the RFQ, such as a DTL – and that users of the (standard) PARMILA code used for DTL simulation should consider upgrades to PARMILA.

In our noon meeting on 21 December, a reference (LA-CP-97-54) on some of the RFQ studies was provided to the team. The main technical problem addressed in this report is the approximation used for space charge computation in the Los Alamos codes that use position along the linac as the independent variable (such as PARMTEQM and PARMILA). Work at Saclay has pointed out a paraxial approximation that should be removed; this and other questions led them to write their own code. Results from less-than-3D and symmetricized space-charge calculations are generally considered less reliable for modern designs, and not necessary with modern computers. Approximated field distributions using transit-time factors are also being replaced by field mesh methods.

At the LANL SNS Division review, it was mentioned that a check of PARMILA in the first SNS DTL tank has now been made using the LANL code PARMELA, designed first for electron beam simulation but now fully capable for ions as well, and thus now misnamed. PARMELA is based on time as the independent variable, uses a field mesh, and has a 3D space charge routine, avoiding these approximations used in PARMILA. A vugraf noted that differences were small, but no evidence was presented.

No mention of joint work between the SNS LANL Linac Beam Dynamics team and their counterparts at Saclay was made at the ASAC meeting. In the LANL SNS Division review, comments about such a collaboration were presented on a vugraf. Two points:

- The vugraf says “dynamics nominally copied from PARMILA”. “Nominally” is not quantitative, and from the above-mentioned knowledge of French work on their simulation codes and also personal contact with them, it can be expected that they will soon, if not already, be improving PARMILA.

- Direct mention of collaboration with Saclay makes it important to be sensitive to nontechnical aspects of the collaboration as well.

• The following are recommended:

  - A detailed comparison of PARMILA with PARMELA is probably the most direct, as they have programming similarities but differences in approximations.

  - Probably significant differences between simulations with PARMILA (PARMELA) and IMPACT have now been noted; plans have already been made by the team to study and resolve these.

  - A element of support should be sought at the project level for continuing code development. This should not be all left to Saclay or other outside groups. This code development would involve analysis of approximations, removal of approximations when “significant” effects are found – or if the underlying physics is thereby basically improved and the underlying basis of the code would be made more sound. Code development will also be needed as precise and quantitative beam analysis techniques are developed under the space charge physics framework discussed in 1.) above.

- Be informed in detail about all on-going and planned work on the PARMILA-class codes at Saclay.

- Include code development in the proposed LDRD/ORNL proposal.

## 8. Beam dynamics sensitivity to cavity tilt errors

This concerns a question raised by Claus Rohde of JLAB at the ASAC meeting. ASAC noted it, but LANL did not. The ASAC report recorded that there might be "the possibility of significant beam dynamics sensitivity to cavity tilt errors". It turns out that "tilt errors" were interpreted by LANL to mean tilts in the electric field distribution of a superconducting module, and it was noted at the LANL SNS Division review that such errors had been found to be insignificant. I, and I think Helen Edwards at the Division review, remembered the question as concerning superconducting module or cryostat alignment tilt errors. Check with Claus Rohde.

## 9. Model BNL group with collaborations, consultants, workshops, papers, …

The ASAC Report states

> "The accumulator ring beam dynamics group has reached out to the SNS partner laboratories for significant contributions. In addition, they have sponsored workshops, engaged consultants, and written extensively about their work. We recommend this as a model for the linac beam dynamics. In particular, an SNS project-wide linac dynamics team is important. It would engage more people, possibly bring in new semi-analytical or simulation approaches for understanding beam halo, and develop staff that will be needed for continued study well through the commissioning phase."

ASAC felt this is crucial to the long-term needs of the project through the commissioning of SNS.

The Division review indicated that this recommendation has been heard and that increased interactions are underway – a HEBT interface workshop, a beam dynamics workshop in November (no proceedings) and another scheduled for Spring 2001, expanding collaboration with Saclay, involving ORNL staff in commissioning studies in mentoring role, plans for interacting with consultants, writing papers for the PAC.

- Recommendations:

  - Engage ORNL staff in really significant ongoing activities. Some of them may be relatively young and inexperienced, but all are not, and those that are can work hard, learn quickly, and make real contributions to pressing problems. Also, commissioning is no trivial task and is no place to relegate young and inexperienced staff for its planning.

  - Involve ORNL staff at LANL. Develop close working relationship also at the leadership level of the person having overall project responsibility for beam dynamics and with the ORNL Linac Group. There was no evidence of this at ASAC meeting or the LANL SNS Division review (granted this is intended primarily as a LANL internal review).

  - Set up a system of internal SNS Linac Technical Reports, and insist that people write full reports. Nothing clarifies thinking like this does. The present culture of documenting only with vugrafs, without even facing text, seems not to constitute good scientific or technological procedure. (Granted that the excessive review climate severely exacerbates this problem. Maybe if reviewers could be referred to good reports they wouldn't have to come so often…?)

## 10. Supporting evidence that the SNS Linac is beam current insensitive

It is important for commissioning that the linac be operable from very small to full peak current. As outlined above, the linac tunes and potential resonance interactions change with current, and it can be expected that matching and other features might have to be readjusted with current. This in itself may not be much trouble with the computer control system. But a design relatively insensitive to current minimizes this retuning and is desirable. It has been stated that the SNS Linac design is current insensitive, but no quantitative data or interpretive discussion has been presented.

## 11. EMQs in DTL

It has probably been a misunderstanding by the LANL team until recently that ASAC is concerned about lack of flexibility in adjusting the whole transverse focusing pattern in the linac up to the end of the DTL if PMQs are used – and not in some ability to rematch the beam, although this is not an insubstantial consideration. The thrust of this concern was also communicated in writing by me to Jim Stovall in early April 2000.

LANL has decided to continue with PMQs and propose a backup position, in which some EMQs would be designed for doing trim matching. The EMQs would be prototyped, and intended for installation during or after commissioning should experience with the actual beam indicate the desirability.

It was also mentioned by Jim Billen in a separate section of the review that the drift tubes in Tanks 1 and 2 are really too short to engineer EMQs for, but that EMQs might be able to be used starting in Tank 3 and certainly in Tanks 4, 5 and 6. There are about 100 drift tubes requiring quads in Tanks 1 and 2, and about 100 in the remaining tanks. Presumably these EMQs could also be substituted later during or after commissioning. Personally I would favor this approach if it is felt to be to late to change now.

It should be noted that shutting the machine down for such installation during commissioning will not be popular. Also, the tanks and supporting utilities should be designed now to minimize the downtime that would be required.

- Recommendations:
    - LANL should formally propose their recommendation to the project.

## 12. Higher order modes – serious beam dynamics calculations

Higher order cavity modes are a problem in electron machines. The APT program (I believe Dominic Chan) evaluated the question for protons, and concluded that the modes start being seen at the 5-10% level in the fields at energies above about 1 GeV, if I remember correctly. A reference should be provided when this question comes up.

## 13. Prototype SRF coupler tests

ASAC applauded collaboration between JLAB and LANL proposed for the prototype SRF coupler tests. Many questions relating to this and to the tests of the production couplers apparently remain – this seems to be a project issue involving costs, schedules and resources more than technical ones.

## 14. Over-voltage test of SRF klystrons

Many klystrons at the high-energy end of the SRF linac must operate at 100% of their rated capacity – not attractive for reliability or future upgrade potential. ASAC suggested that the tubes be tested at ~120% of their rating for acceptance. LANL has learned from vendors that the vendors probably will not agree to tests at 120%, but may agree to 110%. What seems possible, and associated costs, should be formally proposed to the project.

## 15. HVCM headroom

I was involved with the Beam Dynamics team and did not hear this presentation. Risks, costs, and proposed handling of the issue should be clearly and formally communicated to the project, preferably accompanied with a comprehensive written technical report.

## 16. Strategy: (a.) conservative decisions (large aperture), (b.) flexibility (EMQs) and tunability

These were ASAC recommendations on strategy.

### Other

1) Requested a copy of November beam dynamics workshop proceedings. Jim Stovall sent me a note on what occurred – there are no proceedings.

2) I think it is very good if ASAC or other reviewers are asked to spend extra time between meeting at the various labs to directly discuss issues. The ASAC reviews address many concerns at the project level as well as detailed technical issues, and are always very busy.

3) It would seem that the ASAC report should be reviewed at project level and formal interpretation and requests for response sent to the appropriate work package managers, who would then formally respond back to the project before the next ASAC meeting.

4) RAM has been emphasized by David Moncton as a primary project goal, with a very high overall availability goal that translates into about 99% availability for the Linac. ASAC believes this specification requires considerations beyond the usual design and engineering practice used to date. There is no evidence (anywhere in the project) that such consideration is being given, and indeed it is difficult when everything is driven by very strong cost constraints. The project should be formally asked to support or drop this specification.

5) Presentations to review committees should always review the most important specifications, such as the beam loss issue in the linac, and make some comment about the status. It is hard for reviewers who meet only once or twice a year to remember what might have been presented in detail several reviews ago and not mentioned since, and reassuring to them to hear that the most important specifications are always in view.

6) I reiterate the need to find a way to get support for long-term development needed through the commissioning stage, such as "understanding beam halo". The recommendations of the LANL SNS Division review committee for proposing LANL LDRD matched by SNS project funds is excellent. I also found the phrasing of the Review committee's recommendations to be excellent.

(Note 2018: The ASAC tried hard, but essentially almost all of the above was ignored.)

[toc]

## Appendix 5 – Concerning Fast Communication and Possibly Later Formal Publication

The initial draft title of this appendix was "A Diatribe, Polemic, about the abuse of the "review system": what could be done?" Several of us concentrated in 2022 on "what could be done".

The accelerator field communication and publishing situation has changed very much since the 1960's, when everything was quite new. There was some short-papers conference reporting and no particularly dedicated journals. The best expositions were usually "internal" laboratory reports, with access under some restrictions but it was possible – one felt very privileged to be given access. Then came an attempt or two to establish a dedicated journal, and finally success under the strong established umbrella of the APS. But this avenue has proven to have very well-known problems. Journals have no mechanisms for fostering rapid communication which promotes discussion and collaborations. The major part of their slowness, and the major, very well-known problem in general is the "review" system.

The situation is perhaps not so bad as commonly felt. Other fields, many institutions, many authors, also in the accelerator field, have by now found workarounds, even under powerful umbrellas. If this framework could be expanded and supported for general use, a great improvement would be accomplished. Some non-emotional, non-egotistical, non-political improvements might be considered.

The discussion will be *split* into two essentially independent aspects – to have a recognized and well-used *world* mechanism for rapid communication, and the separate aspect of formal publishing.

**The Formal Publishing Aspect**

This sub-section is very long, being developed over many years. But it attempts to offer some suggestion in the spirit of not trying to change anything very much, to not stir up anything emotional of egotistical. It is oriented to the concerns of young, serious, researchers and practitioners.

It is necessary to speak out about the system of "reviewing" author's attempts to express their work and opinions, after observing for so many years its extremely egregious effects. These remarks are a distillation of personal experiences with students and many world-wide CERN-style coffee conversations.

*It is important to highlight and point out that the discussion here is primarily over the narrow, small niche of intense ion beam dynamics with space charge – narrow, but includes design, simulation, theory, analysis and practical application.* Here the problem is particularly bad, often discussed. In other areas of particle accelerator technology, for example engineering, or in other fields, it may not be as bad, but it is widely acknowledged to be a problem, coupled with the challenges of education and publishing in the electronic age. Again, here, a kernel of the problem seems to lie in having a too-small, technically out-dated, not "knowledgeable" or "helpful" (see below) stable of reviewers – a stable which apparently is, perhaps because of no alternative, also now being accessed by journals that until now had a better reputation for "knowledgeable" review.

"Reviewers" are supposed to be qualified as "international experts", whether when requested as reviewers, or self-claimed. However, if the present level of the "reviewers" state-of-the-art – the level of expertise – is being developed to a higher level, being at present even "an international expert (at present)" is not enough. To be unbiased and helpful in the actual responsibility of a "reviewer" – that is, to help an author present his work well – the "reviewer" needs to be "a knowledgeable person" – able to get outside his box and be unbiased and helpful. There can be no real evaluation, discussion, testing of higher level expertise if papers that present new views are just rejected for publication because they do not fit into some perceived, outdated box. Later, in "knowledgeable and helpful" review, in particular, papers claiming to compare something new against the old state-of-the-art can be reviewed with an eye toward whether the comparison has actually been tested or not, etc.

It is not the job of an author to educate a "reviewer". If the "reviewer" can sense that something is being presented that is beyond his own present "state-of-the art", then a review should be done in the frame of being helpful (or, if time for hard self-learning is not available, the "reviewer" should decline). It is the job of the journal (its editorial stable) to require its "review stable" to conform to real standards – first and foremost, to be "knowledgeable and helpful" – plus to a set of rules that prevent misuse, either semantically or other.

It is necessary for the young people to learn at some point that "science" is often not the lofty search for truth, and that if they wish to, or these days must, "publish", they must persevere, learn the pitfalls, and not be demolished by "the review process". The word "demolished" is carefully chosen – in several cases has it been necessary to scrape up the pieces and help someone go on, after an attack that clearly went beyond a helpful review of scientific content, or even non-scientific matters of taste.

Unfortunately having "reviewed" papers has become overly important for young people's careers. The criticism is well known – only one recent reference from the physics area is cited here [82]– it is not just in physics. In "the old days of accelerator technology", we were anyway considered by the pundits of particle physics to be just technicians; "accelerators and beams" was not considered by "physics" as a science in any way. We published among ourselves by unreviewed internal laboratory reports and monographs (and were very grateful when other labs allowed us to have copies of their internal reports), and at conferences. We knew that one has to judge for oneself about the merit of a paper. And we made it possible for the ungrateful ones to do their experiments. Now we have a journal, but with all the problems of the review system, stemming from within the field itself.

Only occasionally, and then very gratefully, is a review helpful in clarifying the intended expression.

Very rarely is a reviewer qualified, or has or takes the time, to do a helpful review.

Very often, there seem to be competitive, political or egotistical motives behind a "review". Even, that there are clearly known to be.

***<u>Especially if the paper proposes something new. Or a new view.</u>***

So often, research progress is stimulated by someone grasping a concept that helps him finally understand the problem he has been struggling with and allows him to break through. But statement of a concept in a paper is stated by "reviewers" to be inappropriate. The review system, mired in tradition, wants re-derivation of the whole background of its own field, the whole background of another, interdisciplinary field if the concept comes from "outside" the immediate field.

Please read again Parts A & B of "Elements of Linear Accelerators, above.

Am author often wants to present something succinctly, without unnecessary padding. Outlined further below, present "reviewer" tendencies work in the opposite direction. All of which then would make the paper too long – bad enough these days when long papers do not get read, and it would be good if new work could be presented for discussion by a short, forceful paper to communicate it and as a challenge for future work.

It is impossible to build up the background using a series of papers, culminating in the new work. Hardly anyone, not to mention "reviewers", would follow the series, read and comprehend the succession, and in each succeeding paper, all the preceding background would again be demanded.

Very often, "reviewer" remarks are given in a condescending and/or insulting way. This is especially destructive for young people who are trying to become "the next generation" and need to learn and express “what everybody knows” in their own and often importantly extended way. I have too much experience in the time and care it takes to get a student re-motivated enough to hope to carry on. In these cases, the fear of competition and territorial defense often become very evident in the “review”.

---

82 "In Referees We Trust?, https://physicstoday.scitation.org/doi/full/10.1063/PT.3.3463, Feb. 2017 (vol. 70, Issue 2), p.44

It is often more sensible to publish outside “the field itself” – in journals related to instrumentation, plasmas, engineering, control theory, mathematics, nonlinear systems, optimization. But that does not help advance “the field itself”, because rarely does anyone read “outside their field”. (It seems even worse – little real reading, comprehension, or thought, not to mention real collaboration, even inside the field itself.)

It is telling that the world-wide International Committee on Future Accelerator (ICFA) publishes primarily outside the (USA-based) dedicated journal and its traditional close neighbors.

Clearly for the next generations at least, something should be done. *For the journals, the above are the responsibility of editorial policy.* Open access without review is clearly acceptable. One must judge an article for one’s self in any case – whether it has been reviewed or not. The idea that an article is correct just because it has been “reviewed” is clearly nonsense.

***Maybe a helpful reminder:***

The article that made us realize that a great advance had been made – the RFQ – was maybe well written in its original Russian, but the translation we had was very poor, without re-derivation of all of the Kapchinsky book, without many examples or figures, *in other words not able to withstand review* – but included a sentence to the effect: “This device allows acceleration of low energy intense ion beams with very little loss of transmission and very little emittance growth.” *One sentence, once comprehended, that changed our world.*

***Non-technical review aspects***

The style requirements of a journal are non-technical but considered appropriate, as usually clearly stated, non-personal.

There are many non-technical aspects of a paper that are clearly non-technical and a matter of taste, *but are now acceptable grounds for rejection* by a “reviewer” and also editor. Journals and editors should not accept rejections based on these aspects – they are too easily used for unethical purposes, and often almost impossible to sensibly respond to.

Flagrant examples include:

• Relation to concepts, as above. Often, the “reviewer” is clearly hardly or not literate in the main subject being reviewed, and of course not in the field from which a valuable concept is found. The introduction of a concept in a paper is to stimulate the reader (perhaps also the “reviewer”) to investigate further as desired or required.

• Amount of theoretical formulae. It is easy for a “reviewer” to insist on more, or fewer, formulae. This is not a scientific condition.

The formal and arrogant style of theoreticians to use few formulas and leave out steps with statements like “it is easy to show”, “clearly”, “obviously”, “easily proved” has been criticized often, but even without these egotistical garnishments, including all the steps would make a paper unacceptably long.

The recent escape from the “smooth approximation” box, that has severely constrained the study of beam dynamics involving resonance modes and beams with space charge, required many more pages of algebra. Although the synopsis of the theory presented in a paper clearly stated this, a “reviewer” rejected the paper on the grounds of not presenting the full theory (while also clearly not at all comprehending the main thrust of the paper). Upon reminder of the situation as already stated in the paper that many pages would be needed, expanded to point out that at least 20 pages of formulae would be needed, the "reviewer" admitted he "had missed that on the first reading", and also came to see the main thrust… A final expanded paper was published, but the authors felt that unnecessary dilutions severely reduced its potential impact.

A style to purposely use a minimum of, or no, formulas is an effective way to present a case without superfluous complication, as caused by obtuse formulas, or formulae with missing steps.

The point is that the number of formulae presented is a matter of purpose or taste, and not an appropriate matter for acceptance or rejection. The author should be competent to decide how many formulae are required to set the basis and stimulate further investigation. The editorial policy is responsible for insisting on this.

• Amount and content of examples. The same applies especially to examples, for instance, the simulation examples chosen to illustrate. A few pointed examples to announce new, original, work are considered not enough – the review system wants every case to be considered. And considered to some ultimate degree – for example, an explanation of zero beam current acceptance, which is obviously larger than acceptance with space charge and therefore clearly covers the concept when space charge is added, is rejected and evidence of the final answer with space charge is required.

The author clearly knows his subject and chooses appropriate examples, which the reader can add to. If the "reviewer" does not have the ability to interpret examples to the main subject, that is his problem. Suggestion that another or added example could specifically help clarify a particular point could be scientifically helpful. But non-scientific requirements for a very broad number of examples, or examples that clearly belong to a future extension of the main subject, are not appropriate grounds for acceptance or rejection, and should not be allowed by editors.

• Amount and choice of references. "Reviewers" often include choice of and amount of references as grounds for rejection. This is non-scientific and should be summarily rejected by journals and editors.

"Reviewers" often reject on the basis that they would have to *read* a reference in order to see the background given in the reference. Knowledgeable readers know they have to do that, and understand that that is the purpose of a reference. A paper cannot have the book-length necessary to train the uninitiated, and such should decline to review.

"Reviewers" object to self-references [83]. If the subject is new, or controversial, and the authors are trying to expose their point of view, self-references are necessary and appropriate. As it is very difficult in the conservative "accelerator physics" to introduce new concepts, authors might try to introduce new ideas piece-wise in a series of articles and not too quickly. This makes it necessary to read references… The series approach has not worked – "reviewers" do not read references, and are not familiar enough with the literature to follow a series (i.e., or even with the previous work of an author).

"Reviewers" often have the temerity, but also the naivety, to suggest references, which are well known to the authors and the initiated to actually be weak papers or irrelevant to the discussion, which of course accentuates his incompetence as a "reviewer" [84].

• Notation. Notation in the accelerator community is not well standardized, although attempts have been made. Many authors egotistically insist on their own notation; papers and even book publications have been accepted by "reviewers" and editors. It is of course appropriate and required that notation be defined and self-contained in a scientific publication. Especially for something new, a new notation, or restructuring of a previous less-clear notation, can be helpful and even necessary. But acceptance or rejection on the basis of some "preferred" notation is not acceptable.

---

83 "Reviewer objection: "7 out of 18 references are self references, indicating the author is not very familiar with work outside of their own. *!!!!!!*

84 For example, the same "reviewer" as in the previous footnote!

• Terminology. The same applies here - Terminology in the accelerator community is not well standardized, although attempts have been made. Many authors egotistically insist on their own terminology, even if it is clearly misleading; papers and even book publications have been accepted by "reviewers" and editors. It is of course appropriate and required that terminology be defined and self-contained in a scientific publication. Especially for something new; a new terminology, or restructuring of a previous less-clear terminology, may be a main point of the manuscript. But acceptance or rejection on the basis of some "preferred" terminology is not acceptable. [85]

• Assessment of reader potential This may be an appropriate acceptance or rejection criterion for a journal like PRL or Nature. But PRAB is supposed to be a journal for the physics (and engineering!) of accelerators and beams, with very many narrow subject fields – even the editor should hardly take it upon himself to reject a publication on the grounds that only certain readers might be interested. And especially not in a niche like resonant interactions of beams with space charge – a clearly active, controversial, much published niche with little coherency. In this case, the ethical grounds for a "reviewer" rejection recommendation [86] are extremely suspect. As a matter of fact, in some cases a dead giveaway, because, based on completely reliable sources, the identity of the "reviewer" is known, and it is known that he and a few other "proprietors" of this niche are colluding in using this and other unworthy methods to stymie competitors and newcomers. A strong indictment but true. In other cases such can only be clearly suspected from the quality of the review statement.

***Technical review aspects***

"Knowledgeable" reviewer comments, questions, requests for clearer exposition or clarification, on actually technical aspects are completely appropriate and welcome.

Hardly known breadth and depth aspects of the subject of resonant interactions of beams with space charge are outlined throughout "Elements of Linear Accelerators". The obvious problem with the common "reviews" of this subject over the present and past decades is that of "the blind studying the elephant". All too obviously, "reviewers" are not well, or even not at all, informed on the subject, certainly not in the direct context of the vast literature on this niche subject – their "state-of-the-art: is inadequate. Thus, they should not accept invitations to review.

A review is not requested to afford "reviewers" to stroke their own egos. They often clearly do not feel that their obligation is to help an author present his views. They do not constructively review a paper on its own merits. They do not read carefully enough to grasp the essential point of the paper, even when clearly stated in the abstract, introduction and conclusion.

---

85 Reviewer" objection to terminology of "FD" to a lattice type: "Is FD a FODO cell? If so use FODO as it is common used in the accelerator community. FD is not."
*[ !!!!!!!! ?????????? Remember, this is a "review" from an "international expert". Is a FODO one cell? Is it really necessary in every paper to explain the fill factor? Is this guy even an "accelerator" person? Have no record of the context now - a response might have been to change to "an alternating gradient lattice" - but this kind of naivety gives no confidence that a knowledgeable review is being received. ]*

86 "The authors gave a very detailed reply to all my comments and objections. I am convinced that their research is rigorous and that the results are correct, thus they achieved an advance in this particular field. However, I believe that this field is very narrow and that it requires a very deep knowledge to fully appreciate that advance and to maybe transform it into a visible gain w.r.t. machine design and performance. The number of potential readers might be very small. Hence I feel that the manuscript does not match the journal's claim regarding impact and accessibility for the broader readership."
*[In other words, PRAB is not the right journal to publish anything requiring deep knowledge!!! Is the "reviewer" insinuating that only he might have such, or admitting he does not?? This is the classic "Chinese rejection letter" – your super paper is too good for our humble journal. **But that an editor would let such a review be forwarded** speaks volumes about the magnitude, depth and breadth of the problem.]*

They raise erroneous, misleading, trivial, preposterous, irrelevant points. Many direct examples from "reviews" or in papers could be listed here (far beyond mine or one I was associated with); they are embarrassing and at first I hesitated to list any. But ok, a few examples might be given in terms of the points already made herein, with a special format – the embarrassing "review" comments are given as footnotes, with a brief outline of needed education in the main text. It would be essentially impossible to educate enough in a response to such a "review".

- with respect to application to linacs:

[87]: Uninformed opinions or observations should not be stated as general guidance. Many operating low-beta linacs, especially RFQs, DTLs, but even long linac designs to several hundred MeV, do or could operate at transverse tune depressions much lower than 0.8, for example to exploit economical advantages. There is no problem to design a linac with transverse tune depression down to 0.4 and even lower, with small beam loss, if equipartitioning is intelligently applied. There are **elements,** including economics involved in selection of the tune depression. In room-temperature ion linacs, there is a direct trade-off between length and amount of rf power, with short length very often emphasized (often without realizing the economic tradeoff); operating at lower transverse tune depression is a strong factor in reducing length. There are differences between ion, electron, room-temperature, superconducting linacs in this respect. Other lattices, e.g. containing nonlinear elements or strongly oscillating tunes, would afford other choices with respect to tune depression. Such studies would be real linac studies.

The following sentences in the review paragraph are grammatically loose, as they pertain only to very small tune depression ~0.8 although this is not emphasized. A practical designer would never place a trajectory inside a stopband long enough for any deleterious result, so statements about "inside the bands of instability" are irrelevant. For linac use, the "growth rates" quoted are irrelevant and the statements are wrong. What is the intended meaning of "and if at the same time T is (sufficiently) above unity …"?

An important technical point: "Temperature", widely understood as a measurement of a thermal quantity, is an inappropriate term for ion linacs or electron linacs at low beta, where the dynamics is without question not thermodynamic (there is no final steady state), nor is it wished to be, nor it is correct to use thermodynamic temperature (i.e. the space charge limit) for design purposes. "T" here refers to the equipartitioning ratio, not temperature. Much time and energy have been wasted on this, including at very recent workshops and conferences, and hanging on to implications to, or use of, "temperature", or the directly inferring nomenclature "T", is at least misleading, and seems to stem also from other reasons.

---

[87] ... for instance, the transverse (y ) tune depression must be below 0.6 (hence the longitudinal one is even significantly smaller according to Fig. 9) to enter into the unstable region of the third order (nonoscillatory) even mode, and even lower for the fourth order (nonoscillatory) even mode. This is hardly the case in any practical design, where one finds the transverse tune depression closer to 0.8. The oscillatory instabilities left of T=1 (see Figs. 11 and 12) have (normalized) growth rates limited to 0.05. This corresponds to the relatively long e -folding time of 20 periods of betatron oscillation (defined without space charge) . The narrow spikes of odd mode instabilities near ½x/½y = 0.5 and ½x/½y = 0.33 are also expected to be harmless. Growth rates inside the bands of instability as well as the width of these bands increase gradually if the transverse tune depression drops below 0.6, and if at the same time T is (sufficiently) above unity (or T sufficiently below unity if the emittance ratio is reversed). The normalized growth rates of 0.25 reached for the transverse tune depression of 0.3 (Fig.12) would result in an e-folding time of four periods of betatron oscillation (defined without space charge). These are also, roughly speaking, the peak values of growth rates we have found for a variety of parameters. ...

[88] - the classic oxymoron. This is much more important, see “Elements…” Sec. 1.7.4, which also exposes many other misuses of the scientific method. How can it be that physics training resists the advantages of an equilibrium?

There is (no) mention that if the beam is or nearly EP that the EP effect kills or makes the resonance effect smaller - e.g. Fig. 2 if EP had no free energy to feed the resonance at kz/kx=0.5 what would happen? EP is a local, instantaneous condition, and will affect a resonance – and tails from other resonances - over a space-charge broadened tune band. This additional "resonance-free" region exists because EP is in effect!

• basic dimensional analysis is useful:

[89] Phase advance by itself has no meaning. The simplest, most direct, most useful for analysis and design expressions for phase advance are the envelope equations, relating emittance = (beamsize)^2*((phase advance)/(unit length)), where dimensional analysis shows how the second term on the right works. This kind of “review” looseness contributes to the confusion about phase advance that is currently beginning to appear as an awareness of how actual linacs are laid out begins to spread.

• on analysis of phenomena, trajectories:

Steady state methods, such as Poincaré plots and other non-self-consistent methods such as test points, in general need to be regarded as providing something like a “roadmap”, and may reveal useful summary characteristics. But can be very misleading, as they do not reveal what is actually going on, like “how to get from point A to B on the roadmap?”, “how does a real beam arrive at point A (how does it acquire the initial conditions of point A)?”. Plotting at homologous points and generalizing from this is very dangerous. The actual dynamics is self-consistent, local (instantaneous).

• on the actual (experimental)/simulation/theory hierarchy:

[90] Ok, but hardly convincingly, usefully, accepted generally; and not at all experimentally ??, or for detailed analysis and design ?? The “reviewer” is clearly only a little bit familiar with the subject, and missed the point of the local state point of view, which was a key emphasis of the paper in question.

• on the problem of equations:

[91] Not familiar with the subject at all. The equations presented as background for the main theme were the standard sequence already used by numerous authors, there was an equation (2), etc., steps were omitted, added or modified steps were emphasized. How can such be dealt with in a reply??

---

88 " Sufficiently large resonance-free regions exist on the stability charts to make non-equipartitioned design work".

89 “ … the phase advances (in particle, envelope, and collective mode) themselves are dimensionless …”

90 “The connection between resonance mechanisms and halo formation in high current transport lines or accelerators has long been established both numerically and analytically.”

91 “I should also add that the effective Hamiltonian reported in Eq. (7), with space-charge potential as in Eq. (1), is not even of the relevant form for the problem at hand, as it would only apply to particle motion within the beam core. Much of the relevant halo particle dynamics, however, happens outside the beam core where the space charge potential behaves asymptotically as 1/r, inconsistent with the form assumed in Eq. (1).

• on the study of collective modes:

[92] The "reviewer" is essentially unfamiliar with the subject, is deluded that only one experimental measurement has been made, and should have declined to "review", but in spite of that he poses himself as somehow an expert. How would it be possible to respond to such nonsense?

If such "reviewer" observations were to be reviewed by other reviewers, the other reviewers would certainly be often amazed, often shocked, find it necessary to reject them. This would make clear the problems that the authors are faced with. Some journals are currently trying to address this problem by trying to set up a review system in which reviewers must act as a committee. This could have some real advantage – it would be a great way to educate reviewers, which is hard for authors to do in each single paper. But the time burden on reviewers would be even more severe. And "everyone knows" that committee decisions are usually driven by non-technical factors and suboptimal.

The method described in Ch. 29, to improve the bunching process and achieve lower longitudinal emittance, is important, so we decided it should be announced via a publication. This afforded a very up-to-date (2022) view of these problems. It was informative that a senior editor was also involved. The process again revealed the gravity of the situation regarding the breadth and depth of current "international expert expertise", with breath-taking naivety and arrogance, but with no objection at the editorial level. The details are too embarrassing to outline here; the situation is alarming.

The paper was then submitted to another respected journal, reviewed and immediately approved, with only the note "ok for publication!" (! original, not added…)

***What could be done?***

The accelerator world is too small, too project-driven, and especially too stove-piped in the traditional "accelerator physics" box to expect a well-focused *multi-disciplinary* (not "multiphysics" hubris) educational or review system. The publication problem can be improved and is anyway not as important as the educational problem. How to achieve the terminology change to something like "accelerator multidisciplinary technology" is worth much study and implementation. A really major problem for the near future is for it to be taken up by multidisciplinary people.

Going on with the review system, good reviewers would be expert and up-to-date, able to comprehend the thrust of a submission, able to perceive its present and future impacts, above conflicts of interest, and with a helpful attitude. But good reviewers are very few, and would be the busiest and most productive people – having little extra time and usually cannot afford to do much extra by way of reviews. So the question is: what can be improved with the handicap of "hobby reviewers" or worse…

---

[92] *"There are contradictions in the scientific method used in the manuscript. In the manuscript authors compare theoretical KV modes growth rate with WB simulations in a strange mix to claim a delicate mechanism: this seems quite an artificial procedure. …. The use of the WB distribution is somehow ambiguous: a KV beam is dominated by the instability of the KV-modes; a WB beam is still a little dominated by KV-like modes, but in addition it has also 4th order component of the force. Why didn't authors use a Gaussian beam? With a Gaussian beam the KV-like modes would be significantly suppressed and the discussion of 2nd/4th coherent-incoherent resonance would be more difficult. …. The subject the manuscript addresses is scientifically interesting, but it is not clear if it has any application: 4th order structure resonance has been experimentally measured at GSI (by LG (Ed.)); coherent 2nd order resonances was never measured by no one. The manuscript is scientifically misleading because of using WB simulations and not comparing them with a WB Vlasov-theory, but rather with a KV Vlasov-theory. Considering that no experimental evidence of the mechanism claimed is provided, I recommend rejection"*

"Field" suggests analogy to a field of earth for crops. Crops cannot grow if the field is blanketed by layers of whatever – layers of uncomposted old crops, layers of plastic… A new plant might sprout, but will be stymied trying to reach the sun. The accelerator beam dynamics field still has many open questions and possibilities for new approaches, but is suffocated.

The problems with the review systems and their clear present and future potential for abuse are impossible to overcome. Ask your colleagues or "knowledgeable" acquaintances directly to give you a helpful and constructive review. "Let the reader beware" is necessary in any case.

It is proposed to just separate the presentation of technical work, into a fast, open framework able to foster the state-of-the-art, and the journal framework as auxiliary, perhaps more archival. Therefore the following discussion below on Fast Communication, in the framework that it is not necessary to change anything at all in the journal system, and that open source can be framed as a complementary, parallel tool with separate advantages and goals.

Fortunately, the new wave is open publishing, with very good and reputable platforms. Savvy people already use arXiv and ResearchGate, etc. to present their work and also as sources. They are the people that one wants to reach anyway.

My opinion is that this is the best way for authors to present their work.

**However, some things might be introduced without controversy within the present journal system.**

**First step – improve some hindrances.**

• Set up denial procedures for non-original, repetitive submissions, e.g. submissions that are just grind-the-handle, use-the-tools, MS level work, (e.g. CST, or to develop another rote design). (That many PhD degrees have been and are being granted for such reflects on the student, but much more on the supervisor. The result is a Myridon, with knowledge of an extremely narrow subject and little else. Such used to be a partial requirement for an apprentice, or M.S. level degree, and the PhD degree required understanding at a level which produced an original new contribution.)

• Set up procedures for "project summaries" submissions – which are informative and appropriate, but which are really not necessary to be in a journal. Conferences papers are more appropriate.

• Or maybe set up an "Announcements" section, to allow notifications that work is in progress at this or that institution, with authors, titles, abstracts only, for anyone interested to contact directly, or important cross-references to the fast communication platform.

This would free up journal space. Granted that there might be less stuff to publish, but might be surprisingly refreshing.

**Second step – identify possible improvements.**

• Define criteria for editors and reviewers

- Remove *completely* the requirement that "Suggested reviewers" "should be *the* international experts in this field, …" Leave the requirement "should not have a conflict of interest with the current work". All, including the blind men investigating the elephant, all those "cited" herein, consider themselves international experts. And blind men automatically have built-in conflict-of-interest of one form or another. And a lot of their severely bounded papers have been "published".

- Use the term "Knowledgeable" as the suggested reviewer requirement. Able to sense the technical content, and be interested in helping authors get their original work communicated. Unfortunately such people are also very rare and very busy.

***I think that this one suggestion – to reorient in terms of "knowledgeable" and follow that up in detail – would be a very fundamental and helpful reform – and can be done within the boundaries of the present review system. Eliminate completely the words "*the international experts*".***

• Define guidelines, rules, for editors and reviewers

- Include specific instructions that non-original, repetitive submissions, e.g. submissions that are just grind-the-handle, use-the-tools, (MS level), work, or that use tools (e.g. simulation programs) known to be inaccurate, below the state-of the-art, should be declined.

- Define realistic paths for presentation of continuing work.
- Define realistic paths for presentation of new, original, innovative work.
- Define realistic paths for handling presentations to "experts" with "very deep knowledge"..

- Short 3-page conference papers are effective for presenting new, continuing work, but turning them into accepted journal papers is not effective – largely because of knowing that a lot of time with poor reviews will have to be wasted and face poor chance of success.

- Set up a forum for monographs

- Not the same as a review article.

- Previously in accelerator technology, these were the carefully and fully written internal laboratory reports. Probably such are rarely written any more – just conference papers or talks with transparencies without accompanying text.

- Transparencies without accompanying text are not acceptable.

- Do not think that super-critical search for weeds is necessary. A great many reviewed papers are not good papers. Let the readers decide – their responsibility anyway.

***Reform of Review System cannot be expected***

That would imply reformed selection of a journal's reviewer stable.

- not possible – egos, no "expert" would dare to rate other "experts", etc.
- Would immediately reduce to a very small stable, then zero because such people would not have time.

Review stable would have to undergo training… - (technically, beyond the editorial framework, which *should* be trained and rigorously observed).

Reviews would have to pass through Editor(s) or a "Review" Review Panel) and then to the Authors. No one would agree to be an editor with such a load, or to be on such a panel.

Jury reviews not practical – no doubt now proven.

Author suggestion of reviewers – better than random from stable more interested in demonstrating their "expertise", could be improved by requiring authors to suggest from a "knowledgeable" stable. But although apparently being tried, authors' suggestions are not used anyway…

***Conferences are also a slow metho***d, only available when a conference is held, one of the authors must attend, and proceedings are often delayed.

**The Fast Communication Aspect**

The idea is how to circumvent the "review" problem entirely - that is the central point. Any change or even change of the journal system is improbable, and any try would stir up all kinds of emotional problems.

As found for particle physics, the accelerator field would benefit very much from the specific development, promotion and continued and improved use of a dedicated, communication-oriented, fast platform. [93]

An internationally organized, visible and focused effort could be made to get people together on a fast communication system for new, innovative, for experts only work, for communicating goals, plans, and progress toward those.

---

93 It was really exciting and useful during the late 1980's - 1990's when chaos and nonlinear systems were breaking out, to check the internet every day. Traveling, so pages of notes in tiny handwriting.

Controversy and perceived competition is avoided if a complementary, parallel, already existing under powerful umbrellas, and *non-competing* tool is available with separate advantages and goals from the journal system. Some non-emotional, non-egotistical, non-political improvements might also be considered specifically for the accelerator field. Fortunately this is completely possible.

For that, arXiv is already being used to some extent as the vehicle, established, impartial, easy to use, was established for exactly this purpose and because of same problems with particle physics publishing. Its use is even promoted by some journals already. arXiv posts would centralize - independent work, preprints, modifications to articles in publishing process, final publications - exactly as it was so wisely set up to do. arXiv is impartial, already international and using it should not introduce rivalries and politics. Everybody would post to arXiv, and then follow whatever route for further publishing or communication. No politics, no delay.

ICFA, as an already established international platform, could be the sponsor of the communication effort and push it worldwide. (Clearly need an international platform – not associated with any one nation.) Kind of an obvious, non-controversial, but not yet thought of, addition to the ICFA charter and mission.

The initial step would be to formally imbed the fast communication approach into the existing ICFA or one of its Panels. Then, through ICFA, enhancement of appropriate arXiv categories. Then the effort to get international accelerator community (both individuals and institutions) on board for the idea that a central "bulletin board" could and would really help to unify, integrate, consolidate, inform and spur new work, possibly even collaborations, in many ways.

Getting the idea implemented would be offered as emphasizing “fast communication”. It should also avoid or at least should help with the problem of an institution's bureaucracy - part of the work in getting it set up would be to try to persuade the institution that timely posting preprints, working papers, promoting open discussion, etc, is good for everyone, and the institution should then follow up by encouraging individuals to formally publish (good for the authors and for the institution's statistics). I think individuals would easily welcome the idea.

The idea avoids any proposal to change anything about the journal and review system - that is the whole merit of the idea! It is not that one gets hard reviews - informed reviews with intent to help getting published are welcome and helpful. Uninformed, and/or not useful, nit-picking reviews are a widespread problem, not just for us. And the process takes far too long - stuff has gotten very stale. The parallel approach is non-competitive for new stuff, stuff still in development, stuff “for experts”. Particle physics and other fields that maybe move faster are able to communicate stuff fast; then decide later about formal publishing, with "review" etc. They had exactly the same problems, and solved them to their satisfaction with the arXiv idea.

The accelerator community would benefit from

- a central forum for fast communication.
- plus a central forum for getting new stuff, stuff still in development, out for discussion and comment.
- plus a central forum for synthesis of progress on current topics, like Grand Challenges of various countries.

A big advantage is that it would not have to start from scratch - use physics community's arXiv and the overall integration forum ICFA. Formal establishment of such would really freshen up the environment - let people be informed, invite timely discussion, objection, find out the extent to which a real teaching effort is unavoidable, etc. – and remove possibly perceived concerned or guilty feelings that “only journal published” articles are legitimate. It seems a straightforward and logical extension of the ICFA mission and forums.

Then whether to eventually formally publish is then up to each author (and institutions) independently.

So again, the idea is how to avoid addressing the "review" problem entirely, and to try to improve integration and fast communication - those are the central points. So, a new idea about how to significantly improve an existing situation, completely impartially, non-politically, non-emotionally, not stirring up any feathers, based on well established procedure in particle physics...

***A point regarding "Archives"***

Nothing electronic is safely archived.

[toc]

# Appendix 6 – CV & Publications

## Curriculum Vitae – Robert A. Jameson

Res. address: (US) 4 Comanche Lane, Los Alamos, NM 87544

Business Address: Institüt Angewandte Physik, Goethe Universität Frankfurt, Max-von-Laue Str. 1, D60438 Frankfurt-am-Main, Germany

E-mail: rajameson@protonmail.com

**Jameson, Robert A.**, was born in New York, USA, on May 3, 1937. Thomas Corners Elementary School, Scotia Jr. High and High School, Scotia, NY.

**Batchelor Degree in Electrical Engineering (EE)** University of Nebraska (1958), USAF 1958-1961, **Master Degree EE** University of Colorado 1962, **PHD EE** University of Colorado August 1965. The dissertation was devoted to the closed-loop control of the rf fields in a linear particle accelerator. During his activity at Los Alamos National Laboratory (1963-1999) R. Jameson was responsible for final installation, turn-on and commissioning of the LAMPF Proton accelerator (1969-1972), and worked as Group Leader of Accelerator Development Group (1972-1980), Division Leader of the Accelerator Technology Division (1978-1987). Dr. Jameson has worked extensively in laboratories in Japan and Germany, as Monbusho Visiting Professorship (1988-1989), Alexander von Humboldt Senior Research Award (1994-1996) (plus 2nd visit award 2009)), and RIKEN Eminent Scientist (2000-2001)). Guest Professor Kyoto University Accelerator Laboratory (Keage and Obaku) (1981-present), Guest Professor KEK/J-Parc (1988-2018), FY2008 JSPS Invitation Fellowship Program for Research in Japan, KEK Visiting Scientist 2010. Member of numerous review committees (most recently the SNS ASAC (2000-2006)). IEEE Centennial Medal - 1984, Fellow, IEEE (1991), Fellow, American Physical Society (1988). Accelerator Facility Team Leader in the International Fusion Materials Irradiation Facility Project, an international program of the IEA (1994-2006). Consultant to Oak Ridge National Laboratory, USA (2000-2006); Guest Professor at the Institute for Applied Physics; Goethe University, Frankfurt, Germany (1993-Present), Guest Scientist Los Alamos National Laboratory (1999-Present), Visiting Scientist RIKEN Radiation Laboratory (2002-2021), Guest Professor Inst. of Modern Physics (IMP), CAS, Lanzhou, China (2004-2018).

His background and main topics of interest are:

1. charged particle beam physics and energy exchange between beam and microwave structures;
2. linear particle accelerator design and technology;
3. accelerator rf systems.

Has more than 200 scientific publications. Served as a member of ICFA (International Committee of Future Accelerators) Working Groups, chaired International Particle Accelerator Conference, US Linear Accelerator Conference, served on numerous review committees,

Meishi 2005:

ROBERT A. JAMESON
Consultant, Accelerator Technology & Management

IFMIF Project Accelerator Team Leader, ORNL
Inst. Appl. Physics, Goethe Uni. Frankfurt
Inst. Phys. & Chem Res. (RIKEN)
Inst. Modern Physics, Lanzhou, China
Uni. Peking, Beijing, China
Retired, Los Alamos National Laboratory, 1963-1999

Inst Angew Physik, Uni Frankfurt, Max-von-Laue Str. 1, D60438
Frankfurt-am-Main, Germany Fax: 496979847407
Email: jameson@riken.jp, RIKEN Tel. 048-467-4397, Fax 048-462-4719

## **Bibliography - Robert A. Jameson**

✚ Registered in ResearchGate (✚ full-text uploaded, ✚ private full-text uploaded)

Conf. on Low Energy Ion Beams, Bath, England, April 14-17, 1980, Inst. of Phys. Conf. Ser. No. 54: Chapter 2, p. 54, April 1980; Los Alamos Scientific Laboratory Report LA-UR-80-1091, 9 April 1980.

✚ R. A. Jameson and R. S. Mills, "On Emittance Growth in Linear Accelerators", Proc. 10th Linear Accelerator Conf., Montauk, New York, September 10-14, 1979, Brookhaven National Laboratory Report BNL-51134, p.231, September 1980; Los Alamos Scientific Laboratory Report LA-UR-79-2541, 12 September 1979.

✚ O. R. Sander, G. N. Minerbo, R. A. Jameson, and D. D. Chamberlin, "Beam Tomography in Two and Four Dimensions", Proc. 10th Linear Accelerator Conf., Montauk, New York, September 10-14, 1979, Brookhaven National Laboratory report BNL-51134, p.314 , September 1980; Los Alamos Scientific Laboratory Report LA-UR-79-2540, 12 September 1979.

✚ G. P. Boicourt and R. A. Jameson, "A Statistical Approach to the Estimation of Beam Spill", Proc. 10th Linear Accelerator Conf., Montauk, New York September 10-14, 1979, Brookhaven National Laboratory report BNL 51134, p. 238, September 1980; Los Alamos Scientific Laboratory Report LA-UR-79-2502, 10 September 1979.

✚ J.T. Ahearne, (R.A. Jameson, ghost), "RFQ Is Alive And Well...", Atom, Vol. 17, No. 4, Los Alamos Scientific Laboratory, July/August 1980.

✚ R. A. Jameson, "Emittance Growth in RF Linacs", Proc. of Heavy Ion Fusion Workshop, Claremont Hotel, Berkeley, CA, October 29-November 9, 1979, Lawrence Berkeley Lab Report LBL10301, September 1980; Los Alamos Scientific Laboratory Report LA-UR-80-309, 28 January 1980.

✚ R. A. Jameson, Chairman, "Summary of Low-b Linac Working Group", Proc. of Heavy Ion Fusion Workshop, Claremont Hotel, Berkeley,CA, October 29-November 9, 1979, Lawrence Berkeley Lab Report LBL-10301, September 1980; Los Alamos Scientific Laboratory Report LA-UR-80-310, 28 January 1980.

✚ R. A. Jameson, "Review of Accelerator Development at LASL Relevant to Intense Neutron Sources", (Invited), Proceedings of ICANS-4 Meeting at National Laboratory for High Energy Physics (KEK), Tsukuba, Japan, October 20-24, 1980, Los Alamos National Laboratory Report LA-UR-80-2724, October 1980. (No copy in Report Library)

✚ "Accelerator Technology Program. Progress report, January-June 1980", E. A. Knapp, R. A. Jameson

✚ G. P. Boicourt, B. R. Chidley, K. R. Crandall, and R. A. Jameson, "Analysis of the Deuteron Distribution Emerging from the FMIT RFQ", Proc. 1981 Particle Accelerator Conf., Washington, DC, March 11-13, 1981, IEEE Trans. Nucl. Sci. 28, p. 3492, June 1981; Los Alamos National Laboratory Report LA-UR-81-527, 18 February 1981.

✚ G. P. Boicourt and R. A. Jameson, "A Study of the Variations of Maximum Beam Size with Quadrupole Gradient Strength in the FMIT Drift-Tube Linac", Proc. 1981 Particle Accelerator Conf., Washington, DC, March 11-13, 1981, IEEE Trans. Nucl. Sci. 28, p. 3504, June 1981; Los Alamos National Laboratory Report LA-UR-81-526, 18 February 1981.

✚ R. A. Jameson, "Beam-Intensity Limitations in Linear Accelerators," (Invited), Proc. 1981 Particle Accelerator Conf., Washington, DC, March 11-13, 1981, lEEE Trans. Nucl. Sci. 28, p. 2408, June 1981; Los Alamos National Laboratory Report LA-UR-81-765, 9 March 1981.

✚ Correction, Jameson, RA; IEEE TRANSACTIONS ON NUCLEAR SCIENCE; 1981; v.28, no.4, p.3665-3665

✚ R. A. Jameson (Compiler), "Radio-Frequency Structure Development for the Los Alamos/NBS Racetrack Microtron", Los Alamos National Laboratory Document LA-UR-83-95, January 21, 1983. (No copy in Report Library)

✚ D. J. Liska and R. A. Jameson, "Particle Accelerator Engineering", University of California Engineering Faculty Meeting, Los Alamos, May 25-27, 1983, Los Alamos National Laboratory document LA-UR-83-1358.

R.O. Bangerter, D. Keefe, R.A. Jameson, "Heavy Ion Fusion Accelerator Research Program Plan for FY84-FY89", Report submitted to DOE; Los Alamos National Laboratory Report LA-UR-83-1717, 13 June 1983. (No copy in Report Library)

R.A. Jameson (participant). "Assessment of the Adequacy of U.S. Accelerator Technology For Department of Energy Missions, WJSA-83-228, report submitted to USDOE-OER-OHENP, prepared by E.T. Gerry, S.A. Mani, W.J. Schafer Associates, Inc. Wakefield MA 01880, 24 June 1983.

✚ R. A. Jameson, "New Linac Technology - For SSC and Beyond", (Invited), 12th International Conference on High-Energy Accelerators, Fermi National Accelerator Laboratory, Batavia, IL., August 11-17, 1983, Los Alamos National Laboratory Report LA-83-2365, 9 August 1983.

✚ "Comments on the 1983 Particle Accelerator Conference", R. A. Jameson, (Conference Chairman, Program Committee Chairman), 1983 Particle Accelerator Conference on Accelerator Engineering & Technology, Santa Fe, New Mexico, March 21-23, 1983, IEEE Trans. Nucl. Sci., Vol. NS-30, No. 4, August 1983; p.1945.

✚ R. A. Jameson, "Ion Source and Ion Accelerator Development at Los Alamos", (Invited), Intl. Ion Engineering Congress - ISIAT '83 and IPAT '83, September 12-16, 1983, Kyoto, Japan, Los Alamos National Laboratory document LA-UR-83-175l, 16 June 1983. (No copy in Report Library)

✚ R. A. Jameson, "New Linear Accelerators", (Invited) INS-Kikuchi Winter School, Jinzai-Kaihatsu Center, Fujiyoshida, Yamanashi 403, Japan, January 29 - February 2, 1984, Los Alamos National Laboratory Report LA-UR-84-25, 5 January 1984. (No copy in Report Library)

(Probably same as: R. A. Jameson, "New Directions in Linear Accelerators", (Invited), Proc. 1984 Linear Accelerator Conference, Darmstadt, Federal Republic of Germany, May 7-11, 1984, GSI-84-11, GSI, Darmstadt, p. 237; & Los Alamos National Laboratory document LA-UR-84-1359, 25 April 1984.)

✚ D. D. Armstrong, W. D. Cornelius, R. A. Jameson, F. O. Purser, and T. P. Wangler, "RFQ Development at Los Alamos", Institute of Nuclear Studies Conference on Heavy Ion Accelerators and Inertial Fusion, University of Tokyo (INS), Tokyo, Japan, January 23-27, 1984; Los Alamos National Laboratory Report LA-UR-84-498, 9 February 1984.

✚ R. A. Jameson, "Introduction to RFQ Session", (Invited), Proc. 1984 Linear Accelerator Conference, Darmstadt, Federal Republic of Germany, May 7-11, 1984, GSI-84-11, GSI, Darmstadt, p. 49; & Los Alamos National Laboratory document LA-UR-84-1375, 26 April 1984.

✚ R. A. Jameson, "New Directions in Linear Accelerators", (Invited), Proc. 1984 Linear Accelerator Conference, Darmstadt, Federal Republic of Germany, May 7-11, 1984, GSI-84-11, GSI, Darmstadt, p. 237; & Los Alamos National Laboratory document LA-UR-84-1359, 25 April 1984.

Probably same as: R. A. Jameson, "New Linear Accelerators", (Invited) INS-Kikuchi Winter School, Jinzai-Kaihatsu Center, Fujiyoshida, Yamanashi 403, Japan, January 29 - February 2, 1984, Los Alamos National Laboratory Report LA-UR-84-25, 5 January 1984. (No copy in Report Library))

✚ R. A. Jameson, "Conference Summary", Proc. 1984 Linear Accelerator Conference, Darmstadt,

Federal Republic of Germany, May 7-Il, 1984, GSI-84-11, GSI, Darmstadt, p. XII, May 1984.

✚ R. A. Jameson, Chairman, "Report of the DOE Ad-Hoc Committee on the Brookhaven National Laboratory National Synchrotron Light Source, 30 May - 1 June 1984", Los Alamos National Laboratory Document No. AT-DO:84-157(Rev.), June 1984.

✚ "Accelerator Technology Program", R. A. Jameson, Status Report, Jan. - Sep. 1983 Los Alamos Scientific Lab., NM. 06/1984; -1.

✚ R. A. Jameson, "Accelerator Technology Working Group Summary", Proceedings of Second Workshop on Laser Acceleration of Particles, University of California, Los Angeles, CA 90024, January 6-l8, 1985, Los Alamos National Laboratory document LA-UR-85-882, January 1985. AIP Conference Proceedings. 07/1985; 130(1):549-559.

✚ R. A. Jameson, "RF Accelerators for Fusion and Strategic Defense", (Invited), Symposium on Lasers and Particle Beams for Fusion and Strategic Defense, University of Rochester, April 18-19, 1985, Fusion Power Associates, Gaithersburg, MD 20879, & Los Alamos National Laboratory document LA-UR-85-3760, 12 November 1985. Journal of Fusion Energy 02/1986; 5(1):23-31.

✚ M. A. Cochran, R. A. Jameson, and C. S. Nicol, "Annotated Bibliography of Accelerator Technology Division Research and Development 1978-1985", September, 1985, Los Alamos National Laboratory document LA-10541-MS, and Addendum, July - December 1985, publ. 1985.

✚ "Accelerator Technology Program. Status report, April-September 1985", R. A. Jameson, S. O. Schriber

✚ "Accelerator technology program. Status report, October 1984-March 1985", R.A. Jameson, S.O. Schriber, 03/1986

✚ "Accelerator Technology Program: Status report, October 1985-March 1986: Volume 1",R. A. Jameson, S. O. Schriber.

✚ R. A. Jameson, "RF Breakdown Limits", (Invited), NATO Advanced Science Institute on High-Brightness Accelerators, Pitlochry, Scotland, July 13-25, 1986, NATO ASI Series B: Physics Vol. 178, Plenum Press, 1988; & Los Alamos National Laboratory document LA-UR-86-2353, 8 July 1986. SEE LA-UR-86-2353 FROM NATO ADA205203.pdf DO NOT UPLOAD, COPYRIGHT – EXTRACTED INSTEAD

✚ R. A. Jameson and D. W. Reid, "RF Power Sources for High-Brightness RF Linacs", (Invited), NATO Advanced Science Institute on High-Brightness Accelerators, Pitlochry, Scotland, July 13-25, 1986, NATO ASI Series B: Physics Vol. 178, Plenum Press, 1988; & Los Alamos National Laboratory document LA-UR-86- 2319, 8 July 1986. SEE LA-UR's-86-2319 FROM NATO ADA205203.pdf DO NOT UPLOAD, COPYRIGHT – EXTRACTED INSTEAD

✚ R. A. Jameson, "High-Brightness Rf Linear Accelerators", (Invited), NATO Advanced Science Institute on High Brightness Accelerators, Pitlochry, Scotland, July 13-25, 1986, NATO ASI Series B: Physics Vol. 178, Plenum Press, 1988; & Los Alamos National Laboratory document LA-UR-86-2243, 26 June 1986. SEE LA-UR's-86-2243 FROM NATO ADA205203.pdf DO NOT UPLOAD, COPYRIGHT – EXTRACTED INSTEAD

✚ R. A. Jameson, "Linacs for Esoteric Applications", (Invited - first open conference talk on the SDI neutral-particle-beam program), 1986 Linear Accelerator Conference, SLAC, Stanford, CA, June 2-6, 1986, SLAC Report 303, Sept. 1986; & Los Alamos National Laboratory document LA-UR-86-1734, 21 May 1986.

R. A. Jameson, “RF Linacs for SDI", (Invited), Ninth Conference on Application of Accelerators in Research and Industry, November 10-12, 1986, Denton, TX, Nuclear Instruments & Methods in Phys. Res., Sec. B, Vol. B24/25, North-Holland, (1987), p. 725; & Los AIamos National Laboratory document LA-UR-86-3642, November 1986. (No copy in Report Library)

R. A. Jameson, “Discussion of High Brightness RF Linear Accelerators", (Invited), Foreign Applied Science Assessment Center (FASAC), January 13, 1987, McLean, VA, Los Alamos National Laboratory document LA-UR-87-69, 12 January 1987.

R. A. Jameson, “High-Brightness $H^-$ Accelerators", (Invited), 1987 Particle Accelerator Conference, Washington, D.C., March 16-19, 1987, IEEE Catalog No. 87CH2387-9, p. 903; & Los Alamos National Laboratory document LA-UR-87-666, 3 March 1987.

R. A. Jameson, “Applications of RF Linacs to Free Electron Lasers and Particle Beams", (Invited), Fusion Power Assoc. Annual meeting, April 8-9, 1987, Pleasanton, CA. Journal of Fusion Energy, Dec 1987, Vol. 6, No. 4, p.329-35; Los Alamos National Laboratory Report LA-UR-88-201, 21 January 1988. (No copy in Report Library) (Protected)

J. E. Leiss, R. H. Abrams, K. W. Ehlers, J. A. Farrell, G. H. Gillespie, R. A. Jameson, D. Keefe, R. K. Parker, "Soviet Exoatmospheric Neutral Particle Beam Research", Foreign Applied Sciences Assessment Center Technical Assessment Center (FASAC) Report (TAR) 3110, Science Applications International Corp, McLean VA., February 1988.

R. A. Jameson, "Beam Energy Variability and Other System Considerations for a Deuteron Linac Materials Research Neutron Source", (Invited), Proc. High Energy Neutron Source for Material Research & Development, Report on JAERI Symposium, Tokyo, Japan, January 12-13, 1989, p. 105, JAERI, March 1989.

R. A. Jameson, (Invited panelist & contributor) "Integration Report III: Soviet Information Sciences", Foreign Applied Sciences Assessment Center Technical Assessment Center (FASAC) Integration Report III, Science Applications International Corp, McLean VA., June 1989.

R. A. Jameson, "RF Linacs for Pion Therapy -- Beyond PIGMI", (Invited), Particle Medicine, Ryushi-sen Igaku, No. 13 1989.7, (c/o Hamamatsu University School of Medicine, Hamamatsu, Japan), July 1989.

R. A. Jameson, "High-Brightness Ion and Electron RF Linear Accelerators", (Invited), Second All-Union Workshop On New Methods Of Particle Acceleration, Nor Amberd, USSR, 10-14 October 1989, & LA-UR-89-2937, 29 August 1989.

R. A. Jameson, "LANTERN -- Los Alamos NeuTrons - Enterprise for Research Needs", Los Alamos National Laboratory Office Memo, AT-DO:89-476, November 20, 1989.

LANTERN-original.pdf, LANTERN DIR-7-11-90.pdf ("LANTERN Presentation to DIR and SMG, 8 June 1990", Los Alamos National Laboratory Office Memo, AT-DO:90- , July 11, 1990), LANTERN still-11-7-90.pdf

R. A. Jameson, "Accelerator-Based Intense Neutron Source For Materials R&D", (Invited), Proc. Second International Symposium on Advanced Nuclear Energy Research -- Evolution by Accelerators, Mito, Japan, January 24-26,1990, p. 34, JAERI, & LA-UR-89-4253, 22 December 1989.

R. A. Jameson, "High Intensity Linear Accelerator Development Topics for Panel Discussion on "Nuclear Energy Research and Accelerators -- Future Prospects"", (Invited), Proc. Second International Symposium on Advanced Nuclear Energy Research -- Evolution by Accelerators, Mito, Japan, January 24-26,1990, p. 163, JAERI, & LA-UR-89-4252, 22 December 1989.

✚ R. A. Jameson, "Considerations for High-Brightness Electron Sources", (Invited), JAERI Free Electron Laser Symposium, Tokyo, Japan, January 29-30, 1990, & LA-UR-90-30, 5 January 1990.

P. Lapostolle, R. A. Jameson, "Accelerators, Linear", chapter for Encyclopedia of Applied Physics, Vol. 1, American Institute of Physics, VCH Publishers, 1990.

✚ R. A. Jameson, "Energy Variable Deuteron Linac For Materials Research Neutron Source", Proc. 1990 European Particle Accelerator Conference, Nice, France, 12-16 June 1990, & LA-UR-90-1858, 25 May 1990.

✚ M. Mizumoto, H. Takeda, T. Nishida, I. Kanno, K. Hasegawa, H. Yasuda, Y. Nakahara, T. Takizuka, H. Akabori, Y. Okumura, M. Sugimoto, H. Shirakata, R.A. Jameson, and Y. Kaneko, "Transmutation of Transuranium Waste with High Energy Proton Induced Spallation Reaction", Proc. 1990 European Particle Accelerator Conference, Nice, France, 12-16 June 1990, p. 346.

✚ "Tribute to the Spirit of Keage", R.A. Jameson, (Commemoration Issue Dedicated to Professor Hidekuni Takekoshi On the Occasion of His Retirement), Citation Bulletin of the Institute for Chemical Research, Kyoto University (1990), 68(2): 185-187 Issue Date 1990-10-31, URL http://hdl.handle.net/2433/77329 KURENAI : Kyoto University Research Information Repository

✚ R.A. Jameson, "Accelerator-Driven Neutron Sources for Materials Development", (Invited), Eleventh International Conference on the Application of Accelerators in Research and Industry, Denton Texas, November 5-8,1990, Proc. NIM B56/57 (1991)982-986; Los Alamos National Laboratory Report LA-UR-____.

✚ R. A. Jameson, "Impressions and Expectations of Research in Japan", (Invited), Researches of Reactor Physics, Issue 40, Japan, 1991; Los Alamos National Laboratory Report LA-UR-91-101, 9 January 1991. (no copy in Report Library)

✚ R. A. Jameson, "Advanced Ion RF Accelerator Applications in the Nuclear Energy Arena", (Invited), 6-7 June 1991 Symposium on High Brightness Beams for Advanced Accelerator Applications,University of Maryland at College Park, College Park, MD. AIP Conf. Proc. NO. 253, Particles and Fields Series 47, ISBN 0-88318-947-X, 1992; also LA-UR-91-2196, 26 June 1991.

✚ R.A. Jameson, Principal Investigator, P.J. Tallerico, W.E. Fox, N. Bultman, T.H. Larkin, R.L. Martineau, S.J. Black, "Scaling and Optimization in High-Intensity Linear Accelerators", Study for the JAERI, LA-CP-91-272, Los Alamos National Laboratory, July 1991. Republished as LA-UR-07-0875, Los Alamos National Laboratory, 2/8/2007. (No copy in Report Library)

✚ R.A. Jameson (Invited), G.P. Lawrence, C.D. Bowman, "Accelerator-Driven Transmutation Technology for Incinerating Radwaste and for Advanced Application to Power Production", LA-UR-91-2687, Los Alamos National Laboratory; for 2nd european conference on accelerators in applied research and technology (ecaart), 3-7 September 1991, Frankfurt-am-Main, Germany, Proceedings - NIM.B.68 (1992) 474-480; Los Alamos National Laboratory Report LA-UR-91-2687, 14 August 1991. (No copy in Report Library)

✚ G.P. Lawrence, R.A. Jameson, S.O. Schriber, "Accelerator Technology for Los Alamos Nuclear Waste Transmutation And Energy Production Concepts", Proc. ICENES '91, Sixth Intl. Conf on Emerging Nuclear Energy Systems, Monterrey, CA, June 19-21, 1991, Fusion Technology, v. 20,.pp. 652-656, Dec. 1991; Los Alamos National Laboratory Report LA-UR-91-2797, 29 August 1991.

✚ "Specialist Meeting on Accelerator-Driven Transmutation Technology for Radwaste and other Applications", 24-28 June 1991, Saltsjöbaden, Stockholm, Sweden, hosted by Swedish National Board for Spent Nuclear Fuel. LA-12205-C, Conference, Los Alamos National Laboratory Report, also SKN Report No. 54, UC-940, Issued November 1991, Compiled & edited by R.A. Jameson, Conference Chairman.

✚ R.A. Jameson, D.P. Rusthoi, “High-Intensity, Low-Beam-Loss Proton Linac Reference Design for Transmutation Technology”, Study for the JAERI, LA-CP-93-178, Los Alamos National Laboratory, 15 July 1993.

✚ R.A. Jameson, “Accelerator-Driven Transmutation of Nuclear Waste and Electrical Power Production", (Invited) 1993 Fusion Power Associates Symposium, ORNL, October 1993; Journal of Fusion Energy, Vol. 12, No. 4, Dec. 1993. Los Alamos National Laboratory Report LA-UR-93-3716, 15 October 1993. (No copy in Report Library) (Protected)

✚ H.A. Thiessen, et. al.. include R.A. Jameson, “Report of the Committee on a TA-53 Upgrade, March 7, 1994”, Los Alamos National Laboratory, LA-UR-94-1924, 7 June 1994.
(No copy in Report Library)

✚ R.A. Jameson, Principal Investigator, et.al., “Deuteron Linac Design Studies for ESNIT/IFMIF”, Study for the JAERI, LA-CP-94-136, Los Alamos National Laboratory, June 1994.

✚ R.A. Jameson, Principal Investigator, et.al., “Review of Studies, and Continuing Work, for JAERI OMEGA and ESNIT/IFMIF Projects”, Study for the JAERI, LA-CP-94-151, Los Alamos National Laboratory, June 1994. LA-UR-07-0878, 2/8/07 (re-publish of , LA-CP-94-151, June 1994).

✚ G.H. Gillespie, B.W. Hill & R.A. Jameson, "Progress in Development of an Accelerator System Modeling Code", BDO-94 Intl. Wksp on Beam Dynamics & Optimization, July 4-8, 1994, St. Petersburg, Russia; Los Alamos National Laboratory Report LA-UR-94-2266, 1 July 1994.. (No copy in Report Library)

✚ G.H. Gillespie, B.W. Hill & R.A. Jameson, "A New Approach to Modeling Linear Accelerator Systems", Proc. 1994 Intl. Conf. on Accelerator-Driven Transmutation Technologies and Applications, 25-29 July, 1994, Las Vegas, Nevada, Los Alamos National Laboratory Report LA-UR-94-2753, 10 August 1994.

✚ B.I. Bondarev, A.P. Durkin, B.P. Murin & R.A. Jameson, "Emittance Growth and Halo Formation in Charge-Dominated Beams", Proc. 1994 Intl. Conf. on Accelerator-Driven Transmutation Technologies and Applications, 25-29 July, 1994, Las Vegas, Nevada, LA-UR-94-2753, Los Alamos National Laboratory.

✚ G.H. Gillespie, B.W. Hill & R.A. Jameson, "A New Tool for Accelerator System Modeling and Analysis", 17th Intl. Linac Conf., 21-26 August 1994, Tsukuba, Japan. Los Alamos National Laboratory Report LA-UR-94-2804, 17 August 1994.

✚ R.A. Jameson, “Self-Consistent Beam Halo Studies & Halo Diagnostic Development in a Continuous Linear Focusing Channel”, LA-UR-94-3753, Los Alamos National Laboratory, 9 November 1994. AIP Proceedings of the 1994 Joint US-CERN-Japan International School on Frontiers of Accelerator Technology, , Maui, Hawaii, USA, 3-9 November 1994, World Scientific, ISBN 981-02-2537-7, pp.530-560.

✚ C. Chen, R.C. Davidson, Q. Qian, R.A. Jameson, “Resonant and Chaotic Phenomena in a Periodically Focused Intense Charged-Particle Beam” (Invited paper for C. Chen), Proc. 10th Intl. Conf on High Power Particle Beams, NTIS, Springfield, VA 22151,(1994), 120-127.

✚ C. Chen & R.A. Jameson, “Self-Consistent Simulation Studies of Periodically Focused Intense Charged-Particle Beams”, Physical Review E, April 1995, PFC/JA-95-9 MIT Plasma Fusion Center

✚ R.A. Jameson, “Report of the International Fusion Materials Irradiation Facility (IFMIF) Conceptual Design Activity (CDA) Accelerator Team Meeting, Dallas, Texas, USA, 30 April 1995”, Los Alamos National Laboratory Report LA-UR-95-1701, 15 May 1995.

✚ B.I. Bondarev, A.P. Durkin, R.A. Jameson, “Space Charge-Dominated Beam Research”, 3rd Intl. Workshop on Beam Dynamics & Optimization, 1-5 July 1996, St. Petersburg, Russia.

✚ R.A. Jameson, "Beam Losses and Beam Halos in Accelerators for New Energy Sources", (Invited), Proc. 7th Intl. Symp. on Heavy Ion Fusion, Princeton, NJ, 6-10 Sept. 1996, Princeton, NJ, Fusion Engineering and Design 32-33 (1996) 149-157; Los Alamos National Laboratory Report LA-UR-96-175, 22 January 1996.

✚ T.E. Shannon,...,R.A. Jameson,...,et. al., “Conceptual Design of the International Fusion Materials Irradiation Facility (IFMIF)”, Proceeding Series of the International Atomic Energy Agency; 1997; p.437-449, 16th Intl. IAEA Fusion Energy Conference, Montreal, Canada, 7-11 October 1996.

✚ “IFMIF - International Fusion Materials Irradiation Facility Conceptual Design Activity, Final Report”, M. Martone, editor, IFMIF-CDA Team (R.A. Jameson, Accelerator Facility Team Leader), ENEA Frascati Report, RT/ERG/FUS/96/11 (December 1996).

✚ “Conceptual Design of the International Fusion Materials Irradiation Facility (IFMIF)”, Shannon, TE; Jameson, RA; Katsuta, H.; Maekawa, H.; Martone, M.; Moeslang, A.; Teplyakov, V.; Rennich, MJ, Journal of Nuclear Materials; Oct. 1998; vol.258-263, pt.A, p.106-12, Proc. 1997 8th Intl Conf on Fusion Reactor Materials, ICFRM. Part 1 (of 2); Oct 26-31 1997; Sendai, Japan

✚ “Accelerator Conceptual Design of the International Fusion Materials Irradiation Facility”, Sugimoto, M.; Jameson, RA; Teplyakov, V.; Berwald, D.; Blind, B.; Bruhwiler, D.; Deitinghoff, H.; Ferdinand, R.; Kinsho, M.; Klein, H.; et. al.; Journal of Nuclear Materials; Oct. 1998; vol.258-263, pt.A, p.367-71; Proceedings of the 1997 8th International Conference on Fusion Reactor Materials, ICFRM. Part 1 (of 2); Oct 26-31 1997; Sendai, Japan (ResearchGate shows Journal of Nuclear Materials 01/1999 ??)

✚ R.A. Jameson, “A Discussion of RFQ Linac Simulation”, Los Alamos National Laboratory Report LA-CP-97-54, September 1997. Re-published as LA-UR-07-0876, 2/8/07.

✚ K. Hasegawa, H. Oguri, Y. Honda, H. Ino, M. Mizumoto, R.A. Jameson, “Beam Dynamics Study of High Intensity Linac for the Neutron Science Project at JAERI”, First Asian Particle Accelerator Conference, March 23-27, 1998, KEK, Tsukuba, Japan.

✚ T. E. Shannon, The University of Tennessee, Knoxville, TN, USA; H. Katsuta, H. Maekawa, JAERI, Tokai-mura, Japan; R. A. Jameson, Los Alamos National Laboratory, Los Alamos, NM, USA; M. Martone, Associazione Euratom - ENEA sulla Fusione, Frascati, Italy; A. Möslang, Euratom - FZK Association, Karlsruhe, Germany; M. J. Rennich, The Oak Ridge National Laboratory, Oak Ridge, TN, USA; V. Teplyakov, Institute for High Energy Physics, Protvino, Moscow Region, RF; “IFMIF (International Fusion Materials Irrradiation Facility): A High Intensity Deuteron Beam Application”, EPAC ‘98.

✚ R.A. Jameson, LANL; H. Klein, IAP; J.M. Lagniel, CEA Saclay; C. Piaszczyk, NGC; M. Sugimoto, JAERI; V. Teplyakov, IHEP, “IFMIF Accelerator Conceptual Design Activities”, AccApp’98, ANS Meeting, Gaitlinburg, Tennessee, September 1998, Los Alamos National Laboratory Report LA-UR-98-2753, 22 June 1998. (No copy in Report Library)

✚ “Suitability and feasibility of the International Fusion Materials Irradiation Facility (IFMIF) for fusion materials studies”, Moslang, A; Ehrlich, K; Shannon, TE; Rennich, MJ; Jameson, RA; Kondo, T; Katsuta, H; Maekawa, H; Martone, M; Teplyakov, V; NUCLEAR FUSION; MAR 2000; v.40, no.3Y, 3, p.619-627; 17th IAEA Fusion Energy Conference; October 19-24, 1998; Yokohama, Japan.

✚ R. A. Jameson, “An Approach to Fundamental Study of Beam Loss Minimization”, AIP Conference Proceedings 480, “Space Charge Dominated Beam Physics for Heavy Ion fusion”, Saitama, Japan, December 1998, Y. K. Batygin, Editor. Workshop on Space Charge Dominated Beam

- "A direct plasma injection system into an RFQ for clean and safe ion implantation", T. Takeuchi,*, T. Katayama, M. Okamura, K. Yano, A. Sakumi, T. Hattori, N. Hayashizaki, R.A. Jameson, Nuclear Instruments and Methods in Physics Research B 188 (2002) 233–237

- "Direct Injection Scheme for RFQ Linac", M. Okamura, T. Takeuchi, T. Katayama and R.A. Jameson, RIKEN, Saitama, Japan, T. Hattori, H. Kashiwagi, TITech, Tokyo, Japan Proceedings, 21st International Conference, *Linac 2002*, Gyeongju, South Korea, August 19-23, 2002, pp91-93.

- "Laser ion source and RFQ Linac for Direct injection scheme"
H. Kashiwagi, M. Okamura, T. Hattori, T. Katayama, R. A. Jameson, R. Becker, A. Schempp, T. Takeuchi, A. Sakumi, N. Hayashizaki, Y. Takahashi, T. Hata, IEEE International Conference on Plasma Science 01/2003;

- "Nd-YAG laser ion source for direct injection scheme", Kashiwagi, H; Hattori, T; Hayashizaki, N; Yamamoto, K; Takahashi, Y; Hata, T; Okamura, M; Jameson, RA; Katayama, T; Mescheryakov, N; RSI, May 2004; v. 75, no.5 PART II, p. 1569-1571; Conf: $10^{th}$ Intl. Conf. On Ion Sources; Sept. 8-13, 2003, Dubna, Russia.

- "Design of ≥ 100 mA C4+ RFQ for Laser Ion Source", R. A. Jameson, M. Okamura, T. Katayama, The Institute of Physical and Chemical Research (RIKEN), Hirosawa 2-1, Wako-shi, Saitama, 351-0198, Japan, RIKEN-AF-AC-43, December 2003, ISSN 1346-2431

- "IFMIF Accelerator Facility", Jameson, RA; Ferdinand, R; Klein, H; Rathke, J; Sredniawski, J; Sugimoto, M; Journal of Nuclear Materials; 1 Aug. 2004; vol.329-333, pt.A, p.193-7; 11th International Conference on Fusion Reactor Materials (ICFRM-11), 7-12 Dec. 2003, Kyoto, Japan. Journal of Nuclear Materials - J NUCL MATER. 01/2004; 329:193-197.

- "A Comparison of High Current Ion Beam Matching From An Ion source To A RFQ By Electrostatic And By Magnetic Lenses", R.Becker, R.A. Jameson, A. Schempp, M. Okamura, A. Sakumi, T. Katayama, H. Kashiwagi, T. Hattori, N. Hayashizaki, K. Yamamoto, Y. Takahashi, T. Hata; EPAC2004, TUPLT024.

- "Matching of a C6+ Ion Beam From A Laser Ion source To A RFQ", R.Becker, R.A. Jameson, A. Schempp, M. Okamura, A. Sakumi, T. Katayama, H. Kashiwagi, T. Hattori, N. Hayashizaki, K. Yamamoto, Y. Takahashi, T. Hata; EPAC 2004, TUPLT025.

- "Nd-YAG laser ion source for direct injection scheme", H. Kashiwagi, T. Hattori, N. Hayashizaki, K. Yamamoto, Y. Takahashi, T. Hata, M. Okamura, R.A. Jameson,, T, Katayama, N. Mescheryakov, Review of Scientific Instruments 05/2004; 75(5):1569-1571.

- "Simulation of RF-Focusing In The RFQ Matching Section by IGUN", R. Becker & R.A. Jameson, ICAP2004, Oral

- "High Current RFQ Using Laser Ion Source", M. Okamura, R. A. Jameson, J. Takano, K. Yamamoto, RIKEN, Saitama, Japan; H. Kashiwagi, JAERI, Gunma, Japan, T. Hattori, N. Hayashizaki, TIT, Tokyo, Japan; A. Schempp, R. Becker, IAP, Goethe-Universität, Frankfurt. Germany; Y. Iwata, NIRS, Chiba, Japan, T. Fujimoto, S. Shibuya, AEC, Chiba, Japan, LINAC2004.

- C. Zhang, Z.Y. Guo, A. Schempp, R. A. Jameson, J. E. Chen,& J. X. Fang, "Low-beam-loss design of a compact, high-current deuteron radio frequency quadrupole accelerator", Physical Review Special Topics - Accelerators and Beams 01/2004; 7(10):100101.

- "Experimental results of DPIS with a new RFQ", M. Okamura, R. A. Jameson, H. Kashiwagi, T. Hattori, N. Hayashizaki, K. Sakakibara, K. Yamamoto, T. Kanesue, Radiation Effects and Defects in Solids 09/2005; 160(10-12):445-449.

- "60 mA Carbon Beam Acceleration with DPIS", M. Okamura, R.A. Jameson, J. Takano, K. Sakakibara, H. Kashiwagi, T. Hattori, N. Hayashizaki, A. Shempp, K. Yamamoto, Y. Iwata, T.

Fujimoto, S. Shibuya, T. Takeuchi, Particle Accelerator Conference, 2005. PAC 2005. Proceedings of the; 06/2005 TPPE031 (OkaPAC05 TPPE031.pdf)

✚ R. A. Jameson, “The IFMIF Accelerator Facility: Space-Charge Physics Design of High Intensity Ion Linacs“, Forum for the Centenary of the Scientific Breakthroughs by Albert Einstein, International Workshop on Heavy Ion Accelerators, October 10 – 17, 2005, Lanzhou, China

✚ “Comparison of Kashiwagi Experimental C6+ vfac Curve to Simulation”, R.A. Jameson, RIKEN Technical Memo, RIKEN-RAJ-20060502

✚ “Acceleration of high current fully-stripped carbon ion beam by direct injection scheme”, Hirotsugu Kashiwagi , …, Masahiro Okamura, R.A.Jameson, et. al., 11th International Conference on Ion Sources, Caen, France, September 2005. Review of Scientific Instruments 03/2006; 77(3):03B305-03B305-4. (This paper is very bad - did not see before publication…)

✚ “Emittance growth as mesh artifact”, R. Becker & R.A. Jameson, Nucl. Instr. & Meth. In Physics Research A 558 (2006) 32-35. ICAP2004, Poster

✚ “Simulation of RF-focusing in the RFQ matching section by IGUN”, R. Becker & R.A. Jameson, Nucl. Instr. & Meth. In Physics Research A, 558 (2006) 205-209; HP-NIS meeting Abingdon 2004 (“rfq-matching.pdf)

✚ “Matching of High Intensity Ion Beams to an RFQ: Comparison of PARMTEQ and IGUN Simulations”, R. Becker, R. A. Jameson, EPAC’2006, TUPLS104.

✚ “Matched and Equipartitioned Dynamics Design for High-intensity Radio Frequency Quadrupole Accelerators”, X. Q. Yan, R. A. Jameson, Y. R. Lu, Z. Y. Guo, J. X. Fang, J. E. Chen, EPAC’2006, THPCH015.

“Recent Progress about DPIS”, M. Okamura, R.A. Jameson (RIKEN) T. Kanesue (Kyushu University), H. Kashiwagi (JAEA/ARTC) A. Kondrashev (ITEP) K. Sakakibara, (RLNR) A. Schempp (IAP) J. Tamura (TIT), EPAC’2006, TUPLS095. (paper not submitted)

✚ “Acceleration of intense beams of highly-charged ions using direct plasma injection scheme”, S. Kondrashev, M. Okamura, R. Jameson, T. Kanasue, H. Kashiwagi, K. Sakakaibara, J. Tamura, HB2006; 01/2006, High Intensity High Brightness Hadron Beams, Tsukuba, Japan (HB2006, Sergei.doc)

✚ “High current carbon beam production with direct plasma injection scheme”, Masahiro Okamura, Hirotsugu Kashiwagi, Kazuhiko Sakakibara, Jumpei Takano, Toshiyuki Hattori, Noriyosu Hayashizaki, Robert A. Jameson, Kazuo Yamamoto, Review of Scientific Instruments 04/2006

“RIKEN Presentation of MO Book to Prof Noyori 20060330.pdf”, RIKEN Library Ref.# 080/JAM/1

✚ “RFQ Designs and Beam-Loss Distributions for IFMIF”, R.A. Jameson, Oak Ridge National Laboratory Report ORNL/TM-2007/001, January 2007. DOI:10.2172/931554, http://info.ornl.gov/sites/publications/files/Pub6207.pdf.

✚ Jameson, Robert A., LA-UR-07-0874, “Progress Toward Scaling And Optimization Criteria For High-Intensity, Low-Beam-Loss Rf Linacs”, 2/8/07 (re-publish of LA-CP-92-221, July 1992).

✚ Jameson, Robert A., LA-UR-07-0875, “Scaling And Optimization In High Intensity Linear Accelerators”, 2/8/07 (re-publish of LA-CP-91-272, July 1991).

✚ Jameson, R.A., “RFQ Design Studies – Investigation (incomplete) of Dependence of Optimization on Codes”, Injector/RFQ Working Group, 1st EU=JA Workshop on IFMIF-EVEDA Accelerator, 7-9 March 2007, Paris.

✚ “Matched and equipartitioned design method for modern high-intensity radio frequency quadrupole accelerators”, X.Q. Yan, R.A. Jameson, et. al., Nuclear Instruments and Methods in Physics Research Section A, Volume 577, Issue 3, 11 July 2007, Pages 402-408

✚ “Direct plasma injection scheme in accelerators (invited)”, M. Okamura, T. Takeuchi, R. A. Jameson, S. Kondrashev, et.al., Invited paper, published as part of the Proceedings of the 12th International Conference on Ion Sources, Jeju, Korea, August 2007. Review of Scientific Instruments 01/2008; 79(2):314-02.

✚ “Beam dynamics design of an RFQ for a planned accelerator, which uses a direct plasma injection scheme”, Zhang Zhou-Li, R.A. Jameson, Zhao Hong-Wei, Liu Yong, Zhang Sheng-Hua, Zhang Cong, Nuclear Instruments and Methods in Physics Research A 592 (2008) 197– 200

✚ “RFQ particle dynamic simulation development”, J. M. Maus_, R. A. Jameson, A. Schempp, N. Müller, U. Bartz, IAP, Pac2009, Vancouver, Canada.

✚ “PteqHI Development and Code Comparing”. J. Maus, R. A. Jameson, A. Schempp, IAP, Frankfurt, Germany, http://accelconf.web.cern.ch/AccelConf/ICAP2009/papers/thpsc031.pdf

✚ “Study of the Beam Pulse Shape by DPIS for High Intensity Heavy Ion Beam Acceleration”, Hirotsugu Kashiwagi, Masahiro Okamura, Toshiyuki Hattori, and Noriyosu Hayashizaki, R. A. Jameson, RSI-ICIS MS# C09557-B-12R, Linac 2009

✚ “Time structure of an accelerated beam using a radio-frequency quadrupole linac with direct plasma injection scheme”, Hirotsugu Kashiwagi, Masahiro Okamura, R A Jameson, Toshiyuki Hattori, Noriyosu Hayashizaki, The Review of scientific instruments 02/2010; 81(2):02B724.

✚ “Influence of projecting electrodes on shunt impedance and flatness of 4-rod radiofrequency quadrupole cavity”, Z. Zhang, H. Zhao, R.A. Jameson, Z. Xu, S. Zhang, Qiangiguang Yu Liziu/High Power Laser and Particle Beams 02/2010; 22(2):430-432, DOI:10.3788/HPLPB20102202.0430

✚ ✂Study on shunt impedance and voltage distribution of 4-rod RFQ cavity”,Zhang Zhou-Li, R. A. Jameson, Zhao Hong-Wei, Xu Zhe, Zhang Sheng-Hu, Zhang Cong, Sun Lie-Peng, Mei Li-Rong, Shen Xiao-Kang, Chinese Physics C (Impact Factor: 0.34). 03/2010; 34(3):398. DOI:10.1088/1674-1137/34/3/018

✚ “A two-meter long RFQ for the direct plasma injection scheme at IMP”, Zhang Zhouli, Zhao Hongwei, Liu Yong, Shi Aimin, Sun Liepeng, Sha Shan, Guo Xiaohong, R.A. Jameson, A Schempp, M. Okamura, LINAC2010, Tsukuba, Japan.

✚ “TESTING OF IMP LIS-RFQ”, Y Liu, Z Zhang, R A Jameson, A Schempp, H Zhao, W Zhan, H Zimmermann, M Vossberg, L Sun, M Okamura, He Yuan, S Sha, X Guo, A Shi, Z Xu, W Yue, Z Wang, X Du, C Zhang, X Lei, TUP043 Proceedings of Linear Accelerator Conference LINAC2010, Tsukuba, Japan

✚ “Development of pteqHI”, J. Maus, R.A. Jameson, A. Schempp, LINAC2010, Tsukuba, Japan

✚ “Beam Acceleration of DPIS RFQ at IMP”, Zhouli Zhang*, Hongwei Zhao, Shan Sha, Yong Liu, He Yuan, Xiaohong Guo, Sun Liepeng, Shi Aimin, IMP, Lanzhou730000, China; M. Okamura, BNL, Upton, New York 11973, U.S.A.; R.A. Jameson, A. Schempp, IAP, Frankfurt am Main, Germany, IPAC 2011, Spain. (IMP DPIS EP RFQ Beam Accel mopc028.pdf - only says EP, no discussion of RFQ itself.)

✚ “Beam dynamics design of a new radio frequency quadrupole for beam-current upgrade of the Japan Proton Accelerator Research Complex linac”, Yasuhiro Kondo, Robert A. Jameson, June 2012, PRSTAB 08/2012; 15(8). DOI:10.1103/PhysRevSTAB.15.080101. (This is a good paper…)

✚ "Design and Simulation of Practical Alternating-Phase-Focused (APF) Linacs – Synthesis and Extension in Tribute to Pioneering Russian AFP Research", R.A. Jameson, Proc. RUPAC2012, St-Petersburg, Russia, MOACH02 & MOACH02-talk

✚ "Practical design of alternating-phase-focused linacs", R. A. Jameson, 2013, arXiv identifier 1404.5176, available at: http://arxiv.org/abs/1404.5176

✚ "Acceleration of Cold Muons", R.A. Jameson, KEK Report 2013-2, April 2013. (Includes method for practical design of alternating-phase-focused linacs).

✚ "Linear Accelerators", Robert Jameson, Joseph Bisognano, and Pierre Lapostolle, *Encyclopedia of Nuclear Physics and its Applications,* First Edition. Edited by Reinhard Stock.
″ 2013 Wiley-VCH Verlag GmbH & Co. KGaA. Published 2013 by Wiley-VCH Verlag GmbH & Co. KGaA.

✚ "Space Charge Induced Collective Modes and Beam Halo in Periodic Channels",
Chao Li, (Q. Qin), 7th Intl. Particle Accelerator Conf., May 8-13, 2016, Busan, Korea, (I did very exensive editing, but was not included as co-author or acknowledged.)

✚ "R.A. Jameson Publications RAJ Pubs", Researchgate, 6/14/2016, **DOI:** 10.13140/RG.2.1.5075.9281

✚ "RFQ Vane Shapes for Efficient Acceleration", Y. Iwashita, Y. Fuwa, R. A. Jameson, LINAC 2016

"On Compensation of an Existing Linac for H- Intrabeam Stripping and Residual Beam Loss". R.A. Jameson, KEK Report 2-17-4 December 2017, A (also ref# 1724004)

✚ "Structure resonances due to space charge in periodic focusing channels", Chao Li & R.A. Jameson, Phys. Rev. Accel. Beams 21, 024204 (2018)

✚ "A new description of space charge induced beam halo", July 2018, Chao Li & R.A. Jameson, ResearchGate

✚ "Interaction between high intensity beam and structure resonances", Chao Li & R.A. Jameson, arXiv 1905.10008 v1, [physics.acc-ph] 24 May 2019.

✚ "Development of a radio frequency quadrupole linac implemented with the equipartitioning beam dynamics scheme", Yasuhiro Kondo, Takatoshi Morishita, Robert A. Jameson, https://doi.org/10.1103/PhysRevAccelBeams.22.120101
(DOI: 10.1103/PhysRevAccelBeams.22.120101)

✚ "Takekoshi_Memorial.pdf", **arXiv**: http://arxiv.org/abs/2008.02718
Title: In Memory of Prof. Hidekuni Takekoshi
Authors: Y. Iwashita, M.Mizumoto, T. Igaki, H. Okamoto, R.A. Jameson
Categories: physics.hist-ph physics.acc-ph
Comments: 23 pages, 18 figures
✚ **ResearchGate**: "Takekoshi_Memorial.docx" Have to download to view

✚ "MINGEI OMIYAGE, collected by R.A. Jameson", 30 November 2020, 822 pages, 82.6MB

✚ "Improved bunching and longitudinal emittance control in an RFQ", R.A. Jameson, B. Yee-Rendon, arXiv 2203.04632 , JINST_058P_1122, accepted for publication 11/28/2022.

✚ "Ilya Mikhailovich Kapchinsky – To the 90th birthday, Collection of memories", by his family and others, in Russian. I translated this to English in 2022 and am working on it editing with Vladimir Skachkov, ITEP and his colleagues. arXiv 2303.00801, ResearchGate.

✚ "Elements of Ion Linear Accelerators + Calm in the Resonances and Other Tales", R.A. Jameson, 2022, updates 2024, perhaps beyond. Posted to arXiv http://arxiv.org/abs/2212.04249, Posted to ResearchGate.

[toc]

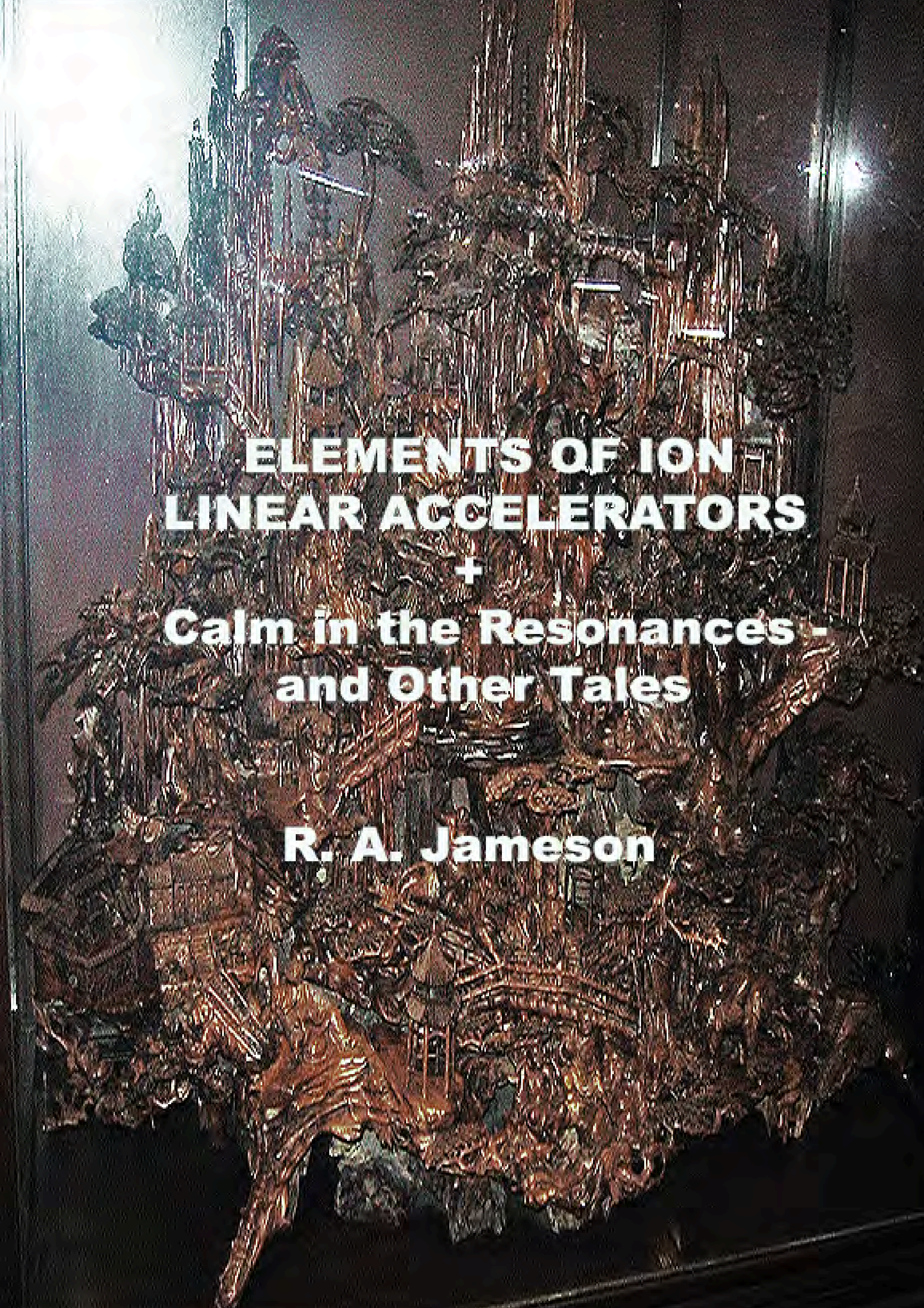

# ELEMENTS OF ION LINEAR ACCELERATORS
# +
# Calm in the Resonances - and Other Tales

R. A. Jameson

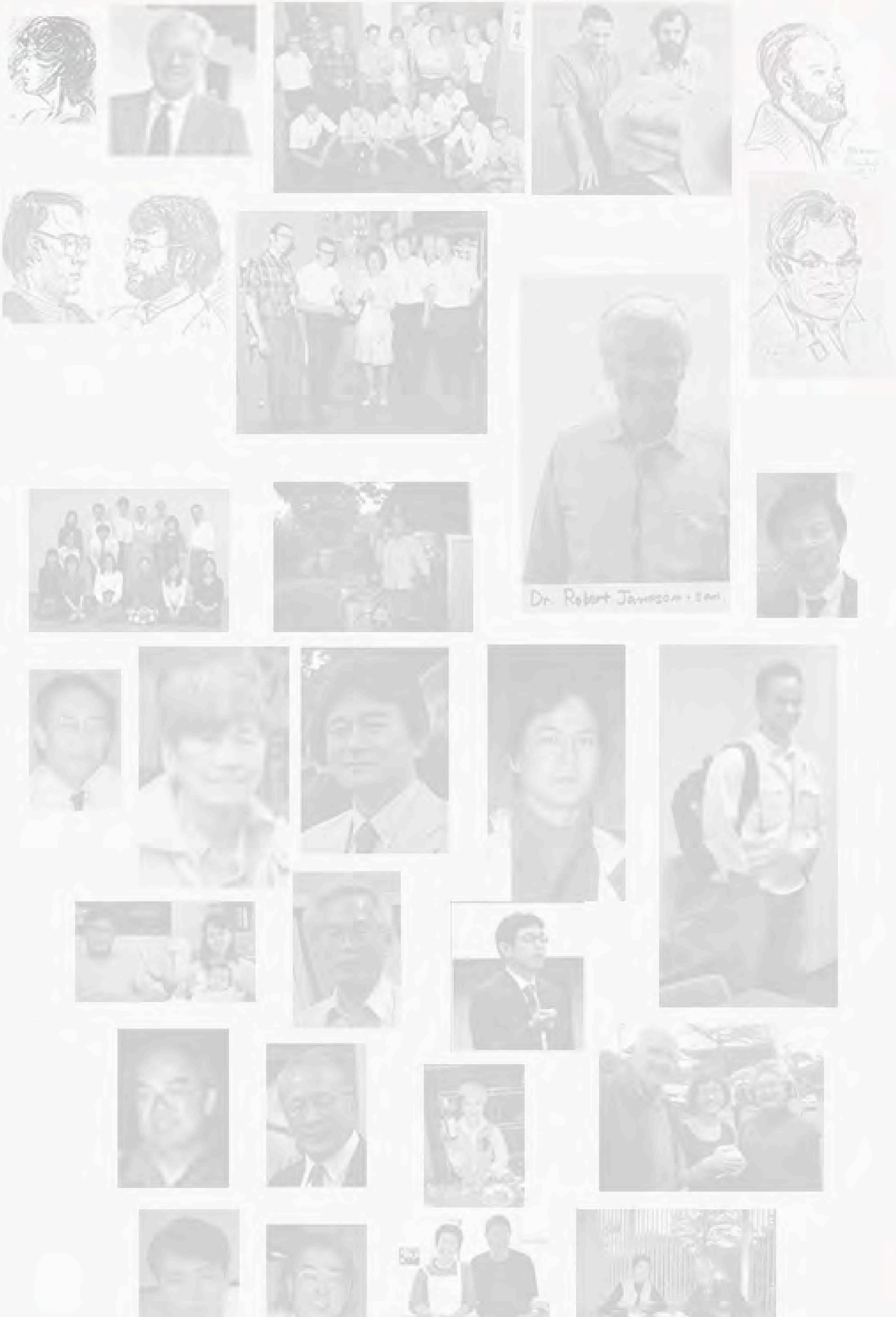
Dr. Robert Jameson · IBM

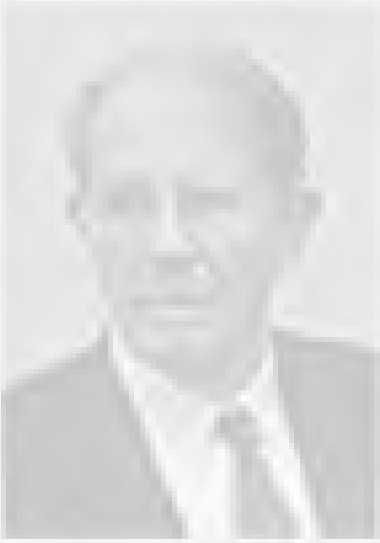
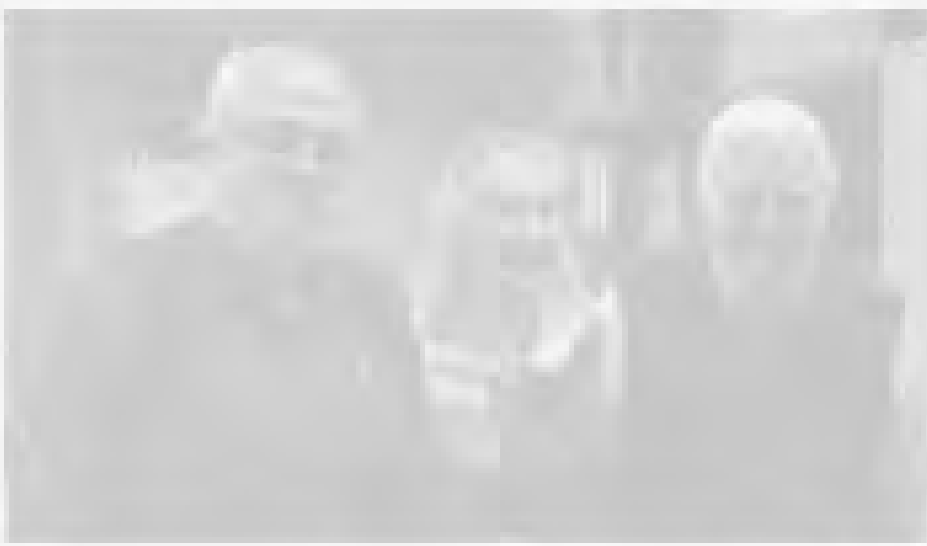
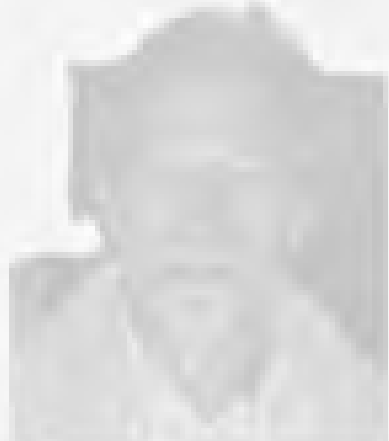
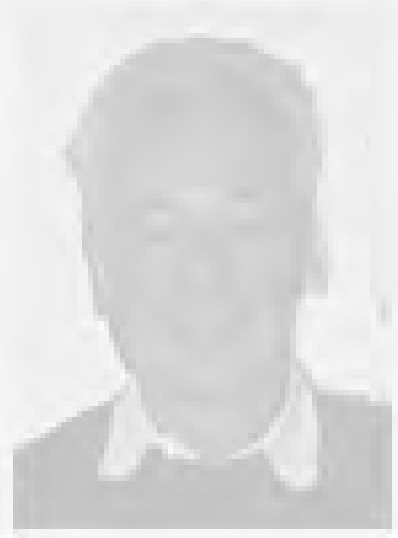
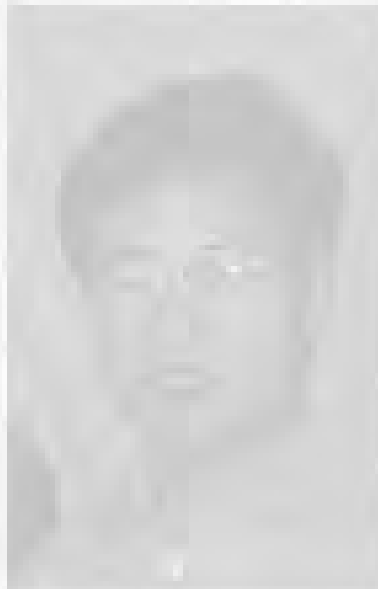
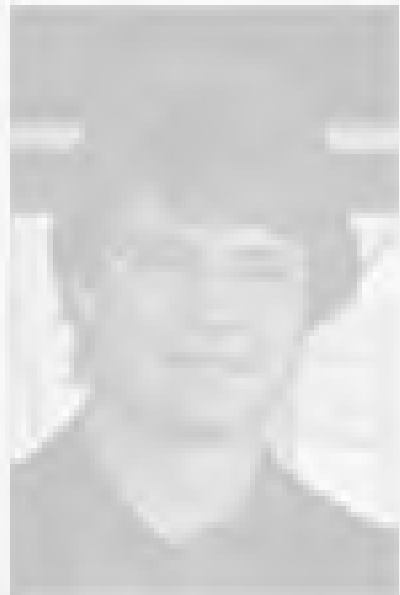
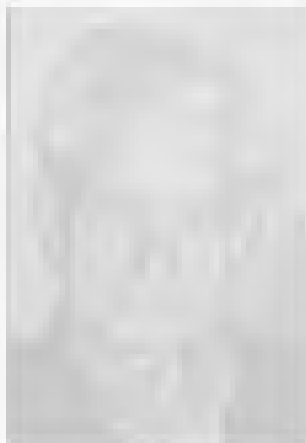
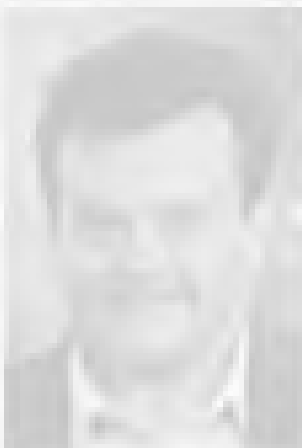
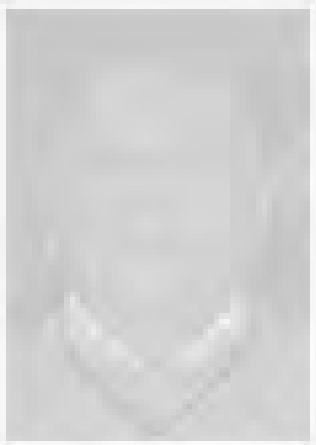
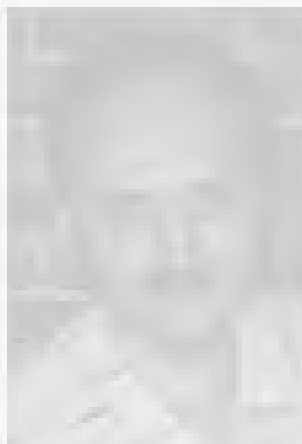
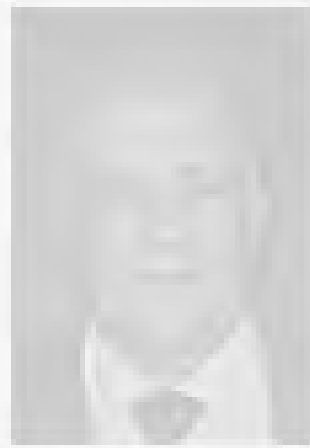
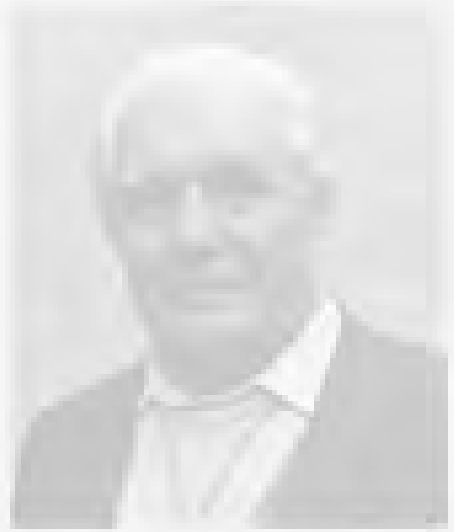
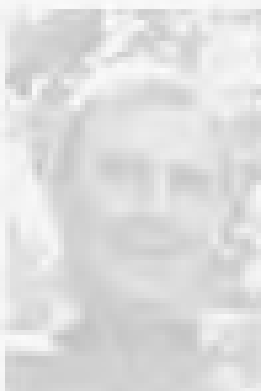
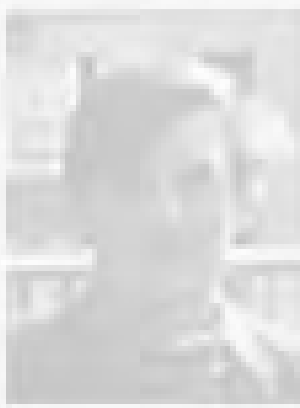
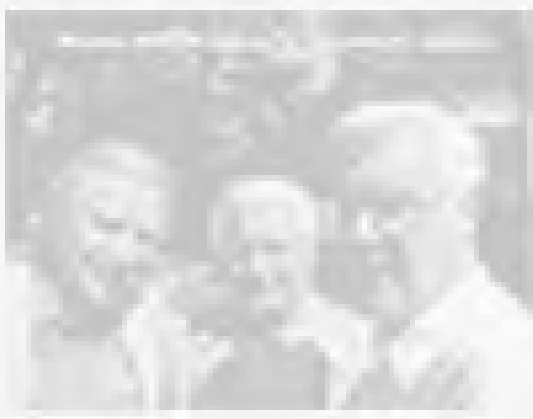
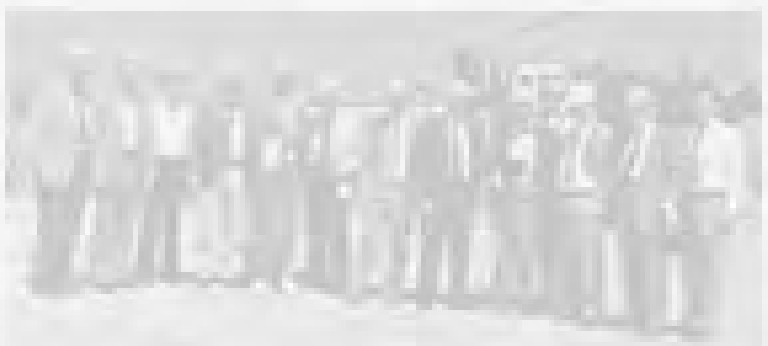
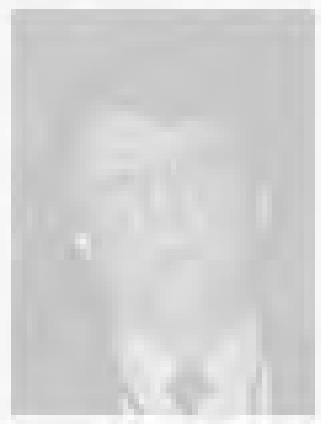
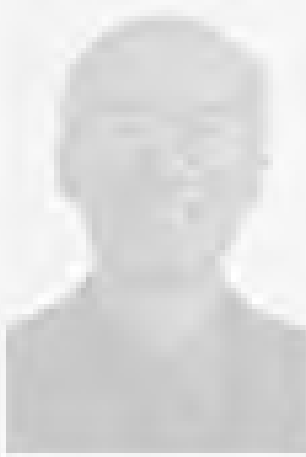
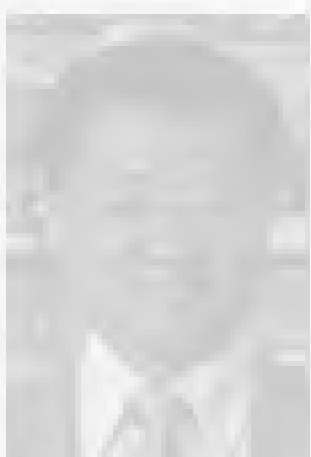
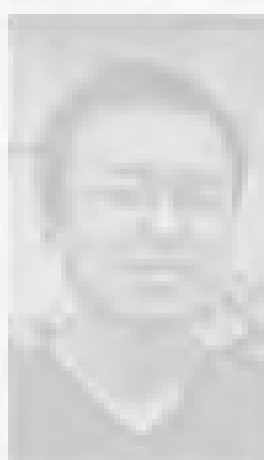
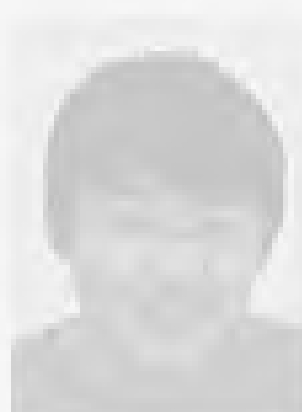
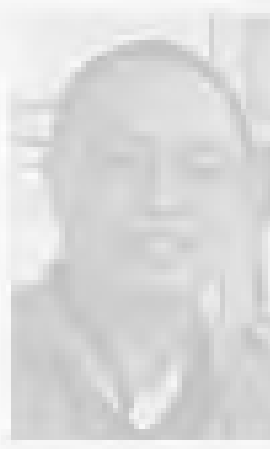
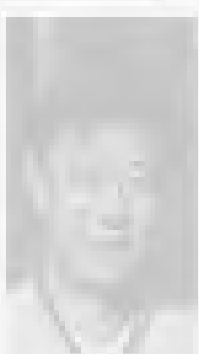
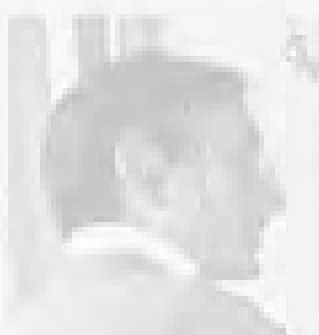
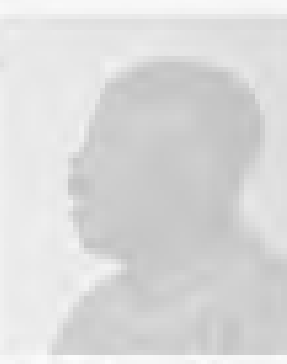
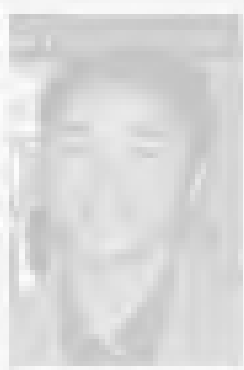
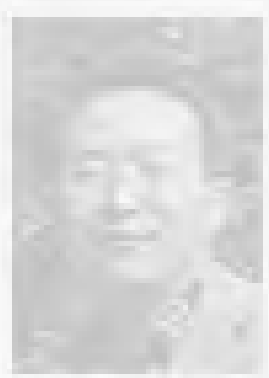
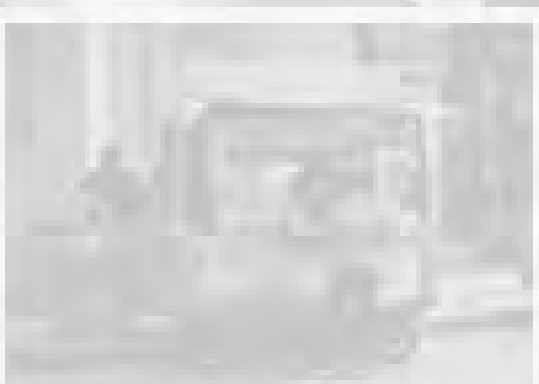
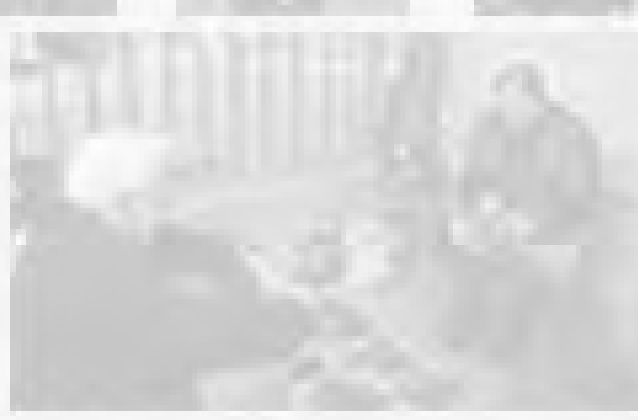